\documentclass[aps,pra,10pt,groupedaddress]{revtex4-1}
\makeatletter

\renewcommand*{\p@subsection}{}

\renewcommand*{\p@subsubsection}{}
\makeatother
\usepackage{amsmath, amsthm, amssymb, amsfonts}
\usepackage{braket}
\usepackage{graphicx}
\usepackage{epstopdf}
\usepackage{makeidx}
\usepackage{bbold}
\usepackage[colorlinks = true,
            linkcolor = blue,
            urlcolor  = blue,
            citecolor = red,
            anchorcolor = blue]{hyperref}
\usepackage{tikz}
\usepackage{calc} 
\tikzset{>=latex} 
 
\newcommand{\beq}{\begin{equation}}
\newcommand{\enq}{\end{equation}}
\newcommand{\bea}{\begin{eqnarray}} 
\newcommand{\ena}{\end{eqnarray}}

\usepackage[T1]{fontenc}

\makeindex
    
\begin{document}
 
\title{The Jaynes-Cummings model and its descendants}
\author{Jonas Larson}
\affiliation{Department of Physics, Stockholm University, AlbaNova
  University Center, SE-106 91 Stockholm, Sweden}
\author{Themistoklis Mavrogordatos}
\affiliation{Department of Physics, Stockholm University, AlbaNova
  University Center, SE-106 91 Stockholm, Sweden}
  \affiliation{ICFO -- Institut de Ci\`{e}ncies Fot\`{o}niques, The Barcelona Institute of Science and Technology, 08860 Castelldefels (Barcelona), Spain}

\date{\today}

\begin{abstract}
The Jaynes-Cummings (JC) model has been at the forefront of quantum optics for almost six decades to date, providing one of the simplest yet intricately nonlinear formulations of light-matter interaction in modern physics. Laying most of the emphasis on the omnipresence of the model across a range of disciplines, this monograph brings up the fundamental generality of its formalism, looking at a wide gamut of applications in specific physical systems among several realms, including atomic physics, quantum optics, solid-state physics and quantum information science. When bringing the various pieces together to assemble our narrative, we have primarily targeted researchers in quantum physics and quantum optics. The monograph also comprises an accessible introduction for graduate students engaged with non-equilibrium quantum phase transitions, quantum computing and simulation, and quantum many-body physics. In that framework, we aim to reveal the common ground between physics and applications scattered across literature and different technological advancements. The exposition guides the reader through a vibrant field interlacing quantum optics and condensed-matter physics. All sections are devoted to the strong interconnection between theory and experiment, historically linked to the development of the various modern research directions stemming from JC physics. This is accompanied by a comprehensive list of references to the key publications that have shaped its evolution since the early 1960s. Finally, we have endeavoured to keep the presentation of such a multi-sided material as concise as possible, interspersing continuous text with various illustrations alongside an economical use of mathematical expressions.
\end{abstract}

\maketitle 
\tableofcontents
\newpage
\vspace*{\fill}
\begin{center}
{\Large ``Get your name in print: Diagonalize a $2\times2$.'' \vspace{1cm}\begin{flushright}
F. Cummings
\end{flushright}}
\end{center}
\vspace*{\fill}
\newpage

\section*{Preface}

I was finishing my Diploma at the Warsaw University in the Division of Quantum Electrodynamics and Statistical Physics of Prof. Iwo Bia{\l}ynicki-Birala in the end of the 1980s. Theoretical Quantum Optics was pioneered then and there by Kazik Rz\k{a}\.{z}ewski and the late Krzysztof W\'{o}dkiewicz, who considered Quantum Optics to be ``applied, non-perturbative QED''. These were the times of simple non-perturbative models, mostly pertaining to single or few body systems. From a theoretical point of view, the challenge was related to ``non-perturbative'' rather than to quantum many-body complexity. Our ``Bible'' at that time was the book by Les Allen and Joseph H. Eberly, {\it Optical Resonance and Two-Level Atoms}~\cite{Allenbook}. The ``Bible'' was in fact primarily dealing with Jaynes-Cummings (JC) model and its variations. ``Serious'' theoreticians in particle or statistical physics somewhat arrogantly did not consider Quantum Optics at all. Even my PhD mentor, the late Fritz Haake, a disciple of the Stuttgart school of Wolfgang Weidlich and Hermann Haken, told me after my PhD exam: ``Maciek, Quantum Optics is not a separate discipline, it is a part of Statistical Physics!''. All of them, in a sense, did not foresee nor understand the enormous potential in the experimental aspect of single-body systems: extreme control of preparation, manipulation and detection of such objects. Clearly, this control, allowing for the boom of experiments and observations, led to the boom of practical Quantum Information Science (realization of qubits, single- and two-party gates), practical Foundation of Quantum Mechanics (observations of certified entanglement and Bell correlations), Quantum Technologies (quantum cryptography and metrology). The same control (especially concerning laser techniques and methods) allowed for unprecedented possibilities of cooling of single and many body systems, observation of Bose Einstein condensation with ultracold alkali atoms, and finally the realization of strongly correlated atomic/molecular systems in traps or optical lattices, where Quantum Optics and Atomic, Molecular and Optical (AMO) physics meets (beats?) Condensed Matter and Quantum Many-Body Physics. 

There were two areas of physics that dealt at that time with this amazing single-body Quantum Mechanics: physics of trapped ions and cavity QED. The first was pioneered by inventors of ion traps, Wolfgang Paul~\cite{PaulNobel} and Hans Dehmelt~\cite{DehmeltNobel}, who obtained Nobel Prize for their work in 1989, together with Norman Ramsey, whose spectroscopic methods made amazing impact on precise detection and measurements in both single and many-body AMO systems. Perhaps the most spectacular experiments of that era concerned the observation of quantum jumps in a laser-driven single trapped ion with three relevant energy levels~\cite{PlenioKnight1998}. These initial works allowed later Dave Wineland to achieve absolutely unprecedented control of single and few-ion systems, realizing single and few-qubit systems, single and two-qubit quantum gates with impressive fidelity, and much more...Wineland obtained the Nobel Prize for his research and achievement in 2012~\cite{WinelandNobel}, together with one of the leaders and pioneers of cavity QED, Serge Haroche~\cite{HarocheNobel}. The 1980s also saw other cavity QED masters, like Daniel Kleppner and the late Herbert Walther. 

Daniel Kleppner and Dave Pritchard were also pioneers of the novel  and powerful laser-cooling techniques that turned out to be essential for the novel precision spectroscopy, realization of BEC, strongly correlated quantum many-body systems in optical lattices, and ultimately single trapped (Rydberg) atoms, which realize in a sense an array of coupled JC models. Laser cooling has also awarded the Nobel Prize to Claude Cohen-Tannoudji~\cite{CTannoudjiNobel}, Steve Chu~\cite{ChuNobel}, and William Phillips~\cite{PhilipsNobel}. Another pioneer of laser cooling was Ted H\"{a}nsch~\cite{HanschNobel}, who got the Nobel Prize in 2005 together with John ``Jan'' Hall~\cite{HallNobel} for precise laser spectroscopy in general, and the frequency comb. The third laureate in 2005 was Roy Glauber, for quantum theory of coherence of light and theory of photon counting~\cite{GlauberNobel}. All these techniques expanded out abilities to control both single and many-body quantum systems. Ted H\"{a}nsch, together with Dan Kleppner, Ike Silveira, and Jook Walraven, belong to major players in the quest for Bose-Einstein condensation, first in spin-polarized hydrogen, and later in alkali atoms. In 2001, Eric Cornell~\cite{CornellNobel}, Carl Wieman~\cite{WiemanNobel} and Wolfgang Ketterle~\cite{KetterleNobel} got the Nobel Prize for BEC of ultracold trapped alkali atoms. 

Very quickly the interests of the AMO community turned from weakly interacting many-body systems, like a standard BEC or weakly interacting Fermi gas that can be described by the Bardeen-Schriefer-Cooper theory, towards strongly-correlated systems, described by various types of Hubbard models. In certain limits, these models reduce to spin systems or related systems of coupled few-level units, {\it i.e.}, systems of coupled generalized JC models. Amazingly, Ted H\"{a}nsch is also behind these developments: after theory proposal of Ignacio Cirac, Peter Zoller and collaborators~\cite{jaksch1998cold}, the seminal experiment on Bose Hubbard models was realized in 2002 in Ted’s group~\cite{greiner2002quantum}. This was the beginning of the new era. In 2005 I was a Vorsitzender of the Quantum Optics division DPG (German Physical Society), preparing the gigantic DPG Tagung in Berlin. I was supposed to incorporate in the already overfull program every laureate of the DPG. During the preparatory meeting, one of the high officials of the DPG told me: ``Herr Lewenstein, don't worry. Quantum Optics will never get the Max Planck Medal - it goes only to Particle and Condensed Matter Physics''. The next day Peter Zoller was awarded...

Of course, nowadays Quantum Optics and AMO Physics are not the only platforms to realize single and coupled JC models. Condensed Matter Physics offers nowadays also enormous possibilities and opportunities: circuit QED ({\it i.e.}, coupled Josephson junctions in microcavities), systems of quantum dots, systems of coupled  Nitrogen vacancy (NV) centers, and more...

The fascinating book by Jonas Larson and Themis Mavrogordatos reflects these equally fascinating developments of Quantum Optics over the last 50-60 years, expanding the previously existing literature on the subject to the new territories and areas. After a short Introduction, Section 2 covers the basic properties of the standard JC model and more: it introduces the models, discusses the JC dynamics, describes driven and open JC models. It goes beyond the Rotating Wave Approximation (RWA), and introduces various kinds of extended JC models. These models under certain conditions turn into single-particle lattice problems. Finally, Section 2 reviews various approximations used for the analysis of JC models. 

Section 3 covers cavity QED, with careful respect to experimentally accessible regimes of parameters. It reports early results on optical bistability and micro-maser, and discusses cavity-induced forces. The rest of the section is relevant for the Quantum Information Science: the authors discuss state preparation, state tomography and quantum information processing.

Section 4 is the first preparatory one towards elaboration on quantum many-body physics - it deals with circuit QED, {\it i.e.}, systems of coupled Josephson qubits in macro-cavities, which correspond to coupled JC models. The section starts with the discussion of the rich physics of Josephson qubits, and then focuses on he (generalized) JC nonlinearity and spectrum revisited in the light of circuit QED.  Finally, control and transfer of quantum information in circuit QED is discussed. 

Section 5 focuses on trapped ions, and covers model Hamiltonians, state preparation and tomography, and quantum information processing. An important closing section on further aspects and perspectives relates trapped ion systems to quantum simulators. First, this is discussed in the trivial case of a single JC model, but then turning to genuine many-body spin chains constructed from arrays of JC models. 

In Section 6, the authors focus on waveguide QED, and start from description of atomic emission in the vicinity of an interface. After revising circuit QED, the authors discuss light-matter interaction in a 1D waveguide: a continuum for correlated photon states, and interaction with matter in nanowire plasmons.

Significant others are discussed in Section 7: nitrogen vacancy centers, strong coupling in photonic crystals, hybrid systems: from nano-mechanics to atomic ensembles. I particularly adore Section 8, devoted to extensions and truly challenging quantum many body systems and problems: Jaynes-Cummings-Hubbard models, many-body cavity, mean-field explorations, critical phenomena for bosons and fermions, and, last but not least, polaritonic chemistry, using the Born-Oppenheimer approximation and deriving molecular JC Hamiltonian. The book ends with conclusions - the next decades...

This is a very complete book: it has over 200 pages and over 1500 references (sic!). This is a XXIst century analogue of the Allen-Eberly book, a new ``Bible'' of Quantum Optics.

\begin{flushright}
Maciej Lewenstein
\end{flushright}

\vspace{2cm}

\section*{Acknowledgments}

Playing such a central role in quantum optics, anyone in the field has or will come across the Jaynes-Cummings model. In the last four to five decades, many of the major discoveries in the field have been somehow linked to this particular model. We have had the pleasure to meet, discuss, and work with several great scientists being active in this development. People who contributed in one way or another to the important discoveries resulting from the Jaynes-Cummings description. In the span of more than half a century, new generations of researchers bring on this legacy. Today, the model finds many applications beyond quantum optics, leading to a more diverse community. It makes the ongoing progress very exciting, and we are most thankful to be part of this, for example by learning from people who belong to other fields.

The person who had an indispensable contribution to this book is the late Stig Stenholm, who was the PhD supervisor for one of us (JL). Without him this book would not have seen the light of day. Stig himself had learned about the model from Willis Lamb in the late 1960s. In other words, he had actively followed its whole development from its infancy through the first Jaynes-Cummings related experiments on cavity quantum electrodynamics (QED) and trapped ions--a history lesson he happily shared. It is hard to imagine a better teacher for learning about the model. Another figure behind some of the most important discoveries related to the model is Peter Knight. Of course, Peter has been an extremely influential researcher in the field. While serving as the PhD opponent for JL's defense, this became clear and he gave tremendous insight into the model, and equally interesting were all the anecdotes about the Jaynes-Cummings history that he revealed during the PhD defense dinner...While Stig was the `academic father' of JL, Howard Carmichael
has inspired TM with his unique vision on quantum optics. Several parts of this book find their origin in his works. Barry Garraway also deserves some extra acknowledgement; together with him JL learned early on about extended Jaynes-Cummings models. TM is grateful to George Fikioris, his mentor at the National Technical University of Athens, for emphasizing the value of rigour in mathematical physics. TM's postdoctoral appointments at University College London helped him appreciate the intricacy of light-matter interaction at the quantum level. Tania Monteiro acquainted TM with the world of optomechanics as a very considerate and astute teacher, while Marzena Szyma\'{n}ska and Eran Ginossar introduced him to circuit QED and the open-systems approach, constantly bringing up how important the link with the experiment is. JL was introduced to many-body theories by Maciej Lewenstein when working on ultracold atomic gases confined in optical cavities. This line of research -- cross-fertilizing quantum many-body theories with quantum optical systems -- is something that has since then formed a central research topic for JL which may actually be reflected in the content of this book.

There is a long list of others that have been significant for us in our personal development as physicists, as well as for the writing of this book. An incomplete list includes: Alexander Altland, Joan Agust\'{i} Bruz\'{o}n, Sibylle Braungardt, Iacopo Carusotto, Peter Domokos, Tobias Donner, Kirsty Dunnett, Stefan Filipp, Vasileios Fragkos, Axel Gagge, Oliver Gould, A. Greentree, Ricardo Guti\'{e}rrez-J\'{a}uregui, Hans Hansson, S\"{o}ren Holst, Elinor Irish, Chaitanya Joshi, Yaron Kedem, Dainius Kilda, Thomas Klein Kvorning, Markus Kowalewski, Dominik Kufel, \r{A}sa Larson, G\"{o}ran Lindblad, Crist\'{o}bal Lled\'{o}, Andrea Maiani, Marcin Malinowski, Andrew Maxwell, Oleg Mitrofanov, Giovanna Morigi, Fernanda Pinheiro, Peter Rabl, Samuel Rudge, Janne Salo, Pil Saugmann, Gleb Siroki, Erik Sj\"{o}qvist, Jon Urrestilla, Tim Wilkinson, Alejandro Zamora Soto.

Finally, JL and TM are grateful for the financial support of the Swedish Research Council (VR), the Wallenberg Academy Fellows program and the EPSRC.

\vspace{3mm}

\underline{For this third substantially revised version}, in particular, we are grateful for the interest and comments of: S. Ashhab, R. Blatt, D. Braak, D. Burgarth, B. Dalton, L. Duan, G. Fux, S. Hughes, C. Lled\'{o}, P. Lodahl, D. Lonigro, P. Meystre, K. M\o lmer, A. J. Omolo, A. Parra-Rodr\'iguez and P. Rice.

\vspace{2.2cm}

\section{Introduction}
\label{sec:intro}
It all started in 1957 when the young student Frederick Cummings knocked on the door of Edwin Jaynes' office asking for a thesis project~\cite{cummings2013reminiscing}. Some two years later, Cummings was told to write things up and publish. However, things got in the way and it wasn't until years later when he was asked to contribute to a special issue in IEEE on ``Quantum electronics'' that the work finally got published. In their highly influential paper of 1963~\cite{jaynes1963comparison}, Jaynes and Cummings (JC) set forth to interpret deviations from Fermi's golden rule\index{Fermi's golden rule} and perturbation theory in the first quantization, with regards to coherent radiation from a microwave resonator (called the {\it beam maser}). As they characteristically note, ``...it is shown that the semiclassical theory, when extended to take into account both the effect of the field on the molecules and the effect of the molecules on the field, reproduces almost quantitatively the same laws of energy exchange and coherence properties as the quantized field theory, even in the limit of one or a few quanta in the field mode. In particular, the semiclassical theory is shown to lead to a prediction of spontaneous emission\index{Spontaneous! emission}, with the same decay rate as given by quantum electrodynamics, described by the Einstein $A$ coefficients\index{Einstein $A$ and $B$ theory}.'' They go on stating that ``Because of this success, and the fact that the correspondence with quantum electrodynamics continues to strengthen as this formalism is applied to a larger group of problems, it is felt that this formalism deserves independent status as a physical theory in its own right, and we suggest it be called the neoclassical theory of electrodynamics\index{Neoclassical! theory}.'' A seemingly brave statement about ones own theory, but now, more than 50 years later, we can confirm that a path of its own was indeed opened. While we can find early tests of the neoclassical theory of radiation\index{Neoclassical theory of radiation} concerning resonance fluorescence \cite{GibbsH1, GibbsH2}, following an attempt to formulate a ``semiquantization to restore agreement with experiment without quantizing the electromagnetic field'' in~\cite{NeoclassicalFailPRL}, these works did not really got the ball rolling. Instead, when the predicted collapse-revival phenomenon~\cite{eberly1980periodic}, proving the field quantization, was verified in 1987 in the group of Herbert Walther~\cite{rempe1987observation} things took off for real. Indeed, since the mid 1980s JC physics has formed the backbone of a plethora of different physical systems, and till this day we see a constant flow of newly developed hybrid systems in which the JC model, despite its simplicity, provides the correct description of their properties.  

Before moving forwards, it is instructive to first move backwards in time to a few decades prior to 1963, where we find a set of works retrospectively related to the JC model. The JC formulation draws an obvious insight from the quantum coherence associated with nuclear magnetic resonance in the language of first quantization, with the solution of Schr\"{o}dinger's equation for a two-level system in a rotating magnetic field, as studied by Rabi in 1937 \cite{Rabi1937}. Its importance to the JC model will become clear later, and devote one section of this monograph to the {\it quantum Rabi model}. The mathematical formulation of the multimode JC Hamiltonian\index{Multimode! Hamiltonian! Jaynes-Cummings} is motivated by certain problems in field theory, such as the interaction of two neutral nonrelativistic fermion fields, $\hat{\psi}_V$ and $\hat{\psi}_N$, and a relativistic scalar field $\hat{\phi} \equiv \hat{A}(\mathbf{r})+\hat{A}^{\dagger}(\mathbf{r})$. Written as an integral over the phase space, the interaction Hamiltonian reads \cite{Lee1954}
\begin{equation}\label{eq:HintFermionBoson}
\hat{H}_{\rm  int}=\int \left[\hat{\psi}_{V}^{\dagger}(\mathbf{r})\hat{\psi}_N(\mathbf{r})\hat{A}(\mathbf{r}) + \hat{\psi}_{V}(\mathbf{r})\hat{\psi}_N ^{\dagger}(\mathbf{r}) \hat{A}^{\dagger}(\mathbf{r})+\delta m_V \hat{\psi}_{V}^{\dagger}(\mathbf{r}) \hat{\psi}_{V}(\mathbf{r})\right]\, d\mathbf{\tau},
\end{equation}
where $g$ is the interaction strength (the last term in eq. \eqref{eq:HintFermionBoson} serves the purpose of mass renormalization). The model was introduced as a method to handle divergences in field theory, a topic that actually led Jaynes to become interested in quantum radiation theory~\cite{cummings2013reminiscing}. When mapped to the fermionic states with occupation number $1$ or $0$ for the two species, the first term of eq. \eqref{eq:HintFermionBoson} generates the (multimode) JC- type interaction term
\begin{equation}\label{eq:simpltwospecies}
\hat{H}_{\rm  int}=\sum_{n=1}^{N}\lambda_{n} (\hat{\sigma}_{-}\hat{a}_{n}^{\dagger} + \hat{\sigma}_{+}\hat{a}_{n}).
\end{equation}
In \cite{StenholmJC}, we find reference to the even earlier work by Friedrichs examining the fate of a single bound state embedded in a continuum \cite{PertContinuum1948}, seemingly disappearing as the coupling constant tends to zero. The extension of Friedrichs' model by Lee consists then in the description of the interaction between a two-state fermionic particle with a continuous bosonic field. 

Also dating back to 1954, as does the reference to the Lee model, is the work of Robert Dicke focusing on the interaction between a molecular gas and an electromagnetic field, resulting in coherent radiation in the dipole approximation~\cite{Dicke1954}. Dicke studied spontaneous emission\index{Spontaneous! emission} from a set of $N$ dipoles, and found that it is possible to enhance the emission rate by a factor $N$. He coined this collective phenomenon {\it superradiance}. 

\begin{figure*}
\centering
\begin{tikzpicture}[]
 
  \newcount\yearOne; \yearOne=1900
  \def\w{15}    
  \def\n{4}     
  \def\lt{0.40} 
  \def\lf{0.36} 
  \def\lo{0.30} 
 
  \def\yearLabel(#1,#2){\node[above] at ({(#1-\yearOne)*\w/\n/10},\lt) {#2};}
  \def\yearArrowLabel(#1,#2,#3,#4){
    \def\xy{{(#1-\yearOne)*\w/\n/10}}; \pgfmathparse{int(#2*100)};
    \ifnum \pgfmathresult<0
      \def\yyp{{(\lt*(0.90+#2))}}; \def\yyw{{(\yyp-\lt*#3)}}
      \draw[<-,thick,black!60!green,align=center] (\xy,\yyp) -- (\xy,\yyw) node[below,black] at (\xy,\yyw) {#4};
    \else
      \def\yyp{{(\lt*(0.10+#2)}}; \def\yyw{{(\yyp+\lt*#3)}}
      \draw[<-,thick,black!25!red,align=center] (\xy,\yyp) -- (\xy,\yyw) node[above,black] at (\xy,\yyw) {#4};
    \fi}
 
  \draw[->,thick] (-\w*0.03,0) -- (\w*1.03,0);
 
  \foreach \tick in {0,1,...,\n}{
    \def\x{{\tick*\w/\n}}
    \def\year{\the\numexpr \yearOne+\tick*10 \relax}
  	\draw[thick] (\x,\lt) -- (\x,-\lt) 
	             node[below] {\year};
 
	\ifnum \tick<\n
	  \draw[thick] ({(\x+\w/\n/2)},0) -- ({(\x+\w/\n/2)},\lf); 
      \foreach \ticko in {1,2,3,4,6,7,8,9}{
        \def\xo{{(\x+\ticko*\w/\n/10)}}
  	    \draw[thick] (\xo,0) -- (\xo,\lo);  
	}\fi
  }
  
    \yearArrowLabel(1901, 1.0,1.0, \footnotesize {\bf Planck}: {\it Blackbody radiation},\\ \footnotesize `energy elements')
        \yearArrowLabel(1905,-2.0,1.2, \footnotesize {\bf Einstein}: {\it Photoelectric effect},\\ \footnotesize `the light quantum')
    \yearArrowLabel(1916, -2.0,1.2, \footnotesize {\bf Einstein}: {\it A and B} \\ \footnotesize {\it coefficients})
  \yearArrowLabel(1913, 1.0,1.0, \footnotesize {\bf Bohr}: {\it space quantization} \\ \footnotesize for atomic stability, \\ \footnotesize {\it indivisible} quantum jump)
  \yearArrowLabel(1937, 1.0,1.5, \footnotesize {\bf Rabi}: disparity in probabilities \\ \footnotesize depending on external field rotation)
  \yearArrowLabel(1922,-4.0,1.2,\footnotesize {\bf Stern-Gerlach} experiment: interaction with quantized momentum)
  \yearArrowLabel(1924.2,1.2,1.3,\footnotesize {\bf De Broglie}: $p\lambda=h$)
  \yearArrowLabel(1925,-1.2,1.2,\footnotesize {\bf Schr\"{o}dinger}: \\ $i\hbar \dot{\psi}=\hat{H}\psi$)
    \yearArrowLabel(1927,4.25,1.2,\footnotesize {\bf Heisenberg}: $\Delta x \Delta p \geq \hbar/2$)
 \yearArrowLabel(1936,-1.2,1.9,\footnotesize {\bf Birkhoff} and {\bf von Neumann}:\\
 \footnotesize `Quantum logic')
\end{tikzpicture}

\noindent\rule[0.5ex]{\linewidth}{2pt}

\begin{tikzpicture}[]
 
  \newcount\yearOne; \yearOne=1940
  \def\w{15}    
  \def\n{4}     
  \def\lt{0.40} 
  \def\lf{0.36} 
  \def\lo{0.30} 
 
  \def\yearLabel(#1,#2){\node[above] at ({(#1-\yearOne)*\w/\n/10},\lt) {#2};}
  \def\yearArrowLabel(#1,#2,#3,#4){
    \def\xy{{(#1-\yearOne)*\w/\n/10}}; \pgfmathparse{int(#2*100)};
    \ifnum \pgfmathresult<0
      \def\yyp{{(\lt*(0.90+#2))}}; \def\yyw{{(\yyp-\lt*#3)}}
      \draw[<-,thick,black!60!green,align=center] (\xy,\yyp) -- (\xy,\yyw) node[below,black] at (\xy,\yyw) {#4};
    \else
      \def\yyp{{(\lt*(0.10+#2)}}; \def\yyw{{(\yyp+\lt*#3)}}
      \draw[<-,thick,black!25!red,align=center] (\xy,\yyp) -- (\xy,\yyw) node[above,black] at (\xy,\yyw) {#4};
    \fi}
 
  \draw[->,thick] (-\w*0.03,0) -- (\w*1.03,0);
 
  \foreach \tick in {0,1,...,\n}{
    \def\x{{\tick*\w/\n}}
    \def\year{\the\numexpr \yearOne+\tick*10 \relax}
  	\draw[thick] (\x,\lt) -- (\x,-\lt) 
	             node[below] {\year};
 
	\ifnum \tick<\n
	  \draw[thick] ({(\x+\w/\n/2)},0) -- ({(\x+\w/\n/2)},\lf); 
      \foreach \ticko in {1,2,3,4,6,7,8,9}{
        \def\xo{{(\x+\ticko*\w/\n/10)}}
  	    \draw[thick] (\xo,0) -- (\xo,\lo);  
	}\fi
  }
 
      \yearArrowLabel(1946, -3.9,1.0, \footnotesize {\bf Ionescu} and {\bf Mihu}: {\it first} H {\it maser})
    \yearArrowLabel(1947, 1.0,1.0, \footnotesize {\bf Lamb} and {\bf Retherford}: {\it shift in $\Delta E$ of H levels,}\\
     \footnotesize `Lamb shift')  )
    \yearArrowLabel(1946, -1.5,1.3, \footnotesize {\bf Bloch} and {\bf Purcell}: {\it Nuclear Magnetic Resonance (NMR)})
    \yearArrowLabel(1953, 5.5,1.0, \footnotesize {\bf Townes}, {\bf Gordon} and {\bf Zeiger}: {\it first ammonia maser})   
        \yearArrowLabel(1954, -4.8,1.0, \footnotesize {\bf Dicke}: {\it Super-radiance} from a molecular gas)  
        \yearArrowLabel(1957, 3.5,1.0, \footnotesize {\bf Townes} and {\bf Schawlow}: {\it optical maser}) 
                       \yearArrowLabel(1961, -6.8,1.0, \footnotesize {\bf Fano}: {\it Coherent scattering} from state coupled to a continuum) 
                        \yearArrowLabel(1963, 1.0,1.0, \footnotesize {\bf Jaynes} and {\bf Cummings}:\\
                       \footnotesize  \underline{\it JC Hamiltonian} for the ammonia maser) 
    \yearArrowLabel(1973, -2.0,1.0, \footnotesize {\bf Hepp} and {\bf Lieb}, {\bf Wang} and {\bf Hioe}:\\
                       \footnotesize  predict phase transition in the {\it Dicke model}) 
    \yearArrowLabel(1973, 1.0,1.0, \footnotesize {\bf Meystre} {\it et al.}:\\
                       \footnotesize  {\it collapse-revival})
     \yearArrowLabel(1980, 1.0,1.0, \footnotesize {\bf Eberly} {\it et al.}:\\
                       \footnotesize  {\it collapse-revival}\\
                       \footnotesize quantum signature)
\end{tikzpicture}

\noindent\rule[0.5ex]{\linewidth}{2pt}

\begin{tikzpicture}[]
 
  \newcount\yearOne; \yearOne=1980
  \def\w{15}    
  \def\n{4}     
  \def\lt{0.40} 
  \def\lf{0.36} 
  \def\lo{0.30} 
 
  \def\yearLabel(#1,#2){\node[above] at ({(#1-\yearOne)*\w/\n/10},\lt) {#2};}
  \def\yearArrowLabel(#1,#2,#3,#4){
    \def\xy{{(#1-\yearOne)*\w/\n/10}}; \pgfmathparse{int(#2*100)};
    \ifnum \pgfmathresult<0
      \def\yyp{{(\lt*(0.90+#2))}}; \def\yyw{{(\yyp-\lt*#3)}}
      \draw[<-,thick,black!60!green,align=center] (\xy,\yyp) -- (\xy,\yyw) node[below,black] at (\xy,\yyw) {#4};
    \else
      \def\yyp{{(\lt*(0.10+#2)}}; \def\yyw{{(\yyp+\lt*#3)}}
      \draw[<-,thick,black!25!red,align=center] (\xy,\yyp) -- (\xy,\yyw) node[above,black] at (\xy,\yyw) {#4};
      
    \fi}
 
  \draw[->,thick] (-\w*0.03,0) -- (\w*1.03,0);
 
  \foreach \tick in {0,1,...,\n}{
    \def\x{{\tick*\w/\n}}
    \def\year{\the\numexpr \yearOne+\tick*10 \relax}
  	\draw[thick] (\x,\lt) -- (\x,-\lt) 
	             node[below] {\year};
 
	\ifnum \tick<\n
	  \draw[thick] ({(\x+\w/\n/2)},0) -- ({(\x+\w/\n/2)},\lf); 
      \foreach \ticko in {1,2,3,4,6,7,8,9}{
        \def\xo{{(\x+\ticko*\w/\n/10)}}
  	    \draw[thick] (\xo,0) -- (\xo,\lo);  
	}\fi
  }
 
      \yearArrowLabel(1985, -2.0,1.0, \footnotesize {\bf Walther} {\it et al.}: {\it Strong-coupling} regime)
       \yearArrowLabel(1987, 1.0,1.0, \footnotesize {\bf Walther} {\it et al.}: {\it JC collapse-revival} demonstrated)
        \yearArrowLabel(1988, -4.0,1.0, \footnotesize {\bf Phoenix} and {\bf Knight}: {\it Entanglement} in the JC model)
          \yearArrowLabel(1995, 3.0,1.0, \footnotesize {\bf Cirac} and {\bf Zoller}: proposal of {\it CNOT gate} with trapped ions)
          \yearArrowLabel(1995, -6.0,1.0, \footnotesize {\bf Wineland} {\it et al.}: experimental demonstration of {\it CNOT gate})
          \yearArrowLabel(1996, -8.8,1.0, \footnotesize {\bf Haroche} {\it et al.}: experiment on the decay of {\it Schr\"{o}dinger cat} state)
          \yearArrowLabel(2004, 2.0,1.0, \footnotesize {\bf Wallraff} {\it et al.}: birth of {\it circuit QED})
          \yearArrowLabel(2005, -2.0,1.0, \footnotesize {\bf Kimble} {\it et al.}: {\it photon blockade} in cavity QED)
          \yearArrowLabel(2006, -4.7,1.0, \footnotesize {\bf Greentree} and {\bf Hartmann}  propose {\it cavity arrays})
          \yearArrowLabel(2007, -6.9,1.0, \footnotesize {\bf Dimer} {\it et al.}  propose {\it Raman transitions} to realize an effective Dicke model)
			\yearArrowLabel(2010,5.0,1.0, \footnotesize {\bf Esslinger} {\it et al.}:  {\it Dicke phase transition}\\
			\footnotesize realized with cavity BEC)
			\yearArrowLabel(2010,8.0,1.0, \footnotesize {\bf Gross} {\it et al.} and {\bf Mooij} {\it et al.} attain the {\it ultra-strong} coupling regime)
			\yearArrowLabel(2012,1.0,1.0, \footnotesize {\bf Braak} solves analytically the {\it Rabi model})
			\yearArrowLabel(2015,-9.6,1.0, \footnotesize {\bf Carmichael} shows that photon blockade \\
			\footnotesize {\it  breaks down by means of a dissipative\index{Dissipative! quantum phase transition} quantum phase transition})
			\yearArrowLabel(2017,-2.6,1.0, \footnotesize {\bf Houck} {\it et al.} couple \\
			\footnotesize {\it 72 superconducting resonators})
\end{tikzpicture}
\caption{{\it A most personal time-line of some of the most important milestones shaping JC physics.} {\bf 1.} At the turn of the 20$^{\rm th}$ century, Planck introduces an energy element $h \nu$ to interpret the spectral energy density of blackbody radiation. {\bf 2.} Based on Planck's theory, in 1905, Einstein proposes that light waves consist of {\it photons} or quanta to interpret the photoelectric effect. {\bf 3.} Bohr introduces {\it space quantization} in 1913, maintaining that electrons in atoms could only exist in certain well-defined, stable orbits. {\bf 4.} Einstein introduces two phenomenological coefficients $A$ and $B$\index{Einstein $A$ and $B$ theory} in 1916, to quantify the rate of spontaneous emission\index{Spontaneous! emission! rate} and stimulated absorption/emission respectively\index{Stimulated! emission}\index{Stimulated! absorption}. {\bf 5.} Gerlach conducts in 1922 the experiment devised by Stern a year earlier, demonstrating that the spatial orientation of angular momentum is quantized. {\bf 6.} de Broglie introduces {\it matter waves} in his 1924 thesis, first experimentally verified by George Paget Thomson's thin metal diffraction experiment three years later.}\label{fig0}
\end{figure*}
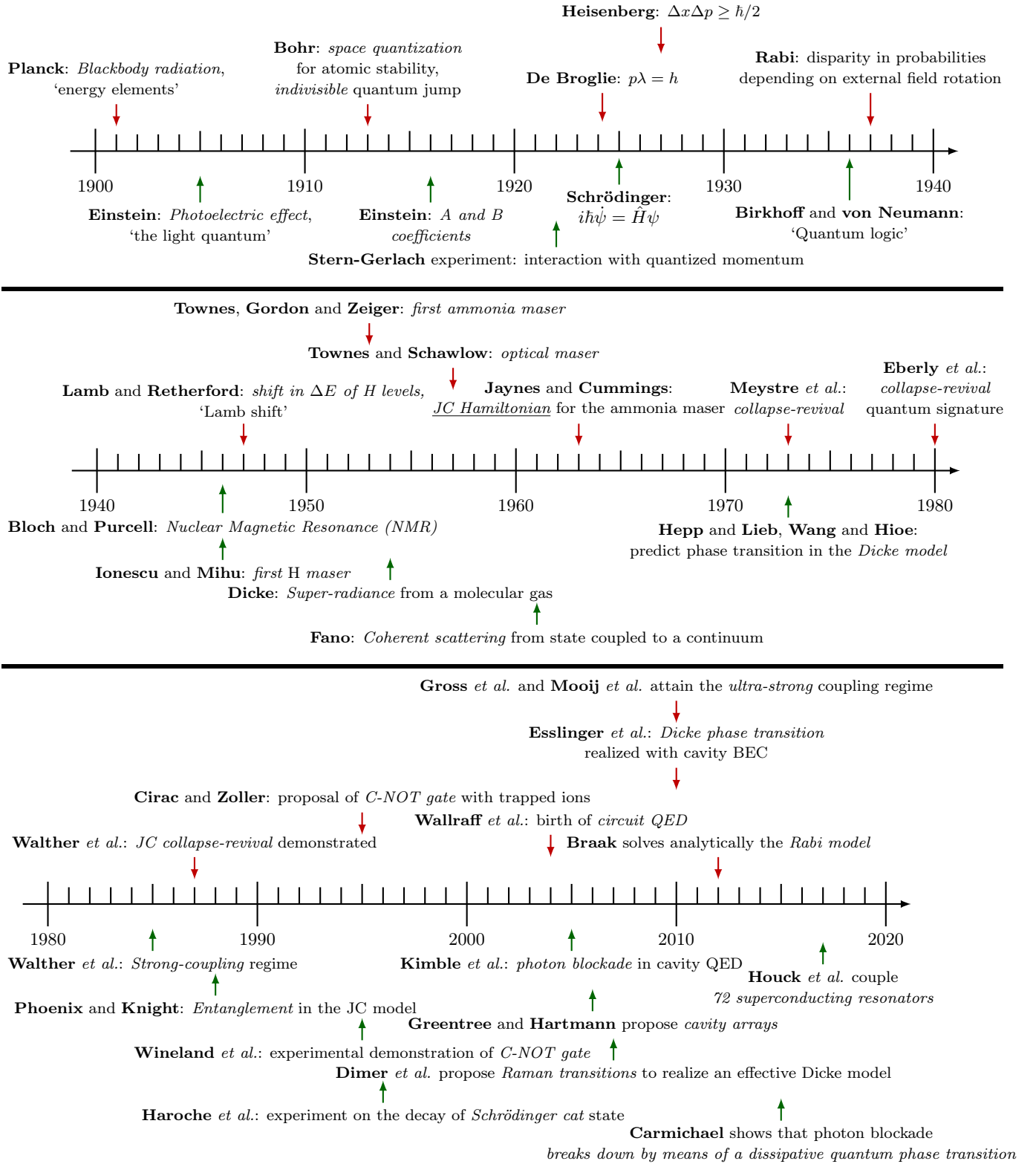

In the early 1960s, the laser entered physics and technology, and optical cavities were designed favoring a discrete mode spectrum; the radiation field, therefore, was coherent enough to sequester only a few atomic levels as candidates for the interaction. This possibility opened the door to the JC formulation. The Hamiltonian of eq. \eqref{eq:simpltwospecies} can be readily reduced to model the interaction of a two-level transition (a `single molecule', as termed by Jaynes and Cummings) with a normal mode arising as a solution to a boundary value problem; a resonant mode in a cavity constitutes a characteristic example. In order to realize the JC model physics experimentally, the quality factor of the cavities needed to be high enough to allow for coherent evolution, and it took until 1985 for the first experiments to reach this so-called {\it strong coupling regime}~\cite{meschede1985one}. However, once this was achieved, the field of cavity quantum electrodynamics (QED) seriously got started~\cite{dutra2005cavity}. Roughly two decades followed with fascinating experiments, ranging from exploring fundamental aspects of quantum mechanics to realizing simple quantum information processing (QIP) schemes, before the next great breakthrough in cavity QED. To not rush too much ahead, we first note that during the 1990s, the JC model found a new arena where it could be experimentally realized -- trapped ion physics~\cite{blockley1992quantum}. These systems turned out extremely versatile and robust, something proved by the group of David Wineland which was the first to demonstrate a $C$-NOT gate in 1995~\cite{monroe1995demonstration}. While in the 1990s JC physics was mainly to be found in the communities of cavity QED and trapped ions, the 2000s saw the birth and rapid growth of circuit QED. In 1999, Nakamura and co-workers had shown, by means of controlling the coherent dynamics of a Cooper-pair box\index{Cooper pair! box}, how macroscopic matter can be used to study quantum phenomena~\cite{nakamura1999coherent}. Five years later such a superconducting q-dot, acting as an artificial two-level atom, was successfully strongly coupled to a transmission line resonator~\cite{Wallraff2004QED}. It was now possible to implement JC physics on an electronic chip. Today, the advance of circuit QED occurs on a massive scale with several experimental groups worldwide. Lately, common to all these three fields, cavity QED, trapped ion physics, and circuit QED, is the trend to push the limits in how far one can go. We briefly mention a few examples, which will be further discussed in later sections: Within cavity QED systems, instead of single atoms, Bose-Einstein condensates formed from tens of thousands of atoms have been trapped and coherently coupled to single cavity modes~\cite{brennecke2007cavity}. Trapped ion physics has, for instance, demonstrated how to construct {\it quantum simulators} of paradigmatic spin systems~\cite{smith2016many}. Interacting spin models have also been experimentally realized in circuit QED~\cite{stockklauser2017strong}, as well as coupling together 72 transmission line resonators on a chip for exploring nonequilibrium phase transitions\index{Nonequilibrium! phase transition} of light~\cite{fitzpatrick2017observation}\index{Phase transition! quantum! out of equilibrium}. In fig.~\ref{fig0} we list, in chronological order, some of the important discoveries/achievements prior and after the 1963 Jaynes-Cummings paper. The reader may discover notable omissions in that list, while it is clear that the timeline only depicts a small selection chosen from hundreds of essential works.

\clearpage
\noindent (cont.) {\bf 7.} In 1925, Schr\"{o}dinger postulates the linear partial differential equation bearing his name, describing the wave function of a quantum-mechanical system. The same year also witnesses the discovery of {\it electron spin} by Uhlenbeck and Goudsmit. One year later, Pauli and Schr\"{o}dinger show that the Rydberg formula for the spectrum of hydrogen follows from the new theory of quantum mechanics. {\bf 8.} In 1927, Heisenberg introduces the {\it uncertainty principle} stating that the more accurately the position of a given particle is determined, the less precisely its momentum can be predicted from initial conditions, and vice versa. The formal inequality written in terms of standard deviations was derived one year latter by Kennard and Weyl. {\bf 9.} In 1936, Birkhoff and von Neumann attempt to reconcile the apparent inconsistency of classical logic with the facts concerning the measurement of complementary variables in quantum mechanics. In a modern interpretation, quantum logic can be regarded as a diagram of classical logic. {\bf 10.} In 1937, Rabi demonstrates a disparity between transition probabilities in a two-state system (of atomic angular momentum in an external rotating magnetic field), depending on the sign of the Land\'{e} factor. {\bf 11.} In 1946, Ionescu and Mihu build and test in Bucharest a precursor of the ammonia maser, a hydrogen-based device operating on stimulated emission\index{Stimulated! emission}. In the same year, Bloch and Purcell demonstrate {\it nuclear magnetic resonance} (NMR) in water and parafine, following the work of Rabi on the magnetic properties of various nuclei.
{\bf 12.} In 1947, Lamb and Retherford measure the small energy shift (termed {\it Lamb shift})\index{Lamb! shift} between the $^2\text{S}_{1/2}$ and $^2\text{P}_{1/2}$ levels of hydrogen, in contradiction to the direct analytical solution of the Dirac equation, thus providing a great stimulus to the development of quantum electrodynamics. {\bf 13.} In 1953, Townes, Gordon and Zeiger build the first ammonia maser in which excited molecules deliver energy to a microwave resonator. Four years later, Schawlow and Townes, demonstrate ``an extension of maser techniques to the infrared and optical region is considered. It is shown that by using a resonant cavity of centimeter dimensions, having many resonant modes, maser oscillation at these wavelengths can be achieved by pumping with reasonable amounts of incoherent light.'' {\bf 14.} In 1954, Dicke demonstrates coherence in spontaneous radiation\index{Spontaneous! emission} emitted from a molecular gas, coining the term {\it super-radiance}. {\bf 15.} In 1961, Fano demonstrates that the interference of a discrete auto-ionized state with a continuum gives rise to asymmetric peaks in the excitation spectra. {\bf 16.} Starting already in 1957-1958 a student project, finally in 1963 Jaynes and Cummings publish their work~\cite{jaynes1963comparison}. {\bf 17.} Hepp and Lieb demonstrate that the Dicke model is critical~\cite{hepp1973superradiant}. {\bf 18.} Eberly {\it et al.} show that the JC model predicts evolution revivals as a result of field quantization~\cite{eberly1980periodic}. About seven years before, in 1973, Meystre and coworkers~\cite{Meystre1973} had calculated the excitation probability and atomic dipole moment under JC evolution for an initial coherent state. {\bf 19.} The strong-coupling regime is reached in the Garching experiments in the group of Walther~\cite{meschede1985one}. {\bf 20.} Two years later, in 1987, the JC collapse-revival pattern is attained by the same group~\cite{rempe1987observation}. {\bf 21.} The first references to entanglement in the JC model is published in a work by Phoenix and Knight~\cite{phoenix1988fluctuations}. {\bf 22.} Cirac and Zoller suggest how to implement a quantum $C$-Not gate with trapped ions following JC physics~\cite{cirac1995quantum}. In the same year, the gate is experimentally demonstrated in the NIST group of Wineland~\cite{monroe1995demonstration}. {\bf 23.} The Paris ENS group led by Haroche captures the decay of a prepared Schr\"{o}dinger cat state from a superposition into a statistical mixture~\cite{brune1996quantum}. {\bf 24.} In 2004, Wallraff and co-workers demonstrate for the first time strong coupling between a q-dot and a transmission line resonator~\cite{Wallraff2004QED}, which marks the beginning of circuit QED. {\bf 25.} Photon blockade, an intrinsic result of the JC nonlinearity, is demonstrated at CalTech in the group of Kimble~\cite{birnbaum2005photon}. {\bf 26.} The following year saw two independent proposals on how to simulate interacting quantum many-body physics with JC physics; Hartmann {\it et al} suggested photonic crystals to construct cavity arrays~\cite{hartmann2006strongly}, while Greentree {\it et al.} considered nitrogen-vacancy centers~\cite{greentree2006quantum}. {\bf 27.} By coherently coupling an atomic BEC to an optical cavity, the ETH group of Esslinger demonstrated the Dicke phase transition, and in the same year both the Delft group of Mooij~\cite{forn2010observation} and the Garching group of Gross~\cite{niemczyk2010circuit} reported the first experiments operating in the ultrastrong coupling regime. {\bf 28.} Braak finds the analytical solution of the quantum Rabi model~\cite{braak2011integrability}. {\bf 29.} In 2015, Carmichael shows that photon blockade in the open driven JC model breaks down by way of a dissipative quantum phase transition\index{Dissipative! quantum phase transition} in zero dimensions, experimentally verified two years later by the group of J. Fink in IST. {\bf 30.} A crucial step towards manufacturing and controlling JC lattices is taken at Princeton through the coupling of 72 driven JC cavities~\cite{fitzpatrick2017observation}.

\noindent\rule[0.5ex]{\linewidth}{1pt}

\vspace{3mm}

The interaction between a linear harmonic oscillator and a two-level atom modelling an atomic transition, a system with paradigmatic nonlinearity, is omnipresent across various fields of modern physics, beyond those mentioned above.  However, the JC model also appears in a wide variety of different and distinct composite systems. For example, in the interaction picture, the JC Hamiltonian can be mapped onto the Dirac equation which is first order in its spatial derivative (contrary to the second-order Schr\"odinger equation)~\cite{rozmej1999dirac}. As such, it has been employed to experimentally simulate relativistic effects like {\it Zitterbewegung}\index{Zitterbewegung}~\cite{gerritsma2010quantum} and the {\it dynamical Casimir effect}~\cite{wilson2011observation}. Furthermore, the JC model has also appeared in recent days in the physics of cold atomic gases, like in the explanation of a {\it quantum Hall effect} in these systems~\cite{goldman2009non}, or the {\it Rydberg blockade} phenomenon~\cite{beterov2014jaynes}. We also encounter JC physics in the study of electron cyclotron motion in graphene~\cite{schliemann2008cyclotron}, or when describing the coherent coupling between a q-dot and spin-waves in a magnet~\cite{tabuchi2015coherent,lachance2017resolving}. 

In this monograph, we will explore notable ramifications stemming from the JC oscillator in modern physics. The Shore and Knight review article titled {\it The Jaynes-Cummings model} dating from 1993~\cite{shore1993jaynes} gives a good summary of the works until that date, but a plethora of results have seen the light since then, as is evident from fig.~\ref{fig0}. Contrary to that reference, the present work also enters into neighbouring areas closely related to the physics of the JC model. The path we take will not go much into details of the various systems and concepts; rather, we aim to visit many of them within our overarching theme, and provide the relevant references for a more in depth reading. At first we introduce the JC model on a purely mathematical level, and discuss general properties of it by placing special emphasis on the role of quantum fluctuations. The prime example is that of {\it collapse-revivals}, but also other instances are also explored like atom-field entanglement, squeezing and the ``thermodynamic'' limits. Much, but far from all, of the work in these subsections can be found in introductory course books in quantum optics, see for example Refs.~\cite{mandel1995optical,scully1999quantum,walls2007quantum,meystre2007elements,ficek2016quantum,klimov2009group,orszag2008quantum,vogel2006quantum,schleich2011quantum,garrison2008quantum}. We then leave aside the more standard aspects that can be also found in more orthodox textbooks, and move on to somewhat more specialist topics namely driven and dissipative JC systems, and in particular the characteristic phenomenon of photon blockade\index{Photon! blockade}. One section is devoted to the quantum Rabi model which has become most relevant in recent years with experiments reaching the ultrastrong coupling regime. The significance of the rotating-wave approximation (RWA) was discussed very early on in the history of the JC model, and analytical approximations to take the counter-rotating terms into account were put forward. The theoretical section also includes a section on various extensions of the JC model. The most important extension is most probably that of considering many atoms, {\it i.e.} the Dicke model. It was long believed that the predicted phase transition (PT) of the Dicke model (or Tavis-Cummings model)~\cite{hepp1973superradiant,wang1973phase} was only of academic interest due to a no-go theorem forbidding it in rather general settings~\cite{rzazewski1975phase,bialynicki1979no,andolina2019cavity}. However, by considering driven nonequilibrium configurations\index{Nonequilibrium! configuration}, it was shown how to circumvent the rule of this theorem~\cite{dimer2007proposed}, and since then it has been demonstrated in different settings~\cite{baumann2010dicke,baden2014realization,lin2011spin}. This theoretical section ends with a rather lengthy discussion on the approximations behind the JC model, something that is usually not found in the literature. In recent years, it has been reported that some of the approximations have far deeper consequences than first thought, namely breaking gauge invariance. The following sections are divided into the various systems in which JC physics can be realized. For both historical and practical reasons we start by considering cavity QED in sec.~\ref{sec:cavQED}. This was the first system to demonstrate JC evolution back in the mid 1980s~\cite{meschede1985one}. Apart from paying attention to experiments that explicitly realize JC physics, we also discuss important related features like optical bistability, while we also place some emphasis on the function of the micromaser. Circuit QED is the topic of sec.~\ref{sec:cirQED}. The similarities to cavity QED are evident, but the ingredients are very different and we thereby devote some time to explain the physics of superconducting circuits. Trapped-ion systems are conceptually different from the previous two; the two-level structure is still represented by internal states of the matter subsystem, but the bosons are not represented by photons but by phonons. In this respect, demonstrating the collapse-revival phenomenon in cavity QED is of much more fundamental interest since it provides evidence for the quantization of the electromagnetic field, while in trapped-ion systems it would only reaffirm that the vibrational energies of an ion are discrete. Nevertheless, trapped-ion systems became one of the most promising platforms for implementing JC physics and QIP. As we have already mentioned above, there is of course a set of other configurations that are well described by a two-level system coupled to a harmonic oscillator, and some of them, especially their hybrid representatives, are briefly acknowledged in sec.~\ref{sec:physreal}, after having focused on light-matter interaction in waveguides in sec.~\ref{sec:waveguideQED} and its links to circuit QED. In recent years, new directions have been laid out in terms of emulating paradigmatic quantum many-body Hamiltonians. Cold-atom systems were the first to break new ground in the direction of {\it quantum simulators}, but nowadays this line of research is also pursued in circuit/cavity QED and trapped-ion systems~\cite{cirac2012goals}. Section~\ref{sec:ext} discusses two types of systems that have come to play an important role in this progress. In the first part, we look into Jaynes-Cummings-Hubbard models which arise from manufacturing on-chip lattices of JC cavities. Effectively, this results in Hubbard-type lattice models. In the second example we focus on the coherent coupling between atomic Bose-Einstein condensates and single cavity modes. Again, these systems often lead to effective interacting quantum-lattice models. And our third example concerns cold molecules coupled to a single or several cavity modes. It has recently been demonstrated how such couplings can influence the `energy landscape' of molecules and thereby alter their behaviour in chemical reactions or in dissociation. In the Conclusions, we give a personal outlook on what we may expect from the coming decades. 


\section{Theoretical aspects}
This section, by far the longest in this monograph, introduces the JC Hamiltonian as a theoretical model. The first two sections go through its analytical solutions and general properties of the model and its corresponding physical quantities, but so far no attention is really payed to the connections to real physical systems. This is left for sec.~\ref{ssec:approx} in which we go through various approximations leading to the JC model. Here we also present a microscopic derivation of the model starting from a minimal coupling Hamiltonian. Incidentally, one of the earliest examples of dressing due to radiation-matter interaction -- worth summarizing on a quantitative level -- appears a few years before the celebrated JC Hamiltonian. In 1961, Fano reports \cite{Fano1961} on the interaction between one discrete state $\phi$ and a continuum of states $\psi_E$, producing the eigenstate 
\begin{equation}\label{eq:Fano}
\Psi_E = a(E) \phi + \int dE^{\prime} b(E^{\prime})\psi(E^{\prime}),
\end{equation}
with 
\begin{equation*}
\begin{aligned}
&|a(E)|^2=\frac{|V(E)|^2}{[E-E_{\phi}-F(E)]^2 + \pi^2 |V(E)|^4}, \hspace{1cm}
 b(E^{\prime})=\frac{V(E^{\prime})}{\pi V(E)} \frac{\sin \Delta}{E-E^{\prime}}-\cos \Delta\, \delta(E-E^{\prime}),
 \end{aligned}
\end{equation*}
where $\Delta \equiv - \arctan[\pi |V(E)|^2/(E-E_{\phi}-F(E))]$. In the above expressions, $E_{\phi}=\braket{\phi|\hat{H}|\phi}$, $V(E) \equiv \braket{\psi(E^{\prime})|\hat{H}|\phi}$ and $F(E)=\mathcal{P}\int dE^{\prime} |V(E^{\prime})|^2/(E-E^{\prime})$ (here $\mathcal{P}$ stands for the principal value). The phase shift due to the interaction determines the matrix element (for a set transmission operator $T$) between an initial state $I$ and the eigenstate $\Psi(E)$, as follows:
\begin{equation}\label{eq:Fanoadmixture}
\braket{\Psi(E)|\hat{T}|I}=\frac{1}{\pi V^{*}(E)} \braket{\Phi|\hat{T}|I}\sin\Delta-\braket{\psi(E)|\hat{T}|I}\cos\Delta,
\end{equation}
with $\Phi \equiv \phi + \mathcal{P} \int dE^{\prime} V(E^{\prime})\psi(E^{\prime})/(E-E^{\prime})$, showing explicitly that the discrete state $\phi$ is modified by a linear combination of states belonging to the continuum. Having now acquainted ourselves with this introductory example, in this section we formalise light-matter interaction in the language of quantum optics alongside the main approximations and simplifications involved in the description. Finally, we discuss some of the numerous extensions of the JC model, like the quantum Rabi model and the Dicke model. 

\subsection{The Jaynes-Cummings model}\label{ssec:JCm}
Throughout this monograph, boson operators of the electromagnetic (or phonon) field are denoted $\hat{a}_i$ and $\hat{a}_i^\dagger$\index{Creation/annihilation operators} where the subscript stand for mode $i$ corresponding to a photon energy $\hbar\omega_i$ (In subsection.~\ref{ssec:mbcQED} we will introduce another set of boson operators $\hat{b}_i$ and $\hat{b}_i^\dagger$ describing matter excitations, and sometimes the subscript $i$ is replaced by $k$ in order to emphasize that it labels a certain momentum $k$.). The {\it Fock states}\index{Fock state}\index{State! Fock} $|n_i\rangle$ ($n_i=0,\,1,\,2,\,...$) are eigenstates of the {\it number operator}\index{Number! operator}\index{Operator! number} $\hat{n}_i\equiv\hat{a}_i^\dagger\hat{a}_i$, {\it i.e.} $\hat{n}_i|n_i\rangle=n_i|n_i\rangle$. Using the boson commutation relation $\left[\hat{a}_i,\hat{a}_j^\dagger\right]=\delta_{ij}$ it further follows that $\hat{a}_i|n_i\rangle=\sqrt{n_i}|n_i-1\rangle$ and $\hat{a}_i^\dagger|n_i\rangle=\sqrt{n_i+1}|n_i+1\rangle$. When a single mode is considered, which is the case most of the time, the subscript will be dropped. Angular momentum operators, or spin operators, are written as $\hat{S}_\alpha$ ($\alpha=x,\,y,\,z$), and they obey the regular commutation relation $\left[\hat{S}_\alpha,\hat{S}_\beta\right]=i\varepsilon_{\alpha\beta\gamma}\hat{S}_\gamma$\index{Angular momentum operator}, where $\alpha$, $\beta$, and $\gamma$ may be $x$, $y$, or $z$ and $\varepsilon_{\alpha\beta\gamma}$ is the antisymmetric Levi-Civita tensor. The total spin is here assumed to be $S$. Most often we are interested in the spin-$1/2$ manifold, {\it i.e.} $S=1/2$, and for this special case we use the more conventional notation $\hat{\sigma}_\alpha$, where
\begin{equation}\label{pauli}
\begin{array}{lll}
\hat{\sigma}_x=\left[\begin{array}{cc}
0 & 1\\ 1 & 0\end{array}\right], & \hat{\sigma}_y=\left[\begin{array}{cc}
0 & -i\\ i & 0\end{array}\right], & \hat{\sigma}_z=\left[\begin{array}{cc}
1 & 0\\ 0 & -1\end{array}\right]
\end{array}
\end{equation}
are the regular Pauli matrices\index{Pauli! matrices}, which will also be refereed to as {\it dipole operators}\index{Dipole! operator}, $\hat\sigma_x$ and $\hat\sigma_y$, and {\it inversion operator}\index{Inversion! operator}\index{Operator! inversion}\index{Operator! dipole}, $\hat\sigma_z$. The Pauli matrices have been expressed in the ``atomic'' ground $|g\rangle$ and excited $|e\rangle$ states. The ``atom'' constitutes a two-level system, and sometimes we also refer to it simply as {\it two-level (two-state) system}\index{Two-level! system} and sometimes also as {\it qubit}\index{Qubit}. The energy difference between the bare atomic states is denoted $\hbar\Omega$. The {\it bare states}\index{Bare states} for a single bosonic mode are defined by the product states as
\begin{equation}\label{bstate}
|e,n\rangle=|e\rangle\otimes|n\rangle,\hspace{1cm}|g,n\rangle=|g\rangle\otimes|n\rangle.
\end{equation}
The generalization to a higher number of modes is straightforward. These states are eigenstates of the {\it excitation operator}\index{Excitation operator} 
\begin{equation}\label{exop}
\hat{N}=\hat{n}+\hat{\sigma}_z/2
\end{equation} 
with eigenvalues $n\pm1/2$ respectively. Thus, $|e,n-1\rangle$ and $|g,n\rangle$ both share the number of ``excitations'' $n-1/2$. 

In the JC model, the two-level system exchanges excitations with the boson mode such that absorption lowers the boson number by one unit, while the two-level system is excited from $|g\rangle$ to $|e\rangle$. The opposite situation (the conjugate process) takes place for emission. The vacuum rate (at resonance) of such {\it Rabi oscillations}\index{Rabi! oscillations} is denoted by $g$ -- the {\it vacuum Rabi coupling}\index{Vacuum Rabi! coupling} or {\it light-matter coupling}\index{Light-matter coupling}. With a simple rotation, either with the unitary $\hat U(\varphi)=\exp(i\hat{\sigma}_z\varphi)$ or $\hat U(\varphi)=\exp(i\hat n\varphi)$, we can always chose $g$ to be real (hence, if not otherwise stated, the atom-field coupling is always taken real.). This is nothing but a gauge transformation\index{Gauge! transformation}, and we will return to this a few times later in the monograph. The JC Hamiltonian now reads (we will keep $\hbar=1$ throughout our discussion apart from places where it might be informative to keep it in the expressions)
\begin{equation}\label{jcham}
\hat{H}_\mathrm{JC}'=\hat H_0+\hat H_\mathrm{int},
\end{equation}
with
\begin{equation}\label{inth}
\hat H_0=\omega\hat{n}+\frac{\Omega}{2}\hat{\sigma}_z,\hspace{1cm}\hat H_\mathrm{int}=g\left(\hat{a}^\dagger\hat{\sigma}_{-}+\hat{\sigma}_{+}\hat{a}\right),
\end{equation}
the {\it free} and {\it interaction Hamiltonians}\index{Interaction! Hamiltonian}\index{Jaynes-Cummings! model}\index{Model! Jaynes-Cummings}, and where the raising/lowering operators $\hat{\sigma}^\pm=\left(\hat{\sigma}_x\pm i\hat{\sigma}_y\right)/2$. In sec.~\ref{ssec:approx}, we will present a microscopic derivation of the JC Hamiltonian~(\ref{jcham}), see also text books like~\cite{mandel1995optical,scully1999quantum,gerry2005introductory}. The number of excitations is manifestly preserved in the JC model. Thus, the JC Hamiltonian possesses a $U(1)$ symmetry\index{$U(1)$ symmetry}
\begin{equation}\label{jcU1}
[\hat{U}(\varphi),\hat{H}_\mathrm{JC}^{\prime}]=0, \hspace{1cm}\hat{U}(\varphi)=\exp\left(-i\varphi\hat{N}\right),\,\,\varphi\in\mathbb{R}.
\end{equation}
It is convenient to turn to an interaction picture\index{Interaction! picture} with respect to $\omega\hat{N}$, {\it i.e.} the Schr\"odinger equation $i\partial_t|\psi\rangle=\hat H'|\psi\rangle$ is transformed by the (time-dependent) unitary operator $\hat U=\exp\left(-i\hat Nt\right)$ leading to a transformed Hamiltonian $\hat H=\hat U\hat H'\hat U^\dagger-i\hat U\partial_t\hat U^\dagger$. Using that $\hat U\hat a\hat U^\dagger=\hat ae^{i\omega t}$ and $\hat U\hat{\sigma}_{-}\hat U^\dagger=\hat{\sigma}_{-}e^{-i\omega t}$~\cite{mandel1995optical} we derive
\begin{equation}\label{jcham2}
\hat{H}_\mathrm{JC}=\frac{\Delta}{2}\hat{\sigma}_z+g\left(\hat{a}^\dagger\hat{\sigma}_{-}+\hat{\sigma}_{+}\hat{a}\right),
\end{equation}
thereby reducing the number of parameters from three to two: the atom-field detuning\index{Atom!-field detuning} $\Delta=\Omega-\omega$ and the light-matter coupling $g$. It is clear that in the bare basis~(\ref{bstate}) the JC Hamiltonian is on block form with $2\times2$ blocks
\begin{equation}\label{block}
\hat H_\mathrm{JC}=\bigotimes_{n}\hat h_n,\hspace{0.4cm}\mathrm{with}\hspace{0.5cm}\hat{h}_n=\left[\begin{array}{cc}
\displaystyle{\frac{\Delta}{2}} & g\sqrt{n+1}\\
g\sqrt{n+1} & -\displaystyle{\frac{\Delta}{2}}\end{array}\right]
\end{equation}
acting on the states $|e,n\rangle$ and $|g,n+1\rangle$. The only bare state which is an eigenstate of the JC Hamiltonian (with $g\neq0$) is $|g,0\rangle$ with an eigenvalue $-\Omega/2$. So there is one block of the Hamiltonian that is simply $1\times1$, while all others are $2\times2$.  Note that the block form results from the continuous $U(1)$ symmetry, hence solving the full problem boils down to diagonalizing a $2\times2$ matrix~\cite{cummings2013reminiscing}.  

\begin{figure}
\includegraphics[width=9cm]{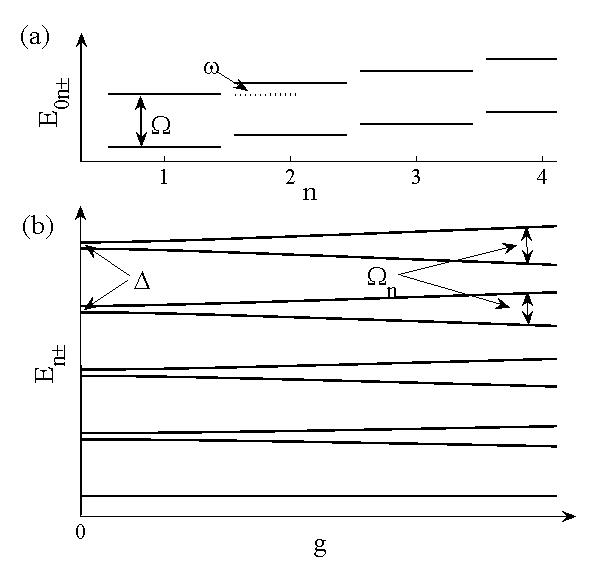} 
\caption{Schematic diagram of the bare {\bf (a)} and dressed {\bf (b)} energies of the JC model. As the atom-field coupling strength increases, the linearity (two equidistant ladders of eigenvalues for $g=0$ in (b)) of the spectrum is lost. This anharmonicity is characteristic for the JC model and plays an important role in many physical systems (see for instance sec.~\ref{sec:cirQED} and \ref{ssec:jch}). The anharmonic spectrum of the JC model is called the {\it Jaynes-Cummings ladder}~\cite{fink2008climbing,kasprzak2010up}. For further larger couplings, the energy levels in (b) begin to cross and build up a much more complex structure.
}
\label{fig1}
\end{figure}

The block form of the JC Hamiltonian implies that it is diagonalized by a unitary $\hat U_\mathrm{JC}=\bigotimes_n\hat U_n$, with 
\begin{equation}\label{jcU}
\hat U_n=\left[
\begin{array}{cc}
\sin(\theta_n/2) & \cos(\theta_n/2)\\
-\cos(\theta_n/2) & \sin(\theta_n/2)
\end{array}\right]
\end{equation}
diagonalizing a given block $\hat h_n$. Hence, the eigenstates\index{Jaynes-Cummings! eigenstates} of $\hat{H}_\mathrm{JC}$ are 
\begin{equation}\label{dstate}
|\psi_{n\pm}\rangle=\sin\left(\frac{\theta_n}{2}\right)|e,n\rangle\pm\cos\left(\frac{\theta_n}{2}\right)|g,n+1\rangle,
\end{equation}
with the angle given by
\begin{equation}
\tan(\theta_n)=\frac{2g\sqrt{n+1}}{\Delta}
\end{equation}
and the corresponding eigenvalues
\begin{equation}\label{eigv}
E_{n\pm}=\pm\Omega_n\equiv\pm\sqrt{\frac{\Delta^2}{4}+g^2(n+1)},
\end{equation}
where we have introduced the {\it Rabi frequencies} $\Omega_n$\index{Rabi! frequency}. The states~(\ref{dstate}) are often referred to as {\it dressed states}\index{Dressed! states} since they can be envisioned as photons "attached" to the atom. In this sense, these states are also the so-called {\it polaritonic states}\index{Polariton!state} or simply {\it polaritons} (see further sec.~\ref{ssec:chem}), which represent composite bosonic quasi-particles~\cite{kasprzak2010up}. The energies are given in the interaction picture, while the true energies become
 \begin{equation}\label{eigv2}
\mathcal{E}_{n\pm}=\omega \left(n\pm\frac{1}{2}\right)\pm\sqrt{\frac{\Delta^2}{4}+g^2(n+1)}.
\end{equation}
At resonance, $\Delta=0$, the angle $\theta_n$ is $n$-independent and the solutions are especially simple; $\cos(\theta/2)=\sin(\theta/2)=1/\sqrt{2}$. In this case we see already from eq.~(\ref{block}) that a so-called {\it Hadamard rotation}\index{Hadamard! transformation} 
\begin{equation}\label{hada}
H_\mathrm{H}=\frac{1}{\sqrt{2}}\left[
\begin{array}{cc}
1 & 1\\
1 & -1
\end{array}\right],
\end{equation}
brings the Hamiltonian into a diagonal form. The basis in which the resonant JC Hamiltonian is diagonal is sometimes referred to as {\it diabatic}~\cite{wang2008quantum}\index{Diabatic! representation}. The eigenstates (here in the interaction picture) are also called {\it dressed states}\index{Dressed! states} from the fact that it is often more relevant to talk about polariton-like particles~\cite{rebic1999large,ridolfo2012photon}\index{Polariton} than atoms and photons separately. Similarly, the eigenvalues are the {\it dressed energies}\index{Dressed! energies}. In fig.~\ref{fig1} both the bare $E_{0n\pm}=\omega n\pm\Omega/2$ and dressed energies (\ref{eigv}) are shown. The presence of the two-level system implies that the harmonic spectrum split into two {\it Jaynes-Cummings ladders}\index{Jaynes-Cummings! ladder}, which for $g=0$ are separated by the energy $\Omega$, as shown in (a). The onset of a $g$ causes a repulsion between the bare levels that are coupled, and the ladders will not be as clear. As a consequence, when a two-level system couples to a boson mode, one finds a splitting of the harmonic energies, which for the lowest state is termed {\it vacuum Rabi splitting}~\cite{sanchez1983theory,agarwal1984vacuum}\index{Vacuum Rabi! splitting} (The vacuum Rabi splitting was one of the early experimental verifications that single atoms can be made to interact coherently with single cavity modes~\cite{bernardot1992vacuum,thompson1992observation}). This shift is related to the {\it Lamb shift}\index{Lamb! shift} occurring in a two-level system interacting with the vacuum of the electromagnetic field~\cite{lamb1947fine}, to which we will return in sec.~\ref{sssec:smapp}.

In the large-detuning limit or {\it dispersive regime}\index{Dispersive! regime} $|\Delta|\gg g\sqrt{\bar{n}}$\index{Large detuning limit} (with $\bar{n}$ the average photon number), it is practical to first perform a {\it Schrieffer-Wolff transformation}\index{Schrieffer-Wolff transformation}\index{Schrieffer-Wolff transformation}~\cite{carbonaro1979canonical,klimov2002effective}
\begin{equation}\label{polaron}
\hat{H}_\mathrm{JC'}=e^{\hat{S}}\hat{H}_\mathrm{JC}e^{-\hat{S}}
\end{equation} 
and then use the operator identity $e^{\hat{A}}\hat{B}e^{-\hat{A}}=\hat{B}+[\hat{A},\hat{B}]+\frac{1}{2!}[\hat{A},[\hat{A},\hat{B}]]+...$ to systematically expand the Hamiltonian such that the linear-in-$g$ term vanishes. The operator $\hat{S}=\frac{g}{\Delta}\left(\hat{\sigma}_{+}\hat{a}-\hat{a}^\dagger\hat{\sigma}\right)$ serves this purpose and one obtains (up to trivial constants)
\begin{equation}\label{largejc}
\hat{H}_\mathrm{JC'}=\frac{\Delta}{2}\hat{\sigma}_z+\frac{2g^2}{\Delta}\hat{n}\hat{\sigma}_z+\mathcal{O}\left(\frac{g^4}{\Delta^3}\right).
\end{equation}  
The above transformation is also referred to as a {\it polaron transformation}~\cite{boissonneault2009dispersive,roy2011phonon,hausinger2010qubit}\index{Polaron! transformation}. For an alternative expansion, the {\it renormalization group} approach has been recently applied to the JC model~\cite{Ilderton2020}. By neglecting higher order terms in~(\ref{largejc}) we arrive at the effective Hamiltonian for dispersive atom-light interaction\index{Dispersive! Hamiltonian}
\begin{equation}\label{effham}
\hat{H}_\mathrm{disp}=\frac{\Delta}{2}\hat{\sigma}_z+\lambda\hat{n}\hat{\sigma}_z,
\end{equation}
where we have introduced the effective atom-field coupling $\lambda=2g^2/\Delta$. It is clear that the first term is just the bare atomic energy, while the second term is the {\it Stark shift}~\cite{alsing1992dynamic}\index{Stark shift} which is proportional to the field intensity $\hat{n}$ and depends on the sign of $\Delta$ (what is called {\it red} or {\it blue} detuned)\index{Red/blue detuning}. To this order, the effective Hamiltonian is diagonal in the bare basis, and in particular the spectrum is linear in the photon number. Next term, $\chi\hat n^2\equiv-8g^4\hat{n}^2/\Delta^3$, in the expansion breaks this linearity and acts as an effective ``interaction term'' for the photons~\cite{imamoglu1997strongly}, see further discussions about this in sec.~\ref{ssec:jch}. The results~(\ref{effham}) and the higher-order terms are interpreted as virtual two, four,...$(2n)$ photon processes respectively; a photon is absorbed/emitted with amplitude $g\sqrt{n}$, `interacts' giving a factor $\sim\pm1/\Delta$, and then re-emits/absorbs the photon giving the contribution $\pm g^2\hat n/\Delta$, and equivalently for the higher order terms. 

We note that the effective Hamiltonian agrees with what one would obtain by expanding the eigenvalues (\ref{eigv}) in the small parameter $g\sqrt{n}/\Delta$. Another systematic way to derive~(\ref{effham}) is via {\it adiabatic elimination}~\cite{grigolini1985basic,sanz2016beyond}\index{Adiabatic! elimination}. Working in the Heisenberg picture\index{Heisenberg! picture}, and assume we can split the system into two subsystems $A$ and $B$, $\hat H=\hat H_A+\hat H_B+\hat H_{AB}$, we have
\begin{equation}\label{adel0}
\begin{array}{l}
\partial_t\hat A=-i\left[\hat A,\hat H_A+\hat H_{AB}\right]=F_A(\hat O_A,\hat O_B),\\ \\
\partial_t\hat B=-i\left[\hat B,\hat H_B+\hat H_{AB}\right]=F_B(\hat O_A,\hat O_B),
\end{array}
\end{equation}
with $\hat A$ and $\hat B$ any operators acting on the respective subsystems, $F_A$ and $F_B$ some functions of operators $\hat O_A$ and $\hat O_B$ that act on the respective subsystems. Assume now that $\hat H_B$ provides the fast time-scale such that $\hat B$ rapidly reaches a steady state\index{Steady state}. We may then set $\partial_t\hat B=0$, and from the identity $F_B(\hat O_A,\hat O_B)=0$ we can solve for the operators $\hat O_B=\hat O_B(\hat O_A)$ and insert these into the first equation
\begin{equation}\label{adel1}
\partial_t\hat A=F_A\Big(\hat O_A,\hat O_B(\hat O_A)\Big)\equiv -i\left[\hat A,\hat H_\mathrm{eff}\right],
\end{equation}
where in the second step we have defined the effective Hamiltonian $\hat H_\mathrm{eff}$ generating the effective time-evolution of subsystem $A$. We may point out that in the general case it is not always possible to find a Hermitian $\hat{H}_\mathrm{eff}$ satisfying the equality above. Furthermore, we will see examples where we apply the adiabatic elimination procedure to open quantum systems.

Letting $A$ be the boson subsystem and $B$ the atomic subsystem we write down the equations of motion~\cite{gardiner2004quantum,gardiner2014quantum}
\begin{equation}\label{adel}
\begin{array}{l}
\partial_t\hat a=-i\left[\hat a,\hat H_\mathrm{JC}\right]=-ig\hat{\sigma}_{-},\\ \\
\partial_t\hat{\sigma}_{-}=-i\Delta\hat{\sigma}_{-}+ig\hat a\hat\sigma_z,\\ \\
\partial_t\hat\sigma_z=2ig\left(\hat a^\dagger\hat{\sigma}_{-}-\hat{\sigma}_{+}\hat a\right).
\end{array}
\end{equation}
Since the fast time scale is set by $\Delta$ we may assume that the atom degrees of freedom follows adiabatically the boson evolution. Thus, we can approximate the atomic operators with their steady state\index{Steady state} values, $\hat{\sigma}_{+}=g\hat a^\dagger\hat\sigma_z/\Delta$ and  $\hat{\sigma}_{-}=g\hat a\hat\sigma_z/\Delta$, and by inserting these into the equations of motion for the field one derives effective dynamical equations for $\hat{a}$ and $\hat{a}^\dagger$, {\it i.e.},
\begin{equation}
\partial_t\hat a=-i\frac{g^2}{\Delta}\hat a\hat\sigma_z,
\end{equation}
and consequently also an effective field Hamiltonian as in eq.~(\ref{effham}) from the relations $\partial_t\hat a=-i\left[\hat a,\hat H_\mathrm{disp}\right]$ and $\partial_t\hat a^\dagger=-i\left[\hat a^\dagger,\hat H_\mathrm{disp}\right]$. The case in which $|\Delta|\gg g\sqrt{\bar{n}}$ is called the {\it adiabatic}~\cite{larson2006validity}\index{Adiabatic! regime} or {\it dispersive regime}\index{Dispersive! regime}, and we will return to it frequently throughout the monograph. Here, the Hamiltonian is quasi-diagonal in the bare basis~(\ref{bstate}). The adiabatic limit, $g\sqrt{\bar{n}}/\Delta\rightarrow0$ is in a sense the opposite of the diabatic limit, $\Delta/g\sqrt{\bar{n}}\rightarrow0$. The adiabatic elimination scheme outlined above for the JC model in the large detuning limit is more general and applicable when there are clear time-scale separations. In sec.~\ref{sssec:dicke} we will consider the situation where we adiabatically eliminate the boson degree of freedom in order to obtain an effective model for the spin degree of freedom, while in sec.~\ref{ssec:rabi} we discuss the {\it Born-Oppenheimer approximation} which again assumes an adiabatic following\index{Adiabatic! following} of the fast-evolving variables.

\begin{figure}
 \includegraphics[width=11cm]{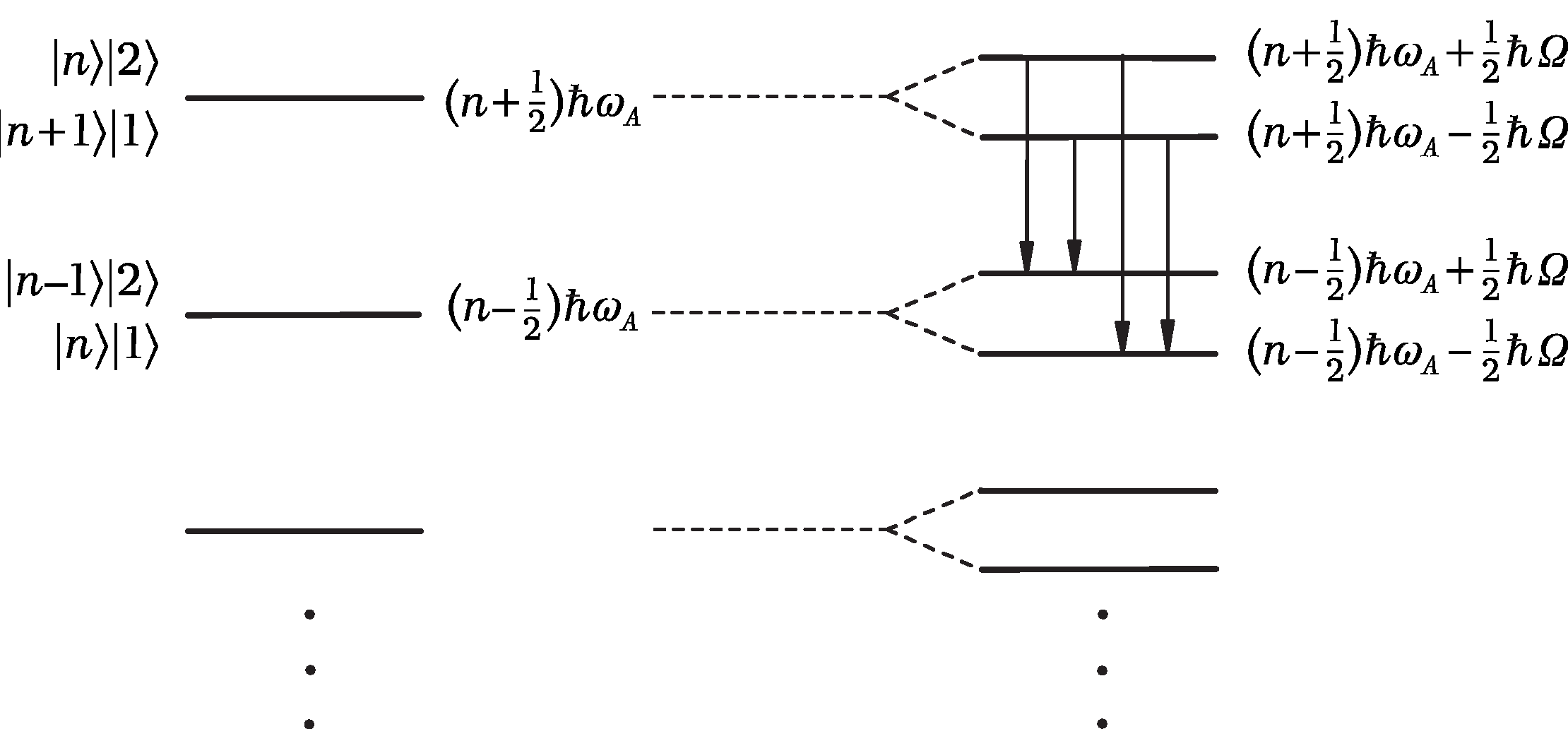}
 \caption{Following up from fig.~\ref{fig1} for resonance fluorescence: {\bf (left)} Degenerate ladder of energy levels for the uncoupled atom-field system; {\bf (right)} Level splitting due to the light-matter interaction. Reading from left to right, the depicted transitions have frequencies $\omega_A$, $\omega_A - \Omega$, $\omega_A + \Omega$ and $\omega_A$ [see eq.~\eqref{eq:populatedeigen}]. The transitions with frequency $\omega_A$ correspond to the central peak of the Mollow triplet\index{Mollow! triplet}, which is enhanced with respect to the side peaks at $\omega_A \pm \Omega$. We remark that the central peak of the Mollow triplet has a subnatural linewidth in the weak-excitation limit~\cite{Rice1988}. Source: Fig. 2.4 of~\cite{BookQO1Carmichael}.}
 \label{fig:dressedResfl}
\end{figure}

We will now explore a connection between the light-matter interaction level splitting in the JC model and the dressed states of resonance fluorescence\index{Resonance fluorescence! dressed states}, aimed at interpreting the celebrated Mollow-triplet\index{Mollow! triplet} spectrum. The Hamiltonian for a resonantly-driven two-level atom is given by
\begin{equation}\label{eq:HsResfl}
 \hat{H}_S=\tfrac{1}{2}\hbar \omega_A \hat{\sigma}_z - d E (e^{-i\omega_A t}\hat{\sigma}_{+} + e^{i\omega_A t}\hat{\sigma}_{-}),
\end{equation}
where $d$ is the projection of the dipole moment $\boldsymbol{d}_{12}$ on the electric field of amplitude $E$. A complementary view of the atomic dynamics is provided by the {\it dressed-states}\index{Dressed! states} formalism, whose application was pioneered by Cohen-Tannoudji and Reynaud~\cite{CTannoudjiReynaud1977}. The formalism is developed around the fully quantized (JC) Hamiltonian
\begin{equation}
 \hat{H}_{S, JC}=\tfrac{1}{2}\hbar \omega_A \hat{\sigma}_z + \hbar \omega_A \hat{a}^{\dagger}\hat{a} + \hbar (\kappa \hat{a}\hat{\sigma}_{+} + \kappa^{*} \hat{a}^{\dagger}\hat{\sigma}_{-}),
\end{equation}
where we must take $\hbar \kappa \braket{\hat{a}}=-d E$ to make the connection with the Hamiltonian of resonance fluorescence\index{Resonance fluorescence! Hamiltonian}. The energy eigenvalues are 
\begin{equation}
 E_{n,\pm}=\left(n+\tfrac{1}{2}\right)\hbar \omega_A \pm \sqrt{n+1} \hbar |\kappa|.
\end{equation}
If the laser field driving the two-level atom is in a coherent state with mean photon number $\overline{n} \gg 1$, we may write 
\begin{equation}
 d E= \hbar |\kappa \braket{\hat{a}}|=\hbar|\kappa|\sqrt{\overline{n}},
\end{equation}
and for all the populated eigenstates,
\begin{equation}\label{eq:populatedeigen}
 E_{n, \pm} \approx \left(n+\tfrac{1}{2}\right)\hbar \omega_A \pm \hbar \left(\frac{d}{\hbar}E\right)=\left(n+\tfrac{1}{2}\right)\hbar \omega_A \pm \tfrac{1}{2}\hbar \Omega,
\end{equation}
where $\Omega$ is the Rabi frequency. Transitions between the eigenstates of the interacting atom-field system identify the frequencies encountered in the solution of the Heisenberg equations of motion under the action of $H_{S}$ in eq.~\eqref{eq:HsResfl}, namely $\omega_A$, $\omega_A + \Omega$ and $\omega_A - \Omega$ as depicted in fig.~\ref{fig:dressedResfl}. If we suppress the term $n\hbar \omega_A$, which distinguishes states in the Fock hierarchy, the four-level structure that remains gives the {\it dressed energies}\index{Dressed! energies} for the atom, $-\tfrac{1}{2}\hbar(\omega_A \mp \Omega)$ and $-\tfrac{1}{2}\hbar(\omega_A \pm \Omega)$ -- these are the {\it quasi}energies\index{Quasi! energy} for the Hamiltonian~\eqref{eq:HsResfl}.
For large $n$, the dressed states approximately factorize as the product of a Fock state for the field and linear combinations of the atomic states $\ket{e}$ and $\ket{g}$, where we neglect the difference between $\ket{n}$ and $\ket{n+1}$~\cite{BookQO1Carmichael}. 

The incoherent spectrum\index{Spectrum! incoherent} is calculated from the expression
\begin{equation}
 S_{\rm incoh}(\omega)=f(\boldsymbol{r})\frac{1}{2\pi} \int_{-\infty}^{\infty}\,d\tau\, e^{i(\omega-\omega_A)\tau}\, \braket{\Delta\hat{\tilde{\sigma}}_{+}(0) \Delta\hat{\tilde{\sigma}}_{-}(\tau)}_{\rm ss},
\end{equation}
where $\Delta \hat{\tilde{\sigma}}_{\mp}\equiv e^{\pm i\omega_A t}(\hat{\sigma}_{\mp}-\braket{\hat{\sigma}_{\mp}}_{\rm ss})$ and 
\begin{equation}
 f(\boldsymbol{r}) \equiv \left(\frac{\omega_A^2 d_{12}}{4\pi \epsilon_0 c^2}\right)^2 \frac{\sin^2 \theta}{r^2}
\end{equation}
is the geometrical factor\index{Geometrical factor} ($\theta$ is the angle between the dipole moment and the position vector with respect to an origin at the location of the atom). In the strong-field limit, incoherent scattering dominates and the spectrum takes the form of a Mollow\index{Mollow! triplet} triplet~\cite{MollowTriplet1969}. The appearance of the triplet is directly related to the four-level structure depicted in fig.~\ref{fig:dressedResfl}. On the other hand, the incoherent part of the resonance-fluorescence spectrum has a subnatural linewidth\index{Linewidth! subnatural} in the weak-field limit, which is due to the squeezing\index{Squeezing} of field fluctuations in phase with the field radiated by the mean induced atomic dipole~\cite{Rice1988}. This instance shows that the atomic emission is nonclassical even for very weak excitation where the Mollow sidepeaks have not yet developed~\cite{BookQO1Carmichael}. Squeezing-induced linewidth narrowing\index{Linewidth! narrowing} is also seen in the transmitted and fluorescent light emanating from a driven optical cavity containing a single resonant two-level atom.

The Mollow spectrum of resonance fluorescence\index{Resonance fluorescence! spectrum}, measured first for atoms, has served as a long-standing figure of merit for the coherent addressing of both semiconductor nanocrystals [discussed in more detail in subsec.~\ref{subsec:phononsc}] and superconducting-circuit qubits~\cite{Muller2007, XXu2007, Flagg2009, Abdumalikov2011}. In a remarkable experiment with a superconducting-circuit qubit and squeezed microwave photons, Toyli and coworkers have extended the observation of resonance fluorescence\index{Resonance fluorescence! squeezed bath} from a two-level system bathed by ordinary vacuum fluctuations to one embedded in a re-engineered bath where vacuum fluctuations are squeezed\index{Squeezed! bath}, thus permitting narrow spectroscopic lines to be observed~\cite{Toyli2016,CarmichaelViewpoint}. Here vacuum fluctuations undergo phase-sensitive amplification and deamplification to produce a set number of photons per mode, and this phase sensitivity dramatically changes the bath-induced radiative decay rates and the spectrum of resonance fluorescence\index{Resonance fluorescence! spectrum}. This development follows a long thread of theoretical investigations initiated by the proposal of Gardiner in 1986~\cite{Gardiner1986, CarLaneWalls1987, DaltonFicek1999}, while the phase-dependent dipole decay was observed for the first time in the circuit QED experiment of~\cite{Murch2013} in 2013.

Finally, we note that the JC model may also be construed as a bipartite atom-field qubit~\cite{OmoloJC}. Writing the JC Hamiltonian as $\hat{H}=\hbar \omega \hat{N} + \hbar g \hat{R}$, with $\hat{N}=\hat{a}^{\dagger}\hat{a} + \hat{\sigma}_{+}\hat{\sigma}_{-}$ and $\hat{R}=\tfrac{1}{2}\beta \hat{\sigma}_z + \hat{a}^{\dagger}\hat{\sigma}_{-} + \hat{a}\hat{\sigma}_{+}$, with $\beta \equiv (\Omega-\omega)/g$, then the action of $\hat{R}$ on the state $\ket{e,n}$ yields $\hat{R} \ket{e, n}=r\ket{\phi}$, with 
\begin{equation}
 \ket{\phi}=c_1 \ket{e,n} + c_2 \ket{g, n+1}, \quad \quad r=\sqrt{\beta^2/4 + n +1}, \quad \quad c_1=\beta/(2r), \quad \quad c_2=\sqrt{n+1}/r.
\end{equation}
On resonance ($\beta=0$), the states $\ket{e,n}$ and $\ket{\phi}=\ket{g, n+1}$ are evidently orthogonal.


\subsection{Jaynes-Cummings dynamics}\label{ssec:JCdyn}
The previous section made clear that the JC model takes a block form in the bare basis (\ref{bstate}). Importantly, this block decomposition derives from number conservation $[\hat{N},\hat{H}_\mathrm{JC}]=0$\index{Number! conservation}. As a result, the Hamiltonian can be easily diagonalized where the corresponding solutions attain simple closed analytic forms~(\ref{dstate}). Despite its simplicity, the JC model generates very intriguing dynamics. The fact that the Rabi frequencies $\Omega_n$\index{Rabi! frequency} depend on the photon number $n$, and moreover that this dependence is nonlinear, imply that the bi-partite atom-field system display complex evolution including phenomena such as; {\it collapse-revivals}, atom-field {\it entanglement}, {\it squeezing}, and the generation of macroscopic superposition field states. It is the aim of the present subsection to give an introduction to these topics. 

\subsubsection{General solution and remarks}\label{sssec:sol}
For a pure state, the general solution to the Schr\"odinger equation is written in the bare basis as
\begin{equation}\label{tstate}
|\psi(t)\rangle=\sum_n\left[c_{en}(t)|e,n\rangle+c_{gn}(t)|g,n\rangle\right].
\end{equation}
It is enough to solve the $2\times2$ problem for a given $N$. In a rotating frame with respect to the term $\Delta\hat\sigma_z/2$, the coefficients obey
\begin{equation}
\begin{array}{l}
\partial_tc_{en}=-ig\sqrt{n+1}e^{i\Delta t}c_{gn+1},\\ \\
\partial_tc_{gn+1}=-ig\sqrt{n+1}e^{-i\Delta t}c_{en}
\end{array}
\end{equation}
with the general solution\index{Time-dependent solution}
\begin{equation}\label{tsol}
\begin{array}{lll}
c_{en}(t) & = & \displaystyle{\left\{c_{en}(0)\left[\cos(\Omega_nt)-\frac{i\Delta}{2\Omega_n}\sin(\Omega_nt)\right]\right.} -\displaystyle{\left.\frac{ig\sqrt{n+1}}{\Omega_n}c_{gn+1}(0)\sin(\Omega_nt)\right\}e^{i\Delta t/2}},\\ \\
c_{gn+1}(t) & = & \displaystyle{\left\{c_{gn+1}(0)\left[\cos(\Omega_nt)+\frac{i\Delta}{2\Omega_n}\sin(\Omega_nt)\right]\right.}-\displaystyle{\left.\frac{ig\sqrt{n+1}}{\Omega_n}c_{en}(0)\sin(\Omega_nt)\right\}e^{-i\Delta t/2}}.
\end{array}
\end{equation}
If the atom is initially excited, $c_{gn}(0)=0$ and $c_{en}(0)=c_n(0)$ with $c_n(0)$ the initial photon amplitudes, the solutions simplify to
\begin{equation}\label{tsol2}
\begin{array}{lll}
c_{en}(t) & = & \displaystyle{c_n(0)\left[\cos(\Omega_nt)-\frac{i\Delta}{2\Omega_n}\sin(\Omega_nt)\right]e^{i\Delta t/2}},\\ \\
c_{gn+1}(t) & = & \displaystyle{-c_n(0)\frac{ig\sqrt{n+1}}{\Omega_n}\sin(\Omega_nt)e^{-i\Delta t/2}.}
\end{array}
\end{equation}
Finally we note that for the resonant case the above expressions relax to 
\begin{equation}\label{tsol3}
\begin{array}{l}
c_{en}(t)=c_n(0)\cos(g\sqrt{n+1}t),\\ \\
c_{gn+1}(t)=-ic_n(0)\sin(g\sqrt{n+1}t).
\end{array}
\end{equation}

In certain cases it might be useful to consider the explicit expression for the JC evolution operator $\hat U_\mathrm{JC}(t)=e^{-i\hat H_\mathrm{JC}t}$. In the interaction picture\index{Interaction! picture}, when $\hat H_\mathrm{JC}=g\left(\hat{a}^\dagger\hat{\sigma}_{-}+\hat{\sigma}_{+}\hat{a}\right)$, it is rather straightforward to derive the evolution operator\index{Evolution operator} by expanding the exponent~\cite{phoenix1988fluctuations}
\begin{equation}
\hat U(t)=\sum_{n=0}^\infty\frac{(-igt)^n}{n!}\left[
\begin{array}{cc}
0 & \hat a^\dagger\\
\hat a & 0
\end{array}\right]^n,
\end{equation}
and use the identities
\begin{equation}
\left[\begin{array}{cc}
0 & \hat a^\dagger\\
\hat a & 0\end{array}\right]^{2n}=\left[
\begin{array}{cc}
\left(\hat a^\dagger\hat a\right)^n & 0\\
0 & \left(\hat a\hat a^\dagger\right)^n
\end{array}\right],\hspace{0.8cm}
\left[\begin{array}{cc}
0 & \hat a^\dagger\\
\hat a & 0\end{array}\right]^{2n+1}=\left[
\begin{array}{cc}
0 & \hat a^\dagger\left(\hat a^\dagger\hat a\right)^n \\
\hat a\left(\hat a\hat a^\dagger\right)^n & 0
\end{array}\right].
\end{equation}
One then finds the interaction picture JC evolution operator\index{Jaynes-Cummings! evolution operator}
\begin{equation}
\hat U_\mathrm{JC}(t)=\left[
\begin{array}{cc}
\cos\left(gt\sqrt{\hat a\hat a^\dagger}\right) & -i\hat a\sqrt{\hat a^\dagger\hat a}\sin\left(gt\sqrt{\hat a^\dagger\hat a}\right)\\ 
-i\hat a^\dagger\sqrt{\hat a\hat a^\dagger}\sin\left(gt\sqrt{\hat a\hat a^\dagger}\right) & \cos\left(gt\sqrt{\hat a^\dagger\hat a}\right)
\end{array}\right].
\end{equation}
This can be further reshuffled such that it is written on the
form~\cite{scully1999quantum,gerry2005introductory} 
\begin{equation}\label{evop}
\hat{U}_\mathrm{JC}(t)=\left[\begin{array}{cc}
\cos(g\sqrt{\hat{n}+1}t) & \displaystyle{-i\frac{\sin(g\sqrt{\hat{n}+1}t)}{\sqrt{\hat{n}+1}}\hat{a}}\\ \\
\displaystyle{-i\hat{a}^\dagger\frac{\sin(g\sqrt{\hat{n}+1}t)}{\sqrt{\hat{n}+1}}} & \cos(g\sqrt{\hat{n}}t)
\end{array}\right].
\end{equation}

Experiments in cavity QED typically rely on atoms traversing a resonator one after the other~\cite{raimond2006exploring}. The finite probability for the atoms to exit the cavity in their ground states implies that the atoms pump energy into the resonator. This is the idea of the so-called {\it micromaser}~\cite{krause1986quantum,scully1991micromaser,walther1992experiments,an1994microlaser,berman1994cavity,englert2002elements}\index{Micromaser} which will be considered in sec.~\ref{sssec:micro}. Even without cavity losses, the growth of the field amplitude $\langle\hat{n}\rangle$ typically terminates. To understand this `field freezing' we note that the atomic velocity determines the effective interaction time $t_\mathrm{f}$. Given $g$, $\Delta$, and $t_\mathrm{f}$, if there is an $n$ such that $\Omega_nt_\mathrm{f}=\nu\pi$ ($\nu\in N$) the atom will leave the cavity in its excited state provided the field state is in the precise Fock state $|n\rangle$. These $|n\rangle$'s are called {\it trapping states}~\cite{meystre1988very,slosser1989harmonic}\index{Trapping states}, and will be discussed in much more detail in sec.~\ref{ssec:cqedearly}. Note that the frozen population does not mean that the atom is not interacting with the field; rather, it is performing exactly $\nu$ Rabi oscillations while traversing the cavity. Trapping states was observed in the micromaser in the group of Walther~\cite{weidinger1999trapping}. For a sufficiently narrow initial distribution $P(n)$, the micromaser field will normally approach one particular trapping state and can thereby be employed for state preparation of photon number states~\cite{varcoe2000preparing}. Another type of trapping effect in the JC model was found by Cirac and Sanches-Soto ~\cite{cirac1990population}, who showed that for special initial atom-field states, no population swapping between the two atomic states $|e\rangle$ and $|g\rangle$ took place, whence the internal atomic population was instead trapped.

We note that in the dispersive regime\index{Dispersive! regime}, the atomic states $|g\rangle$ and $|e\rangle$ are disconnected and the general solutions~(\ref{tsol}) are approximated as
\begin{equation}\label{dispcoef}
\begin{array}{l}
c_{en}(t)=c_{en}(0)\exp\left(-i\lambda(n+1)t\right),\\ \\
c_{gn}(t)=c_{gn-1}(0)\exp\left(i\lambda nt\right),
\end{array}
\end{equation}
where we have turned into an interaction picture with respect to the term $\Delta\hat\sigma_z/2$. As demonstrated below, these expressions allow for closed analytical expressions of various expectation values for certain initial states.

\begin{figure}
\includegraphics[width=10cm]{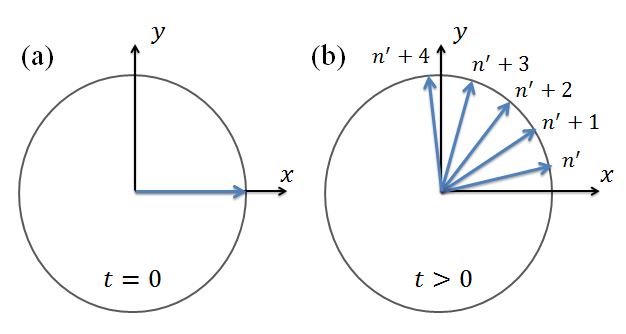} 
\caption{Schematic representation of the atomic state (not normalized) on the Bloch sphere where the circle represents the equator and the blue arrows the atomic states for different photon numbers $n'$. Initially the atomic state points in the positive $x$-direction {\bf (a)}. After some time $t$ the precession of the state vector depends on the field intensity $n$ {\bf (b)}.}
\label{fig2}
\end{figure}

Even if the large detuning Hamiltonian (\ref{effham}) is diagonal in the bare basis, if the initial state is a linear superposition of bare states the time evolved state will in general be an entangled atom-field state. This derives from the $n$-dependence of the phase factors of (\ref{dispcoef}), which means that the atomic two-level state is `rotated' (in a spin sense) differently depending on the photon number $n$. More precisely, if the atom is initially pointing in the spin $x$-direction ({\it e.g.}, $(|g\rangle+|e\rangle)/\sqrt{2}$), the time evolved atom-field state can be expressed as 
\begin{equation}\label{dispev}
|\psi(t)\rangle=\frac{1}{\sqrt{2}}\sum_ne^{i\lambda nt}\left(c_{n-1}(0)|g\rangle+e^{-i\lambda(2n+1)t}c_n(0)|e\rangle\right)|n\rangle.
\end{equation} 
Thus, for given $n$ and assuming $c_n(0)\approx c_{n-1}(0)$ the atomic state has been rotated an angle $\lambda nt$ along the equator on the Bloch sphere. Figure~\ref{fig2} pictures the time-evolution of the atomic state (not properly normalized) on the Bloch sphere. Ideally, if the state of the atom could be determined it would imply that information about the photon number $n$ has been extracted (without measuring directly on the field). This is the idea behind {\it quantum non-demolition measurements} (QND)~\cite{poizat1994characterization,grangier1998quantum}\index{Non-demolition measurement} and {\it interaction-free measurements}~\cite{elitzur1993quantum}\index{Interaction-free measurement}. In order for such schemes to be experimentally feasible, the parameter $\lambda$ needs to be larger than the transition linewidths\index{Linewidth! transition} of both the atom and the photons. This is termed the {\it number-split regime}~\cite{gambetta2006qubit,schuster2007resolving,johnson2010quantum,heeres2015cavity}\index{Number-split regime}. JC type of QND measurements have been experimentally implemented numerous times, something we will come back to in sec.~\ref{sec:cavQED} and subsec.~\ref{ssec:JCspectrcQED}. 

A special interesting case of~(\ref{dispev}) is when the field is in an initial {\it coherent state}\index{Coherent state}\index{State! coherent}
\begin{equation}\label{cohe}
|\alpha\rangle=\sum_nc_n|n\rangle\equiv e^{-|\alpha|^2/2}\sum_n\frac{\alpha^n}{\sqrt{n!}}|n\rangle.
\end{equation}
As will be discussed further in sec.~\ref{sssec:crsubsec}, coherent states were introduced as quantum states most closely following classical states, {\it e.g.} they are {\it minimum uncertainty states}\index{Minimum uncertainty states} meaning that they satisfy the Heisenberg uncertainty relation. Another property is that they are eigenstates of the annihilation operator, {\it i.e.} $\hat a|\alpha\rangle=\alpha|\alpha\rangle$, where the amplitude $\alpha$ is in general a complex number~\cite{glauber1963coherent,gazeau2009coherent}. From this is directly follows that the average photon number $\bar n\equiv\langle\hat n\rangle=|\alpha|^2$. The coherent state can be constructed by acting with a {\it displacement operator}\index{Displacement! operator} on the vacuum state~\cite{glauber1963coherent,mandel1995optical,Scully1997a}
\begin{equation}\label{dispcoh}
|\alpha\rangle=\hat D(\alpha)|0\rangle\equiv e^{\alpha\hat a^\dagger-\alpha^*\hat a}|0\rangle.
\end{equation}
By using the expansion~(\ref{cohe}) of the coherent state in terms of Fock states it is readily derived that the atom-field states evolves into~\cite{brune1992manipulation,guo1996generation}
\begin{equation}\label{cat1}
|\psi(t)\rangle=\frac{1}{\sqrt{2}}\left(|\alpha e^{i\lambda t}\rangle|g\rangle+e^{-i\lambda t}|\alpha e^{-i\lambda t}\rangle|e\rangle\right)
\end{equation} 
under evolution with the dispersive JC Hamiltonian~(\ref{effham}). A useful formula for coherent states is~\cite{mandel1995optical}
\begin{equation}\label{coover}
N\equiv\langle\beta|\alpha\rangle=e^{-\frac{1}{2}\left(|\beta|^2+|\alpha|^2-2\beta^*\alpha\right)}\hspace{0.3cm}\rightarrow\hspace{0.3cm}|\langle\beta|\alpha\rangle|^2=e^{-|\alpha-\beta|^2},
\end{equation}
which measures how `dislike' two coherent states are. In particular, we see that the further apart the states are (increasing $|\alpha-\beta|$), the smaller the overlap, which implies a larger entanglement between the atom and the field in the state~(\ref{cat1}). We will come back to this type of states, called {\it Schr\"odinger cats} or simply {\it cat states}\index{Cat state}\index{State! Cat}, repeatedly in this monograph and especially in secs.~\ref{ssec:cqedstateprep} and \ref{ssec:ionpreptom}. A more general single mode cat state is given by
\begin{equation}\label{catstate}
|\psi_\mathrm{cat}\rangle=\frac{1}{\sqrt{N}}(|\alpha\rangle+e^{i\phi}|\beta\rangle,
\end{equation}
where $|\alpha\rangle$ and $|\beta\rangle$ are both coherent states, $N$ a normalization constant,  and $\phi$ some phase. In order to count as a proper cat we should have that the overlap~(\ref{coover}) is small. The non-classical properties of such cat states will be discussed further in sec.~\ref{sssec:openjc}.

For a solution of the JC model in the Heisenberg picture we refer to ref.~\cite{scully1999quantum}. An alternative way of treating the JC system~\cite{stenholm1981bargmann} is by introducing the {\it Bargmann representation}~\cite{bargmann1961hilbert}\index{Bergmann representation}, {\it i.e.} \begin{equation}\label{barg}
\hat{a}^\dagger\rightarrow z, \hspace{1cm}\hat{a}\rightarrow\partial_z.
\end{equation} 
For a further discussion on the Bargmann representation and its application to the quantum Rabi model see sec.~\ref{sssec:rabiint}. The Bargmann approach has also been employed for analyzing the spectrum of the Tavis-Cummings model\index{Tavis-Cummings model} discussed in sec.~\ref{sssec:dicke}, see also~\cite{Mohamadian2019}.). In a similar way, the problem can alternatively be tackled in the quadrature representation\index{Quadrature representation}~\cite{larson2007dynamics}
\begin{equation}\label{quad}
\begin{array}{l}
\hat{x}=\displaystyle{\frac{1}{\sqrt{2}}\left(\hat{a}^\dagger+\hat{a}\right)},\\ \\
\hat{p}=\displaystyle{\frac{i}{\sqrt{2}}\left(\hat{a}^\dagger-\hat{a}\right)}.
\end{array}
\end{equation}
Here, $\hat{x}$ and $\hat{p}$ are conjugate variables, $[\hat{x},\hat{p}]=1$. In the quadrature representation, the JC Hamiltonian~(\ref{jcham2}) in the interaction picture becomes
\begin{equation}\label{jchamquad}
\hat H_\mathrm{JC}=\frac{\Delta}{2}\hat\sigma_z+\frac{g}{\sqrt{2}}\left(\hat x\hat\sigma_x+\hat p\hat\sigma_y\right).
\end{equation}
As already mentioned in the introduction, the linearity in $\hat p$ connects the JC model with the Dirac equation, more precisely the {\it Dirac oscillator} in 1D~\cite{DOJC1999,bermudez2007exact}. We will return to this mapping to a relativistic equation when discussing trapped ion physics in sec.~\ref{ssec:ionfur}. Thanks to this, some grasp of relativistic quantum mechanics may provide meaningful insight into the JC model. As an example, the JC model has been shown to support an $SO(4)$ algebra\index{$SO(4)$ algebra} much like the Dirac oscillator~\cite{Song2020}\index{Dirac! oscillator}. The JC Hamiltonian has also been shown to be {\it supersymmetric}, allowing an alternative method for its treatment~\cite{andreev1989supersymmetry}. The supersymmetric property\index{Supersymmetry} of the JC without the RWA, {\it i.e.} the quantum Rabi model of sec.~\ref{ssec:rabi}, was further analyzed  in~\cite{tomka2015supersymmetry}, while it has also been discussed in terms of generalized JC models (see sec.~\ref{ssec:exjc})
~\cite{maldonado2021underlying}.

Since the Schr\"odinger evolution is unitary, any initial pure atom-field state will remain pure for all times. It is often, however, interesting to consider initial mixed states, for example evolution of thermal field states. To describe these situations we introduce the density operator $\hat{\rho}(t)$~\cite{nielsen2010quantum,sakurai1995modern} for the combined atom-field system. Its dynamical behaviour is ruled by the Liouvillian\index{Liouvillian! equation}
\begin{equation}
\partial_t\hat{\rho}(t)=i[\hat{\rho},\hat{H}_\mathrm{JC}]\hspace{0.4cm}\Leftrightarrow\hspace{0.4cm}\hat{\rho}(t)=\hat{U}_\mathrm{JC}(t)\hat{\rho}(0)\hat{U}_\mathrm{JC}^{-1}(t),
\end{equation}
where the evolution operator is $\hat{U}_\mathrm{JC}(t)=\exp(-i\hat{H}_\mathrm{JC}t)$\index{Jaynes-Cummings! evolution operator}. Remember that in the resonant case, the interaction picture evolution operator was given in eq.~(\ref{evop}). Since any mixed state can always be written as a sum of pure density operators, {\it i.e.} $\hat{\rho}(t)=\sum_jC_j|\varphi_j(t)\rangle\langle\varphi_j(t)|$\index{Density operator}, it follows from the linearity of the evolution operator that the general solution (\ref{tsol}) can provide $\hat{\rho}(t)$\index{Field! density operator} for any initial state. The {\it reduced density operator} $\hat{\rho}_\mathrm{a}(t)$\index{Atom! density operator} ($\hat{\rho}_\mathrm{f}(t)$) of the atom (field) is obtained by tracing the full density operator over the field (atom) degrees of freedom;\index{Reduced! density operator}
\begin{equation}
\hat{\rho}_\mathrm{a}(t)=\mathrm{Tr}_\mathrm{f}[\hat{\rho}(t)],\hspace{1cm}\hat{\rho}_\mathrm{f}(t)=\mathrm{Tr}_\mathrm{a}[\hat{\rho}(t)].
\end{equation}
Note that this is a general result for any $\hat{\rho}(t)$, mixed or pure. If the full density operator is pure, $\hat{\rho}(t)=|\psi(t)\rangle\langle\psi(t)|$, the reduced density operators $\hat{\rho}_\mathrm{a,f}(t)$ need not be pure. In fact, as we will see in sec.~\ref{sssec:ent} the reduced states\index{Reduced! state} are in general mixed, and the evolution of the reduced states separately is not unitary. An example of this non-unitary\index{Non-unitary} sub-space evolution is the collapse-revival phenomenon that will be investigated below in sec.~\ref{sssec:semclas}. 

Expectations of some observable $\hat{A}$, often encountered in the bibliography and numerous applications, are given by\index{Pure state}\index{Mixed state}\index{State! pure}\index{State! mixed}
\begin{equation}
\begin{array}{llll}
\langle\hat{A}\rangle=\langle\psi(t)|\hat{A}|\psi(t)\rangle, &&& \mathrm{pure\,\,state},\\ \\
\langle\hat{A}\rangle=\mathrm{Tr}[\hat{A}\hat{\rho}(t)], &&& \mathrm{mixed\,\,state}.
\end{array}
\end{equation}
For the special case of a density matrix corresponding to a pure state, the latter expression reduces to the former. In terms of quantum information processing there are a few multi-qubit states of special interest, both from a fundamental viewpoint but also in terms of applications. Let us mention some of them (using the standard notation $|0\rangle$ and $|1\rangle$) that will be recurring in our monograph. The {\it Bell} or {\it EPR states} (Einstein-Podolsky-Rosen) \index{Bell! state}\index{EPR state}\index{State! EPR}\index{State! Bell}
\begin{equation}\label{eprstate}
|{\rm EPR}\rangle_\pm=\frac{1}{\sqrt{2}}(|1,0\rangle\pm|0,1\rangle).
\end{equation}
These are maximally entangled two-qubit states (the reduced states will be maximally mixed) that extremize the violation of the {\it Bell inequality}~\cite{nielsen2010quantum}. The {\it GHZ state} (Greenberger-Horne-Zeilinger)\index{GHZ state} which for three qubits read\index{State! GHZ}
\begin{equation}\label{ghzstate}
|{\rm GHZ}\rangle=\frac{1}{\sqrt{2}}(|0,0,0\rangle+|1,1,1\rangle).
\end{equation}
The state can be generalized to any number of qubits. The reduced state for two qubits, {\it i.e.} when we trace out one qubit, becomes
\begin{equation}\label{trghz}
\hat\rho_{12}^\mathrm{GHZ}=\frac{1}{2}(|0,0\rangle\langle0,0|+|1,1\rangle\langle1,1|),
\end{equation}
which shows classical correlations between the two qubits, but no entanglement. Application of the GHZ state involves ruling out {\it local hidden variable theories}\index{Local! hidden variable}, similar as for the Bell experiments but without satisfying an inequality; such an experiment obeys an equality~\cite{nielsen2010quantum,stenholm2005quantum,jaeger2007quantum}. Another three-qubit state of interest is the W state\index{W state}\index{State! W} 
\begin{equation}\label{wstate}
|W\rangle=\frac{1}{\sqrt{3}}(|0,0,1\rangle+|0,1,0\rangle+|1,0,0\rangle).
\end{equation}
Like the GHZ state, the W state is {\it non-biseperable}, {\it i.e.} all three qubits display multi-partite entanglement. However, as one qubit is traced out, the remaining two qubits
\begin{equation}\label{trw}
\hat\rho_{12}^\mathrm{W}=\frac{1}{3}(|{\rm EPR}\rangle_{+\,+}\!\langle {\rm EPR}|+|0,0\rangle\langle0,0|)
\end{equation}
 are entangled for the W state, contrary to the state (\ref{trghz}). The two three-qubit states, (\ref{ghzstate}) and (\ref{wstate}), cannot be transformed into one another by simple single qubit operations, and they thereby belong to two different classes of entangled three-qubit states~\cite{nielsen2010quantum,stenholm2005quantum,jaeger2007quantum}. Finally we encounter the {\it Dicke states}\index{Dicke! state}\index{State! Dicke}~\cite{lucke2014detecting} defined as
 \begin{equation}\label{dickestates}
\hat{S}^2|s,m\rangle=s(s+1)|s,m\rangle,\hspace{0.7cm}\hat{S}_z|s,m\rangle=m|s,m\rangle,
\end{equation}
where $\hat S_\alpha$ are the spin angular momentum operators, see further sec.~\ref{sssec:dicke}. Here $s$ is the total spin for the state. The Dicke states are entangled multi-partite qubit states, {\it e.g.} the $s=0$ singlet $|0,0\rangle$ is an EPR state, the $s=1$ triplet $|1,0\rangle$ is also an EPR state, while for three qubits the $W$ state $|W\rangle=(|011\rangle+|101\rangle+|110\rangle)/\sqrt{3}$ belongs to the $s=3/2$ spin sector. 

There are several physical quantities that will concern us in this work -- the most important ones are listed below:
\begin{enumerate}

\item {\it Atomic inversion}\index{Atomic! inversion}
\begin{equation}\label{inv}
W(t)=\langle\hat{\sigma}_z\rangle
\end{equation}
gives the population difference between the two atomic states $|e\rangle$ and $|g\rangle$, and the explains why $\hat\sigma_z$ is called the inversion operator. Within the bare basis, the inversion contains no information about the coherences. $W(t)$ is, however, often easily measured experimentally~\cite{raimond2006exploring}. Furthermore, rotations of the atomic two-level system allow for other spin directions to be measured.

\item The {\it Bloch vector}\index{Bloch! vector}
\begin{equation}\label{bloch}
\mathbf{R}(t)=\left(R_x(t),R_y(t),R_z(t)\right)=\left(\langle\hat{\sigma}_x\rangle,\langle\hat{\sigma}_y\rangle,\langle\hat{\sigma}_z\rangle\right),
\end{equation}
that parametrizes the atomic state 
\begin{equation}\label{bloch2}
\hat{\rho}_\mathrm{a}(t)=\frac{1}{2}\left[1+R_x(t)\hat{\sigma}_x+R_y(t)\hat{\sigma}_y+R_z(t)\hat{\sigma}_z\right],
\end{equation} 
and its length $R(t)=|\mathbf{R}(t)|$ is a measure of the atomic-state {\it purity}~\cite{gea1992new}\index{Purity}, {\it i.e.} it indicates the amount of ``mixedness''. Note that $R_z(t)=W(t)$ and that $R_x(t)$ (for real atom-field coupling $g$), which is the expectation of the dipole operator $\hat\sigma_x$, is often referred to as the {\it atomic dipole value}. In particular, $R_x(t)$ measures how ``aligned'' the atom is with the field. In subsecs.~\ref{sssec:semclas} and \ref{sssec:ent} we will be  particularly interested in $R_x(t)$ when we analyze the behavior of the bi-partite system for large amplitude fields. All equations that govern the evolution of the state $\hat{\rho}_\mathrm{a}(t)$ can equally well be recast into a form dictating the evolution of the Bloch vector, {\it i.e.}
\begin{equation}\label{blocheq}
\partial_t\mathbf{R}(t)=\mathbf{MR}(t),
\end{equation}
where $\mathbf{M}$ is a $3\times3$ matrix. These are called {\it optical Bloch equations}\index{Bloch! equations}~\cite{allen1987optical}, and they can be generalized to any dimensions~\cite{mathisen2018liouvillian}. Furthermore, the form of the optical equations is general such that it could include non-unitary time-evolution ({\it e.g.} for unitary evolution the matrix $\mathbf{M}$ is anti-symmetric). For open systems, {\it i.e.}, when $\hat{\rho}_\mathrm{a}(t)$ is the reduced state of a larger system, non homogeneous Bloch equations may arise: $\partial_t\mathbf{R}(t)=\mathbf{MR}(t) + \mathbf{b}$, with $\mathbf{b}$ a three-component column vector--see eq.~\eqref{eq:BlochNH}.  

\item The {\it von Neumann entropy}\index{Entanglement! von Neumann entropy}\index{von Neumann! entropy}
\begin{equation}\label{vN}
S_\mathrm{vN}(t)=-\mathrm{Tr}\left[\hat{\rho}(t)\log_2\hat{\rho}(t)\right],
\end{equation}
defined for some general state $\hat\rho(t)$~\cite{wehrl1978general,amico2008entanglement}. In the basis where $\hat{\rho}(t)$ is diagonal with eigenvalues $\lambda_n(t)$, the above expressions should be understood as $S_\mathrm{vN}(t)=-\sum_n\lambda_n(t)\log_2\lambda_n(t)$. Whenever $\hat{\rho}(t)$ is pure, the entropy vanishes, $S_\mathrm{vN}(t)=0$. While, if the state is maximally mixed $S_\mathrm{vN}=\log_2 N$ with $N$ being the Hilbert space dimension. 
The eigenvalues of $\hat{\rho}_\mathrm{a}(t)$ are $\lambda_\pm(t)=\frac{1}{2}\pm\frac{1}{2}\sqrt{R_x^2(t)+R_y^2(t)+R_z^2(t)}$. In sec.~\ref{sssec:ent} we will use the von Neumann entropy in order to quantify the amount of atom-field entanglement. We finally note that the {\it purity}\index{Purity} \begin{equation}\label{purity}
P(t)=\mathrm{Tr}[\hat{\rho}^2(t)]
\end{equation}
(or equivalently to the {\it linear entropy}\index{Linear entropy} or {\it Tsallis entropy}\index{Tsallis entropy}, $S_\mathrm{L}(t)=1-P(t)$), which measures the amount of `mixedness' of the state, can be directly related to the Bloch vector for the atom, as $P(t)=\left(1+R^2(t)\right)/2$. A generalization of the purity are the so-called {\it R\'enyi entropies}\index{R\'enyi! entropy} defied as $S^{(n)}(t)=\mathrm{Tr}\left[\hat\rho^n(t)\right]/(1-n)$~\cite{amico2008entanglement}. Clearly, $S^{(2)}(t)=-P(t)$, but we also have $\lim_{n\rightarrow1}S^{(n)}(t)=S_\mathrm{vN}(t)$. 

\item The {\it Mandel $Q$-parameter}~\cite{mandel1982squeezed,mandel1995optical}\index{Mandel $Q$-parameter}
\begin{equation}\label{qpara}
Q(t)=\frac{\langle\hat{n}^2\rangle-\langle\hat{n}\rangle^2}{\langle\hat{n}\rangle}-1,
\end{equation}
which is indicative of the uncertainty in the photon number, {\it i.e.} whether the photon distribution $P(n)$ is sub- ($Q<0$) or super-Poissonian\index{Super-Poissonian} ($Q>0$). For a Fock state $|n\rangle$, the uncertainty in the photon number vanishes and $Q=-1$. For a coherent state (Poissonian photon distribution) $Q=0$. In general, $Q<0$, {\it i.e.} sub-Poissonian\index{Sub-Poissonian} statistics, is taken as a smoking gun of ``non-clasicallity''~\cite{davidovich1996sub}. 
\item {\it Second-order correlation function} 
\begin{equation}\label{corr2}
g^{(2)}(\tau)=\frac{\langle\hat{a}^{\dagger}(0)\hat a^\dagger(\tau)\hat{a}(\tau)\hat a(0)\rangle}{\langle\hat{n}\rangle^2}.
\end{equation}
The zero-delay value $g^{(2)}(0)$ is directly related to the $Q$-parameter, {\it e.g.} a state with sub-Poissonian\index{Sub-Poissonian} statistics is characterized by $g^{(2)}(0)<1$. The function $g^{(2)}(\tau)$ indicates how correlated two photons are in time~\cite{mandel1995optical,scully1999quantum} and is typically measured with a Hanbury Brown and Twiss setup\index{Hanbury Brown and Twiss! technique/interferometer} [see also sec.~\ref{subsec:WPCorrelator}]. If the light source is classical, for example thermal, one that has $g^{(2)}(\tau)\leq g^{(2)}(0)$. This instance corresponds to the so-called {\it photon bunching} and implies that photons from the source prefer to be bunched rather than randomly spread. If the source is quantum, like a single two-level atom, $g^{(2)}(\tau)>g^{(2)}(0)$, and the light is said to be {\it anti-bunched}~\cite{carmichael1985photon,mandel1995optical,scully1999quantum}. Anti-bunched light is different from random as it is more evenly spread, for random light every now and then two consecutively detected photons will appear very close in time. In sec.~\ref{ssec:lightforce} we will also mention higher-order correlation functions when discussing antibunching of photons pairs, while in sec.~\ref{sssec:JCzerodim} we will point to the link between nonclassical statistics and photon blockade.

\item The {\it quadrature variances}
\begin{equation}\label{quadvar}
\begin{array}{l}
\Delta x(t)=\langle\hat{x}^2\rangle-\langle\hat{x}\rangle^2,\\ \\
\Delta p(t)=\langle\hat{p}^2\rangle-\langle\hat{p}\rangle^2,
\end{array}
\end{equation}
where we $\hat{x}$ and $\hat{p}$ are defined in eq.~(\ref{quad}). With a simple rotation\index{Quadrature operators}\index{Operator! quadrature}
\begin{equation}\label{quadvar2}
\begin{array}{l}
x_\phi=x\cos\phi+p\sin\phi,\\ \\
p_\phi=-x\sin\phi+p\cos\phi
\end{array}
\end{equation}
the variances $\Delta x_\phi$ and $\Delta p_\phi$ for the angle $\phi$ are equivalently obtained. Since $\Delta x_{\phi+\pi/2}=\Delta p_\phi$, squeezing of the photon field implies a variance $\Delta x_\phi^2<1/2$  for some $\phi\in[0,2\pi)$~\cite{loudon1987squeezed,mandel1995optical}. 

\item The {\it state fidelity}\index{State fidelity}\index{Fidelity}
\begin{equation}\label{fidel}
F(\rho_1,\rho_2)=\left(\mathrm{Tr}\left[\sqrt{\sqrt{\hat{\rho}_1(t)}\,\hat{\rho}_2(t)\sqrt{\hat{\rho}_1(t)}}\right]\right)^2
\end{equation}
between two states $\hat\rho_1$ and $\hat\rho_2$. For pure states it identifies $F_{\Psi_1,\Psi_2}=|\langle\Psi_1|\Psi_2\rangle|^2$, and thus it is the overlap between the two states such that if the two states are (physically) the same $F=1$, and if they orthogonal $F=0$. There exists, of course, other measures for how alike two states are, like the {\it trace distance}\index{Trace distance}, $T(\rho_1,\rho_2)=\frac{1}{2}\mathrm{Tr}\left[\sqrt{\left(\hat\rho_1-\hat\rho_2\right)^\dagger\left(\hat\rho_1-\hat\rho_2\right)}\right]$~\cite{nielsen2010quantum}. We will, however, consider the fidelity whenever we compare different states and not the trace distance. We note that the fidelity is sometimes defined without the square of the R.H.S. and we will sometimes show the non-squared fidelity. 
\end{enumerate}

All information of the time-evolved state is contained in $\hat{\rho}(t)$. However, in order to picture the state and get an intuition for its properties it is convenient to introduce\index{Distribution! phase-space} {\it phase space distributions}~\cite{walls2008quantum,schleich2011quantum,gardiner2004quantum}. More precisely, phase space distributions are employed for representing quantum mechanical states in a phase space, similar to how they are used in classical mechanics. There are many ways how these distributions can be defined~\cite{agarwal1970calculus}, but among the most common ones are: the Glauber-Sudarshan $P$-function~\cite{sudersham1963,glauber1963coherent}, the $Q$-function~\cite{husimi1940some}, and the Wigner distribution\index{Distribution! Wigner}~\cite{wigner1932quantum}. We will now focus on the last two. The $Q$-function for a single (spinless) particle in one dimension is defined as
\begin{equation}\label{husimi}
Q(\alpha)\equiv\frac{1}{\pi}\langle\alpha|\hat{\rho}|\alpha\rangle,
\end{equation}
where $\hat{\rho}$ is the state of the particle and $|\alpha\rangle$ is a coherent state with amplitude $\alpha$, see eq.~(\ref{cohe}), and hence $Q(\alpha)$ is a function of two variables -- the real and imaginary part of $\alpha$. For a particle with spin plus a boson field, the $Q$ function is defined as $Q(z,\alpha)=\frac{1}{\pi^2(1+|z|^2)}\langle z,\alpha|\hat{\rho}|z,\alpha\rangle$, where $|z, \alpha\rangle \equiv |z\rangle \otimes |\alpha\rangle$ is the product of a spin coherent state with amplitude $z$~\cite{arecchi1972atomic} and the bosonic coherent state $|alpha\rangle$. The Wigner distribution (not to be confused with the atomic inversion (\ref{inv})) is defined accordingly
\begin{equation}\label{wigner}
W(x,p)=\frac{1}{\pi}\int\,dy\,\langle x-y/2|\hat{\rho}|x+y/2\rangle e^{ipy}.
\end{equation}
The Wigner distribution\index{Distribution! Wigner} is not positive definite and is therefore not a proper probability distribution. However, its marginal distributions\index{Marginal distributions}\index{Distribution! marginal}, $P(x)=\int dp\,W(x,p)$ and $P(p)=\int dx\,W(x,p)$, give the correct probability distributions for the quadratures $x$ and $p$. The $Q$-function can be obtained from $W(x,p)$ via a Gaussian average\index{Gaussian! average}~\cite{rajagopal1983classical}
\begin{equation}\label{Qwig}
Q(\bar{x},\bar{p})=\frac{1}{\pi}\int\,dxdp\,e^{-(x-\bar{x})^2-(p-\bar{p})^2}W(x,p).
\end{equation}
This `smoothening' of the Wigner function implies that $Q(\alpha)\geq0$. This is indeed the case, since the quantity $\pi Q(\alpha)$ is the diagonal matrix element of the density operator taken with respect to the coherent state $\ket{\alpha}$. Hence, it is a probability in the strict sense, namely the probability for observing the coherent state $\ket{\alpha}$ [see Ch. 4 of ~\cite{BookQO1Carmichael}]. 

In the dispersive regime, closed-form analytical expressions of $W(x,p,t)$ and $Q(\alpha,t)$ for the JC model are straightforward to derive for certain initial field states. For the simplest non-trivial example, we consider the initial product state $|\psi(0)\rangle=\left(|e\rangle|\beta\rangle+|g\rangle|\beta\rangle\right)/\sqrt{2}$, {\it i.e.} the atom in an equal superposition of its bare states and the field in a coherent state with an amplitude $\beta$ taken to be real. Time evolution, according to eq.~(\ref{dispcoef}), gives $|\psi(t)\rangle=\left(|e\rangle|\beta e^{-i\lambda t}\rangle+|g\rangle|\beta e^{i\lambda t}\rangle\right)/\sqrt{2}$~\cite{guo1996generation}. After tracing out the atomic degrees of freedom the field is in general in a mixed state (apart from the times $\lambda t=n\pi$, $n\in N$). One then obtains the phase space distributions\index{Distribution! phase-space}
\begin{equation}\label{dispdist}
\begin{array}{cll}
Q(\alpha,t) & = & \displaystyle{\frac{1}{2\pi}\left(e^{-\left|\alpha-\beta e^{-i\lambda t}\right|^2}+e^{-\left|\alpha-\beta e^{i\lambda t}\right|^2}\right)},\\ \\
W(x,p,t) & = & \displaystyle{\frac{1}{2\pi}\left(e^{-\left(x-\sqrt{2}\beta\cos(\lambda t)\right)^2-\left(p+\sqrt{2}\beta\sin(\lambda t)\right)^2}\right.}\displaystyle{\left.+e^{-\left(x-\sqrt{2}\beta\cos(\lambda t)\right)^2-\left(p-\sqrt{2}\beta\sin(\lambda t)\right)^2}\right)}.
\end{array}
\end{equation}
Thus, both $Q(\alpha,t)$ and $W(x,p,t)$ are given by two Gaussian distributions\index{Gaussian! distribution} counter propagating around the phase space origin. At $\lambda t=\nu\pi$ the two Gaussians completely overlap, while at $\lambda t=\nu\pi/2$ they are maximally separated. These results hinge on `linearizing' the time-evolution operator, which results in the Gaussian evolution. The properties and the evolution of the field phase space distributions for more general JC interactions have been extensively studied over the last decades, see for example refs.~\cite{gea1991atom,eiselt1991quasiprobability,buvzek1992superpositions,garraway1992quantum,orl1995dynamical,de1996nonlinear,buvzek1997cavity,larson2007dynamics}. 

We make here a brief comment on the application of phase-space approaches to the JC model. In quantum optics, the quantized electromagnetic field is often treated using phase-space methods, where the bosonic mode annihilation and creation operators are represented by complex-number phase space variables. According to the prescription, the density operator is associated with a distribution function of these variables. {\it Fokker--Planck equations}\index{Fokker--Planck equation} for the distribution function are then obtained, and either used to determine quantities of direct experimental interest or in order to develop {\it Langevin equations}\index{Heisenberg-Langevin equations} for stochastic versions of the phase-space variables from which experimental quantities are extracted as stochastic averages (see {\it e.g.}, \cite{BookQO1Carmichael}). 

Phase-space methods have also been employed in the study of atomic systems, with the atomic spin operators mapped to complex-number phase space variables. Haken, Risken and Weidlich introduced a direct extension of the Glauber-Sudarshan $P$ representation to represent atomic states, when developing their theory for the laser~\cite{Haken1967, Weidlich1967B, Hakenbook}. In addition, {\it atomic coherent states} (or {\it coherent spin states})\index{Atomic coherent states}~\cite{Radcliffe1971, Arecchi1972} have been introduced for a collection of $N$ atoms, particularly convenient for situations preserving the total angular momentum where Dicke states\index{Dicke! state} are used as basis states. In the limit $N \to \infty$, they may be formally connected to the coherent states we find in the harmonic oscillator. This corresponds to the radius of the sphere a spin is mapped to becoming infinite to yield the flat phase space for bosons. The associated $Q$ representation has been used in the study of superfluorescence~\cite{Lee1984} as well as in the description of thermalization in the Dicke model~\cite{altland2012quantum}. The atomic spin operators however satisfy the standard angular momentum commutation rules rather than the commutation rules for bosonic annihilation and creation operators; there exist in fact numerous mappings, both quadratic and non-quadratic -- see the Schwinger spin-boson mapping discussed in subsec.~\ref{sssec:strans}. Since the Hilbert space for bosons is infinite, but the one for a spin $S$ is not, one needs to restrict the number of bosons available in the system (through particle conservation) to accomplish an exact mapping.

Although the phase-space methods where the fermionic operators are represented directly by phase-space variables are yet to find a meaningful application, the anti-commutation rules these operators obey point to the possibility of using Grassmann variables\index{Grassmann! variables}, which have similar anti-commutation properties. Nevertheless, despite the impact of the seminal work by Cahill and Glauber~\cite{Cahill1999} and a few other notable applications~\cite{Plimak2001, Anastopoulos2000}, the use of phase space methods in quantum optics employed to treat fermionic systems through a direct representation of fermionic annihilation and creation operators by Grassmann phase space variables\index{Grassmann! variables} is rather rare. The analysis of Dalton and collaborators~\cite{Dalton2013} shows that phase-space methods using a positive type distribution function, involving complex-number variables for the cavity mode alongside Grassmann variables\index{Grassmann! variables} for the two-level atom, can be used to study the Jaynes–Cummings model, where quantities of experimental interest are related to quantum correlation functions -- expectation values of normally-ordered products of bosonic and fermionic operators.

The Grassmann phase-space theory\index{Grassmann! theory} allows the derivation of complex-variable equations governing the evolution (over time or temperature) of quantum correlation functions describing the position probabilities for atoms of opposite spins in a single or a couple of Cooper pairs\index{Cooper pair}. These correlators may be measured using Bragg spectroscopy. The theory is also applicable to simple Fermion systems, such as those described by the Fermi-Hubbard model. Functional Fokker--Planck equations\index{Fokker--Planck equation} are derived after establishing correspondence rules, followed by the equivalent Ito stochastic field equations. The required correlators are then computed in the Fourier space as stochastic averages of products of Grassmann stochastic momentum fields\index{Grassmann! fields}.

With the phase-space methods already enjoying some popularity in the treatment of laser-like systems, the theme of revival for the population inversion of the atomic states in the JC model in the presence of cavity losses is revisited in the 1990s. {\it Quasi}probability distributions are given in~\cite{Eiselt1991}, while a phase-space distribution\index{Phase-space distribution}\index{Phase-space distribution}\index{Distribution! phase-space} function is employed in~\cite{chough1996nonlinear} to visualize the departure from the perfect reconstruction of the quantum state in the driven JC model with dissipation absent. Typical examples of how the phase-space distributions evolve will be presented in the subsections that follow. In particular, we will see how the phase-space dynamics of the JC model can be related to the collapse-revival phenomenon, squeezing, and atom-field entanglement.

\begin{figure}
\includegraphics[width=10cm]{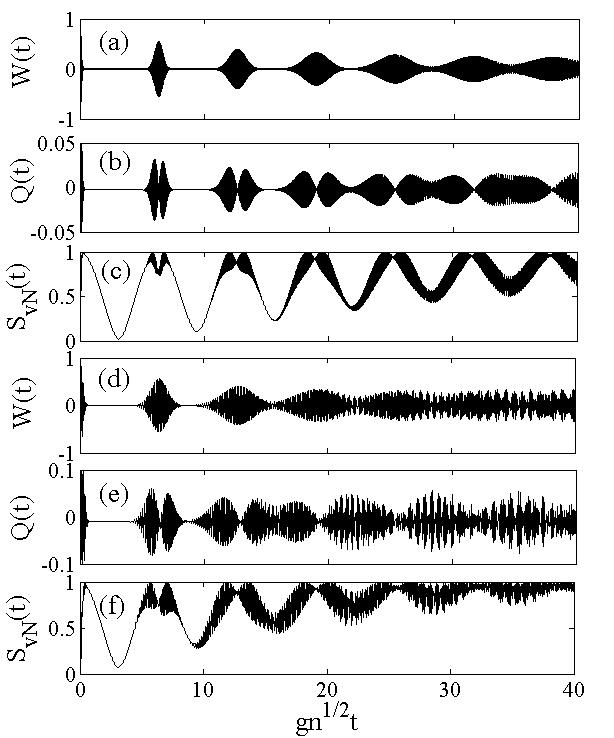} 
\caption{Time evolution of various physical quantities at resonance ($\Delta=0$); atomic inversion $W(t)$\index{Atomic! inversion}\index{Mandel $Q$-parameter} in {\bf (a)} and {\bf (d)}, the Mandel $Q$-parameter in {\bf (b)} and {\bf (e)}, and the von Neumann entropy $S_\mathrm{vN}(t)$\index{Entanglement! von Neumann entropy}\index{von Neumann! entropy} in {\bf (c)} and {\bf (f)}. Since the initial state is pure, the entropies of the atom and the field are identical. More precisely, the atom is initially excited and the field initially in a coherent state with amplitude $\alpha=10$ (a)-(c), and $\alpha=5$ (d)-(f). The time axis has been scaled with $g\sqrt{\bar{n}}$ in order to help comparing the two cases; $\bar{n}=100$ upper plots and $\bar{n}=25$ lower plots. It is clear that revivals get more distinct the larger the field amplitude is. A close-up of the atomic inversion and entropy production in the collapse region can be seen in fig.~\ref{classlim} below.}
\label{fig3}
\end{figure}


\subsubsection{Collapse-Revival}\label{sssec:crsubsec}\index{Collapse-revival}
For any initial state with an uncertainty in the photon number, $\Delta n^2=\langle\hat{n}^2\rangle-\langle\hat{n}\rangle^2>0$, one expects to find a collapse of any observable $\langle\hat{A}\rangle$ as time progresses~\cite{cummings1965stimulated}. Considering one of the simplest examples, the atomic inversion at resonance becomes\index{Atomic! inversion}
\begin{equation}\label{inv2}
W(t)=\sum_n|c_n(0)|^2\cos(2g\sqrt{n+1}t),
\end{equation}
where the coefficients $c_n(0)$ are determined from the initial state of the cavity field and the two-level atom. It follows from the above expression that such a collapse is an interference effect between the various terms in the expansion. Thus, it is a result of quantum fluctuations of the radiation field. Signatures of a collapse does not, however, prove that the radiation field is quantized, {\it i.e.} a collapse could in principle emerge from fluctuations of a classical electromagnetic field~\cite{scully1999quantum}. What if the different terms in the sum~(\ref{inv2}) return back in phase at some later time $t_\mathrm{R}$? The square root $n$-dependence of the Rabi frequency forbids in general a complete rephasing whenever the sum contains many non-zero terms. Nevertheless, a partial rephasing might well occur. This is the phenomenon of quantum revivals in the JC model~\cite{robinett2004quantum}. Contrary to the collapse, the occurrence of revivals in the JC model is a direct result of the field quantization.  

To the best of our knowledge, the first mention of collapse and revival in the JC model appeared in 1973 in a Note to {\it Lettere al Nuovo Cimento} titled {\it Destruction of Coherence by Scattering of Radiation on Atoms}~\cite{Meystre1973}. Therein, Meystre and coworkers consider the elastic scattering of radiation by a two-level atom inside a Fabry-P\'{e}rot resonator\index{Fabry-P\'erot! cavity/resonator}. Quoting their results directly, for the atom initially in its excited state $\ket{+}$ and a single cavity mode prepared in a coherent state $\ket{\alpha}$ -- the initial conditions required for the application of eq.~\eqref{inv2} -- the probability $P_{+}(t)$ of finding the atom in the excited state $\ket{+}$ upon evolution with the JC Hamiltonian [see also eq.~\eqref{eq:meanph3}] is [note that $W(t)=2P_{+}(t)-1$]
\begin{equation}
P_{+}(t)=\tfrac{1}{2} \exp(-|\tilde{\alpha}|^2) \sum_{n=0}^{\infty} \frac{|\tilde{\alpha}|^{2n}}{n!}[1+\cos(\sqrt{n}x)].
\end{equation}
The authors also calculate the atomic dipole moment as
\begin{equation}
 \braket{\mathcal{D}}(t)=\exp(-|\tilde{\alpha}|^2)\sum_{n=0}^{\infty} \frac{|\tilde{\alpha}|^{2n+1}}{\sqrt{n!(n+1)!}}\cos(\sqrt{n}x)\,\sin(\sqrt{n}x) \{i \exp[i(\omega t-\phi(\tilde{\alpha}))]\braket{+|\mathcal{D}|-} + \text{c.c.}\},
\end{equation}
with $x\equiv g t\,\sqrt{p}$ the dimensionless time and $\tilde{\alpha}\equiv \alpha/\sqrt{p}$ (the atom is assumed to have no permanent dipole moment). The index $p$ enumerates the cavity mode and scales the coupling constant -- in the present case $p=1$. Two years later, the same authors extended their analysis to include a set of initial conditions for the intracavity field~\cite{Meystre1975}. 

Subsequent explorations of the revival date back to the 1980s in works by Eberly, Milburn, Stenholm among others~\cite{eberly1980periodic,narozhny1981coherence,buck1981exactly,yoo1981non,stenholm1981bargmann,knight1982quantum,milburn1984interaction,barnett1986dissipation,puri1986collapse,filipowicz1986quantum,wright1989collapse}. In terms of the electrodynamic field, revivals were first demonstrated experimentally in the micromaser by the group of H. Walther~\cite{rempe1987observation}. Later, revivals as a proof of the quantization of the field has also been verified in the group of S. Haroche~\cite{brune1996quantum}, and beautifully demonstrated in trapped ion systems by Wineland and co-workers~\cite{meekhof1996generation}. The mechanism behind the collapse-revival evolution is well understood and can be found in most text books on quantum optics, see for example~\cite{walls2008quantum,schleich2011quantum,scully1999quantum,gerry2005introductory,meystre2007elements,orszag2008quantum,klimov2009group}. The idea relies on assuming that the relative width $\delta n=\Delta n/\bar n$, with $\bar n=\langle\hat n\rangle$, of the photon distribution $P(n)$ vanishes as $\bar n\rightarrow\infty$. This is certainly true for coherent and squeezed states. The collapse time occurs when the phases $\Omega_nt$ of the different $n$-dependent components of the evolved state are spread out over an interval of length unity, see fig.~\ref{fig2} (b). This leads to the collapse time\index{Collapse time}
\begin{equation}
\left(\Omega_{\bar n+\Delta n}-\Omega_{\bar n-\Delta n}\right)t_\mathrm{c}=1\hspace{0.5cm}\Rightarrow\hspace{0.5cm}t_\mathrm{c}=\frac{1}{\Omega_{\bar n+\Delta n}-\Omega_{\bar n-\Delta n}},
\end{equation}
since $\left(\Omega_{\bar n+\Delta n}-\Omega_{\bar n-\Delta n}\right)$ can be taken as an estimate of the spread in the Rabi frequencies. For a coherent state, $\Delta n=\sqrt{\bar n}$, and for $\bar n\gg1$ we can expand the square root in the Rabi frequencies to obtain~\cite{scully1999quantum,fleischhauer1993revivals}
\begin{equation}\label{ctime}
t_\mathrm{c}=\frac{1}{2g}\sqrt{1+\frac{\Delta^2}{4g^2\bar{n}}}.
\end{equation}
Similarly, the revival should occur when the involved components return in phase\index{Revival time}, {\it i.e.} 
\begin{equation}\label{rtime}
\left(\Omega_{\bar n}-\Omega_{\bar n-1}\right)t_\mathrm{r}=1\hspace{0.5cm}\Rightarrow\hspace{0.5cm}t_\mathrm{r}=\frac{2\pi m}{\Omega_{\bar n}-\Omega_{\bar n-1}}\approx\frac{2\pi m\sqrt{\bar{n}}}{g}\sqrt{1+\frac{\Delta^2}{4g^2\bar{n}}}.
\end{equation}
This time, $\left(\Omega_{\bar n}-\Omega_{\bar n-1}\right)$ is an estimate for the phase difference of two neighbouring terms in the sum~(\ref{inv2}). Again, we assumed a coherent state with $\bar n\gg1$, and $m=1,\,2,\,\dots$ denotes the first revival, second revival and so on. For $g\bar{n}\gg|\Delta|$ (for example at resonance, $\Delta=0$), both $t_\mathrm{c}$ and $t_\mathrm{r}$ scale as $\sim g^{-1}$. However, while the collapse time is independent on the mean photon number $\bar{n}$ in this limit, the revival is delayed for increasing photon numbers. The latter $\sqrt{\bar{n}}$-dependence implies that revivals become more and more difficult to observe when the field amplitude $\bar{n}$ grows big. This regime is sometimes referred to as the {\it classical limit of the JC model}\index{Jaynes-Cummings! classical limit} which will be considered in more detail in sec.~\ref{sssec:semclas}. For now we note that the more ``classical'' a field is, {\it i.e.} $\bar n$ is very large, the harder it is to see the revivals which we have argued is a purely quantum phenomenon.

The derivations leading to the expressions~(\ref{ctime}) and~(\ref{rtime}) rely on a relatively sharply peaked distribution $P(n)$. Indeed, when $\bar{n}\gg1$ the revivals become more complete as the distribution narrows. This may occur, for example, for initial squeezed field states depending on the squeezing parameter\index{Squeezing! parameter}~\cite{milburn1984interaction,satyanarayana1989ringing,rekdal2004preparation}. Another interesting localized field state is the {\it binomial state}\index{Binomial state}~\cite{stoler1985binomial}
\begin{equation}
    |p,N\rangle=\sum_{n=0}^N\left[\frac{N!}{(N-n)!n!}p^n(1-p)^{N-n}\right]^{1/2}|n\rangle,
\end{equation}
which has been thoroughly studied by Joshi {\it et al.} in terms of the JC model and the occurrence of the collapse-revivals pattern~\cite{joshi1987effects,joshi1989effects,joshi1989effects2}. 

It is the `sharpness' of the distributions $P(n)$ that causes the collapse time $t_\mathrm{c}$ to be $\bar{n}$-independent. For initial thermal field states\index{Thermal state}\index{State! thermal}~\cite{mandel1995optical}
\begin{equation}\label{thermal}
P(n)=\frac{1}{1+\bar{n}}\left(\frac{\bar{n}}{1+\bar{n}}\right)^n,
\end{equation}
with $\bar{n}=\left[\exp(\omega/T)-1\right]^{-1}$ the thermal photon number and $T$ the temperature (the Boltzmann constant $k_\mathrm{B}$ has been set to unity), one finds instead $t_\mathrm{c}=\Omega_{\bar{n}}/2$~\cite{knight1982quantum,barnett1986dissipation}. In general, the large (thermal) fluctuations in a thermal state weaken or completely destroy the revival signatures~\cite{chumakov1993analytical}. The thermal state is far from pure (indeed, it is the state with maximum entropy given $\bar{n}$). In fact, the presence of revivals in the JC model does not crucially depend on the field purity and off-diagonal coherence. The effect of such mixedness has been analyzed~\cite{vidiella1992interaction} and it was found that the absence of coherence may particularly delay and suppress the revivals. This does not, however, contradict the claim that JC revivals result from the quantum graininess of the electromagnetic field. The importance of the initial atomic state has also been studied~\cite{joshi1989effects3,zhou1992effect,zaheer1989phase}. In ref.~\cite{joshi1989effects3} it was demonstrated that the revivals may build up a double peak structure, while in~\cite{zaheer1989phase} it was shown that the revivals may be greatly suppressed for certain atomic initial states (this phenomenon will be discussed further in the next subsection).

\begin{figure}
\includegraphics[width=10cm]{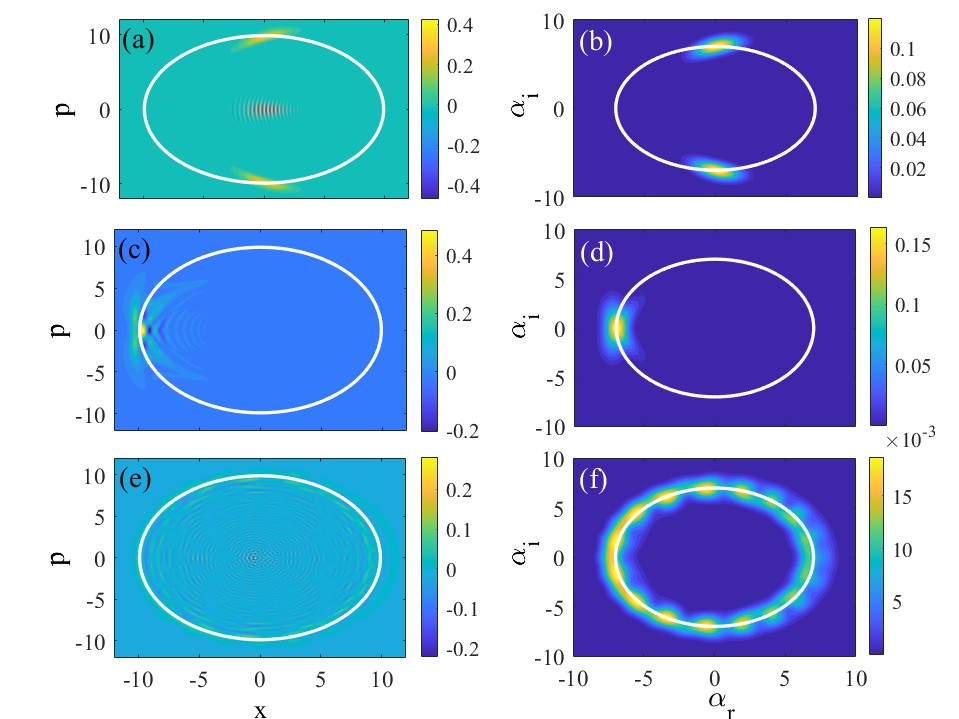} 
\caption{The plots of the left column display examples of the Wigner distribution\index{Wigner! distribution}\index{Distribution! Wigner} at different times, while the right column plots show the corresponding \index{Q distribution}\index{Distribution! Q} $Q$-function. In {\bf (a)} and {\bf (b)}, $t=t_\mathrm{r}/2$ ({\it i.e.} in the collapse region), in {\bf (c)} and {\bf (d)} $t=t_\mathrm{r}$, and in {\bf (e)} and {\bf (f)} $t=500$ (in dimensionless units). The initial state $|\psi(0)\rangle=|\beta\rangle|e\rangle$ is the atom in its excited state and the field in a coherent state with the amplitude $\beta=7$. The detuning $\Delta=0$ and $g=1$. The ``classical'' phase-space trajectory is indicated by the white circle.}
\label{fig4}
\end{figure}

In fig.~\ref{fig3}, we depict the atomic inversion $W(t)$ in frames (a) and (d), and the Mandel $Q$-parameter\index{Mandel $Q$-parameter} in frames (b) and (e) for the atom initially in the $|e\rangle$ state and an initial coherent field state $|\beta\rangle$ with $|\beta|^2=100$ (a) and (b) and $|\beta|^2=25$ (d) and (e). On resonance ($\Delta=0$) the inversion oscillates around its zero long-time mean. Both the collapse and the revivals are clearly visible. For the scaled time $g\sqrt{\bar{n}}t$, the revivals (but not the collapses) occur approximately at the same instant in the two examples depicted here. The larger the amplitude $\beta$, the clearer the collapse-revival structure is. In frames (d) and (e), for long times we observe a slower variation within the Rabi oscillations. These are the so-called {\it super-revivals}~\cite{gora1993superstructures,moya1998long,fang2000super}. Collapse and revivals are beautifully explained in the phase-space representation~\cite{robinett2004quantum} (see sec.~\ref{sssec:semclas} and \ref{fig3}). During the collapse period, for example, the phase-space distribution\index{Phase-space distribution} is split into localized sub-distributions. In particular, super-revivals can be understood as ``interference of different revivals''. More precisely, the width of the revival periods grows for larger revivals, and finally two consecutive revivals will ``overlap''~\cite{gora1993superstructures,moya1998long,fang2000super}. The phenomenon of super-revivals can be made more apparent for initial sub-Poissonian\index{Sub-Poissonian} photon distributions (as compared to the Poissonian\index{Poissonian} ones of fig.~\ref{fig3}. For completeness, we end this subsection by noting that {\it fractional revivals}\index{Fractional! revivals}~\cite{averbukh1989fractional,robinett2004quantum} which correspond to temporal formation of symmetric phase-space distributions have been discussed in terms of the JC model~\cite{averbukh1992fractional}. For a more thorough comparison of revivals deriving from anharmonicities in the potentials and those of the JC model we refer to ref.~\cite{wang2008quantum} (see fig.~\ref{fig4}).

In~\cite{fleischhauer1993revivals}, the sum of the atomic inversion\index{Inversion! atomic}\index{Atomic! inversion} is replaced by an infinite sum of integrals via the Poisson summation formula\index{Poisson! summation formula}; each revival is then associated with one term of this infinite series of integrals, calculated via the stationary phase method, while in~\cite{KlimovChumakov1999} the inversion sum is replaced by a sum of two integrals for an initial thermal field applying the Abel-Plana formula\index{Abel-Plana formula}. The fist term represents the initial collapse and the second term gives the revival as a quantum correction; this rationale has been extended to account for a thermal coherent state\index{Thermal state!coherent}\index{State! thermal!coherent} [a mixed state with $\hat{\rho}(\alpha,\beta)=(1-e^{-\hbar\beta\omega})\,\hat{D}(\alpha) e^{-\hbar \beta \omega a^{\dagger}a}\hat{D}^{\dagger}(\alpha)$, where $\beta=1/(k_B T)$ and $D(\alpha)|0\rangle=|\alpha\rangle$] at low temperature in~\cite{Azuma2010}. Novel asymptotic formulae for the atomic inversion were identified in~\cite{Karatsuba_2009}, relying on the functional equation of the Jacobi theta function. It has been recently demonstrated that the atomic inversion in the JC model on resonance has an exact representation as an integral over the Hankel contour. Using the integral representation of the Bessel function of half-integer order, the inversion sum is replaced by a contour integral in the complex plane which is evaluated through the saddle-point method\index{Saddle-point! method} for a highly excited initial coherent state of the cavity field. The saddle points identified from a differential equation, the solutions of which are trajectories in the complex plane--the Lambert W-function\index{Lambert W-function} which is pivotal in the manifestation of collapse and revivals. The method carries over as an asymptotic approximation for the non-resonant case~\cite{pavlik2023inside}.   

The excited-state population in the multimode JC model satisfies an integrodifferential equation~\cite{Nemet2019}\index{Integrodifferential equation}, and the kinklike behavior in the profile of the population~\cite{Parker1987} is explained by recursively solving a series of equations for sequential time intervals, while sending the number of modes to infinity. The number of modes required for the numerical results to agree with the exact solution increases with the light-matter coupling strength~\cite{Hasebe2022}. Finally, we mention that collapse and revival has been studied for the dissipative\index{Dissipative! JC model! multimode} multimode JC model~\cite{Islam2022}, and the numerical results were compared against the experimental observations of~\cite{QuantumRabi1996}. 

Collapse and revivals in the light-matter interaction described by the JC model have been recently discussed in the pedagogical account of~\cite{LaPierre2022}. 


\subsubsection{Semiclassical regime and the classical limit}\label{sssec:semclas}\index{Classical limit}\index{Semiclassical! regime}
The meaning of a ``classical limit'' in the JC model is ambiguous as the atom (two-level system) has no classical counterpart (In sec.~\ref{sssec:dicke}, when discussing the Dicke model, we will see that there exists, on the contrary, a proper classical limit.). The electromagnetic field, on the other hand, has a natural classical limit in terms of coherent states first introduced by Glauber~\cite{glauber1963coherent}. The classical limit here corresponds to letting the coherent state amplitude $\alpha$ grow to infinity~\cite{zhang1990coherent}. The relative number fluctuations $\delta n\equiv\Delta n/\bar{n}=1/\sqrt{\bar{n}}$ tend to zero in the large amplitude limit. In phase space, a coherent state is a symmetric Gaussian\index{State! Gaussian}\index{Gaussian! state}, see eq.~(\ref{dispdist}), which furthermore is a minimum uncertainty state, meaning that it is maximally localized according to the Heisenberg uncertainty principle. If the Hamiltonian is purely bosonic and quadratic in terms of these operators (in any number of degrees of freedom) it can be solved via Bogoliubov transformation\index{Bogoliubov transformation}~\cite{weedbrook2012gaussian}. The linearity of the Bogoliubov transformation implies that a coherent state will stay coherent under evolution of the quadratic Hamiltonian, and in particular it will follow the classical phase space trajectories. In field theories, a coherent-state {\it ansatz} is often referred to as a mean-field approximation\index{Mean-field approximation}. For a normally ordered Hamiltonian $\hat{H}(\hat{a}_\mathrm{i},\hat{a}_\mathrm{i}^\dagger)$, the corresponding mean-field energy functional is simply $H[\alpha]=\hat{H}(\alpha,\alpha^*)$. In such an approach for the JC model, any quantum correlations between the atom and the field are neglected even when the two-level system is treated at a full quantum level. This can only be valid for times long before the collapse has set in. In order to explain the evolution for longer times it is necessary to include quantum fluctuations and correlations. This was the topic of the seminal works by Gea-Banacloche, and Knight and co-workers~\cite{gea1990collapse,gea1991atom,buvzek1992schrodinger,phoenix1991comment}. Before recapitulating their results we discuss this observation in somewhat more general terms, and argue that the ``classical limit'' for the JC model does not amount to simply letting $\alpha\rightarrow\infty$, but some additional care is needed.

\begin{figure}
\includegraphics[width=10cm]{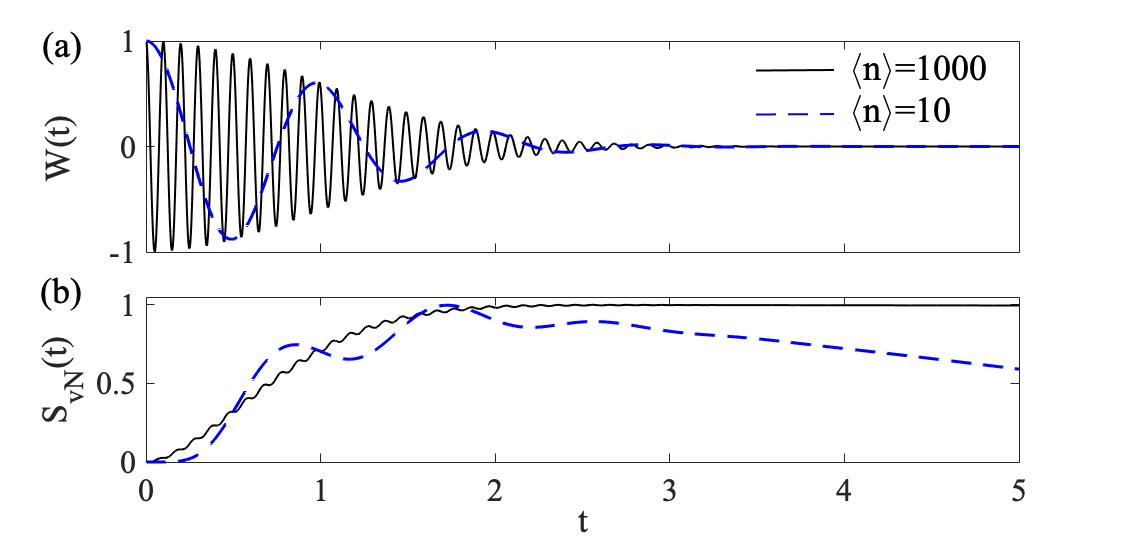} 
\caption{The time evolution in the JC model for scales of the collapse time $t_\mathrm{c}$. The two curves both correspond to initial coherent state of the field with very different amplitudes, $\alpha=\sqrt{10}$ (dashed blue) and $\alpha=\sqrt{1000}$ (solid black). In both plots, $g=1/2$ and $\Delta=0$, such that $t_\mathrm{c}=2$. In the upper plot {\bf (a)} we display the atomic inversion with the atom initially in its excited state $|e\rangle$, and in the lower plot {\bf (b)} we give the corresponding von Neumann entropies. What is evident from the figure is that the collapse, as well as the entanglement production, depends solely on the coupling $g$ and not on the field amplitude $|\alpha|$. Hence, the degree of ``classicality'' of the field is not important for entanglement formation. In the main text we suggest that the classical limit of the JC model should therefore be taken as $|\alpha|\rightarrow\infty$ while $g|\alpha|$ is kept constant. One clear distinction between the two cases is the drop of entanglement within the collapse region; this occurs much earlier for smaller $|\alpha|$. This follows from the atom-field disentangling at $t_\mathrm{r}/2\sim|\alpha|$ which is explained in the main text.}
\label{classlim}
\end{figure}

We saw in the previous section that the collapse derives from the fluctuations of the field around the mean-field background. These fluctuations build up the entanglement between the field and the atom; as demonstrated in fig.~\ref{fig2} the atomic state possesses information about the photon distribution. We will aim to provide an instructive explanation for how the entanglement forms when discussing the results of refs.~\cite{gea1990collapse,gea1991atom}. For now we note that given that the field is initially in a coherent state and the atom in one of its internal states $|g\rangle$ or $|e\rangle$, the time-scale for the entanglement generation between the two subsystems agrees with the collapse time~(\ref{ctime}). What is striking is that, provided that $\bar n$ is large, this time is independent of the coherent-state amplitude, and depends solely on the inverse coupling, $g^{-1}$. Thus, the combined atom-field system gets entangled on the same time-scale irrespective of how ``classical'' the field is, {\it i.e.}, the size of the coherent state amplitude $\alpha$. We wish, somehow, that the classical limit of the JC model should reproduce the {\it Rabi model}\index{Rabi! model}\index{Rabi! model} -- a two-level system driven by a monochromatic electromagnetic field~\cite{allen1975optical}
\begin{equation}\label{crabi0}
i\partial_t|\varphi(t)\rangle=\left(\frac{\Omega}{2}\hat\sigma_z+I\cos(\omega t)\hat\sigma_x\right)|\varphi(t)\rangle,
\end{equation}
with the field intensity $I^2$ proportional to the field amplitude $|\alpha|^2$.  After imposing the RWA (see secs.~\ref{ssec:rabi} and~\ref{sssec:rwasub}) we derive the {\it Rabi model}\index{Rabi! model}
\begin{equation}\label{crabi}
i\partial_t|\varphi(t)\rangle=\left[\begin{array}{cc}
\displaystyle{\frac{\Delta}{2}} & g\alpha\\
g\alpha^* & -\displaystyle{\frac{\Delta}{2}}
\end{array}\right]|\varphi(t)\rangle,
\end{equation}
where $|\varphi(t)\rangle=[a_e(t)\,\,\,a_g(t)]^T$ is the state vector for the atom. The Rabi model~(\ref{crabi}) should be viewed as the semiclassical analogue of the JC model - the quantum properties of the electromagnetic field are disregarded completely making the field classical.

In fig.~\ref{classlim} we show how the JC collapse occurs in relation to the formation of atom-field entanglement formation. We display two examples with very different field amplitudes $\alpha$, and what is clear is that it is indeed $g^{-1}$ determining the relevant time-scale, and not how ``classical'' the field is. What distinguishes the two curves is the Rabi frequency that scales as $\Omega_n=g\sqrt{n}=g|\alpha|$, {\it i.e.} the larger $|\alpha|$ the more Rabi oscillations take place before the collapse and the build-up of entanglement. As a result of this observation, to compare the semiclassical Rabi model to the JC model, one should not simply look at the large $\alpha$-limit, but take this limit while keeping the Rabi frequency $\Omega_n$, see eq.~(\ref{eigv}), fixed. Thus, the classical limit of the JC model\index{Jaynes-Cummings! classical limit} is thereby defined as an initial coherent field state $|\alpha\rangle$ and take the limit $|\alpha|\rightarrow\infty$ while keeping $|g\alpha|$ fixed. In this limit, the collapse time $t_\mathrm{c}$ goes to infinity at the same time as the Rabi oscillation frequency stays intact. There is a physical motivation behind this. By placing the atom inside a cavity, it can couple more strongly to the light field compared to the atom in free space. To compensate for the small $g$ in free space we need a light field of strong intensity. As a side remark, in secs.~\ref{ssec:rabi} and~\ref{sssec:dicke} we will discuss another type of classical limit for the quantum Rabi model, that does not rely on the state of the electromagnetic field but only on the system parameters. In that limit one lets the spectrum of the harmonic oscillator collapse into a continuum (as for a classical harmonic oscillator).  

\begin{figure}
\includegraphics[width=10cm]{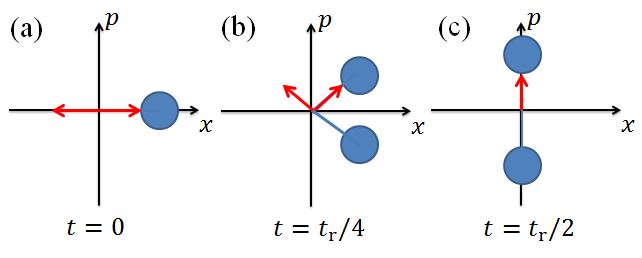} 
\caption{A graphic explanation of the large field JC evolution. Initially {\bf (a)} the atomic state (red arrows) can be decomposed into a component aligned or anti-aligned with the field (blue circle). As time goes, the two atomic components follows the fields and therefore are no longer pointing in opposite directions {\bf (b)}. At half the revival time {\bf (c)}, however, the two atom components coincide and consequently disentangle from the field.}
\label{fig5}
\end{figure}

To return to the results of refs.~\cite{gea1990collapse,gea1991atom,buvzek1992schrodinger,phoenix1991comment}, let $\alpha$ be real and large compared to $\Delta/g$, the semiclassical Rabi Hamiltonian then simply becomes $\hat H_\mathrm{cR}=g\alpha\hat{\sigma}_x$. The {\it dipole states}\index{Dipole! states}\index{State! dipole}
\begin{equation}\label{dipol}
|+\rangle=\frac{1}{\sqrt{2}}\left[\begin{array}{c}
1 \\ 1\end{array}\right],\hspace{1cm}
|-\rangle=\frac{1}{\sqrt{2}}\left[\begin{array}{c}
1 \\ -1\end{array}\right]
\end{equation}
are eigenstates of the Hamiltonian. In the language of spins, the state $|\pm\rangle$ points to the positive/negative $x$-(spin)direction. As time progresses, $|+\rangle$ will rotate anti-clockwise around the Bloch sphere along the equator, while $|-\rangle$ rotates clockwise. After quarter rotation both states will point in the positive $y$-direction. This will have interesting consequences as we show next. For finite $\alpha$, quantum fluctuations in the field will always be present. Assuming that $\bar{n}=|\alpha|^2\gg1$, the photon distribution $P(n)$ is then rather localized around its mean $\bar n$, and we may thereby expand the Rabi frequencies $\Omega_n$\index{Rabi! frequency} around $\bar{n}$, which at resonance gives~\cite{gea1991atom}
\begin{equation}\label{expans}
\Omega_n=g\sqrt{n+1}\approx g\sqrt{\bar{n}}+\frac{g}{2}\frac{n-\bar{n}}{\sqrt{\bar{n}}}-\frac{g}{8}\frac{(n-\bar{n})^2}{\bar{n}^{3/2}}+...\,.
\end{equation}
The first term is just an overall phase shift, the second term induces a phase $\mp gt/2\sqrt{\bar{n}}$ upon the coherent states given the atomic states $|\pm\rangle$, while the third nonlinear term induces a deformation of the Gaussian (coherent) field state\index{Gaussian! state}. Thus, within the linear regime ({\it i.e.} neglecting quadratic and higher order terms) the atom-field interaction of an initial state $|\alpha\rangle|\pm\rangle$ will cause a circulation of the field state around the phase space origin and a rotation of the atomic state along the equator~\cite{gea1990collapse,gea1991atom}. This type of evolution is schematically presented in fig.~\ref{fig5}. Initially, any atomic state (not entangled with the field) can be decomposed as $|\varphi(0)\rangle=a_+|+\rangle+a_-|-\rangle$. Thus, the atom has one component aligned and one component ant-aligned with the field. The field is initially in a coherent state with some amplitude $\alpha$, {\it i.e.} $|\psi(0)\rangle=a_+|+\rangle|\alpha\rangle+a_-|-\rangle|\alpha\rangle$ (see fig.~\ref{fig5} (a)). At a later time, fig.~\ref{fig5} (b), the field hat split up into two parts and the atom is `following' the field rotation. At half the revival time\index{Revival time} $t_\mathrm{r}/2$ (recall that the revival time $t_\mathrm{r}\sim\sqrt{\bar n}/g$, see eq.~(\ref{rtime})), fig.~\ref{fig5} (c), the field is in a coherent-state superposition $|\theta(t_\mathrm{r}/2)\rangle\propto(a_+|i\alpha\rangle+a_-|-i\alpha\rangle)$, as was already demonstrated in fig.~\ref{fig4} (a) and (b) which showed the Wigner and $Q$ phase-space distributions\index{Q distribution}\index{Distribution! Q}, and the atom has disentangled from the field. For $|a_+|=|a_-|$ the field is {\it deterministically} prepared in a cat state\index{Schr\"odinger cat! states}\index{Cat state}~\cite{gerry1997quantum}, {\it i.e.} a macroscopic superposition state~(\ref{catstate})\index{Cat state}. For initially mixed atomic states, but still not initially entangled with the field, the atom can again be prepared in a pure state~\cite{orszag1992preparation}.  

This disentangling is an astonishing property of the dynamics, since this effect does not depend on the initial atomic state $|\varphi(0)\rangle$, and at $t=t_\mathrm{r}/2$ the atom is always `prepared' in a given state. It reflects the non-unitary evolution of the reduced atomic state $\hat{\rho}_\mathrm{a}(t)$~\cite{knoll1995distance}. The characteristic evolution is not relying on a coherent-state {\it ansatz} for the field as long as the distribution $P(n)$ is narrow and peaked around some large $\bar n$. The disentangling effect may in fact be enhanced for initial squeezed states~\cite{rekdal2004preparation}. Complete disentanglement at $t_\mathrm{r}/2$ occurs only in the large field limit $\bar{n}\rightarrow\infty$, which on the other hand from eq.~(\ref{rtime}) implies that it is impractical for any experimental realization since the rival time grows with $\bar{n}$. The natural question is then how entangled the atom and field are at $t_\mathrm{r}/2$, {\it i.e.} how much effect does the quadratic and higher order terms in the expansion (\ref{expans}) have. In fig.~\ref{fig3} (c) and (f) the von Neumann entropy is shown for $\bar{n}=100$ and $\bar{n}=25$ respectively. In this case when the initial state is a pure atom-field product state, $S_\mathrm{vN}(t)$ is a direct measure of the amount of entanglement; $S_\mathrm{vN}=\log_2(N)$ for maximal entanglement and $S_\mathrm{vN}=0$ for no entanglement. Already for $\bar{n}=20$ the disentanglement is very evident. At later times $\tau$, such that $\tau g(n-\bar n)^2/8\bar n^{3/2}\sim\pi$, the small terms in the expansion (\ref{expans}) can no longer be neglected and during the second collapse period the disentanglement is greatly reduced.

Having understood the semiclassical evolution of the JC model, we can give a qualitative explanation for the connection between the collapse and entanglement formation. In the linear regime, where we truncate the expansion~(\ref{expans}) after the second term, in phase space the initial coherent state splits and the two components encircle the origin. After some time, the two components will approximately be orthogonal and the two subsystems are then approximately maximally entangled.
A coherent-state component close to the origin will move slower than a coherent-state component further away from the origin. However, the overlap between the two components will not depend much on $\alpha$, since the width of the coherent state increases with $\alpha$ in the same way as the velocity of rotation. An overlap which is roughly zero between the two field components means that the reduced states $\hat\rho_\mathrm{a}$ and $\hat\rho_\mathrm{f}$ {\it have lost their phase coherence} -- this instance marks the collapse. 

\begin{figure}
\includegraphics[width=10cm]{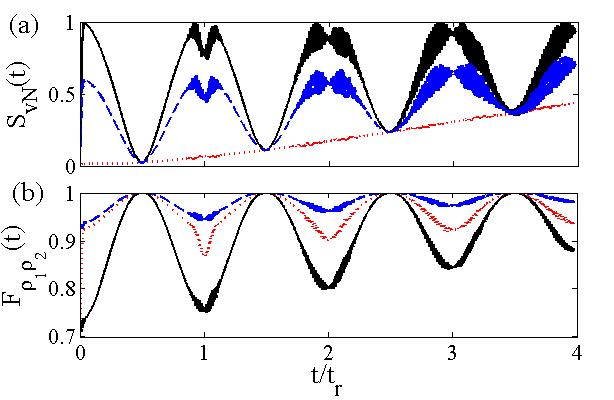} 
\caption{Time evolution for the von Neumann entropy\index{Entanglement! von Neumann entropy}\index{von Neumann! entropy} {\bf (a)} and the atomic fidelity {\bf (b)} for three different initial states; $|\psi_1(0)\rangle=|e\rangle|\alpha\rangle=\frac{1}{\sqrt{2}}\left(|+\rangle+|-\rangle\right)|\alpha\rangle$ (black solid line in (a)), $|\psi_2(0)\rangle=\frac{1}{\sqrt{2}}\left(|e\rangle+|g\rangle\right)|\alpha\rangle=|+\rangle|\alpha\rangle$ (red dotted line in (a)), and $|\psi_3(0)\rangle=\frac{1}{\sqrt{2}}\left(|e\rangle+e^{i\pi/4}|g\rangle\right)|\alpha\rangle=\frac{1}{2}\left[\left(1+e^{i\pi/4}\right)|+\rangle+\left(1-e^{i\pi/4}\right)|-\rangle\right]|\alpha\rangle$ (blue dashed line in (a)), and $\alpha=10$. In the lower plot; $F_{\rho_1\rho_2}(t)$ (black solid line), $F_{\rho_1\rho_3}(t)$ (red dotted line), and $F_{\rho_2\rho_3}(t)$ (blue dashed line). It is particularly noted that the atom-field disentangling at $t=t_\mathrm{r}/2$ does not seem to depend on the initial atomic state. More striking is the perfect fidelity\index{State fidelity} at times $t_\mathrm{r}/2,\,\,3t_\mathrm{r}/2,\,...$, regardless whether the atom-field entanglement is large or not.
}
\label{fig6}
\end{figure}

We end this subsection by considering the importance of the initial atomic state. This topic was briefly discussed in~\cite{joshi1989effects3,zhou1992effect}. As explained above and in fig.~\ref{fig5}, at $t_\mathrm{r}/2,\,\,3t_\mathrm{r}/2,\,...$ the atom disentangled from the intracavity field in the limit of a large field amplitude. The atomic state at these instants is independent of the initial state. However, for other times the atomic state will strongly depend on the initial state. In fig.~\ref{fig6} (a) we give three examples of $S_\mathrm{vN}(t)$. If the atom starts in its excited state (black line) the atom and field build up an appreciable entanglement, while if the atom starts in the `dipole state' $|+\rangle$ (which is aligned with the field) the atom-field entanglement remains low for short times. A linear combination of these two limiting cases (blue line) falls in between the two other results. The three lines indeed coincide at the expected disentanglement times. In principle, $S_\mathrm{vN}(t)$ cannot reveal if the atomic states at those times are independent of the initial conditions. This is indeed the case, as is verified in fig.~\ref{fig6} (b) which displays the three different state fidelities~(\ref{fidel})~\cite{jozsa1994fidelity}.
The fidelity is extremely close to unity at the disentanglement times even when the atom is still entangled with the field. At very long times the fidelity approaches unity, which is due to the fact that the atomic state becomes maximally mixed.

\subsubsection{Entanglement}\label{sssec:ent}
The particularly simple, yet non-trivial, JC Hamiltonian has served as a model formulation for numerous studies on quantum correlations between different parts of a composite system, characteristically exemplified by {\it entanglement}\index{Entanglement}. Understanding the very nature of entanglement, as the source of many quantum information processing protocols, constitutes by now a research field in its own right~\cite{horodecki2009quantum,nielsen2010quantum}. We will not attempt to review all the different properties of quantum correlations (let alone multi-partite entanglement), but will instead focus on those ones conspicuously related to the JC model.

The earliest study relating quantum correlations to the JC model dates back to the first half of the 1980s~\cite{aravind1984two}, {\it i.e.} before the rise of {\it quantum information science}. Aravind and Hirschfelder especially looked at the purity of the atom as the bi-partite system evolved. The first introduction of the concept of {\it entanglement} in the JC model seems to appear in the early 1990s~\cite{phoenix1991establishment}. However, it should be noted that many of the conclusions of Ref~\cite{phoenix1991establishment} had already been put into action some years earlier in~\cite{phoenix1988fluctuations} when the von Neumann entropy of the atom and field sub-systems was thoroughly studied. Since the beginning of the 1990s, numerous papers have discussed entanglement properties in the JC model in one or another way, see for example Refs.~\cite{gea1991atom,gea1993jaynes,guo1997preparation,furuichi1999entanglement,furuichi2001entanglement,bose2001subsystem,son2002entanglement,scheel2003hot,boukobza2005entropy,ghosh2007effects,akhtarshenas2007negativity}. In the coming section, when discussing specific physical systems, we will return in more detail to many of the schemes devised for generating entangled states (see, for example, subsecs.~\ref{ssec:cqedstateprep} and~\ref{ssec:qipi}).

For pure initial states $|\psi(0)\rangle$, the von Neumann entropy (\ref{vN})\index{von Neumann! entropy} is a good measure of entanglement between any two subsystems with reduced density operators $\hat{\rho}_\mathrm{A}(t)$ and $\hat{\rho}_\mathrm{B}(t)$~\cite{nielsen2010quantum}. Moreover, for such states Araki and Lieb have proved that $S_\mathrm{vN}(\hat{\rho}_\mathrm{A},t)=S_\mathrm{vN}(\hat{\rho}_\mathrm{B},t)$, {\it i.e.} the entropy for either of the reduced states must be equal~\cite{araki2002entropy}. This result is expected; subsystem $A$ is as entangled with subsystem $B$ as $B$ is entangled with $A$. In sec.~\ref{sssec:sol}, the von Neumann entropy $S_\mathrm{vN}(t)$, the purity $P(t)$, and the linear entropy $S_\mathrm{L}(t)$ were all expressed in terms of the length of the Bloch vector $\mathbf{R}(t)$. Thus, provided that the system is initialized in a pure state, and that the evolution is unitary, it follows that measuring the three components of the Bloch vector, $\langle\hat{\sigma}_x\rangle$, $\langle\hat{\sigma}_y\rangle$, and $\langle\hat{\sigma}_z\rangle$ (which naturally determines the full atomic state), the intrinsic atom-field entanglement is accessible. Figure~\ref{fig3} has already illustrated two examples of the von Neumann entropy for an initial state $|\psi(0)\rangle=|e\rangle|\alpha\rangle$ with $\alpha=10$ (c) and $\alpha=5$ (f). During the collapse, the atom and field become almost maximally entangled due to the mechanism explained in the previous subsection.   

Characterizing quantum correlations for mixed states is a much more subtle problem~\cite{vedral1997quantifying,vidal2001computable,Nha2004Ent,horodecki2009quantum}. For initially pure states, we saw that there is a one-to-one correspondence between the purity $P(t)$ and the entropy $S_\mathrm{vN}(t)$. Clearly, this cannot be true for more general states, since one cannot say whether the mixedness of the reduced state stems from entanglement with the other subsystem or from mixedness of the full system state. For two qubits, the {\it concurrence}\index{Concurrence}\index{Entanglement! Concurrence}~\cite{wootters1998entanglement},
\begin{equation}\label{conc}
\mathcal{C}(t)\equiv\mathrm{max}\left\{0,\,\lambda_1(t)-\lambda_2(t)-\lambda_3(t)-\lambda_4(t)\right\},
\end{equation}
where the $\lambda_i(t)$'s are eigenvalues (in decreasing order) of
\begin{equation}
 \hat{R}(t)\equiv\sqrt{\sqrt{\hat{\rho}(t)}\hat{\rho}'(t)\sqrt{\hat{\rho}(t)}}\quad \text{with} \quad\hat{\rho}'(t)=\left(\hat{\sigma}_y\otimes\hat{\sigma}_y\right)\hat{\rho}^*(t)\left(\hat{\sigma}_y\otimes\hat{\sigma}_y\right),
\end{equation}
is a necessary and sufficient entanglement measure. This then is only a good measure for the $\hat{N}=1$ special case of the JC model, {\it i.e.} a maximum of one photon. For more general bi-partite systems, the {\it logarithmic negativity}~\cite{vidal2001computable}\index{Logarithmic negativity}\index{Entanglement! Logarithmic negativity} can be used as a sufficient condition for entanglement (in $2\times3$ it is also a necessary condition). The logarithmic negativity is also a {\it monotone}~\cite{plenio2005logarithmic}\index{Entanglement! monotone}, meaning that it can be used a measure of entanglement. It is defined as
\begin{equation}\label{logneg}
E_\mathrm{N}(t)\equiv \log_2(2\mathcal{N}(t)+1),
\end{equation}
where
\begin{equation}
\mathcal{N}(t)\equiv\sum_i\frac{|\lambda_i(t)|-\lambda_i(t)}{2}
\end{equation}
is the {\it negativity}\index{Negativity}\index{Entanglement! negativity} and $\lambda_i(t)$ is the $i$'th eigenvalue of $\hat{\rho}^{\Gamma_\mathrm{A}}(t)$ which is the partial transpose with respect to subsystem $A$ of the state $\hat{\rho}(t)$. For the JC system, the partial transpose with respect to the atom is
\begin{equation}
\hat{\rho}^{\Gamma_\mathrm{a}}(t)=\left[
\begin{array}{cc}
\hat{\rho}_{ee}^T(t) & \hat{\rho}_{eg}^T(t)\\
\hat{\rho}_{ge}^T(t) & \hat{\rho}_{gg}^T(t)
\end{array}\right],
\end{equation}
with $\hat{\rho}_{ij}(t)=\langle i|\hat{\rho}(t)|j\rangle$ ($i,\,j=e,\,g$), and $T$ stands for the regular matrix transpose. The negativity, or logarithmic negativity, has been used to study how atom-field entanglement builds up in the JC model, both for initial pure field state but mixed atomic states~\cite{akhtarshenas2007negativity,chen2011entanglement}, and also when both bi-partite subsystems are initially mixed~\cite{scheel2003hot}. In ref.~\cite{scheel2003hot} it was explored how `hot' -- thereby mixed -- the initial state can be in order to build up entanglement. By `hot' we mean that the temperature of the effective thermal field state is high. It was found in particular that entanglement might survive even for very hot fields, given certain initial mixed atomic states. There exist many other entropic measures, {\it e.g.} recently the {\it quantum relative entropy}\index{Quantum! relative entropy}\index{Entropy! quantum relative} was considered in terms of characterizing the flow of information between the subsystems of the JC model~\cite{megier2021entropic}, it is defined for two states $\hat\rho$ and $\hat\varrho$ as $S(\hat\rho,\hat\varrho)=\mathrm{Tr}\left[\hat\rho\log\hat\rho-\hat\rho\log\hat\varrho\right]$, and the {\it Wigner-Yanase skew information}\index{Wigner!-Yanase skew information} has also been considered for the JC model~\cite{dai2020atomic}, which is defined for a state $\hat\rho$ and an observable $\hat O$ as $I(\hat\rho,\hat O)=-\mathrm{Tr}\left[[\sqrt{\hat\rho},\hat O]^2\right]/2$. The {\it Wehrl entropy}\index{Wehrl entropy}\index{Entropy! Wehrl}, $S_W=-\frac{1}{\pi}\int Q(\alpha)\log Q(\alpha)d^2\alpha$, with $Q(\alpha)$\index{$Q$-function} the $Q$-function~(\ref{husimi})\index{Q distribution}\index{Distribution! Q}, was early on studied for the JC evolution~\cite{paul1995dynamical}. 

\begin{figure}
\includegraphics[width=10cm]{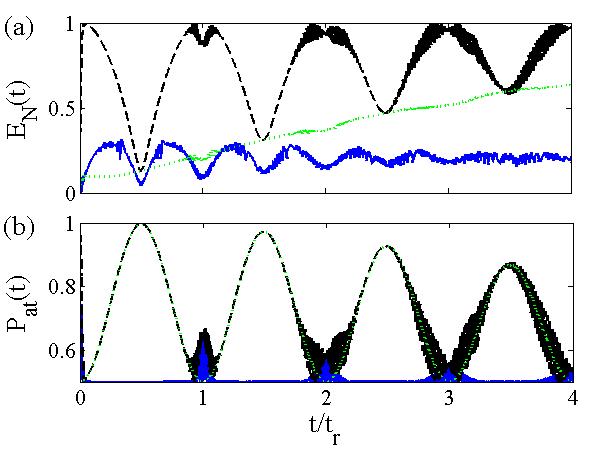} 
\caption{Logarithmic negativity {\bf (a)} and atomic purity {\bf (b)} for various initial atom-field states; $\hat{\rho}_\mathrm{I}(0)$ (black dashed line), $\hat{\rho}_\mathrm{II}(0)$ (blue solid line), and $\hat{\rho}_\mathrm{III}(0)$ and $\hat{\rho}_\mathrm{IV}(0)$ (green dotted line)  see eq.~(\ref{inst}). The two initial states $\hat{\rho}_\mathrm{III}(0)$ and $\hat{\rho}_\mathrm{IV}(0)$ give identical evolution for the two plotted quantities. The coherent-state field amplitude is $\alpha=10$.
}
\label{fig7}
\end{figure}

To give an example of entanglement formation in the JC model for mixed states, in fig.~\ref{fig7} we plot the logarithmic negativity and atomic purity for the initial states
\begin{equation}\label{inst}
\begin{array}{l}
\displaystyle{\hat{\rho}_\mathrm{I}(0)=|\alpha\rangle\langle\alpha|\otimes|e\rangle\langle e|} ,\\ \\
\displaystyle{\hat{\rho}_\mathrm{II}(0)=|e\rangle\langle e|\otimes\sum_n |c_n|^2|n\rangle\langle n|}\\ \\
\displaystyle{\hat{\rho}_\mathrm{III}(0)=\frac{1}{2}\left(|e\rangle+e^{i}|g\rangle\right)\left\langle e|+e^i\langle g|\right)\otimes\sum_n |c_n|^2|n\rangle\langle n|} \\ \\
\displaystyle{\hat{\rho}_\mathrm{IV}(0)=\frac{1}{2}\left(|e\rangle\langle e|+|g\rangle\langle g|\right)\otimes\sum_n |c_n|^2|n\rangle\langle n|}, 
\end{array}
\end{equation}
where the coefficients $c_n$ are given in eq.~(\ref{cohe}) for a coherent state of amplitude $\alpha=10$. The two last cases, $\hat{\rho}_\mathrm{III}(0)$ and $\hat{\rho}_\mathrm{IV}(0)$ yield identical negativities and atomic purities. The first case, $\hat{\rho}_\mathrm{I}(0)$, represents a pure state, and the logarithmic negativity resembles the von Neumann entropy of fig.~\ref{fig6}. The blue and green curves of fig.~\ref{fig7}, corresponding to initially mixed states, demonstrate how mixedness and entanglement are not necessary one and the same thing.

In recent years, an interesting aspect about entanglement among qubits has been pointed out by Yu and Eberly~\cite{yu2009sudden}; two initially entangled qubits interacting with independent ``reservoirs'' can suddenly become disentangled. This phenomenon has been named {\it entanglement sudden death} (ESD)\index{Entanglement! sudden! birth}\index{Entanglement! sudden! death}, and similarly the entanglement can be suddenly reestablished ({\it entanglement sudden birth}~\cite{ficek2008delayed} (ESB)). A proper entanglement measure for the two qubits was given by the concurrence defined in eq.~(\ref{conc}). Note that, similar to the negativity, the concurrence is defined in terms of the density matrix and is therefore valid for pure as well as for mixed states. From the definition of $\mathcal{C}(t)$, it is clear that ESD/ESB occurs whenever the quantity $\lambda_1(t)-\lambda_2(t)-\lambda_3(t)-\lambda_4(t)$ changes sign. While the ESD effects have been demonstrated with polarized photons~\cite{almeida2007environment}, the early theoretical investigations ware often devoted to JC-like systems~\cite{yonacc2006sudden}. More precisely, the model systems comprise two decoupled JC Hamiltonians where the two atoms are initially entangled. Thus, the two fields serve the function of independent reservoirs. Various types of initial entangled states have been considered~\cite{man2008conditions}, alongside decoherece effects~\cite{sainz2007entanglement,abdel2011multi} which demonstrate that ESD can survive Markovian information loss to an external reservoir. Generalizations to other JC setups have been considered in numerous works, see for example refs.~\cite{yonacc2007pairwise,cui2007study,qiang2008control,zhang2009atomic,zhang2009controlling,sadiek2019manipulating}.

Independently, Vedral with Henderson and Ollivier with Zurek have demonstrated that there exist quantum correlations beyond classical correlations but still for disentangled states~\cite{ollivier2001quantum,henderson2001classical}. More precisely, for disentangled mixed states there may still exist correlations with no classical interpretation, {\it i.e.}, for a situation where correlations are due to quantum physical effects and classical physics would predict no correlations. Such {\it quantum discord}\index{Quantum! discord} has been proven an actual asset for quantum computing~\cite{datta2008quantum}. Some aspects of the quantum discord have been discussed in terms of JC type systems~\cite{wang2011classical}. In ref.~\cite{fanchini2010non}, it was specifically shown that in the time periods of ESD ({\it i.e.}, for vanishing qubit entanglement) the quantum discord may be non-zero.

The generation of atom-field entanglement in the JC model naturally creates the possibility to carry out quantum logic gate operations~\cite{nielsen2010quantum} (see further subsecs.~\ref{ssec:cqedQI} and \ref{ssec:qipi} for more detailed discussions of specific quantum information processing schemes). There exist either single qubit gates applied to the atom, or two-qubit gates addressing the atom and the field~\cite{domokos1995simple,rauschenbeutel1999coherent}, or equally two-qubit gates addressing two field modes mediated by an atom~\cite{zubairy2003cavity,larson2004dynamics,biswas2004quantum}, as well as two-qubit gates addressing two atoms interacting with the same field mode~\cite{imamog1999quantum,zheng2000efficient,jane2001quantum,pachos2002quantum,zheng2002quantum,yang2003possible,song2005quantum,feng2007scheme}. The general idea for entangling two qubits, field qubits or atomic qubits, is to let the parts interact via an {\it ancilla}\index{Ancilla! system} subsystem. For two atomic qubits, the ancilla is the cavity mode (sometimes also referred to as a {\it quantum bus}\index{Quantum! bus} shuffling information between the two qubits), while for the field and qubits, an auxiliary atom may serve as the ancilla. As demonstrated in the previous section, a field in the vacuum state may still interact with an atom and this interaction, in particular, causes an energy shift which may be utilized for performing conditional {\it phase-gates}\index{Quantum! phase gate}~\cite{imamog1999quantum,zheng2000efficient,yang2003possible}. Even for a general cavity state~\cite{zhu2005geometric} (including a thermal field state~\cite{zheng2002quantum}) are qubit gate operations feasible. Most of the above schemes are not prone to cavity decay since the field mode is only virtually excited by the atoms. On the other hand, they are sensitive to actual gate times. A way to circumvent such errors, {\it adiabatic quantum computing} has been proposed in~\cite{farhi2000quantum}. Here the evolution is adiabatic and does not rely on operation times as long as the adiabatic constraint is met. Adiabatic logic gates have also been discussed for the JC model, see for example~\cite{recati2002holonomic,zheng2004unconventional,zheng2005nongeometric,zhu2005geometric,lacour2006arbitrary,song2007quantum,solinas2003semiconductor,xue2007simple}. These adiabatic implementations often rely on the existence of {\it dark states}\index{Dark! state}\index{State! dark}, {\it i.e.} instantaneous eigenstates of the time-dependent Hamiltonian with vanishing eigenvalues,
\begin{equation}\label{darkstate}
\hat{H}(t)|\psi(t)\rangle=0.
\end{equation}
In the adiabatic limit, the state evolves from $|\psi(-\infty)\rangle$ to $|\psi(+\infty)\rangle$, while the fact that the state is dark implies suppression of photon losses. Other adiabatic schemes rely on the system geometry by taking advantage of the {\it Berry phase}\index{Berry phase}~\cite{berry1984quantal,bohm2003geometric}.

The JC model -- an exemplary realizable model of supersymmetric~\cite{andreev1989supersymmetry} quantum mechanics -- exhibits a nontrivial behavior in the course of an adiabatic evolution of the system parameters. For a cyclic variation of the atom-field detuning $\Delta$ and the dipole coupling constant $\mu=|\mu| e^{i\phi}$ (so that $\dot{\Delta}/\Delta$ and $\dot{\mu}/\mu$ are the smallest characteristic frequencies in the system evolution), Andreev and collaborators~\cite{andreevJETP}, using the WKB approximation\index{WKB approximation}, found that the initial state
\begin{equation}
 \ket{\psi}=\ket{N,-},
\end{equation}
transforms to
\begin{equation}
 \ket{\psi^{\prime}}=\cos\theta\,\ket{N,+} + \sin\theta\,\ket{N-1,-}, 
\end{equation}
where
\begin{equation}
 \theta=\int^{t} \frac{\Delta \dot{\phi}}{4\Omega}\,d\tau, \quad\quad \Omega \equiv \sqrt{4|\mu|^2 N + \Delta^2},
\end{equation}
has the meaning of a topological phase. In the limit $N=0$ one obtains $\theta=0$, {\it i.e.}, the vacuum state does not acquire a topological phase. For a large $N$ (which is an integral of motion), $\theta \sim 1/\sqrt{N} \to 0$, {\it i.e.}, the effect is of a purely quantum nature and disappears in the classical limit $N\to \infty$. For a sinusoidal variation of the detuning, the phase and the amplitude of the coupling constant, with $\Delta(t)= \pm \epsilon \cos(\gamma_R t)$, $\phi=\sin(\gamma_R t)$, $2\sqrt{N} |\mu|(t)=\epsilon \sin(\gamma_R t)$ (with $\gamma_R \ll |\Delta|$,$|\mu|$), one finds a large Berry phase\index{Berry phase}, namely $\theta=\pm\pi/4$. 

The authors of~\cite{andreevJETP} further propose that the accumulated geometric phase\index{Geometric phase} may be measured by a Ramsey interferometer\index{Ramsey! interferometry}. In the course of a measurement a wavefunction of the form $\ket{\psi}=\alpha_0 \ket{a} + \beta_0 \ket{b}$ is transformed into the wavefunction $\ket{\psi^{\prime}}=\tilde{\alpha}_0 \ket{a} + \tilde{\beta}_0 \ket{b}$, where $\tilde{\alpha}_0, \tilde{\beta}_0$ are the modified amplitudes and $\phi$ is the phase of a reference oscillator. Ramsey's arrangement enables the determination of the quantity $|\tilde{\alpha}_0 \tilde{\beta}_0^{*}|\cos(\theta + \phi - \omega t_0)$, where $\theta={\rm arg}(\tilde{\alpha}_0 \tilde{\beta}_0^{*})={\rm arg}(\alpha_0 \beta_0^*)$, and $\omega t_0$ is linked to the transit time between the two fields. To make the connection, the authors note that the two atomic states (ground and excited) play the role of the interferometer arms, while the resonator is to be prepared in a Fock state with $N$ photons~\cite{krause1987state}. A generalized kinematic approach for the computation of the geometric phases\index{Geometric phase} acquired in both unitary and dissipative JC models\index{Dissipative! JC model} has been recently developed in~\cite{Viotti2022}.


\subsubsection{Squeezing}\label{sssec:squeez}\index{Squeezing}
Squeezing~\cite{loudon1987squeezed,drummond2004quantum} of the electromagnetic-field fluctuations is defined as $\Delta x_\phi^2<1/2$ for some angle $0\leq\phi<2\pi$, where the variance $\Delta x_\phi$ was introduced in (\ref{quadvar2}). Thus, squeezing means suppressing quantum fluctuations of the field below the semiclassical limit of a coherent state for which $\Delta x_\phi^2=1/2,\,\,\forall\phi$. In this respect, squeezing is a quantum signature intimately related to antibunching\index{Antibunching} and sub-Poissonian\index{Sub-Poissonian} photon statistics~\cite{mandel1982squeezed,peverina1984relations}. As such, it has naturally been the subject of intense research in quantum optics in general and also within the realm of JC physics. Is is also interesting to note that the generation of fluctuation squeezing has been very recently associated with a self-gravitating quantum field (such as an axion\index{Axion}), an instance which may be used to probe whether dark matter\index{Dark! matter} can exhibit observable quantum characteristics~\cite{KoppFragkos2022}. 

The first demonstration of squeezing in the JC model is due to Meystre and Zubairy in the early 1980s~\cite{meystre1982squeezed}. They studied in particular the amount of squeezing as a function of the field amplitude for initial coherent states. Knight continued this analysis in~\cite{knight1986quantum}, while Kuklinski {\it et al.} performed a thorough numerical analysis to demonstrate that the amount of squeezing for coherent states can become substantial only for large initial field amplitudes~\cite{kuklinski1988strong} (see also ref.~\cite{hillery1989squeezing} for another early reference on the same topic). The question regarding higher-order squeezing in the JC model was addressed by Gerry and coworkers in ref.~\cite{gerry1988squeezing},who showed that it may indeed appear but not for general initial atomic states. Aravind {\it et al.} considered various initial atomic states (not only $|e\rangle$ or $|g\rangle$ as in the aforementioned references) and maintained that squeezing can be improved for certain initial conditions~\cite{aravind1988influence}. Using the large-field amplitude expansion~(\ref{expans}), valid for moderate times $t$, Gea-Banacloche provided an analytical expression for the maximum amount of squeezing in~\cite{gea1991atom,woods1993squeezing}. Subsequently, Gerry {\it et al.} proposed projective atomic measurements in order to improve the amount of squeezing~\cite{gerry1997squeezing}.

\begin{figure}
\includegraphics[width=10cm]{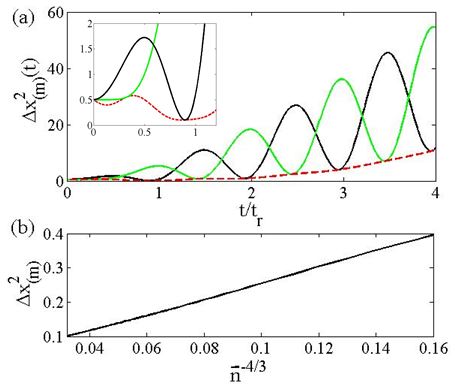} 
\caption{The plot in {\bf (a)} shows the squared variances of the $x$ (black solid line) and $p$ (green solid line) quadratures, together with the maximum squeezing $\Delta x_{(m)}^2(t)$ (red dashed line). The inset is the same but for short times. The state is initially $|\psi(0)\rangle=|\alpha\rangle|+\rangle$ with $\alpha=10$. The plot in {\bf (b)} gives the maximum attainable squeezing, {\it i.e.} $\min_t\Delta x_{(m)}^2(t)$, for the same type of initial state but varying the amplitude. It is plotted towards $\bar{n}^{-4/3}$ in order to reproduce a straight line. The rest of the parameters are $g=1$ and $\Delta=0$.
}
\label{fig8}
\end{figure}

Figure~\ref{fig4} illustrated the electromagnetic field phase-space distributions\index{Distribution! phase-space}\index{Phase-space distribution} at various times when the atom was initially excited and the field was in a coherent state. According to fig.~\ref{fig5}, we know that the splitting of the phase space distribution can be understood by expressing the atomic state in terms of the two atomic dipole states $|\pm\rangle$. It is clear from fig.~\ref{fig5} (a) and (b) that for large squeezing in the $x_0$ quadrature one should not start with the atom in $|e\rangle$ or $|g\rangle$ but instead in either of $|\pm\rangle$. If the atom is initialized in say $|+\rangle$, in the large field limit the distribution will not split up but will stay localized for time-scales $t\sim t_\mathrm{r}$. At $t~\sim t_\mathrm{r}/2$ squeezing in the $p_0$ quadrature is obtained, while for $t~\sim t_\mathrm{r}$ the squeezing occurs in the $x_0$ quadrature instead. The tilt of the distributions at say $t_\mathrm{r}/2$ implies that maximum squeezing is not exactly obtained for $\phi=\pi/2$. Figure~\ref{fig8} (a) displays $\Delta x^2(t)$ (black line), $\Delta p^2(t)$ (green line), and $\Delta x_{(m)}^2(t)$ (red line), where the maximum squeezing is defined as
\begin{equation}\label{minsq}
\Delta x_{(m)}^2=\min_{\phi\in[0,2\pi)}\Delta x_\phi^2(t).
\end{equation} 
With a slight shift (see inset), maximum squeezing at the times $t=t_\mathrm{r}/2,\,t_\mathrm{r},\,3t_\mathrm{r}/2,\,...$ occurs approximately along either the $x$- or the $p$-direction. The squeezing angle that satisfies eq.~(\ref{minsq}) is to first order independent of the detuning $\Delta$ and grows as $\phi=\pi g\bar{n}$ with the average photon number, {\it i.e.} it does not share the $\sqrt{\bar{n}}$-dependence of the Rabi frequency $\Omega_n$ [see eq.~(\ref{eigv})]. As demonstrated in ref.~\cite{woods1993squeezing}, after a certain time $t_\mathrm{max}\sim\bar{n}^{3/4}/g$, squeezing is no longer present. The maximum attainable squeezing as a function of the average photon number is shown in fig.~\ref{fig8} (b). As for what happens at $t_\mathrm{max}$, this amount of squeezing shows a fractional power dependence~\cite{woods1993squeezing}.

Atomic squeezing in the JC model has also been considered in numerous works~\cite{li1989squeezing,hu1989dynamical,xie1996relationship,ramon1998collective,fang2000entropy}. The Heisenberg uncertainty relation reads $\Delta\hat{\sigma}_x^2\Delta\hat{\sigma}_y^2\geq\frac{1}{16}\langle\hat{\sigma}_z\rangle^2$. Spin squeezing is usually defined as for the boson field, namely if $\Delta\hat{\sigma}_i^2<\frac{1}{4}\langle\hat{\sigma}_z\rangle^2$ for either $i=x,\,y$ the state is said to be squeezed. Such definition directly generates some issues since $\langle\hat{\sigma}_z\rangle$ may vanish for states that are supposedly `squeezed', or even {\it coherent spin states}\index{Atomic! coherent states} may appear to be squeezed if the basis is appropriately selected~\cite{kitagawa1993squeezed}. In order to overcome such subtlety one has to assume that the spin has a non-zero mean $\langle\hat{\sigma}_z\rangle$. Other ideas to circumvent this issue rely on entropic properties and has been applied to the JC model~\cite{fang2000entropy}. This entropic approach led to somewhat contradicting results when compared to the predictions of the Heisenberg uncertainty relation. More precisely, the definition relying on atomic entropy may signal the presence of squeezing, while the definition based on the uncertainty relation predicts a vanishing squeezing. Comparing atomic and field squeezing in the JC model showed that there is indeed some correlation between the two; following the above definition it may happen, however, that the boson field is squeezed while the atomic state is not~\cite{xie1996relationship}. 


\subsection{Driven and open Jaynes-Cummings physics}
\label{ssec:drjc}
Historically, JC physics has played an important role in the coherent control of quantum systems and the early explorations of quantum information processing (QIP). Many of the breakthroughs are indeed related to results obtained from operating deep in the quantum regime where the system becomes extremely sensitive to dissipation and decoherence. As a result, the effect of losses in the JC model has been extensively studied. The driven JC model is also of great relevance for experiments of multiphoton quantum nonlinear optics\index{Multiphoton! quantum optics}, comprising an active research field. In this subsection we review some of the most important aspects of the open driven JC model. 

\subsubsection{Field or atom driving}\label{sssec:drivejc}
A classical drive (or pump) of the cavity field adds a term to the standard JC Hamiltonian~\cite{alsing1991spontaneous,raimond2006exploring}, which becomes\index{Jaynes-Cummings! driven}\index{Model! driven Jaynes-Cummings}
\begin{equation}\label{fdrive}
\hat{H}_\mathrm{fJC}'=\omega\hat{n}+\!\frac{\Omega}{2}\hat{\sigma}_z+g\!\left(\hat{a}^\dagger\hat{\sigma}_{-}+\hat{\sigma}_{+}\hat{a}\right)+\eta\!\left(\hat{a}^\dagger e^{i\omega_\mathrm{p}t}\!+\!\hat{a}e^{-i\omega_\mathrm{p}t}\right)\!,
\end{equation}
where $\omega_\mathrm{p}$\index{Driven! Jaynes-Cummings model} is the drive frequency and $\eta$ the drive amplitude. A standard procedure when analyzing time-periodic Hamiltonians is to employ {\it Floquet theory}~\cite{shirley1965solution}\index{Floquet theory}, but for most purposes involving the model~(\ref{fdrive}) this would be equivalent to killing a fly with an elephant gun since the Hamiltonian becomes time-independent in a rotating frame (in sec.~\ref{sssec:dicke} we will briefly discuss examples where the time-dependence of the periodic drive cannot be absorbed into a rotating frame\index{Rotating frame}). Namely, in the interaction picture with respect to $\omega_\mathrm{p}\hat{N}$ we have
\begin{equation}\label{fielddrive}
\hat{H}_\mathrm{fJC}=\delta_\mathrm{p}\hat{n}+\!\frac{\Delta_\mathrm{p}}{2}\hat{\sigma}_z+g\!\left(\hat{a}^\dagger\hat{\sigma}_{-}+\hat{\sigma}_{+}\hat{a}\right)+\eta\!\left(\hat{a}^\dagger+\hat{a}\right)\!,
\end{equation}
with the field-pump detuning $\delta_\mathrm{p}=\omega-\omega_\mathrm{p}$ and the atom-pump detuning $\Delta_\mathrm{p}=\Omega-\omega_\mathrm{p}$\index{Atom!-pump detuning}\index{Field-pump detuning}.

\noindent Driving the atom means that the two states $|e\rangle$ and $|g\rangle$ are coupled via an external field in the Hamiltonian
\begin{equation}
\hat{H}_\mathrm{aJC}'\!=\!\omega\hat{n}\!+\!\frac{\Omega}{2}\hat{\sigma}_z\!+\!g\!\left(\hat{a}^\dagger\hat{\sigma}_{-}\!+\!\hat{\sigma}_{+}\hat{a}\right)\!+\!\eta\!\left(\hat{\sigma}_{+} e^{i\omega_\mathrm{p}t}\!+\!\hat{\sigma}_{-}e^{-i\omega_\mathrm{p}t}\right)\!,
\end{equation}
which in the interaction picture becomes
\begin{equation}\label{atomdrive}
\hat{H}_\mathrm{aJC}=\delta_\mathrm{p}\hat{n}+\!\frac{\Delta_\mathrm{p}}{2}\hat{\sigma}_z+g\!\left(\hat{a}^\dagger\hat{\sigma}_{-}+\hat{\sigma}_{+}\hat{a}\right)+\eta\hat{\sigma}_x,
\end{equation}
with the same meaning of the two detunings. The last two terms of the above expression can be combined and written as $g\left[\left(\hat{a}^\dagger+\eta/g\right)\hat{\sigma}_{-}+\hat{\sigma}_{+}\left(\hat{a}+\eta/g\right)\right]$. Thus, the pump term `shifts' the field by $\eta/g$. Using the properties of the displacement operators\index{Displacement! operator}~\cite{glauber1963coherent}
\begin{equation}\label{dispop}
\hat{D}(\nu)=e^{\nu\hat{a}^\dagger-\nu^*\hat{a}},
\end{equation}
namely $\hat{D}(\nu)\hat{a}\hat{D}^\dagger(\nu)=\hat{a}+\nu^*$ and $\hat{D}(\nu)\hat{a}^\dagger\hat{D}^\dagger(\nu)=\hat{a}^\dagger+\nu$, it follows that $\hat{H}_\mathrm{fJC}$ and $\hat{H}_\mathrm{aJC}$ are unitary equivalent~\cite{alsing1991spontaneous}. Thus, the properties of the two possible driven JC models are most similar. The {\it quasi}energies and eigenstates of the resonantly-driven Hamiltonian~\eqref{fdrive} ($\delta_p=\Delta_p=0$), commonly known as the {\it dressed JC eigenstates}\index{Dressed JC eigenstates}, were derived in~\cite{alsing1992dynamic}. The {\it quasi}energy spectrum comprises a ground level $e_0=0$ and the doublets
\begin{equation}\label{eq:doubletsdrivenJC}
 e_{n,\pm}=\pm \sqrt{n}\,\hbar g [1-(2\eta/g)^2]^{3/4}, \quad \quad n=1,2,\ldots
\end{equation}
The splittings collapse\index{Spectral! collapse} to zero at $\eta=g/2$; beyond that point, the spectrum is continuous\index{Continuous spectrum}. The singularity locates a critical point\index{Critical! point}, which is the organizing center of a second-order dissipative quantum phase transition~\cite{CarmichaelPRX}\index{Phase transition! dissipative! quantum! second-order}\index{Dissipative! quantum phase transition}. We will have more to say on this topic in subsec~\ref{sssec:JCzerodim}.  

It is clear that for the driven systems, $\hat{N}$ is no longer a constant of motion, {\it i.e.}, the continuous $U(1)$ symmetry is broken. As a result, there is no reason to expect that the driven JC model is integrable in the general case of non-zero detunings (see, however, sec.~\ref{sssec:rabiint}). However, in the dispersive regime, as for the regular JC model, a Schrieffer-Wolff transformation\index{Schrieffer-Wolff transformation}~\cite{carbonaro1979canonical,klimov2002effective} diagonalizes the spin part of the Hamiltonian~\cite{bishop2010response}
\begin{equation}\label{drivedis}
\hat{H}_\mathrm{fJC'}=\delta_\mathrm{p}\hat{n}+\left(\frac{\Delta_\mathrm{p}}{2}+\frac{g^2\hat{n}}{\Delta_\mathrm{p}}\right)\hat{\sigma}_z+\eta\left(\hat{a}^\dagger+\hat{a}\right).
\end{equation}
When applying the Schrieffer-Wolff transformation to the driven JC model, care must be taken since the transformation does not normally commute with the drive terms~\cite{klimov2003master}. An alternative semiclassical approach in order to derive an effective model for the low-energy states relies on the so-called {\it Born-Oppenheimer approximation}\index{Born-Oppenheimer approximation} (BOA)~\cite{atkins1997molecular}. The idea of the BOA is that the spin precesses on a much faster timescale than the field, and can thereby be assumed to adiabatically follow the field  ({\it i.e.}, the spin degree of freedom can be adiabatically eliminated). For the JC model, the BOA approach was first considered in~\cite{larson2007dynamics,larson2012absence}, and it consists of finding the {\it adiabatic eigenstates}\index{Eigenstate! adiabatic} which diagonalize the two-level structure of the Hamiltonian. A more thorough analysis of the {\it cavity} BOA was presented in~\cite{flick2017cavity}, which we will return to in sec.~\ref{ssec:chem} when we consider molecules (instead of atoms) coupled to quantized cavity fields. The corresponding diagonalization matrix $\hat{U}(\hat{a},\hat{a}^\dagger)$ will typically not commute with neither $\hat{n}$ nor the pump term. As will be shown in sec.~\ref{ssec:chem}, this gives rise to {\it non-adiabatic corrections} (also referred to as {\it synthetic gauge potentials}~\cite{bohm2003geometric})\index{Gauge! potential! synthetic} which are typically neglected in the BOA. In the quadrature representation~(\ref{quad}) we see that the effect of the pumping is just to shift the harmonic oscillators by the term $\sqrt{2}\eta\hat{x}$. The BOA for the driven JC model has been considered in ref.~\cite{peano2010quasienergy} in order to study both static and dynamical properties. In secs.~\ref{ssec:rabi} and~\ref{sssec:dicke} we will describe in some detail how the BOA works when applied to the quantum Rabi and Dicke models respectively. 
 
In general, most of the explorations of the driven JC model have concentrated on dynamical aspects, especially on how an initial state evolves when the system is driven~\cite{moya1998long,jyotsna1993jaynes}. In particular, collapses and revivals of the mean photon number occur at a much longer time scale than the revival time for the two-state inversion~\cite{chough1996nonlinear}. The inherent nonlinearity of the JC model also implies that when the system is driven on resonance the amount of energy pumped into the system remains finite~\cite{larson2013chaos}. This nonlinearity gives rise to the onset {\it optical bistability} in certain regimes of operation~\cite{bonifacio1978optical,drummond1980quantum,gardiner2004quantum,armen2006low,walls2008quantum} at the single atom level~\cite{savage1988single,alsing1991spontaneous,kilin1991single,bishop2010response,peano2010dynamical, CarmichaelPRX}. Optical bistability, see sec.~\ref{sssec:optbis}, manifests itself as a hysteresis effect when the pump amplitude is adiabatically ramped up or down. In cavity QED conditions it has been observed for the first time in the group of Kimble where, however, the nonlinearity due to the presence of a single atom was not strong enough to trigger hysteresis~\cite{rempe1991optical}. We will briefly revisit bistability in its many-atom version in subsec.~\ref{ssec:quantumglcavQED}.

By driving the JC model it is also possible to create bespoke interactions such as that of the {\it anti-JC}\index{Anti-Jaynes-Cummings model}\index{Model! anti-Jaynes-Cummings}~\cite{solano2003strong,ballester2012quantum} (see sec.~\ref{sec:ion} and next section on the quantum Rabi model). To see this we consider the atom-driven Hamiltonian~(\ref{atomdrive}), and take $\Delta_\mathrm{p}=0$. We then turn to an interaction picture with respect to $\delta_\mathrm{p}\hat n+\eta\hat\sigma_x$, giving us
\begin{equation}\label{drham}
\hat H_\mathrm{int}=\frac{g}{2}\left(|+\rangle\langle+|+|-\rangle\langle-|+e^{2i\eta t}|+\rangle\langle-|-e^{-2i\eta t}|-\rangle\langle+|\right)\hat ae^{-i\delta_\mathrm{p}t}+\text{h.c.},
\end{equation}
where we have expressed the Hamiltonian in the atomic dipole basis\index{State! dipole}~(\ref{dipol}). Provided that $\eta\gg g,\delta_\mathrm{p}$ we can perform a RWA and drop the rapidly oscillating terms to arrive at
\begin{equation}\label{solanoeff}
\hat H_\mathrm{int}=\frac{g}{2}\left(\hat a e^{-i\delta_\mathrm{p}t}+\hat a^\dagger e^{i\delta_\mathrm{p}t}\right)\hat\sigma_x.
\end{equation}
We will have reasons to come back to this effective Hamiltonian in later sections. In sec.~\ref{ssec:cqedstateprep}, it is shown how this type of effective Hamiltonian can be utilized to prepare exotic states like Schr\"odinger cat and entangled states, and in sec.~\ref{ssec:cqedQI} it will be deployed for QIP in cavity QED setups.

Let us pause here to link the characteristic collapses and revivals in the JC model to a nonlinear oscillator behaviour for its driven version, following closely the analysis of~\cite{chough1996nonlinear} (reinstating $\hbar$). We consider a two-state atom driven on resonance by an external field and coupled to a resonant cavity mode. The Hamiltonian in the interaction picture reads
\begin{equation}
\hat{H}=i g (\hat{\sigma}_{+}\hat{a}-\hat{\sigma}_{-}\hat{a}^{\dagger}) + i\eta (\hat{\sigma}_{+}-\hat{\sigma}_{-}),
\end{equation}
where $\eta$ is the real amplitude of the external field. Introducing the displacement\index{Operator! displacement}\index{Displacement! operator} 
\begin{equation}\label{eq:disptran}
\hat{D}^{\dagger}(\alpha) \hat{a} \hat{D}(\alpha)=\hat{a} + \alpha, \quad \quad \alpha\equiv\eta/g,    
\end{equation}
with $\hat{D}(\alpha) \equiv \exp[\alpha(\hat{a}-\hat{a}^{\dagger})]$, we obtain
\begin{equation}
\hat{\tilde{H}} \equiv \hat{D}^{\dagger}(\alpha) \hat{H} \hat{D}(\alpha)= i\hbar g (\hat{\sigma}_{+}\hat{a}-\hat{\sigma}_{-}\hat{a}^{\dagger}).
\end{equation}
We then solve the corresponding Schr\"{o}dinger equation,
\begin{equation}\label{eq:HtildenonlinJC}
 i\hbar \frac{d}{dt} \ket{\tilde{\psi}}=\hat{\tilde{H}}   \ket{\tilde{\psi}},\quad\quad  \ket{\tilde{\psi}} \equiv \hat{D}^{\dagger}(\alpha) \ket{\psi},
\end{equation}
with initial state $\ket{\psi(0)}=\ket{0}\ket{+}$, transformed to $\ket{\tilde{\psi}(0)}=\hat{D}(\alpha)\ket{\psi(0)}=\ket{\alpha}\ket{+}$ (where $\ket{\alpha}=\hat{D}(\alpha)\ket{0}$ is a coherent state of amplitude $\alpha$). If we denote $P_{+}$ the occupation probability of the state $\ket{+}$, then eq. \eqref{eq:HtildenonlinJC} preserves the sum $\braket{\tilde{\psi}|\hat{a}^{\dagger}\hat{a}|\tilde{\psi}} + P_{+}$. Hence, from eq. \eqref{eq:disptran} the average photon number can be written as
\begin{equation}\label{eq:meanph1}
  \braket{\hat{a}^{\dagger}\hat{a}} \equiv \braket{\psi|\hat{a}^{\dagger}\hat{a}|\psi}=\braket{\tilde{\psi}|(\hat{a}^{\dagger}-\alpha)(\hat{a}-\alpha)|\tilde{\psi}}=2\alpha\left(\alpha-\langle\tilde{\psi}|\hat{a}|\tilde{\psi}\rangle\right) +1 - P_{+}.  
\end{equation}
The standard JC model yields the solution
\begin{equation}\label{eq:meanph2}
   \braket{\tilde{\psi}|\hat{a}|\tilde{\psi}}=\sum_{n=1}^{\infty} C_n C_{n-1} \left\{\frac{\sqrt{n+1}+\sqrt{n}}{2} \cos[(\sqrt{n+1}-\sqrt{n})gt] - \frac{\sqrt{n+1}-\sqrt{n}}{2} \cos[(\sqrt{n+1}+\sqrt{n})gt] \right\} 
\end{equation}
and
\begin{equation}\label{eq:meanph3}
P_{+}=\frac{1}{2} \left[1  + \sum_{n=1}^{\infty} C_{n-1}^2\cos(2\sqrt{n}\, gt) \right],   
\end{equation}
where the constants 
\begin{equation}\label{eq:meanph4}
C_n = e^{-\alpha^2/2}\frac{\alpha^n}{\sqrt{n!}}    
\end{equation}
are the familiar expansion coefficients of the coherent state $\ket{\alpha}$ in the $\ket{n}$ representation. 

Like in sec.~\ref{sssec:crsubsec}, we now expand the difference-frequency term about the mean photon number as
\begin{equation}\label{eq:seriesa}
\sqrt{n+1}-\sqrt{n}=\frac{1}{2\alpha}-\frac{2k+1}{\alpha^3}+\ldots,    
\end{equation}
where $k=n-\alpha^2$ is an integer. Keeping only the first term in this series, effecting the approximations $(\sqrt{n+1}+\sqrt{n})/2 \approx \sqrt{n}$, $C_n C_{n-1} \approx C_n^2$, and replacing $P_{+}$ in eq. \eqref{eq:meanph3} by its time average, yields 
\begin{equation}\label{eq:mainosc}
\braket{\hat{a}^{\dagger}\hat{a}} \approx 2\alpha^2\{1-\cos[gt/(2\alpha)]\} + 1/2,    
\end{equation}
with a period of oscillation given by
\begin{equation}
2g T_{\rm inv}=4\pi \alpha,    
\end{equation}
where $T_{\rm inv}$ is the revival time for the two-state inversion. Keeping the second term in the series \eqref{eq:seriesa} accounts for the collapse and revival of the oscillation defined by eq. \eqref{eq:mainosc}. The collapse takes place when the frequency component located one standard deviation away [coefficient $C_n^2(k=\alpha)$] dephases from the central component [coefficient $C_n^2(k=0)$]. The collapse time is then\index{Collapse time}
\begin{equation}
g T_{\rm c}=4\pi \alpha^2\Rightarrow T_{\rm c}=\frac{4\pi\eta^2}{g^3}.    
\end{equation}
At the photon number revival all frequency components return back to phase, which occurs at\index{Revival time}
\begin{equation}
gT_{\rm r}=8\pi \alpha^3\Rightarrow T_{\rm r}=\frac{8\pi\eta^3}{g^4}. 
\end{equation}
Compared to the corresponding results for the nondriven JC model, eqs.~(\ref{ctime}) and~(\ref{rtime}), we see how the collapse/revival times now scale with different powers of the coupling strength $g$. 

\begin{figure}
\includegraphics[width=12cm]{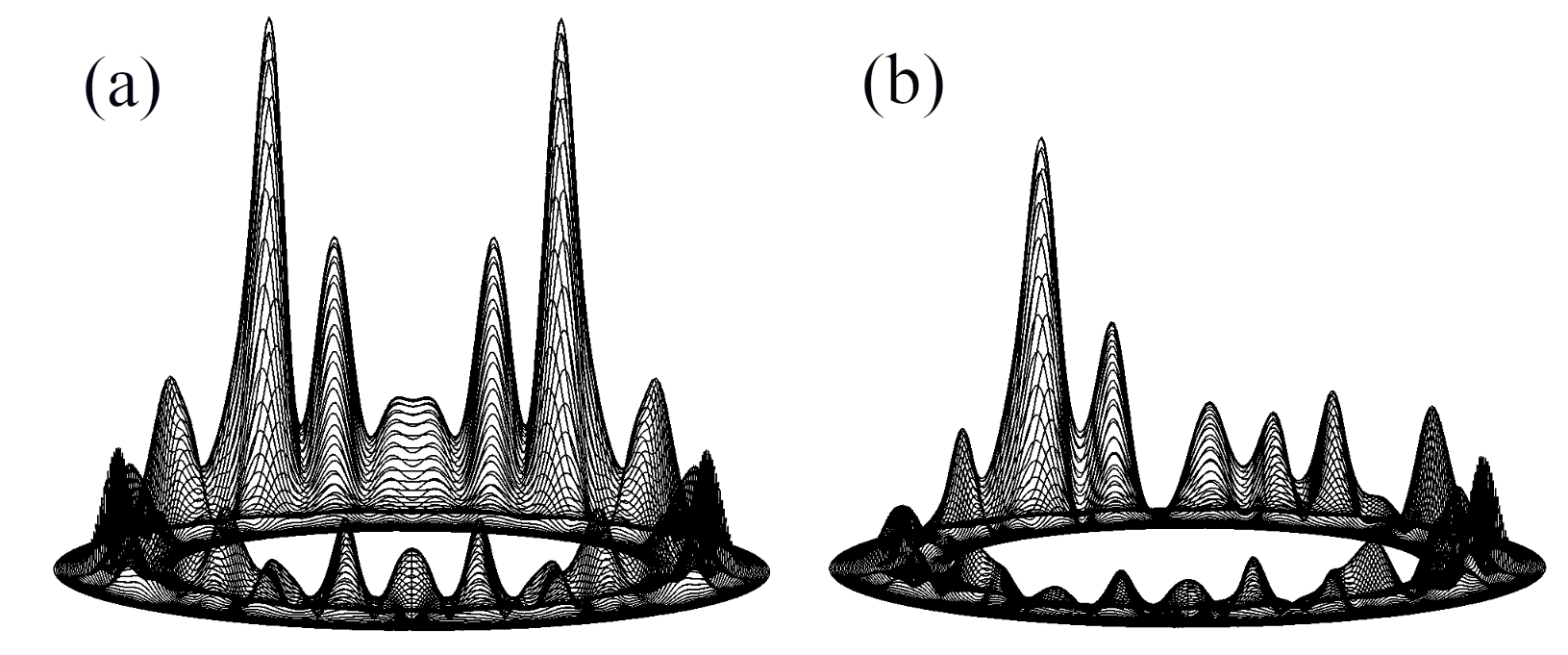} 
\caption{The {\it quasi}probability distribution $Q(\tilde{z})$ (in the high excitation approximation) calculated for the time when the photon number attains its maximum in the course of the first revival, for $\alpha=15$ and an initial Gaussian distribution\index{Gaussian! distribution} peaking at the front center of the plot. {\bf (a)} The symmetric distribution obtained from the sum over the two orthogonal anharmonic oscillator modes. {\bf (b)} The contribution from the $\hat{A}_u$ mode, rotating clockwise (while the distribution corresponding to $\hat{A}_l$ rotates anti-clockwise). (Source: Fig. 3  of~\cite{chough1996nonlinear}). Reproduced with permission from the author and the APS.}
\label{fig:Qanh}
\end{figure}

As pointed out in~\cite{chough1996nonlinear}, the solution for the mean photon number given by eqs. \eqref{eq:meanph1}-\eqref{eq:meanph4} and the associated wavepacket dynamics (see~\cite{robinett2004quantum} for a review), suggest an underlying anharmonic oscillator for the electromagnetic field instead of the two-state oscillator for which the JC model is most commonly known\index{Nonlinear oscillator model}. To clarify this point, we begin with the eigenstates of the Hamiltonian $\hat{\tilde{H}} + \hbar\omega_0 (\hat{a}^{\dagger}\hat{a} + \hat{\sigma}_z/2 + 1/2)$, which are the ground state $\ket{G}=\ket{0}\ket{-}$, with energy $E_g=0$, and the excited-state doublets
\begin{equation}
\begin{aligned}
\ket{u_n}&=(\ket{n-1}\ket{+}-i\ket{n}\ket{-})/\sqrt{2},\\
\ket{l_n}&=(\ket{n-1}\ket{+}+i\ket{n}\ket{-})/\sqrt{2},
\end{aligned}    
\end{equation}
for positive integers $n$ and with energies $E_{u_n}=n\hbar \omega_0 + \hbar \sqrt{n} g$ and $E_{l_n}=n\hbar \omega_0 - \hbar \sqrt{n} g$. Considered on their own, each of these two energy ladders\index{Jaynes-Cummings! ladder}, both referred to the same ground state, defines an anharmonic oscillator, where the anharmonicity enters as an energy-dependent frequency shift\index{Energy-dependent frequency shift}. For highly excited states ($n \gg 1$), this shift is inversely proportional to the square root of the energy,
\begin{equation}\label{eq:sqrtnshift}
(E_{u_n, l_n} - E_{u_{n-1},l_{n-1}})/\hbar - \omega_0 \approx \pm g/(2\sqrt{n}).   
\end{equation}
To construct quantized anharmonic oscillators, we rewrite the Hamiltonian of eq. \eqref{eq:HtildenonlinJC} in the energy representation,
\begin{equation}\label{eq:Htran1}
\tilde{H}=\hbar g \sum_{n=1}^{\infty} \sqrt{n} (|u_n\rangle \langle u_n| - |l_n\rangle \langle l_n|), \end{equation}
and we introduce the ladder operators
\begin{equation}
\hat{A}_{\lambda}=|g\rangle \langle \lambda_1| + \sum_{n=2}^{\infty}\sqrt{n}  |\lambda_{n-1}\rangle \langle \lambda_n|,   
\end{equation}
with $\lambda=u,l$. The two ladders share the same ground state, as reflected in the commutator
\begin{equation}
[\hat{A}_{\lambda}, \hat{A}_{\lambda}^{\dagger}]=\hat{P}_{\lambda} + |g\rangle \langle g|, \quad \quad \hat{P}_{\lambda}=\sum_{n=1}^{\infty}|\lambda_n\rangle \langle \lambda_n|.   
\end{equation}
The Hamiltonian of eq. \eqref{eq:Htran1} can now be written in terms of the newly defined ladder operators, and, in particular, in terms of the square root of the number operator which count the excitations on the two JC ladders:
\begin{equation}
\tilde{H}=\hbar g \left(\sqrt{\hat{A}^{\dagger}_{u}\hat{A}_{u}}-\sqrt{\hat{A}^{\dagger}_{l}\hat{A}_{l}}\right)    
\end{equation}
It is interesting to note that $\hat{A}^{\dagger}_{\lambda}$ and $\hat{A}_{\lambda}$ create and annihilate entangled excitations of the electromagnetic field coupled to the two-state atom, rather than single electromagnetic quanta. The system state can then be written as a sum of two orthogonal states of almost orthogonal anharmonic oscillator modes:
\begin{equation}
\ket{\tilde{\psi}(t)}=[\ket{\tilde{U}(t)} + \ket{\tilde{L}(t)}]/\sqrt{2},    
\end{equation}
with 
\begin{equation}
\ket{\tilde{\Lambda}(t)}=\sqrt{2} \exp\left[\mp i gt \sqrt{\hat{A}^{\dagger}_{\lambda}\hat{A}_{\lambda}}\right]\hat{P}_{\lambda} \ket{\alpha}\ket{+},   
\end{equation}
for $\Lambda=U,L$. For highly excited states, the two anharmonic oscillator modes can be considered orthogonal, and $\hat{a}=\hat{A}_u + \hat{A}_l$. The coherent states of $\hat{A}_{\lambda}$, denoted by $\ket{\tilde{z}_{\lambda}}_{\lambda}$, are states of the electromagnetic field and the two-state atom. To demonstrate the potential of this new model to produce a reconstruction of the quantum state during revival, in fig.~\ref{fig:Qanh} we plot the phase-space distribution\index{Phase-space distribution} function [frame (a)] 
\begin{equation}
Q(\tilde{z})=\frac{1}{2}\left(|\braket{\tilde{U}|\tilde{z}}_u|^2 + |\braket{\tilde{L}|\tilde{z}}_l|^2\right),
\end{equation}
at the time of the first revival, isolating explicitly the contribution of the first term in the sum, in frame (b). This is an exemplary manifestation of dispersion in the JC model; the Gaussian wave-packets\index{Gaussian! wavepacket} spread over time interfering with one another after running around a circle in phase space. The distribution depicted in frame (a) is indistinguishable from the exact $Q$ function\index{Q distribution}\index{Distribution! Q} of the intracavity field on the scale of the figure. The partial reconstruction, which lacks the symmetry of the initial Gaussian state\index{Gaussian! state}, is primarily a result of neglecting higher-order terms when making the approximation \eqref{eq:seriesa} to estimate the revival time. These terms also restrict the revival of the photon number oscillation to about a third of its initial magnitude (see fig. 1 of~\cite{chough1996nonlinear}).

The excited states of the JC Hamiltonian can be accessed by driving the cavity mode coupled to the two-level atom with broadband chaotic light. For a sufficiently large dipole coupling strength the spectrum of the scattered light exhibits peaks at the many resonance frequencies of the JC Hamiltonian. A similar multipeaked spectrum is observed in transmission. The spectrum of the light reflected from the cavity shows corresponding absorption dips superimposed onto the broadband input~\cite{TianProc90}.


\subsubsection{Applying the open systems formalism}\label{sssec:openjc}
In typical quantum optical experiments, the characteristic frequencies of the system and their baths/reservoirs are often such that the mechanism behind losses, either {\it dissipation}\index{Dissipation} (energy loss) or {\it decoherence} (loss of quantum coherence)\index{Decoherence}, can be assumed to be {\it Markovian}~\cite{breuer2002theory}. Even if defining {\it non-Markovianity} in quantum mechanical systems is far from trivial~\cite{breuer2009measure}, in general terms Markovianity can be thought of as if any information about the system that has been extracted by the bath will {\it not} `flow' back into the system. The {\it Markovian approximation}\index{Markovian approximation} thereby relies on rapid decay of correlations within the bath compared to any time-scale of the system evolution. To treat such Markovian system-bath evolution for the JC model one typically considers a {\it master equation}\index{Master equation} approach (Schr\"odinger picture) or a {\it Heisenberg-Langevin approach} (Heisenberg picture). Which method that is preferable depends on the question asked, but in terms of number of works, the former seems somewhat more popular in the JC literature. However, when analyzing the light leaking the cavity the {\it input-output theory}\index{Input-output! theory}, as an application of the Heisenberg-Langevin method\index{Heisenberg-Langevin equations}, has been developed~\cite{walls2007quantum}. We note here in passing that the {\it transient spectrum}\index{Spectrum!transient} in a cavity QED setup is found by generating a sequence of control pulses, which induce non-Markovian coherence revivals, with an envelope that can be determined by a phase-sensitive measurement of the output field~\cite{McIntyre2022}. Of significance is also the so-called {\it time-resolved physical spectrum}\index{Spectrum!time-resolved, physical}, which brings forward the role of causality when measuring the transient magnitude of characteristic cavity QED features, like the Rabi doublet~\cite{Yamaguchi2022}.  

\begin{figure}
\includegraphics[width=8cm]{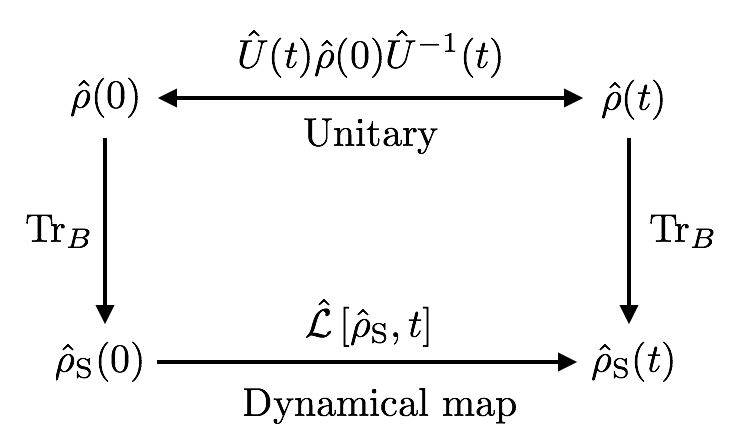} 
\caption{The general idea behind the Lindblad master equation\index{Master equation}. The system state $\hat{\rho}(t)$ is obtained by tracing out the bath degrees of freedom from the full system$+$bath density operator $\hat{\rho}_\mathrm{tot}(t)$ (vertical arrows). Instead of solving the full (unitary) dynamics of the system$+$bath, the {\it dynamical map} $\hat{\mathcal{L}}[\hat{\rho},t]$ evolves the system state (normally non-unitarily). Losses due to the bath implies that the time-evolution under the dynamical map is irreversible and the set of Liouvillian operators $\hat{\mathcal{L}}$ therefore forms a semi-group and not a group as for the unitary time-evolution operators $\hat{U}$~\cite{lindblad1976generators}. In the figure, the irreversibility can be seen as stemming from the partial trace over the bath.  
}
\label{figdyn}
\end{figure}

In order to derive a master equation\index{Master equation} for the system state $\hat{\rho}(t)$, the system-bath coupling is assumed `weak' (in the sense that the reservoir is only slightly affected by the system). In the resulting {\it Born approximation}\index{Born approximation}, system-bath entanglement is neglected. The `size' of the bath is also assumed to be much larger than that of the system such that its state is minimally affected by the system. The final approximation before arriving at a master equation\index{Master equation} on the {\it Lindblad form}~\cite{lindblad1976generators} a RWA is imposed (in this community it is called {\it secular approximation}\index{Secular equation}). The general Lindblad master equation\index{Master equation} reads
\begin{equation}\label{lindblad}
\partial_t\hat{\rho}(t)=\hat{\mathcal{L}}\hat{\rho}(t)=i\left[\hat{\rho}(t),\hat{H}\right]+\sum_{j=1}^J\hat{\mathcal{D}}[\hat{L}_j]\hat{\rho}(t),
\end{equation}
where the {\it super-operator}\index{Super-operator} $\hat{\mathcal{L}}$ is the {\it Liouvillian}\index{Liouvillian! super-operator} and the {\it Lindblad super-operator}\index{Lindblad! super-operator}
\begin{equation}
\hat{\mathcal{D}}[\hat{L}]\hat{\rho}=\hat{L}\hat{\rho}\hat{L}^\dagger-\frac{1}{2}\left(\hat{L}^\dagger\hat{L}\hat{\rho}+\hat{\rho}\hat{L}^\dagger\hat{L}\right),
\end{equation}
with {\it jump operator}\index{Jump operator} $\hat{L}_j$, in principle arbitrary operators acting on the system's degrees of freedom. In a strict mathematical sense, $\sum_{j=1}^J\hat{L}_j^\dagger\hat{L}_j$ should be a bounded operator in order to preserve positivity and trace norm of $\hat{\rho}$~\cite{lindblad1976generators}. This constraint is in general not crucial, and we will indeed consider unbounded operators. $\hat{H}$ in (\ref{lindblad}) is the system Hamiltonian (more precisely, the Hamiltonian part should also contain a {\it Lamb shift}\index{Lamb! shift} term~\cite{breuer2002theory}, which, however, is usually negligible), and thus, the term $\hat{\mathcal{D}}[\hat{L}_j]$ accounts for irreversible dissipation and decoherence. For a single qubit, for example, described by the Hamiltonian $\hat H_\mathrm{qu}=\frac{\Omega}{2}\hat\sigma_z$, coupled to a zero temperature bath of harmonic oscillators we would find the master equation\index{Master equation}
\begin{equation}\label{openqubit}
\partial_t\hat\rho_\mathrm{qu}(t)=i\left[\hat\rho_\mathrm{qu}(t),\frac{\Omega}{2}\hat\sigma_z\right]+\gamma\left(2\hat{\sigma}_{-}\hat\rho_\mathrm{qu}(t)\hat{\sigma}_{+}-\hat{\sigma}_{+}\hat{\sigma}_{-}\hat\rho_\mathrm{qu}(t)-\hat\rho_\mathrm{qu}(t)\hat{\sigma}_{+}\hat{\sigma}_{-}\right).
\end{equation}
Using the parametrization~(\ref{bloch2}) for the density operator and substituting into the above equation we derive the corresponding Bloch equations\index{Bloch! equations}~\cite{breuer2002theory},
\begin{equation}\label{eq:BlochNH}
\begin{array}{l}
\displaystyle{\frac{d}{dt}R_x(t)=-\frac{\gamma}{2}R_x(t)},\\ \\
\displaystyle{\frac{d}{dt}R_y(t)=-\frac{\gamma}{2}R_y(t)},\\ \\
\displaystyle{\frac{d}{dt}R_z(t)=-\gamma R_z(t)}+\gamma.
\end{array}
\end{equation}
A schematic picture of the irreversible time-evolution generated by $\hat{\mathcal{L}}$ is presented in fig.~\ref{figdyn}. 

In the limit $g\ll\Omega,\,\omega$, cavity losses are typically modeled by the phenomenological master equation\index{Master equation}~\cite{barnett1986dissipation,quang1991quantum,alsing1991spontaneous,eiselt1991quasiprobability,daeubler1992analytic,gea1993jaynes,van1997exact,de1999dissipative}\index{Open Jaynes-Cummings model}\index{Model! open Jaynes-Cummings}
\begin{equation}\label{master1}
\partial_t\hat{\rho}(t)= i\left[\hat{\rho}(t),\hat{H}_\mathrm{JC}\right]+\displaystyle{\frac{\kappa}{2}n_\mathrm{th}\left(2\hat{a}^\dagger\hat{\rho}(t)\hat{a}-\hat{a}\hat{a}^\dagger\hat{\rho}(t)-\hat{\rho}(t)\hat{a}\hat{a}^\dagger\right)}+\displaystyle{\frac{\kappa}{2}(n_\mathrm{th}+1)\left(2\hat{a}\hat{\rho}(t)\hat{a}^\dagger-\hat{a}^\dagger\hat{a}\hat{\rho}(t)-\hat{\rho}(t)\hat{a}^\dagger\hat{a}\right)},
\end{equation}
where the reservoir is assumed to be thermal (at temperature $T$)\index{Thermal state}\index{State! thermal} with an average number of photons $n_\mathrm{th}=1/\left(e^{\omega/T}-1\right)$, ($k_\mathrm{B}=1$). For optical photons, one can often assume $T=0$, {\it i.e.} $n_\mathrm{th}=0$ and only the last term survives (no thermal photons enter the cavity). In this case, the state $\hat{\rho}=|g,0\rangle\langle g,0|$ is a steady state\index{Steady state} of the master equation\index{Master equation}. Moreover, since $\hat{L}|g,0\rangle=0$, or $\hat{\mathcal{L}}\hat\rho=0$, the state is called a {\it dark state}\index{Dark! state}\index{State! dark}~\cite{diehl2008quantum,bardyn2013topology} or {\it decoherence-free state}\index{Decoherence-free! state}\index{State! decoherence-free}~\cite{lidar1998decoherence}, {\it i.e.} it is transparent for the decoherence. In the long time limit of the JC model coupled to a zero temperature photon bath, the system state will approach $|g,0\rangle$ regardless of the initial condition. This reflects the irreversible evolution; the steady state\index{Steady state!unique} is unique such that it is impossible to reverse time and regain the initial state.

As a demonstrative example of how the bath affects the time-evolution we consider a cat state~(\ref{catstate})\index{Cat state}\index{Cat state} which, of course, is extremely sensitive to decoherence. Due to the {\it fluctuation-dissipation theorem}, we expect photon losses to inevitably imply some sort of decoherence (fluctuations)~\cite{mandel1995optical}. According to eq.~(\ref{master1}), a harmonic oscillator coupled to a zero temperature boson bath evolves as
\begin{equation}
\partial_t\hat\rho(t) = -i\left[\hat\rho(t),\omega\hat a^\dagger\hat a\right] + 
\kappa\left(2\hat a\hat\rho(t)\hat a^\dagger - \hat a^\dagger\hat a\hat\rho(t) - 
\hat\rho\hat a^\dagger\hat a\right).
\end{equation}
For an initial {\it even cat state} (see sec.~\ref{sssec:subscat})
\begin{equation}\label{eq:evencats}
|\psi\rangle=\frac{1}{\sqrt{N}}(|\alpha\rangle+|-\alpha\rangle),
\end{equation}
with $N=\sqrt{2\left(1+\mathrm{Re}\left[e^{i\phi}\langle\alpha|-\alpha\rangle\right]\right)}$ the normalization constant, the density matrix is
\begin{equation}
\hat\rho(0) = \frac{1}{N^2}\left(|\alpha\rangle\langle\alpha| + |-\alpha\rangle\langle-\alpha| + 
|\alpha\rangle\langle-\alpha|+|-\alpha\rangle\langle\alpha|\right).
\end{equation}
The time-evolved state becomes~\cite{walls1985effect,phoenix1990wave}
\begin{equation}\label{catdec}
\hat\rho(t)=\frac{1}{N(t)} \big( |\alpha(t)\rangle \langle \alpha(t)| + 
|-\alpha(t)\rangle \langle - \alpha(t)|+ C(t)|\alpha(t)\rangle \langle - \alpha(t)|  + C(t)| -\alpha(t) \rangle \langle \alpha(t)| \big),
\end{equation}
where $\alpha(t)=\alpha e^{-i\omega t-\kappa t}$, and the factor multiplying the coherence terms reads 
\begin{equation}\label{cohfac}
C(t)=\exp\left(-2|\alpha|^2(1-e^{-2\kappa t})\right)
\end{equation}
if we assume $\alpha$ to be real. For $\kappa t \ll 1$, we have that the coherence terms 
vanish as $C(t)\approx e^{-2\kappa|\alpha|^2t}$. By noting that $2|\alpha|^2$ is the 
``distance'' between the two coherent states that form the cat state we see how the 
cat becomes extremely fragile to decoherence the larger the cat is (the cat state size 
is the number of bosons $n=|\alpha|^2$). Thus, after an exponentially short time the 
cat has turned into a statistical mixture\index{Statistical mixture}
\begin{equation}\label{catmix}
\hat\rho=\frac{1}{N^2}\left(|\alpha\rangle\langle\alpha|+|-\alpha\rangle\langle-\alpha|\right).
\end{equation}

An insight to these results is provided by a particularly illuminating application of quantum-trajectory theory\index{Quantum! trajectories}, which has to do with extracting photon-counting records from a damped coherent state~\cite{OpenQuantCh4}. Working in the interaction picture, a damped resonator is described by the jump operator\index{Jump operator} $\hat{J}=\sqrt{2\kappa}\,\hat{a}$ and the non-Hermitian Hamiltonian\index{Non-Hermitian! Hamiltonian} $\hat{H}=-i\hbar\kappa \hat{a}^{\dagger}\hat{a}$. The initial conditional state is $\ket{\overline{\psi}_{\rm REC}(0)}=\ket{\alpha}$, while we adopt the {\it ansatz}
\begin{equation}
 \ket{\overline{\psi}_{\rm REC}(t)}=A(t)\ket{\alpha(t)},
\end{equation}
with
\begin{equation}
 \ket{\alpha(t)}=\exp[\alpha(t)a^{\dagger}-\alpha^{*}(t)a]\ket{0}.
\end{equation}
The norm $\braket{\overline{\psi}_{\rm REC}(t)|\overline{\psi}_{\rm REC}(t)}=|A(t)|^2$ yields the record probability density. By solving the trajectory equations, we solve not only the master equation\index{Master equation} but a photon counting problem as well. 

We consider a record with $n$ counts up to time $t$. The evolution of the conditional state consists of a sequence of jumps at the ordered sequence of count times $t_1, t_2, \ldots t_k$, with a continuous evolution between these jumps. A jump at time $t_k$ preserves the {\it ansatz} while it changes the norm as 
\begin{equation}
 A(t_k) \to \sqrt{2\kappa}\,\alpha(t_k)A(t_k).
\end{equation}
The time evolution between the jumps must satisfy the equation of motion
\begin{equation*}\label{eq:eom}
\frac{d\ket{\overline{\psi}_{\rm REC}(t)}}{dt}=-(\kappa \hat{a}^{\dagger}\hat{a}) \ket{\overline{\psi}_{\rm REC}(t)}.
\end{equation*}
It preserves the {\it ansatz} too provided $\alpha(t)$ and $A(t)$ satisfy a set of equations. To determine them we substitute the conditional state in both sides of~\eqref{eq:eom}, 
\begin{equation}
 \frac{dA(t)}{dt}\ket{\alpha(t)} + A(t) \left(\frac{d\alpha(t)}{dt}\hat{a}^{\dagger}-\frac{d\alpha^{*}(t)}{dt}\alpha(t)\right)\ket{\alpha(t)}=-\kappa \hat{a}^{\dagger}\alpha(t) A(t) \ket{\alpha(t)}.
\end{equation}
Equating the two sides yields
\begin{equation}
\frac{d\alpha(t)}{dt}=-\kappa \alpha(t) \quad \text{and} \quad \frac{1}{A(t)}\frac{dA(t)}{dt}= \frac{d\alpha^{*}(t)}{dt}\alpha(t)=(-\kappa |\alpha(t)|^2)=\frac{d}{dt}\left(\tfrac{1}{2}|\alpha(t)|^2\right),
\end{equation}
with solutions $\alpha(t)=\alpha \exp(-\kappa t)$ and $A(t)=A(t_k)\exp[-\tfrac{1}{2}|\alpha|^2 (e^{-2\kappa t_k}-e^{-2\kappa t})]$, for $t_k \leq t \leq t_{k+1}$. Putting the pieces together we arrive an analytical expression for the conditional state:
\begin{equation}\label{eq:condstateD}
\ket{\overline{\psi}_{\rm REC}(t)}=(\sqrt{2\kappa}\alpha e^{-\kappa t_n})\ldots (\sqrt{2\kappa}\alpha e^{-\kappa t_1})\exp[-\tfrac{1}{2}|\alpha|^2(1-e^{-2\kappa t})]\ket{\alpha e^{-\kappa t}},
\end{equation}
from where we obtain the {\it record probability}\index{Record probability}
\begin{equation}
 \braket{\overline{\psi}_{\rm REC}(t)|\overline{\psi}_{\rm REC}(t)}=(2\kappa|\alpha|^2 e^{-2\kappa t_n})\ldots (2\kappa|\alpha|^2 e^{-2\kappa t_1})\exp[-|\alpha|^2(1-e^{-2\kappa t})].
\end{equation}
This result is expected from the semiclassical treatment of photoelectric emission for a classical field of a decaying intensity (in photon flux units)  $2\kappa |\alpha|^2 \exp(-2\kappa t)$: the exponential factor $\exp[-|\alpha|^2(1-e^{-2\kappa t})]$ is the {\it null measurement probability}\index{Null measurement probability} in a time span up to $t$ -- precisely the factor of eq.~\eqref{cohfac} -- while $2\kappa |\alpha|^2 e^{-2\kappa t_k}$ is the probability density for one count at the time $t_k$. From the record probability we obtain the photon count distribution\index{Distribution! photon count}, {\it i.e.}, the probability of $n$ counts in time $T$, summing (integrating) over all possible counting times. Using 
\begin{equation}
(2\kappa |\alpha|^2)^n \int_{0}^{T}dt_n\int_{0}^{t_n}dt_{n-1} \cdots \int_{0}^{t_2}dt_1\,e^{-2\kappa t_n}\ldots e^{-2\kappa t_1}=\frac{\left[2\kappa|\alpha|^2\int_0^T e^{-2\kappa t^{\prime}}dt^{\prime}\right]^n}{n!} 
\end{equation}
to calculate the sum, we arrive at the familiar Poisson distribution\index{Distribution! Poisson}\index{Poisson! distribution},
\begin{equation}
 P(n,T)=\frac{[|\alpha|^2(1-e^{-2\kappa T})]^n}{n!}\, \exp[-|\alpha|^2(1-e^{-2\kappa T})].
\end{equation}
Now, since the evolution of the unnormalized conditional state $\ket{\overline{\psi}_{\rm REC}(t)}$ is linear, we can extend eq.~\eqref{eq:condstateD} to an arbitrary coherent-state superposition, such as the even cat of eq.~\eqref{eq:evencats}, written as
\begin{equation}
 \ket{\overline{\psi}_{\rm REC, even\,cat}(0)}=\frac{1}{\sqrt{2}} \frac{\ket{\alpha} + \ket{-\alpha}}{\sqrt{1 + \exp(-2|\alpha|^2)}}.
\end{equation}
The two pats of the superposition differ only in the action of the jump operator, on the basis of the different signs of the coherent-state amplitudes. Finally, we obtain
\begin{equation}
\ket{\overline{\psi}_{\rm REC, even\,cat}(t)}=(\sqrt{2\kappa}\alpha e^{-\kappa t_n})\ldots (\sqrt{2\kappa}\alpha e^{-\kappa t_1})\exp[-\tfrac{1}{2}|\alpha|^2(1-e^{-2\kappa t})]\frac{1}{\sqrt{2}} \frac{\ket{\alpha e^{-\kappa t}} + (-1)^n \ket{-\alpha e^{-\kappa t}}}{\sqrt{1 + \exp(-2|\alpha|^2)}}.
\end{equation}
From the expression we see that ``preserving the coherence between the two pieces of the superposition  is a matter of tracking every last photon transferred to the environment\index{Cat state}\index{Schr\"odinger cat! states!decoherence}. It is not so much that the prepared cat `dies'. It simply leaves the resonator to exist in the output field; it is necessary to know what part is on the inside and what on the outside (down to the level of one photon) if we are to continue to have access to the coherence of the prepared cat.''~\cite{OpenQuantCh4} We will revisit quantum trajectories when dealing with the strong-coupling limit in subsec.~\ref{sssec:JCzerodim}, as a method which is instrumental for assessing the importance of quantum fluctuations. There, expectedly, we give up on the convenience of an analytical solution for most cases of interest.  

\begin{figure}
\includegraphics[width=14cm]{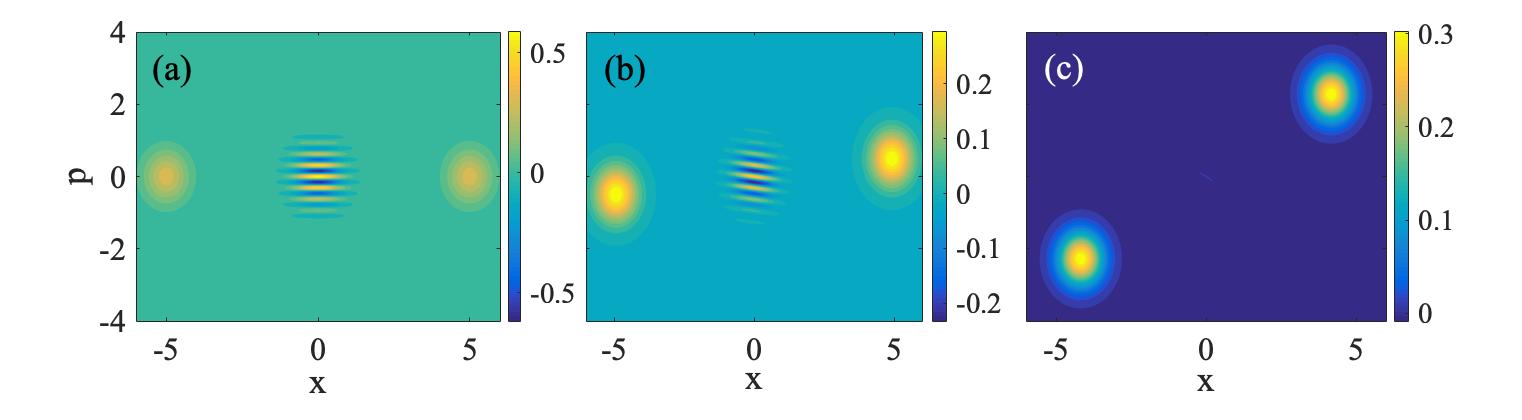} 
\caption{Time shots of the Wigner function~(\ref{wigcateq}) for the decaying cat state. In {\bf (a)}, $t=0$ such that the cat is in a perfect superposition even cat with $\alpha=5$. The coherent superposition is manifested in the interference fringes at the center which also display negative values. As time progresses, the distribution starts to rotate with the angular frequency $\omega$ (in this example $\omega=10$), but decoherence sets in due to the non-zero decay rate $\kappa=1$. In principle, dissipation also kicks in implying that the two blobs approach the origin. However, on these time-scales this is a minor effect compared to the decoherence. In {\bf (b)} $t=0.01$ and we already see how the interference pattern is weaker, and the negative values smaller. Finally in {\bf (c)}, for $t=0.05$, almost all interference is lost and the cat has decayed into a statistical mixture, while the field amplitude is approximately intact. } 
\label{catfig}   
\end{figure}

In phase space, a cat state consists of two well separated blobs, as already shown in fig.~\ref{fig4} (a) and (b). The Wigner function for the state\index{Wigner! function}~(\ref{catdec}) becomes~\cite{gerry1997quantum}
\begin{equation}\label{wigcateq}
W(x,p)=\frac{1}{N_C(t)}\left[e^{-2|x-ip-\alpha(t)|^2}+e^{-2|x-ip+\alpha(t)|^2}+2C(t)e^{-2|\alpha(t)|^2}\cos\left(4\mathrm{Im}[(x-ip)\alpha(t)]\right)\right],
\end{equation}
where $N_C(t)$ is a normalization constant. The term, proportional to $C(t)$, displays the coherences of the cat seen in the oscillations of the cosine-function. In fig.~\ref{catfig} we show the Wigner function at three different times. The first plot gives the initial even cat with a fully developed interference pattern, {\it i.e.} $C(t=0)=1$. In (b) the interference term has been decreased by the factor $C(t=0.01)\approx0.4$, and in (c) $C(t=0.05)\approx0.014$. We also note how the cat rotates around the origin with the frequency $\omega$. The dissipation occurs on a much longer time-scale than that of the decoherence. For example, in (b) $e^{-\kappa t}=0.99$ and in (c) $e^{-\kappa t}=0.95$, to be compared to the factor $C(t)$. In this example $\alpha(0)=5$, meaning that the decoherence takes place on a time-scale 50 ($=2|\alpha|^2$) times that of dissipation! This rapid transition into a statistical mixture is the idea behind {\it einselection}\index{Einselection} and {\it pointer states}\index{Pointer state}\index{State! pointer}~\cite{zurek2003decoherence}. It has been beautifully demonstrated by Haroche and co-workers in a cavity QED experiment~\cite{brune1992manipulation}, which will be further discussed in sec.~\ref{sssec:subscat}. For a similar analysis as the one above, but for finite temperature baths we refer to~\cite{kim1992schrodinger}. 

Effects of the photon loss have been thoroughly  studied in the past in terms of cat states, but it has also been shown how the reservoir induced decoherence suppresses the revivals\index{Quantum! revivals}~\cite{puri1986collapse,barnett1986dissipation,puri1987finite,eiselt1991quasiprobability,quang1991quantum,barnett2007damped}. In the weak-damping limit\index{Damping! weak, limit}, $n^2\kappa\ll\sqrt{n+1}g$, the atomic inversion\index{Atomic! inversion}~(\ref{inv2}) is modified as
\begin{equation}\label{inv3}
W(t)=\exp\left(-\frac{\kappa t}{2}\right)\left[\sum_{n=0}^{\infty}|c_n(0)|^2e^{-n\kappa t}\cos(2g\sqrt{n+1}t)
+\sum_{n=0}^{\infty}\sum_{l=n}^\infty\frac{[2(l-n)]!}{2^{2(l-n)}[(l-n)!]^2}\left(1-e^{-\kappa t}\right)^{l-n}|c_l(0)|^2\right]-1.
\end{equation}
What is clear is that the JC revivals will be exponentially sensitive to photon losses. In ref.~\cite{puri1987finite} Puri and Agarwal, apart from analyzing the atomic inversion, also studied the influence of damping on the second order correlation function~(\ref{corr2}). One sort of decoherence, not stemming from coupling to a photon bath, is phase noise, {\it i.e.} the atom-field coupling is subject to stochastic fluctuations $g\rightarrow g\, e^{i\theta(t)}$ where $\theta(t)$ is a stochastic variable. Such noise results also in an exponential suppression of the revivals~\cite{joshi1994jaynes,joshi1995effects}. The same type of behaviour of the inversion is also found in the JC model with quenched disorder~\cite{Ghoshal2020}. By quenched disorder we mean that in an experiment some parameter, here the coupling $g$, varies from one experimental run to the next, and as one average over realizations one finds a damping of the revivals. An alternative to the standard Lindblad form is to follow phase space approaches as considered in Refs.~\cite{eiselt1991quasiprobability,daeubler1992analytic}, and especially it was shown how to derive a Fokker--Planck equation\index{Fokker--Planck equation} for the phase-space distributions\index{Phase-space distribution}.

In the dispersive regime~(\ref{effham}) the Hamiltonian is diagonal in the number basis and it is easy to solve eq.~(\ref{master1}) for initial coherent field states and arbitrary atomic states~\cite{de1999dissipative}. It should be emphasized that performing the adiabatic elimination\index{Adiabatic! elimination} in the presence of a bath normally results in different Lindblad operators $\hat{\mathcal{D}}[\hat{L}]$ since the Schrieffer-Wolff transformation\index{Schrieffer-Wolff transformation}~(\ref{polaron}) may not commute with the $\hat{L}$'s~\cite{klimov2003master,boissonneault2009dispersive}. Assuming the above form of the Lindblad operators, since the dispersive JC Hamiltonian only accounts for a phase shift of the coherent state amplitudes, the field states (for the atom in either of the two bare states) remain coherent with an exponentially decaying amplitude~\cite{scully1991micromaser,walls2008quantum,schleich2011quantum,gardiner2004quantum}. Expectedly, cavity losses have also been shown to suppress squeezing~\cite{buvzek1992superpositions}. More surprising is that photon losses may actually favour entanglement between atoms collectively coupled to a boson mode~\cite{plenio1999cavity}. Quantum correlations between the atom and the field can also arise from collective coupling to the same environment. In ~\cite{Badveli2020}, the time-evolution of the negativity, the mutual information, and the discord was analyzed for a non-interacting system, {\it i.e.} the coupling $g=0$, coupled to a common reservoir of harmonic oscillators. As long as the photon loss rate was small compared to the spontaneous emission\index{Spontaneous! emission! rate} rate of the atom, all three measures indicated non-classical correlations, even in the Markovian limit.

Atomic loses, described by the eq.~(\ref{openqubit}), are phenomenologically mimicked by adding a term $\hat{\mathcal{D}}[\hat{\sigma}_{-}]$ to the master equation\index{Master equation}~\cite{scully1991micromaser,walls2008quantum,schleich2011quantum,gardiner2004quantum}. Pure atom dephasing is obtained instead by for example $\hat{\mathcal{D}}[\hat{\sigma}_z]$. Both dissipation~\cite{alsing1991spontaneous,joshi1994jaynes,armen2006low,boissonneault2008nonlinear} and dephasing~\cite{clemens2000nonclassical,wilson2002quantum,obada2004entanglement,larson2008rabi,rendell2003revivals,gonzalez2009effect,auffeves2010controlling} have been the subject of interest. In ref.~\cite{rendell2003revivals}, for example, the entanglement evolution of an initial Bell state\index{Bell! state} ({\it e.g.}, $|\psi\rangle=(|g,1\rangle+|e,0\rangle)/\sqrt{2}$) was studied in the case of dephasing. A similar study was performed in~\cite{obada2004entanglement}, but for unentangled initial atom-field states. Most of these works, however, discuss how the cavity spectrum is affected by atom dephasing. 

One solvable version of the open JC model arises when we consider the jump operator $\hat L=\hat H_\mathrm{JC}$~\cite{moya1993intrinsic}
\begin{equation}\label{intr}
\partial_t\hat{\rho}(t)=i\left[\hat{\rho}(t),\hat{H}_\mathrm{JC}\right]-\kappa\left[\left[\hat{\rho}(t),\hat{H}_\mathrm{JC}\right],\hat H_\mathrm{JC}\right].
\end{equation}
Such a Lindblad equation was first proposed in the general case by Milburn in order to mimic {\it intrinsic decoherence}\index{Intrinsic decoherence}~\cite{milburn1991intrinsic}. Intrinsic decoherence, as arising from coupling to a gravitational field, had been speculated whether it could explain the collapse of the wave function upon a measurement. It is clear that the effect of the double commutator is to kill off all non-diagonal terms of the density operator in the energy eigenbasis. Generally, a double commutation in the equations of motion implies some sort of diffusion. The general solution $\hat\rho(t)$ of~(\ref{intr}) was derived in~\cite{moya1993intrinsic} using a method of super-operators. It is expressed as an infinite sum and slightly complicated, but the expression of the inversion~(\ref{inv}) is rather simple. Instead of the expression~(\ref{inv2}), valid at the resonant case for the JC model, one finds\index{Atomic! inversion}
\begin{equation}\label{inv2dec}
W(t)=\sum_n|c_n(0)|^2e^{-2\kappa(n+1)g^2t}\cos(2g\sqrt{n+1}t).
\end{equation}
Thus, instances of coherence, as seen for example in the revivals, are exponentially suppressed by the presence of decoherence.

The above explorations are only valid in the weak coupling regime. Increasing the coupling $g$ means that at some point the ground state will be the dressed state $|\psi_{0-}\rangle$. For even larger $g$, the excitation number of the ground state increases even further. It is clear that the master equations\index{Master equation} discussed above cannot reproduce a realistic situation; at zero temperature the system will not relax to its ground state but instead reproduce unphysical results~\cite{werlang2008rabi}. This issue was first discussed by Carmichael and Walls~\cite{carmichael1973master}. In the advent of circuit QED, where the coupling $g$ can become rather big (see sec.~\ref{sec:cirQED}), this topic is even more relevant and hence has been discussed in numerous works~\cite{cresser1992thermal,briegel1993quantum,scala2007microscopic,beaudoin2011dissipation,ridolfo2012photon,agarwal2013dissipation,nathan2020universal}. In order to derive a more accurate Markovian master equation\index{Master equation} for the JC model, the operators $\hat{L}_j$ are often taken to be projection operators onto the dressed states. Another way to circumvent problems like this is to instead leave the Schr\"odinger picture and work in the Heisenberg picture, {\it i.e.} solve the Heisenberg-Langevin equations\index{Heisenberg-Langevin equations} which for the JC model read
\begin{equation}\label{langeq}
\begin{array}{l}
\partial_t\hat{a}(t)=-ig\hat{\sigma}_{-}(t)-\kappa\hat{a}(t)+\sqrt{2\kappa}\hat{a}_\mathrm{in}(t),\\ \\
\partial_t\hat{\sigma}_{-}(t)=i\Delta\hat{\sigma}_{-}(t)+ig\hat{a}(t)\hat{\sigma}_z(t)-\displaystyle{\frac{\gamma}{2}\hat{\sigma}_{-}}+\sqrt{\gamma}\hat{\sigma}_z\hat{f}_\mathrm{in}(t),
\end{array}
\end{equation}
where $\hat{a}_\mathrm{in}(t)$ and $\hat{f}_\mathrm{in}(t)$ are the field and atomic input (Langevin) noise terms~\cite{walls2008quantum,gardiner1985input}. The properties of the bath determine the characteristics of the noise terms. For a thermal bath, for example, they are taken to have vanishing means, $\langle\hat{a}_\mathrm{in}(t)\rangle=\langle\hat{f}_\mathrm{in}(t)\rangle=0$, and to be $\delta$-correlated, $\langle\hat{a}_\mathrm{in}(t)\hat{a}_\mathrm{in}^\dagger(t')\rangle=\delta(t-t')$ and $\langle\hat{f}_\mathrm{in}(t)\hat{f}_\mathrm{in}^\dagger(t')\rangle=\delta(t-t')$. This approach has been frequently employed to the JC model, see for example ~\cite{lawande1994stochastic,cirac1997quantum,mabuchi1998retroactive,law2000jaynes,sarovar2005high,an2009quantum,de2009extracavity,ridolfo2012photon}. Like with the Schr\"odinger vs. the Heisenberg representations, which one is the most convenient one depends strongly on the problem being studied. For example, in terms of the JC model, studying say statistics like correlators of the photon field might be more easily done with the Langevin input-output formalism\index{Input-output! formalism} (as we already mentioned above), while if one is interested in entanglement properties it might be more practical to work with the master equation\index{Master equation}.

Above we introduced the decoherence-free states\index{Decoherence-free! state}\index{State! decoherence-free} as those obeying $\hat{\mathcal{D}}[\hat{\rho}_\mathrm{df}]=0$. If, furthermore $\left[\hat{L}_j,\hat{H}\right]=0$ $\forall j$, it implies that an initial state will typically decay into one of the decoherence free states. The set of states $\hat{\rho}_\mathrm{df}$ is called {\it decoherence free subspace}\index{Decoherence-free! subspace}~\cite{lidar1998decoherence}. Once the system has relaxed into this subspace, the states are protected by dissipation. If these states contain interesting properties, like entanglement, squeezing, or are topological, they become experimentally very attractive. Thus, given the Hamiltonian, if the loss channels $\hat{L}_j$ can be experimentally engineered and controlled this presents a robust method for state preparation~\cite{diehl2008quantum,verstraete2009quantum,barreiro2011open}. These ideas have also been considered in terms of the JC model or extensions of it~\cite{kastoryano2011dissipative,busch2011cooling,reiter2012driving,sweke2013dissipative}. Here, spontaneous emission\index{Spontaneous! emission} of the atoms are acting as a resource by projecting the system state down onto an entangled two-atom state. The method can be extended also to ensembles of atoms~\cite{cho2011optical,muller2011simulating}

Decoherence arises also from quantum measurements - any measurement induces a backaction\index{Backaction! measurement-induced} on the system~\cite{wiseman2010quantum}. Among the first examples of such effects in terms of the JC model was the analysis of the {\it micromaser}\index{Micromaser}~\cite{krause1986quantum,meystre1988very,scully1991micromaser}. In the micromaser (see sec.~\ref{sssec:micro}), excited single atoms traverse a resonator one-by-one and resonantly interact with one of the cavity modes~\cite{berman1994cavity}. The states of the atoms when exiting the interaction region is not recorded, {\it i.e.} the atomic degrees of freedom are traced out, and only the properties of the cavity field is considered. Most generally, tracing out the atoms leaves the field in a mixed state. A somewhat related situation emerges when the state of the atom is repeatedly measured. This has been considered as a method for implementing a {\it quantum Zeno effect}\index{Quantum! Zeno effect}~\cite{Misra1977, sakurai1995modern} which then can slow down the field evolution~\cite{larson2011anomalous,lizuain2010zeno}. The quantum Zeno effect of the field in a microwave cavity, induced by atomic measurements, has been observed in the group of Haroche~\cite{bernu2008freezing}. They studied a pumped cavity, where they interrupted the coherent pumping, performing non-demolition measurements by sending atoms through the resonator. Likewise, by performing non-demolition measurements of the field the evolution of the atom can be frozen~\cite{gagen1992quantum,helmer2009quantum}. It has also been suggested how the Zeno effect could help in protecting entanglement shared between two atoms~\cite{maniscalco2007protecting}. One of the earliest studies of backaction\index{Backaction} from field measurements was by Imoto and co-workers~\cite{imoto1990microscopic}. Recently, the backaction\index{Backaction} from either the atom~\cite{clerk2007using} or the field~\cite{gambetta2006qubit,gambetta2008quantum,hausinger2008dissipative,boissonneault2012back} has been addressed. As a remark, deriving microscopic equations of motion for the system state including measurements backaction typically leads to master equations\index{Master equation} which are not in a Lindblad form~\cite{gagen1992quantum,gambetta2008quantum,woolley2010continuous}. In general, the idea that measurements can explain emergence of chaos in quantum systems and the occurrence of classical behaviour~\cite{zurek2003decoherence} has also been considered in terms of the JC model~\cite{everitt2009quantum}. Continuous measurements prohibit the oscillation collapse, maintaining Rabi oscillations over long periods. An efficient and experimentally feasible method to realize the Zeno effect\index{Quantum! Zeno effect} has been proposed in~\cite{Zeno2022} for a dissipative two-level system\index{Dissipative! two-level system} interacting with a single-mode field. The method relies on the dressed-state dynamics in the open JC model, yielding five orders of magnitude longer decay times for the H atom, compared to that of the bare-state model.

From a more general perspective, Minganti and coauthors have explored the fundamental properties on first-order dissipative quantum phase transitions\index{Dissipative! quantum phase transition} as well as second-order transitions associated with symmetry breaking~\cite{Minganti2018}. The density matrix solving a Lindblad master equation\index{Master equation}, with the spectrum and eigenvectors of the Liouvillian defined from $\mathcal{L}\hat{\rho}_{i}=\lambda_i \hat{\rho}_{i}$, can be written in the form
\begin{equation}
 \hat{\rho}(t)=\hat{\rho}_{\rm ss} + \sum_{i\neq 0}c_i (0) e^{\lambda_i t}\hat{\rho}_i,
\end{equation}
where ${\rm tr}(\hat{\rho}_{\rm ss})=1$ and ${\rm tr}(\hat{\rho}_{i})=0$. The eigenvalues, satisfying ${\rm Re}[\lambda_i] \leq 0$, are ordered such that $|{\rm Re}[\lambda_0]| < |{\rm Re}[\lambda_1]|<\cdots <|{\rm Re}[\lambda_n]|$, with $\lambda_0=0$ corresponding to the eigenvector which comprises the steady state\index{Steady state}. The next eigenvalue in this sequence, defines the Liouvillian gap~\index{Liouvillian! gap} via $\lambda=|{\rm Re}[\lambda_1]|$, which determines the slowest relaxation rate in the long-time limit. 

When there is only one real Liouvillian eigenvalue, $\lambda_i$, the corresponding density $\hat{\rho}_{i}$ can be diagonalized with the spectral decomposition~\cite{Rivas2011}\index{Spectral! decomposition} 
\begin{equation}
 \hat{\rho}_{i}=\sum_{n} p_n^{(i)} |\psi_{n}^{(i)} \rangle  \langle \psi_n^{(i)}|,
\end{equation}
where $\braket{\psi_{n}^{(i)}|\psi_{m}^{(i)}}=\delta_{n,m}$ and all coefficients $p_n^{(i)}$ must be real; some must be positive and others negative since ${\rm tr}(\rho_i)=0$. One may order the coefficients with respect to an integer $M$ such that $p_{n}^{(i)}>0$ for $n \leq M$, and $p_{n}^{(i)}<0$ for $n > M$. We can then explicitly write
\begin{equation}\label{eq:spectrdecom}
 \hat{\rho}_{i}=c (\hat{\rho}_{i}^{+} -  \hat{\rho}_{i}^{-}), \quad \quad \hat{\rho}_{i}^{+}=\sum_{n \leq M}p_{n}^{(i)} |\psi_{n}^{(i)} \rangle  \langle \psi_n^{(i)}|, \quad \quad \hat{\rho}_{i}^{-}=-\sum_{n > M}p_{n}^{(i)} |\psi_{n}^{(i)} \rangle  \langle \psi_n^{(i)}|,
\end{equation}
where the coefficients $p_n$ have been normalized to ensure that ${\rm tr [\hat{\rho}_{i}^{+}]}={\rm tr [\hat{\rho}_{i}^{-}]}=1$, which renders $\hat{\rho}_{i}^{\pm}$ density matrices. This means that a state of the form $\hat{\rho}(0)=\hat{\rho}_{\rm ss} + c_1 \hat{\rho}_i$ will evolve in time as~\cite{Macieszczak2016} 
\begin{equation}
 \hat{\rho}(t)=\hat{\rho}_{\rm ss} + c_1 e^{\lambda_i t} (\hat{\rho}_{i}^{+} -  \hat{\rho}_{i}^{-}).
\end{equation}
For a straightforward extension to a complex eigenvalue, see sec.II B of~\cite{Minganti2018}. 

In the thermodynamic limit\index{Thermodynamic limit}, the Liouvillian gap $\lambda=|{\rm Re}[\lambda_1]|$ closes when a characteristic parameter of the Liouvillian, called $\zeta$, assumes a critical value $\zeta_c$. When the scale parameter is taken to infinity, $N \to \infty$, a dissipative quantum transition between two phases is characterized by the nonanalytical behavior exhibited by an observable $\hat{O}$ which is independent of the parameter $\zeta$ which approaches a critical value $\zeta_c$. Formally, there is a phase transition of order $m$ if
\begin{equation}
 \lim_{\zeta \to \zeta_c} \left|\frac{\partial^{m}}{\partial \zeta^{m}}\lim_{m\to \infty} {\rm tr}[\hat{\rho}_{\rm ss}(\zeta, N) \hat{O}]\right|=+\infty.
\end{equation}
Since the observable $\hat{O}$ does not depend on the parameter $\zeta$, the discontinuity evinced by the above limit is attributed to $\hat{\rho}_{\rm ss}(\zeta, N\to \infty)$ and associated with a level crossing in the Liouvillian spectrum~\cite{Kato1995}. Now, as $\hat{\rho}_{\rm ss}$ is associated with the eigenvalue $\lambda_0=0$, the phase transition should coincide with the closure of the Liouvillian gap~\cite{Kessler2012, Horstmann2013}. The thermodynamic limit\index{Thermodynamic limit} is then associated with the emergence of multiple steady states\index{Steady state!multiple}. 

Following the analysis of~\cite{Minganti2018}, we discuss in some detail the occurrence of criticality in a system with discrete $Z_2$ symmetry, represented by the super-operator $\mathcal{Z}_2=\hat{Z}_2 \cdot \hat{Z}_2$ which admits the eigenvalues $\pm 1$. For $\zeta < \zeta_c$, there exists a unique steady state\index{Steady state!unique} $\hat{\rho}_{\rm ss}$ corresponding to the eigenvalue $\lambda_0=0$, with $\mathcal{Z}_2 \hat{\rho}_{\rm ss}=\hat{\rho}_{\rm ss}$. For $\zeta \geq \zeta_c$, a phase transition accompanied by spontaneous symmetry breaking\index{Spontaneous! symmetry breaking} takes place identifying two solutions, $\hat{\rho}_0$ and $\hat{\rho}_1$ (with $\lambda_0=\lambda_1=0$) which belong to two different symmetry sectors. These two solutions are orthogonal, given that $\mathcal{Z}_2$ is Hermitian:
\begin{equation}
 \braket{\hat{\rho}_0|\hat{\rho}_1}=\braket{\mathcal{Z}_2 \hat{\rho}_0|\hat{\rho}_1}=\braket{\hat{\rho}_0|\mathcal{Z}_2 \hat{\rho}_1}=-\braket{\hat{\rho}_0|\hat{\rho}_1}.
\end{equation}
Since $\lambda_1$ is real and of degeneracy one, the matrix $\hat{\rho}_1$ is Hermitian. We infer that the density matrices 
\begin{equation}\label{eq:rhopm}
 \hat{\rho}^{\pm}\equiv \frac{\hat{\rho}_0 \pm \hat{\rho}_1}{{\rm tr}[\hat{\rho}_0]}
\end{equation}
represent steady states\index{Steady state} of the master equation\index{Master equation} breaking the symmetry, since $\mathcal{Z}_2 \hat{\rho}^{\pm}=\hat{\rho}^{\mp}$. From eq.~\eqref{eq:rhopm} it follows that $\hat{\rho}^{+}$ and $\hat{\rho}^{-}$ are as well orthogonal, whence we can write
\begin{equation}
 \hat{\rho}_0 \propto \hat{\rho}^{+} + \hat{\rho}^{-}, \quad \quad \hat{\rho}_1 \propto \hat{\rho}^{+} - \hat{\rho}^{-}
\end{equation}
and conclude that the two symmetry-broken states originate from the spectral decomposition\index{Spectral! decomposition}  of $\hat{\rho}_1$, see eq.~\eqref{eq:spectrdecom}. For a finite-system size with a unique steady state\index{Steady state!unique}, we have 
\begin{equation}
 \hat{\rho}_{\rm ss}(\zeta \geq \zeta_c, N) \approx \frac{1}{2}[\hat{\rho}_1^{+}(\zeta_c,N) + \hat{\rho}_1^{-}(\zeta_c,N)],
\end{equation}
with ${\rm tr}[\hat{\rho}_1^{+}(\zeta_c,N)]={\rm tr}[\hat{\rho}_1^{-}(\zeta_c,N)]=1$. Since we are dealing with a second-order phase transition, we must make sure that the unique steady state\index{Steady state!unique} for $\zeta=\zeta_c^{-}$ coincides with both the symmetry-broken steady states for $\zeta=\zeta_c^{+}$, {\it i.e.}, $\hat{\rho}_{\rm ss}(\zeta_c^{-})=\hat{\rho}^{+}(\zeta_c^+)=\hat{\rho}^{-}(\zeta_c^+)$. It readily follows that $\hat{\rho}_1(\zeta_c)=0$, which means that a second-order phase transition is characterized by the coalescence of two eigenvectors of the Liouvillian~\cite{Minganti2018}.  

In addition to the well-known criticality of the single-photon driven Kerr oscillator\index{Kerr! medium}\index{Kerr! effect}, a comprehensive experimental and theoretical analysis of both first- and second-order quantum phase transitions occurring in a superconducting two-photon driven Kerr resonator\index{Two-photon! driven Kerr oscillator} has been recently published in~\cite{beaulieu2023observation}. The breakdown of photon blockade on resonance constitutes a characteristic second-order phase transition for the open driven JC oscillator~\cite{Curtis2021}, which we will discuss in sec.~\ref{sssec:JCzerodim}, while a discrete-$Z_2$ symmetry breaking underlies the Dicke phase transition, which we meet in sec.~\ref{sssec:dicke}.


\subsubsection{Quantum fluctuations and criticality: photon blockade and its breakdown}
\label{sssec:JCzerodim}
The coherently-driven JC model is a characteristic example of a nonlinear oscillator which exhibits {\it photon blockade}\index{Photon! blockade}~\cite{imamoglu1997strongly}. Like Coulomb or Rydberg blockade, the photon blockade\index{Photon! blockade} arises due to an anharmonicity induced by interaction. The mechanism is most clearly understood from the JC spectrum~(\ref{eigv2}). Let us for brevity consider the resonant JC model, {\it i.e.} $\Delta=\omega-\Omega=0$. The ground state of the JC model is the state $|g,0\rangle$ with an energy $-\omega/2$. The first excited states are, according to eq.~(\ref{dstate}), $|\psi_{0\pm}\rangle=(|e,0\rangle\pm|g.1\rangle)/\sqrt{2}$, with corresponding energies $\mathcal{E}_{1\pm}=\frac{\omega}{2}\pm g$. Thus, the energies needed to excite the JC system from the ground state to any of these are $\delta\mathcal{E}_1=\omega\pm g$. If we drive the JC oscillator with a coherent field of frequency $\omega_\mathrm{dr}=\omega-g$ and the system is initially in its ground state, then a photon of the pump will be excited to put the system in its first excited state $|\psi_{0-}\rangle$. Next pair of states in the JC ladder\index{Jaynes-Cummings! ladder} are $|\psi_{1\pm}\rangle=(|e,1\rangle\pm|g.2\rangle)/\sqrt{2}$, and with energies $\mathcal{E}_{2\pm}/\hbar=\frac{3\omega}{2}\pm g\sqrt{2}$. To excite the system to any of these states we need an energy (in this section we reinstate $\hbar$ for clarity) $\mathcal{E}_{2}/\hbar=\omega\pm g(\sqrt{2}\mp1)$. Thus, the frequency of the pump is off resonant with any transition and transitions to higher states are suppressed -- the one-photon blockade.
The phenomenon has been experimentally verified in the group of Jeff Kimble for a single trapped atom in an optical resonator~\cite{birnbaum2005photon,dayan2008photon}. The idea can be generalized to multiphoton processes\index{Multiphoton! process}, and the two-photon blockade\index{Photon! blockade}\index{Two-photon!blockade} has also been experimentally demonstrated~\cite{hamsen2017two}. The blockade is depleted but may persist also in the dispersive regime~\cite{hoffman2011dispersive}, and by adding a Stark shift term it is possible to reestablish strong nonlinearity~\cite{Tang2019}. The strongly-dispersive JC Hamiltonian with coherent driving of the two-level atom has been employed in~\cite{MunozNP2014} to assess the process of $N$-photon emission upon the generation of maximally entangled polaritons\index{Polariton} of the type $(\ket{g,0} \pm \ket{e,N})\sqrt{2}$. The coherent drive dresses the two-level atom, opening up a Mollow ladder\index{Mollow! ladder} where the cavity Purcell-enhances the $N$-photon transitions from the $\ket{g}$ to the $\ket{e}$ states. Also in circuit QED (see sec.~\ref{sec:cirQED}) the blockade has been experimentally analyzed in the groups of Wallraff~\cite{lang2011observation} and Houck~\cite{hoffman2011dispersive}.  

The result of photon blockade is that a single photon (with the appropriate frequency) can be transferred from the pump field to the JC oscillator, but a second photon is prohibited from being transferred. However, after some time the photon will decay and the JC system will be reset in its ground state, opening up the possibility for another photon to be transmitted. A detector measuring the decayed photons will see single photons separated in time, {\it i.e.} they will show antibunching\index{Antibunching} [see eq.~(\ref{corr2}]. Since the anharmonicity of the JC spectrum scales as $\sim g$ one expects strong antibunching, {\it i.e.} efficient photon blockade, in the strong coupling regime where $g$ exceeds the line broadening set by $\kappa$. It thereby came as a surprise that perfect antibunching can also be achieved for very weak anharmonicities (when the size of $g$ only a few percent of $\kappa$)~\cite{liew2010single}. The mechanism behind this phenomenon, which has been termed {\it unconventional photon blockade}\index{Photon! blockade! unconventional}, has a different origin than for the regular photon blockade as it derives from destructive interference for injecting two photons into the cavity~\cite{bamba2011origin}. The unconventional photon blockade\index{Photon! blockade} has been experimentally verified by coupling a q-dot to a two-mode {\it micropillar cavity}~\cite{snijders2018observation}. An effective non-Hermitian Hamiltonian has been derived in~\cite{ZhangDuan23} to model photon blockade with a phenomenologically introduced two-photon dissipation\index{Two-photon!dissipation}.

Conventional photon blockade\index{Photon! blockade} breaks down as an out-of-equilibrium process that involves fundamentally both inputs and outputs\index{Nonequilibrium phase transition}\index{Phase transition! quantum! out of equilibrium}. The latter send information into the environment measuring the quantum state of the open system, while the former impose some sort of boundary constraints. This dynamical effect takes place by means of a dissipative quantum phase transition\index{Dissipative! quantum phase transition}\index{Driven! dissipative quantum phase transition}. The transition is first-order\index{Phase transition! dissipative! quantum! first-order} apart from one critical point\index{Critical! point} in the phase space set by the drive, where the phases of the system are characterized by continuous solutions\index{Phase transition! dissipative! quantum! second-order} \cite{CarmichaelPRX}. A similar dissipative quantum phase transition will be discussed in sec.~\ref{ssec:cqedearly} in terms of the micromaser response. 
\begin{figure}[!ht]
\begin{center}
\includegraphics[width=0.45\textwidth]{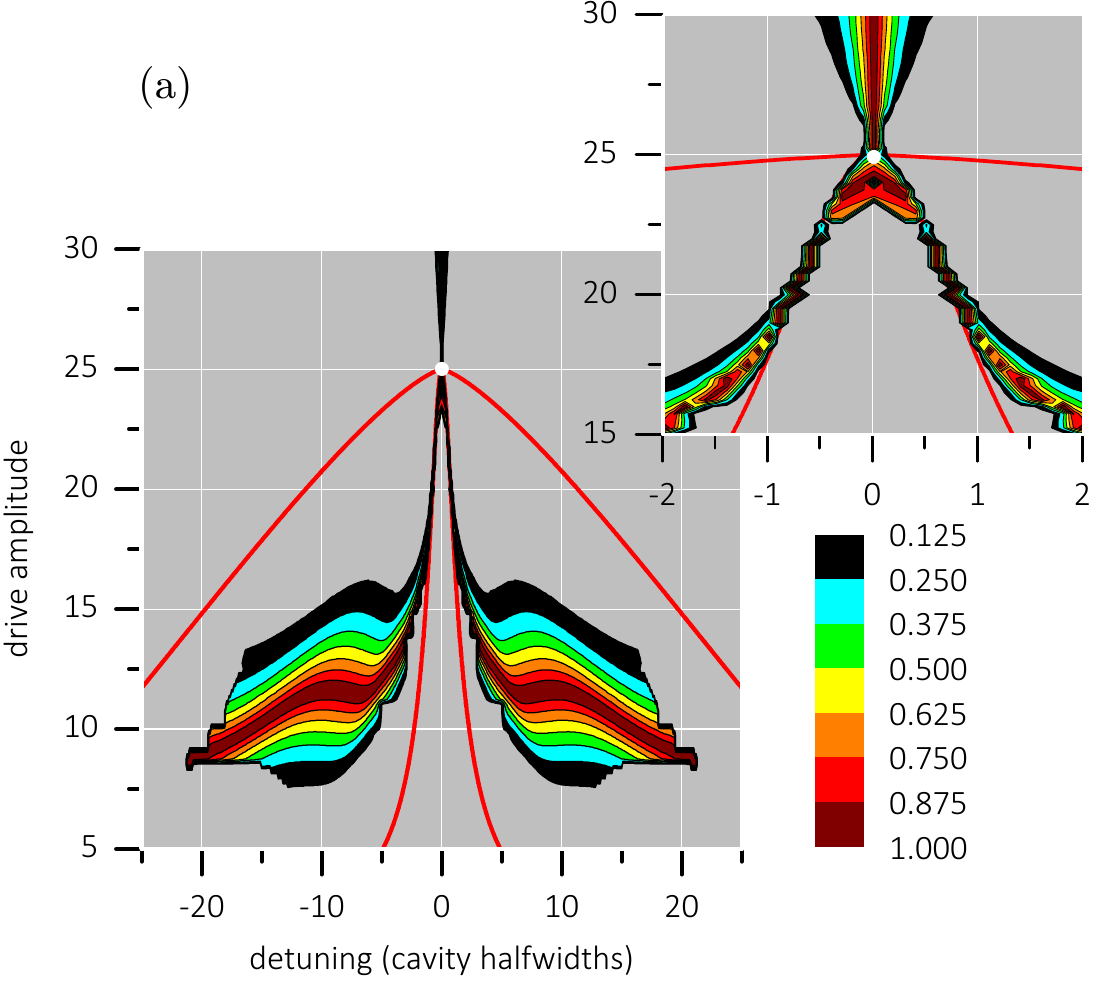}
\includegraphics[width=0.45\textwidth]{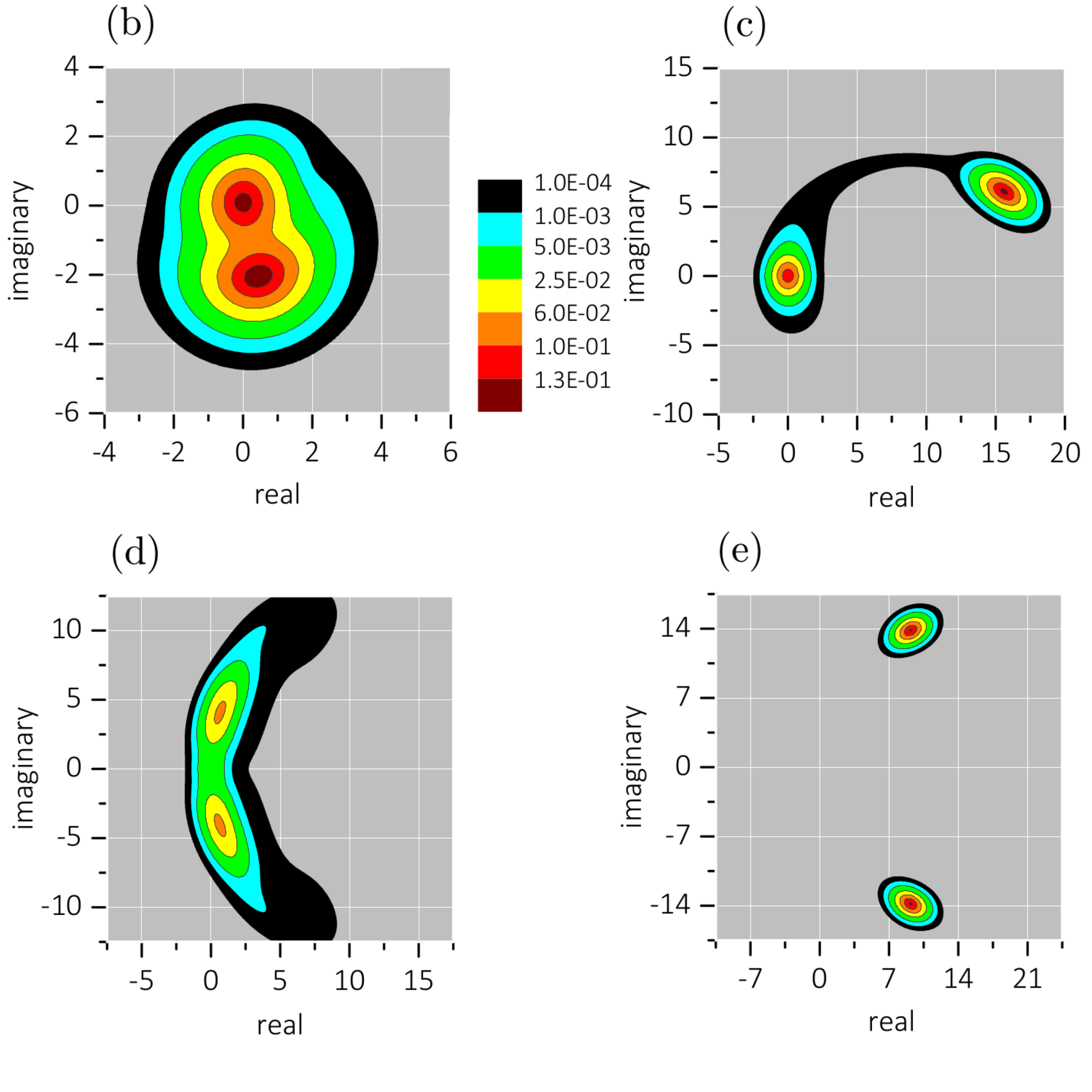}
\end{center}
\caption{{\bf Amplitude and phase bimodality in zero dimensions} (for a single atom coupled to a cavity mode and with $n_{\rm sat, wc}=0$). {\bf (a)} Region of drive detuning-amplitude plane within which the steady-state Q-function\index{$Q$-function}\index{Q distribution}\index{Distribution! Q} of the intracavity-field distribution is bimodal\index{Distribution! bimodal}. Contours of $r=1-|h_1-h_2|/(h_1+h_2)$ are shown, with $h_1$ and $h_2$ the peak heights. Red lines demark the region of bistability according to the neoclassical equations\index{Neoclassical! equations} of motion\index{Neoclassical theory of radiation}. {\bf (b)}-{\bf (e)} Sample $Q$-functions plotted, respectively, for $(\eta/\kappa, \delta_p/\kappa)=(10,16); (18,1.1); (25,0); (30,0)$, respectively. Source: Fig. 2 of~\cite{CarmichaelPRX}. Reproduced with permission from the author.}
\label{fig:PhaseDiagramBist}
\end{figure}

For a cavity resonant with the two-state atom ($\Omega=\omega$) and coherently-driven with a detuning $\delta_p=\omega_p-\omega$, the Hamiltonian in the interaction picture reads
\begin{equation}\label{HJCDw}
\hat{H}^{\rm int}_{\rm JC}=-\hbar \delta_p (\hat{a}^{\dagger}\hat{a} + \hat{\sigma}_{+}\hat{\sigma}_{-})+ i\hbar g(\hat{a}^{\dagger}\hat{\sigma}_{-}-\hat{a}\hat{\sigma}_{+})+i\hbar \eta(\hat{a}^{\dagger}-\hat{a}).
\end{equation}
Resonant excitation of multiphoton transitions\index{Multiphoton! transition} in the JC model induces an additional semiclassical Rabi spitting, the so-called\index{Dressing of the dressed states} ``dressing of the dressed states''~\cite{TianCarmichael,BookQO2Carmichael}, arising from multiphoton blockade\index{Photon! blockade}\index{Multiphoton! blockade} in the strong-coupling regime ($g \gg 2\kappa, \gamma$). Shamailov and collaborators showed that this behavior is revealed in the first and second-order correlations of the quasi-elastically scattered radiation~\cite{Shamailov2010}. In particular, for the two-photon case ($|\delta_p|/g \approx 1/\sqrt{2}$), they resorted to a minimal model comprising only the first four dressed states of the JC Hamiltonian, namely,
\begin{subequations}\label{eq:4levels}
\begin{align}
 &\ket{0}\equiv\ket{0,g}, \\
 &\ket{1}\equiv \frac{1}{\sqrt{2}} (\ket{1,g} - \ket{0,e}), \\
 & \ket{2}\equiv \frac{1}{\sqrt{2}} (\ket{1,g} + \ket{0,e}), \\
 &  \ket{3}\equiv \frac{1}{\sqrt{2}} (\ket{2,g} - \ket{1,e}). 
\end{align}
\end{subequations}
Following a perturbative treatment which relies on expanding in powers of $\eta/g \ll 1$, they calculated the second-order correlation function\index{Second-order correlation function} of the light emitted by the cavity in the forwards direction ($\rightarrow$) applying the quantum regression formula\index{Quantum! regression formula},
\begin{equation}\label{eq:g2blockade}
 g^{(2)}_{\rm ss, \rightarrow}(\tau)=1 + e^{-\gamma |\tau|}[c_1 \cos(2\Omega_r \tau) + c_2 \sin(2\Omega_r |\tau|) + c_3 e^{-\gamma |\tau|} + c_4 \cos(\nu \tau)],
\end{equation}
where the four coefficients $c_1, c_2, c_3, c_4$ are functions of the effective two-photon Rabi frequency\index{Two-photon!Rabi frequency} $\Omega_r=2\sqrt{2}\eta^2/g$ and the dissipation rates (for $\gamma=2\kappa$). It turns out that bunching and antibunching\index{Antibunching} both occur depending on the driving strength. Weak excitation yields strong photon bunching, while $g^{(2)}_{\rm ss, \rightarrow}(0)$ decreases with increasing saturation\index{Saturation} of the $\ket{0} \to \ket{3}$ transition, reaching eventually its minimum value $16/25$. The last term on the RHS of eq.~\eqref{eq:g2blockade} is a rapid oscillation at frequency $\nu \approx 2g$ (plus perturbative corrections of order $\eta^2/g$ and higher). This oscillation is a quantum beat\index{Quantum! beat} which arises when the states $\ket{1}$ and $\ket{2}$ are prepared as a superposition,
\begin{equation}
 \ket{3} \to \sqrt{\frac{2}{3}} \left(\frac{\sqrt{2}-1}{2} \ket{1} + \frac{\sqrt{2}+1}{2} \ket{2} \right), 
\end{equation}
following the emission of a post-steady-state photon leaving the system in a mixed state. The two-photon dressing\index{Two-photon!dressing} of the dressed states appears as a semiclassical Rabi oscillation with frequency $2\Omega_r \ll \nu$ induced on top of the quantum beat (see fig. 5 of~\cite{Shamailov2010}). Photon bunching and fast oscillations at frequency $2g$ were observed in the intensity correlation function of the light transmitted from an optical microcavity strongly coupled to a gated quantum dot when driving the two-photon JC resonance\index{Two-photon!resonance}, in the experiment conducted by Najer and collaborators~\cite{Najer2019}.

In the absence of spontaneous emission\index{Spontaneous! emission} but with photon loss present, at a rate $2\kappa$, the evolution of the reduced density operator $\hat{\rho}$ is described by the Lindblad master equation\index{Master equation}
\begin{equation}\label{MEZDim}
\frac{d\hat{\rho}}{dt}=\frac{1}{i\hbar}\left[\hat{H}^{\rm int}_{\rm JC}, \hat{\rho}\right] + \kappa (2\hat{a}\hat{\rho}\hat{a}^{\dagger}-\hat{a}^{\dagger}\hat{a}\hat{\rho}-\hat{\rho}\hat{a}^{\dagger}\hat{a}).
\end{equation}
In connection with the anhamronic oscillator we introduced in sec.~\ref{sssec:drivejc}, the ME \eqref{MEZDim} on resonance ($\delta_p=0$) can be approximated by the following equation involving the $\hat{A}_u$ (or equally the $\hat{A}_l$) ladder operator:
\begin{equation}
 \frac{d\hat{\rho}}{dt}=\frac{1}{i\hbar}[\hat{H}_u, \hat{\rho}] + \kappa (2\hat{A}_u^{\dagger}\hat{\rho}\hat{A}_u - \hat{A}_u^{\dagger}\hat{A}_u \hat{\rho}-\hat{\rho}\hat{A}_u^{\dagger}\hat{A}_u),   
\end{equation}
where
\begin{equation}
\hat{H}_u \equiv \hbar \omega_0 \hat{A}_u^{\dagger}\hat{A}_u + \hbar g \sqrt{\hat{A}_u^{\dagger}\hat{A}_u} + i\hbar\eta(\hat{A}_u^{\dagger}-\hat{A}_u),    
\end{equation}
and we assume a small damping $\kappa \ll g$. At high excitation, this model reproduces the semiclassical result (see chap. 16 of~\cite{BookQO2Carmichael})
\begin{equation}
\braket{\hat{a}^{\dagger}\hat{a}}_{\rm ss} \approx \left|\frac{\eta}{\kappa + i g /(2\sqrt{\braket{\hat{a}^{\dagger}\hat{a}}_{\rm ss}})} \right|^2 \hspace{0.3cm}\Rightarrow\hspace{0.3cm} \braket{\hat{a}^{\dagger}\hat{a}}_{\rm ss} \approx (\eta/\kappa)^2-[g/(2\kappa)]^2.    
\end{equation}

The aforementioned phase transitions are organized by quantum fluctuations underlying complex-amplitude and phase bistability, portrayed by means of a {\it quasi}-probability distribution function for the intracavity field (in this case the $Q$-function) in fig. \ref{fig:PhaseDiagramBist}. The participation of quantum metastable states is quantified through the coefficient $r=1-|h_1-h_2|/(h_1+h_2)$, where $h_1$ and $h_2$ are the peak-heights of the {\it quasi}-probability  distribution. Complex-amplitude bistability gives its place to phase bistability upon approaching the critical point\index{Critical! point} $(\eta/\kappa, \delta_p/\kappa)=(g/2,0)=(25,0)$ and along the line $(\eta/\kappa, 0)$, with $\eta \geq g/2$. As we have already mentioned in subsec.~\ref{sssec:drivejc}, on resonance ($\delta_p=0$), the quasi-energy spectrum comprises the ground-state {\it quasi}energy ($e_0=0$) and the doublets $e_{n,\pm}$ of eq.~\eqref{eq:doubletsdrivenJC}, collapsing to a continuous spectrum at the critical point $\eta=g/2$\index{Continuous spectrum}. Making a notable extension, the driven JC model has been studied in connection to a charged Dirac particle subject to an external electromagnetic field. The transformation to a reference frame where the electric or the magnetic field vanishes is mapped to the scanning of a coherent drive strength across the critical point of the JC second-order phase transition~\cite{Dirac}\index{Phase transition! dissipative! quantum! second-order}.

The form of the coherently-driven JC model justifies the observation that for weak driving fields we expect the spectrum to consist of the JC ladder plus a perturbative level shift depending on the drive amplitude (of order $ (2\eta/g)^2$). In the opposite extreme, for strong driving fields, the roles of the interactions are reversed and the JC interaction becomes the perturbation. The dominant interaction becomes that with the external field, which has a continuous spectrum since it is formally a potential energy proportional to the position (or the momentum) of a harmonic oscillator. Hence, we anticipate a transition from a discrete to a continuous spectrum to occur at the critical point\index{Critical! point} $2\eta/g=1$, as argued in chap. 16 of~\cite{BookQO2Carmichael}. At the same time, the semiclassical analysis identifies a symmetry-breaking transition passing through threshold in the limit of zero system size: while in absorptive bistability the phases of the field amplitude are fixed by the driving field, when the limit $\gamma \to 0$ is taken {\it a priori}, the phases of the {\it individual} field states do not reflect the phase of the external field (the same happens for the atomic states on the Bloch sphere). Above threshold, the induced dipole moment changes under the influence of the intracavity field while at the same time radiates into the cavity mode  whence the intracavity field.  This phenomenon was termed {\it spontaneous dressed-state polarization}\index{Spontaneous! dressed-state polarization} by Alsing and Carmichael~\cite{alsing1991spontaneous}. It can be described by means of an analogy with a magnetic moment moving under the influence of a time-varying magnetic field, dynamically coupled to the magnetic moment so that the pair either align or anti-align with each other in steady state\index{Spontaneous! dressed-state polarization}\index{Steady state}. In particular, the Bloch equations can be written in the form\index{Bloch! equations}
\begin{equation}
\dot{\boldsymbol{\sigma}}=\boldsymbol{B} \times \boldsymbol{\sigma},    
\end{equation}
with 
\begin{equation}
 \boldsymbol{\sigma}=(v_x, v_y,m), \quad \quad  \boldsymbol{B}=2g(-z_y,z_x,0),  
\end{equation}
where $z=e^{i\omega_0 t} \braket{\hat{a}}$, $v=2 e^{i\omega_0 t}\braket{\hat{\sigma}_{-}}$, and $m=2\braket{\hat{\sigma}_z}$. Then, $\boldsymbol{\sigma}$ is stationary if it either aligns or anti-aligns with the field $\boldsymbol{B}$, which is not a prescribed but a dynamical field. Hence, the new radiation states must be determined self-consistently.

Let us here mention the conventional optical bistability (further discussed in sec.~\ref{sssec:optbis}) in order to appreciate the origin of the zero system-size limit. In the presence of spontaneous emission\index{Spontaneous! emission} (at rate $\gamma$), the steady-state solution\index{Steady state!solution} of the Maxwell-Bloch equations\index{Maxwell-Bloch! equations} for the intracavity amplitude on resonance reads
\begin{equation}\label{MBres}
\alpha_{\rm ss}=-i \eta \left[\kappa + \frac{2g^2}{\gamma(1+8g^2 |\alpha_{\rm ss}|^2/\gamma^2)}\right]^{-1},
\end{equation}
suggesting that the nonlinearity can no longer be treated as a negligible perturbation for photon numbers $|\alpha|^2 \sim n_{\rm sat, wc} \equiv \gamma^2/(8g^2)$. The corresponding ``thermodynamic limit'' is a {\it weak-coupling (wc) limit}\index{Thermodynamic limit! weak-coupling}, since taking $g$ to zero implies that both $n_{\rm sat, wc}$ and the mode volume $V$ tend to infinity. This is precisely the scale parameter for absorptive optical bistability\index{Optical! bistability}. Single-atom bistability was demonstrated by obtaining numerical solutions for the quantum-mechanical density operator in the late 1980s \cite{savage1988single}. On the other hand, the mean-field photon number obtained from the steady-state solution\index{Steady state!solution} of the neoclassical equations\index{Neoclassical! equations} on resonance ($\delta_p=0$), reading
\begin{equation}
\frac{|\alpha_{\rm ss}|^2}{n_{\rm sat, sc}}\left[\frac{|\alpha_{\rm ss}|^2}{n_{\rm sat, sc}} + 1 - \left(\frac{2\eta}{g}\right)^2\right]=0,
\end{equation}
with $n_{\rm sat, sc}=g^2/(2\kappa)^2$, indicates a {\it strong-coupling (sc) limit}, where the scale parameter increases with the coupling strength\index{Strong-coupling ``thermodynamic limit''}\index{Thermodynamic limit! strong-coupling}. In the regime of photon blockade\index{Photon! blockade}, we find a constant disparity between the semiclassical predictions and the full quantum treatment as the ``thermodynamic limit'' (with $n_{\rm sat, sc} \to \infty$) is approached [see fig. 4(a) of~\cite{CarmichaelPRX}]. Increasing the spontaneous emission\index{Spontaneous! emission! rate} rate from zero to a finite value provides the link between the weak and strong-coupling limits defined above, bringing about significant changes in the response of the JC oscillator as depicted in the drive-detuning-amplitude plane (see {\it e.g.}, fig. 7 of~\cite{CarmichaelPRX}). On the experimental side, and in particular in circuit QED (see also section \ref{sec:cirQED}), the first-order dissipative quantum phase transition\index{Dissipative! quantum phase transition} of the driven JC model was evidenced via recorded trajectories, {\it quasi}-probability distribution functions, and vacuum Rabi spectra, for a setup using a driven {\it coplanar waveguide resonator} coupled to up to three transmon qubits (see section~\ref{sec:cirQED})~\cite{fink2017observation}. In a circuit QED setup and for the generalized JC model (see sec. \ref{ssec:JCspectrcQED}), Sett and coworkers have recently reported on bistable switching with a characteristic dwell time as high as 6 seconds between a bright coherent state with $\approx 8 \times 10^3$ intra-cavity photons and the vacuum state with equal probability, for $g/\kappa=287$~\cite{Sett2022}. 

From a more formal point of view, the critical theory for the breakdown of photon blockade in terms of the spontaneous breaking\index{Spontaneous! symmetry breaking} of an anti-unitary PT symmetry has been very recently developed in~\cite{Curtis2021}. In a generalized Jaynes-Cummings-Rabi model with three control parameters, the breakdown of photon blockade\index{Photon! blockade!breakdown} has been unified with the dissipative extension of the Dicke quantum phase transition\index{Dicke! phase transition}\index{Dissipative! quantum phase transition} (see sec.~\ref{sssec:dicke}) through a previously unreported phase of the Dicke model~\cite{DickePhotonBlockade2018}. A dynamical multilevel atom-cavity blockade and its breakdown transition in time were identified in~\cite{Clark2022}. As with optical bistability, atoms initially block transmission by effectively detuning the cavity mode from the coherent driving field. The interacting system, however, eventually reaches an uncoupled state via a critical runaway process -- substantiated by off resonant circular polarized light -- which gives rise to maximum transmission.  

In concluding this entire subsection, we note that the critical response of the driven JC model on resonance has been intimately tied to the inception of quantum trajectory theory \index{Quantum! trajectories} in the beginning of the 1990s (see sec. 5 of~\cite{alsing1991spontaneous}). This early application of quantum trajectories on macroscopic phase switching targeted a stochastic process in Hilbert space conditioned on a record of scattering events; concurrently, quantum trajectories were introduced in~\cite{Dalibard1992Traj, Dum1992Traj} in somewhat different ways. In~\cite{Dalibard1992Traj}, the algorithm for computing the source-field correlation function from quantum trajectories is presented in the frame of what the authors term a Monte Carlo Wavefunction (MCWF) approach\index{Monte Carlo!wavefunction}. Three years back in time, in 1989, Carmichael and coworkers had calculated the waiting-time distribution\index{Distribution! of waiting times} between photoelectric counts (see also the perspective offered by~\cite{Brandes2008}) in resonance fluorescence\index{Resonance fluorescence! waiting-time distribution} based on the Kelley-Kleiner\index{Kelley-Kleiner theory of photoelectric detection} theory of photoelectric detection~\cite{KelleyKleiner1964} (as well as on modified Bloch equations). In their report~\cite{CarmichaelSingh1989}, they conclude that ``...photon emission may be viewed as a realized process underlying observed photoelectron counting sequences. The detector simply records an emitted photon with probability $\eta^{\prime}$, or fails to record it with probability $(1-\eta^{\prime})$. From this point of view the atomic state reduction associated with each photoelectric count is caused by the irreversible interaction of the atom with the multimode vacuum\index{Multimode! vacuum}, not by the detection process itself...''.   

Quantum trajectories realize records that disentangle the system and environment without discarding all correlations. The environment is not merely traced out, as is done in the derivation of the Lindblad master equation. The obtained records describe the environment in terms of time series of real numbers. In that sense, the description is not complete in capturing the full Schr\"odinger state; the trajectories reveal an aspect of the physical reality, and they do so in a way that allows us to talk about the system plus environment complex with reference to its correlated parts. Despite being incomplete in the sense we pointed out, they are complete in the sense that every scattered photon is accounted for in the making of the record, which is then able to answer any question we might ask about the photon counting statistics. However, the scattered field has other properties like a wave amplitude and spectrum, in addition to the particle number. Considering the possibility of devising an unraveling which is able to reveal these attributes, we encounter a recurring theme in the work of Bohr, namely the aspect of {\it complementarity}\index{Complementarity/contextuality}: since the photon is indivisible whenever it is irreversibly exchanged, then each quantum causes one and only one happening. It follows that it is impossible to make records of different kind in parallel~\cite{CarmichaelQJRev}. In sec.~\ref{subsec:WPCorrelator}, we will meet an unraveling based on the dual wave/particle nature of light: homodyne measurements\index{Homodyne detection} of the field quadrature (wave aspect) are performed, conditioned on a photon click (particle aspect) registered by a detector~\cite{CarmichaelFosterChapter}.  

Let us turn to a particular example of one of the earliest quantum trajectory unravelings. The formalism developed in~\cite{alsing1991spontaneous} separated the spontaneous emission\index{Spontaneous! emission} events that cause switching between excitation paths formed by the two JC ladders {\it at high excitation with mean photon number $\overline{n} \gg 1$} (with states denoted by $\ket{u}=(1/\sqrt{2})(\ket{+} + i \ket{-})$ - upper and $\ket{l}=(1/\sqrt{2})(\ket{+} - i \ket{-})$ - lower) from the independent evolution along each path that takes place between the jumps. For that purpose, the time-dependent system density operator (two-state atom plus cavity mode) is written as a Dyson expansion\index{Dyson expansion} of the form\index{Master Equation!unraveling} 
\begin{equation}
\begin{aligned}
\hat{\rho}(t)&=\sum_{k=0}^{\infty}\int_{t_0}^{t}dt_k \int_{t_0}^{t_k}dt_{k-1}\cdots \int_{t_0}^{t_2}dt_1 \exp{[(\mathcal{L}-\mathcal{S})(t-t_k)]}\mathcal{S}\exp{[(\mathcal{L}-\mathcal{S})(t_k-t_{k-1})]}\mathcal{S}\cdots\\
&\quad\quad\cdots \mathcal{S}\exp{[(\mathcal{L}-\mathcal{S})(t_2-t_1)]}\mathcal{S} \exp{[(\mathcal{L}-\mathcal{S})(t_1-t_0)]}\hat{\rho}(t_0).
\end{aligned}
\end{equation}
where $\mathcal{S}$ is the super-operator which describes the collapse of the quantum state at the emission times $t_1, t_2, \ldots,t_k$ (with $t_0 \leq t_1 \leq t_2\leq \ldots t_k \leq t$). The propagator\index{Propagator} $\exp{[(\mathcal{L}-\mathcal{S})(t_k-t_{k-1})]}$ describes the evolution between the emission times $t_{k-1}$ and $t_k$. For a given sequence of spontaneous emission times\index{Spontaneous! emission! times}, the quantum trajectory is conditioned on all earlier emissions and is given by the conditional density operator (denoted by the subscript $c$), 
\begin{equation}
\hat{\rho}_c(t)=\begin{cases}
\displaystyle\frac{\exp{[(\mathcal{L}-\mathcal{S})(t-t_{k-1})]}\hat{\rho}_c(t_{k-1})}{{\rm tr}[\exp{[(\mathcal{L}-\mathcal{S})(t-t_{k-1})]}\hat{\rho}_c(t_{k-1})]}, \quad \quad t_{k-1} \leq t < t_k \\ \\
\displaystyle\frac{\exp{\mathcal{S}[(\mathcal{L}-\mathcal{S})(t_k-t_{k-1})]}\hat{\rho}_c(t_{k-1})}{{\rm tr}[\mathcal{S}\exp{[(\mathcal{L}-\mathcal{S})(t_k-t_{k-1})]}\hat{\rho}_c(t_{k-1})]}, \quad \quad t=t_k
\end{cases}
\end{equation}
For the resonantly driven JC oscillator, in the presence of photon decay at a rate $2\kappa$ and spontaneous emission\index{Spontaneous! emission! rate} at rate $\gamma$ (with $g \gg 2\kappa, \gamma$), the switching transitions are defined through the action of the super-operators for any system operator $\hat{O}$,
\begin{equation}\label{superopJC}
\begin{array}{c}
\mathcal{S}\hat{O}=\tfrac{1}{4}\gamma(\hat{d}_{-}\hat{O}\hat{d}_{+} + \hat{d}_{+}\hat{O}\hat{d}_{-}),\\ \\
(\mathcal{L}-\mathcal{S})\hat{O}=-i g/(2\sqrt{\overline{n}}) \frac{1}{2} \left(\hat{d}_z [\hat{a}^{\dagger}\hat{a}, \hat{O}]+ [\hat{a}^{\dagger}\hat{a}, \hat{O}] \hat{d}_z\right) + \eta[\hat{a}^{\dagger}-\hat{a}, \hat{O}]+ \kappa (2\hat{a}\hat{O}\hat{a}^{\dagger} - \hat{a}^{\dagger}\hat{a} \hat{O}- \hat{O}\hat{a}^{\dagger}\hat{a}) + (\gamma/4) (\hat{d}_z \hat{O} \hat{d}_z - 2 \hat{O}),
\end{array}
\end{equation}
where $\hat{d}_{+}=|u\rangle \langle l|=\hat{d}_{-}^{\dagger}$ is the operator responsible for the switching between the two JC ladders\index{Jaynes-Cummings! ladder}, while $\hat{d}_z=|u\rangle\langle u| - |l\rangle\langle l|$ leaves the excitation path unaltered. In eq. \eqref{superopJC}, $\pm g/(2 \sqrt{\overline{n}})$ are the field detunings from the upper($+$) and the lower($-$) paths, introduced in eq.~\eqref{eq:sqrtnshift} (see also fig. 1 of~\cite{alsing1991spontaneous}). At the end of 1992, quantum trajectory simulations of the two-state behavior of the JC oscillator response, when excited near one of the vacuum Rabi resonances\index{Vacuum Rabi! resonance}, were reported in~\cite{TianCarmichael}. The Stark splitting is demonstrated via a simulation of the optical spectrum using a cascaded-system formalism\index{Cascaded! systems} (source plus scanning interferometer)\index{Cascaded! systems}. The authors characteristically remark in the last paragraph of their report: ``The quantum trajectory approach allows us to visualize the dynamics of a photoemissive source free from the straightjacket of classical diffusion models. It is particularly useful in cavity quantum electrodynamics where the standard diffusion models break down.'' In 2004, the phase bimodality of the resonantly driven JC model was used as a characteristic case to exemplify contextual entanglement based on quantum trajectory unravelings of the open system dynamics~\cite{Nha2004Ent}. Therein, it was directly verified that ``the degree of entanglement in an open quantum system varies according to how information in the environment is read''. It's also worth mentioning that the quantum trajectory formalism was used in the unraveling of the master equation\index{Master equation!cascaded systems}\index{Cascaded! systems} for cascaded quantum open systems~\cite{Carmichael1993QOpen} in 1993, simultaneously with Gardiner's quantum Langevin equations\index{Heisenberg-Langevin equations}~\cite{Gardiner1993}. A recent report~\cite{Kiilerich2019} by Kiilerich and M{\o}lmer presents a related derivation (based on the input-output formalism\index{Input-output! formalism}) and application of a master equation\index{Master equation} where the input and output pulses are treated as single-oscillator modes that both couple to a local system in a cascaded fashion. 

\subsubsection{Old quantum theory revisited}
\label{subsubsec:oldquantum}

The master equation\index{Master equation!resonance fluorescence} of resonance fluorescence\index{Resonance fluorescence! master equation} describes the quantum state of a two-level atom alone, after tracing over every mode of the radiation field. To provide a consistent description of thermal jumps upwards, which cannot reasonably be identified with a detector ``click'', Chough and Carmichael proposed the idea to raise one mode of the field to the same status as the atom by including it, together with its interaction with the atom, in the system Hamiltonian~\cite{Thermaljumps2001}. All other modes are to be treated as a reservoir, as usual, and their interaction with the atom is to be described by quantum jumps. Let us first connect the problem to the `old quantum theory' by considering a two-state atom in thermal equilibrium with Planck radiation at temperature $T$\index{Blackbody radiation}. In the Einstein $A$ and $B$ theory\index{Einstein $A$ and $B$ theory} photons are exchanged between the atom and the radiation field as the atom jumps randomly between its two stationary states. The jump rates follow a prescription that includes spontaneous emission\index{Spontaneous! emission}, stimulated emission\index{Stimulated! emission! rate}, and absorption\index{Stimulated! absorption! rate}, with rates
\begin{align}
 \Gamma_{\rm down}&=A + B\, \sigma(\omega_0),\\
 \Gamma_{\rm up}&=B\, \sigma(\omega_0),
\end{align}
where
\begin{equation}
 \sigma(\omega_0)=\overline{n}(\omega_0)\hbar \omega_0 [\rho(\omega_0)/V]
\end{equation}
is the energy density of the radiation field at the resonance frequency $\omega_0$ of the atom, with average photon number per mode
\begin{equation}
 \overline{n}(\omega_0)=[e^{\hbar \omega_0/(k_B T)}-1]^{-1}
\end{equation}
and mode density, in volume $V$,
\begin{equation}
 \rho(\omega_0)=\frac{\omega_0^2 V}{\pi^2 c^3}.
\end{equation}
The Einstein $A$ and $B$ coefficients should satisfy
\begin{equation}
 \frac{B}{A}=\frac{\pi^2 c^3}{\hbar \omega_0^3}
\end{equation}
in order for the two-state atom to be brought into thermal equilibrium with the radiation field. With the help of the above relation we can write
\begin{subequations}
\begin{align}
 \Gamma_{\rm down}&=A[\overline{n}(\omega_0) + 1],\label{eq:einst1} \\
 \Gamma_{\rm up}&=A\overline{n}(\omega_0).\label{eq:einst2}
\end{align}
\end{subequations}
Fermi's golden rule assigns a value to the coefficient $A$ as
\begin{equation}
 A=2\pi \sum_{\lambda}\int d\Omega \, \rho(\omega_0)|\kappa_{\lambda,\hat{\boldsymbol{n}}}(\omega_0)|^2,
\end{equation}
where
\begin{equation}\label{eq:magnkappa}
 |\kappa_{\lambda,\hat{\boldsymbol{n}}}(\omega_0)|=\sqrt{\frac{\omega_0}{2\pi \epsilon_0 V}}|\hat{e}_{\lambda,\hat{\boldsymbol{n}}}\cdot \boldsymbol{d}_{eg}|
\end{equation}
is the dipole coupling strength to a mode of the radiation field with polarization $\lambda$ and direction of propagation specified by the unit vector $\hat{\boldsymbol{n}}$. The polarization vector is $\hat{e}_{\lambda,\hat{\boldsymbol{n}}}$, while $\boldsymbol{d}_{eg}$ is the atomic dipole matrix element.
\begin{figure}
 \includegraphics[width=10cm]{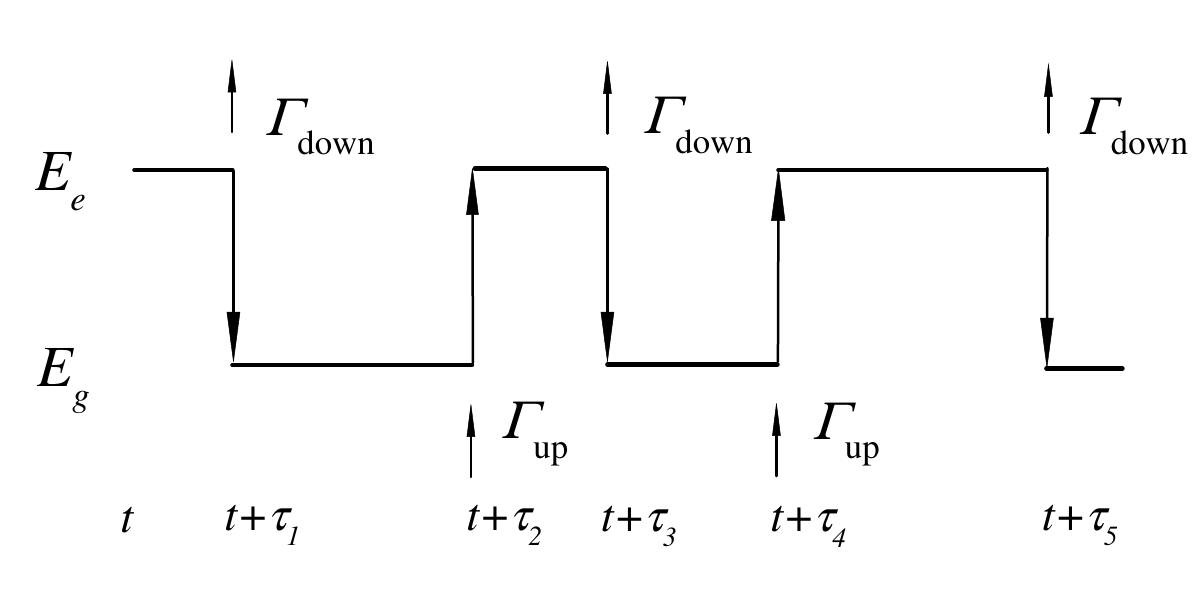}
 \caption{Sample realization of the Einstein stochastic process, associated with a record of certain jump times and jump types. Source:Fig. 2 of~\cite{Thermaljumps2001}.}
 \label{fig:Einsteinprocessfig}
\end{figure}

The Einstein theory defines a stochastic process, one that may be visualized in terms of quantum jumps whose occurrences unfold randomly in time. An individual realization of such a process is depicted in fig.~\ref{fig:Einsteinprocessfig}. Each realization of this process is associated  with a record of two jump types and a sequence of jump times. As a matter of fact, although Bohr has himself put forward the notion of a quantum jump to explain the association of stationary-state energy differences with electromagnetic wave frequencies in his proposed model of the hydrogen atom, he became quite dissatisfied with the idea in the concrete form it acquired in Einstein's theory. The notion of the light quantum, in particular, troubled him the most. Bohr maintained that since so many optical phenomena rely on the continuity of coherent waves, the wave nature of light could simply not be dismissed. He was then unavoidably drawn towards some kind of merging of the two ideas~\cite{CarmichaelTalk2000}. 

What has come to be known as the Bohr-Kramers-Slater (BKS) proposal~\cite{BKS1924, Slater1924}\index{BKS proposal} posits that during the residence times in a stationary state, an atom {\it is not inactive} in its interaction with the electromagnetic field; rather, it acts through a coherent dipole radiator -- a ``virtual oscillator -- constantly radiating an electromagnetic wave of frequency $\omega_0=(E_e-E_g)/\hbar$. This wave is either in phase or out of phase with the external radiation at frequency $\omega_0$ depending on whether the residence is in the stationary state with energy $E_e$ or $E_g$. This way, there is an energy transfer either from the dipole to the electromagnetic field or in the reverse direction, depending on the stationary state -- this is dictated by the laws of classical electrodynamics. Wave-based physics motivated Einstein~\cite{Einstein1916, Einstein1917} to include stimulated emission\index{Stimulated! emission! jump} jumps (in the present nomenclature) together with absorption jumps. In~\cite{Einstein1917} we read: ''If a Planck resonator is located in a radiation field, the energy of the resonator is changed through the work done on the resonator by the electromagnetic field of the radiation; this work can be negative or positive depending on the phases of the resonator and the oscillating field. We correspondingly introduce the following hypothesis. ...'' The goal was to eliminate the quanta emitted and absorbed at the times of jumps although still permitting the atom to jump. The proposal thus attempted to retain yet keep separate two incompatible mechanisms for energy exchange: a wave mechanism for the change of material state energies (continuous) and a particle mechanism for the change of material state energies (discontinuous). The proposal foundered on the violation of energy conservation at the level of individual quantum events, contradicting the findings of Compton scattering experiments~\cite{Bothe1925,Compton1925}.

An essential omission from the BKS proposal is the label on the state vector, ${\rm REC}$ (compare figs. 1 and 12 of~\cite{CarmichaelTalk2000}). Through this label, the state of the incoming light field is allowed to depend on the history of the data record -- the detection events that have already taken place. At the time of each event, an operator representing the light field acts on the state vector to annihilate a light particle, and in doing so updates the state of the incoming light to be consistent with the obtained record of photoelectric counts. In this way, correlations at the level of individual quantum events are taken into account. Let us here contrast the Einstein $A$ and $B$ theory to a quantum trajectory description of a coherent field of amplitude $\mathcal{E}$, resonantly exciting the two-state atom. Due to the induced coherence, it is necessary that the system state be a superposition of the stationary states $\ket{E_e}$ and $\ket{E_g}$, which we denote precisely by $\ket{\psi_{\rm REC}(t)}$. As the BKS theory suggests, there is a coherent interaction with the electromagnetic field between the quantum jumps. This is accounted for by a continuous evolution under the Schr\"{o}dinger equation
\begin{equation}
 \frac{d\ket{\overline{\psi}_{\rm REC}}}{dt}=\frac{1}{i\hbar}\hat{H}_B \ket{\overline{\psi}_{\rm REC}},
\end{equation}
with non-Hermitian Hamiltonian
\begin{equation}
 \hat{H}_B=\tfrac{1}{2}\hbar(\omega_0 - i \Gamma_{\rm down})|E_e \rangle \langle E_e| - \tfrac{1}{2}\hbar(\omega_0 + i \Gamma_{\rm up}) |E_g \rangle \langle E_g| + i\hbar \mathcal{E}\left(e^{i\omega_0 t} |E_g \rangle \langle E_e| - e^{-i\omega_0 t} |E_e \rangle \langle E_g|\right),
\end{equation}
in the dipole and rotating wave approximations. The quantum jumps are governed by the probabilistic rules of the Einstein $A$ and $B$ theory (first linked to Schr\"{o}dinger's equation by Dirac~\cite{Dirac1926, Dirac1927}), generalized here to account for the fact that the system is not definitely in a particular stationary state at any time. There are jumps
\begin{align}
 &\ket{\overline{\psi}_{\rm REC}} \xrightarrow[]{\Gamma_{\rm down}} (|E_g \rangle \langle E_e|)\ket{\overline{\psi}_{\rm REC}}, \\
  &\ket{\overline{\psi}_{\rm REC}} \xrightarrow[]{\Gamma_{\rm up}} (|E_e \rangle \langle E_g|)\ket{\overline{\psi}_{\rm REC}},
\end{align}
with rates
\begin{align}
 R_{\rm down}&=\Gamma_{\rm down}\left|\braket{E_e|\overline{\psi}_{\rm REC}}\right|^2, \\
 R_{\rm up}&=\Gamma_{\rm up}\left|\braket{E_g|\overline{\psi}_{\rm REC}}\right|^2.
\end{align}
For a sufficiently large drive amplitude in excess of the spontaneous emission rate\index{Spontaneous! emission! rate}, the dominant mechanism for evolution between the stationary states is a coherent Rabi oscillation -- a coherent nonperturbative evolution that only became accessible with the invention of the laser. This nonperturbative coherence was not anticipated by the BKS proposal.
\begin{figure}
 \includegraphics[width=16cm]{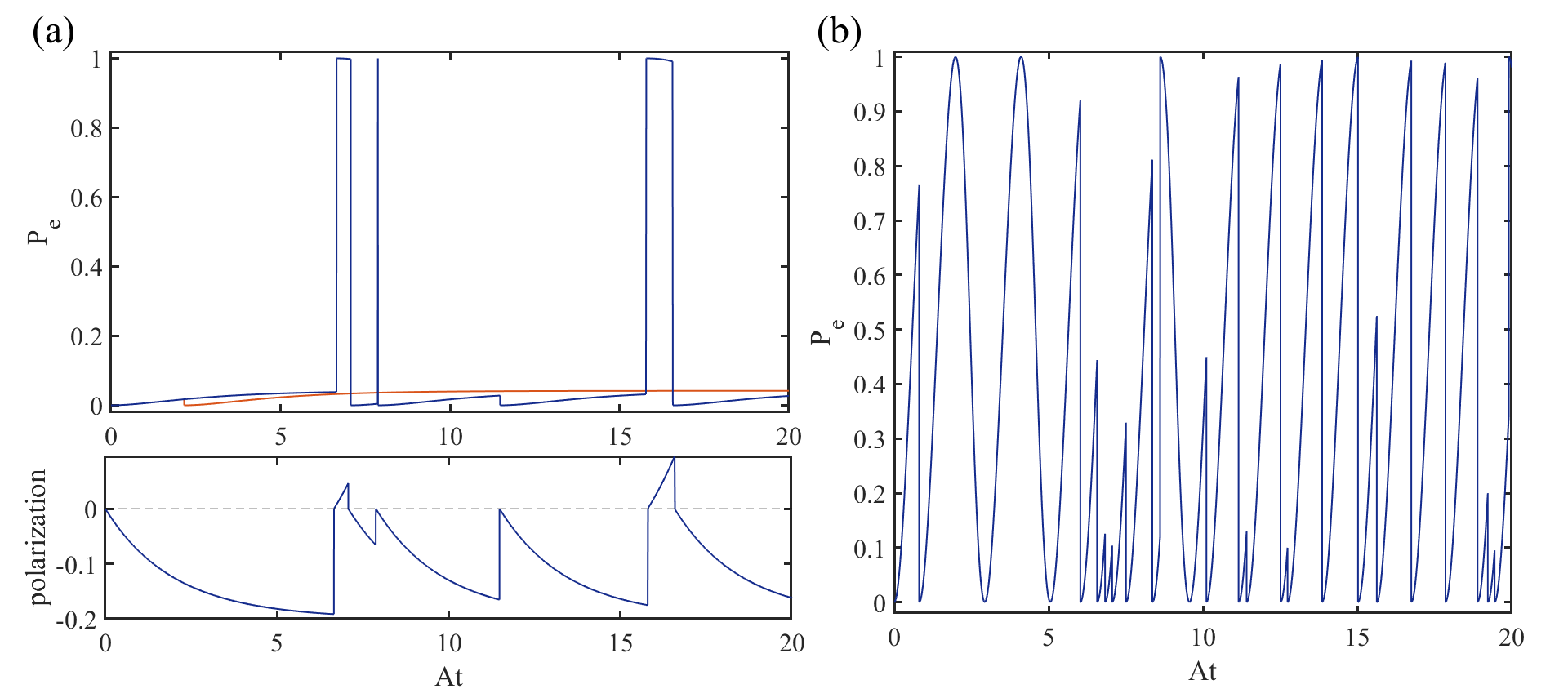}
  \includegraphics[width=17cm]{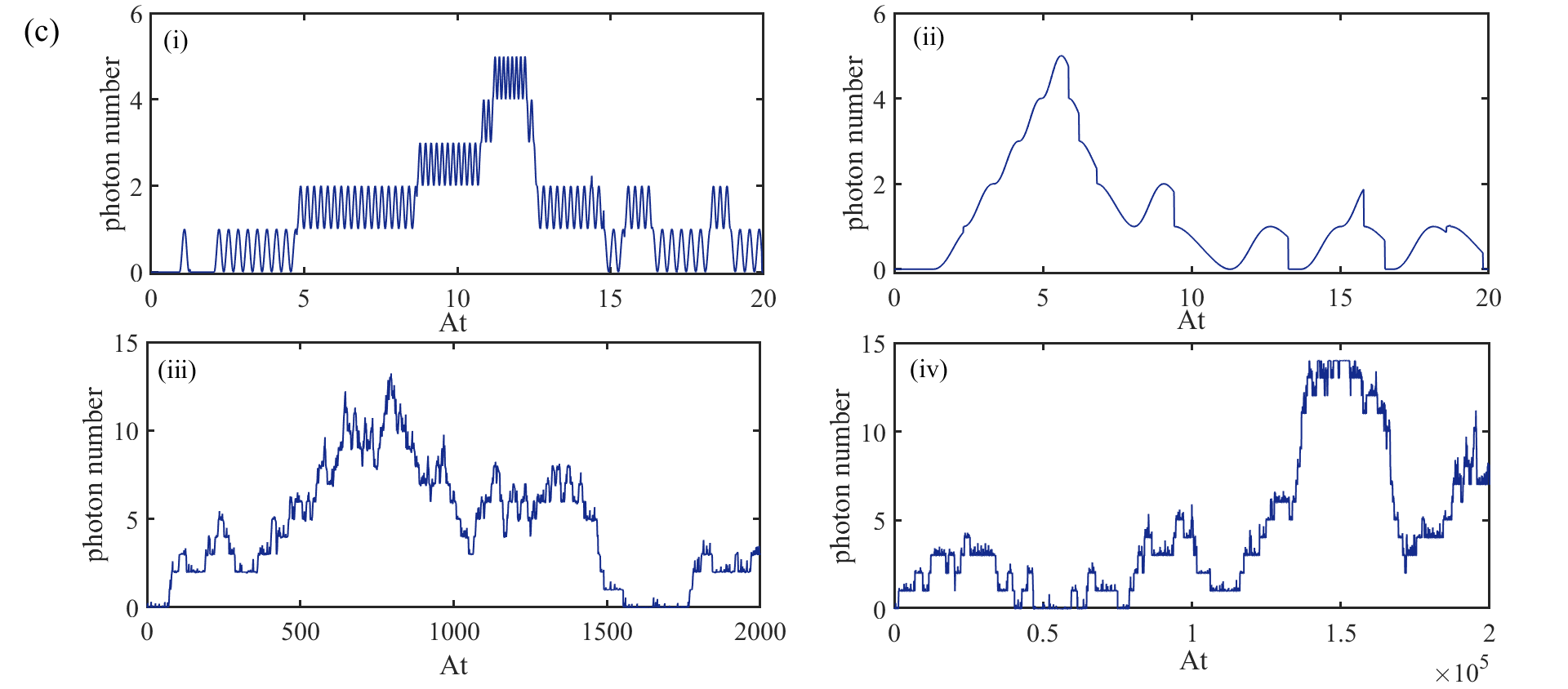}
  \caption{{\bf (a, b)} Sample quantum trajectories obtained by a Monte Carlo algorithm, with both thermal jumps and induced coherence (external driving by a coherent field $\mathcal{E}$). $P_e$ is the probability to find the atom in the excited state. In {\bf (a)} $\mathcal{E}/A=0.1$, while in {\bf (b)} $\mathcal{E}/A=1.5$. In both cases we take $\overline{n}(\omega_0)=0.25$. In (a), the trajectory in orange is plotted for $\overline{n}(\omega_0)=0$ (ordinary resonance fluorescence). {\bf (c)} Sample quantum trajectories for the photon number expectation of a resonant single-field mode included as part of the system. We take $|\kappa_{\lambda,\hat{\boldsymbol{n}}}(\omega)|/A=10, 1.0, 0.1$ and $0.01$ in (i), (ii), (iii) and (iv), respectively. In all subframes of (c), $\overline{n}(\omega_0)=1$.}
  \label{fig:thermaljumpsfig}
\end{figure}

Figure~\ref{fig:thermaljumpsfig} illustrates sample realizations of this stochastic process. In (a) the coherent excitation is relatively weak, and the overall form of the evolution remains close to that predicted by the Bohr-Einstein quantum jumps. In addition of the switching of the energy there is, however, a weak induced coherence carried along by the continuous evolution between jumps as a nonvanishing polarization amplitude, reminding us of the BKS virtual oscillator. In (b), the coherent excitation is much stronger, and the dominant mechanism for the evolution between stationary states is a coherent Rabi oscillation. This is specifically a form of evolution predicted by the Schr\"{o}dinger equation, hence it could not have been anticipated by the BKS proposal. 

At sufficiently low temperatures, when $\overline{n}(\omega_0) \ll 1$, the jump record that labels the state can be mostly made by detectors monitoring the scattered light. Almost all jumps will be down-jumps at a rate equal to $A$; they may be identified with emitted photons counted as isolated excitations of the vacuum. While this may be sufficient for numerous applications of quantum optics, thermal jumps cannot be set aside from a fundamental point of view. The authors of~\cite{Thermaljumps2001} appeal to the language of old quantum mechanics and show that these jumps are consistent with the Schr\"{o}dinger equation in the weak-coupling limit. 

The Hamiltonian for a two-state atom interacting with the radiation field of a thermal environment is
\begin{equation}\label{eq:Hammanymodes}
 \hat{H}=\tfrac{1}{2}\hbar \omega_0 |E_e \rangle \langle E_e| - \tfrac{1}{2}\hbar \omega_0 |E_g \rangle \langle E_g| + \sum_{\lambda^{\prime}, \hat{\boldsymbol{n}}^{\prime},\omega^{\prime}}\hbar \omega^{\prime} \hat{r}^{\dagger}_{\lambda^{\prime}, \hat{\boldsymbol{n}}^{\prime},\omega^{\prime}} \hat{r}_{\lambda^{\prime}, \hat{\boldsymbol{n}}^{\prime},\omega^{\prime}} + \sum_{\lambda^{\prime}, \hat{\boldsymbol{n}}^{\prime},\omega^{\prime}}\hbar \left[\kappa_{\lambda^{\prime}, \hat{\boldsymbol{n}}^{\prime}}(\omega^{\prime})|E_e \rangle \langle E_g|\hat{r}_{\lambda^{\prime}, \hat{\boldsymbol{n}}^{\prime},\omega^{\prime}} + \textrm{H.c.} \right],
\end{equation}
where $\hat{r}_{\lambda^{\prime}, \hat{\boldsymbol{n}}^{\prime},\omega^{\prime}}$ and $\hat{r}^{\dagger}_{\lambda^{\prime}, \hat{\boldsymbol{n}}^{\prime},\omega^{\prime}}$ are creation and annihilation operators for the field mode with polarization $\lambda$, frequency $\omega$, propagating along the direction $\hat{\boldsymbol{n}}$; $\kappa_{\lambda, \hat{\boldsymbol{n}}}(\omega)$ is the mode coupling whose magnitude is defined in eq.~\eqref{eq:magnkappa}. The stochastic process we described above can be formally built around the master equation~\index{Master equation} derived from the Hamiltonian~\eqref{eq:Hammanymodes}. This master equation describes the quantum state of the atom alone after tracing out all modes of the radiation field. The proposal of~\cite{Thermaljumps2001} consists in raising one mode of that field to the same status as the atom by including it and its interaction with the atom in the system Hamiltonian -- reminiscent of a JC coupling but in the absence of a cavity. All other modes are to be treated as a bath like we did previously; their interaction with the atom gives rise to the quantum jumps. The stochastic process remains the same, but with the non-Hermitian Hamiltonian $\hat{H}_B$ replaced by
\begin{equation}
 \hat{H}_B=\tfrac{1}{2}\hbar(\omega_0 - i \Gamma_{\rm down})|E_e \rangle \langle E_e| - \tfrac{1}{2}\hbar(\omega_0 + i \Gamma_{\rm up}) |E_g \rangle \langle E_g| + \hbar \omega \hat{r}^{\dagger}_{\lambda, \hat{\boldsymbol{n}},\omega}\hat{r}_{\lambda, \hat{\boldsymbol{n}},\omega} + \hbar \left[\kappa_{\lambda, \hat{\boldsymbol{n}}}(\omega)|E_e \rangle \langle E_g|\hat{r}_{\lambda, \hat{\boldsymbol{n}},\omega} + \textrm{H.c.} \right].
\end{equation}

Removing one mode from the reservoir of infinite modes does not affect the overall jump rates for the atom. What changes here is that we can follow the evolution of an explicit Hilbert space vector for the selected mode, one entangling this mode with the two-state atom. The natural question to ask, now, is whether this mode experiences quantum jumps. 

Figure~\ref{fig:thermaljumpsfig} (c) shows individual realizations for the selected-mode photon number expectation with decreasing coupling strengths in frames (i)-(iv) and assuming resonance with the atom. When the coupling is strong compared to the Einstein coefficient $A$, coherent Rabi oscillations are observed. There are discontinuous changes which, in the case of strong coupling, are a direct manifestation of the quantum jumps for the atom. The monotonous sequence of jumps ``up'' and ``down'' is interrupted by repeated up-jumps which can transfer many energy quanta to the field mode. At an intermediate coupling strength [frame (ii)], Rabi oscillations are still present, but when the coupling becomes weak these disappear. Instead, a new kind of evolution sets in and the jumps proceed at a rate far less than the total jump rate of the atom. Their rate decreases with the square of the coupling constant [frames (iii) to (iv)].   

To demonstrate the self-consistency of thermal trajectories, we appeal to the coherent evolution based on Schr\"{o}dinger equation. Let us denote the number of energy quanta shared between the atom and field mode at any time $t$ by $n_{\omega}+1$. We denote the time of the very last jump of the atom by $t_k$ and let $t_{k+1}$ be the time of the jump that is to occur next. In between these times, for $t_k < t < t_{k+1}$, the entangled state of the atom and field mode may be expanded as 
\begin{equation}
 \ket{\overline{\psi}_{\rm REC}(t)}=\overline{C}_{e|i}(t)\ket{E_e}\ket{n_{\omega}} + \overline{C}_{g|i}(t)\ket{E_g}\ket{n_{\omega}+1},
\end{equation}
with initial conditions $\overline{C}_{e|i}(0)=\delta_{e,i}$ and $\overline{C}_{g|i}(0)=\delta_{g,i}$ [where $i=e$ ($i=g$) for an up (down) jump at time $t_k$]. The equations of motion for the conditional state amplitudes are 
\begin{subequations}
\begin{align}
 \frac{d \tilde{C}_{e|i}}{dt}&=-\tfrac{1}{2}(\Gamma_{\rm down}-i\Delta \omega)\tilde{C}_{e|i}-i|\kappa_{\lambda,\hat{\boldsymbol{n}}}(\omega)|\sqrt{n_{\omega}+1}\,\tilde{C}_{g|i},\label{eq:eomjmumps1}\\
 \frac{d \tilde{C}_{g|i}}{dt}&=-\tfrac{1}{2}(\Gamma_{\rm up}+i\Delta \omega)\tilde{C}_{g|i}-i|\kappa_{\lambda,\hat{\boldsymbol{n}}}(\omega)|\sqrt{n_{\omega}+1}\,\tilde{C}_{e|i},\label{eq:eomjmumps2}
\end{align}
\end{subequations}
where we have defined
\begin{subequations}
\begin{align}
 \overline{C}_{e|i}(t)&\equiv e^{-i(n_{\omega}+\tfrac{1}{2})\omega t}e^{i\tfrac{1}{2}\phi_{\hat{\boldsymbol{n}}}(\omega)}\tilde{C}_{e|i}(t), \\
  \overline{C}_{g|i}(t)&\equiv e^{-i(n_{\omega}+\tfrac{1}{2})\omega t}e^{-i\tfrac{1}{2}\phi_{\hat{\boldsymbol{n}}}(\omega)}\tilde{C}_{g|i}(t).
\end{align}
\end{subequations}

We consider now the case of an up-jump at $t_k$, such that the initial amplitudes are set to $\overline{C}_{e|i}(0)=1$ and $\overline{C}_{g|i}(0)=0$. When $|\kappa_{\lambda,\hat{\boldsymbol{n}}}(\omega)|/A \ll 1$, we will have $\overline{C}_{e|i}(t) \approx 1$ and $\overline{C}_{g|i}(t) \sim |\kappa_{\lambda,\hat{\boldsymbol{n}}}(\omega)|$, with a strong likelihood that the next jump of the atom will be a down-jump. By the same argument, it is very likely that the down-jump is followed by another up-jump, thus forming a telegraph signal. Because, however, of the non-negligible amplitude excited by the coupling of the atom to the selected mode [with $\overline{C}_{g|i}(t) \sim |\kappa_{\lambda,\hat{\boldsymbol{n}}}(\omega)|$ or $\overline{C}_{e|i}(t) \sim |\kappa_{\lambda,\hat{\boldsymbol{n}}}(\omega)|$] there is always a small probability that a jump will occur to break the alternating sequence. Two up-jumps or two down-jumps may occur in a row, as we can observe in fig.~\ref{fig:thermaljumpsfig}(c). In frames (iii) and (iv) the presence of the small amplitude underlying the anomalous jump mechanism can be seen as a perturbation on top of the developing smooth curve. From a physical point of view, each of the anomalous events represents the scattering of a photon between one of the many field modes and the mode selected to be viewed. For example, two up-jumps occur in a row because in the interval between them the energy absorbed on the first jump is transferred to the selected mode. The scattering record resolves this transfer at the time of the second up-jump. 

The equations of motion~\eqref{eq:eomjmumps1} and \eqref{eq:eomjmumps2} can be used to calculate the rates for the unlikely jumps. Multiplying both sides by the complex conjugate of the probability amplitudes, we obtain the following set of four equations:
\begin{subequations}
 \begin{align}
  \frac{d|\tilde{C}_{e|i}|^2}{dt}&=-\Gamma_{\rm down}|\tilde{C}_{e|i}|^2-2|\kappa_{\lambda,\hat{\boldsymbol{n}}}(\omega)|\sqrt{n_{\omega}+1}\,{\rm Im}(\tilde{C}_{e|i}\tilde{C}^{*}_{g|i}), \\
   \frac{d|\tilde{C}_{g|i}|^2}{dt}&=-\Gamma_{\rm up}|\tilde{C}_{g|i}|^2+2|\kappa_{\lambda,\hat{\boldsymbol{n}}}(\omega)|\sqrt{n_{\omega}+1}\,{\rm Im}(\tilde{C}_{e|i}\tilde{C}^{*}_{g|i}), \\
   \frac{d\, {\rm Re}(\tilde{C}_{e|i}\tilde{C}^{*}_{g|i})}{dt}&=-\tfrac{1}{2}(\Gamma_{\rm down} + \Gamma_{\rm up}) {\rm Re}(\tilde{C}_{e|i}\tilde{C}^{*}_{g|i})-\Delta\omega\,{\rm Im}(\tilde{C}_{e|i}\tilde{C}^{*}_{g|i}),\\
    \frac{d\, {\rm Im}(\tilde{C}_{e|i}\tilde{C}^{*}_{g|i})}{dt}&=-\tfrac{1}{2}(\Gamma_{\rm down} + \Gamma_{\rm up}) {\rm Im}(\tilde{C}_{e|i}\tilde{C}^{*}_{g|i})+\Delta\omega\,{\rm Re}(\tilde{C}_{e|i}\tilde{C}^{*}_{g|i}) + |\kappa_{\lambda,\hat{\boldsymbol{n}}}(\omega)|\sqrt{n_{\omega}+1}\left(|\tilde{C}_{e|i}|^2-|\tilde{C}_{g|i}|^2\right).
 \end{align}
\end{subequations}
Starting from the time $t_k$ of the last jump, we define the quantities
\begin{equation}
 W_e \equiv \int_{t_k}^{\infty} dt |\tilde{C}_{e|i}|^2, \quad W_g \equiv \int_{t_k}^{\infty} dt |\tilde{C}_{g|i}|^2, \quad U \equiv {\rm Re}\left[\int_{t_k}^{\infty} dt\, \tilde{C}_{e|i}\tilde{C}^{*}_{g|i}\right], \quad V \equiv {\rm Im}\left[\int_{t_k}^{\infty} dt\, \tilde{C}_{e|i}\tilde{C}^{*}_{g|i}\right].
\end{equation}
The products $\Gamma_{\rm up}W_g$ and $\Gamma_{\rm down} W_e$ are then the probabilities that the unlikely jump will occur, given that $i$ is $e$ or $g$, respectively. Integrating then the equations of motion over time and setting the amplitudes to zero at infinity (see also~\cite{Cohdecoh1997}) we obtain:
\begin{subequations}\label{eq:timelimiteqs}
 \begin{align}
-\delta_{e,i}&=-\Gamma_{\rm down}W_e - 2 |\kappa_{\lambda,\hat{\boldsymbol{n}}}(\omega)| \sqrt{n_{\omega}+1}\,V,\\
-\delta_{g,i}&=-\Gamma_{\rm up}W_g + 2 |\kappa_{\lambda,\hat{\boldsymbol{n}}}(\omega)| \sqrt{n_{\omega}+1}\,V,\\
0&=-\tfrac{1}{2}(\Gamma_{\rm down} + \Gamma_{\rm up}) U - \Delta \omega\, V, \\
0&=-\tfrac{1}{2}(\Gamma_{\rm down} + \Gamma_{\rm up}) V + \Delta \omega\, U  + |\kappa_{\lambda,\hat{\boldsymbol{n}}}(\omega)| \sqrt{n_{\omega}+1}(W_e-W_g), 
 \end{align}
\end{subequations}
whence
\begin{subequations}
 \begin{align}
\Gamma_{\rm up} W_g|_{i=e}&=\frac{\tfrac{1}{2}(\Gamma_{\rm down}+\Gamma_{\rm up})/\pi}{[\tfrac{1}{2}(\Gamma_{\rm down}+\Gamma_{\rm up})]^2 + (\Delta \omega)^2} \frac{2\pi  |\kappa_{\lambda,\hat{\boldsymbol{n}}}(\omega)|^2}{\Gamma_{\rm down}}(n_{\omega}+1),\label{eq:probun1} \\
\Gamma_{\rm down} W_e|_{i=g}&=\frac{\tfrac{1}{2}(\Gamma_{\rm down}+\Gamma_{\rm up})/\pi}{[\tfrac{1}{2}(\Gamma_{\rm down}+\Gamma_{\rm up})]^2 + (\Delta \omega)^2} \frac{2\pi  |\kappa_{\lambda,\hat{\boldsymbol{n}}}(\omega)|^2}{\Gamma_{\rm up}}(n_{\omega}+1), \label{eq:probun2}
 \end{align}
\end{subequations}
where the set of eqs.~\eqref{eq:timelimiteqs} is solved to lowest order in the coupling strength.  

Equations~\eqref{eq:probun1} and~\eqref{eq:probun2} determine the probability for the improbable jump to occur after any preparation of the initial state $i$. To obtain the jump rates for the photon number we must multiply these expressions by the rate at which the state $i$ is prepared -- the jump rates for the atom. Then, eq.~\eqref{eq:probun1} is multiplied by $\Gamma_{\rm up}p_{e}^{\rm eq}$, and eq.~\eqref{eq:probun2} by $\Gamma_{\rm down}p_{g}^{\rm eq}$, where $p_{g,e}^{\rm eq}$ are the state occupation probabilities in thermal equilibrium. In addition, we set $n_{\omega}=N_{\omega}$ in eq.~\eqref{eq:probun1} and $n_{\omega}+1=N_{\omega}$ in eq.~\eqref{eq:probun2}, where $N_{\omega}$ is the photon number expectation plotted in fig.~\ref{fig:thermaljumpsfig}(c). The photon jump rates then read:
\begin{subequations}\label{eq:probunph}
 \begin{align}
\gamma_{N_{\omega}}^{\rm up}&=\frac{\tfrac{1}{2}(\Gamma_{\rm down}+\Gamma_{\rm up})/\pi}{[\tfrac{1}{2}(\Gamma_{\rm down}+\Gamma_{\rm up})]^2 + (\Delta \omega)^2} 2\pi  |\kappa_{\lambda,\hat{\boldsymbol{n}}}(\omega)|^2 (N_{\omega}+1)p_{e}^{\rm eq},\label{eq:probunph1} \\
\gamma_{N_{\omega}}^{\rm down}&=\frac{\tfrac{1}{2}(\Gamma_{\rm down}+\Gamma_{\rm up})/\pi}{[\tfrac{1}{2}(\Gamma_{\rm down}+\Gamma_{\rm up})]^2 + (\Delta \omega)^2} 2\pi  |\kappa_{\lambda,\hat{\boldsymbol{n}}}(\omega)|^2 N_{\omega}p_{g}^{\rm eq}. \label{eq:probunph2}
 \end{align}
\end{subequations}
Using the principle of detailed balance\index{Detailed balance} after setting up rate equations based on the expressions~\eqref{eq:probunph} and solving in the steady state\index{Steady state}, we obtain
\begin{equation}
 p_{N_{\omega}+1}^{\rm eq}\gamma_{N_{\omega}}^{\rm down} (N_{\omega}+1)= p_{N_{\omega}}^{\rm eq}\gamma_{N_{\omega}}^{\rm up} N_{\omega},
\end{equation}
whence 
\begin{equation}
 p_{N_{\omega}+1}^{\rm eq}/p_{N_{\omega}}^{\rm eq}=p_e^{\rm eq}/p_g^{\rm eq},
\end{equation}
After normalizing we obtain the equilibrium probability to find $N_{\omega}$ photons in the selected mode:
\begin{equation}
 p_{N_{\omega}}^{\rm eq}=\left(1 - \frac{p_e^{\rm eq}}{p_g^{\rm eq}} \right) \left( \frac{p_e^{\rm eq}}{p_g^{\rm eq}} \right)^{N_{\omega}}=\frac{1}{\overline{n}(\omega_0)+1} \left[\frac{\overline{n}(\omega_0)}{\overline{n}(\omega_0)+1} \right]^{N_{\omega}},
\end{equation}
where we have used the principle of detailed balance\index{Detailed balance} for the atom $p_e^{\rm eq}/p_g^{\rm eq}=\Gamma_{\rm up}/\Gamma_{\rm down}$ and the Einstein relations~\eqref{eq:einst1} and~\eqref{eq:einst2}. This is a Bose-Einstein distribution\index{Bose-Einstein! distribution}\index{Distribution! Bose-Einstein} with average photon number $\overline{N_{\omega}}=\overline{n}(\omega_0)$. We note that the particular mode coupling strength $|\kappa_{\lambda,\hat{\boldsymbol{n}}}(\omega)|$ affects only the rate of approach to equilibrium. In reality, of course, each mode of the electromagnetic field couples to a large number of two-level atoms, and most strongly to the ones with which it is nearly resonant. For that situation, then, we expect to find $\overline{N_{\omega}}=\overline{n}(\omega)$, in agreement with the Planck radiation formula\index{Planck!radiation formula}. 

The last requirement is to demonstrate that the jump rates are self-consistent. That is, if we sum the rates~\eqref{eq:probunph1} and~\eqref{eq:probunph2} over all frequencies, directions of propagation and polarization states, we should recover the jump rates initially assumed for the atom. Indeed, summing~\eqref{eq:probunph1} over all $\lambda, \hat{\boldsymbol{n}}, \omega$, with $N_{\omega}$ replaced by $\overline{n}(\omega_0)$ and neglecting the frequency dependence of the density of states and that of the dipole coupling constant (Lorentzian lineshape), returns $\Gamma_{\rm down}p_{e}^{\rm eq}$, while summing~\eqref{eq:probunph2} returns $\Gamma_{\rm up}p_{g}^{\rm eq}$. Hence, the net jump rates are in accord with the Einstein relations as well as with Fermi's golden rule\index{Fermi's golden rule}. 

Coming now to the electronic transitions of an emitter, we find that their rates become
dependent on the overlap between vibrational configurations in the initial and final states, which are generally displaced from one another. The dynamical influence of the electromagnetic reservoir depends on the vibrational coupling between the atom and its environment, leading to the formation of manifolds corresponding to the ground and excited electronic configurations~\cite{Maybook2011}. The Franck-Condon principle\index{Franck-Condon! principle} can be then incorporated into a nonadditive master equation\index{Master Equation!nonadditive} -- respecting the nonperturbative nature of the vibrational coupling -- in a so-called {\it collective coordinate representation}\index{Collective coordinate representation} of the vibrational environment~\cite{Maguire2019}. 


\subsection{Beyond the rotating-wave approximation: the quantum Rabi model}\label{ssec:rabi}\index{Quantum! Rabi model}\index{Model! quantum Rabi}
In order to derive the JC model from, say, a minimal-coupling {\it ansatz}, one must impose numerous approximations: {\bf 1.} {\it Dipole approximation}\index{Dipole! approximation} -- the `size' of the atoms is assumed small on the electromagnetic wavelength scale such that we  disregard any spatial dependence. {\bf 2.} {\it Single-mode approximation}\index{Single-mode approximation} -- for the atom to couple to a single mode a cavity with a very high quality factor $Q$ is requested, meaning that the linewidths\index{Linewidth} are small in comparison to the separation of the mode frequencies. {\bf 3.} Two-level approximation (TLA)\index{Two-level! approximation}-- we can limit the analysis to only two electronic levels of the atom due to selection rules and since other transitions are assumed far detuned. {\bf 4.} {\it Neglecting the diamagnetic self-energy} -- from the minimal coupling Hamiltonian a term scaling as the squared vector potential\index{Vector potential} appears, but since it is proportional to the electric charge $e$ divided by the speed of light $c$ it can in most circumstances be neglected (see further discussions in sec.~\ref{sssec:dicke}). {\bf 5.} {\it The rotating-wave approximation (RWA)}\index{Rotating-wave approximation} -- the terms responsible for the virtual exchange\index{Virtual processes} of photons are not taken into account. These approximations will be all discussed in greater depth in sec.~\ref{ssec:approx}.

It should be clear that these approximations are interconnected, and that the relative size of the coupling $g$ is essential for their validity. Contrary to early beliefs, nowadays we know that one of the approximations may break down while the another one may still hold. For example, a common understanding was that in the ultrastrong coupling regime the TLA will break down. This turns out to not always be the case, especially with the discovery of the flux qubit\index{Qubit! flux}~\cite{niemczyk2010circuit,forn2010observation,yoshihara2017superconducting} where experiments agree with the JC model only when the counter rotating terms are included. In most recent times, the breakdown of the TLA has instead led to the definition of a new coupling regime, the {\it extreme strong coupling regime} occurring when $g$ is a few tenths of the cavity frequency $\omega$~\cite{Ashida2021}. The validity of the TLA in the JC model was given a thorough discussion in~\cite{frasca2003modern}. Since then, new facets of the TLA have emerged in terms of {\it gauge invariance}\index{Gauge! invariance}~\cite{de2018cavity,de2018breakdown,stokes2019gauge,di2019resolution}. This issue of the TLA will be discussed in some detail in sec.~\ref{sssec:tla}. Despite this, the fact that the RWA can break down before the other approximations as $g$ is increased implies that we need to understand the physics of the quantum Rabi model, {\it i.e.} the JC model without imposing the RWA. 

To set the stage, we start by introducing the different parameter regimes of interest~\cite{casanova2010deep,rossatto2017spectral,forn2019ultrastrong,kockum2019ultrastrong,settineri2019gauge,le2020theoretical, Ashida2021, Lamata2020}. In the table~\ref{regimetable} we give the parameters defining the five most commonly considered regimes. How they are defined may vary between different works, {\it e.g.} here we compare the coupling strength $g$ to the bare photon frequency $\omega$ and the bare photon loss rate $\kappa$, and not to the spontaneous emission rate\index{Spontaneous! emission! rate} $\gamma$ and the bare atomic frequency $\Omega$. Strictly speaking, this is a simplified picture, and in general, the dynamical evolution is also dictated by these two rates, $\Omega$ and $\gamma$. For example, if $\gamma>g$ it would not make much sense to talk about strong/ultrastrong/deep/extreme strong coupling regimes\index{Ultrastrong coupling regime}\index{Deep strong coupling regime}\index{Deep strong coupling regime}\index{Extreme strong coupling regime}; it should be understood that $\gamma$ is small for these cases. Furthermore, the validity of the RWA is not solely depending on the ratio $g/\omega$, as suggested by the table, but also on $\Omega$. State-of-the-art experiments in circuit QED, see section~\ref{sec:cirQED}, have reached the ultrastrong coupling regime~\cite{niemczyk2010circuit,forn2010observation,forn2019ultrastrong,le2020theoretical}, with $g/\omega\sim0.1$, and more recently also the deep strong coupling regime~\cite{yoshihara2017superconducting}. The deep strong coupling regime with $g/\omega\sim2$ has also been demonstrated in plasmonic nanoparticle crystals~\cite{mueller2020deep}. One explanation why it is easier to reach these regimes more easily with circuit QED compared to cavity QED is because the effective coupling $g$ scales differently with respect to the {\it fine structure constant}\index{Fine structure constant}; proportional and inversely proportional, respectively, to $\sqrt{\alpha}$~\cite{devoret2007circuit}. 

\begin{table}[h!]
  \centering

\begin{tabular}{|c|c|p{5cm}|}
\hline
Regime & Parameters & Description\\
  \hline \hline
  {\it Weak coupling} & $g/\omega\ll1$ \& $g<\kappa$ & The coherent time-scale $g^{-1}$ is the longest -- the system evolution is dominated by losses.  \\
  \hline
  {\it Strong coupling} & $g/\omega\ll1$ \& $g>\kappa$ & The evolution for time-scales $g^{-1}$ is predominantly unitary. The RWA is valid.  \\ 
  \hline
  {\it Ultrastrong coupling} &  $g/\omega\sim0.1$ \& $g>\kappa$ & The RWA breaks down. Perturbation theory still captures the corrections. The TLA breaks down.\\
  \hline
  {\it Deep strong coupling} & $g/\omega\gtrsim1$ \& $g>\kappa$ & The properties of the lower energy states of the model changes qualitatively, {\it e.g.} the expectation $\langle\hat n\rangle$ for the ground state is no longer approximately zero. The TLA breaks down.  \\
  \hline
  {\it \,\,Extreme strong coupling\,\,} & $\,\,g/\omega\gtrsim10$ \& $g\gg\kappa\,\,$ & Perturbative results hold in the polaron basis\index{Polaron! basis} such that the light-matter degrees of freedom decouple. The TLA breaks down.  \\
  \hline
  \end{tabular}
  \caption{Definition of the different regimes of the quantum Rabi model. }
  \label{regimetable}
  \end{table}

In the JC model, the RWA removes terms describing simultaneous excitation of the atom and creation of one photon or atomic de-excitation accompanied with annihilation of one photon~\cite{scully1999quantum,gerry2005introductory}. Including these {\it counter rotating terms} (CRT)\index{Counter rotating terms} gives the {\it quantum Rabi model}\index{Quantum! Rabi Hamiltonian}~\cite{rabi1936process,rabi1937space}
\begin{equation}\label{rabiham}
\hat{H}_\mathrm{R}=\omega\hat{n}+\frac{\Omega}{2}\hat{\sigma}_z+g\left(\hat{a}+\hat{a}^\dagger\right)\hat{\sigma}_x.
\end{equation}
Contrary to the JC model Hamiltonian, $\hat H_\mathrm{R}$ does not preserve the number of excitations $\hat N$ of eq.~(\ref{exop}). Nevertheless, the quantum Rabi model possesses a $\mathbb{Z}_2$ parity symmetry characterized by $\left(\hat a,\hat a^\dagger,\hat\sigma_x,\hat\sigma_y,\hat\sigma_z\right)\leftrightarrow\left(-\hat a,-\hat a^\dagger,-\hat\sigma_x,-\hat\sigma_y,\hat\sigma_z\right)$, see also eq.~(\ref{parity}). As will be discussed in more detail in secs.~\ref{sssec:rabiint} and~\ref{sssec:dicke}, this symmetry is of great importance for the characteristics of the model. A consequence of the $\mathbb{Z}_2$ symmetry is that the eigenstates and eigenvalues
\begin{equation}\label{rabieigstates}
|\psi_{n\pm}\rangle,\hspace{1.2cm}E_{n\pm},
\end{equation}
can be labeled by a discrete quantum number $n=0,1,2,\dots,$ and a quantum number $j=\pm$ for the two parities~\cite{casanova2010deep}. It may prove practical to split the Hamiltonian as 
\begin{equation}\label{counterterms}
\hat H_\mathrm{R}=\hat H_\mathrm{JC}+\hat V_\mathrm{CR},\hspace{1,6cm}\hat V_\mathrm{CR}=g\left(\hat a^\dagger\hat{\sigma}_{+}+\hat a\hat{\sigma}_{-}\right),
\end{equation}
where the first term is the JC Hamiltonian~(\ref{jcham}), and the second term comprises the counter rotating terms. Note further that
\begin{equation}\label{crcom}
\left[\hat V_\mathrm{CR},\hat H_\mathrm{JC}\right]=g(\omega+\Omega)\left(\hat a\hat{\sigma}_{-}-\hat a^\dagger\hat{\sigma}_{+}\right)+g^2\left(\hat a^{\dagger 2}-\hat a^2\right)\hat\sigma_z,
\end{equation}
meaning that the additional term $V_\mathrm{CR}$ of the quantum Rabi Hamiltonian provides a non-trivial contribution.

To build a deeper understanding of the quantum Rabi model it is practical to go to the quadrature representation\index{Quadrature representation}~(\ref{quad}), which gives~\cite{larson2007dynamics,ashhab2010qubit}
\begin{equation}\label{rabiham2}
\hat{H}_\mathrm{R}=\omega\left(\frac{\hat{p}^2}{2}+\frac{\hat{x}^2}{2}\right)+\frac{\Omega}{2}\hat{\sigma}_z+\sqrt{2}g\hat{x}\hat{\sigma}_x.
\end{equation}
Expressed in this form it is readily seen that the quantum Rabi model is what in the molecular physics community is called the $E\times\beta$ Jahn-Teller model\index{$E\times\beta$ Jahn-Teller model}\index{Model! $E\times\beta$ Jahn-Teller}~\cite{grosso,hines2004entanglement} (the splitting term $\sim\hat{\sigma}_z$ is in molecular physics due to a spin-orbit coupling). In molecular group theoretical language, the Latin capital letter signifies the presence of a two-level system, and the greek letter signifies the coupling to a single vibrational mode. In sec.~\ref{sssec:multi} we will discuss the $E\times\varepsilon$ Jahn-Teller model\index{$E\times\varepsilon$ Jahn-Teller model}\index{Model! $E\times\varepsilon$ Jahn-Teller} which is a two-level system coupled to two degenerate harmonic oscillators, see eq.~(\ref{jt}).  

\begin{figure}
\includegraphics[width=10cm]{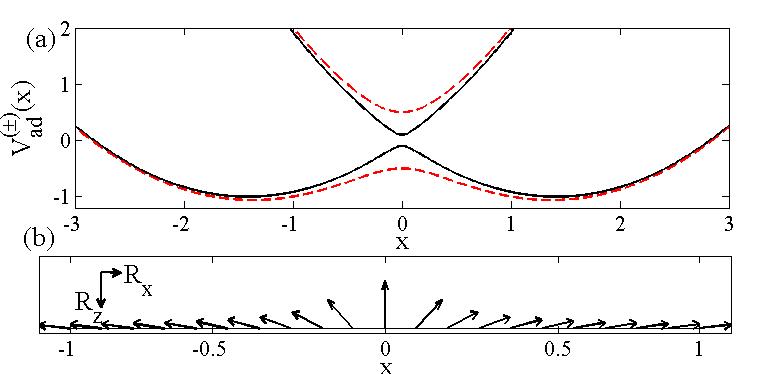} 
\caption{The adiabatic potential curves of the quantum Rabi model {\bf (a)}, and the Bloch vector\index{Bloch! vector} of the lower adiabatic state {\bf (b)}. The parameters are $\omega=g=1$ and $\Omega=1$ (red dashed curve) and $\Omega=0.2$ (black solid curve). The gap at $x=0$ equals $\Omega$. The lower plot gives the Bloch vector ($R_y\equiv0$) for the adiabatic state corresponding to the lower adiabatic potential\index{Adiabatic! potential}. At $x=0$, the atom is in the state $|g\rangle$, and the further away the $x$-component of the atomic state starts to dominate over the $z$-component. Hence, as the field amplitude grows (increasing $x$) the atoms aligns more with the field. }
\label{fig9}
\end{figure}

As explained above, the BOA\index{Born-Oppenheimer approximation} consists in diagonalizing the Hamiltonian with respect to the atomic degrees of freedom and neglect the off-diagonal terms. The matrix diagonalizing the atomic part, fixing the gauge, takes the simple form (recall the unitary (\ref{jcU})
\begin{equation}\label{adbas}
\hat{U}_\mathrm{ad}(\hat{x})=\left[
\begin{array}{cc}
\sin(\theta/2) & \cos(\theta/2)\\
-\cos(\theta/2) & \sin(\theta/2)
\end{array}\right],
\end{equation}
with $\tan(\theta)=\sqrt{8}g\hat{x}/\Omega$. The transformed Hamiltonian becomes
\begin{equation}\label{boaham}
\hat{H}_\mathrm{R}'=\omega\left[\frac{\left(\hat{p}-\hat{A}\right)^2}{2}+\frac{\hat{x}^2}{2}\right]+\sqrt{\frac{\Omega^2}{4}+2g^2\hat{x}^2}\hat{\sigma}_z,
\end{equation}
where the {\it synthetic gauge potential}\index{Synthetic! gauge potential}\index{Gauge! potential} 
\begin{equation}\label{syngauge}
\hat{A}=i\hat{U}_\mathrm{ad}^{-1}\partial_{\hat{x}}\hat{U}_\mathrm{ad}=i\partial_{\hat{x}}\theta\hat{\sigma}_x.
\end{equation}
We will occasionally return to these synthetic gauge potentials\index{Gauge! potential! synthetic}, and especially discuss them in more detail in sec.~\ref{sssec:mfmQED}. In the BOA, $\partial_{\hat{x}}\theta\rightarrow0$ meaning that adiabaticity is expected either when $\Omega\gg|g\hat{x}|$ or the opposite $\Omega\ll|g\hat{x}|$. The {\it non-adiabatic coupling}\index{Non-adiabatic! coupling} $\partial_{\hat{x}}\theta$ which gives the off-diagonal elements of the transformed Hamiltonian is small for smooth $\theta$. To suppress the off-diagonal terms further, a second BOA can be applied such that these terms instead go as $\partial_{\hat{x}}^2\theta$ which was the subject of ref.~\cite{zhi2013dynamics} applied to the quantum Rabi model. The gauge freedom lies in the choice of overall phase of the adiabatic eigenstates which forms the matrix $\hat{U}_\mathrm{ad}(\hat x)$~\cite{bohm2003geometric}. In the limit when $\Omega=0$, the quantum Rabi model is given by two decoupled (displaced) Harmonic oscillators~\cite{irish2005dynamics,ashhab2013superradiance}. A non-zero $\Omega$ couples the two oscillators and opens up a gap between the two `potentials'~\cite{larson2007dynamics,atkins1997molecular}. In fig.~\ref{fig9} (a), the two {\it adiabatic potential curves}\index{Adiabatic! potential! curve} 
\begin{equation}\label{adpot}
V_\mathrm{ad}^{(\pm)}(x)=\omega\frac{x^2}{2}\pm\sqrt{\frac{\Omega^2}{4}+2g^2x^2}
\end{equation}
are shown. An interesting observation is that in the BOA we assume that the spin degree of freedom follows adiabatically the boson field, which is similar to the idea of adiabatic elimination\index{Adiabatic! elimination} discussed at the end of sec.~\ref{ssec:JCm}. In the adiabatic elimination we find an effective model by assuming that the fast degrees of freedom reach their steady states\index{Steady state} instantaneously. Upon eliminating the spin in this way one finds the effective boson Hamiltonian
\begin{equation}\label{adbos}
\hat H_\mathrm{Bos}=\omega\frac{\hat p^2}{2}+V_\mathrm{ad}^{(-)}(x),
\end{equation}
which is exactly what one finds within the BOA.

From the expression (\ref{adpot}) for the adiabatic potentials, it follows that for $|g|<\sqrt{\omega\Omega}/2$ the lower potential has a single minimum at $x=0$, while for $|g|>\sqrt{\omega\Omega}/2$ a double-well structure of the potential is formed, which also defines the {\it deep strong coupling regime}\index{Deep strong coupling regime}~\cite{casanova2010deep,ying2015ground,rossatto2017spectral}, see tab.~\ref{regimetable}. The potential barrier separating the two minima is given by $\delta_\mathrm{bar}=\frac{g^2}{\omega}+\frac{\omega\Omega^2}{16g^2}-\frac{\Omega}{2}$. This qualitative change of the potential is expected to have drastic changes for the ground state properties~\cite{ashhab2013superradiance,ermann2020jaynes}. As a finite system, it was shown that the double-well formation is indeed not representing a true $\mathbb{Z}_2$ symmetry breaking, but the ground state is non-degenerate and possessing an even parity~\cite{lo1998there}. However, in the so-called {\it classical limit}\index{Classical limit}\index{Rabi! classical limit} $\omega/\Omega\rightarrow0$~\cite{bakemeier2012quantum,larson2017some}, Plenio and co-workers showed~\cite{hwang2015quantum} that the transition bear all types of characteristics expected for a phase transition (PT), {\it i.e.} a discontinuity in the derivative of the ground state energy and universal critical exponents\index{Critical! exponents} (here agreeing with those of the Dicke phase transition). We will return to this PT in sec.~\ref{sssec:poormodel}. As will be discussed in some detail both in sec.~\ref{sssec:dicke} and sec.~\ref{sssec:dia}, reaching the critical coupling in circuit and cavity QED setups is experimentally challenging. However, for trapped ions, see sec.~\ref{ssec:ionham}, the situation is very different since the coupling strength can be controlled externally.  And as a result, this type of phase transition in the quantum Rabi model has been experimentally observed using a singly trapped $^{171}\mathrm{Yb}^+$ ion within a quasi 1D trap~\cite{Cai2021}. At the expected critical point\index{Critical! point}, they saw a rapid increase in both the number of phonons and internal ion excitations. The appearing PT structure in the classical limit can be understood from the adiabatic potentials by noting that for vanishing $\omega$ the barrier $\delta_\mathrm{bar}$ becomes large meaning that the even ground and first excited odd states become approximately degenerate. They further showed universality in terms of excitation production ({\it Kibble-Zurek mechanism})\index{Kibble-Zurek mechanism} as the system is driven through the critical point, see also a more recent work~\cite{srivastava2020scaling}. We will further discuss the same type of symmetry breaking in sec.~\ref{sssec:dicke} when the Dicke quantum phase transition is revised. In the corresponding symmetry broken phase, Irish and Gea-Banacloche studied the Josephson-type oscillations\index{Josephson! oscillations} appearing in the Rabi double-well structure~\cite{irish2013oscillator}. A similar analysis has also been performed for the emerging double-well potential in the Dicke model~\cite{chen2007quantum}.

To derive the JC model from the quantum Rabi model, one turns to the rotating frame with respect to the first two bare terms, for which the interaction contains the JC terms $\hat{a}^\dagger\hat{\sigma}_{-}e^{-i(\omega-\Omega)t}$ and $\hat{\sigma}_{+}\hat{a}e^{i(\omega-\Omega)t}$, and the CRTs\index{Counter rotating terms} $\hat{a}^\dagger\hat{\sigma}_{+}e^{-i(\omega+\Omega)t}$ and $\hat{a}\hat{\sigma}_{-}e^{i(\omega+\Omega)t}$. We normally assume that $|\Delta|\ll\omega,\,\Omega$ meaning that the CRTs are oscillating rapidly compared to the JC terms. If, furthermore, $g\ll\omega$ we can at short or moderate time scales neglect contributions from the CRTs. For long time-scales, however, this is not necessarily true~\cite{larson2012absence} as we demonstrate in sec.~\ref{sssec:rwasub}. In the 1980s and 1990s, experimental realizations of JC physics was mainly to be found in cavity QED with single atoms coupled to single resonator modes, either microwave or optical, or in trapped ion physics. In the former (see section~\ref{sec:cavQED}), the system could be made to operate in the {\it strong coupling regime} defined as $g\sqrt{\bar{n}}>\kappa,\,\gamma$ where $\kappa$ and $\gamma$ are, respectively, the cavity and atom loss rates. This allowed for coherent quantum dynamics to be explored. However, the coupling $g$ was always several orders of magnitude smaller than the photon frequency $\omega$ and for the experiments one needed not to worry too much about the validity of this approximation. With the advent of circuit QED and artificially manufactured atoms coupled to transmission line resonators, the coupling could approach the photon frequency~\cite{forn2010observation,niemczyk2010circuit,yoshihara2017superconducting}. In other words, the new physics arising from the CRTs is no longer only of academic interest but also of practical nature (see also sec.~\ref{sssec:dicke} on the Dicke phase transition). Also, we note that in cavity QED there are a few proposals how one can construct effective models which would simulate the deep strong coupling regime by starting from driven JC type systems ~\cite{solano2003strong,dimer2007proposed,ballester2012quantum,grimsmo2013cavity,munoz2020simulating}. One experimental work constructed a circuit QED system, operating in the strong coupling regime, and performed a {\it digital quantum simulation} in order to explore the deep strong coupling regime~\cite{langford2017experimentally}. Recently, another interesting arena where the quantum Rabi model plays an important role is for {\it synthetic gauge fields} in cold atom physics~\cite{dalibard2011colloquium}. By Raman dressing harmonically confined ultracold atoms it is possible to derive an effective model where the atomic velocity is coupled to the internal atomic degrees of freedom, {\it i.e.} a {\it spin-orbit coupling}\index{Spin-orbit coupling}. For two dimensions, any type of spin-orbit coupling can be achieved from combinations of {\it Rashba} and {\it Dresselhaus} couplings, and the quantum Rabi model results from an equal strength of the two~\cite{lin2011spin}. The emergence of the double-well structure of the quantum Rabi model when approaching the deep strong coupling regime was probed in a dilute gas of $^{87}$Rb atoms~\cite{lin2011spin}. It has also been suggested that this symmetry breaking is a realization of the Dicke phase transition~\cite{zhang2013tunable} (see sec.~\ref{sssec:dicke}). The results of ref.~\cite{lin2011spin} can be understood by neglecting atom-atom interaction, but it is clear that when interactions become more important the model cannot be effectively described in terms of a quantum Rabi model. This also applies to the more recent experiments on spin-orbit coupled Fermi gases~\cite{wang2012spin,cheuk2012spin}.

\begin{figure}
\includegraphics[width=10cm]{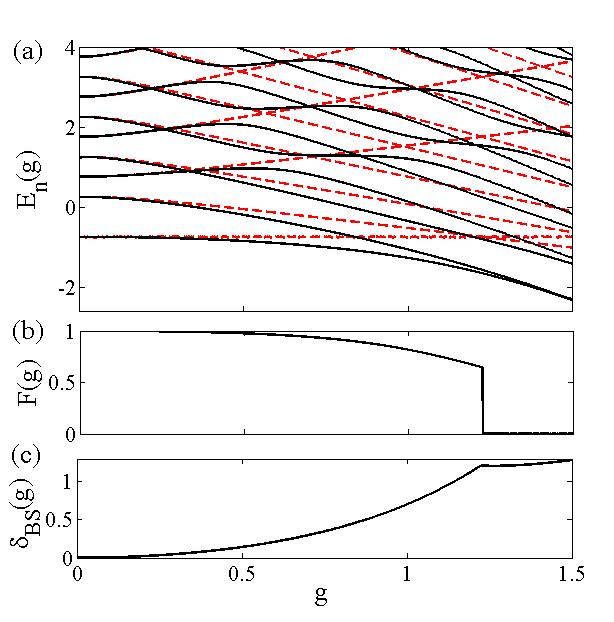} 
\caption{The low-lying energies of the Rabi (black solid lines) vs. the JC (red dashed lines) models as a function of the atom-field coupling {\bf (a)}, the fidelity~(\ref{fidel}) of the ground states of the two models {\bf (b)}, and the scaled Bloch-Siegert shift of the ground state {\bf (c)}. The first points for when the two solid lines cross have been termed {\it Juddian points}\index{Juddian! points}~\cite{braak2011integrability}. Below these, the perturbative Hamiltonian~(\ref{bsham}) quantitatively predicts the correct energies~\cite{rossatto2017spectral}. The field and atom frequencies are in all examples $\omega=1$ and $\Omega=1.5$ respectively.}
\label{fig10}
\end{figure}


\subsubsection{Effect of counter rotating terms, the ultrastrong coupling regime}\label{sssec:rwaeff}
Before discussing more subtle consequences stemming from the additional CRTs~\cite{le2020theoretical}\index{Counter rotating terms} in the following subsections, a direct effect from these terms is to shift the eigenenergies~\cite{kockum2019ultrastrong}. These are the so-called {\it Bloch-Siegert shifts}\index{Bloch!-Siegert shift}~\cite{bloch1940magnetic}. An estimate of these shifts can be achieved perturbatively for small couplings $\Gamma=g/(\omega+\Omega)\ll1$. Transforming the Hamiltonian~(\ref{rabiham}) with the unitary
\begin{equation}\label{unit}
\hat U=e^{\Gamma\left(\hat a\hat{\sigma}_{-}-\hat a^\dagger\hat{\sigma}_{+}+\frac{g}{2\omega}(\hat a^2-\hat a^{\dagger 2})\hat \sigma_z\right)},
\end{equation}
and expanding to second order in $\Gamma$ one finds~\cite{rossatto2017spectral}
\begin{equation}\label{bsham}
\hat H_\mathrm{R}'\approx\omega_\mathrm{BS}\hat\sigma_z\hat a^\dagger\hat a+\frac{\omega_\mathrm{BS}}{2}\hat\sigma_z-\frac{\omega_\mathrm{BS}}{2}+\hat H_\mathrm{JC}.
\end{equation}
Here the $\omega_\mathrm{BS}=g\Gamma$ is the estimate of the Bloch-Siegert shift. Not surprisingly, the exponent of the unitary~(\ref{unit}) bears similarities with the commutator~(\ref{crcom}). In fig.~\ref{fig10} (a) we plot the low lying energies of the JC and quantum Rabi models as functions of the atom-field coupling. In the figure, the visible effects sets in after $g\sim0.2$. In order to get a better feeling for the size of the Bloch-Siegert shifts, in (c) we plot the relative shift $\delta_{BS}(g)=(E_\mathrm{0,JC}(g)-E_\mathrm{0,R}(g))/E_\mathrm{0,JC}(g)$ where $E_\mathrm{0,JC}(g)$ and $E_\mathrm{0,R}(g)$ are the ground state energies of the JC and quantum Rabi model respectively. The sudden change in the shift occurs when the ground state of the JC model switch from $|0,g\rangle$ to the dressed state $|\psi_{0,-}\rangle$. Properties of the spectrum for the quantum Rabi model was discussed in~\cite{graham1984two,zaheer1989photon,zhang2011analytical,qing2012spectrum}. Zaheer and Zubairy, for instance, used a path integral approach in order to study the shifts in both energy and photon number arising due to the CRTs~\cite{zaheer1989photon}, while Law focused on the shift of the vacuum field due to strong atom-field coupling~\cite{law2013vacuum}.

From fig.~\ref{fig9} (a) we can directly conclude that in the deep ultrastrong coupling regime, {\it i.e.} when the lower potential attains a double-well shape, the ground state of the field should be in an even Schr\"odinger cat state\index{Cat state}~\cite{irish2005dynamics,ashhab2010qubit,ashhab2013superradiance}. Moreover, the crossover from one global to two global minima leads to a hysteresis effect~\cite{shore1973ground}. Further characteristics of the ground state of the quantum Rabi model has been considered in numerous works~\cite{shore1973ground,graham1984two,chen1989variational,fessatidis2002moments,liu2007lower,xiao2009ground,hwang2010variational,ashhab2010qubit,zhang2011analytical,hirokawa2012absence}. The ground state is often compared to that of the JC model, and one interesting aspect is the build-up of photons in the ground state~\cite{zhang2011analytical}, {\it i.e.} how $\langle\hat{n}\rangle$ scales with $g$. For very large $g$, it follows from eq.~(\ref{dstate}) for the JC and using the BOA for the quantum Rabi model that $\langle\hat{n}\rangle\sim g^2$ in both models but the proportionality constant is twice as big for the quantum Rabi model. This results from `equal contribution' from the JC terms and CRTs in the deep ultrastrong coupling regime (the ground state energy for the two models both scale as $\sim-g$). Already at the transition point $|g|=\sqrt{\omega\Omega}/2$ are the two terms, JC and CRT, equally important (see also sec.~\ref{sssec:dicke}). A direct measure how `similar' the two ground states are is given by the state fidelity\index{State! fidelity}~(\ref{fidel}) $F=|\langle\psi_\mathrm{0JC}|\psi_\mathrm{0R}\rangle|^2$ which is displayed in fig.~\ref{fig10} (b). As for the Bloch-Siegert shift, the fidelity changes drastically when the ground state of the JC model goes from $|0,g\rangle$ to $|\psi_{0,-}\rangle$. Another important property is the amount of squeezing of the ground state which has been thoroughly investigated in Refs.~\cite{chen1989variational,hwang2010variational,ashhab2010qubit} and how the CRTs affect the phase space distributions~\cite{phoenix1991effect}. In particular, from the BOA [defined above as neglecting $\hat{A}$ in eq.~(\ref{boaham})], it is clear that field squeezing is only possible in the $p$-quadrature since the width of the lower adiabatic potential $V_\mathrm{ad}^{(-)}(x)$ is larger than {\it diabatic} (bare) potential $V_\mathrm{dia}(x)=\omega x^2/2$. Field and dipole squeezing were also considered in the evolving cases in~\cite{lais1990squeezing,xie1995numerical}. Phoenix argued that interference between different terms, CRT and JC terms, may become important at large time-scales for certain initial atomic states~\cite{phoenix1989counter}. The long time properties with and without CRTs will be further discussed in the following subsections. The influence of the CRTs were also considered in terms of collapse-revivals~\cite{zaheer1988atom,bonci1991quantum,larson2007dynamics} (where the first corrections appear as a rapid superimposed oscillation on top of the JC collapse-revival pattern~\cite{larson2007dynamics,he2014absence}, and see also~\cite{huang2017manipulating} where it was proposed how the importance of the counter-rotating terms can be enhanced by a periodic driving of the two-level system), the photon blockade~\cite{ridolfo2012photon}, and on atom-field entanglement~\cite{fang1996quantum}. Entanglement properties between two~\cite{jing2009breakdown,chen2010entanglement,lee2013ground} and three~\cite{wang2010one} atoms have also been explored with the CRTs present. In Refs.~\cite{jing2009breakdown,chen2010entanglement}, the effect of the CRTs in terms of ESD\index{Entanglement! sudden! death} was considered and it was demonstrated that the ESD may qualitatively change by taking the CRTs into account. In~\cite{wang2010one} it was discussed how to perform robust and fast preparation of GHZ states, see eq.~(\ref{ghzstate})\index{GHZ state}\index{GHZ state}, {\it i.e.} a state of the form
\begin{equation}\label{ghz}
|{\rm GHZ}\rangle=\frac{1}{\sqrt{3}}\left(|g,g,g\rangle+|e,e,e\rangle\right), 
\end{equation}
in a three-qubit-resonator system operating in the ultrastrong coupling regime. How cavity losses affect the entanglement between two atoms (qubits) has as well been addressed~\cite{sabin2010dynamics,ma2012entanglement,altintas2013dissipative}. It has in particular been shown that for the steady state\index{Steady state} solution of the corresponding master equation\index{Master equation}, entanglement shared between the qubits can indeed survive. One objective with increasing the atom-field coupling is that the characteristic time-scales decrease which is desirable for any QIP. With this in mind, proposals how to implement two-qubit gates have been put forward~\cite{romero2012ultrafast,wang2012validity,lu2012two}. For another circumstance involving the counter-rotating terms, it has recently been argued that a uniformly accelerated atom in Minkowski space-time emits entangled photon pairs in a squeezed state \index{Squeezing}, akin to the entanglement of the Minkowski vacuum. The entanglement \index{Entanglement! Unruh, Hawking and Cherenkov radiation} is formed between Minkowski modes which are dominantly in opposite causal wedges of the space-time. A similar emission of photon pairs takes place when a two-state atom is held above the event horizon of a black hole \cite{Unruh2022}.


\subsubsection{Analytical approximations}
There have been numerous proposals how to find accurate solutions of the JC model beyond the RWA. We have witnessed an especial increase of such suggestions during the last decade since the realization of circuit QED which has increased the coupling to the field frequency ratio by a couple of orders of magnitude. Naturally, the idea is to find solutions that are valid in other parameter regimes where the regular RWA breaks down.

In any adiabatic approximation, the relevant time-scales should be identified. For example, the regular RWA assumes a weak coupling and small detuning which makes the 'rotating terms' to evolve on a much shorter time-scale than the 'counter-rotating terms'. As we see from fig.~\ref{fig10}(a), when the coupling is large the BOA suggests that the bare states are not good approximations for the low-energy physics. Irish identified a better basis of states for performing a RWA in this strong coupling regime, namely considering {\it displaced Fock states}\index{Displaced Fock states}\index{State! displaced Fock} of the boson mode~\cite{irish2007generalized}. Such a {\it generalized rotating-wave approximation}\index{Generalized rotating-wave approximation} has proven very powerful as it gives analytical expressions for both the energies as well as the (generalized) dressed states. A related scheme for identifying the relevant time-scales for carrying out an effective RWA was considered in ref.~\cite{wu1997jaynes} in the realm of trapped ion physics. Improving the generalized RWA to be valid in larger parameter regimes has been considered in refs.~\cite{gan2010dynamics,zhang2011analytical}. More recently, similar ideas as the one behind the generalized RWA were employed in order to improve the BOA~\cite{li2021generalized}. Like in~\cite{irish2007generalized}, it was found that this generalized BOA could much more correctly capture the eigenenergies in large parameter regimes.

Another popular approach is to consider corrections to the JC results in a perturbative fashion. Following standard perturbation techniques where the quantum Rabi model is written as (\ref{counterterms}), with the unperturbed Hamiltonian being the JC one and the perturbation $\hat{V}_\mathrm{CR}$ the counter-rotating terms was employed in ref.~\cite{sornborger2004superconducting}. Perzeverzev and Bittner performed a similar analysis without considering the full perturbation $\hat{V}_\mathrm{CR}$; rather, certain terms of it that arose after a unitary transformation~\cite{pereverzev2006exactly}. Phoenix developed a technique relying on taking corrections from non-commuting operators systematically into account~\cite{phoenix1989counter} and used this to calculate the atomic inversion. Yu {\it et al.} made use of the {\it polaron transformation}\index{Polaron! transformation}~\cite{Mahan,wagner1986unitary}
\begin{equation}\label{polaron2} 
\hat{U}_p=\exp\left[\sqrt{2}g\hat{\sigma}_z\left(\hat{a}^\dagger-\hat{a}\right)\right]
\end{equation}
to rewrite the Hamiltonian in a new {\it polaron basis}\index{Polaron! basis} which was then expanded in the boson operators~\cite{yu2012analytical} (in some communities the polaron transformation~(\ref{polaron2}) has also been called a {\it Lang-Firsov transformation}~\cite{lang1963kinetic}\index{Lang-Firsov transformation}). The same polaron transformation was also utilized in ref.~\cite{hausinger2010qubit}, to be followed by performing a perturbation theory in the bare atomic Hamiltonian. In this way, the results were valid for negative detunings $\Delta$. Another perturbative method was employed in ref.~\cite{he2012first} by first writing the Schr\"odinger equation as a recursion equation and then systematically increasing the number of terms in the recursion. To lowest order, the JC results were regained. The same kind of approach was also used in ref.~\cite{liu2009generalized} where the authors gave an expression (given in terms of a determinant of a $4\times4$ matrix) for the energies, and also in~\cite{chen2011exact}, where the recursive formula was derived from a coherent-state {\it ansatz}. Using a $t$-expansion method~\cite{horn1984t}, Fessatidis {\it et al.} studied low energy properties of the quantum Rabi model~\cite{fessatidis2002moments}.

Other methods that have been employed include to use the Bargmann representation~(\ref{barg})\index{Bargmann! representation} and from there on derive a polynomial equation~\cite{kocc2002quasi} or a continued fraction~\cite{braak2013continued}. A continued fraction approach was also considered in ref.~\cite{ziegler2012short}, but it resulted from different, algebraic, methods. An advantage with analytical approximations relying on continued fractions is that convergence is often fast compared to different expansion techniques. Several works have also analyzed the quantum Rabi model by means of variational methods~\cite{chen1989variational,stolze1990quality,bishop1999variational,hwang2010variational}. In ref.~\cite{chen1989variational} the polaron transformation (\ref{polaron2}) was utilized, and the vacuum was used as an {\it ansatz} state. Using a variational approach based on the {\it coupled cluster method}, Bishop {\it et al.} studied the formation of the double-well structure as explained above in terms of the BOA and discussed if this could be identified as a second-order quantum phase transition\index{Phase transition! Dicke}.  Hwang {\it et al.} constructed a semiclassical theory similar to the BOA to motivate an {\it ansatz} state consisting of two superimposed coherent states~\cite{hwang2010variational}. This allowed them to obtain good estimates for the amount of squeezing. The semiclassical approximation valid for large field amplitudes and explained above in sec.~\ref{sssec:semclas} for the JC model has also been modified to the quantum Rabi model in~\cite{finney1994quasiclassical}.


\subsubsection{Integrability of the quantum Rabi model}\label{sssec:rabiint}
The remarkable simplicity of the quantum Rabi model has spurred intense activity in trying to determine its analytical solutions. At first, it is not clear that they should exist, {\it i.e.} whether the model is integrable or not. We have seen that numerous analytical approximations exist which are valid in certain parameter regimes. Nevertheless, at large times these approximations and even numerical diagonalization may fail to predict accurate results. This is a more practical motivation why an analytic solution is desirable. Another reason, of more fundamental and broad interest, is that understanding these aspects of the quantum Rabi model could shed light on ideas revolving around {\it quantum integrability}\index{Quantum! integrability}~\cite{caux2011remarks}. In classical physics, the concept of integrability is well defined in terms of number of degrees of freedom relative to the numbers of conserved quantities of motion~\cite{arnol1989mathematical,gutzwiller1990chaos}. Similar definitions for quantum systems directly lead to several issues~\cite{caux2011remarks}. Nevertheless, the general consensus seems to be that the more constants of motion that a quantum system supports the more likely it is to be integrable.

The JC and quantum Rabi models couple spin and motional degrees of freedom (`motional' refers here to the boson mode). A symmetry operation should then transform both degrees of freedom. In the JC model, the conserved total number of excitations generates a continuous $U(1)$ symmetry, the individual number of excitations of the atom or the field are not conserved. In the quantum Rabi model it is clear that the counter-rotating terms do not preserve the excitation number. However, the adiabatic potentials of fig.~\ref{fig10} (a) suggests that there is a preserved parity where $x\leftrightarrow-x$. Remember how the BOA is constructed, {\it i.e.} the spin degree of freedom is adiabatically slaved to the field, it follows that the parity transformation of the field alone is not a symmetry. In fig.~\ref{fig10} (b) we show how the spin of the adiabatic states corresponding to the lower adiabatic potential $V_\mathrm{ad}^{(-)}(x)$ depends on the variable $x$. By flipping the sign of $x$ we also need to make a $\pi$-rotation of the spin around $\hat{\sigma}_z$ in order to have a full symmetry operation. Thus, the quantum Rabi model supports the $\mathbb{Z}_2$ parity symmetry~\cite{casanova2010deep,larson2007dynamics,bina2012solvable,larson2008wave}
\begin{equation}\label{parity}
\hat{\Pi}=e^{i\pi\left(\hat{n}+\hat{\sigma}_z/2\right)}.
\end{equation}
Recently it has also been shown that for certain parameter conditions, the quantum Rabi model also has a `hidden symmetry'~\cite{gardas2013new} as will be discussed further below. 

The presence of the discrete $\mathbb{Z}_2$ symmetry does not automatically mean that the quantum Rabi model is expected to be solvable. To demonstrate the importance of symmetries it has been shown that the breaking of the JC continuous $U(1)$ symmetry by including the counter rotating terms can have great influence on the system properties and also on physical observables in the weak coupling regime~\cite{larson2012absence,larson2013rotating}. The reason for this is that the topology of the 'energy landscapes' of the JC and quantum Rabi models are different and in the adiabatic limit when time-scales are very long the topology can indeed be manifested in for example {\it geometric phases}\index{Geometric phase}. Performing the RWA or not may also alter the universality classes of related critical models as demonstrated in ~\cite{zueco2009qubit,altintas2013dissipative}. Related to these observations is also the qualitative change in the corresponding classical equations of motion between the JC and quantum Rabi models. 
If a coherent-state {\it ansatz} for both the field and the atom (in terms of spin coherent states\index{Spin! coherent state}\index{State! spin coherent}) is assumed, one can assign the quantum Rabi model a corresponding classical model~\cite{graham1984two,larson2013integrability}. The corresponding classical equations of motion are chaotic, {\it i.e.} the solutions show an exponential sensitivity to perturbations in their initial conditions (characterized by a positive {\it Lyapunov exponent}\index{Lyapunov exponent} $\lambda$~\cite{arnol1989mathematical,gutzwiller1990chaos}). Indeed, the equations of motion for the Dicke model are identical in form where it is since long known that relaxing the RWA leads to classical chaos\index{Chaos!classical}~\cite{milonni1983chaos}. Kujawski has shown, however, that for certain parameter values one can find periodic solutions of the semiclassical quantum Rabi model~\cite{kujawski1988exact}. The existence of periodic solutions in classically chaotic systems typically results in {\it quantum scars}~\cite{heller1984bound}\index{Quantum! scar} in the corresponding quantum systems. This has not been studied in terms of the quantum Rabi model, but instead for the highly related Dicke model~\cite{Villasenor2020,sinha2021fingerprint,pilatowsky2021ubiquitous} to be discussed in sec.~\ref{sssec:dicke}. Another important feature of integrability and chaos is the {\it level statistics}~\cite{haake2010quantum}\index{Level! statistics}. Integrable models\index{Integrable model} typically exhibit {\it Poissonian level statistics}, and it has shown that the quantum Rabi model does not obey this property~\cite{graham1984two,larson2013integrability}.
 
The above argumentation does not rule out the solvability (or integrability) of the quantum Rabi model. The standard approach in order to study the solvability of the quantum Rabi model is to rewrite the Schr\"odinger equation in the Bargmann representation~(\ref{barg})~\cite{bargmann1961hilbert}. The Schr\"odinger equation is either written as a single second-order differential equation or, alternatively, as two coupled first-order differential equations. For certain parameters, the emerging differential equation can be written in a known form where the solutions are the {\it confluent Heun functions}~\cite{zhong2013analytical}. However, it  is more common to expand the solutions in a series of other known functions and the coefficients are given as a recurrence relation. Reik {\it et al.} argued that such expansions may terminate, which opens up the possibility of closed-form solutions of the quantum Rabi model~\cite{reik1986exact,reik1987exact}. Kus and Lewenstein were the first to use this method to identify a set of isolated solutions~\cite{kus1986exact}. There is an equivalence between the quantum Rabi model and the $E\times\beta$ Jahn-Teller model\index{$E\times\beta$ Jahn-Teller model}\index{Model! $E\times\beta$ Jahn-Teller}\index{Jahn-Teller model} originating from molecular physics~\cite{reik1987exact,reik1994generalized,szopa1997canonical,kocc2002quasi}, and the solutions found in~\cite{kus1986exact} are the so-called {\it Juddian solutions}\index{Juddian! solution} known from the $E\times\varepsilon$ Jahn-Teller model~\cite{judd1979exact}. The properties of these solutions, plus the discovery of possible other solutions, were discussed in~\cite{maciejewski2013full,wang2015comment}.  

A breakthrough in the history of solvability of the quantum Rabi model came with ref.~\cite{braak2011integrability}. Braak conjectured that the quantum Rabi model is indeed solvable and that the spectrum is given by the zeros of transcendental functions $G_{\pm}(x)$, where $x=(E+g^2)/\omega$ is the {\it spectral parameter}\index{Spectral! parameter} and $\pm$ denotes the parity quantum number. $G_{\pm}(x)$ are linear combinations of confluent Heun functions. He also discussed the driven quantum Rabi model, which has become known as the {\it asymmetric quantum Rabi model},\index{Asymmetric quantum Rabi model}\index{Model! asymmetric quantum Rabi} 
\begin{equation}\label{rabiham4}
\hat{H}_\mathrm{asR}=\omega\hat{n}+\frac{\Omega}{2}\hat{\sigma}_z+g\left(\hat{a}+\hat{a}^\dagger\right)\hat{\sigma}_x+\frac{\epsilon}{2}\hat\sigma_x,
\end{equation}
and claimed that it was also solvable despite lacking the $\mathbb{Z}_2$ parity symmetry characteristic of the quantum Rabi model (Level crossings appear in the eigenspectrum of the asymmetric quantum Rabi model for certain values of the drive parameter.). Hence, it would be an example of a quantum solvable, but not integrable model (according to the suggested definition of quantum integrability given in~\cite{braak2011integrability}). Chen {\it et al.} derived the same analytical expression which generates the spectrum not using the Bargmann method but a `Bogoliubov operator method'~\cite{chen2012exact}\index{Bogoliubov transformation}. Moroz came to the conclusion~\cite{moroz2012spectrum,moroz2013solvability} that the quantum Rabi model is not {\it exactly solvable} but it is instead an example of a {\it quasi-exactly solvable model}~\cite{turbiner1987spectral,bender1995quasi} (see also Refs.~\cite{debergh2001quasi,kocc2002quasi}). He claimed that the existence of parity degenerate Juddian exact isolated analytic solutions is a direct consequence of the quasi-exact solvability (for those energies  corresponding to the eigenvalues of the displaced harmonic oscillator,  {\it i.e.}, for an integer $x$). On the other hand, for generic values of the parameters $\omega, \Omega,g$ no quasi-exact solutions exist and the regular spectrum is given by the zeros of $G_\pm(x)$, where $x$ is not an integer, while the eigenstates are not polynomials but Heun functions.

To rephrase, the main hindrance in solving the quantum Rabi model is the failure of any finite {\it ansatz} for the wave function, as is possible for the harmonic oscillator or the hydrogen atom; in~\cite{zhong2013analytical} we read that the eigenstates of the model must be expressed through transcendental functions beyond the hypergeometric class, the aforementioned confluent Heun functions, and cannot be written as any finite linear combination of states with fixed photon number. We pause here to discuss some further consequences of breaking the $U(1)$ symmetry of the JC model down to $\mathbb{Z}_2$. In the quantum Rabi model, one can naturally divide the spectrum into a regular and an exceptional part, the latter comprising the degenerate solutions which are at the same time quasi-exact -- termed {\it Juddian} solutions\index{Juddian! solution}. The spectral graph consists of two intersecting ladders, each with infinitely many rungs. On the other hand, in the JC model we have infinitely many intersecting ladders each with two rungs. This is the central difference between the two models. The energy distance between JC-doublets grows with both $g$ and $n$, eventually leading to level crossings in the ground state. In the quantum Rabi model, however, the levels are roughly equidistant for a fixed parity, for all couplings and energies if the cavity frequency is not too large. This qualitative feature can be read off from the pole structure of $G_{\pm}(x)$ whose zeros determine the eigenenergies. The poles are located at integer values of $x$, the quasi-exact energies, which indicates that these solutions are usually doubly degenerate. In this sense and only, one may call this a proper solution of a restricted problem with regard to the spectrum. To extract other important features, such as the form of the ground state or the content of squeezing, approximate methods are better suited. The same also applies to systems amenable to the Bethe {\it ansatz}\index{Bethe {\it ansatz}}. The Bethe {\it ansatz} yields only the spectrum and its qualitative features but not the exact eigenstates. In~\cite{Braak2013}, Braak introduced a complex-valued generalization of $G_{\pm}(x)$. Through a suitable choice of an additional complex parameter $z$, a much better numerical control over the the high-energy part of the spectrum can be achieved. Braak used this generalization to dispel the doubts that some of the zeros of $G_{\pm}(x)$ are not physical, as Maciejewski and collaborators had maintained in~\cite{Maciejewski2012A, Maciejewski2012B}. Subsequently, the latter authors proceeded to claim that ``for an integer value of the spectral parameter\index{Spectral! parameter} $x$, in addition to the finite number of the classical Juddian states there exist infinitely many possible eigenstates'' in~\cite{Maciejewski2014}. The non-degenerate exceptional solutions have been discussed by Wakayama and Yamasaki in~\cite{Wakayama_2014, sym11101259}. For each integer $x=N$,  one may derive a (transcendental) function $G(N,g,\Omega)$ whose zeros in the parameters $g$ and $\Omega$ give the conditions for the existence of a non-degenerate exceptional solution with $E=N-g^2$ $(\omega=1)$. This is analogous to the constraint polynomials in $g$ and $\Omega$ for the Juddian solutions reported in~\cite{kus1986exact}. The solutions in question, however, are not quasi-exact, for the eigenstates are not polynomials.  

When discussing the question of integrability, attention should be drawn to the distinction between discrete and continuous degrees of freedom. The situation is simple if one has only one type in the system under consideration. The Hubbard model and spin chains, for example, only have discrete degrees of freedom, whereas the Lieb-Liniger model only has continuous degrees of freedom; in the latter, we are dealing with the quantization of a classical model, in contrast to the Hubbard and the Rabi model. The quantum Rabi model, on the other hand, has one continuous and one discrete degree of freedom, which complicates the situation. Braak has introduced a notion of integrability based on a ``fine-grained'' labeling of states, which is also applicable to the known integrable models (see~\cite{reyes2021remarks}). In the Hubbard model, for instance, it is the set of conserved rapidities\index{Rapidity} associated with integer ``quantum numbers'' appearing in the Bethe {\it ansatz}\index{Bethe {\it ansatz}} which allow a fine-grained labeling of eigenstates. What is essential is to have one quantum number for each degree of freedom, which is the case for the quantum Rabi model but not for the Dicke model. It becomes increasingly clear that to understand integrability on the quantum level, one has to look at a somewhat arcane mathematical concept, the {\it analytic commutant}\index{Commutant} of the Hamiltonian~\footnote{D. Braak, private communication}.

Determining whether an operator $\hat{Y}$, satisfying $[\hat{Y}, \hat{H}]=0$, is independent of the Hamiltonian $\hat{H}$ or not has been the major obstacle to define integrability for quantum systems~\cite{caux2011remarks}. If there is a spectral degeneracy, then a ``symmetry operator'' $\hat{Y}$ exists, belonging to the commutant\index{Commutant} of the Hamiltonian, which is not generated by $\hat{H}$ itself; such a degeneracy, however, should not be necessary for the existence of a non-trivial symmetry generator. Every quantum system seems to have a complete set of mutually commuting operators, the projectors $\hat{P}_{\lambda_i}$ onto the eigenvectors of $\hat{H}$ with eigenvalue $\lambda_i$. The projectors generate the commutant of $\hat{H}$ provided that it is self-adjoint and possesses a pure point spectrum.  These operators, however, cannot play a role analogous to that of the $N$ functions $L_i$ on a $2N$-dimensional phase space with the property $\{L_i, L_j\}=0$ (where $\{\cdot, \cdot\}$ is the Poisson bracket\index{Poisson! bracket}), functions whose existence renders the system integrable in classical mechanics (in the Liouvillian sense). For classically integrable systems, the functions $L_i$ must be independent from the Hamiltonian $H=L_1$ and from each other. On the other hand, the operators $\hat{P}_{\lambda_i}$ exist for every self-adjoint operator $\hat{H}$ and generate its commutant, however they are not all independent from $\hat{H}$.  In fact, $\hat{P}_{\lambda_i}$ can be written as a bounded function of $\hat{H}$ itself: $\hat{P}_{\lambda_i} =\chi_{\lambda_i}(\hat{H})$, where $\chi_{\lambda_i}(\lambda)$ is the characteristic function on any interval containing only  $\lambda_i$.

``Locality'', a notion usually invoked to discern ``true'' conserved quantities from these projectors in many-body systems, is ill-defined for systems with only few degrees of freedom such as the JC model and its relatives,  where ``spatial distance'' lacks physical meaning. Therefore, a modified notion of independence is warranted. A proposal in this direction was made in~\cite{reyes2021remarks} (see our discussion below on the asymmetric QRM). Still, there is no general consensus on the proper definition of integrability in quantum mechanics~\cite{caux2011remarks}. For a discussion on the connection between integrability and quantum thermalization in the driven Rabi model\index{Rabi! model!driven} see~\cite{Larson_2013}.

Let us briefly visit the unified approach to nonlinear Rabi models based on the analysis of~\cite{Duan2022}. The generators of the su(1,1) group satisfy the commutation relations
\begin{equation}
 \left[\hat{K}_0, \hat{K}_{\pm}\right]=\pm \hat{K}_{\pm}, \quad \quad \left[\hat{K}_{+}, \hat{K}_{-}\right]=-2\hat{K}_0.
\end{equation}
The corresponding Casimir operator\index{Casimir operator} can be written as
\begin{equation}
 \hat{C}=\hat{K}_0^2 - \tfrac{1}{2} \left(\hat{K}_{+}\hat{K}_{-} + \hat{K}_{-}\hat{K}_{+}\right),
\end{equation}
which commutes with all elements of the su(1,1) algebra. The basis state $\ket{k,m}$ can be chosen such that the following relations are satisfied:
\begin{equation}
\begin{aligned}
 &\hat{K}_0\ket{k,m}=(k+m)\ket{k,m}, \quad \quad \hat{K}_+\ket{k,m}=\sqrt{(m+1)(m+2k)}\ket{k,m+1},\\
 &\hat{K}_-\ket{k,m}=\sqrt{m(m+2k-1)}\ket{k,m-1}, \quad\quad \hat{C}\ket{k,m}=k(k-1) \ket{k,m}, 
 \end{aligned}
\end{equation}
where $m=0,1,2,\ldots$ and $k$ is the Bargmann index\index{Bargmann! index}, separating different irreducible representations. All states can be obtained from the lowest $\ket{k,0}$ through successive actions of the raising operator $\hat{K}_{+}$ according to
\begin{equation}
 \ket{k,m}=\sqrt{\frac{\Gamma(2k)}{m! \Gamma(2k+m)}}\,\hat{K}_{+}^{m}\ket{k,0}.
\end{equation}

Within the su(1,1) algebra, a general nonlinear Rabi model has the Hamiltonian
\begin{equation}\label{eq:HnonlinR}
 \hat{H}_{nl, R}= \tfrac{1}{2} \Omega\hat{\sigma}_z + \omega \hat{K}_0 + g \hat{\sigma}_x \left(\hat{K}_{+} + \hat{K}_{-}\right).
\end{equation}
Now, the parity operator can be defined as $\hat{\Pi} \equiv -\hat{\sigma}_z \otimes \hat{T}$, with $\hat{T} \equiv \exp[i\pi (\hat{K}_0-k)]$. We may easily deduce that:
\begin{equation}
 \hat{\Pi} \hat{\sigma}_z \hat{\Pi}^{\dagger}=\hat{\sigma}_z, \quad \hat{\Pi} \hat{\sigma}_x \hat{\Pi}^{\dagger}=-\hat{\sigma}_x, \quad \hat{\Pi} \hat{K}_0 \hat{\Pi}^{\dagger}=\hat{K}_0, \quad \hat{\Pi} \hat{K}_{\pm} \hat{\Pi}^{\dagger}=-\hat{K}_{\pm},
\end{equation}
whence $\hat{\Pi} \hat{H}_{nl, R} \hat{\Pi}^{\dagger}=\hat{H}_{nl, R}$ and $[\hat{H}_{nl, R}, \hat{\Pi}]=0$. Since the parity operator has eigenvalues $\Pi=\pm 1$, we can separate the entire Hilbert space into two subspaces with even and odd parities respectively. Unlike the linear quantum Rabi model, the Hamiltonian~\eqref{eq:HnonlinR} also commutes with the Casimir operator $\hat{C}$, which separates the entire Hilbert space into different subspaces distinguished by the Bargmann index $k$~\cite{Duan2022}. Each type of nonlinear Rabi model is associated with a particular Bargmann index. For instance, in the two-photon quantum Rabi model\index{Two-photon!quantum Rabi model}~\cite{Emary2002, Duan2016} we select: $\hat{K}_0=\frac{1}{2}(\hat{a}^{\dagger}\hat{a} + \frac{1}{2})$, $\hat{K}_{+}=\frac{1}{2}(\hat{a}^{\dagger})^2$, $\hat{K}_{-}=\hat{K}_{+}^{\dagger}$. In the Fock-state basis $\ket{n}_a$, the basis state $\ket{k,m}$ can be written as $\ket{k,m}=\ket{2(m+k-\frac{1}{4})}_{a}$; therefore, the number of bosons is odd or even for $k=\frac{3}{4}$ or $k=\frac{1}{4}$, respectively.  

\begin{figure}
 \includegraphics[width=\textwidth]{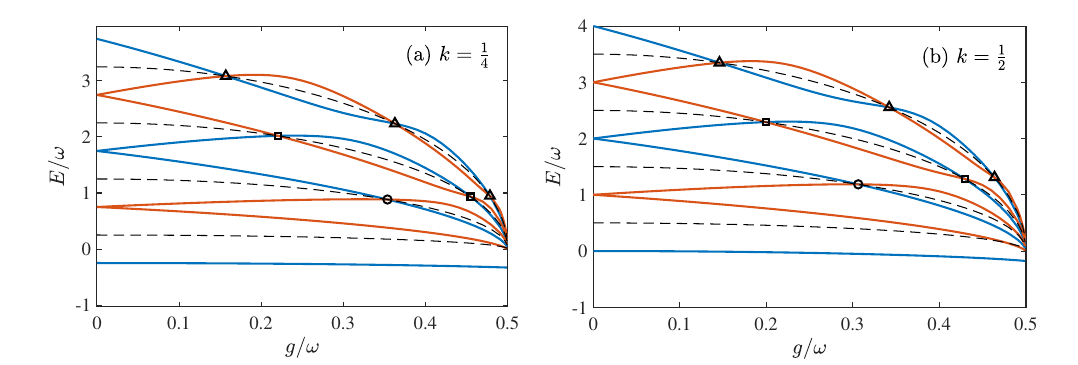}
 \caption{Energy spectrum as a function of the coupling strength $g$ at $\Omega=\omega=1$ for $k=\frac{1}{4}$ in {\bf (a)} and $k=\frac{1}{2}$ in {\bf (b)}. The blue (red) solid lines correspond to the even (odd) parity. The dashed lines refer to the baseline energies $E=\sqrt{\omega^2-4g^2}(k+M)$. The circle, square and triangle correspond to the exact isolated solutions with $M = 1, 2$ and $3$, respectively, at the level crossings between different parities, for which the eigenstates can be reduced to a closed form. Source: Fig. 1 of~\cite{Duan2022}; reproduced with permission from the author.}
 \label{fig:DuanFigNJP}
\end{figure}

In the basis of the $\hat{\sigma}_{x}$ eigenstates $\ket{\pm}$ (with $\hat{\sigma}_x \ket{\pm}=\pm \ket{\pm}$), the Hamiltonian $\hat{H}_{nl, R}$ can be written in a matrix form as
\begin{equation}
 \hat{H}_{nl, R}=\begin{pmatrix}
                  \omega \hat{K}_0 + g (\hat{K}_{+} + \hat{K}_{-}) && -\tfrac{1}{2}\epsilon\\
                  -\tfrac{1}{2} \epsilon && \omega \hat{K}_0 - g (\hat{K}_{+} + \hat{K}_{-})
                 \end{pmatrix}.
\end{equation}
Since the diagonal elements are linear combinations of the su(1,1) algebra generators, one may easily attain its eigenvalues and eigenstates, which satisfy
\begin{equation}
 [\omega \hat{K}_0 \pm g (\hat{K}_{+} + \hat{K}_{-})] \ket{k,m}_{\pm}= \sqrt{\omega^2-4g^2}\,(k+m) \ket{k,m}_{\pm},
\end{equation}
with 
\begin{equation}
 \ket{k,m}_{\pm}=\hat{S}(\mp r) \ket{k,m}, \quad \quad \hat{S}(\mp r)=\exp\left[\mp \frac{r}{2} (\hat{K}_{+} - \hat{K}_{-}) \right],
\end{equation}
where $\hat{S}(\mp r)$ is a (generalized) squeezing operator. When $g \to \omega/2$, a spectral collapse\index{Spectral! collapse} takes place and the model is no longer self-adjoint~\cite{Braak2023} -- compare with the collapse of the quasi-energy spectrum occurring in the driven JC model that we met in subsec.~\ref{sssec:drivejc}. The collapse is visible in the energy spectrum plotted for $k=\frac{1}{2}, \frac{1}{4}$ in fig.~\ref{fig:DuanFigNJP}.

The spectrum of the two-photon quantum Rabi model\index{Two-photon!quantum Rabi model}, with Hamiltonian
\begin{equation}
 \hat{H}_{R, 2p}=\omega \hat{a}^{\dagger}\hat{a} + (\hat{a}^2 + \hat{a}^{\dagger 2})\hat{\sigma}_z + \Delta \hat{\sigma}_z,
\end{equation}
(scaled with $g$ set to unity) undergoes a dramatic change when $\omega$ approaches the critical value $\omega_c=2$ from above. For $\omega<2$, the Hamiltonian is no longer bounded from below, while the spectrum contains a discrete and a continuous part exactly at $\omega=2$. The generalized spectral determinants of this model for $\omega>2$ exhibit different properties depending on their derivation, with respect to their poles and the possibility of their analytical determination. Chen's $G$-function stands out since the pole distance yields the average distance of energy levels, as well as the quasi-exact Juddian points of the spectrum, in the same way as the corresponding $G$-function of the linear quantum Rabi model~\cite{Braak2023}.

As we mentioned, Braak's result also applies for the {\it asymmetric (anisotropic) quantum Rabi model}\index{Quantum! Rabi model! asymmetric} in which the $\mathbb{Z}_2$ symmetry is explicitly broken, but still a solution is found. This naturally opens up for the question whether such a model possesses a {\it hidden symmetry}\index{Hidden symmetry} that could explain its solvability. In fact, for certain drive amplitudes $\epsilon=\mathbb{Z}/2$, the energy levels of the asymmetric quantum Rabi model display level crossings which typically signal some symmetry. This has been numerically explored in~\cite{ashhab2020attempt} with the conclusions that if there exists a hidden symmetry it is not a traditional symmetry since it seems to depend on system parameters. The level crossings appear as conical intersections\index{Conical intersection} in the energy landscape when parametrized as a function of $g$ and $\epsilon$~\cite{batchelor2015energy,li2020conical}. These conical intersections are different from those to be discussed in sec.~\ref{sssec:multi} (see fig.~\ref{fig11}); the present ones appear in parameter space while those of the next section show up in variable space. The quest for understanding this hidden symmetry has been pushed forward by Batchelor and others~\cite{li2015algebraic,li2020hidden,mangazeev2021hidden,reyes2021remarks,lu2021hidden,reyes2021degeneracy}. In~\cite{mangazeev2021hidden,reyes2021degeneracy}, the hidden symmetry was looked for via an operator series expansion at $\epsilon=\mathbb{Z}/2$, and a set of coupled recurrence relations were derived, when solved gives the desired symmetry $\hat J_\epsilon$ that commutes with the Hamiltonian. For the $\epsilon=0$ case, the method was shown to render a symmetry $\hat J_0$ identical to the parity symmetry~(\ref{parity}), while for $\epsilon=1/2$ the symmetry is
\begin{equation}
\hat J_{1/2}=2\,e^{i\pi\hat a^\dagger\hat a}\left[
\begin{array}{cc}
\frac{\Delta}{2g} & g-\hat a\\
g+\hat a^\dagger & \frac{\Delta}{2g}
\end{array}\right].
\end{equation}
For higher values $\epsilon$, the expressions for the symmetries become cumbersome. These operators are in general not parity operators since $\hat J_\epsilon^2\neq1$, and the authors postulated that one has $\hat J_\epsilon^2=p(\hat H_\mathrm{asR})$ with $p(x)$ a polynomial of $x$. This conjecture has been proven in~\cite{reyes2021remarks} and entails that $\hat{J}_\epsilon$ and $\hat{H}_{\mathrm{asR}}$ are the two \emph{independent} generators of the \emph{analytic} commutant\index{Commutant} of $H_{\mathrm{asR}}$, because $\hat{J}_\epsilon$ is an analytic function of $\hat{a}$ and $\hat{a}^\dagger$ but not an analytic function of $\hat{H}_{\mathrm{asR}}$ itself. On the other hand, the projectors $\hat{P}_{\lambda_j}$ we mentioned above do not belong to the analytic commutant because they are continuous but not analytic functions of $\hat{H}_{\mathrm{asR}}$. 

The properties of the ground state and the excitation gap for the {\it anisotropic Rabi model}\index{Quantum! Rabi model! anisotropic}, with anisotropy parameter $\lambda$, were assessed in~\cite{YingZJ2022}, where it is reported that a hidden symmetry breaking accompanies the conventional quantum phase transition at the boundary $g_c^{(\lambda)}=2/(1+|\lambda|) g_c$, where $g_c=\sqrt{\Omega \omega}/2$ is the transition point of the ordinary quantum Rabi model. Tuning the frequency up reveals a fine structure of the phase diagram, while the additional boundaries involve a parity change [see fig. 1 of~\cite{YingZJ2022}]. The entanglement entropy in the {\it asymmetric quantum Rabi model}\index{Quantum! Rabi model! asymmetric} -- where the two-level atom is statically driven -- shows a number of resonance valleys. This feature signifies an efficient coupling of the relevant quantum states, essentially determined by the energy spectrum of the system~\cite{Shi2022}.

At the beginning of this section, we listed the approximations behind the JC model, and that by relaxing the RWA we derived the quantum Rabi model~(\ref{rabiham}). If we are to include the diamagnetic\index{Diamagnetic term} self-energy term\index{Self-!energy term} the model becomes 
\begin{equation}\label{rabiham3}
\hat{H}_\mathrm{anR}=\omega\hat{n}+\frac{\Omega}{2}\hat{\sigma}_z+g\left(\hat{a}+\hat{a}^\dagger\right)\hat{\sigma}_x+\mu\left(\hat a+\hat a^\dagger\right)^2,
\end{equation}
where the new coupling constant is $\mu$. In sec~\ref{sssec:dicke} we will discuss in more detail the implications of the self-energy term when one considers the deep strong coupling regime. For now we may notice that the self energy term can be transformed away by application of the {\it squeezing operator}\index{Squeezing! operator}\index{Operator! squeezing}
\begin{equation}\label{sqop}
\hat S(z)=\exp\left[\frac{1}{2}\left(z^*\hat a^2-z\hat a^{\dagger 2}\right)\right],
\end{equation}
with a properly chosen squeezing parameter. The transformed Hamiltonian takes the form of a quantum Rabi model with renormalized parameters~\cite{feranchuk2020exact}. An alternative way to transform the Hamiltonian such that the self-energy term vanishes is to note that the bare field Hamiltonian $\hat H_\mathrm{f}=\hat a^\dagger\hat a+\mu\left(\hat a+\hat a^\dagger\right)^2$ is quadratic and can be diagonalized by a Bogoliubov transformation~\cite{de2018breakdown}\index{Bogoliubov transformation}. The interesting observation is that the transformation does not change the structure of the interaction term, and the resulting Hamiltonian has the desired form. We thereby conclude that the solvability of the model survives the addition of a self-energy term. 

We note that since the work of Braak, the same approach has been applied in order to find energies analytically for the $N=3$ atom quantum Rabi model~\cite{braak2013solution} ({\it i.e.} Dicke model with three atoms), for the two atom quantum Rabi model with additional dipole couplings between the atoms~\cite{peng2013regular}, and special solutions containing at most a single photon for the multi-mode and multi qubit quantum Rabi model
~\cite{peng2021one}. The case of different strengths between the JC's and the counter-rotating coupling terms, 
\begin{equation}\label{anRabi}
\hat{H}_\mathrm{anR}=\omega\hat{n}+\frac{\Omega}{2}\hat{\sigma}_z+g_\mathrm{jc}\left(\hat{\sigma}_{+}\hat{a}+\hat{a}^\dagger\hat{\sigma}_{-}\right)+g_\mathrm{ajc}\left(\hat{\sigma}_{-}\hat{a}+\hat{a}^\dagger\hat{\sigma}_{+}\right)
\end{equation}
is called the {\it anisotropic quantum Rabi model}\index{Anisotropic! quantum Rabi model}\index{Model! anisotropic quantum Rabi}, and has been thoroughly analyzed~\cite{tomka2013exceptional,xie2014anisotropic,zhang2015analytical,joshi2016quantum,liu2017universal,zhang2017analytical,wang2018quantum, Chen2021}. Using both circularly polarized light and a circularly polarized dipole transition, the counter rotating term vanishes, {\it i.e.} $g_\mathrm{ajc}=0$ and the RWA is the correct picture~\cite{xie2014anisotropic}. In ref.~\cite{mahmoodian2019chiral}, the author considered such a scenario but with two degenerate oppositely circularly polarized modes in such a way that particular symmetry was not conserved (just like in the quantum Rabi model), but instead a chiral $U(1)$-symmetry emerged\index{$U(1)$ symmetry}\index{Chiral symmetry}. Leaving such a two mode model, we note that whenever both coupling terms, $g_\mathrm{jc}$ and $g_\mathrm{ajc}$, are non-zero, the continuous $U(1)$ symmetry of the $JC$ model has been broken down to a $\mathbb{Z}_2$ symmetry. We will return to this type of Hamiltonian in section~\ref{ssec:ionham} when we discuss trapped ion realizations of the JC model. It turns out that in those systems it is rather straight forward to control the two coupling strengths independently. The anisotropic quantum Rabi model can be unitarily transformed to a {\it parametrically driven JC model}\index{Parametrically driven Jaynes-Cummings model}\index{Model! parametrically driven Jyanes-Cummings} (see the previous sec.~\ref{sssec:drivejc} for driven JC models) 
\begin{equation}
\hat H_\mathrm{pdJC}=\delta_p\hat a^\dagger\hat a+\frac{\Delta_p}{2}\hat\sigma_z+g\left(\hat a^\dagger\hat{\sigma}_{-}+\hat{\sigma}_{+}\hat a\right)+\eta\left(\hat a^{\dagger 2}+\hat a^2\right),
\end{equation}
as shown in ~\cite{Gutierrez2021}, and where $\delta_p=\omega-\omega_p$ and $\Delta_p=\Omega-\omega_p$ with $\omega_p$ the drive frequency. The unitary is again the squeezing operator~(\ref{sqop}), {\it i.e.} up to a constant $\hat H_\mathrm{pdJC}=\hat S^\dagger(z)\hat{H}_\mathrm{anR}\hat S(z)$. The squeezing parameter obeys $\cosh(z)=g_\mathrm{jc}/g$ and $\sinh(z)=g_\mathrm{ajc}/g$. Finally, the solutions of~\cite{braak2011integrability} have also been employed in order to extract various correlation functions for different coupling strengths, ranging from the JC regime to the deep strong coupling regime~\cite{wolf2013dynamical}. They were also used to analyze the aforementioned conical intersections~\cite{batchelor2015energy}. In particular, these conical intersections coincide with the Juddian points\index{Juddian! solution}, see fig.~\ref{fig10}.\\ \\

The Rabi model and its extension called the {\it Rabi-Stark model}\index{Rabi-Stark model}\index{Model! Rabi-Stark} is described by the Hamiltonian~\cite{eckle2017generalization,xie2019quantum,xie2020first,xie2019exact} 
\begin{equation}\label{eq:RabiStarkHam}
 H_R=\left(\Delta/2+U\hat{a}^{\dagger}\hat{a}\right)\hat{\sigma}_z + \omega\hat{a}^{\dagger}\hat{a} + g(\hat{a}+\hat{a}^{\dagger})\hat{\sigma}_x,
\end{equation}
(where $U$ is the nonlinear coupling strength) has been reported to belong to different universality  classes~\cite{Chen2020}. Braak's method for solving the quantum Rabi model can be applied also to this model~\cite{eckle2017generalization}. The Rabi-Stark Hamiltonian\index{Rabi-Stark model}\index{Model! Rabi-Stark}~\eqref{eq:RabiStarkHam} can be mapped to an effective JC Hamiltonian. When the Stark coupling is zero, the resulting Hamiltonian aligns to that of the quantum Rabi model. We remark here that applying a perturbation theory that omits weak multiphoton processes\index{Multiphoton! process}, analytical solutions for the Rabi model can be obtained through a mapping to an effective JC model selecting a proper variation parameter~\cite{yu2012analytical}.  

In addition now to the usual JC term, the interaction between a plasma mode and a qubit contains a term which is quadratic in the oscillator variables, as has already been derived in~\cite{bertet2005dephasing}. The Hamiltonian of the canonical two-photon Rabi model includes the additional interaction term $g^{\prime}(\hat{a}^2 + \hat{a}^{\dagger 2})\hat{\sigma}_x$, arising in a superconducting circuit implementation where a nonlinear resonator is inductively coupled to a flux qubit\index{Qubit! flux}~\cite{felicetti2018two}. The model exhibits spectral collapse\index{Spectral! collapse} in a transition from a linear to an inverted oscillator with the critical point\index{Critical! point} being analogous to a free particle~\cite{rico2020spectral}, while in~\cite{Ying2021} we find a detailed analysis on the symmetry properties of the ground state. Finally, an infinite family of exact solutions of the two-photon Rabi model at energy level crossings outside the Juddian class was reported in~\cite{Maciejewski2019}. 

The notion of a time-dependent resonance condition in a generalized Rabi model was introduced in~\cite{Grimaudo2018}, when studying the exact quantum dynamics of a single spin-1/2 in a generic time-dependent classical magnetic field (see also~\cite{Belousov2021}). It was demonstrated that the time evolution of the transition probability between the two eigenstates of the atomic inversion may exhibit a regime of distorted oscillations, or may also display a monotonic behavior. An integrable model of two interacting qubits, which can be exactly and reduced to two independent single-spin quantum Rabi models via a unitary transformation has recently been introduced in~\cite{Grimaudo2023}, where the spin-spin coupling plays the role of the transverse field. Moreover, the free energy difference between the quantum and classical descriptions of a Hamiltonian has been used to decipher the so-called {\it quantumness} for the quantum to classical transition in the generalized Rabi model~\cite{zhuang2023quantumness}. 

A two-level system coupled to a linear harmonic oscillator and embedded in a viscous fluid is described by the dissipative quantum Rabi model\index{Quantum! Rabi model!dissipative}\index{Model! quantum Rabi!dissipative} for which, in the Ohmic regime, a BKT quantum phase transition\index{Phase transition! quantum! BKT} occurs by varying the coupling strength even for a very low dissipation rate~\cite{DeFilippis2023}. Shortly before, it had been shown that thermalization always occurs in the open quantum Rabi model when the two degrees of freedom are coupled to the same heat bath, but only for equal-temperature baths in the case of individual couplings~\cite{LiuFan2023}. Finally, it has been recently reported that eliminating the spin degree of freedom in the closed and open quantum Rabi, as well as in the Dicke model, leads to an effective harmonic oscillator whose eigenstates are squeezed\index{Squeezing} with respect to the physical harmonic oscillator. This formulation leads to time-independent uncertainties in the quadratures~\cite{gietka2023unique}.


\subsection{Extended Jaynes-Cummings models}\label{ssec:exjc}
Naturally, the list of extensions of the JC model can be made very long. As will be discussed in this section, extended JC models may exhibit characteristics very different from the regular JC model. Moreover, many of these extensions are not only of a theoretical interest but are also of relevance for modern experimental investigations (see the following sections). This section has been divided into five subsections, but it is understood that some works to be discussed belongs to more than a single sub-category and that the list of references is far longer than presented here. 

\subsubsection{Kerr medium and intensity dependent or multiphoton couplings}\label{sssec:kerr}
Typically, eliminating some degrees of freedom results in nonlinear terms in the emerging effective models. For the Jaynes-Cummings model, the elimination could be an additional boson field~\cite{buvzek1990dynamics,buvzek1991emission,joshi1992dynamical}, {\it e.g.}, a {\it Kerr medium}\index{Kerr! medium}\index{Kerr! effect}, or largely detuned atomic levels~\cite{alsing1987collapse}. The simplest way to describe a Kerr medium coupled to the cavity field would be to add the terms $\omega_\mathrm{k}\hat{b}^\dagger\hat{b}+\Gamma\left(\hat{a}^\dagger\hat{b}+\hat{b}^\dagger\hat{a}\right)$, where the bosonic operators $\hat{b}^\dagger$ and $\hat{b}$ create and annihilate Kerr excitations respectively. If the characteristic frequency $\omega_\mathrm{k}$ is large compared to other frequencies, the Kerr medium can be adiabatically eliminated to generate an effective photon-photon interaction term. Assuming also general interaction terms (in the RWA), the nonlinear JC model becomes
\begin{equation}\label{kerr}
\hat{H}_\mathrm{nlJC}=\omega\hat{n}+\frac{\Omega}{2}\hat{\sigma}_z\!+\!\chi\hat{a}^\dagger\hat{a}^\dagger\hat{a}\hat{a}+g\left[f(\hat{a},\hat{a}^\dagger)\,\hat{a}^\dagger\hat{\sigma}_{-}\!+\text{h.c.}\right]\!,
\end{equation}
where $\chi$ is the strength of the Kerr induced nonlinearity\index{Kerr! nonlinearity}, and $f(\hat{a},\hat{a}^\dagger)$ is a function of the field boson operators. If $f(\hat{a},\hat{a}^\dagger)=f(\hat{n})$ (normally $f(\hat{n})=\sqrt{\hat{n}}$ like in the {\it Buck-Sukumar model}~\cite{buck1981exactly})\index{Buck-Sukumar model}\index{Model! Buck-Sukumar} we say that the coupling is {\it intensity dependent}, while if $f(\hat{a},\hat{a}^\dagger)=\hat{a}^j\left.\hat{a}^\dagger\right.^k$ for some integers $j$ and $k$ (normally either $j$ or $k$ is zero) we say that the interaction is of the {\it multiphoton} type\index{Multiphoton! interaction}. The case of an intensity dependent coupling $f(\hat a,\hat a^\dagger)=\hat n$ is often termed a $\chi^{(2)}$-nonlinearity\index{$\chi^{(2)}$-nonlinearity}, to be compared to the term $\chi\hat{a}^\dagger\hat{a}^\dagger\hat{a}\hat{a}$ which gives a  $\chi^{(3)}$-nonlinearity\index{$\chi^{(3)}$-nonlinearity}. Since the RWA is assumed, like in the JC model, the number of excitations (modulo the number of photon exchange) is preserved which generates a continuous $U(1)$ symmetry. Due to this, the model $\hat{H}_\mathrm{nlJC}$ is analytically solvable.

The first analytical solution for the two-photon case stems from 1981~\cite{sukumar1981multi} and shortly afterwards several authors considered also the intensity dependent case, writing down the general solution~\cite{buck1981exactly,singh1982field}. For the solution with a general $\hat{f}(\hat{a},\hat{a}^\dagger)$ see~\cite{kochetov1987exactly}. While solving these models could be straightforward generalizations of how the JC models is solved, there have been other powerful methods relying on algebraic approaches~\cite{chaichian1990quantum,yang1997unified}. With the analytic solutions, expressions for the collapse and revival times have been presented for various initial field states~\cite{singh1982field,alsing1987collapse,chaichian1990quantum,vcrnugelj1994properties}. Also the field squeezing has been studied in the case of intensity dependent couplings~\cite{shumovsky1987squeezing,buzek1989jaynes,buvzek1989light,vcrnugelj1994properties} and it was found that there exist solutions where the amount of squeezing evolves periodically. Squeezing in the two-photon JC model\index{Two-photon! Jaynes-Cummings model}\index{Model! two-photon Jaynes-Cummings} was first studied by Gerry~\cite{gerry1988two} and he particularly considered the amount of squeezing as a function of the field intensity. Entanglement properties for the intensity dependent coupling was addressed in ref.~\cite{guo2011entropy} and for the multiphoton model\index{Multiphoton! model} in ref.~\cite{abdel2002entanglement}. A solution of the two-photon JC model including photon losses was given in~\cite{puri1988coherent} and it was argued that the two-photon JC model is more sensitive to decoherence than the regular JC model.

Effects of the Kerr medium\index{Kerr! medium} was a central topic in the 1990s with a series of works. The phase space evolution was considered in~\cite{werner1991q,joshi1992dynamical}. The collapse and revival times were estimated in the large field limit in~\cite{joshi1992dynamical,obada1998phase}, and also other field properties have been analyzed~\cite{buvzek1990dynamics}. One interesting observation is that there exists a `critical' detuning $\Delta_c$ for which the collapse is greatly suppressed by constructive interference~\cite{gora1992nonlinear}. In the same work it was shown how superstructures of the observables appear in the long time regime. Several works have discussed entanglement in the Kerr JC model\index{Kerr! Jaynes-Cummings model}\index{Model! Kerr Jaynes-Cummings} and many of these concluded that a large Kerr nonlinearity\index{Kerr! nonlinearity} reduces the atom-field entanglement~\cite{fang1995properties,abdel2002entanglement,yu2010partial,guo2011entropy}. Tara {\it et al.} demonstrated that in the largely detuned case the Kerr medium can actually be utilized for the preparation of field Schr\"odinger cat states~(\ref{catstate})\index{Cat state}\index{Cat state}~\cite{tara1993production}. In more recent years, a set of works considered external driving of the qubit in order to engineer the nonlinearity to realize desired forms. It was shown that if the driving is chosen in certain ways it is possible to achieve large nonlinearities, and also higher order nonlinearities, like terms scaling as $\sim\hat n^3$~\cite{wang2021photon}. The same method can be utilized in order to cancel existing Kerr nonlinearities\index{Kerr! nonlinearity}, which could compromise various state-preparation schemes.


\subsubsection{Multimode and multi-level atoms}\label{sssec:multi}
The number of possibilities to theoretically extend the Jaynes-Cummings model to include additional bosonic modes or more internal atomic levels is enormous. Practically, isolating certain transitions, which should also be allowed by selection rules, is, of course, not trivial. While there exist numerous models with considerably more levels or modes considered, here we will mainly focus on bimodal and three-level atom\index{Three-level atom} scenarios with the exception of the spin-boson model of eq.~(\ref{spinboson}). A discussion about the case of infinite number of modes will be presented in sec.~\ref{sssec:smapp}, where we discuss in some depth the single-mode approximation. For an early review on the topic of a few number of modes we refer to ref.~\cite{yoo1985dynamical}, and for more recent but less comprehensive one, to~\cite{messina2003interaction}. 

Among the simplest extensions is to keep the two-level structure of the atom and let it couple to two boson modes~\cite{benivegna1994new,napoli1996dressed,gerry1997generation,wildfeuer2003generation,larson2006scheme}\index{Bimodal! Jaynes-Cummings model}\index{Model! bimodal Jaynes-Cummings}
\begin{equation}\label{2mode}
\hat{H}_\mathrm{2m} = \omega_1\hat{n}_a+\omega_2\hat{n}_b+\frac{\Omega}{2}\hat\sigma_z+\left(g_a\hat{a}^\dagger\hat{\sigma}_{-}+g_a^*\hat{\sigma}_{+}\hat{a}\right)+\left(g_b\hat{b}^\dagger\hat{\sigma}_{-}+g_b^*\hat{\sigma}_{+}\hat{b}\right),
\end{equation}
with the two modes now labeled $a$ and $b$. As shown in~\cite{benivegna1994new}, by making use of a symmetry operator that includes both boson modes the above model can be transformed into a regular JC model plus a decoupled field. This transformation is especially simple for $\omega_a=\omega_b$ and when both couplings $g_a$ and $g_b$ are real~\cite{mattinson2001adiabatic,larson2006scheme}, it amounts to introducing the new Boson operators $\hat{A}=\left(g_a\hat{a}+g_b\hat{b}\right)/\sqrt{g_a^2+g_b^2}$ and $\hat{B}=\left(g_b\hat{a}-g_a\hat{b}\right)/\sqrt{g_a^2+g_b^2}$. Hence, the problem reduces to the one solved in sec.~\ref{ssec:JCm}, now with the coupling $\sqrt{g_a^2+g_b^2}$. The original boson modes do still, however, show non-trivial dynamics and in particular they build up mutual entanglement~\cite{wildfeuer2003generation}. In particular, by applying the semiclassical approach described in subsec~\ref{sssec:semclas} it is possible to prepare the two field states in so-called {\it entangled coherent states}\index{Entangled coherent states}\index{State! entangled coherent}~\cite{larson2006scheme}. These types of states can also be generated in the dispersive regime~\cite{gerry1997generation} as outlined in sec.~\ref{sssec:sol}. The populations of the individual modes in the present model was discussed in~\cite{benivegna1996quantum} and it was shown that photons tend to oscillate between the two modes. The method of introducing new boson operators can be readily generalized to any number of modes as long as they share the frequencies $\omega_k=\omega$~\cite{seke1985extended,lo1997canonical,wickenbrock2013collective,alaeian2019creating,hung2021quantum}. However, in the more realistic situation with different $\omega_k$'s the new modes will couple and the model does not allow for a simple solution, see the discussion on the spin-boson model below.

\begin{figure}
\includegraphics[width=7cm]{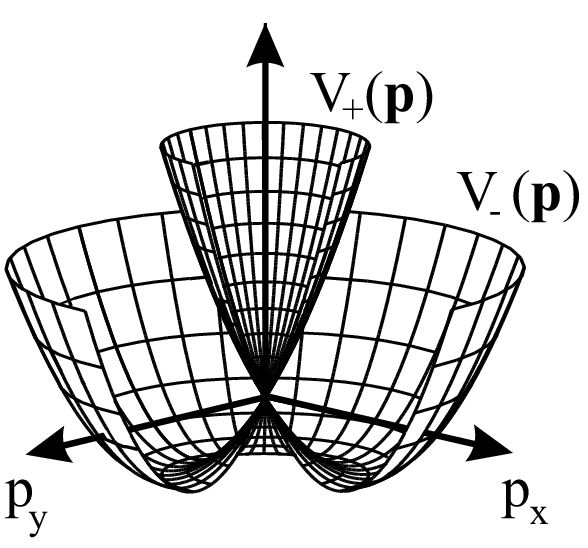} 
\caption{The two adiabatic potential surfaces\index{Adiabatic! potential! surfaces} of the conjugated $E\times\varepsilon$ Jahn-Teller model, eq.~(\ref{jt}), for $\Omega=0$. The lower surface has the familiar sombrero shape, and for $(p_x,p_y)=(0,0)$ the two surfaces touch in a conical intersection\index{Conical intersection}.}
\label{fig11}
\end{figure}

The phase dependence of the couplings $g_a$ and $g_b$ in eq.~\ref{2mode} turns out to have rather drastic consequences on the system evolution. In sec.~\ref{ssec:JCm} we argued that we can always choose the light-matter coupling real via a simple gauge transformation\index{Gauge! transformation}. This only works as long as we have imposed the RWA. When the RWA is relaxed this situation is more subtle; let us assume $g_a\in\mathbb{R}$, $g_b\in\mathbb{I}$ and $|g_b|=g_a\equiv g$ and $\omega_a=\omega_b=\omega$, and we express the Hamiltonian  in the quadrature representation (where we have done a transformation $\hat{x}\leftrightarrow\hat{p}_x$ and the same for the $b$ mode $\hat{y}\leftrightarrow\hat{p}_y$)~\cite{larson2008jahn}
\begin{equation}\label{jt}
\hat{H}_{E\times\varepsilon} = \displaystyle{\omega\left(\frac{\hat{p}_x^2}{2}+\frac{\hat{p}_y^2}{2}+\frac{\hat{x}^2+\hat{y}^2}{2}\right)+\frac{\Omega}{2}\hat{\sigma}_z} +\sqrt{2}g\left(\hat{p}_x\hat{\sigma}_x+\hat{p}_y\hat{\sigma}_y\right).
\end{equation}
This is a `conjugate' $E\times\varepsilon$ {\it Jahn-Teller model}\index{$E\times\varepsilon$ Jahn-Teller model}\index{Model! $E\times\varepsilon$ Jahn-Teller} known from molecular physics~\cite{longuet1958studies,wagner1986unitary}. The fact that $g_b$ was purely imaginary manifests in the fact that the $y$ mode couples to the $\hat\sigma_y$ dipole operator. For a general complex coupling, the coupling would include also the $\hat\sigma_x$ dipole operator. There is no simple rotation of the Pauli operators such that the two modes couple to the same dipole operators, {\it i.e.} we do not have the gauge freedom to chose both coupling amplitudes real.

In the BOA we diagonalize the internal degrees of freedom giving the adiabatic potential surfaces (`momentum dependent') $V_\pm(p_x,p_y)=\omega\frac{p_x^2+p_y^2}{2}+\pm\sqrt{\frac{\Omega^2}{4}+2g^2\left(p_x^2+p_y^2\right)}$. For $\Omega=0$, at $(p_x,p_y)=(0,0)$ the two surfaces become degenerate in a {\it Dirac cone}\index{Dirac! cone}, see fig.~\ref{fig11}. For $g>\sqrt{\omega\Omega}/2$ the lower potential builds up a sombrero shape and the minima is along a circle with non-zero `momentum' $|{\bf p}|$. This is the mechanism behind the {\it Jahn-Teller effect}\index{Jahn-Teller effect} which says that the ground state configuration is distorted from the most symmetric one ($(p_x,p_y)=(0,0)$). Not surprisingly, in sec.~\ref{sssec:dicke} this coupling defines the critical point\index{Critical! point} for the Dicke quantum phase transition\index{Phase transition! Dicke}. The linear dispersion in the vicinity of the Dirac cone implies that for small $p_x$ and $p_y$ the dynamics is effectively relativistic, which also holds true for the quantum Rabi model~\cite{bermudez2007exact,lamata2007dirac,larson2010travelling}. In molecular physics, the quadratures are interchanged such that the adiabatic potentials are instead $V_\pm(x,y)$ and the degeneracy is called a conical intersection\index{Conical intersection}. Encircling the conical intersection (or equivalently the Dirac cone) gives rise to a non-trivial {\it Berry phase}\index{Berry phase}~\cite{longuet1958studies}, which has been studied in terms of cavity QED in~\cite{larson2008jahn} and can be understood in the language of {\it Mead-Berry gauge theory}\index{Mead-Berry gauge theory}~\cite{mead1992geometric,baer2006beyond}. This Berry phase has been very recently observed experimentally in a trapped ion setup~\cite{Valahu2022} (see further sec.~\ref{sec:ion} on trapped ions). An initial Gaussian state\index{Gaussian! state} was prepared at the minima of the lower adiabatic potential of fig.~\ref{fig11}, and it was then let free to evolve. This causes the state to spread along the minima of the potential; eventually it will populate the entire region of the minima, and a self-interference occurs. The Berry phase manifests itself in the emerging interference patter. The dynamics in the vicinity of the Dirac cone -- the conical intersection -- was also explored experimentally, once more in a trapped-ion system, in ref.~\cite{Whitlow2022}. The emphasis was placed on the breakdown of the BOA. The type of coupling between the two-level system and the `momentum' is in the condensed matter community called (up to a unitary spin rotation) a {\it Rashba} or {\it Dresselhaus spin-orbit coupling}\index{Spin-orbit coupling}\index{Rashba spin-orbit coupling}\index{Dresselhaus spin-orbit coupling}~\cite{bychkov1984oscillatory}. We note that the Hamiltonian can be rewritten as~\cite{larson2009effective} 
\begin{equation}\label{jt2}
\hat{H}_{E\times\varepsilon} = \displaystyle{\omega\left(\frac{\left(\hat{p}_x-\hat{A}_x\right)^2}{2}+\frac{\left(\hat{p}_y-\hat{A}_y\right)^2}{2}+\frac{\hat{x}^2+\hat{y}^2}{2}\right)}+\displaystyle{\frac{\Omega}{2}\hat{\sigma}_z-\frac{g^2}{\omega}},
\end{equation}
where the {\it synthetic gauge potential}\index{Synthetic! gauge potential}\index{Gauge! potential! synthetic} $\left(\hat{\phi},\hat{A}_x,\hat{A}_y\right)=\left(\frac{\Omega}{2}\hat{\sigma}_z,\frac{\sqrt{2}g}{\omega}\hat{\sigma}_x,\frac{\sqrt{2}g}{\omega}\hat{\sigma}_y\right)$. Since $[\hat{A}_x,\hat{A}_y]\neq0$ the gauge structure of this model is {\it non-Abelian}\index{Non-Abelian} (see further discussions in sec.~\ref{sssec:mfmQED}). This inherent (spin-orbit generated) gauge structure (different from the Mead-Berry gauge theory mentioned above~\cite{larson2013anomalous}) can be used to understand complex dynamics like {\it intrinsic spin} and {\it anomalous Hall effect}\index{Anomalous Hall effect}\index{intrinsic spin Hall effect}~\cite{larson2010analog}. As a final remark on the bimodal two-level system, if, say, both $g_a$ and $g_b$ are real, the Dirac cone is not present and the Hall effects would vanish and so would the non-trivial Berry phase.

\begin{figure}
\includegraphics[width=11cm]{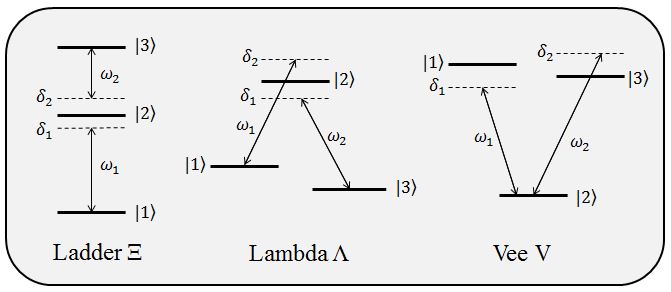} 
\caption{Possible three-level atom-light interaction configurations. A coupling between $|1\rangle$ and $|3\rangle$ is always assumed to be zero.}
\label{fig12}
\end{figure}

Turning to the three-level atoms\index{Three-level atom}\index{Atom! three-level}\index{Atom! Lambda} we have three main configurations; {\it Ladder} Xi ($\Xi$), {\it Lambda} ($\Lambda$), and {\it Vee} $V$\index{Ladder! atom}\index{Lambda! atom}\index{Vee atom}\index{$\Lambda$ atom}\index{$\Xi$ atom}\index{$V$ atom}. Like the two-level atoms can be viewed as qubits, the three-level atom can be considered as {\it qutrits}\index{Qutrit}, {\it i.e.} quantum dots with three internal logical states. The transitions are schematically pictured in fig.~\ref{fig12}, and within the RWA their respective Hamiltonians are
\begin{equation}\label{ladder}
\hat{H}_\Xi =  \displaystyle{\omega_1\hat{n}_a+\omega_2\hat{n}_b+\sum_{i=1}^3E_i\hat{\sigma}_{ii}+g_a\left(\hat{a}^\dagger\hat{\sigma}_{12}+\hat{\sigma}_{21}\hat{a}\right)} +g_b\left(\hat{b}^\dagger\hat{\sigma}_{23}+\hat{\sigma}_{32}\hat{b}\right),
\end{equation}
\begin{equation}\label{lambda}
\hat{H}_\Lambda= \displaystyle{\omega_1\hat{n}_a+\omega_2\hat{n}_b+\sum_{i=1}^3E_i\hat{\sigma}_{ii}+g_a\left(\hat{a}^\dagger\hat{\sigma}_{12}+\hat{\sigma}_{21}\hat{a}\right)}+g_b\left(\hat{b}^\dagger\hat{\sigma}_{32}+\hat{\sigma}_{23}\hat{b}\right),
\end{equation}
and
\begin{equation}\label{vee}
\hat{H}_{\it V} = \displaystyle{\omega_1\hat{n}_a+\omega_2\hat{n}_b+\sum_{i=1}^3E_i\hat{\sigma}_{ii}+g_a\left(\hat{a}^\dagger\hat{\sigma}_{21}+\hat{\sigma}_{12}\hat{a}\right)}+g_b\left(\hat{b}^\dagger\hat{\sigma}_{32}+\hat{\sigma}_{23}\hat{b}\right).
\end{equation}
Here, $E_i$ is the energy of atomic level $|i\rangle$ ($i=1,\,2,\,3$) and $\hat{\sigma}_{ij}=|i\rangle\langle j|$ are the atomic operators. In the absence of spontaneous emission\index{Spontaneous! emission} or other decay channels, the difference of the three models is in the indices $ij$ of the $\hat{\sigma}_{ij}$ operators, telling if a photon is absorbed or emitted. In the RWA all models have a continuous $U(1)$ symmetry characterizing conservation of excitations. As shown in fig.~\ref{fig12}, by turning to an interaction picture with respect to the bare energies, and properly shifting the zero energy, the Hamiltonians can be expressed in four relevant parameters, the detunings $\delta_1$ and $\delta_2$ and the couplings $g_a$ and $g_b$. As for the case of the JC model, within the RWA and in the interaction picture, the problem reduces to the manipulation of a $3\times3$ matrix, and the analytical solutions are readily obtainable. With a coupling between $|1\rangle$ and $|3\rangle$, however, a general analytic solution (valid for any couplings) is not attainable. Note further that we can consider the single-mode case by putting $\hat{a}=\hat{b}$ and $\omega_a=\omega_b=\omega/2$. We will briefly return to the single-mode three-level systems in sec.~\ref{ssec:focklattice}.

Occasionally, when discussing three-level atoms\index{Three-level atom}, it is convenient to introduce the {\it Gell-Mann matrices}\index{Gell-Mann matrices}~\cite{georgi2018lie}
\begin{equation}\label{gellmann}
\begin{array}{llll}
\hat\lambda^{(1)}=\left[
\begin{array}{rrr}
0 & 1 & 0\\
1 & 0 & 0\\
0 & 0 & 0
\end{array}\right], & \hat\lambda^{(2)}=\left[
\begin{array}{rrr}
0 & -i & 0\\
i & 0 & 0\\
0 & 0 & 0
\end{array}\right], & 
\hat\lambda_3=\left[
\begin{array}{rrr}
1 & 0 & 0\\
0 & -1 & 0\\
0 & 0 & 0
\end{array}\right], &\hat \lambda^{(4)}=\left[
\begin{array}{rrr}
0 & 0 & 1\\
0 & 0 & 0\\
1 & 0 & 0
\end{array}\right],\\ \\
\hat\lambda^{(5)}=\left[
\begin{array}{rrr}
0 & 0 & -i\\
0 & 0 & 0\\
i & 0 & 0
\end{array}\right], &\hat \lambda^{(6)}=\left[
\begin{array}{rrr}
0 & 0 & 0\\
0 & 0 & 1\\
0 & 1 & 0
\end{array}\right], &
\hat\lambda^{(7)}=\left[
\begin{array}{rrr}
0 & 0 & 0\\
0 & 0 & -i\\
0 & i & 0
\end{array}\right], & \hat\lambda^{(8)}=\displaystyle{\frac{1}{\sqrt{3}}}\left[
\begin{array}{rrr}
1 & 0 & 0\\
0 & 1 & 0\\
0 & 0 & -2
\end{array}\right]
\end{array}
\end{equation}
instead of the common notation $\hat{\sigma}_{ij}=|i\rangle\langle j|$. This is especially true when we do not work in the RWA. Like for the Pauli matrices~(\ref{pauli}), which are generators of the $SU(2)$ algebra\index{$SU(2)$ algebra}, the Gell-Mann matrices\index{Gell-Mann matrices} are generators of the $SU(3)$ algebra\index{$SU(3)$ algebra} and furthermore, the matrices are traceless, $\mathrm{Tr}\left[\hat\lambda^{(\alpha)}\right]=0$, and mutually orthogonal $\mathrm{Tr}\left[\hat\lambda^{(\alpha)}\hat\lambda^{(\beta)}\right]=2\delta_{\alpha\beta}$. The $\hat\lambda^{(\alpha)}$'s span all $3\times3$ matrices, and like for qubits~(\ref{bloch2}), it is possible to introduce a Bloch vector $\mathbf{R}=(R_1,R_2,\dots,R_8)$ also for qutrits~\cite{kimura2003bloch,bertlmann2008bloch}
\begin{equation}
\hat\rho=\frac{1}{3}\left(1+\sum_{i=1}^8R_i\hat\lambda^{(i)}\right).
\end{equation}
Both the Bloch vector and Gell-Mann matrices\index{Gell-Mann matrices} can be generalized to higher dimensions, {\it i.e.} for {\it qudits}\index{Qudit} with $d$ internal states. 

Already in the previous subsection, we mentioned that multiphoton transitions\index{Multiphoton! transition} arise from adiabatic eliminations of virtual intermediate atomic levels. More precisely, in the regime when $|\delta_1|\gg g_a\sqrt{n_a}$ and $|\delta_2|\gg g_b\sqrt{n_b}$ are both fulfilled, the atomic level $|2\rangle$ can be eliminated. Such procedures have been taken by numerous authors, either following the regular adiabatic elimination scheme~\cite{buck1981exactly,puri1988quantum,phoenix1990periodicity,gerry1990dynamics,ashraf1990theory,cardimona1991quantum} or using the bimodal version of the Schrieffer-Wolff transformation~\cite{alexanian1995unitary,wu1996effective,wu1997effective}. The effective atom-field interaction term for, say, the $\Lambda$ system becomes (assuming $g_a,\,g_b\in\mathbb{R}$)
\begin{equation}\label{lambdaeff}
\hat{V}_\Lambda=\frac{g_ag_b}{2}\left(\frac{1}{\delta_1}+\frac{1}{\delta_2}\right)\left(\hat{a}^\dagger\hat{b}\hat{\sigma}_{23}+\hat{b}^\dagger\hat{a}\hat{\sigma}_{32}\right).
\end{equation}
An interesting observation by Wu {\it et al.} was that the Schrieffer-Wolff transformation\index{Schrieffer-Wolff transformation} may actually be applied to decouple the three-level system into a 2+1 level system regardless of parameter values ({\it i.e.} no perturbation expansion is assumed), both for the $\Lambda$~\cite{wu1996effective} and the $\Xi$ system~\cite{wu1997effective}. An important extension of the $\Lambda$ configuration is the so-called {\it tripod atom}, in which there are three (instead of two) lower states and one excited. One of the main reasons why the tripod atom is of particular interest is because the corresponding Hamiltonian hosts two dark states\index{Dark! state}\index{State! dark} (instead of one dark state for the $\Lambda$ atom), {\it i.e.} eigenstates composed as linear combinations of only the lower atomic states and with zero eigenvalues. In the subspace of these two dark states, it is possible to explore effects deriving from non-Abelian\index{Non-Abelian} evolution~\cite{unanyan1999laser}, which can be helpful for QIP. For the three-level systems and within the elimination framework, the collapse-revival\index{Collapse-revival} phenomenon has been studied~\cite{gerry1990dynamics}, and it has been shown that the particular coherence between the two initial field states may affect the collapse/revival times in these systems~\cite{lai1991dynamics}. The elimination also gives rise Stark shifts of the two bare levels $|1\rangle$ and $|3\rangle$, and it has been shown that these (dynamical) terms can severely alter the collapse-revival dynamics~\cite{gou1990dynamics,lai1991dynamics} and also the trapping states in the $\Lambda$ system~\cite{agarwal1993coherent} is normally affected by the Stark shifts~\cite{li1995influences}. When the two-photon transition in the $\Lambda$ (or $V$) system contains a single mode we note that in the resonant case the Rabi frequencies $\Omega_n\propto n$, {\it i.e.} the JC nonlinear $\sqrt{n}$-dependence is lost~\cite{buck1981exactly}. As a result, the revivals will be periodic and ``perfect''~\cite{puri1988quantum,phoenix1990periodicity}. Adding a nonlinearity in terms of a Kerr medium\index{Kerr! medium} will destroy this periodic collapse-revival pattern~\cite{abdel2000influence}. In the single mode $\Xi$ model, the elimination gives an effective coupling term containing the two-photon processes $\hat{a}^2$ and $\left.\hat{a}^\dagger\right.^2$. The time-evolution operator is then containing a squeezing part\index{Squeezing}, which was analyzed in~\cite{villas2003squeezing} where large field squeezing was indeed demonstrated. Non-classical properties of the fields in the bimodal $\Lambda$ model has also been considered within the adiabatic elimination~\cite{joshi1990characteristics,abdalla1990dynamics}. For example, a `beating' effect was found where field bunching/antibunching was alternating between the two modes. Closer to to our days, $N$ identical, V-shaped three-level atoms\index{Three-level atom} coupled to a dissipative cavity\index{Dissipative! resonator} in the thermodynamic limit\index{Thermodynamic limit} ($N\to \infty$) were considered in~\cite{LinRui2022} to ascertain the stabilization of a continuous family of dark and nearly dark excited many-body states\index{Dark! state}\index{State! dark}. 

As already stated, the three Hamiltonians (\ref{ladder}), (\ref{lambda}), and (\ref{vee}) can be analytically diagonalized~\cite{bogolubov1984two,li1984generalized,bogolubov1985photon,li1985quantum,li1987nonresonant,li1989nonresonant,mahran1990squeezing,ashraf1994cavity}. The collapse-revival evolution has been discussed in several works, see for example~\cite{li1984generalized,li1985quantum,cardimona1990effect,cardimona1991quantum}. In the $V$ system it was found that for specific initial conditions the characteristic collapse-revival pattern can be completely absent~\cite{cardimona1990effect}. In the $\Lambda$ system super-revivals have been demonstrated~\cite{cardimona1991quantum}. Also squeezing effects, both dipole and field, in the $\Xi$ and the $\Lambda$ systems have been analyzed~\cite{li1989nonresonant,mahran1990squeezing,poizat1992nondegenerate,ashraf1994cavity}. It was found that a direct correlation between field and dipole squeezing did not exist. The driven three-level systems have been considered in terms of state preparation of Fock states~\cite{nha2000single} and squeezed states\index{Squeezed! state}\index{State! squeezed}~\cite{ficek1995fluorescence}.

In the bimodal setups, the three sub-systems are coupled which may lead to interesting entanglement properties~\cite{abdel2000influence,ikram2002engineering,abdel2003engineering,larson2004dynamics,li2007enhancement}. The idea is often to use the atom as an ancilla state which mediates the effective interaction between the two field modes~\cite{larson2004dynamics,li2007enhancement}. Alternatively, one introduces an effective logic gate between the two modes~\cite{zubairy2003cavity,abdel2003engineering,larson2004dynamics,biswas2004quantum,jian2006scheme}. Genuine tripartite entanglement generation has also been demonstrated, like the preparation of GHZ\index{GHZ state}\index{State! GHZ} or W states, eqs.~(\ref{ghzstate})\index{W state}\index{State! W} and (\ref{wstate}) respectively~\cite{ikram2002engineering,abdel2003engineering,biswas2004preparation}. It has also been shown that Kerr media\index{Kerr! medium}~\cite{abdel2003quantum,abdel2003engineering} or external pumping~\cite{zhou2006single} may help in achieving large entanglement in these systems. In sec.~\ref{sssec:dicke} we continue discussing entanglement properties but there the field is the effective `generator' of entanglement between multi-level atoms. 

Without going into any details, we mention that generalizations to multi-level atoms have been addressed in numerous works, see among others~\cite{buck1984solution,abdel1987n,liu1988properties,rebic2002polariton,abdel2007interaction,lu2008continuous}. One recent observation is that multilevel atoms enhance the effective atom-light coupling strength $g$~\cite{gross2018near}. Nonclassical optical switching from strong single-photon blockade\index{Single-photon!blockade} to two-photon bundles alongside super-Poissonian\index{Super-Poissonian} photon
emission have been recently reported for the spin-1 JC model~\cite{Jing2022}. In sec.~\ref{sssec:dicke}, we discuss the phenomenon of superradiance\index{Superradiance} which is an effect arising when $N$ identical two-level emitters radiate collectively. The emission of radiation is enhanced by a factor $\sqrt{N}$ compared to $N$ independent emitters. The same enhanced coupling can be achieved by generalizing $V$-atoms to atoms with a single lower state and $N$ quasi-degenerate excited states $|e_i\rangle$~\cite{tufarelli2021single}. Thus, the atom-light interaction takes the form
\begin{equation}\label{mmcoup}
\hat V=\sum_{i=1}^Ng_i\left(\hat a|e_i\rangle\langle g|+\hat a^\dagger|g\rangle\langle e_i|\right).
\end{equation}
To see how the $\sqrt{N}$-dependence comes about we introduce the collective excited state $|e\rangle=\sum_ig_i|e_i\rangle/G$, with $G^2=\sum_ig_i$. With this collective state interaction term takes the regular JC-form, with an effective coupling $g_\mathrm{eff}$ enhanced by $\sqrt{N}$. It might seem hard to achieve large $N$-values in realistic systems, but $N$ can be made on the order of hundred in modern emitters~\cite{tufarelli2021single}. 

\begin{figure}
\includegraphics[width=7cm]{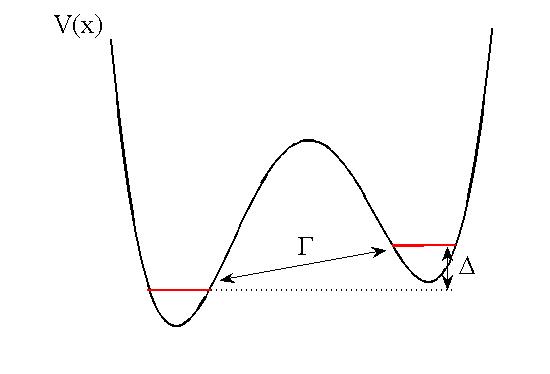} 
\caption{Schematic picture of the emergent two-level structure in the spin-boson model; the dynamics is projected onto two well separated energy levels (red solid lines) of a double-well potential $V(x)$ with $\Delta$ the energy mismatch and $\Gamma$ the tunneling amplitude through the barrier.}
\label{fig13}
\end{figure}

The two-level bimodal case discussed above finds applications especially in cavity/circuit QED and trapped ion physics. Generalizing this further, {\it e.g.}, by coupling a two-level system to an infinite number of bosons, yields the {\it spin-boson model}\index{Spin-boson model}\index{Model! spin-boson}~\cite{leggett1987dynamics,weiss2012quantum}
\begin{equation}\label{spinboson}
\hat{H}_\mathrm{SB}=\sum_i\omega_i\hat{n}_i+\frac{\eta}{2}\hat{\sigma}_z-\frac{\Omega}{2}\hat{\sigma}_x+\frac{\hat{\sigma}_z}{2}\sum_ig_i\left(\hat{a}_i^\dagger+\hat{a}_i\right).
\end{equation}
In the language of the JC model, the spins have been rotated $\hat{\sigma}_x\leftrightarrow\hat{\sigma}_z$ and the RWA is not applied. Furthermore, a (time-independent) pumping term proportional to $\eta$ is included. This is a paradigmatic model describing open system dynamics of a two-level system and finds numerous applications in various fields of physics. The Lindblad master equation (\ref{lindblad}) was derived in the RWA and Born-Markov approximation, while for the spin-boson model one is normally not imposing any of these approximations\index{Master equation}. The breakdown of such approximations frequently appears in solid-state systems where the characteristic time-scales of the bath and the system are rather different from optical systems where a Lindblad approach is often justified. In the solid-state community, $\hat{\sigma}_z$ is seen as the `position' of the particle and $\hat{\sigma}_x$ drives tunneling of the particle. This picture is given in fig.~\ref{fig13}. We make a {\it two-mode approximation} of a double-well system, where one quasi-bound state is kept in either of the two wells, the energy offset of the double-well is given by $\eta$ and $\Omega$ determines the tunneling amplitude. Now, the particle is coupled to a boson bath, which induces particle decoherence. 

The properties of the spin-boson model is largely determined by the {\it spectral function}\index{Spectral! function} defined as~\cite{Mahan}
\begin{equation}\label{spectral}
J(\omega)=\pi\sum_ig_i^2\delta(\omega-\omega_i).
\end{equation}
The spectral function is assumed algebraic
\begin{equation}\label{specfun}
J(\omega)=2\pi\alpha\omega^sf(\omega,\omega_c),\hspace{1.3cm}s>-1,
\end{equation}
where $\alpha$ determines the strength of the interaction, and $f(\omega,\omega_c)$ is a cutoff function\index{Frequency cutoff} such that frequencies $\omega>\omega_c$ are greatly suppressed. Note that the properties of the system should not depend on the details of $f$. The generic features of the model can be divided into the regimes\index{Ohmic bath}\index{Sub-ohmic bath}\index{Super-ohmic bath}
\begin{equation}
\begin{array}{lll}
-1<s<0, & \hspace{1cm} & \mathrm{inverse},\\
0<s<1, & \hspace{1cm} & \mathrm{sub-Ohmic},\\ 
s=1, & \hspace{1cm} & \mathrm{Ohmic},\\ 
1<s, & \hspace{1cm} & \mathrm{super-Ohmic}.
\end{array}
\end{equation}
For a phonon bath one often takes $s=3$, for an impurity bath $s=5$, and for $1/f$-noise\index{$1/f$-noise} $0<s<1$. How to control the type of dissipation ({\it i.e.} $s$) in an atomic setting was discussed in ref.~\cite{recati2005atomic}.

Without the coupling to the boson modes, the particle will show Rabi oscillations between the two wells, while if it couples to the bath of boson modes they will tend to dephase the oscillations. It has long been known that the Ohmic case is quantum critical where for a critical coupling $\alpha_c$ there is a zero temperature {\it Kosterlitz-Thouless transition}\index{Kosterlitz-Thouless transition} from a {\it localized}\index{Localized phase} to a {\it delocalized}\index{Delocalized! phase} (or incoherent to coherent) phase~\cite{spohn1985quantum,leggett1987dynamics,weiss2012quantum}. In the localized phase, tunneling is suppressed by dephasing while in the delocalized phase tunneling is still present despite the coupling to the bath. Note how the localized phase resembles a {\it self-trapping} effect common in many-body double-well systems\index{Self-!trapping}. In the delocalized phase, the particle can be seen as dressed with a cloud of oscillator excitations. This picture is especially clear by transforming the Hamiltonian with its corresponding polaron transformation\index{Polaron! transformation}~\cite{chin2006coherent}. It was long believed that the delocalized phase would not survive in the sub-Ohmic regime~\cite{leggett1987dynamics}, which was proven wrong by using infinitesimal transformation similar to a renormalization procedure~\cite{kehrein1996spin}. Indeed, for small coupling strengths $\alpha$ the zero temperature delocalized phase survives also in sub-Ohmic regime~\cite{shnirman2002noise,bulla2003numerical,vojta2005quantum}. However, it is here a proper second-order {\it quantum phase transition}\index{Phase transition! quantum}~\cite{sachdev2007quantum}. The nature of this transition was long debated, while renormalization methods\index{Renormalization! group} suggested that the transition was not ``classical'', {\it i.e.} the universality properties of the quantum phase transition cannot be ascribed to a classical phase transition\index{Phase transition! classical}~\cite{vojta2005quantum}. Other methods indicated that the renormalization results were false~\cite{winter2009quantum,alvermann2009sparse}, and finally Vojta could resolve the erroneous results in the application of the renormalization group technique~\cite{vojta2012numerical}. In the super-Ohmic case the delocalized phase is not stable even at zero temperature. Thus, as a function of the `dimension' $s$ there is a line of second-order critical couplings $\alpha_c(s)$ which terminates in a Kosterlitz-Thouless transition (cross-over) at $s=1$. Interestingly, a similar behaviour was also demonstrated in the {\it Kondo problem}\index{Kondo problem}~\cite{fritz2004phase}. The transition survives for finite temperature $T<T_*$ in both the sub-Ohmic and Ohmic cases~\cite{chin2006coherent}. Using a variational approach, similar to the BOA where corrections are self-consistently taken into account, it was also found that the finite temperature system may show a ``re-entrant'' behaviour where the system goes as localized-delocalized-localized\index{Re-entrance transition} as $T$ is increased~\cite{chin2006coherent}. Typical for quantum phase transitions\index{Phase transition! quantum} or cross-overs is that the entanglement rapidly grows in the vicinity of the transition point, which has also been shown for the spin-boson model in the sub-Ohmic case~\cite{le2007entanglement,hur2008entanglement}.

When describing the dissipation (decoherence) of a two-level system, the spin-boson model can also be combined with the JC (or quantum Rabi) model, {\it i.e.} the two-level atom of the JC model is coupled to a bath of bosons~\cite{wilson2002quantum}
\begin{equation}\label{rabisb}
\hat{H}_\mathrm{RSB} = \displaystyle{\omega\hat{n}+\sum_i\omega_i\hat{n}_i+\tfrac{\Omega}{2}\hat{\sigma}_z+\sqrt{2}g\hat{\sigma}_x\left(\hat{a}^\dagger+\hat{a}\right)},\displaystyle{+\hat{\sigma}_z\sum_i\lambda_i\left(\hat{b}_i^\dagger+\hat{b}_i\right)},
\end{equation}
where now $\lambda_i$ are the bath coupling strengths. Thus, the atom in the quantum Rabi model (\ref{rabisb}) is subject to non-Markovian dissipation\index{Non-Markovian dissipation}~\cite{wenderoth2021non} and thereby this type of approach to atomic dephasing goes beyond that of the Markovian Lindblad approach discussed in sec.~\ref{sssec:openjc}. For a derivation of a Born-Markov master equation\index{Master equation} with the Hamiltonian (\ref{rabisb}) as a starting point, see~\cite{roy2011phonon,roy2011influence,roy2012anomalous}. It has been demonstrated, however, that in solid state systems, some experimental features cannot be described in the framework of a Markovian bath, that non-Markovian effects can really be important~\cite{kaer2010non}. Comparing to the regular spin-boson model (\ref{spinboson}) it follows that the tunneling is accompanied by interaction with the quantized boson field of the quantum Rabi model. Recently, Henriet {\it et al}. considered the Ohmic case and showed how the Rabi oscillations of this system dies out due to the bath~\cite{henriet2014quantum}, and they further demonstrated how a (polariton) steady state\index{State! polariton}\index{Steady state}\index{Polariton} solution with a single excitation can be achieved when an AC driving is included. The decay of Rabi oscillations, but now within the RWA of the Rabi coupling and for a super-Ohmic coupling $J(\omega)\sim\omega^3$ (characterizing coupling to phonons) was also analyzed in~\cite{zhu2005excitonic} making use of the polaron transformation (with respect to the bath), see also the work by Wilson and Imamo{\u{g}}lu~\cite{wilson2002quantum} where a field drive was also considered. The same work also addressed collapse-revival structures and found that the revivals indeed vanishes for strong couplings $\alpha$. However, it has since then been shown that the non-Markovian coupling to a phonon bath can also stabilize the collapse-revival pattern, {\it e.g.} making it more regular~\cite{carmele2013stabilization}. As will be discussed in further detail in sec.~\ref{sssec:smapp}, the polaron transformation in this system renormalizes the JC atom-field coupling~\cite{wai2003rabi,larson2008rabi} and thereby alters the Rabi frequency\index{Rabi! frequency}. In the transformed basis it is, under certain conditions, motivated to perform a RWA between the two-level system and the bath modes (similar to the generalized RWA for the JC as discussed by Irish~\cite{irish2007generalized}) under which analytical expressions can be obtained for various quantities like collapse and revival times~\cite{larson2008rabi}. The coherence has also been explored in terms of the Mandel $Q$-parameter\index{$Q$-parameter} (\ref{qpara}) in the presence of a phonon bath, {\it i.e.} $J(\omega)\sim\omega^3$, and it was argued that the atom-field detuning can be a handle to achieve sub-Poissonian\index{Sub-Poissonian} field statistics~\cite{harouni2009influence}. Heat engines with non-Markovian thermal operations can be experimentally realized by using a qubit coupled to Markovian thermal baths via auxiliary bosonic modes. The qubit is coupled to the bosonic modes with a JC-type interaction~\cite{Ptaszynski2022}. 


\subsubsection{Time-dependent and adiabatic Jaynes-Cummings models}~\label{sssec:timeJCsec}
Explicit time-dependence of a system means some sort of external `driving'. For the JC model this could for example be a varying pumping term. In an interaction picture, this could give rise to time-dependent detunings. Also the atom-field coupling $g$ can be effectively time-dependent by considering multiphoton transitions\index{Multiphoton! transition} where one (or more) of the transitions are driven by external (classical) time-dependent fields. Naturally there are numerous other ways to construct an explicit time-dependence, which will depend on the particular physical system at hand. Yet another possibility in cavity QED is that an effective time-dependence emerges in a semiclassical\index{Semiclassical! approximation} picture where the center-of-mass motion of the atom is treated classically. In such a model, where the time-dependence derives due to inherent evolution, one replaces the atomic velocity and position with their classical values, $\hat{p}/m\rightarrow v$ and $\hat{x}\rightarrow x_0+vt$~\cite{schlicher1989jaynes}. It should be clear that when doing such an approximation it is assumed that the atomic kinetic energy $mv^2/2$ is considerably larger than the JC interaction energy $E_n(x)=\sqrt{\frac{\Delta^2}{2}+g^2(x)(n+1)}$ (here $x=\langle\hat{x}\rangle$ and for simplicity we consider the 1D case). In the next subsection we will address the situation when this semiclassical approximation breaks down, and one must quantize also the atomic motion. In sec.~\ref{sssec:velapp} we will return to this situation and briefly discuss the validity of this replacement.

In the interaction picture, the time-dependent JC model leads to the Schr\"odinger equation
\begin{equation}
i\frac{d}{dt}|\psi(t)\rangle=\left[\frac{\Delta(t)}{2}\hat{\sigma}_z+g(t)\left(\hat{a}^\dagger\hat{\sigma}_{-}+\hat{\sigma}_{+}\hat{a}\right)\right]|\psi(t)\rangle.
\end{equation}
Expressed in the bare basis, we again regain the two-level block structure (\ref{block})
\begin{equation}\label{2level}
i\frac{d}{dt}\left[
\begin{array}{c}
c_{ne}(t)\\ c_{ng}(t)\end{array}\right]=
\left[\begin{array}{cc}
\displaystyle{\frac{\Delta(t)}{2}} & g(t)\sqrt{n+1} \\
g(t)\sqrt{n+1} & -\displaystyle{\frac{\Delta(t)}{2}}
\end{array}\right]\left[
\begin{array}{c}
c_{ne}(t)\\ c_{ng}(t)\end{array}\right].
\end{equation}
Thus, the time-dependent JC problem takes the form of a general time-dependent two-level Schr\"odinger equation~\cite{larson2003adiabatic}. There are a few (limiting) cases that can be solved directly; ($i$) $\Delta(t)$ and $g(t)$ both constant, ($ii$) $\Delta(t)\equiv0$, ($iii$) $g(t)\equiv0$, and ($iv)$ $\Delta(t)\propto g(t)$. For example, for case ($ii$) and the initial condition $c_{ne}(0)=c_n$ and $c_{ng}(0)=0$ (that is the atom is initially excited and the field in the state characterized by the amplitudes $c_n$) we have
\begin{equation}\label{trivial}
\begin{array}{l}
c_{ne}(t)=\cos(A(t))c_n,\\ \\
c_{ng}(t)=\sin(A(t))c_{n+1},
\end{array}
\end{equation}
where $A(t)=\int_0^tg(t')dt'$ is the `coupling-area'\index{Coupling area}. Despite its simplicity, apart from the examples listed above, very few exactly solvable models of the two-level time-dependent Schr\"odinger equation are known~\cite{nikitin1984theory,kyoseva2006coherent}. The most famous solvable models are: Landau-Zener\index{Landau-Zener model}\index{Model! Landau-Zener}~\cite{landau1932,zener1932}, Rosen-Zener\index{Rosen-Zener model}\index{Model! Rosen-Zener}~\cite{rosen1932double}, Demkov-Kunike\index{Demkov-Kunike model}\index{Model! Demkov-Kunike}~\cite{demkov1969}, Nikitin\index{Nikitin model}\index{Model! Nikitin}~\cite{nikitin1984theory}, Allen-Eberly\index{Allen-Eberly model}\index{Model! Allen-Eberly}~\cite{allen1975optical}, and Carrol-Hioe\index{Carrol-Hioe model}\index{Model! Carrol-Hioe}~\cite{carroll1986new}. Having access to analytical solutions implies that the parameter dependence on, {\it e.g.}, non-adiabatic excitations, is readily obtained. 

The Landau-Zener problem has been discussed in terms of the JC model, for preparation of non-classical field states~\cite{larson2003adiabatic,larson2004cavity} (also the Demkov-Kunike model was considered in these references) and for general properties of the field~\cite{keeling2008collapse}. Adiabatic properties in cavity QED as the atom traverses the cavity perpendicularly was considered in~\cite{larson2003photon}, where the Rosen-Zener model was employed in order to mimic the Gaussian perpendicular mode profile\index{Gaussian! mode! profile}. Dasgupta also considered the Rosen-Zener model and explored the effect of time-dependence on the Rabi oscillations and on the collapse-revival structure, and it was especially demonstrated how the Rabi frequency changes smoothly in the adiabatic regime~\cite{dasgupta1999analytically}. The effect on collapse-revivals stemming from a time-dependent coupling has also been analyzed in terms of the Nikitin model~\cite{prants1992jaynes}. 

We will now focus on the influence of a linear sweep in the Rabi model following closely ref.~\cite{ashhab2023controlling} and revisiting some part of our discussion in sec.~\ref{ssec:rabi} pertinent to the Born-Oppenheimer approximation. An equivalent writing of the Hamiltonian in eq.~\eqref{rabiham} is
\begin{equation}\label{eq:RabiHambias}
 \hat{H}_R=-\frac{\Delta}{2}\hat{\sigma}_x- \frac{\epsilon}{2}\hat{\sigma}_z + \omega \hat{a}^{\dagger}\hat{a} + g \hat{\sigma}_z(\hat{a} + \hat{a}^{\dagger}),
\end{equation}
where $\epsilon$ is now the bias parameter\index{Rabi! model!bias} (it is more common to reserve the Pauli operator $\hat{\sigma}_z$ for the bare qubit energy). For the moment we keep $\epsilon=0$. When $g^2 \ll \omega\Delta/4$, the system is in the normal phase and the eigenstates are approximately,
\begin{equation}\label{eq:eigenstatesRabi1}
 |\rightarrow,n \rangle=\tfrac{1}{\sqrt{2}}(|\uparrow \rangle + |\downarrow \rangle) \otimes |n\rangle, \quad \quad |\leftarrow,n \rangle=\tfrac{1}{\sqrt{2}}(|\uparrow \rangle - |\downarrow \rangle) \otimes |n\rangle,
\end{equation}
where $|\leftarrow\rangle$ and $|\rightarrow\rangle$ are eigenstates of $\hat{\sigma}_x$, while the ground state is $|\rightarrow,0\rangle$. In the opposite limit, when $g^2 \gg \omega\Delta/4$, the system is in the superradiant state, and the low-lying states read
\begin{equation}\label{eq:eigenstatesRabi2}
 |+,n \rangle=\tfrac{1}{\sqrt{2}}(|\uparrow \rangle \otimes \hat{D}(\alpha)|n\rangle + |\downarrow \rangle \otimes \hat{D}(-\alpha)|n\rangle), \quad \quad |-,n \rangle=\tfrac{1}{\sqrt{2}}(|\uparrow \rangle \otimes \hat{D}(\alpha)|n\rangle - |\downarrow \rangle \otimes \hat{D}(-\alpha)|n\rangle),
\end{equation}
where $\hat{D}(\alpha)=\exp(\alpha \hat{a}^{\dagger}-\alpha^{*}\hat{a})$ is the familiar displacement operator\index{Displacement! operator}, and $\alpha\equiv g/\omega$. The ``ground state'' is now $|+,0\rangle$. We remark here that the normal to superradiant phase transition\index{Phase transition! normal-superradiant} in the quantum Rabi model has been recently studied in~\cite{Yang_2023}, by means of two successive diagonalizations, one pertaining to the the two-level system and the other to cavity mode in a truncated Fock space. In this procedure, the two branches of eigenstates are decoupled under the Born-Oppenheimer approximation\index{Born-Oppenheimer approximation}, as we discussed in sec.~\ref{sssec:drivejc}. As the coupling strength traverses its critical value, the photon population statistics is approximately Poissonian, and subsequently conforming to the random matrix distribution of a Gaussian unitary ensemble\index{Gaussian! unitary ensemble}.

In circuit QED, finite-duration quenches can be readily realized, where one of the system parameters is swept through the critical point that separates the normal and superradiant phases. In fact, the role of parameter fluctuations on the superradiance phase transition for a configuration where a large number of superconducting qubits are coupled to a single cavity mode has already been investigated in~\cite{Ashhab2017}, with emphasis on typical superconducting architectures. A natural choice would be to set $g=vt$, where $v$ is the sweep rate and $t$ is the time variable with $t\in[0,T]$; provided that $vT \gg \sqrt{\Delta \omega}/2$, the system is brought to the superradiant phase at the final time. This approach, however, would also lead to a divergence of the common observables in the system, such as the photon number, one can alternatively tune $\Delta$. Superconduncting qubits with a tuneable gap have been realized in a number of experiments~\cite{Paauw2009, Zhu2010CohOp}. The sweep from a normal to a superradiant phase is accomplished with the Hamiltonian
\begin{equation}\label{eq:RabiHambiasSweep}
 \hat{H}_R=-\frac{\Delta_i-vt}{2}\hat{\sigma}_x + \omega \hat{a}^{\dagger}\hat{a} + g \hat{\sigma}_z(\hat{a} + \hat{a}^{\dagger}),
\end{equation}
where the final time is $T=(\Delta_i-\Delta_f)/v \approx \Delta_i/v$, and the initial and final values of the gap are selected such as to satisfy $\Delta_i \gg 4g^2/\omega$ and $\Delta_f \ll 4g^2/\omega$, respectively. The asymptotic expressions for the lowest-energy eigenstates are independent of $\Delta$ in the limits $\Delta\to 0$ and $\Delta\to \infty$ whence no physical divergences occur in the final state of the system. A related experimental investigation reported on multiphoton transitions\index{Multiphoton! transition} between two macroscopic quantum mechanical superposition states created by two opposite circulating currents in a superconducting loop~\cite{Saito2004}. Fock states were generated at the end of the parameter sweep.
\begin{figure}
\begin{center}
 \includegraphics[width=0.6\textwidth]{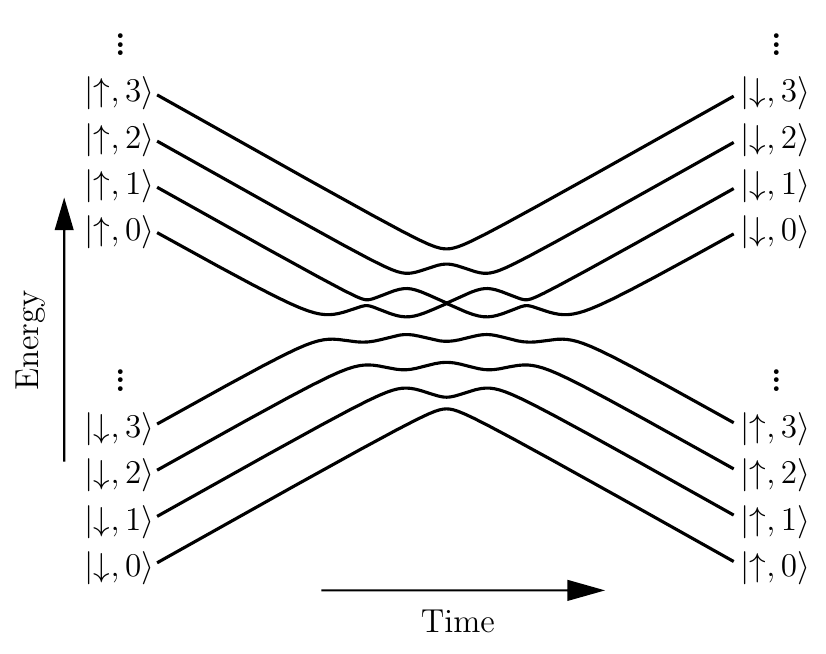}
\end{center}
\caption{Schematic diagram of eight representative energy levels in a problem of Landau-Zener sweep for a qubit coupled to a harmonic oscillator. The energy level ladders extend to $\infty$, as indicated by the vertical dots. Source: Fig. 5 of~\cite{ashhab2023controlling}. Reproduced with permission from the author.}
\label{fig:LZsweep}
\end{figure}
  
The so-called semiclassical limit -- the analogue of a large number of two-level systems coupled to an oscillator -- is realized for a single qubit when $\Delta \gtrsim 10^3$~\cite{ashhab2013superradiance}. In that limit, one encounters a sharp boundary separating the normal and superradiant phases. If the coupling strength is much larger than the cavity frequency, then an abrupt change is expected during the sweep since the transition point is located as $\Delta/\omega=(2g/\omega)^2$. The initial state is taken to be the ground state $|\rightarrow, 0\rangle$, in general the easiest state to prepare in an experiment. We note that the ground state is symmetric with respect to the parity operator, and the system remains in the parity-symmetric subspace of the Hilbert space throughout its evolution.

For an adiabatic evolution ($v \to 0$) the system remains in the ground state, while in the fast-sweep limit ($v \to \infty$), the final-state probabilities are given by the overlaps between the initial and the final energy eigenstates. In the limits $\Delta_i \to \infty$ and $\Delta_f=0$, these states are given exactly by eqs.~\eqref{eq:eigenstatesRabi1} and~\eqref{eq:eigenstatesRabi2}, respectively. The probability amplitudes in the final states then read [see eq.~\eqref{cohe}]
\begin{equation}\label{eq:asymptprob}
 P(n)=|\langle n|\hat{D}(g/\omega)|0\rangle|^2=e^{-\langle n \rangle}\frac{\langle n \rangle^n}{n!}, \quad \quad \text{with}\quad \langle n \rangle=(g/\omega)^2,
\end{equation}
peaking at $n=\langle n \rangle$. The probabilities for intermediate values of $v$ have been numerically calculated in~\cite{ashhab2023controlling}. For relatively small values of $g/\omega$ all the probabilities $P(\pm,n)$ monotonically increase to their asymptotic values given by eq.~\eqref{eq:asymptprob}, while for large values of $g/\omega$, where the highest occupation in the fast-sweep limit leads to a high intracavity photon number, intermediate peaks appear in $P(\pm,n)$ when plotted against $v/\omega^2$. 

Let us now reinstate the bias $\epsilon$ in eq.~\eqref{eq:RabiHambias}, which is not to be swept from a large negative to a large positive value:
\begin{equation}\label{eq:RabiHambiasLZ}
 \hat{H}_R=-\frac{\Delta}{2}\hat{\sigma}_x- \frac{vt}{2}\hat{\sigma}_z + \omega \hat{a}^{\dagger}\hat{a} + g \hat{\sigma}_z(\hat{a} + \hat{a}^{\dagger}).
\end{equation}
The first two terms in eq.~\eqref{eq:RabiHambiasLZ} describe the Landau-Zener problem\index{Model! Landau-Zener}, while the other two terms involve the harmonic oscillator, which is the accessible and controllable part of the system. For the analysis of the dynamics it is more meaningful to use a correlated basis, in which the photon number $\hat{n}$ is $n_{\uparrow}=(\hat{a}^{\dagger}+g/\omega)(\hat{a}+g/\omega)$ when the qubit is in the excited state ($\uparrow$), and $n_{\downarrow}=(\hat{a}^{\dagger}-g/\omega)(\hat{a}-g/\omega)$ when the qubit is in the ground state ($\downarrow$). 

We gain an intuitive understanding by first setting $\Delta=0$. As $t$ varies from $-\infty$ to $\infty$, level crossings occur at $vt=m \omega$, with $m$ integer. Energy levels corresponding to the states $|\downarrow, n\rangle$ and $|\uparrow,n+m\rangle$ intersect for any $m$ and $n=0,1,\ldots$, producing a mesh of energy level crossings. Reinstating $\Delta$ turns all the energy level crossings into avoided crossings with gaps determined by the model parameters. If the initial state at $t\to -\infty$  is the ground state $|\downarrow, 0\rangle$, the above explanation accounts for the probabilities of the final energy eigenstates so long as $\omega$ is comparable to $\Delta$~\cite{Brundobler1993, Malla2021}. A sequence of independent LZ transitions through the avoided crossings one by one determines the probabilities:
\begin{equation}\label{eq:LZP1}
\begin{aligned}
 &P(\uparrow, 0)=1-\exp[-\pi \Delta_0^2 /(2v)], \quad P(\uparrow, 1)=\exp[-\pi \Delta_0^2 /(2v)]\{1-\exp[-\pi \Delta_1^2 /(2v)]\},\\
 &P(\uparrow, 2)=\exp[-\pi \Delta_0^2 /(2v)]\exp[-\pi \Delta_1^2 /(2v)]\{1-\exp[-\pi \Delta_2^2 /(2v)], \ldots,
 \end{aligned}
\end{equation}
while 
\begin{equation}\label{eq:LZP2}
 P(\downarrow, 0)=\exp[-\pi (\sum_n \Delta_n^2) /(2v)]=\exp[-\pi \Delta^2 /(2v)]\quad \text{and} \quad P(\downarrow,k)=0 \quad \text{for} \quad k \geq 1.
\end{equation}
The gaps at the different avoided crossings, schematically depicted in fig.~\ref{fig:LZsweep}, read
\begin{equation}
 \Delta_n=\frac{1}{\sqrt{n!}} \left(\frac{2g}{\omega}\right)^n \exp[-2(g/\omega)^2] \Delta.
\end{equation}
We note that $\sum_{n=0}^{\infty}\Delta_n^2=\Delta$. The latter follows from the completeness of the photon-number basis states and ensures that all probabilities in eqs.~\eqref{eq:LZP1}, \eqref{eq:LZP2} sum up to one. In the limit $g/\omega \ll 1$, the probability of creating any amount of photons in the oscillator remains very small for all values of the sweep rate $v/\Delta^2$, in contrast to the limit of high values of $g/\omega$, where one can prepare a Fock state with a chosen $n$ in a deterministic fashion~\cite{ashhab2023controlling}. 

While analytical solutions are always practical, the two-level problem of eq.~(\ref{2level}) can be easily integrated for any other time-dependent functions $\Delta(t)$ and $g(t)$. The most studied case is the one with a constant detuning and a periodic coupling, $g(t)=g_0\sin(vt)$~\cite{schlicher1989jaynes,wilkens1992spectrum,wilkens1992spectrum,ren1992spontaneous,prants1992jaynes,fang1998effects,xie2009dynamic}. It was introduced by Schlicher as a model for an atom traversing the cavity along the standing wave mode~\cite{schlicher1989jaynes}. Wilkens and Meystre extended this model to the maser system (see sec.~\ref{sssec:micro}) and studied the effects on the maser emission spectrum~\cite{wilkens1992spectrum}. Atom-field entanglement and the preparation of Schr\"odinger cat states (\ref{catstate})\index{Cat state} (in the large $n$-limit~\cite{gea1991atom}) was discussed in~\cite{fang1998effects}. In particular, they assumed zero detuning situation of eq.~(\ref{trivial}) for which the level crossings (the instantaneous {\it adiabatic energies} are $E_\mathrm{ad}^{(n)}(t)=\pm g_0\sin(vt)\sqrt{n+1}$), induced by the periodic coupling, are not avoided and the alternating sign of $g(t)$ then acts as an effective time-reversal operation. Joshi and Lawande instead considered a constant detuning $\Delta$ and a coupling $g(t)=vt$ which would represent a well localized atom in a slowly varying mode profile which justifies a linearization around the atomic position~\cite{joshi1993generalized}. Due to the changing amplitude of the coupling, the revival time was altered.

As will become evident, the adiabatic regime, valid when $\langle\partial_t\hat{H}\rangle_{ij}\ll(E_{i+1}(t)-E_{i}(t))$\index{Adiabatic! criteria}~\cite{ballentine1998quantum}, is of great interest for JC physics as it has applications in {\it adiabatic QIP}. Most of the schemes relay on {\it adiabatic passage}~\cite{bergmann1998coherent,vitanov2001laser,shore2011manipulating}. Let us consider a $\Lambda$ atom\index{$\Lambda$ atom}\index{Atom! Lambda} in a bimodal cavity, see fig.~\ref{fig12}. For a given excitation sector $N$, the relevant three bare states (or diabatic states) are $\left\{|1,n_a,n_b\rangle,\,|2,n_a-1,n_b\rangle,\,|3,n_a-1,n_b+1\rangle\right\}$ where the ket represents; atomic state, number of photons in mode $a$, and number of photons in mode $b$. For simplicity we assume that $\delta_1=\delta_2=\delta$ (=constant) such that there is only one effective detuning. The time-dependent Schr\"odinger equation becomes
\begin{equation}\label{stirap}
i\displaystyle{\frac{d}{dt}}\left[\!\!\begin{array}{c}
c_{N1}(t)\\ c_{N2}(t)\\ c_{N3}(t)\end{array}\right]=
\left[\!\begin{array}{ccc}
0 & g_1(t)\sqrt{n_a} & 0\\
g_1(t)\sqrt{n_a} & \delta & g_2(t)\sqrt{n_b+1}\\
0 & g_2(t)\sqrt{n_b+1} & 0
\end{array}\right]\left[\begin{array}{c}
c_{N1}(t)\\ c_{N2}(t)\\ c_{N3}(t)\end{array}\right].
\end{equation} 
With the couplings both real, the adiabatic eigenstates\index{Adiabatic! state}\index{State! adiabatic} become particularly simple
\begin{equation}
\begin{array}{lllll}
\left[\begin{array}{c}
\sin\phi\sin\theta\\
\cos\phi\\
\sin\phi\cos\theta\end{array}\right], & &
\left[\begin{array}{c}
\cos\theta\\
0\\
-\sin\theta\end{array}\right], & & 
\left[\begin{array}{c}
\cos\phi\sin\theta\\
-\sin\phi\\
\cos\phi\cos\theta\end{array}\right],
\end{array}
\end{equation}
where the two angles are defined as
\begin{equation}
\begin{array}{lll}
\tan\theta=\frac{g_1(t)\sqrt{n_a}}{g_2(t)\sqrt{n_b+1}}, & & 
\tan2\phi=\frac{2\sqrt{g_1^2(t)n_a+g_2^2(t)(n_b+1)}}{\delta}.
\end{array}
\end{equation}
The corresponding adiabatic eigenstates\index{Adiabatic! eigenenergies} are
\begin{equation}
\begin{array}{l}
E_{N+}(t)=\frac{1}{2}\left(\delta+\sqrt{\delta^2+4g_1^2(t)n_a+4g_2^2(t)(n_b+1)}\right),\\ 
E_{0N}(t)=0,\\
E_{N-}(t)=\frac{1}{2}\left(\delta-\sqrt{\delta^2+4g_1^2(t)n_a+4g_2^2(t)(n_b+1)}\right).
\end{array}
\end{equation} 
The middle adiabatic eigenstate
\begin{equation}
|\psi_\mathrm{D}(t)\rangle=\cos\theta|1,n_a,n_n\rangle-\sin\theta|3,n_a-1,n_b+1\rangle
\end{equation}
has a zero adiabatic eigenvalue and it does not contain the excited bare atomic level $|2\rangle$. Thus, when the system occupies this state in an adiabatic following\index{Adiabatic! following}, and the system occupies this state it means that losses due to atomic spontaneous emission\index{Spontaneous! emission} of the excited $|2\rangle$ state is greatly suppressed. This state is therefore called a {\it dark state}~\cite{vitanov2001laser}\index{Dark! state}\index{State! dark} introduced in eq.~(\ref{darkstate}). Note that the dark state\index{Dark! state}\index{State! dark} does not require a dynamical phase factor when the system evolves adiabatically (a geometric phase\index{Geometric phase} factor is still possible). Note the similarity between this dark state and the one discussed in sec.~\ref{sssec:openjc}; here the dark state is an eigenstate of the Hamiltonian with zero eigenvalue, while in sec.~\ref{sssec:openjc} it was an eigenstate of the Lindblad operator with zero eigenvalue. 

\begin{figure}
\includegraphics[width=9cm]{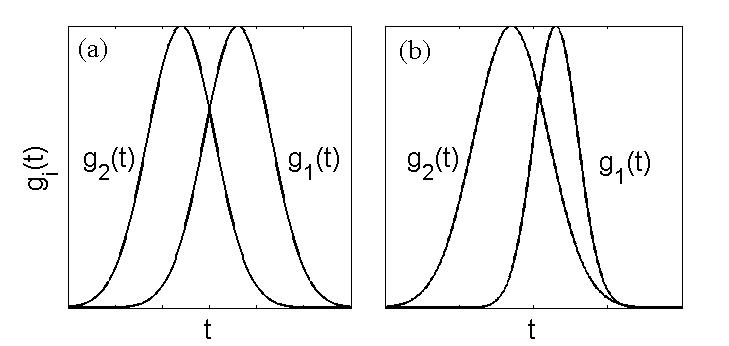} 
\caption{The pulse sequence in STIRAP {\bf (a)} and $f$-STIRAP\index{$f$-STIRAP} {\bf (b)}. The counter-intuitive sequence implies that $\lim_{t\rightarrow-\infty}G(t)=0$, and $\lim_{t\rightarrow+\infty}G^{-1}(t)=0$ for STIRAP and $\lim_{t\rightarrow+\infty}G^{-1}(t)=\mathrm{const.}$ for $f$-STIRAP.}
\label{fig14}
\end{figure}

While the adiabatic passage is not restricted to vanishing detuning, in order to demonstrate the general idea we here let $\delta=0$. This implies that $\cos\phi=\sin\phi=1/\sqrt{2}$. It follows that the adiabatic eigenstates are fully determined by the ratio $G(t)\equiv g_1(t)\sqrt{n_a}/g_2(t)\sqrt{n_b+1}$. Thus, if we assume that $\lim_{t\rightarrow-\infty}G(t)=0$ the system is initially in the state $|1,n_a,n_b\rangle$, and provided the evolution is adiabatic, if $\lim_{t\rightarrow+\infty}G^{-1}(t)=0$ the final state will be $|3,n_a-1,n_b+1\rangle$. Thus, by carefully choosing the time-dependence of the couplings $g_1(t)$ and $g_2(t)$ it is possible to adiabatically transfer population from the states $|1,n_a,n_b\rangle|$ to the states $|3,n_a-1,n_b+1\rangle$. A possible choice for the couplings $g_i(t)$ is pictured in fig.~\ref{fig14} (a); first the coupling between the bare atomic states $|2\rangle$ and $|3\rangle$ is smoothly turned on, followed by the coupling between the bare states $|2\rangle$ and $|1\rangle$. The pulse sequence is therefore applied counterintuitively. Furthermore, as long as the evolution remains adiabatic, the intermediate state will never be populated - population will be transfered between two states even though there is no direct coupling between them and the intermediate state coupling them together is never populated. This is the so-called {\it Stimulated Raman Adiabatic Passage} (STIRAP)\index{STIRAP} scheme frequently used in atomic and molecular physics~\cite{vitanov2001laser,shore2011manipulating}. An alternative version of the method is the $f$-STIRAP (fractional STIRAP)\index{Fractional! STIRAP} where the initial condition of $G(t)$ is the same but $\lim_{t\rightarrow+\infty}G^{-1}(t)$ attains a non-zero constant value. This leaves the final state in a coherent superposition of bare/diabatic states~\cite{vitanov1999creation}. The typical pulse sequence is shown in fig.~\ref{fig14} (b). Another method to prepare coherent superposition of diabatic states is to couple the intermediate state to several states all with the same mode profile $g_2(t)$ (not necessarily with the same strength nor same phase if it is complex). 

\begin{figure}
\includegraphics[width=7cm]{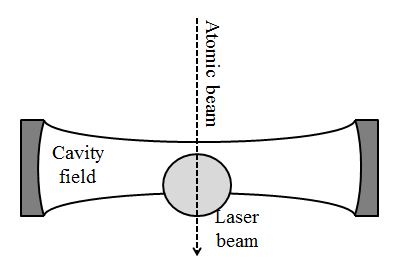} 
\caption{The STIRAP scheme for Fock state preparation. Every single atom adiabatically transfers one photon from the laser beam to the cavity field as it traverses the setup.}
\label{fig15}
\end{figure}

Note that the STIRAP results above do not rely on a bimodal JC model, for example one or both boson fields could be replaced with classical fields with a varying amplitude (which of course was the case for the original development of these coherent-state preparation methods as they grew out from atomic and molecular physics). In the first proposal for employing adiabatic methods in cavity QED, one field was that of the fundamental cavity mode and the other a classical field~\cite{parkins1993synthesis}. The idea is that atoms prepared in the state $|1\rangle$ traverse a Fabry-P\'erot cavity\index{Fabry-P\'erot! cavity/resonator} transversely and simultaneously couple to a classical laser field. The idea is sketched in fig.~\ref{fig15}, a $\Lambda$ atom has non-vanishing dipole transitions to the cavity model between levels $|2\rangle$ and $|3\rangle$ and non-vanishing dipole transition to the laser beam between $|1\rangle$ and $|2\rangle$. The atom is initially in state $|1\rangle$ and the atomic trajectory is such that the effective time-dependence of the couplings appears counterintuitive and as long as the motion is adiabatic the atom will pass one photon from the laser beam into the cavity as it passes through the two fields~\cite{kuhn1999controlled}. This Fock state preparation method has been experimentally realized in the group of Rempe where single photons were created~\cite{hennrich2000vacuum}. Creating superpositions of bare states implies that the atom and cavity field are typically entangled. The present scheme with one classical and one quantized field has been suggested for such entanglement creation~\cite{amniat2004atom}. Moreover, if the atom couples to two perpendicularly polarized cavity modes, ``+'' and ``-'', it follows that the photon will end up in a maximally entangled EPR-state\index{EPR state}\index{State! EPR}~(\ref{eprstate})\index{EPR state} of the type $\frac{1}{\sqrt{2}}\left(|1_+,0\rangle+|0,1_-\rangle\right)$ with $1_+$ and $1_-$ being the single photon labeling of the two modes~\cite{lange2000dynamic}. By considering $N$ identical atoms, simultaneously performing an adiabatic passage, Deng {\it et al}. showed how the atoms can be prepared in a multi-qubit W state\index{W state}\index{State! W}~\cite{deng2006generation}\index{W state}, {\it i.e.} the photon excitation is shared equally among the atoms
\begin{equation}\label{werner} |\psi\rangle=\frac{1}{\sqrt{N}}\left(|e,g,g,...,g\rangle+|g,e,g,...,g\rangle+...+|g,...,g,e\rangle\right).
\end{equation}
Two atoms can also become entangled by letting them traverse the setup along different trajectories. By carefully choosing the propagation paths, the resulting dark state\index{Dark! state}\index{State! dark} can generate an atomic EPR state~\cite{amniat2005decoherence}. It is also possible to achieve {\it state transfer}\index{Adiabatic! state! transfer} where a quantum state of the atoms is passed over to the cavity~\cite{parkins1995quantum}.

The STIRAP scheme combining two modes with a $\Lambda$ atom can only transfer a single photon per pulse sequence. Mattinson {\it et al.} showed how state transfer of arbitrary field states can be realized via adiabatic passage  using instead a two-level atom~\cite{mattinson2001adiabatic}. The time-dependent bimodal Hamiltonian is given by\index{Bimodal! Jaynes-Cummings model}\index{Model! bimodal Jaynes-Cummings}
\begin{equation}
\hat{H}_\mathrm{2m}= \omega\hat{n}_a+\omega\hat{n}_2+\frac{\Delta}{2}\hat{\sigma}_z +\left(g_1\hat{a}^\dagger+g_2\hat{b}^\dagger\right)\hat{\sigma}_{-}+\hat{\sigma}_{+}\left(g_1\hat{a}+g_2\hat{b}\right).
\end{equation}
By introducing a new pair of boson operators
\begin{equation}\label{bostr}
\begin{array}{l}
\hat{A}=K^{-1}\left(g_2\hat{a}-g_1\hat{b}\right),\\
\hat{B}=K^{-1}\left(g_1\hat{a}+g_2\hat{b}\right),
\end{array}
\end{equation}
where $K=\sqrt{g_1^2+g_2^2}$, the Hamiltonian can be rewritten in the interaction picture as
\begin{equation}\label{2mham}
\hat{H}_\mathrm{2m}=\frac{\Delta}{2}\hat{\sigma}_z+K\left(\hat{B}^\dagger\hat{\sigma}_{-}+\hat{\sigma}_{+}\hat{B}\right).
\end{equation}
Since $\left[\hat{A},\hat{B}^\dagger\right]=0$ there is a set of dark states\index{Dark! state}\index{State! dark}
\begin{equation}
|\psi_\mathrm{D}^{(n)}\rangle=\frac{1}{\sqrt{n!}}\left(\hat{A}^\dagger\right)^n|g,0,0\rangle,
\end{equation}
with $|g,0,0\rangle$ the atom in its ground state and both modes in vacuum. The eigenvalues are constant but not strictly zero, which, however, is just a matter of an overall energy shift. Now, by choosing the time-dependence of $g_1$ and $g_2$ such that $\lim_{t\rightarrow-\infty}\hat{A}^\dagger=\hat{a}^\dagger$ and $\lim_{t\rightarrow+\infty}\hat{A}^\dagger=-\hat{b}^\dagger$ it follows that any field can be adiabatically transferred from the occupied mode $a$ to the empty mode $b$. This idea has been generalized to transfer states between spatially separated cavities by only virtually populating excited intermediate states and to create entangled coherent states\index{Entangled coherent states}\index{State! entangled coherent}~\cite{larson2005cavity} (see further sec.~\ref{ssec:cqedstateprep}). More recently, in ref.~\cite{chen2021shortcuts} the preparation of entangled coherent states has also been explored in terms of the mechanism of {\it Shortcut to adiabaticity}\index{Shortcut to adiabaticity}~\cite{torrontegui2013shortcuts}. This is not relying on adiabatic following\index{Adiabatic! following}, but to tailor the explicit time-dependence such that the evolution seems adiabatic by suppressing the non-adiabatic coupling terms. 
 
Using dark states\index{Dark! state}\index{State! dark} and adiabatic evolution is advantageous when it comes to QIP\index{QIP} schemes since ideally states subject to losses are only virtually populated and the logic gates do not depend on particular operational times~\cite{lukin2000entanglement,pachos2002quantum,biswas2004preparation,song2007entangled,chen2007generation,ye2008deterministic,lu2008dispersive,zhou2009quantum}. Lukin {\it et al.} presented a scheme with multi-level atoms interacting with a cavity and a classical field, and by ramping up the classical field amplitude an initial product state evolved into a highly entangled one~\cite{lukin2000entanglement}. The same setup could also be used for adiabatic state transfer between atoms and cavity fields. Pachos and Walther showed how {\it universal quantum computing}\index{Universal quantum computing} ({\it i.e.} combination of a two-qubit conditional logic gate and single qubit rotations) can be realized by individually addressing two atoms in a cavity~\cite{pachos2002quantum}. Also entangling spatially separated atoms, placed in cavities that are connected via an optical fiber, is possible by utilizing adiabatic methods~\cite{chen2007generation,song2007entangled,ye2008deterministic,lu2008dispersive}. Biswas and Agarwal discussed STIRAP methods to prepare both multi-qubit atomic W and GHZ states\index{GHZ state}\index{W state}\index{State! GHZ}\index{State! W}, {\it e.g.} states of the form $|\psi\rangle=\frac{1}{\sqrt{N}}\left(|g,g,...,g\rangle+|e,e,...,e\rangle\right)$. M\o lmer {\it et al.} considered a pumped gas of Rydberg atoms with a strong dipole-dipole interactions and derived an effective model Hamiltonian on the form a driven JC system~\cite{moller2008quantum}. It was demonstrated that using adiabatic passage this system could be used for entanglement generation among a large number of atoms. Multi-partite entanglement generation has also been addressed in~\cite{goto2004multiqubit}, and in particular the possibilities to perform conditional multi-qubit gates.


\subsubsection{Quantized atomic motion}\label{sssec:qatmo}
In the traditional cavity QED realization of JC physics, the atom is typically ``flying'', {\it i.e.} it is traversing the cavity with a preassigned velocity given by some distribution $P(v)$. The velocity can, to a large extent, be controlled by Doppler shifted lasers that prepare the atom in its desired internal states. While in most early cavity QED experiments, the atomic kinetic energy is often much larger than any other energy scales, later works considered laser cooled atoms such that this is no longer valid. Then, the atomic motion must be treated quantum mechanically and will thereby add extra degrees of freedom to the problem, {\it i.e.} the cavity photons couple to both the internal atomic states $|g\rangle$ and $|e\rangle$, and to the external atomic states $|\mathbf{x}\rangle$. In particular, the spatial dependence of the coupling $g(\mathbf{x})$ implies that the atom experiences a light-force (proportional to $\mathbf{\nabla}g(\mathbf{x})$, see sec.~\ref{ssec:lightforce}) which will act differently on the internal atomic states. Furthermore, the effective atom-light coupling will be proportional to the overlap of the atomic density with the mode profile, and thus, in an adiabatic picture the instantaneous Rabi frequency will typically vary in time. This coupled dynamics may indeed lead to many novel phenomena as explained in this subsection.

In 1D, the extended JC model becomes~\cite{meystre1989atomic,englert1991reflecting,haroche1991trapping,scully1996induced}
\begin{equation}\label{qm}
\hat{H}_\mathrm{vJC}=\frac{\hat{p}_z^2}{2m}+\frac{\Delta}{2}\hat{\sigma}_z+g(z)\left(\hat{a}^\dagger\hat{\sigma}_{-}+\hat{\sigma}_{-}\hat{a}\right),
\end{equation}
where $\hat{p}_z$ and $\hat{z}$ are the center-of-mass atomic momentum and position, respectively, and $m$ is the bare atomic mass. It should be understood that the first term multiplies the unit matrix $\mathbb{I}$. A general state $|\Psi(t)\rangle$ can be written in the bare basis as, following the notation of eq.~(\ref{tstate}),
\begin{equation}\label{zstate}
|\Psi(t)\rangle=\sum_n\left[c_{en}(z,t)|e,n\rangle+c_{gn}(z,t)|g,n\rangle\right],
\end{equation}
where $c_{en}(z,t)$ and $c_{gn}(z,t)$ are the (spatial) wave functions given the internal atomic states $|g\rangle$, $|e\rangle$ and the photon Fock state $|n\rangle$.

It is informative to go to the dressed state basis~(\ref{dstate}), by transforming the Hamiltonian by (\ref{jcU}). In other communities, the dressed states -- which now depend on the variable $z$ -- are called {\it adiabatic states}\index{Adiabatic! state}~\cite{larson2020conical}. Modifying the approach of sec.~\ref{ssec:JCm} to the case of a spatially varying coupling strength $g=g(z)$, the Hamiltonian (\ref{qm}) in the adiabatic basis reads
\begin{equation}
\hat{H}_\mathrm{vJC}'=\frac{\left(\hat{p}_z-\hat A(z)\right)^2}{2m}+\sqrt{\frac{\Delta^2}{4}+g^2(\hat n+1)}\hat{\sigma}_z,
\end{equation}
with the synthetic gauge potential\index{Synthetic! gauge potential}\index{Gauge! potential! synthetic}~\cite{larson2006validity}
\begin{equation}\label{jcgaugepot}
\hat A(z)\equiv\frac{1}{2}\frac{\partial\theta}{\partial z}\hat\sigma_y=\frac{1}{2}\frac{\Delta\sqrt{n+1}}{\Delta^2+4g^2(z)(n+1)}\frac{\partial g(z)}{\partial z}\hat\sigma_y.
\end{equation} 
Note how this is in line with the BOA approach to the quantized Rabi model discussed in sec.~\ref{ssec:rabi}, and especially eq.~(\ref{boaham}). Like before, the gauge potential\index{Gauge! potential} $\hat A(z)$ provides the non-adiabatic corrections\index{Non-adiabatic! correction}, which in the BOA are neglected and the Hamiltonian is diagonal in both the atomic and field degrees of freedom. It is important to appreciate the differences between the transformation of this section and that of sec.~\ref{ssec:rabi}; the variables giving rise to the non-adiabatic corrections are the position and momentum of the atom here, while for the quantum Rabi model it was the field quadratures of the electromagnetic field. In sec.~\ref{ssec:chem} we consider molecules coupled to cavity fields and there, just like here, an additional continuous degree of freedom emerges from the nuclear motion of the molecule.

At resonance, $\Delta=0$, the gauge potential\index{Gauge! potential} vanishes, which yields the two decoupled expressions
\begin{equation}
\hat{H}_\mathrm{vJC}^\pm=\frac{\hat{p}_z^2}{2m}\pm g(z)\sqrt{\hat{n}+1}
\end{equation}
and thus, the problem becomes one of 1D scattering against a potential $V_\pm^{(n)}(z)=\pm g(z)\sqrt{n+1}$. For $g(z)>0$, the lower diabatic potential $V_-^{(n)}(z)$ is attractive, while the upper one is repulsive. In the dispersive regime, $|\Delta|\gg|g(z)|\sqrt{n+1}$, we instead find that the gauge potential diminishes since $\hat A(z)\sim\Delta^{-1}$ according to eq~(\ref{jcgaugepot}), which again  decouples the system (to lowest order, see sec.~\ref{ssec:JCm}) and the effective model Hamiltonian becomes
\begin{equation}
\hat{H}_\mathrm{vJC'}=\frac{\Delta}{2}\hat{\sigma}_z+U(z)(\hat{n}+1)\hat{\sigma}_z,
\end{equation}
with $U(z)=g^2(z)/\Delta$. What we have done is simply an adiabatic elimination as explained in the first section~\ref{ssec:JCm}, but things complicate slightly since the Schrieffer-Wolff transformation will not commute with the kinetic energy operator. Depending on the sign of the detuning, the bare atomic states $|g\rangle$ and $|e\rangle$ experience either an attractive or repulsive potential $U(z)(\hat{n}+1)$. These two limiting cases, $|\Delta|\ll |g(x)|\sqrt{n+1}$ or $|\Delta|\gg |g(x)|\sqrt{n+1}$, represent the diabatic and adiabatic limits\index{Diabatic! basis}\index{Adiabatic! basis} (see sec.~\ref{ssec:JCm}). The intermediate regime where non-adiabatic corrections are substantial has been analyzed in ref.~\cite{larson2006validity}. We note in particular that the adiabatic elimination of the excited atomic state $|e\rangle$, see eq.~(\ref{adel}), will no longer solely depend on the condition $|\Delta|\gg |g(x)|\sqrt{n+1}$, but also on the size of the non-adiabatic corrections arising from the non-zero $\hat A(z)$. More precisely, if the coupling $g(z)$ varies rapidly relative to the atomic velocity, these variations can induce non-adiabatic transitions between the internal atomic states.  

In a Fabry-P\'erot cavity\index{Fabry-P\'erot! cavity/resonator}, the fundamental TEM$_{00}$ mode\index{TEM$_{nm}$ mode} shape can be approximated with~\cite{kollar2015adjustable}.
\begin{equation}\label{fpprof}
g(x,y,z)=\left\{\begin{array}{lll}
g_0\cos(kz)\exp\left(-\frac{x^2+y^2}{\sigma^2}\right), & & z\in[-l,l],\\
0, & & \mathrm{otherwise}
\end{array}\right.,
\end{equation}
where $k$ ($=2\pi/\lambda$) is the longitudinal wave number of the cavity mode, $\sigma$ is the characteristic mode width, and the cavity length $L=2l$. Thus, if the atom traverses the cavity transversely the mode is assumed Gaussian\index{Gaussian! mode}, while if it traverses it longitudinally it is periodic. Naturally, the two cases show very different characteristics. Controlling the velocity in a given direction can also be done by varying the angle at which the atom enters the cavity, for example if the incident angle is slightly shifted from perpendicular the dynamics could be assumed ``classical' in the perpendicular direction and quantized along the longitudinal direction.

For a quantized transverse motion, the problem was first addressed in Refs.~\cite{englert1991reflecting,haroche1991trapping}. Englert {\it et al.} studied the scattering problem at zero detuning and by approximating the mode function by a mesa function ({\it i.e.} box-shaped potential)~\cite{englert1991reflecting}. They considered both how the additional degree of freedom affects the Rabi oscillations and the possibility to reflect atoms from the light field. Haroche {\it et al}. asked the question whether the light field generated from vacuum would be strong enough to trap single atoms~\cite{haroche1991trapping}. The conclusion was that ideally this should be possible. Later,  the groups of Rempe and Kimble independently demonstrated atomic trapping in cavity fields on the single photon level~\cite{pinkse2000trapping,hood2000atom} (for further details see section~\ref{sec:cavQED}). The results of~\cite{englert1991reflecting} have been extended to study the maser problem (at zero detuning) with quantized center-of-mass motion~\cite{scully1996induced,meyer1997quantum,schroder1997quantum}. Scully and co-workers showed how the stimulated emission\index{Stimulated! emission! probability} probability is greatly affected by the atomic motion, which in return alters the maser spectrum. They termed this quantized maser system the {\it mazer}\index{Micromaser}\index{Mazer} with the ``z'' indicating quantized motion in the $z$ direction. In their analysis they considered both a mesa function mode profile and a P\"oschl-Teller potential\index{P\"oschl-Teller potential} ($V(x)=\mathrm{sech}^2(x)$~\cite{landau1958quantum}) mode profile. Typical for quantum scattering problems, they found tunneling resonances (see fig.~\ref{microm} in sec.~\ref{sssec:micro}) for certain atomic velocities where the transmission coefficient is very large. It was argued that these resonances could be used for preparing atomic beams with well defined velocities~\cite{loffler1998velocity}. The effects of quantum fluctuations in the atomic velocity were addressed for a Gaussian mode profile\index{Gaussian! mode! profile}~\cite{larson2009quantum}. The tunneling resonances can be greatly suppressed by the fluctuations, while in the adiabatic regime an {\it echo effect}\index{Quantum! echo} was found which leaves the maser field unchanged. The characteristic tunneling time was considered in~\cite{arun2011tunneling}. The maser action for non-zero detuning was first discussed in ref.~\cite{bastin2003detuning} in the case of a mesa function (the problem is analytically solvable in this case). Since its introduction, the maser action has been generalized to various setups; bimodal cavities~\cite{arun2000maser,agarwal2000resonant}, two-photon transitions~\cite{zhang1999quantum,zhi1999quantum}, $\Lambda$ atoms~\cite{zhang1998quantum,zhi2000spectrum}, and $V$ atoms~\cite{zhi2001atomic,ai2005influence}. Effects of quantized motion in the STIRAP scheme was analyzed in~\cite{jonasthesis}, and it was particularly found that the adiabatic passage breaks down for sufficiently slow atoms and then a series of tunneling resonances appears.

\begin{figure}
\includegraphics[width=10cm]{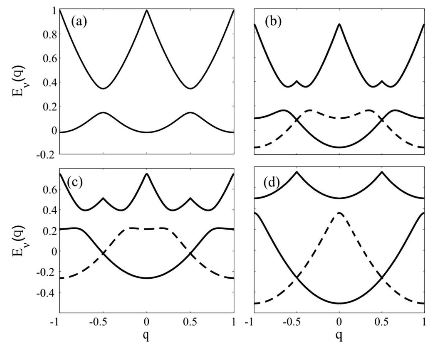} 
\caption{Spectrum\index{Band spectrum} of the Hamiltonian (\ref{2mat}) for a given excitation number and for different (positive) detunings $\Delta$, in growing order from {\bf (a)}--{\bf (d)}. The solid and dashed lines represents different parities, {\it i.e.}, eigenstates to the operator $\hat{\tau}$ with different eigenvalues.}
\label{fig16}
\end{figure}

For longitudinal atomic passages the coupling is taken as $g(x)=g_0\cos(kz)$, and, by scaling the length by $k^{-1}$ and energies by the recoil energy $E_\mathrm{r}=\frac{\hbar^2k^2}{2m}$\index{Recoil energy}, the Hamiltonian in these new dimensionless parameters/variables becomes
\begin{equation}\label{2mat}
\hat{H}_\mathrm{vJC}=-\frac{d^2}{dz^2}+\frac{\Delta}{2}\hat{\sigma}_z+g_0\cos(z)\left(\hat{a}^\dagger\hat{\sigma}_{-}+\hat{\sigma}_{+}\hat{a}\right).
\end{equation}  
Written as $2\cos(z)=e^{iz}+e^{-iz}$, it follows that the atom-field interaction terms contains four terms where every absorption/emission of a cavity photon is accomplished by a momentum kick/recoil changing the center-of-mass momentum by $\pm 1$. This underlines the coupling of internal and external degrees of freedom which led to the notion {\it well-dressed states}\index{Well-dressed states}~\cite{vernooy1997well}, where the idea is that the internal atomic states are not only ``dressed'' by photons but also by the momentum states deriving from the quantized motion. Since the Hamiltonian is periodic, if we let the cavity length $L\rightarrow\infty$, the energy spectrum $E_\nu(q)$ has a band-structure\index{Energy band spectrum} characterized by a {\it quasi-momentum} $q$\index{Quasi! momentum} and a {\it band index} $\nu$~\cite{ashcroftmermin}. More precisely, the translation operator $\hat{T}=e^{\pm i2\pi\hat{p}_z}$ commutes with $\hat{H}_\mathrm{vJC}$, and one therefore expects the first {\it Brillouin zone}\index{Brillouin zone} to span quasi-momentum $q\in[-1/2,+1/2)$. However, there is an additional translational symmetry\index{Translational! symmetry}
\begin{equation}
\hat{\tau}=\hat{\sigma}_ze^{\pm i\pi\hat{p}_z}.
\end{equation}
The action of $\hat{\tau}$ simultaneously flips the sign of $g(z)$ and of $\hat{\sigma}_{\pm}$. As a result, the actual first Brillouin zone is better defined for $q\in[-1,+1)$~\cite{ren1995spontaneous}. Note further that $\hat{\tau}^2=\hat{T}$ and that the additional $\hat{\tau}$ symmetry separates the Hamiltonian into two blocks of momentum states~\cite{larson2005effective} (apart from the $2\times2$ block structure deriving from the preserved total number of excitations). In other words, $\hat{\tau}$ has a $\mathbb{Z}_2$\index{$\mathbb{Z}_2$ symmetry} structure where the basis can be labeled by their ``parity''. In the diabatic limit $\Delta=0$, the effective light-induced potential is $V_\mathrm{d}(z)\sim\cos(z)$, while in the adiabatic limit, $|\Delta|\gg|g(z)|\sqrt{n}$, it is of the form $V_\mathrm{ad}(z)\sim\cos^2(z)$. In either case, the Hamiltonian is of the general {\it Mathieu} form~\cite{zwillinger1998handbook} corresponding to the {\it Mathieu equation}\index{Mathieu equation}, and the spectrum has the ``common'' structure as shown in fig.~\ref{fig16} (a) for the $\Delta=0$ case (remember that since the Brillouin zone\index{Brillouin zone} is twice the size of the regular Mathieu equation, we usually have $-1/2\leq q\leq+1/2$). In the intermediate regime, $|\Delta|\sim|g(z)|\sqrt{n}$, the spectrum attains rather peculiar shapes as demonstrated in fig.~\ref{fig16} (b)-(d). 

The properties of the dynamics are significantly affected by the shape of the atomic center-of-mass wavepacket $\psi(z,t)$~\cite{vaglica1995jaynes,chough2002dynamics,leach2004cavity,larson2005effective}. If $\psi(z,t)$ is broad on a length scale $k^{-1}$, the atom `feels' the periodicity of the potential, and the momentum distribution is typically much narrower than the size of the Brillouin zone\index{Brillouin zone}, meaning that the atom can be assigned an approximate {\it quasi}momentum $q_\mathrm{cm}$. This regime was studied in detail in ref.~\cite{larson2005effective} where the concept of an {\it effective mass}\index{Effective! mass}, $m_\mathrm{eff}(q)=\left(\frac{\partial^2 E_\nu(q)}{\partial q^2}\right)^{-1}$, and of the {\it group velocity}\index{Group velocity}, $v_g(q)=\frac{\partial E_\nu(q)}{\partial q}$, were introduced. Since these quantities depend on the photon number $n$, it was demonstrated that by measuring the traversal times of the atoms, it is possible to extract the photon number via non-demolition measurements~\cite{larson2009cavity}\index{Non-demolition measurement}. Another non-demolition scheme was proposed in order to measure the atomic velocity via detection of the field quadratures~\cite{quadt1995measurement}. The effective mass also provides an estimate for how fast the wavepacket spreads in the periodic potential. The spreading
has also been studied by other means~\cite{vaglica1995jaynes,chough2002dynamics}, for example in the {\it Raman-Nath approximation}\index{Raman-Nath approximation}. Since the potential couples certain momentum states, $|q_0+N\rangle$ ($N\in\mathbb{Z}$, and $q_0$ is some initial momentum eigenstate), in this basis the kinetic energy operator is simply $(q_0+N)^2$. The Raman-Nath approximation consists in neglecting the $q_0^2$, and by this analytical approximate expressions can be obtained~\cite{holland1991quantum,herkommer1992quantum}. The coupling of discrete atomic momentum states implies that in the momentum representation the system mimics a tight-binding lattice model~\cite{wang2014superradiance}. For a wavepacket width on the length scale of the periodicity, the atom no longer experiences an effective periodic potential~\cite{wilkens1992spontaneous,vaglica1995jaynes,leach2004cavity}. Here the overlap between the atomic wavepacket and the mode function determines the effective atom-field coupling, and variations in this quantity imply among other phenomena the decay of the collapse-revival feature~\cite{vaglica1995jaynes}. One possibility to confine the atom within the length of a single period could be to introduce an external trapping potential~\cite{leach2004cavity}. In sec.~\ref{sssec:ent} we discussed aspects of atom-field entanglement. In the present case there is the additional quantum degree of freedom of the atomic motion. As we have argued, all three degrees of freedom (internal and external atomic and field) become correlated. These correlations are of course affected by decoherence. Doherty {\it et al.}, showed that cavity loss not only affects the correlations but also induces an atomic diffusion mechanism~\cite{doherty1998motional}. 

Early studies of cavity QED with quantized atomic motion consisted in considering deflection of atomic beams incident transversely on a standing wave mode profile~\cite{meystre1989atomic}. Since absorption/emission of photons induces transverse (in the direction of the beam) momentum kicks the beam will split up according to the number of performed Rabi oscillations, which was explicitly analyzed in the Raman-Nath approximation in~\cite{meystre1989atomic} (a focusing effect is also possible in the dispersive regime~\cite{averbukh1994quantum}). It naturally follows that for large field amplitudes the atom can perform more oscillations during the effective interaction time, and it is therefore possible to conduct quantum non-demolition measurements of the field amplitude~\cite{holland1991quantum}. Following this work, it has been shown how to extract both the photon distribution~\cite{herkommer1992quantum}, and to do full field tomography\index{Tomography measurement} (see further sec.~\ref{ssec:cqedtomo}) by initializing the internal atomic states to coherent superpositions of $|g\rangle$ and $|e\rangle$~\cite{freyberger1994probing}.

\begin{figure}
\includegraphics[width=9cm]{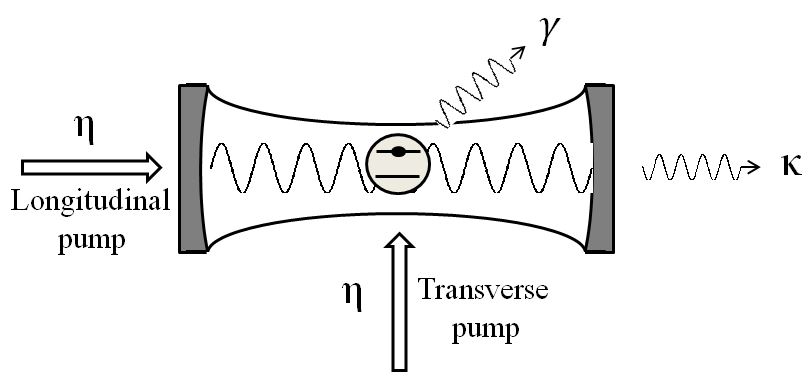} 
\caption{The system setup for cavity cooling schemes. An atom (gray circle), either confined in an external trap or freely moving within the cavity, couples to a cavity mode and its excited state $|e\rangle$ has a spontaneous emission decay rate\index{Spontaneous! emission! rate} $\gamma$. The photon decay rate is $\kappa$, and either the cavity is pumped via one of its mirrors or the atom is directly pumped by an external laser.  
}
\label{fig18}
\end{figure}

We now turn to the configuration depicted in fig.~\ref{fig18}: a single two-level atom, with spontaneous emission rate\index{Spontaneous! emission! rate} $\gamma$, is located in a leaky cavity with decay rate $\kappa$; in addition there is a pump, either parallel to the cavity axis or in the transverse direction and exciting the atom. The master equation (from the weak to the strong-coupling regime, see sec.~\ref{sssec:openjc}) reads~\cite{domokos2003mechanical}
\begin{equation}
\partial_t\hat{\rho}(t)=i\left[\hat{\rho}(t),\hat{H}_\mathrm{c}\right]+\hat{\mathcal{D}}\left[\kappa\hat{a}\right]\hat{\rho}(t)+\hat{\mathcal{D}}\left[\gamma\hat{A}\right]\hat{\rho}(t).
\end{equation}
The Hamiltonian is
\begin{equation}
\hat{H}_\mathrm{c} = \displaystyle{-\mathbf{\nabla}^2+V_\mathrm{ext}(\mathbf{x})+\frac{\Delta_\mathrm{p}}{2}\hat{\sigma}_z+\delta_\mathrm{p}\hat{n}} +g(\mathbf{x})\left(\hat{a}^\dagger\hat{\sigma}_{-}+\hat{\sigma}_{+}\hat{a}\right)+\hat{H}_\mathrm{pump},
\end{equation}
where $V_\mathrm{ext}(\mathbf{x})$ is an external atomic potential, $\Delta_\mathrm{p}=\Omega-\omega_\mathrm{p}$ is the atom-pump detuning, $\delta_\mathrm{p}=\omega-\omega_\mathrm{p}$ is the field-pump detuning, and
\begin{equation}\label{pumpterm}
\begin{array}{lll}
\hat{H}_\mathrm{pump} & = & \left\{\begin{array}{lll}
\eta\left(\hat{a}^\dagger+\hat{a}\right), & & \mathrm{longitudinal\,\,pump},\\
\eta\left(\hat{\sigma}_{+}+\hat{\sigma}\right), & & \mathrm{transverse\,\,pump},
\end{array}\right.
\end{array}
\end{equation}
is the external pump term of either the field or the atoms. The Lindblad super-operators\index{Lindblad! super-operator} are
\begin{equation}
\hat{\mathcal{D}}\left[\kappa\hat{a}\right]\hat{\rho}(t)=\kappa\left(\hat{a}\hat{\rho}\hat{a}^\dagger-\frac{1}{2}\hat{n}\hat{\rho}-\frac{1}{2}\hat{\rho}\hat{n}\right)
\end{equation}
and
\begin{equation}\label{eq:Dterm}
\hat{\mathcal{D}}\left[\gamma\hat{\sigma}_{-}\right]\hat{\rho}(t)=  \displaystyle{\gamma\left(\int d^2\mathbf{u}\,N(\mathbf{u})\hat{\sigma}_{-}e^{-ik_\mathrm{A}\mathbf{u\cdot x}}\hat{\rho}e^{ik_\mathrm{A}\mathbf{u\cdot x}}\hat{\sigma}_{+}\right.} \displaystyle{\left.-\frac{1}{2}\hat{\sigma}_{+}\hat{\sigma}_{-}\hat{\rho}-\frac{1}{2}\hat{\rho}\,\hat{\sigma}_{+}\hat{\sigma}_{-}\right)},
\end{equation}
with $\mathbf{u}$ the unit vector in the direction of the emitted photon, $N(\mathbf{u})$ is the direction distribution, and $k_\mathrm{A}$ is the wave number of the emitted photon ({\it i.e.} $k_\mathrm{A}=\Omega/c$) -- compare with eq.~\eqref{eq:MEion} for a driven and damped two-level ion. As we mentioned above, the spatial dependence of $g(\mathbf{x})$ generates a light-induced force felt by the atom~\cite{domokos2003mechanical,black2005collective}. The dissipation channels will, on the one hand, give rise to atomic diffusion~\cite{doherty1998motional}, and on the other also to effective atomic {\it friction forces}~\cite{horak1997cavity,hechenblaikner1998cooling,domokos2001semiclassical,domokos2003mechanical}, which can be utilized for atomic cooling. 

Let us first consider the situation of no external trapping potential, and discuss the trapping case later. Free space laser cooling relies on atomic spontaneous emission\index{Spontaneous! emission} to convert atomic kinetic energy into photon energy. In free space, since the emitted photons are unlikely to be reabsorbed by the atoms they irreversibly remove kinetic energy from the system~\cite{metcalf1999laser}. Loosely speaking, the idea of {\it cavity cooling}\index{Cavity! cooling} is that photon losses of the cavity can serve as an irreversible loss mechanism instead of atomic spontaneous emission\index{Spontaneous! emission}. The lower temperature limit $T_\mathrm{D}$ of regular {\it Doppler cooling}\index{Doppler! cooling} is limited by the atomic linewidth\index{Linewidth! atomic}, {\it i.e.} $k_\mathrm{B}T_\mathrm{D}\sim\hbar\gamma$ where $k_\mathrm{B}$ is the Boltzmann's constant. The cavity linewidth\index{Linewidth! cavity} $\kappa$ sets the lower limit for the corresponding cavity cooling and it can be considerably lower than that of Doppler cooling~\cite{horak1997cavity}. 

Without a trap, cavity cooling derives from the atom-field backaction\index{Backaction}; the motion of the atom within the field alters the cavity field which in return acts back on the propagation of the atom. If the fastest time-scale in the problem is determined by the photon decay rate $\kappa$, the cavity field follows the atomic evolution adiabatically. In this bad cavity limit\index{Bad-cavity limit}, the mechanism of the cavity cooling becomes the same as for regular Doppler cooling~\cite{horak1997cavity,domokos2003mechanical}. For smaller $\kappa$ there are non-adiabatic contributions which can become an asset for the cooling, and the cooling is more reminiscent of {\it Sisyphus cooling}\index{Sisyphus cooling}~\cite{horak1997cavity}. Classically, the cooling mechanism can be understood from the retarded field evolution, {\it i.e.} there is a finite time for the field to `respond' to the atomic motion. For an atom passing through a field maximum, the field has responded to the passage when the atom is already on its way down the slope. Another way of seeing the process is that the photons leaking through the cavity have somewhat higher energy than those of the pump and the missing energy is provided by looking at the atomic kinetic energy. The atomic evolution, including friction and atomic diffusion has been studied in phase space by deriving a Fokker--Planck equation\index{Fokker--Planck equation}  which generates the evolution of the Wigner function\index{Wigner! function}~\cite{domokos2001semiclassical}. Quantum corrections to this approach were considered in~\cite{vukics2005cavity}, which was particularly motivated by the fact that the atomic state can be altered already for very weak cavity fields containing a few photons. In the good cavity regime it has been explained that atomic pumping is preferable over cavity pumping. A characteristic property of cavity cooling is that it can be independent of internal atomic states and thereby it is possible to preserve atomic coherence throughout the process~\cite{vuletic2000laser,vuletic2001three}. In particular, in the large-detuning limit, the excited atomic level can be adiabatically eliminated~\cite{domokos2001semiclassical}. Note that the elimination results in an additional decay channel which induces decoherence of the cavity. With the cooling depending on internal atomic states, it is also possible to cool molecules with complex electronic states~\cite{vuletic2000laser,lev2008prospects}. By extending the setup to a multimode cavity\index{Multimode! cavity} it has even been suggested how external, internal electronic, vibrational, and rotational molecular degrees of freedom\index{Rotational! degree of freedom} can be cooled~\cite{morigi2007cavity}. A completely different type of cavity cooling is {\it self-organization cooling}\index{Self-!organization cooling}~\cite{domokos2002collective,asboth2005self}. Here a large number of atoms are transversely pumped by a standing wave field in the large detuning limit such that photons from the pump laser are scattered upon the atoms into the cavity. For a certain pump amplitude, the atoms {\it self-organize} in 'checkerboard pattern' in the square lattice formed by the cavity and pump fields. While passing to this organized phase, the atomic kinetic energy is highly reduced.  

In the case of a harmonic trapping potential, the cavity cooling modifies the {\it sideband cooling}\index{Sideband cooling} developed in trapped ion physics~\cite{diedrich1989laser} (see sec.~\ref{ssec:ionham}). Again, standard sideband cooling build on spontaneous emission\index{Spontaneous! emission}, while in cavity assisted sideband cooling the irreversible mechanism stems from photon losses~\cite{vuletic2001three}. Morigi and co-workers derived effective rate equations between motional states by assuming the {\it Lamb-Dicke regime}\index{Lamb!-Dicke! regime} (see sec.~\ref{ssec:ionham}, and in particular eq.~\ref{ldregime}) and then expand in the {\it Lamb-Dicke parameter}\index{Lamb!-Dicke! parameter} $\eta$~\cite{zippilli2005cooling,zippilli2005mechanical,bienert2012cooling}. Both types of pumping were analyzed, longitudinal cavity~\cite{zippilli2005cooling,zippilli2005mechanical} and transverse atomic~\cite{bienert2012cooling}. They identified both heating and cooling parameter regimes, with possible cooling down to average phonons much less than unity. Also prospects of cavity cooling of atoms residing in an optical lattice have been analyzed~\cite{murr2006three}, as well as cavity feedback cooling~\cite{steck2006feedback}. 

Experimentally, cavity cooling has been demonstrated in numerous groups~\cite{chan2003observation,maunz2004cavity,nussmann2005vacuum,boozer2006cooling,leibrandt2009cavity,wolke2012cavity}. Collective cooling below the Doppler limit in a gas containing millions of atoms was observed in~\cite{chan2003observation}. The cavity was here vertical and the cloud of atoms fell freely within the cavity while simultaneously being perpendicularly pumped. Other collective cavity cooling has been seen in a Bose-Einstein condensate\index{Bose-Einstein! condensate} of a density $10^{14}$ cm$^{-1}$~\cite{wolke2012cavity}. Cooling of the condensate down to the sub-recoil regime was achieved, and also cavity heating was seen. Sideband cooling has also been verified for both atoms~\cite{boozer2006cooling} and ions~\cite{leibrandt2009cavity}. In the case of an atom, motional ground state cooling was reached. Single atom cavity cooling was also considered in~\cite{maunz2004cavity,nussmann2005vacuum}. The cavity cooling was found to be five times as efficient as for free space cooling and atoms could, thanks to the low temperatures, be trapped within the cavity up to 17 seconds.


\subsubsection{The Dicke and Tavis-Cummings models}\label{sssec:dicke}
The extension of the JC model that has deserved the most attention is without doubt the one by Dicke \cite{dicke1954coherence,garraway2011dicke,kirton2019introduction,roses2020dicke} or Tavis-Cummings (TC)~\cite{tavis1968exact}, namely to consider $N$ ($>1$) two-level atoms rather than a single one. The generalisation is straightforward,
\begin{equation}\label{dickeham}
\hat{H}_D=\omega\hat{n}+\sum_{i=1}^N\left[\frac{\Omega}{2}\hat{\sigma}_z^{(i)}+g\left(\hat{a}^\dagger+\hat{a}\right)\hat{\sigma}_x^{(i)}\right]
\end{equation} 
for the Dicke model, and
\begin{equation}\label{tcham}
\hat{H}_{TC}=\omega\hat{n}+\sum_{i=1}^N\left[\frac{\Omega}{2}\hat{\sigma}_z^{(i)}+g\left(\hat{a}^\dagger\hat{\sigma}_{(i)}^-+\hat{\sigma}_{(i)}^+\hat{a}\right)\right]
\end{equation}
for the TC model. Here the sub/superscript $`(i)'$ stands for atom number $i$, and consequently operators with different $i$'s commute. From the definitions above we note that the TC model is identical to the Dicke one within the RWA, and hence, the Dicke model is the corresponding many-atom version of the quantum Rabi model, while the TC model is the extension of the JC model. Accordningly, the TC model is integrable~\cite{bogoliubov1996exact}. In particular, it can be solved within a Bethe {\it ansatz}\index{Bethe {\it ansatz}}, but as is often the case with such solutions, extracting physical quantities are not always trivial, and the Bethe {\it ansatz} solutions are not capable of capturing dynamical correlations. The TC model is also within the class of models studied by Gaudin\index{Gaudin model}\index{Model! Gaudin}~\cite{gaudin1976diagonalisation}, which has turned out important for more general aspects of integrable models. The influence of the neglected counter rotating terms in the Tavis-Cummings model\index{Tavis-Cummings model} was studied in ref.~\cite{seke1991many,li2012collective}.

While we define the two models according to eqs.~(\ref{dickeham}) and (\ref{tcham}), in the literature, it is common that the Dicke model is referred to for $N\gg1$ and for few atoms, $1<N<10$, one talks instead about the TC model. It is convenient to introduce the total spin operators\index{Spin! operator}\index{Operator! spin}
\begin{equation}\label{spinop}
\hat{S}_\alpha=\sum_{i=1}^N\hat{\sigma}_\alpha^{(i)},\hspace{1cm}\alpha=x,\,y,\,z
\end{equation}
which obey the standard commutation relations $\left[\hat{S}_\alpha,\hat{S}_\beta\right]=i\varepsilon_{\alpha\beta\gamma}\hat{S}_\gamma$ and $\left[\hat{S}_\alpha,\hat{S}^2\right]=0$, where $\hat{S}^2=\hat{S}_x^2+\hat{S}_y^2+\hat{S}_z^2$. Since $\left[\hat{S}^2,\hat{H}_D\right]=\left[\hat{S}^2,\hat{H}_{TC}\right]=0$, the eigenstates of the two Hamiltonians can be characterised by their total angular momentum quantum number $s$. As always, we chose a basis $\left\{|s,m\rangle\right\}$ ($-s\leq m\leq +s$) where $\hat{S}_z$ is diagonal, {\it i.e.}
$\hat{S}^2|s,m\rangle=s(s+1)|s,m\rangle$ and $\hat{S}_z|s,m\rangle=m|s,m\rangle.$ The $|s,m\rangle$'s are the so-called Dicke states introduced in (\ref{dickestates})\index{Dicke! state}, and which have shown to host interesting entanglement properties~\cite{toth2007detection,guhne2009entanglement}. For the multi-particle system we can interpret
\begin{equation}
j_z=\langle\hat S_z\rangle
\end{equation}
as the `magnitization'. We note that $s$ is bounded by $N/2$ and since one is often interested in the spin sector containing the lowest energy state it is enough to consider the $s=N/2$ Dicke states. This, of course, decreases the atomic Hilbert space dimension from exponential, $\mathcal{D}=2^N$, to linear, $\mathcal{D}=2N+1$, in the particle number. It also implies that the models (\ref{dickeham}) and (\ref{tcham}) can be seen as a bosonic mode coupled to a spin-$s$ particle.  

As an alternative to the collective spin operators of eq.~(\ref{spinop}), we can consider a unitary transformation into a new spin basis. For example we call the spin variable
\begin{equation}
\hat\sigma_\alpha^{(B)}=\frac{1}{\sqrt{N}}\sum_{i=1}^N\hat\sigma_\alpha^{(i)},
\end{equation}
for the {\it bright spin operator}\index{Bright! spin operator}\index{Operator! bright spin}, while the remaining $N-1$ operators $\hat\sigma_\alpha^{(D_\mu)}$ are called {\it Dark spin operators}\index{Dark! spin operators}\index{Operator! dark spin}~\cite{agarwal1998spectroscopy,vetter2016single,ribeiro2018polariton}. The TC Hamiltonian, in the new spin basis, becomes
\begin{equation}\label{tcham2}
\hat{H}_{TC}=\omega\hat{n}+\frac{\Omega}{2}\hat{\sigma}_z^{(B)}+g\sqrt{N}\left(\hat{a}^\dagger\hat{\sigma}_{(B)}^-+\hat{\sigma}_{(B)}^+\hat{a}\right)+\frac{\Omega}{2}\sum_{\mu=1}^{N-1}\hat\sigma_z^{(D_\mu)}.
\end{equation}
Thus, we have decoupled the system into two subspaces, the `bright' and the `dark'\index{Bright/dark subspace}. The idea is the same as the approach for the bimodal JC model when we instead transformed the boson modes~(\ref{bostr}). In the bright subspace the Hamiltonian is a renormalized JC one, with $g\rightarrow g\sqrt{N}$, and in the dark subspace the Hamiltonian is simply the bare spin energies. We will see next the effect of this coupling renormalization on the radiation emission. Since the TC model written on the above form identifies a JC model, we note that in order to reach a strong light-matter coupling one can consider a set of identical atoms. The bright state\index{Bright! state}, $|B\rangle=\hat\sigma_{(B)}^+|0\rangle$, is symmetric under permutations of the atomic index $i$, which is not true for the dark states\index{Dark! state}, $|D_\mu\rangle=\hat\sigma_{(D_\mu)}^+|0\rangle$. 
 
Dicke originally introduced his model in order to understand how many densely-spaced identical atoms collectively spontaneously radiate\index{Spontaneous! radiation}. It should be clear that any deviation from $N$ independently radiating dipoles derives from the common coupling to the same radiation mode. Looking at the matrix element corresponding to the spontaneous emission\index{Spontaneous! emission} we have the rate of photon emission
\begin{equation}
\begin{array}{lll}
\Gamma_{\rm ph.em.} & \propto & \sum_\psi|\langle\psi|\hat{a}^\dagger \hat{S}_{(i)}^-|s,m\rangle|0\rangle|^2\\ \\
& = & (s+m)(s-m+1).
\end{array}
\end{equation}
Here, $|0\rangle$ is the vacuum state of the field mode and the sum is over all possible states $|\psi\rangle$. Assuming all $N\gg1$ atoms to be initially excited ($m=s=N/2$) we find $\Gamma_{\rm ph.em.}\propto N$ as for independent atoms. If instead $m=0$, {\it i.e.}, a gas of partially excited atoms, we have $\Gamma_{\rm ph.em.}\propto N^2$ implying a large enhancement of the emission due to the cooperative effect. This phenomenon, called {\it superradiance}\index{Superradiance}~\cite{scully2009super}, derives from quantum correlations among the initialised atomic state~\cite{wiegner2011quantum,duan2001long}  and has been experimentally observed in various systems, see for example the reviews~ \cite{gross1982superradiance,ficek2002entangled}. There is also a {\it subradiance}\index{Subradiance} effect where instead a decreased emission rate among the $N$ two-level systems is found~\cite{dicke1954coherence,scully2015single}. We will expand more on this subject in sec.~\ref{subsec:generalDickeext} where we discuss collective spontaneous emission\index{Spontaneous! emission} from an array of atoms using an application of quantum-trajectory theory. 
 
In a recent extension, the authors of~\cite{Orioli2022} model atom-light interaction in the dipole approximation by the multilevel JC Hamiltonian, to study the collective decay dynamics of atoms with a generic multilevel structure coupled to two light modes of different polarization inside a cavity. They find that dark states\index{Dark! state} with appreciable entanglement arise after superradiant decay, due to the destructive interference between different internal transitions. Such Dicke dark states are highly entangled and could be of interest for various QIP schemes. 
 
\begin{figure}
\includegraphics[width=10cm]{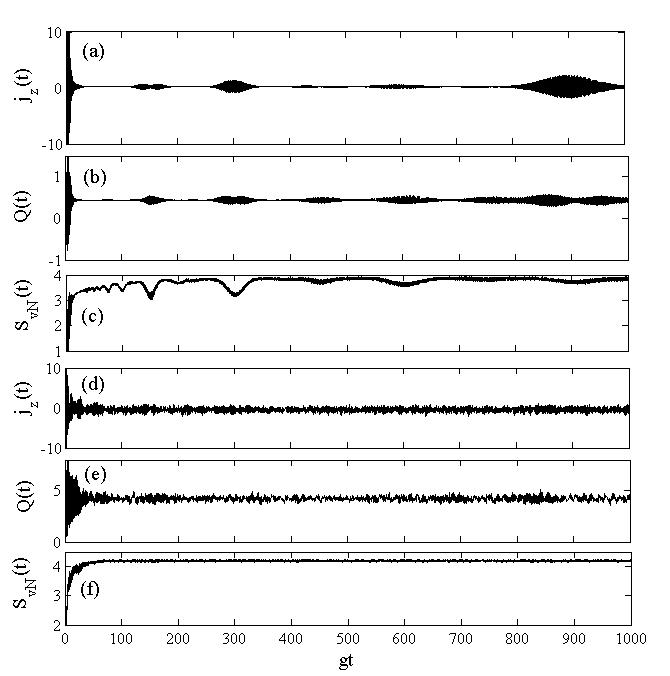}  
\caption{Same as fig.~\ref{fig3} but for the Dicke model. The initial state is a coherent field state with $\bar n=100$ and a spin state $|s,m\rangle$ with $s=m=20$. In the upper three plots the coupling $g=0.1$ implying that applying the RWA is justified. In this case we also envision clear revivals, however not as prominent as for the JC model displayed in fig.~\ref{fig3}. This results from the additional anharmonicity emerging due to the multiple number of atoms in the Dicke model. In the lower three plots $g=0.5$ and the nonlinearity is so strong that we do not encounter any revivals. Here, contrary to the case with $g=0.1$, the evolution is dominated by chaos and the vanishing of fluctuations is typical for quantum thermalization\index{Quantum! thermalization}~\cite{altland2012quantum,larson2013chaos}. In either cases, $g=0.1$ or $g=0.5$, the two sub-systems become almost maximally entangled. This is certainly expected in the chaotic case as the evolution is ergodic\index{Ergodic}. The dimensionless other parameters are $\omega=\Omega=1$. } 
\label{figdickeevolve}  
\end{figure}
 
The multi-atom (qubit) character of the Dicke model naturally opens up for entanglement generation among the atoms~\cite{guo2002scheme,tessier2003entanglement,hou2004decoherence}; entanglement deriving from common interaction to the same boson mode. In the large detuning limit an effective atomic model is obtained from elimination of the cavity field, and it has been demonstrated that it can be utilized for dynamical generation of multi-qubit entangled GHZ\index{GHZ state}\index{State! GHZ} (\ref{ghzstate}) and W (\ref{wstate})\index{W state}\index{State! W} states~\cite{guo2002scheme}. The two-atom case was considered in~\cite{tessier2003entanglement}. By tracing over the field degrees of freedom the two-atomic state is generally mixed and entanglement measures such as purity or von Neumann entropy (\ref{vN}) are no longer valid and instead the concurrence or logarithmic negativity (\ref{logneg}) is applicable. The evolution of atom-field entanglement as the coupling is increased was the subject of ref.~\cite{hou2004decoherence}. As discussed below, beyond a certain coupling the Dicke model becomes chaotic which is reflected in strong entanglement among the constituent parts. This is, indeed, expected since in this regime the eigenenstates are ergodic (see also paragraph on thermalization) meaning that information should be shared among all parts. However, these references all apply the rotating-wave approximation, {\it i.e.} they consider the Tavis-Cummings model\index{Tavis-Cummings model}. Entanglement properties for the Dicke model has also been thoroughly explored~\cite{schneider2002entanglement,latorre2005entanglement,ficek2008delayed,chen2008numerically,larson2010circuit,joshi2015cavity}. In Refs.~\cite{schneider2002entanglement,larson2010circuit,joshi2015cavity} the effect of decoherence was a main focus; can entanglement survive certain kinds of decoherence? In \cite{schneider2002entanglement} the effective coupling among the atoms is `incoherent', the coupling is accompanied by noise which tend to decohere the steady state\index{Steady state}, but still it was shown that sustainable atom-atom entanglement is possible. Related to dynamical atom-atom entanglement generation is the idea of realizing atom-atom logic gates. Similar to the schemes discussed in sec.~\ref{sssec:multi} for logic operations between different photon modes it is also possible to construct gates between the atoms. In ref.~\cite{zheng2004unconventional} a proposal for a two-qubit phase gate is presented. The novelty of the method relies on combining the atom-field interaction with an atomic driving and by carefully choosing the parameters the cavity mode is only virtually populated and thereby prone to photon losses. In terms of QIP an interesting aspect is that superradiance can be a source of an undesirable effective decoherence~\cite{yavuz2014superradiance}. In particular, fault-tolerant error correction is not capable of handling such non-local noise.

In 2006, Hillery and Zubairy provided a class of inequalities, arising from uncertainty relations, whose violation indicates the presence of entanglement in two-mode systems (with photon annihilation operators $\hat{a}$ and $\hat{b}$)~\cite{HilleryZubairy2006}. They conclude that a state is entangled if
\begin{equation}\label{eq:ZubHill}
 |\braket{\hat{a}^m (\hat{b}^{\dagger})^n}|^2 > \braket{(\hat{a}^{\dagger})^m \hat{a}^m (\hat{b}^{\dagger})^n \hat{b}^n}.
\end{equation}
To form a representation of the su(1,1) algebra, they also defined the variable 
\begin{equation}
 \hat{K}(\phi)=e^{i\phi}\hat{a}^{\dagger}\hat{b}^{\dagger} + e^{-i\phi}\hat{a} \hat{b}
\end{equation}
and proposed that states whose uncertainty in one of the variables $\hat{K}(\phi=0)$ or $\hat{K}(\phi=\pi/2)$ is close to one of the un-normalizable eigenstates of the corresponding operators will be entangled. In principle, $\hat{K}(\phi=0)$ and $\hat{K}(\phi=\pi/2)$ are proportional to quadratures that can be determined by means of homodyne measurements\index{Homodyne detection}. In their recent report, Cheng and Zubairy set out to assess the usefulness of criteria like that given by eq.~\eqref{eq:ZubHill} -- conditions which are, in general, sufficient but not necessary for entanglement -- for the detection of entanglement in $N$-qubit systems. They then proceed to study entanglement in states of total angular momentum and apply the developed criteria to the Tavis-Cummings model\index{Tavis-Cummings model}. They find that for any $N$, when the atoms start in the ground state with only one photon present in the cavity mode, the resulting state is always entangled, except for a discrete set of times~\cite{ChengGB2022}.

\begin{figure}
\includegraphics[width=10cm]{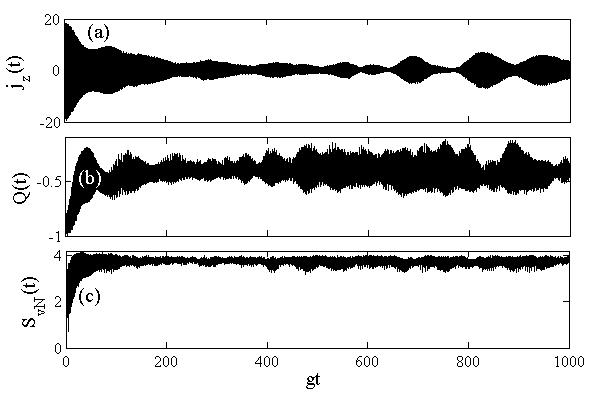}  
\caption{Time evolution for an initial Fock state $|150\rangle$ and a fully polarized spin state $|s,m\rangle$ with $s=-m=40$ for the Dicke model~(\ref{dickeham2}). The coupling $g=0.01$ (note that the coupling term of the Hamiltonian is scaled by $1/\sqrt{N}$) which suggests validity of the RWA even for such a large initial Fock state. Nevertheless, we find especially for the `magnetization' $j_z(t)$ [plotted in {\bf (a)}] that a set of collapse-revivals emerges\index{Collapse-revival}, in stark contrast to the case of the JC model which exhibits perfect Rabi oscillations [see eq.~(\ref{tsol3})]. The Mandel $Q$-parameter\index{$Q$-parameter} plotted in {\bf (b)} displays that the field stays sub-Poissonian\index{Sub-Poissonian} throughout. The approximate plateau $\approx 4$ of the von Neumann entropy\index{von Neumann! entropy} in {\bf (c)} should be compared to the value $\approx4.3$ for a maximally entangled atom-field state. Hence, the two subsystems are almost perfectly correlated. The dimensionless frequencies are $\omega=\Omega=1$.} 
\label{figdickeevolve2}  
\end{figure}

In the theory of multi-partite entanglement it is a known fact that spin squeezing\index{Spin! squeezing} can be an indicator (witness) of entanglement\index{Entanglement! witness}~\cite{sorensen2001entanglement}. This idea has also been applied to discuss entanglement in the Dicke model~\cite{song2009spin}. Early on, the squeezing of the field quadratures (\ref{quadvar}) for the Tavis-Cummings model\index{Tavis-Cummings model} was analyzed and it was concluded that a higher excitation of the initial atomic state tends to produce a higher amount of squeezing~\cite{butler1986squeezed}. These results were further elaborated on in refs.~\cite{li1990squeezing,retamal1997squeezing} by looking at higher order squeezing and deriving analytical approximations for the quadrature squeezing. As for the JC and quantum Rabi models, one may expect the greatest squeezing in the collapse regime between revivals. In fig.~\ref{figdickeevolve} (a)--(c) we show the evolution of a few quantities for an initial coherent field state. It is particularly seen that during the revival there is a drop in the atom-field entanglement which suggests that also squeezing decreases at these times. Unlike the JC model, where an initial de-excited atom interacting with a single Fock field state produces trivial Rabi oscillations, in the Tavis-Cummings model\index{Tavis-Cummings model} one obtains a complex evolution; typical quantities, like for example the expected photon number $\langle\hat n\rangle$, show a series of fractional and full revivals~\cite{chumakov1996dicke}. This is demonstrated in fig.~\ref{figdickeevolve2} where the spin is initially in its bare ground state and the field in a highly excited Fock state ($\bar n=150$). Quantitative conditions for the validity of the Tavis-Cummings model\index{Tavis-Cummings model}, taking into consideration the cascaded interaction\index{Cascaded! interaction} of the photons with all consecutive emitters, have been recently derived in~\cite{Blaha2022}.

Just like for the phenomenon of superradiance, the collapse-revival structure depends on the inherent atomic correlations for the initial state~\cite{ramon1998collective}. In particular, for initial Dicke states $|s,m\rangle$, the same type of derivation for the revival time, assuming $\bar n\gg1$, as briefly outlined in sec.~\ref{sssec:crsubsec} results in 
\begin{equation}
t_r=\frac{2\pi}{g}\sqrt{\bar{n}+m}.
\end{equation} 
In fig.~\ref{figdickeevolve}, we present numerical results for the evolution of some quantities of the Dicke model. In the first plots the coupling $g$ is weak such that RWA is applicable and as a result we encounter clear collapse-revival structures, just like in fig.~\ref{fig3} showing the same for the JC model. 

\begin{figure}
\includegraphics[width=10cm]{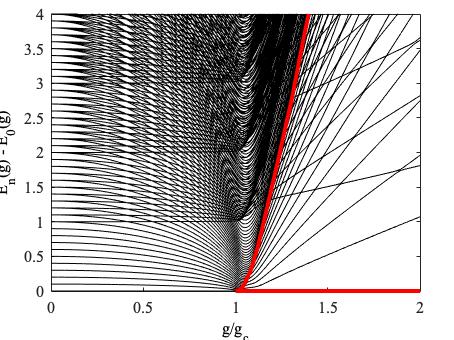}  
\caption{Lower part of the (shifted) spectrum of the Dicke model for $\Omega=1$ and $\omega=0.1$. Here, $g_c=\sqrt{\Omega\omega}/2$ marks the onset of a double-well structure of the lowest BOA potential\index{Adiabatic! potential}, and hence it also gives the critical coupling for the Dicke PT\index{Dicke! phase transition}\index{Phase transition! Dicke}. As explained in the main text, below $g_c$ the adiabatic theory predicts that the spectrum derives from a set of decoupled oscillators, shifted from each other in energy by $m\Omega$ and with a frequency $\approx\omega$. This is indeed what is found as seen for small $g$'s. Above $g_c$ we instead expect the spectrum to be formed from decoupled double-wells. Again this is what is seen, for example in the appearance of approximate double degenerate eigenvalues (even and odd parities); each pair of energies merge after a certain $g$. The thick red line give this for the lowest double-well (the same happens for the higher double-wells but we have not indicated it with further red lines). Focusing on one single double-well it is seen how the number of quasi-degenerate states with energies below the potential barrier increases for larger $g$. This is natural since the barrier (\ref{barhight}) grows when the coupling $g$ is increased. While the structure of the spectrum shown in the figure can be understood from the BOA, it is an approximation and the non-adiabatic corrections is not strictly zero. The non-zero non-adiabatic couplings\index{Non-adiabatic! coupling} manifest in the avoided crossings between the curves. Visible in the figure is also that the density of states $\nu(E)$, {\it i.e.}, the number of states given the energy $E$, changes along the red line. Discontinuities in $\nu(E)$ are related to excited state PTs\index{Excited state phase transition}\index{Phase transition! excited-state} and was studied in ref.~\cite{brandes2013excited} for the Dicke model. Naturally, the results of this plot are obtained for a finite system (N=50) and one must think of the density of states as coarse grained. In the thermodynamic limit the corresponding thick red curve would give a critical coupling for such excited state PT. } 
\label{figdickespec}  
\end{figure}

Contrary to the JC and quantum Rabi models, both the Dicke and the TC models support a direct classical limit by letting the field amplitude and the spin $s$ grow to infinity~\cite{lieb1973classical,zhang1990coherent,perelomov1977generalized}. However, in order for this thermodynamic limit\index{Thermodynamic limit} to render interesting non-trivial results the coupling should be rescaled as $g\rightarrow g/\sqrt{N}$. This scaling can be understood by following the BOA approach applied to the quantum Rabi model in sec.~\ref{ssec:rabi}. If we scale the coupling  $g\rightarrow g/\sqrt{N}$, the adiabatic potential functions\index{Adiabatic! potential! curve} for the Dicke model are 
\begin{equation}
V_\mathrm{ad}^{(m)}(x)=\omega\frac{x^2}{2}+m\sqrt{\frac{\Omega^2}{4}+2\frac{g^2}{N}x^2},
\end{equation}
with $-N/2\leq m\leq N/2$~\cite{graham1984two}, and for $m=-N/2$ (lowest energy potential) we find that the quadrature $x$ minimising the potential $x\sim\sqrt{N}$ (provided that $g>\sqrt{\Omega\omega}/2$). Thus, after this rescaling every term of the Hamiltonian scales as $\mathcal{O}(N)$, and the rescaled Dicke model has 
\begin{equation}\label{dickeham2}
\hat{H}_D=\omega\hat{n}+\frac{\Omega}{2}\hat{S}_z+\frac{g}{\sqrt{N}}\left(\hat{a}^\dagger+\hat{a}\right)\hat{S}_x.
\end{equation} 
This makes also physical sense; normally the thermodynamic limit\index{Thermodynamic limit} consists in letting the system size grow to infinity while keeping a density fixed. In the Dicke model the density would be the number of atoms $N$ divided by the effective mode volume $V$, but the light-matter coupling $g$ scales with $1/\sqrt{V}$~\cite{raimond2006exploring} and the above scaling enters naturally. Recalling sec.~\ref{ssec:rabi} on the quantum Rabi model and especially fig.~\ref{fig9} showing the adiabatic potential curves we find that the corresponding BOA potentials ($m<0$) for the Dicke model also changes character for $g_c=\sqrt{\Omega\omega}/2$. For $g\gg g_c$ we find that within the BOA the lower part of the spectrum should be characterized by a set of decoupled double-well potentials, where the lowest double-well has a barrier height
\begin{equation}\label{barhight}
\delta_\mathrm{bar}=N\left(\frac{g^2}{4\omega}+\frac{\omega\Omega^2}{16g^2}-\frac{\Omega}{4}\right).
\end{equation}
On the other hand, for $g<g_c$ we expect instead that a set of decoupled single potential wells shifted in energy by $m\Omega$ to determine the energies. We demonstrate this in fig.~\ref{figdickespec}, which gives the lowest energies $E_n(g)=\varepsilon_n(g)-\varepsilon_0(g)$ of the Dicke model. This spectrum structure has been pointed out using the BOA in ref.~\cite{relano2017approximated}. 

Opposite to the quantum Rabi model, the transition from a single potential minimum to two minima is critical in the thermodynamic limit $N\rightarrow\infty$ independent of the fraction $\omega/\Omega$. Remember that the quantum Rabi model is critical in the classical limit\index{Rabi! classical limit} $\omega/\Omega\rightarrow0$~\cite{bakemeier2012quantum,hwang2015quantum,larson2017some,puebla2020universal,garbe2020phase}. Considering the barrier separation of eq.~(\ref{barhight}) it follows that the height diverges as $N\rightarrow\infty$ such that the tunneling rate between the two wells vanishes. Similarly, in the classical limit the distance between the two wells diverges which again suppresses the tunneling. One can think of the classical limit $\omega/\Omega\rightarrow0$ as sort of thermodynamic limit\index{Thermodynamic limit} just like $N\rightarrow\infty$ for the Dicke model. In the former the harmonic oscillator is considered classical with a continuous spectrum, while in the latter the atomic degree of freedom is the classical variable. In either case, in the symmetry broken phase a macroscopic number of bare states will get populated in the ground state. The corresponding symmetry breaking, whether it is the quantum Rabi or the Dicke model, is thereby represented by populating one of the two wells. This second order PT was first pointed out by Hepp, Lieb, Wang and Hioe back in 1973~\cite{hepp1973superradiant,wang1973phase} (interestingly, the corresponding instability of the ground state was predicted some years earlier by Mallory~\cite{mallory1969solution}). Naturally, they did not consider the BOA approach presented above, but used other (but highly related) mean-field methods. In particular, Wang and Hioe calculated the partition function for the TC model using the transfer matrix method and a coherent-state {\it ansatz} for the field which they argued should be accurate in the thermodynamic limit\index{Thermodynamic limit}. This assumption was later proven correct when it was demonstrated that the transition survives also at zero temperature $T=0$~\cite{hillery1985semiclassical}, and more recently it was revisited and proven with the help of {\it Wick's theorem}~\cite{gammelmark2011phase}. As a finite $T$ transition it is a classical PT. The corresponding derivation for the Dicke model which includes the counter rotating terms was carried out the same year~\cite{hioe1973phase,carmichael1973higher}. The critical coupling, for $T=0$, differ by a factor two between the Dicke and the TC model, \begin{equation}\label{dickecrit}
g_c=\left\{
\begin{array}{ll}
\displaystyle\frac{\sqrt{\Omega\omega}}{2},\hspace{0.5cm}\mathrm{Dicke},\\ \\
\sqrt{\Omega\omega},\hspace{0.5cm}\mathrm{TC}.
\end{array}\right.
\end{equation}
It is interesting to note that the RWA totally breaks down for such large couplings - we are in the deep strong coupling regime (see sec.~\ref{ssec:rabi}) - and it may seem natural that the Dicke model and not the TC models should have been explored first. The two phases separated by the critical coupling $g_c$ have become known as the {\it normal} and {\it superradiant phase}\index{Superradiant! phase}, and consequently the Dicke PT is often referred to as the {\it normal-superradiant PT}\index{Normal-superradiant phase transition}\index{Phase transition! normal-superradiant}. At the mean-field level, like the BOA approach presented above, we may directly specify the ground states in the two phases for a finite $N\gg1$. In the normal phase the adiabatic potential has a global minimum at $x=0$ implying that the field is in vacuum and all atoms in their ground states, {\it i.e.} $|\psi_0\rangle=|s,-s\rangle|0\rangle$. In the superradiant phase the effective potential $V_\mathrm{ad}^{(m=-s)}(x)$ has built up a double-well structure and the ground state is the even solution which is here a Schr\"odingier cat\index{Cat state}\index{Schr\"odinger cat! states}
\begin{equation}\label{dickecat}
|\Psi_\mathrm{SR}\rangle=\frac{1}{\sqrt{\mathcal{N}}}\left(|\alpha\rangle|\chi_+\rangle+|-\alpha\rangle|\chi_-\rangle\right),
\end{equation}
where 
\begin{equation}\label{sramp}
\alpha=\sqrt{N}\sqrt{\frac{g^2}{2\omega^2}\left(1-\frac{g_c^4}{g^4}\right)}
\end{equation}
is the coherent state amplitude scaling as $\sqrt{N}$, $|\chi_\pm\rangle$ is the spin state with corresponding `Bloch vector' $R_\pm\propto\left(\pm\frac{g^2}{\omega}\sqrt{1-\frac{g_c^4}{g^4}},0,-\frac{\Omega}{2}\right)$ (see fig.~\ref{fig9}) on the spin phase space, and $\mathcal{N}$ is a normalization constant which approaches $2$ in the thermodynamic limit\index{Thermodynamic limit}. We note that $\alpha$ is real meaning that the quadrature $p$ vanishes and at the critical point\index{Critical! point} $g=g_c$ the field is in vacuum and all atoms are in their lower state pointing towards the south pole of the sphere forming the spin phase space. Naturally, in the thermodynamic limit $N\rightarrow\infty$ a proper second order PT emerges and then in the superradiant symmetry broken phase one of the two states in the linear combination (\ref{dickecat}) survives. Thus, the cat state\index{Cat state} becomes exceedingly fragile to any fluctuations. By a proper PT we mean that the derivative $d^2E_0(g)/dg^2$, where $E_0$ is the ground state energy, becomes discontinuous at the critical point\index{Critical! point} $g=g_c$ and moreover that the properties in the vicinity of $g_c$ can be uniformly described with a set of critical exponents\index{Critical! exponents}. However, as discussed above, letting the size of the spin go to infinity, {\it i.e.} $s\rightarrow\infty$, is not the only possibility to achieve such singularity. Another limit we already mentioned in sec.~\ref{ssec:rabi} is the `classical limit' $\Omega/\omega\rightarrow\infty$ while keeping $N$ and $\omega\Omega$ fixed~\cite{hwang2015quantum}. Tricriticality and the occurrence of exotic phases have been reported for a Dicke triangle where three cavities, each containing an ensemble of three-level atoms\index{Three-level atom}, are connected to each other by means of an artificial magnetic field. Such a field can be achieved through temporal modulation of the photon-hopping strength on each cavity~\cite{Cheng2022}. 

\begin{figure}
\includegraphics[width=10cm]{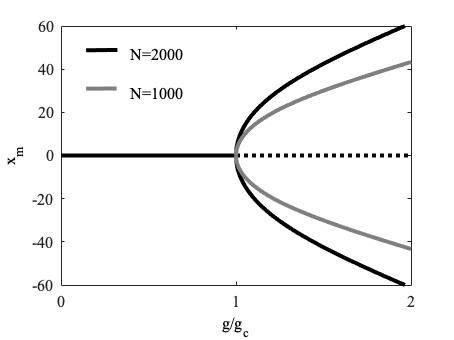} 
\caption{The Dicke pitchfork bifurcation resulting from the classical Dicke Hamiltonian (\ref{clasdicke}). Solid lines give the stable fix points and the dashed one the unstable one. The crossover from stable to unstable of the $x_m=0$ fix point mark the onset of the double-well structure of the adiabatic potential $V_\mathrm{ad}^{(-s)}(x)$; two new global minima emerges while the previous minimum turns into a local maximum. It is also seen that the transition becomes more dramatic as the atom number $N$ is increased.
}
\label{bifurcationDicke} 
\end{figure}

\begin{figure}
\includegraphics[width=10cm]{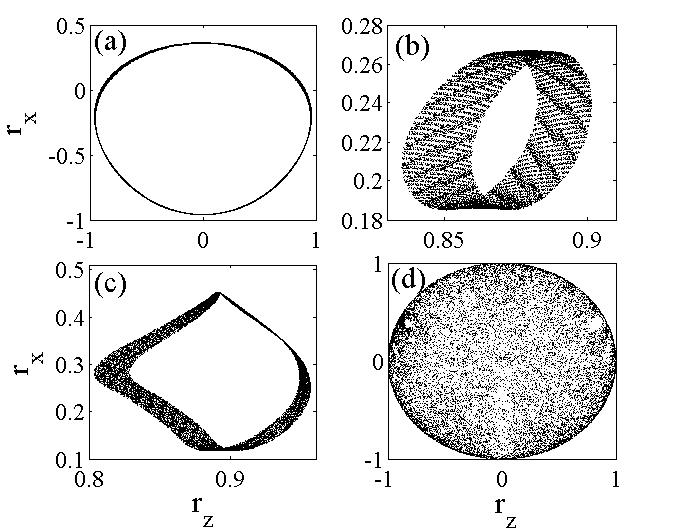} 
\caption{Poincar\'e sections for $g=0.2,\,0.48,\,0.5,\,0.52$ in {\bf (a)}--{\bf (d)}, respectively, on the plane $\psi=0$. The variables depicted are the scaled (with total spin) `Bloch vector' components $r_x$ and $r_z$. The transition from regular (a) to fully chaotic (d) evolution is clearly visible. The frequencies $\omega=\Omega=1$ and the initial condition in all for plots is $(s(t),\psi(0),\phi(0),\theta(0))=(10,0,0,\pi/2)$. Thus, the sections are plotted for a single trajectory. It is often common to consider several randomly picked initial states keeping some conserved quantity fixed (normally the energy for Hamiltonian systems), but the plots presented here still clearly serve their purpose by demonstrating the onset of chaos.
}
\label{figpoin} 
\end{figure}

Before analyzing quantum corrections to the mean-field results presented in the previous section we discuss dynamical aspects of the classical version of the Dicke model. The classical analogue of the Dicke model can be obtained by writing down the functional $H(z,\alpha)=\langle z,\alpha|\hat{H}_D|\alpha,z\rangle$ with $|z\rangle$ a spin coherent state\index{Spin! coherent state}\index{State! spin coherent}~\cite{radcliffe1971some} and $|\alpha\rangle$ a regular boson coherent state~(\ref{cohe})\index{Coherent state}\index{State! coherent}. Parametrizing the amplitudes as $z=e^{i\phi}\tan(\theta/2)$ and $\alpha=\sqrt{Ns}e^{i\varphi}$ with the canonical variables obeying the Poisson brackets\index{Poisson! bracket} $\left\{s,\varphi\right\}=\left\{\cos\theta,\phi\right\}=1$, the classical Hamiltonian becomes~\cite{altland2012quantum}
\begin{equation}\label{clasdicke}
H_{cD}=\omega s+\frac{\Omega}{2}\cos\theta+2g\sqrt{s}\sin\theta\cos\phi\cos\varphi.
\end{equation}
It should be noted, as already pointed out in sec.~\ref{sssec:rabiint}, that the classical Hamiltonian (\ref{clasdicke}) is identical to the one of the quantum Rabi model. Hamilton's equations become
\begin{equation}\label{dickeeom}
\begin{array}{l}
\dot s=-2g\sqrt{s}\sin\varphi\sin\theta\cos\phi,\\ \\
\displaystyle{\dot\varphi=-\omega-\frac{g}{2\sqrt{s}}\cos\varphi\sin\theta\cos\phi,}\\ \\
\displaystyle{\dot\phi=\frac{\Omega}{2}-2g\sqrt{s}\cos\varphi\cot\theta\cos\phi,}\\ \\
\dot\theta=2g\sqrt{s}\cos\varphi\sin\phi.
\end{array}
\end{equation}
The Dicke PT\index{Dicke! phase transition}\index{Phase transition! Dicke} in the classical model is manifested as a {\it pitchfork bifurcation}\index{Pitchfork bifurcation}, {\it i.e.} for $g=g_c$ the classical steady state\index{Steady state!classical} solution representing the normal phase becomes unstable and two new solutions (superradiant phase) emerge~\cite{milonni1983chaos,muller1991classical,larson2011photonic,bhaseen2012dynamics}, see fig.~\ref{bifurcationDicke} showing the mean-field quadrature $x$ as a function of $g$. This bifurcation is readily understood from the mean-field BOA Hamiltonian 
\begin{equation}\label{mfdicke}
H_{MF}(x,p)=\omega\left(\frac{p^2}{2}+\frac{x^2}{2}\right)+V_\mathrm{ad}^{(-N/2)}(x),
\end{equation}
where the adiabatic potential\index{Adiabatic! potential}
\begin{equation}\label{dickeadpot}
V_\mathrm{ad}^{(m)}(x)=m\sqrt{\frac{\Omega^2}{4}+2\frac{g^2}{N}x^2},\hspace{1.2cm}-\frac{N}{2}\leq m\leq\frac{N}{2},
\end{equation}
(here not defined with the harmonic $x^2$ term) and the steady state\index{Steady state} represents the minima $(x_m,p_m)$ of $H_{MF}(x,p)$; $(x_m,p_m)=(0,0)$ for $g<g_c$ and $(x_m,p_m)=\left(\sqrt{(N/2)g^2\left(1-g_c^4/g^4\right)}/\omega,0\right)$ for $g>g_c$. The magnetization (steady state for the spin $S_z$ component) becomes
\begin{equation}
S_{zm}=\left\{
\begin{array}{lll}
-N/2,& & g<g_c,\\
-N\left[2(1+g^4/g_c^4)\right]^{-1/2}, & & g>g_c
\end{array}\right.
\end{equation}
A notable property of the classical Dicke model is that it displays chaos. For weak couplings $g$ the dynamics is almost perfectly regular, while for increasing $g$ it becomes more and more chaotic to show full blown chaos deep in the superradiant phase~\cite{milonni1983chaos,muller1991classical,furuya1998quantum,emary2003quantum,emary2003chaos,larson2011photonic}. One signature of classical chaos\index{Chaos!classical} is the structure of the {\it Poincar\'e sections}\index{Poincar\'e! section}~\cite{haake2010quantum}. The Poincar\'e section shows the classical solution in one hyperplane in phase space. Four such examples for the classical Dicke model (\ref{clasdicke}) are displayed in fig.~\ref{figpoin}, above and below the bifurcation at $g_c=\sqrt{\omega\Omega}/2$. The scaled `Bloch vector' components $r_x=\langle\hat S_x\rangle/(N/2)$ and $r_z=\langle\hat S_z\rangle/(N/2)$ are shown in the plane $\psi=0$. In (a), far below $g_c$, the dynamics is regular, still below but close to the bifurcation point (b) the solutions still evolve on a hyper torus, for $g=g_c$ (c) the torus has began to significantly deform in accordance with the {\it KAM-theorem}\index{KAM-theorem}~\cite{tabor1989chaos}, and finally in (d) for $g>g_c$ the Poincar\'e section shows full chaos. In a recent work~\cite{Juarez2020}, the semiclassical equations of motion~(\ref{dickeeom}) were simulated experimentally in a classical system comprising two coupled $LC$ circuits. The authors were able to explore a whole range of properties of these equations; transition from regular to chaotic motion, the Dicke PT, the {\it Lyapunov exponent}\index{Lyapunov exponent} $\lambda_\mathrm{L}$, and the so-called {\it out-of-time-order-correlator}\index{Out-of-time-order-correlator (OTOC)} (OTOC)~\cite{maldacena2016remarks,fan2017out,swingle2018unscrambling} (see further discussions below on the OTOC and chaos in the Dicke model). Both the Dicke and the Jaynes-Cummings-Hubbard Hamiltonians can be construed as idealised models of a quantum battery\index{Quantum! battery}. A numerical investigation of the charging properties of both these models was carried out in~\cite{Hogan2022}. The interaction of a three-level atom\index{Three-level atom} with the electromagnetic field of a cavity in the presence of a laser field has been recently discussed in terms of two quantum quantum-battery\index{Quantum! battery} setups. In the first, a single three-level atom interacts sequentially with many cavities, each in a thermal state, while in the second a stream of atoms\index{Three-level atom} in a thermal state sequentially interacts with a single cavity mode initially prepared in a thermal state at the same temperature as the atoms~\cite{beleno2023laser}.

The counter-rotating terms, present in the Dicke model, break the conservation of excitations in the TC model and thereby lower the symmetry from a continuous $U(1)$ to a discrete $\mathbb{Z}_2$. Thus, by artificially adding such terms to the TC model and by controlling their strengths one can explore the transition from integrable to non-integrable, which was done in ref.~\cite{fox1987systematic} where classical chaos was studied.

\begin{figure}
\includegraphics[width=10cm]{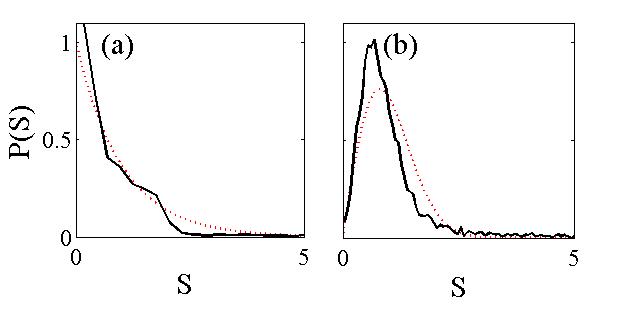} 
\caption{Two examples of the distribution $P(S)$ for the Dicke model (black solid lines) in the normal phase {\bf (a)} and the superradiant phase {\bf (b)}. For the sake of comparison, the Poisson and Wigner-Dyson ($\alpha=1$) distributions\index{Poisson! distribution} are shown as red dotted curves. The level repulsion in the superradiant phase is clearly seen. The parameters are $\omega=\Omega=1$, $N=100$ atoms, while $g=0.2$ in (a) and $g=2$ in (b).}  
\label{figleveldist}  
\end{figure}

Naturally, the Dicke model (\ref{dickeham2}) displays also signatures of `quantum chaos'~\cite{graham1984quantum,emary2003quantum,emary2003chaos,perez2011excited}. Quantum chaos\index{Quantum! chaos}\index{Quantum! chaos} is often analyzed in terms of the {\it level statistics}\index{Level! statistics}~\cite{haake2010quantum}. The information about level statistics is contained in the probability distribution $P(S)$ where $S$ is the energy difference between neighbouring eiegenenergies; $S_n=\varepsilon_{n+1}-\varepsilon_n$. Characteristic for chaotic systems is {\it level repulsion}\index{Level! repulsion}, the few symmetries in the Hamiltonian implies that energy levels do not in general cross but are avoided. This is also manifested in the distribution $P(S)$ which for chaotic systems takes the generic {\it Wigner-Dyson}\index{Wigner!-Dyson distribution}\index{Distribution! Wigner-Dyson} form~\cite{haake2010quantum}
\begin{equation}\label{WD}
P_\mathrm{WD}(S)\propto S^\alpha e^{-\pi S^2/4},\hspace{1cm}\left\{
\begin{array}{lll}
\alpha=1, & \hspace{0.2cm} & {\rm orthogonal}\\
\alpha=2, & \hspace{0.2cm} & {\rm unitary}\\
\alpha=4, & \hspace{0.2cm} & {\rm symplectic}
\end{array}\right..
\end{equation}
The three universality classes\index{Universality! class} , {\it orthogonal}\index{Orthogonal symmetry class}, {\it unitary}\index{Unitary symmetry class}, and {\it symplectic}\index{Symplectic symmetry class}, are determined from the symmetries of the Hamiltonian. In particular, the larger the exponent $\alpha$ we note that the level repulsion is more pronounced. For the Dicke model we expect the level statistics $P_\mathrm{WD}(S)=\pi S\exp\left(-\pi S^2/4\right)/2$, {\it i.e.}, belonging to the orthogonal class. For integrable models, on the other hand, the generic level structure obeys instead a Poissonian distribution\index{Distribution! Poisson}\index{Poissonian! level statistics}, 
\begin{equation}\label{pos}
P_\mathrm{P}(S)=e^{-S},
\end{equation} 
which for the Dicke model should apply deep in the normal phase. Properties of the eigenvalues for the quantum Rabi model ({\it i.e.} the special case of the Dicke model with $s=1/2$) were first discussed by Graham and H{\"o}hnerbach in~\cite{graham1984quantum}. As was already mentioned in sec.~\ref{sssec:rabiint}, the quantum Rabi model does not show neither Poissonian nor Wigner-Dyson statistics~\cite{larson2013integrability}. The larger the spin $s$, the better is the agreement with the generic forms (\ref{WD}) and (\ref{pos}) which has been numerically demonstrated in the works by Emary and Brandes~\cite{emary2003quantum,emary2003chaos}. As examples of the level statistics for the Dicke model we display examples of $P(S)$ in fig.~\ref{figleveldist} in the normal as well as in the superradiant phases. The agreement with the generic distributions $P_\mathrm{P}(S)$ and $P_\mathrm{WD}(S)$ is not perfect, but the indication of level repulsion is clearly illustrated. One reason why the agreement is not better is because the spectrum has not been {\it unfolded}. Unfolding the spectrum means that secular variations are omitted by subtracting the smooth part of the spectrum~\cite{haake2010quantum}. Instead, in the figure we only consider energies away from the edges of the spectrum since the low and high energy states are known to be non-ergodic~\cite{santos2010onset}.  
   
\begin{figure}
 \includegraphics[width=0.495\textwidth]{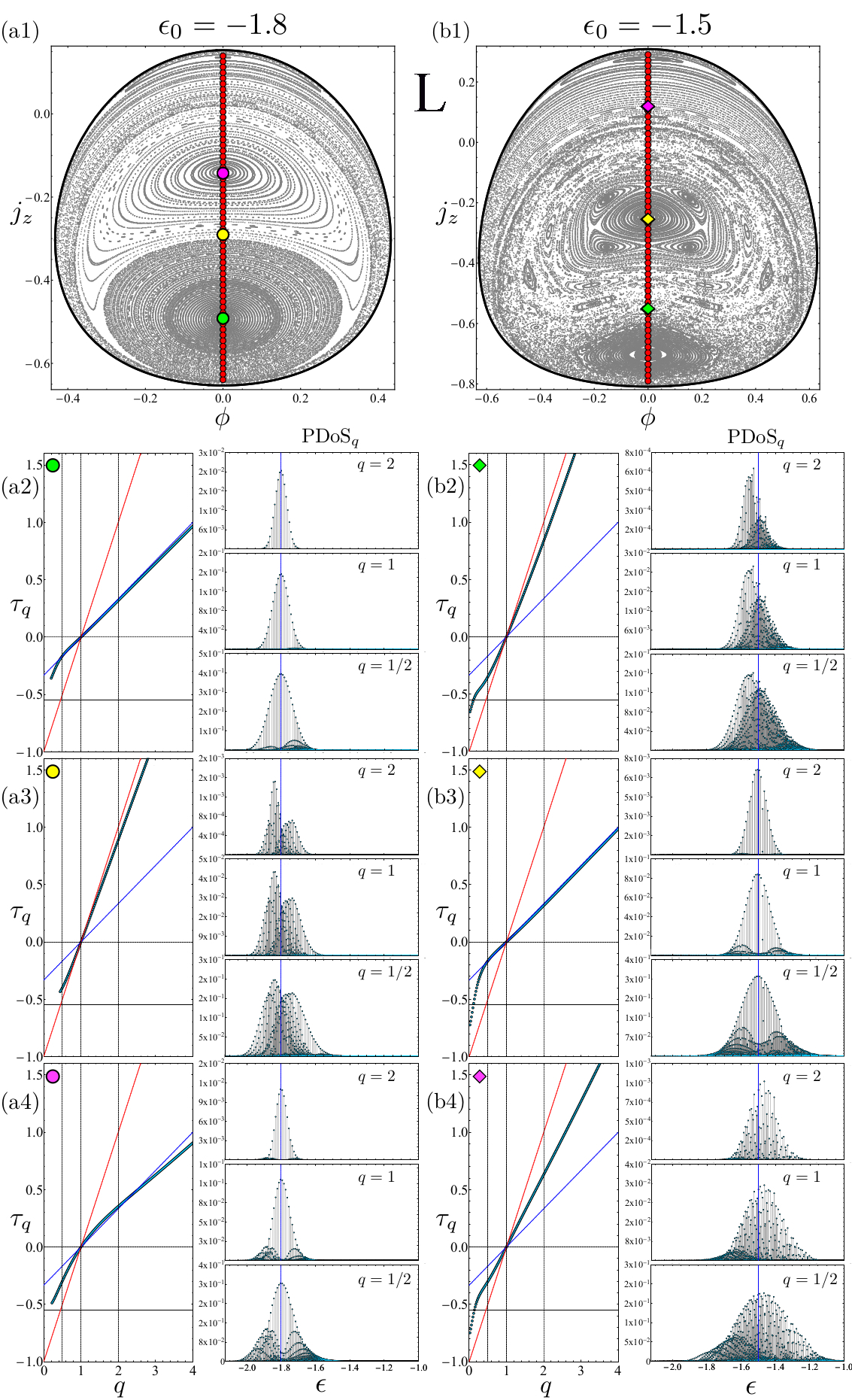}
  \includegraphics[width=0.495\textwidth]{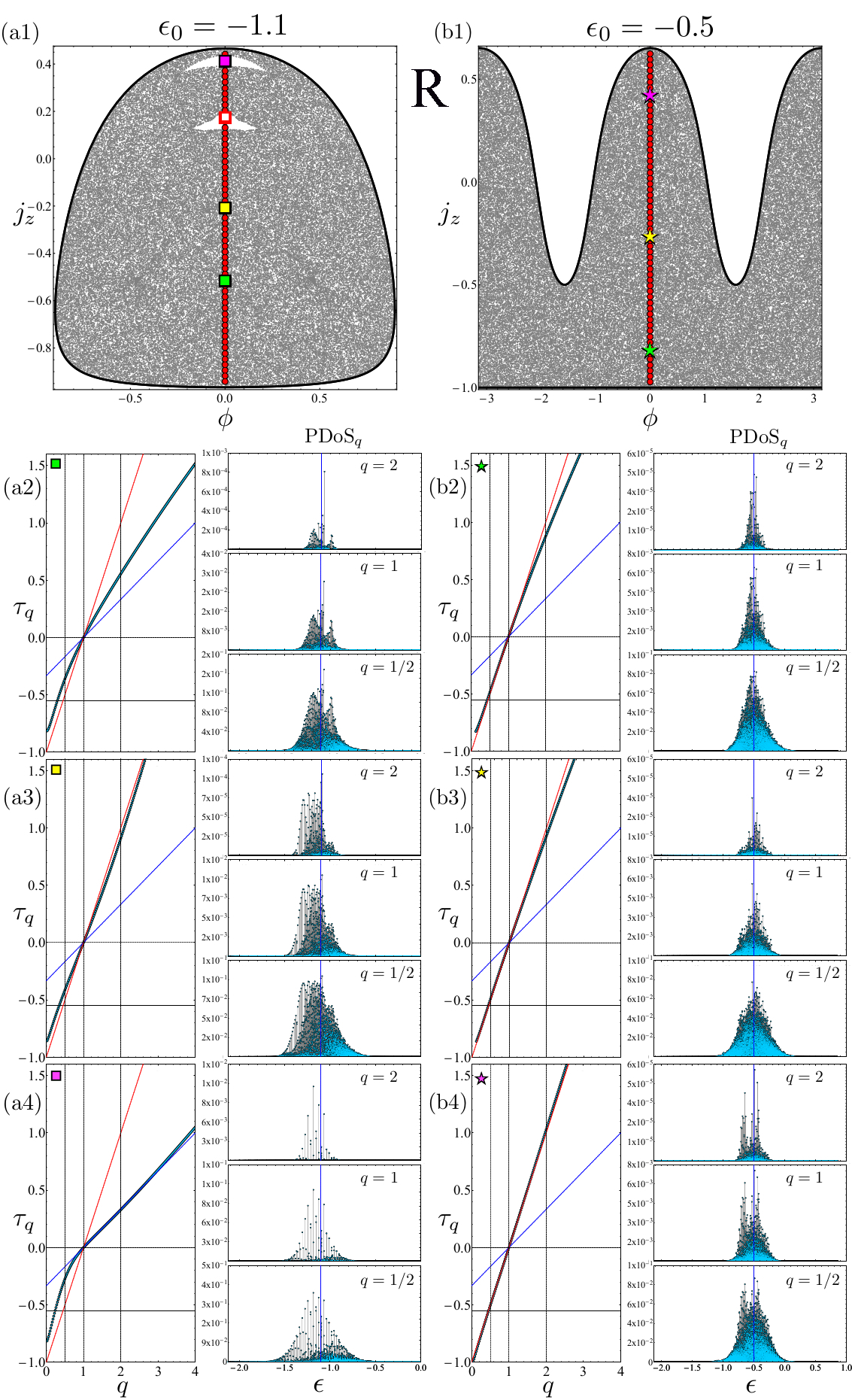}
  \caption{{\it Multifractality under the quantum-classical correspondence in the totally symmetric subspace.} ({\bf L}eft panel) Top row: Poincar\'{e} surface of sections\index{Poincar\'e! section} with regular structures in phase space. ({\bf R}ight panel) Top row: Poincar\'{e} surface of sections with predominant chaos. The rest of the rows in both panels depict the mass exponents $\tau_q$ and the distributions PDo$S_q(\epsilon)$ as a function of $q$. Numerical results are plotted by dotted cyan lines, while the ergodic (regular) limits of $D_q$ are indicated by red (blue). Source: Figs. 1 and 2 of~\cite{bastarracheamagnani2023quantum}.}
  \label{fig:Dickechaos}
\end{figure}

\begin{figure}
\includegraphics[width=10cm]{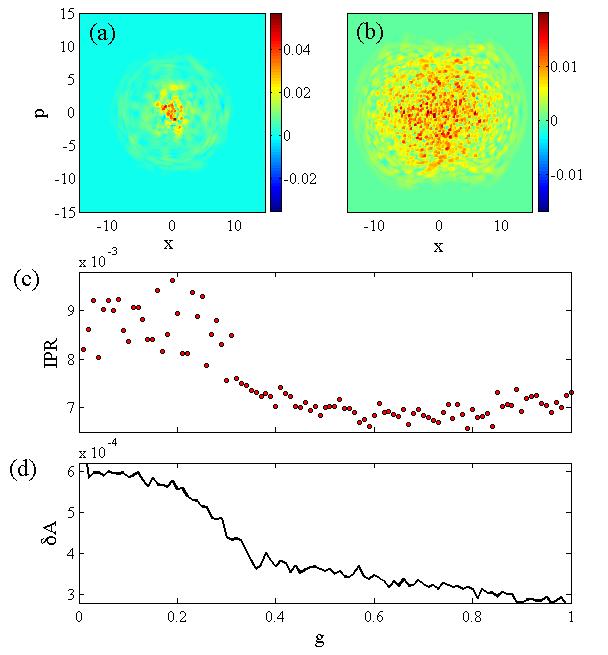} 
\caption{The upper two plots, in frames {\bf (a)} and {\bf (b)}, display the Wigner functions\index{Wigner! function} after a time $t_f=1000$ for an initial state $|\alpha\rangle|s,-s\rangle$ with $\alpha=\sqrt{39}$ and $s=40$. In (a) the atom-field coupling $g=0.1$ (normal phase) while in (b) $g=1$ (superradiant phase). By comparing the two plots it is seen that the one with stronger coupling shows a more vivid structure with finer patterns typical for chaotic evolution. It has also a more evenly spread out distribution which reflects ergodicity\index{Ergodicity}, another outcome of chaos. In {\bf (c)}, the inverse partition ration (\ref{ipr}) is presented as a function of the $g$, and the remaining parameters are the same as for the other plots. Even though the IPR fluctuates much in the normal phase, it is clear that it is higher in this phase than in the superradiant one. These fluctuations are partly due to numerics since the $xp$-grid has been discretized for the numerical simulations. The drop in the IPR occurs before the critical coupling, for $\omega=\Omega=1$ used here, $g=1/2$. However, one must keep in mind that these results are for evolution of some excited state, and nothing prevents chaos to emerge for smaller $g$'s as long as the dynamics takes place far from the fix point $(x_m,p_m)=(0,0)$. Indeed, classical chaos in the Dicke model has been shown to appear also for $g>g_c$~\cite{altland2012quantum}. The same kind of transition is found also in the variance of the autocorrelation function (\ref{auto}) plotted in {\bf (d)}. By increasing the system size (larger $\alpha$ and $s$) the variance decreases (checked numerically) which is expected since the state should thermalize in this limit. } 
\label{figdickethermal}  
\end{figure}  

The statistics contained in the function $P(S)$ does not fully characterize the properties of the spectrum. Another quantity is the density of states $\nu(\varepsilon)$, {\it i.e.} $\nu(\varepsilon)d\varepsilon$ is the number of states within the energy interval $[\varepsilon,\varepsilon+d\varepsilon]$, see fig.~\ref{figdickespec}. This quantity has been thoroughly explored for the Dicke model in ref.~\cite{brandes2013excited}. Within the thermodynamic limit\index{Thermodynamic limit}, $\nu(\varepsilon)$ can be found exactly and it shows a singularity which is identified as a PT for excited states -- a so-called {\it excited-state phase transition}\index{Phase transition! excited-state}\index{Excited state phase transition}~\cite{Cejnar2021, Swan2021}. A related aspect was considered in~\cite{bastidas2012nonequilibrium}, where the periodically driven Dicke model was demonstrated to show phase transition-like behaviour in its effective steady states\index{Steady state!effective} (time-averaged).  

Models like the Tavis-Cummings\index{Tavis-Cummings model} and the Lipkin-Meshkov-Glick belong to the class of the so-called integrable models of Gaudin type~\cite{Skrypnyk_2022}\index{Model! Gaudin type}. Integrability\index{Integrable model} in these models relies on the existence of a matrix, defined on the tensor product of three Hilbert spaces, which satisfies the Young-Baxter equation\index{Young-Baxter equation}~\cite{Yang1967, Baxter1978}. A new representation of the Gaudin algebra in the form of a Laurent-series expansion\index{Laurent series} in terms of the complex parameter on which the Lax-matrix coefficients depend\index{Lax matrix}. A physical generalized Hamiltonian can be constructed from the commuting operator-valued coefficients of the above series~\cite{kurlov2023generalization}.

For excitation energies above a certain threshold in the superradiant phase of the Dicke model, both the semiclassical dynamics are chaotic and the quantum eigenspectrum exhibits level repulsion. By employing the eigenstate expectation values and the distributions of the off-diagonal elements of the number of photons and the number of excited atoms one may verify the diagonal and off-diagonal eigenstate thermalization hypothesis\index{Eigenstate! thermalization hypothesis}, thus ensuring thermalization. The authors of~\cite{Villasenor2023} propose a so-called ``efficient basis''~\index{Dicke! model! efficient basis} with a clear advantage over the widespread employed Fock basis when investigating the unbounded spectrum of the Dicke model, yielding access to a larger number of converged states.

The notion of {\it multifractality}\index{Multifractality} was introduced in 1974 to describe the statistical properties of turbulent flow~\cite{mandelbrot_1974}. It signifies the presence of an infinite set of critical exponents determining the scaling of the moments generated by a given distribution. In quantum mechanics, multifractality is intertwined with the concepts of localization, ergodicity and chaos, and has been recently employed as a local measure of chaos for the kicked top model\index{Model! kicked top}~\cite{WangQian2021}. Its presence signifies that the wave function is extended but effectively restricted to a portion of the Hilbert space, owing to the different scaling of each component. The partition function of the system scales as~\cite{bastarracheamagnani2023quantum}.
\begin{equation}
 Z(q)=\sum_{k=1}^{N}p_k^q \sim N^{-\tau_q},
\end{equation}
with $Z(1)=1$, whence $\tau_{q=0}=0$. The measure is multifractal when the {\it mass exponents} $\tau_q$\index{Mass exponent} are nonlinear functions of $q$, introducing the R\'{e}nyi dimension~\cite{Hentschel1983, Hasley1986}\index{R\'enyi! dimension} $D_q$ such that $\tau_q=D_q(q-1)$. Multifractality is also assessed by means of the so-called general participation local density of states PDo$S_q(\epsilon) \equiv \sum_{k}|\langle E_k|\boldsymbol{z}_0 \rangle|^{2q} \delta(\epsilon-\epsilon_k)$, where $|E_k\rangle$ are the eigenstates of the Dicke Hamiltonian, $\epsilon_k=2 E_k/N$ are the scaled energies (with $N$ being the total number of two-level systems) and $|\boldsymbol{z}_0 \rangle$ is the coherent state in phase space. 

As we saw from eq.~\eqref{clasdicke}, under the quantum-classical correspondence, the Dicke Hamiltonian [see eq.~\eqref{dickeham}]
\begin{equation}
 \hat{H}_D=\omega \hat{a}^{\dagger}\hat{a} + \omega_0 \hat{J}_z + \frac{\Omega}{\sqrt{N}}(\hat{a}^{\dagger}+\hat{a})\hat{J}_x,
\end{equation}
is mapped to the classical Hamiltonian
\begin{equation}
 h_D(\boldsymbol{z})=\tfrac{1}{2}(x^2 + p^2) + \omega_0 j_z + \Omega\sqrt{(N/2)^2-j_z^2}\,x\cos\phi,
\end{equation}
where $\beta=\sqrt{N/2}(x+ip)$, $w=\tan(\theta/2)e^{i\phi}$, $j_z=-(N/2)\cos\theta$ and $\phi=\tan^{-1}(j_y/j_x)$. The distributions plotted in fig.~\ref{fig:Dickechaos} reveal signatures of multifractality hidden in the coherent states $|\boldsymbol{z}\rangle=|\beta\rangle \otimes |w\rangle$ for different values of $q$.

Chaos, non-integrability, and ergodicity are all often taken as prerequisites for a closed quantum system to thermalize (for a discussion on integrability of the Dicke model see~\cite{batchelor2015integrability,he2015exact,babelon2007bethe}). Following the above discussion, thermalization is expected to occur also in the Dicke model. The OTOC\index{Out-of-time-order-correlator (OTOC)} has become a standard tool in exploring quantum chaotic systems. Imagine two operators $\hat V$ and $\hat W$, where we time evolve one of them, {\it i.e.} $\hat W(t)=\exp(i\hat H t)\hat W(0)\exp(-i\hat H t)$. These can, for example, be local operators with support initially on disconnected locations; $[\hat V(0),\hat W(0)]=0$. The time-evolution typically implies that the support of $\hat W(t)$ increases and we may find $[\hat V(0),\hat W(t)]\neq0$. The OTOC measures such an overlap and is defined as~\cite{swingle2018unscrambling}
\begin{equation}\label{otoc}
\mathcal{F}(t)=\langle\hat W^\dagger(t)\hat V^\dagger\hat W(t)\hat V\rangle.
\end{equation}
It relates to the above commutator via the identity $\mathrm{Re}1-[\mathcal{F}(t)]=\langle [\hat V^\dagger,\hat W^\dagger(t)][\hat W(t),\hat V]\rangle/2$, and for short times (and for `fast scramblers')\index{Scrambling} one can define the (quantum) Lyapunov exponent $\mathcal{F}(t)=1+\varepsilon e^{\lambda_\mathrm{L}t}+\dots$. The OTOC thus provides insight into how fast local information spreads in the system, the so-called {\it scrambling}\index{Scrambling! information}\index{Scrambling}, and hence the thermalization\index{Quantum! thermalization} of the system. The OTOC for different variations of the Dicke model in an open quantum system setting has been calculated in~\cite{tiwari2023quantum}, focusing on the link between the superradiant phase transition of the Dicke model and the onset of quantum chaos\index{Chaos!quantum}.

The exponential growth of the fidelity OTOC in the beginning of the time evolution of a Loschmidt echo\index{Loschmidt echo} signal cannot be attributed to a classical unstable point in the general case for the two-mode Dicke model\index{Model! Dicke!two-mode}. Such a model exhibits a normal to superradiant quantum phase transition with spontaneous breaking either of a discrete or continuous symmetry\index{Spontaneous! symmetry breaking}~\cite{Kirkova_2023}. 

Based on the fact that an inverted harmonic oscillator may exhibit the behavior of thermal energy emission, in close analogy to the Hawking radiation emitted by black holes, Tian and collaborators proposed the use of a trapped ion to realize a concrete analogue-gravity system and check the upper bound of the associated Lyapunov exponent, $\lambda_L \leq 2\pi T/\hbar$~\cite{TianZ2022}. They found that the effective Hawking temperature\index{Hawking temperature} of the trapped-ion-inverse-oscillator indeed matches this upper bound for the speed of scrambling\index{Scrambling}.  

The unitary evolution of quantum systems, and hence linearity of quantum mechanics, has since long been puzzling researchers; how can chaotic motion emerge from it? For example, the fidelity~(\ref{fidel}) between two pure states is constant in time, and there is no obvious quantum version of the {\it butterfly effect}\index{Butterfly effect}~\cite{haake2010quantum}. Likewise, the correlator $[\hat V(t),\hat W(t)]$ is also constant in time, and this explains why one looks at the out-of-time ordered correlator for the OTOC. We may note that the exponential growth, characteristic for the butterfly effect and the presence of a positive Lyapunov exponent, can also appear using different approaches. For example, if losses are added to the otherwise closed system, unitarity is lost and the fidelity~(\ref{fidel}) between two initially nearby states behaves as $F(t)=\exp(-\lambda_Lt)$~\cite{bhattacharya2000continuous}. Alternatively, the overlap $|\langle\psi_0(t)|\psi_\varepsilon(t)\rangle|^2$, where the two states are initially the same but $|\psi_0(t)\rangle$ has been evolved with an unperturbed Hamiltonian and $|\psi_\varepsilon(t)\rangle$ has been evolved with an unperturbed Hamiltonian, shows also an exponential decay with the Lyapunov exponent~\cite{peres2006quantum}.

As a simple, but still experimentally relevant, non-integrable system, the Dicke model has attracted much attention in recent years regarding OTOC and related issues~\cite{safavi2018verification,lewis2019unifying,chavez2019quantum,alavirad2019scrambling,sun2020out,hu2020out}. In exploring properties of the evolution of complex quantum systems one is usually interested in generic and universal features, and one expects the OTOC $\mathcal{F}(t)$ to not depend strongly on the particular choice of $\hat V$ and $\hat W$. For the Dicke model different choices have been considered; $\hat V=\hat W=\hat a^\dagger\hat a$~\cite{sun2020out,hu2020out}, $\hat V=\hat p$ and $\hat W=\hat q$~\cite{chavez2019quantum}, and $\hat V=\hat\rho(0)$, $\hat V=\hat\sigma_i$ and $\hat W=\hat\sigma_j$ or $\hat V=\hat W=\hat x$~\cite{alavirad2019scrambling}, and $\hat W=\exp(i\delta\phi\hat x)$~\cite{lewis2019unifying}. In the latter case, the $\hat V$ operator is a projection onto the initial state of the system, and for pure states the OTOC becomes $\mathcal{F}(t)=|\langle\psi(0)|e^{-i\hat Ht}e^{i\delta\phi\hat x}e^{i\hat H t}|\psi(0)\rangle|^2$. This quantity was termed the {\it fidelity-OTOC}, and for small $\delta\phi\ll1$ it approximates $\mathcal{F}(t)\approx1-\delta\phi^2\mathrm{var}\left[\hat x(t)\right]$, {\it i.e.} it measures the variance of the $x$-quadrature. The Lyapunov exponent $\lambda_L$ for the Dicke model was studied in ref.~\cite{alavirad2019scrambling} in the whole of the phase diagram, and it was found that the model is the most chaotic in the vicinity of the critical point\index{Critical! point}. By using an expansion similar to the Holstein-Primakoff method described below, and a {\it Majorana fermion representation}\index{Majorana fermion! representation}, analytical expressions for the OTOC's could be given. From this, $\lambda_L$ can be extracted. In ref.~\cite{chavez2019quantum} the Lyapunov exponent defined via the OTOC was compared to the one of the corresponding classical Dicke model~(\ref{clasdicke}). It was found that $\lambda_L$ agrees with the classical one in the chaotic regime of the model. The same conclusion was found by studying the aforementioned fidelity-OTOC~\cite{lewis2019unifying}, and links between the OTOC, quantum thermalization, and scrambling\index{Scrambling} was found. The OTOC has also been analyzed in terms of the anisotropic Dicke model [see eq.~(\ref{anDicke})]. By time-averaging the OTOC it was shown in~\cite{sun2020out} that the OTOC captures the critical behaviour at both zero and finite temperatures, and furthermore the time average shows universal features, {\it i.e.} it scales as $(g-g_c)^\beta$ for some exponent $\beta$. 

Normally, quantum thermalization is connected to systems possessing a large number of degrees of freedom~\cite{d2015quantum}. However, as properties of quantum chaotic systems can often be understood from random matrix theory\index{Random matrix theory}~\cite{haake2010quantum} it should be clear that `sufficiently large' systems can also be achieved in a system with few degrees of freedom as long as the Hilbert space is large. For the Dicke model, thermalization can only emerge in the thermodynamic limit\index{Thermodynamic limit}, spin $S\rightarrow\infty$. In ref.~\cite{altland2012quantum} thermalization in the Dicke model was considered, and especially the role played by quantum fluctuations was analyzed. In particular, in phase space the dynamics is governed by a Fokker--Planck equation\index{Fokker--Planck equation} consisting of an $\hbar$-independent `classical drift part' and an $\hbar$-dependent part giving rise to quantum corrections. In this phase space representation (particularly for the $Q$-function) of quantum mechanics the Heisenberg uncertainty manifests itself in preventing structures beyond the $\hbar$-scale to form. In fig.~\ref{figdickethermal} we give a few snap-shot examples of the Wigner function for the Dicke model in the normal (a) and superradiant (b) regimes. Ergodicity\index{Ergodicity} is evident in the superradiant case as expected. The seemingly chaotic structure in (b) is an outcome of chaos and signature of thermalization. A measure of how localized, {\it i.e.} ergodic the state is, is given by the {\it inverse partition ratio}\index{Inverse partition ratio} which for the Dicke model we define here as
\begin{equation}\label{ipr}
\text{IPR}=\int dx dp\,W^2(x,p),
\end{equation}
where $W(x,p)$ is the Wigner function for the boson mode. In analog with the linear entropy for density operators introduced in sec.~\ref{sssec:sol}, the IPR is sometimes also referred to as a linear entropy~\cite{manfredi2000entropy}. Note that $\text{IPR}\leq1$, and a large value of the IPR implies a localized distribution, while a small value represents a smeared out distribution as expected in the ergodic regime. We see in (c) that a clear drop in the IPR occurs slightly before the critical coupling. As a final example of the transition from regular to thermalizing evolution we consider the {\it autocorrelation function}
\begin{equation}\label{auto}
A(t)=|\langle\psi(t)|\psi(0)\rangle|.
\end{equation}
If the system thermalizes it follows that any expectation should settle. Thus, we look at the variance $\delta A$ of $A(t)$ in fig.~\ref{figdickethermal} (d). It displays the same trend as the IPR; a drop in the variance is seen as we approach the superradiant (chaotic) regime. 

A generalized Dicke model describes the interaction of a quantized light field with a two-level atomic ensemble coupled by microwave fields in optical cavities. Multicritical points\index{Critical! point! multi-critical} for the superradiant quantum phase transition have been shown to occur, which can be effectively manipulated by tuning system parameters~\cite{Hu_2023}.

The phenomenon of {\it quantum scars}\index{Quantum! scar}\index{Scarring}~\cite{heller1984bound} is also due to localization in phase space. These are eigenstates of a chaotic Hamiltonian $\hat H$ which are not uniformly spread in phase space. Instead they concentrate around unstable classical periodic orbits of the corresponding classical Hamiltonian $H$. Thus, chaotic signatures of the classical model find their way towards the quantum eigenstates. It hasn't until recently been understood that scarred states\index{State! scarred} occur also in many-body models; it was for long believed that the many degrees of freedom make scarred states exponentially unlikely. Actually, scarred states are very common in the Dicke model~\cite{pilatowsky2021ubiquitous,pilatowsky2021quantum}. In fact, it has been claimed that all states of the Dicke model are scarred to some degree; typically an eigenstate populates only half of the accessible phase space~\cite{pilatowsky2021ubiquitous}. This is striking since one would expect most states to be ergodic. However, the scarring in the Dicke model is not able to prevent thermalization.

The above explanation of the Dicke PT hinges on classical/mean-field arguments, and we only mentioned that it also survives on a quantum level. A classical PT is driven by thermal fluctuations while a quantum PT occurs at zero temperature, $T=0$ and results from quantum fluctuations~\cite{sachdev2007quantum}. The first discussion about the Dicke PT at zero temperature dates back to a work by Hillery {\it et al.}~\cite{hillery1985semiclassical}. However, the zero-temperature Dicke PT should not be called a quantum PT in the above sense since in the thermodynamic limit\index{Thermodynamic limit} quantum fluctuations are negligible. In particular, the number of degrees of freedom in the Dicke model is still two in the thermodynamic limit, {\it i.e.} it is independent of the system size. Nevertheless, the ground-state energy $E_0(g)$ shows a kink at the critical point\index{Critical! point}, which is a characteristic of quantum PTs~\cite{sachdev2007quantum}. An elegant approach to study the zero temperature Dicke PT is to apply the {\it Holstein-Primakoff method}\index{Holstein-Primakoff! method} (HP)~\cite{holstein1940field}. Even though the HP method had been previously applied to related topics~\cite{ressayre1975holstein,persico1975coherence}, Emary and Brandes were the first to demonstrate its power when employed to the study of the $T=0$ Dicke PT~\cite{emary2003chaos}. The HP transformation maps spin to boson degrees of freedom according to 
\begin{equation}\label{hpeq}
\begin{array}{l}
\hat{S}^-=\sqrt{2s-\hat{b}^\dagger\hat{b}}\hat{b},\\ \\
\hat{S}^+=\hat{b}^\dagger\sqrt{2s-\hat{b}^\dagger\hat{b}},\\ \\
\hat{S}_z=\hat{b}^\dagger\hat{b}-s.
 \end{array}
 \end{equation}
The 'bosonized' Dicke Hamiltonian becomes (up to trivial constants)
 \begin{equation}\label{hbdicke}
 \hat{H}_\mathrm{D}=\omega\hat{n}+\frac{\Omega}{2}\hat{b}^\dagger\hat{b}+g\left(\hat{a}^\dagger+\hat{a}\right)\left(\hat{b}^\dagger\sqrt{1-\frac{\hat{b}^\dagger\hat{b}}{2s}}+\sqrt{1-\frac{\hat{b}^\dagger\hat{b}}{2s}}\hat{b}\!\right),
 \end{equation}
 where we used $N=2s$. The idea is now to expand the fields around their mean-field solutions,
 \begin{equation}\label{linj}
 \begin{array}{l}
 \hat{a}\rightarrow\sqrt{\alpha}+\hat{\delta}_a,\\ \\
 \hat{b}\rightarrow\sqrt{\beta}+\hat{\delta}_b.
 \end{array}
 \end{equation}
Here, $\hat{\delta}_a$ and $\hat{\delta}_b$ are the quantum fluctuations on top of the mean-field solutions $\alpha$ and $\beta$ (note that in the symmetry broken superradiant phase we pick one of the two possible solutions). This approach of linearizing around the mean-field solutions yields a low-order expansion which captures universal critical properties; when studying the dynamical evolution, however, it will fail to reproduce the correct properties after relatively short times. In the normal phase $\alpha=\beta=0$ and we set $\hat{b}^\dagger\hat{b}/2s=0$; the effective model is simply
 \begin{equation}\label{hbdickeN}
 \hat{H}_\mathrm{eff}^{(N)}=\omega\hat\delta_a^\dagger\hat\delta_a+\frac{\Omega}{2}\hat\delta_b^\dagger\hat\delta_b+g\left(\hat\delta_a^\dagger+\hat\delta_a\right)\left(\hat\delta_b^\dagger+\hat\delta_b\right).
 \end{equation}
As a quadratic Hamiltonian this can be readily diagonalised by a {\it Bogoliubov transformation}\index{Bogoliubov transformation}~\cite{assa1994interacting}. Denoting the new bosonic modes in which the Hamiltonian is diagonal as $\hat c_1$ and $\hat c_2$, we have~\cite{emary2003chaos}
\begin{equation}\label{hbdickeN2}
 \hat{H}_\mathrm{eff}^{(N)}=\epsilon_+^{(N)}\hat c_1^\dagger\hat c_1+\epsilon_-^{(N)}\hat c_2^\dagger\hat c_2+C_N,
\end{equation}
where $C_N$ is a constant, and for the two excitation modes in the normal phase ($g<g_c$) we obtain
\begin{equation}\label{exN}
 \epsilon_\pm^{(N)2}=\frac{1}{2}\left[\omega^2+\frac{\Omega^2}{4}\pm\sqrt{\left(\omega^2-\frac{\Omega^2}{4}\right)^2+4g^2\omega\Omega}\right].
\end{equation}
In the superradiant phase we have $\alpha,\,\beta\neq0$ as evident from eq.~(\ref{sramp}), and one expands around the corresponding values. In doing so, the effective low-energy Hamiltonian in the superradiant phase becomes
\begin{equation}\label{hbdickeSR2}
 \hat{H}_\mathrm{eff}^{(SR)}=\epsilon_+^{(SR)}\hat d_1^\dagger\hat d_1+\epsilon_-^{(SR)}\hat d_2^\dagger\hat d_2+C_{SR},
 \end{equation}
where $\hat d_{1,2}$ represent the boson operators for the new Bogoliubov modes, and again $C_{SR}$ is a constant, and for the two excitation modes in the superradiant phase ($g>g_c$) one obtains
\begin{equation}\label{exSR}
 \epsilon_\pm^{(SR)2}=\frac{1}{2}\left[\omega^2+\frac{\Omega^2}{4\mu^2}\pm\sqrt{\left(\omega^2-\frac{\Omega^2}{4\mu^2}\right)^2+g^2\omega\Omega}\right],
\end{equation}
with $\mu=g_c^2/g^2$. At $g=g_c$ the two modes corresponding to the the normal and the superradiant phase meet, while in particular $\epsilon_-^{(N)}(g=g_c)=\epsilon_-^{(SR)}(g=g_c)=0$ as expected at the critical point\index{Critical! point}.  

\begin{figure}
\includegraphics[width=10cm]{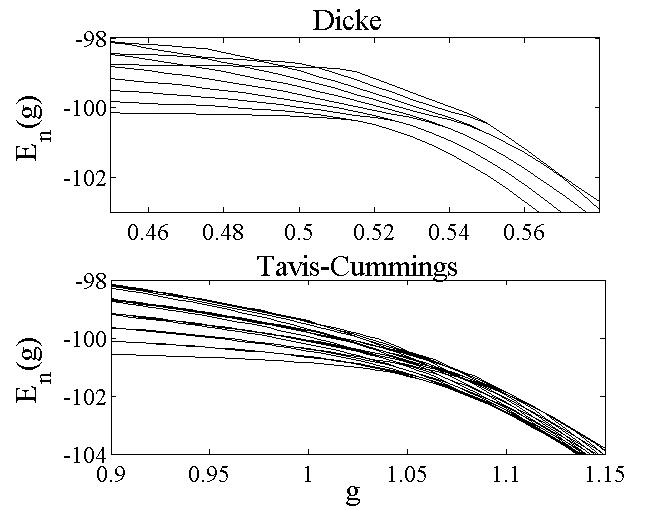} 
\caption{Spectrum of the lowest (10 -- Dicke and 20 -- TC) energies for the Dicke and TC models. The critical couplings are $g_c=1/2$ (Dicke) and $g_c=1$ (TC), and for the plots $N=200$ particles. In the superradiant phase, $g>g_c$, the Dicke spectrum is (approximately for finite $N$) doubly degenerate reflecting the $\mathbb{Z}_2$ symmetry and the TC spectrum shows instead a a quasi continuous band of energies (only in the thermodynamic limit will the spectrum become exactly continuous). These two cases exemplifies the cases of massive Higgs excitation and massless Goldstone excitations respectively.   
} 
\label{figdickespec2}   
\end{figure}

\begin{figure}
\includegraphics[width=10cm]{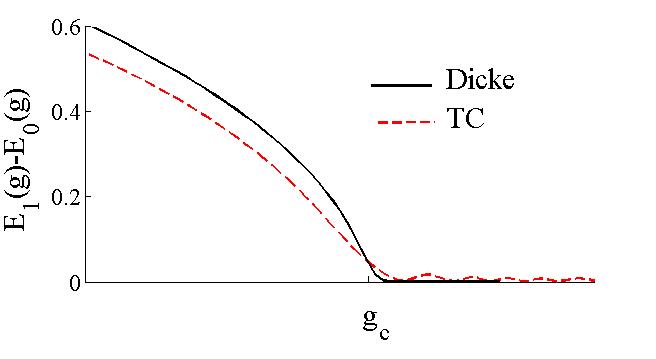} 
\caption{Comparing the gap-closening between the first excited $E_1(g)$ and ground state $E_0(g)$ energies for the Dicke and the TC models. The parameters are the same as in fig.~\ref{figdickespec2}. Naturally, with a finite size system, $N=200$, the exponents cannot be fully extracted, but the difference between the two models is still evident - softer closing of the TC model. How the gap scales with the system size $N$ were explored in detail in refs.~\cite{vidal2006finite,liu2009large}, and see also~\cite{bastarrachea2011numerical}.} 
\label{figdickegap}  
\end{figure}

From the resulting energies it is possible to extract the critical exponent $\nu=1/2$ for the gap closing at $g=g_c$ (see a further discussion in sec.~\ref{ssec:mbcQED} where the open/driven Dicke model is considered). Vidal and Dusuel generalised the above HP results valid in the thermodynamic limit to finite systems, {\it i.e.} studying finite scaling effects~\cite{vidal2006finite}. They concluded, among others, that the value of the exponent for the Dicke model agrees with those of the {\it Lipkin-Meshkov-Glick model}\index{Lipkin-Meshkov-Glick! model}\index{Model! Lipkin-Meshkov-Glick}~\cite{lipkin1965validity}
\begin{equation}\label{lmg}
\hat H_\mathrm{LMG}=\omega\hat S_z+\lambda\hat S_x^2.
\end{equation}
In fact, the link to the Lipkin-Meshkov-Glick model\index{Lipkin-Meshkov-Glick! model} becomes apparent when noting that the two are identical when the boson field of the Dicke model has been eliminated~\cite{latorre2005entanglement,morrison2008dynamical,larson2010circuit,li2021collective}. The physics of the $T=0$ PTs of the Dicke and TC models are fundamentally different as the two models support different symmetries, just like the JC and quantum Rabi models. The Dicke model has the $\mathbb{Z}_2$ parity symmetry\index{$\mathbb{Z}_2$ symmetry}~(\ref{parity}) modified to spin-$s$ ($\hat{\sigma}_z\rightarrow\hat{S}_z$), while the TC model supports the $U(1)$ symmetry\index{$U(1)$ symmetry} of eq.~(\ref{jcU1}). Naturally, the adiabatic elimination of the boson field in the TC model generates another Lipkin-Meshkov-Glick model which indeed supports a continuous symmetry, see eq.~(\ref{lmgU}). A direct consequence of the different symmetries is that excitations in the symmetry broken superradiant phase is given by a massive {\it Higgs mode}\index{Higgs mode} in the Dicke model and a gapless {\it Goldstone mode}\index{Goldstone mode} in the TC model. This qualitative difference between the two models can have far reaching consequences. For example, photon losses destabilize the superradiant phase in the TC model such that only the normal phase survives~\cite{larson2017some}. Actually, for the {\it anisotropic Dicke model}\index{Anisotropic! Dicke model}, 
\begin{equation}\label{anDicke}
\hat{H}_\mathrm{anD}=\omega\hat{n}+\frac{\Omega}{2}\hat{S}_z+\frac{g_\mathrm{jc}}{\sqrt{N}}\left(\hat S^+\hat{a}+\hat{a}^\dagger\hat S^-\right)+\frac{g_\mathrm{ajc}}{\sqrt{N}}\left(\hat S^-\hat{a}+\hat{a}^\dagger\hat S^+\right),
\end{equation}
the superradiant phase does not survive photon losses even for non-zero $g_\mathrm{ajc}$ provided $g_\mathrm{jc}$ is large enough~\cite{soriente2018dissipation,stitely2020nonlinear}. The Higgs mode, or {\it roton mode}\index{Roton mode}, can be understood from the double-well structure of the effective (adiabatic) potential $V_\mathrm{ad}^{(m)}(x)$, and has been experimentally observed~\cite{mottl2012roton}. A corresponding effective potential, or action, for the TC model has the form of a `Mexican hat', $V_\mathrm{ad}^{(m)}(x,p)=\omega\left(\frac{p^2}{2}+\frac{x^2}{2}\right)+m\sqrt{\frac{\Omega^2}{4}+\frac{g^2}{2N}\left(p^2+x^2\right)}$; it does not cost any energy to displace the state (order parameter) along the minimum (azimuthal angle) of the hat potential. This discrepancy between the two models is evident in the spectrum of the low energies, as illustrated in fig.~\ref{figdickespec2} (for further numerical results on the criticality of the Dicke model, see ref.~\cite{bastarrachea2011numerical}). After passing the critical value for the coupling, {\it i.e.} entering the superradiant phase, the spectrum becomes doubly degenerate (in the thermodynamic limit\index{Thermodynamic limit}) for the low lying energies in the Dicke model. The corresponding two (symmetry broken) ground states are the ones of the left/right minima of the double-well potential. Note that physically this symmetry breaking is directly reflected in the phase of the field, in the left minima the phase is $\pi$ and in the right it is $0$. Indeed, this symmetry breaking has been experimentally verified by measuring the phase (relative and not absolute) of the field~\cite{baumann2011exploring}, which we discuss further in sec.~\ref{ssec:mbcQED}. It has been suggested that for a small perturbation that breaks the $\mathbb{Z}_2$ symmetry the system will prefer one of the two wells and upon measurement of the state it is possible to get an indirect handle of the perturbation, {\it i.e.} the symmetry breaking mechanism can by utilised for quantum sensing~\cite{ivanov2015spontaneous}. In the TC model, on the other hand, in the symmetry broken phase a large number of energies cluster together to form a quasi continuous band, see fig.~\ref{figdickespec2}. The larger number $N$ of particles, the higher density of states at low energies one would get. Physically, the Mexican hat shaped 'potential' implies that in the TC model the field phase of the superradiant state can take any value in the range $[0,2\pi)$. Also the universality classes of the transitions in the two models are different, with in particular the Dicke model PT is of the {\it Ising class}\index{Ising! symmetry class}. We already mentioned that the exponent for the gap closing in the Dicke model is $1/2$. This is also shown in fig.~\ref{figdickegap}. The figure also displays the gap for the TC model and a `weaker' closening is found in this case. The distinction of universality classes\index{Universality! class} between a model with discrete or continuous symmetries was also found in the Lipkin-Meshkov-Glick model~\cite{dusuel2005continuous}. The critical exponent for the {\it fidelity susceptibility}\index{Fidelity! susceptibility}, {\it i.e.} how sensitive the properties of the system ground state are to variations in the coupling, for the Dicke model has also been calculated~\cite{liu2009large,castanos2012universal} and again it agreed with the one of the Lipkin-Meshkov-Glick model. One further aspect of the spectral properties that can be qualitatively understood from the adiabatic potentials $V_\mathrm{ad}^{(m)}(x)$ of eq.~(\ref{dickeadpot}) is that the potential barrier of the double-well structure sets some `critical' energy $E_c$ such that eigenstates with energies below $E_c$ can show symmetry breaking, while those above $E_c$ will be of definite $\mathbb{Z}_2$ parity symmetry (remember discussion of fig.~\ref{figdickespec}). This observation was explored in a similar but slightly different approach, also semiclassical, in ref.~\cite{puebla2013excited}.  

More recently, the deep strong-coupling analysis of~\cite{HuZhiguo2023} reveals a quantum critical point, anticipated by mean-field theory, at which it is found that the Ising-Rabi lattice model\index{Model!Ising-Rabi} belongs to the universality class of the $XYZ$ model\index{Model! Heisenberg! $XYZ$}. Beyond the mean-field regime in 1D, the $XYZ$ model exhibits a far richer phase diagram than, say, the transverse-field Ising model, including both BKT\index{Phase transition! quantum! BKT} and $\mathcal{Z}_2$-symmetry breaking transitions~\cite{Sela2011}. Hence, the $XYZ$ model opens up interesting prospects in the field of quantum simulators. 

As the effective expanded HP Hamiltonians for the Dicke model in the two phases are both quadratic it follows that the ground states are Gaussian\index{Gaussian! state} and it is thereby possible to analytically determine properties of the two phases like squeezing and entanglement in the vicinity of the critical point\index{Critical! point}
~\cite{emary2003chaos,lambert2004entanglement}. These two properties are known to be highly interrelated for Dicke states~(\ref{dickestates})\index{Dicke! state}, squeezing may act as an {\it entanglement witness}\index{Entanglement! witness}~\cite{wang2003spin,messikh2003spin,toth2007optimal}. Typical, and also so for the Dicke model, is that the amplitudes of these properties blow up at the critical point. The same holds true for other properties like the fidelity susceptibility~\cite{liu2009large,mumford2014impurity} and field quadrature squeezing~\cite{mumford2014impurity}. In particular, at the critical point the ground state $p$-quadrature gets infinitely squeezed.

For many applications, it is desirable to simplify the description based on the Tavis-Cummings Hamiltonian\index{Tavis-Cummings model}, towards an analytical and more intuitive description of the atom-light interaction which is directly applicable  to existing experiments. Such experiments often deal with nearly classical light beams containing many photons, to which for most purposes a coherent state may be assigned. In this case it is possible to derive simpler and analytical expressions for the light-atom and light-mediated and entanglement-generating spin-spin interaction, as is done in~\cite{Zeyang2022}.
     
The Dicke model is one of the simplest examples -- one collective spin coupled to a bosonic degree of freedom -- of a critical quantum system. As such, it has been the focus of numerous theoretical works that we have met in the previous paragraphs. However, a most important and relevant contribution to the whole topic of the Dicke PT was published only a couple of years after the first prediction of a PT in the Dicke model. Rzazewski and co-workers pointed out that under the standard assumptions, the PT is an artefact of neglecting the {\it diamagnetic self-energy} term\index{Diamagnetic term}\index{Electromagnetic self-energy} of the EM-field~\cite{rzazewski1975phase}. With the scenario of $N$ atoms confined in a `single-mode' cavity, the minimal coupling Hamiltonan
\begin{equation}\label{mincoup}
\hat H_\mathrm{mc}=\omega\hat a^\dagger\hat a+\frac{1}{2m}\sum_{n=1}^N\left(\hat p_n-\frac{e}{c}\hat A(0)\right)^2+V(\hat x_n),  
\end{equation}   
where $\hat A(0)\propto\left(\hat a^\dagger+\hat a\right)$ is the vector potential\index{Vector potential} and we have made the dipole approximation. The bare Hamiltonian $\hat H_n=\frac{\hat p_n^2}{2m}+V(\hat x_n)$ describes the atomic dipole. The Hamiltonian~(\ref{mincoup}) arises when the field is quantized in the {\it Coulomb gauge}\index{Coulomb gauge}\index{Gauge! Coulomb}~\cite{cohen1997photons}. Opening the square, and imposing the TLA, we derive the Hamiltonian~\cite{Scully1997a}
\begin{equation}
\hat H_{A^2}=\hat H_D+\mu\left(\hat a^\dagger+\hat a\right)^2,
\end{equation}
where $\hat H_D$ is given in eq.~(\ref{dickeham2}). The new contribution to the Hamiltonian is the so-called diamagnetic term\index{Diamagnetic term}, with the coupling $\mu$ scaling as the inverse of the mode volume. In the mean-field picture\index{Mean-field approximation} discussed above, the adiabatic potentials now become
\begin{equation}\label{adpot2}
V_\mathrm{ad}^{(m)}(x)=\omega\frac{x^2}{2}+m\sqrt{\frac{\Omega^2}{4}+2\frac{g^2}{N}x^2}+2\mu x^2.
\end{equation}
Clearly, the self-energy renormalizes the bare photon frequency $\omega$ and in particular it counteracts the formation of the double-well structure. The main result of ref.~\cite{rzazewski1975phase} was to show that under realistic assumptions (the atomic density is very dilute~\cite{mallory1975superradiant} and the lower atomic state $|g\rangle$ an electronic ground state), the {\it Thomas-Reiche-Kuhn sum rule}\index{Sum rule!Thomas-Reiche-Kuhn}~\cite{reiche1925zahl} prevents the appearance of a PT. Hence, the coupling can never reach the critical value for the rescaled frequency; $g_c=\sqrt{\Omega(\omega+4\mu)}/2$. Some years later, the authors strengthened their results by formulating a {\it no-go theorem}\index{Dicke! no-go theorem} for the Dicke PT~\cite{bialynicki1979no}. In particular, they showed that by disregarding the self-energy term the Hamiltonian is not gauge invariant\index{Gauge! invariance}, see also refs.~\cite{andolina2019cavity,andolina2021no} for a recent discussion on the no-go theorem and the importance of gauge invariance. In sec.~\ref{sssec:dia} we will return to this discussion when we analyze the effects of the diamagnetic term\index{Diamagnetic term} more qualitatively. Further, long after the ref.~\cite{bialynicki1979no} it has been understood that gauge invariance in these light-matter models is more subtle and already truncating the matter Hilbert space to few levels can violate gauge invariance as we will demonstrate in sec.~\ref{sssec:tla}. Nevertheless, the results reported in~\cite{rzazewski1975phase,bialynicki1979no} suggest that a cavity QED experimental realization of the Dicke PT employing a set of two-level atoms is out of reach. However, the accuracy of the starting Hamiltonian~(\ref{mincoup}) has been recently questioned. Keeling pointed out~\cite{keeling2007coulomb} that dipole-dipole interactions among the atoms may qualitatively alter the conclusions of~\cite{rzazewski1975phase} (see also~\cite{al2002several}). Domokos {\it et al.} went even further to explore the role of the atoms in the quantization of the EM-field~\cite{vukics2012adequacy,vukics2014elimination}; a more compete picture of the composite system of atoms and field, in the situation of strong coupling, is obtained via a systematic field quantization taking the atoms into account. Indeed, it was found that a PT may be possible, even if it might be difficult to reach in realistic experiments. An addition to this debate was made by Rabl and others; when the TLA is imposed, gauge invariance\index{Gauge! invariance} is usually lost. For example, the quantum Rabi model derived as above in the Coulomb gauge does not agree with the one derived in the {\it dipole gauge}\index{Dipole! gauge}~\cite{de2018breakdown} (see a more detailed discussion about gauge invariance in the next sec.~\ref{sssec:tla}). In particular, in the dipole gauge the application of the Thomas-Reiche-Kuhn sum rule does not necessarily rule out the presence of a superradiant instability. We also note that the debate on whether the Dicke PT is accessible or not has also spurred activity in the circuit QED community in which the parameter relations appearing in the Thomas-Reiche-Kuhn sum rule are modified~\cite{nataf2010no,viehmann2011superradiant}. A somewhat similar effect may arise if one considers the multimode Dicke model\index{Multimode! Dicke model}. The presence of the highly detuned modes will induce an effective interaction among the two-level systems which can be written as $D\hat S_x^2$ with $D\sim g^2$~\cite{jaako2016ultrastrong}. As one writes the Dicke Hamiltonian, including this additional term, in the Holstein-Primakoff repersentation~(\ref{hpeq}) one finds a term $ND\left(\hat b+\hat b^\dagger\right)^2/4$ which has the same form as the diamagnetic term\index{Diamagnetic term}. Its effect will serve to counteract the normal-superradiant PT and disentangle the two subsystems. However, since the term describes an interaction, the two-level system will be in a highly entangled state when the field disentangles from matter. Closer to our days, a different approach in realizing the Dicke PT, overcoming the problem with the self-energy diamagnetic term, has been presented. It consists of introducing a squeezing term $-\xi\left(\hat a^\dagger+\hat a\right)^2$ in the Hamiltonian offsetting the diamagnetic term. This idea was put into practice in an $N\!M\!R$ (nuclear magnetic resonance)\index{Nuclear! magnetic resonance} experiment by demonstrating the appearance of a Schr\"odinger cat state~(\ref{dickecat}) in the superradiant phase~\cite{chen2021experimental}. 
 
Despite the new results indicating that a Dicke PT may not be ruled out by sum rules or a no-go theorem, it remains the case that experimentally one must reach the deep strong coupling regime which to date is experimentally quite challenging. Therefore, alternatives have been proposed for how a {\it driven Dicke PT} might be realized~\cite{dimer2007proposed}. The idea is to drive an atomic transition and thereby control the coupling strength with an external classical field. In ref.~\cite{dimer2007proposed} a Raman-coupled $\Lambda$ scheme is considered (see fig.~\ref{fig12}). Considering the dispersive regime and after eliminating the excited atomic state, the coupling is given by eq.~(\ref{lambdaeff}). In the case where one classical field drives one transition while the other one is driven by the cavity field, this term becomes
\begin{equation}\label{ramnd}
\hat V_D=\frac{g\eta}{\Delta}\left(\hat a^\dagger+\hat a\right)\hat S_x,
\end{equation} 
with $g$ the cavity vacuum Rabi coupling, $\eta$ the amplitude of the driving field, and $\Delta$ the atom-field detuning. Note also that the coefficient in front of the term $\hat S_z$ is also controllable in this setup since one considers a rotating frame. By tuning up $\eta$, the effective atom-field coupling can be made very large, while the diamagnetic self-energy\index{Self-!energy term} is unaltered and the transition is indeed within reach. The proposal has been experimentally demonstrated~\cite{baden2014realization}. In particular, combining one cavity mode and two external lasers two Raman transitions were driven. After elimination of the excited level (see sec.~\ref{sssec:multi}) an anisotropic Dicke type\index{Anisotropic! Dicke model} of model is obtained. As in~(\ref{anDicke}), anisotropy means that the coupling strengths of the rotating and counter-rotating terms may differ. However, by appropriately selecting the external laser Rabi frequencies, isotropy can be restored. In addition to the Raman coupling, the two levels are Stark shifted which is represented by a term $\delta\hat a^\dagger\hat a\hat S_z/N$, realizing the so-called {\it Dicke-Stark model}\index{Dicke-Stark model}\index{Model! Dicke-Stark}~\cite{Mu2020}. Such a term does not break the $\mathbb{Z}_2$ parity symmetry of the Dicke model and the criticality survives this modification~\cite{bhaseen2012dynamics}, but the critical coupling is now shifted. At the mean-field level, including photon losses at a rate $\kappa$ and spontaneous emission\index{Spontaneous! emission} of the atoms at a rate $\Gamma$, the new critical coupling becomes~\cite{xie2021quantum}
\begin{equation}\label{critcoupnew}
g_c=\frac{1}{2}\sqrt{\frac{\left[\kappa^2+(\omega-\delta/2)^2\right]\left(\Gamma^2+\Omega^2\right)}{\Omega(\omega-\delta/2)}}.
\end{equation}
From this expression we also see how the photon loss $\kappa$ pushes the critical coupling to larger values. This abides by our intuition as in this case the driving must win over photon loss to the surroundings. In the experiment of ref.~\cite{baden2014realization}, around $2\times 10^5$ Rubidium atoms were loaded into a cavity and the amplitude of the output cavity was measured. By ramping up the amplitude of the drive lasers a rapid increase appeared for some critical value. The increased photon count was argued to mark the transition from the normal to the superradiant phase. An alternative approach to realize a dynamical Dicke PT is to consider a `pumped' atomic condensate confined within a cavity~\cite{nagy2010dicke}. This setup will be discussed in more detail in sec.~\ref{ssec:mbcQED}. Here we only note that the external pump scatters photons into the cavity via the atoms which thereby experience photon recoils; this, in turn, excites collective vibrational modes of the condensate. The phonon operators describing these vibrations can be mapped into spin operators via the {\it Schwinger spin-boson mapping}\index{Schwinger spin-boson mapping}~\cite{sakurai1995modern} and one ends up with an effective model identical to the Dicke one with the additional aforementioned Stark shift term (see the derivation in sec.~\ref{ssec:mbcQED}). Properties of the Dicke PT have been thoroughly analyzed experimentally in the BEC setup, especially symmetry breaking~\cite{baumann2011exploring} and critical exponents~\cite{brennecke2013real}. Common to both these driven models is the fact that photon losses cannot be overlooked, as eq.~(\ref{critcoupnew}) for the critical coupling indicates. 
Since the Dicke model is fully connected (every spin couples collectively to the same boson mode), one cannot talk about a characteristic (correlation) length scale which, for critical models, defines an exponent $\nu$. Nevertheless, one can still introduce universal exponents and, interestingly, it is found that the openness of the system affects their values. For example, the average photon number scales as 
\begin{equation}\label{dcrit}
n\sim|g-g_c|^{-\beta},
\end{equation}
with $\beta=1/2$ for the closed and $\beta=1$ for the open Dicke model~\cite{nagy2010dicke,nagy2011critical,oztop2012excitations,nagy2015nonequilibrium,hwang2018dissipative}. Another exponent studied is the one determining the singularity of the fidelity susceptibility\index{Fidelity susceptibility}~\cite{gu2008fidelity,cozzini2007quantum}
\begin{equation}
\chi(g)=-\left.\frac{1}{2}\frac{d^2}{dg^2}\langle\psi_0(g)|\psi_0(g+\delta)\rangle\right|_{\delta=0},
\end{equation} 
where $|\psi_0(g)\rangle$ is the ground state conditioned on the coupling $g$. The fidelity susceptibility method for determining dynamical critical exponents in the Dicke model with $\chi\sim|g-g_c|^\alpha$ in the vicinity of the critical coupling and with $\alpha=-2$ for $g<g_c$, and $\alpha=-1/2$ for $g>g_c$ has been used in~\cite{liu2009large,mumford2015dicke}. 

We note that the driven Dicke PT can also be established in spin-orbit coupled atomic condensates as was recently observed in the lab~\cite{hamner2014dicke}, and in optomechanical cavities~\cite{mumford2015dicke} as well as in the atomic Josephson effect modified with impurity atoms~\cite{mumford2014impurity}. It is also possible to include an additional pump term (\ref{pumpterm}) to the Dicke model and still preserve the criticality as long as the extra term obeys the parity symmetry~\cite{zou2013quantum,mumford2015dicke,zhu2015quantum}. In ref.~\cite{mumford2015dicke} some critical exponents\index{Critical! exponents} of the PT in the pumped Dicke model were extracted and they were found to be identical to the regular Dicke model. However, qualitative differences between the two models arise away from the universal critical regime. A driven model may also entail an explicit time-dependence, and it has been demonstrated that the physics of the Dicke model with a sinusoidal coupling $g(t)=g_0\sin(\omega_dt)$ is extremely rich~\cite{bastidas2012nonequilibrium}. In this case, one cannot talk anymore about PTs of the ground state. Nevertheless, by turning to a Floquet basis\index{Floquet theory} it is possible to define an effective time-independent Hamiltonian. In ref.~\cite{bastidas2012nonequilibrium} this Hamiltonian was considered and a whole set of `excited state' PTs was identified. In a recent report, Boneberg and coworkers find that the presence of local spin decay leads to stronger quantum correlations and ``stabilizes'' an entangled nonequilibrium superradiant phase\index{Nonequilibrium! phase! superradiant} in the open Dicke model~\cite{Boneberg2022}.
 
 \begin{figure}[ht]
\begin{center}
\includegraphics[width=10cm]{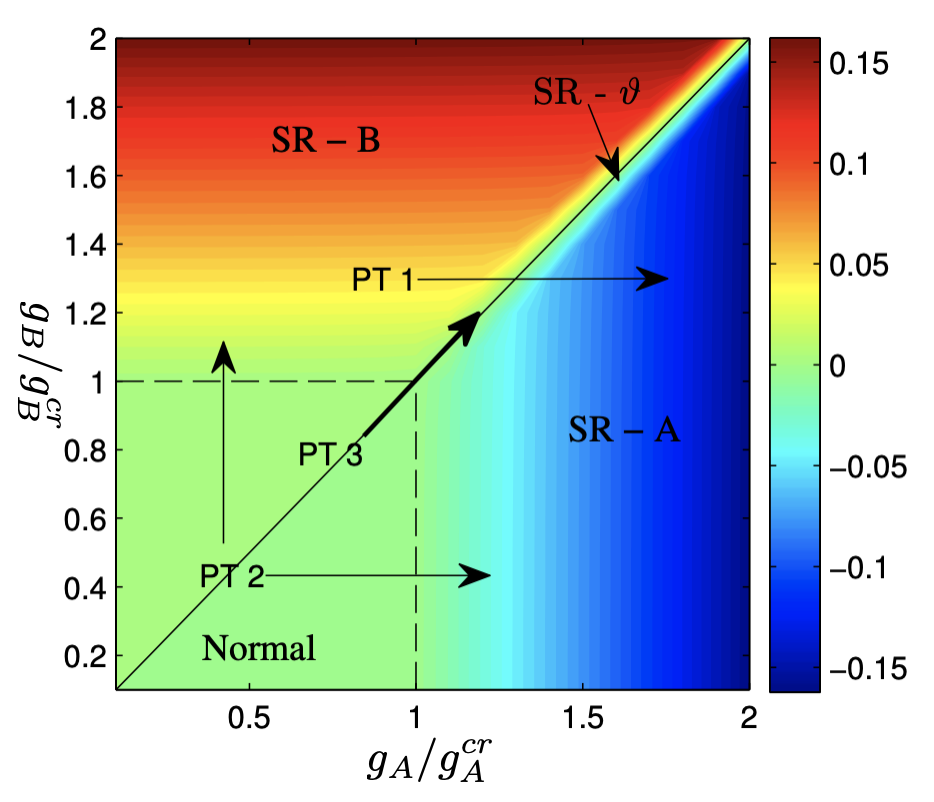}
\caption{Phase diagram for the $SU(3)$-Dicke model~(\ref{su3dicke}) for equal detuning of the cavities {\it i.e.}, $\omega_A = \omega_B$. We have plotted the difference between the average photon numbers in the modes B and A, {\it i.e.}, $\langle \hat b^\dagger \hat b \rangle - \langle \hat.a^\dagger \hat a \rangle$ calculated in the ground state computed by exact diagonalisation for $N = 100$. The vertical and horizontal dotted lines indicate $g_A = g_A^{cr}$ and $g_B = g_B^{cr}$. We have indicated the three possible phase transitions that can be accessed by tuning the coupling. PT 2 and PT 3 are second order, while PT 1 is a first-order phase transition\index{Phase transition! first-order}.} \label{fig:phasediag}
\end{center}
\end{figure}

\begin{figure}[ht]
\begin{center}
\includegraphics[width=7cm]{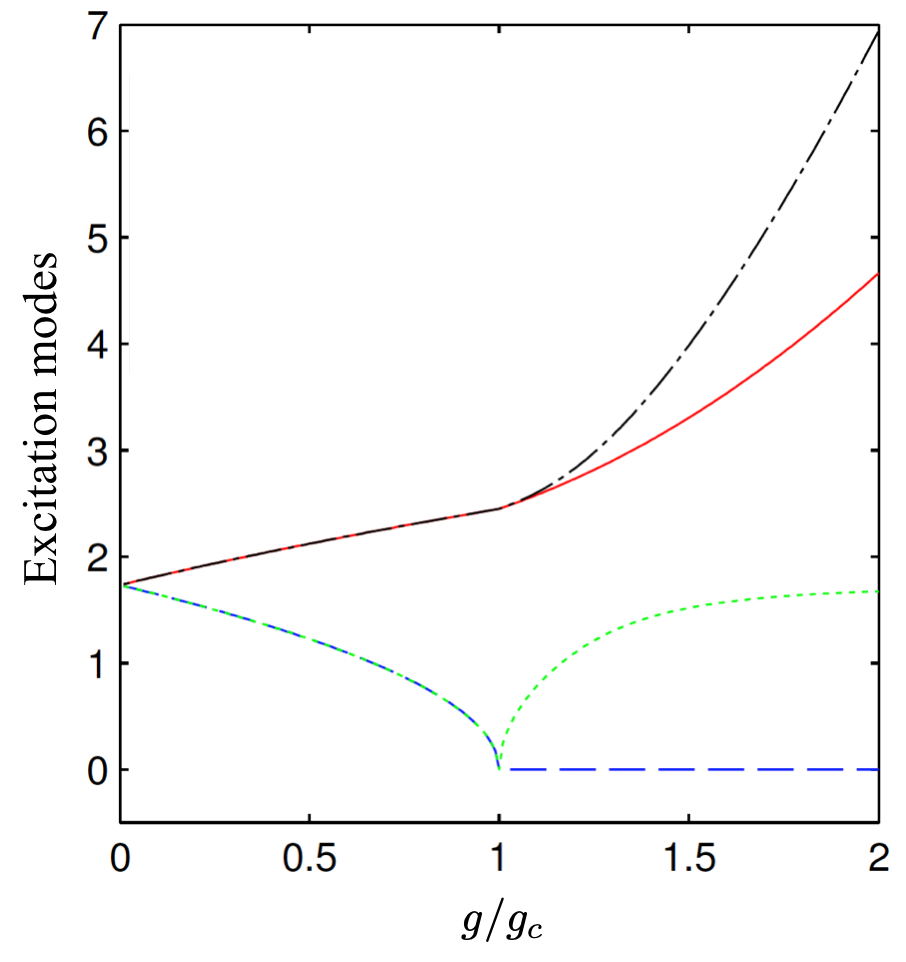}
\caption{The four excitation modes of the isotropic $SU(3)$ Dicke model~(\ref{su3dicke}) at resonance. In the normal phase, $g<g_c$, we see two gapped modes whose gaps close at the critical point $g=g_c$. In the superrdiant phase, $g>g_c$, one mode opens up (Higgs mode\index{Higgs mode}) a gap while the other remains gapless (Goldsone modes\index{Goldstone mode}). } \label{excfig}
\end{center}
\end{figure}

Other studies of criticality of extended Dicke models have been considered. For example, in a bimodal setup with a combination of a $\hat\sigma_x$ and $\hat\sigma_y$ coupling to the respective modes, as discussed in sec.~\ref{sssec:multi} (see fig.~\ref{fig11}), the system hosts a continuous $U(1)$ symmetry~\cite{emary2004phase,ivanov2013simulation,fan2014hidden} and the low lying excitations in the symmetry broken phase are of the soft/gapless Goldstone type\index{Goldstone mode}. Breaking down the continuous symmetry to a discrete $\mathbb{Z}_2$ symmetry can easily be achieved by considering different coupling strengths~\cite{hayn2011phase} and the excitations become Higgs-like\index{Higgs mode}, {\it i.e.} gapped. The Goldstone mode also appears in the so-called $SU(3)$ Dicke model\index{Generalized Dicke model}\index{$SU(3)$ Dicke model}\index{Model! generalized Dicke}\index{Model! $SU(3)$ Dicke}~\cite{leonard2017monitoring,cordero2021quantum,lopez2021quantum}. Here the two boson modes couple to $SU(3)$ pseudo-spins, {\it e.g.} three level atoms in any of the configurations of fig.~\ref{fig12}. The matter degrees of freedom are then better expressed in terms of the eight Gell-Mann matrices\index{Gell-Mann matrices} $\hat\lambda_\alpha$~(\ref{gellmann}), and it is clear that the variations of light-matter couplings are greater than for the $SU(2)$ Dicke model. Let us consider one version, that is relevant for the experiment~\cite{leonard2017monitoring} that will be discussed further in sec.~\ref{ssec:mbcQED}, namely 
\begin{equation}\label{su3dicke}
\hat{H}_{su3} = \omega_A\hat{n}_A+\omega_B\hat{n}_B+\Omega\hat{\Lambda}_8\displaystyle{+\frac{g_A}{\sqrt{N}}\left(\hat{a}+\hat{a}^\dagger\right)\hat{\Lambda}_4+\frac{g_B}{\sqrt{N}}\left(\hat{b}+\hat{b}^\dagger\right)\hat{\Lambda}_6},
\end{equation}
where $\hat\Lambda_\alpha=\sum_{i=1}^N\hat\lambda_\alpha^{(i)}$ ($\alpha=1,2,\dots,8$) are the collective Gell-Mann matrices\index{Gell-Mann matrices}, see eq.~(\ref{gellmann}).  Recall that the $SU(3)$ group supports three $SU(2)$ subgroups; $\{\hat{\Lambda}_1,\hat{\Lambda}_2,\hat{\Lambda}_3\}$, $\{\hat{\Lambda}_4,\hat{\Lambda}_5,\hat{X}\}$, and $\{\hat{\Lambda}_6,\hat{\Lambda}_7,\hat{Y}\}$ where $\hat{X}$ and $\hat{Y}$ are linear combinations of $\hat{\Lambda}_3$ and $\hat{\Lambda}_8$, and hence, $\hat H_{su3}$ is not a simple generalization of an $SU(2)$ Dicke model. Nevertheless, we find two $\mathbb{Z}_2$ parity symmetries defined by the unitaries
\begin{equation}\label{su2sym1}
\hat{\Pi}_A=e^{i\pi\left(\hat{n}_A-\hat{\Lambda}_3/2+\sqrt{3}\hat{\Lambda}_8/2\right)},\hspace{1.6cm}
\hat{\Pi}_B=e^{i\pi\left(\hat{n}_B+\hat{\Lambda}_3/2+\sqrt{3}\hat{\Lambda}_8/2\right)}.
\end{equation}
We may also note that a generalized Dicke model supporting a $\mathbb{Z}_3$ symmetry\index{$\mathbb{Z}_3$ symmetry} was recently analyzed~\cite{sedov2020chiral}, and the presence of a continuous symmetry broken phase for the $SU(3)$ case has also been demonstrated in the bimodal quantum Rabi model~\cite{zhang2020quantum}. More interesting than the two discrete symmetries (\ref{su2sym1}) is the fact that in the isotropic case ($\omega_A=\omega_B$ and $\lambda_A=\lambda_B$) the model also hosts a continuous $U(1)$ symmetry\index{$U(1)$ symmetry}~\cite{leonard2017monitoring} represented by the operator
\begin{equation}\label{su3sym2}
\hat{U}_\vartheta=e^{i\vartheta\left(\hat{n}_A-\hat{n}_B-\hat{\Lambda}_2\right)},
\end{equation}
{\it i.e.}, $\hat{U}_\vartheta$ commutes with $\hat{H}_{su3}$ in the isotropic case. We note that $\hat{U}_\vartheta$ transforms $\hat{\Lambda}_4$ and $\hat{\Lambda}_6$ as
\begin{equation}
\left[\begin{array}{c}
\hat{\Lambda}'_4\\
\hat{\Lambda}'_6\end{array}\right]=\left[\begin{array}{cc}
\cos\vartheta & -\sin\vartheta\\
\sin\vartheta & \cos\vartheta\end{array}\right]
\left[\begin{array}{c}
\hat{\Lambda}_4\\
\hat{\Lambda}_6\end{array}\right],
\end{equation}
while $\hat{\Lambda}_8$ is not affected by $\hat{U}_\vartheta$. Likewise, the transformation governed by $\hat{U}_\vartheta$ rotates the $A$-mode quadratures $\hat{x}$ and $\hat{p}_x$ clockwise by an angle $\vartheta$, and the $\hat{x}$ and $\hat{p}_x$ quadratures of the $B$-mode anti-clockwise. The tri-modal version of the $SU(3)$ Dicke model would include an additional $\mathbb{Z}_2$ symmetry reflecting a sign change in $\hat{\Lambda}_1$. In addition, the $U(1)$ symmetry discussed above would be replaced by a $U(2)$ symmetry in that case. The phase diagram of the $SU(3)$ Dicke model can be analyzed in much the same fashion as for the regular Dicke model, {\it i.e.} in terms of the BOA or mean-field, or employing the HP method generalized to $SU(3)$ spins. The phase diagram is shown in fig.~\ref{fig:phasediag} in the $g_Ag_B$-plane for the isotropic case which supports the continuous $U(1)$ symmetry. For small enough coupling the system is in the symmetric normal phase, and upon increasing the couplings across the PT the system has three possibilities by breaking any of the symmetries. If $g_A>g_B$ the symmetry $\hat\Pi_A$ is spontaneously broken\index{Spontaneous! symmetry breaking} and the $A$ mode becomes populated while the $B$ remains in vacuum, and the revers holds for $g_B>g_A$ (in the figure these are marked by the PT2 arrows). For $g_A=g_B$ all three symmetries are broken and both modes get populated, however with a random fraction (PT3 arrow in the figure). We have marked these three superradiant phases by SR-A, SR-B, and SR-$\vartheta$. The transition (PT1 in the figure) between two symmetry broken phases SR-A and SR-B is first order. The breaking of the $\mathbb{Z}_2$ symmetries upon entering the superradiant phases SR-A or SR-B follow the same universality classes as the regular Dicke, {\it i.e.} the critical exponent $\beta=1/2$ as in eq.~(\ref{dcrit}). In the phase SR-$\vartheta$ both modes get populated and it is the spontaneously chosen parameter\index{Spontaneous! choosing} $\vartheta$ that determines how much each one of them gets populated. A mean-field as well as a HP analysis predicts that the critical exponent $\beta=1/2$ also for this transition. In the SR-$\vartheta$ phase, where the $U(1)$ symmetry has been spontaneously broken\index{Spontaneous! symmetry breaking}, the spectrum is gapless as mentioned above. Within the HP formalism it is possible to find the excitation modes analytically to linear order as done for the regular Dicke model in eqs.~(\ref{exN}) and (\ref{exSR}). In the present model one has four boson modes; $\hat a$ and $\hat b$, plus two, $\hat c$ and $\hat d$, coming from the HP transformation of the $\hat\Lambda_\alpha$ operators. We do not present the derivation here (the derivation follows the same logic as for the Dicke model, see discussion above and ref.~\cite{emary2003chaos}), but state the results for the isotropic case ($\omega_A=\omega_B=\omega$ and $g_A=g_B=g$). In the normal phase one finds two doubly degenerate modes
\begin{equation}
\tilde{\epsilon}_{\pm}^{(N)2} = \displaystyle{\frac{1}{2}\left\{\omega^2 + \frac{\Omega^2}{8\mu^2}\left(1+\mu\right)^2 \pm \sqrt{\left[\frac{\Omega^2}{8\mu^2}\left(1+\mu\right)^2-\omega^2\right]^2 +\frac{\Omega}{\mu}2(1+\mu)^2 \omega g^2}\right\}}, 
\end{equation}
while in the $U(1)$ broken superradiant phase one finds three massive (Higgs) modes
\begin{equation}
\begin{array}{lll}
\epsilon_{0}^{(SR)} & = &  \displaystyle{\sqrt{\omega^2+\frac{\Omega^2}{16\mu^2}(1+\mu)^2}},\\ \\
\epsilon_{\pm}^{(SR)} & = &  \displaystyle{\frac{1}{2}\left[\frac{\Omega^2}{4\mu^2}+\omega^2\pm\sqrt{\left(\frac{\Omega^2}{4\mu^2}-\omega^2\right)^2+\omega^2\Omega^2}\right]},
\end{array}
\end{equation}
and one massless Goldstone mode (recall that $\mu=g_c^2/g^2$). The excitations, as a function of $g$, are shown in fig~\ref{excfig}.

The non-commutativity between the Gell-Mann matrices\index{Gell-Mann matrices} implies an underlying non-Abelian\index{Non-Abelian} (synthetic) gauge structure, similar to that discussed for the bimodal quantum Rabi model in eq.~(\ref{jt2}). Let us elaborate a bit further along these lines. Defining the force as $\hat{\mathbf{F}}=\ddot{\mathbf{r}}/2$~\cite{nikolic2005transverse}, where the `dot' represents time derivative and $\mathbf{r}=(\hat{x},\hat{y})$, we derive 
\begin{equation}\label{force}
\hat{\mathbf{F}}=\frac{1}{2}\left[\hat{H}_{su3},[\hat{\mathbf{r}},\hat{H}_{su3}]\right]=\frac{1}{2}\left(\omega^2\hat{x}-\Omega \tilde g \sqrt{3}\tilde g\hat{\Lambda}_7-\hat{p}_y\tilde g^2\hat{\Lambda}_2,\,\omega^2\hat{y}-\Omega \tilde g \sqrt{3}\tilde g\hat{\Lambda}_5+\hat{p}_x\tilde g^2\hat{\Lambda}_2\right),
\end{equation}
with $\tilde g=g/\sqrt{N/2}$. Now the first contributions are simply those of the harmonic potential (in the dimensionless scaling of the quadrature operators). The last terms are reminiscent of a {\it Lorentz force} arising from a magnetic field in the perpendicular $z$-direction; $\hat{\mathbf{F}}_\mathrm{Lor}=\tilde g^2\left(\hat{\mathbf{p}}\times\mathbf{e}_z\right)\hat{\Lambda}_2$. The second term represents instead a synthetic electric force $\hat{\mathbf{F}}_\mathrm{el}=\hat{\mathbf{E}}$ (with the synthetic ``charge'' being unity). To make this view more transparent we introduce the synthetic gauge potential\index{Synthetic! gauge potential}\index{Gauge! potential! synthetic}
\begin{equation}
\hat{\mathbf{A}}=(\hat{\Phi},\hat{A}_x,\hat{A}_y,\hat{A}_z)=\left(\Omega \hat{\Lambda}_8,\frac{\tilde g}{\omega}\hat{\Lambda}_4,\frac{\tilde g}{\omega}\hat{\Lambda}_6,0\right).
\end{equation}
The non-commutativity between $\hat{A}_0$, $\hat{A}_x$, and $\hat{A}_y$ demonstrates that the gauge potentials are non-Abelian\index{Gauge! potential! non-Abelian}. Following Wong~\cite{wong1970field}, the field tensor
\begin{equation}
\hat{F}=\left[
\begin{array}{cccc}
0 & \hat{E}_x & \hat{E}_y & \hat{E}_z\\
-\hat{E}_x & 0 & \hat{B}_z & -\hat{B}_y\\
-\hat{E}_y & -\hat{B}_z & 0 & \hat{B}_x\\
-\hat{E}_z & \hat{B}_y & -\hat{B}_x & 0
\end{array}\right]
\end{equation}
is given by
\begin{equation}\label{tensor}
\hat{F}_{\alpha\beta}=\partial_\alpha\hat{A}_\beta-\partial_\beta\hat{A}_\alpha-i\left[\hat{A}_\alpha,\hat{A}_\beta\right].
\end{equation}
Since $\hat{\mathbf{A}}$ is constant ($\partial_\alpha\hat{\mathbf{A}}=\mathbf{0}$), the first two terms of (\ref{tensor}) vanish and the only contribution derives from the non-Abelian\index{Non-Abelian! gauge field}\index{Synthetic! gauge field} property of the gauge potential\index{Gauge! potential}. The synthetic electric field becomes
\begin{equation}
\hat{\mathbf{E}}=\Omega \sqrt{3}\tilde g\left(\hat{\Lambda}_7,\,\hat{\Lambda}_5,\,0\right),
\end{equation}
and the magnetic field
\begin{equation}
\hat{\mathbf{B}}=\tilde g^2\left(0,\,0,\,\hat{\Lambda}_2\right).
\end{equation}
It is clear that these results are in agreement with the equations of motion (\ref{force}). Thus, we can interpret the forces acting upon the quasi-particle (represented in our model by the electromagnetic fields of the cavities) as a classical force from the ``potential'', and a synthetic electric and magnetic force. For $\Omega=0$ the synthetic electric field vanishes and the particle moves in a perpendicular synthetic magnetic field. As a result, one encounters the anomalous Hall effect\index{Anomalous Hall effect} discussed in sec.~\ref{sssec:multi} for the two-mode quantum Rabi model~(\ref{jt2}) 
~\cite{larson2013anomalous,larson2010analog}. As soon as $\Omega\neq0$ the synthetic electric field distorts the motion driven by the magnetic Lorentz force and the Hall effect is rapidly destroyed.  

Another model extensively studied is that of two spatially separated spins $\hat S_\alpha^{(1)}$ and $\hat S_\alpha^{(2)}$ coupled to a common boson mode~\cite{joshi2015cavity,zhu2020finite}. Criticality in such a model is not different from the regular Dicke model, however, since the spins are assumed to be separated in space we can treat them as two entities, and entanglement properties between the two spins can be studied in the presence of photon losses~\cite{joshi2015cavity}. In particular, entanglement may survive losses. In~\cite{zhu2020finite}, the two spins were assumed to be driven by the fields $\pm\hat S_x^{(i)}$ in opposite directions. This breaks down the $\mathbb{Z}_2$ symmetry, but if the two fields have equal amplitudes a pseudo $\mathbb{Z}_2$ symmetry arises; the previous symmetry of the Dicke model accompanied with $\hat S_\alpha^{1}\leftrightarrow\hat S_\alpha^{2)}$. The resulting phase diagram contains {\it tri-critical points}\index{Critical! point! tri-critical} in which the normal phase merges with two superradiant phases. The generalization to quantized atomic motion has also been explored~\cite{larson2009dilute}. In the regular Dicke model, the PT is a competition between the bare energies and the atom-field coupling energy, and by including also atomic motion as in~\cite{larson2009dilute} it becomes an interplay between three energy contributions instead (the above two plus the atomic kinetic energy). This greatly affects the phase diagram; in particular, the normal-to-superradiant transition may become first order. In~\cite{larson2009dilute} the atomic motion is quantized taking into account quantum fluctuations. A simplification could be to consider the atoms as point-like fixed points, but spatially distributed in the cavity such that the coupling strength becomes $g\rightarrow g_i$ with $g_i$ a random variable with the subscript $i$ marking atom $i$~\cite{buvzek2005instability,bazhenov2020superradiant}. It was found that this modification could lead to a series of new transitions between ground states with abrupt changes in their atom-field entanglement. Randomness naturally appears in a multimode cavity\index{Multimode! cavity} where the effective coupling to the various modes varies due to the specific mode profiles and also the different frequencies~\cite{gopalakrishnan2011frustration,strack2011dicke,buchhold2013dicke}. After eliminating the bosonic degrees of freedom one obtains a transverse-field Ising model
\begin{equation}\label{effis}
\hat H_\mathrm{Is}=\frac{\Omega}{2}\sum_{i=1}^N\hat\sigma_z^{(i)}-\frac{1}{2}\sum_{j,k=1}^NJ_{jk}\hat\sigma_x^{(j)}\hat\sigma_x^{(k)}.
\end{equation}   
To a good approximation, $J_{jk}$ is a random variable, and above a critical strength of disorder at zero temperature, a {\it spin glass phase}\index{Spin! glass} emerges~\cite{strack2011dicke}. The transition to a glass phase, {\it i.e.}, increasing the strength of randomness, could be achieved by increasing the number of cavity modes coupled to the atoms~\cite{gopalakrishnan2011frustration}. The appearance of the glass phase relies on the fact that the atom-field coupling is not too strong; in that case the system transforms instead into a traditional superradiant phase. Additional insight into the disordered Dicke model was gained by realizing that in the thermodynamic limit\index{Thermodynamic limit} it is equivalent to a quantum {\it Hopfield network}\index{Hopfield! network}\index{Model! Hopfield network}~\cite{rotondo2015dicke,fiorelli2020signatures,carollo2020proof}. This allows, in particular, for predictions about the system ground state. In the framework of {\it Keldysh theory}\index{Keldysh! theory}, these results were extended to account for photon loses and it was shown that the glass-phase physics also survives in this case~\cite{buchhold2013dicke}. However, just as for the Dicke model, the universality class is modified due to the photon losses. Quasi disorder can also be achieved by combining two incommensurate optical lattices, as has been realized experimentally for cold atoms in two laser fields~\cite{roati2008anderson}. In ref.~\cite{habibian2012bose} it was demonstrated that a {\it Bose glass phase}\index{Bose glass} may emerge if one of the two lattices is replaced by a cavity standing wave mode. A Bose glass is an insulating compressible phase. The main difference to~\cite{roati2008anderson} is that here the disorder is relevant only in the superradiant phase, {\it i.e.} the coupling has to exceed the critical one. In terms of realizing Dicke physics by mapping phonon modes of an atomic condensate to effective spin degrees of freedom (as mentioned above and more discussed in sec.~\ref{ssec:mbcQED}) the multimode Dicke model\index{Multimode! Dicke model} may also lead to {\it frustration}\index{Frustration}~\cite{gopalakrishnan2009emergent}. Critical exponents for the multimode Dicke model have been extracted by utilizing the HP approach, see eq.~(\ref{hpeq}), and especially criticality of the atom-atom entanglement (concurrence)\index{Concurrence}~\cite{tolkunov2007quantum}. Introducing disorder into the game may have the effect of enhancing quantum fluctuations such that the critical behaviour is no longer captured by simple mean-field predictions. The question of a proper quantum normal-superradiant PT was also addressed in~\cite{gagge2020superradiance} by considering an array of connected cavities (see sec.~\ref{ssec:jch}). In this scenario, the quantum fluctuations grow linearly with the system size and cannot be disregarded in the thermodynamic limit\index{Thermodynamic limit}, see also ref.~\cite{kurcz2014hybrid}. The extension to multi-level atoms, instead of two-level atoms, also alters the criticality of the model as shown in~\cite{xu2020multicriticality}.
 
 Let us end this rather long section on the Dicke model on a recent topic that has gained much attention, namely {\it time-crystals}~\cite{wilczek2012quantum,sacha2017time}. These are nonequilibrium phases of matter\index{Nonequilibrium! phase! Floquet system} in periodically driven (Floquet) systems. The state breaks the discrete time-transnational symmetry of the driven Hamiltonian and oscillates with a multiple of the drive period. Periodic doubling of the states were found in {\it many-body localized}\index{Many-body! localization} systems~\cite{khemani2019brief}. These are disordered interacting quantum systems that prevents thermalization, and under periodic driving they do not heat to infinite temperatures which is crucial for robust time-crystals. In analyzing periodically driven systems, it is practical to turn to the {\it Floquet picture}~\cite{bayfield1999quantum}. With $T$ the period of the drive, the time-evolution operator\index{Operator! time-evolution} from $t_0$ to a generic later time can be decomposed as 
\begin{equation}\label{fevo}
\hat U(t+nT,t_0)=\hat U(t,t_0)\hat U^n(T,t_0),
\end{equation}
with $t\in[t_0,t_0+T]$, and furthermore, we can define the (Hermitian) {\it Floquet Hamiltonian} $\hat H_\mathrm{F}$ via the relation
\begin{equation}
\hat U(T,t_0)=e^{-i\hat H_\mathrm{F}T}.
\end{equation}
The eigenstates of this operator,
\begin{equation}
\hat U(T,t_0)|u_j(t_0)\rangle=e^{-i\epsilon_jT}|u_j(t_0)\rangle,    
\end{equation}
form a complete basis, and the exponents $\epsilon_j$ are called {\it quasi energies}\index{Quasi! energy} which are clearly defined modulo $2\pi$. A solution to the time-dependent Schr\"odinger equation can be written as
\begin{equation}
|\psi(t)\rangle=e^{-i\epsilon_j(t-t_0)}|u_j(t)\rangle,\hspace{1.2cm}|u_j(t+T)\rangle=|u_j(t)\rangle.
\end{equation}
A general solution can be expanded as
\begin{equation}
|\psi(t)\rangle=\sum_jC_je^{-i\epsilon_j(t-t_0)}|u_j(t)\rangle,\hspace{1.2cm}|u_j(t+T)\rangle=|u_j(t)\rangle,
\end{equation}
such that the coefficients $C_j$ determine the initial state $|\psi(t_0)\rangle$. Note the similarities between the Floquet theory\index{Floquet theory} and the Bloch theory for spatially periodic Hamiltonians.  

To achieve the formation of time crystals in non-disorder systems, Russomanno {\it et al.} considered a periodically $\delta$-kicked Lipkin-Meshkov-Glick model\index{Lipkin-Meshkov-Glick! model}\index{Model! Lipkin-Meshkov-Glick}~\cite{russomanno2017floquet}. What is important to note here is that the model is `infinite range', {\it i.e.} every spin interacts with every other spin, and that it supports a discrete $\mathbb{Z}_2$ symmetry. It was found that robust time-crystal dynamical phases can indeed exist in such systems. Having understood that infinite range interactions together with a $\mathbb{Z}_2$ symmetry can stabilize time-crystal phases, many people studied also the Dicke model~\cite{gong2018discrete,kessler2019emergent,zhu2019dicke,cosme2019time}. In particular, the question whether these states survive photon losses is a most relevant one~\cite{gong2018discrete}. At the mean-field level, bu studying the truncated Heisenberg equations, it was shown that time-crystal states exist for the driven Dicke model. At the quantum level, the discrete time-crystal states eventually decay, but their lifetimes may be long on experimental time-scales. The effects of other `imperfections' were addressed in~\cite{zhu2019dicke}, like dipole interactions among the atoms and non-uniform light-matter couplings. For ferromagnetic spin-spin couplings\index{Ferromagnetic! spin-spin coupling} the time-crystal state could survive. In refs.~\cite{kessler2019emergent,cosme2019time} the possible time-crystal states in the setup of transversely pumped Bose-Einstein condensates were considered. It was found~\cite{cosme2019time} that the time-crystal states were realizations of so-called {\it prethermal states}\index{Prethermalization} which are quasi-stationary many-body states that can survive over very long times~\cite{berges2004prethermalization,gring2012relaxation}. Most recently, experimental evidence of a time-crystalline phase was realized in such experiments which could be described by a periodically driven $SU(3)$ Dicke model~\cite{skulte2021driven,kongkhambut2021realization}. These pumped condensate systems will be discussed in great detail in sec.~\ref{ssec:mbcQED}. Kongkhambut and collaborators have very recently reported on the observation of spontaneous breaking of a continuous time translation symmetry\index{Spontaneous! symmetry breaking} in an atomic BEC inside a high-finesse optical cavity~\cite{Kongkhambut2022}. Since the observed limit cycles are robust against temporal perturbations, they evince the realization of a {\it continuous time crystal}\index{Time crystal! continuous}. At the same time, in the periodically-kicked JC model\index{Model! Jaynes-Cummings! periodically kicked}, chiral symmetry conserved by commensurate kicks preserves the phase of an initial quantum state against phase fluctuations during the dynamical evolution, in contrast to broken chiral symmetry (for incommensurate kicks) which erases the initial phase, as reported in~\cite{Qin2023}.

We close this section out by a brief account on the relevance of the Dicke and Tavis-Cummings models\index{Tavis-Cummings model} in the rapidly expanding field of polaritonic chemistry. Microcavity polaritons\index{Polariton} may be used to control chemical reactions with electromagnetic fields, which has been a long-standing goal in chemistry. Modifications in the reaction rates for chemical processes due to strong coupling between reactant molecular vibrations and the cavity vacuum have been reported in~\cite{Ahn2023}. The reaction-rate constants were extracted from evolving cavity-transmission spectra; a strong cavity frequency dependence was noted that closely followed the reactant infrared absorption spectrum. At the same time, it has been found that manipulating the light-matter coupling strength and cavity frequencies may either inhibit or accelerate the hydrogen-bond dissociation rate of water dimer by altering the vibrational dissociation channels~\cite{YuQi2023}. Additional enhancements of a dissociation probability can be expected for a cavity with thermal excitations or for multimode cavities\index{Multimode! cavity}~\cite{Triana2023}. 

A common feature of emitter-ensemble models developed to incorporate the details relevant to chemical reactivity (such as molecular vibrations along reaction coordinates) along with collective effects is the presence of dark states\index{Dark! state}\index{State! dark} we have met several times so far, which do not couple to the field of the cavity or an external field. In a recent report, Davidsson and Kowalewksi demonstrated that increasing the rate of pure dephasing induces a stronger coupling between the dark and polariton\index{Polariton} states~\cite{Davidsson2023}. The behaviour of a CO molecule together with a varying number of resonantly-coupled two-level emitters, which have their own individual dephasing operators, was studied via a quantum-trajectory\index{Quantum! trajectories} unraveling of the master equation\index{Master equation! unraveling}. 

\subsubsection{Generalizing the Dicke model to an extended medium}
\label{subsec:generalDickeext}

There is an equivalence between generalized Dicke models and interacting quantum models. Rom\'{a}n-Roche and coworkers present an analytical solution to quantum long-range models\index{Quantum!long-range model}~\cite{Mukamel2008, defenu2021longrange} in a lattice of $N$ sites to confirm that the mean-field solution is exact~\cite{Mori2010, Mori2012}. The system is described by a Hamiltonian of the form
\begin{equation}\label{eq:stronglongrange}
 \hat{H}_{\rm lr}=-\sum_{ij}^{N}J_{ij}\hat{C}_i \hat{C}_j,
\end{equation}
where $\hat{C}_i$ is a local Hermitian operator acting on site $i$. They consider systems with power-law decaying interactions of the form $J_{ij}=\Gamma \tilde{J}(\boldsymbol{r}_{ij})/\tilde{N}$, where $\tilde{N}=\sum_{i}\tilde{J}_{ij}$ is a renormalization factor independent of $j$ (following the Kac prescription) that restores extensivity and ensures a well-defined thermodynamic limit\index{Thermodynamic limit}, $\Gamma$ is the interaction strength and 
\begin{equation}
 \tilde{J}(\boldsymbol{r}_{ij})=\begin{cases}
                                 b \quad \text{if}\quad \boldsymbol{r}_{ij}=\boldsymbol{0}\\
                                 |\boldsymbol{r}_{ij}|^{-\alpha} \quad \text{otherwise}
                                \end{cases}
\end{equation}
The constant $\alpha$ sets the range of interactions compared to the dimensionality of the lattice $d$. The absence of additivity for $\alpha < d$ entails the emergence of particular statistical and dynamical phenomena in contrast to the ordinary short-range models, such as {\it quasi}stationary states, a negative specific heat or different phase diagrams when the system is studied within two different ensembles~\cite{Cohen2012}. The authors of~\cite{romanroche2023exact} introduce a generalized Hubbard-Stratonovich transformation~\index{Transformation!Hubbard-Stratonovich} to produce a closed expression for the free energy o the system at any temperature.

To demonstrate the method, it is helpful to first consider the classical Ising model, with
\begin{equation}
H_{\rm cl, Is}=h \sum_{i}^{N}s_i - \sum_{ij}^{N} J_{ij} s_i s_j,
\end{equation}
where $s_i$ is a discrete variable. Following the diagonalization of the interaction matrix as $J=\Lambda D \Lambda^{T}$, we can write the coupling term as $\sum_k D_k(\sum_i \Lambda_{ik} s_i)^2$, where $D_k$ are the eigenvalues of the interaction matrix. Consequently, the coupling is expressed as a sum of interaction terms which are quadratic in $\sum_i \Lambda_{ik} s_i$. These quadratic interactions are then eliminated via the Hubbard-Stratonovich transformation\index{Transformation!Hubbard-Stratonovich}, based on the following relation for the partition function involving Gaussian integrals\index{Gaussian! integral}:
\begin{equation}
 Z=\sum_{s_i}e^{-\beta H_{\rm cl, Is}} \propto \int du_k \sum_{s_i} \exp\bigg[-\beta\bigg(h \sum_i s_i + \sum_k u_k^2/D_k - 2 \sum_{ik}\Lambda_{ik}s_i u_k\bigg)\bigg],
\end{equation}
where $u_k$ are real auxiliary variables. Subject to some conditions imposed on the eigenspectrum of the $J_{ij}$ matrix, the integral over $u_k$ can be evaluated within the saddle-point approximation\index{Saddle-point! approximation} in the large $N$ limit. 

To extend the application of the Hubbard-Stratonovich transformation\index{Transformation!Hubbard-Stratonovich} into quantum models, one needs to reframe the partition function in terms of commuting quantities. Rom\'{a}n-Roche and his collaborators draw an equivalence between some quantum long-range models and a cavity QED model, namely the generalized Dicke model\index{Dicke! model!generalized} with
\begin{equation}\label{eq:genDicke}
 \hat{H}_{\rm gen.\,Dicke}=\sum_{k=0}^{M-1}\omega_k \hat{a}_k^{\dagger}\hat{a}_k + \hat{H}_0 - \sum_{k,i}(\hat{a}_k + \hat{a}_k^{\dagger})\frac{\lambda_{ik}}{\sqrt{N}}\hat{C}_i,
\end{equation}
where $\hat{H}_0$ is an exactly solvable Hamiltonian pertaining to the ``matter'' degrees of freedom, and $\hat{C}_i$ is the local Hermitian operator which couples the site $i$ to the bosonic modes, satisfying $[\hat{a}_{k}, \hat{a}_{k^{\prime}}^{\dagger}]=\delta_{k,k^{\prime}}$, with strength $\lambda_{ik}$. In the thermodynamic limit\index{Thermodynamic limit} ($N\to \infty$), integrating out the electromagnetic field modes yields an exact effective description for the matter degrees of freedom~\cite{RomanRoche2021, RomanRocheZueco2022} with a Hamiltonian
\begin{equation}\label{eq:HeffDickematter}
 \hat{H}_m^{\rm eff}=\hat{H}_0 - \sum_{ij}^{N}\sum_{k=0}^{M-1}\frac{\lambda_{ik}\lambda_{jk}}{N \omega_k} \hat{C}_i \hat{C}_j.
\end{equation}
The exact mapping is limited to the thermodynamic limit\index{Thermodynamic limit} as well as to a number of modes $M$ satisfying $\lim_{N\to\infty}M/N=0$. Equation~\eqref{eq:HeffDickematter} defines a quantum model with interaction coupling coefficients $(J_{\rm eff})_{ij}=\sum_{k=0}^{M-1}\lambda_{ik}\lambda_{kj}/(N\omega_k)$. For systems with translational invariance, the interaction matrix can be diagonalized in the Fourier space and the large-$N$ behaviour of the eigenvalues can be obtained analytically, while in the general case the model for an arbitrary interaction matrix $J$ is tractable if $\lim_{N\to \infty}(1/N)\sum_{k=0}^{M-1} D_k=\lim_{N\to \infty} (1/N){\rm Tr}(J)=0$. 

The authors of~\cite{romanroche2023exact} show that every strong long range model ($\alpha<d$) can be cast in the form of eq.~\eqref{eq:HeffDickematter}. For an arbitrary extensive model with $J_{ij}=\sum_{k=0}^{N-1} \Lambda_{ik} D_k \Lambda_{jk}$, the number $M$ of nonzero eigenvalues of $J$ (bounded by construction) scales with the size of the matrix, $N$, while for a model with a power-law decay of interactions and periodic boundary conditions, the number of nonzero eigenvalues (whence the number of nonzero modes) depends on the decay rate of the interaction, $\alpha$. For $\alpha<d$, only a small fraction of modes have a nonzero eigenvalue, therefore the condition $\lim_{N\to\infty}M/N=0$ is satisfied (see also fig. 2 of~\cite{romanroche2023exact}). Having now mapped the strong long-range quantum model of eq.~\eqref{eq:stronglongrange} to the generalized Dicke model of eq.~\eqref{eq:genDicke}, the solution follows the steps outlined by Wang and Hioe~\cite{wang1973phase, hioe1973phase}. For $N\to \infty$, the trace over the cavity modes is replaced by a set of complex Gaussian integrals\index{Gaussian! integral} with the mode operators giving their place to complex fields in the expression
\begin{equation}
 Z=\int \prod_{k=0}^{M-1}\frac{d^2\alpha_k}{\pi} {\rm Tr}_{\rm m}\left\{\exp\left[-\beta \left(\sum_{k=0}^{M-1}\omega_k |\alpha_k|^2 + \hat{H}_0 + \sum_{k,i}\frac{2\lambda_{ik}x_k}{\sqrt{N}}\hat{C}_i\right)\right]\right\},
\end{equation}
where the trace is taken over the ``matter'' degrees of freedom and we set $\alpha_k=x_k  + i y_k$. This expression for the partition function draws the parallel between the effective theory employing auxiliary bosonic modes and the standard Hubbard-Stratonovich transformation\index{Transformation!Hubbard-Stratonovich} employing auxiliary classical fields. The integral over $y_k$ yields an unimportant constant, while for the integral over the real parts, we perform a scale change of variables $u_k^2=x_k^2/N$ to define
\begin{equation}
 Z_m[u_k] \equiv Z_m (u_0,\ldots,u_{M-1})={\rm Tr}_m \bigg\{\exp\bigg[-\beta\bigg(\hat{H}_0 + \sum_{k,i}\frac{2\lambda_{ik}x_k}{\sqrt{N}}\hat{C}_i\bigg)\bigg]\bigg\}\quad \text{and}\quad f_m=\ln(Z_m[u_k])/N,
\end{equation}
thereby recasting the partition function in the form 
\begin{equation}
 Z=\int \prod_{k=0}^{M-1}\sqrt{\frac{N}{\pi \omega_k}}\, du_k\, \exp(N \phi[u_k]), \quad \text{with} \quad \phi[u_k]=-\beta \sum_{k=0}^{M-1}\omega_k u_k^2 + f_m[u_k].
\end{equation}
Since the exponent has an explicit linear dependence on $N$, the zero-order saddle-point approximation allows one to express the partition function in terms of the integrand at $\phi[\overline{u}_k]={\rm max}_{\{u_k\}}\phi[u_k]$. The resulting multivariate maximization problem is simplified by considering solutions which are homogeneous in the lattice -- these homogeneous solutions are proven to be the global maxima (see Appendices of~\cite{romanroche2023exact}). Second-order corrections to the zero-order term are negligible, whence the {\it exact expression for the partition function of strong long-range models} is $Z=\int \prod_{k=0}^{M-1}\sqrt{N/(\pi \omega_k)}\, du_k\, \exp(N \phi[\overline{u}_k])$.  

Carrying on with the theme of developing a multimode theory\index{Multimode! theory} for assessing collective behaviour in a quantum optical model, though marking a departure from the thermodynamic limit\index{Thermodynamic limit} and the relevance of mean-field dynamics, we will now examine spontaneous emission\index{Spontaneous! emission} from a line of atoms -- the simplest example of an extended medium -- briefly discussing an effective boson approximation for many atoms. The frequency and angular dependence of the spontaneous emission spectrum\index{Spontaneous! emission! spectrum} of two excited atoms alongside the time-dependent emission intensity as a function of the ratio of their radiative widths, detuning and Coulomb interaction is investigated in~\cite{Lehmberg1970, Lehmberg1970II, SmirnovJETP}. To determine the angular distribution of the photons emitted from a line of several atoms we will resort to a paradigmatic application of quantum trajectory theory\index{Quantum! trajectories} closely following~\cite{Clemens2003}. In fact, Dicke's discussion of the angular correlation of successive photons (see p. 108 of~\cite{dicke1954coherence}) is formulated based on the language of quantum trajectories, since he considers the detection of each emitted photon in the far field and constructs the state resulting after the emission of the first $s-1$ photons. The state in question is conditioned on the sequence of directions $\boldsymbol{k}_1, \ldots \boldsymbol{k}_{s-1}$ of those emissions. This construction corresponds to the density operator unraveling of Carmichael and Kim~\cite{CarmichaelKim2000}.  

Clemens and coworkers consider a system of $N$ identical two-state atoms located at positions $r_i$, $i=1,2,\ldots,N$ whose atomic dipole moments are all aligned along the direction $\hat{d}$. The atomic density operator obeys the Lehmberg-Agarwal\index{Master equation!Lehmberg-Agarwal} master equation~\cite{Lehmberg1970II} in the interaction picture (assuming the standard electric-dipole, Born-Markov and rotating wave approximations) 
\begin{equation}\label{eq:MEatomsline}
 \dot{\rho}=-i\sum_{i\neq j=1}^{N}\Delta_{ij}[\hat{\sigma}_{i+}\hat{\sigma}_{j-},\rho] + \tfrac{1}{2}\sum_{i,j=1}^{N}\gamma_{ij} (2\hat{\sigma}_{i+}\rho \hat{\sigma}_{i+} - \hat{\sigma}_{i+}\hat{\sigma}_{j-}\rho-\rho\hat{\sigma}_{i+}\hat{\sigma}_{j-}),
\end{equation}
with 
\begin{equation}
 \Delta_{ij}=\tfrac{3}{4}\gamma\left\{-[1-(\hat{d}\cdot\hat{r}_{ij})^2]\frac{\cos\xi_{ij}}{\xi_{ij}} + [1-3(\hat{d}\cdot\hat{r}_{ij})^2]\left(\frac{\sin\xi_{ij}}{\xi_{ij}^2} + \frac{\cos\xi_{ij}}{\xi_{ij}^3}\right)\right\}
\end{equation}
accounting for dipole-dipole interactions, and
\begin{equation}
 \gamma_{ij}=\tfrac{3}{2}\gamma\left\{[1-(\hat{d}\cdot\hat{r}_{ij})^2]\frac{\sin\xi_{ij}}{\xi_{ij}} + [1-3(\hat{d}\cdot\hat{r}_{ij})^2]\left(\frac{\cos\xi_{ij}}{\xi_{ij}^2} - \frac{\sin\xi_{ij}}{\xi_{ij}^3}\right)\right\}
\end{equation}
accounting for collective emission. In the above expressions, $\gamma$ is the Einstein coefficient $A$\index{Einstein $A$ and $B$ theory}, $\lambda_0$ is the resonant wavelength, $\xi_{ij}=k_0 r_{ij}=2\pi r_{ij}/\lambda_0$ and $\boldsymbol{r}_{ij}=r_{ij}\hat{r}_{ij}=\boldsymbol{r}_i-\boldsymbol{r}_j$. The pseudospin operators obey the commutation relations
\begin{equation}
 [\hat{\sigma}_{i+}, \hat{\sigma}_{j-}]=\delta_{ij}\hat{\sigma}_{iz}, \quad  [\hat{\sigma}_{i\pm}, \hat{\sigma}_{jz}]=\mp 2\delta_{ij}\hat{\sigma}_{i \pm}.
\end{equation}
We note that owing to the Markov approximation, the master equation~\eqref{eq:MEatomsline} does not fully account for propagation effects; the size of the atomic ensemble is restricted such that a time needed for a photon to traverse it must be much shorter than the characteristic timescale of collective radiative decay. 

The master equation is unraveled into quantum trajectories based on direct photodetection\index{Master Equation!unraveling}. We decompose the density matrix as a sum over pure states,
\begin{equation}
 \rho(t)=\sum_{\rm REC}P_{\rm REC}|\psi_c(t)\rangle \langle\psi_c(t)|,
\end{equation}
where ${\rm REC}$ is the record attributed to a particular sequence of photon emissions up to time $t$, and $P_{\rm REC}=\langle \overline{\psi}_c(t)|\overline{\psi}_c(t)\rangle$ is the probability density for that record to occur. The states $|\psi_c(t)\rangle$ and $|\overline{\psi}_c(t)\rangle$ are the normalized and un-normalized states, respectively, of the atoms, conditioned on the occurrence of the record ${\rm REC}$. The time evolution of $|\overline{\psi}_c(t)\rangle$ is governed by a non-Hermitian Hamiltonian interrupted by jumps generated by the times of photon emissions. For a general master equation\index{Master equation!  unraveling} in Lindblad form
\begin{equation}
 \dot{\rho}=\frac{1}{i\hbar}[\hat{H},\rho] + \sum_{i}(2\hat{O}_i \rho \hat{O}_i^{\dagger} - \hat{O}_i^{\dagger}\hat{O}_i \rho -\rho \hat{O}_i^{\dagger}\hat{O}_i),
\end{equation}
the non-Hermitian Hamiltonian is
\begin{equation}
 \hat{H}_B=\hat{H} - i\hbar \sum_{i}\hat{O}_i^{\dagger}\hat{O}_i,
\end{equation}
and the jumps are generated by the set of operators $\hat{O}_i$. We will outline two ways of identifying the jump operators, both casting the master equation in Lindblad form\index{Master equation}. In the first one, we define~\cite{CarmichaelKim2000}
\begin{equation}\label{eq:SopP}
 \hat{S}(\theta,\phi)=\sqrt{\gamma D(\theta,\phi)\,d\Omega}\,\sum_{j=1}^{N}e^{-ik_0 \hat{R}(\theta,\phi)\cdot \boldsymbol{r}_j}\,\hat{\sigma_{j-}},
\end{equation}
operators which apply when a photon is detected in the far field withing the elemental solid angle $d\Omega$ and in the direction $\hat{R}(\theta,\phi)$. The quantity
\begin{equation}
 D(\theta,\phi)=\frac{3}{8\pi}\{1-[\hat{d}\cdot \hat{R}(\theta,\phi)]^2\}
\end{equation}
is the dipole radiation pattern from a single isolated atom. For the ensemble of $N$ atoms, an additional angular dependence enters through the propagation phase factors in eq.~\eqref{eq:SopP}. When diagonalized in terms of these operators, the master equation~\eqref{eq:MEatomsline} is recast in the form
\begin{equation}
 \dot{\rho}=-i\sum_{i\neq j=1}^{N}\Delta_{ij}[\hat{\sigma}_{i+}\hat{\sigma}_{j-},\rho] + \tfrac{1}{2} \int [2\hat{S}(\theta,\phi) \rho \hat{S}^{\dagger}(\theta,\phi)-\hat{S}^{\dagger}(\theta,\phi)\hat{S}(\theta,\phi)\rho - \rho \hat{S}^{\dagger}(\theta,\phi)\hat{S}(\theta,\phi)],
\end{equation}
where the jump operators have a clear physical interpretation in terms of the detection of outgoing photons in the direction $\hat{R}(\theta,\phi)$. 

An alternative identification is provided by diagonalizing the matrix of coefficients $(\gamma_{ij})$~\cite{Ressayre1976, Ressayre1977}. The physical meaning of these operators stems from an expansion of the collective atomic polarization in source modes. Since the matrix $(\gamma_{ij})$ is a real and symmetric, it can be diagonalized by an orthogonal transformation of the form $(\gamma_{ij})=\boldsymbol{B}^{T} \boldsymbol{\Lambda} \boldsymbol{B}$, where $\boldsymbol{\Lambda}={\rm diag}(\lambda_1, \ldots,\lambda_N)$ is a diagonal matrix containing the eigenvalues of $(\gamma_{ij})$ with the corresponding normalized eigenvectors populating the columns of matrix $\boldsymbol{B}^{T}=(b_{ij})^{T}$, which are denoted by $\boldsymbol{b}_l=(b_{l1},\ldots,b_{lN})^{T}$. The master equation in this unraveling\index{Master equation! unraveling} assumes the form
\begin{equation}
 \dot{\rho}=-i\sum_{i\neq j=1}^{N}\Delta_{ij}[\hat{\sigma}_{i+}\hat{\sigma}_{j-},\rho] + \tfrac{1}{2}\sum_{l=1}^{N} (2 \hat{J}_{l}\rho \hat{J}_{l}^{\dagger} - \hat{J}_{l}^{\dagger}\hat{J}_{l}\rho - \rho \hat{J}_{l}^{\dagger}\hat{J}_{l}),
\end{equation}
where the source-mode collective jump operators are
\begin{equation}
 \hat{J}_l=\sqrt{\lambda_l}\, \boldsymbol{b}_l^T \hat{\boldsymbol{\Sigma}}, \quad  \hat{J}_l^{\dagger}=\sqrt{\lambda_{l}}\,\hat{\boldsymbol{\Sigma}}^{\dagger}\boldsymbol{b}_l,\quad \hat{\boldsymbol{\Sigma}} \equiv (\hat{\sigma}_{1-},\ldots,\hat{\sigma}_{N-})^{T}, \quad \hat{\boldsymbol{\Sigma}}^{\dagger} \equiv (\hat{\sigma}_{1+},\ldots,\hat{\sigma}_{N+}).
\end{equation}
Armed with these two unravelings, our aim is to trace the evolution of the angular distribution for photon emission and its potential deviation from the dipole radiation pattern -- which is observed for the first emitted photon -- to a directed distribution suggested by theories of superradiance; for this we need to calculate the angular distribution for the $k$th emitted photon, in other words the probability for that photon to be emitted within the solid angle $d\Omega$ in direction $(\theta, \phi)$.

Combining elements from both unravelings, the authors of~\cite{Clemens2003} adopt source-mode super-operators for the first $k-1$ emissions and the jump operator~\eqref{eq:SopP} only for the emission of the $k$th photon. Subsequently, they sum over a finite number $N$ of alternate series in the record on each emission. The explicit expression for the time evolution of the conditional wavefunction then reads
\begin{equation}
 |\overline{\psi}_{l_1,t_1;\ldots l_{k-1}, t_{k-1};\theta,\phi,t_k}\rangle=\hat{S}(\theta,\phi)\hat{B}(t_k-t_{k-1})\hat{J}_{l_{k-1}}\hat{B}(t_{k-1}-t_{k-2})\ldots \hat{J}_{l_1}\hat{B}(t_1)|\{+\}\rangle, 
\end{equation}
from which we calculate the probability as a sum over records covering all permutations of source modes and emission times for the first $k-1$ emissions,
\begin{equation}\label{eq:angdistqtr}
 P_k(\theta,\phi)\,d\Omega=\sum_{l_1=1}^{N}\ldots\sum_{l_{k-1}}^{N}\int_0^{\infty}dt_1\ldots\int_{t_{k-1}}^{\infty}dt_k\, \langle \overline{\psi}_{l_1,t_1;\ldots l_{k-1}, t_{k-1};\theta,\phi,t_k}|\overline{\psi}_{l_1,t_1;\ldots l_{k-1}, t_{k-1};\theta,\phi,t_k}\rangle, 
\end{equation}
where
$\hat{B}(\tau) \equiv \exp(-i\hat{H}_B \tau/ \hbar)$ and $|\{+\}\rangle$ denotes the state with all $N$ atoms excited. The non-Hermitian Hamiltonian governing the evolution between emissions is
\begin{equation}
 \hat{H}_B=\hbar \sum_{i\neq j=1}^{N}\Delta_{ij} \hat{\sigma}_{i+}\hat{\sigma}_{j-} - i\hbar \tfrac{1}{2} \sum_{l=1}^{N} \hat{J}_{l}^{\dagger}\hat{J}_{l}.
\end{equation}
It proves convenient to specialize to a line of atoms located along the $z$-axis, with position vectors $\boldsymbol{r}_j=(j-1)(s\lambda_0) \hat{z}$, $j=1,\ldots,N$, where $s$ is the interatomic separation in units of the resonant wavelength. The atomic dipole moments are all aligned perpendicular to the $z$-axis; due to the cylindrical symmetry the phase factors in eq.~\eqref{eq:SopP} are independent of $\phi$, and only the dipole radiation pattern exhibits a dependence on the azimuth. Integrating over the azimuthal angle we replace $\hat{S}(\theta,\phi)$ with
\begin{equation}\label{eq:newStheta}
 \hat{S}(\theta)=\sqrt{\gamma D(\theta) \sin\theta\, d\theta}\, \sum_{j=1}^{N}e^{-i2\pi(j-1)s\cos\theta}\hat{\sigma}_{j-}, \quad D(\theta)=\tfrac{3}{4}\left(1-\tfrac{1}{2}\sin^2\theta\right).
\end{equation}
From eq.~\eqref{eq:angdistqtr}, the angular distribution of the first photon emission is
\begin{equation}
 P_1(\theta)\sin\theta \,d\theta=\int_{0}^{\infty}dt_1 \langle \overline{\psi}_{\theta,t_1}|\overline{\psi}_{\theta,t_1} \rangle, \quad |\overline{\psi}_{\theta,t_1} \rangle =\hat{S}(\theta)\hat{B}(t_1)|\{+\}\rangle=e^{-(N\gamma/2)t_1}\hat{S}(\theta)|\{+\}\rangle.
\end{equation}
Substituting for $\hat{S}(\theta)$ from eq.~\eqref{eq:newStheta} we note that the interference terms have zero expectation in the excited state irrespective of the number of atoms. Therefore, $P_1(\theta)=D(\theta)$, {\it i.e.}, the first photon is always emitted following the dipole radiation pattern. Our aim now is to determine $P_k (\theta)$ for $k>1$.

The case of two atoms is special. Lehmberg obtained the dipole radiation pattern after solving the master equation\index{Master equation} in a standard way~\cite{Lehmberg1970II}, in an apparent contradiction with the directional correlations noted by Dicke~\cite{dicke1954coherence}. Let us now see what insight the quantum trajectory theory is able to offer in that observation. Appealing once more to eq.~\eqref{eq:angdistqtr}, we find that the angular distribution of the second emitted photon from the linear array of $N$ atoms is given by a sum over the $N$ possible jumps corresponding to the emission of the first photon, as
\begin{equation}\label{eq:P2theta}
\begin{aligned}
 P_2(\theta)\sin\theta\,d\theta=\sum_{l_1=1}^{N}\int_0^{\infty}dt_1 \int_{t_1}^{\infty}dt_2\,\langle \overline{\psi}_{l_1,t_1; \theta,t_2}| \overline{\psi}_{l_1,t_1; \theta,t_2} \rangle, \quad | \overline{\psi}_{l_1,t_1; \theta,t_2} \rangle&=\hat{S}(\theta)\hat{B}(t_2-t_1)\hat{J}_{l_1}\hat{B}(t_1)|\{+\}\rangle\\
 &=e^{-\gamma t_1}\hat{S}(\theta)\hat{B}(t_2-t_1)\hat{J}_{l_1}|\{+\}\rangle.
 \end{aligned}
\end{equation}
The state produced after the first jump,
\begin{equation}
 |l_1 \rangle \equiv \hat{J}_{l_1}|\{+\}\rangle=\sqrt{\lambda_{l_1}}\,\boldsymbol{b}_{l_1}^{T}\,\hat{\boldsymbol{\Sigma}}|\{+\}\rangle,
\end{equation}
is an eigenstate of the operator $\sum_{l=1}^{N} \hat{J}_{l}^{\dagger}\hat{J}_{l}$, with
\begin{equation}
 \left(\sum_{l=1}^{N} \hat{J}_{l}^{\dagger}\hat{J}_{l}\right) |l_1 \rangle=[(N-2)\gamma + \lambda_{l_1}] |l_1 \rangle.
\end{equation}
We now specify to a pair of atoms setting $N=2$; we write
\begin{equation}
 (\gamma_{ij})=\begin{pmatrix}
                \gamma & \Gamma \\
                \Gamma & \gamma
               \end{pmatrix},
\text{ with eigenvalues and eigenvectors } \lambda_{1,2}=\gamma \pm \Gamma, \quad \boldsymbol{b}_{1,2}=\tfrac{1}{\sqrt{2}} (1, \pm 1)^{T}
\end{equation}
and jump operators
\begin{equation}
 \hat{J}_1=\sqrt{\gamma+\Gamma}\, \frac{\hat{\sigma}_{1-} + \hat{\sigma}_{2-}}{\sqrt{2}}, \quad  \hat{J}_2=\sqrt{\gamma-\Gamma}\, \frac{\hat{\sigma}_{1-} - \hat{\sigma}_{2-}}{\sqrt{2}}.
\end{equation}
In the two-atom case only, the state $|l_1\rangle$ is an eigenstate of the dipole-dipole interaction Hamiltonian $\hbar \sum_{i\neq j=1}^{N}\Delta_{ij} \hat{\sigma}_{i+}\hat{\sigma}_{j-}$ with eigenvalues $\pm \hbar \Delta_{12}$, whence also of the non-Hermitian Hamiltonian $\hat{H}_B$, with eigenvalues $-i\hbar\lambda_{l_1}/2 \pm \hbar \Delta_{12}$. Computing now the evolution between jumps and substituting in eq.~\eqref{eq:P2theta}, we obtain the angular distribution of the second emitted photon as
\begin{equation}
 P_2(\theta)=D(\theta)\tfrac{1}{2} \sum_{l_1=1}^{2} \lambda_{l_1}^{-1} \langle l_1 |\hat{\sigma}_{1+}\hat{\sigma}_{1-} + \hat{\sigma}_{2+}\hat{\sigma}_{2-} + e^{-i\zeta}\hat{\sigma}_{1+}\hat{\sigma}_{2-} + e^{i\zeta}\hat{\sigma}_{2+}\hat{\sigma}_{1-}|\l_1 \rangle,
\end{equation}
with $\zeta \equiv 2\pi s \cos\theta$. Although the interference terms play a role in individual trajectories, allowing for the correlation noted by Dicke, their contribution cancels in the sum over records and we recover the dipole radiation pattern according to Lehmberg. Analytical expressions for the stationary density matrix of and system observables pertaining to the fluorescent emission from two interacting nonidentical dipoles under coherent driving have been recently obtained in~\cite{VivasViana2021}. For a three-atom array, on the other hand, Clemens and coworkers find that the nine individual trajectories do not cancel in the sum and there is a distortion of the dipole radiation pattern. The most characteristic property is the enhanced directionality of the third emitted photon, with an increased probability for emission near the axis of the atomic emission~\cite{Clemens2003}. In numerical simulations, just before the time $t_k$ when the $k$-jump is executed, the Monte Carlo average\index{Monte Carlo!average} of the norm $\langle \overline{\psi}_c(t_k)|\hat{S}^{\dagger}(\theta)\hat{S}(\theta)|\overline{\psi}_c(t_k)\rangle$ determines the angular distribution of the $k$th emitted photon. Numerical results for $N\approx 10$ atoms showed a correlation between increased directionality and longer waiting times, in contrast to what is expected from superradiance. 

To take the calculations significantly beyond $N\approx 10$ it is clear that some approximation is warranted. First, we note that a relationship between the collective operators $\hat{S}(\theta)$ and $\hat{J}_l$ can be established due to the fact that both are expansions over the single-atom operators $\hat{\sigma}_{i-}$, $i=1,2,\ldots,N$. We write the phase factors in vector form as
\begin{equation}\label{eq:Pivector}
 \boldsymbol{\Pi}(\theta) \equiv \begin{pmatrix}
                                  \pi_1(\theta)\\
                                  \vdots\\
                                  \pi_N(\theta)
                                 \end{pmatrix}\equiv
                                 \begin{pmatrix}
                                  1\\
                                  e^{-i2\pi s\cos\theta}\\
                                  \vdots\\
                                  e^{-i2\pi(N-1)s \cos\theta}
                                 \end{pmatrix}
\text{ such that } \hat{S}(\theta)=\sqrt{\gamma D(\theta)d\theta}\,\boldsymbol{\Pi}^{T}(\theta)\hat{\boldsymbol{\Sigma}}.
\end{equation}
We also collect the source-mode annihilation and creation operators together in row and column vectors, defining
\begin{equation}\label{eq:Jvector}
 \hat{\boldsymbol{J}} \equiv \begin{pmatrix}
                              \hat{J}_1\\
                              \vdots\\
                              \hat{J}_N
                             \end{pmatrix}=\sqrt{\boldsymbol{\Lambda}} \boldsymbol{B} \hat{\boldsymbol{\Sigma}}, \quad\quad \hat{\boldsymbol{J}}^{\dagger} \equiv (\hat{J}_1^{\dagger},\ldots, \hat{J}_N^{\dagger})=\hat{\boldsymbol{\Sigma}}^{\dagger}\boldsymbol{B}^{T}\sqrt{\boldsymbol{\Lambda}}.
\end{equation}
Equation~\eqref{eq:Jvector} can be inverted and since $\boldsymbol{B}$ is an orthogonal matrix we have
\begin{equation}
 \hat{\boldsymbol{\Sigma}}=\boldsymbol{B}^{T} \sqrt{\mathbf{\Lambda}^{-1}}\,\hat{\boldsymbol{J}}, \quad \quad  \hat{\boldsymbol{\Sigma}}^{\dagger}=\hat{\boldsymbol{J}}^{\dagger}\sqrt{\mathbf{\Lambda}^{-1}}\boldsymbol{B}
\end{equation}
and then substituting for $\hat{\boldsymbol{\Sigma}}$ in eq.~\eqref{eq:Pivector} we arrive at the expansion
\begin{equation}\label{eq:connectionJS}
 \hat{S}(\theta)=\sqrt{\gamma D(\theta) \sin\theta \, d\theta}\, \boldsymbol{\Xi}^{T}(\theta) \sqrt{\mathbf{\Lambda}^{-1}}\,\hat{\boldsymbol{J}}, \quad \text{ with } \boldsymbol{\Xi}(\theta) \equiv (\xi_1(\theta),\ldots,\xi_N(\theta))^{T}=\boldsymbol{B} \boldsymbol{\Pi}(\theta).
\end{equation}
We note that $\boldsymbol{\Xi}^{T}(\theta)\boldsymbol{\Xi}(\theta)=\boldsymbol{\Pi}^{\dagger}(\theta)\mathbf{B}^{T}\mathbf{B} \boldsymbol{\Pi}(\theta)=N$ due to the orthogonality of $\boldsymbol{B}$. We also observe that the coefficients $\xi_i(\theta)$ featuring in the expansion~\eqref{eq:connectionJS} can be expressed as off-diagonal matrix elements of $\hat{S}(\theta)$ taken with respect to the vacuum state and the one-quantum collective excitations $|l\rangle \equiv (J_l^{\dagger}/\sqrt{\lambda_l})|\{-\}\rangle$:
\begin{equation}
 \frac{\langle\{-\}|\hat{S}(\theta)\hat{\boldsymbol{J}}^{\dagger}\sqrt{\mathbf{\Lambda}^{-1}}|\{-\}\rangle}{\sqrt{\gamma D(\theta)\sin\theta\,d\theta}}=\boldsymbol{\Xi}^T(\theta) \boldsymbol{B} \langle\{-\}|\hat{\boldsymbol{\Sigma}}\hat{\boldsymbol{\Sigma}}^{\dagger}|\{-\}\rangle=\boldsymbol{\Xi}^T(\theta),
\end{equation}
where we have used 
\begin{equation}\label{eq:sumrule1}
 \langle \{-\}|\hat{\boldsymbol{\Sigma}}\hat{\boldsymbol{\Sigma}}^{\dagger}|\{-\}\rangle=\boldsymbol{I}_N
\end{equation}
and the orthogonality of $\boldsymbol{B}$. Hence, the expansion coefficients read
\begin{equation}
 \xi_l(\theta)=\frac{\langle\{-\}|\hat{S}(\theta)|l\rangle}{\sqrt{\gamma D(\theta)\sin\theta\,d\theta}}.
\end{equation}
From the above formula, the angular distribution of the emission originating from the one-quantum excitation $|l\rangle$ can be expressed in terms of $\xi_l(\theta)$. When the dipole-dipole interactions ($\propto\Delta_{ij}$) are neglected, $|l \rangle$ is an eigenstate of $\hat{H}_B$, with
\begin{equation}
 \hat{H}_B|l\rangle=-i\hbar \tfrac{1}{2}\hat{\boldsymbol{J}}^{\dagger}\hat{\boldsymbol{J}}|l\rangle=-i\hbar \tfrac{1}{2}\lambda_l |l\rangle.
\end{equation}
We deduce that $\hat{B}(t_1)=\exp(-\lambda_l t_1/2)|l\rangle$, hence the angular distribution of emission with initial state $|l\rangle$ is
\begin{equation}\label{eq:Qltheta}
 Q_l(\theta)\sin\theta\,d\theta=\int_0^{t_1}dt_1 \langle l|\hat{B}^{\dagger}(t_1)[\hat{S}^{\dagger}(\theta)\hat{S}(\theta)]\hat{B}(t_1)|l\rangle=(\gamma/\lambda_l)D(\theta) |\xi_l(\theta)|^2 \sin\theta\,d\theta.
\end{equation}
A weighted sum of these distributions returns the dipole radiation pattern:
\begin{equation}\label{eq:sumrule2}
 \sum_{l=1}^{N}\lambda_l Q_l(\theta)=\gamma D(\theta) \sum_{l=1}^N |\xi_l(\theta)|^2=N\gamma D(\theta),
\end{equation}
from where we see that $|\xi_l(\theta)|^2$ indicates the difference between the emission from the collective excitation $|l\rangle$ and the dipole radiation pattern. Integrating~\eqref{eq:sumrule2} over $\theta$ yields the sum rule $\sum_{l=1}^{N}\lambda_l=N\gamma$.  

Regarding now the emission properties of collective excitations near the fully excited state $|\{+\}\rangle$, we have already found that the first photon is emitted according to the dipole distribution; considered in terms of the source modes this follows from the\index{Sum rule!Collective spontaneous emission}\index{Spontaneous! emission! collective} sum rules~\eqref{eq:sumrule1} and~\eqref{eq:sumrule2}. To make this point clear we write
\begin{equation}\label{eq:SdagSQ}
 \frac{\langle\{+\}|\hat{S}^{\dagger}(\theta)\hat{S}(\theta)|\{+\}\rangle}{\sin\theta\,d\theta}=\gamma D(\theta)\langle\{+\}|\hat{\boldsymbol{J}}^{\dagger}\sqrt{\boldsymbol{\Lambda}^{-1}}\boldsymbol{\Xi}^{*}(\theta)\boldsymbol{\Xi}^{T}(\theta)\sqrt{\boldsymbol{\Lambda}^{-1}}\hat{\boldsymbol{J}}|\{+\}\rangle=\gamma D(\theta)\sum_{l=1}^{N}|\xi_l(\theta)|^2=\sum_{l=1}^N \lambda_l Q_{l}(\theta),
\end{equation}
where we have used $\langle\{+\}|\hat{\boldsymbol{J}}^{\dagger}\hat{\boldsymbol{J}}|\{+\}\rangle=\boldsymbol{\Lambda}$. Without loss of generality, we assume that the first photon is emitted from the source mode $l_1$. We would now like to determine the angular distribution of the second emitted photon given the first emission from source mode $l_1$, with $l_1=\hat{J}_{l_1}|\{+\}\rangle$. We write
\begin{equation}\label{eq:condprobP2}
 \frac{P_2^{l_1}(\theta)\sin\theta\,d\theta}{\lambda_{l_1}/(N\gamma)}=\lambda_{l_1}^{-1} \int_{0}^{\infty} d\tau_2 \langle l_1|\hat{B}^{\dagger}(\tau)[\hat{S}^{\dagger}(\theta)\hat{S}(\theta)]\hat{B}(\tau)|l_1\rangle,
\end{equation}
where $\tau=t_2-t_1$. The quantity $P_2^{l_1}(\theta)\sin\theta\,d\theta$ is the joint probability\index{Joint probability} for the first photon to be emitted by mode $l_1$ {\it and} the second to be emitted in the direction $\theta$, while the denominator, $\lambda_l/(N\gamma)$ is the probability that the first photon is emitted by mode $l_1$. It is not straightforward to evaluate the integral in the expression~\eqref{eq:condprobP2} since $|l_1\rangle$ is not an eigenstate of $\hat{H}_B$, even with dipole-dipole interactions neglected. We are able though to determined the emission rate for small $\tau_2$. The conditional emission rate per unit solid angle is
\begin{equation}
 \begin{aligned}
R_2^{l_1}(\theta) &\equiv \frac{\langle \{+\}|\hat{J}_{l_1}^{\dagger}[\hat{S}^{\dagger}(\theta)\hat{S}(\theta)]\hat{J}_{l_1}|\{+\}\rangle}{\lambda_{l_1}\sin\theta\,d\theta}=\gamma D(\theta)\sum_{i,j=1}^{N}b_{l_1 i}b_{l_1 j} \sum_{n,m=1}^{N}\pi_n^{*}(\theta)\pi_m(\theta) \langle\{+\}| \hat{\sigma}_{i+}\hat{\sigma}_{n+}\hat{\sigma}_{m-}\hat{\sigma}_{j-}|\{+\}\rangle\\
&=\gamma D(\theta)\left[\left(\sum_{i,n=1}^{N} b_{l,i}^2 |\pi_n (\theta)|^2 -  \sum_{i=1}^{N} b_{l,i}^2 |\pi_i (\theta)|^2 \right) + \left(\sum_{i,n=1}^{N} b_{l_1 i}\pi_{i}(\theta)b_{l_1 n}\pi_n^{*}(\theta) - \sum_{i=1}^{N}b_{l,i}^2|\pi_i(\theta)|^2 \right)\right].
 \end{aligned}
\end{equation}
Every sum apart from the third is trivially constant since $|\pi_i(\theta)|^2=1$, and we arrive at the result
\begin{equation}\label{eq:shorttimesD}
 R_2^{l_1}(\theta)=\gamma D(\theta)(N-2 + |\xi_{l_1}(\theta)|^2).
\end{equation}
Equation~\eqref{eq:shorttimesD} is a generalization of the stimulated emission enhancement\index{Stimulated! emission! enhancement} factor noted by Dicke~\cite{dicke1954coherence}. Dicke remarked that following the first emission in direction $\hat{\boldsymbol{k}}_1$, the rate for the second emission is enhanced by a factor of two over that for spontaneous emission\index{Spontaneous! emission! independent atoms} from independent atoms. To interpret eq.~\eqref{eq:shorttimesD}, we observe that the rate $R_2^{l_1}(\theta)$ may be divided into a factor $(N-1)\gamma D(\theta)$ accounting for emission from $N-1$ independent atoms, plus an ``enhancement factor'' equal to $\gamma D(\theta)(|\xi_{l_1}(\theta)|^2-1)$ whose angular distribution is that of the source mode from which the first atom was emitted. In fact, the enhancement may be either positive or negative. Since the integral of either $D(\theta)$ or $Q_{l_1}(\theta)$ is unity, we have
\begin{equation}\label{eq:SupersubM}
 \int_{0}^{\pi}\sin\theta\,d\theta \gamma D(\theta)(|\xi_{l_1}(\theta)|^2-1))=\lambda_{l_1}-\gamma,
\end{equation}
where we have used eq.~\eqref{eq:Qltheta}. For super-radiant source modes ($\lambda_{l_1} > \gamma$), eq.~\eqref{eq:SupersubM} indicates a positive enhancement, whereas for subradiant\index{Subradiance} modes ($\lambda_{l_1} < \gamma$) an inhibition occurs. 

While we have so far identified the directionality of certain collective excitations of the atoms, nothing has been said about the dynamical process, about the evolution between the jumps and the sequence of states visited as the spontaneous emission\index{Spontaneous! emission} proceeds. We will now formulate an approximation to address these limitations. This approximation\index{Boson approximation} stems from the observation that the source-mode operators act like boson operators near the fully excited state $|\{+\}\rangle$ of near the ground state $|\{-\}\rangle$. This follows from the fact that the sets of operators $\sqrt{\boldsymbol{\lambda}^{-1}}\,\hat{\boldsymbol{J}}$ and $\hat{\boldsymbol{J}}^{\dagger}\,\sqrt{\boldsymbol{\lambda}^{-1}}$ are formed from orthogonal eigenvectors of the symmetric matrix $(\gamma_{ij})$. Using the commutation relation $[\hat{\sigma}_{i-}, \hat{\sigma}_{j+}]=\delta_{ij}\hat{\sigma}_{iz}$, we can evaluate the commutator
\begin{equation}
 (\lambda_{l}\lambda_{l^{\prime}})^{-1/2}[\hat{J}^{\dagger}_{l}, \hat{J}_{l^{\prime}}]=\sum_{i=1}^{N}b_i^{l}b_i^{l^{\prime}} \hat{\sigma}_{iz}.
\end{equation}
Acting with this commutator on states $|\{+\}\rangle$ and $|\{-\}\rangle$ gives
\begin{equation}
  (\lambda_{l}\lambda_{l^{\prime}})^{-1/2}[\hat{J}^{\dagger}_{l}, \hat{J}_{l^{\prime}}]|\{\pm\}\rangle=\pm \boldsymbol{b}_l^{T} \boldsymbol{b}_{l^{\prime}}|\{\pm\}\rangle=\pm \delta_{l l^{\prime}}|\{\pm\}\rangle. 
\end{equation}
This expression shows that $\hat{J}^{\dagger}_{l}$ and $\hat{J}_{l^{\prime}}$ act like the annihilation (creation) and creation (annihilation) operators of independent boson modes in state $|\{+\}\rangle$ ($|\{-\}\rangle$). More generally, if the commutator above acts on a state that is $n\ll N$ excitations below $|\{+\}\rangle$ or $n\ll N$ excitations above $|\{-\}\rangle$, the result will be accurate to within a correction of order $n/N$. For large $N$ and excitations close to the fully-excited state, we can propose the following boson approximation:
\begin{equation}\label{eq:bosonapp1}
 \hat{J}_{l} \to \sqrt{\lambda_l}\,\hat{a}^{\dagger}_{l}, \quad \text{with} \quad [\hat{a}_l, \hat{a}^{\dagger}_{l^{\prime}}]=\delta_{l, l^{\prime}}.
\end{equation}
It follows that the boson creation operators $\hat{a}_l^{\dagger}$ create delocalized holes within a population of many excited atoms. Conversely, close to the ground state, Clements and coauthors~\cite{Clemens2003} propose
\begin{equation}\label{eq:bosonapp2}
  \hat{J}^{\dagger}_{l} \to \sqrt{\lambda_l}\,\hat{b}^{\dagger}_{l}, \quad \text{with} \quad [\hat{b}_l, \hat{b}^{\dagger}_{l^{\prime}}]=\delta_{l, l^{\prime}}.
\end{equation}
The boson operators $\hat{b}_l^{\dagger}$ create delocalized excitations\index{Delocalized! excitations} within a population of many atoms which are unexcited. We note that, without the boson approximation, the source-mode operators only define a simplifying algebra in the single-mode limit or in the Dicke limit. In the latter ($s\to 0$), there is one superradiant eigenvalue $\lambda_N=N\gamma$ while the rest of the eigenvalues are zero. $\hat{J}_N$ and $\hat{J}^{\dagger}_N$ are angular-momentum operators and the emission follows a straightforward sequence of jumps between the angular-momentum states $|J,M\rangle$, with $J=N/2$ and $M=-N/2,\ldots,N/2$. The conditional state following the $k$th jump is $|\psi_c(t_k^{+})\rangle=|N/2, N/2-k\rangle$, and the evolution between jumps brings a trivial rescaling of the norm. 

The boson approximation~\eqref{eq:bosonapp1} enormously simplifies the algebra\index{Boson approximation} in the general multimode case\index{Multimode! Boson approximation}. It returns us to a situation where the operator algebra is simple and the book-keeping of the conditional state evolution is straightforward. The conditional state after the $k$th emission is 
\begin{equation}
 |\psi_{l_1,t_1;\ldots;l_k,t_k}\rangle=\prod_{l=1}^{N}|n_l\rangle \equiv |\{n_l\}_k\rangle,
\end{equation}
where $\sum_{l=1}^{N}n_l=k$ and $|\{n_l\}_k\rangle$ is a multimode Fock state. The occupation numbers $n_l$, $l=1,\ldots,N$ count the holes created in the population of $N$ initially excited atoms. It follows that the conditional emission rate per unit solid angle for the $(k+1)$th emitted photon is
\begin{equation}\label{eq:Rbosonapp}
 R_{k+1}^{l_1,\ldots,l_k}(\theta)=\gamma D(\theta) \langle\{n_l\}_k|\hat{\boldsymbol{A}}^{\dagger}\sqrt{\boldsymbol{\Lambda}^{-1}} \boldsymbol{\Xi}^{*}(\theta)\boldsymbol{\Xi}^{T}(\theta)\sqrt{\boldsymbol{\Lambda}^{-1}}\hat{\boldsymbol{A}}|\{n_l\}_k\rangle=\gamma D(\theta) \sum_{l=1}^{N} |\xi_l(\theta)|^2 (n_l +1).
\end{equation}
In the above expression we have defined
\begin{equation}
 \hat{\boldsymbol{A}}\equiv \begin{pmatrix}
                             \sqrt{\lambda_1} \hat{a}_1^{\dagger}\\
                             \vdots\\
                             \sqrt{\lambda_N} \hat{a}_N^{\dagger}
                            \end{pmatrix}
,\quad \quad  \hat{\boldsymbol{A}}^{\dagger}\equiv (\sqrt{\lambda_1}\hat{a}_1,\ldots,\sqrt{\lambda_N}\hat{a}_N).
\end{equation}
We find that after a first emission from source mode $l_1$, the angular distribution for the second emission is
\begin{equation}
 R_2^{l_1}(\theta)=\gamma D(\theta) (N + |\xi_{l_1}(\theta)|^2),
\end{equation}
a rate which differs from the exact result [eq.~\eqref{eq:shorttimesD}] by $\mathcal{O}(1/N)$. In place of eq.~\eqref{eq:SupersubM} it yields the ``enhancement factor''
\begin{equation}
 \int_{0}^{\pi}\sin\theta\,d\theta \gamma D(\theta)(|\xi_{l_1}(\theta)|^2+1))=\lambda_{l_1}+\gamma.
\end{equation}
When considering the distribution of the one-quantum excitation, we note that the approximation~\eqref{eq:bosonapp2} reproduces the exact result of eq.~\eqref{eq:Qltheta}. In the boson approximation, the distributions $Q_l(\theta)$, $l=1,2,\ldots,N$ characterize quite generally the angular emission properties of the source modes. Combining eqs.~\eqref{eq:Qltheta} and eq.~\eqref{eq:Rbosonapp} we arrive at a generalization of eq.~\eqref{eq:SdagSQ},
\begin{equation}\label{eq:condemrate}
  R_{k+1}^{l_1,\ldots,l_k}(\theta)=\sum_{l=1}^{N}\lambda_l Q_l(\theta)(n_l+1),
\end{equation}
underpinning the bosonic stimulation from the hole occupation numbers $n_l$.

We are finally in position to consider the dynamical process that produces directed emission. Equation~\eqref{eq:condemrate} gives the conditional emission rate per unit solid angle, summed over source modes, describing a stochastic evolution similar to the mode competition\index{Mode! competition} in a laser. From eq.~\eqref{eq:Rbosonapp} we find that at time $t$ the mean-directed emission rate per unit solid angle is given by 
\begin{equation}\label{eq:stochint}
 \frac{\langle\hat{S}^{\dagger}(\theta)\hat{S}(\theta)\rangle\,(t)}{\sin\theta\,d\theta}=\overline{\langle\{n_l\}_t|\hat{\boldsymbol{A}}^{\dagger}\sqrt{\boldsymbol{\Lambda}^{-1}} \boldsymbol{\Xi}^{*}(\theta)\boldsymbol{\Xi}^{T}(\theta)\sqrt{\boldsymbol{\Lambda}^{-1}}\hat{\boldsymbol{A}}|\{n_l\}_t\rangle}=\gamma D(\theta)\sum_{l=1}^{N}|\xi_i(\theta)|^2 e^{\lambda_l t},
\end{equation}
where $\{n_{l}\}_{t}$ is the set of hole occupation numbers at time $t$, while the overbar takes the stochastic average over $\{n_l\}_{t}$. To a good approximation, the superradiant eigenvalues, $2Ns$ in total, are determined by the dipole radiation distribution and the interatomic spacing $s$, according to the formula (for $2Ns$ integer)
\begin{equation}
 \lambda_l/\gamma=(1/s)D(\theta_l),\quad \text{with} \quad \cos\theta_l=\begin{cases}
                                                                         (q-1)/(2Ns), \,\,q=1,3,\ldots\\
                                                                         (q-2)/(2Ns),
                                                                         \,\,q=2,4,\ldots
                                                                        \end{cases}, \quad l=N-2Ns+q \quad \text{and} \quad q=1,\ldots,2Ns
\end{equation}
The intensity distribution given by eq.~\eqref{eq:stochint} is proportional to the dipole distribution $D(\theta)$ for $t=0$ (since $\sum_{l=1}^{N} |\xi_l (\theta)|^2=N$), and for long times approaches the distribution $Q_N(\theta)$ of the source mode with the largest eigenvalue. Monte Carlo averages show that as long as the amplification continues, the different amplification rates for different source modes eventually concentrate the emission into a cone near the line axis where the atoms are located. There is a hole right on the axis, though, where the emission is dominated by the subradiant modes\index{Subradiance}. Emission from individual source modes can be associated with the ``rays''\index{Rays!in collective atomic emission} of~\cite{ErnstStehle1968}. The boson approximation, however, does not account for energy depletion as the excitation in the atomic ensemble runs out, without which it is impossible to describe a superradiant pulse. Moreover, no appreciable directionality will develop if the $N$ initial quanta are emitted before the onset of sufficient amplification.  

It is thus necessary to modify the simple boson approximation in order to make a reliable description of the development of directionality, by enforcing energy depletion. This is achieved by altering the source-mode jump rates. In the boson approximation, the jump rates after the $k$th photon emission are given by the substitution
\begin{equation}\label{eq:jumpbapp}
 \langle \hat{J}_l^{\dagger}\hat{J}_l \rangle \to \lambda_l \langle \{n_{l^{\prime}}\}_k |\hat{a}_l \hat{a}_l^{\dagger}|  \{n_{l^{\prime}}\}_k\rangle=\lambda_l (n_l+1), \quad \text{with} \quad l=1,\ldots,N.
\end{equation}
Summing over these rates we obtain the net rate of eq.~\eqref{eq:condemrate}, integrated over the angle $\theta$. The jumps here are performed by the hole creation operators $\hat{a}_l^{\dagger}$. Let us consider the single-mode limit, where there is one nonzero eigenvalue $\lambda_N$ with jump operator
\begin{equation}
 \hat{J}_N=\sqrt{\lambda_N/N}\,\hat{J}_{-}, \quad \text{with} \quad \hat{J}_{-}=\sum_{j=1}^{N}\hat{\sigma}_{j-}.
\end{equation}
We now adopt the Schwinger representation\index{Schwinger-boson mapping}~\cite{Sakurai1994}, writing the Dicke collective operators as
\begin{equation}
 \hat{J}_{-}=\hat{a}^{\dagger}\hat{b}, \quad \hat{J}_{+}=\hat{a}\hat{b}^{\dagger}, \quad \hat{J}_{z}=\hat{b}^{\dagger}\hat{b}-\hat{a}^{\dagger}\hat{a},
\end{equation}
with $\hat{a}^{\dagger}\hat{a} + \hat{b}^{\dagger}\hat{b}=N$ and $\hat{a}, \hat{b}$ independent boson modes. Here $\hat{a}^{\dagger}$ creates an atom in the ground state while $\hat{b}$ annihilates an excited-state atom. The conditional state after $k$ emissions is then represented as the two-boson number state $|k\rangle_a|N-k\rangle_b$ and the exact jump rate is calculated as
\begin{equation}
 \langle \hat{J}_N^{\dagger}\hat{J}_N \rangle=\lambda_N N^{-1}\langle (\hat{a}\hat{a}^{\dagger})(\hat{b}^{\dagger}\hat{b}) \rangle=\lambda_N (1-k/N)(n_N+1).
\end{equation}
The boson approximation overlooks then the depletion factor in the jump rate $(1-k/N)$ through replacing $|N-k\rangle_b$ by $|N\rangle_b$. 

To account for energy depletion in the multimode case, we replace the jump operator $\hat{a}_l^{\dagger}$ by $\hat{a}_l^{\dagger} \hat{b}$; the number of atoms remaining in the excited state is $\hat{b}^{\dagger}\hat{b}=N-\sum_{l=1}^{N}\hat{a}_l^{\dagger}\hat{a}_l$. The conditional state after $k$ emissions is then written as 
\begin{equation}
 |\psi_{l_1,t_1;\ldots;l_k,t_k}\rangle=|\{n_l\}_k\rangle_a |N-k\rangle_b,
\end{equation}
and eq.~\eqref{eq:jumpbapp} is replaced by
\begin{equation}\label{eq:jumpbapp2}
 \langle \hat{J}_l^{\dagger}\hat{J}_l \rangle \to \lambda_l(1-k/N)(n_l+1).
\end{equation}
This way, every photon emission reduces the future emission rate of all source modes, whence the modes are not independent. The approximation becomes exact in the single-mode limit. While it accounts for a superradiant pulse, this modification to the boson approximation fails to produce a directed subradiant emission in the tail of the pulse. Another somewhat {\it ad hoc} (yet exact in the single-mode limit) modification is called for, recognizing that the emission process develops in two phases, an initial phase where stimulated creation\index{Stimulated! creation} dominates, and a second phase at the end of the pulse where there is a decay of coherence due to the hole creation in the amplification phase. After half of the photons have been emitted, the jump operator $\hat{a}_l^{\dagger} \hat{b}$ should change to $\hat{a}_l^{\dagger} \hat{b}_l$ where we adopt the conditional state (for $k\geq N/2$)  
\begin{equation}
 |\psi_{l_1,t_1;\ldots;l_k,t_k}\rangle=|\{n_l\}_k\rangle_a |\{m_l\}_k\rangle_b, \quad \text{with} \quad \{m_l\}_k\equiv 2 \{n_l\}_{N/2}-\{n_l\}_k. 
\end{equation}
The jump rates for the second half of the pulse read $\langle \hat{J}_l^{\dagger}\hat{J}_l \to \lambda_l m_l(n_l+1)/(2m_l^{\rm (max)})$, which allows independent emission from the source modes. To achieve this independence, one replaces the state $|\{n_l\}_{N/2}\rangle_a |N/2\rangle_b$ reached after $N/2$ emissions by the state $|\{n_l\}_{N/2}\rangle_a |\{n_l\}_{N/2}\rangle_b$, effectively equating the number $n_l$ of excited state holes created in mode $l$ in the course of the pulse growth with the number of grounds-state excitations which are to be annihilated during its decay. With this modification, the directed subradiant emission is reproduced at the tail of the pulse (see Figs. 12-14 of~\cite{Clemens2003}).

\subsubsection{``Poor man's'' models}\label{sssec:poormodel}
Universality\index{Universality} in physics postulates that certain quantities may not depend on the details of a system, but instead on global properties like dimensions and symmetries~\cite{cardy1996scaling,goldenfeld2018lectures}. This leads, for example, to the classification of continuous phase transitions\index{Phase transition! continuous} into {\it universality classes}\index{Universality! class} characterized by their critical exponents\index{Critical! exponents}. In the Ginzburg-Landau\index{Ginzburg-Landau theory} mean-field description of critical phenomena, an {\it order parameter}\index{Order parameter} $\phi(\lambda)$ is introduced to distinguish the {\it normal} phase from the {\it symmetry broken} phase\index{Symmetry breaking}
\begin{equation}
\phi(\lambda)=\left\{
\begin{array}{lll}
0,&\hspace{1cm}&\lambda\leq\lambda_c,\\
\neq0,&\hspace{1cm}&\lambda>\lambda_c,
\end{array}\right.
\end{equation}
where $\lambda$ is some system-coupling parameter and $\lambda_c$ is the critical coupling which separates the two phases. In the Ginzburg-Landau paradigm, the order parameter is local, {\it i.e.} a local measurement of $\phi$ can reveal the system state. At the critical point\index{Critical! point}, the system shows scale invariance meaning that it becomes correlated on all length scales and in particular the characteristic length $\xi$ diverges with an exponent $\nu$, {\it i.e.} $\xi\sim|\lambda-\lambda_c|^{-\nu}$. These results rely on the model being spatially dependent, like for example a lattice model as discussed in section~\ref{sec:ext}. The models discussed so far, JC, quantum Rabi, TC and Dicke, lack such spatial dependencies and instead they are what is called {\it fully connected models}~\cite{botet1982size,hwang2015quantum,larson2017some}\index{Fully connected models}. Nevertheless, much of the properties of critical phenomena of, for example, lattice models can be translated to fully connected models, the counterpart of a characteristic length can be introduced as a {\it coherence length}\index{Coherence length}~\cite{botet1983large}. A characteristic of these models is that in the thermodynamic limit\index{Thermodynamic limit} many properties, like critical exponents, can be extracted from simple mean-field theory. This typically implies that quantum fluctuations become negligible, and the analysis simplifies considerably~\cite{larson2017some}. It is interesting to note, however, that certain quantum properties like entanglement may survive in the thermodynamic limit~\cite{buvzek2005instability}.

In sec.~\ref{sssec:dicke} we pointed out how the Lipkin-Meshkov-Glick model~(\ref{lmg}) emerges as one adiabatically eliminates the boson field from the Dicke model. These two models share the same critical behaviour, and hence if one is interested in the universal critical properties of the Dicke model it is sufficient to study the simpler Lipkin-Meshkov-Glick model. Such methods of reducing the number of degrees of freedom in order to find effective models is rather general~\cite{altland2010condensed}, and adiabatic elimination is one version where retardation effects between the two subsystems is disregarded. Eliminating the boson field from the TC model results in the Lipkin-Meshkov-Glick model
\begin{equation}\label{lmgU}
\hat H_\mathrm{LMG}=\omega\hat S_z+\lambda\left(\hat S_x^2+\hat S_y^2\right),
\end{equation}
which is clearly symmetric with respect to the rotations $U(\varphi)=\exp\left(-i\hat S_z\varphi\right)$, corresponding to the $U(1)$ symmetry of particle conservation in the TC model. If, instead, the spin degree of freedom is adiabatically eliminated we saw in sec.~\ref{ssec:rabi} that the effective boson Hamiltonian $\hat H_\mathrm{Bos}$ of ~(\ref{adbos}) becomes identical to the one obtained in the BOA. That is, at the critical point\index{Critical! point} of the Dicke model, the potential transitions from a single to a double-well shape. Also in this case, the Hamiltonian $\hat H_\mathrm{Bos}$ captures the correct critical exponents.  The inclusion of an anharmonic term in the LMG Hamiltonian gives rise to an excited-state quantum phase transition, which has been recently analyzed in terms of the time evolution of the survival probability and the local density of states after a quantum quench, as well as on the Loschmidt echoes\index{Loschmidt echo} and the microcanonical OTOC~\cite{KhaloufRivera2023}. 

We will now focus on another property of universality which may be helpful in studying critical properties of quantum optical models. We call this {\it poor man's models} and we actually already saw an example of it in discussing criticality of the quantum Rabi and the Dicke models. The thermodynamic limit\index{Thermodynamic limit} of the Dicke model~(\ref{dickeham2}) consists in letting the spin $S\rightarrow\infty$, which results in two nonlinearly coupled harmonic oscillators as seen from the HP mapping and the transformed Hamiltonian~(\ref{hbdicke}). The Hilbert space is thus $\mathcal{H}=\mathcal{H}_A\otimes\mathcal{H}_B$, with $\mathcal{H}_{A,B}$ the harmonic oscillator Hilbert spaces. In the normal phase, the ground state is the vacuum (in the thermodynamic limit), and upon crossing the critical point the number of bosons $n_A$ and $n_B$ grows as $|g-g_c|^{1/2}$. Thus, in the very vicinity of the critical point only a few Fock states get populated in the ground state, and one may truncate the infinite dimensional Hilbert space to one of a finite dimension. This is demonstrated in the quantum Rabi model~\cite{bakemeier2012quantum,ashhab2013superradiance,hwang2015quantum,puebla2016excited,puebla2017probing,puebla2020universal}. The spin-$S$ of the Dicke model is replaced by a spin-$1/2$ system, and to achieve the non-analyticity at the critical point the thermodynamic limit\index{Thermodynamic limit} has to be redefined. More precisely, to achieve criticality, the thermodynamic limit\index{Thermodynamic limit} is taken as the classical limit $\omega/\Omega\rightarrow0$, meaning that the level spacing of the oscillator becomes infinitely small in comparison to the energy splitting of the two-level system (the classical oscillator would have a continuous spectrum). The result is that the boson mode gets rapidly highly populated when entering the superradiant phase. The critical behaviour of both the Dicke and quantum Rabi models is described by the same universality classes, while the criticality of the open versions of these systems is the same~\cite{larson2017some,hwang2018dissipative}.

A less known example of a poor man's model is obtained when the boson field of the Dicke model is replaced by a single spin-1/2 particle~\cite{mumford2014impurity}. This is equivalent to truncating the boson degrees of freedom to the lowest two states $|0\rangle$ and $|1\rangle$. Calling the pseudo-spin operators $\hat\tau_\alpha$ we obtain
\begin{equation}\label{csm}
\hat H_\mathrm{cs}=N\omega\hat\tau_z+\frac{\Omega}{2}\hat S_z+g\hat\tau_x\hat S_x,
\end{equation}
where the factor $N$ has been included in order to make all terms extensive scaling like $\mathcal{O}(N)$. Thus, we do not redefine the thermodynamic limit\index{Thermodynamic limit} like we did for the quantum Rabi vs. Dicke model PT, but introduce $N$ and keep the former thermodynamic limit\index{Thermodynamic limit} as $N$ (and thereby $S$) going to infinity. The model~(\ref{csm}) has been studied in the past under various names; {\it central spin model}~\cite{mermin1991can,levine2000linear,garmon2011density}\index{Central spin model}\index{Model! central spin}, {\it spin star model}~\cite{breuer2004non}\index{Spin! star model}\index{Model! spin star}, $SU(2)$ {\it Jaynes-Cummings model}\index{$SU(2)$ Jaynes-Cummings model}\index{Model! $SU(2)$ Jaynes-Cummings}~\cite{ellinas1995motion}, {\it Gaudin model}\index{Model! Gaudin}~\cite{gaudin1976diagonalisation}\index{Gaudin model}. Gaudin was the first to study the model, however in a more general setting, in order to explore integrable models. Mermin considered it as a most simple example of a critical model, while Breuer {\it et al}. were interested in non-Markovian effects in a system of a spin-1/2 coupled to an environment of other spin-1/2 particles similar to the spin-boson model~(\ref{spinboson}). For the TC or anisotropic Dicke model\index{Anisotropic! Dicke model} one would again find a central spin model, but now also with a coupling $g\hat\tau_y\hat S_y$. Similar to our discussion above, the poor man's models give the same critical properties as the full models. The scheme can, of course, be applied to other models like multi-mode or multi-level atoms. In ref.~\cite{kurcz2014hybrid}, the authors considered an array of quantum Rabi like models (see sec.~\ref{ssec:jch}) and could show that model could be described by a poor man's model having the shape of a transverse field Ising model. A related model considering multi-level atoms was discussed in Ref.~\cite{gagge2020superradiance}, and here the corresponding poor man's model comprises a chain of spin-1/2 particles and $SU(3)$ pseudo spins. 

\subsubsection{Connection to generalized spin-boson models}

Lonigro and collaborators have considered a class of generalized spin-boson models (GSBs) to account for non-normalizable coupling functions between a quantum mechanical system and a structured boson environment~\cite{lonigro2022generalized, lill2023}. A commonly encountered variant of the GSBs (which, for a monochromatic field, reduces to the Jaynes-Cummings model) can be written in the form
\begin{equation}\label{eq:f1}
\hat{H}=\hat{H}_{\rm free} + \lambda[\hat{\sigma}_{+} \hat{a}(f) + \hat{\sigma}_{-}\hat{a}^{\dagger}(f)],
\end{equation}
where $\hat{H}_{\rm free}$ is the free energy of the system and the field. The creation and annihilation operators are explicit functions of the form factor $f$\index{Form factor}, meaning that when the atom switches from the excited to the ground state a boson is created with wavefunction $f$. 

The so-called {\it ultraviolet (UV) problem} for the multi-atomic version of Eq.~\eqref{eq:f1},
\begin{equation}\label{eq:fN}
\hat{H}_{f_1,\cdots,f_N}=\hat{H}_{\rm free} + \sum_{j=1}^{N}[\hat{\sigma}_{j+} \hat{a}(f_j) + \hat{\sigma}_{j-}\hat{a}^{\dagger}(f_j)],
\end{equation} 
where the form factors may exhibit {\it mild} UV divergences, is discussed in~\cite{lonigro2023selfadjointness}. The form factors are subject to the weak normalization constraint
\begin{equation}\label{eq:condW}
\int_{\Omega} \frac{|f_j(k)|^2}{\omega(k)} d\Omega < \infty, \quad j=1,\ldots,N,
\end{equation}
in which $\omega(k)$ is the dispersion relation for the bosonic mode, and $\Omega$ is the reciprocal space for a given geometry. The Hamiltonian of Eq.~\eqref{eq:fN}, written as $\hat{H}_{f_1,\cdots,f_N}=\hat{H}_{\rm free} + \hat{A} + \hat{A}^{\dagger}$, admits the following tri-diagonal decomposition in the domain formed by the tensor product of the spins and the boson field (the same as that of $\hat{H}_{\rm free}$):
\begin{equation}\label{eq:blocktrid}
	\hat{H}_{f_1,\dots,f_N}\simeq\left[\begin{array}{c|c|c|c|c}
		\hat{H}_0 & \hat{A}_{0,1}^\dag& \phantom{\ddots}& \phantom{\ddots}&\\\hline 
		\hat{A}_{0,1} & \hat{H}_1 & \hat{A}_{1,2}^\dag & \phantom{\ddots} &\\\hline  
		& \hat{A}_{1,2} & \hat{H}_2 & \hat{A}_{2,3}^\dag &\phantom{\ddots} \\\hline
		& & \hat{A}_{2,3} & \hat{H}_3 & \ddots\\\hline 
		&&& \ddots& \ddots
	\end{array}\right],
\end{equation}
with the number of blocks scaling linearly with the number of spins, $N$. The case where $f_1,\ldots,f_N$ are only subject the condition~\eqref{eq:condW}, whence $\hat{A}_{n,n+1}, \hat{A}^{\dagger}_{n-1,n}$ are not well defined unbounded operators in the Hilbert space of the spin-field states, is extensively treated in~\cite{lonigro2023selfadjointness}.  

Moreover, the validity of the quantum regression formula beyond the Born-Markov approximation has been assessed in~\cite{LonigroPRA2022} for a class of GSB models. In the latter, a multi-level generalization of the familiar amplitude-damping evolution arises, while a particular choice of coupling functions leads to semi-group dynamics at all times. The presence of a flat form factor generates {\it strong} UV divergences, which require a renormalization of the atomic excitation energy. The less stringent condition
\begin{equation}
\int_{\Omega} \frac{|f(k)|^2}{\omega^2(k)}\,d\Omega < \infty
\end{equation}
is then to be upheld.   


\subsection{Extended Jaynes-Cummings models turned into single particle lattice problems}\label{ssec:focklattice}\index{Fock-state lattice}
As explained in sec.~\ref{ssec:JCdyn}, the JC dynamics can be construed as a set of Rabi oscillating two-level systems, characterized by a detuning $\Delta$ and a coupling $g\sqrt{n+1}$. Focusing on one such two-level system, population will oscillate between the two bare states $|e,n\rangle$ and $|g,n+1\rangle$ according to eq.~(\ref{tsol}). Similar oscillatory evolution appears for a single particle in a double-well potential\index{Double-well potential} -- if the particle is initialized in one of the wells it will tunnel through the separating potential barrier, and for non-zero times there will be a finite probability to find the particle in the second well. The tunneling amplitude between the two wells sets an effective coupling  $J$, and the energy offset between the wells gives the effective detuning $\delta$. In the {\it two-mode approximation}\index{Two-mode approximation}, valid when the detuning is smaller than the tunneling amplitude, we include single site localized states for the two wells (if we extend the system to the one with infinite wells/lattice sites, this approximation is the so-called {\it single-band approximation}\index{Single-band approximation}~\cite{Bloch2008a}). Within the two-mode approximation, the resulting Hamiltonian is identical to the JC one~(\ref{block}) after identifying $J\rightarrow g\sqrt{n+1}$ and $\delta\rightarrow\Delta$. Alternatively, we may picture the bare states~(\ref{bstate}) as states representing single sites in some lattice -- a {\it Fock state lattice}\index{Fock-state lattice}~\cite{wang2016mesoscopic, Cai2020, Saugmann2022, DengDong2022}. Thus, the properties of the models are identical to those of a single particle hopping on specific lattices. We will therefore use the term {\it hopping} instead of population transfer or similar. In particular, in this section we will demonstrate that thinking of extended JC models in this way will give much new insight, and furthermore provide a link to the physics of lattice models. It should be clear, though, that these lattice models are not translationally invariant\index{Translational! invariance} since the tunneling strengths become site dependent. Such varying tunneling amplitudes can be visualized as a {\it strain} in the lattice~\cite{Cai2020}.  

\subsubsection{Fock-state lattices of single-mode models}

In the present and the following subsection we discuss a set of extended JC models and their resulting Fock-state lattices\index{Fock-state lattice}. These and some more are summarized in table~\ref{focklat}. 

\begin{figure}
\includegraphics[width=16.2cm]{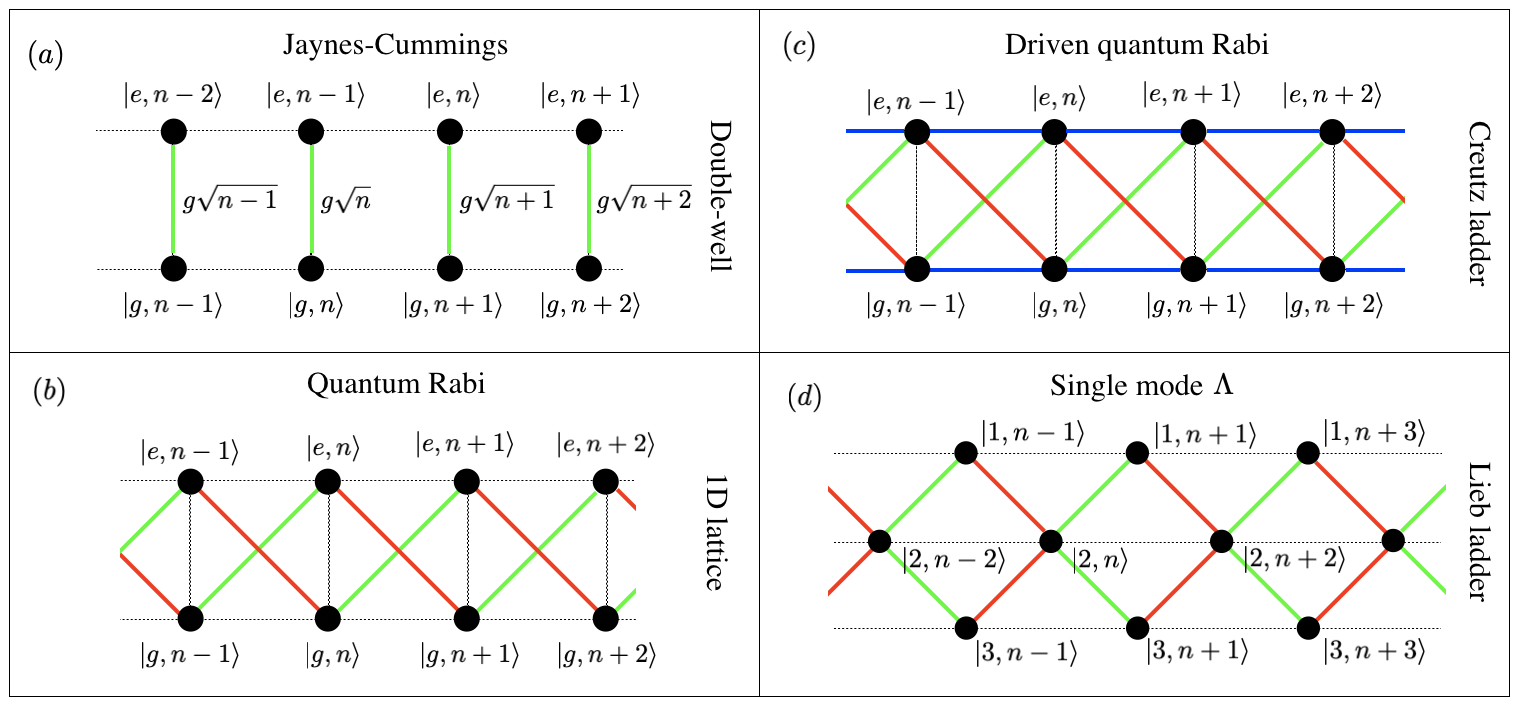} 
\caption{The JC Fock-state lattice {\bf (a)}: The Fock states $|e,n\rangle$ ($|g,n\rangle$) denote the states along the upper (lower) leg of a one dimensional ladder chain. The two-level block form of the JC Hamiltonian in the Fock (bare) basis implies non-zero couplings only along the rungs of the ladder, marked by solid thick green lines in the figure. The strengths of the couplings are included in the plot in order to remind that the lattice is not translationally invariant. In frame {\bf (b)}, we display the Fock state lattice corresponding to the quantum Rabi model. The JC coupling terms are now combined with the counter-rotating terms (solid thick red lines), and each lattice sites couples to two neighbouring sites and the lattice becomes two copies (one for each parity sector of the quantum Rabi model) of 1D lattices. Field driving couples the two parity sectors, as depicted in {\bf (c)} by the thick blue lines along the legs of the ladder, and one finds the so-called Creutz ladder\index{Creutz ladder}\index{Model! Creutz} lattice. In this scenario there are competing nearest and next nearest neighbour coupling terms. Finally, in {\bf (d)} we show the lattice of the three-level $\Lambda$ system. The lower internal states $|1\rangle$ and $|3\rangle$ couple to one other internal state $|2\rangle$, {\it i.e.} they have two nearest neighbours, while the $|2\rangle$ states have four nearest neighbours. This is the characteristic of a Lieb lattice\index{Lieb! lattice}. The dark state of the $\Lambda$ setup translates into a flat `band' for the Fock state lattice. In all plots, green lines represents JC type of couplings, and red lines counter-rotating (anti-JC) coupling terms.} 
\label{figJClat}  
\end{figure}

For the JC model the Fock-state lattice will take the form of a 1D {\it ladder lattice}\index{Ladder! lattice}. The two legs of the ladder represent the two internal states $|g\rangle$ and $|e\rangle$, while the rungs of the ladder can be denoted the JC excitations $n$, see fig.~\ref{figJClat} (a). The interaction Hamiltonian $\hat H_\mathrm{int}$ of eq.~(\ref{inth}) causes hopping along the rungs of the ladder with an amplitude $g\sqrt{n}$, but no hopping along the ladder legs. The free Hamiltonian $\hat H_0$ is diagonal in the Fock/bare basis and its influence will solely be on shifting the single site energies, but it will {\bf not} induce any hopping. Needless to say, the free Hamiltonian will affect the system properties, but despite this, for the main part of this section we will focus on the hopping in the lattice, {\it i.e.} the lattice geometry and hence only consider $\hat H_\mathrm{int}$. In fig.~\ref{figJClat} (a), giving the Fock lattice for the JC model, we write out the hopping amplitudes to emphasize that the lattice is not translationally invariant. This is true for all lattices considered in this section, and as a result there is no quasi momentum serving as a good quantum number. Nevertheless, many properties of the corresponding translationally invariant lattices are found also in these Fock state lattices.

The first step beyond the JC model is to include the counter-rotating terms\index{Counter rotating terms}~(\ref{counterterms}) and thereby explore the quantum Rabi model~(\ref{rabiham}), with the interaction Hamiltonian given by $\hat H_\mathrm{int}=g\left(\hat a^\dagger+\hat a\right)\hat\sigma_x$. Any change in boson number is accompanied by a flip in the internal states, and you have two possibilities (unless the state is in vacuum); add or subtract a boson. Thus, a given Fock state, say $|e,n\rangle$ has two neighbours $|g,n\pm1\rangle$, resulting in two disconnected 1D lattices, as depicted in fig.~\ref{figJClat} (b). The two lattices correspond to the two parity sectors of the quantum Rabi model. As such we still represent the lattice as a ladder system with diagonal hopping between the legs. 

Rather general, for finite spins, each boson mode gives rise to one extra dimension in the Fock state lattice. However, a continuous symmetry constrains the lattice and removes one dimension. So for the JC model we have a single boson mode, but also one $U(1)$ symmetry, and thus the lattice becomes zero dimensional -- decoupled double-wells. While in the quantum Rabi model, the $U(1)$ is broken down to a $\mathbb{Z}_2$ parity symmetry, such that the lattice is one dimensional (the $\mathbb{Z}_2$ symmetry decouples the two legs of the ladder). Since the Hilbert space of the boson mode is infinite dimensional, the Fock state lattices become also infinite. However, there is still an edge since the Fock states have a lower bound, $\hat a|0\rangle=0$. To get an upper bound one must truncate the Hilbert space. Alternatively, one can replace the boson degree of freedom by spin-$S$ particle, {\it i.e.} $\hat a^\dagger\rightarrow\hat S^+$ and $\hat a\rightarrow\hat S^-$ with $\hat S^\pm$ the raising/lowering operators for the spin. The Fock states are the {\it spin states} $|S,m\rangle$ where $\hat S^\pm|S,m\rangle=\sqrt{S(S+1)-m(m\pm1)}|S,m\pm1\rangle$. As a side remark, if this method is applied to the JC model one obtains the so-called {\it central spin model}~\cite{mermin1991can,breuer2004non} to be discussed further below in sec.~\ref{sssec:strans}.

Neither the JC nor the quantum Rabi models render specifically exotic lattices. However, by coupling the two parity sectors in the quantum Rabi model with a field drive (breaking the $\mathbb{Z}_2$ symmetry), {\it i.e.} considering the interaction Hamiltonian $\hat H_\mathrm{int}=g\left(\hat a^\dagger+\hat a\right)\hat\sigma_x+\eta\left(\hat a^\dagger+\hat a\right)$, results in more interesting features. This configuration is presented in fig.~\ref{figJClat} (c), from which we see that hopping now occurs along the legs of the ladder. Driving the qubit generates hopping along the rungs of the ladder. The figure makes clear that hopping occurs between nearest neighbours as well as between next nearest neighbours. Such competing hopping can give rise to novel phenomena, and this specific lattice geometry is that of a {\it Creutz ladder}~\cite{creutz1999end}. Special to the Creutz ladder is the occurrence of a flat energy band ({\it i.e.} dispersionless band which is independent of the quasi momentum) for certain hoppings. Even though our Fock lattice is not translationally invariant, it is easy to see that some remnants of the flat band survive. By diagonalizing $\hat H_\mathrm{int}$ we find the energies $E_\pm(x)=\sqrt{2}(\eta\pm g)\hat x$, where $\hat x=\left(\hat a^\dagger+\hat a\right)/\sqrt{2}$. Thus, for $g=\eta$ we find infinitely-many degenerate states with zero energy. It is important to remember that this degeneracy is lifted if we include the bare Hamiltonian $\hat H_0$, and the full Hamiltonian is of course bounded from below contrary to $\hat H_\mathrm{int}$.  

\begin{table}[h!]
  \centering

\begin{tabular}{|c|c|c|}
\hline
{\bf Model} & {\bf Interaction Hamiltonian} & {\bf Lattice type}  \\
  \hline \hline
Jaynes-Cummings & $\hat H_\mathrm{int}=g\left(\hat a^\dagger\hat{\sigma}_{-}+\hat a\hat{\sigma}_{+}\right)$ & Double-well, fig.~\ref{figJClat} (a)  \\
  \hline
  $N$ atom Tavis-Cummings\index{Tavis-Cummings model}\index{Model! Tavis-Cummings} & $\hat H_\mathrm{int}=g\left(\hat a^\dagger\hat S^-+\hat a\hat S^+\right)$ & $N$ potential-well  \\
  \hline
  Quantum Rabi & $\hat H_\mathrm{int}=g\left(\hat a^\dagger+\hat a\right)\hat\sigma_x$ & 1D chain, fig.~\ref{figJClat} (b)   \\
  \hline
  $N$ atom Dicke & $\hat H_\mathrm{int}=g\left(\hat a^\dagger+\hat a\right)\hat S_x$ & Square lattice  \\
   \hline
  Central spin model\index{Central spin model}\index{Model! central spin} & $\hat H_\mathrm{int}=g_x\hat\sigma_x\hat S_x+g_y\hat\sigma_y\hat S_y$ & Finite SSH chain  \\
  \hline
 Anisotropic quantum Rabi\index{Anisotropic! quantum Rabi model}\index{Model! anisotropic quantum Rabi} & $\hat H_\mathrm{int}=g_\mathrm{jc}\left(\hat{\sigma}_{+}\hat{a}+\hat{a}^\dagger\hat{\sigma}_{-}\right)+g_\mathrm{ajc}\left(\hat{\sigma}_{-}\hat{a}+\hat{a}^\dagger\hat{\sigma}_{+}\right)$ & Infinite SSH chain\\
  \hline
  Driven quantum Rabi\index{Driven! quantum Rabi model}\index{Model! driven quantum Rabi} & $\hat H_\mathrm{int}=g\left(\hat a^\dagger+\hat a\right)\hat\sigma_x+\eta\left(\hat a^\dagger+\hat a\right)$ & Creutz ladder, fig.~\ref{figJClat} (c)   \\
  \hline
  Single mode $\Lambda$ & $\hat H_\mathrm{int}=g\left(\hat a^\dagger+\hat a\right)\left(\hat\lambda^{(1)}+\hat\lambda^{(6)}\right)$ & Lieb ladder\index{Lieb! model}, fig.~\ref{figJClat} (d)   \\
  \hline
  Two-mode JC\index{Bimodal! Jaynes-Cummings model}\index{Model! bimodal Jaynes-Cummings} & $\hat H_\mathrm{int}=g_a\left(\hat a^\dagger\hat{\sigma}_{-}+\hat a\hat{\sigma}_{+}\right)+g_b\left(\hat b^\dagger\hat{\sigma}_{-}+\hat b\hat{\sigma}_{+}\right)$ & SSH chain, fig.~\ref{figJClat2} (a)  \\
  \hline
  Two-mode detuned JC & $\hat H_\mathrm{int}=t\left(\hat a^\dagger\hat b+\hat b^\dagger\hat a\right)$ & CDEL chain, fig.~\ref{figJClat2} (b)  \\
  \hline
  Two-mode quantum Rabi\index{Bimodal! quantum Rabi model}\index{Model! bimodal quantum Rabi} & $\hat H_\mathrm{int}=g\left[\left(\hat a^\dagger+\hat a\right)+\left(\hat b^\dagger+\hat b\right)\right]\hat\sigma_x$ & (Layered) square lattice   \\
  \hline
   Two mode $\Lambda$ & $\hat H_\mathrm{int}=g_a\left(\hat a^\dagger+\hat a\right)\hat\lambda^{(1)}+g_b\left(\hat b^\dagger+\hat b\right)\hat\lambda^{(6)}$ & 2D Lieb lattice   \\
  \hline
  Three-mode JC & $\hat H_\mathrm{int}=g_a\left(\hat a^\dagger\hat{\sigma}_{-}+\hat a\hat{\sigma}_{+}\right)+g_b\left(\hat b^\dagger\hat{\sigma}_{-}+\hat b\hat{\sigma}_{+}\right)+g_c\left(\hat c^\dagger\hat{\sigma}_{-}+\hat c\hat{\sigma}_{+}\right)$ & Hexagonal lattice, fig.~\ref{figJClat2} (c)  \\
  \hline
  Three-mode detuned JC & $\hat H_\mathrm{int}=t\left(\hat a^\dagger\hat b\, e^{i\varphi}+\hat b^\dagger\hat c+\hat a^\dagger\hat c+\text{h.c.}\right)$ & Triangular, fig.~\ref{figJClat2} (d)  \\
  \hline
  Three-mode tripod\index{Tripod! model}\index{Model! tripod} & $\hat H_\mathrm{int}=g_a\left(\hat a^\dagger+\hat a\right)\hat\sigma_{12}+g_b\left(\hat b^\dagger+\hat b\right)\hat\sigma_{13}+g_c\left(\hat c^\dagger+\hat c\right)\hat\sigma_{14}+\text{h.c.}$ & Perovskite (Lieb) lattice\index{Perovskite! lattice}   \\
  \hline
   Three-mode quantum Rabi & $\hat H_\mathrm{int}=g\left[\left(\hat a^\dagger+\hat a\right)+\left(\hat b^\dagger+\hat b\right)+\left(\hat c^\dagger+\hat c\right)\right]\hat\sigma_x$ & Cubic lattice   \\
  \hline
    \end{tabular}
  \caption{List of various extended JC models, their interaction Hamiltonians and the corresponding Fock state lattices. In the table the operators $\hat\lambda^{(\alpha)}$ are the Gell-Mann matrices~(\ref{gellmann})\index{Gell-Mann matrices} and $\hat\sigma_{ij}=|i\rangle\langle j|$.}
  \label{focklat}
  \end{table}
  
Another zero-energy degeneracy is found for the $\Lambda$ atom of eq.~(\ref{lambda}). The $\Lambda$ setup supports a zero-energy (dark) eigenstate\index{Dark! state}\index{State! dark}, also beyond the RWA. For a single mode the interaction Hamiltonian $\hat H_\mathrm{int}=g\left(\hat a^\dagger+\hat a\right)\left(\hat\lambda^{(1)}+\hat\lambda^{(6)}\right)$, and the zero energy eigenstate is $|\psi_0\rangle=(|1,\phi\rangle-|3,\phi\rangle)/\sqrt{2}$ for an arbitrary field state $|\phi\rangle$. The specific structure of the $\Lambda$ system appears in the {\it Lieb lattice}~\cite{lieb1989two}, see fig.~\ref{figJClat} (d). In a Lieb lattice, each unit cell contains three sites; the `central' site couples to the other two sites, while the `outer' sites only couple to the center one. A destructive interference effect occurs, in which the propagation of particles in the lattice is hindered. This is manifested in a flat (dispersionless) energy band\index{Flat energy band}. While we do not have a band spectrum in the Fock state lattice (due to the absence of translational invariance), the massive $E=0$ degeneracy survives. The corresponding eigenstates can always be constructed/superpositioned such that they are localized within the lattice. 

\begin{figure}
\includegraphics[width=14cm]{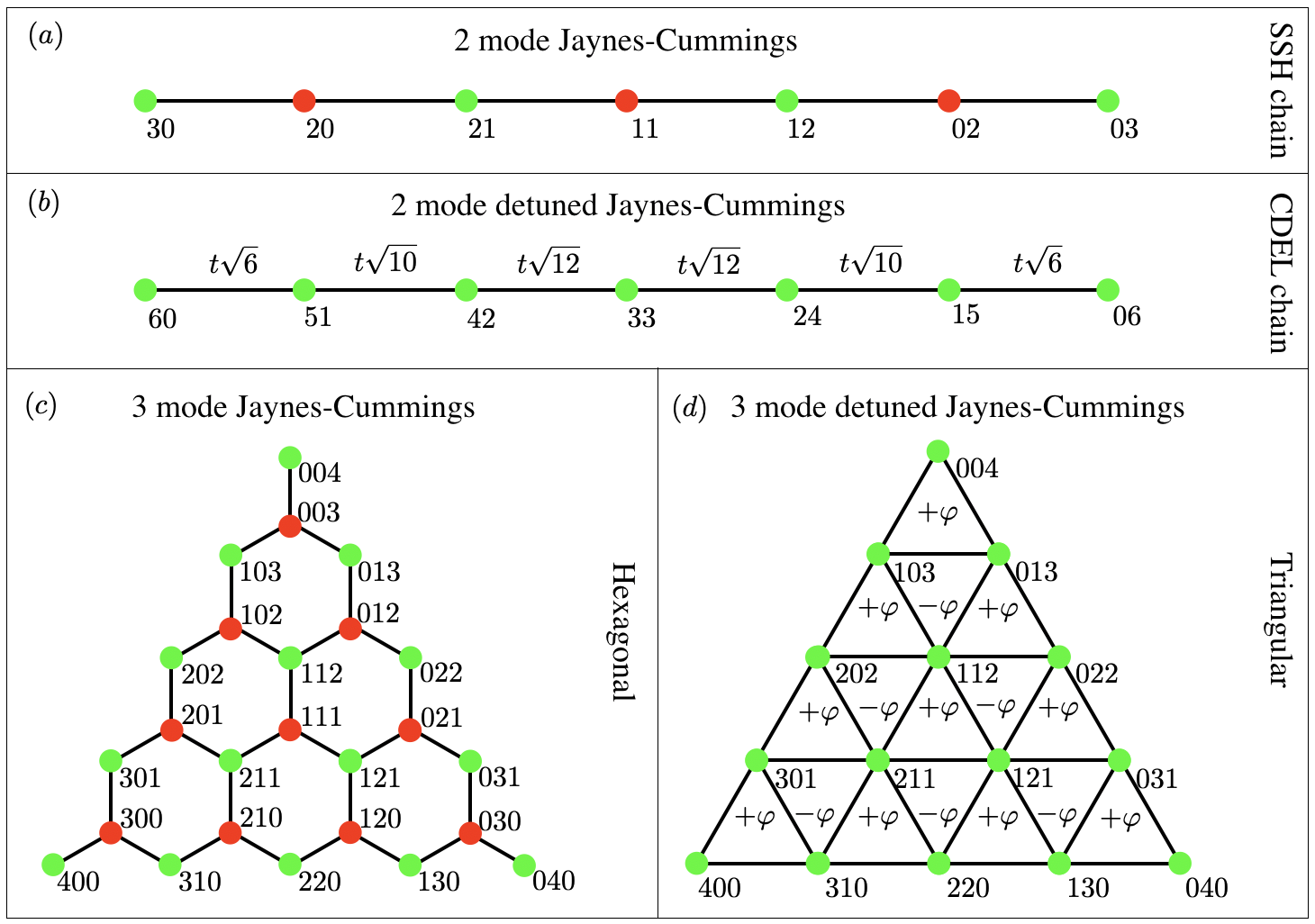} 
\caption{A few Fock-state lattices for multimode models. The red and green dots mark sites with internal atomic states $|e\rangle$ and $|g\rangle$ respectively. Furthermore, the sites are labeled by the corresponding photon numbers of the modes. In {\bf (a)} we present the two-mode JC model, where we allow for varying strengths for the two light-matter couplings; for equal strengths we find a regular 1D chain, while unequal strengths result in a SSH chain\index{SSH model}\index{Model! SSH} type of lattice. If the excited internal states $|e\rangle$ are adiabatically eliminated in the large detuning regime we find the chain of {\bf (b)}. In this plot we write out the coupling strengths, which are exactly those of the CDEL chain explored in terms of perfect state transfer in quantum information processing. The three-mode JC model realizes a hexagonal lattice as shown in {\bf (c)}. A hexagonal lattice\index{Hexagonal lattice} is formed by two connected triangular lattices (green and red sites), and upon eliminating one we find a triangular lattice\index{Triangular lattice} shown in {\bf (d)}. For the triangular lattice in (d) we have included an alternating synthetic magnetic flux through the plaquettes which emerges if one of the coupling terms between the three boson modes becomes complex, see main text.} 
\label{figJClat2}  
\end{figure}  

\subsubsection{Fock-state lattices of multimode models}
The natural way to introduce higher dimensional lattices is to include more bosonic modes. Recall that the dimension of the lattice is typically given by the number of boson modes minus the number of continuous symmetries. For example, to have a 1D lattice with the $U(1)$ symmetry representing number (excitation) conservation present we must consider two boson modes, {\it e.g.} the bimodal JC model described by the Hamiltonian~(\ref{2mode}). Let us parametrise the coupling amplitudes with the angle $\theta$; $g_a=g\cos\theta$ and $g_b=g\sin\theta$, such that $\theta=\pi/4$ gives the `balanced' bimodal JC model and $\theta=0$, for example, reproduces the regular JC model. As pointed out below eq.~(\ref{2mode}), the bimodal JC model is readily diagonalized by introducing the collective boson operators $\hat A\propto g_a\hat a+g_b\hat b$ and $\hat B\propto g_a\hat a-g_b\hat b$. In the transformed basis the problem relaxes to a regular JC model and a decoupled harmonic oscillator. Nevertheless, in the original Fock basis, defined by the $\hat a$ and $\hat b$ modes, the system properties are non-trivial~\cite{larson2006scheme}. The resulting Fock state lattice, shown in fig.~\ref{figJClat2} (a), takes the structure of a {\it Su-Schrieffer-Heeger} (SSH) model\index{Su-Schrieffer-Heeger model}\index{Model! Su-Schrieffer-Heeger}~\cite{su1979solitons}. The SSH model describes a single particle tunneling between nearest neighbours in 1D. What is different compared to a simple 1D tight-binding lattice is that the tunneling strengths alternate between even and odd bonds, to define inter vs. intra site tunneling.

Next up in the row is the three-mode JC model. Again, it is preferably diagonalized by introducing collective boson operators like for the bimodal JC model. By decoupling three modes instead of two, the degeneracy is higher, and moreover we find a 2D Fock state lattice having {\it hexagonal} structure shown in fig.~\ref{figJClat2} (c). Like in the above example, the (conserved) excitation number $\hat N$ determines the lattice size, {\it e.g.} in the example of the figure $\hat N=4$. The properties of hexagonal lattices (also called {\it honeycomb lattices}\index{Honeycomb lattices}) have been extensively studied as these are the lattices of graphene~\cite{larson2020conical}. The most striking feature of the spectrum is the {\it conical intersection}\index{Conical intersection} (a point degeneracy) of the lowest energy bands at the Brillouin edge, giving rise to relativistic properties of the electrons in graphene. Since the translational symmetry is broken in the Fock state lattice of fig.~\ref{figJClat2} (c), the corresponding spectrum does not have the band structure as for a regular hexagonal lattice. A reconfigurable valley-platform scheme based on a honeycomb lattice of JC emitters, which can be implemented by cavity- or circuit-QED lattice cells has been proposed in~\cite{DongJ2023}, aiming at realizing a tuneable topological quantum router.   

\begin{figure}
\includegraphics[width=4cm]{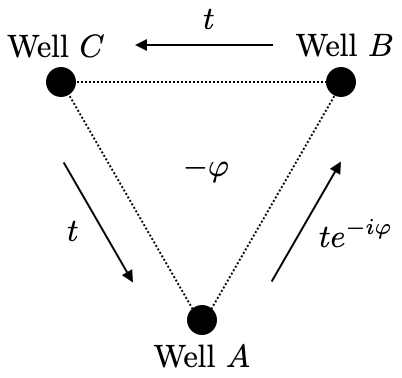} 
\caption{Schematic of the triple-well\index{Triple-well} system described by the interaction Hamiltonian~(\ref{tremod}). The three wells are marked by the filled black circles and the tunneling amplitudes between the wells are $|t|$. A boson hopping around the triple well once counter-clockwise experiences a synthetic magnetic field $-\varphi$. A gauge transformation\index{Gauge! transformation} can alter the phase factors of the three tunneling amplitudes, {\it e.g.} the phase factor $\exp(-i\varphi)$ could be put between any other two wells. However, the summed phases (equals the magnetic flux) of all three tunneling amplitudes is gauge invariant\index{Gauge! invariance}. } 
\label{tripfig}  
\end{figure}

Another interesting model derives as we adiabatically eliminate the internal state $|e\rangle$. Following the elimination procedure described in sec.~\ref{ssec:JCm} we find the interaction Hamiltonian for three modes
\begin{equation}\label{tremod}
\hat H_\mathrm{int}=t\left[\left(\hat a^\dagger\hat b\,e^{i\varphi}+\hat b^\dagger\hat a\,e^{-i\varphi}\right)+\left(\hat b^\dagger\hat c+\hat c^\dagger\hat b\right)+\left(\hat a^\dagger\hat c+\hat c^\dagger\hat a\right)\right],
\end{equation}
with $t=2g^2/\Delta$, where we assume that all three light-matter coupling amplitudes are $g$ and the corresponding detunings $\Delta$ are also identical. We have, however, introduced a non-trivial phase factor $\exp(i\varphi)$ for the first term of the interaction Hamiltonian. This factor does not automatically follow from considering a complex $g$ for one of the coupling terms, since this could be made real by a gauge transformation\index{Gauge! transformation} as explained in sec.~\ref{ssec:JCm}. However, it is possible to generate the phase factor by {\it Floquet driving}, {\it i.e.} periodically drive the system and after averaging over the rapid drive one finds a complex coupling~\cite{wang2016mesoscopic} (see discussion around eq.~(\ref{fevo}) above). The Hamiltonian of eq.~(\ref{tremod}) describes non-interacting bosons in a triple-well system as schematically depicted in fig.~\ref{tripfig}. The number of bosons in a given site $\alpha$ is $n_\alpha$ and a single boson can tunnel between wells with an amplitude $|t|$. The tunneling from the well $A$ to the well $B$ is accompanied by a phase $\exp(-i\varphi)$, and upon encircling the well counter-clockwise results in a phase shift of the wave function by $-\varphi$. This can be viewed as a synthetic magnetic flux\index{Synthetic! magnetic flux} of strength $-\varphi$ through the triple-well~\cite{larson2020conical}. Since the flux is a gauge invariant\index{Gauge! invariance} quantity, how the phase factors are divided among the three tunneling terms is irrelevant as long as they sum up to $-\varphi$, {\it e.g.} we could equally well consider the symmetric case with all three tunneling amplitudes equal to $t\exp(-i\varphi/3)$.

The complex tunneling $t\exp(-i\varphi)$ is translated into Fock-state lattice tunneling amplitudes that are also complex. In particular, the resulting synthetic magnetic field in the Fock state lattice becomes {\it staggered} -- alternating between $\pm\varphi$ between neighbouring lattice plaquettes. The spectrum, and hence the system properties, all rely on this magnetic flux as will be demonstrated in the next two sections. By considering more complicated (number dependent) complex tunneling amplitudes, other fluxes can be achieved, as for example a constant magnetic flux.

Turning now to a very recent proposal, a circuit QED system with photons and superconducting qubits on a hyperbolic lattice\index{Hyperbolic lattice} has been studied in~\cite{Bienias2022}, following the demonstration of hyperbolic lattices, lattices in an effective curved space which cannot be isometrically embedded~\cite{Kollar2019}. The photons were modeled by a tight-binding Hamiltonian, with a spectrum bounded from below and above, while the qubits were approximated as two-level systems. The bound-state wavefunction for a qubit at position $z_i$ on the Poincar\'{e} disk\index{Poincar\'e! disk} in a finite hyperbolic lattice, with an energy in the vicinity of the lower bound state, is
\begin{equation}
 \ket{\psi_B} \propto \left(\hat{\sigma}_{+}^{(i)} - \int \frac{d^2 z}{(1-|z|^2)^2}u(z) \hat{a}^{\dagger}(z)\right)\ket{\downarrow, 0},
\end{equation}
where the complex amplitude $u(z)$ is proportional to the continuum approximation of the photon Green function, evaluated at the lower bound-state energy. The continuum approximation enables the analytical quantification of the size of the single-particle bound state, and it is found that on a hyperbolic lattice correlations are truncated by the curvature radius. In~\cite{Bienias2022} we also read that despite the fact that the boundary effects have a noticeable impact on the photonic density of states, the spectral density\index{Spectral! density} is well described by the continuum theory.

\subsubsection{Fractal spectra}
The spectrum for a single particle hopping between neighbouring sites on a square lattice in the presence of a perpendicular constant magnetic field displays a fractal structure~\cite{hofstadter1976energy}\index{Fractal spectrum}. This is known as the {\it Hofstadter butterfly}\index{Hofstadter butterfly} and it can be generalized to other lattice geometries and dimensions. The intuition for the emerging fractal structure is the following. For zero flux, the unit cell of the square lattice contains a single site, and in the tight-binding-approximation\index{Tight-binding approximation} we have a single energy band. If the flux is a multiple of $2\pi$, a particle hopping around a loop in the lattice will return to its initial state, and the system is `transparent' to the magnetic field. If, say $\varphi=\pi$, the particle returns to its initial state if it hops around, for example, two plaquettes. The unit cell is thereby doubled, and it contains two sites. As a result, for $\varphi=n\pi$, with $n$ an integer, we have now two energy bands. If $\varphi=n\pi/m$ for integers $n$ and $m$ the system is still periodic, but the unit cell size is changed. We can still assign a quasi momentum to the eigenstates, but the size of the Brillouin zone\index{Brillouin zone} and the number of bands will depend on the integers $n$ and $m$. Irrational fluxes $\varphi/\pi$ will result in models with broken translational symmetry, and the spectrum is then discrete. 

\begin{figure}
\includegraphics[width=10cm]{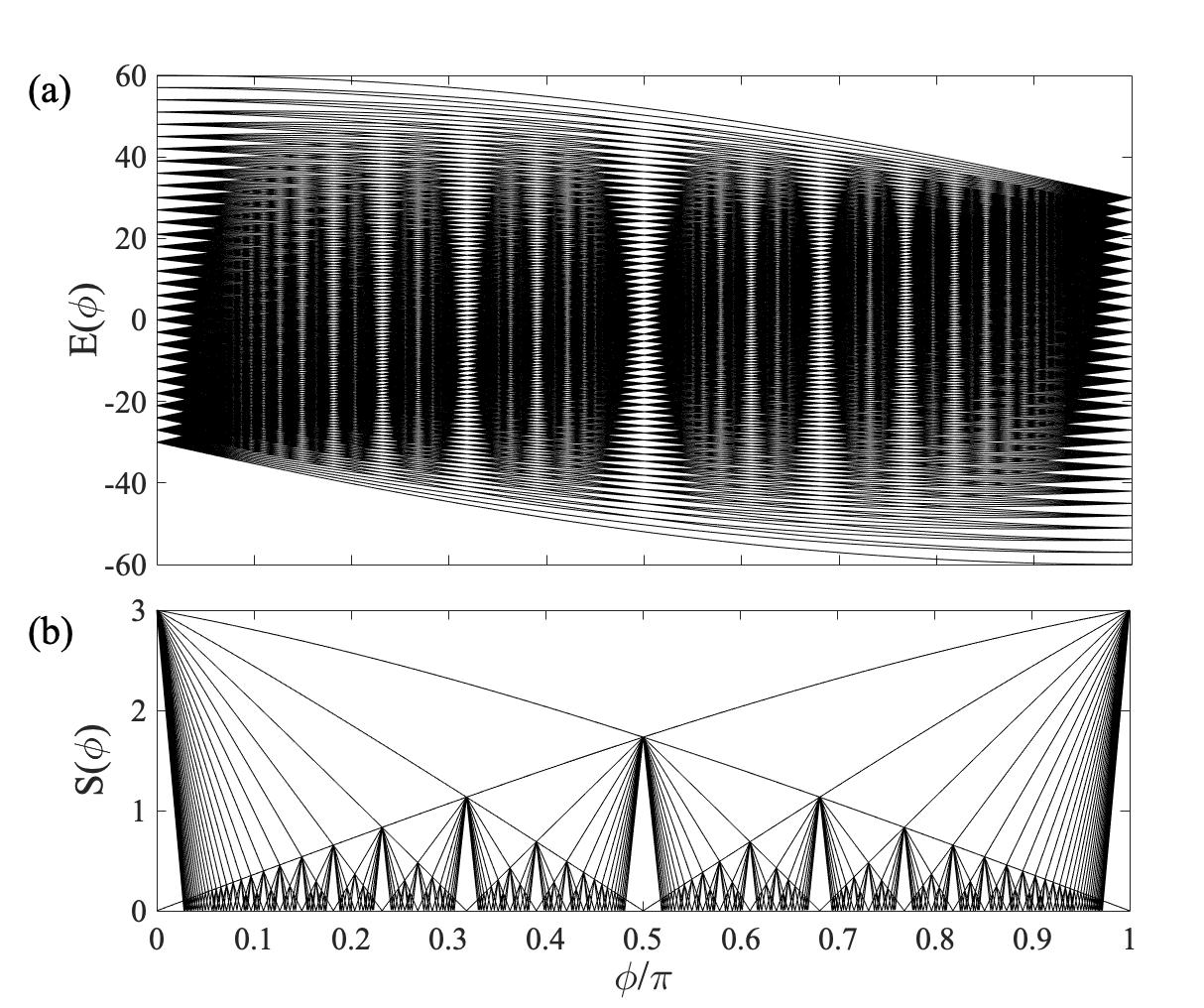} 
\caption{The spectrum {\bf (a)} of the Hamiltonian~(\ref{tremod}) as a function of the phase $\varphi$ and for $N=30$ bosons. For certain phases, multiple energies cluster together. To explore the properties of this clustering we also show how the energy differences $S=E_{n+1}-E_n$ varies with $\varphi$ {\bf (b)}. The plot reveals a fractal structure which is hidden in the spectrum. For the phases $\varphi=n\pi$ and $\varphi=n\pi/2$ ($n$ and integer) there is a single time-scale in the system and the evolution displays perfect revivals for these phases. The dimensionless tunneling amplitude $t=1$. 
} 
\label{fracfig}  
\end{figure}  

It is interesting to explore whether the flux in the Fock-state lattices can also give rise to a fractal structure. Since there is no translational symmetry, one would expect that the spectrum should not be fractal, {\it i.e.} one cannot use the same argument as for the Hofstadter butterfly which relies on the size of the unit cells. As we will see, despite this one finds a fractal spectrum for the Hamiltonian~(\ref{tremod}). The underlying mechanism, however, derives from another property. Let us give an example of the spectrum in fig.~\ref{fracfig} (a), where we display the energies as a function of the phase $\varphi$. Clearly, there is a clustering of energies for certain phases. Like in the previous section, when we explored spectral properties of the Dicke model, it is convenient to consider the energy difference $S_n=E_{n+1}-E_n$ between consecutive energies. The resulting energy differences $S_n$ are presented in fig.~\ref{fracfig} (b). The fractal structure now becomes evident. For larger particle numbers $N$ there will occur more steep lines around $\varphi=0,\,\pi$.

To see how this pattern comes about we rewrite the Hamiltonian on the quadratic form
\begin{equation}\label{3m}
\frac{\hat H_\mathrm{int}}{t}=\left[\hat a^\dagger,\,\hat b^\dagger,\,\hat c^\dagger\right]
\left[\begin{array}{ccc}
0 & e^{i\varphi} & 1\\
e^{-i\varphi} & 0 & 1\\
1 & 1 & 0
\end{array}\right]\left[
\begin{array}{c}
\hat a\\ \hat b\\ \hat c\end{array}\right].
\end{equation}
The characteristic equation for the $3\times3$ matrix is $\lambda^3-3\lambda-2\cos\varphi=0$, and the three roots
\begin{equation}
    \lambda_k  = 2\cos\left(\frac{\varphi-2\cdot\pi\cdot k}{3}\right)\quad, \quad k\in\left\{0,1,2\right\}
\end{equation}
 Assume we can order them as $\lambda_0(\varphi)\leq\lambda_1(\varphi)\leq\lambda_2(\varphi)$. With $n_k$ non-negative integers obeying $n_0+n_1+n_2=N$, the spectrum is
\begin{equation}
 E_\mathbf{n}(\varphi)=\lambda_0n_0+\lambda_1n_1+\lambda_2n_2=\lambda_0N+(\lambda_1-\lambda_0)n_1+(\lambda_2-\lambda_1)n_2.
 \end{equation}
 The first term on the R.H.S. is an overall constant, {\it i.e.} irrelevant for the evolution of observables. Thus, multiples of the two energies $\delta\varepsilon_1=\lambda_1-\lambda_0$ and  $\delta\varepsilon_2=\lambda_2-\lambda_1$ determine the characteristic time-scales of the system. In particular, for $\varphi=\pi/2$ we find $\delta\varepsilon_1=\delta\varepsilon_2$ and the spectrum is thereby equidistant. For this special case we find a perfect revival\index{Perfect revival} for $T_r=2\pi/\delta\varepsilon_1$. This is not the only phase for which revivals may occur. For a fraction $\delta\varepsilon_1/\delta\varepsilon_2=n/m$, with $n$ and $m$ positive integers, we find a clustering of energies, and for sufficiently long evolution times it is possible to get a perfect revival. The locations of the peak-like features of fig.~\ref{fracfig} (b) are determined by exactly these fractions, while their `widths' are set by $N$. 

It comes natural to ask whether the phase $\varphi$ is crucial for the fractal structure, or would it work with any parameter dependence of the eigenvalues $\lambda_k$. Taking, for example, real tunneling terms between the three $a$, $b$, and $c$ modes, but letting their amplitudes vary, one cannot in general fulfill the $\delta\varepsilon_1/\delta\varepsilon_2=n/m$ condition. In other words, the fractal structure is in general not obtained upon varying some parameter. For higher number of modes, there are more constraints to be fulfilled. As a result, the clear fractal structure is lost, even though some of its features may survive. 

\begin{figure}
\includegraphics[width=10cm]{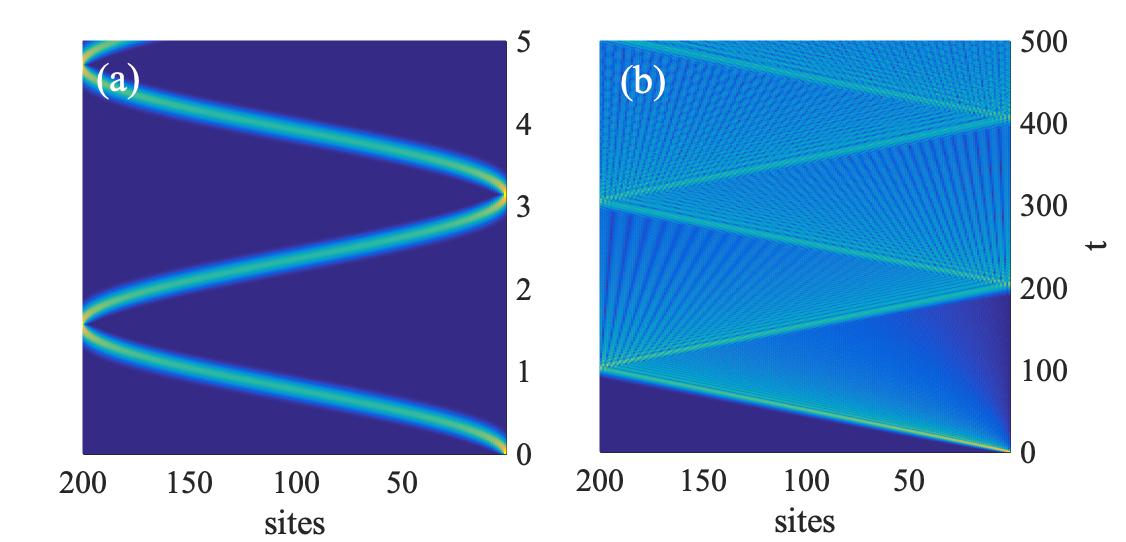} 
\caption{Demonstration of the perfect transfer of a Fock state from one end of the lattice to the other end in the CDEL model {\bf (a)}. Plotted are the probability densities within the lattice. Due to the varying tunneling amplitude in the lattice, the wave packet moves slower towards the edges of the lattice and spreading is hindered. If we consider instead a lattice with constant tunneling amplitudes $t$ we recovered a very different evolution {\bf (b)}. The wave packet spreads while traversing the lattice, and as it hits the edge forward and backward components scatter to give rise to an interference pattern. Due to the square roots of the tunneling amplitudes in the CDEL lattice, the characteristic time-scales become much shorter as is evident by comparing the $y$-axes labeling. In both examples the tunneling amplitude $t=1$ (to not be confused with the time on the $y$ axis), and the lattice is 200 sites long, {\it i.e.} the initial state is $|200,0\rangle$.  
} 
\label{trans1fig}  
\end{figure}

\subsubsection{State transfer and edge states}\label{sssec:strans}
1D chains are particularly interesting for their ability to sustain transfer of quantum states between spatially separated locations. In the seminal work~\cite{christandl2004perfect}, Christandl, Datta, Ekert, and Landahl (CDEL)\index{Christandl-Datta-Ekert-Landahl model}\index{Model! Christandl-Datta-Ekert-Landahl} showed how a perfect state transfer could be accomplished in a tight-binding 1D lattice by allowing for varying tunneling amplitudes. In a spatial lattice the desired type of coupling strengths is not trivially realized, but in the Fock state lattice of the dispersive bimodal JC model it comes automatically. Thus, we consider the interaction Hamiltonian $\hat H_\mathrm{int}=t\left(\hat a^\dagger\hat b+\hat b^\dagger\hat a\right)$, which is a simple version of the {\it Hopfield model}\index{Hopfield! model}\index{Model! Hopfield}~\cite{hopfield1958theory}. For a given photon number $N$ we can identify a spin of amplitude $S=N/2$ and rewrite the interaction Hamiltonian using the Schwinger spin-boson mapping\index{Schwinger spin-boson mapping}~\cite{sakurai1995modern}, {\it i.e.} $\hat H_\mathrm{int}=2t\hat S_x$. It is clear that the spectrum is equidistant, and a perfect revival results regardless of the initial state. Thus, starting with, say, a Fock state at one end of the lattice, {\it i.e.} $|N,0\rangle$, after the revival time\index{Revival time} $T_r=2\pi/t$ the system reappears in the initial state $|N,0\rangle$. During this time, the wave packet has actually traversed the lattice back and forth twice. For each quarter of the revival time, {\it i.e.} $\tau_{j=1,2,3,4}=jT_r/4$, the system populates a Fock state $|N\rangle$ in either of the modes. The phase factor of this state is $(-i)^{4(N-1)\tau_j/T_r}$, and hence, it depends on the photon number $N$. Note, however, that the revival time $T_r$ is independent of system size $N$, so it would take any initial Fock state the same amount of time to be transferred from the $a$ to the $b$ mode. This implies that an initial coherent state $|\alpha,0\rangle$ will be transferred to $|0,-i\alpha\rangle$ after a time $T_r/4$. In fig.~\ref{trans1fig} (a) we demonstrate the perfect state transfer for an initial Fock state in one of the modes. We depict the density of the various lattice sites as time progresses. With the chosen parameters the revival time $T_r=2\pi$, but already at $t=\pi$ we have a perfect revival, but here there is an overall phase factor $-1$. To compare the evolution with a regular 1D lattice with constant tunneling amplitudes we also show the corresponding density for such a lattice in (b). Here, the curvature of the energy dispersion causes the wave packet to spread out and we find a complicated interference pattern emerging after the wave packet hits the first edge. 

\begin{figure}
\includegraphics[width=9cm]{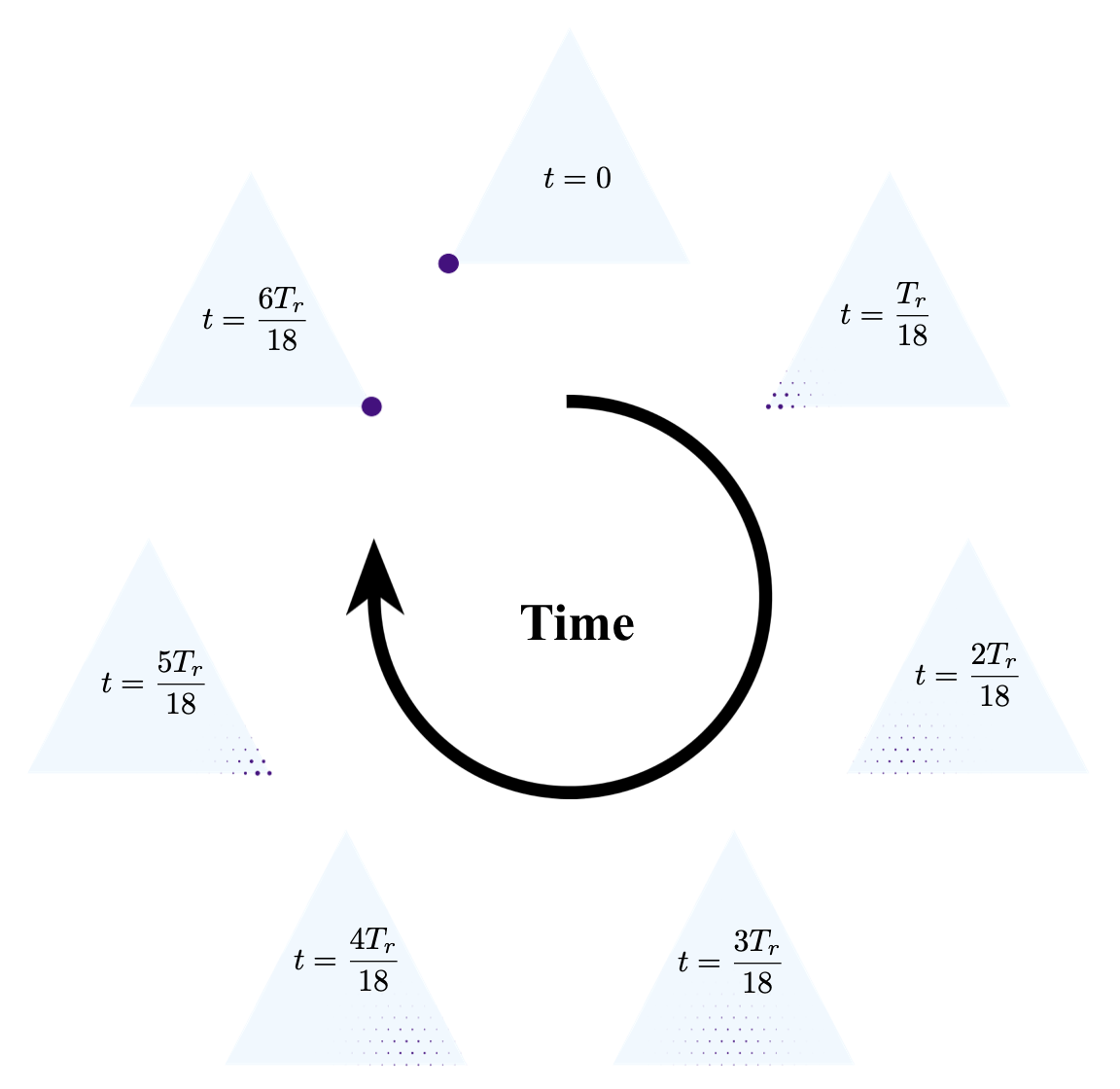} 
\caption{Snapshots of the population in the different lattice sites (time goes clockwise); the sizes of the filled dark blue circles represent the amount of population, and the light blue triangle represents the triangular lattice. Initially the system is in a Fock state $|N,0,0\rangle$ (in this example $N=20$) and after $T_r/3$ the state has evolved into the Fock state $|0,N,0\rangle$ by mainly populating the lower edge of the triangular lattice during the transition between the two Fock states. After $2T_r/3$ the state will be $|0,0,N\rangle$ and so on.  
} 
\label{trans2fig}  
\end{figure}

Perfect revivals are also found in the three-mode detuned JC model\index{Three-mode Jaynes-Cummings model}. In the spectra depicted in fig.~\ref{fracfig} (a), and for $\varphi=\pi/2$ we saw that the energies are equally spaced. For this phase the Hamiltonian supports a chiral symmetry, {\it i.e.} there exists a unitary operator that anti-commutes with the Hamiltonian. The unitary operator is $\hat U_\mathrm{C}=\hat K\exp\left(i\pi\hat n_c\right)$ where $\hat K$ stands for complex conjugation. A chiral symmetry\index{Chiral symmetry} is characterized by a symmetric spectrum around $E=0$, which we can also observe in fig.~\ref{fracfig} (a). The chirality has consequences for how the system evolves in time, {\it e.g.} we can expect an asymmetric evolution in ``left'' and ``right''. It turns out that a state starting out in a corner of the triangular lattice, {\it i.e.} with two of the modes in the vacuum state, will predominantly follow the edges either clockwise or anti-clockwise depending on the sign $\varphi=\pm\pi/2$, as shown in fig.~\ref{trans2fig}. Note that the states along the three edges of the triangular lattice have one of the modes unoccupied, see fig.~\ref{figJClat2} (d). 

\begin{figure}
\includegraphics[width=10cm]{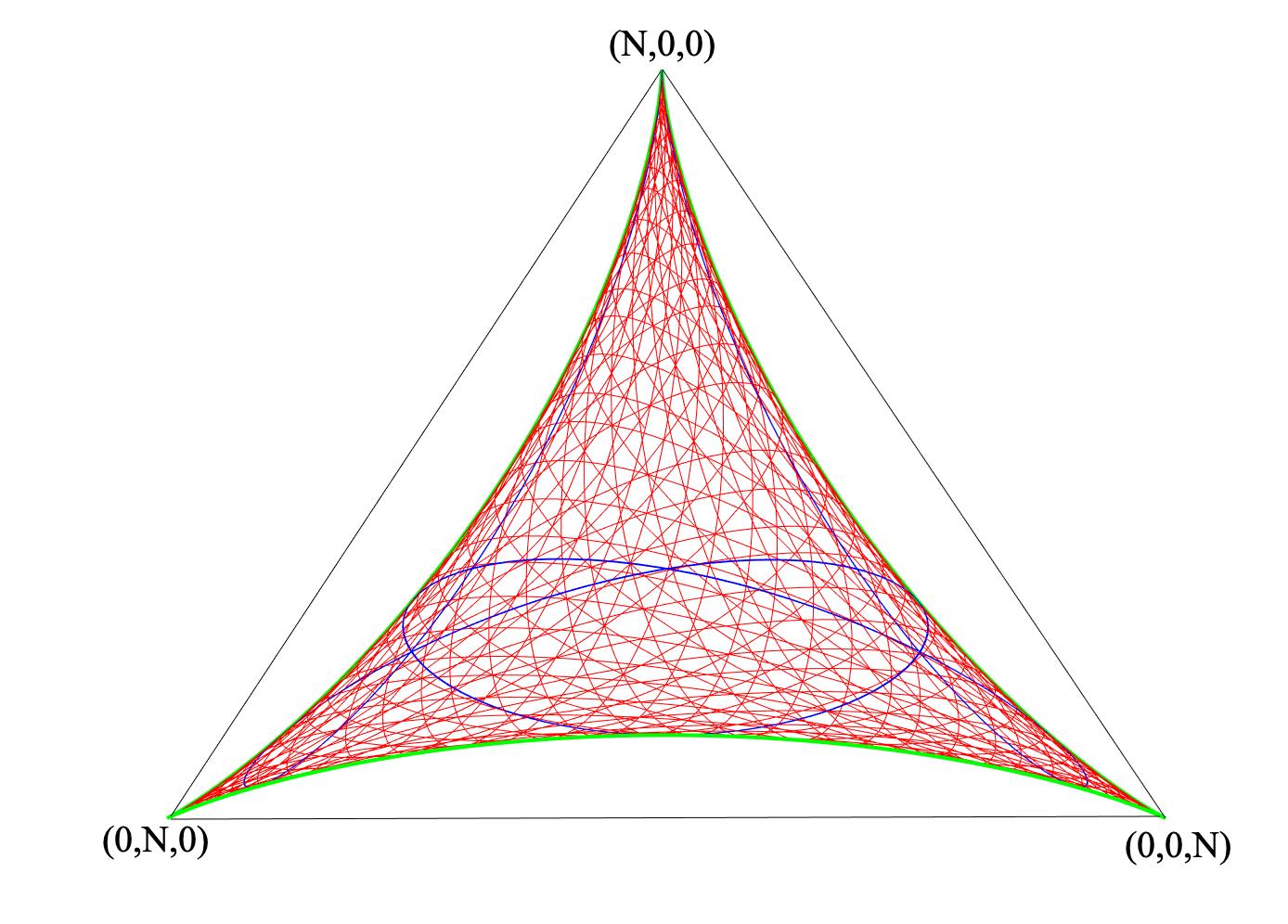} 
\caption{Examples of three trajectories $N(t)$ in the triangular lattice for the dispersive three-mode JC model; green $\phi=\pi/2$ gives the edge states with perfect revivals, blue $\phi\approx\pi/3$ with approximately perfect revivals, and red $\phi=$random with no revivals. 
} 
\label{trans2fig2}  
\end{figure}

To explore the evolution in more detail we introduce the {\it occupation vector} $\bar N(t)=\frac{1}{N}\left(\langle\hat n_a\rangle_t,\langle\hat n_b\rangle_t,\langle\hat n_c\rangle_t\right)$, {\it i.e.} the scaled expectation value of the photon number in any of the three modes. Since the number $N$ is preserved, the vector $|\bar N(t)|\leq1$, and it lives in a 2D plane. Furthermore, if we focus on states initialized in one corner, then $\bar N(t)$ will be independent of $N$ meaning that any initial state of the form, say, $|\phi,0,0\rangle$ will render the same trajectory $\bar N(t)$. Thus, we may take $|\phi\rangle$ to be Gaussian\index{Gaussian! state} and then, since the Hamiltonian~(\ref{3m}) is quadratic, we can obtain $\bar N(t)$ from solving the corresponding classical (Hamilton's) equations of motion
\begin{equation}
\begin{array}{lll}
\dot x_a & = & -p_b-p_c,\\
\dot x_b & = & -p_a-\cos(\varphi)p_c-\sin(\varphi)x_c,\\
\dot x_c & = & -p_a-\cos(\varphi)p_b+\sin(\varphi)x_b,\\
\dot p_a & = & x_b+x_c,\\
\dot p_b & = & x_a+\cos(\varphi)x_c-\sin(\varphi)p_c,\\
\dot p_c & = & x_a+\cos(\varphi)x_b+\sin(\varphi)p_b,
\end{array}
\end{equation}
such that $\langle\hat n_\alpha\rangle_t=\left(p_\alpha^2(t)+x_\alpha^2(t)\right)/2$, and we have assumed the tunneling amplitude $t=1$. Three examples of $N(t)$ trajectory in the 2D plane are displayed in fig.~\ref{trans2fig2}. The green line gives the trajectory for $\varphi=\pi/2$ and it follows the edges of the lattice. The blue line is for the second clustering phase around $\phi\approx\pi/3$ (see fig.~\ref{fracfig}), and we see also a rivival in this time even though the trajectory is more irregular. Finally the red line represents a trajectory for a random phase $\varphi$. 

A paradigmatic model when it comes to discussing topology in quantum lattice systems is the SSH model\index{Model! Su-Schrieffer-Heeger}\index{Model! SSH}\index{SSH model}\index{Su-Schrieffer-Heeger model}~\cite{su1979solitons,batra2019understanding}. The SSH Hamiltonian 
\begin{equation}
\hat H_\mathrm{SSH}=\sum_{i=1}^N\left(v\hat c_{i,1}^\dagger\hat c_{i,2}+w\hat c_{i,2}^\dagger\hat c_{i+1,1}+\text{h.c.}\right)
\end{equation}
describes a single particle moving in a tight-binding 1D {\it superlattice}\index{Superlattice}. By superlattice we mean that the unit cell contains more than a single lattice site, in this case each unit cell hosts two sites, and $v$ is the amplitude for inter tunneling amplitude for hopping between sites within the cell, and w is the intra tunneling amplitude. Thus, $\hat c_{i,j}^\dagger$ ($\hat c_{i,j}$) creates (annihilates) a particle at inter-site $j=1,2$ in cell $i$. For simplicity, the two tunneling amplitudes are taken to be real and positive. The number of sites $N$ is even. For periodic boundary conditions, the Hamiltonian is diagonalized by the Fourier transform, $\hat c_{i,j}=\frac{1}{\sqrt{N}}\sum_k\hat c_{k,j}e^{ika}$, with $k$ the quasi momentum\index{Quasi! momentum}. Since there are two sites per unit cell, the {\it Bloch Hamiltonian}\index{Bloch! Hamiltonian} $h(k)$ becomes a $2\times2$ matrix, where the eigenvalues $\varepsilon_\pm(k)$ of $h(k)$ give the two energy bands. The Bloch Hamiltonian can be expressed in terms of a Bloch vector as~\cite{batra2019understanding}
\begin{equation}
h(k)=\mathbf{h}(k)\cdot \bar\sigma,
\end{equation}
where $\mathbf{h}(k)=(h_x(k),h_y(k),h_z(k))$ and $\bar\sigma=(\hat\sigma_x,\hat\sigma_y,\hat\sigma_z)$, with
\begin{equation}
h_x(k)=v+w\cos(ka),\hspace{1cm}h_y=w\sin(ka),\hspace{1cm}h_z(k)=0.
\end{equation}
The eigenvalues are $\varepsilon_\pm(k)=\pm\sqrt{v^2+w^2+2vw\cos(ka)}$, with the corresponding eigenstates $|\theta_\pm(k)\rangle=\left[\pm e^{-\phi(k)}\,\,\,1\right]^T/\sqrt{2}$, with $\tan(\phi(k))=h_y(k)/h_x(k)$. The Bloch vector $\mathbf{h}(k)$ lives in the $xy$-plane and for $k:\,\,0\rightarrow2\pi$ it makes a circle of radius w around the point $(x,y)=(v,0)$. For $w<v$, the origin $(0,0)$ lies outside the resulting circle, while for $w>v$ the circle winds around the origin (one says that the {\it winding number}\index{Winding number} equals one). The limiting case with $v=w$ marks a special point. As long as $v\neq w$ the two energy bands $\varepsilon_\pm(k)$ are separated, while for $v=w$ the bands become degenerate at the Brillouin edge, {\it i.e.} an energy gap closes as the two tunneling strengths become equal. Upon crossing this point, the {\it Zak phase}\index{Zak phase} $\gamma=i\oint\langle\theta_\pm(k)|\nabla_k|\theta_\pm(k)\rangle dk$ swaps from $\gamma=0$ to $\gamma=\pm\pi$. A non-zero Zak phase marks a non-trivial topology, and in higher dimensions it is replaced by the {\it Chern number}\index{Chern number}. The {\it bulk-boundary correspondence}~\cite{hasan2010colloquium} connects properties of the internal bulk with those of the boundary, namely the Chern number tells us how many {\it edge states}\index{Edge state} we have. Thus, it is a result pertaining to finite systems where strict translational invariance is broken. An edge state $|\phi_{l,r}\rangle$ is one which is exponentially localized to the left/right edge of the system. For the SSH model, if the system is finite one finds that the spectrum is symmetric around $E=0$ (a result of chiral symmetry)\index{Chiral symmetry}, but more importantly, when $\gamma=\pm\pi$ there exists two eigenstates with $E=0$ and these are localized to the edges. 
For the Fock-state lattices there are no Zak phases nor Chern numbers\index{Chern number}, but we may still ask if there exists something like edge states with zero energy~\cite{Cai2020}. The problem with some of the 1D models listed in table~\ref{focklat} is that they do not support two edges since the Hilbert space is infinite. The two-mode JC model, with a corresponding SSH type of Fock state lattice, has two edges owing to the constraint coming from the preserved number of excitations. Properties of the localized zero-energy eigenestates (solitons\index{Soliton}) were explored in~\cite{Cai2020}. Those states are not exponentially localized, as is typical for edge states. A superconducting quantum circuit simulator of the 2D SSH lattice has been recently proposed in~\cite{Li2022SSH} to investigate the higher-order topological phase transitions induced by continuously varying magnetic field.

\begin{figure}
\includegraphics[width=10cm]{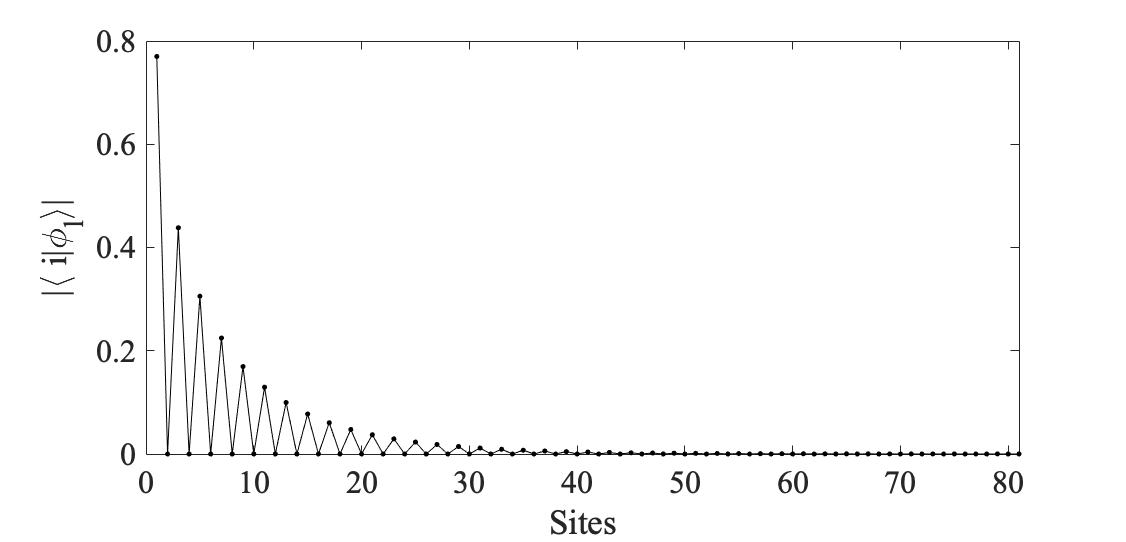} 
\caption{The probability amplitude $|\langle i|\phi_l\rangle|$ to populate site $i$ for the $E=0$ edge eigenstate. To a very good approximation the state is exponentially localized with the probability distribution $P_i\propto\left(\frac{g_\mathrm{r}}{g_\mathrm{cr}}\right)^i$, where we assume $g_\mathrm{r}<g_\mathrm{cr}$. Thus, the ratio $g_\mathrm{r}/g_\mathrm{cr}$ determines the exponential decay rate. In the shown example,  $g_\mathrm{r}/g_\mathrm{cr}=0.8$, and the spin $S=40$ giving 81 sites in each 1D chain.
} 
\label{edgefig}  
\end{figure} 

Exponentially localized edge states may still occur in the Fock-state lattices, and to demonstrate this we introduce the {\it central spin model}\index{Central spin model}\index{Model! central spin}~\cite{mermin1991can,breuer2004non}, which can be seen as a {\it poor man's quantum Rabi model} (see previous section). The central spin model is obtained from the quantum Rabi model by replacing the boson degree of freedom by a spin degree of freedom, {\it i.e.} $\left(\hat a+\hat a^\dagger\right)\rightarrow\hat S_x$ and $\hat a^\dagger\hat a\rightarrow\hat S_z$ where $\hat S_\alpha$ are the $SU(2)$ spin operators for a spin-$S$ particle. Thus, the infinite dimensional Hilbert space of the boson has become a finite Hilbert space with dimension $2S+1$. The central spin model has been studied as a simple model showing critical behaviour, and expectedly, in the thermodynamic limit\index{Thermodynamic limit} the model shows the same critical universal behaviour as the Dicke and quantum Rabi models. To achieve the SSH type of lattice we need to consider the central spin model with both $\hat\sigma_x\hat S_x$ and $\hat\sigma_y\hat S_y$ coupling terms, hence the interaction Hamiltonian is
\begin{equation}
\hat H_\mathrm{int}=g_x\hat\sigma_x\hat S_x+g_y\hat\sigma_y\hat S_y=(g_x+g_y)\left(\hat{\sigma}_{+}\hat S^-+\hat{\sigma}_{-}\hat S^+\right)+(g_x-g_y)\left(\hat{\sigma}_{+}\hat S^++\hat{\sigma}_{-}\hat S^-\right).
\end{equation}
This is reproducing two decoupled finite 1D chains of lengths $2S+1$; each chain belongs to one parity sector\index{Parity symmetry}. In the second step we have written the model in terms of the rotating and counter rotating terms in analogy to the anisotropic quantum Rabi model\index{Anisotropic! quantum Rabi model}\index{Model! anisotropic quantum Rabi}, see eq.~(\ref{anRabi}), which also results in an SSH model (however infinite). Due to the $\mathbb{Z}_2$ parity symmetry the spectrum is doubly degenerate, and in particular there are two $E=0$ eigenstates $|\phi_{l,r}\rangle$ that are exponentially localized to one edge of either of the two chains. These states have the additional property that $\hat\sigma_z|\phi_{l,r}\rangle=\pm|\phi_{l,r}\rangle$, {\it i.e.} in the Fock state lattice only every second site will be populated. An example of the edge state is shown in fig.~\ref{edgefig}, and we find the same exponential decay away from the edge as well as that every second site is populated, just like in the SSH model. 


\subsection{Review of the approximations underlying the JC model}\label{ssec:approx}
The JC model, as we have pointed out, is the simplest model to describe light-matter interaction on a quantum level. Despite its simplicity it manages to capture and explain many effects observed in a wide variety of quantum optical systems. Of course, such experimental systems have been tailored for exploring clean and simple quantum phenomena -- these are typically not systems that arise naturally. The technology needed for realizing JC-type experimental systems developed over the span of several decades, which is something we will discuss further in the following sections. It may seem paradoxical that much of the interest today is not focusing on exploring clean systems that realize JC physics, but instead on pushing the boundaries such that the JC description is no longer valid -- some of the underlying approximations become inapplicable. The prime example was presented in sec.~\ref{ssec:rabi}, namely the breakdown of the RWA. This happens as the light-matter coupling $g$ becomes comparable to $\omega$, and the ground state of the Dicke model\index{Dicke! model}\index{Model! Dicke} becomes superradiant, see subsec.~\ref{sssec:dicke}. We also saw examples how new phenomena stem from coupling of the atomic motion to the other degrees of freedom in subsec.~\ref{sssec:qatmo}. Thus, territories beyond the JC regime are now about to be explored with new impetus. At the same time, more fundamental questions are asked, {\it e.g.} regarding gauge invariance\index{Gauge! invariance} and symmetries. The point of this section is to briefly summarize the approximations leading to the JC model. A number of them have already been discussed in some detail in previous sections, and we thereby will not repeat everything here.

We started this long section by introducing the JC Hamiltonian~(\ref{jcham}) as a mathematical model, from which we analyzed several phenomena. To shed light on the underlying approximations let us follow a derivation starting from a microscopic theory. Our starting point is the minimal-coupling Hamiltonian resulting from field quantization in the {\it Coulomb gauge}\index{Coulomb gauge}~\cite{walls2007quantum,schleich2011quantum,de2018breakdown,feranchuk2016physical}. The general Hamiltonian for an atom (in one dimension) acting as an electric dipole\index{Electric dipole} in an EM-field is (in units where the speed of light is set to $c=1$)
\begin{equation}\label{coulham}
\hat H_\mathrm{C}=\hat H_\mathrm{I}+\hat H_\mathrm{f}=\frac{(\hat p-q\hat A(\hat x))^2}{2m}+V(\hat x)+\hat H_\mathrm{f},
\end{equation}
where $m$ is the dipole mass, $q$ the electronic charge, $\hat A(\hat x)$ is the vector potential\index{Vector potential}, $V(\hat x)$ the dipole potential ({\it i.e.} typically a Coulomb potential for an atom), and finally $\hat H_\mathrm{f}$ is the free-field Hamiltonian. We note that the minimal-coupling Hamiltonian~(\ref{coulham}) is unitary equivalent, via a {\it Power-Zienau-Woolley transformation}\index{Power-Zienau-Woolley transformation}~\cite{power1959coulomb,cohen1997photons}
\begin{equation}
\hat U_\mathrm{PZW}=e^{-iq\hat A\hat x}.
\end{equation}
This gauge transformation\index{Gauge! transformation} displaces the momentum $\hat p\rightarrow\hat p+q\hat A$, which removes the coupling of $\hat p$ to the vector potential\index{Vector potential}. Instead one obtains the free-matter Hamiltonian plus a dipole interaction term; 
\begin{equation}
\hat H_\mathrm{I}'=\hat U_\mathrm{PZW}\hat H_\mathrm{I}\hat U_\mathrm{PZW}^\dagger=\hat H_0-q\hat x\hat E(\hat x),
\end{equation} 
since $\partial_t\hat U_\mathrm{PZW}$ is non-zero~\cite{scully1999quantum,schleich2011quantum}. Thus, the minimal-coupling Hamiltonian is identical to the one for an electric dipole\index{Electric dipole} $-qx$ in an electric field $\hat E(\hat x)$.

To proceed we impose our first approximations; the single-mode approximation and the dipole approximation. Hence, $\hat H_\mathrm=\omega\hat a^\dagger\hat a$, and $\hat A(\hat x)=\hat A(0)\equiv\mathcal A_0\left(\hat a+\hat a^\dagger\right)$, while we diagonalize the dipole Hamiltonian $\hat H_\mathrm{d}\equiv=\frac{\hat p^2}{2m}+V(\hat x)=\sum_j\omega_j|\varphi_j\rangle\langle\varphi_j|$ with $\omega_j$ and $|\varphi_j\rangle$ the respective eigenenergies and eigenstates. When evaluating the square in~(\ref{coulham}) we get the interaction term 
\begin{equation}\label{intte}
\hat H_\mathrm{int}=-\frac{q}{m}\mathcal{A}_0\left(\hat a+\hat a^\dagger\right)\hat p,
\end{equation}
and the diamagnetic self-energy term\index{Self-!energy term}
\begin{equation}\label{set}
\hat H_\mathrm{dia}=\frac{q^2\mathcal{A}_0^2}{2m}\left(\hat a+\hat a^\dagger\right)^2.
\end{equation}
The TLA consists in projecting the Hamiltonian onto the relevant states $|g\rangle\equiv|\varphi_i\rangle$ and $|e\rangle\equiv|\varphi_j\rangle$, {\it i.e.}
\begin{equation}
\hat H_\mathrm{C}=\omega\hat a^\dagger\hat a+\omega_i|g\rangle\langle g|+\omega_j|e\rangle\langle e|+\sum_{i,j=g,e}g_{ij}\left(\hat a+\hat a^\dagger\right)|i\rangle\langle j|+\frac{q^2\mathcal{A}_0^2}{2m}\left(\hat a+\hat a^\dagger\right)^2,
\end{equation}
where the effective atom-light coupling $g_{ij}=-\frac{q}{m}\mathcal{A}_0\langle i|\hat p|j\rangle$. Note that in evaluating the matrix element of $\hat p$ it is convenient to first transform into a matrix element of $\hat r$ by using the identity $\hat p=-im[\hat r,\hat H]$~\cite{sakurai1995modern}. Furthermore, we note that $g_{ij}=0$ for $i=j$, and after introducing the Pauli matrices~(\ref{pauli}) and shifting the overall energy we obtain
\begin{equation}\label{coulham2}
\hat H_\mathrm{C}=\omega\hat a^\dagger\hat a+\frac{\Omega}{2}\hat\sigma_z+g\left(\hat a+\hat a^\dagger\right)\hat\sigma_x+\mu\left(\hat a+\hat a^\dagger\right)^2,
\end{equation}
with $\mu=q^2\mathcal{A}_0^2/2m$ and $g=-q\mathcal{A}_0/m$. Neglecting the last term, diamagnetic term\index{Diamagnetic term}, and imposing the RWA results in the desired JC Hamiltonian~(\ref{jcham}), while only neglecting the last term yields the quantum Rabi model\index{Quantum! Rabi model}~(\ref{rabiham}).

\subsubsection{Electric-dipole approximation}\index{Electric dipole! approximation}\label{eda}
There is an obvious reason why the dipole approximation has been very sparsely analyzed in the literature on JC-like models. It simply is a very good approximation in most cases, whether the electromagnetic field is quantized or not (the approximation does not really rely on the quantization). Discussions regarding the approximation in the semiclassical case can be found in several textbooks on quantum mechanics, for example~\cite{landau1958quantum,sakurai1995modern,ballentine1998quantum,gasiorowicz2007quantum,shankar2012principles}. 

Inside a resonator, the single-mode vector potential\index{Vector potential} assumes the form 
\begin{equation}
A(x,t)\propto\left(\hat a^\dagger+\hat a\right)\cos(kx-\omega t),
\end{equation}
with $k=2\pi/\lambda$ the wave vector. For $kx\ll1$ we may expand the cosine to first order, $\cos(kx-\omega t)\approx\cos(\omega t)$. In order for this to be true, the size of the atom should be much less than the wavelength $\lambda$. For light atoms, we may estimate the atomic size by $R_\mathrm{atom}\sim a_0/Z$ where $a_0$ is the {\it Bohr radius}\index{Bohr radius} and $Z$ the charge. Combining this with the fact that the photon energy $\hbar\omega$ should roughly equal the transition energy $Z^2e^2/a_0$ in the atom, one finds an estimate 
\begin{equation}
\frac{R_\mathrm{atom}}{\lambda}\sim\frac{Ze^2}{c\hbar}\approx\frac{Z}{137}\ll1.
\end{equation}
Thus, the wavelength is typically two orders of magnitude larger than the size of the atom. One may argue that in the microwave regime, see tab.~\ref{expdatatable}, when atomic Rydberg states\index{Rydberg state} are used, the sizes of the atoms become very large. The size scales as $\sim n^2$ with $n$ the principle quantum number. However, the transition energy between the atomic states falls off like $n^{-3}$ meaning that the wavelength actually scales as $n^3$ and the approximation is actually even better.  

\begin{figure}
\includegraphics[width=10cm]{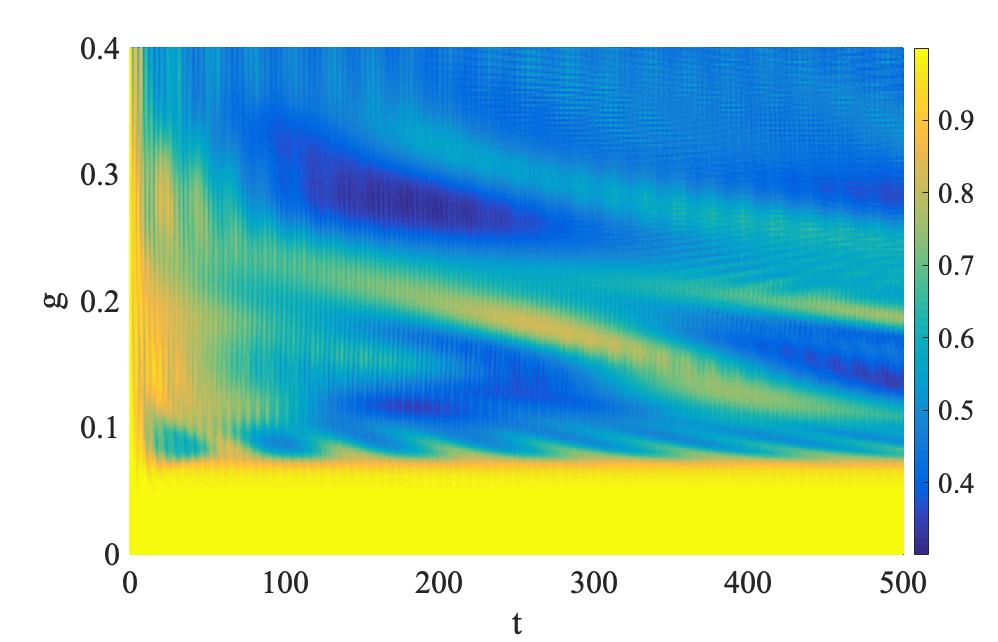} 
\caption{The purity $P=\mathrm{Tr}\left[\hat\rho_\mathrm{af}^2(t)\right]$ for the reduced state of the fundamental mode and the two-level atom, {\it i.e.} the second mode has been integrated out. The fundamental mode is resonant with the atomic transition, $\omega=\Omega=1$, while the second mode is largely detuned, $\omega_2=2\omega$. Shown is the time-evolution of the purity for different coupling strengths $g$ and for an initial product state with the atom excited, the fundamental mode in a coherent state and the second mode in vacuum; $|\psi(0)\rangle=|\alpha\rangle|0\rangle|e\rangle$, with $\alpha=\sqrt{39}$. Expectedly, for short times, $t\lesssim10$ the purity is close to unity, {\it i.e.} the dephasing due to the second mode is negligible. What is most striking is the rapid drop in the purity as the coupling $g\gtrsim0.08$. }
\label{2modpur}
\end{figure}

\subsubsection{Single-mode approximation}\label{sssec:smapp}
The multimode quantum Rabi model\index{Multimode! quantum Rabi model} pertaining to a two-level system coupled to an infinite set of quantized resonator modes is described by the Hamiltonian  
\begin{equation}\label{mrabiham}
\hat{H}_\mathrm{multiR}=\sum_m\omega_m\hat{n}_m+\frac{\Omega}{2}\hat{\sigma}_z+\sum_mg_0\sqrt{m+1}\left(\hat{a}_m+\hat{a}^\dagger_m\right)\hat{\sigma}_x,
\end{equation}
where for a perfect lossless cavity the frequencies 
\begin{equation}\label{freqm}
\omega_m=\frac{\pi c}{L}(m+1),\hspace{1.4cm}m=0,\,1,\,2,\,\dots
\end{equation}
with $c$ the speed of light in vacuum.  Here we have used the fact that the electric field amplitude scales as $E\propto\sqrt{\omega_m}\propto\sqrt{m+1}$, whence the light-matter coupling between the $m$'th mode and the atom scales as~\cite{sundaresan2015beyond,gely2017convergence} \begin{equation}\label{gm}
g_m=g_0\sqrt{m+1}.
\end{equation} 
To give a first simple idea on how the presence of off-resonant modes affects the single-mode physics, we take the two-mode quantum Rabi model, {\it i.e.} we include only the first two terms in the sum of~(\ref{mrabiham}). If the additional mode is far detuned from the fundamental mode, to lowest order it will not participate in the evolution. However, for long enough time-scales the additional mode will inevitably get entangled with the qubit and the other mode. When this happens, the purity~(\ref{purity})\index{Purity} of the reduced density state $\hat\rho_\mathrm{af}$ will drop below unity. We show how this manifests in fig.~\ref{2modpur} as the coupling is varied. What is somewhat surprising is that the second mode stays disentangled from the rest for very long time-scales provided the coupling is relatively small. But beyond a certain value of the coupling strength, this disentanglement behaviour is suddenly lost and already at moderate times the second mode cannot be neglected. 

Both the single-mode approximation discussed in this section and the TLA discussed in the next section rely on truncating the Hilbert space. As we will see, in either case the dimensional reductions lead to conceptual problems not often discussed in the literature on JC type of models. The Hilbert space truncation\index{Hilbert space truncation} turns out to be problematic in terms of deriving a gauge invariant\index{Gauge! invariance} effective model, which for the single-mode approximation results in divergences in QED. In particular, including an infinite number of modes in QED is known to result in divergences and the necessity of renormalization\index{Renormalization}~\cite{peskin2018introduction}. 

Upon eliminating the off-resonant modes, a single mode gives rise to an energy shift of the bare atomic levels
\begin{equation}\label{vsh}
\gamma_m\approx\frac{2g_m^2}{\omega_m}\hat\sigma_z=\frac{2g_0}{\omega_0}\hat\sigma_z,
\end{equation}
which summed together gives the Lamb shift\index{Lamb! shift} $\gamma$. However, this diverges in the limit of infinite modes; 
\begin{equation}\label{gammalam}
\gamma=\sum_m\gamma_m\rightarrow\infty
\end{equation}
as $m\rightarrow\infty$. Thus, in such perturbative approach the complication of divergences, characteristic of QED, survives~\cite{houck2008controlling,filipp2011multimode,gely2017convergence}. Actually, the spontaneous emission\index{Spontaneous! emission! rate} rate of an atom in free space will also diverge unless an ultraviolet cutoff\index{Frequency cutoff! ultraviolet} is introduced~\cite{malekakhlagh2017cutoff}. Another inconsistency resulting from including infinite number of modes is found in the multimode Dicke model\index{Multimode! Dicke model}, {\it i.e.} consider the Dicke model~(\ref{dickeham}) for an infinite number of modes. This accounts to take the Hamiltonian~(\ref{mrabiham}) and replace the spin-$1/2$ Puali operators with general spin-$S$ operators; $\hat\sigma_\alpha\rightarrow\hat S_\alpha$. We discussed in some detail the normal-superradiant PT\index{Normal-superradiant phase transition}\index{Phase transition! normal-superradiant} of the Dicke model\index{Dicke! model}\index{Model! Dicke} in sec.~\ref{sssec:dicke}. In particular, there is a critical coupling $g_c=\sqrt{\omega\Omega}/2$, see eq.~(\ref{dickecrit}), beyond which the ground state becomes macroscopically populated, {\it i.e.} $\langle\hat n\rangle\gg0$ in the limit of very large spins $S$. In the multimode Dicke model\index{Multimode! Dicke model} the critical point\index{Critical! point} for which the vacuum ground state becomes unstable can be derived analytically in the same way as for the single mode Dicke model using the Holstein-Primakoff mapping\index{Holstein-Primakoff! mapping} (\ref{hbdicke}) to recast the problem into one quadratic in the boson operators~\cite{tolkunov2007quantum}. The result is given by the solution of the equation
\begin{equation}
4\sum_m\frac{g_m^2}{\omega_m}=\Omega,
\end{equation}
which for the single-mode case reproduces the expression~(\ref{dickecrit}). Using eqs.~(\ref{freqm}) and~(\ref{gm}) one has $g_c^{(M)}=g_c^{(1)}/M$ for $M$ modes, and hence for an infinite set of modes $g_c\rightarrow0$. Thus, placing the atoms inside the (perfect) resonator populates the modes macroscopically! In the next section~\ref{sssec:dia} we will see that this problem also depends on neglecting the diamagnetic term\index{Diamagnetic term} which shifts the value of the critical coupling $g_c$. In fact, Malekakhlagh {\it et al.} pointed out that the divergences result from considering a gauge non-invariant model~\cite{malekakhlagh2017cutoff}. More particularly, the inclusion of the diamagnetic term\index{Diamagnetic term} $\hat A(0)$ is essential in order to find divergence-free quantities. It was shown, in a circuit QED configuration, that in a gauge-invariant\index{Gauge! invariance} approach the couplings scale as $g_M=g_0/\sqrt{M}$ for large $M$'s, which makes the sums finite as $M\rightarrow\infty$. The results of ref.~\cite{malekakhlagh2017cutoff} suggest that, as in the TLA, the problem of divergences originates from giving up gauge invariance\index{Gauge! invariance}. Another conceptual issue arising in the single-mode quantum Rabi model is that it allows for superluminal signaling, {\it i.e.} breakdown of causality~\cite{munoz2018resolution}. Causality is restored by introducing the set of neglected modes. 

For a long time, the problem of a diverging Lamb shift\index{Lamb! shift} was not discussed. However, when reaching large values of $g_m$ with the birth of circuit QED (section~\ref{sec:cirQED}) the issue was taken up more seriously. Nigg {\it et al.} suggested that to overcome the divergences, the system, qubit plus electromagnetic field, should be quantized simultaneously -- in the spirit of a``{\it blackbox quantization}''~\cite{nigg2012black,solgun2014blackbox}. The emerging low-energy effective model is free from divergences and expressed in terms of a nonlinear polaronic mode $\hat a_p$. In such quadrature representation~(\ref{quad}) of this mode, the resulting Hamiltonian contains even powers of $\hat x\propto\left(\hat a_p+\hat a_p^\dagger\right)$ up to sixth order. 

A drawback of the blackbox quantization\index{Blackbox quantization} is that the effective model does not have the structure of a quantized Rabi model with a spin-$1/2$ coupled to a boson mode. This motivated the authors of~\cite{gely2017convergence} to take a different approach based on renormalizing the system parameters as the number $M$ of modes was increased, {\it i.e.} $g_m\rightarrow g_m^{(M)}$ and $\Omega\rightarrow\Omega^{(M)}$. The effective model is obtained by the corresponding quantum Rabi model when $M\rightarrow\infty$. The authors showed that when the renormalization is handled with care, the system parameters approach a non-zero finite value which is in agreement with the blackbox quantization~\cite{nigg2012black}. Importantly, the final Hamiltonian has the desired Rabi form, and the renormalized atomic transition frequency and light-matter coupling strength remain on the same order as those of the original single-mode approximation model. A corresponding renormalization scheme, or black-box quantization, as discussed in terms of circuit QED seem to be missing for a cavity QED setting. 

Let us for now leave the issue of divergences and return to the multimode quantum Rabi model\index{Multimode! quantum Rabi model}~(\ref{mrabiham}). In fact, this model has already been discussed in sec.~\ref{sssec:multi} in terms of the spin-boson model\index{Spin-boson model} [see eq.~(\ref{spinboson})], for which the spectral density\index{Spectral! density} of the boson modes was given by~(\ref{specfun}). It was found that the exponent $s$ of the spectral density determines the characteristics of the model. In a cavity, the spectral density is not expressed as a continuous function $\sim\omega^s$, but the spectral function is {\it structured} according to eq.~(\ref{freqm}). We can get some estimates by looking at typical experimental data. The mode separation\index{Mode! separation} $\Delta\omega=\pi c/L$. With typical values  $L\sim25$ mm for a Fabry-P\'erot microwave\index{Fabry-P\'erot! cavity/resonator} cavity~\cite{bernu2008freezing} and  $L\sim10-100$ $\mu$m for optical Fabry-P\'erot cavities~\cite{hood2001characterization,maunz2004cavity}, we find $\Delta\omega\sim40$ GHz and $\Delta\omega\sim20$ THz respectively. Thus, this is a lower scale of the detuning from the omitted modes, which should be compared to corresponding light-matter couplings, $g<1$ MHz and $g\sim100$ MHz, see tab.~\ref{expdatatable}. From the solutions~(\ref{dstate}) we find that in the perturbative regime the population of the neglected mode should scale as $4g^2/\Delta\omega^2$ which is $\sim1\times10^{-9}$. This simple estimate shows that in typical cavity QED settings, either operating in the microwave or optical regime (see next section), the single-mode approximation is justified. The mode structure~(\ref{freqm}) is similar in circuit QED~\cite{egger2013multimode} (see section~\ref{sec:cirQED}), {\it e.g.} the resonator length $L\sim1$ cm~\cite{Blais2004QED}. Here, on the other hand, the light-matter coupling $g$ can be substantially larger in comparison to the photon frequency $\omega$ such that the system operates in the ultrastrong coupling regime\index{Ultrastrong coupling regime}~\cite{niemczyk2010circuit,forn2010observation,bosman2017multi,forn2019ultrastrong,le2020theoretical}. The single-mode approximation is then put to a test, and the off-resonant modes may become important~\cite{houck2008controlling,gely2017convergence,bosman2017multi}. 

The spontaneous emission\index{Spontaneous! emission} of a two-level system in such a structured environment generated by the resonator was experimentally studied in a circuit QED setting, and due to the strong coupling the far detuned modes of the transmission line had to be taken into account in order to explain the experimental results~\cite{houck2008controlling}. In ref.~\cite{sundaresan2015beyond} (see also~\cite{sletten2019resolving}), a multimode circuit QED system\index{Multimode! circuit QED system} was studied experimentally by tuning the q-dot frequency $\Omega$ through a series of avoided level crossings corresponding to the different modes frequencies. It was found that the results could not be explained within a set of simple single-mode quantum Rabi models, but several modes needed to be simultaneously included in order to get good agreement between theory and experiment.

For configurations involving more than a single two-level system, the additional detuned modes do not only result in virtual energy shifts but also induce effective couplings between the two-level systems~\cite{filipp2011multimode,jaako2016ultrastrong}. This was already noted in sec.~\ref{sssec:dicke} when discussing the Dicke model and how it results in the Lipkin-Meshkov-Glick model~(\ref{lmg})\index{Lipkin-Meshkov-Glick! model}\index{Model! Lipkin-Meshkov-Glick} upon eliminating the boson degrees of freedom~\cite{latorre2005entanglement}. For two qubits, and assuming the RWA, this induces an $XY$ type of coupling~\cite{zheng2000efficient,blais2007quantum}, {\it i.e.} 
\begin{equation}\label{excoup}
\hat H_{XY}=J\left(\hat\sigma_1^+\hat\sigma_2^-+\hat\sigma_2^+\hat\sigma_1^-\right)=J\left(\hat\sigma_x^{(1)}\hat\sigma_x^{(2)}+\hat\sigma_y^{(1)}\hat\sigma_y^{(2)}\right)
\end{equation} 
for an effective exchange coupling strength $J$ scaling like the Lambda shift\index{Lambda! shift}~(\ref{gammalam}). In the next section we will come back to this when discussing entanglement generation in cavity QED setups, see eq.~(\ref{atat}). For two identical qubits, the interaction term~(\ref{excoup}) couples the bare states of the system and in particular, the eigenstates of $\hat H_{XY}$ are the EPR states\index{EPR state}\index{State! EPR} $|EPR\rangle=(|e,g\rangle\pm|g,e\rangle)/\sqrt{2}$. This coupling implies an avoided crossing between the bare states, which would not be present within the single-mode approximation. This has been experimentally observed in a circuit QED setting~\cite{filipp2011multimode}, which gives an estimate of the coupling strength $J$. 

\begin{figure}
\includegraphics[width=10cm]{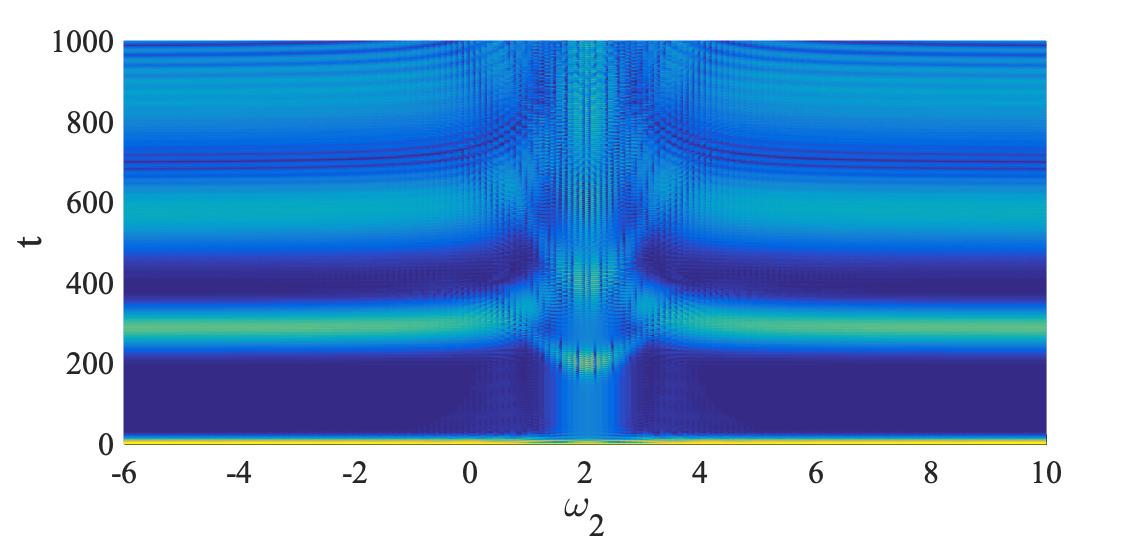} 
\caption{The absolute value of the atomic inversion\index{Atomic! inversion} $|W(t)|=|\langle\hat\sigma_{ee}\rangle-\langle\hat\sigma_{gg}\rangle|$ of the extended JC model~(\ref{3ham}) including one extra internal level $|a\rangle$. On the $x$-axis we vary the energy $\omega_2$ of the third level, and as it becomes far detuned from the resonant transition, the inversion displays the regular collapse-revival structure as first shown in fig.~\ref{fig3} (a). In the regime $0<\omega<4$ the collapse revival pattern is greatly altered. Here, the third level gets populated and interferences between the involved transitions causes new structures to emerge. The remaining parameters are $\omega=\Omega=1$ and $g=0.1$, and the initial state is $|\psi(0)\rangle=|\alpha\rangle|e\rangle$, with $\alpha=\sqrt{50}$.
} 
\label{twolevapp}  
\end{figure} 

\subsubsection{Two-level approximation}\label{sssec:tla}
For a long time, a common belief in the community was that whenever the coupling $g$ becomes large enough such that the RWA breaks down, so will the TLA\index{Two-level! approximation}. Of course, this assumption must depend on details of the system under investigation. Normally, the neglected levels are assumed to be far-off-resonance compared to the corresponding light-matter coupling. Thus, in this case these levels will only be virtually populated. The effect, to lowest order, will be Stark shifts of the two levels forming the JC dimer~\cite{moya1991power,villanueva2020effect}. It may happen that the RWA is not justified in such a configuration, but a TLA is. Eventually, as $g$ is increased the TLA must however break down~\cite{Ashida2021}. Let us first demonstrate how the virtual energy shifts come about, and then argue that the TLA will break down in the extreme-coupling regime\index{Extreme coupling regime}. We will also show that other issues arise due to the TLA, namely the breakdown of gauge invariance\index{Gauge! invariance}. 

To give an example of the Stark-shift effect we take a most simple extension of the JC model to include a third internal level $|a\rangle$ with a dipole coupling between this level and the level $|e\rangle$ with a coupling strength $g$ the same as for the JC transition between $|g\rangle$ and $|e\rangle$. For simplicity we assume the RWA, and the Hamiltonian is given by
\begin{equation}\label{3ham}
\hat H_\mathrm{JC+1}=\omega\hat n+\Omega\hat\sigma_{ee}+\omega_2\hat\sigma_{aa}+g\left(\hat a\sigma_{eg}+\hat a^\dagger\hat\sigma_{ge}\right)+g\left(\hat a\sigma_{ea}+\hat a^\dagger\hat\sigma_{ae}\right),
\end{equation}
where we use the terminology of sec.~\ref{sssec:multi}, {\it i.e.} $\hat\sigma_{\alpha\beta}=|\alpha\rangle\langle\beta|$. For the observable we take the atomic inversion $|W(t)|=|\langle\hat\sigma_{ee}\rangle-\langle\hat\sigma_{gg}\rangle|$~(\ref{inv}), which we know shows clear signatures of the collapse-revivals\index{Collapse-revival}, like for example in fig.~\ref{fig3} (a), given that the initial state is localized in Fock space, {\it e.g.} a coherent state. In fig.~\ref{twolevapp} we give an example of the inversion as a function of the third atomic level energy $\omega_2$; for $\omega_2=\Omega=1$ the third level is resonant with the field and the weights of the levels $|e\rangle$ and $|a\rangle$ are identical. Far from the resonance the importance of the third level is diminished. For $\omega_2=2\Omega$ some interferences between the involved bare states take place which greatly affect the inversion. 

In the vicinity of $\omega_2=2\Omega$ it is not possible to explain the effect of the third levels by simple dispersive shifts\index{Dispersive! shift} of the bare JC energies. However, further away from this regime the characteristics of the system are qualitatively intact and the effect of the virtually populated levels can be treated perturbatively. In ref.~\cite{villanueva2020effect}, the size $\chi$ of such Stark-shift term\index{Dispersive! Hamiltonian}
\begin{equation}
\hat H_{\mathrm{eff}JC}=\hat H_{JC}+\chi\hat n\hat\sigma_z
\end{equation}
was derived. The authors considered $K$ ancilla levels with energies $\omega_k$, and dipole coupled to $|e\rangle$ with strengths $\eta_k$. To adiabatically eliminate\index{Adiabatic! elimination} the ancilla levels, the Schrieffer-Wolff transformation $\exp(\hat S)$\index{Schrieffer-Wolff transformation} was applied [see eq.~(\ref{polaron})], with 
\begin{equation}\label{starkjc}
\hat S=\sum_{k=1}^K\xi_k\left(\hat a|k\rangle\langle e|-\hat a^\dagger|e\rangle\langle e|\right).
\end{equation}
Like in sec.~\ref{ssec:JCm}, the parameters $\xi_k$ are chosen such that the transformed Hamiltonian does not contain direct coupling terms to linear order $\mathcal{O}(\xi_k)$. With $\Delta_k=\omega_k-\omega$, the Stark-shift\index{Stark shift} amplitude becomes $\chi=2\sum_{k=1}^K\frac{\eta_k^2}{\Delta+\Delta_k}$, where the RWA\index{Rotating-wave approximation} has been implemented by assuming $\omega_k\gg\omega$. The Stark shifted JC model is still analytically solvable, with the effect of shifting the detuning $\Delta\rightarrow\Delta+\chi(2n+1)$. The result of the Stark-shift term was analyzed in terms of the resonance fluorescence spectrum\index{Resonance fluorescence! spectrum}
\begin{equation}
S(\nu)=\mathrm{Re}\left[\int_0^\infty e^{-i\nu\tau}\langle\hat{\sigma}_{+}(t+\tau)\hat{\sigma}_{-}(t)\rangle d\tau\right].
\end{equation}
Expectedly, for $\chi$ on the order of the coupling $g$ the spectrum is changed considerably~\cite{villanueva2020effect}.

The above analysis is phenomenological, no actual physical system is considered. {\it Ab initio} studies of the TLA problem in cavity QED systems seem to be lacking. In circuit QED, the questions was raised in ref.~\cite{manucharyan2017resilience}, by considering up to 100 levels of the q-dot. The results from numerical simulations, either truncating to two levels or including all 100 levels were compared. It was found that details could become important, {\it e.g.} while the quantum Rabi model correctly captured the characteristics for the Cooper pair box\index{Cooper pair! box}, it failed for the {\it fluxion} q-dot (see section~\ref{sec:cirQED} for various types of q-dots in circuit QED) in the deep strong coupling regime\index{Deep strong coupling regime}, $g\gg\omega$. A more recent {\it ab initio} study was also presented in ref.~\cite{kaur2020absence} for the {\it Cooper pair box}\index{Cooper pair! box} and the {\it transmon qubit}\index{Transmon qubit}. In their study, also the single mode approximation was not taken into account and they focused on the QPT of the spin-boson model~(\ref{spinboson}) discussed in sec.~\ref{sssec:multi}. Without imposing the TLA, they derived a modified spin-boson model, and they solved it numerically with a renormalization group\index{Renormalization! group} method. Complementing their numerical results with analytical approximate results (similar to the no-go theorem\index{No-go theorem} of the Dicke PT discussed in sec.~\ref{sssec:dicke}) they demonstrated that the TLA breaks down for the transmon in the ultrastrong coupling regime. As a result, there is no QPT for the transmon realization of the spin-boson model, unless it is possible to greatly increase the nonlinearity that would reestablish the validity of the TLA. 

In ref.~\cite{Ashida2021}, Ashida, Imamoglu, and Demler considered a rather generic interacting light-matter model and introduced a new regime; extreme strong coupling regime\index{Extreme strong coupling regime} in which the TLA breaks down. Hence, this regime is defined by the breakdown of the TLA, stated as ``{\it ...we conclude that
effective models derived by level truncations, such as tight-binding models or the quantum Rabi model, must inevitably break down when $g$ becomes sufficiently large}''. The crucial observation in~\cite{Ashida2021} is that in the polaron basis\index{Polaron! basis}, {\it i.e.}, after unitary transforming the Hamiltonian with an operator $\hat U$ similar in form as~(\ref{polaron}), the matter and light degrees of freedom decouple as $g\rightarrow\infty$. Importantly, they do not impose any TLA at this stage, but consider a general situation with an electron in some potential $V(x)$. In the polaron frame, the transformed Hamiltonian reads
\begin{equation}\label{polham}
\hat H_\mathrm{Pol}=\frac{\hat p^2}{2m_\mathrm{eff}}+V(x+\xi_g\hat\pi)+\Omega_g\hat b^\dagger\hat b,
\end{equation}
where the renormalized frequency $\Omega_g=\sqrt{\omega^2+2g^2}$, the effective electron mass $m_\mathrm{eff}=m\left[1+2(g/\omega)^2\right]$, $\hat\pi=i\left(\hat b^\dagger-\hat b\right)$, $\xi_g=g/\sqrt{m}\Omega_g^{3/2}$, and the Bogoliubov transformed\index{Bogoliubov transformation} boson operators are expressed in terms of the old ones by
\begin{equation}
\hat b=\frac{1}{2}\left(\sqrt{\frac{\Omega_g}{\omega}}+\sqrt{\frac{\omega}{\Omega_g}}\right)\hat a+\frac{1}{2}\left(\sqrt{\frac{\Omega_g}{\omega}}-\sqrt{\frac{\omega}{\Omega_g}}\right)\hat a^\dagger.
\end{equation}
As $g\rightarrow\infty$, the two degrees of freedom decouple since $\xi_g\rightarrow0$, and the wave function in this polaron picture is a simple product, {\it e.g.} the ground state assumes the form $|\Psi_\mathrm{Pol}\rangle\approx|\psi_{U}\rangle|0\rangle_{\Omega_g}$. The electron wave function $|\psi_{U}\rangle$ will be well localized in real space since the effective mass\index{Effective! mass} scales as $\sim g^2$.  Going back to the original basis, the matter and light states will be entangled, so the polaron transformation is a convenient basis in which the problem simplifies. To arrive at this result, the diamagnetic self-energy term\index{Diamagnetic term}~(\ref{set}) must be included. Note that the above Hamiltonian~(\ref{polham}) does not contain any TLA (within the TLA the polaron transformation has also been employed for decoupling the degrees of freedom~\cite{kurcz2014hybrid}). The tight localization of $|\psi_{U}\rangle$ for large $g$ values implies the occupation of large-momentum components and that in the original basis $|\psi_{U}\rangle$ many bare-energy eigenstates will be populated. Thus, for some large $g$ the TLA must break down as mentioned above.

In sec.~\ref{sssec:drivejc} we discussed the driven JC model, and in particular it was shown how a coherent drive could be used to engineer certain Hamiltonians, see eq.~(\ref{drham}). However, periodic driving of a system typically lead to heating provided the spectrum of the system is sufficiently nonlinear. Thus, we can imagine a breakdown of the TLA for driven JC systems. In~\cite{lescanne2019escape}, a periodically driven transmon q-dot coupled to a transmission line (see section~\ref{sec:cirQED}) was experimentally studied and it was found that the TLA broke down as unbound q-dot states were coupled.

Let us discuss a problem of the TLA of more fundamental character~\cite{barton1974frequency,settineri2019gauge} (see also the review article~\cite{stokes2020implications}). To derive the Hamiltonian~(\ref{coulham2}), given in the Coulomb gauge, we projected the full dipole Hilbert space $\{|\varphi_i\rangle\}$ onto a truncated subspace\index{Truncated Hilbert space};
\begin{equation}
\hat H_\mathrm{C}=\hat P\left(\hat H_\mathrm{f}+\hat H_\mathrm{d}+\hat H_\mathrm{I}+\hat H_\mathrm{SE}\right)\hat P,
\end{equation}
where $\hat P=|g\rangle\langle g|+|e\rangle\langle e|$. The projection implies that the resulting Hamiltonian is not gauge invariant\index{Gauge! invariance}, {\it i.e.} the gauge choice made for the original Hamiltonian will affect the form of the effective Hamiltonian achieved after the projection~\cite{de2018breakdown,stokes2019gauge,di2019resolution,le2020theoretical}. To demonstrate this, we consider the {\it dipole gauge}, in which the vector potential\index{Vector potential} appearing in the canonical momentum is `boosted away' by the unitary transformation\index{Gauge! transformation} $\hat U_\mathrm{D}=\exp\left(-iq\hat x\hat A\right)$, {\it i.e.} 
\begin{equation}
\hat H_\mathrm{D}=\hat U_\mathrm{D}\hat H_\mathrm{C}\hat U_\mathrm{D}^\dagger=\frac{\hat p^2}{2m}+\tilde V(\hat x)+\omega\hat a^\dagger\hat a+i\omega q\mathcal{A}_0\left(\hat a^\dagger-\hat a\right)\hat x,
\end{equation}
with the transformed potential $\tilde V(\hat x)=V(\hat x)+q^2\mathcal{A}_0^2\omega\hat x^2$. It follows that $\hat P\hat H_\mathrm{D}\hat P$ is not unitary-equivalent to the corresponding Hamiltonian in the Coulomb gauge, $\hat P\hat H_\mathrm{C}\hat P$, since in general $\left[\hat P,\hat U_\mathrm{D}\right]\neq0$. Whether the projector will commute with the unitary $\hat U_\mathrm{D}$ or not, depends on the shape of the dipole eigenstates $|\varphi_i\rangle$, and thereby on the potential $V(\hat x)$. This boils down to the explicit expressions for matrix elements $\langle\varphi_i|\hat p|\varphi_j\rangle$ and $\langle\varphi_i|\hat x|\varphi_j\rangle$ appearing in the derivation of the JC model~(\ref{coulham2}). For $V(\hat x)\propto\hat x^2$ the two gauges are identical, but for, say, a double-well potential they will be different~\cite{de2018breakdown}. It is possible to introduce a more general gauge that is parametrized by some real parameter $\alpha$, and which includes both the dipole and the Coulomb\index{Coulomb gauge} choices by picking $\alpha=0$ or $\alpha\neq 0$ respectively~\cite{stokes2019gauge}. The resulting projected Hamiltonian, which now is $\alpha$-dependent, becomes the anisotropic quantum Rabi model\index{Anisotropic! quantum Rabi model}~(\ref{anRabi}). The gauge setting fixes the coupling strengths $g_\mathrm{jc}(\alpha)$ and $g_\mathrm{ajc}(\alpha)$. An interesting observation is that there exists a particular $\alpha=\alpha_\mathrm{jc}$, the {\it Jaynes-Cummings gauge}\index{Jaynes-Cummings! gauge}~\cite{savasta2020gauge}, for which the derived effective Hamiltonian has the form of a JC model~\cite{stokes2019gauge}. Thus, within this gauge the $U(1)$ symmetry\index{$U(1)$ symmetry} is restored and the model is solvable. However, the renormalized system parameters, like the photon frequency, will now depend on the regular atom-light coupling $g$ obtained in the Coulomb gauge.

From the above discussion it transpires that gauge invariance\index{Gauge! invariance} is lost in the ultrastrong coupling regime\index{Ultrastrong coupling regime}~\cite{de2018breakdown,stokes2019gauge,stokes2020gauge}. In fact, Stokes and Nazir conclude that ``gauge non-invariance'' is a necessity of the TLA~\cite{stokes2020gauge}. This, of course, is rather unsatisfying since we want our physical observables to be gauge invariant -- that there is one ``correct'' (effective) model that describes reality. In a series of papers~\cite{di2019resolution,garziano2020gauge,savasta2020gauge} by Nori and co-workers it was demonstrated how gauge invariance could be reestablished in the quantum Rabi model by performing the TLA in such a way that a unique quantum Rabi model derives. The origin of the problem lies in that gauge transformation\index{Gauge! transformation} $\psi(x)\rightarrow e^{iq\theta(x)}\psi(x)$ is non-local, and in principle $\hat Pe^{iq\theta(x)}\hat P\neq e^{iq\theta(\hat Px\hat P)}$ since $\hat P=|g\rangle\langle g|+|e\rangle\langle e|$ is not the identity operator. Nori {\it et al.} argue that $\hat P$ should be used as the identity since the two-level system lives in the truncated space~\cite{savasta2020gauge}, and then they derive the {\it gauge invariant quantum Rabi model}\index{Gauge! invariant quantum Rabi model} 
\begin{equation}\label{girabi}
\hat H_\mathrm{giR}=\omega\hat a^\dagger\hat a+\frac{\Omega}{2}\left\{\hat\sigma_z\cos\left[2\eta\left(\hat a+\hat a^\dagger\right)\right]+\hat\sigma_x\sin\left[2\eta\left(\hat a+\hat a^\dagger\right)\right]\right\},
\end{equation}
where $\eta=g/\omega$~\cite{settineri2019gauge}. A comparison of the spectrum of the model~(\ref{girabi}) and of the quantum Rabi model (\ref{rabiham}) was given in ref.~\cite{di2019resolution}, and the two agreed even for very large coupling strengths $g$. They extended their analysis of `gauge invariance' of the quantum Rabi model to open systems in~\cite{salmon2021gauge}. The issue here is whether measurable quantities, like the cavity emission spectra, will be gauge invariant. To further the perspective, the authors of~\cite{salmon2021gauge} conclude that the nature of the system-bath coupling is also of importance. It should be related to the quadrature coupling to the external fields -- distinguishing for example vector potential\index{Vector potential} coupling from electric-field coupling -- and the observables to ensure a gauge-invariant master equation\index{Master equation}, going beyond a simple phenomenological system-bath interaction. The limits of phenomenological dissipative JC models\index{Dissipative! JC model! phenomenological} are put into test by the analysis of~\cite{Franke2019} introducing a second quantization scheme based on {\it quasi}normal modes\index{Quasi! normal modes} with complex eigenfrequencies. The creation and annihilation operators for photonic or plasmonic resonances are derived from the vector-valued {\it quasi}normal mode eigenfunctions -- satisfying Helmholtz equation in a structured dielectric medium with an appropriate boundary condition -- subject to a symmetrization procedure to bypass the non-Hermitian nature of the problem. The proposed scheme enables the construction of Fock states, required for addressing problems in multiplasmon or multiphoton quantum optics\index{Multiphoton! quantum optics}. Shortly afterwards, significant differences in the quantum master equation\index{Master equation} and input-output relations\index{Input-output! theory} between the {\it quasi}normal-mode quantum model and phenomenological dissipative JC models were identified in~\cite{Franke2020}. Light-matter interaction in an inhomogeneous dispersive and absorbing dielectric is a special example in a general theory of gauge invariance under material and photonic subspace truncation in arbitrary media developed in~\cite{Gustin2022}. We finally note that, since the system-bath interaction is crucial in the formulation of the open-system dynamics, gauge ambiguities may influence the calculation of photodetection probabilities\index{Photoelectric detection} since the canonical momentum for the field is gauge-dependent~\cite{Healy1980, Power1980}. A resolution of this issue is achieved in~\cite{settineri2019gauge} with reference to Glauber's formula for the photodetection probability as well as to Fermi's golden rule\index{Fermi's golden rule} for the excitation rate of two-level sensors. For instance, when a system is prepared in the initial energy eigenstate $\ket{j_C}$, the detection rate at a frequency $\omega=\omega_{j,k}=\omega_j - \omega_k$ for a single transition in the Coulomb gauge is proportional to 
\begin{equation}
W_{j,k}=|\braket{k_C|\hat{\mathcal{P}}|j_C}|^2, 
\end{equation}
with $\hat{\mathcal{P}}=i(\hat{a}-\hat{a}^{\dagger})$. In contrast, in the dipole gauge one instead has
\begin{equation}
 W_{j,k}=\left|\braket{k_D|i(\hat{a}-\hat{a}^{\dagger})-2\eta \hat{\sigma}_x|j_D}\right|^2.
\end{equation}
The differences from the (incorrect) expression $W_{j,k}^{\prime}=\left|\braket{k_D|i(\hat{a}-\hat{a}^{\dagger})|j_D}\right|^2$ are evident already for $\eta \sim 0.1$ (see fig. 2 of~\cite{settineri2019gauge}). A generalized master equation\index{Master equation} in the dressed basis was used in~\cite{Mercurio2022} for the calculation of emission rates and spectra in the Rabi model under incoherent excitation, from the weak to the deep strong-coupling regime. Different metrics quantifying the gauge performance can be introduced depending on the observable of interest. The optimal gauge choice is generally mode dependent, meaning that a different gauge is needed for each cavity mode, while the degree of light-matter entanglement is not in correlation with that choice~\cite{Arwas2023}.

As discussed in sec.~\ref{sssec:openjc}, upon reaching the ultrastrong coupling regime, the phenomenological master equation\index{Master equation}~(\ref{master1}) mimicking photon losses breaks down~\cite{carmichael1973master,werlang2008rabi}, and more involved Lindblad equations must be considered. Furthermore, Stokes and Nazir argue that in controlling which gauge-invariant observables are used to define a material system, the choice of gauge affects the balance between the material system's localization and its electromagnetic dressing, which has an imprint to the route from the effective models derived down to photodetection theory~\cite{stokes2020implications}\index{Photoelectric detection}. Furthermore, in their recent work, Hughes and collaborators have shown that a correspondence between a classical and a quantum mechanical dissipative Hopfield model\index{Hopfield! model}\index{Model! Hopfield}\index{Dissipative! Hopfield model} in the ultrastrong coupling regime can be established only when the quantum models respect gauge invariance and employ the appropriate bath-coupling operators in the generalized master equation\index{Master equation}~\cite{hughes2023reconciling}. 

\begin{figure}
\includegraphics[width=10cm]{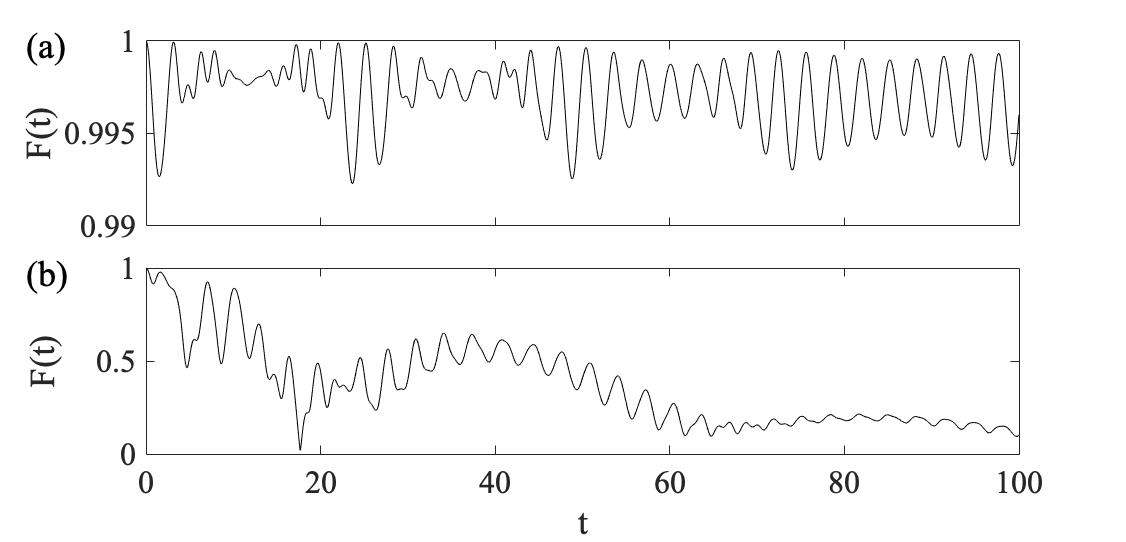} 
\caption{The fidelity~(\ref{fidel}) between the states $|\psi_\mathrm{JC}(t)\rangle$ and $|\psi_\mathrm{R}(t)\rangle$, where the respective states are time-evolved using either the resonant JC Hamiltonian~(\ref{jcham}) or the resonant quantum Rabi Hamiltonian~(\ref{rabiham}). The initial state is $|\psi(0)\rangle=|\alpha\rangle|e\rangle$, with $\alpha=\sqrt{50}$. In the plot of frame {\bf (a)} $\omega/g=0.018$, which is a typical experimental value in circuit QED~\cite{schuster2007resolving}, while in plot (b) we have $\omega/g=0.1$ corresponding to the circuit QED experiment operating beyond the ultrastrong coupling regime~\cite{niemczyk2010circuit}. For these time-scales we see that the fidelity states above 99$\%$ with $g=0.018\omega$. In both figures we consider the resonant case $\omega=\Omega=1$. For longer times the fidelity will drop, but these are time-scales well beyond the photon decay rate, and hence the RWA is a rather good approximation in this case. In {\bf (b)}, however, the fidelity drops rapidly and the RWA is not justified here.} 
\label{rwafid}  
\end{figure} 

\subsubsection{Rotating-wave approximation}\label{sssec:rwasub}
The breakdown of the RWA\index{Rotating-wave approximation} was discussed in some detail in sec.~\ref{sssec:rwaeff} when we compared the quantum Rabi model with the JC model. We will not repeat that discussion here, but will summarize a few results of that section. In particular, the spectra for the two models were shown in fig.~\ref{fig10} (a). The energy shifts arising due to the counter rotating terms are called the {\it Bloch-Siegert shifts}\index{Bloch!-Siegert shift}, and it becomes apparent that in the ultrastrong coupling regime\index{Ultrastrong coupling regime} these terms can no longer be neglected, as was also experimentally observed in~\cite{forn2010observation}. The same figure also showed the fidelity\index{State! fidelity}~(\ref{fidel}) $F=|\langle\psi_\mathrm{0JC}|\psi_\mathrm{0R}\rangle^2|$ between the two respective ground states. The fidelity drops smoothly for increasing $g$, as expected. At the critical point\index{Critical! point} where the system enters the superradiant\index{Superradiant} state, see sec.\ref{sssec:dicke}, the fidelity drops abruptly to zero. This is a result of the different symmetries of the two models; a $U(1)$ excitation symmetry for the JC model and a $\mathbb{Z}_2$ parity symmetry of the quantum Rabi model\index{Quantum! Rabi model}. 

Given the quantum Rabi Hamiltonian, the justification of the RWA depends solely on the amplitudes of the system parameters $\omega$, $\Omega$, and $g$, but not on other real physical quantities like,  say, the quality factor of the cavity or electronic states of the atom. In this sense, in order to understand the implications of the RWA we do not need to consider some real physical system and perform some {\it ab initio} calculation like for the TLA. To complement the numerical results of fig.~\ref{fig10}, we again look at the fidelity, but this time between the time-evolved states $|\psi_\mathrm{JC}(t)\rangle$ and $|\psi_\mathrm{R}(t)\rangle$, $F(t)=|\langle\psi_\mathrm{JC}(t)|\psi_\mathrm{R}(t)\rangle|$\index{State! fidelity}, where the two states have been time-evolved from the same initial coherent state $|\alpha\rangle|e\rangle$ by either the JC Hamiltonian~(\ref{jcham}) or the quantum Rabi Hamiltonian~(\ref{rabiham}) respectively. For short times $t\ll1$ we can approximate the fidelity by employing the {\it split-operator approximation} which splits the exponential and neglects corrections beyond linear terms in $t$,
\begin{equation}
\begin{array}{lll}
F(t) & = & \left|\langle\alpha,e|\exp\left(i\hat H_\mathrm{JC}t\right)\exp\left(-i(\hat H_\mathrm{JC}+\hat V_\mathrm{CR})t\right)|\alpha,e\rangle\right|\approx\left|\langle\alpha,e|\exp\left(-i\hat V_\mathrm{CR}t\right)|\alpha,e\rangle\right|\\ \\
& = & 1-g^2|\alpha|^2t^2/2+\mathcal{O}(t^4),
\end{array}
\end{equation}
where $\hat V_\mathrm{CR}$ is the counter rotating interaction~(\ref{counterterms}), and we have Taylor expanded the exponential in the last step. This result is expected as an example of the non-exponential decay of quantum states~\cite{sakurai1995modern}, and it is precisely the quadratic $t$-dependence underlying the mechanism of the quantum Zeno effect\index{Quantum! Zeno effect} that we briefly discussed in sec.~\ref{sssec:openjc}. Numerically one finds that in the strong coupling regime with $g\ll\omega$ the squared time-dependence of the fidelity survives for long times $t$ which demonstrates that the RWA is doomed to break down after sufficiently long times. Numerical results of the fidelity for two different coupling strengths $g$, corresponding to different experimental realizations, are displayed in fig.~\ref{rwafid}. The upper plot represents a situation in the strong coupling regime, {\it i.e.} $g\ll\omega$, while for the lower plot $g$ is such that the system operates in the ultrastrong coupling regime. In the ultrastrong coupling regime the fidelity drops rapidly on the time-scale of a few tenths of Rabi oscillations $(g\alpha)^{-1}$. In the example we have used $\alpha=\sqrt{50}$, and the decay of the fidelity depends on the magnitude of $\alpha$; the larger it is the faster the drop.  

Burgarth and coworkers~\cite{BurgathRWA2023, Burgarth2024tamingrotatingwave} consider the evolution of a Fock state $\ket{n}$ under the Rabi and JC Hamiltonian to assess the validity of the RWA for short times via the parameter $\sqrt{n}\,(g/\omega)$. They find that the maximal error $\epsilon_n$ the RWA incurs in a time interval up to $\pi/\omega$ is constrained by the non-perturbative bounds:
\begin{equation}
 5 \frac{g}{\omega}\sqrt{n+3}  \geq \epsilon_n \geq \frac{1}{6} - \frac{1}{216 n} \left(\frac{\omega}{g}\right)^2 - \frac{7}{12n},
\end{equation}
which shows that the approximation deteriorates for large photon numbers. Furthermore, corrections to the RWA have been recently obtained in~\cite{wang2023rotating} using renormalized perturbation theory.  

\begin{figure}
\includegraphics[width=10cm]{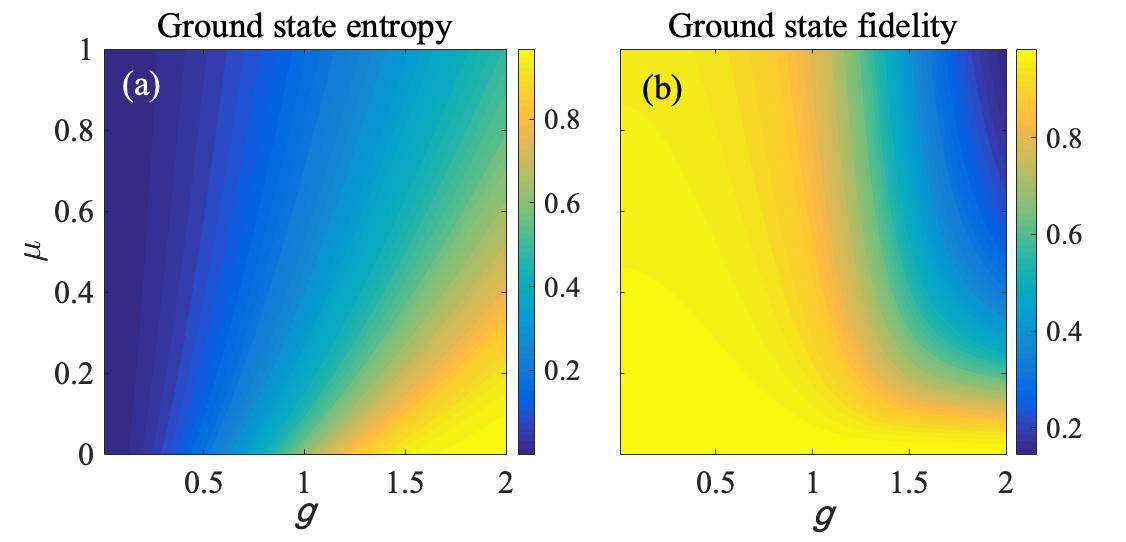} 
\caption{The von Neumann entropy\index{von Neumann! entropy}~(\ref{vN}) for the ground state {\bf (a)} and the fidelity $F(\mu)=|\langle\psi_{0+}(\mu)|\psi_{0+}(0)\rangle|$ {\bf (b)} as functions of the light-matter coupling $g$ and the diamagnetic coupling $\mu$. Along the horizontal line $\mu=0$ we regain the quantum Rabi model, {\it e.g.} the atom-field entanglement takes off around the critical coupling $g_c=\sqrt{\omega\Omega}/2$, and naturally here $F(0)=1$. We note that the presence of the diamagnetic term\index{Diamagnetic term}, $\mu\neq0$, suppresses the entanglement, as well as greatly reduce the fidelity. The drastic fidelity drop occurs in the regime where the adiabatic potential of the quantum Rabi model possesses two global minima, while the adiabatic potential of the Hamiltonian $\hat H_\mathrm{C}$ only supports a single global minimum. In both figures we used $\omega=\Omega=1$.} 
\label{diamag}  
\end{figure} 

\subsubsection{Neglecting the self-energy diamagnetic term}\label{sssec:dia}
Section~\ref{sssec:tla} made clear that the issue of gauge invariance becomes complicated in the truncated Hilbert space. This poses the question if there are preferable gauges in different physical contexts~\cite{viehmann2011superradiant,ciuti2012comment,jaako2016ultrastrong}. We will not continue this discussion here but take the minimal coupling Hamiltonian $\hat H_\mathrm{C}(\mu)$ of eq.~(\ref{coulham}) as our starting point. We note that the model obtained by keeping the diamagnetic term\index{Diamagnetic term} is sometimes referred to as the {\it Hopfield model}\index{Hopfield! model}~\cite{nataf2010vacuum,lolli2015ancillary,garziano2020gauge} dating back to the seminal work~\cite{hopfield1958theory}. We assume the dipole approximation such that we can write $\hat A(0)=\mathcal A_0\left(\hat a+\hat a^\dagger\right)$. The amplitude reads
\begin{equation}
\mathcal A_0=\left(\frac{\omega}{2V\epsilon_0}\right)^{1/2},
\end{equation}
where $V$ is the effective mode volume~\cite{Scully1997a,gerry2005introductory}. Naturally, the Hamiltonian $\hat H_\mathrm{C}(\mu)$ depends on the diamagnetic parameter $\mu$, {\it i.e.} its eigenstates $|\psi_{n\pm}(\mu)\rangle$ will depend on $\mu$ such that $|\psi_{n\pm}(0)\rangle$ reproduces the eigenstates~(\ref{rabieigstates}) of the quantum Rabi model. The light-matter interaction term and the diamagnetic self-energy terms are given by eqs.~(\ref{intte}) and (\ref{set}) respectively. In particular, the interaction term can be written $\hat H_\mathrm{int}=\Omega d\mathcal{A}_0\left(\hat a^\dagger+\hat a\right)\hat\sigma_x$, where $d=q\langle g|\hat r|e\rangle$ is the dipole moment between the two electronic states. Adding the bare Hamiltonians, the interaction and diamagnetic terms\index{Diamagnetic term} one finds the full Hamiltonian~(\ref{coulham2}). It was pointed out early on that omitting the diamagnetic terms results in a Hamiltonian that is not gauge invariant~\cite{bialynicki1979no,schafer2020relevance}.

The diamagnetic term, being proportional to $\hat x^2$ in the quadrature representation, has the effect of squeezing the field (for $g=0$ the field ground state to be a squeezed vacuum). With this in mind we transform the Hamiltonian~(\ref{coulham2}) with the squeezing operator $\hat S(z)$\index{Squeezing! operator}, eq.~(\ref{sqop}), {\it e.g.}~\cite{mandel1995optical}
\begin{equation}
\hat b=\hat S(z)\hat a\hat S^\dagger(z)=\cosh(z)\hat a+\sinh(z)\hat a^\dagger,
\end{equation}
which holds provided $z\in\mathbb{R}$. If we chose $z=\frac{1}{4}\log\left(\frac{\omega}{\omega+2\mu}\right)$, and rewrite the Hamiltonian in the new boson operator $\hat b$ and $\hat b^\dagger$ one finds 
\begin{equation}
\hat H_\mathrm{C}=\sqrt{\omega(\omega+2\mu)}\hat b^\dagger\hat b+\frac{\Omega}{2}\hat\sigma_z+g\left(\frac{\omega}{\omega+2\mu}\right)^{1/4}\left(b^\dagger+\hat b\right)\hat\sigma_x.
\end{equation}
Thus, the Hamiltonian~(\ref{coulham2}) has been rewritten in the form of the quantum Rabi Hamiltonian~(\ref{rabiham}), with a rescaled photon frequency and light-matter coupling. With this mapping we can identify the critical coupling $g_c=\sqrt{(\omega+2\mu)\Omega}/2$ for which the ground state of the field can no longer be approximated by a Gaussian\index{Gaussian! state}, and the atom-light subsystems can become strongly entangled. The Thomas-Reiche-Kuhn sum rule\index{Sum rule!Thomas-Reiche-Kuhn} imposes constraints on which the values the couplings $g$ and $\mu$ can attain as discussed in sec.~\ref{sssec:dicke}. Thus, the two are in general not independent and one of them cannot be tuned without affecting the other. This was the reason for formulating the no-go theorem for the Dicke normal-superradiant PT\index{Dicke! phase transition}\index{Phase transition! Dicke} in cavity QED setups -- under rather general situations it is impossible to reach the superradiant regime~\cite{nataf2010no}. The specific values $g$ and $\mu$ are determined by details of the system in mind, {\it e.g.} whether one has cavity/circuit QED systems in mind, which type of two-level systems and so on~\cite{malekakhlagh2016origin}. In most cases $\mu$ is an order of magnitude smaller than $g$ and can be safely neglected.

To qualitatively explore how the presence of the diamagnetic term\index{Diamagnetic term} affects the system properties we will treat $g$ and $\mu$ as independent quantities and subject them to a variation over large parameter regimes~\cite{sakhi2019effect}. It is instructive to start from the BOA\index{Born-Oppenheimer approximation}~\cite{larson2007dynamics} and the adiabatic potentials\index{Adiabatic! potential}~(\ref{adpot}) of the quantum Rabi model. As we discussed in sec.~\ref{ssec:rabi} for the quantum Rabi model, in the deep strong coupling regime, $g>\sqrt{\omega\Omega}/2$, the lower adiabatic potential $V_\mathrm{ad}^{(-)}(x)$ has a double-well shape (equivalent to the adiabatic potentials of the Dicke model, see eq.~(\ref{adpot2})). The dynamics in the adiabatic regime ($\Omega \ll \omega$) of the quantum Rabi model have been recently considered in~\cite{Castanos_Cervantes_2023}. The same transition for $\hat H_\mathrm{C}$ occurs now for the aforementioned critical coupling $g_c=\sqrt{(\omega+2\mu)\Omega}/2$. In the thermodynamic limit\index{Thermodynamic limit} ($\omega/\Omega\rightarrow 0$ for the quantum Rabi model, and $N\rightarrow\infty$ for the Dicke model), the appearance of the double-well structure for increasing $g$ manifests as a continuous PT, see sec.~\ref{sssec:dicke}. The ground state (for a finite system where no symmetry breaking has occurred) is then a highly entangled cat state~(\ref{dickecat})~\cite{buvzek2005instability}. Now, the diamagnetic term\index{Diamagnetic term} counteracts the formation of a double-well structure~\cite{rzkazewski2006comment}. However, for nonequilibrium driven systems the effective\index{Nonequilibrium! driven system} light-matter coupling $g$ can indeed be tuned independently from $\mu$ as will be discussed further in sec.~\ref{sssec:dicke}. This motivates a study of $\hat H_\mathrm{C}$ treating $g$ and $\mu$ as independent. 

We restrict our analysis to properties of the ground state $|\psi_{0+}(\mu)\rangle$ (recall that the eigenstates of the quantum Rabi model are labelled with respect to their parity, in this case the + subscript), and how these depend on the diamagnetic amplitude $\mu$ as if it would be a free parameter. In fig.~\ref{diamag} we display both the ground-state fidelity $F(\mu)=|\langle\psi_{0+}(\mu)|\psi_{0+}(0)\rangle|$ and the ground-state von Neumann entropy $S_\mathrm{vN}(\mu)$ in the $g\mu$-plane (see~\cite{sakhi2019effect} for similar results). As we mentioned in the paragraph above, the diamagnetic term\index{Diamagnetic term} tries to prevent the formation of an entangled cat state. This effect has been further analyzed in~\cite{garcia2015light} for the spin-boson model\index{Spin-boson model}\index{Model! spin-boson}~(\ref{spinboson}). In the present model this decoupling is seen in the figure as the decreased entropy for growing values of $\mu$. With $\mu=0$ and $g>\sqrt{\omega\Omega}/2$, the ground state phase is superradiant (in the thermodynamic limit), but as $\mu$ is increased the system will return into a normal phase since as some point $g<g_c=\sqrt{(\omega+2\mu)\Omega}/2$ (recall that the field will now be a squeezed vacuum, rather than a non-squeezed vacuum as for the quantum Rabi model).  This reappearance of the normal phase for large couplings $g$ is evident in fig.~\ref{diamag} (b), ascertained by the low fidelity of the two ground states for sufficiently large $\mu$'s. 

The issue of detecting, or estimating, the amplitude of $\mu$ has been addressed in~\cite{tufarelli2015signatures,garcia2015light,rossi2017probing}. The analysis in these references starts from the Dicke model in the thermodynamic limit\index{Thermodynamic limit}, which after imposing the Holstein-Primakoff transformation\index{Holstein-Primakoff! transformation} read
\begin{equation}
\hat H_\mathrm{D}=\omega\hat a^\dagger\hat a+\Omega\hat b^\dagger\hat b+g\left(\hat a^\dagger+\hat a\right)\left(\hat b^\dagger+\hat b\right)+\mu\left(\hat a^\dagger+\hat a\right)^2,
\end{equation}
which can be diagonalized via a {\it Bogoliubov transformation}\index{Bogoliubov transformation}~\cite{assa1994interacting}. Hence, when diagonalized, the Hamiltonian simply becomes $\hat H_\mathrm{D}=\omega_c\hat c^\dagger\hat c+\omega_d\hat d^\dagger\hat d$ for the new boson operators $\hat c$ and $\hat d$. The normal-superradiant PT\index{Dicke! phase transition}\index{Phase transition! Dicke} manifests itself as one of the frequencies $\omega_c$ or $\omega_d$ becoming negative. As shown above, the diamagnetic term\index{Diamagnetic term} shifts the value of the critical coupling. In this linearized Dicke model\index{Linearized! Dicke model}, the diamagnetic amplitude, according to the Thomas-Reiche-Kuhn sum rule, takes the value~\cite{tufarelli2015signatures}
\begin{equation}
\mu_\mathrm{TRK}=\frac{g^2}{\Omega}.
\end{equation}
Given this value, it is impossible to reach the superradiant phase by increasing $g$ (the no-go theorem)\index{Dicke! no-go theorem}. In ref.~\cite{rossi2017probing} the authors used quantum estimation theory to study the effect of the diamagnetic term\index{Diamagnetic term}, especially how its presence influences the so-called {\it quantum Fisher information}\index{Quantum! Fisher information}. In particular, they concluded that the diamagnetic term could, if optimal schemes are employed, be detected at coupling amplitudes $g$ a few tenths of the mode frequency $\omega$.   

 \subsubsection{Neglecting the kinetic energy term}\label{sssec:velapp}
We may take two limits in which the kinetic energy term $\hat T=\frac{\hat p^2}{2m}$ can be disregarded or at least handled effectively; the ultracold regime with very small kinetic energies, and the semiclassical regime in which the kinetic energy is much higher than other involved energies and the atom can be seen as a point particle subject to a time-varying atom-light coupling $g(t)$. Let us start with the first case where the atomic velocity $v_{at}$ is small enough such that $E_{kin}=\frac{mv_{at}^2}{2}$ is smaller than the other energies involved. We saw in sec.~\ref{sssec:qatmo} that a plethora of new phenomena emerge when this is not the case, {\it i.e.} when $E_{kin}$ is comparable to say $g_0\sqrt{\bar n}$. Even though we briefly discussed the approximation in that section, we will complement it with some further observations here. Of course, given the JC model (\ref{qm}), when we include quantized motion\index{Quantized atomic motion} and a spatially varying coupling $g(z)$, the presence of non-commuting terms implies that quantum fluctuations may start to play a role, like for example in the maser problem. We will return to this in sec.~\ref{ssec:mbcQED} when discussing the coupling of atomic Bose-Einstein condensates to optical resonators. This applies even when we neglect $\hat T$, due to the extension of the atoms, determined by the coefficients $c_{gn}(z,t)$ and $c_{en}(z,t)$. The overlap of these with the coupling $g(z)$ will affect the evolution e.g., when determining the Rabi frequency. If, say, the atom is delocalized compared to the variations of $g(z)$, this amounts to an effective spatial averaging of the coupling. 

As shown in sec.~\ref{sssec:qatmo}, the inclusion of the kinetic energy term implies that when we transform to the adiabatic basis a gauge potential $\hat A(z)$\index{Gauge! potential} appears in the form of eq.~(\ref{jcgaugepot}). This term vanishes in the BOA\index{Born-Oppenheimer approximation} where we neglect corrections arising from the commutator $[\hat z,\hat p_z]$. A non-zero $\hat A(z)$ causes the occurrence of a non-adiabatic transition. This effect was studied in great detail in ref.~\cite{larson2006validity}. Loosely speaking, a large detuning $\Delta$ and a smooth coupling $g(z)$ reduce the non-adiabatic transitions. The smoothness is assessed with respect to the velocity of the atom -- a fast moving atom will see a rapidly varying $g(z)$. One manifestation of the non-adiabatic transitions can be explored in the large detuning limit $|\Delta|\gg g\sqrt{\bar n}$, the JC model is conveniently written as~(\ref{largejc}). Transitions between the bare atomic states $|g\rangle$ and $|e\rangle$ are suppressed by powers of $g/\Delta$. In particular, from eq.~(\ref{tsol2}) we see that for an atom initialized in one of its internal states, the other internal state will never get more populated than the initial one, multiplied by $(g/\Delta)^2$. This is no longer true if the kinetic term is included since then the off-diagonal $\hat A(z)$ will induce non-adiabatic transitions between the internal atomic (adiabatic) states\index{State! adiabatic}\index{Adiabatic! state}. 

\begin{figure}
\includegraphics[width=10cm]{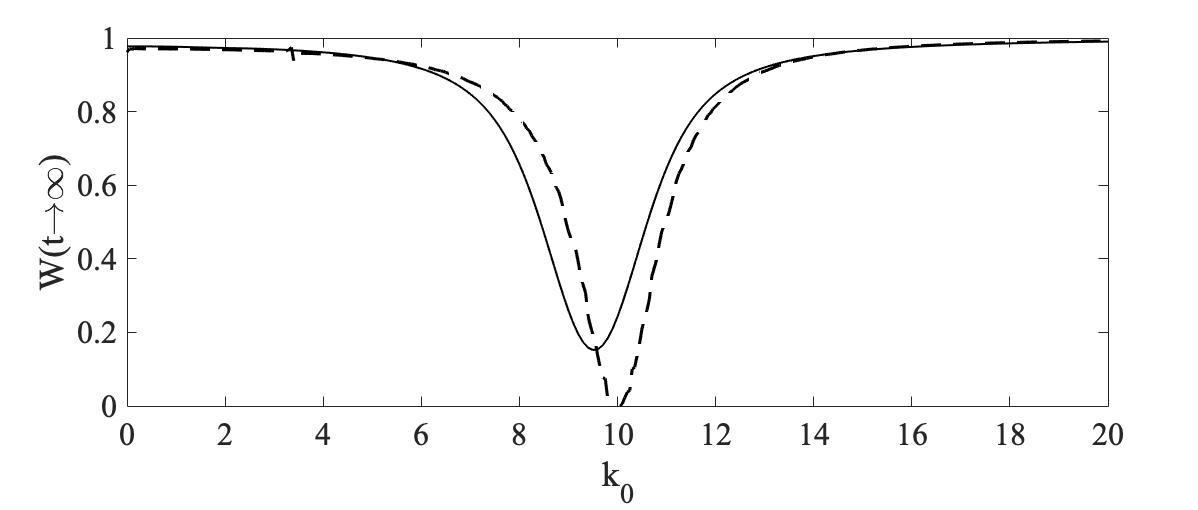} 
\caption{Solid line: The asymptotic value of the atomic inversion\index{Atomic! inversion} W as function of the wave number (bare velocity) $k_0$ for the initial Gaussian atomic wavepacket\index{Gaussian! wavepacket}~(\ref{inast}) numerically propagated in time for the Hamiltonian~(\ref{Mham}). We consider the large detuning scenarion with $\Delta=10g_0=10$ (and the other dimensionless parameters $m=\sigma_0=k=1$). For a frozen atom, $k_0=0$, the inversion stays close to 1. As the wave packet starts to propagate within the standing wave, $k_0\neq0$, the internal atomic state $|g\rangle$ gets more populated due to non-adiabatic transitions induced by the spatially varying coupling $g(z)$. For $k_0\sim10$ ($=\Delta$), the two internal atomic levels are roughly equally populated in the asymptotic limit $t\rightarrow\infty$. Interestingly, for even faster wave packets, the population transfer is dropped again, and for very fast atoms we are back to almost no transfer. This is due to a resonance occurring as explained in the main text. To verify this more quantitatively we show the corresponding inversion W (dashed line) obtained from the propagation of the semiclassical Hamiltonian (\ref{scMham}) after identifying $k_0=v$, and average $W(t)$ over time. As expected, the same dip is found, however slightly shifted.} \label{nonadQ}   
\end{figure} 

To demonstrate the emergence of non-adiabatic transitions between internal atomic states, inspired by the mode profile~(\ref{fpprof}) we consider the Hamiltonian
\begin{equation}\label{Mham}
\hat H_\mathrm{M}=\frac{\hat p_z^2}{2m}+\frac{\Delta}{2}\hat\sigma_z+g_0\cos(k\hat{z})\hat\sigma_x.
\end{equation}
This Hamiltonian is that of a two-level atom in classical EM-field~\cite{allen1975optical}. Without the internal structure it is simply a {\it Mathieu Hamiltonian}\index{Mathieu equation}~\cite{zwillinger1998handbook} that we will return to occasionally in our monograph, see for example section~\ref{sec:cirQED}. As an initial atomic state (note that since the EM field is not quantized, the involved degrees of freedom comprise the internal atomic states and the atomic center-of-mass motion), we take a Gaussian wavepacket\index{Gaussian! wavepacket}
\begin{equation}\label{inast}
|\Psi\rangle=\left(\frac{1}{\sqrt{\pi}\sigma_0}\right)^{1/2}e^{-\frac{(z-z_0)^2}{2\sigma_0^2}}e^{-ik_0z}|e\rangle.
\end{equation}
Thus, the atom is initially excited, and its center-of-mass is a Gaussian with width $\sigma_0$, centered around $z_0$, and with an average momentum $p_z=k_0$ ($\hbar=1$). As discussed in sec.~\ref{sssec:qatmo}, as time evolves this Gaussian wavepacket will start to spread, determined by some effective mass\index{Effective! mass} $m_\mathrm{eff}(k_0)$, and propagate along $z$ with a velocity set by the group velocity\index{Group velocity} $v_g(k_0)$~\cite{larson2005effective}. We expect the non-adiabatic transitions to set in as the velocity $k_0$ increases. This is also found numerically as depicted in fig.~\ref{nonadQ}, which shows the asymptotic value of the atomic inversion~(\ref{inv}), $W(t\rightarrow\infty)$. For $k_0=0$, where the non-adiabatic corrections vanish, we have in this case a very small population transfer between the internal atomic states, given that $\Delta=10g_0$. As we ramp up $k_0$, more and more population is transferred between the internal states. The maximum is reached for a $k_0$ of the order of $\Delta$, while for $k_0\gg\Delta$ we return to a situation with suppressed population transfer. This comes about due to the Doppler shift\index{Doppler! shift} of the atom, implying that the transition $|g\rangle\leftrightarrow|e\rangle$ becomes far-off resonant. In the semiclassical treatment of atomic motion, sec.~\ref{sssec:timeJCsec}, where $\hat p_z\rightarrow mv$ and $\hat z\rightarrow vt$, this resonance condition follows rather straightforwardly. The semiclassical Hamiltonian corresponding to~(\ref{Mham}) is~\cite{schlicher1989jaynes} (setting $k=1$) 
\begin{equation}\label{scMham}
\hat H_\mathrm{scM}=\frac{\Delta}{2}\hat\sigma_z+g_0\cos(vt)\hat\sigma_x.
\end{equation}
This Hamiltonian can be interpreted as that of a two-level system driven by a classical field with a frequency $v$. Within the RWA\index{Rotating-wave approximation}, perfect Rabi oscillations are found for $v=\Delta$. The counter-rotating terms give rise to Bloch-Siegert shifts\index{Bloch!-Siegert shift} and imperfect Rabi oscillations\index{Rabi! oscillations} even at resonant driving. This explains the behaviour of fig.~\ref{nonadQ} with the clear drop of population of the initial atomic state. We complement the result (solid line) with the corresponding result obtained instead using the semiclassical Hamiltonian~(\ref{scMham}) plotted with a dashed line. We point out that the semiclassical result of $W(t)$ keeps oscillating indefinitely, and we therefore time-average $W(t)$ in this case. 

\begin{figure}
\includegraphics[width=10cm]{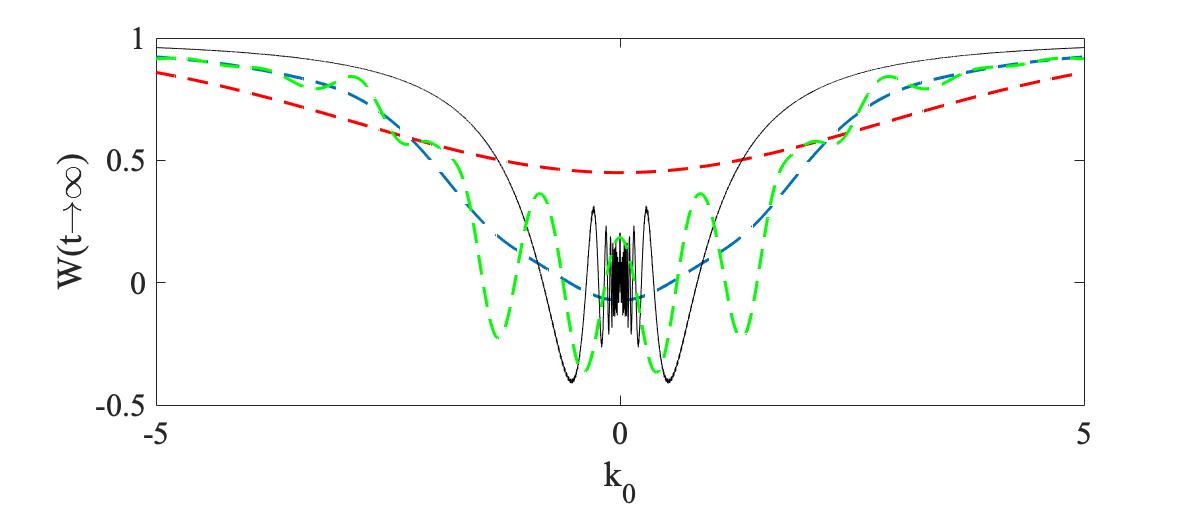} 
\caption{Same as fig.~\ref{nonadQ}, but this time for the resonant situation $\Delta=0$, and we vary the widths of the initial Gaussian wave-packets\index{Gaussian! wavepacket}~(\ref{inast}); $\sigma_0=0.2$ (dashed red line), $\sigma_0=1$ (dashed blue line), and $\sigma_0=5$ (dashed green line). The solid black line gives the semiclassical result. In the semiclassical result, where quantum fluctuations in the atomic motion are disregarded, we see a `beating effect' between the Rabi oscillations and the period of the drive, leading to rapid variations in the time averaged inversion for small $k_0=v$. These are lost as quantum fluctuations are taken into account. Here, on the other hand, for broad initial wave packets we see an oscillating inversion as a function of $k_0$.} 
\label{nonadQ2}   
\end{figure} 

In the semiclassical approximation of sec.~\ref{sssec:timeJCsec} the particle is considered point-like, and thus, the uncertainties in $x$ and $p$ are neglected. Whenever the uncertainty $\Delta x$ is comparable to the wave length $\lambda=2\pi/k$ of the mode, this may greatly influence the characteristics of any physical quantities~\cite{larson2009quantum}. Let us give an example where these quantum uncertainties in the atomic motion cause the semiclassical approximation to break down. For these we consider $\Delta=0$, and from Subsec.~\ref{ssec:JCm} we note that the eigenstates (\ref{dstate}) do not depend on the coupling $g$ in this case. This results from the fact that a simple Hadamard rotation\index{Hadamard! transformation}~(\ref{hada}) diagonalizes the Hamiltonian. Furthermore, the problem becomes analytically solvable even for a time-dependent coupling $g(t)$ as given in eq.~(\ref{trivial}). What is found is that only the area $A(t)=\int_0^tg(t')dt'$ matters, but not the actual shape of $g(t)$, {\it e.g.} how smooth it is. This was discussed in ref.~\cite{larson2006validity}; the $\Delta=0$ limit shows some characteristics of adiabatic following\index{Adiabatic! following}. With $g(t)=g_0\cos(vt)$ we can find the atomic inversion for an initial state $|e\rangle$ evolving under the Hamiltonian~(\ref{scMham}),
\begin{equation}
W_\mathrm{scM}(t)=\cos\left(\frac{2g_0}{v}\sin(vt)\right)=J_0\left(\frac{2g_0}{v}\right)+2\sum_{k=1}^\infty J_{2k}\left(\frac{2g_0}{v}\right)\cos(2kvt),
\end{equation}
where $J_\alpha(x)$ are the Bessel functions\index{Bessel function}. The period of $W_\mathrm{scM}(t)$ is $T=\pi/v$, such that the time average simply becomes
\begin{equation}\label{semiinv}
W=\frac{1}{T}\int_0^TW_\mathrm{scM}(t)dt=J_0\left(\frac{2g_0}{v}\right).
\end{equation}
 The Bessel functions oscillate around zero and decay for increasing argument $x$. Thus, as $v\rightarrow0$ the inversion W will oscillate rapidly around zero. This is depicted in fig.~\ref{nonadQ2} by the solid black line. For large values of $v$ we obtain $W\rightarrow1$. 
 
When we quantize the center-of-mass motion we consider the Hamiltonian~(\ref{Mham}). For $\Delta=0$ we again can decouple the two equations via a Hadamard transformation~(\ref{hada}) to obtain Mathieu equations~\cite{zwillinger1998handbook}. The width $\sigma_0$ of the wavepacket~(\ref{inast}) will influence the evolution as discussed in sec.~\ref{sssec:qatmo}; initially narrow packets will more rapidly broaden. When the width is comparable or larger than the wavelength of the standing wave, an inherent dephasing of Rabi oscillations will set in and, as seen in fig.~\ref{nonadQ}, the atomic inversion $W(t)$ approaches a fixed value for large times. We show this value as a function of $k_0$ and different widths $\sigma_0$ in fig.~\ref{nonadQ2}. Compared to the previous fig.~\ref{nonadQ}, the resonance transition now occurs for $k_0=\Delta\equiv0$. Interestingly, as the initial width $\sigma_0$ is comparable to the wave-length $\lambda=2\pi$ the asymptotic inversion displays an oscillatory dependence on initial velocity $k_0$. However, this dependence does not resemble that of the semiclassical inversion~(\ref{semiinv}) which stems from a beating between two time-scales set by the pump and transition frequencies. 
 
In which regimes do experiments operate? In the {\it ENS} experiments conducted by the group of S. Haroche, rubidium atoms were sent perpendicularly through a Fabry-P\'erot\index{Fabry-P\'erot! cavity/resonator} microwave cavity with a mode waist around 6 mm (see next section, section~\ref{sec:cavQED}). The velocities typically ranged between $v=100$ m/s to $v=700$ m/s~\cite{brune1996quantum,gleyzes2007quantum}. As we see from tab.~\ref{expdatatable}, the coupling in these experiments $g/2\pi=51$ kHz. With these experimental values we find a lower kinetic energy $E_\mathrm{kin}\approx5$meV, and an interaction energy $E_\mathrm{int}\equiv\hbar g\approx1.3$ neV. Thus, even for non-zero fields, inside a cavity with a few hundred photons, the kinetic energy is at least four orders of magnitude larger than the interaction energy. If the atoms are within the $1$m/s regime and the coupling is increased somewhat like in optical Fabry-P\'erot cavity experiments (see tab.~\ref{expdatatable}) we cannot treat the atomic motion classically any longer. As will be discussed in the next section, this was experimentally demonstrated in the group of J. Kimble, where they studied a regime with $E_\mathrm{kin}\approx0.4$neV and $E_\mathrm{int}\equiv\hbar g\approx457$ neV~\cite{hood2000atom}. This will also be thoroughly discussed in sec.~\ref{ssec:mbcQED} where we consider ultracold atoms or Bose-Einstein condensates trapped inside optical resonators.

\subsubsection{Neglecting losses}
 Any experiment exploring quantum phenomena will inevitably be subject to losses and decoherence. In cavity QED experiments (sec.~\ref{sec:cavQED}) the photon and qubit lifetimes typically range from ms in the microwave regime to $\mu$s in the optical regime as indicated in tab.~\ref{expdatatable}, while in circuit QED (section~\ref{sec:cirQED}) the corresponding lifetimes are typically some order of magnitude shorter, see tab.~\ref{table:parameters}. For trapped ions (section~\ref{sec:ion}), the coherence times for single qubits can be much longer as recently demonstrated for a trapped $^{171}$Yb ion achieving coherence times up to an hour~\cite{wang2017single,wang2020single}! To reach such extreme long-lived coherence the qubit must be encoded in metastable Zeeman levels, and the motional state of the ion is effectively decoupled from the internal degrees of freedom. In cavity and circuit QED, the photon degrees of freedom often play a more active role, and since we cannot disregard losses in these subsystems, we are left with much shorter lifetimes. 

Thus, neglecting losses typically boils down to comparing time-scales, see tabs.~\ref{expdatatable} and \ref{table:parameters}. Originating from NMR (nuclear magnetic resonance)\index{Nuclear! magnetic resonance} terminology and the processing of spin-1/2 systems, the effect of losses is often divided into two time-scales $T_1$ and $T_2$. Nowadays one sees this notion more spread and shortly, $T_1$ characterizes the time for dissipation or relaxation like photon losses, while $T_2$ sets the time for dephasing or decoherence like the transition from a superposition state into a statistical mixture. The two relaxation processes are related via the fluctuation-dissipation theorem~\cite{mandel1995optical}, and one has in general that $T_2<T_1$. An example of this was presented in sec.~\ref{sssec:openjc} when we discussed the evolution of the cat state going from a quantum superposition into a statistical mixture~(\ref{catdec}). In this example $T_1\sim\kappa^{-1}$ and $T_2\sim(\kappa|\alpha|^2)^{-1}$, and thus $T_2=T_1/|\alpha|^2$. Let us explore the fragility of the cat-state preparation somewhat more.

\begin{figure}
\includegraphics[width=10cm]{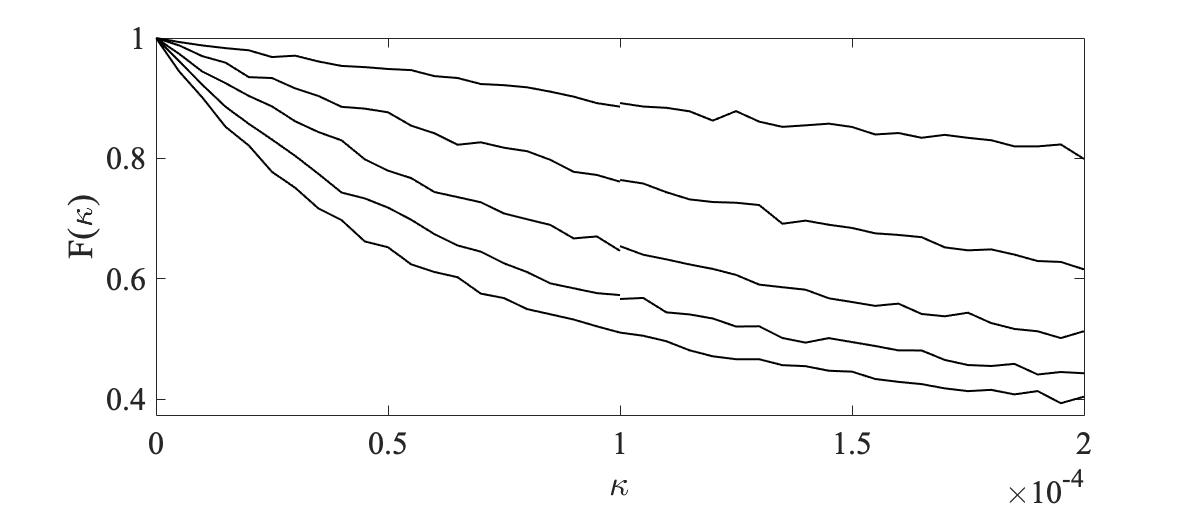} 
\caption{Demonstration of the sensitivity of the JC cat state preparation due to photon losses. For an initial coherent state $|\alpha\rangle$ and the atom in say $|e\rangle$, the JC evolution predicts that the field is approximately prepared in a cat state after half the revival time~(\ref{rtime}), see fig.~\ref{fig4} (a). The state preparation of the cat gets more accurate the larger the field amplitude $\alpha$, but simultaneously, the more fragile it will be towards photon losses (or any other source of decoherence). In this plot, we present the fidelity~(\ref{fidel}) between the state with zero photon loss and with the state $\hat\rho_\kappa(t_\mathrm{R}/2)$ with a non-zero decay rate $\kappa$. The different curves corresponds to initial coherent state amplitudes $|\alpha|^2=10,\,20,\,30,\,40,\,50$ (from top to bottom), and we consider the resonant case $\omega=\Omega$ with $g=0.1$ in all cases. Furthermore, we assume a zero temperature photon bath, {\it i.e.} $n_\mathrm{th}=0$ in the master equation~(\ref{master1})\index{Master equation}. The figure agrees with the discussion in sec.~\ref{sssec:sol} about exponential sensitivity of the cat coherence with respect to $\kappa|\alpha|^2$. } 
\label{fidcat}   
\end{figure} 

We saw in fig.~\ref{fig4} (a) and (b) how an initial coherent state $|\alpha\rangle$, and the qubit in, say, $|e\rangle$, evolves under the JC Hamiltonian into a cat state after half the revival time, $t_\mathrm{R}/2$~(\ref{rtime})\index{Revival time}. This scheme for preparing cat states\index{Cat state}, as described in sec.~\ref{sssec:sol}, has been implemented in the cavity QED setting~\cite{auffeves2003entanglement}. As explained, the negativity of the Wigner function of fig.~\ref{fig4} (a), seen at the origin of the $xp$-plane, is the result of the quantum superposition of coherent states. These distinguish a proper superposition from a statistical mixture, and these features will first die out in the presence of photon losses. Let us fix the interaction time to half the revival time, where the cat structure is the most pronounced, and the atom approximately decouples from the field (see sec.~\ref{sssec:sol}). The larger the coherent-state amplitude $\alpha$, the closer the state will be to a cat composed of two coherent states out of phase by $\pi$, but the more fragile it will be to decoherence. To demonstrate this we calculate the state fidelity~(\ref{fidel})\index{State! fidelity} between the two states $\hat\rho_0(t_\mathrm{R}/2)$ and $\hat\rho_\kappa(t_\mathrm{R}/2)$ representing, respectively, the states obtained with either a photon decay rate $\kappa=0$ or $\kappa\neq0$. Thus, we initialize a state $|\alpha\rangle|e\rangle$ and let it evolve according to the Lindblad equation\index{Lindblad! equation}~(\ref{master1}) with $n_\mathrm{th}=0$. The resulting fidelity, as a function of $\kappa$ and different amplitudes $\alpha$, is shown in fig.~\ref{fidcat}. It is clear that to reach a reasonable fidelity $F>0.9$ we need a $\kappa\sim10^{-4}g$, which is in general not reached in neither cavity nor circuit QED settings. In addition, we have not taken qubit relaxation and dephasing into account in this simulation which would decrease the fidelity even further. All in all, this example demonstrates how losses cannot be disregarded for the preparation of the highly non-classical cat states, and that producing larger cats becomes extremely difficult. 

Simple gate operations can, however, be performed on shorter time scales, typically $g^{-1}$ and hence a factor $1/\sqrt{\bar n}$ shorter than what is required for the cat-state preparation. The gate operations are often implemented via external driving which makes it possible to go to much shorter times of the order of ns. Over these short times, a lifetime of a few ms for either the qubit or the photon is very long, and as will be discussed in more detail in the next sections it is actually possible today to achieve entanglement with fidelities well above 90$\%$ using two-qubit gates~\cite{yamaguchi2002quantum,leibfried2003experimental,benhelm2008towards,stojanovic2012quantum,barends2014superconducting,puri2016high,krinner2020demonstration}. 
   

\section{Cavity QED}\label{sec:cavQED}
Over the past decades we observe an ever growing interest in ways of modifying the radiative properties of atoms by modifying the boundary condition for the electromagnetic field. Edward Purcell was actually the one to initiate the area of modern physics now known by the name of cavity quantum electrodynamics (cavity QED), by predicting that the spontaneous emission\index{Spontaneous! emission! rate} rate for a nuclear magnetic moment transition should be modified by a confining resonator~\cite{purcell1946spontaneous}. No cavity was involved in his proposal but the basic idea is the same: coupling a radiative transition to an oscillator on resonance opens up a new loss channel to significantly increase the overall spontaneous decay rate of that transition. This is due to the change in the radiation mode functions (hence the change of the mode coupling strengths to the two-level transition) originating from the presence of the cavity\index{Purcell enhancement}.

\begin{figure}
\includegraphics[width=10cm]{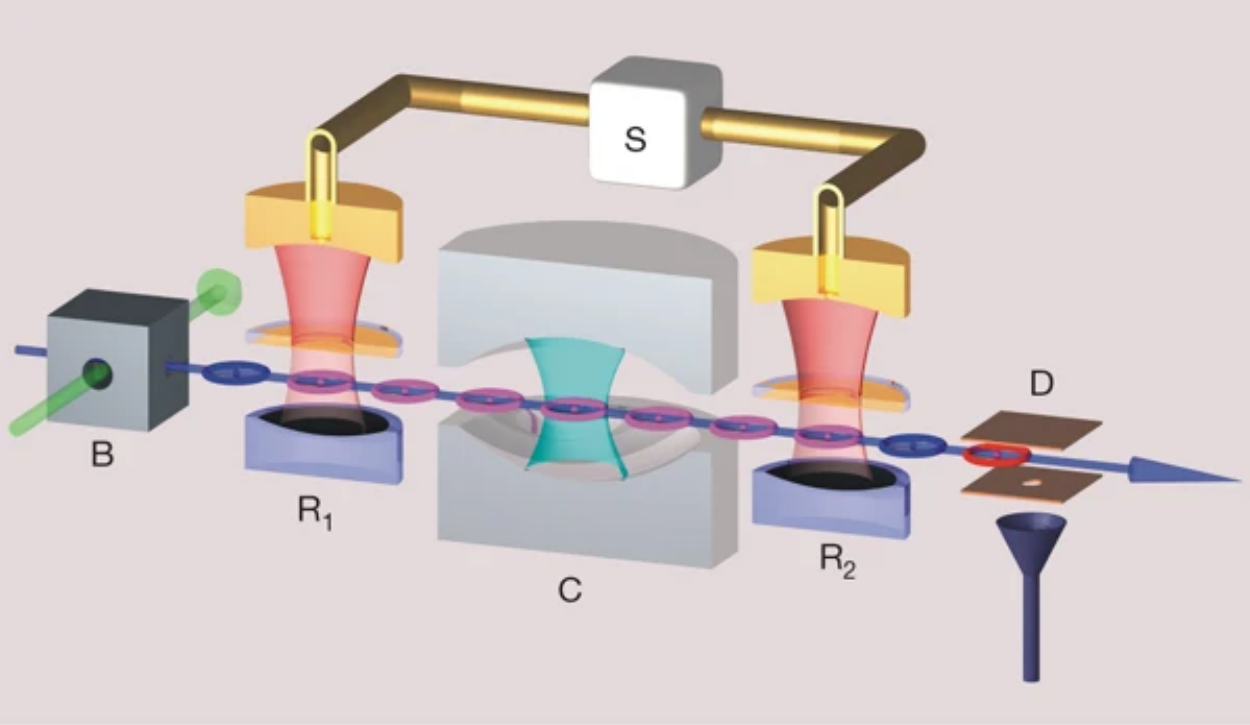} 
\caption{Typical setup for cavity QED experiments. Atoms (marked by circles representing the highly excited Rydberg states\index{Rydberg state}) pass the setup from left to right. The first step B is a `velocity-selector'; only atoms with the desired velocity is excited to the target internal electronic state. Thus, the velocity-selector acts also as state preparation of the internal atomic state. The process relies on Doppler shifts, only atoms with correct velocity are resonant with the transition. In earlier experiments, {\it Fizeau velocity-selector} with pairs of rotating plates were used instead. The two cavities $\mathrm{R}_1$ and $\mathrm{R}_2$ rotates the atomic two-level system. For example performing a $\pi/2$-pulse\index{$\pi/2$-pulse} for the implementation of a Hadamard gate\index{Hadamard! gate}. With $\mathrm{R}_1$ and $\mathrm{R}_2$ atomic Ramsey interference\index{Ramsey! interferometry} experiments can be performed. It should be mentioned that these two cavities have low $Q$-values and are classically driven such that their quantum nature is negligible. The high-$Q$ cavity\index{High-$Q$ cavity} is the middle one, C. The velocity of the atoms sets the effective interaction time $\tau$. D is the atomic detector, which together with $\mathrm{R}_2$ is capable to measure in any basis. The `central' S controls the sources $\mathrm{S}_1$ and $\mathrm{S}_2$. Reproduced with permission from Springer Nature.} 
\label{cavityQEDsetup}   
\end{figure}

The JC model appeared for the first time in 1963 as the simplest model to describe quantum coherent evolution of single atoms confined in optical resonators~\cite{jaynes1963comparison}. For an early demonstration of the $\sqrt{n}$ splitting in the energy levels see also~\cite{Harry1963}. Naturally, the physics of cavity QED predates the JC model. As we mentioned above, in 1946 Edward Purcell predicted that the spontaneous emission rate\index{Spontaneous! emission! rate} for a nuclear magnetic moment transition should be modified by a confining resonator~\cite{purcell1946spontaneous} due to the change in the radiation mode functions (hence the change of the mode coupling strengths to the two-level transition) originating from the presence of the cavity\index{Purcell enhancement}. A few years later the first maser operation was demonstrated~\cite{gordon1954molecular,gordon1955maser}, and Schawlow and Townes suggested that a Fabry-P\'erot resonator\index{Fabry-P\'erot! cavity/resonator} could be used for the maser action in the infrared regime~\cite{schawlow1958infrared}. The first laser was realized in 1960~\cite{thmaiman} spurring a lot of theoretical as well as experimental research, for instance the fully quantum model of Jaynes and Cummings~\cite{jaynes1963comparison}. The next major step connected with the JC model came with achieving the strong coupling regime\index{Strong coupling regime}~\cite{meschede1985one,brune1987realization} where the vacuum Rabi frequency\index{Vacuum Rabi! frequency} $g$ exceeds both the photon and atomic loss rates, $\kappa$ and $\gamma$ respectively. By reaching this regime it was possible to demonstrate the collapse-revival structure of the JC model, and by this to provide evidence for the quantization of the electromagnetic field~\cite{rempe1987observation}. The early pioneering experiments came mainly from the Garching group of the late Herbert Walther and the Paris group of Serge Haroche. Both of them operated in the microwave regime which became possible by exciting the atoms to highly-excited Rydberg states\index{Rydberg state}. The large dipole moments, scaling as $\sim n^2$ with $n$ being the principle quantum number (typically $\sim50$ for the microwave regime) of the corresponding electronic level, of these transitions are crucial for reaching strong -coupling conditions. Indeed, the high finesse of the cavities and the long life times of the Rydberg state implied that the JC model could be tested in the laboratory (the involved assumptions leading to the JC model was discussed in the previous sec.~\ref{ssec:approx}, see also ref.~\cite{feranchuk2016physical}). In the optical regime, the group of Jeff Kimble demonstrated the strong-coupling regime by measuring the JC mode splitting\index{Jaynes-Cummings! mode splitting}~\cite{raizen1989normal}. The smaller mode volume of the optical resonators (compare the cavity length $L=45$ $\mu$m for the Kimble and $L\sim3$ cm for the Haroche experiments) implies a factor of 1000-fold larger atom-field coupling. However, the higher frequency and the smaller principle quantum numbers $n$ make the decay rates $\kappa$ and $\gamma$ much higher. Typically, the experiments working in the microwave regime have reached a higher {\it cooperativity factor}\index{Cooperativity}~\cite{kimble1998strong}
\begin{equation}\label{coop}
C=\frac{4g^2}{\kappa\gamma},
\end{equation}
mainly due to the very long lifetimes (small $\gamma$'s) of the highly-excited Rydberg states\index{Rydberg state}. Table~\ref{expdatatable} summarises a few experimental parameters from four of the leading experimental groups. A schematic picture of the experimental setup is given in fig.~\ref{cavityQEDsetup} (reproduced from the ENS experiments~\cite{gleyzes2007quantum}). 

\begin{table}[h!]
  \centering

\begin{tabular}{|c|c|c|c|c|c|}
\hline
{\bf Group} & $g/2\pi$ & $\omega/2\pi\approx\Omega/2\pi$ & $\kappa/2\pi$ & $\gamma/2\pi$ & $\tau$ \\
  \hline \hline
  Haroche~\cite{gleyzes2007quantum} & 50 kHz & 50 GHz & 200 Hz & 5 Hz & 2.5 $\mu$s \\
  \hline
  Walther~\cite{brattke2001generation} & 7 kHz & 21 GHz & 0.5 Hz & 400 Hz & 80 $\mu$s \\ 
  \hline
  Kimble~\cite{ye1999trapping} & 30 MHz & 352 THz & 4 MHz & 2.6 MHz & 0.05 $\mu$s \\
  \hline
  Rempe~\cite{maunz2005normal} & 16 MHz & 385 THz & 1.4 MHz & 3 MHz & 3 $\mu$s \\
  \hline
  \end{tabular}
  \caption{Typical experimental parameters. Here, as before, $g$ is the effective atom-field coupling strength, $\omega$ the photon frequency\index{Photon! frequency}, $\Omega$ the atomic transition frequency\index{Atomic! transition frequency}, $\kappa$ the photon decay rate, and $\gamma$ is the spontaneous\index{Photon! decay rate} emission rate\index{Spontaneous! emission! rate} of the excited atomic level. $\tau$ is the effective interaction time. Note that the first two sets of parameters take place in the microwaves while the last two in the optical frequencies.}
  \label{expdatatable}
  \end{table}

\subsection{Early results and predictions}\label{ssec:cqedearly}
Two of the main directions of research within cavity QED during the 80's were {\it optical bistability}\index{Optical! bistability}~\cite{bonifacio1978photon,lugiato1984theory} and the {\it micromaser}~\cite{krause1986quantum,filipowicz1986theory}. 

\begin{figure}
\includegraphics[width=10cm]{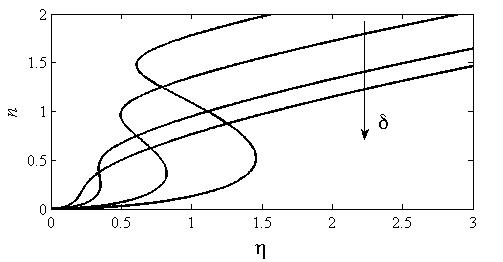} 
\caption{The mean-field results\index{Mean-field approximation} for the steady state photon number~(\ref{mfbistability}), as a function of the pumping amplitude. As the cavity-pump detuning $\delta$ is raised (according to the arrow, with respective values $\delta=-3,\,-2,\,-1,\,-0.5$) the bistable regime is lost. For $\delta<-1$, the photon intensity builds up the familiar `S'-shaped hysteresis curve which implies sudden jumps in the output light field intensity as the pumping amplitude is changed~\cite{rempe1991optical}. The remaining dimensionless parameters are $\kappa=\chi=1$.} 
\label{bistability}   
\end{figure} 

\subsubsection{Optical bistability}\label{sssec:optbis}
In sec.~\ref{sssec:JCzerodim}, when discussing the photon blockade\index{Photon! blockade} phenomenon, we saw a first example of how the inherent JC nonlinearity manifests. The photon blockade occurs for low photon numbers where quantum fluctuations significantly influence the response. At larger photon numbers, drastic changes due to the nonlinearity may also appear at the mean-field level where quantum fluctuations are negligible. This is the case of optical bistability. Due to the inherent nonlinearity of the JC model it becomes natural to look for bistability in cavity QED systems. Here we restrict the analysis to the dispersive optical bistability occurring for a large detuning, and leave out the situation of bistability in the resonant case ({\it absorptive bistability})\index{Absorptive bistability}, see for example~\cite{drummond1981quantum,carmichael1985photon,ben1986intrinsic,savage1988single}. The effective single atom Hamiltonian valid in the dispersive regime was given in eq.~(\ref{effham}). However, at this level the model is linear and in order to see any bistability one has to include higher-order terms. These terms describe multiphoton exchange interactions\index{Multiphoton! interaction! exchange}, among which the simplest is described by the Hamiltonian
\begin{equation}\label{kerrham}
\hat H_0=\delta\hat a^\dagger\hat a+\chi\hat a^\dagger\hat a^\dagger\hat a\hat a+i\eta\left(\hat a^\dagger-\hat a\right),
\end{equation}
obtained after adiabatic elimination\index{Adiabatic! elimination} of the atoms, and where $\delta$ is the detuning between the drive and cavity-mode frequencies. We need to include a driving term (proportional to $\eta$) in order to obtain a nonvanishing field amplitude. The second term is the {\it Kerr term}\index{Kerr! term} with the strength $\chi$ being proportional to the number of atoms, {\it i.e.} increasing the atom number leads to a stronger nonlinear effect. The full master equation\index{Master equation} for the cavity field follows from eq.~(\ref{master1}) and in the case of a zero-temperature bath ($n_\mathrm{th}=0$) becomes
\begin{equation}\label{masterbist1}
\partial_t\hat{\rho}(t) = i\!\left[\hat{\rho}(t),\hat{H}_0\right]\!
+\displaystyle{\frac{\kappa}{2}\left(2\hat{a}\hat{\rho}(t)\hat{a}^\dagger-\hat{a}^\dagger\hat{a}\hat{\rho}(t)-\hat{\rho}(t)\hat{a}^\dagger\hat{a}\right)}.
\end{equation}
The traditional procedure to explore optical bistability is to reformulate the problem in terms of a Fokker--Planck equation\index{Fokker--Planck equation} for the Glauber-Sudarshan $P(\alpha)$-distribution\index{Glauber-Sudarshan $P(\alpha)$-distribution}\index{Distribution! Glauber-Sudarshan $P$}~\cite{drummond1980quantum,gardiner2004quantum,walls2007quantum}
\begin{equation}\label{peq}
\displaystyle{\frac{\partial}{\partial t}P(\alpha)} = \displaystyle{\left[\frac{\partial}{\partial\alpha}\left(\frac{\kappa}{2}\alpha-2\chi\alpha^2\alpha^*-\eta\right)-\chi\frac{\partial^2}{\partial\alpha^2}\alpha^2\right.}
\displaystyle{\left.\!\frac{\partial}{\partial\alpha^*}\!\left(\frac{\kappa}{2}\alpha^*\!-2\chi\alpha^{*2}\alpha\!-\!\eta\right)\!-\chi\frac{\partial^2}{\partial\alpha^{*2}}\alpha^{*2}\right]\!\!P(\alpha)},
\end{equation} 
where we have assumed $\delta=0$. For a finite temperature bath the additional term $\kappa n_\mathrm{th}\frac{\partial^2}{\partial\alpha\partial\alpha^*}$ appears which results in a diffusion stemming from thermal bath fluctuations. The terms proportional to the first-order derivatives represent the `drift' of $P(\alpha)$, while the other two terms are the `diffusion' terms causing a spreading of $P(\alpha)$. From the Fokker--Planck equation\index{Fokker--Planck equation} one can formulate a stochastic equation for the amplitudes $\alpha$ and $\alpha^*$~\cite{drummond1980quantum}. Instead of following this full analysis including fluctuations we demonstrate the emergence of bistability at a mean-field level\index{Mean-field approximation} by writing down the equations of motion for $\alpha$ and $\alpha^*$, and explore the expectation values $\alpha=\langle\hat a\rangle$ and $\alpha^*=\langle\hat a^\dagger\rangle$. The mean-field approximation consists in factorizing the expectations, noting that the Langevin terms vanish when considering average values. The resulting equations for the $c$-numbers become
\begin{equation}
\frac{\partial}{\partial t}\alpha=-i\delta\alpha-2i\chi\alpha^*\alpha^2-\frac{\kappa}{2}\alpha+\eta,
\end{equation}
and the equation for $\alpha^*$ is obtained from complex conjugation. The steady state\index{Steady state} photon number ($n=|\alpha|^2$) is
\begin{equation}\label{mfbistability}
n=\frac{\eta^2}{\frac{\kappa^2}{4}+(\delta+2\chi n)^2}.
\end{equation}
Obeying a third-order equation, the photon number can show hysteresis upon varying the system parameters, as shown in fig.~\ref{bistability}. The manifestation of the bistable behaviour will be seen in the output photon field; for weak pumping a small photon intensity is established and by slowly increasing $\eta$ the intensity increases moderately until the curve bends backwards where the photon number makes a sudden jump to higher values. After the jump, the pump intensity can be lowered such that the photon number follows the upper branch and one recovers the hysteresis effect. This was first demonstrated in the group of J. Kimble at CalTech for a moderate number of atoms $3\leq N\leq65$, and in particular bistability was observed for $N>15$~\cite{rempe1991optical}. Corrections to the mean-field results can be obtained from linearizing the full stochastic equations~\cite{drummond1980quantum,walls2007quantum}. More recently, a full quantum analysis of optical bistability was presented for only a few atoms~\cite{dombi2013optical}. 

In ref.~\cite{harshawardhan1996controlling} it was shown that by considering a pump for the atoms, the nonlinear effect is greatly enhanced such that the bistability threshold becomes much lower. More specifically, $\Lambda$-atoms (see fig.~\ref{fig12}) were considered in which the cavity mode couples to the $|1\rangle\leftrightarrow|2\rangle$ transition and the pump, acting as a control parameter, is coupled to the $|2\rangle\leftrightarrow|3\rangle$ transition. The scheme is similar to that of {\it electromagnetically induced transparency}\index{Transparency! electromagnetically-induced} (EIT)~\cite{fleischhauer2005electromagnetically}, where typically the $\Lambda$ configuration is considered. It had already been shown that such EIT setup could be used to generate strong nonlinear Kerr effects\index{Kerr! effect}~\cite{imamo?lu1989lasers,schmidt1996giant}. In the EIT, the pump field can control the transmission of the light pulses through the medium, and, in particular, the light propagation can be considerably slowed down or even stopped. This knowledge led Lukin and co-workers to explore the possibility to enhance the photon lifetime inside an optical resonator~\cite{lukin1998intracavity}. The predicted cavity linewidth\index{Linewidth! cavity! narrowing} narrowing was soon afterwards experimentally demonstrated in ref.~\cite{wang2000cavity} by considering a vapor of Rb atoms and where a narrowing by a factor of $15$ was found. The effect was later displayed for a single trapped Rb atom inside the resonator. The scaling of the linewidth\index{Linewidth! scaling} with the atom number $N$ was explored, and experimental data agreed well with the predicted $N^{-1}$ scaling~\cite{mucke2010electromagnetically}. The low-photon regime, where the photon blockade may set in as explained in sec.~\ref{sssec:JCzerodim}, has also been analyzed in terms of the EIT setup~\cite{werner1999photon}. It was shown the blockade may survive, albeit a great enhancement was not found. 

\begin{figure}
\includegraphics[width=10cm]{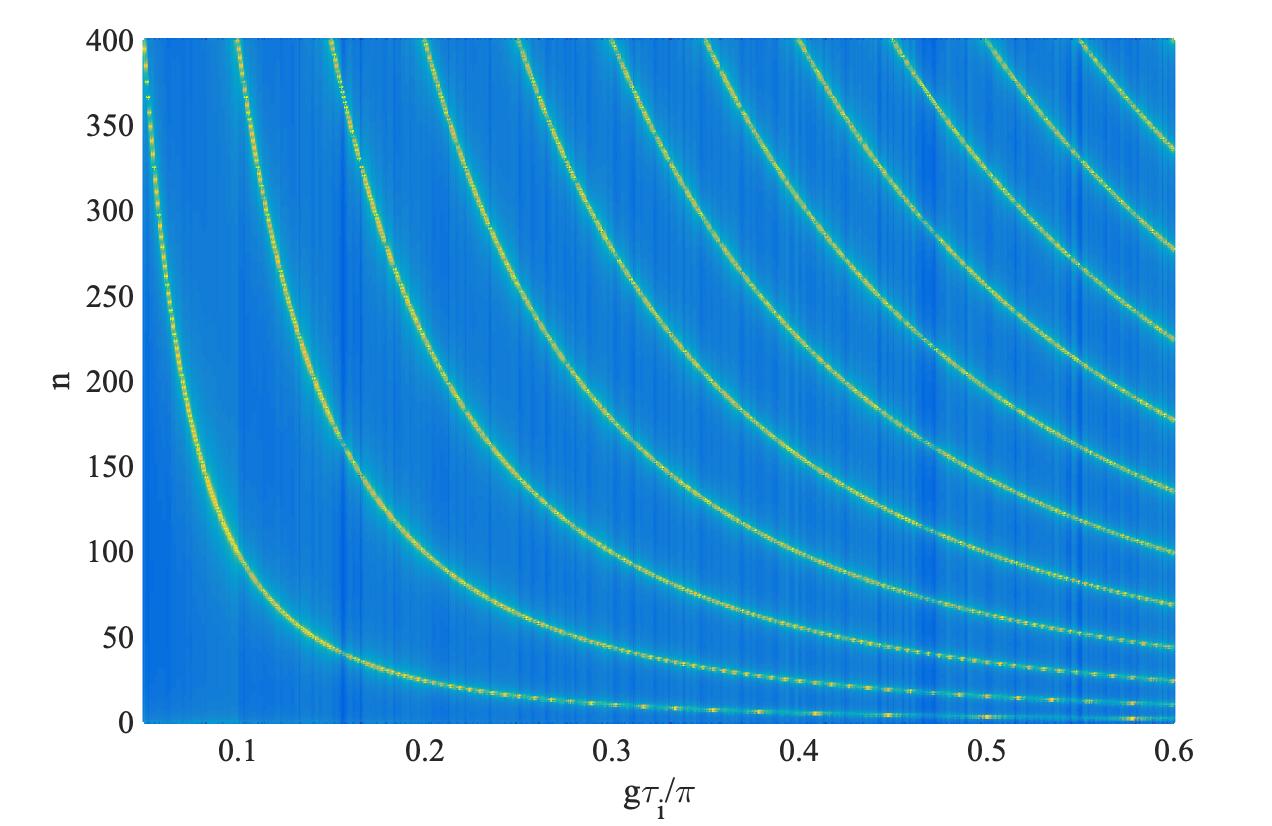} 
\caption{The steady-state photon number distribution~(\ref{pdist0}) as a function of the interaction time $g\tau/\pi$. Given $g\tau/\pi$, for the majority of photon numbers $n$ we have $P_n(g\tau)\approx0$. However, for certain values of $n$, the distribution blows up due to the vanishing of the denominator $1-\cos^2(g\tau\sqrt{n+1})$. These mark the so-called trapping states; for a trapping state a traversing atom completes an integer number of Rabi oscillations during the interaction time $\tau$, and thereby exits the cavity in its excited state. } 
\label{micromnoloss}   
\end{figure} 

\subsubsection{The micromaser}\label{sssec:micro}
In the micromaser\index{Micromaser} a monochromatic beam of excited two-level atoms traverses a microwave resonator, with rate $r$, one at a time. In comparison to a regular laser, where energy is pumped into the resonator by driving a medium inside the resonator, in the micromaser the energy increase is due to photon emission from the two-level atoms~\cite{meschede1985one,rempe1987observation}. We are interested in the cavity-field properties, and not in the atoms leaving the interaction region, {\it i.e.} we trace over those degrees of freedom. During the traversal of a single atom, the atom and field systems interact via the JC Hamiltonian~(\ref{jcham}) for, say, a time $\tau$. Taking photon losses into account implies that one solves the full master equation~(\ref{master1})\index{Master equation}.  If, however, $\tau$ is short in comparison to the photon lifetime\index{Photon! lifetime} $\kappa^{-1}$ we may neglect dissipation and decoherence during the time an atom spends inside the cavity. Nevertheless, in between the passage of two consecutive atoms, the photon field experiences a coupling to a thermal reservoir. Thus, if we consider the passage of a single atom and a relaxation time $t_r$ until the next atom enters the cavity, denoting the full atom-field state by $\hat\rho$ and the JC time-evolution operator\index{Time-evolution operator}\index{Operator! time-evolution} by $\hat U_\mathrm{JC}(t)$, the field state $\hat\rho_f$\index{Reduced! density operator} evolves as~\cite{filipowicz1986theory}
\begin{equation}
\hat\rho_f(t_r+\tau+t_0)=\exp(\hat Lt_r)\mathrm{Tr}_a\left[\hat U_\mathrm{JC}(\tau)\hat\rho(t_0)\hat U_\mathrm{JC}^\dagger(\tau)\right].
\end{equation}
Here, $t_0$ is the time when the atom enters the cavity, $\mathrm{Tr}_a$ is the trace over the atomic degree of freedom, and $\hat L$ is the Lindblad super-operator\index{Lindblad! super-operator} [see eq.~(\ref{master1})]
\begin{equation}
\hat L\hat\rho_f = \displaystyle{\frac{\kappa}{2}(n_\mathrm{th}+1)\left(2\hat a\hat\rho_f\hat a^\dagger-\hat a^\dagger\hat a\hat\rho_f-\hat\rho_f\hat a^\dagger\hat a\right)} + \displaystyle{\frac{\kappa}{2}n_\mathrm{th}\left(2\hat a^\dagger\hat\rho_f\hat a-\hat a\hat a^\dagger\hat\rho_f-\hat\rho_f\hat a\hat a^\dagger\right)}.
\end{equation}
With this approach it is possible to analytically solve for the steady state\index{Steady state} photon distribution $P_n=\hat\rho_{nn}=\langle n|\hat\rho_f|n\rangle$~\cite{filipowicz1986theory,lugiato1987connection}. However, let us instead write down a full master equation\index{Master equation} of the field and analyze the behaviour with that approach. Assuming the coarse-graining\index{Coarse-graining} described above, and utilizing techniques from laser theory, it is possible to derive the Lindblad master equation\index{Lindblad! master equation}~\cite{krause1986quantum,lugiato1987connection}
\begin{equation}
\begin{aligned}
\displaystyle{\frac{d}{dt}\hat\rho_{n,n+k}(t)} & = -r\left[1-\cos(g\tau\sqrt{n+1})\cos(g\tau\sqrt{n+1+k})\right]\hat\rho_{n,n+k}+r\sin(g\tau\sqrt{n})\sin(g\tau\sqrt{n+k})\hat\rho_{n-1,n-1+k}\\ 
&  \displaystyle{-\frac{\kappa}{2}(n_\mathrm{th}+1)\left[\left(2n+k\right)\hat\rho_{n,n+k}+\kappa\sqrt{(n+1)(n+1+k)}\hat\rho_{n+1,n+k+1}\right]}\\
&  \displaystyle{-\frac{\kappa}{2}n_\mathrm{th}\left[(2n+k+2)\hat\rho_{n,n+k}-2\sqrt{n(n+k)}\hat\rho_{n-1,n-1+k}\right],}
\end{aligned}
\end{equation}
with $n_\mathrm{th}$ the number of thermal photons of the bath with the relevant frequency, and we assume the resonant scenario where $\omega=\Omega$. To get the photon distribution $P_n$ one lets $k=0$. Clearly, the first two terms proportional to the atomic injection rate $r$ give the unitary JC-evolution. 

Some insight will be gained by first considering the case of no photon losses, {\it i.e.} $\kappa=0$. The steady-state photon distribution is in this case
\begin{equation}\label{pdist0}
P_n=P_0\prod_{l=1}^n\frac{\sin^2(g\tau\sqrt{l})}{1-\cos^2(g\tau\sqrt{l+1})},
\end{equation}
with $P_0$ a normalization factor determined from $\sum_nP_n=1$. If
\begin{equation}\label{trapc}
g\tau\sqrt{n+1}=k\pi,\hspace{1cm}k\in\mathbb{N}
\end{equation}
for a given interaction time $g\tau$ and some $n$, the denominator in (\ref{pdist0}) becomes zero and the distribution blows up. Of course, $n$ is restricted to integer values, and the condition might not be exactly met. When the field occupies such a Fock state\index{Fock state}\index{State! Fock}, every atom passing through the cavity will have sufficient time to complete exactly $k$ Rabi cycles. Thus, the atom leaves the cavity in its excited state without changing the photon number of the field. Such peculiar states are called {\it trapping states}~\cite{meystre1988very}, and their presence has been experimentally verified~\cite{weidinger1999trapping}. In fig.~\ref{micromnoloss} we display the photon distribution for a range of interaction times $g\tau$. These trapping states are plotted with bright lines. Even in the ideal situation of no photon losses, the trapping states are not perfectly stable since the condition~(\ref{trapc}) is never exactly met. If we assume that the cavity field is initially in vacuum, what will happen is that as atoms traverse the cavity the photon number will increase, but it will have difficulties to populate states with $n$'s larger than the first trapping state~\cite{weidinger1999trapping}. However, since the trapping state is not perfect, if the pumping of atoms is continuous, higher photon states will eventually get populated until the evolution gets hindered by the next trapping state. As a final remark, since the trapping states infinitely continue to appear with increasing $n$ values, we clearly have $\langle\hat n\rangle=\sum_nnP_n=\infty$. So, without losses the steady state contains an infinite number of photons!

\begin{figure}
\includegraphics[width=10cm]{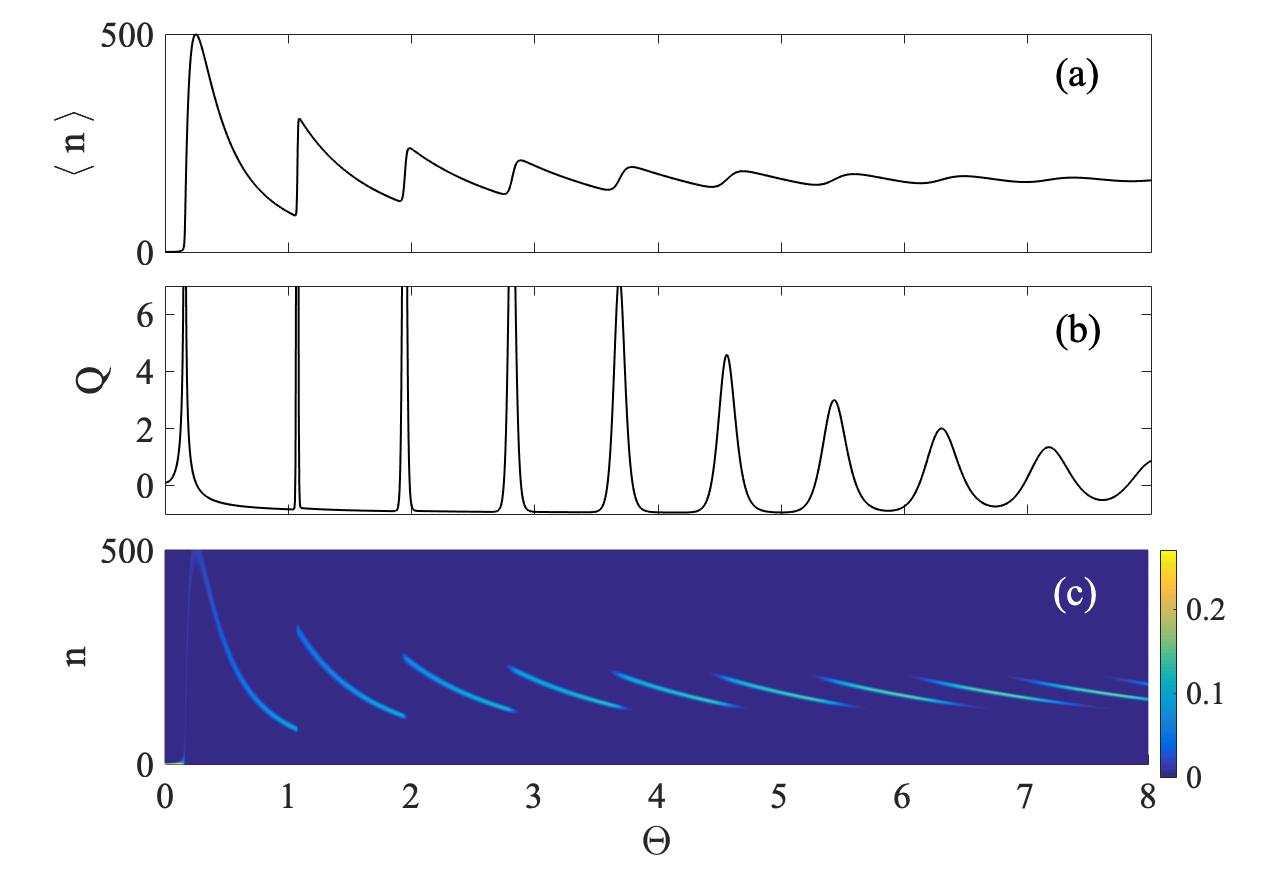} 
\caption{The upper plot of frame {\bf (a)} shows the number of photons $\langle\hat n\rangle$ in the micromaser, while the middle plot {\bf (b)} shows the Mandel $Q$-parameter\index{Mandel $Q$-parameter}, both as a function of the scaled interaction time $\Theta$~(\ref{intt}). Finally, the lower plot {\bf (c)} gives the photon distribution~(\ref{pdist1}). As a first observation we note that the trapping states of fig.~\ref{micromnoloss} manifest also in the lossy case; as the interaction time $\Theta$ is varied, the photon distribution jumps between different trapping states. At these jumps, the distribution gets super-Poissonian\index{Super-Poissonian}, while in between it may become sub-Poissonian\index{Sub-Poissonian}, {\it i.e.} $Q<0$. The dimensionless parameters are $\kappa=0.1$,\index{Sub-Poissonian} $r=50$, and $n_\mathrm{th}=0.1$. } 
\label{microm}   
\end{figure} 

We now turn to the more realistic situation of a non-zero $\kappa$. The photon distribution is then given by~\cite{lugiato1987connection}
\begin{equation}\label{pdist1}
P_n=P_0\prod_{l=1}^n\frac{n_\mathrm{th}\kappa+\sin^2(g\tau\sqrt{l})/l}{\kappa(n_\mathrm{th}+1)}.
\end{equation}
Since $\kappa\neq0$, the denominator is non-vanishing for all parameters, and one may ask whether the occurrence of trapping states is an artifact from neglecting photon losses. An answer is given in fig.~\ref{microm}; in (a) we present the average photon number $\langle\hat n\rangle$, in (b) the Mandel $Q$-parameter, see eq.~(\ref{qpara}), and in (c) the photon distribution~(\ref{pdist1}). The quantities are presented as functions of the scaled interaction time
\begin{equation}\label{intt}
\Theta=\frac{g\tau}{2\pi}\sqrt{\frac{r}{\kappa}}.
\end{equation}
Let us try to summarize the features revealed in this figure. For small $\Theta$ the field is approximately in vacuum until it reaches a critical $\Theta_c$ after which the amplitude $\langle\hat n\rangle$ rapidly increases. Simultaneously the $Q$-parameter blows up. This `micromaser threshold'\index{Micromaser! threshold} corresponds to the lasing threshold in a laser as the pump amplitude is increased~\cite{graham1970laserlight}. Like for the laser, this is a realization of a nonequilibrium continuous phase transition\index{Phase transition! out of equilibrium}\index{Nonequilibrium! phase transition! continuous}, where the field amplitude $\langle\hat n\rangle$ serves as the order parameter\index{Order parameter}, and the thermodynamic limit\index{Thermodynamic limit} is identified as $N_\mathrm{ex}=r/\kappa$ going to infinity~\cite{filipowicz1986theory}. The maximum value of $\langle\hat n\rangle$ approximately coincides with the trapping state value for the given $g\tau$, see fig.~\ref{micromnoloss}. The amplitude is then decreased as the interaction time $\Theta$ grows. In particular, $\langle\hat n\rangle$ follows closely the trajectory of the trapping states\index{Trapping states}\index{State! trapping}. However, this ends abruptly for $\Theta\approx1$, where the amplitude suddenly jumps to a larger value. This time $\langle\hat n\rangle$ follows the second branch of trapping states. At the transition point $\Theta\approx1$ the photon distribution, shown in (c), displays two peaks around the corresponding trapping states. The sudden jumps continue for larger values of $\Theta$, but they progressively get smoother, while in frame (c) we see that the double-peak structure of $P_n$ survives for longer intervals. Finally the jumps are completely smeared out, and here the distribution $P_n$ is composed of several local maxima. For large periods of time between jumps, the photon distribution is sub-Poissonian\index{Sub-Poissonian} with $Q<0$~\cite{rempe1990sub}. In ref.~\cite{filipowicz1986theory} the sudden changes in $\langle\hat n\rangle$ were analyzed within the framework of a heuristic Fokker--Planck equation\index{Fokker--Planck equation} which was derived in the thermodynamic limit\index{Thermodynamic limit} where the photon number was taken as a continuous variable. By ascribing an effective potential to the problem and identifying its minima, the jumps can be construed as first-order phase transitions\index{Phase transition! first-order}~\cite{filipowicz1986theory,bergou1989role,raimond2006exploring}. Much later, the phase transitions in the micromaser were examined from a quantum-trajectory approach by deriving statistics for the quantum jumps~\cite{garrahan2011quantum}. By varying $\Theta$ it was suggested that the micromaser field amplitude should exhibit a hysteresis bahaviour, something which was later experimentally demonstrated~\cite{benson1994quantum}. 

\begin{figure}
\includegraphics[width=10cm]{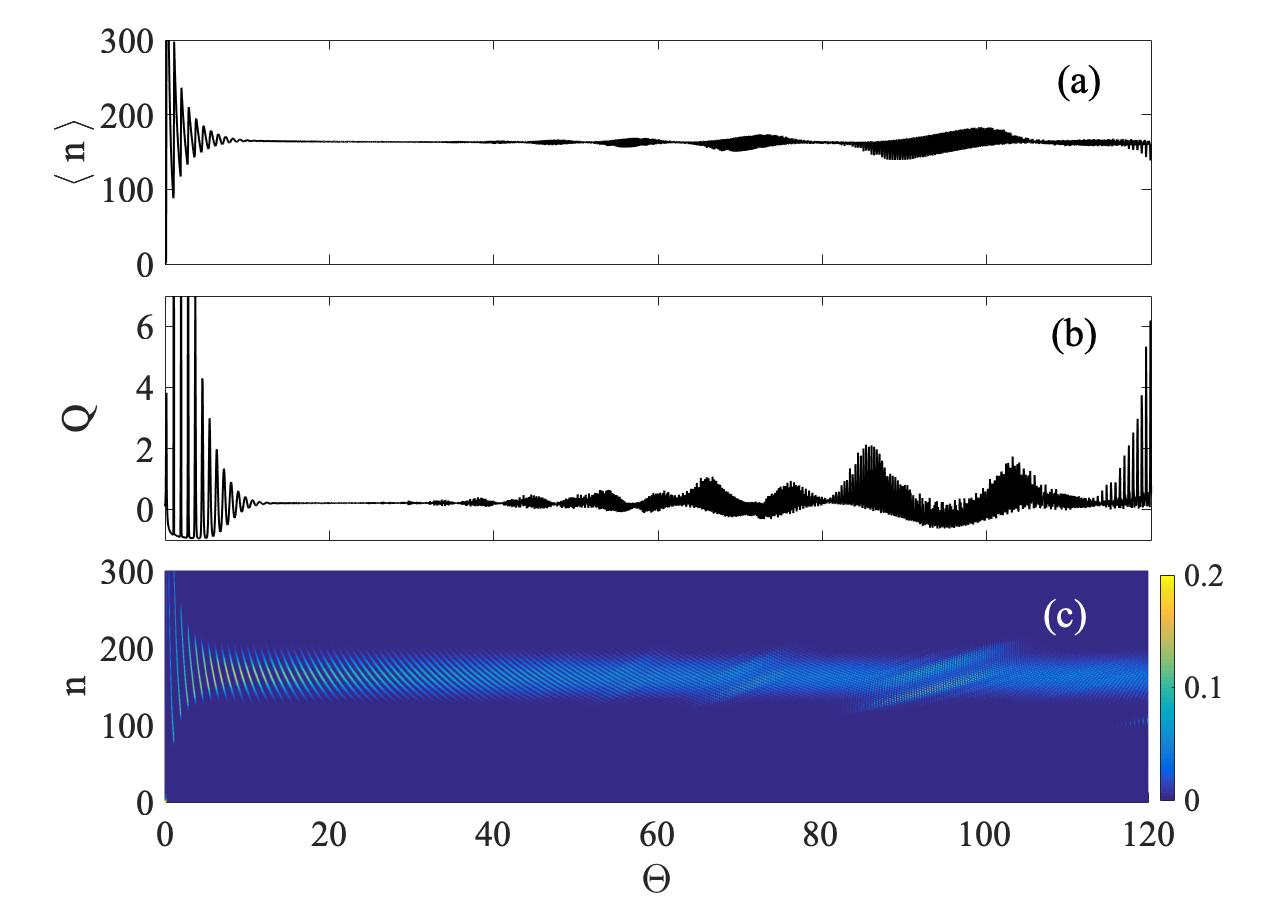} 
\caption{Same as fig.~\ref{microm} but for longer interaction times $\Theta$. After sufficiently long times, new structures of the photon distribution emerge resulting in a revival-like bahaviour of the field amplitude $\langle\hat n\rangle$ and sub-Poissonian\index{Sub-Poissonian} photon statistics. } 
\label{microm2}   
\end{figure} 

For longer periods, when the appearance of sudden changes of $\langle\hat n\rangle$ vanishes, the steady-state distribution~(\ref{pdist1}) is not exactly Poissonian which one could expect from maser and laser theory~\cite{filipowicz1986theory}. When studying the trapping states of the photon distribution~(\ref{pdist0}) for the loss-less case, one finds new patterns developing for large interaction times. These manifest also in the distribution~(\ref{pdist1}) for the lossy case as shown in fig.~\ref{microm2}. The parameters are exactly the same as for fig.~\ref{microm}, but now larger $\Theta$ values are used. The period of jumps due to hysteresis is followed by a `collapse' period, before the field starts to oscillate again~\cite{wright1989collapse}. This pattern is reminiscent of the JC revivals demonstrated in fig.~\ref{fig3}. However, the $\tau$-dependence of the probabilities $P_n$ is more complex than that of, say, the JC atomic inversion~(\ref{inv2}). During the revivals, the photon distribution periodically becomes sub-Poissonian\index{Sub-Poissonian} as the system swaps between different trapping states. 

The spectrum of the micromaser was analyzed in a series of papers in the early 1990s~\cite{scully1991micromaser,vogel1993calculation,quang1993calculation}. Here the spectrum is defined as the Fourier transform of the electromagnetic field $\langle E(t)\rangle$. With the field state $\hat\rho_{n,m}(t)$ given in the Fock representation, the quantity of interest is 
\begin{equation}
\langle E(t)\rangle\sim\sum_n\sqrt{n+1}\hat\rho_{n,n+1}(t).
\end{equation}
Via the {\it ansatz}
\begin{equation}
\hat\rho_{n,n+1}(t)=e^{-D_n(t)}\hat\rho_{n,n+1}(0),
\end{equation}
it is possible, using detailed balance\index{Detailed balance}, to derive a first-order differential equation for the micromaser linewidth\index{Linewidth! micromaser}\index{Micromaser! linewidth} $D_n(t)$~\cite{scully1991micromaser}. By expanding around the photon mean $\bar n=\langle\hat n\rangle$ (with the same arguments as in sec.~\ref{sssec:crsubsec} when we discussed the JC collapse-revivals) one finds~\cite{scully1991micromaser}
\begin{equation}
D=4r\sin^2\left(\frac{g\tau}{4\sqrt{\bar n}}\right)+\frac{\gamma(2n_\mathrm{th}+1)}{4\bar n}.
\end{equation}
For large photon numbers, $g\tau/4\sqrt{\bar n}\ll1$, the micromaser linewidth agrees with the Schawlow-Townes linewidth\index{Schawlow-Townes! linewidth}\index{Linewidth! Schawlow-Townes} for the maser~\cite{scully1967quantum}. 

Since the pioneering works on the micromaser in the second half of the 1980s, there have been numerous extensions and generalizations. For example, the `two-photon micromaser'\index{Two-photon!micromaser}~\cite{brune1987theory,davidovich1987quantum,ashraf1990theory}, effects of a non-vanishing detuning~\cite{jonaslic}, and the maser problem~\cite{scully1996induced,meyer1997quantum,loffler1997quantum,schroder1997quantum}  already discussed in sec.~\ref{sssec:qatmo}. In the maser, ultracold atoms are considered such that the atomic kinetic energy $\hat p^2/2m$ is similar in size to the interaction energy $g\bar n$. When the atomic motion is quantized, the properties of the micromaser field can be greatly altered, and in the extreme case the atoms can be reflected by the field~\cite{englert1991reflecting}. 

Another aspect that has been explored is the role played by atomic coherence, {\it i.e.} whether the incoming atoms are initialized in their upper excited state or in some other superposition of their two internal states. In the micromaser problem discussed above, when the atoms are prepared in their excited state $|e\rangle$, the steady state\index{Steady state} of the photon field has no preferred phase. At the micromaser threshold\index{Micromaser! threshold}, this phase is spontaneously broken\index{Spontaneous! symmetry breaking! phase}. However, if the atoms are in a superposition state, the field phase is not spontaneously broken but is instead inherited from the phase of the atoms~\cite{krause1986quantum}. A somewhat related topic is the entanglement generated between two successive atoms. This was considered in ref.~\cite{phoenix1993non} for two atoms traversing a lossless cavity. The focus was whether such photon generated atom-atom entanglement could be used to demonstrate violations of Bell inequalities\index{Bell! inequality}. 

\subsection{Cavity-induced atomic forces}\label{ssec:lightforce}
In sec.~\ref{sssec:qatmo} we discussed the forces experienced by an ultracold atom via interaction with a cavity mode. By utilizing the photon dissipation, the cavity could be used in order to cool atoms down below the sub-Doppler limit. Due to the strong atom-light coupling, one can trap single atoms with very weak fields~\cite{miller2005trapped} (for numerical simulations of the atomic motion see~\cite{doherty2000trapping}). Theoretically it was envisioned that atoms could be trapped solely by the cavity vacuum~\cite{haroche1991trapping,schon2003trapping}, which, however, has to date not been achieved experimentally. Much of the efforts in experimentally exploring these aspects of `ultracold' cavity QED has been pursued by the Kimble group at CalTech. By studying the fluorescence of single atoms trapped inside optical cavities, they could determine the lifetime of the atom in the trap to be around $30$ ms~\cite{ye1999trapping}. This experiment was followed up by one where an atom could be trapped for $10$ms in a cavity field composed of roughly one photon~\cite{hood2000atom}. This work introduced the `atomic microscope'; by analyzing the light transmitted through the cavity, information about the atomic motion could be gained. This idea was employed in the Rempe group, where atoms could again be trapped by single photons~\cite{pinkse2000trapping}. The information gained from the detected light was used to drive the cavity with a feedback loop, and in this way it was shown that the trapping time could be increased by 30$\%$~\cite{fischer2002feedback}.

The idea to use the transmitted cavity photons as a non-destructive probe of the atoms naturally predates the trapping experiments mentioned above. As the atom enters the cavity, its presence alters the index of refraction, which will be reflected in the output cavity field. By letting cold single atoms freely fall through the cavity region, a change in the transmitted light intensity is observed~\cite{mabuchi1996real,hood1998real,munstermann1999dynamics}.   

The presence of an atom has an even more dramatic effect when the structures of the JC spectrum are resolved. In these experiments, one end mirror is pumped with a monochromatic classical field, and the intensity of the light emitted through the other mirror is detected. By varying the frequency of the pump field it is possible to probe the JC spectrum. The presence of a two-level atom splits the spectrum into two `JC ladders'\index{Jaynes-Cummings! ladder} as seen in fig.~\ref{fig1}. In an early experiment, this type of splitting due to the atom was demonstrated~\cite{raizen1989normal,thompson1992observation}. However, the number of atoms inside the resonator was more than one and as a result the splitting results in many `ladders'. To observe the JC splitting one must make sure that only a single atom interacts with the cavity field. This was achieved by laser cooling and dipole trapping an atom inside the cavity~\cite{maunz2004cavity}. A next step would be to resolve the JC excitations with single photons. The JC spectrum~(\ref{eigv2}) is anharmonic, {\it i.e.} the energy difference between nearby eigenenergies varies according to the position in the spectral ladder. The anharmonicity gives rise to the photon blockade effect as discussed in sec.~\ref{sssec:JCzerodim}. The single photon blockade effect was first demonstrated in~\cite{birnbaum2005photon} by trapping a Cesium atom, strongly coupled to the cavity mode, inside an optical resonator. Their data were improved in a work appearing shortly afterwards~\cite{dayan2008photon}. As discussed in sec.~\ref{sssec:JCzerodim}, the photon blockade manifests in photon antibunching which is described by the second-order correlation function~(\ref{corr2}). Higher-order photon blockade\index{Photon! blockade} means that $n$ photons can be injected into the cavity, but the $(n+1)$'th photon is hindered. In an $n$-photon blockade scenario one expects to see $n+1$ photon antibunching, but $n$ photon bunching. To realize such a blockade, multiphoton transitions\index{Multiphoton! transition} in the atom must be addressed. These are, however, typically much weaker which puts challenges on experimental verification. In order to demonstrate pairwise antibunching, Rempe and co-workers measured the third order correlation function $g^{(3)}(\tau_1,\tau_2)=\langle\hat n\cdot\hat n(\tau_1)\cdot\hat n(\tau_1+\tau_2)\rangle/\langle\hat n\rangle^3$~\cite{hamsen2017two}. A different type of photon blockade, not deriving from the anhormonicity of the JC spectrum, was predicted~\cite{liew2010single}. The effect comes about due to destructive interference, and results in strong antibunching~\cite{bamba2011origin}. This unconventional photon blockade\index{Photon! blockade! unconventional} was recently experimentally demonstrated in two q-dot experiments~\cite{snijders2018observation,vaneph2018observation}.

Antibunching\index{Antibunching} results from a quantum mechanical light source. If the source is classical, the emitted light obeys the Cauchy-Schwartz inequality $g^{(2)}(\tau)\geq g^{(2)}(0)$~\cite{mandel1995optical}. The light leaking from the cavity containing a single two-level atom is thereby anti-bunched. However, if the number of atoms increases the source becomes more classical and the bunching effect should cease. The transition from antibunching to bunching as the number of trapped atoms was increased was explored in~\cite{hennrich2005transition}. It was found that already for two atoms, a bunching peak in the $g^{(2)}(\tau)$ function develops inside the antibunching minimum. The experimental results could be explained by considering a model of independent two-level emitters ~\cite{carmichael1978intensity}. The spectrum of JC `molecule' excited by a series of ultrashort pulses has been very recently reported to evince a quantum Mollow {\it quadruplet}\index{Mollow! quadruplet} at high excitation~\cite{Allcock2022}, `` quantizing the semiclassical Mollow triplet into a coherent superposition of a large number of transitions between rungs of the ladder''.

\subsection{State preparation}\label{ssec:cqedstateprep}
Reaching the strong coupling regime in cavity QED together with refining the experiments (better control over system parameters, preparation and detection, and increased photon lifetimes), enabled the experimental exploration of the quantum nature of the electromagnetic field. Field states of particular interest are photon Fock states\index{Fock state}\index{State! Fock}~\cite{brattke2001generation,santos2001conditional,bertet2002generating,brown2003deterministic,kuhn2010cavity}, squeezed states~\cite{carmichael1985photon,wodkiewicz1987squeezing}, cat states (\ref{catstate})\index{Cat state}~\cite{brune1992manipulation,brune1996observing,guo1996generation,davidovich1996mesoscopic,deleglise2008reconstruction}, and entangled states~\cite{raimond2001manipulating}.\\

\subsubsection{Fock states} 
The appearance of trapping states in the micromaser~\cite{weidinger1999trapping}, discussed in the previous subsection, results in sub-Poissonian\index{Sub-Poissonian} photon distributions\index{Distribution! sub-Poissonian} as experimentally observed in the Garching group~\cite{rempe1990observation}. In an ideal scenario it is possible to make use of the micromaser trapping states to prepare high fidelity Fock states. This has been demonstrated for the $n=1$ state, but with a rather low success rate of 85$\%$~\cite{brattke2001generation}. Photon losses and parameter fluctuations make the fidelity of such schemes go down. A more straight forward method to prepare photon number states is to initialize the cavity field in vacuum and then via atomic state reduction determine the number of photons in the cavity~\cite{krause1987state}. More precisely, if the internal state of every atom exiting the cavity is recorded, and provided the atoms entered the cavity in their excited state the photon number will equal the number of atoms found in their lower state. This idea has been used to prepare both $n=1$ and $n=2$ photon number states~\cite{varcoe2000preparing}. If the interaction time $\tau$ can be controlled for every atom, it is also possible to fix it such that every atom performs half a Rabi cycle\index{Rabi! cycle} inside the cavity, thereby leaving behind exactly one additional photon in the cavity~\cite{krause1989preparation}. Such a method has also been generalized to prepare desired superpositions of Fock states~\cite{hofheinz2009synthesizing}. The ENS group of Haroche used instead a Raman\index{Raman! coupling} scheme in which the two-photon\index{Two-photon!scattering} process involved scattering of one photon from a populated source mode into a target mode~\cite{bertet2002generating}. A main source of errors was spontaneous emission\index{Spontaneous! emission} from atoms prepared in their excited state, and the success rate for the $n=2$ state was determined to be 37$\%$. 

A long-standing goal has been the `on-demand' generation of photon number states~\cite{brattke2001generation}. By this it is meant that a given Fock state should be generated at a given time. For practical purposes, this is typically what is asked for -- we do not want to wait until the moment when we happen to have the correct Fock state. If one has the possibility to control the occurrence of the atom one can, as mentioned above, vary the atomic velocity such that it emits one photon into the cavity. More robust methods rely on adiabatic transfer~\cite{domokos1998photon,kuhn2002deterministic,larson2003adiabatic}. In ref.~\cite{kuhn2002deterministic}, as we explained in sec.~\ref{sssec:timeJCsec} and especially in fig.~\ref{fig15}. the STIRAP\index{STIRAP} was employed to transfer single photons from a classical laser field to the cavity field. An alternative way is to adiabatically tune the detuning while the atom flies through the cavity, such that the atomic transition frequency $\Omega$ is at first far detuned from the photon frequency $\omega$ but while in the cavity they become resonant $\Delta=\omega-\Omega=0$~\cite{domokos1998photon,larson2003adiabatic}. The Kimble group trapped single cesium atoms inside the cavity for up to 3 seconds. With an external laser they could drive a Raman transition\index{Raman! transition} which created a single photon in the cavity~\cite{mckeever2004deterministic}. A second laser took the atom back to its initial state and the process could be repeated. For a single trapped ion they could produce more than thousand cavity photons. Similarly, the Rempe group used an optical dipole trap that held a $^{85}$Rb atom trapped inside a resonator for up to 30 seconds, and during this time it could generate almost 300 000 photons~\cite{hijlkema2007single}. A recent report~\cite{Walker2020} analyzed the indistinguishability of the generated photons, related to how coherent the generation is. In a quantum network scheme, see sec.~\ref{ssec:cqedQI} below, the emitted photons serve as information carriers with a high demand on quantum coherence. However, decoherence during the Raman transition would harm such coherences. In their experiment a $ ^{40}$Ca$^+$ ion was trapped inside an optical resonator, and two types of Raman schemes were considered for the photon generation. They found an efficient generation of photons, but when employing the {\it Hong-Ou-Mandel interferometer}\index{Hong-Ou-Mandel! interference}~\cite{hong1987measurement} it was shown that imperfections cause the photon indistinguishability to drop. Concurrently, the dynamics of retrieval of single photons emitted by a solid-state quantum memory\index{Quantum! memory} as a function of the stored excitation and of the properties of the memory were analyzed in~\cite{Schmit2021}, accounting for a mismatch between the group velocities\index{Group velocity} and wavenumbers of the input/output photons and the read/write pulses. 

\begin{figure}
\includegraphics[width=12cm]{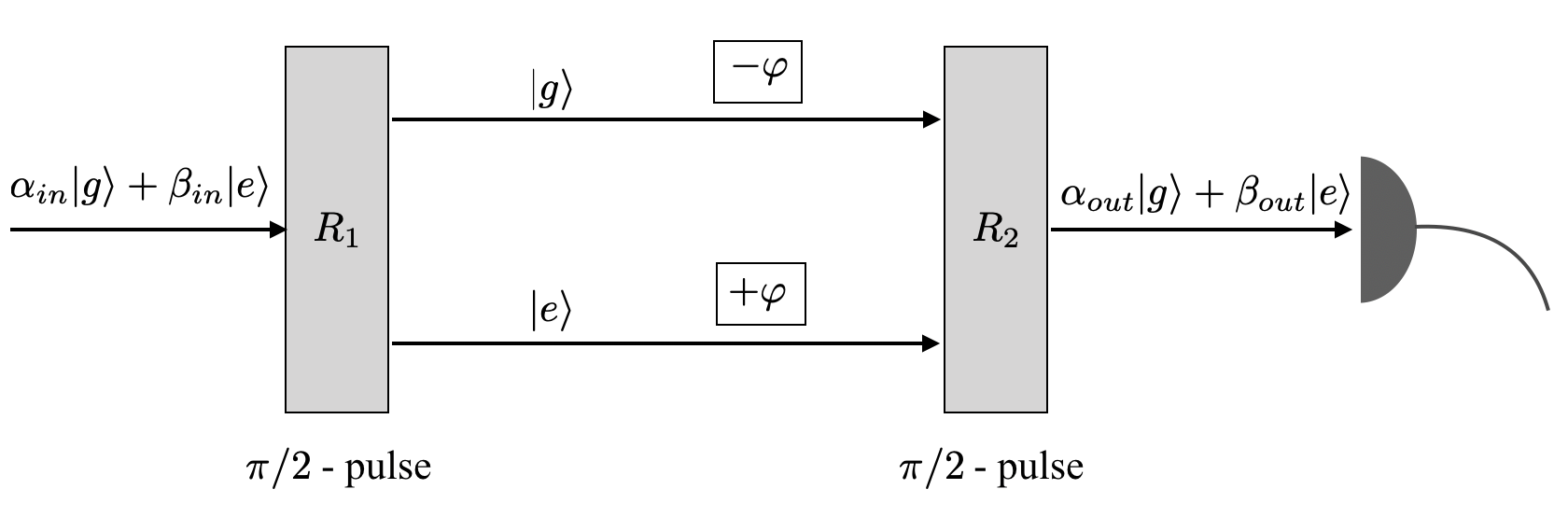} 
\caption{Sketch of the Ramsey interferometer\index{Ramsey! interferometry}. An initial atomic state enters the first Ramsey-field which realizes a $\pi/2$-pulse. The atom traverses the interaction region where the state acquires a phase  $\pm\varphi$ depending on its internal state. After the interaction the atoms experiences a second $\pi/2$ Ramsey pulse\index{Ramsey! pulse} before its internal state is being detected. The Ramsey pulses play the same role as the beam-splitters in a Mach-Zehnder interferometer\index{Mach-Zehnder interferometer}.} 
\label{ramsey}   
\end{figure}

A method for Fock-state preparation that has proven very efficient~\cite{geremia2006deterministic,brattke2001generation,sayrin2011real} relies on quantum feedback control~\cite{wiseman2009quantum}. To understand the idea of the feedback scheme we first need to explain the basics of the Ramsey interferometer\index{Ramsey! interferometry}~\cite{bransden2003physics,raimond2006exploring}. 

The Ramsey interferometer\index{Ramsey! interferometry} is the matter analog of a Mach-Zehnder interferometer\index{Mach-Zehnder interferometer} for light, see fig.~\ref{ramsey}. The beam-splitters are replaced by Ramsey fields which apply $\pi/2$-pulses\index{$\pi/2$-pulse} on the internal atomic states. The two paths in the Mach-Zehnder apparatus are represented by the two internal states of the atom as it traverses the cavity. The interaction between the atom and the cavity field is assumed to be dispersive such that the cavity field does not induce any population transfer between the atomic state, but only causes a light shift of the atomic levels. This action is described by the dispersive JC-Hamiltonian~(\ref{effham}). If we assume that the atom is initialized in the ground state $|g\rangle$, and notice that the $\pi/2$-pulse is nothing but a Hadamard transformation\index{Hadamard! transformation}, it is straightforward to find the output state\index{Output state}\index{State! output}
\begin{equation}\label{outf}
|\Psi_\mathrm{out}\rangle=\sum_{n=0}^\infty\left[c_n\cos(\varphi n)|n\rangle|g\rangle+c_n\sin(\varphi n)|n\rangle|e\rangle\right],
\end{equation}
where the $c_n$'s are the amplitudes of the initial photon Fock states. For say an initial Fock state $|n\rangle$, the probability to detect the internal atomic state $|g\rangle$ becomes $P_g=\cos^2(\varphi n)$, which clearly oscillates when varying the accumulated phase shift $\varphi$. Looking at the atomic state on the Bloch sphere, initially after the first Ramsey pulse\index{Ramsey! pulse} it points to the equator in the negative $x$-direction. After the second Ramsey pulse it has been rotated along the equator by an angle $\varphi n$. For a general initial photon state the output state is~(\ref{outf}), {\it i.e.} an entangled state and hence an atomic projective measurement\index{Projective measurement} will affect the state of the field. When the internal atomic state is detected, the photon distribution is modified accordingly. If, for example, the atom is found in the $|g\rangle$ state, the photon distribution becomes
\begin{equation}\label{filter}
P(n)\propto\cos^2(\varphi n)|c_n|^2.
\end{equation}
The photon distribution has been modified by the {\it filter function} $\cos^2(\varphi n)$. If the process is repeated, for every atomic measurement the distribution will be adjusted by the corresponding filter function~\cite{larson2004cavity}. This method, using quantum non-demolition measurements\index{Non-demolition measurement}, has been used to prepare Fock states from initial coherent field states\index{State! coherent}\index{Coherent state}~\cite{guerlin2007progressive}. For sufficiently many filter functions, a single Fock state will eventually survive. It is, however, {\it a priori} not known which is the Fock state that will finally survive. Quantum trajectory\index{Quantum! trajectories} simulations of the excited-state population of the Jaynes-Cummings system subject to weak continuous probing were given in~\cite{Andersen2022}. Andersen and M\o lmer discuss quantum non-demolition measurements\index{Non-demolition measurement} which do not project the system on the eigenstate of the corresponding observable but rather on a superposition state persistently evolving within a given subspace.

We return to the feedback method for preparing Fock states on-demand. Assume that the target state is $|\psi_t\rangle=|n\rangle$, and the initial state is a known coherent state. After the application of the first pulse, the entangled state is given by~(\ref{outf}), while the atomic projective measurement will collapse the field state by multiplying the amplitudes $c_n$ with the corresponding filter function ($\cos(\varphi n)$ or $\sin(\varphi n)$), which results in a state $|\psi_c\rangle$. Hence, after the detection the field state is still known. This information is fed into a classical drive that injects a coherent state into the cavity, effectively realizing the displacement operator $\hat D(\alpha)=\exp\left(\alpha\hat a^\dagger-\alpha^*\hat a\right)$. Now, the amplitude $\alpha$ is chosen such that the state fidelity\index{State! fidelity}~(\ref{fidel}) $F=|\langle n|\psi_c\rangle|$ is minimized~\cite{sayrin2011real}. By repeating these steps with new atoms and measurements, the field state will approach the desired Fock state. In a real experiment photons dissipate and this should be taken into account in the feedback. In fact, the feedback injection of the displacement operator can be utilized in order to stabilize the desired Fock state; it is possible to correct for a photon loss to regain the desired Fock state.\\

\subsubsection{Schr\"odinger cat states}\label{sssec:subscat}
In cavity QED, a cat state (\ref{catstate})\index{Cat state} (or simply cat state) is thought of as the superposition of two well distinguished states, typically two coherent states with opposite phase and with a large amplitude $|\alpha|^2\gg0$~\cite{raimond2006exploring}, {\it i.e.}
\begin{equation}\label{cats}
|\psi_\mathrm{cat}\rangle=\frac{1}{N}\left(|\alpha\rangle+e^{i\phi}|-\alpha\rangle\right),
\end{equation} 
with the normalization $N=\sqrt{2\left(1+\mathrm{Re}\left[e^{i\phi}\langle\alpha|-\alpha\rangle\right]\right)}$. There are certain values of the phase $\phi$ that naturally appear in state preparations: $\phi=0$ {\it even Schr\"odinger cat} whose photon distribution only contains even Fock states, $\phi=\pi$ {\it odd Schr\"odinger cat}\index{Odd Schr\"odinger cat state}\index{State! odd Schr\"odinger cat}\index{Even Schr\"odinger cat state}\index{State! even Schr\"odinger cat} whose photon distribution only contains odd Fock states, and $\phi=\pi/2$ {\it Yurke-Stoller state}\index{Yurke-Stoller state}\index{State! Yurke-Stoller}~\cite{yurke1986generating}. The interest in cat states in cavity QED has largely been driven by fundamental questions of quantum mechanics, but these states can also be useful for applications in QIP~\cite{gilchrist2004schrodinger,ourjoumtsev2006generating,vlastakis2013deterministically} and spectroscopy~\cite{kira2011quantum}.

What sparked this interest in cavity QED was the fact that the dispersive JC Hamiltonian~(\ref{effham}) generates cat states~\cite{savage1990macroscopic,brune1992manipulation,guo1996generation}. Initializing the atom with the first Ramsey pulse\index{Ramsey! pulse} $R_1$ in the superposition state $|\psi_\mathrm{at}\rangle=(|g\rangle+|e\rangle)/\sqrt{2}$, and the field in a coherent state $|\psi_\mathrm{fi}\rangle=|\alpha\rangle$, we saw that in the dispersive regime this state evolves into that of eq.~(\ref{cat1}). The application of a second Ramsey pulse $R_2$ followed by a projective atomic measurement\index{Projective measurement} will leave the field in the cat state 
\begin{equation}\label{cat2}
|\psi_\mathrm{cat}\rangle=\frac{1}{N}\left(|\alpha e^{i\lambda t}\rangle\pm e^{i\lambda t}|\alpha e^{-i\lambda t}\rangle\right),
\end{equation} 
and by choosing the interaction time $\lambda t=\pi/2$ we obtain the Yurke-Stoler cat
\begin{equation}\label{cat3}
|\psi_\mathrm{cat}\rangle=\frac{1}{N}\left(|i\alpha\rangle\pm i|-i\alpha\rangle\right).
\end{equation} 
The two coherent states\index{Coherent state}\index{State! coherent} are maximally separated in phase space, and for $|\alpha|^2\gg0$ we have that $\langle-i\alpha|i\alpha\rangle\approx0$. In general, the overlap will depend on the angle $\lambda t$, and for $\lambda t=\pi$ the atom will disentangle from the field which is left in a simple coherent state. Actually the overlap $\mathcal{O}(t)=\langle\alpha e^{-i\lambda t}|\alpha e^{i\lambda t}\rangle$ determines the amount of atom-field entanglement. In the seminal experimental paper~\cite{brune1996observing}, this type of Ramsey interferometer\index{Ramsey! interferometry} was explored in detail. In their experiment, the probability $P_g$ to detect the atom in the $|g\rangle$ state after the second Ramsey pulse\index{Ramsey! pulse} was determined by repeating the experiment sufficiently many times. The probability $P_g$ was found to oscillate as a function of the frequency of the applied Ramsey pulses. In particular, if $\mathcal{O}(t)\approx1$ there is roughly no atom-field entanglement and the Ramsey interference fringes were almost perfect (not taking into account losses and other imperfections). However, as the overlap $\mathcal{O}(t)$ decreased, by considering larger interaction times, the fringe contrast went down. The smaller $\mathcal{O}(t)$ is, the more entangled are the two subfields, and the more information is contained in the field state on which interference path the atom followed. If $\mathcal{O}(t)=0$, a measurement of the field state would reveal whether the atom passed the cavity in the upper or lower state, and the fringe contrast is completely lost in this case. In such a situation the atom-field complex is in a maximally entangled EPR-like state\index{EPR state}\index{State! EPR} (\ref{eprstate}). As we see in fig.~\ref{fig4} (a), the Wigner function\index{Wigner! function} of a cat state displays fine structures in between the two blobs. Quite unexpectedly, the Wigner function for a cat state may posses sub-Planck\index{Planck!scale} structures, {\it i.e.} features on a scale smaller than the phase-space area $\hbar^2$, and it was proposed that the above type of Ramsey experiment should be able to reveal these features~\cite{toscano2006sub}. It is interesting to note that the effective Hamiltonian~(\ref{solanoeff}), which was derived for a strongly driven JC model, also generates a cat state~\cite{solano2003strong} 
\begin{equation}\label{cat4}
|\Psi_\mathrm{cat}\rangle=\frac{1}{\sqrt{2}}\left(|\alpha\rangle|+\rangle+|-\alpha\rangle|-\rangle\right),
\end{equation} 
where $|\pm\rangle$ are the atomic dipole states~(\ref{dipol}), $\alpha=g\left(e^{i\delta_\mathrm{p}t}-1\right))/2\delta_\mathrm{p}$ (for $\delta_\mathrm{p}=0$ we have $\alpha=-igt/2$), and the initial state was taken as $|0\rangle|g\rangle$. As for the dispersive regime, an atomic projective measurement in the basis $\{|g\rangle,\,|e\rangle\}$ will prepare the field in a cat state.

As already discussed in sec.~\ref{sssec:openjc}, a cat state is extremely sensitive to decoherence. In particular, an initial even cat state will evolve into the state~(\ref{catdec}) when coupled to a zero-temperature bath. From this we concluded that the cat becomes a statistical mixture on a timescale $\sim 1/(\kappa|\alpha|^2)$, which may be orders of magnitude smaller than the decay rate $1/\kappa$ provided the amplitude satisfies $|\alpha|\gg1$. The decay into a statistical mixture is an irreversible process, and an insight into it is obtained via `echo spectroscopy'~\cite{morigi2002measuring} (see also ref.~\cite{meunier2005rabi} for another experiment utilizing echo dynamics in order to induce revivals in the Rabi oscillations\index{Echo measurement}\index{Revivals} ). After a given time $\tau$, the time-evolution is reversed such that at $2\tau$ the system should return to its initial state provided there were no losses. How close the evolved state $|\psi(2\tau)\rangle$ is to the initial state $|\psi(0)\rangle$ says how important was the effect of the losses. For the aforementioned cat experiment, time was not reversed but instead a second atom was injected into the cavity which could `partly' reverse time~\cite{davidovich1996mesoscopic} and the resulting Ramsey fringes\index{Ramsey! interferometry! fringes} for the second atom revealed information about the conversion of a true cat into a statistical mixture~\cite{brune1996observing}, or as stated in~\cite{raimond2001manipulating} "decoherence caught in the act".

A drawback with the dispersive generation of cat states in cavity QED is that the parameter $\lambda=2g^2/\Delta$ is rather small (since we demand $|\Delta|\ggg\sqrt{\bar n}$), and the time needed for preparing a cat state becomes too long in order to sustain quantum coherence. Due to this, in ref.~\cite{brune1996observing} the field amplitude was limited to $|\alpha|^2\approx9.5$. In circuit QED (see next section), the dispersive regime was again used and in such systems one could reach cat states with $|\alpha|^2\approx100$~\cite{vlastakis2013deterministically}. Alternatively, we saw in sec.~\ref{sssec:semclas} that for initial coherent states\index{Coherent state}\index{State! coherent} $|\alpha\rangle$ with a large amplitude, at half the revival time the atom approximately disentangles from the field, and the field is prepared in a cat state~\cite{gea1991atom,buvzek1992schrodinger}. The mechanism for this was explained in fig.~\ref{fig5}, and a numerical example of the prepared cat state was shown in fig.~\ref{fig4} (a) and (b). The ENS group of Haroche employed this idea to experimentally prepare cat states with amplitudes $|\alpha|^2\approx36$~\cite{auffeves2003entanglement}.

As for bosons, it is possible to form coherent states also for spins~\cite{arecchi1972atomic,gazeau2009coherent}. The spin coherent states (also called $SU(2)$ coherent states) can be formed with a displacement operator, like for bosons in eq.~(\ref{dispcoh}), by acting with a displacement operator on the `vacuum',
\begin{equation}
|\theta,\phi\rangle=e^{i\phi\hat S_z}e^{i\theta\hat S_x}|s,-s\rangle,
\end{equation}
where the collective spin operators\index{Collective spin operator}\index{Operator! collective spin} were defined in eq.~(\ref{spinop}), and the spin states below that equation. Note that, given the spin $s$, a spin coherent state is defined by the polar and azimuthal angles. It is possible to define phase-space distributions\index{Distribution! phase-space}\index{Phase-space distribution} also for spins, e.g. the function $Q(\theta,\phi)\propto\langle\theta,\phi|\hat\rho|\theta,\phi\rangle$, which lives on a sphere with radius $s$. An atomic cat state is thus a superposition state formed from two well separated spin coherent states, and where $s\gg1$. In ref.~\cite{agarwal1997atomic} it was demonstrated how it is possible to create atomic cat states starting from the Tavis-Cummings Hamiltonian\index{Tavis-Cummings model}\index{Model! Tavis-Cummings}~(\ref{tcham}). The authors therein consider the open Tavis-Cummings model where the photon mode was coupled to a thermal bath. As we have already mentioned in sec.~\ref{sssec:dicke}, when eliminating the boson degree of freedom, we end up with an open Lipkin-Meshkov-Glick model\index{Lipkin-Meshkov-Glick! model}\index{Model! Lipkin-Meshkov-Glick} that effectively describes the atoms. In this way, the authors were able to demonstrate the preparation of cat states. In general, if the cavity field is prepared in a superposition state $|\psi_\mathrm{fi}\rangle=(|0\rangle+e^{i\varphi}|1\rangle)/\sqrt{2}$ and the ensemble of atoms in their ground state $|s,-s\rangle$, when the system evolves with the dispersive Tavis-Cummings Hamiltonian, it is straightforward to show that the atoms develop a cat state~\cite{gerry1997generation}.

\subsubsection{Entangled states} 
Aspects of entanglement in the JC model was already discussed in sec.~\ref{sssec:ent}. We have studied how the quantum correlations of the atom-field state evolve in time, both in terms of the von Neumann entropy\index{von Neumann! entropy}~(\ref{vN}) as shown in fig.~\ref{fig3} (c) and (f), and in terms of the logarithmic negativity in fig.~\ref{fig7} (a). The first reference to quantum correlations between the two subsystems dates back to the work of Phoenix and Knight~\cite{phoenix1988fluctuations} from 1988, even if it took them three years more before they actually used the word {\it entanglement}~\cite{phoenix1991establishment}. The first experiments exploring entanglement started in 1997, in the group of S. Haroche. 

The simplest entangled state is generated by preparing the state $|e,0\rangle$ and then letting the atom interact with the cavity mode for half a Rabi cycle, {\it i.e.} for an interaction time $t_\mathrm{f}=\pi/4g$. The atom-field complex will be prepared in a maximally EPR-like state\index{EPR state}\index{State! EPR}
\begin{equation}\label{afent1}
|\psi(t_\mathrm{f})\rangle=\frac{1}{\sqrt{2}}\left(|e,0\rangle-i|g,1\rangle\right).
\end{equation}
By adjusting the second Ramsey pulse\index{Ramsey! pulse}, the atom can be measured in any desirable basis, and depending on the outcome the photon mode will collapse into the corresponding superposition state of vacuum and the single-photon Fock state\index{Single-photon!state}\index{Fock state}~\cite{maitre1997quantum}. Note how the atomic measurement is spatially separated from the cavity and that it is the entanglement causing the non-local collapse of the photon field. If we initialize the state $|\psi\rangle=(c_g|g\rangle+c_e|e\rangle)|0\rangle$ and let $t_\mathrm{f}=\pi/2g$ the final state becomes
\begin{equation}
|\psi(t_\mathrm{f})\rangle=\left(c_g|0\rangle-ic_e|1\rangle\right)|g\rangle.
\end{equation}
The information about the atomic state, carried by the amplitudes $c_g$ and $c_e$ has in this case been transferred to the photon field. A second atom, prepared in the state $|g\rangle$ and traversing the cavity with the same velocity will exit the cavity in the state $c_g|g\rangle+c_e|e\rangle$, {\it i.e.} identical to that of the first atom before it entered the cavity. The state has been swapped between the two atoms by using the cavity mode as an intermediate quantum memory\index{Quantum! memory}~\cite{maitre1997quantum}. The scheme can be combined by varying the initial atomic states and the interaction times~\cite{phoenix1993non}. For example, if the state~(\ref{afent1}) is prepared and the second atom, initialized in $|g\rangle$, makes a full Rabi cycle\index{Rabi! cycle} we end up with the atom-atom EPR state~(\ref{eprstate})\index{EPR state}
\begin{equation}\label{afent2}
|{\rm EPR}_-\rangle=\frac{1}{\sqrt{2}}\left(|e_1,g_2\rangle-|g_1,e_2\rangle\right)|0\rangle
\end{equation}
for the two atoms. Using this procedure, entanglement between two atoms, separated by approximately one centimeter, was experimentally demonstrated with a state purity of 63$\%$~\cite{hagley1997generation}. The low fidelity could be ascribed to experimental imperfections like losses, and once these were taken into account a very good agreement with the experiment was found. In their later work they also extended the experiment to measure a Bell `signal', but the low fidelity did not allow for violation of the Bell inequality~\cite{raimond2001manipulating}. The idea presented above with consecutively passing atoms through a cavity can be generalised to more particles, as was done in ref.~\cite{rauschenbeutel2000step}. For an initially empty cavity, the first atom subject to a $\pi/2$-pulse and the second atom to a $\pi$-pulse, the resulting three-partite entangled state is a GHZ state~(\ref{ghzstate}). 

Following the theoretical proposal by Zheng and Guo~\cite{zheng2000efficient}, a different method for entangling a pair of atoms was implemented experimentally in ref.~\cite{osnaghi2001coherent}. The idea is to let the photon mode mediate an effective atom-atom interaction in the dispersive regime, see eq.~(\ref{effham}) for the single atom effective model. When discussing the Dicke model in sec.~\ref{sssec:dicke} we mentioned that if the boson field is adiabatically eliminated one finds the Lipkin-Meshkov-Glick model\index{Lipkin-Meshkov-Glick! model}\index{Model! Lipkin-Meshkov-Glick}, which is an Ising-type of model with infinite-range interaction\index{Infinite-range interaction}. For simply two atoms one obtains the dispersive atom-atom Hamiltonian~\cite{zheng2000efficient}
\begin{equation}\label{atat}
\hat H_\mathrm{aa}=\frac{g^2}{\Delta}\left(\hat\sigma_1^+\hat\sigma_1^-+\hat\sigma_2^+\hat\sigma_2^-+\hat\sigma_1^+\hat\sigma_2^-+\hat\sigma_2^+\hat\sigma_1^-\right),
\end{equation}
where we have assumed the cavity mode to be in the vacuum state. If the two atoms are initialized in the state $|e_1\rangle|g_2\rangle$, and the interaction time is taken as $gt=\pi/4$, the bipartite atomic state will become (up to an overall phase factor)
\begin{equation}
|\psi_\mathrm{EPR}\rangle=\frac{1}{\sqrt{2}}\left(|e_1\rangle|g_2\rangle-i|g_1\rangle|e_2\rangle\right).
\end{equation}
The scheme does not rely on the existence of the vacuum state for the intracavity field (it can also be realized for a thermal state), which makes it very robust against photon losses. The atom-atom entanglement generation stems from what has been termed `cavity-assisted collisions'. A difficulty in the experiment was to avoid having more than two atoms inside the cavity at once, and another source for errors was detector efficiency. Nevertheless, clear quantum correlations between the two atoms were reported~\cite{osnaghi2001coherent}. 

One type of methods for preparing entangled states, which turn out very robust and efficient, employs state collapse~\cite{duan2003efficient,sorensen2003probabilistic,chen2015carving}. Typically, one starts with two or more atoms, initially unentangled, and lets them interact with a common photon mode. Subsequently a selective photonic detection\index{Selective measurement} is carried out, and depending on the measurement outcome the atoms will collapse into some entangled state. The state preparation is not deterministic, but depends on the particular result of the measurement. To exemplify the idea we consider the Duan and Kimble scheme~\cite{duan2003efficient}. The set-up is the following: (i) Two identical cavities, both supporting two degenerate cavity modes $h$ and $v$ with different polarizations. (ii) Each cavity contains a tripod atom, {\it i.e.} a four-level atom with one excited electronic state $|e\rangle$ and three lower states $|0\rangle,\,|g\rangle,\,|1\rangle$, with $|0\rangle$ and $|1\rangle$ degenerate. The atoms are prepared in the states $|e\rangle$. (iii) Each atom can emit either a $h$-photon via the transition $|e\rangle\rightarrow|0\rangle$ or a $v$-photon via $|e\rangle\rightarrow|1\rangle$. (iv) The photons escaping the cavities are merged in a beam-splitter\index{Beam-splitter} and then detected by two detectors. Due to the presence of the beam-splitter it is not possible to determine which cavity the photon came from, and if a photon is detected in each cavity we know with certainty that the atoms are in the maximally entangled EPR state (\ref{eprstate}) (to prepare this particular Bell state\index{Bell! state} one also needs a polarization rotation of one cavity photon). Note that the success relies on that one photon is detected in each cavity, so on average only half of the experimental runs will generate the desired state (it was shown that the method could be made deterministic though~\cite{browne2003robust}). In the experiment~\cite{casabone2013heralded} the scheme was somewhat modified to consider two trapped ions inside a single cavity, but two detectors were used and a single click in each of them collapsed the two atoms into the entangled state. The measured fidelity of the prepared state was around 92$\%$. To achieve higher robustness, one can deploy driven-dissipative systems (see further details in sec.~\ref{ssec:ionfur}). The idea is that the drive and loss are monitored in such a way that they balance one another and the resulting steady state\index{Steady state} possesses non-classical properties~\cite{barreiro2011open}. Kastoryano {\it et al}. employed such a concept to propose how to maximally entangle two atoms inside an optical cavity~\cite{kastoryano2011dissipative}. The dissipation-induced preparation of entangled states has not been experimentally demonstrated in cavity QED so far, but it has been implemented in both circuit QED~\cite{shankar2013autonomously} (see section~\ref{sec:cirQED}) and in trapped ions~\cite{barreiro2011open,lin2013dissipative} (see section~\ref{sec:ion}). 

Adiabatic following\index{Adiabatic! following}, like STIRAP\index{STIRAP} (see sec.~\ref{sssec:timeJCsec}), is another approach to the robust preparation of entangled states~\cite{pachos2002quantum,biswas2004preparation,song2007entangled,chen2007generation}. Adiabatic methods are preferable due to their robustness against parameter fluctuations and timing. However, they are by default time costly, and losses may become a problem. It has been suggested to circumvent this by applying the idea of a {\it shortcut to adiabaticity}\index{Shortcut to adiabaticity}~\cite{del2013shortcuts}, which aims at finding the optimal time scheme (time-dependent Hamiltonian) for taking some initial state to a desired final state in as short time as possible and with a high fidelity. Two approaches have been developed through the years, one based on {\it Lewis-Riesenfeld invariants}\index{Lewis-Riesenfeld invariants} and one termed {\it transitionless quantum  driving}. It turns out that they are highly connected~\cite{chen2011lewis}, and both have been applied for entanglement state preparation in cavity QED systems, see for example refs.~\cite{lu2014shortcuts,chen2014efficient,chen2016fast,chen2017deterministic}. Turning now to circuit QED, an eight-fold decrease in the adiabatic evolution time of a single lossy mode has been experimentally demonstrated in~\cite{Yin2022}, aiming at the design of fast open-system protocols. The concept of shortcuts to adiabaticity has recently motivated a proposal to accelerate the measurement on a qubit longitudinally coupled to a cavity~\cite{Lopez2022}. The proposed setup employs an asymmetric SQUID threaded by an external magnetic flux. 

Adding a counteradiabatic driving term to the JC Hamiltonian, whose form judiciously involves the dressed eigenstates, allows for the generation of arbitrary Fock states of the bosonic mode, as well as coherent quantum superpositions of a Schr\"{o}dinger cat-like states~\cite{Abah2020}. The results were assessed against the non-classicality of photon-added states, after their paradigmatic conditional generation in spontaneous parametric down conversion\index{Spontaneous! down conversion}~\cite{Zavatta2004}. A single-photon--added coherent state\index{Coherent state!single-photon--added}\index{Single-photon!--added coherent state} in the result of the ``most elementary amplification process of classical light fields by a single quantum of excitation...they offer the opportunity to closely follow the smooth transition between the particle-like and the wavelike behavior of light.'' An extension of the above shortcut to adiabaticity\index{Shortcut to adiabaticity} has been recently presented in~\cite{YangMan_2023}, applied to the anti-JC model. 

By trapping the atoms/ions inside the resonators, new perspectives open up, not only by allowing for longer interaction times. Some of these experiments are motivated by quantum networks, see sec.~\ref{ssec:cqedQI}. In such networks, the trapping of atoms/ions inside the resonators serves as an essential ingredient. Rempe and co-workers prepared entangled atom-atom states in a probabilistic manner following theoretical proposals~\cite{chen2015carving}. Starting with two trapped $^{87}$Rb atoms prepared in the unentangled state $(|g_1\rangle+|e_1\rangle)(|g_2\rangle+|e_2\rangle)/2$, the method, like for the Duan/Kimble scheme~\cite{duan2003efficient}, relies on the scattering of coherent light pulses off the cavity with certain polarizations followed by their detection. The light will couple differently to the atoms depending on their polarizations, and by performing a polarization-dependent projective measurement\index{Projective measurement} on the light field afterwards, collapses the atomic state into an EPR state\index{EPR state}\index{State! EPR}~(\ref{eprstate}). Using this method, Duan and Kimble were able to prepare all four different EPR states with up to $\sim90\%$ fidelity~\cite{welte2017cavity}. Unlike Rempe, the Blatt group does not work with dipole trapped atoms, but rather with $^{40}$Ca$^+$ ions held in the cavity with the help of a Paul trap. In the experiment of~\cite{stute2012tunable} they maximally entangled, with $97\%$ fidelity, a two-level ion with a single photon possessing two different polarizations. This was achieved by preparing the ion in an ancilla electronic state\index{Ancilla! state}\index{State! ancilla}, and then employing two Raman transitions that transfer the internal ionic state into two different hyperfine electronic states. In the process a vertically or horizontally polarized photon is transmitted into the cavity, very much in the same vain as in ref.~\cite{duan2003efficient}. However, unlike in that reference, here the two internal ionic states were entangled with the polarization states of the photon.  

Instead of entangling massive objects like two atoms, it is, of course, also possible to consider entanglement between photons of different cavity modes. By varying the atom-field detuning while the atom interacts with the cavity field, Haroche and co-workers prepared the two-mode entangled state $|\psi\rangle(|1,0\rangle+e^{i\phi}|0,1\rangle)\sqrt{2}$~\cite{rauschenbeutel2001controlled}. {\it Entangled coherent states} have been thoroughly studied since the 1990s. This two-mode entangled state has the expression {\it entangled coherent states}\index{Entangled coherent states}\index{State! entangled coherent},
\begin{equation}
|\psi_\mathrm{2cat}\rangle=\frac{1}{\sqrt{N}}\left(|\alpha\rangle|\beta\rangle+e^{i\phi}|-\alpha\rangle|-\beta\rangle\right),
\end{equation}
formed from coherent states $|\pm\alpha\rangle$ and $|\pm\beta\rangle$~\cite{chai1992two,van2001entangled,sanders2012review}. Such states find applications in a variety of QIP schemes, like teleportation~\cite{van2001entangled} and metrology~\cite{joo2011quantum}, and they are examples of {\it entangled non-orthogonal states}. In most studies $\alpha=\beta$ and $e^{i\phi}=\pm1$. The generation of cat states, either in the dispersive regime as described in sec.~\ref{sssec:sol} and in particular eq.~(\ref{cat1}) or in the resonant case pictured in fig.~\ref{fig5}, can be generalized to more modes as the bimodal JC Hamiltonian~(\ref{2mode}). For example, consider the dispersive two-mode model given by the Hamiltonian
\begin{equation}
\hat H_\mathrm{2m}=\lambda\left(\hat a^\dagger\hat a-\hat b^\dagger\hat b\right)\hat\sigma_z,
\end{equation}
which upon evolving a state $|\alpha\rangle|\beta\rangle|\pm\rangle$ generates an entangled coherent state~\cite{gerry1997generation,guo1997preparation}. The idea was also extended to more than two modes in~\cite{wang2001multipartite}, and to spatially separated cavities in~\cite{gerry1996proposal,yang2013generating}. Note that the above Hamiltonian $\hat H_\mathrm{2m}$ does not contain `cross-terms' which describe scattering of photons from one mode to the other. If no selection rule prohibits this, or different detunings are considered, the dispersive Hamiltonian, obtained from eq.~(\ref{2mham}), reads
\begin{equation}\label{2mham1}
\hat H_\mathrm{2m}=\frac{\Delta}{2}\hat\sigma_z+\frac{g^2}{\Delta}\hat A^\dagger\hat A\sigma_z,
\end{equation}
with $g=\sqrt{g_1^2+g_2^2}$ and we have introduced two new boson operators as
\begin{equation}
\left[\begin{array}{c}
\hat A\\ \hat B\end{array}\right]=U\left[\begin{array}{c}
\hat a\\ \hat b\end{array}\right],\hspace{1cm}U=\left[
\begin{array}{cc}
\cos\theta & \sin\theta\\
-\sin\theta & \cos\theta
\end{array}\right]
\end{equation}
with $\cos\theta=g_1/g$. If we label the original two modes (represented by $\hat a$ and $\hat b$) by subscripts 1 and 2 and the transformed two modes (represented by $\hat A$ and $\hat B$) by I and II, then for the displacement operators~(\ref{dispcoh}) we have
\begin{equation}
\hat D_1(\alpha)\hat D_2(\beta)=\hat D_\mathrm{I}(\mu)\hat D_\mathrm{II}(\nu),
\end{equation}
where 
\begin{equation}
\left[\begin{array}{c}
\mu\\ \nu\end{array}\right]=\left[\begin{array}{cc}
\cos\theta & \sin\theta\\
-\sin\theta & \cos\theta
\end{array}\right]\left[\begin{array}{c}
\alpha\\ \beta\end{array}\right].
\end{equation}
Thus, the product of two coherent states remains a product of coherent states under a rotation. In other words, even when the cross terms are taken into account the dispersive bimodal Hamiltonian\index{Bimodal! Jaynes-Cummings model}\index{Model! bimodal Jaynes-Cummings}~(\ref{2mham1}) will generate entangled coherent states~\cite{larson2006scheme}. In fact, even at zero detuning, the Hamiltonian 
\begin{equation}\label{2mham2}
\hat H_\mathrm{2m}=g\left(\hat A^\dagger\hat{\sigma}_{-}+\hat A\sigma^+\right)
\end{equation}
can be utilized in order to prepare entangled cat-like states following the idea presented in fig.~\ref{fig5}. However, the method works better for entangled squeezed states instead of coherent states~\cite{larson2006scheme}. 

\begin{figure}
\includegraphics[width=5cm]{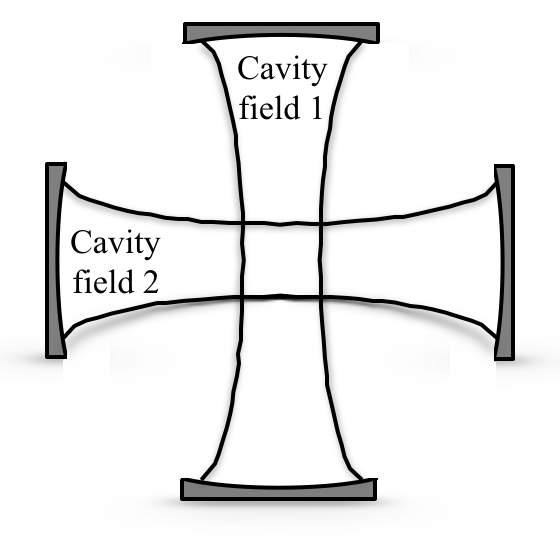} 
\caption{Sketch of the crossed cavity setup. In the middle the two cavity fields overlap, and this is where the atom traverse the fields perpendicularly to the cavity plane. The mode waists are assumed Gaussian\index{Gaussian! mode}, and the two symmetry axes are slightly displaced such that for an adiabatic evolution an atom in its ground state $|g\rangle$ induces the STIRAP process of sec.~\ref{sssec:timeJCsec}, which transfer the state of the first cavity into the initially empty second cavity. With the help of an atomic ancilla state\index{Ancilla! state}\index{State! ancilla}, as explained in the main text, it is possible to prepare the two cavity modes in an entangled coherent state~(\ref{entcat}).} 
\label{threecav}   
\end{figure}

In sec.~\ref{sssec:timeJCsec} we presented a method, relying on the STIRAP\index{STIRAP}, for state transfer between cavity modes belonging to different cavities~\cite{mattinson2001adiabatic}. Such `crossed-cavity' setups have been recently realized at ETH~\cite{leonard2017supersolid} (see further sec.~\ref{ssec:mbcQED}). By letting a two-level atom adiabatically traverse the two cavities in the regime where the fields overlap, a many-body STIRAP may take place, and one can transfer an arbitrary state from one cavity into the second cavity, e.g.
\begin{equation}
|\alpha\rangle|0\rangle|g\rangle\xrightarrow[\mathrm{STIRAP}]{}|0\rangle|-\alpha\rangle|g\rangle.
\end{equation}
Let us demonstrate the entangled coherent state preparation by considering the example of fig.~\ref{threecav}. The two cavity fields spatially overlap, but the first cavity is shifted slightly downwards (recall the STIRAP scheme presented in sec.~\ref{sssec:timeJCsec}). This cavity is initialized in a coherent state $|2\alpha\rangle$, while the second is in a vacuum state. To prepare the desired state we introduce an ancilla state $|q\rangle$ of the atom (a trick commonly used for entanglement state preparation/logic gates, as for example the Cirac-Zoller gate~\cite{cirac1995quantum} discussed in sec.~\ref{ssec:qipi}), which is not interacting with the cavity fields. Given that the atom is prepared in $|g\rangle$, as it traverses the cavities it will transfer the coherent state of the first cavity into the second cavity. And by using the ancilla we realize the multimode STIRAP\index{Multimode! STIRAP}~\cite{larson2005cavity} 
\begin{equation}\label{catgen}
|2\alpha,0\rangle\frac{1}{\sqrt{2}}\left(|g\rangle+|q\rangle\right)+\xrightarrow[\mathrm{STIRAP}]{}\frac{1}{\sqrt{2}}\left(|0,-2\alpha\rangle|g\rangle+|2\alpha,0\rangle|q\rangle\right).
\end{equation}
Now, perform a projective atomic measurement\index{Projective measurement} in the $(|g\rangle\pm|q\rangle)/\sqrt{2}$ basis, and displace the coherent states by $\hat D(-\alpha)$ in the first cavity and by $\hat D(-\alpha)$ for the second cavity. We then end up with the entangled coherent state\index{Entangled coherent states}\index{State! entangled coherent}
\begin{equation}\label{entcat}
|\psi_\mathrm{2cat}\rangle=\frac{1}{\sqrt{N}}\left(|-\alpha,-\alpha\rangle\pm|\alpha,\alpha\rangle\right).
\end{equation}

Let us end this subsection by noting that there exist, of course, many other interesting types of entangled states, which have implications in QIP and have been discussed in terms of cavity QED. To mention the most important ones we note: atomic {\it cluster states}~\cite{zou2005schemes,cho2005generation}, N$00$N states\index{N$00$N state}~\cite{nikoghosyan2012generation,merkel2010generation}, atomic Dicke states\index{Dicke! state}\index{State! Dicke}~\cite{hong2002quasideterministic,stockton2004deterministic,haas2014entangled}, multi-atom GHZ states\index{GHZ state}\index{State! GHZ}~\cite{Zhao2021}, multimode entangled coherent states\index{Multimode! state! coherent entangled}~\cite{wang2001multipartite}, and GHZ and W entangled coherent states\index{W state}\index{W state}~\cite{jeong2006greenberger}. We close this subsection out by drawing attention to a recent proposal for generating Gottesman-Kitaev-Preskill (GKP) states in the optical domain, which is applicable to several platforms of cavity QED~\cite{Hastrup2022}. These states are of significance for attaining fault-tolerant optical continuous-variable quantum computing. 

\subsection{State tomography}\label{ssec:cqedtomo}
With the second Ramsey pulse\index{Ramsey! pulse} $R_2$ of fig.~\ref{ramsey}, followed by a projective measurement (typically state ionization -- the atom is ionized if in state $|e\rangle$ but not in state $|g\rangle$) in the bare basis, the atomic state can be measured in any desired basis. By repeated measurements, the atomic state $\hat\rho_\mathrm{at}$ can be determined. Measuring the state $\hat\rho_\mathrm{field}$ is, however, more tricky. Nevertheless, it is of most interest to get full information about the state of the electromagnetic field. For example, non-classical features, like that of a cat state (\ref{catstate}), can be revealed by measuring the negative values of the Wigner function~(\ref{wigner}). The difference between a proper cat formed as a superposition of two coherent states~(\ref{cats}), and that of a statistical mixture~(\ref{catmix}) can be seen in the `interference fringes' of the Wigner function, as shown in fig.~\ref{fig4} (a). Since the Wigner function contains the full information about this state~\cite{moyal1949quantum}, it is equally sufficient to know $W(x,p)$ as it is to know $\hat\rho_\mathrm{field}$. The seminal work by Vogel and Risken~\cite{vogel1989determination} showed how there is a one-to-one mapping between the Wigner function (or any phase space distribution\index{Distribution! phase-space}) and the field quadratures\index{Field! quadrature}~(\ref{quad}). If the quadratures can be measured in an arbitrary direction, {\it i.e.} measuring $\hat X_\theta=\cos(\theta)\hat x-\sin(\theta)\hat p$, the full Wigner function can be reconstructed. The quadrature $\hat X_\theta$ can indeed be measured via {\it homodyne detection}\index{Homodyne detection}~\cite{collett1987quantum,welsch1999ii}.

For cavity QED (and also for trapped ions, see sec.~\ref{ssec:ionpreptom}), Lutterbach and Davidovich~\cite{lutterbach1997method} demonstrated that by a suitable transformation of the field before performing the Ramsey interferometry\index{Ramsey! interferometry}, as described in fig.~\ref{ramsey}, it is possible to extract the Wigner function by only measuring the atomic inversion~(\ref{inv}). We start by noting that the Wigner function can be expressed in the form~\cite{cahill1969density}
\begin{equation}\label{wigexp}
W(\alpha)=2\mathrm{Tr}\left[\hat{D}^{-1}(\alpha)\hat{\rho}\hat{D}(\alpha)e^{-i\pi\hat{n}}\right].
\end{equation}
As a side note, this identity together with the expression~(\ref{q2}) below can be used to derive eq.~(\ref{Qwig}) which connects the $Q$-function with the Wigner function. We can regard (\ref{wigexp}) as originating from the state $\hat{D}^{-1}(\alpha)\hat{\rho}\hat{D}(\alpha)$, which evolves under the action of the Hamiltonian $H=\omega\hat n$ for a time $t=\pi/\omega$. Both these steps are experimentally easy to implement; the displacement of the field $\hat D(\alpha)$ is achieved by driving the cavity with a classical source, and the Hamiltonian is simply the dispersive JC model~(\ref{effham}). Thus, we start by displacing the field and then send an atom through the cavity and let it interact for the correct time. The important contribution of ref.~\cite{lutterbach1997method} was to realize that, given the appropriate initial atomic state, the atomic inversion~(\ref{inv}) becomes
\begin{equation}
\langle\hat\sigma_z\rangle=W(-\alpha)/2,
\end{equation}
and hence by simply repeating the measurements for different displacements $\hat{D}^{-1}(\alpha)\hat{\rho}\hat{D}(\alpha)$ the full Wigner function can be mapped out. This method for extracting the Wigner function was first implemented in the ENS group for detecting negativity of the Wigner function for the $n=1$ Fock state~\cite{nogues2000measurement}, and for measuring the full Wigner function for the vacuum and the $n=1$ Fock state\index{State! Fock}\index{Fock state}~\cite{bertet2002direct}. Some years later the same group extended the study to Fock states up to four photons and also to even and odd cat states with an average photon number $\bar n=3.5$~\cite{deleglise2008reconstruction}. Moreover, also the decay of the cat into a statistical mixture, as visualized in fig.~\ref{catfig}, was experimentally explored in that reference. In other words, the coherence factor could be measured~(\cite{bertet2002direct}). This method has also been used in circuit QED~\ref{sec:cirQED} to determine the Wigner function for non-classical field states~\cite{hofheinz2009synthesizing}. In circuit QED, the group of Schoelkopf used a different, but related method, in order to measure the $Q$-function\index{$Q$-function}\index{Q distribution}\index{Distribution! Q}~(\ref{husimi}). We first note that the $Q$-function can be interpreted as the overlap between the displaced state with vacuum; 
\begin{equation}\label{q2}
Q(\alpha)=|\langle0|\hat D(-\alpha)|\psi\rangle|^2/\pi=P(n=0)/\pi.
\end{equation}
Hence, if the state is first displaced by $\hat D(-\alpha)$, and then the probability of finding the cavity in the vacuum state is measured, one determines the $Q$-function\index{Q distribution}\index{Distribution! Q} describing the state. In order to obtain $P(n=0)$ we recall fig.~\ref{fig2} which demonstrates how the qubit is rotated on the Bloch sphere for different Fock states. By adjusting the Ramsey pulse\index{Ramsey! pulse}, the qubit can then be measured in the desired basis in order to find the desired probability.

\subsection{Quantum information processing}\label{ssec:cqedQI} 
The photon mode may act as a qubit when it is limited to include two Fock states, typically the $|0\rangle$ and $|1\rangle$ states. The simplest two-qubit gate within a cavity QED setting is thereby composed of the atom and the two lowest Fock states of the cavity mode. Among the first proposals for such a gate was the one presented in ref.~\cite{domokos1995simple}. This idea relies on utilizing the dressed states of the JC model and assuming an adiabatic evolution in the sense that the spatial variations of the mode coupling $g(z)$ do not induce transitions among the dressed (adiabatic) states. In sec.~\ref{sssec:qatmo} we will discuss the physics of the JC model for cold atoms that takes quantized motion into account. In the adiabatic regime, {\it i.e.} where the Born-Oppenheimer approximation is valid, the JC energies can be parametrized by the atomic center-of-mass position $z$, $E_{n\pm}(z)$. By applying a $\pi$-pulse to the atom, between the correct dressed states, as the atom traverses the cavity it is possible to realize a CNOT gate in which the atom swaps state given that there is one photon in the cavity, while it remains unaltered if the cavity mode is in the vacuum state.  

The first cavity QED experimental demonstration of a two-qubit gate used a different scheme in order to perform a phase gate~\cite{rauschenbeutel1999coherent}. A two-qubit ($\pi$)phase gate is defined by the action $|i,j\rangle\rightarrow(-1)^{\delta_{i1}\delta_{j1}}|i,j\rangle$, {\it i.e.} the state changes sign provided both qubits are in the state $|1\rangle$~\cite{nielsen2010quantum}. It is straightforward to show that the phase gate, combined with a Hadamard gate realizes a CNOT gate, and thereby the phase gate is universal. To realize the desired phase gate, an atomic ancilla state\index{Ancilla! state}\index{State! ancilla} $|q\rangle$ was introduced. We already saw how ancilla states\index{Ancilla! state} can be employed in eq.~(\ref{catgen}) above for the generation of entangled states. Such an extension of the Hilbert space is often common when devising logic gate operations, with the paradigmatic example set by the Cirac-Zoller gate~\cite{cirac1995quantum} to be discussed in sec.~\ref{ssec:qipi}. We mention here in passing that the close relation of quantum trajectories\index{Quantum! trajectories} to a {\it positive operator valued measurement (POVM)}\index{Positive operator valued measurement (POVM)}, arising by the interaction of the measured system with an ancilla (via CNOT or SWAP gates) and a subsequent projective measurement of the ancilla, is explored in~\cite{BrunPOVM}, where a connection to the formalism of {\it consistent histories}\index{Consistent (coherent) histories} is also attempted -- for the equivalence between quantum trajectories and consistent sets see~\cite{Diosi1995, Brun1997, Brun2000}. For the cavity QED phase gate, the logic states $|0\rangle$ and $|1\rangle$ of the atom consisted of the ancilla $|q\rangle$ and the JC $|g\rangle$ state. The ancilla does not feel the presence of the cavity mode and passes the cavity unaffected. Nevertheless, a Ramsey field couples the $|q\rangle$ and $|g\rangle$ states such that single qubit gates can be implemented. The timing, {\it i.e.} the atomic velocity, is chosen such that the state $|1,g\rangle$ performs a full Rabi cycle meaning that $|1,g\rangle\rightarrow-|1,g\rangle$, while $|0,g\rangle$ again is invariant. Summing up, as the atom passes the cavity, all logic states remain the same except $|1,g\rangle$ which changes sign, whence the phase gate is realized. In~\cite{rauschenbeutel1999coherent}, a range of phase gates not limited to the $\pi$ phase were experimentally demonstrated.

As discussed above, by considering two two-level atoms simultaneously coupled to the photon mode it was possible to engineer an effective atom-atom interaction of the type~(\ref{atat}). This could be employed for entanglement generation, but it may equally well be used for performing two-qubit gates between the two atoms~\cite{zheng2000efficient}. A similar scheme, {\it i.e.} adiabatically eliminating\index{Adiabatic! elimination} the cavity field, was conceived by Imamoglu {\it et al}. in ref.~\cite{imamog1999quantum}. However, these authors had in mind two $\Lambda$ atoms where the excited level was eliminated under the large detuning assumption, and the resulting effective Hamiltonian takes the same form as that of eq.~(\ref{atat}), whence it can be implemented to realize, for example, CNOT gates. 

Another interesting scheme for implementing an atom-atom logic gate relies on the consideration of a driven system, see sec.~\ref{sssec:drivejc}. For a strong atomic drive which is resonant with the pump, $\Delta_\mathrm{p}=\Omega-\omega_\mathrm{p}=0$, we derived an effective Hamiltonian in eq.~(\ref{solanoeff}). The time-evolution operator of this Hamiltonian, in the interaction picture with respect to the drive, can be written as~\cite{sorensen2000entanglement,zheng2002quantum}
\begin{equation}
\hat U(t)=e^{-iA(t)\hat S_x^2}e^{-iB(t)\hat a\hat S_x}e^{-iC(t)\hat a^\dagger\hat S_x},
\end{equation}
where $\hat S_x=\hat\sigma_x^{(1)}+\hat\sigma_x^{(2)}$ is the collective spin operator, and the coefficients
\begin{equation}
A(t)=\frac{g^2}{4\delta}\left[t+\frac{1}{i\delta}\left(e^{-i\delta t}-1\right)\right],\hspace{0.7cm}B(t)=\frac{g}{2i\delta}\left(e^{i\delta t}-1\right),\hspace{0.7cm}C(t)=-\frac{g}{2i\delta}\left(e^{-i\delta t}-1\right),
\end{equation}
and recall that for a resonant drive, the detuning $\delta=\omega-\Omega$ becomes the regular light-atom detuning.  Assume now that the detuning is taken such that $\delta t=2\pi$. Upon returning to the original frame, the evolution operator becomes
\begin{equation}\label{evop2}
\hat U(t)=e^{-\left(\eta\hat S_x+g^2\hat S_x^2/4\delta\right)t}.
\end{equation}
By tuning the parameters such that $g^2t/2\delta=\pi$ and $\eta t=(2k+1/2)\pi$ ($k$ an integer, and recall that $\eta$ is the pump amplitude), the atoms are subject to a phase gate in the $|\pm\rangle$-basis~\cite{zheng2002quantum}. The significance of the effective Hamiltonian~(\ref{solanoeff}) becomes evident by noticing that it forms the backbone for a plethora of theoretical proposals, {\it e.g.} for state teleportation~\cite{jin2005teleportation} and for implementing the Deutsch-Jozsa algorithm~\cite{yang2006simple} or the Grover search algorithm~\cite{deng2005simple}. Another application of the model is found for realizing {\it entanglement concentration}\index{Entanglement! concentration}\cite{bennett1996concentrating}, {\it i.e.} use a set of $N$ partially entangled qubit pairs in order to prepare $M$ ($<N$) maximally entangled Bell pairs~\cite{yang2005entanglement}.

The main advantage of the scheme above lies in the fact that the evolution operator~(\ref{evop2}) is independent of the photonic operators, and thereby does not rely on the actual state of the field. As a consequence, the scheme is resilient to photon losses, and, in particular, the cavity field could be taken as a thermal state~\cite{zheng2003generation}. This observation has been used in many more proposals, as for example for {\it quantum dense coding}~\cite{ye2005scheme}, in which the idea is to utilize entanglement in order to transfer two bits of information with a single qubit, and for teleportation of atomic states~\cite{jin2005teleportation}.

Teleporting\index{Teleportation} atomic states between distant atoms was also considered in~\cite{bose1999proposal}. Their idea relies on initializing the atoms in two spatially separated cavities. Photons escaping the cavities pass through a beam-splitter before being detected, and as such a detection is incapable of determining where the photon came from. Thus, the detection induces the correlation between the two subsystems, and by correctly monitoring the interactions, the atomic state from say atom $A$ can be transmitted to atom $B$.  

Transferring states between two atoms, as in the scheme above, has not been demonstrated experimentally. However, the group of Kimble has realized an adiabatic state transfer back-and-forth between an atom and the photon mode~\cite{boozer2007reversible}. This can be seen as a first step towards QIP with a {\it quantum network}, in which it should be possible to both store and transfer quantum information between different nodes~\cite{cirac1997quantum,kimble2008quantum,reiserer2015cavity}. This is an approach to QIP pursued by the Rempe group where they store information in ions trapped inside resonators, and photons transmit the information. In an early experiment~\cite{wilk2007single}, they used a cavity supporting two modes and started from the state $|0,c\rangle$ where $|c\rangle$ is an ancilla state\index{Ancilla! state}. They then drove a STIRAP transition from this state into an entangled state $|\psi\rangle=(|1_+,g\rangle-|1_-,e\rangle)/\sqrt{2}$, with $1_\pm$ representing an $n=1$ photon state with different circular polarizations. A second STIRAP\index{STIRAP} was applied where the atom was taken back to its initial state $|c\rangle$ by populating the second photon mode as $|\psi\rangle=(|1_+,1_-\rangle-|1_-,1_+\rangle)\sqrt{2}$, {\it i.e.} the two photon modes were prepared in a Bell state\index{Bell! state}. We may regard the second step as transferring the atomic state onto the second photon mode.  Some five years later, the same group expanded on the aforementioned idea by demonstrating how two modes of a quantum network could talk to one another~\cite{ritter2012elementary}. The node in their setup is a Rb atom in an optical resonator, and the two nodes were placed in two laboratories separated by $21$m. Using the same type of Raman schemes as in their first experiment, they could transfer an atomic state from node $A$ to node $B$, and cause maximal entanglement between the two atoms (fidelity of 98$\%$) via the transmitted photon. An architecture for realizing a cavity-mediated quantum gate between any two qubits applied to Rydberg atoms as well as to multiple chains of trapped ions has been recently proposed in~\cite{Ramette2022} to expand the capabilities of Hamiltonian simulations and enable more robust high-dimensional error-correcting schemes.

Finally, coming to the second law of thermodynamics, a theoretical description together with an experimental realization of an autonomous Maxwell's demon\index{Maxwell's demon} where information is employed to transfer heat from a cold qubit to a hot harmonic oscillator (microwave cavity) has recently been reported in~\cite{Santos2020}. In this cavity QED setup, the evolution of the joint qubit-demon-cavity system is unitary such that its entropy remains constant in time. Under this reversible process, the actions performed by the demon can be schematically split into a readout part and a feedback mechanism. For an up-to-date and comprehensive review on quantum information processing with cavities, we refer the reader to~\cite{Meher2022}. 

\subsection{Quantum fluctuations and coherence in the weak-excitation limit}
\label{ssec:quantumglcavQED}

There is a very extensive literature devoted to the topic of quantum fluctuations for many atoms inside a cavity in the small-noise limit\index{Small-noise limit}, mainly during the 1970s and the 1980s when optical bistability\index{Optical! bistability} was widely studied. The large number of two-state atoms, $N$, is commonly taken as the system-size parameter in the construction of a linearized Fokker--Planck equation\index{Fokker--Planck equation!linearized} to account for the fluctuations. A comprehensive review of the field was given by Lugiato~\cite{LugiatoReview}, with a focus on the so-called ``hybrid electro-optical systems'' exhibiting a bistable response. At about the same time, a linearized theory of quantum fluctuations\index{Linearized! theory of quantum fluctuations} for absorptive bistability without adiabatic elimination was developed in~\cite{Carmichael1986nonadiabatic}, demonstrating nonclassical features such as squeezing\index{Squeezing} and photon antibunching\index{Antibunching}. These features are strongest in the regime of weak excitation. In the small-noise limit of many-atom cavity QED, the spectrum of the transmitted light consists of two parts, one of the usual Lorentzian form and another one which is a squared Lorentzian (see also~\cite{Rice1988} for a discussion on the role of squeezing in inducing a {\it spectral hole} when operating in the good-cavity limit).

In this last section on cavity QED, we turn to an alternative to the linearized treatment of fluctuations, applied to the weak-excitation limit in cavity QED. The derived results pertain as well to the single-atom system, $N=1$, as was the case in sec.~\ref{ssec:drjc}. Here, however, the scattering processes destroying the coherence of the quantum evolution are not in principle capable of interrupting the regression of a typical fluctuation. This preserves the purity of the quantum state to lowest order in the drive strength. In that framework, we closely follow the analysis of chap. 16 of~\cite{BookQO2Carmichael} and adopt the notation used therein. Our starting point is the master equation (ME)\index{Master equation! optical bistability}\index{Optical! bistability! Master equation} for optical bistability in the interaction picture\index{Interaction! picture}, involving $N$ two-state atoms coupled to a cavity mode driven by coherent field of amplitude $\overline{\mathcal{E}}_0$, as well as a surrounding bath with zero thermal occupation,
\begin{equation}\label{eq:MEQED}
\begin{aligned}
\frac{d\hat{\tilde{\rho}}}{dt}=\tilde{\mathcal{L}}\hat{\tilde{\rho}}&=-i\frac{1}{2}\omega_A [\hat{J}_z, \tilde{\rho}]-i\omega_C[\hat{a}^{\dagger}\hat{a}, \hat{\tilde{\rho}}] +g[\hat{a}^{\dagger}\hat{J}_{-}-\hat{a} \hat{J}_{+},\hat{\tilde{\rho}}]-i[\overline{\mathcal{E}}_0 \hat{a}^{\dagger}+\overline{\mathcal{E}}_0^{*}\hat{a}, \hat{\tilde{\rho}}] \\
&+\frac{\gamma}{2}\left(\sum_{j=1}^{N}2\hat{\sigma}_{j-}\hat{\tilde{\rho}} \hat{\sigma}_{j+}-\frac{1}{2}\hat{J}_{z}\hat{\tilde{\rho}} - \frac{1}{2}\hat{\tilde{\rho}} \hat{J}_{z} -N\hat{\tilde{\rho}}\right) + \frac{\gamma_p}{2} \left(\sum_{j=1}^{N} \hat{\sigma}_{jz}\hat{\tilde{\rho}} \hat{\sigma}_{jz} - N\hat{\tilde{\rho}}\right)+\kappa(2\hat{a}\hat{\tilde{\rho}} \hat{a}^{\dagger}-\hat{a}^{\dagger}\hat{a}\hat{\tilde{\rho}}-\hat{\tilde{\rho}} \hat{a}^{\dagger}\hat{a}),
\end{aligned} 
\end{equation} 
where $\hat{J}_{\pm}, \hat{J}_z$ are collective atomic operators,
\begin{equation}
  \hat{J}_{\pm} \equiv \sum_{j=1}^{N}\hat{\sigma}_{j\pm}, \quad \quad \hat{J}_z \equiv \sum_{j=1}^{N}\hat{\sigma}_{jz},  
\end{equation}
satisfying the familiar angular momentum commutation relations
\begin{equation}
[\hat{J}_{+}, \hat{J}_{-}]=\hat{J}_z, \quad \quad [\hat{J}_{\pm}, \hat{J}_z]=\mp 2\hat{J}_{\pm}.    
\end{equation}
We limit our treatment now to the regime of weak-excitation. A {\it pure-state factorization}\index{Pure-state factorization}, consistent with a truncation of matrix-element equations based on an expansion in powers of $|\overline{\mathcal{E}}_0|$ follows from the ME~\eqref{eq:MEQED} with $\gamma_p=0$, by making the approximation
\begin{equation}\label{eq:MEfactform}
\tilde{\mathcal{L}} \approx g[\hat{a}^{\dagger}\hat{J}_{-}-\hat{a}\hat{J}_{+},\cdot]-i[\overline{\mathcal{E}}_0 \hat{a}^{\dagger}, \cdot] - \frac{\gamma}{4}[\hat{J}_z + N, \cdot]_{+} - \kappa[\hat{a}^{\dagger}\hat{a}, \cdot]_{+},
\end{equation}
where $[\, , \,]_{+}$ denotes the anti-commutator. We have also dropped the term $\overline{\mathcal{E}}_0^{*} \hat{a}$ as it only contributes at higher order in $|\overline{\mathcal{E}}_0|$. Upon substitution of the factorized form $\hat{\tilde{\rho}}(t) \approx |\tilde{\psi}(t)\rangle \langle \tilde{\psi}(t)|$, we find that the pure state $\ket{\tilde{\psi}}$ satisfies the Schr\"{o}dinger equation 
\begin{equation}\label{eq:SchrEqfact}
\frac{d\ket{\tilde{\psi}}}{dt}=\left[g(\hat{a}^{\dagger}\hat{J}_{-}-\hat{a}\hat{J}_{+})-i\overline{\mathcal{E}}_0 \hat{a}^{\dagger}-\frac{\gamma}{4}(\hat{J}_z+N)-\kappa \hat{a}^{\dagger}\hat{a}\right] \ket{\tilde{\psi}}.
\end{equation}
For weak excitation we can adopt the two-quanta truncation, accounting as well for the possibility that two atoms are simultaneously excited. A factorization can then be proposed with pure-state expansion (the Fock states $\{ \cdot\}_a$ pertain to the cavity field)
\begin{equation}\label{eq:purestexp}
\ket{\tilde{\psi}(t)}=\ket{0}^{(N)}\ket{0}_{a} + \tilde{\alpha}(t)\ket{0}^{(N)}\ket{1}_{a} + \tilde{\beta}(t)\ket{1}^{(N)}\ket{0}_{a}+\tilde{\eta}(t)\ket{0}^{(N)}\ket{2}_{a} + \tilde{\zeta}(t)\ket{1}^{(N)}\ket{1}_{a} + \tilde{\theta}(t)\ket{2}^{(N)}\ket{0}_{a},
\end{equation}
for a basis
\begin{equation}\label{eq:basisfact}
\left\{\ket{0}^{(N)}\ket{0}_{a},
\ket{0}^{(N)}\ket{1}_{a},
\ket{1}^{(N)}\ket{0}_{a},
\ket{0}^{(N)}\ket{2}_{a},
\ket{1}^{(N)}\ket{1}_{a},
\ket{2}^{(N)}\ket{0}_{a}\right\}
\end{equation}
where the collective atomic states ($\ket{1}_j$ and $\ket{2}_j$ are the ground and excited states, respectively, for the atom $j$)
\begin{subequations}\label{eq:atomicstates}
\begin{align}
&\ket{0}^{(N)} \equiv \prod_{j=1}^{N} \ket{1}_j=\ket{1,N/2, -N/2}, \\
&\ket{1}^{(N)} \equiv \frac{1}{\sqrt{N}} J_{+}\ket{0}^{(N)}=\ket{1,N/2, -N/2+1}, \\
&\ket{2}^{(N)} \equiv \sqrt{\frac{2}{N-1}} J_{+}\ket{1}^{(N)}=\ket{1,N/2, -N/2+2},
\end{align}
\end{subequations}
with $\ket{\lambda, J, M}$ ($M=-J, -J+1 \ldots, J$) a Dicke state (of degeneracy $\lambda$)\index{Dicke! state}\index{State! Dicke}. Substituting in eq.~\eqref{eq:SchrEqfact}, we obtain the following equations of motion for the state amplitudes:
\begin{subequations}\label{eq:eomstateampl}
\begin{align}
&\dot{\tilde{\alpha}}=-\kappa \tilde{\alpha} + \sqrt{N} g \tilde{\beta} -i \overline{\mathcal{E}}_0, \label{eq:eomstateamplA}\\
&\dot{\tilde{\beta}}=-\frac{\gamma}{2}\tilde{\beta} - \sqrt{N}g \tilde{\alpha},\label{eq:eomstateamplB} \\
&\dot{\tilde{\eta}}=-2\kappa\tilde{\eta} + \sqrt{2} \sqrt{N} g \tilde{\zeta} -i\sqrt{2}\overline{\mathcal{E}}_0 \tilde{\alpha},\\
&\dot{\tilde{\zeta}}=-(\kappa+\gamma/2)\tilde{\zeta} -\sqrt{2}\sqrt{N} g \tilde{\eta} + \sqrt{2}\sqrt{N-1} g \tilde{\theta} - i \overline{\mathcal{E}}_0 \tilde{\beta}, \\
&\dot{\tilde{\theta}}=-\gamma \tilde{\theta} - \sqrt{2} \sqrt{N-1} g \tilde{\zeta},
\end{align}
\end{subequations}
first reported in~\cite{Carmichael1991}.  

The pure-state factorization can be readily applied to the calculation of the intensity correlation function of the forwards-scattered field\index{Second-order correlation function}. We begin with the general expression
\begin{equation}\label{eq:g2general}
 g^{(2)}_{\rightarrow}(\tau)=\frac{\braket{(\hat{\tilde{a}}^{\dagger}\hat{\tilde{a}})(\tau)}_{\hat{\tilde{\rho}}(0)=\hat{\tilde{\rho}}^{\prime}_{\rm ss}}}{\braket{\hat{\tilde{a}}^{\dagger}\hat{\tilde{a}}}_{\rm ss}},
\end{equation}
with
\begin{equation}
\hat{\tilde{\rho}}_{\rm ss}^{\prime} \equiv \frac{\hat{\tilde{a}} \hat{\tilde{\rho}}_{\rm ss} \hat{\tilde{a}}^{\dagger}}{{\rm Tr}(\hat{\tilde{a}} \hat{\tilde{\rho}}_{\rm ss} \hat{\tilde{a}}^{\dagger})},
\end{equation}
the state resulting after the emission of a first photon from the cavity. Using the pure-state factorization $\hat{\tilde{\rho}}(t) \approx |\tilde{\psi}(t)\rangle \langle \tilde{\psi}(t)|$, we can write
\begin{equation}\label{eq:g2fact}
g^{(2)}_{\rightarrow}(\tau)=\frac{\braket{\tilde{\psi}(\tau)|\hat{\tilde{a}}^{\dagger}\hat{\tilde{a}}|\tilde{\psi}(\tau)}}{\braket{\tilde{\psi}_{\rm ss}|\hat{\tilde{a}}^{\dagger}\hat{\tilde{a}}|\tilde{\psi}_{\rm ss}}},
\end{equation}
with
\begin{equation}\label{eq:psionephoton}
 \ket{\tilde{\psi}(\tau)}=\ket{0}^{(N)}\ket{0}_a + \tilde{\alpha}(\tau) \ket{0}^{(N)}\ket{1}_a + \tilde{\beta}(\tau)\ket{1}^{(N)}\ket{0}_a.
\end{equation}
The initial conditions $\tilde{\alpha}(0)$ and $\tilde{\beta}(0)$ are defined by 
\begin{equation}\label{eq:incondpsi}
  \ket{\tilde{\psi}(0)} \equiv \frac{\hat{\tilde{a}}\ket{\tilde{\psi}_{\rm ss}}}{\sqrt{\braket{\tilde{\psi}_{\rm ss}|\hat{\tilde{a}}^{\dagger}\hat{\tilde{a}}|\tilde{\psi}_{\rm ss}}}}\, e^{-i\,\text{arg}\left(_{a}\bra{0}^{(N)}\braket{0|\tilde{a}|\tilde{\psi}_{\rm ss}}\right)},
\end{equation}
where $\ket{\tilde{\psi}_{\rm ss}}$ is the stationary state vector obtained from the steady-steady solutions of the eqs.~\eqref{eq:eomstateampl}. Combining eqs. \eqref{eq:psionephoton}, \eqref{eq:g2fact} and the solutions to \eqref{eq:eomstateampl}, yields the simple expression
\begin{equation}
 g^{(2)}_{\rightarrow}(\tau)=\frac{|\tilde{\alpha}(\tau)|^2}{|\tilde{\alpha}_{\rm ss}|^2},
\end{equation}
where $\tilde{\alpha}(\tau)$ is the conditional one-photon amplitude satisfying the coupled equations \eqref{eq:eomstateamplA} and \eqref{eq:eomstateamplB}, with initial conditions 
\begin{equation}
 \tilde{\alpha}(0)=\frac{\sqrt{2}\tilde{\eta}_{\rm ss}}{\tilde{\alpha}_{\rm ss}}, \quad \quad \tilde{\beta}(0)=\frac{\tilde{\zeta}_{\rm ss}}{\tilde{\alpha}_{\rm ss}}.
\end{equation}
Following through with the remaining calculations, we finally arrive at the second-order correlation function of forwards scattering in the weak excitation limit, which is a perfect square -- the square of the one-photon amplitude conditioned upon the detection of a forwards-scattered photon at $\tau=0$:
\begin{equation}\label{eq:g2finalQEDequal}
{g_{\rightarrow}^{(2)}(\tau)=\Bigg\{1-2C_1 \frac{\xi}{1+\xi}\frac{2C}{1+2C-2C_1 \xi/(\xi+1)} e^{-\frac{1}{2}(\kappa+\gamma/2)\tau}\left[\cos(G\tau) + \frac{1}{2}\frac{\kappa+\gamma/2}{G} \sin(G\tau)\right]\Bigg\}^2},
\end{equation}
where $\xi=2\kappa/\gamma$, $C=Ng^2/(\kappa\gamma)$ is the cooperativity\index{Cooperativity}, $C_1=C/N$ and $G=\sqrt{Ng^2-(1/4)(\kappa-\gamma/2)^2}$. For a single atom ($N=1$) we can calculate in that scheme the intensity correlation function of side-scattered light, as
\begin{equation}\label{eq:g2side}
  g^{(2)}_{\rm side}(\tau)=\left\{1-e^{-\frac{1}{2}(\kappa+\gamma/2)\tau} 
\left[\cos(g^{\prime}\tau) + \frac{\kappa-\gamma^{\prime}/2}{\kappa+\gamma/2}\frac{\frac{1}{2}(\kappa + \gamma/2)}{g^{\prime}} \sin(g^{\prime}\tau)\right]\right\}^2,
\end{equation}
where $\gamma^{\prime}=\gamma(1+2C_1)$ is the {\it cavity-enhanced emission rate}\index{Cavity!-enhanced emission rate} and $g^{\prime}=G(N=1)$. Finally, the expression given by eq. \eqref{eq:g2finalQEDequal} can be shown to violate the Schwartz inequality\index{Schwartz inequality}
\begin{equation}\label{eq:Schwartzineq}
|g^{(2)}(\tau)-1| \leq g^{(2)}(0)-1 \geq 0,
 \end{equation}
for a single atom, $\xi=1$ and $g/\kappa=1.85$ (see e.g., fig. 16.1 (b) of~\cite{BookQO2Carmichael}). A violation of the upper bound of \eqref{eq:Schwartzineq} has also been observed for many atoms in the experiment of Mielke and coworkers~\cite{Mielke1998}. Photon antibunching advocates for the discrete nature of light (particles), contrasted with amplitude squeezing which speaks of the continuous (waves). The tension between particles and waves was demonstrated by the experiments of Foster and coworkers~\cite{Foster2000A,Foster2000B}, which combine the measurement strategies used to observe nonclassical behaviors of light~\cite{CarmichaelTalk2000}. Incidentally, the report~\cite{Foster2000A} coincided with the centenary of the first statement in public of the law for the spectrum of blackbody radiation\index{Blackbody radiation} by Planck~\cite{Planck1900}.

When spatial effects are taken into account, we must use a direct product state basis since the atoms are no longer indistinguishable. The resulting system of equations is not closed -- instead, we encounter a dependence on an infinite hierarchy of collective variables. The final result in the weak-excitation limit reads
\begin{equation}\label{eq:unequalcouplg2}
g_{\rightarrow}^{(2)}(\tau)=\Bigg\{1- \frac{[1+\xi(1+C)]S-2C}{1+(1+\xi/2)S} e^{-\frac{1}{2}(\kappa+\gamma/2)\tau}\left[\cos(G\tau) + \frac{1}{2}\frac{\kappa+\gamma/2}{G} \sin(G\tau)\right]\Bigg\}^2,
\end{equation}
with
\begin{equation}
 S \equiv \sum_{j=1}^{N} \frac{2C_{1j}}{1+\xi(1+C)-2C_{1j}\xi},
\end{equation}
where
\begin{equation}
 2C_{1j} \equiv 2 \frac{g^2 (\boldsymbol{r}_j)}{\gamma\kappa}
\end{equation}
is the spontaneous emission\index{Spontaneous! emission! enhancement} enhancement factor for an atom at the position $\boldsymbol{r}_j$. In the expression of eq.~\eqref{eq:unequalcouplg2}, which has quantitative differences with respect to its equal-coupling strengths counterpart--eq~\eqref{eq:g2finalQEDequal}, the many-atom Rabi frequency\index{Rabi! frequency!many-atom} is now defined as
\begin{equation}
G \equiv \sqrt{N_{\rm eff}g_{\rm max}^2 -\frac{1}{4}(\kappa-\gamma/2)^2},
\end{equation}
where 
\begin{equation}
2C \equiv 2 \frac{\sum_{j=1}^{N} g^2(\boldsymbol{r}_j)}{\gamma\kappa}=2\frac{N_{\rm eff} g_{\rm max}^2}{\gamma\kappa}.
\end{equation}
Rempe and collaborators derived eq.~\eqref{eq:unequalcouplg2} to attain a better fit of the theory to experimental results, considering as well an average over an ensemble of atomic configurations~\cite{rempe1991optical}. These results give us a flavour of a nonperturbative treatment in QED. In the next section, we will concentrate on the attainment of strong-coupling conditions in scalable setups allowing the study and control of light-matter interaction at the quantum level in unprecedented detail. We will also revisit the theme of second-order coherence in the weak-excitation limit in sec.~\ref{ssec:plasmons}, when dealing with the interaction between a surface plasmon and a two-state atom. 

\subsection{The wave-particle correlator: extending the Hanbury Brown and Twiss technique}
\label{subsec:WPCorrelator}

\begin{figure}
 \includegraphics[width=0.62\textwidth]{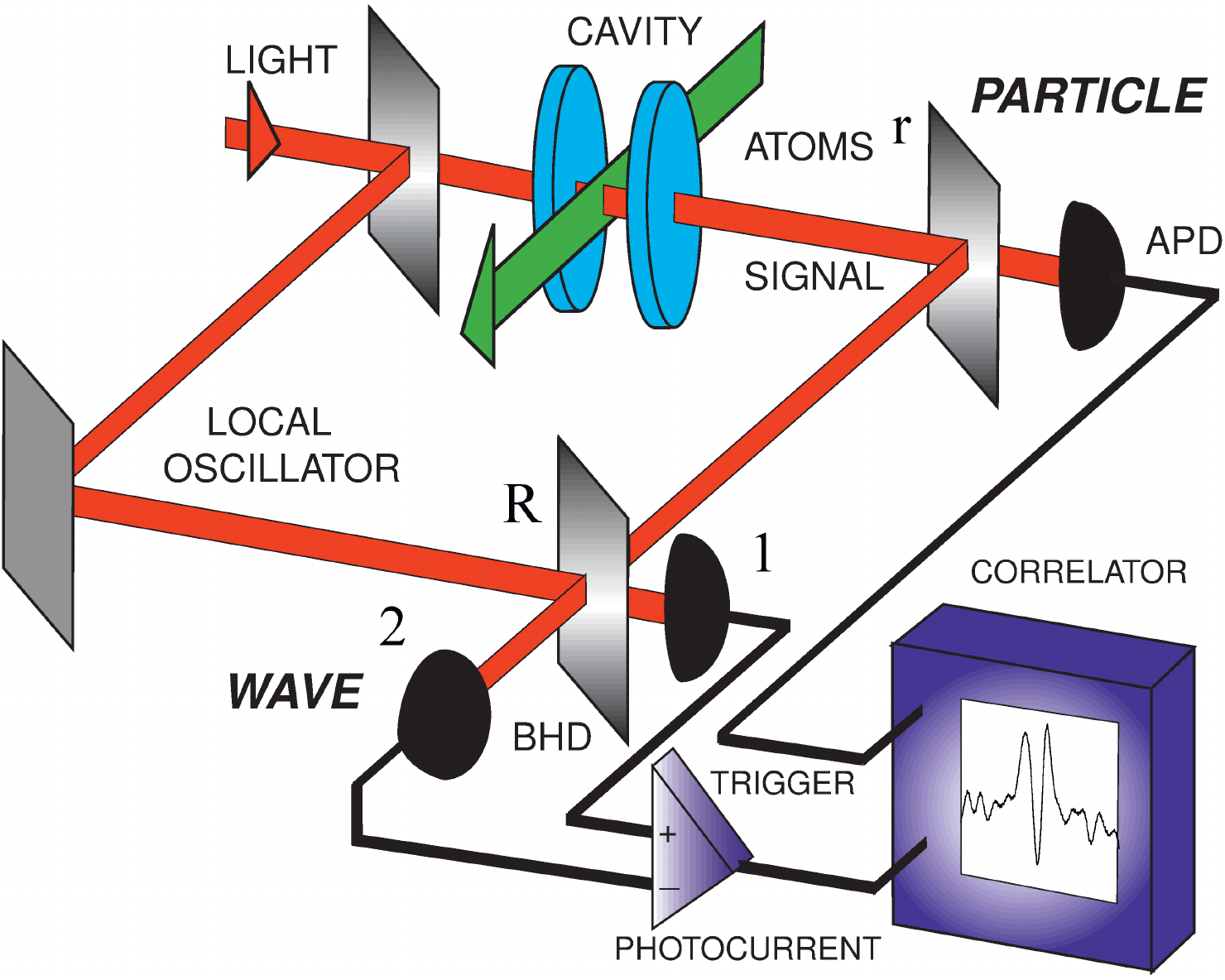}
 \includegraphics[width=0.37\textwidth]{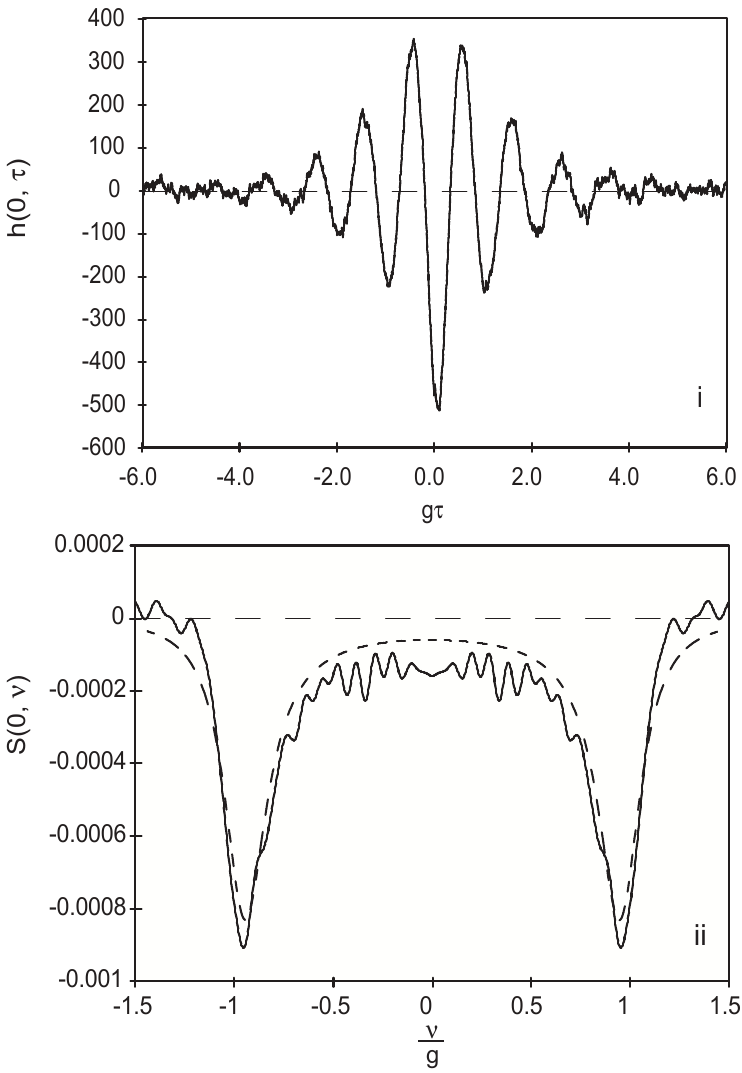}
  \caption{{\bf (left)}{\it Experimental realization of the wave-particle correlator}\index{Wave-particle correlator}. Light of wavelength $\lambda=780\,$nm emanating from a Ti:sapphire laser enters a Mach-Zehnder interferometer\index{Mach-Zehnder interferometer}, driving a cavity QED system in one arm and providing a local oscillator (LO) for the balanced homodyne detector (BHD) in the other arm. The BHD has strong common mode suppression allowing a phase-sensitive measurement of the field quadrature. A fraction of the signal leaving the cavity generates photodetections at the avalanche photodiode (APD)\index{Avalanche photodiode (APD)}, which trigger a digital oscilloscope to record in coincidence the BHD photocurrent. A lock cycle maintains the cavity on resonance, chopping between a higher power auxiliary lock beam and the weak drive. A beam splitter directs a fraction $r=0.15$ of the signal to the particle detector, represented in the figure by a single APD\index{Avalanche photodiode (APD)}. The experiment used a pair of APDs behind a 50/50 beam splitter. The photon detections at the APDs which trigger the digital oscilloscope to record the BHD photocurrent can alternatively be used to obtain the intensity correlation function. The remaining part $1-r=0.85$ of the signal is directed to the wave detector, provided by the BHD, where it is mixed with the LO--whose relative phase is controlled with a piezoelectric transducer--and detected with fast photodiodes. The photocurrent is ac amplified by $65\,$dB, $70\,$MHz low-pass filtered, and sampled and averaged with a fast digital oscilloscope at $2$Gs/s. Source: Fig. 1 of~\cite{Foster2000}. {\bf (right)} (i.) Wave-particle correlation $h_{0^{\circ}}(\tau)$ for very
low intensity excitation calculated from the quantum trajectory
implementation of the conditioned homodyne photocurrent\index{Homodyne detection! photocurrent}. The
following parameters were used:
$(g,\kappa,\gamma,\Gamma)$/$(2/\pi)$ = $(38.0, 8.7, 3.0, 100)$
MHz, $X=2.99\times10^{-4}$, $r=0.5$ and $N_{\rm starts}$ =
$55000$. (ii.) Spectrum of squeezing calculated from the cosine
Fourier transform of $h_{0^{\circ}}(\tau)$. The dashed line is the
spectrum of squeezing calculated directly from the master equation\index{Master equation} with the quantum regression formula\index{Quantum! regression formula}. Source: Fig.3 of~\cite{Reiner2001}. Reproduced with permission from the APS and the authors.}
  \label{fig:WPcorrfig}
\end{figure}

We will now focus on an approach which detects the the fluctuations of a nonclassical electromagnetic field by correlating its intensity and amplitude. The technique introduced by Hanbury-Brown and Twiss\index{Hanbury Brown and Twiss! technique/interferometer} of intensity cross-correlating light marks the beginning of the systematic study of its quantum fluctuations~\cite{Brown1956, BrownHTwissI}. The scheme made a conditional measurement\index{Conditional measurement} by collecting data on the signal of a photon count that identifies times when a fluctuation is in progress. Carmichael and coworkers proposed an extension by developing a new way to measure the squeezing spectrum through a third-order correlation function of the field. The proposed setup correlates the detection of a photon with the record of the output of the photocurrent from a balanced homodyne detector\index{Homodyne detection! balanced}~\cite{CarmichaelFosterChapter}. This method is a variant of the intensity correlation measurements that have shown non-classical features of the electromagnetic field. Figure~\ref{fig:WPcorrfig}(left) shows the apparatus that measures the correlation between the amplitude and the intensity of the electromagnetic field. The particle (photon), representing a fraction $r$ of the signal, produces a trigger `click' in an avalanche photodiode (APD)\index{Avalanche photodiode (APD)} and conditionally the wave (electromagnetic field amplitude) gets recorded in the photocurrent output of a balanced homodyne detector (BHD) to which the remaining fraction $1-r$ is directed. 

The observations of Ref.~\cite{Foster2000} consider a conditioned field evolution described by a third order correlation function of an electromagnetic field mode $\hat b$ (signal) that has a non-zero, steady state\index{Steady state} average $\langle \hat{b} \rangle = \lambda$. The measured correlation is (with $:\,\,:$ denoting normal ordering) 
\begin{equation} \label{eq:prop}
h_\theta(\tau)=\frac{\langle:\!(\hat b^\dagger\hat b)(0)\hat
Q_\theta(\tau)\!:\rangle}{\sqrt{\eta}\lambda\langle\hat
b^\dagger\hat b\rangle},
\end{equation}
where $\hat Q_\theta\equiv(\hat b {\rm{exp}}(-i\theta)+\hat b^\dagger {\rm{exp}}(i\theta))/2$ is the quadrature amplitude  selected by the local oscillator phase $\theta$ and $\eta$ is the coupling efficiency into the BHD. Equation~\eqref{eq:prop} can be generalized to any source~\cite{GiantViolations2000}.

When the source field is small and nonclassical, its fluctuations--a manifestation of the uncertainty principle--dominate over its steady-state amplitude. It is precisely these fluctuations we are interested in, and we separate them from the mean field writing
\begin{equation}
\hat b=\lambda {\rm{exp}}(i\theta)+\Delta\hat b.
\end{equation}
The correlation function can be rewritten in terms of the field
fluctuations. In the limit of Gaussian fluctuations\index{Gaussian! fluctuations} we may neglect
third-order moments of the field fluctuations and the correlation
function is~\cite{Foster2000}
\begin{equation}\label{eq:cfun}
h_{0^{\circ}}(\tau)=1 + 2\frac{\langle:\!\Delta\hat
Q_{0^{\circ}}(0) \Delta\hat
Q_{0^{\circ}}(\tau)\!:\rangle}{\lambda^{2}+ \langle\Delta\hat
b^\dagger\Delta\hat b\rangle}+\xi(\tau), 
\end{equation}
where $\Delta\hat Q_{0^{\circ}}\equiv(\Delta\hat b +\Delta\hat
b^\dagger)/2$ and $\xi(\tau)$ is the residual shot noise\index{Shot noise} with the
following correlation function\index{Shot noise!correlation function}
\begin{equation}\label{eq:shotcorr}
\overline{\xi(0)\xi(\tau)}=\frac{\Gamma}{16\eta N_s|\langle\hat
b\rangle|^2}{\rm{exp}}(-\Gamma\tau),
\end{equation}
where $\Gamma$ is the BHD bandwidth. It is apparent from eq.~\eqref{eq:shotcorr} that once the conditioned field is normalized
for a fixed $|\langle\hat{b}\rangle|^2$ then the shot noise,
$\xi(t)$, is determined by the detection bandwidth and the number
of samples averaged. Therefore one must average over many start
``clicks" to obtain a good signal to noise ratio.

The spectrum of squeezing\index{Squeezing! spectrum} is the cosine transform of the normalized correlation function,
\begin{equation}\label{eq:spectrum}
S(\theta,\nu)=4F\int_0^\infty
d\tau\cos(2\pi\nu\tau)[\overline{h_\theta(\tau)}-1],
\end{equation}
where $F=2\kappa\langle\hat b^\dagger\hat b\rangle$ is the photon
flux into the correlator and $\overline{h_\theta(\tau)}$ is the
average of $h_\theta(\tau)$ with respect to $\xi(\tau)$. This
average is independent of the fraction of light, $1-r$, sent to
the BHD.

The wave-particle correlator so defined has two practical advantages over a standard squeezing measurement. First, the fluctuations in time
reveal the entire spectrum of squeezing up to a frequency set by
the detector bandwidth. The second is that the BHD efficiency only
enters through the residual shot noise in eq.~\eqref{eq:cfun}.
This shot noise\index{Shot noise} can be made to vanish by averaging over many trigger starts, which implies that the spectrum of squeezing is not degraded by imperfect detector efficiencies and propagation errors which beset the standard squeezing measurement. 

Here we revisit our discussion in sec.~\ref{sssec:JCzerodim}, where we saw how spontaneous dressed-state polarization\index{Spontaneous! dressed-state polarization} in the open JC model driven on resonance can be treated by expanding the solution of the master equation\index{Master equation} as a generalized sum over \textit{quantum trajectories}\index{Quantum! trajectories}, a term coined in 1991 by Alsing and Carmichael in~\cite{SponDressState} to describe a formalism ''related to the theory of continuous quantum measurement"~\cite{CarmichaelSingh1989}. We are only interested in those unravelings which correspond directly to an experimental measurement. Such a measurement probes the system by opening channels between the system and the environment. Excitations into these channels are represented by the following super-operator:
\begin{eqnarray}\label{eq:super}
{\mathcal
S}_i\hat\rho=\hat{k}(\hat{s_i},\hat{s_i}^{\dag})\hat\rho\hat{k}^{\dag}
(\hat{s_i},\hat{s_i}^{\dag}), 
\end{eqnarray}
where ${\hat{k}}$ is a function of the creation and annihilation
operators, ${\hat{s}_i^{\dag}}$ and ${\hat{s}_i}$, for the
$i^{\rm th}$ system operator coupled to the environment. The
probability for the system to decay through one of these channels
depends on the occupation of that particular mode,
\begin{equation}\label{eq:collapseprob}
P_i={\rm Tr}({\mathcal S}_{\it i}\hat\rho)\,dt.
\end{equation}
Quantum trajectories also provide a picture of what is happening
to the system between measurements. Equation~\eqref{eq:super}
describes how measurements are performed on the system. To see
what happens between measurements, subtract eq.~\eqref{eq:super} from
the RHS of the Lindblad master equation\index{Master equation} governing the unconditional dynamical evolution, $\mathcal{L}\rho$:
\begin{equation}\label{eq:prop}
({\mathcal L}-{\sum_i{\mathcal
S}_i})\hat\rho=\frac{1}{i\hbar}(\hat{H}_{\rm sys}\hat\rho -
\hat\rho\hat{H}_{\rm sys}^{\dagger}) 
\end{equation}

If we assume that the system state is initially pure, then
eqs.~\eqref{eq:super} and~\eqref{eq:prop} guarantee that the conditional
density operator can be written in the factorized form
$\rho=|\psi\rangle\langle\psi|$. Therefore eq.~\eqref{eq:prop}
implies that a propagator\index{Propagator} exists for the system wavefunction
between measurements. This propagator is a modified, non-unitary,
Hamiltonian with a real and imaginary part, $\hat{H}_{\rm
sys}=\hat{H}_R+i\hat{H}_I$, where $\hat{H}_R$ and $\hat{H}_I$ are Hermitian. Below we present the quantum trajectory algorithm\index{Monte Carlo! algorithm} which propagates the system wavefunction forward in time.

\boxed{1.}  Choose an initial state for the system.

\boxed{2.} Calculate the probability for the system to decay through each of the output channels defined from eq.~\eqref{eq:collapseprob}.
\begin{equation}\label{eq:expprob}
P_i=\langle\hat{k}^{\dagger}(\hat{s}_i)\hat{k}(\hat{s}_i)\rangle
dt 
\end{equation}

\boxed{3.} Associate a uniformly distributed random number between $0$ and $0$ with each loss channel in step 2. If the emission probability along the $i^{\rm th}$ channel is greater than its corresponding random number then collapse the system wavefunction,
\begin{equation}\label{eq:eqcollapse}
|\psi(t+dt)\rangle=\hat{k}(\hat{s}_i)|\psi(t)\rangle
\end{equation}
In the unlikely event of multiple collapses, choose only one
collapse via another random number.

\boxed{4.} If the probability for loss through each channel is less than
the corresponding random numbers, then propagate the system wavefunction with the effective Hamiltonian from eq.~\eqref{eq:prop},
\begin{equation}\label{eq:eqprop}
|\psi(t+dt)\rangle=\left(1-\frac{\hat{H}_{\rm sys}
}{i\hbar}dt\right)|\psi(t)\rangle 
\end{equation}

\boxed{5.}  Normalize the system wavefunction and repeat from step 2.

\noindent These steps form the basis for the generation of quantum trajectories\index{Quantum! trajectories}. A master equation\index{Master equation} for the reduced density operator $\hat\rho$ for the case of $N$-atoms identically coupled to a single cavity mode reads [see also eq.~\eqref{eq:MEQED} in sec.~\ref{ssec:quantumglcavQED}]
\begin{equation}\label{eq:rhodef}
\dot{\hat{\rho}}\equiv{\mathcal{L}}{\hat{\rho}}={\mathcal
E}[\hat{a}^\dagger -\hat{a},\hat{\rho}]+ g[\hat{a}^\dagger \hat{J}_- - \hat{a} \hat{J}_+,\hat{\rho}] + \kappa(2 \hat{a} \hat{\rho}
\hat{a}^\dagger - \hat{a}^\dagger \hat{a} \hat{\rho} - \hat{\rho}
                                 \hat{a}^\dagger \hat{a})+ \gamma/2 \sum_{j=1}^{N}
                 (2 \hat{\sigma}^j_- \hat{\rho} \hat{\sigma}^j_+ -
                 \hat{\sigma}^j_+  \hat{\sigma}^j_- \hat{\rho}  -
                  \hat{\rho} \hat{\sigma}^j_+ \hat{\sigma}^j_-),
\end{equation}
where $\hat{J}_{\pm} = \sum_j \sigma^j_{\pm}$ are the collective
raising and lowering operators for the atoms. In order to unravel eq.~\eqref{eq:rhodef} with the wave-particle correlator we need first to add the BHD to the system. The BHD consists of a local oscillator which is modeled by a driven, single mode coherent field.  We write the master equation\index{Master equation} for the system plus local oscillator mode as~\cite{Reiner2001}
\begin{equation}\label{eq:rhodeflo}
\begin{aligned}
\dot{\hat{\rho}}=&{\mathcal
E}[\hat{a}^\dagger -\hat{a},\hat{\rho}]+
                     g[\hat{a}^\dagger \hat{J}_- - \hat{a} \hat{J}_+,\hat{\rho}] + \kappa(2 \hat{a} \hat{\rho}
\hat{a}^\dagger - \hat{a}^\dagger \hat{a} \hat{\rho} - \hat{\rho}
                                 \hat{a}^\dagger \hat{a}) + \kappa_{LO}(2 \hat{c} \hat{\rho}
\hat{c}^\dagger - \hat{c}^\dagger \hat{c} \hat{\rho} - \hat{\rho}
\hat{c}^\dagger \hat{c}) + \kappa_{LO}\chi[\hat{c}^\dagger -
\hat{c},\hat{\rho}]\\
&+ \gamma/2
\sum_{j=1}^N
                 (2 \hat{\sigma}^j_- \hat{\rho} \hat{\sigma}^j_+ -
                 \hat{\sigma}^j_+  \hat{\sigma}^j_- \hat{\rho}  -
                  \hat{\rho} \hat{\sigma}^j_+ \hat{\sigma}^j_-),
                  \end{aligned}
\end{equation}
where $\hat{c}^\dagger$ and $\hat{c}$ represent the raising and
lowering operators for the local oscillator mode. The coherent
field occupies a mode whose strength we denote by $\chi$. Under a
BHD scheme the beam splitter combines the signal and local
oscillator fields equally $(R=1/2)$, as in fig.~\ref{fig:WPcorrfig} (left), to give the following total measured fields at photodetectors 1 and 2:
\begin{equation}\label{eq:hom1}
\hat{\mathcal{E}}_{\rm BHD^{1,2}}=\pm i\sqrt{\kappa_{\rm
LO}}\hat{c}+\sqrt{\kappa(1-r)}\hat{a} 
\end{equation}
while the measured field at the photon counting detector gives
\begin{equation}\label{eq:count1}
\hat{\mathcal{E}}_{\rm count}=\sqrt{2\kappa r}\hat{a}.
\end{equation}

We assume that the signal mode does not affect the local
oscillator mode. This implies that the density matrix separates
into a piece which corresponds to the cavity QED system and a
piece which corresponds to the local oscillator,
\begin{equation} \label{eq:separate}
\hat\rho=\hat\rho_{s}|\chi\rangle\langle\chi|.
\end{equation}
We then define the local oscillator flux in terms of the strength of
the local oscillator mode and its cavity decay rate:
\begin{equation}\label{eq:fdef}
f=\kappa_{\rm LO}|\chi|^2. 
\end{equation}
From the action of the various super-operators we construct the master equation\index{Master equation} terms which correspond to eqs.~\eqref{eq:hom1}--\eqref{eq:count1} and spontaneous emission\index{Spontaneous! emission} events:
\begin{equation}\label{eq:homm}
{\mathcal S}_{\rm BHD^{1,2}}\hat\rho_{s} =
(\pm\sqrt{f}e^{i\theta}+\sqrt{2\kappa(1-r)}\hat{a})
\times\hat{\rho}_{s}(\pm\sqrt{f}e^{-i\theta}+\sqrt{2\kappa(1-r)}\hat{a}^{\dagger}),
\end{equation}
\begin{equation} \label{eq:countm}
{\mathcal S}_{\rm count}\hat{\rho}_{s} = 2\kappa
r\hat{a}\hat{\rho}_{s}\hat{a}^{\dagger},
\end{equation}
\begin{equation} \label{eq:spont}
{\mathcal S}_{\rm spont}\hat{\rho}_{s} =
\gamma\hat\sigma_{-}\hat\rho_{s}\hat\sigma_{+}.
\end{equation}
After subtracting eqs.~\eqref{eq:homm}-\eqref{eq:spont} from
eq.~\eqref{eq:rhodeflo}, we arrive at an expression which corresponds
to eq.~\eqref{eq:prop}:
\begin{equation}\label{eq:propm}
\begin{aligned}
({\mathcal{L}}-{\mathcal{S}}_{\rm BHD^{1,2}}-{\mathcal{S}}_{\rm
count}-{\mathcal{S}_{\rm spont}})\hat{\rho}_{s} = &{\mathcal
E}[\hat{a}^\dagger -\hat{a},\hat{\rho}_{s}] +
g[\hat{a}^\dagger \hat{S}_- - \hat{a} \hat{S}_+,\hat{\rho}_{s}] -
\kappa(\hat{a}^\dagger \hat{a} \hat{\rho}_{s} +
\hat{\rho}_{s}\hat{a}^\dagger \hat{a})\\
&-\gamma/2 \sum_{j=1}^N
(\hat{\sigma}^j_+ \hat{\sigma}^j_- \hat{\rho}_{s} + \hat{\rho}_{s}
\hat{\sigma}^j_+ \hat{\sigma}^j_-) - f\hat\rho_{s}.
\end{aligned}
 \end{equation}
Equation~\eqref{eq:propm} provides us with a modified Hamiltonian which
propagates the system's wavefunction between measurements, while eqs. \eqref{eq:homm} and~\eqref{eq:countm} govern how ``clicks" at the BHD and APD disrupt that evolution. There are two types of detections, not including spontaneous emissions\index{Spontaneous! emission}, which occur under the conditional BHD measurement. The first comes from eq.~\eqref{eq:countm} which corresponds to detections at the APD\index{Avalanche photodiode (APD)}. These collapses occur with a probability of $2\kappa r\langle\hat{a}^{\dagger}\hat{a}\rangle dt$ and they produce a significant change in the wavefunction. The second type of collapse corresponds to the ``clicks" at the  BHD. The strength of the local oscillator tells us that there are many BHD ``clicks" within a time interval set by $\kappa^{-1}$. Equation~\eqref{eq:eqcollapse} shows that the effect of each of these ``clicks" on the system wave function is almost negligible. This implies that a straightforward trajectory simulation with the Monte-Carlo algorithm described above in eqs.~\eqref{eq:expprob}-\eqref{eq:eqprop}
will not provide an efficient method for modeling the
wave-particle correlator. Instead we may associate a photocurrent
with the detections registered by the BHD. The superposition of
the cavity field with the local oscillator field in eq.~\eqref{eq:hom1} implies that the BHD photocurrent and the system wavefunction evolution are not independent. With the principles outlined in secs. 8.4, 9.2, and 9.4 of~\cite{OpenSystemsBook1993} we combine eqs.~\eqref{eq:homm}-\eqref{eq:propm} to simultaneously calculate this photocurrent from the BHD setup and then evolve a corresponding stochastic Schr\"{o}dinger equation forward in time. This analysis leads to the following expressions for the difference in photocurrents between photodetectors 1 and 2 in fig.~\ref{fig:WPcorrfig}, and wavefunction propagation between photodetections at the APD\index{Avalanche photodiode (APD)} and spontaneous emissions\index{Spontaneous! emission} out the sides of the cavity:
\begin{equation}\label{eq:current}
di=-\Gamma(i\,dt -\sqrt{8\kappa(1-r)}\langle\hat{a}_\theta\rangle_c dt + dW_t)
\end{equation}
\begin{equation}\label{eq:stochschr}
d|\bar{\psi}\rangle_c = \left[\frac{H_{\rm
sys}}{i\hbar}dt+\sqrt{2\kappa(1-r)}\hat{a}{\rm{exp}}(-i\theta)(\sqrt{8\kappa(1-r)}\langle\hat{a}_\theta\rangle_c dt +
dW_t)\right]|\bar{\psi}\rangle_c, 
\end{equation}
where $\Gamma$ is the BHD bandwidth and $dW_t$ is the same Wiener
noise increment in both equations.

This BHD difference current, $i(t)$, along with a set of start
times $\{t_j\}$, form a stochastic measurement record. The source
quasimode $\hat a$ is in a quantum state $|\psi_{\rm
REC}(t)\rangle$ conditioned on the past record. We simulate
eqs.~\eqref{eq:current}-\eqref{eq:stochschr} on a computer. By sampling an
ongoing realization of $i(t)$ for many ``start'' times (APD
detections realized concurrently) we may calculate the following
averaged photocurrent,
\begin{equation}\label{eq:signal}
{\mathcal{H}(\tau)}=\frac1{N_s}\sum_{j=1}^{N_s}i(t_j+\tau).
\end{equation}
A connection exists between the averaged photocurrent and the
wave-particle correlation function~\cite{GiantViolations2000}. In the
limit of large bandwidth this connection reads
\begin{equation}\label{eq:conversion}
h_{\theta}(\tau)=\frac{\mathcal{H}(\tau)}{|\langle\hat{a}\rangle|\sqrt{8\kappa(1-r)}}.
\end{equation}

The dual nature of the measurement process provides one of the many strengths of the conditional field measurement. It allows the use of quantum trajectory theory to unravel the master equation\index{Master equation! unraveling} in two distinct ways. A numerical simulation of eqs.~\eqref{eq:current}-\eqref{eq:stochschr} reproduces the experimental results for the averaged photocurrent. We can understand some of the mechanisms which lead to this averaged result by simulating the algorithm described in eqs.~\eqref{eq:expprob}-\eqref{eq:eqprop} and replacing eq.~\eqref{eq:eqprop} with eq.~\eqref{eq:stochschr}. Setting $r\approx1$ reduces the conditional homodyne simulation to a cleaner photocounting simulation. This leads to an understanding of the dynamical processes which create the spectrum of squeezing. Figure~\ref{fig:WPcorrfig} (right frames) shows the equivalence of the third-order correlation function $h_{0^o}(\tau)$ and the spectrum of squeezing for a cavity QED system in the strong-coupling regime. The spectrum of squeezing has also been calculated via the quantum regression formula\index{Quantum! regression formula}. A large positive peak in the spectrum of squeezing appears at zero frequency when one increases the strength of the driving field to the case when there is about one tenth of a photon in the cavity and two atoms interacting with the mode. Spontaneous emission degrades the squeezing signal due to the fact that the average value of the conditioned field is much larger than the steady-state field, a process visualized by the quantum trajectories of~\cite{Reiner2001}. 

Denisov, Castro-Beltran and Carmichael explored the time-reversal properties of the particle-wave correlations. They found that while the particle-particle correlation function is necessarily time symmetric, the wave-particle correlation function may be time asymmetric for non-Gaussian fluctuations\index{Gaussian! fluctuations}. This time asymmetry indicates a breakdown of detailed balance\index{Detailed balance}~\cite{Denisov2002, CarmichaelFosterChapter}.


\section{Circuit QED}\label{sec:cirQED}

In the late 1990s, artificial atoms based on Josephson junctions\index{Josephson! junction} were proposed as candidates for the realization of a quantum bit and its scalable extensions, following the proposal of a macroscopic order parameter -- related to the density and common phase of Bose-condensed Cooper pairs\index{Cooper pair! Bose-condensation} of electrons -- some ten years earlier~\cite{Eckern1984}. Central to the idea was the expectation that the relevant subspace would comprise the ground and first excited states of this artificial atom, while the sufficient anharmonicity of the energy spectrum would prevent leakage of the encoded information out of the relevant subspace, facilitating control at a quantum level. This proposal was soon followed by the experimental demonstration of quantum coherence in a superconducting qubit\index{Superconducting qubit} in 1999 by Y. Nakamura and coworkers, then at NEC in Tokyo. With this significant accomplishment, it was only a matter of time for the branch of quantum electrodynamics based on superconducting qubits to gain greater popularity. A perspective offered in~\cite{schoelkopf2008wiring} about a decade later, evidenced the rapid development of the emergent field, poised to explore the ``limits of coherence'' in solid-state implementations of quantum systems, while the coherence time of superconducting qubits had increased by a factor of almost a thousand at the time of writing. Remarkable developments in control and measurement techniques, connectivity, qubit architecture, and coherence performance have been comprehensively summarized in the recent review of~\cite{Mamgain2023}, focusing on the performance of small-scale quantum processors based on superconducting qubits.

In this section, we reinstate $\hbar$ in keeping with the relevant literature. In doing so, we emphasize the distinction between eigen-energies and operating frequencies (typically in the GHz range). We will now briefly demonstrate how to quantize circuits built from superconducting elements, and obtain the basic Hamiltonians used to assess quantum fluctuations in circuit QED; see also~\cite{garciaripoll2022}.   

\subsection{From the Cooper pair box to the transmon qubit: the generalized JC model}
\label{ssec:evolcQED}

The first realization of an ``island'' hosting a coherent superposition of charge states differing by a pair of electrons, the {\it Cooper pair}\index{Cooper pair}, was demonstrated in 1998~\cite{FirstCooperpair}. Measurements of individual Cooper pairs were performed via coupling to a single electron transistor. Shortly afterwards, quantum oscillations were observed in a single voltage-biased Cooper pair box\index{Cooper pair! box}, as a result of the superposition between two charge states~\cite{Nakamura1999}. The mesoscopic island is connected to a large superconducting reservoir via a Josephson junction with energy $E_{J}$ and capacitance $C_{J}$. The energy $E_{J}=\Phi_0 I_{c}/(2\pi)$, with $\Phi_0=h/(2e)$ the magnetic flux quantum, depends on the superconducting energy gap via the critical supercurrent $I_{c}$ flowing through the junction~\cite{Tunneling1963}. 

Replacing the singe Josephson junction by a SQUID\index{SQUID} (Superconducting Quantum Interference Device) configuration makes the effective Josephson energy $E_{J}$ tunable by an external magnetic field as $E_{J}=E_{J0}|\cos(\pi \Phi/\Phi_0)|$, where $\Phi$ is the external magnetic flux through the area formed by a pair of Josephson junctions \cite{Blais2004QED}. The CPB is also connected to a gate capacitor ($C_{g}$), which is in turn directly linked to a gate voltage $V_{g}$ inducing charge on its electrode. Hence, by modifying the flux $\Phi$ in the loop formed by the pair of junctions alongside the gate voltage $V_{g}$, we can gain control over the effective fields acting on the qubit. The total box capacitance reads $C_{\Sigma}=C_{J}+C_{g}$. If the energy gain for two electrons associated with the formation of a Cooper pair is larger than both the charging energy $E_{c}=e^2/(2C_{\Sigma})$\index{Charging energy} and thermal energy, we can use the number of Cooper pairs\index{Cooper pair} on the island, $N$, to form a basis for the Hamiltonian describing the coherent dynamical evolution~\cite{Blais2004QED, KochTransmon2007},
\begin{equation}\label{CPBH}
\hat{H}_{Q}=4E_{c}\sum_{N}\left[(N-n_{g})^2 \ket{N}\bra{N}-\frac{E_{J}}{2}\left(\ket{N+1}\bra{N}+{\rm H.c.}\right)\right]=4E_{C}(\hat{n}-n_{g})^2-E_{J}\cos\hat{\phi},
\end{equation}
with $n_{g}=C_{g}V_{g}/(2e) + Q_{r}/(2e)$ a control parameter representing the gate charge injected into the island by the voltage source (where we have taken into account an offset charge $Q_{r}$ induced by the environment). In eq. \eqref{CPBH}, $\hat{n}$ and $\hat{\phi}$ are the new conjugate operators, denoting the number of Cooper pairs\index{Cooper pair} transferred to and from the island and the gauge-invariant phase difference across the junction, with $\hat{n}=-i \partial/\partial \hat{\phi}$, satisfying the commutation relation $[\hat{\phi}, \hat{n}]=i$.

Based on \eqref{CPBH} we can now recast the corresponding stationary Schr\"{o}dinger equation in the phase basis:
\begin{equation}\label{eqphase}
\left[4E_{C} \left(-i\frac{d}{d\phi}-n_{g}\right)^2 -E_{J} \cos{\phi}\right]\psi(\phi)=E \psi(\phi),
\end{equation}
with the periodic boundary condition $\psi(\phi)=\psi(\phi+2\pi)$. Equation~\eqref{eqphase} takes the form of a Mathieu equation\index{Mathieu equation} with the substitution $g(x)=e^{2in_{g}x}\psi(2x)$, yielding
\begin{equation}\label{Mathieu}
\frac{d^2 g(x)}{dx^2} + \left(\frac{E}{E_{C}}+\frac{E_{J}}{E_{C}}\cos(2x)\right)g(x)=0,
\end{equation}
which admits the well-known Mathieu solutions, depending on the ratio $E_{J}/E_{C}$ and the offset charge $n_{g}$ (for a more detailed discussion see Appendix B of~\cite{KochTransmon2007}).

We now move to the limiting case of a {\it charge qubit}\index{Qubit! charge}, where only the charge degree of freedom is relevant. For $4E_{C} \gg E_{J}$ and $0 \leq n_{g} \leq 1$ we can restrict our description to a pair of adjacent charge states, reducing the Hamiltonian of eq. \eqref{CPBH} to a $2 \times 2$ matrix
\begin{equation}\label{CPBHred}
\hat{H}_{Q}^{r}=-2 E_{C} (1-2n_{g})\hat{\tilde{\sigma}}_{z} -\frac{1}{2}E_{J} \hat{\tilde{\sigma}}_{x},
\end{equation}
mapping the CPB to a pseudospin-$1/2$ particle with effective fields along the $x$ and $z$ axis. Transforming now to a basis in which $\hat{H}_{Q}^{r}$ is diagonal, we obtain the eigenvalues $\lambda_{1,2}=\pm (1/2)\hbar\Omega_{\rm cq}$, with
\begin{equation}\label{energysplit}
\hbar \Omega_{\rm cq}=\sqrt{E_{J}^2 + [4E_{C}(1-2n_{g})]^2}
\end{equation}
the energy splitting for the qubit in the new representation. The operators transform according to the relation 
\begin{equation}\label{transform}
\left[\begin{array}{c}
\hat{\tilde{\sigma}}_{x} \\ \hat{\tilde{\sigma}}_{z} \end{array}\right]=
\left[\begin{array}{cc}
\cos\theta & \sin\theta \\ -\sin \theta & \cos \theta
\end{array}\right]
\left[\begin{array}{c}
\hat{{\sigma}}_{x} \\ \hat{\sigma}_{z} \end{array}\right],
\end{equation}
where $\theta=\arctan\{E_{J}/[4E_{C}(1-2n_{g})]\}$. The interaction with the microwave field is taken into account with a term of the form $\hbar g \hat{\tilde{\sigma}}_z (\hat{a}+\hat{a}^{\dagger})$, where $g$ is the coupling strength with an expression depending upon the circuit design and the type of coupling, {\it capacitive} or {\it inductive} \cite{MWPhotonics}.   

A similar procedure is followed for the {\it flux qubit}\index{Qubit! flux}, comprising a superconducting loop interrupted by a number of Josephson junctions, responsible for creating an effective potential energy profile. In this setup, Cooper pairs\index{Cooper pair} flow continuously through the loop, in contrast to the charge qubit\index{Qubit! charge} where individual pairs are tunneling across the junction. The fact that the wavefunction must be a single-valued function along a loop requires an integer number of flux quanta that are allowed to penetrate the superconducting ring. As the persistent external bias produces a steady flux through the loop, clockwise and counter-clockwise currents are developed to counteract the non-integer number of flux quanta. Allowing also for tunneling between the two stable solutions (indicated by the minima of the potential energy) we arrive at the Hamiltonian (see also \cite{FluxQubitRep} for the demonstration of coherent dynamics)
\begin{equation}\label{fluxQH}
\hat{\tilde{H}}_{\rm fq}=\frac{1}{2}\varepsilon_{B} \hat{\tilde{\sigma}}_{z} +\frac{1}{2} \hbar \Delta \hat{\tilde{\sigma}}_{x},
\end{equation}
where $\varepsilon_{B}=2 I_{p}\Phi_{0}(\phi_{p}-1/2)$ is the magnetic energy of the qubit, with $I_p$ the persistent current and $\phi_p=\Phi/\Phi_0$ the magnetic frustration, and $\Delta$ is the tunneling rate between the two minima. Diagonalizing \eqref{fluxQH} we obtain $\hat{H}_{\rm fq}=(1/2)\hbar\omega_{\rm fq} \hat{\sigma}_{z}$, with $\hbar\Omega_{\rm fq}=\sqrt{4(\phi_{p}-1/2)^2 I_p^2 \Phi_0^2 + \hbar^2 \Delta^2}$. A simple intuitive correspondence between the circuit Hamiltonian of a Josephson-junction flux qubit\index{Qubit! flux} coupled to an LC oscillator, and the quantum Rabi Hamiltonian has been established in~\cite{YoshiharaAshhab2022}. 

\subsubsection{Engineering the coupling strength}
\label{sssec:couplingstrength}

The demonstrated coherence based on the Josephson effect, led subsequently to the incorporation of a Cooper pair box\index{Cooper pair! box} (CPB), comprising an island with a macroscopic number of conduction electrons, into a microwave resonator, giving birth to the field of circuit QED. Keeping the notation introduced for the artificial two-level atom in the charge regime, we now include the interaction with the microwave cavity mode, to obtain the Hamiltonian \cite{Blais2004QED} (in the transformed qubit frame)
\begin{equation}\label{FullH}
H=\hbar \omega_{r} \hat{a}^{\dagger}\hat{a} + \frac{1}{2}\hbar \Omega_{\rm cq} \hat{\sigma}_{z} - 2e \frac{C_{g}}{C_{\Sigma}}\sqrt{\frac{\hbar \omega_{r}}{2C_{r}}}(\hat{a}+\hat{a}^{\dagger}) [1-2n_{g}-\cos\theta \hat{\sigma}_{z} +\sin\theta \hat{\sigma}_{x}],
\end{equation}
where $C_{r}$ is the capacitance of the resonator. For $n_{g}=1/2$ (and analogously for $\phi_p=1/2$ in a charge qubit\index{Qubit! charge}), called the {\it degeneracy point}, leading to $\theta=\pi/2$, the Hamiltonian of eq. \eqref{FullH} in the RWA assumes the JC form [see eq. \eqref{jcham}] for a transverse coupling with strength
\begin{equation}\label{coupling}
g=\frac{\beta e}{\hbar} \sqrt{\frac{\hbar \omega_{r}}{2C_{r}}},
\end{equation}
in which $\beta\equiv C_{g}/C_{\Sigma}$. Therefore, {\it the voltage biased} CPB as well as the flux qubit\index{Qubit! flux} inside a microwave resonator can be mapped to the problem of a two-level atom inside a cavity discussed in the previous section. We note here that suitably designed circuits supporting a strong {\it longitudinal coupling} term to the microwave field have nowadays become increasingly interesting for fast quantum non-demolition qubit readout as well as for the generation of nonclassical states in the resonator (see~\cite{MWPhotonics} and references therein). For an experimentally realistic coupling of $\beta \sim 0.1$, owing to a highly polarizable artificial ``atom'', the CPB heralded a vacuum Rabi splitting three orders of magnitude larger than the corresponding atomic microwave experiments of cavity QED \cite{Wallraff2004QED, Blais2004QED, RaimondRev}. With a particular scheme coupling the artificial atom to the magnetic field of the resonator, the coupling strength can be shown to be proportional to the inverse square root of the fine-structure constant, and, consequently, significantly enhanced \cite{LargeCC}. What remained to be addressed was the issue of decoherence. Experiments had already shown that a simple adjustment of the quantum circuit design led to significantly larger dephasing times (see~\cite{LesHouches} for an overview of the field in 2003). The dependence of the dephasing time $T_2$ on gate charge suggested that low frequency fluctuations were the main source of dephasing away from the degeneracy point $n_{g}=1/2$ \cite{Bladh2005}. 

\subsubsection{Mitigating dispersion and decoherence}

The decisive step towards that direction was taken in 2007 with the introduction of the {\it transmon qubit}\index{Transmon qubit}, comprising two superconducting islands coupled via two Josephson junctions, to allow for the tuning of the Josephson energy, but isolated from the rest of the system \cite{KochTransmon2007}. The transmon qubit is very closely related to the CPB, but operates in a significantly larger ratio of $E_{J}/E_{C}$. The transmon emerged as an ideal candidate for mitigating charge noise and increasing the qubit-photon strength, while still retaining the amount of aharmonicity required for quantum manipulations. While it mimics the architecture of a CPB, it is operated in the regime $E_{J} \gg E_{C}$. This is achieved by means of an additional large capacitance $C_{B}$ between the two superconductors forming the Josephson junctions, making the charging energy $E_{c}=e^2/(2C_{\Sigma})$ (with $C_{\Sigma}=C_{J}+C_{B}+C_{g}$) significantly smaller than the Josephson energy. Smoothly increasing the ratio $E_{J}/E_{C}$ maps the CPB into the transmon. This is in contrast to the alternative scalable design of the {\it phase qubit}\index{Phase qubit} operating in a regime of $E_{J}\gg E_{C}$ \cite{Martinis2002}, with large-area Josephson junctions (the Josephson energy scales proportionally with the area of the tunnel junction), with a ``washboard potential'' (where the phases $\phi$ and $\phi+2\pi$ are distinct). 

The sensitivity of the transmon to charge noise is related to the differential charge dispersion $\partial E_{ij}/\partial n_{g}$, with $E_{ij}$ the separation between the eigenenergies $E_{i}$ and $E_{j}$. One can approximate the dependence of the eigenenergies $E_{m} (n_{g})$ ($m=0,1,2\ldots$) of the Mathieu equation \eqref{Mathieu} on the offset charge by a cosine term in the limit of large $E_{J}/E_{C}$. We can further show using WKB (Wentzel-Kramers-Brillouin) methods\index{WKB approximation} that the peak-to-peak value for the charge dispersion follows the law
\begin{equation}\label{chargedisp}
\epsilon_{m} \equiv E_{m}(n_{g}=1/2)-E_{m}(n_{g}=0)\simeq (-1)^{m}E_{C} \frac{2^{4m+5}}{m!} \frac{2}{\pi} \left(\frac{E_{J}}{2E_{C}}\right)^{\frac{m}{2}+\frac{3}{4}}\, e^{-\sqrt{8E_{J}/E_{C}}}, 
\end{equation} 
an approximation bettering with increasing $E_{J}/E_{C}\gg 1$. The crucial term in this result is the presence of the exponential decrease which will eventually dominate over the power law in the limit of very large $E_{J}/E_{C}$. In that very limit, the dispersion can be considered flat with respect to $n_{g}$, with $E_{1}-E_{0}\equiv E_{01}\sim \sqrt{8E_{J}E_{C}}$, the so-called {\it plasma energy}. The challenge presented here is the operation in a regime with significantly improved charge-noise sensitivity with respect to the CPB, yet a sufficiently large anharmonicity, quantified by $\alpha_{r}\equiv (E_{12}-E_{01})/E_{01} \approx - (8E_{J}/E_{C})^{-1/2}$. 
\begin{figure}
    \centering
        \includegraphics[width=0.5\columnwidth]{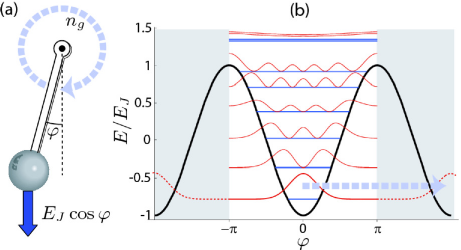}
    \caption{{\bf (a)} Mechanical rotor analogy for the transmon. The transmon Hamiltonian can be understood as a charged quantum rotor in a constant magnetic field $\sim n_g$. For large $E_J/E_C$, there is a significant ``gravitational'' pull on the pendulum and the system typically remains in the vicinity of $\varphi=0$. Only tunneling events between adjacent cosine wells ({\it i.e.}, a full $2\pi$ rotor movement) will acquire an Aharonov-Bohm\index{Aharonov-Bohm phase} like phase due to $n_g$. The tunneling probability decreases exponentially with $E_J/E_C$, explaining the exponential decrease of the charge dispersion. {\bf (b)} Cosine potential (black solid line) with corresponding eigenenergies and squared moduli of the eigenfunctions. Reproduced with permission from the APS.
\label{fig:rotor}}
\end{figure}
In order to get a grasp of the physical picture we map the transmon to the charged quantum rotor with mass $m$ and charge $q$, moving in the presence of a homogeneous magnetic field, as depicted in fig. \ref{fig:rotor}. The motion of the rotor is restricted to the $x-y$ plane, perpendicular to the magnetic field, and with the homogeneous gravitational field pointing at the $x$-direction. The corresponding potential energy acquires a term proportional to $\cos\phi$, with the angle $\phi$ identified as the phase difference across the Josephson junction. On the other hand, the angular momentum, pointing towards the $z$-axis, is expressed as $\hat{L}_{z}=-i\hbar\, \partial/\partial \hat{\phi}$, playing the role of the qubit pair number operator $\hat{n}$. The Hamiltonian governing the rotor's motion reads
\begin{equation}\label{rotorH}
\hat{H}_{\rm rot}=\frac{\hat{L}_{z}^2}{2ml^2}-mgl\cos\hat{\phi}.
\end{equation}
In addition to these interactions we represent the magnetic field by the vector potential\index{Vector potential} (we omit the operator symbol on top of the Cartesian co-ordinates) $\hat{\mathbf{A}}=(1/2)B_{0}(-y, x, 0)$ giving rise to the canonical momentum $\hat{\mathbf{P}}=\hat{\mathbf{p}}+q\hat{\mathbf{A}}$, thereby writing $\hat{L}_{z}\rightarrow \hat{L}_{z}-q (\mathbf{r}\times \hat{\mathbf{A}})\cdot \hat{z}=\hat{L}_{z}-(1/2) q B_0 l^2$. We can therefore identify $n_{g}$ with the term $(1/2) q B_0 l^2$. The $2\pi$ periodicity of the eigenstates of \eqref{rotorH} yields discrete values for the angular momentum, establishing the mapping to the well-defined eigenvalues of the charge operator $\hat{n} \leftrightarrow \hat{L}_{z}/\hbar$. For the energies we identify $E_{J} \leftrightarrow mgl$ and $E_{C} \leftrightarrow \hbar^2/(8ml^2)$. 

In the regime $E_{J}/E_{C} \gg 1$ the gravitational potential energy dominates, favoring small oscillations about $\phi=0$. Neglecting then the periodic boundary condition we can eliminate $n_{g}$ via a gauge transformation\index{Gauge! transformation}, we expand the $\cos\phi$ term up to the fourth order to yield the potential energy $-E_{J} + E_{J}\phi^2/2 -E_{J}\phi^4/24$, and recast $H_{\rm rot}$ to the form
\begin{equation}\label{HDuff}
\hat{H}_{D}=\sqrt{8E_{J}E_{C}}\, (\hat{b}^{\dagger}\hat{b}+1/2)-E_{J}-(E_{J}/12)(\hat{b}^{\dagger}+\hat{b})^4.
\end{equation}
With this perturbative expansion the phase variable $\phi$ becomes non-compact and the Hamiltonian is mapped to that of a Duffing oscillator\index{Duffing oscillator}~\cite{nayfeh2008nonlinear}. On the other hand, the periodic boundary condition $\psi(\phi+2\pi)=\psi(\phi)$ has a physical meaning, as it is associated with {\it instanton}\index{Instanton} tunneling events through the cosine potential barrier. The offset charge $n_{g}$ is associated to a rare $2\pi$ rotation of the rotor, which brings us to the nonvanishing charge dispersion since it is associated with the Aharonov-Bohm effect via the characteristic role of the vector potential\index{Vector potential} in quantum mechanics.   

The transmon has been incorporated to a superconducting transmission line, coupled with a resonant mode with an antinode at the center of the resonator. For a resonator capacitance $C_{r} \gg C_{\Sigma}$, we can write the Hamiltonian of the coupled system in the basis of the uncoupled transmon states $\ket{l}$ as
\begin{equation}\label{HamTr}
\hat{H}=\hbar\sum_{m}\omega_{n}\ket{m}\bra{m} +\hbar \omega_{r} \hat{a}^{\dagger} \hat{a} + \hbar \sum_{l,m}g_{lm}\ket{l}\bra{m}(\hat{a}+\hat{a}^{\dagger}),
\end{equation}
with coupling constants
\begin{equation}\label{couplingmn}
\hbar g_{lm}\equiv 2e\beta \sqrt{\frac{\hbar \omega_{r}}{2C_{r}}} \braket{l|\hat{n}|m},
\end{equation}
where the number operator $\hat{n}$ assumes the asymptotic bosonic form $\hat{n}=-(i/\sqrt{2})[E_{J}/(8E_{c})]^{1/4}(\hat{b}-\hat{b}^{\dagger})$. In the limit $E_{J}/E_{C} \to \infty$, we can write for the matrix elements of the number operator 
\begin{equation}\label{matrixel}
|\braket{m+k|\hat{n}|m}|\approx \sqrt{\frac{m+1}{2}}\left(\frac{E_{J}}{8E_{C}}\right)\delta_{k1}.
\end{equation}
Eliminating now the fast-rotating terms, we arrive at the generalized JC Hamiltonian,
\begin{equation}\label{generalRWA}
\hat{H}_{\rm GJC}=\hbar\sum_{m}\omega_{n}\ket{m}\bra{m} +\hbar \omega_{r} \hat{a}^{\dagger} \hat{a}+\left[\hbar \sum_{m}g_{m,m+1}\ket{m}\bra{m+1}\hat{a}^{\dagger} + {\rm H.c.}\right].
\end{equation}
The coupling energies $g_{mn}$ grow with increasing an already large $E_{J}/E_{C}$, allowing for utilizing the transmon as an actual qubit. The transmon can be interpreted as a harmonic
oscillator in the charge basis, with its quadratic potential centered at $n_{g}$. Despite the displacement of $n_{g}$, the frequency of the harmonic oscillator remains unaltered, mitigating any possible dephasing. On the other hand, driving the transmon on resonance induces transitions between the different oscillator states, leading to a highly polarizable configuration that remains moderately anharmonic. 
\begin{figure}
\begin{center}
\includegraphics[width=0.45\textwidth]{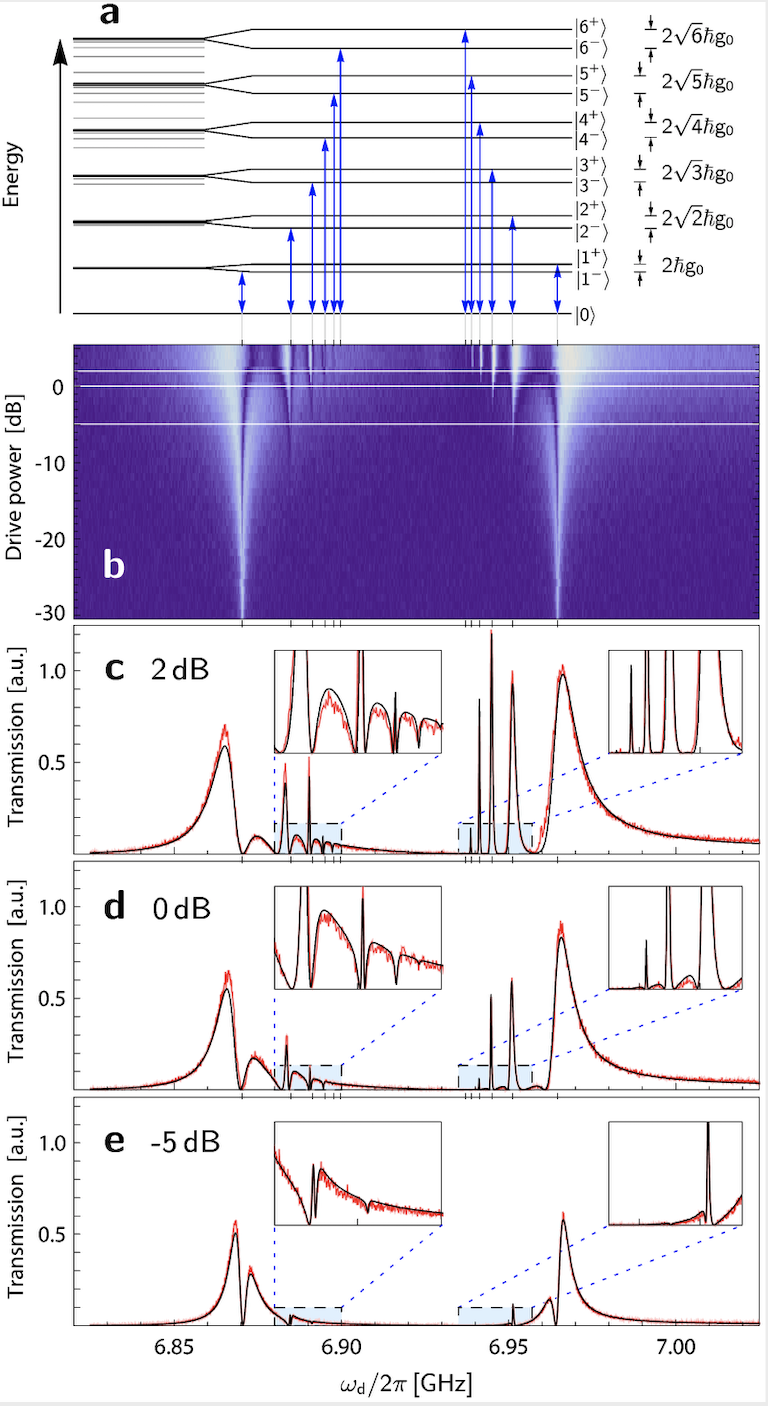}
\end{center}
\caption{{\bf (a)} The $\sqrt{n}$ spectrum of the extended resonant JC oscillator. All levels are shown to scale in the left part of the diagram: black lines represent the levels $\ket{n,\pm} \simeq (\ket{n,0}\pm \ket{n-1,1})/\sqrt{2}$ with only small contributions from higher transmon states (for $j>1$); grey lines represent the levels  with large contributions from higher transmon states. In the right part of the diagram, the $\sqrt{n}$ scaling of the splitting between the $\ket{n,\pm}$ states is exaggerated for clarity, and the transitions observed in plots {\bf (b)-(e)} are indicated at the $x$-coordinate, $E_{n\pm}/(n\hbar)$ of their $n$-photon transition frequency from the ground state. {\bf (b)} Measured intensity ($A^2$, heterodyne amplitude squared)\index{Heterodyne detection} in color scale as a function of drive frequency and power. The multiphoton transitions\index{Multiphoton! transition} shown in {\bf (a)} are observed at their calculated positions. {\bf (c)-(e)} Examples of cuts for constant power, at the values indicated in {\bf (b)} (results from the master equation \eqref{ME} in black; experimental results in red), demonstrating excellent agreement between theory and experiment, which is reinforced in the enlarged insets. Good agreement is found over the full range in drive power for a single set of parameters (Source: Fig. 3 of~\cite{Bishop2008}). Reproduced with permission from Springer Nature.}
\label{fig:JCspectr}
\end{figure}
Like the CPB, the transmon is typically operated in the dispersive regime ($|E_{01}-E_{r}| \gg \hbar g_{01}$, $|E_{01}(1+\alpha_{r})-E_{r}| \gg \hbar g_{01}$) for qubit readout and control. Taking into account virtual transitions through the various transmon states with a canonical transformation, similar to the Schrieffer-Wolff/polaron transformation~(\ref{polaron}), the effective Hamiltonian in the linear dispersive regime reads
\begin{equation}\label{EffHDisp}
\hat{H}_{\rm disp,l}=\hbar[(\omega_{01}+\chi_{01})\ket{1}\bra{1} + \hat{a}^{\dagger}\hat{a}(\omega_{r}-\chi_{01}\ket{0}\bra{0}+ \chi_{01}\ket{1}\bra{1}-\chi_{12}\ket{1}\bra{1})],
\end{equation}
which can be rewritten as
\begin{equation}\label{EffDisp2}
\hat{H}_{\rm disp,l}=\frac{1}{2}\hbar \omega_{01}^{\prime}\, \hat{\sigma}_{z} + \hbar (\omega_{r}^{\prime} + \chi \hat{\sigma}_{z})\hat{a}^{\dagger}\hat{a},
\end{equation}
where the ``qubit'' and resonator frequencies ($\omega_{01}$ and $\omega_{r}$ respectively) are renormalized by the partial dispersive shifts $\chi_{ij}=g_{ij}^2/(\omega_{ij}-\omega_{r})$, with $\omega_{ij}=\omega_{j}-\omega_{i}=E_{ij}/\hbar$, as $\omega_{01}^{\prime}=\omega_{01}+\chi_{01}$ and $\omega_{r}^{\prime}=\omega_{r}-\chi_{12}/2$, while $\chi=\chi_{01}-\chi_{12}/2$.  

Introducing a coherent classical drive with amplitude $\eta$ at frequency $\omega_d$, see eq.~(\ref{fdrive}), we write $\hat{H}_{\rm GJC,d}=\hat{H}_{\rm GJC}+\eta (e^{-i\omega_d t}\hat{a}^{\dagger}+e^{i\omega_d t}\hat{a})$, such that the Markovian master equation\index{Master equation} for the transmon reads \cite{Bishop2008}
\begin{equation}\label{ME}
\frac{d\hat{\rho}}{dt}=\frac{1}{i\hbar}[\hat{H}_{\rm GJC,d},\hat{\rho}] + 2\kappa\mathcal{D}[\hat{a}]\hat{\rho}+\gamma \mathcal{D}\left[\sum_{m}\alpha_m \ket{m}\bra{m+1}\right]\hat{\rho}+\gamma_{\phi}\mathcal{D}\left[\sum_{m}\beta_m \ket{m}\bra{m}\right]\hat{\rho},
\end{equation}
where $\mathcal{D}[\hat{A}]\hat{\rho}$ is the standard Lindblad super-operator\index{Lindblad! super-operator} $\mathcal{D}[\hat{A}]\equiv (1/2)([\hat{A}\hat{\rho}, \hat{A}^{\dagger}]+[\hat{A}, \hat{\rho}\hat{A}^{\dagger}])$. In the ultrastrong coupling regime (see sec. \ref{sssec:couplingstrength}), the form of eq. \eqref{ME} becomes inappropriate, as one cannot consider two reservoirs coupled to the qubit and cavity independently to produce additive dissipation terms. Instead, the derivation of the master equation\index{Master equation} must now take into consideration the qubit-resonator coupling, recasting the description into the dressed-state picture \cite{Beaudoin}.

\subsection{The (generalized) JC nonlinearity and spectrum revisited in the light of circuit QED}
\label{ssec:JCspectrcQED}

The rapid advancement of well-controlled mesoscopic vibrational systems reappraised the role of quantum fluctuations in the dynamical response of nonlinear oscillators (for an extensive coverage see~\cite{DykmanBook}), of which the JC model occupies a prominent position. Let us return at this point to the JC spectrum, for a two-level system with resonance frequency $\omega_{q}$  coupled to a cavity with resonance frequency $\omega_{r}=\omega_{q}$ with the excited-state doublets (\ref{dstate}). On resonance ($\omega_r=\omega_q=\omega_0$) the excited-state doublets take the form
\begin{equation}\label{doublets}
\begin{aligned}
\ket{\psi_{n+}}&=\frac{1}{\sqrt{2}}(\ket{+}\ket{n}+i\ket{-}\ket{n+1}),\\
\ket{\psi_{n-}}&=\frac{1}{\sqrt{2}}(\ket{+}\ket{n}-i\ket{-}\ket{n+1}),
\end{aligned}
\end{equation}
with corresponding resonant energies
\begin{equation}\label{doubletenergies}
\begin{aligned}
E_{\rm n+}&=\left(n+\frac{1}{2}\right)\hbar\omega_{0} + \sqrt{n+1}\hbar g,\\
E_{\rm n-}&=\left(n+\frac{1}{2}\right)\hbar\omega_{0} - \sqrt{n+1}\hbar g,
\end{aligned}
\end{equation}
and $E_{G}=-(1/2)\hbar \omega_{0}$ the ground state energy. For high levels of excitation, the detunings between a driving field, which is resonant with the uncoupled atom and cavity mode, are  given by $\pm (\sqrt{n}-\sqrt{n-1})g \simeq \pm g/(2\sqrt{n})$ and $\pm (\sqrt{n}+\sqrt{n-1})g \simeq \pm 2g\sqrt{n}$. This brings into play a detuning which depends on the level of excitation, whose importance depends on the strength of the coupling relative to the dissipation (see fig. 1 of~\cite{SponDressState}). For strong coupling, this mechanism forms the basis of the dispersive transition producing high excitation along alternate near-resonant paths with increasing drive strength [see sec. 2 of~\cite{SponDressState} and eq. (24) of~\cite{CarmichaelPRX}]. It is also the mechanism lying behind the photon blockade effect~\cite{hoffman2011dispersive,viehmann2011superradiant,ridolfo2012photon,fink2017observation}.

We now consider a coherently-driven resonant cavity mode with a driving field of amplitude $\varepsilon_{d}$ tuned to the frequency of the lower Rabi resonance rather than at the cavity resonance frequency $\omega_{r}$. Assuming that the drive strength is not too large, we can consider an effective two-level system comprising the ground state $\ket{G}=\ket{-}\ket{0}$ and the excited state $\ket{\psi_{1-}}=\ket{-}\ket{1}$, making the approximation 
\begin{equation}\label{approx2l}
\begin{aligned}
\sqrt{2}(i\hat{a},\hat{\sigma}_{-}) \rightarrow \hat{\Sigma}_{-} \equiv \ket{G}\bra{\psi_{1-}}, \\
\sqrt{2}(-i\hat{a}^{\dagger},\hat{\sigma}_{+}) \rightarrow \hat{\Sigma}_{+} \equiv \ket{\psi_{1-}}\bra{G},
\end{aligned}
\end{equation}
modifying the master equation~\eqref{ME}\index{Master equation} to the one we find in resonance fluorescence\index{Resonance fluorescence! master equation}\index{Master equation!resonance fluorescence} \cite{TianCarmichael}
\begin{equation}\label{ME2level}
\begin{aligned}
\frac{d\hat{\rho}}{dt}&=-i\left[\left(\frac{1}{2}\omega_{0}-g\right)\hat{\Sigma}_{+}\hat{\Sigma}_{-}-\frac{1}{2}\omega_{0}\hat{\Sigma}_{-}\hat{\Sigma}_{+}, \hat{\rho}\right]+\frac{\eta}{\sqrt{2}}\left[e^{-i(\omega_{0}-g)t}\hat{\Sigma}_{+}-e^{i(\omega_{0}-g)t}\hat{\Sigma}_{-}, \hat{\rho}\right]\\
&+\frac{1}{2}[\kappa+(\gamma+\gamma_{\phi})/2]\left(2\hat{\Sigma}_{-} \hat{\rho} \hat{{\Sigma}}_{+}-\hat{\Sigma}_{+}\hat{\Sigma}_{-}\hat{\rho} -\hat{\rho}\hat{\Sigma}_{+}\hat{\Sigma}_{-}\right).
\end{aligned}
\end{equation}
We denote by $\Delta=\omega_{0}-g-\omega_{d}$ the detuning between the drive frequency and the lower vacuum Rabi resonance frequency, as in~\cite{Bishop2008}. In the two-level subspace, the system evolution can be described by the semiclassical Bloch oscillations\index{Bloch! oscillations} for the three components of the reduced system density matrix $\hat{\rho}=(1+x \hat{\Sigma}_{x} + y \hat{\Sigma}_{y} + z\hat{\Sigma}_{z})/2$ [with $\hat{\Sigma}_{x}=(1/2)(\hat{\Sigma}_{+}+\hat{\Sigma}_{-})$, $\hat{\Sigma}_{y}=[1/(2i)](\hat{\Sigma}_{+}-\hat{\Sigma}_{-})$ and $\hat{\Sigma}_{z}=2\hat{\Sigma}_{+}\hat{\Sigma}_{-}-1$], resulting in the corresponding Bloch equations
\begin{equation}\label{eq2level}
\begin{aligned}
\dot{x}&=-x/T_{2}^{\prime}-\Delta y, \\
\dot{y}&=\Delta x -y/T_{2}^{\prime}-\Omega z,\\
\dot{z}&=\Omega y-(z+1)/T_{1}^{\prime}, 
\end{aligned}
\end{equation}
where $\Omega=\sqrt{2} \varepsilon_{d}$ is the effective drive strength, and $T_{1}^{\prime}=(\kappa+\gamma/2)^{-1}$, $T_{2}^{\prime}=\{[\gamma+2(\gamma_{\phi}+\kappa)]/4\}^{-1}$ are the effective relaxation and dephasing times. The steady-state solution\index{Steady state!solution} of the system \eqref{eq2level} yields the heterodyne amplitude\index{Heterodyne detection}
\begin{equation}
A=2V_{0} \left|\braket{a}\right|=\frac{V_{0}T_{2}^{\prime}\Omega\sqrt{(\Delta^2 {T_{2}^{\prime}}^2+1)/2}}{\Delta^2 {T_{2}^{\prime}}^2 + T_{1}^{\prime}T_{2}^{\prime}\Omega^2 +1}.
\end{equation}
The above expression predicts a crossover from the linear regime of low drive strengths, with a Lorentzian response, to the regime where the nonlinearity is responsible for the splitting of the main peak for stronger driving, occurring at $\Omega^2=T_{1}^{\prime}T_{2}^{\prime}$. The separation of the two emerging peaks scales as $2{T_{2}^{\prime}}^{-1} \sqrt{T_{1}^{\prime}T_{2}^{\prime}\Omega^2-1}$. This behavior compares well to the experiment for moderate drive strengths~\cite{Bishop2008}, yet fails to account for the developing asymmetry due to the occupation of higher levels in the extended JC ladder. The corresponding detunings with respect to the ``dressed'' states are not equal the lower vacuum Rabi resonance. For such an effective two-level system the effects of dynamic Stark splitting have been investigated by producing the optical spectrum, which demonstrated an asymmetric Mollow triplet\index{Mollow! triplet} varying with drive strength \cite{TianCarmichael}. The quantum-trajectory approach captures the asymmetry due to the unequal detuning from the state $\ket{2, L}$ for transitions out of the ``dressed'' states, as well as due to the unequal matrix elements between the states $\ket{\psi_{2-}}$ and $\ket{\psi_{1-}}$ for the operators $\hat{a}$ and $\hat{\sigma}_{-}$. The $\sqrt{n}$ spectrum with the associated multiphoton transitions\index{Multiphoton! transition}, revealed by heterodyne detection\index{Heterodyne detection}, is depicted in fig. \ref{fig:JCspectr}, where the full ladder is shown, following a prior observation of the JC interaction in circuit QED involving states with up to two photons \cite{Fink2008}. Vacuum Rabi mode splitting with one, two, and three qubits evidenced the $\sqrt{N}$ coupling in the Tavis-Cummings Hamiltonian\index{Model! Tavis-Cummings}\index{Tavis-Cummings model} modelling the experiment of~\cite{fink2009dressed}. A second-order effective JC Hamiltonian for the region close to a spectroscopically-evidenced level anticrossing, under a coherent two-photon\index{Two-photon!driving} driving of the two-state atom is derived using a Dyson-series approach and reveals the role of the symmetry in the qubit potential, establishing selection rules similar to those governing electric dipole transitions\index{Electric dipole! transition}~\cite{deppe2008two}. When probing the JC ladder with spin circuit QED, additional transitions have been observed in the vacuum Rabi splitting spectrum due to the involvement of higher excited states~\cite{Bonsen2022}.

The physical system considered in~\cite{ballester2012quantum} consists of a superconducting qubit strongly coupled to a microwave resonator mode. Working at the qubit degeneracy point, the Hamiltonian is given by
\begin{eqnarray}\label{eq:HamilDiag}
{\cal \hat{H}} =  \frac{\hbar \omega_q}{2} \hat{\sigma}_z + \hbar  \omega \hat{a}^\dagger \hat{a}  -\hbar  g    \hat{\sigma}_x  (\hat{a} + \hat{a}^\dagger),
\end{eqnarray}
where $\omega_q$, $\omega$ here denote the qubit and cavity frequencies, and $g$ stands for the coupling strength. Likewise, $\hat{a}$($\hat{a}^\dagger$) represent the annihilation(creation) operators of the photon field mode, while $\hat{\sigma}_x = \hat{\sigma}_{+} + \hat{\sigma}_{-} = |+ \rangle \langle -|+ |- \rangle \langle +|$, $\hat{\sigma}_z = |+ \rangle \langle +|-|- \rangle \langle -|$, with ${\ket{-},\ket{+}}$ the ground and excited eigenstates of the two-state atom. In a typical circuit QED implementation, this Hamiltonian can be simplified further by applying the RWA (see sec.~\ref{sssec:rwasub}). If $\{|\omega-\omega_q|, g\} \ll\omega+\omega_q$, then it can be effectively approximated as
\begin{equation}\label{eq:HamilRWA}
{\cal \hat{H}} = \frac{\hbar \omega_q}{2} \hat{\sigma}_z +\hbar  \omega \hat{a}^\dagger \hat{a} -\hbar  g (\hat{\sigma}_{+} \hat{a} + \hat{\sigma}_{-} \hat{a}^\dagger),
\end{equation}
in direct correspondence to the JC Hamiltonian of quantum optics. In the proposal of~\cite{ballester2012quantum}, the two-state atom is orthogonally driven by two classical fields. The Hamiltonian of the full system reads
\begin{equation}\label{eq:HamilDriv}
  {\cal \hat{H}} =  \frac{\hbar \omega_q}{2} \hat{\sigma}_z +\hbar  \omega \hat{a}^\dagger \hat{a} -\hbar  g (\hat{\sigma}_{+} \hat{a} + \hat{\sigma}_{-} \hat{a}^{\dagger})- \hbar \Omega_1  ( e^{i \omega_1 t} \hat{\sigma} + e^{-i \omega_1 t} \hat{\sigma}_{+}) - \hbar  \Omega_2 ( e^{i \omega_2 t} \hat{\sigma}_{-} + e^{-i \omega_2 t} \hat{\sigma}_{+}), 
\end{equation}
where $\Omega_j$, $\omega_j$ denote the amplitude and frequency of the $j$-th drive field. In order to obtain the expression of eq. \eqref{eq:HamilDriv}, we have implicitly assumed that the RWA applies not only to the atom-resonator interaction, but equally to both drive terms. Next, we will write \eqref{eq:HamilDriv} in the reference frame rotating with the frequency of the first drive field, $\omega_1$, 
\begin{equation}
{\cal \hat{H}}^{L_1} =\hbar  \frac{\omega_q-\omega_1}{2} \hat{\sigma}_z +\hbar  (\omega-\omega_1) \hat{a}^{\dagger} \hat{a} - \hbar g  (\hat{\sigma}_{+} \hat{a} +\hat{\sigma}_{-} \hat{a}^{\dagger}) - \hbar \Omega_1  ( \hat{\sigma}_{-} + \hat{\sigma}_{+}) - \hbar  \Omega_2   ( e^{i (\omega_2-\omega_1) t} \hat{\sigma}_{-} + e^{-i (\omega_2-\omega_1) t} \hat{\sigma}_{+}).
\end{equation}
This allows us to transform the original first driving term into a time independent one ${\cal \hat{H}}_0^{L_1} = - \hbar \Omega_1  (\hat{\sigma}_{-} + \hat{\sigma}_{+})$, leaving the excitation number unchanged. The authors of~\cite{ballester2012quantum} assume this to be the most significant term and treat the others perturbatively by going into the interaction picture with respect to ${\cal \hat{H}}_0^{L_1} $, ${\cal \hat{H}}^{I} (t) = e^{i {\cal \hat{H}}_{0}^{L_1} t/\hbar} ({\cal \hat{H}}^{L_1}   - {\cal \hat{H}}_0^{L_1})    e^{-i {\cal \hat{H}}_{0}^{L_1} t/\hbar } $. Using the rotated spin basis $\ket{\tilde{\pm}} = (\ket{-} \pm \ket{+})/\sqrt2$, we have
\begin{equation}
\begin{aligned}
{\cal \hat{H}}^{I} (t) &=   -\hbar  \frac{\omega_q-\omega_1}{2}    ( e^{-i 2  \Omega_1  t}  |\tilde{+} \rangle \langle \tilde{-}|    +     {\rm H.c.})   + \hbar (\omega-\omega_1) \hat{a}^{\dagger} \hat{a}\\
  & - \frac{\hbar g }{2} \left( \left\{  |\tilde{+} \rangle \langle \tilde{+}| - |\tilde{+} \rangle \langle \tilde{+}| + e^{-i 2  \Omega_1 t} |\tilde{+} \rangle \langle \tilde{-}| \right. \right. 
  -  \left.\left. e^{i 2  \Omega_1  t}|\tilde{-} \rangle \langle \tilde{+}|  \right\} \hat{a} + {\rm H.c.} \right) \\
  &-  \frac{\hbar \Omega_2 }{2} \left(   \left\{  |\tilde{+} \rangle \langle \tilde{+}| - |\tilde{+} \rangle \langle \tilde{+}| - e^{-i 2  \Omega_1  t} |\tilde{+} \rangle \langle \tilde{-}|  \right. \right. 
  \left. \left. + e^{i 2 \Omega_1  t}|\tilde{-} \rangle \langle \tilde{+}|  \right\} e^{i (\omega_2-\omega_1) t} + {\rm H.c.} \right).
  \end{aligned}
\end{equation}
By tuning the parameters of the external driving fields as $\omega_1-\omega_2=2 \Omega_1 ,$ we can choose the resonant terms in this time-dependent Hamiltonian. Then, in case of having a relatively strong first driving, $ \Omega_1$, we can approximate the expression above by a time-independent effective Hamiltonian as
\begin{equation}\label{eq:HamilEff}
{\cal \hat{H}}_{\rm eff}
=  \hbar (\omega-\omega_1) \hat{a}^{\dagger} \hat{a} + \frac{\hbar\Omega_2}{2} \hat{\sigma}_z   -  \frac{\hbar g}{2} \hat{\sigma}_x (\hat{a}+\hat{a}^{\dagger}).  
\end{equation}
We note the resemblance between the original Hamiltonian of eq. \eqref{eq:HamilDiag} and the effective Hamiltonian of \eqref{eq:HamilEff}. While the value of the coupling $g$ is fixed in eq. \eqref{eq:HamilEff}, we can still tailor the effective parameters by tuning the amplitude and frequency of the coherent drive fields. If we can reach values such that $\Omega_2  \sim (\omega-\omega_1) \sim g/2$, the dynamics of the original system will simulate that of a two-state atom coupled to the resonator with a relative interaction strength beyond the SC regime \textemdash{ideally} in the ultrastrong coupling (USC)/deep strong coupling regimes (DSC). The coupling regime reproduced through the effective Hamiltonian \eqref{eq:HamilEff} can be quantified by the ratio $g_{\rm eff} / \omega_{\rm eff}$, where $g_{\rm eff} \equiv g/2$ and $ \omega_{\rm eff}\equiv\omega-\omega_1$.

Coming now to superconducting flux qubits\index{Qubit! flux}, the macroscopic nature of the persistent current states allows
strong coupling to the other elements of the circuit. Strong anharmonicity enables the separation of two lowest energy levels from the rest of the ladder. In the setup of~\cite{yoshihara2017superconducting}, the flux qubit is coupled to an LC oscillator via a tunable inductance $L_{\rm c}$. A tunable coupler is also employed, consisting of four parallel Josephson junctions\index{Josephson! junction}. As a consequence of the inductive coupling a flux qubit and the LC oscillator via Josephson junctions, circuits were realized with $g/\omega_{r} >1$ and $g/\omega_{q} \gg 1$, prompting  ground-state-based entangled-pair generation. The Hamiltonian of the LC oscillator\index{LC oscillator} can be written as $\hat{\mathcal{H}}_{\rm o} = \hbar \omega_{\rm o}(\hat{a}^{\dagger}\hat{a} + 1/2)$, where $\omega_{\rm o} = 1/\sqrt{(L_0+L_{\rm qc})C}$ is the resonance frequency, $L_{\rm 0}$ is the inductance of the superconducting lead, $L_{\rm qc}(\simeq L_{\rm c})$ is the inductance across the qubit and coupler, $C$ is the capacitance,  and $\hat{a}$ $(\hat{a}^{\dagger})$ are the usual annihilation (creation) operators. The freedom of choosing $L_0$ for large zero-point fluctuations\index{Zero-point! fluctuations} in the current, $I_{\rm zpf}$, is one of the advantages of lumped-element LC oscillators over coplanar-waveguide resonators for the experiment. In addition, a lumped-element LC oscillator has only one resonant mode. Such a setup readily realizes the Rabi model \textemdash{one} of the simplest possible quantum models of qubit-oscillator systems \textemdash{with} no additional energy levels lying in the range of interest.

The coupling Hamiltonian can be written as $\hat{\mathcal{H}}_{\rm c}=\hbar g \hat{\sigma}_z (\hat{a} + \hat{a}^{\dagger})$,
where $\hbar g=MI_{\rm p}I_{\rm zpf}$ (where $I_{\rm p}$ is the maximum persistent current in the flux qubit\index{Qubit! flux}) is the coupling energy and $M (\simeq L_{\rm c})$ is the mutual inductance between the qubit and the LC oscillator. Importantly, a Josephson-junction\index{Josephson! junction} circuit is used as a large inductive coupler, which together with the large $I_{\rm p}$ and $I_{\rm zpf}$ allows us to achieve deep strong coupling. The total Hamiltonian of the circuit is then given by
\begin{equation}
\hat{\mathcal{H}}_{\rm total} = -\frac{\hbar}{2}(\Delta \hat{\sigma}_x + \varepsilon \hat{\sigma}_z) + \hbar \omega_{\rm o}(\hat{a}^{\dagger}\hat{a} + \frac{1}{2}) + \hbar g\hat{\sigma}_z(\hat{a} + \hat{a}^{\dagger}).
\end{equation}
Nonlinearities in the coupler circuit lead to higher-order terms in $(\hat{a} + \hat{a}^{\dagger})$.
The leading-order term can be written as $\propto \hbar g(\hat{a} + \hat{a}^{\dagger})^2$ \textemdash{the} familiar $A^2$ term. This contribution can be eliminated from $\hat{\mathcal{H}}_{\rm total}$ by means of a variable transformation.

We will close out this discussion on the nonlinearity of radiation-matter interaction accessible to circuit QED by a brief mention to the experiment of~\cite{mlynek2014observation}, characterizing the subradiant decay of a two-qubit ensemble. We summarize below some of the key findings reported therein, verifying the prediction on what Dicke termed {\it single-atom superradiance} in 1954~\cite{Dicke1954}:

\noindent {\bf (i)} The power emitted for two qubits prepared in their excited state is twice as large as in the single-qubit case. The deviation from the average single-qubit level is stronger at initial times and recedes later on.\\
{\bf (ii)} The two-qubit collective decay rate is initially smaller and then accelerates to exceed the single-qubit decay rate, an effect attributed to the phase synchronization due to the interaction between the two emitters. This manifests the presence of entanglement in the only allowed decay channel formed between the state $\ket{ee}$ and the ``bright'' state $\ket{B} \equiv (1/\sqrt{2})(\ket{eg} + \ket{ge})$. Owing now to the correlations, the decay rate from $\ket{B}$ to the ground state of the ensemble, $\ket{gg}$, is twice that of the single decay rate of the states $\ket{eg}$ or $\ket{ge}$. \\
{\bf (iii)} Superradiant decay is not subject to a purely exponential law, as its rate is not always proportional to the number of excitations in the system. 
We will explore interactions of the like in more detail when we reach the section on waveguide QED. Finally, concerning the study of collective phenomena, the so-called {\it no-go theorem}\index{Dicke! no-go theorem} for the occurrence of criticality at equilibrium in the Dicke Hamiltonian of cavity QED was argued to also hold for the microscopic description of circuit QED with a minimal coupling in~\cite{viehmann2011superradiant}. This demonstration followed as a response to the claim, made in~\cite{nataf2010no}, that a superradiant phase transition should be in principle possible in a circuit QED system (as opposed to cavity QED) by tailoring the ratio $E_J/E_C$. 
\begin{table*}[ht]
\begin{center}
\resizebox{\textwidth}{!}{\begin{tabular}{|l|l|c|c||c|c|c|c|}\hline
{\bf parameter}&{\bf symbol}&{\bf 3D optical}&{\bf 3D microwave}&{\bf Blais}~\cite{Blais2004QED}&{\bf Schoelkopf}~\cite{Bishop2008}&{\bf Martinis}~\cite{Martinis2016}&{\bf Yoshihara}~\cite{yoshihara2017superconducting} \\ \hline %
resonance/transition frequency&$\omega_{r}/ 2\pi$, $\Omega/ 2 \pi$&$350 \, \rm{THz}$&$51 \, \rm{GHz}$& $10 \, \rm{GHz}$&$6.3\, \rm{GHz}$&$6.78\, \rm{GHz}$, $5.4\, \rm{GHz}$&$5.71\, \rm{GHz}$, $0.44\, \rm{GHz}$\\ \hline %
vacuum Rabi frequency&$g/\pi$, $g/\omega_r$&$220 \, \rm{MHz}$, $3 \times 10^{-7}$&$47 \, \rm{kHz}$, $1 \times 10^{-7}$&$100 \, \rm{MHz}$, $5\times 10^{-3}$&$347 \, \rm{MHz}$, $2.75\times 10^{-2}$&$174 \, \rm{MHz}$, $1.3\times 10^{-2}$&$15.26 \, \rm{GHz}$, $1.34$\\ \hline %
transition dipole&$d/e a_0$&$\sim 1$&$1 \times 10^3$&$ 2\times 10^4$&$\sim 10^4$ &$\sim 10^4$ & $\sim 10^6$ \footnote{magnetic dipole moment $m/\mu_B$ (where $\mu_B$ is the Bohr magneton)}\\ \hline%
cavity lifetime&$1/\kappa$&$10 \, \rm{ns}$ & $1 \, \rm{ms}$&$160 \, \rm{ns}$&$530 \, \rm{ns}$ & $37 \, \rm{ns} $& $\sim 100\, \rm{ns}$\\ \hline %
atom lifetime &$T_1\equiv 1/\gamma$&$ 61\,\rm{ns}$&$30 \, \rm{ms}$&$ 2 \, \rm{\mu s}$&$ 1.7 \, \rm{\mu s}$& $20\rm{\mu s} \leq T_1 \leq 40\rm{\mu s} $ & $\sim 600\,\rm{ns}$\\ \hline %
atom transit time&$t_{\rm transit}$&$\ge 50 \, \rm{\mu s}$&$100 \, \rm{\mu s}$&$\infty$&$\infty$&$\infty$&$\infty$\\ \hline %
critical atom number&$N_0=2\gamma\kappa/g^2$&$6 \times 10^{-3}$&$3\times10^{-6}$&$\leq 6 \times 10^{-5}$ &$ \leq 2 \times 10^{-6} $& $\sim 2 \times 10^{-5}$& $\leq 2 \times 10^{-8}$\\ \hline %
critical photon number&$m_0=\gamma^2/2g^2$&$3\times10^{-4}$&$3\times10^{-8}$&$\leq 1 \times 10^{-6}$& $ \leq 1.5 \times 10^{-7} $& $\sim 7 \times 10^{-9}$& $6 \times 10^{-10}$\\ \hline %
\# of vacuum Rabi flops&$n_{\rm Rabi}= 2g/(\kappa+\gamma)$ &$\sim 10$& $\sim 5$ &$\sim 10^2$&$ 8.8 \times 10^2$&$20$& $\geq 6.5 \times 10^3$ \\ \hline %
\end{tabular}}
\end{center}
\caption{Key rates and QED parameters for optical~\cite{hood2000atom} and microwave~\cite{Raimond2001} atomic systems using 3D cavities, compared against one-dimensional superconducting circuits in the columns following the double vertical line, showing the possibility for attaining the strong cavity QED limit ($n_{\rm Rabi} \gg 1$). The coupling strength over the resonance frequency takes the form $g/\omega=(L/r)\sqrt{2\alpha/\pi}$, where $\alpha \approx 1/137$ is the fine-structure constant, $L$ is the length and $r$ is the transverse extent of the microwave resonator \cite{schoelkopf2008wiring}. To maximize the coupling strength the atom must fill the transverse direction ($L/r \sim 1$).}
\label{table:parameters}
\end{table*}

Let us now shift our focus to the dispersive regime of QED, which proves advantageous for quantum non-demolition measurements\index{Non-demolition measurement}~\cite{poizat1994characterization,grangier1998quantum,Blais2004QED}, see sec.~\ref{sssec:sol}. Following the polaron transformation\index{Polaron! transformation}~(\ref{polaron}), the JC Hamiltonian can be recast in the form~\cite{BishopJC,Carbonaro}
\begin{equation}\label{JCDisp}
\hat{H}_{\rm disp}=\hbar \omega_c \hat{a}^{\dagger}\hat{a} + \frac{1}{2}(\omega_c-\hat{\Delta})\hat{\sigma}_{z},
\end{equation} 
where we have introduced a new operator ($\delta \equiv \omega_q-\omega_c$)
\begin{equation}\label{Deltaop}
\hat{\Delta}=\left(\delta^2 + 4g^2 \hat{N}\right)^{1/2},
\end{equation}
depending on the operator of system excitations, $\hat{N}=\hat{a}^{\dagger}\hat{a}+\hat{\sigma}_{+}\hat{\sigma}_{-}$. In the dispersive regime, where $n_{\rm sat, disp}=[\delta/(2g)]^2 \gg 1$, we can expand the square root in \eqref{Deltaop} in powers of $\hat{N}$. The first-order term yields the linear dispersive Hamiltonian with an {\it ac Stark shift}\index{Stark shift} (see \textit{e.g.}, eq. (12) of~\cite{Blais2004QED}, generalized to eq. (3.8) of~\cite{KochTransmon2007} for a transmon qubit\index{Transmon qubit}). Keeping now the second-order term, when the nonlinearity is not negligible, brings us to the dynamic behaviour of the quantum Duffing oscillator\index{Duffing oscillator}, giving rise to a two-photon transition\index{Two-photon!transition} term $\hbar\tilde{\chi} {{\hat{a}}^{\dagger\, 2}} {\hat{a}}^2$ in the Hamiltonian, where we have assumed that the excitation of the qubit is negligible such that $\braket{\hat{\sigma}_z}\approx \hat{\sigma}_z\approx -1$. As we saw in sec.~\ref{ssec:cqedearly}, nonlinearity manifests itself as complex-amplitude bistability, evidenced by the solution of a nonlinear Fokker--Planck equation\index{Fokker--Planck equation} for the positive $P$-representation~(\ref{peq}). See also~\cite{DrummondWallsDuffing} for a discussion on the dispersive optical bistability in the Duffing oscillator\index{Duffing oscillator}. Following the predictions of~\cite{Blais2004QED}, the strong dispersive regime was accessed in the experiment of~\cite{schuster2007resolving}, where the qubit transition energy was resolved into a separate spectral line for each photon number state of the microwave field. The intensity of each line in the spectrum is a measure of the probability of finding the corresponding Fock state in the cavity, without having to absorb the photon. Extending the well-known results from the theory of resonance fluorescence\index{Resonance fluorescence! dressing of dressed states} and the dressing of the dressed states in cavity QED\index{Dressing of the dressed states}, a generalized JC Hamiltonian in its dispersive limit with a coherent drive was employed to model the AC Stark effect\index{Stark shift} and the Mollow triplet\index{Mollow! triplet} spectrum, measured experimentally in the circuit QED setup of~\cite{baur2009measurement}. 

For a cavity with a mean photon number $\overline{n} \gg n_{\rm sat, disp}$, recent experiments reveal the importance of terms that do not conserve the excitation number in the system Hamiltonian (the so-called {\it non-RWA terms}) allowing for resonances occurring between ladders produced by RWA terms \cite{Martinis2016}. This requires a modification of eq. \eqref{generalRWA} together with the inclusion of charge-like states which are not bound by the Josephson cosine-potential term\index{Josephson! oscillations}. Terms outside the RWA must also be retained to account for a Bloch-Siegert shift\index{Bloch!-Siegert shift}, visible when measuring the response of a flux qubit\index{Qubit! flux} embedded in the coplanar waveguide resonator. In the latter configuration, super-harmonic resonances occur for high drive amplitudes at points where the ratio $\omega_{q}/\omega_{r}$ is a natural number larger than unity \cite{FluxSHR}. Some key rates and parameters for circuit QED systems discussed in this section are presented in Table \ref{table:parameters}, compared against earlier configurations employing 3D cavities. With these results in hand, we revisit the overarching topic of light-matter interaction \textemdash{extended} to a continuum of modes inside a waveguide \textemdash{in} section~\ref{sec:waveguideQED}. 

Operating in the strong dispersive regime has allowed the characterization of mechanical quantum states, recently reported in~\cite{vonLupke2022}. The parity of the prepared phonon Fock states is determined by means of spectroscopy and Ramsey measurements\index{Ramsey! interferometry}. 

\subsection{Quantization in the ultrastrong coupling regime}

In the ultrastrong light-matter interaction regime, the coupling between a system and its environment ought to include counter-rotating terms, while it is important whether the coupling is mediated by an electric or a magnetic field -- capacitive or inductive circuits~\cite{Bamba2014}. For a semi-infinite transmission line coupled to a resonator where the inductor is in series with a CPB, the system-environment coupling Hamiltonian contains a frequency-dependent bare loss rate scaling as $\sim 1/|1-i\zeta(\omega)|$, where $\zeta(\omega)=\omega Z_T C_c$; $Z_T$ is the characteristic impedance of the transmission line and $C_c$ is the (modulated) contact capacitance. We focus on the general case of {\it mixed linear coupling} and consider a circuit comprising a nonlinear network of degrees of freedom $\boldsymbol{\phi}$ linearly coupled through a capacitor $C_g$ and an inductor $L_g$ to a transmission line at one end -- see fig. 1 of~\cite{ParraRodriguez_2018}. Given that the network contains a nonlinear potential in flux variables -- as is the case in the presence of Josephson junctions\index{Josephson! junction} -- it is convenient to select the flux variables as the set of position-like coordinates. The Lagrangian of this circuit can be written in terms of a discrete set of flux variables describing the network, collected in the column vector $\boldsymbol{\phi}$, and a flux field $\Phi(x,t)$,
\begin{equation}\label{eq:LagrcQED}
\begin{aligned}
L&=\frac{1}{2}\dot{\boldsymbol{\phi}}^T \mathsf{A}\dot{\boldsymbol{\phi}}-\frac{1}{2}\boldsymbol{\phi}^T \mathsf{B}^{-1}\boldsymbol{\phi}-V(\boldsymbol{\phi})+\int_{0}^{L}dx\,\left[\frac{c}{2}\dot{\Phi}(x,t)^2-\frac{1}{2l}(\Phi'(x,t))^2\right]\\
 &+C_g\left(\frac{\dot{\Phi}(0,t)^2}{2}-\dot{\boldsymbol{\phi}}^T\boldsymbol{a} \dot{\Phi}(0,t)\right)-\frac{1}{L_g}\left(\frac{\Phi(0,t)^2}{2}-\boldsymbol{\phi}^T\boldsymbol{b}\Phi(0,t)\right),
 \end{aligned}
\end{equation}
where $\mathsf{A}=\tilde{\mathsf{A}}+C_g \boldsymbol{a}\boldsymbol{a}^T$ and $\mathsf{B}^{-1}=\tilde{\mathsf{B}}^{-1}+\boldsymbol{b}\boldsymbol{b}^T/L_g$. Here, $\tilde{\mathsf{A}}$ and $\tilde{\mathsf{B}}^{-1}$ are the capacitance and inductance submatrices of the network, respectively, while $\boldsymbol{a}$ and $\boldsymbol{b}$ are coupling vectors to the network from the transmission line with finite norm. No specific description of the network in terms of branch or node flux variables is assumed; nevertheless, the network must be non-trivially connected to the common ground in order for current to circulate through $C_g$ and $L_g$. The limits $C_g\rightarrow0$ and $L_g\rightarrow\infty$ disconnect the transmission line from the network at the position $x=0^{+}$. The corresponding Euler-Lagrange equations of motion read:
\begin{eqnarray}
c\ddot{\Phi}(x,t)&=&\frac{\Phi''(x,t)}{l}\label{eq:EulLag_TL_LCcoup_Network11},\\
\frac{\Phi'(0,t)}{l}&=&C_g\left(\ddot{\Phi}(0,t)-\boldsymbol{a}^T\ddot{\boldsymbol{\phi}}\right)+\frac{1}{L_g}\left({\Phi}(0,t)-\boldsymbol{b}^T{\boldsymbol{\phi}}\right),\label{eq:EulLag_TL_LCcoup_Network12}\\
 \mathsf{A}\ddot{\boldsymbol{\phi}}+\mathsf{B}^{-1}{\boldsymbol{\phi}}&=&C_g\boldsymbol{a}\ddot{\Phi}(0,t)+\frac{1}{L_g}\boldsymbol{b}\Phi(0,t)-\frac{\partial V(\boldsymbol{\phi})}{\partial \boldsymbol{\phi}}\label{eq:EulLag_TL_LCcoup_Network13}.
\end{eqnarray}
For simplicity we assume that the transmission line has finite length $L$. A textbook analysis would introduce a normal-mode decomposition of the flux field in basis functions, as $\Phi(x,t)=\sum_n \Phi_n(t)u_n(x)$. There is an issue in this case, however, since the coupling with the network at the endpoint $x=0^{+}$ involves the second derivative with respect to time of the flux field. Even if all the network variables were set to zero, we would still have a boundary condition that involving the separation constant, according to eq.~\eqref{eq:EulLag_TL_LCcoup_Network12}. Furthermore, setting the network variables to zero would be inconsistent, since the transmission line is a source term in equation~\eqref{eq:EulLag_TL_LCcoup_Network13}. To address the issue, the authors of~\cite{ParraRodriguez_2018} assume a separation constant that is dependent upon the boundary condition, by introducing a length parameter $\alpha$, the value of which is set by an optimization procedure; namely, they require that there be no coupling amongst the transmission line modes in the Hamiltonian representation. 

Following this prescription, the field equations for the transmission line yield the homogeneous eigenvalue problem:
\begin{align}
\ddot{\Phi}_n(t)&=-\omega_n^2\Phi_n(t),\label{eq:EVP_TL_LCcoup_Network_eq0}\\
u_n''(x)&=-k_n^2u_n(x),\label{eq:EVP_TL_LCcoup_Network_eq1}\\
u_{n}'(0)&=-k_n^2\alpha u_n(0)+\frac{1}{\beta}u_n(0),\label{eq:EVP_TL_LCcoup_Network_eq2}\\
u_n(L)&=0,\label{eq:EVP_TL_LCcoup_Network_eq3}
\end{align}
where $\omega_n^2=k_n^2/lc$, and the authors have assumed without loss of generality a short to ground boundary condition at $x=L$. For fixed parameters $\alpha$ and $\beta$, the system of equations~\eqref{eq:EVP_TL_LCcoup_Network_eq1} -- \eqref{eq:EVP_TL_LCcoup_Network_eq3} defines a generalized eigenvalue problem; the corresponding generalized eigenfunctions satisfy the following orthogonality conditions:
\begin{align}
 \langle u_n,u_m\rangle_{\alpha}&=c\left(\int_{0}^{L}dx\, u_n(x) u_m(x)+  \alpha u_n(0) u_m(0)\right)=N_{\alpha}\delta_{nm},\label{eq:TL_LCcoup_Network_ortho_1}\\
\langle u_n,u_m\rangle_{1/\beta}&=\frac{1}{l}\left(\int_{0}^{L}dx\, u_n'(x) u_m'(x)+ \frac{1}{\beta} u_n(0) u_m(0)\right)=\omega_n^2 N_{\alpha}\delta_{nm},\label{eq:TL_LCcoup_Network_ortho_2}
\end{align}
where $N_{\alpha}$ is a normalization constant [F]. Expanding in this basis, we write the Hamiltonian derived from the Lagrangian~\eqref{eq:LagrcQED} as
\begin{equation}
H=\frac{1}{2}{\boldsymbol{P}}^T\mathsf{C}^{-1}{\boldsymbol{P}}+\frac{1}{2}\boldsymbol{X}^T \mathsf{L}^{-1}\boldsymbol{X}+V(\boldsymbol{\phi}),\label{eq:Ham_LCcoup_Network}
\end{equation}
with the vector of fluxes 
\begin{equation}
  \boldsymbol{X}=
  \begin{pmatrix}
    \boldsymbol{\phi}\\
    \boldsymbol{\Phi}
  \end{pmatrix},
\end{equation}
and the conjugate charge variables to the fluxes, $\boldsymbol{P}=\partial L/\partial \boldsymbol{X}\equiv(\boldsymbol{q}^T,\boldsymbol{Q}^T)^T$. The inverse capacitance matrix is 
\begin{equation}
	\mathsf{C}^{-1}= \begin{pmatrix}
	\mathsf{A}^{-1}+ \frac{C_g^2 |\boldsymbol{u}|^2}{D} \mathsf{A}^{-1}\boldsymbol{a}\boldsymbol{a}^T \mathsf{A}^{-1}&   \frac{C_g}{D} \mathsf{A}^{-1} \boldsymbol{a}\boldsymbol{u}^T\\
	 \frac{C_g}{D} \boldsymbol{u}\boldsymbol{a}^T\mathsf{A}^{-1}& \frac{1}{N_{\alpha}}\mathbb{1}+ \frac{1}{|\boldsymbol{u}|^2}\left( \frac{1}{D}- \frac{1}{N_{\alpha}}\right) \boldsymbol{u} \boldsymbol{u}^T
	\end{pmatrix},\nonumber 
\end{equation}
and the inverse inductance matrix reads
\begin{equation}
 \mathsf{L}^{-1}=\begin{pmatrix}
	\mathsf{B}^{-1}& - \boldsymbol{b}\boldsymbol{u}^T/L_g\\
	- \boldsymbol{u}\boldsymbol{b}^{T}/L_g & N_{\alpha} (\omega_n^2) + e\boldsymbol{u} \boldsymbol{u}^T
	\end{pmatrix},
\end{equation}
with $D= N_{\alpha}+|\boldsymbol{u}|^2(d- C_g^2 \boldsymbol{a}^T \mathsf{A}^{-1}\boldsymbol{a})$. In the above expressions, $\boldsymbol{u}\equiv (u_0(0), u_1(0),...u_n(0),\ldots)^T$ is the normalizable coupling vector, and we define $d \equiv C_g-c \alpha$ and $e\equiv 1/L_g - 1/\beta l$; $\mathbb{1}$ is the identity matrix of infinite dimension and $(\omega_n^2)=\mathrm{diag}(\omega_0^2,\omega_1^2,...)$ is the diagonal matrix of squared eigenfrequencies. The quantity $\boldsymbol{u}^T\left[N_\alpha \mathbb{1}+d\boldsymbol{u}\boldsymbol{u}^T\right]^{-1}\boldsymbol{u}=1/C_g$ is finite unless $C_g$ is zero.

We can now impose the requirement that there be no mode-mode coupling for the transmission line. We select the parameters $\alpha$ and $\beta$ to satisfy the equations $D=N_{\alpha}$ and $e=0$, to remove the dependence from the couplings in the diagonal elements. 
\begin{align}
\alpha&=\frac{C_g(1-C_g\boldsymbol{a}^T \mathsf{A}^{-1}\boldsymbol{a})}{c},\label{eq:TL_LCcoup_Network_alpha_fix}\\
\beta&=L_g/l.\label{eq:TL_LCcoup_Network_beta_fix}
\end{align}
The frequencies $\omega_n$ are determined from the solution of the eigenvalue problem~\eqref{eq:EVP_TL_LCcoup_Network_eq1}--~\eqref{eq:EVP_TL_LCcoup_Network_eq3}) with the above specified values for $\alpha$ and $\beta$. After this step is completed, we can write the final Hamiltonian as
\begin{equation}
\begin{aligned}
H=&\frac{1}{2}\boldsymbol{q}^T(\mathsf{A}^{-1}+ \frac{C_g^2}{\alpha c} \mathsf{A}^{-1}\boldsymbol{a}\boldsymbol{a}^T \mathsf{A}^{-1})\boldsymbol{q}+\frac{1}{2}\boldsymbol{\phi}^T\mathsf{B}^{-1}\boldsymbol{\phi}+V(\boldsymbol{\phi})+\sum_n \frac{Q_n^2}{2N_{\alpha}}+\frac{N_{\alpha}\omega_n^2\Phi_n^2}{2}\\
&+\frac{C_g}{N_{\alpha}} (\boldsymbol{q}^T\mathsf{A}^{-1} \boldsymbol{a})\sum_n Q_n u_n(0)-\frac{1}{L_g} (\boldsymbol{\phi}^T\boldsymbol{b})\sum_n \Phi_n u_n(0),\label{eq:Ham_LCcoup_Network2}
\end{aligned}
\end{equation}
where we have used the normalization $|\boldsymbol{u}|^2=N_{\alpha}/(\alpha c)$. The process of quantization is completed by promoting the conjugate variables to operators satisfying the commutation relation $[\hat{X}_i,\hat{P}_j]=i\hbar \delta_{ij}$. The resulting quantized Hamiltonian in terms of annihilation and creation operators, related to flux and charge variables by $\hat{\Phi}_n=i\sqrt{\hbar/(2\omega_n N_\alpha)}(\hat{a}_n-\hat{a}_n^{\dagger})$ and $\hat{Q}_n=\sqrt{\hbar\omega_n N_\alpha/2}(\hat{a}_n+\hat{a}_n^{\dagger})$,
\begin{equation}
\begin{aligned}
\hat{H}=&\frac{1}{2}\hat{\boldsymbol{q}}^T(\mathsf{A}^{-1}+ \frac{C_g^2}{\alpha c} \mathsf{A}^{-1}\boldsymbol{a}\boldsymbol{a}^T \mathsf{A}^{-1})\hat{\boldsymbol{q}}+\frac{1}{2}\hat{\boldsymbol{\phi}}^T\mathsf{B}^{-1}\hat{\boldsymbol{\phi}}+V(\hat{\boldsymbol{\phi}})+\sum_n \hbar\omega_n \hat{a}_n^\dagger \hat{a}_n\\
&+C_g\sqrt{\frac{\hbar}{2 N_\alpha}} (\hat{\boldsymbol{q}}^T\mathsf{A}^{-1} \boldsymbol{a})\sum_n (\hat{a}_n+\hat{a}_n^\dagger)\sqrt{\omega_n} u_n(0)\\
&-\frac{i}{L_g}\sqrt{\frac{\hbar}{2 N_\alpha}} (\hat{\boldsymbol{\phi}}^T\boldsymbol{b})\sum_n (\hat{a}_n-\hat{a}_n^{\dagger}) \frac{u_n(0)}{\sqrt{\omega_n}}.
\end{aligned}
\end{equation}
This Hamiltonian is as exact as the starting Lagrangian of eq.~\eqref{eq:LagrcQED}. From its form we deduce that the (capacitive) coupling constants $\tilde{g}_n\propto \sqrt{\omega_n}u_n(0)$ do not grow without bound. It has further been proven in~\cite{ParraRodriguez_2018} that the large $n$ behaviour of $u_n(0)$ is $1/n$, while $\omega_n$ asymptotically grow as $n$. It then follows that $\tilde{g}_n\sim n^{-1/2}$, whence there is no need for the introduction of an ultraviolet cutoff\index{Frequency cutoff! ultraviolet} extrinsic to the model of eq.~\eqref{eq:LagrcQED}. Instead, the correct choice of modes to expand in has provided us with a natural length scale -- intrinsic to the model -- that corresponds to an intrinsic ultraviolet cutoff (see also ch. 3 of~\cite{AParraThesis}). In the absence of any cutoff, the coupling rate to mode number $n$ scales as $\sqrt{n}$ [$g_n=g_0 \sqrt{2n+1}$] and leads to the divergences found in typical multimode extensions\index{Multimode! quantum Rabi model! extension} of the quantum Rabi model. A renormalization of the bare atomic parameters on the basis of the above discussion, however, alleviates the divergence of the Lamb shift\index{Lamb! shift}~\cite{Gely2017} (see also the effective Rabi model derived in~\cite{malekakhlagh2016origin}).

In fact, the transition from the USC to the DSC regime presents its own intricacies. The experimentally obtained (and numerically fitted via the quantum Rabi model with optimized parameters) spectrum of a circuit-QED system undergoes multiple qualitative transformations in this intermediate regime, such that multiple coupling regimes can be identified, each possessing its own unique spectral features. The different spectral transformations may be associated with crossings between energy level differences as well as with changes in the symmetries of the energy eigenstates~\cite{Yoshihara2017}. In the recent work of Ao and collaborators, the frequency $\Delta$ of a flux qubit\index{Qubit! flux} coupled to a $\lambda/4$ coplanar wavequide resonator is renormalized by the multiple modes, to define the relative Lamb shift\index{Lamb! shift} $(\Delta-\Delta_0)/\Delta_0$ in the DSC regime as
\begin{equation}\label{eq:cutoffDelta}
 \Delta=\Delta_0\exp\left\{-2\left(\frac{g_1}{\omega_1}\right)^2 \sum_{k=0}^{\infty}\frac{1}{(2k+1)\left[ 1 + \frac{(2k+1)^2}{n_{\rm cutoff}^2}\right]}\right\},
\end{equation}
where $n_{\rm cutoff}\equiv\omega_{\rm cutoff}/\omega_1$. The natural cutoff frequency $\omega_{\rm cutoff}$\index{Frequency cutoff} is the ratio of the characteristic impedance of the waveguide and the linear inductance of the Josephson junction\index{Josephson! junction} coupling the flux qubit\index{Qubit! flux} to the waveguide resonator~\cite{ashhab2023highfrequency, malekakhlagh2017cutoff}. The renormalization remains finite even in the case of the strong coupling to an infinite number of modes~\cite{Ao2023}. For $n_{\rm cutoff} \gg 1$, the sum in eq.~\eqref{eq:cutoffDelta} is approximated by $0.635 + 0.5\,\log{n_{\rm cutoff}}$~\cite{ashhab2023highfrequency}. The slow logarithmic dependence of the sum, which is $\sim 1$, entails that the Lamb shift\index{Lamb! shift} does not vary appreciably as a function of $n_{\rm cutoff}$ in comparison to other parameters in the Hamiltonian. In a more general configuration, the ground-state energy shift of a single dipole owing to its coupling to the electromagnetic vacuum in a confined geometry has been recently calculated in~\cite{Rocio2023}. The observable energy difference contains contributions accounting for the purely electrostatic corrections and the genuine vacuum shifts from the transverse modes and the cavity mode.

\subsection{Control and transfer of quantum information in circuit QED}
\label{ssec:contrtransQI}

The remarkable coherence properties of artificial atoms alongside strong coupling, discussed in sec. \ref{ssec:evolcQED}, enhance the storage and transfer of quantum information. The concept of a quantum bus comprising a resonator coupled to several entangled qubits, facilitated by dispersive quantum non-demolition\index{Non-demolition measurement} readout, has been tied to the inception of circuit QED (see~\cite{Devoret1169} and section VIII of~\cite{Blais2004QED}). Current circuit QED implementations have also proved promising for entanglement generation and quantum information transfer. The generation of entangled photon states in a deterministic fashion across two spatially separated microwave resonators $A$ and $B$ was accomplished by manipulating the photon states with a pair of superconducting phase qubits in~\cite{wang2011deterministic}. The realization of multiphoton entangled states\index{Multiphoton! state! entangled} of a two-mode field, called N$00$N states\index{N$00$N state}\index{State! N$00$N}:
\begin{equation*}
 \ket{\psi}=\frac{1}{\sqrt{2}}(\ket{N}_A \ket{0}_B + \ket{0}_A \ket{N}_B),
\end{equation*}
was used to benchmark the process. The two-resonator photon states are then characterized using bipartite Wigner tomography \textemdash{a} special means to distinguish entanglement from an incoherent ensemble. Using the Josephson mixer\index{Josephson! mixer} \textemdash{a} superconducting circuit parametrically coupling two resonators of different frequencies $\omega_A$ and $\omega_B$ via a pump field of a frequency equal to their sum, $\omega_A + \omega_B$ spatially separated two-mode squeezed states were realized in the experiment of~\cite{flurin2012generating}. 

The dynamical evolution of a QED system consisting of a two-state atom coupled by two-photon interactions\index{Two-photon!interaction} to two degenerate radiation field modes of an ideal resonator is discussed in~\cite{Migliore2022}, with the radiation field initially prepared in N$00$N and generalized related states. The evolution crucially depends on the parity of $N$ (see also~\cite{ANapoli_1997}).  An effective perturbative Hamiltonian for a superconducting $\Delta$-type qutrit\index{Qutrit} (simplified to a $\Lambda$-type) coupled to two microwave resonators in the dispersive regime has been derived in~\cite{QiShifan2023} to model the generation of bipartite entangled states, such as Bell\index{Bell! state} and N$00$N states\index{N$00$N state}. The scheme takes advantage of the Stark shift\index{Stark shift} in the qutrit transition frequency, which depends on the number of excitations. Furthermore, for a circuit QED system described by the dispersive Hamiltonian\index{Dispersive! Hamiltonian} $\hat{H}/\hbar=-(\chi/2)\hat{a}^{\dagger}\hat{a}\hat{\sigma}_z$ [see eqs.~\eqref{2mham1} and~\eqref{JCDisp}] and subject to a large cavity displacement $\alpha_0$, a technique called {\it echoed conditional displacement}\index{Displacement! conditional, echoed} allows for operations enacted in a duration of $1/(\chi \alpha_0)$ instead of $1/\chi$ to protect the quantum interference of cat states against decoherence~\cite{PanXiaozhou2023}. Single-step schemes for realizing quantum swap gates and generating two-mode entangled coherent states in circuit QED are discussed in~\cite{sym14091951}.

In a recent setup, two cavities each containing a superconducting qubit have been coupled via a third superconducting qubit in a fashion precluding energy exchange between the qubit system and the cavities, as well as the interaction between the two cavities mediated by the coupling qubit \cite{NJPEntanglementT}. Building upon the propagation of multiphoton wavepackets\index{Multiphoton! wavepacket} and the generation of entanglement between stationary and travelling quantum states [see \textit{e.g.}, \cite{cat}], a very recent experiment reports on the efficient deterministic entanglement of two remote transmon qubits through a driven microwave two-photon transition\index{Two-photon!transition} \cite{PRLremoteEnt}. The properties of a quantum many-body localized (MBL) phase\index{MBL phase}\index{Many-body! localization}, arising due to the interplay of interactions and strong disorder between superconducting qubits, have been experimentally assessed in~\cite{chiaro2019direct}. A tomographic reconstruction of single and two-qubit density matrices was employed for the determination of spatial and temporal entanglement growth between the localized sites. 

In circuit QED, it is possible to decouple the qubit from and the classical part of the cavity field, as has been recently shown in~\cite{QubitCloaking2022}. This effect has interesting consequences such as preventing the vacuum Rabi oscillations from collapsing when a driven cavity is filled with photons, or accelerating dispersive qubit readout. It is feasible thanks to the fact that the cavity and the qubit can be addressed by microwave drives via two different ports. The simple intuition is that a driven cavity develops a field that acts as an effective drive for the qubit~\cite{SponDressState, AlsingCardimona1992}. An additional external cancellation tone can be sent to the qubit to exactly destructively interfere with this effective drive. To see this, we first add a general cavity drive $i\eta(t)(\hat a^\dag - \hat a)$ to the Rabi Hamiltonian, eq.~\eqref{rabiham}. Moving to a displaced frame via the unitary $\hat D(\alpha_t) = \exp(\alpha_t \hat a^\dag - \alpha_t^* \hat a)$, the displaced Hamiltonian $\hat{\mathcal H }_D = \hat D^\dag(\alpha_t)\hat{\mathcal H} \hat D(\alpha_t) -i \hat D^\dag(\alpha_t) \dot{\hat D}(\alpha_t)$ takes the form
\begin{equation}
\hat{\mathcal H}_D(t) = \frac{\hbar \omega_q}{2}\hat \sigma_z + \hbar \omega \hat a^\dag \hat a - \hbar g \hat \sigma_x(\hat a + \hat a^\dag) - \hbar g (\alpha_t + \alpha_t^*)\hat \sigma_x,
\end{equation}
where the complex amplitude has been chosen as $\alpha_t = \int_0^t \eta(\tau) e^{i\omega(\tau - t)} d\tau$ to exactly cancel the cavity drive. Due to the Rabi coupling, this drive is effectively passed to the qubit. Now it becomes evident that adding a cancellation tone of the form $\hbar g(\alpha_t + \alpha_t^*)\hat \sigma_x$, which is not affected by the displacement transformation, will leave the system devoid of drive fields in the displaced frame. Cavity losses at a rate $\kappa$ can be accounted for by replacing $\omega_r \to \omega_r -i\kappa/2$ in the expression for $\alpha_t$, while the result remains unchanged under qubit energy decay or dephasing. The emerging picture is that, from the point of view of the qubit, the cavity is not driven, and only vacuum fluctuations participate in the dynamics. On the other hand, from the point of view of the cavity, the dynamics between the cavity and the qubit develop on top of a classical field $\alpha_t$, which is insensitive to the presence of the qubit. The effect in question has been termed {\it qubit cloaking}\index{Qubit! cloaking}, which can be implemented to improve qubit readout\index{Qubit! readout}~\cite{munozarias2023qubit}.

Multimode cavities consist a novel paradigm of superconducting architecture, promoting lifespans to greatly exceed $10$ms, a marked improvement over their circuit QED counterparts such as transmon qubits\index{Qubit! transmon}~\cite{Chakram2021}. Their unique topology, encountering its own architectural challenges, reduces transpiled\index{Transpilation} coupling distances between qubits compared to the conventional ``sea-of-qubits'' 2D lattice~\cite{stein2023multimode}. Microwave cavity mode control has been so far accomplished via two well-established universal instruction sets: (i) the selective number-dependence arbitrary phase (SNAP) protocol\index{SNAP! protocol}, where displacement operators $\hat{D}(\alpha)$ are applied to the cavity field, and the SNAP gate\index{SNAP! gate} is defined as 
\begin{equation}
 \hat{S}(\phi_1,\ldots,\phi_n)=\sum_n e^{i\phi_n}|n\rangle \langle n|,
\end{equation}
with $|n\rangle$ the Fock states of the cavity field; (ii) the echoed conditional displacement (ECD) gate\index{ECD gate}, defined as
\begin{equation}
 {\rm ECD}(\alpha)=\hat{D}(\alpha/2)|e\rangle \langle g| + \hat{D}(-\alpha/2)|g\rangle \langle e|,
\end{equation}
where $|e\rangle , |g\rangle$, are the excited and ground states of the qubit\index{Qubit}, respectively~\cite{Eickbusch2022}. Owing to the slow
dispersive coupling between the transmon qubits\index{Qubit! transmon} and the cavity
mode, however, as well as to the relatively low coherence time of the transmons, attaining high-fidelity universal intermode control on the
microwave modes remains challenging~\cite{Chakram2021}. Multimode superconducting resonators, on the other hand, have a natural virtual topology capable of storing multiple qubits in one physical device well aligned with quantum simulation problems and exhibiting much longer lifespans in comparison to other planar superconducting hardware~\cite{stein2023multimode}. 

Finally, a circuit QED setup can be used to investigate the nature of quantum jumps. The jumps have been found to be coherent, continuous\index{Quantum! trajectories} and reversible, giving warning signs of their imminent occurrence~\cite{Minev2019}. This is one among the most recent in the line of experiments that follow a long investigation of the coherent and stochastic evolution in quantum trajectories since the proposal of a double-resonance scheme by Dehmelt (see \textit{e.g.}, \cite{KimblePRL} and references therein) allowing for a direct visualization of quantum jumps\index{Quantum! jumps}. 


\section{Trapped ions}\label{sec:ion}
The first demonstrations of single trapped ions date back to the early 1980s~\cite{neuhauser1980localized,wineland1981spectroscopy}. Some years later, ground-state cooling was achieved in the group of Wineland~\cite{diedrich1989laser}. Realizing that these systems could be used for simulating JC physics~\cite{blockley1992quantum} greatly boosted this field. Thus, ion physics constituted an alternative platform for studying phenomena otherwise known from cavity QED. Following this insight, trapped ion physics soon became one of the first playgrounds for quantum-information processing both theoretically and experimentally, which naturally sparked additional interest in these systems. For summaries of the field we refer to the numerous reviews that have been published during the last couple of decades; general concepts are outlined in~\cite{wineland1997experimental,leibfried2003quantum}, discussions on trapped ion quantum information processing are reviewed in Refs.~\cite{steane1997ion,esteve2004quantum,haffner2008quantum}, and the more modern topic trapped-ion quantum simulators is considered in~\cite{johanning2009quantum,blatt2012quantum,monroe2021programmable}. 


\subsection{Model Hamiltonians}\label{ssec:ionham}
\begin{figure}
\includegraphics[width=10cm]{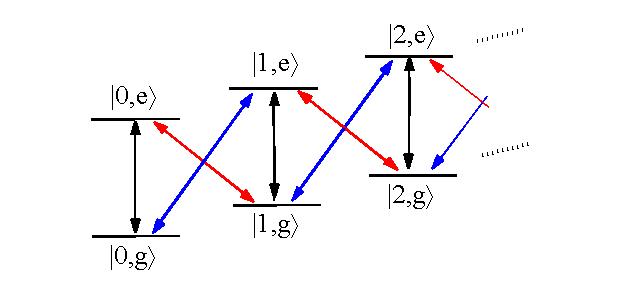} 
\caption{The fundamental transitions in the trapped ion scenario: ($i$) ``Carrier transition'' (black arrows) couples internal atomic states without affecting the external motional state, ($ii$) ``red sideband transiton'' (red arrows) is the traditional JC coupling, while ($iii$) ``blue sideband transition'' (blue arrows) is the aJC coupling.}
\label{fig19}
\end{figure}

In this subsection we give a brief derivation of the JC-type Hamiltonians arising in trapped ion physics. We do not, however, go into any details of the realizations of the trapping potentials. Detailed discussions on the physics of the traps can be found in the reviews~\cite{wineland1997experimental,leibfried2003quantum}. Furthermore, for simplicity we consider the one dimensional situation, noting the generalization to higher dimensions is straightforward. Thus, from now on we assume that the ion is trapped in a harmonic potential
\begin{equation}
\hat{H}_\mathrm{m}=\omega\left(\frac{\hat{p}^2}{2}+\frac{\hat{x}^2}{2}\right),
\end{equation}
with $\omega$ the trapping frequency, and $\hat{p}$ and $\hat{x}$ the ion's center-of-mass momentum and position, respectively. $\hat{H}_\mathrm{m}$ therefore describes the ion's motional degrees of freedom. Note that since the trap couples to the electric charge, it is not dependent upon the internal atomic states (contrary to atomic traps). We continue by assuming the {\it two-level approximation}\index{Two-level! approximation} (TLA), {\it i.e.} we consider only two internal atomic states, separated in energy by $\Omega$, $|g\rangle$ and $|e\rangle$. The internal ionic energy is then
\begin{equation}
\hat{H}_\mathrm{in}=\frac{\Omega}{2}\hat{\sigma}_z.
\end{equation}
The idea now is to dress the ion with a laser field, creating a spatially varying profile. Inducing electronic transitions between the two internal levels is accompanied by changes in the ionic momentum as described in sec.~\ref{sssec:qatmo}. The effective atom-light coupling for a running wave in the dipole approximation\index{Dipole! approximation} becomes
\begin{equation}
\hat{V}_\mathrm{c}=\mu\hat{\sigma}_x\left(e^{-i(k\hat{x}-\omega_\mathrm{p}t+\varphi)}+e^{i(k\hat{x}-\omega_\mathrm{p}t+\varphi)}\right),
\end{equation}
where the effective coupling is $\mu$, $k$ is the laser wave number, $\omega_\mathrm{p}$ the laser frequency, and $\varphi$ is a frequency shift which sets the effective position of the trap relative to the laser. Turning first to an interaction picture with respect to $\hat{H}_\mathrm{in}$ gives
\begin{equation}
\hat{V}_\mathrm{c1}=\!\mu\!\left(\hat{\sigma}_{+}e^{i\Omega t}\!+\!\hat{\sigma}_{-}e^{-i\Omega t}\right)\!\!\left(e^{-i(k\hat{x}-\omega_\mathrm{p}t+\varphi)}\!+\!e^{i(k\hat{x}-\omega_\mathrm{p}t+\varphi)}\right)\!.
\end{equation}
We now perform the first RWA\index{Rotating-wave approximation} by assuming that $|\delta|\equiv|\omega_\mathrm{p}-\Omega|\ll\Omega$ and neglecting the fast rotating terms,
\begin{equation}
\hat{V}_\mathrm{c1}=\!\mu\left(\hat{\sigma}_{-}e^{-i(k\hat{x}-\delta t+\varphi)}+\hat{\sigma}_{+}e^{i(k\hat{x}-\delta t+\varphi)}\right).
\end{equation}
Before deriving the final Hamiltonian we introduce the {\it Lamb-Dicke parameter}\index{Lamb!-Dicke! parameter}
\begin{equation}\label{lambdicke}
\eta=k\sqrt{\frac{1}{2\omega}}
\end{equation}
and turn to bosonic creation/annihilation operators\index{Creation/annihilation operators}\index{Operator! creation/annihilation} $\hat{a}^\dagger/\hat{a}$. Transforming further to a rotating frame with respect to $\hat{H}_\mathrm{m}$ we obtain the interaction term 
\begin{equation}
\hat{V}_\mathrm{c12}= \mu\left\{\hat{\sigma}_{-}\exp\left[-i\left(\eta\left(\hat{a}^\dagger e^{i\omega t}+\hat{a}e^{-i\omega t}\right)-\delta t+\varphi\right)\right]+\hat{\sigma}_{+}\exp\left[i\left(\eta\left(\hat{a}^\dagger e^{i\omega t}+\hat{a}e^{-i\omega t}\right)-\delta t+\varphi\right)\right]\right\}.
\end{equation}
To understand the interaction we expand the exponent in the Lamb-Dicke parameter, {\it i.e.},
\begin{equation}
\exp\left[i\left(\eta\left(\hat{a}^\dagger e^{i\omega t}+\hat{a}e^{-i\omega t}\right)-\delta t+\varphi\right)\right]=e^{i(\varphi-\delta t)}\sum_m\frac{(i\eta)^m}{m!}\left(\hat{a}^\dagger e^{i\omega t}+\hat{a}e^{-i\omega t}\right)^m.
\end{equation}
The various terms in the expansion describe $m$-phonon transitions where the amplitude of these scales as $\eta^m$. Thus, the different terms realize versions of nonlinear JC interactions~\cite{vogel1995nonlinear} as discussed in sec.~\ref{sssec:kerr}. The advantage of using trapped ion systems to achieve these types of interaction is that the terms involved need not be weak, which is often the case for multiphoton transitions\index{Multiphoton! transition}. For the different transitions there is a corresponding effective resonance condition $\delta+l\omega$ with $l$ an integer. Thus, if $|\delta+l\omega|$ is large compared to other characteristic frequencies a second RWA\index{Rotating-wave approximation} can be applied to single out the dominant transitions between the states $|g,n\rangle$ and $|e,n+l\rangle$. The effective coupling between these states is
\begin{equation}
g=\mu|\langle n|e^{i\eta\left(\hat{a}+\hat{a}^\dagger\right)}|n+l\rangle|=\mu e^{-\eta^2/2}\eta^{|l|}\sqrt{\frac{n_<!}{n_>!}}L_{n_<}^{|l|}(\eta^2),
\end{equation}
where $n_<$ and $n_>$ are the lesser and greater of $n$ and $n+l$, and $L_n^\alpha(x)$ is the generalized Laguerre polynomial. The exponential scaling of the coupling is the so-called {\it Huang-Reys factor}~\cite{huang1950theory}. It follows that the validity of the second RWA depends on $|l|$. In most practical settings, $\eta$ is a small parameter and one may expect a truncation of the series expansion. Thus, we define the {\it Lamb-Dicke regime}\index{Lamb!-Dicke! regime} by the condition
\begin{equation}\label{ldregime}
\eta\sqrt{\langle(\hat{a}+\hat{a}^\dagger)^2\rangle}\ll1.
\end{equation}
In this regime, i.e when the spatial variation of the laser is weak on the trap length scale $l_0=\omega^{-1/2}$, single phonon transitions will dominate. We thus have three types of transitions,
\begin{equation}
\left\{\begin{array}{lll}
l=-1, & \hspace{1cm} & \text{Red--sideband}\,\,\mathrm{transition},\\
l=0, & \hspace{1cm} & \mathrm{Carrier}\,\,\mathrm{transition},\\
l=+1, & \hspace{1cm} & \text{Blue--sideband}\,\,\mathrm{transition}.
\end{array}\right.
\end{equation}
The three coupling processes are schematically shown in fig.~\ref{fig19}. 

Now the corresponding Hamiltonians are the JC for the red-sideband transitions\index{Red! sideband transitions}
\begin{equation}
\hat{H}_\mathrm{JC}=\frac{\delta_\mathrm{jc}}{2}\hat{\sigma}_z+g\left(\hat{a}^\dagger\hat{\sigma}_{-}e^{-i\varphi}+\hat{\sigma}_{+}\hat{a}e^{i\varphi}\right),
\end{equation}
the ``classical Rabi'' model for the carrier transitions\index{Carrier transitions}\index{Rabi! model}\index{Model! Rabi}
\begin{equation}
\hat{H}_\mathrm{car}=\frac{\delta}{2}\hat{\sigma}_z+g\left(\hat{\sigma}_{+}e^{i\varphi}+\hat{\sigma}_{-}e^{-i\varphi}\right),
\end{equation}
and finally the {\it anti-JC} (aJC)\index{Anti-Jaynes-Cummings model}\index{Model! anti-Jaynes-Cummings} for the blue-sideband transitions\index{Blue! sideband transitions}
\begin{equation}\label{aJC}
\hat{H}_\mathrm{aJC}=\frac{\delta_\mathrm{ajc}}{2}\hat{\sigma}_z+g\left(\hat{a}^\dagger\hat{\sigma}_{+}e^{i\varphi}+\hat{\sigma}_{-}\hat{a}e^{-i\varphi}\right).
\end{equation}
Here, $\delta_\mathrm{jc}$ and $\delta_\mathrm{ajc}$ are the corresponding detunings of the driven transitions. Separately, all three models are readily diagonalizable. The aJC does not preserve the total number of excitations\index{Number! conservation} $\hat{N}=\hat{n}+\frac{\hat{\sigma}_z}{2}$ which is a constant of motion for the JC model, see eq.~(\ref{jcU1}). However, since $\hat{H}_\mathrm{aJC}$ can be transformed into the form of $\hat{H}_\mathrm{JC}$ via application of the operator $\hat{\sigma}_x$ and renaming $\delta_\mathrm{ajc}$ and $\varphi$, it directly follows that the aJC model supports another continuous $U(1)$ symmetry represented by the constant of motion $\hat{\Delta}=\hat{n}-\frac{\hat{\sigma}_z}{2}$ ($=\hat{\sigma}_x\hat{N}\hat{\sigma}_x$). Thus, while in the JC model the total number of excitations is preserved, in the aJC it is instead the difference between bosonic (photonic or here phononic) and atomic excitations that remains constant. 

The phase $\varphi$ determines the position of the trap relative to the standing-wave laser field. The importance of $\varphi$ has been addressed by Wu and Yang in~\cite{wu1997jaynes}, who demonstrated that regardless of $\varphi$, the trapped ion Hamiltonian can be mapped onto a JC one in the Lamb-Dicke regime\index{Lamb!-Dicke! regime}. The mapping is made such that in the transformed basis the internal atomic states are not the original bare ones. 

It is clear from the above derivation that in the trapped-ion setting, by means of an appropriate driving, the effective model in the Lamb-Dicke regime can be recast as a combination of $\hat{H}_\mathrm{JC}$, $\hat{H}_\mathrm{car}$, and $\hat{H}_\mathrm{aJC}$. For example, with $\varphi=0$ the JC and aJC can be combined to give the quantum Rabi model $\hat{H}_\mathrm{R}$ discussed in sec.~\ref{ssec:rabi}. However, it is also possible to derive effective anisotropic quantum Rabi model\index{Anisotropic! quantum Rabi model}~(\ref{anRabi}) where the strengths of the JC interaction terms and the counter-rotating terms are not equal. The first discussion for this model seems to date back to the early 1970s in a work by Hioe on the Dicke phase transition~\cite{hioe1973phase}. Note that this was long before the trapped-ion Hamiltonian had been born, and therefore the field attracted more of an academic type of interest back then. Much more recently, the ground-state energy of the anisotropic quantum Rabi model has been analyzed~\cite{shen2013ground} where an analytic approximation was derived by making use of the polaron transformation~(\ref{polaron}). Collapse-revivals of this model were discussed in ref.~\cite{rodriguez2005combining}. Also, the approach followed by Braak to obtain the solution of the quantum Rabi model~\cite{braak2011integrability} (see sec.~\ref{sssec:rabiint}) has been generalized to the anisotropic quantum Rabi model~\cite{xie2014anisotropic,zhang2015analytical}. 

The physics beyond the Lamb-Dicke regime contains ``multi-phonon'' transitions and the multiple time-scales typically involved may result in rather complex dynamics. However, a large parameter $\eta$ could be considered for speeding up operations which is desirable in terms of quantum information processing~\cite{wei2002quantum,garcia2003speed}. It has further been demonstrated that a quantum interference phenomenon in the system evolution can lead to regular evolution with pronounced revivals and super-revivals beyond the Lamb-Dicke and RWA assumptions~\cite{wei2004engineering,wang2008quantum}. Moya-Cessa also noticed an equivalence between the trapped-ion Hamiltonian~\cite{moya2000unitary,moya2003family}
\begin{equation}\label{zerodet}
\hat{H}=\omega\hat{n}+\frac{\delta}{2}\hat{\sigma}_z+\mu\left(\hat{\sigma}_{+}e^{i\eta\hat{x}}+\hat{\sigma}_{-}e^{-i\eta\hat{x}}\right)
\end{equation}
and the driven quantum Rabi model\index{Driven! quantum Rabi model}. The idea is to appreciate that $\hat{D}(i\eta)=e^{i\eta\hat{x}}$ is a representation of the displacement operator~\eqref{dispop}. If this operation can be ``redone'' by means of a reverse displacement, the coupling term becomes a drive of the two-level atom. But such a transformation will in turn displace the vibrational harmonic oscillator. The unitary operator accomplishing the displacement in question reads
\begin{equation}
\hat{T}=\frac{1}{\sqrt{2}}\left[
\begin{array}{cc} 
\hat{D}^\dagger(\beta) & -\hat{D}^\dagger(\beta)\\
\hat{D}(\beta) & \hat{D}(\beta)
\end{array}\right],
\end{equation}
with $\beta=-i\eta/2$. The transformed Hamiltonian $\hat{H}'=\hat{T}^\dagger\hat{H}\hat{T}$ becomes
\begin{equation}
\hat{H}'=\omega\hat{n}+\mu\hat{\sigma}_z+\frac{i\omega\eta}{2}\left(\hat{a}+\hat{a}^\dagger\right)\hat{\sigma}_x-\frac{\delta}{\omega\eta}\hat{\sigma}_x.
\end{equation}
As it stands, the trapped-ion system should be a platform for realizing the ultrastrong or deep strong coupling regime of the quantum Rabi model. However, in deriving (\ref{zerodet}) we have already assumed one instance of the RWA, and this might not be justified in the regime under consideration. 

Carrying on with this approach to touch upon entangled bosonic state generation for quantum computation, we briefly focus on a single spin-boson system, which we have met several times so far, where now the spin (two-level atom) is subject to a set number of $N_r$ rotations~\cite{Casanova2018, PueblaCasanova2019}. The driven Hamiltonian with linear coupling terms,
\begin{equation}
 \hat{H}_{SB}=\omega \hat{n} + g(\hat{a} + \hat{a}^{\dagger})\hat{\sigma}_{x} + \sum_{j=0}^{N_r}\frac{\varepsilon_j}{2} [\cos(\Delta_j t + \phi_j)\hat{\sigma}_z + \sin(\Delta_j t + \phi_j)\hat{\sigma}_y],
\end{equation}
can be transformed via the spin-dependent displacement operator $\hat{T}=\hat{T}(\beta)$ to
\begin{equation}
 \hat{H}_{\rm disp}=T^{\dagger}(-g/\omega) \hat{H}_{SB} \hat{T}(-g/\omega)=\omega \hat{n} + \sum_{j=0}^{N_r}\frac{\varepsilon_j}{2} \left[\hat{\sigma}_{+}e^{(2g/\omega)(\hat{a}-\hat{a}^{\dagger})}e^{-i(\Delta_j t + \phi_j)} + {\rm h.c.}\right],
\end{equation}
which, under the RWA and in the Lamb-Dicke regime, assumes the form of an $n^{\rm th}$ order quantum Rabi\index{Quantum! Rabi model! n-th order} model~\cite{Casanova2018}:
\begin{equation}\label{eq:HSBLD}
 \hat{H}_{SB}^{\prime}=\sum_{j=0}^{N_r}\frac{\varepsilon_j}{2} [\hat{\sigma}_{+} e^{-i\phi_j}\hat{f}(\Delta_j)+{\rm h.c.}], \quad \text{where} \quad \hat{f}(\Delta_j)=\displaystyle\begin{cases}
\displaystyle\frac{(2g/\omega)^m}{m!}(-\hat{a}^{\dagger})^m, \quad \Delta_j=+m\omega,\\
\displaystyle\frac{(2g/\omega)^m}{m!}(\hat{a})^m, \quad \Delta_j=-m\omega,\\
1-\frac{(2g/\omega)^2}{2}-(2g/\omega)^2\hat{n}, \quad \Delta_j=0.                                                                                                           \end{cases}
\end{equation}
Two spin-boson subsystems (1 and 2) each with the Hamiltonian of eq.~\eqref{eq:HSBLD}, coupled by an interaction term of the form $\lambda \hat{\sigma}^{z}_{1} \hat{\sigma}^{z}_{2}$, have been considered in~\cite{mcaleese2023engineering} as a means to achieve a synthesis of both entangled coherent states\index{Entangled coherent states}\index{State! entangled coherent} and N$00$N states\index{N$00$N state}. The generation of these states requires a nonlinear interaction between two subsystems -- a cross-Kerr interaction and a non-linear beam splitter interaction, respectively. To that aim, the analysis of~\cite{mcaleese2023engineering} demonstrates that various classes of effective bosonic interaction Hamiltonians can be engineered, mediated by the coupled spins. Time evolution proceeds via the Magnus expansion\index{Magnus expansion}~\cite{Magnus1954} of the time-ordered time propagator, having as an argument an interaction Hamiltonian with renormalized parameters.   

The master equation\index{Master equation} for a single two-level ion interacting with laser light and damped by spontaneous emission\index{Spontaneous! emission! rate} with rate $\gamma$ is~\cite{Cirac1992, Cirac1993Fock}
\begin{equation}\label{eq:MEion}
 \frac{d\hat{\rho}}{dt}=-i\left[\nu \hat{a}^{\dagger}\hat{a} + \frac{\Delta}{2}\hat{\sigma}_z-\frac{\Omega}{2}(\hat{\sigma}_{+} + \hat{\sigma}_{-})\sin[\eta(\hat{a} + \hat{a}^{\dagger})],\hat{\rho}\right] + \frac{\gamma}{2}(2\hat{\sigma}_{-}\overline{\rho}\hat{\sigma}_{+} - \hat{\sigma}_{+}\hat{\sigma}_{-}\hat{\rho} - \hat{\rho}\hat{\sigma}_{+}\hat{\sigma}_{-}),
\end{equation}
in which $\nu$ is the frequency of the trap, $\Delta$ is the detuning of the two-level transition with respect to the laser frequency, $\Omega$ is the laser Rabi frequency, and $\eta$ is the Lamb-Dicke parameter~\eqref{lambdicke}. In~\eqref{eq:MEion}, the term $\overline{\rho}$ accounts for the momentum transfer linked to the spontaneous emission\index{Spontaneous! emission} of a photon,
\begin{equation}\label{eq:Wrad}
 \overline{\rho}=\frac{1}{2}\int_{-1}^{1}du\, \mathcal{W}(u)\, e^{i\eta(\hat{a} + \hat{a}^{\dagger})u} \, \hat{\rho}\, e^{-i\eta(\hat{a} + \hat{a}^{\dagger})u},
\end{equation}
where $\mathcal{W}(u)$ is the angular distribution\index{Distribution! angular} of the spontaneously emitted photons [see also eq.~\eqref{eq:Dterm}]. For an ion cooled to its lowest quantum state (for $\nu>\gamma$) and localized to a region small compared to the wavelength of the cooling radiation, we expand eqs.~\eqref{eq:MEion} and~\eqref{eq:Wrad} to first order in $\eta \ll 1$, to obtain
\begin{equation}\label{eq:MEionapprox}
 \frac{d\hat{\rho}}{dt}=-i\left[\nu \hat{a}^{\dagger}\hat{a} + \frac{\Delta}{2}\hat{\sigma}_z-\frac{\Omega}{2}\eta(\hat{\sigma}_{+} + \hat{\sigma}_{-})(\hat{a} + \hat{a}^{\dagger}),\hat{\rho}\right] + \frac{\gamma}{2}(2\hat{\sigma}_{-}\hat{\rho}\hat{\sigma}_{+} - \hat{\sigma}_{+}\hat{\sigma}_{-}\hat{\rho} - \hat{\rho}\hat{\sigma}_{+}\hat{\sigma}_{-}).
\end{equation}
The rotating-wave approximation\index{Rotating-wave approximation}, rendering the above to the familiar JC master equation\index{Master equation} with light-matter coupling strength $\eta \Omega/2 \to g$, is consistent with the sideband cooling limit. Moreover, in this mapping the only source of decoherence is the coupling of the two-level ion to the modes of the vacuum radiation field; the field mode is undamped ($\kappa=0$). At the same time, the strong-coupling regime ($g \gg \gamma$) is readily attained by adjusting the laser Rabi frequency $\Omega$. A scheme for preparing coherent squeezed states for the motional degrees of freedom in an ion trap was proposed in~\cite{Cirac1993}, relying on the multimode excitation\index{Multimode! excitation} of the trapped ion by standing and traveling-wave fields. 

Spin-dependent optical parametric drives, applied at twice the motional frequency of a trapped-ion ensemble, generate a coordinate transformation of the collective ionic motion in phase space, actuating displacement forces that are nonlinear in the spin operators, under the Lamb-Dicke approximation. This procedure forms the basis of a recently proposed protocol for generating high-order spin-dependent Hamiltonians, such as the many-body stabilizers\index{Stabilizer Hamiltonian}~\cite{Katz2023}.

\subsection{State preparation and tomography}\label{ssec:ionpreptom}
The trapped-ion setup allows for similar operations as for those of cavity/circuit QED, and in some respects even more in terms of driving the different transitions (fig.~\ref{fig19}). As a result, this is a good setting for performing state preparation and also tomography on the ionic state. Naturally, state preparation has been focusing on non-classical states\index{Non-classical state}\index{State! non-classical} such as
entangled~\cite{kneer1998preparation}, Fock\index{Fock state}\index{State! Fock}~\cite{cirac1993preparation,meekhof1996generation,roos1999quantum,myatt2000decoherence}, squeezed\index{Squeezed! state}\index{State! squeezed}~\cite{meekhof1996generation,gou1997generation,Cirac1993}, and Schr\"odinger cats\index{Cat state}\index{Schr\"odinger cat! states}~\cite{monroe1996schrodinger,de1996even,gerry1997generation,zheng1998preparation,myatt2000decoherence}.

One of the earliest demonstrations of state preparation in these systems was the experimental study of the quantum Zeno effect\index{Quantum! Zeno effect}~\cite{itano1990quantum}. Here a $V$-ion was employed with one of the transitions driven by a strong field. Fluorescence indicated that the ionic transition was being driven. If, on the other hand, the ion were prepared in the excited state of the other transition, the 'absence' of fluorescence would act as a Zeno measurement\index{Zeno measurement}, preventing the ion from decaying in the lower state (We should mention that the conclusions of these results led to much discussion following their appearance, see for example~\cite{ballentine1991comment}). A much more recent study used the fact that the strength of the coupling between internal and vibrational degrees of freedom can be tuned, and the group of Davidson could thereby implement a scheme that could alternate between weak and strong measurements~\cite{pan2020weak}. {\it Weak measurements}\index{Weak measurement} affect a system minimally, coming with the price that the information gain is very small~\cite{aharonov1988result}. Thus, the measurement does not `collapse' the state into an eigenstate as in a von Neumann measurement\index{von Neumann! measurement}. 

The idea of preparing cat states~(\ref{catstate}) is similar to the scheme followed in cavity/circuit QED: apply a state dependent (internal state of the ion) displacement $\hat{D}(\beta)$ to the vibrational state, let the system evolve and then reverse the displacement via Ramsey measurements of the atom; the interference visibility reflects the coherence of the cat. This technique was early on applied in the Wineland group~\cite{monroe1996schrodinger}, and later modified in the same group by engineering the decoherence of the cat~\cite{myatt2000decoherence}. Decoherence was enginered by randomly varying the trap frequency $\omega(t)$. In the same work~\cite{myatt2000decoherence}, the coherence pertaining to the superposition of two Fock states $|n_1\rangle$ and $|n_2\rangle$ was also explored. As for the cat states, the distance $|n_1-n_2|^2$ determines how fragile the superposition is to decoherence.

In ref.~\cite{meekhof1996generation} successive applications blue/red/carrier transitions were implemented in order to prepare the Fock state $|2,e\rangle$ and the decay of Rabi oscillations was measured. In the same work, the collapse-revival structure (see sec.~\ref{sssec:crsubsec}) was seen, alongside some evidence of a squeezed vacuum state. Decoherence of vibrational states was studied in~\cite{turchette2000decoherence} by `engineering' a system-reservoir coupling; applying random electric fields was employed to mimic a high temperature reservoir, random changes in the trap frequency simulates phase decoherence, and laser cooling a zero temperature bath. The Blatt group also used the blue sideband transition, following a ground-state cooling of the ion, to prepare the $n=1$ Fock state; as a result, up to 30 Rabi oscillations were recorded~\cite{roos1999quantum}.

In terms of state tomography numerous theoretical proposals have been put forward, see for example Refs.~\cite{wallentowitz1995reconstruction,bardroff1996endoscopy,leibfried1996experimental,poyatos1996motion,lutterbach1997method,kneer1998preparation}. Lutterbach and Davidovich~\cite{lutterbach1997method} used the representation~(\ref{wigexp}) for the Wigner function to propose its extraction for the vibrational state of the ion. The action of the displacement operators $\hat{D}(\alpha)$ can be implemented as we discussed above, and the action of $e^{i\pi\hat{n}}$ is readily realized by operating in the dispersive regime and controlling the interaction time such that $g^2t/\delta=\pi$. By changing the phase in a Ramsey interference scheme\index{Ramsey! interferometry}, they demonstrated that the average inversion is (\ref{inv}) $\langle\hat{\sigma}_z\rangle=W(-\alpha)$, with $\alpha$ directly related to the Ramsey phase. Wallentowitz and Vogel~\cite{wallentowitz1995reconstruction} suggested a different approach where the ion couples to its vibrational quadrature $\hat{x}$ [see eq.~(\ref{quadvar2})] with the help of an additional laser. The inversion (\ref{inv}) can be expressed in terms of a ``diagonal'' and ``off-diagonal'' part (referred to in~\cite{wallentowitz1995reconstruction} as {\it incoherent} and {\it coherent}, respectively). Whether the ion is prepared in a pure state or a statistical mixture (achieved with an external laser coupled to an ancilla state\index{Ancilla! state}) it is possible to measure the diagonal/off-diagonal contributions to the inversion separately. These two measured quantities can then be used to construct the full density operator $\hat{\rho}_\mathrm{f}$ for the ionic vibrational state.

On the experimental side, the Wineland group extracted the $Q$ and Wigner functions\index{$Q$-function}\index{Wigner! function} both for a $N=1$ Fock state, a coherent superposition of two Fock states $N=0$  and $N=2$, and a coherent state~\cite{leibfried1996experimental}. In their method, off-diagonal elements of the density operator, {\it i.e.} $\rho_{nm}=\langle n|\hat{\rho}_\mathrm{f}|m\rangle$ for $n\neq m$, could be extracted by making use of an effective Fourier transform, shifting the state along a circle in phase space by means of a displacement operation.  An experimental demonstration of low-depth {\it amplitude estimation}\index{Amplitude estimation} onto a state-of-the-art trapped ion quantum computer, where two fans of individually addressable beams illuminate a chain of individually imaged ions, is reported in~\cite{GiurgicaTiron2022}.

Thus far we have only discussed tomography of the ion's vibrational state. In the following subsection we mention some results obtained from extracting the internal ionic state.

\subsection{Quantum information processing}\label{ssec:qipi}
There is no doubt that the realization of the $C$-NOT gate between the internal and external vibrational states of a trapped ion~\cite{monroe1995demonstration} was one of the most important reasons why the field of trapped ion systems attracted so much attention from the late 1990s onwards~\cite{steane1997ion,haffner2008quantum,blatt2008entangled,singer2010colloquium}. This resulted in that the field rapidly became one of the more promising and advanced candidates for quantum information processing. The $C$-NOT gate was first implemented in the Wineland group using $^9$Be ions, following the ideas of Cirac and Zoller~\cite{cirac1995quantum}. Later, the Blatt group extended the early experiments and demonstrated the `Cirac-Zoller' gate~\cite{schmidt2003realization} between the internal states of two $^{40}$Ca ions.

\begin{figure}
\includegraphics[width=10cm]{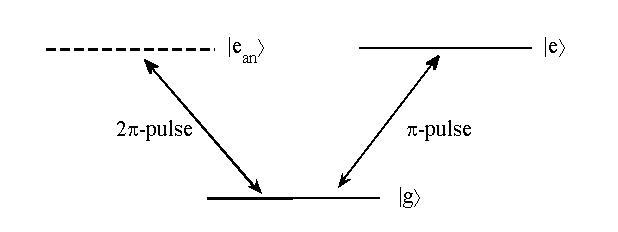} 
\caption{The laser-ion coupling scheme in the Cirac-Zoller gate: ($i$) a $\pi$-pulse between the ground state $|g\rangle$ and the excited state $|e\rangle$ is applied to ion $m$ which de-excite the ion, if it is initially excited, and thereby leaves one phonon to the common vibrational mode. ($ii$) a $2\pi$-pulse between the ground state $|g\rangle$ and the ancilla state\index{Ancilla! state} $|e_\mathrm{an}\rangle$ is applied to ion $n$ causing a sign change provided ion $n$ is in the state $|e\rangle$ and the collective vibrational state is in $|1\rangle$. ($iii$) another $\pi$-pulse is imposed on ion $m$ to redo the excitation swapping of ($i$). }
\label{fig20}
\end{figure}

The crucial building blocks of the Cirac-Zoller gate are, ($i$) couple internal states of different ions to a collective center-of-mass vibrational mode of the ions which are aligned in a a quasi-1D trap (thus the vibrational mode is used to indirectly make the different ions interact with each other) and ($ii$) use an auxiliary ancilla state\index{Ancilla! state} to implement a phase gate. The ions are aligned in a highly anisotropic trap ($\omega_x\ll\omega_y,\,\omega_z$) with a spatial separation allowing for single-ion addressing with external lasers. By driving the red sideband a JC-type interaction is created. Initially the atoms are cooled to the vibrational ground state. First a $\pi$-pulse is applied to one of the atoms which leaves the system unaltered if the ion is in the lower electronic state $|g\rangle$, while it excites the center-of-mass mode from $|0\rangle$ to $|1\rangle$ if the ion starts out in the excited state $|e\rangle$. A second JC interaction links the internal to the vibrational states of another ion. This time we have a  $2\pi$-pulse coupling the ground state $|g\rangle$ to an excited ancilla state\index{Ancilla! state} $|e_\mathrm{an}\rangle$, which has the effect of shifting the sign of the state $|g,1\rangle$ (the states $|e,1\rangle$ and $|e,0\rangle$ are transparent to the pulse). In the last step, another $\pi$-pulse is employed to the first ion to de-excite the vibrational mode. The net result of the three-pulse sequence is to shift the sign of the state $|e,e,0\rangle$ while all other states ($|g,g,0\rangle$, $|g,e,0\rangle$, and $|e,g,0\rangle$) are unchanged. Note that the Cirac-Zoller scheme relies on a precise timing required to achieve the correct pulse areas, and the vibrational mode is excited during the gate operation. In this respect it is different from the adiabatic proposals making use of dark states discussed in the end of sec.~\ref{sssec:ent}.

The first logic gate was experimentally demonstrated, as mentioned above, in the Wineland group~\cite{monroe1995demonstration}. Here the $c$-NOT gate\index{$c$-NOT gate}\index{Quantum! logic gate} operated between the internal electronic states and the vibrational state of a single trapped ion; the transition in question was $|e,1\rangle\rightarrow|g,1\rangle$, while the remaining three computational states were left unchanged. The group extended their first demonstration to also prepare two-ion entangled states of the form $|\psi\rangle=\frac{1}{5}\left(3|e,g\rangle-4|g,e\rangle\right)$ with a 70 $\%$ fidelity~\cite{turchette1998deterministic}.

\begin{figure}
\includegraphics[width=10cm]{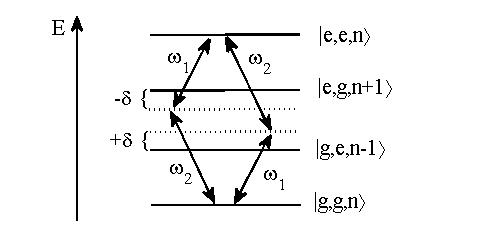} 
\caption{Coupling scheme between the bare states in the M\o lmer-S\o rensen entangling gate. By choosing the two detunings equal in size and opposite in sign a destructive interference between the two paths makes the gate independent on the number $n$ of phonons. Furthermore, the large detunings $\delta$ prevents the intermediate states to be populated. }
\label{fig21}
\end{figure}

To circumvent possible errors arising from phonon dissipation/decoherence in the Cirac-Zoller gate\index{Cirac-Zoller gate}, M\o lmer and S\o rensen\index{M\o lmer and S\o rensen gate} proposed another protocol to prepare entangled ion states which does not rely on the particular vibrational state of the ions~\cite{sorensen1999quantum,molmer1999multiparticle}. The bare states coupled in the M\o lmer-S\o rensen gate are depicted in fig.~\ref{fig21}. This instance sidesteps the reliance on several qubits and inevitably many atoms, obviating the necessity to perform gate operations among them. Two classical lasers, with frequencies $\omega_1$ and $\omega_2$ respectively, illuminate two of the ions via the red and blue sidebands, coupling the state $|g,g,n\rangle$ with $|e,g,n+1\rangle$ and $|g,e,n-1\rangle$ and these latter states with the state $|e,e,n\rangle$. While the transition $|g,g,n\rangle \leftrightarrow|e,e,n\rangle$ is resonant, the intermediate states are highly detuned and consequently will only be virtually populated. After adiabatic elimination\index{Adiabatic! elimination} of these states, an effective coupling between the remaining two states is obtained and, due to the different signs in the two detunings, the $n$-dependence of the coupling drops out. Thus, the effective Rabi frequency between the two coupled states will not depend on the center-of-mass vibrational state of the two ions, {\it i.e.} the coupling is insensitive to phonon losses and does not rely on ground state cooling of the vibrational state. Applying a $\pi/2$-pulse implies, for example, $|g,g,n\rangle\rightarrow\frac{1}{\sqrt{2}}\left(|g,g\rangle-i|e,e\rangle\right)|n\rangle$. Owing to some Stark shifts in the states $|e,g\rangle$ and $|g,e\rangle$, the prepared Bell state\index{Bell! state} will actually acquire an $n$-dependent phase and thereby become entangled with the phonons. This Stark shift\index{Stark shift} can be `redone' in an echo-type procedure by shifting  the sign of the detuning $\delta$ halfway through the duration of the $\pi/2$-pulse. In a recent proposal~\cite{Andrade2022}, the M\o lmer-S\o rensen scheme has been extended to simulate purely three-spin dynamics via tailored first- and second-order spin-motion couplings. 

Several more theoretical proposals on how to perform quantum information processing without relying on the vibrational state have followed, see for example~\cite{milburn2000ion,poyatos1998quantum}. In ref.~\cite{milburn2000ion}, Milburn {\it et al.} again introduce an ancilla electronic state\index{Ancilla! state} which is only virtually populated, and by combining a Stark-shift\index{Stark shift} operation with two STIRAP\index{STIRAP} processes (see sec.~\ref{sssec:timeJCsec}) a $c$-phase gate is achievable. The STIRAPs are by definition adiabatic, requiring long operation times and an alternative non-adiabatic scheme (requiring shorter operation times) was also presented in~\cite{milburn2000ion}. To speed up the Cirac-Zoller gate, it has been suggested that by driving the carrier transitions the energy levels can be Stark-shifted such that transitions can be made resonant thereby lowering the overall time needed for the gate implementation~\cite{jonathan2000fast,steane2000speed}.  Childs and Chuang considered a different situation where the ancilla electronic ion state\index{Ancilla! state} is not used in the gate operation~\cite{childs2000universal}. All the schemes discussed above depend crucially on the interaction time. Fluctuations in the interaction time unavoidably lead to gate errors. To overcome this, adiabatic proposals have been put forward~\cite{duan2001geometric,zhu2003unconventional}. Duan {\it et al.} consider geometric phases\index{Geometric phase} in the tripod system, {\it i.e.} three lower electronic states coupled to one excited state (direct generalization of the $\Lambda$-configuration discussed in sec.~\ref{sssec:multi}). The tripod system\index{Tripod! system} supports two dark states\index{Dark! state}\index{State! dark} and within this subspace (resulting from adiabatic elimination of the bright states)~\cite{unanyan1999laser}, the geometric phase\index{Geometric phase} is non-Abelian\index{Non-Abelian} and can be utilized for entanglement operations.   

While the above theoretical proposals are intended to improve the quantum information processing, they can be difficult to implement experimentally. One proposal that has been frequently implemented, however, is the M\o lmer-S\o rensen gate~\cite{sackett2000experimental,leibfried2003experimental,leibfried2004toward,benhelm2008towards}. In ref.~\cite{sackett2000experimental}, 2-qubit and 4-qubit entangled states were generated, with 83 $\%$ and 57 $\%$ fidelities respectively, and in~\cite{leibfried2004toward} a 3-qubit GHZ state\index{GHZ state}\index{State! GHZ}~(\ref{ghzstate}) was prepared with 89 $\%$ fidelity. In ref.~\cite{benhelm2008towards}, the 2-qubit entanglement preparation fidelity was as high as 99.3 $\%$, and in~\cite{leibfried2003experimental} 97 $\%$ (single-qubit gates\index{Qubit! gate} have reached fidelities as high as 99.9999 $\%$ at room temperature which is way beyond what is needed for fault-tolerant computing\index{Fault-tolerant computing}~\cite{harty2014high}). Similar ideas have also been extended to experimentally realize a {\it Toffoli gate}\index{Toffoli gate} ({\it i.e.} a $cc$-NOT gate - three qubits with two control and one target qubit)~\cite{monz2009realization} (see Refs.~\cite{wang2001multibit,yang2005efficient} for theory proposals), the {\it Deutsch-Josza algorithm} ({\it i.e.} determining whether a quantum-computer implemented function is ``constant'' or ``balanced'')~\cite{gulde2003implementation}, and {\it quantum teleportation} between two ions~\cite{barrett2004deterministic}.

Naturally, the goal is to extend the present schemes for quantum-information processing to include controlled gate implementation between several qubits. That is to be able to address and couple any pair of qubits\index{Qubit! gate} in a system consisting of many qubits. This is the problem of {\it scalability} which is crucial for any practical quantum computing scheme~\cite{loss1998quantum}. One of the early protocols addressing this issue was devised by Cirac and Zoller~\cite{cirac2000scalable}. The idea is to load single ions (``target ions'') in separated traps located in a two-dimensional plane. An additional (or many additional) ion (`head ion'') is in a trap that can be moved `on-top' of the plane, and in particular, it can be brought close to any of the target ions on the plane. Applying a state-dependent force to the two interacting ions, the two states $|g\rangle$ and $|e\rangle$ experience different Coulomb-induced energy shifts, and a $c$-phase gate is implemented between the head and the target ion. By letting the head ion interact with another target ion, in principle any 2-qubit gate between the two target ions can be implemented. In the scheme of ref.~\cite{kielpinski2002architecture}, decoeherence-free states of two ions~\cite{kielpinski2001decoherence} is utilized. Again, the entangling operation is achieved via state dependent Coulomb interaction, and the two target atoms are adiabatically/coherently moved to come close to each other. It has also been analyzed whether photons could serve as {\it quantum buses}\index{Quantum! bus} which mediate  effective interaction between far-separated ions~\cite{duan2004scalable}. Indeed, the ion-photon entanglement required for such a scheme has been observed experimentally with 97 $\%$ fidelity~\cite{blinov2004observation}. The use of cavities to generate entanglement between macroscopically separated ions is another possibility~\cite{simon2003robust,browne2003robust}, and single ions have been coherently coupled to single cavity modes~\cite{mundt2002coupling,keller2004continuous,steiner2013single}. Such setups may serve as `on-demand' sources for single-photon generation\index{Single-photon!source}~\cite{keller2004continuous}. In the more recent experiment of ref.~\cite{steiner2013single}, with the use of an optical-fiber cavity the strong-coupling regime between the ion and the field was reached. The extension to coupling crystals of ions to cavity modes has also been demonstrated~\cite{herskind2009realization}. The collective coupling, see sec.~\ref{sssec:dicke}, between the ions and the cavity mode was shown to belong to the strong-coupling regime. Closer to our days, a high-quality deterministic single-photon source\index{Single-photon!source}, operating in the ultrastrong coupling regime and emitting two single photons with an arbitrary time separation in a one-excitation process, has been recently proposed in~\cite{PengJie2023}. The one-photon solutions to the two-qubit quantum Rabi and JC models can be used to implement the proposal via two consecutive adiabatic evolution processes. The system returns to the initial state of the next period automatically following the photon emission. 
 
Even if the scalability discussed along the lines of the previous paragraph has not been thoroughly explored experimentally, steps towards controlling many-body entangled states have been taken~\cite{sackett2000experimental,haffner2005scalable,leibfried2005creation,lanyon2011universal,monz201114}. For measuring such many qubit states tomography methods are needed. Just like in state tomography of the vibrational state discussed in sec.~\ref{ssec:ionpreptom}, here one also measures expectations of certain observables to work backwards in order to reconstruct the (entangled) qubit state. One of the early demonstrations of qubit tomography\index{Qubit! tomography} for Bell states\index{Bell! state} was presented in~\cite{roos2004bell}; the full $4\times4$ density operator was determined and especially the decay of coherence was measured. It was later experimentally established that the decay rate scales as the square of the number of qubits~\cite{monz201114} in agreement with superradiance\index{Superradiance}, see subsec~\ref{sssec:dicke}. The larger the number of qubits, the more experimental preparations and measurements need to be repeated in order to perform reliable tomography on the full state. For example, for the eight-qubit tomography\index{State! tomography} of ref.~\cite{haffner2005scalable} around 600 000 repetitions were performed. The fidelity of the prepared eight-qubit W state\index{$W$-state}\index{State! $W$} [see eq.~(\ref{werner})] was found to be 72 $\%$. In the same work, W states were also prepared for three-seven ionic qubits. Also the multi-qubit GHZ state [see eq.~(\ref{ghz})] has been experimentally prepared for up to six qubits in the Wineland group~\cite{leibfried2005creation} and for up to 14 qubits in the Blatt group~\cite{monz201114}. For the largest number states, both groups measured a fidelity of around 50 $\%$. Note that the M\o lmer-S\o rensen gate discussed above can be directly generalized for the preparation of many qubit GHZ states.

We end this subsection by noticing that quantum walks\index{Quantum! walk}, which are related to quantum-information processing, have also been demonstrated in the context of trapped ions~\cite{schmitz2009quantum,zahringer2010realization}. Following the theoretical scheme of~\cite{travaglione2002implementing}, Schmitz {\it et al.} implemented a three-step quantum walk~\cite{schmitz2009quantum}. The idea is that the ion acts as the (quantum) coin and by preparing the ion in the linear combination states of ``head'' and ``tail'' $(|H\rangle\pm|T\rangle)/\sqrt{2}$ (with $|H\rangle$ and $|T\rangle$ two internal electronic states of the ion) the coherent vibrational state for the ion is displaced along a circle in phase space (see fig.~\ref{fig5}). Already after three steps a clear difference from a classical random walk was shown. By tailoring the vibrational-internal ion interaction to contain both a JC and a anti-JC contribution, the quantum walk has also been carried out along a line ($x$-quadrature\index{$x$-quadrature}, see eq.~(\ref{quadvar})) instead of a circle~\cite{zahringer2010realization}. The vibrational distribution $P(x)$\index{Distribution! vibrational} could be extracted between every step and up to 23 steps were analysed. An additional feature was also explored by using a ``three-sided'' coin; stay still, go left, or go right. This was achieved by using two trapped ions, and the resulting distribution $P(x)$ displayed different interference structures compared to the one using only a regular coin.

\subsection{Further aspects and perspectives}\label{ssec:ionfur}
One of the more interesting prospects for trapped-ion physics is the possibility to perform quantum simulations~\cite{johanning2009quantum,blatt2012quantum,schneider2012experimental}. While not proper quantum simulators in their original meaning, {\it i.e.} simulating quantum systems that are not tractable on classical computers~\cite{feynman1982simulating}, the Blatt group has been able to experimentally realize effects analogous to {\it Zitterbewegung}\index{Zitterbewegung}~\cite{gerritsma2010quantum,roos2011quantum} and the {\it Klein paradox}\index{Klein paradox}~\cite{gerritsma2011quantum} following theoretical proposals by Lamata {\it et al.}~\cite{lamata2007dirac,bermudez2008nonrelativistic,lamata2011relativistic}. By simultaneously resonantly driving the blue and red sidebands, the Hamiltonian takes the general form
\begin{equation}
\hat{H}_\mathrm{Dirac}=\frac{\Omega}{2}\hat{\sigma}_z+\mu\left(\hat{\sigma}_{+}e^{i\varphi_+}+\hat{\sigma}_{-}e^{-i\varphi_+}\right)\left(\hat{a}^\dagger e^{i\varphi_-}+\hat{a}e^{i\varphi_-}\right)
\end{equation}
in the Lamb-Dicke regime\index{Lamb!-Dicke! regime}. Now, if we chose the frequencies $\varphi_+=0$ and $\varphi_-=\pi/2$ we obtain the Dirac Hamiltonain in $1+1$ dimensions,
\begin{equation}\label{diracham}
\hat{H}_\mathrm{Dirac}=\mu\sqrt{2}\hat{\sigma}_x\hat{p}+\frac{\Omega}{2}\hat{\sigma}_z.
\end{equation}
Note that the ``counter-rotating terms'' are present due to driving the blue sideband, while the RWA has been imposed. Since the transitions are driven resonantly, the pump-photon detunings are zero and the bare photon energy part vanishes. This makes the resulting Hamiltonian linear in the ``momentum'' quadrature $\hat{p}$, thereby mimicking the relativistic model. Zitterbewegung\index{Zitterbewegung} is manifested in a ``trembling'' motion of a particle with a localized wave function (arising from the spin-orbit type coupling $\hat{p}\hat{\sigma}_x$ where internal and external degrees of freedom are coupled). Due to the light mass of the electron, observing the effect in its original setting seems experimentally unrealistic or at least very challenging. In the trapped-ion system, the amplitude of oscillations can be made considerably larger; as a result, these oscillations have been observed in~\cite{gerritsma2010quantum} (Zitterbewegung has also been observed in an ultracold trapped gas of atoms with the help of a laser induced spin-orbit coupling~\cite{qu2013observation}. The Hamiltonian describing this system, ignoring atom-atom interaction, is indeed identical to the driven quantum Rabi model\index{Quantum! Rabi model}). For simulating Klein tunneling, {\it {\it i.e.}} transmittance of a particle through a potential barrier by turning into its own antiparticle, a spatially dependent potential $\phi(x)$ should be added to the Dirac Hamiltonian\index{Dirac! Hamiltonian}~(\ref{diracham}). In ref.~\cite{gerritsma2011quantum}, the potential $\phi(x)\sim x$ was achieved by considering a second ion that is coupled to the ``position'' quadrature $\hat{x}$. The Hamiltonian is then similar to the Landau-Zener model (see sec.~\ref{sssec:timeJCsec}) and a transition between the two states (corresponding particle/anti-particle) occurs due to the presence of non-adiabatic transitions during the curve-crossing. Tunneling was observed by measuring the vibrational density $|\psi(x,t)|^2$ at different time-shots~\cite{gerritsma2011quantum} (Like Zitterbewegung, also Klein tunneling has been experimentally realized in an ultracold atomic gas propagating in an optical lattice, and again the effective model describing this system has the form of some generalized driven quantum Rabi model~\cite{salger2011klein}.). It has further been proposed how to simulate other relativistic models~\cite{bermudez2007exact,tenev2013proposal}, and even quantum field theory models or cosmological models~\cite{alsing2005ion,schutzhold2007analogue,casanova2012quantum,noh2012quantum}, while the $(2+1)$ Dirac oscillator has been recently studied in the context of the $\kappa$-deformed Poincar\'{e}-Hopf algebra\index{Poincar\'e! -Hopf algebra} in~\cite{Uhdre2022}, where we read that its Hamiltonian is non-Hermitian but has real eigenvalues. The linear dispersions when two modes are considered results in a Dirac cone as depicted in fig.~\ref{fig11}. Such point degeneracies typically result in non-trivial topologies and, in particular, in a corresponding Mead-Berry gauge theory. When the degeneracies occur in real space rather than in momentum space, as for the Dirac cones, they are termed {\it conical intersections}\index{Conical intersection}~\cite{LarsonConical}. The Berry phase acquired upon encircling a conical intersection was experimentally studied in a system of a trapped $^{171}$Yb$^+$ ion system~\cite{Valahu2022}. In a related study, the dynamical consequences of a conical intersection were also assessed in the experiment of~\cite{Whitlow2022}.

To realize true quantum simulators the system has to be scaled up to operate in a many-body configuration. An elegant scheme mimicking spin models was put forward by Porras and Cirac~\cite{porras2004effective,deng2005effective}. The spin degree of freedom is mapped onto the internal electronic states $|g\rangle$ and $|e\rangle$, and the collective vibrational modes mediate effective spin-spin interactions. The electronic configuration of the ion is such that the computational basis $|\uparrow\rangle=|e\rangle$ and $|\downarrow\rangle=|g\rangle$ comprises two metastable states Raman-coupled\index{Raman! coupling} via an auxiliary state in a $\Lambda$-configuration\index{$\Lambda$-configuration} (see fig.~\ref{fig12}). One of the states, $|\downarrow\rangle$ is dark\index{Dark! state}\index{State! dark} and does not experience a light-induced force by the lasers, while $|\uparrow\rangle$ couples linearly to the collective modes: $\sim\left(\hat{a}_{\alpha,i}^\dagger+\hat{a}_{\alpha,i}\right)$ where $\alpha$ is a direction index and $i$ the site index. Applying a polaron transformation\index{Polaron! transformation}, eq.~(\ref{polaron}), an effective spin-spin interaction $\sum_{\alpha,i,j}J_{i,j}^\alpha\hat{\sigma}_i^\alpha\hat{\sigma}_j^\alpha$ appears in the transformed Hamiltonian. Here, $\hat{\sigma}_i^\alpha$ is the Pauli $\alpha$ ($=x,\,y,\,z$) matrix\index{Pauli! matrices} acting on the internal states $|\uparrow\rangle$ and $|\downarrow\rangle$ of ion $i$. Since the {\it normal modes} (collective phonon modes) depend on the positions of the ions, also the coupling $J_{i,j}^\alpha$ will typically depend on the distance between the ions as $|q_i-q_j|^{-\nu}$ with $0\leq\nu\leq3$ ($\nu=0$ -- infinite range\index{Infinite-range interaction}, $\nu=1$ -- Coulomb, $\nu=2$ -- monopole-dipole, and $\nu=3$ -- dipole-dipole\index{Dipole!-dipole interaction}). If the ions are trapped in individual traps and the fluctuations around the equilibrium positions $\langle\hat{q}_i\rangle$ are small, to lowest order we can replace the position dependence of the coupling $J_{i,j}^\alpha$ with these equilibrium positions, so-called {\it stiff mode limit}. By adjusting the lasers coupling the internal and external degrees of freedom and the trapping potentials a variety of coupling terms can be achieved. For the case when all three $J_{i,j}^\alpha$'s ($\alpha=x,\,y,\,z$) are different, the long-range {\it Heisenberg $XY\!Z$ model}\index{Heisenberg! model! $XYZ$}\index{Model! Heisenberg! $XYZ$} in an external field can be realized. Making two or all three couplings equal, $XXZ$ or $XXX$ Heisenberg models\index{Heisenberg! model}\index{Model! Heisenberg} in longitudinal or transverse fields can be constructed. By setting two of the couplings to zero, the longitudinal or transverse-field {\it Ising model}\index{Ising! model}\index{Model! Ising}\index{Transverse-field Ising model}\index{Model! transverse field Ising} is instead realized. The sign of the coupling indicates whether ferromagnetic\index{Ferromagnetic! exchange interaction} ($J_{i,j}^\alpha<0$) or anti-ferromagnetic\index{Anti-ferromagnetic exchange interaction} ($J_{i,j}^\alpha>0$) order is favored; if the coupling is negative the spins tend to align in the same direction. In higher dimensions (two and three), these paradigmatic spin models are in general non-integrable\index{Integrable/non-integrable model} (especially with long range interaction)~\cite{assa1994interacting}. In one dimension, also, many of the trapped-ion chains are non-integrable.  Whether the simulated model is integrable or not is important, not only because its properties are qualitatively different. If the aim is to perform quantum simulations one is not expected to be able to reproduce the same results on a classical computer. Along these lines, thermalization, {\it i.e.} how a closed quantum system evolves at long time scales, was considered in refs.~\cite{gong2013prethermalization,Kaplan2020}. In the theoretical work~\cite{gong2013prethermalization} the authors considered a trapped-ion realization of a Heisenberg $XY$\index{Model! Heisenberg! $XY$}\index{Heisenberg! model! $XY$} chain. In~\cite{Kaplan2020}, the Monroe group implemented a realization of a transverse-field Ising model with up to eight $^{171}$Yb$^+$ ions. As we discussed above, in the trapped-ion realization of the spin chains, the range of the interaction is tunable from short to infinite range (recall that the infinite-range transverse-field Ising model becomes the Lipkin-Meshkov-Glick model)\index{Lipkin-Meshkov-Glick! model}\index{Model! Lipkin-Meshkov-Glick}. In their experiment, temporal fluctuations were measured after a quench and for different exponents of the power-law interaction strength. As expected from our understanding of thermalization theories, like the {\it eigenstate thermalization hypothesis}\index{Eigenstate! thermalization hypothesis}~\cite{d2015quantum}, it was found that the long-time fluctuations decrease with the system size. 

The XXZ central spin model\index{Model! central spin! XXZ}\index{Central spin model! XXZ} [see also our discussion in secs.~\ref{sssec:poormodel} and~\ref{sssec:dicke}], which can be construed as a single spin-half particle coupled to $N$ bath spins, has been found to exhibit a normal-to-superradiant phase transition in the limit where the ratio of the transition frequency of the central spin to that of the bath spins and the number of bath spins tend to infinity~\cite{ShaoLei2023}. 

Related to the concept of thermalization is that of {\it information scrambling}\index{Scrambling! information}~\cite{swingle2016measuring}. Scrambling in a 10 $^{40}$Ba$^+$ ion setup has been measured via the out-of-time ordered correlator (\ref{otoc})~\cite{Joshi2020}. Once again a transverse Ising model was realized, and in particular, the components $\hat V=\hat\sigma_1^z$ and $\hat W(t)=\hat\sigma_j^x(t)$ were recorded and the OTOC was extracted. The wavefront obtained from the operator spread is a measure of the scrambling. The OTOC was also studied in a different setup comprising more than 100 ions in a 2D Coulomb crystal~\cite{garttner2017measuring}. Here a fully-connected system was considered, {\it i.e.} a Lipkin-Meshkov-Glick model, and the OTOC could in this case, by using a Loschmidt echo\index{Loschmidt echo} mechanism, be transformed to rather simple observables, like the total $\langle\hat S_x\rangle$. The inclusion of an anharmonic term in the LMG Hamiltonian gives rise to an excited-state quantum phase transition\index{Phase transition! excited-state}, which has been recently analyzed in terms of the time evolution of the survival probability and the local density of states after a quantum quench, as well as on the Loschmidt echoes and the microcanonical OTOC~\cite{KhaloufRivera2023}. 

The capability to emulate dynamical phases out of equilibrium in optical cavities without real Cooper pairs\index{Cooper pair! BCS superconductor} in a BCS superconductor demonstrates that programmable simulators are able to overcome many challenges faced by traditional approaches. This allows to engineer unconventional superconductors and to probe beyond mean-field effects, and for increasing coherence time in quantum sensing~\cite{young2023observing}. 

Many-body evolution has been considered in terms of time crystals\index{Time crystal}~\cite{zhang2017observation,Kyprianidis2021}. We briefly discussed the concept of time-crystals at the end of sec.~\ref{sssec:dicke} when we considered the Dicke model. As mentioned there, (discrete) time crystals are nonequilibrium states\index{Nonequilibrium! state} occurring in periodically driven systems. These are typically found in either systems of infinite range interactions, or in many-body localized (MBL)\index{Many-body! localization} systems. MBL prevents the system from rapidly heating up during the driving. Such a MBL time crystal was for the first time realized in the Monroe group for an Ising-like system~\cite{zhang2017observation}. In particular, the time translational symmetry breaking was verified by the Fourier transform of the measured average $\langle\hat\sigma_i^x\rangle$, which showed clear subharmonics. The same group also demonstrated time-crystalline order in a clean system~\cite{Kyprianidis2021}, where heating was prohibited by very rapid Floquet driving. The nonequilibrium state appears in a prethemral phase\index{Prethermal state}\index{State! prethermal}, and the more rapid is the driving, the more stable is the prethermal phase. 

The theoretical scheme~\cite{gong2013prethermalization} was implemented to two $^{25}$Mg ions in the group of Schaetz~\cite{friedenauer2008simulating}. A crossover from a paramagnetic\index{Paramagnetic state}\index{State! paramagnetic} to a ferromagnetic\index{Ferromagnetic! state}\index{State! ferromagnetic} state was observed and a Bell state\index{Bell! state} was prepared with 88$\%$ fidelity. The Monroe group coupled three $^{171}$Yb ions and in this way studied {\it frustration}\index{Frustration}~\cite{kim2010quantum}. The three atoms were held in a linear trap with ion 1 and 2, and ion 2 and 3 coupled by $J_1$, and ion 1 and 3 by $J_2$. By considering different signs of $J_1$ and $J_2$, the system is frustrated since there is a competition between ferro- and anti-ferromagnetic ordering. The Monroe group has further extended their research to a chain of nine ions to study signals of an emerging quantum phase transition\index{Phase transition! quantum}~\cite{islam2011onset}. By extending the system size from two all the way up to nine ions they identified a sharper transition. In a finite Ising chain, the system can pass through a {\it Devil's staircase}\index{Devil's staircase} as the field is varied~\cite{hauke2010complete}. In such a staircase, the magnetization $M=\sum_i\langle\hat{\sigma}_i^z\rangle$ (where the expectation is taken for the ground state) attains a steplike structure as a function of the field. The larger the system is, the more steps appear, and in the thermodynamic limit\index{Thermodynamic limit} the staircase becomes `complete', meaning that the magnetization can attain any fraction $n/m$ for some integers $n$ and $m$. This behavior has been experimentally explored in~\cite{richerme2013quantum} for a 10-ion chain. Frustration\index{Frustration}, which substantiates various glassy phases of matter, was explored in a chain of up to 16 ions by tuning the ratio between interaction $J_{i,j}^x$ and an effective magnetic field strength $B$ ({\it i.e.} a term $B\sum_i\hat{\sigma}_i^z$ added to the Hamiltonian)~\cite{islam2013emergence}. In this experiment, the Ising interaction was long range with an exponent $\nu\approx1$. By slowly lowering the field amplitude $B$, a crossover from a paramagnetic to an anti-ferromagnetic state was observed. Frustration was detected by mapping out the structure function (the Fourier transform of the {\it two-point correlator}\index{Two-point correlator} $C_{i,j}=\langle\hat{\sigma}_i^x\hat{\sigma}_j^x\rangle-\langle\hat{\sigma}_i^x\rangle\langle\hat{\sigma}_j^x\rangle$) which gives an indication of the density of states (a high degree of degeneracy of the low-lying energy states is typical for frustrated systems). The Blatt group explored various models, Ising, $XY$, and $XY\!Z$, for up to six ions and prepared GHZ states\index{GHZ state}\index{State! GHZ}~(\ref{ghz})~\cite{lanyon2011universal}. More recently, the 32-site Ising model with both a longitudinal and transverse field has been explored in an experiment which combines $^{171}$Yb$^+$ and $^{138}$Ba$^+$ ions~\cite{chertkov2021holographic}. The transverse field was periodically applied ({\it kicked Ising model}) leading to chaotic evolution. Note that 32 qubits is already hard to simulate classically. With the possibility of realizing a universal set of gates, such systems could be important for future realisations of {\it digital quantum simulators}\index{Digital quantum simulator}~\cite{buluta2009quantum,hauke2012can}. A step towards simulating computationally intractable systems was presented in~\cite{britton2012engineered}. Around 300 $^9$Be$^+$ ions were laser cooled and kept in a 2D Penning trap. The ions organize in a triangular lattice structure, and in the experiment the interaction exponent $\nu$ was varied across the range $0\lesssim\nu\lesssim1.4$. Coherent evolution in the system, following from the Ising interaction, was probed by measuring the precession of the total magnetization $M=\sum_i\langle\hat{\sigma}_i^z\rangle$. The experimental data could be reproduced well by means of numerical mean-field calculations\index{Mean-field approximation}. While this experiment did not directly demonstrate strongly-correlated (highly entangled) quantum many-body states, it benchmarked the realization of coherent many-body dynamics. For a model where two sub-wavelength--size ensembles of spins interact with a single quantized radiation field with opposite strengths, an exotic superradiant phase with ferromagnetic ordering\index{Ferromagnetic! ordering} has been recently predicted, in which multistabilities generate non-stationary behaviour in the thermodynamic limit~\cite{mivehvar2023unconventional}. 

In a 1D lattice, the {\it Jordan-Wigner transformation}\index{Jordan-Wigner transformation}~\cite{sachdev2007quantum} maps spin to fermion degrees of freedom or vice versa. For a lattice with $N$ sites we explicitly write down
\begin{equation}
\begin{array}{lllll}
\hat\sigma_j^+=e^{-i\pi\sum_{j=1}^{N-1}\hat c_j^\dagger\hat c_j}\hat c_j^\dagger, & & \hat\sigma_j^-=e^{+i\pi\sum_{j=1}^{N-1}\hat c_j^\dagger\hat c_j}\hat c_j, & & \hat\sigma_{zj}=2\hat c_j^\dagger\hat c_j-1,
\end{array}
\end{equation}
where the subscript $j$ denotes the site, and $\hat c_j/\hat c_j^\dagger$ are the annihilation/creation operators of a spinless fermion at site $j$. This mapping allows for simulations of fermionic systems with trapped ions possessing a pseudo spin-$1/2$ degree of freedom~\cite{casanova2011quantum,hauke2013quantum,garcia2017digital,casanova2011quantum,casanova2012quantum,mezzacapo2012digital,havlivcek2017operator,perez2021quantum}. The general idea is to derive effective models of the trapped-ion system which in return describe known Hamiltonians of interacting fermions, {\it e.g.} the {\it Holstein model}\index{Holstein model}\index{Model! Holstein}, the {\it Agassi model}\index{Agassi model}\index{Model! Agassi}, or lattice gauge models\index{Lattice gauge model}. Following the first digital simulation of the Schwinger model~\cite{Martinez2016}, Nguyen and coworkers experimentally realize the time dynamics of the lattice Schwinger model within its purely fermionic representation~\cite{Nguyen2022}, underlying the importance of applying symmetry-preserving error-suppression schemes.

Engineering effective-spin models with trapped-ion systems hinges on internal electronic states of the atoms, while the motional states of the ions play an indirect role as mediators of an effective interaction between internal states of separated ions. Non-trivial Hamiltonians may also emerge for trapped ``spin-less'' ions, {\it i.e.} ions with no effective internal degrees of freedom~\cite{schiffer1993phase,dubin1993theory}. Here, an interplay between the Coulomb interaction energy and the potential energy brings about a second-order phase transition\index{Phase transition! second-order} in the crystal structure of the trapped ions. Consequently, the transition is not a direct result of a JC type of interaction, and we will not give a detailed description of the theory here. In a quasi 1D setup the transition appears between a {\it linear} and a {\it zigzag} crystalline structure. In the linear phase the potential energy dominates and the ions align, while in the zigzag phase the Coulomb energy dominates and the ions form a zigzag pattern which increases the distance between them but this comes at the price of a higher potential energy. In 2D and 3D the crystalline structures are more complicated~\cite{schiffer1993phase}. We are dealing here with a second-order quantum phase transition\index{Phase transition! second-order! quantum}, which was analyzed in terms of a mean-field Ginzburg-Landau theory\index{Ginzburg-Landau theory} in ref.~\cite{fishman2008structural}. The first observation of the transition is due to the late Herbert Walther whose group trapped $^{24}$Mg$^+$ ions in a ring trap and saw signatures of the different crystal structures~\cite{waki1992observation}. As a continuous phase transition, Del Campo {\it et al.} applied the {\it Kibble-Zurek mechanism}\index{Kibble-Zurek mechanism} to study non-adiabatic excitations in the crystal chain as the system is driven through the critical point\index{Critical! point}~\cite{del2010structural}. Related to these results, the same Kibble-Zurek physics was also examined experimentally in a chain of 16 $^{40}$Ca$^+$ ions~\cite{ulm2013observation}. By recording the fluorescent radiation, one can detect topological excitations/defects, which here appear as domain walls along the chain. Extending the spin chains to include also the motional degrees of freedom has been proposed for the generation of novel features~\cite{porras2012quantum}. In particular, this work considers a monitored coupling to realise a lattice version of coupled Jahn-Teller\index{Jahn-Teller model}\index{Model! Jahn-Teller} systems, as discussed in sec.~\ref{sssec:multi}. At a critical coupling the system displays a second-order quantum phase transition\index{Phase transition! second-order! quantum} into a superradiant phase\index{Superradiant! phase} in analogy with the Dicke PT, see sec.~\ref{sssec:dicke}. Both the Dicke PT\index{Dicke! phase transition}\index{Phase transition! Dicke}~\cite{safavi2018verification} and the PT accompanying the quantum Rabi model~\cite{Cai2021} (see sec.~\ref{ssec:rabi}), have been observed in trapped-ion settings. In the former, a 2D Coulomb crystal of around 70 $^9$Be$^+$ ions was studied as the system was quenched through the normal-superradiant PT\index{Normal-superradiant phase transition}\index{Dicke! phase transition}\index{Phase transition! Dicke}\index{Phase transition! normal-superradiant}, while in the latter a quasi 1D trapped $^{171}\mathrm{Yb}$ ions was used~\cite{Cai2021}. They could identify a sharp increase in the phonon number as the effective coupling exceeded the critical value. Another possibility where the collective vibrational modes play a significant role was discussed in~\cite{porras2008mesoscopic} where the realization of spin-boson models was demonstrated (see sec.~\ref{sssec:multi}).  

As we have already discussed in sec.~\ref{sssec:openjc}, by engineering the Lindblad operators $\hat{L}_i$ the system state can relax down to a decoherence-free subspace. In a first step towards such decoherence-induced state preparation, 4 qubits were prepared in a multi-qubit GHZ state\index{GHZ state}\index{State! GHZ}~(\ref{ghz}) and exposed to decoherence in a controlled way~\cite{barreiro2010experimental}. Also, the group of Wineland explored dissipation for the preparation of two ions in a maximally entangled state~\cite{lin2013dissipative}, and more recently a W state~(\ref{wstate}) was dissipatively prepared reaching a fidelity of 98$\%$~\cite{cole2021dissipative}. By varying the decoherence rate, entangled multi-qubit states can be prepared, while the ESD/ESB effect\index{Entanglement! sudden! death}, which was discussed in sec.~\ref{sssec:ent}, may be observed. In ref.~\cite{barreiro2011open} an ancilla ion was employed for the preparation of two and four-qubit GHZ states. The ancilla\index{Ancilla! state}\index{State! ancilla} ion was entangled with the remaining qubits, and via the decay of the ancilla into one of the states in the computational basis the remaining qubits ended up in the desired state. The fidelities reached about 80$\%$ for a bipartite entangled state and 55$\%$ for a four-qubit multipartite entangled state.  

In the previous subsection we mentioned that ions can be trapped inside optical resonators and coherently couple to the photonic cavity field. Another recent advancement has been the confinement of an ion within an atomic condensate~\cite{zipkes2010trapped,schmid2010dynamics}. One interesting aspect of these systems is the possibility to simulate {\it polaron physics} or {\it impurity models}, {\it i.e.} the ion (`impurity') gets dressed by the surrounding bosons. In the Denschlag experiment~\cite{schmid2010dynamics}, by extracting the fluorescence emitted from the ion, the density profile of the condensate was obtained. The K\"oho experiment~\cite{zipkes2010trapped} demonstrated sympathetic cooling\index{Sympathetic cooling} of the ion from a few tens of Kelvin down to the sub-Kelvin regime. The ion temperature, obtained via the collected fluorescent signal, was contrasted with the atom loss in the condensate, originating from the scattering occurring between the ion and neutral condensed atoms.

As the measurement rate of an open quantum system is varied, a purification phase transition is predicted to emerge at a critical point\index{Critical! point}. A direct experimental observation of a pure and a mixed phase associated with a measurement-induced purification transition in a trapped-ion quantum computer has been reported in~\cite{Noel2022}. This goes hand in hand with the finding that a dissipative quantum phase transition\index{Dissipative! quantum phase transition}\index{Phase transition! dissipative} between a ferromagnetic ordered phase\index{Ferromagnetic! ordering} and a paramagnetic disordered phase emerges for long-range systems as a function of measurement probabilities~\cite{Sierant2022dissipativefloquet}. On the implementation side, reliable measurements of multi-qubit correlation operators are aided by the fault-tolerant parity measurement scheme which was demonstrated in~\cite{Hilder2022}, while the authors of~\cite{FossFeig2022} report that the entanglement structure of an infinite system -- in particular the half-chain entanglement spectrum -- is conveniently encoded within a small register of ``bond qubits''\index{Qubit! bond}, revealing the spatial structure of certain tensor network\index{Tensor network! states} states. 
 
Turning now to a coherent interaction with an external drive, the requirements needed for a pulse sequence to decouple a spin system, subject to two classes of controllable operations, from an external field without altering the intended dynamics have been recently established in~\cite{Morong2022}. The experimental demonstration of this technique for a trapped-ion system shows a significant improvement of coherence for practical applications, while a quantum simulation for an exactly solvable paradigm for long-range interacting spins has also been engineered. A disordered phase with frustrated correlations has also been observed in a trapped-ion spin chain~\cite{LeiFeng2022}, upon preparing states with long-range spin order spanning over the system size of up to 23 spins.

In the cold trapped-ion experimental implementation of Hunanyan, Koch and collaborators~\cite{hunanyan2023periodic, KochHunanyan2023} simulating the periodic quantum Rabi model, the two-state system is represented by two Bloch bands of cold atoms in an optical lattice, and the bosonic mode by oscillations in a superimposed optical dipole trap potential. Deviations from the predictions of the conventional quantum Rabi model are observed at the edge of the Brillouin zone\index{Brillouin zone}. Concurrently, the observation of the destructive interference caused by a geometric phase\index{Geometric phase} during the dynamical evolution of a wavepacket around a conical intersection was reported in~\cite{Valahu2023}. This is achieved by engineering the $E \otimes e$ Jahn-Teller Hamiltonian\index{Model! Jahn-Teller} in a trapped-ion quantum simulator, employing a mixed-qudit-boson encoding where both electronic and motional degrees of freedom for the ion are involved.

 
\section{Waveguide QED}\label{sec:waveguideQED}
 
This section revolves around the quantum transport of strongly correlated photons within the framework of waveguide QED; here, one dimensional bosons in a radiation channel interact with two-state atoms. The standard JC treatment is extended to a continuum of modes, supported by many experimental platforms over the last two decades. Let us begin with a more ``static'' formulation of the problem where an atom is radiatively coupled to the vacuum modes close to an interface formed by two different materials. 

\subsection{Atomic emission in the vicinity of an interface}

The master equation\index{Master equation} and the response-function formulations were used for the development of a QED theory of spontaneous emission\index{Spontaneous! emission} in the presence of dielectrics and conductors in 1975~\cite{Agarwal1975IV} following experimental results reported by Drexhage~\cite{Drexhage1974}, as well as the work of Carniglia and Mandel on the quantization of evanescent waves~\cite{CarnigliaMandel1971}. As an alternative to first-order perturbation theory, we can formulate a description of the problem from the master equation\index{Master equation} for the reduced density operator of the atomic system in the presence of a dielectric or a conducting surface. The average polarization obeys
\begin{equation}\label{eq:modifiedBloch}
 \frac{d\braket{\hat{S}^{-}}}{dt} + [i(\omega+ \Omega) + \gamma]\braket{\hat{S}^{-}}=0,
\end{equation}
with
\begin{subequations}\label{eq:paramsGB}
 \begin{align}
\gamma(\boldsymbol{b}, \omega)&=\sum_{\alpha, \beta} d_{\alpha} d_{\beta}\mathcal{E}_{\alpha \beta}(\boldsymbol{b},\boldsymbol{b},\omega)=\sum_{\alpha \beta} d_{\alpha} d_{\beta} \coth(\beta \omega/2) {\rm Im}[\chi_{\alpha\beta EE}(\boldsymbol{b},\boldsymbol{b},\omega)], \label{eq:paramsGBa}\\
\Omega(\boldsymbol{b}, \omega)&=-\frac{1}{\pi} P \int_{0}^{\infty} d\omega_0 \gamma(\boldsymbol{b}, \omega)[(\omega_0-\omega)^{-1} - (\omega_0+\omega)^{-1}] \equiv\delta E^{+}-\delta E^{-}, \label{eq:paramsGBb}
 \end{align}
\end{subequations}
where 
\begin{subequations}\label{eq:CorrChi}
 \begin{align}
  \mathcal{E}_{\alpha\beta}(\boldsymbol{r}_1, \boldsymbol{r}_2,\tau)&\equiv\frac{1}{2}\braket{\{E_{\alpha}(\boldsymbol{r}_1, \tau), E_{\alpha}(\boldsymbol{r}_2, 0)\}},\\
  {\rm Im} [\chi_{\alpha\beta EE}(\boldsymbol{r}_1, \boldsymbol{r}_2,\tau)]&\equiv \frac{1}{2}\braket{[E_{\alpha}(\boldsymbol{r}_1, \tau), E_{\alpha}(\boldsymbol{r}_2, 0)]},
 \end{align}
\end{subequations}
[$\chi_{\alpha\beta EE}(\boldsymbol{r}_1, \boldsymbol{r}_2,\tau)$ is the susceptibility tensor defined for the electric field~\cite{Agarwal1975I} and $\{A,B\}=AB+BA$ denotes the anti-commutator] and $d_{\alpha}$ is the $\alpha$ component of the dipole moment operator, $\hbar \omega$ is the energy separation between the bare atomic levels, and $\boldsymbol{b}$ denotes the position of the atom. In eq.~\eqref{eq:modifiedBloch}, $\Omega(\boldsymbol{b},\omega)$ is the effective shift in the energy separation of the two levels, equal to the difference between the Lamb shifts\index{Lamb! shift} in the excited and ground state energies, while $\gamma$ represents the width of the excited state. This formalism is used in~\cite{Agarwal1975IV} for the calculation of the atomic widths and shifts in the presence of a dielectric with a $\epsilon_0(\omega)$ occupying the semi-infinite region $-\infty < z \leq 0$ while the atom is located along the $z$-axis at $\boldsymbol{b}=(0, 0, b)$ in free space ($\epsilon=1$). The results depend on the orientation of the dipole relative to the interface [parallel ($\parallel$) or perpendicular ($\perp$); in the former case, the dipole is assumed to be randomly oriented in the $x-y$ plane]. The final expressions for the damping coefficients,
\begin{subequations}
  \begin{align}
\gamma_{\parallel}(b,\omega)-\gamma^{(0)}&=\frac{3}{4}\gamma^{(0)} {\rm Re} \int_{0}^{\infty}\frac{\kappa d\kappa}{\mu} e^{i\mu x} \left(\kappa^2 - \frac{2\mu \kappa^2}{\epsilon_0 \mu + \mu_0}-2(\mu-\mu_0)^2(\epsilon_0-1)^{-1} \right), \\
\gamma_{\perp}(b,\omega)-\gamma^{(0)}&=\frac{3}{2}\gamma^{(0)} {\rm Re} \int_{0}^{\infty}\frac{\kappa^3 d\kappa}{\mu} e^{i\mu x} \frac{\mu \epsilon_0-\mu_0}{\mu \epsilon_0 + \mu_0},
  \end{align}
\end{subequations}
where $\mu^2 \equiv 1- \kappa^2$, $\mu_0^2=\epsilon_0-\kappa^2$, $x\equiv 2\omega b /c$ and $\gamma^{(0)}$ is half of the Einstein $A$ coefficient\index{Einstein $A$ and $B$ theory} (for a random orientation of the dipole moment in free space), are to be substituted in eqs.~\eqref{eq:paramsGBa}, \eqref{eq:paramsGBb} to yield the surface-dependent energy shifts. The general result for the decay rates is a damped oscillatory dependence on $x$, while, if the atom is embedded inside the dielectric, one obtains $\gamma_{\perp, \parallel} \sim \exp(-5x/4)$ [see also the variation of atomic energy levels near a metal surface in the Auger process~\cite{Hagstrum1954}].

Moving on to the 1990s, numerous studies were devoted to the modification of the electromagnetic field surrounding an atom located close to a surface. Such a proximity alters the radiation properties of the atom. In particular, the presence of a surface changes the natural lifetime and the energy spacing of the atomic levels, as well as the spontaneous emission\index{Spontaneous! emission! distribution} radiation distribution\index{Distribution! radiation pattern} pattern. The quantum relaxation equation for the average of an arbitrary atomic observable $\hat{S}$ is expressed in terms of classical damping rates\index{Damping! rate, modification} $\Gamma_q$, as~\cite{Courtois1996}
\begin{equation}\label{eq:classicalrate}
 \frac{d\braket{\hat{S}}}{dt}=-\frac{1}{2}\sum_{m}\Gamma_{m}\braket{\hat{d}_{m}^{+}\hat{d}_{m}^{-}\hat{S} + \hat{S} \hat{d}_{m}^{+}\hat{d}_{m}^{-}-2\hat{d}_{m}^{+} \hat{S} \hat{d}_{m}^{-}},
\end{equation}
where the interaction Hamiltonian in the rotating-wave approximation is $\hat{V}_{\rm int}=-D \sum_{m=-1}^{1}[(\hat{\boldsymbol{d}}^{+}\cdot \boldsymbol{u}_m) \hat{\boldsymbol{E}}_{m}^{+} + (\hat{\boldsymbol{d}}^{-}\cdot \boldsymbol{u}_m) \hat{\boldsymbol{E}}_{m}^{-}]$. Here, $D$ is the scalar attributed to the electric dipole moment\index{Electric dipole! moment} of the atomic transition and $\hat{E}_{m}^{+}$ is the positive-frequency component of the electric-field operator. eq.~\eqref{eq:classicalrate} holds for a general geometry and state of the electromagnetic field. The effective Hamiltonian corresponding to the level shifts can be derived within the framework of a theory employing the method of images for dielectrics~\cite{Zhou1995}; this is essentially the viewpoint of radiation reaction embodied in the correlation functions of the electric field at the position of the atom [see eq.~\eqref{eq:CorrChi}] involving only {\it classical} electrodynamics. 

For one of the most studied realizations of a coherent mirror for atomic de Broglie waves, the recoil effects due to spontaneous emission\index{Spontaneous! emission} in the vicinity of a vacuum-dielectric interface were investigated in~\cite{Henkel1998} to provide a dynamical description of the process through a generalized optical Bloch equation. Based on this report by Henkel and Courtois, the notation of which we adopt for a while, we first identify the general features of such an equation in the presence of an interface. In free space, the master equation\index{Master equation} describing the interaction of a single multilevel atom with a coherent field proceeds in two steps\index{Master equation}. In the first step, one considers the evolution of the total system density matrix comprised by the atom and the electromagnetic field. In the framework of nonrelativistic QED and within the electric-dipole approximation\index{Electric dipole! approximation}, the equation governing the evolution relies on the atom-field Hamiltonian
\begin{equation} \label{eq:Htot}
\hat{H}=\hat{H}_{0}+\hat{H}_{R}+\hat{V}_{AL}+\hat{V}_{AR}. 
\end{equation}
The first term on the right-hand side of eq.~\eqref{eq:Htot} is the atomic Hamiltonian accounting for the sum of the internal and kinetic energies for the bare atom:
\begin{equation}\label{eq:Ha}
\hat{H}_{0}=\frac{\hat{\boldsymbol{P}}^{2}}{2M}+\frac{\hbar \omega _{0}}{2}\hat{S}_z,
\end{equation}
where $\hat{\boldsymbol{P}}$ is the atomic momentum operator, $M$ is the atomic mass, and $\hat{S}_z$ is the usual inversion operator formed by the difference of the projection operators onto the ground and excited states, respectively. The second term is the free Hamiltonian of the quantized electromagnetic field in the Coulomb gauge\index{Coulomb gauge}, while $\hat{V}_{AL}$ is the time-dependent Hamiltonian
\begin{equation}\label{eq:Vav}
\hat{V}_{AL}=- \hat{\boldsymbol{D}}\cdot {\bf \mathcal{E}}_{L}\left( {\bf R},t\right)  
\end{equation}
accounting for the interaction of the atomic dipole $\hat{\boldsymbol{D}}$ with the coherent laser field. The last term
\begin{equation}\label{eq:Var}
\hat{V}_{AR}=-\hat{\boldsymbol{D}}\cdot \boldsymbol{E}(\boldsymbol{R}) 
\end{equation}
accounts for the coupling between the atom and the reservoir vacuum field $\boldsymbol{E}(\boldsymbol{R})$. We note that in eq.~\eqref{eq:Htot} both fields ${\bf \mathcal{E}}_{L}\left( {\bf R},t\right) $ and $\boldsymbol{E}(\boldsymbol{R})$ are evaluated at the location of the atomic center-of-mass ${\bf R}$.

Subsequently, the master equation\index{Master equation} for the atomic density matrix $\hat{\rho}$ is obtained after applying second order perturbation theory for the atom-reservoir interaction, and tracing out the reservoir degrees of freedom. This procedure yields a dynamical-evolution equation in which the reservoir field has two distinct contributions. The first one, associated with an effective Hamiltonian, describes the energy shifts to the atomic levels stemming from their coupling to the vacuum field (the well-known Lamb shifts\index{Lamb! shift}). These are conventionally incorporated into the definition of $\hat{H}_{0}$, yielding the actual atomic Hamiltonian $\hat{H}_{A,\infty }$. The second contribution, $\dot{\hat{\rho}}_{{\rm relax},\infty}$, represents the dissipation of the atomic system due to its coupling with the reservoir. Finally, the free-space time-evolution of the atomic density matrix assumes the form 
\begin{equation}\label{eq:equaBlochfs}
\dot{\hat{\rho}}={\cal L}_{\infty }\,\hat{\rho},
\end{equation}
with
\begin{equation}\label{eq:Liouvillian}
{\cal L}_{\infty }\,\hat{\rho} =\frac{1}{i\hbar }\left[ \hat{H}_{A,\infty }+\hat{V}_{AL},\hat{\rho}
\right] +\dot{\hat{\rho}}_{{\rm relax},\infty },  
\end{equation}
and where we have written down explicitly the free-space Liouville super-operator ${\cal L}_{\infty}$.

After delineating the standard procedure, we now consider an atom placed in the vicinity of a vacuum-dielectric
interface\index{Vacuum-dielectric interface}. Similarly to~\cite{Agarwal1975IV}, we might wonder what are the modifications to the master equation~\eqref{eq:equaBlochfs}\index{Master equation} induced by the abutting dielectric medium? First, since new boundary conditions have now come into play, the modes of the fields $\boldsymbol{\mathcal{E}}_{L}\left( {\bf r},t\right)$ as well as those of the quantized electromagnetic vacuum field are altered and may become evanescent. It is clear that this does not affect the operators $\hat{H}_{0}$, $\hat{H}_{R}$, and $\hat{V}_{AL}$, all of which retain their free-space form. In contrast, the structure of the reservoir markedly changes. The contribution of $\hat{V}_{AR}$ to the atomic dynamics through the induced energy level shifts and the spontaneous emission\index{Spontaneous! emission! rate} rate is therefore expected to be different from the familiar free-space results. Moreover, owing to the instantaneous Coulomb interaction between the atomic and dielectric charges, we anticipate an additional electrostatic contribution $\hat{H}_{\rm es}$ to the energy level shifts. $\hat{H}_{\rm es}$ accounts for the London-Van der Waals interaction of the instantaneous atomic dipole with its image placed inside the dielectric medium (higher multipoles may be neglected provided the atomic radius is much smaller than the distance between the atom and the dielectric surface). Denoting by $\Delta \hat{H}_{A}$ and $\dot{\hat{\rho}}_{{\rm relax},\,{\rm int}}$ the modifications to the Hamiltonian and the dissipative parts\index{Dissipative! part} of the atomic density matrix evolution due to the presence of the interface, we obtain the following form of the generalized optical Bloch equations:
\begin{equation}\label{eq:equaBlochgen}
\dot{\hat{\rho}}={\cal L}_{\infty }\,\hat{\rho} +{\cal L}_{\rm int}\,\hat{\rho}
\end{equation}
where 
\begin{equation}\label{eq:modifeqbloch}
{\cal L}_{\rm int}\,\hat{\rho} =\frac{1}{i\hbar }\left[ \Delta \hat{H}_{A},\hat{\rho} \right] +
\dot{\hat{\rho}}_{{\rm relax},\, {\rm int}}  
\end{equation}
captures the impact of the interface on the atomic dynamics in its totality. We note that ${\cal L}_{\rm int}\,\hat{\rho}$ tends to zero when the atom is far from the dielectric surface. Since the expressions for the atomic level shifts close to a vacuum-dielectric interface have been given in~\cite{Agarwal1975IV, Courtois1996}, the emphasis should be placed on the dissipative contribution\index{Dissipative! part} to eq.~\eqref{eq:modifeqbloch}. Following the adiabatic elimination of the excited state, this equation is transformed into a Fokker--Planck-type equation\index{Fokker--Planck equation} for the phase-space distribution\index{Distribution! phase-space}\index{Phase-space distribution} function of the atomic ground state. The time-evolution of the Wigner representation $W({\bf r},{\bf p},t)$ of the ground state atomic density matrix defined as
\begin{equation}\label{eq:defWigner}
W({\bf r},{\bf p},t)=\frac{1}{(2\pi \hbar )^{3}}\int \!d^{3}s\,
\rho_{gg}({\bf r};{\bf s})
\exp {(-}i{\bf p}\cdot{{\bf s}/\hbar)},
\end{equation}
with
\begin{equation}
\rho_{gg} ({\bf r};{\bf s})\equiv \left\langle {\bf r}+{\textstyle \frac{1}{2}}%
{\bf s}|\rho_{gg} |{\bf r}-{\textstyle \frac{1}{2}}{\bf s}\right\rangle,
\end{equation}
involves quantities inside the integral that exhibit an ${\bf s}$ dependence. Hence, it connects $\partial _{t}W({\bf r}, {\bf p},t)$ to the distribution with displaced momentum, $W({\bf r},{\bf p}+\delta {\bf p},t)$, which is reminiscent of the atomic recoil during a photon absorption or emission process, \textit{i.e.}, $\left| \delta {\bf p}\right| \approx \hbar k$.

\subsection{Circuit QED revisited}
\label{ssec:QEDrevisited}

Let us now spend a little time recalling  and summarizing some key results from our earlier section on circuit QED. A typical implementation of circuit QED is using a transmission-line resonator whose electric fields are coupled to a superconducting charge qubit\index{Qubit! charge}. A central superconducting wire running between two ground planes defines the transmission line. Gaps in the wire, placed an integer number of half-wavelengths (a few centimetres) apart, are the ``mirrors'' employed to form a cavity, which is the microwave version of the Fabry-P\'{e}rot geometry\index{Fabry-P\'erot! geometry} used in optics. The size and shape of the gaps control the rate at which photons enter and leave the cavity, and the entire structure can be made using conventional microelectronic fabrication techniques. Such superconducting transmission lines have been extensively studied in the past. But recent experiments at temperatures close to absolute zero, where they are used as detectors for astrophysics, have shown that photons can make up to a million bounces before being lost. This means that the losses are remarkably low \textemdash{a} gigahertz photon travels back and forth a total distance of several kilometres before being lost.

The artificial atom \textemdash{an} isolated Josephson junction\index{Josephson! junction} \textemdash{is} placed between the wire and the ground planes, at or near an antinode of the standing wave of the voltage on the line, so it couples to the electrical fields of the transmission line. Exciting the atom corresponds to transporting one or a few pairs of bound electrons (Cooper pairs\index{Cooper pair}) from one electrode of the junction to the other. This means that the dipole moment of this artificial atom is very large, often more than four orders of magnitude greater than the typical value for an electronic transition of a real atom. Because the atom size and shape are adjustable, the dipole coupling can also be engineered by having the atom essentially fill the transverse dimension of the cavity, which means that the vacuum Rabi frequency \textemdash{often} expressed as a fraction of the photon frequency  \textemdash{approaches} a maximum value of a few per cent set by the fine-structure constant. In comparison, the values obtained so far using real atoms in either optical or microwave cavities are typically much smaller \textemdash{of} the order of one part in $10^6$. The considerably extended interactions achievable in circuit QED readily allow the attainment of the strong coupling limit of cavity QED. Another key advantage of circuit QED is that it avoids the difficulties of cooling and trapping the atom, since the artificial atom can be fabricated at exactly the desired location inside the resonator.

The rest of this section will be devoted to the collective effects that emerge in waveguide QED with artificial atoms with emphasis on photon-mediated interactions \textemdash{a} paradigmatic experiment reported in~\cite{vanLoo2013}. Our narrative follows closely the results reported by Lalumi\`{e}re and coworkers in~\cite{Lalumiere2013}. We consider an ensemble of $N$ artificial atoms, each comprising $M$ levels. They are dipole coupled to a 1D transmission line. The electromagnetic field in the transmission line can be described by the multimode Hamiltonian\index{Multimode! Hamiltonian}
\begin{equation}\label{eq:fullHfield}
\hat{H}_\mathrm{F}=\int_0^\infty d\omega \hbar \omega \left[\hat{a}^{\dagger}_\mathrm{R}(\omega) \hat{a}_\mathrm{R}(\omega)+\hat{a}^{\dagger}_\mathrm{L}(\omega) \hat{a}_\mathrm{L}(\omega) \right],
\end{equation}
where $\hat{a}^{\dagger}_{\mathrm{R}(\mathrm{L})}(\omega)$ creates right- (left-) moving excitations at frequency $\omega$ along the transmission line. The Hamiltonian for the artificial atoms is
\begin{equation}\label{eq:fullHatoms}
\hat{H}_\mathrm{A} = \sum_{j=0}^{N-1} \sum_{m=0}^{M-1} E_{mj} \ket{m_j} \bra{m_j},
\end{equation} 
where $E_{mj}$ is the energy of the $m$th state of the $j$th atom. The interaction Hamiltonian between the  electric field along the line and the electric dipole\index{Electric dipole} for the free artificial atoms can be written as
\begin{equation}\label{eq:fullHinteraction}
\hat{H}_\mathrm{I} = \sum_{j=0}^{N-1}\sum_{m=0}^{M-1}  \hbar g_j \sqrt{m+1} \left(\hat{\Xi}_j+\hat{\Xi}_j^\dagger \right)\hat{\sigma}_x^{mj}.
\end{equation}
In this expression, the operator $\hat{\Xi}_j$ is related to the electric field at the location $x_j$ of the $j$th artificial atom,
\begin{equation}\label{eq:define_xi}
\hat{\Xi}_j \equiv -i\int_0^\infty d\omega \sqrt{\omega} \left[ \hat{a}_\mathrm{L} (\omega) e^{-i \omega x_j/v_g}+\hat{a}_\mathrm{R}(\omega) e^{i \omega x_j/v_g}\right],
\end{equation}
with $v_g$ the group velocity\index{Group velocity} in the transmission line. We define
\begin{equation}
\hat{\sigma}_x^{mj}=\hat{\sigma}_-^{mj} + \hat{\sigma}_+^{mj},
\end{equation}
with
\begin{equation}
\hat{\sigma}_-^{mj}=\ket{m_j}\bra{(m+1)_j}=\left(\hat{\sigma}_+^{mj} \right)^\dagger,
\end{equation}
the lowering operator for the $(m+1)$th state of the $j$th atom. The interaction only involves transitions between adjacent states of the atoms, which is a valid approximation for the transmon superconducting qubit\index{Qubit! transmon} behaving as a weakly nonlinear oscillator~\cite{Koch2007}. Finally, in eq. \eqref{eq:fullHinteraction}, $g_j$ is the dimensionless coupling strength between atom $j$ and the field. 

The effective master equation\index{Master equation} for the artificial atoms after tracing out the field degrees of freedom can be expressed\index{Master equation} as~\cite{Lehmberg1970,Ficek2002,LeKien2005Tr,Gonzalez2011,Zueco2012,Chang2012} 
\begin{equation}\label{eq:mainMasterEquation}
\dot{\hat{\rho}}=-\frac{i}{\hbar}\left[\hat{H},\hat{\rho}\right]+
\sum_{mj,nk}
\gamma_{mj,nk}\left[\hat{\sigma}_-^{mj}\hat{\rho}\hat{\sigma}_+^{nk}-  \frac{1}{2}\left\{ \hat{\sigma}_+^{nk} \hat{\sigma}_-^{mj},\hat{\rho} \right\}\right],
\end{equation}
with the Hamiltonian 
\begin{equation}\label{eq:Heffwqed}
\hat{H} = \hat{H}_\mathrm{A}+\hbar\sum_{mj} d_{mj}(t)\hat{\sigma}_x^{mj}+\hbar\sum_{mj, nk} J_{mj,nk} \hat{\sigma}_-^{mj}\hat{\sigma}_+^{nk}.
\end{equation}
This effective Hamiltonian contains a drive term with a time-dependent amplitude $d_{mj}(t)$. For input coherent states incoming from the left (right) and of frequency $\omega_d$, phase $\theta_{\text{L(R)}}$ and radiation power $P_{\text{L(R)}}$ it can be shown that
\begin{equation}\label{eq:drive_amplitude_full_Main}
d_{mj}(t)=-2\sqrt{\frac{\gamma_{mj,mj}}{2}} \left( \sqrt{\frac{P_\mathrm{L}}{\hbar\omega_{mj}}}\sin\left[ \omega_d (t+t_j+\theta_\mathrm{L})\right]+\sqrt{\frac{P_\mathrm{R}}{\hbar\omega_{mj}}}\sin \left[\omega_d (t-t_j+\theta_\mathrm{R}) \right]\right)
\end{equation}
with $t_j=x_j/v_g$ defining the retarded time due to the propagation of the signal with group velocity\index{Group velocity} $v_g$.

\begin{figure}
\begin{center}
\includegraphics[width=11cm]{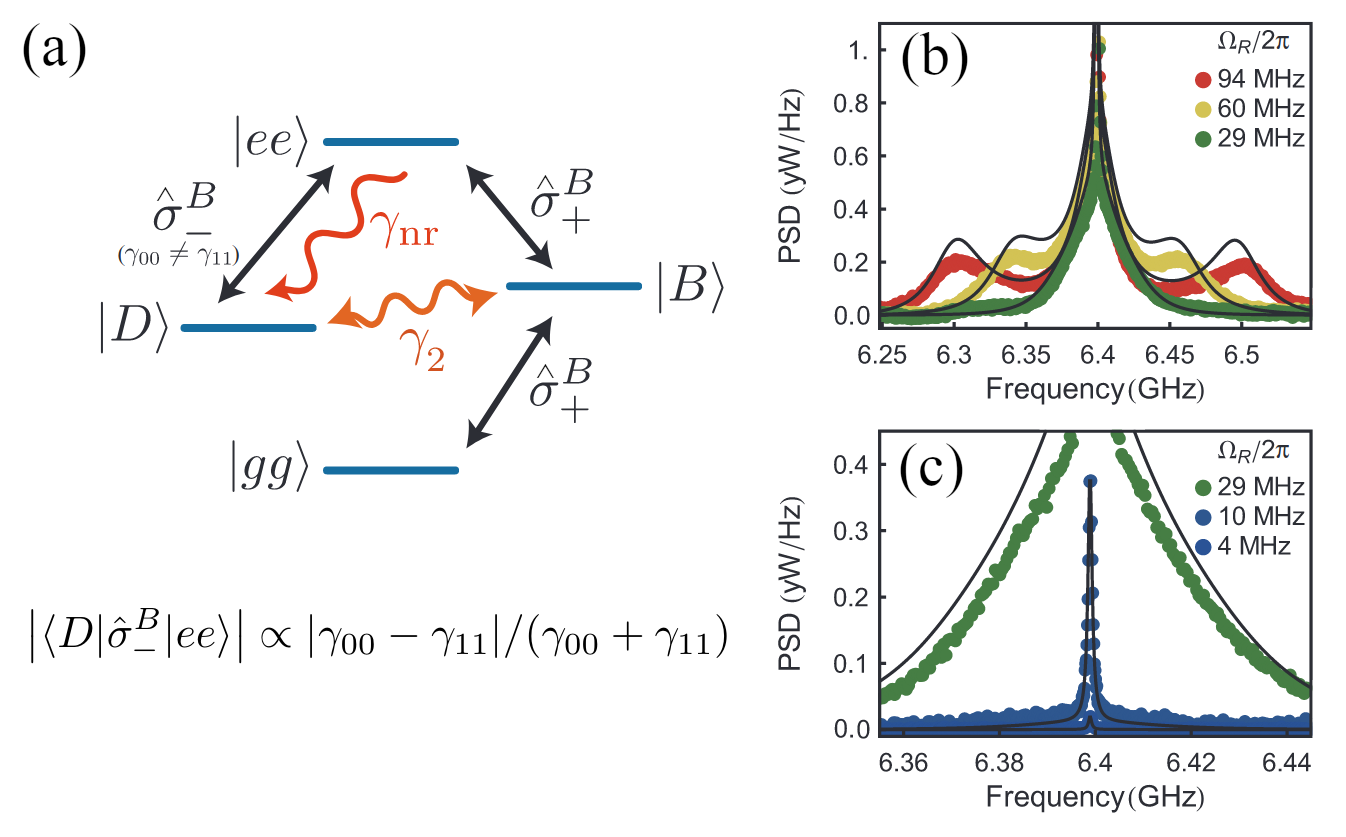}
\end{center}
\caption{{\it Photon mediated interaction between distant transmon qubits.\index{Qubit! transmon}} {\bf (a)} Schematic energy level diagram for a pair of two-state atoms in a transmission line. The dark state\index{Dark! state}\index{State! dark} $\ket{D}$ cannot be driven directly from the ground state $\ket{gg}$. If the two-state atoms are identical, it can however be excited by nonradiative relaxation $\gamma_{\mathrm{nr}}$ from $\ket{ee}$. In the situation where the two-state atom relaxation rates are different ($\gamma_{00}\neq\gamma_{11}$), it can be excited indirectly from $\ket{B}$ via $\ket{ee}$. Dephasing ($\gamma_2$) can also cause transition between $\ket{B}$ and $\ket{D}$. {\bf (b-c)} Power spectral density\index{Spectral! density} (in reflection) for $d \sim \lambda_0$ at the indicated Rabi frequencies $\Omega_R$. The two atoms are driven at the same phase. At high drive powers when the Mollow triplet appears, the spectra are qualitatively similar to those obtained for a single two-state system but with a linewidth\index{Linewidth} which is twice as large \textemdash{a} signature of superradiance [frame {\bf (b)}]. At low drive powers the system behaves as a single two-state system with ground state $\ket{gg}$ and excited state $\ket{B}$ [frame {\bf (c)}]. Sources:~\cite{Lalumiere2013, vanLoo2013}. Reprinted with permission from Science and the APS.}
\label{fig:photonmed}
\end{figure}

We note that the Hamiltonian of eq.~\eqref{eq:Heffwqed} is Hermitian since $J_{mj,nk}=J_{nk,mj}^*$. In obtaining this expression, we have used the rotating-wave approximation, dropped small non-positive terms in the dissipators and incorporated the Lamb shifts\index{Lamb! shift} into the definition of $\hat{H}_\mathrm{A}$. As we can see from eq.~\eqref{eq:mainMasterEquation}, the effect of the interaction with the transmission-line field amounts eventually to generating an atomic dissipation at the rate
\begin{equation}\label{eq:relaxationRate}
\gamma_{mj,nk}= 2\pi g_k g_j \sqrt{(m+1)(n+1)} 
 \left(\chi_{mjk}+\chi_{nkj}^*\right).
\end{equation}
with 
\begin{equation}
\chi_{mjk}=\omega_{mj} e^{i \omega_{mj} t_{kj}}, \quad \hbar\omega_{mj}=E_{m+1,j}-E_{mj}
\end{equation}
and $t_{kj}= |x_k - x_j|/v_g$ the time required for the propagation of the signal from atom $k$ to atom $j$. For $j = k$, eq.~\eqref{eq:relaxationRate} yields the standard relaxation rates of the atoms. For $j \neq k$, however, this corresponds to a correlated decay. 

The last term of eq.~\eqref{eq:Heffwqed} is an exchange interaction between the atoms \textemdash{mediated} by virtual excitations along the line \textemdash{with} amplitude
\begin{align}
\label{eq:exchangeInteraction}
J_{mj,nk}
=& -i \pi g_k g_j \sqrt{(m+1)(n+1)}\left( \chi_{mjk}-\chi_{nkj}^*\right).
\end{align}

For the special case of a pair of levels in two atoms tuned to resonance, $\omega_{mj}=\omega_{nk}$, the expressions for $\gamma_{mj,nk}$ and $J_{mj,nk}$ can be cast to a simpler form~\cite{LeKien2005Tr, Gonzalez2011,Zueco2012}:
\begin{equation}\label{eq:SimplerGamma}
\gamma_{mj,nk}= 4\pi g_k g_j \omega_{mj} \sqrt{(m+1)(n+1)}
 \cos \left(\omega_{mj}t_{kj}\right),
\end{equation}
and
\begin{equation}\label{eq:SimplerJ}
J_{mj,nk}=2 \pi g_k g_j \omega_{mj}  \sqrt{(m+1)(n+1)}\sin \left(\omega_{mj}t_{kj}\right).
\end{equation}
The latter expression explicitly demonstrates that the magnitude of these two quantities exhibits an oscillatory dependence on the interatomic separation.

Specializing to a pair of two-level atoms ($N=M=2$) in a frame rotating with the drive frequency $\omega_d$, we get
\begin{equation}\label{eq:reduced_me_22}
\dot{\hat{\rho}}=-\frac{i}{\hbar}\left[\hat{H},\hat{\rho}\right]+\sum_{jk}\gamma_{jk}\left[\hat{\sigma}_-^{j}\hat{\rho} \hat{\sigma}_+^{k}-  \frac{1}{2}\left\{ \hat{\sigma}_+^{k} \hat{\sigma}_-^{j},\hat{\rho} \right\}\right],
\end{equation}
where
\begin{equation}\label{eq:Heff:22}
\hat{H}/\hbar= \sum_{j}  \Delta_{j}\ket{e_j}\bra{e_j}+\sum_{j} \left(  \epsilon_{j} \sigma_+^{j}+\text{h.c.}\right)+ J (\hat{\sigma}_-^{1}\hat{\sigma}_+^{2} + \hat{\sigma}_+^{1}\hat{\sigma}_-^{2}),
\end{equation}
with $\ket{e_j}$ the excited state of two-state atom $j$, $\Delta_{j} = \omega_{0j} - \omega_d$, $J=J_{0j,0k}$ and $\gamma_{jk}=\gamma_{0j,0k}+\gamma^{j}_{\mathrm{nr}}\delta_{jk}$. The rate  $\gamma^{j}_{\mathrm{nr}}$ represents nonradiative decay of two-state atom $j$. In practice, it is easy to realize a situation where the two-state atom decay will be dominated by emission into the line, that is $\gamma_{jj}\gg \gamma^{j}_{\mathrm{nr}}$. Satisfying this inequality in the open-line configuration takes us to the strong-coupling regime~\cite{Abdumalikov2011}. Assuming that the two-state atoms are driven from the left only, eq.~\eqref{eq:drive_amplitude_full_Main} for the drive amplitude can now be recast into the simpler form
\begin{equation}
\epsilon_{j}= -i \sqrt{\frac{\gamma_{0j,0j}\omega_d}{2\omega_{0j}}}  \braket{a_{\text{in}}^\mathrm{L}}
e^{-i \omega_d t_j}.
\end{equation}

To account for the correlated decay described by the last term in eq.~\eqref{eq:reduced_me_22}, it is helpful to move to a basis that diagonalizes the dissipation matrix with components $\gamma_{j,k}$. This casts the last term of eq.~\eqref{eq:reduced_me_22} to the common form:
\begin{equation}
\sum_{\mu =B,D} \Gamma_\mu \mathcal D \left[ \hat{\sigma}_-^\mu \right]\hat{\rho},
\end{equation}
where $\mathcal D[\hat{x}] \hat{\rho} \equiv \hat{x} \hat{\rho} \hat{x}^\dagger-\left\{\hat{x}^\dagger \hat{x},\hat{\rho} \right\}/2$ is the standard dissipator acting on the dressed lowering operators
\begin{equation}\label{eq:sigmaHL}
\hat{\sigma}_{-}^{\mu}=\frac{\left(\Gamma_{\mu}-\gamma_{11} \right) \hat{\sigma}_-^0 + \gamma_{01}^*\hat{\sigma}_-^1}{\sqrt{\left(\Gamma_{\mu}-\gamma_{11} \right)^2+|\gamma_{01}|^2}},
\end{equation}
where $\mu=B,D$ (and the subscripts $B$ and $D$ refer to bright and dark respectively). The correlated decay rates emerge then as
\begin{equation}
\Gamma_{B/D}=\frac{\gamma_{00}+\gamma_{11}}{2}\pm \sqrt{\left(\frac{\gamma_{00}-\gamma_{11}}{2}\right)^2+|\gamma_{01}|^2}.
\end{equation}

Due to the dependence on the spatial separation of the two-state atoms, both the correlated two-state atom decay $\Gamma_{\mu}$ and exchange interaction $J$ can be adjusted via a modification of the two-state atom transition frequency. As a characteristic example we consider a pair of two-state atoms tuned in resonance at a frequency $\omega_{0}$, whose associated wavelength $\lambda_{0}$ is equal to $d = \lambda_{0} / 2$. To simplify the discussion, we let $\gamma_{\mathrm{nr}} \equiv \gamma_{\mathrm{nr}}^0 \sim \gamma_{\mathrm{nr}}^1$. In the strong-coupling regime, $\gamma^{j}_{\mathrm{nr}}\ll \gamma_{00},\gamma_{11}$, and the nonradiative relaxation rate can be treated as a small perturbation. The above assumption that the nonradiative rates are equivalent for both two-state atoms will therefore not affect the results much. 

With this simplification and the choice $d = \lambda_{0} / 2$, the off-diagonal decay rate $\gamma_{01}$ defined in eq.~\eqref{eq:relaxationRate} can be written as
\begin{equation}
\gamma_{01} = \pm \sqrt{(\gamma_{00}-\gamma_{\mathrm{nr}}) (\gamma_{11}-\gamma_{\mathrm{nr}})}.
\end{equation}
This leads to
\begin{equation}\label{eq:GammaDarkBrightLambdaOver2}
\Gamma_{D}= \gamma_{\mathrm{nr}}
\ll
\Gamma_{B}=\gamma_{00}+\gamma_{11}-\gamma_{\mathrm{nr}}.
\end{equation}
In other words, for $d = \lambda_{0} / 2$ the state $\ket{D}$ defined by $\sigma_-^{D} \ket{D}  = 0$ is dark as its decay rate is purely nonradiative. On the other hand, the state $\ket{B}$ defined by $\hat{\sigma}_-^{B} \ket{B} = 0$ is bright.  This setting corresponds, respectively, to subradiance and superradiance~\cite{LeKien2005}. Moreover, for this half-wavelength tuning, the exchange interaction is absent with $J=0$. The fact that the states $\ket{B}$ and $\ket{D}$ are bright and dark, respectively, can also be ascertained from the Hamiltonian of eq. \eqref{eq:Heff:22}. Indeed, by inverting eq.~\eqref{eq:sigmaHL}, it is possible to rewrite the driving term in eq.~\eqref{eq:Heff:22} as
\begin{equation}
\sum_{\mu = B,D}\hbar\left(\epsilon_\mu \hat{\sigma}_+^\mu+\text{h.c.}\right).
\end{equation}
For $|\Delta_j|/\omega_0 \ll 1$, which is easily satisfied, the drive amplitudes can be written as
\begin{equation}\label{eq:epsilon:LambdaOver2}
\begin{split}
\epsilon_{D} & \approx 0,
\\
\epsilon_{B} & \approx -i  \langle a_{\text{in}}^\mathrm{L} \rangle e^{-i \omega_d t_0} \sqrt{\frac{\gamma_{00}+ \gamma_{11}}{2} -\gamma_{\mathrm{nr}}}.
\end{split}
\end{equation}

It is important to point out that, since $\left[\hat{\sigma}^{B/D}_\pm,\hat{\sigma}^{B/D}_\mp\right]\neq 0$ and $\left[\hat{\sigma}^{B/D}_\pm,\hat{\sigma}^{D/B}_\mp\right] \neq 0$, the state $\ket{D}$ will not be completely dark in practice and especially not in the presence of finite nonradiative decay $\gamma_{\mathrm{nr}}$. Indeed, as shown in fig.~\ref{fig:photonmed} (a), the joint excited state $\ket{ee}$ can be reached from $\ket{B}$ by driving with $\hat{\sigma}_+^B$. From this state, $\ket{D}$ can be populated with the action of $\hat{\sigma}_-^B$ when $\gamma_{00} \neq \gamma_{11}$. This is because the matrix element of $\hat{\sigma}_-^B$ taken between the states $\ket{ee}$ and $\ket{D}$ is proportional to the asymmetry $(\gamma_{00} - \gamma_{11})/(\gamma_{00} + \gamma_{11})$. The dark state\index{Dark! state}\index{State! dark} can also be populated by nonradiative decay. While nonradiative relaxation cannot be controlled in this system, the indirect driving of $\ket{D}$ from $\ket{ee}$ can be tuned by controlling the asymmetry between $\gamma_{00}$ and $\gamma_{11}$, something that can be done by tuning the two-state atom frequency. Experimental results obtained for a separation $d=\lambda_0$ are depicted in Figs.~\ref{fig:photonmed} (b-c), where superradiance is in evidence against a background qualitatively similar to resonance fluorescence. 

\subsection{Light-matter interaction in a 1D waveguide: a continuum for correlated photon states}
\label{ssec:lighmatter1Dcorr}

Spontaneous emission of a cesium atom interacting with guided and radiation modes in the vicinity of a subwavelength-diameter optical waveguide was studied in~\cite{LeKien2005} shortly before the demonstration of strongly correlated photon transport in the photonic analogue of the Kondo effect. In the next few pages we will discuss in some length the main directions set by the remarkable work of Shen and Fan~\cite{Shen2007}. In their theoretical setup, a two-state atom is weakly coupled to a one-dimensional broadband continuum in which the photons propagate in either direction. When the resonance energy of the two-level system is away from the cutoff frequency of the dispersion relation\index{Frequency cutoff}, the paradigmatic Hamiltonian modeling the interaction with a continuum of modes is
\begin{equation}\label{eq:Hwgtla}
 \begin{aligned}
  \hat{H}&=\hat{H}_{\rm field} + \hat{H}_{\rm int} + \hat{H}_{\rm atom}\\
  &=\int dx \left\{-i v_g \hat{a}_{R}^{\dagger}(x)\frac{\partial}{\partial
x} \hat{a}_{R}(x)
+ i v_g 0 \hat{a}_{L}^{\dagger}(x)\frac{\partial}{\partial x} \hat{a}_{L}(x)\right. \\
&\quad \left.+ \overline{V} \delta(x) \left(\hat{a}^{\dagger}_{R}(x) \hat{\sigma}_{-} +
\hat{a}_{R}(x) \hat{\sigma}_{+}
+ \hat{a}^{\dagger}_{L}(x) \hat{\sigma}_{-} + \hat{a}_{L}(x) \hat{\sigma}_{+}\right)\right\} \\
&\quad +E_{e} \hat{c}^{\dagger}_{e} \hat{c}_{e}+ E_{g} \hat{c}^{\dagger}_{g} \hat{c}_{g}.
 \end{aligned}
\end{equation}

As in sec. \ref{ssec:QEDrevisited}, the above Hamiltonian accounts for the propagation of photons with group velocity\index{Group velocity} $v_g$, associated with the bosonic operators $\hat{a}_{R}^{\dagger}(x)$($\hat{a}_{L}^{\dagger}(x)$) creating a right (left) propagating photon at the position $x$ along the waveguide. In eq.~\eqref{eq:Hwgtla}, $\overline{V}$ is the light-matter coupling constant, $\hat{c}^{\dagger}_{g}$($\hat{c}^{\dagger}_{e}$) is the creation operator of the ground (excited) state of the atom, $\hat{\sigma}_{+}=\hat{c}^{\dagger}_{e} \hat{c}_{g}$($\hat{\sigma}_{-}=\hat{c}^{\dagger}_{g} \hat{c}_{e}$) is the atomic raising (lowering) ladder operator in the Schwinger representation, satisfying $\hat{\sigma}_{+}|n,-\rangle =|n,+\rangle$ and $\hat{\sigma_{+}}|n,+\rangle =0$, where $|n, \pm\rangle$ describes the state of the composite system with $n$ photons and the atom in the excited ($+$) or ground ($-$) state. As usual, $E_{e}-E_{g} (\equiv\hbar \Omega)$ is the atomic transition energy.

By employing the following transformation $\hat{a}^{\dagger}_e (x)\equiv \frac{1}{\sqrt{2}}(\hat{a}^{\dagger}_{R}(x)+ \hat{a}^{\dagger}_{L}(-x))$, $\hat{a}^{\dagger}_o (x)\equiv
\frac{1}{\sqrt{2}}(\hat{a}^{\dagger}_{R}(x)- \hat{a}^{\dagger}_{L}(-x))$, the original Hamiltonian is transformed into two decoupled ``single-mode'' parts, $\hat{H}=\hat{H}_e + \hat{H}_o$, where
\begin{align}
\hat{H}_e &= \int dx (-i) v_g \hat{a}_{e}^{\dagger}(x)\frac{\partial}{\partial x}
\hat{a}_{e}(x) + \int dx V \delta(x)\left(\hat{a}^{\dagger}_{e}(x)
\hat{\sigma}_{-} + \hat{a}_{e}(x) \hat{\sigma}_{+}\right)+E_{e} \hat{c}^{\dagger}_{e} \hat{c}_{e}+
E_{g} \hat{c}^{\dagger}_{g}
\hat{c}_{g}\notag\\
\hat{H}_o &= \int dx (-i) v_g \hat{a}_{o}^{\dagger}(x)\frac{\partial}{\partial x}
\hat{a}_{o}(x).
\end{align}

In the above decomposition, $\hat{H}_o$ is an interaction-free single-mode Hamiltonian, while $\hat{H}_e$ describes a non-trivial one-mode interacting model with coupling strength $V\equiv\sqrt{2}\, \overline{V}$. The component $\hat{H}_e$ is identical in form to the $s$-$d$ model [see sec. II of~\cite{Anderson1961}], which describes the $S$-wave scattering of electrons by a magnetic impurity in three dimensions. In this description, however, instead of fermionic operators assigned to electrons, we have bosonic operators describing photons.

Simplifying the notation by setting $v_g$ and $\hbar$ equal to unity, and omitting the subscript $e$ in $\hat{a}^{\dagger}_e$, the one-photon eigenstate of $\hat{H}_e$ assumes the form $|k\rangle \equiv \int dx [e^{i k x} \left(\theta(-x)+ t_k \theta(x)\right)\hat{a}^{\dagger}(x) + e_k \hat{\sigma}_{+}] |0, -\rangle$~\cite{ShenFan2005}, where 
\begin{equation}
t_k \equiv\frac{k - \Omega - i \Gamma/2}{k -\Omega + i \Gamma/2}, \quad \Gamma\equiv V^2
\end{equation}
is the transmission amplitude of unit magnitude, and $e_k = \frac{\sqrt{\Gamma}}{k-\Omega+i \Gamma/2}$ is the excitation amplitude. Single-photon resonance occurs when the energy $k$ is close to the transition energy $\Omega$ of the two-state atom. 

To construct the scattering matrix, an additional \index{Two-photon! bound state}%
\emph{two-photon bound state}\index{Two-photon!bound state} is required. We first describe the general features of the scattering problem. The two-photon Hilbert spaces of the ``\emph{in}'' (prior to scattering) and ``\emph{out}'' (following scattering) states coincide with those of the free photons. They consist of totally symmetric functions of the photon coordinates, $x_1$ and $x_2$. Such a Hilbert space is spanned by the complete basis $\{|S_{k,p}\rangle: k\leq p\}$ defined by
\begin{equation}
\langle x_1, x_2|S_{k, p}\rangle \equiv \frac{1}{2\pi}\frac{1}{\sqrt{2}}\left(e^{i k x_1} e^{i p x_2} +e^{i k x_2} e^{i p x_1}\right)= \frac{\sqrt{2}}{2\pi} e^{i E x_c}\cos\left(\Delta x\right), 
\end{equation}
where $E=k+p$ is the total energy of the photon pair, $x_c \equiv 1/2(x_1 + x_2)$, $x\equiv x_1 - x_2$, and $\Delta \equiv (k - E/2) = 1/2 (k-p) \leq 0$. An alternative choice of basis $\{|A_{k,p}\rangle: k\leq p\}$ spanning the same Hilbert space is defined as
\begin{equation}
\langle x_1, x_2|A_{k,p}\rangle \equiv \frac{1}{2\pi}\frac{1}{\sqrt{2}}\,\mbox{sgn}(x)\left(e^{i k x_1} e^{i p x_2} -e^{i k x_2} e^{i p x_1}\right)
= \frac{\sqrt{2}i}{2\pi}\,\mbox{sgn}(x) \, e^{i E x_c}\sin\left(\Delta x\right)
\end{equation}
where $\mbox{sgn}(x)\equiv \theta(x)-\theta(-x)$ is the sign function.  We point out that, while both $\{|S_{k,p}\rangle: k\leq p\}$ and $\{|A_{k,p}\rangle: k\leq p\}$ are complete, an arbitrary linear combination of the form $\{ a_{k, p} |S_{k,p}\rangle + b_{k, p} |A_{k,p}\rangle: k\leq p\}$ may not be. 

The transport properties of two photons in the presence of the two-state atom are described by the $S$-matrix (denoted by $\mathbf{S}$) providing us with a mapping of the Hilbert space for the {\it in} to the Hilbert space for the {\it out} states: $|\mbox{out}\rangle = \mathbf{S} |\mbox{in}\rangle$. In particular, the matrix element of the $S$-matrix, $\langle S_{k, p}|\mathbf{S}|S_{k', p'}\rangle$, yields the transition amplitude for the process under consideration. The $S$-matrix for the two-photon scattering\index{Two-photon!scattering} can be diagonalized as 
\begin{equation}\label{eq:smatrix}
\mathbf{S}\equiv\sum_{k< p}  t_k t_p |W_{k,p}\rangle \langle W_{k, p}| + \sum_{E} t_E |B_E\rangle\langle B_E|, 
\end{equation}
with
\begin{equation}\label{eq:statesdef}
|W_{k,p}\rangle \equiv \frac{1}{\sqrt{(k-p)^2 +\Gamma^2}}\left[(k-p)|S_{k,p}\rangle + i \Gamma |A_{k,p}\rangle\right],\hspace{0.5cm}
\langle x_1, x_2 |B_E\rangle \equiv \frac{\sqrt{\Gamma}}{\sqrt{4 \pi}} e^{i E x_c -\Gamma |x|/2}, \hspace{0.5cm} t_E \equiv \frac{E-2\Omega-2 i \Gamma}{E-2\Omega+ 2 i \Gamma}.
\end{equation}

As we keep following~\cite{Shen2007}, we present here a short proof that $|W_{k,p}\rangle$ and $|B_{E}\rangle$ are eigenstates of the scattering matrix. A two-photon eigenstate\index{Two-photon!eigenstate} of the component $\hat{H}_e$ has the general form:
\begin{equation}
|\Phi\rangle \equiv\left(\int dx_1 dx_2 \, g(x_1, x_2) \hat{a}^{\dagger}(x_1) c^{\dagger}(x_2) + \int dx \, e(x) \hat{a}^{\dagger}(x) \hat{\sigma}_{+}\right)|0, -\rangle,
\end{equation}
where $e(x)$ is the probability amplitude of finding the atom in its excited state. Since we are dealing with boson statistics, the wavefunction satisfies $g(x_1, x_2) = +g(x_2, x_1)$ [moreover, $g(x_1, x_2)$ is continuous on the line $x_1 = x_2$]. From $\hat{H}_e |\Phi\rangle = E |\Phi\rangle$, we obtain the following system of coupled equations:
\begin{equation}
    \begin{array}{c}
\displaystyle{\left(-i \frac{\partial}{\partial x_1}  -i \frac{\partial}{\partial x_2} - E\right) g(x_1, x_2) + \frac{V}{2} \left(e(x_1) \delta(x_2) + e(x_2) \delta(x_1)\right) =0,} \\ \\
\displaystyle{\left(-i \frac{\partial}{\partial x} - E + \Omega\right) e(x) + V \left(g(0, x) + g(x,0)\right) = 0},    
\end{array}
\end{equation}
where $g(0, x) = g(x, 0) \equiv 1/2 \times(g(0^-, x) + g(0^+, x))$. The functions 
$g(x_1, x_2)$ and $e(x)$ are piecewise continuous. From the from of the $\delta$-functions we find that the interactions take place along the two coordinate axes $x_1=0$ and $x_2=0$. From the above system of equations we obtain the following conditions on the boundary between quadrants II ($x_1< 0 < x_2$) and III ($x_1, x_2 < 0$):
\begin{equation}\label{eq:BC1}
\begin{array}{c}
\displaystyle{-i \left[g(x_1, 0^+) - g(x_1, 0^-)\right]  +\frac{V}{2} e(x_1)=0,}\\ \\
\displaystyle{\left(-i\frac{\partial}{\partial x_1}-(E-\Omega)\right) e(x_1) +  V\left[g(x_1, 0^+) + g(x_1, 0^-)\right]=0},
\end{array}
\end{equation}
while on the boundary of quadrants II ($x_1< 0 < x_2$) and I ($0 < x_1, x_2$) we have:
\begin{equation}\label{eq:BC2}
\begin{array}{c}
\displaystyle{-i \left[g(0^+, x_2) - g(0^-, x_2)\right] +\frac{V}{2} e(x_2)=0,}\\ \\
\displaystyle{\left(-i\frac{\partial}{\partial x_2}-(E-\Omega)\right) e(x_2) +  V\left[g(0^+, x_2) + g(0^-, x_2)\right]=0}.
\end{array}
\end{equation}
These boundary conditions must be supplemented by a further continuity condition
\begin{equation}\label{eq:BC3}
e(0^-) = e(0^+),
\end{equation}
to ensure self-consistency. Invoking once more the boson symmetry, we only need to consider the half space $x_1 \leq x_2$. For $x_1 < x_2 <0$ we assume the form 
\begin{equation*}
 g(x_1, x_2) =B_3 e^{i k x_1 + i p x_2} + A_3 e^{i p x_1 + i k x_2}
\end{equation*}
Then, using eqs.~\eqref{eq:BC1}-\eqref{eq:BC3}, we obtain
\begin{equation*}
 g(x_1, x_2) = t_k t_p (B_3 e^{i k x_1 + i p x_2} + A_3 e^{i p x_1 + i k x_2})
\end{equation*}
for $0< x_1 <x_2$, provided
\begin{equation*}
B_3/A_3 = (k - p - i \Gamma)/(k - p + i \Gamma)
\end{equation*}
as dictated by the continuity condition for the amplitude $e(x)$. Hence, in the \emph{full} quadrant III, the incoming state, $|W_{k,p}\rangle$ given by [see eq.~\eqref{eq:statesdef}]
\begin{align}
\langle x_1, x_2|W_{k,p}\rangle & = \left(A_3 e^{i k x_1 + i p x_2} + B_3 e^{i p x_1 + i k x_2}\right) \theta(x_1 - x_2)+ \left(B_3 e^{i k x_1 + i p x_2} + A_3 e^{i p x_1 + i k x_2}\right) \theta(x_2 - x_1)\notag\\
&\propto (k-p)\langle x_1, x_2 |S_{k, p}\rangle + i \Gamma \langle x_1, x_2|A_{k, p}\rangle,
\end{align}
is an eigenstate of the $S$-matrix with eigenvalue $t_k t_p$. A crucial step in the analysis of the scattering and transport properties beyond the single-photon\index{Single-photon!excitation} case is the realization that the set $\{|W_{k,p}\rangle: k < p\}$ does not comprise a complete set of basis of the free two-photon Hilbert space. Instead, there exists \emph{one additional eigenstate} of the $S$-matrix, denoted by $|B_{E}\rangle$ and defined by eq.~\eqref{eq:statesdef}). In order to see that $|B_{E}\rangle$ is as well an eigenstate of the $S$-matrix, we assume $g(x_1, x_2) = e^{i E x_c} e^{- \Gamma |x|/2}$ in quadrant III. Employing once more eqs.~\eqref{eq:BC1}-\eqref{eq:BC3}), we obtain $g(x_1, x_2) = t_E e^{i E x_c} e^{- \Gamma |x|/2}$ in quadrant I. This bound state features in the calculation of an exact expression for the ground-state energy in the Anderson model\index{Anderson! model}~\cite{Kawakami1981}. We then conclude that the set of eigenstates $\{|W_{k,p}, |B_{E}\rangle\}$ forms a complete and orthonormal basis that spans the free two-photon Hilbert space. We finally note that the two-photon bound state\index{Two-photon!bound state} described by $|B_{E}\rangle$, with a spatial extent $1/\Gamma$, is attributed to a single composite particle with energy $k+p$. This effective particle is not decomposed when passing through the two-state atom; as a result, the atom is capable of manipulating composite particles  made of photons. 
From our discussion above, we can then write down the momenta distribution of the out-going state $\langle S_{k_2, p_2} |\mbox{out}\rangle$ for an arbitrary incoming state of $|\mbox{in}\rangle = |S_{k_1, p_1}\rangle$: 
\begin{equation}\label{eq:smatrixelement}
\langle S_{k_2, p_2}|\mathbf{S}|S_{k_1, p_1}\rangle =  t_{k_1}t_{p_1}\delta(\Delta_1 -\Delta_2)\delta(E_1 - E_2) + t_{k_1} t_{p_1}\delta(\Delta_1 +\Delta_2)\delta(E_1 - E_2)+ B\delta(E_1 -E_2),
\end{equation}
where the first two terms on the right-hand side describe a direct momentum exchange of the incident particles. The third term, with complex amplitude
\begin{equation}
B(E_1, \Delta_1, \Delta_2)=\frac{16 i \Gamma^2}{\pi}\frac{E_1-2\Omega + i\Gamma}{\left[4\Delta_1^2 -(E_1 - 2\Omega + i\Gamma)^2\right] \left[4\Delta_2^2 -(E_1 - 2\Omega + i\Gamma)^2\right]}
\end{equation}
represents the background fluorescence in the presence of two-photon scattering\index{Two-photon!scattering}. When $\Delta_1 \neq \Delta_2$, $|B(E_1, \Delta_1, \Delta_2)|^2$ is the probability density of finding the outgoing photon pair in the $(E_1, \Delta_2)$ state when the incoming pair is in the $(E_1, \Delta_1)$ state. 

\index{Background fluorescence}%
Background fluorescence differs substantially from ordinary resonance fluorescence\index{Resonance fluorescence! vs. background fluorescence}. In the current two-photon case, the background fluorescence results from the fact that the momentum of each photon \emph{is not individually conserved}. As a result, the interaction with the two-state atom redistributes the momenta of the photon pair over a continuous range obeying the constraint of total energy and momentum conservation. We also note that the poles of $B$ at $k_{1,2}=p_{1,2}=\Omega -i\Gamma/2$ signify that either of the photons has energy approximately equal to $\Omega$ ($\Gamma \ll \Omega$). Hence, the background fluorescence comes about as the result of inelastic photon scattering off a composite particle formed when the two-state atom has already absorbed the other photon from the incident pair.  

Let us briefly comment here on the main properties of the background fluorescence amplitude, as depicted in fig. 2 of~\cite{Shen2007}. We first note that $|B(E, \Delta_1, \Delta_2)|^2$ is an even function of $E-2\Omega$. When $|E-2\Omega|\leq \Gamma$, there is a single peak centered at $\Delta_1=\Delta_2=0$. The height of the peak reaches its maximum value at $E=2\Omega$, gradually decreasing as $|E-2\Omega|$ grows. When $|E-2\Omega| = \Gamma$, the top of the peak flattens. When $|E-2\Omega| > \Gamma$, the amplitude function has four peaks centered at $(\pm\sqrt{(E-2\Omega)^2-\Gamma^2}/2, \pm\sqrt{(E-2\Omega)^2-\Gamma^2}/2)$, respectively. For any $E$ and $\Delta_1$, the locations of the maxima in $|B(E, \Delta_1, \Delta_2)|^2$ are independent of $\Delta_1$. In contrast, the $\delta$-functions in the $S$-matrix are centered along the $\Delta_1=\Delta_2$ line [see eq.~\eqref{eq:smatrixelement}]. 
 
Background fluorescence may also indicate an effective spatial interaction between the two photons. For an incoming state $|\mbox{in}\rangle = |S_{E_1, \Delta_1}\rangle$, the outgoing state is written as 
\begin{equation}
\langle x_c, x |\mbox{out}\rangle= 
e^{i E_1 x_c}\frac{\sqrt{2}}{2\pi}\left(t_{k_1} t_{p_1} \cos\left(\Delta_1 x\right)-\frac{4\Gamma^2}{4\Delta_1^2 -(E_1-2\Omega +  i\Gamma)^2} e^{i (E_1-2\Omega) |x|/2 -\Gamma |x|/2}\right) \equiv e^{i E_1 x_c} \langle x|\phi\rangle,
\end{equation}
with $\langle x |\phi\rangle$ the wavefunction in the relative coordinate $x (\equiv x_1-x_2)$. From the second term on the right-hand side we deduce that the deviation of the out-state wavefunctions from the interaction-free case is more pronounced when $\Delta_1 \simeq \pm (E_1/2 -\Omega)$, \textit{i.e.}, when at least one of the incident photons is close to resonance. A positive (negative) deviation from the interaction-free case means that the two photons are bunched (anti-bunched) after scattering. Therefore, the hyperbola $4\Delta_1^2 -(E_1-2\Omega)^2 = \Gamma^2$ indicating where such a deviation is zero, separates the regimes of photon bunching and antibunching. The deviation is maximal at $E_1 -2\Omega=\Delta_1=0$, when both incident photons resonant with the two-state atom. In this case, the two photons form a bound state after being scattered off the two-level system; the bound state has a spatial extent set by $1/\Gamma$. When $E_1 -2\Omega$ is kept at zero, the height of the peak at $x=0$ decreases with increasing $|\Delta_1|$, while for $\Delta_1=-\sqrt{3}\Gamma/2$ the peak at $x=0$ disappears. Both bunching and antibunching occur at other non-resonant values of $E_1$ and $\Delta_1$, but the effects are generally weaker. Consequently, resonance conditions generate an effective repulsion or attraction between the scattered photon pair.

\begin{figure*}
\begin{center}
\includegraphics[width=10cm]{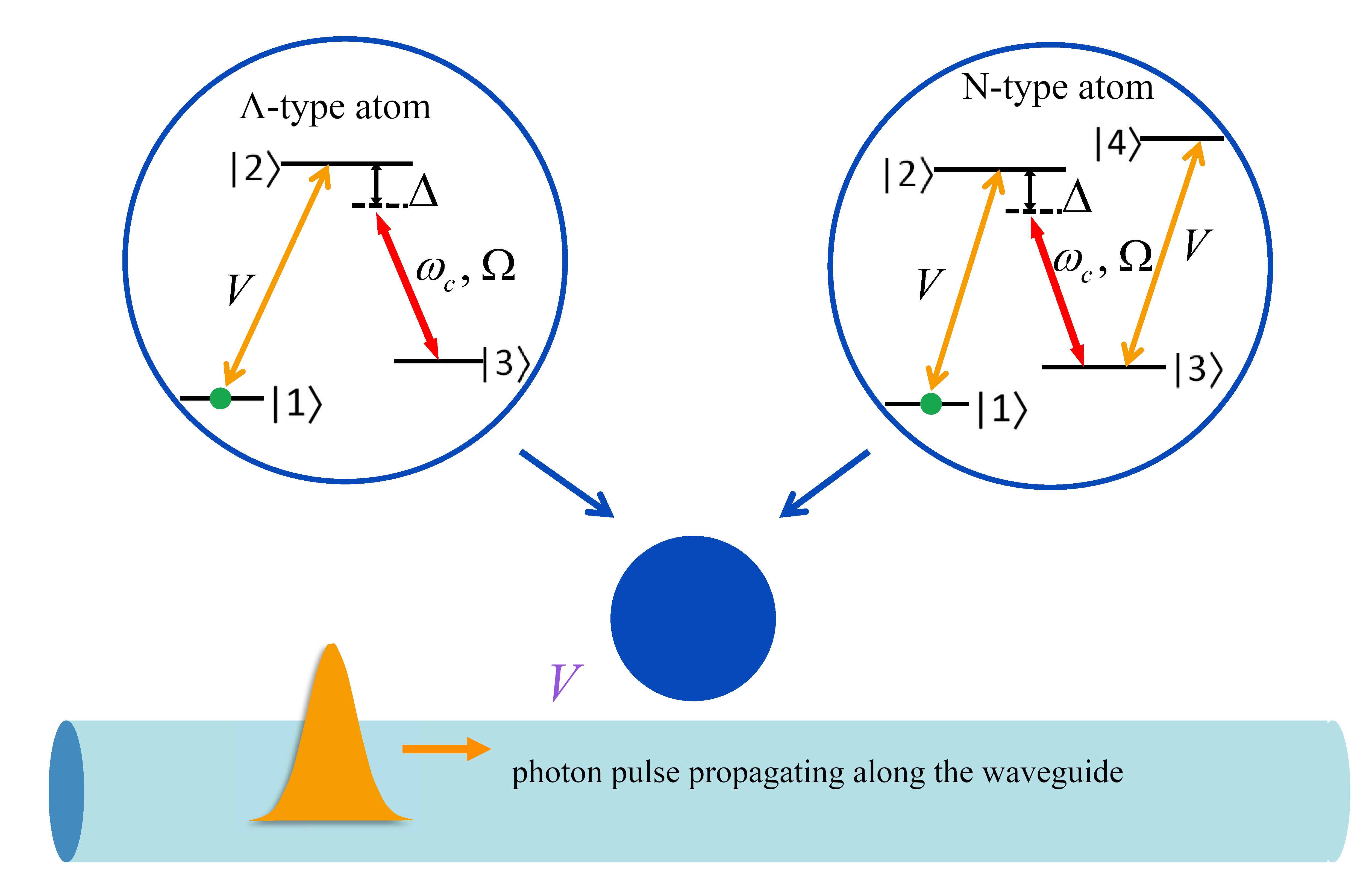}
\end{center}
\caption{{\it Schematic representation of the composite atom-waveguide system.} A $\Lambda$-type three-level system (left) next to an $N$-type four-level system (right). The photons (yellow wavepacket) in a 1D waveguide are coupled to the composite atom (blue). The transitions $|1\rangle\leftrightarrow|2\rangle$ and $|3\rangle\leftrightarrow|4\rangle$ are coupled to the waveguide modes with strength $V$. The transition $|2\rangle\leftrightarrow|3\rangle$ is driven by a semiclassical control field with frequency $\omega_c$, detuning $\Delta$. The control field yields a Rabi frequency $\Omega$. Source: Fig. 1 of~\cite{Zheng2012}. Reproduced with permission from the APS.}
\label{fig:schematicwvgQED}
\end{figure*}

Following the investigation of the fundamental light-matter interaction in a continuum of modes, we move on to the use of a waveguide-QED system to generate strongly-correlated photons through coupling to a more complicated local quantum system, \textemdash{e.g.}, a three-level or four-level atom (3LS or 4LS) as depicted in fig. \ref{fig:schematicwvgQED}. It has been reported that two-photon correlation\index{Two-photon!correlation} is much stronger in a waveguide with a driven three-level atom\index{Three-level atom} than a two-level atom~\cite{Roy2011}. To probe the strong photon-photon correlation mediated by the multilevel system, we address photonic transport, number statistics and second-order coherence associated with the correlated photon states, guided by the analysis of~\cite{Zheng2012}. Focusing for brevity on the $N$-type system (4LS), we write down the components of the Hamiltonian as
\begin{eqnarray}
&&\hat{H}_{\rm atom}^{(N)} = \sum_{j=2}^{4}\hbar\Big(\epsilon_{j}-\frac{i\Gamma_{j}}{2}\Big)|j\rangle\langle j|+\frac{\hbar\Omega}{2}\Big(|2\rangle\langle3|+{\rm H.c.}\Big), \\
&&\hat{H}_{\rm int}^{(N)} = \int dx\,\hbar V\delta(x)\Big\{ [\hat{a}_{R}^{\dagger}(x)+\hat{a}_{L}^{\dagger}(x)](|1\rangle\langle2|+|3\rangle\langle4|)+{\rm H.c.}\Big\}.
\end{eqnarray}
Here, the energy reference is the energy of the ground state $|1\rangle$, and $\epsilon_2=\omega_{21}$,
$\epsilon_3=\epsilon_2-\Delta$, and $\epsilon_4=\epsilon_3+\omega_{43}$, where $\omega_{21}$
and $\omega_{43}$ are the $|1\rangle\leftrightarrow|2\rangle$, and $|3\rangle\leftrightarrow|4\rangle$
transition frequencies, respectively. In the spirit of the quantum jump picture~\cite{Carmichael1993, BookQO2Carmichael}, we include an imaginary term in the energy level to model the spontaneous emission\index{Spontaneous! emission} of the excited states to modes other than the photonic continuum, at rate $\Gamma_{j}$. The spontaneous emission rate to the 1D waveguide continuum is given by $\Gamma=2V^{2}/c$ as dictated by Fermi's golden rule\index{Fermi's golden rule}. Applying the decoupling transformation which transforms the right and left to even and odd modes, as done in~\cite{Shen2007}, yields
\begin{equation}
\hat{H}_{\rm int}^{(N)}=\int dx\hbar\overline{V}\delta(x)\left\{\hat{a}_{e}^{\dagger}(x)\left(|1\rangle\langle2|+|3\rangle\langle4|\right)+{\rm H.c.}\right\},
\end{equation}
where $\overline{V}=\sqrt{2}V$. The applicability of the rotating-wave approximation is ensured by the fact that $\hbar\Gamma\ll \hbar\omega_{21}$, a condition which is upheld in several exemplary experiments~\cite{Astafiev2010, Abdumalikov2010,Claudon2010,Bleuse2011,Hoi2011}.

Hereinafter, we concentrate on solving for the scattering eigenstates in the even space. Since $[\hat{H},\,\hat{n}_{e}+\hat{n}_{\text{atom}}]=[\hat{H},\,\hat{n}_{o}]=0$ for the number operators $\hat{n}_{e/o}\equiv\int dx\;\hat{a}_{e/o}^{\dagger}(x)\hat{a}_{e/o}(x)$ and the atomic excitation $\hat{n}_{\text{atom}}$, the total number of excitations in both the even and odd spaces are separately conserved. Therefore, a general $n$-excitation state in the even space ($n=n_{e}+n_{\text{atom}}$) is given by
\begin{equation}
\begin{aligned}
|\Psi_{n}^{(N)}\rangle_{e}&=  \bigg[\int dx^{n}\;g^{(n)}(x)\;\hat{a}_{e}^{\dagger}(x_{1})\cdots\hat{a}_{e}^{\dagger}(x_{n}) + \int dx^{n-1} \sum_{j=2,3}f_{j}^{(n)}(x)\;\hat{S}_{1j}^{+} \;\hat{a}_{e}^{\dagger}(x_{1})\cdots\hat{a}_{e}^{\dagger}(x_{n-1})+ \\
&\int dx^{n-2}f_{4}^{(n)}(x)\;\hat{S}_{14}^{+}\;\hat{a}_{e}^{\dagger}(x_{1})\cdots\hat{a}_{e}^{\dagger}(x_{n-2})\bigg]|\emptyset,1\rangle, 
\end{aligned}
\end{equation}
where $|\emptyset,1\rangle$ is the zero-photon state with the atom in the ground state $|1\rangle$, and $\hat{S}_{ij}^{+} \equiv |j\rangle\langle i|$. The scattering eigenstates are constructed by imposing the open boundary condition we met in~\cite{Shen2007} that $g^{(n)}(x)$ is a free plane wave propagating far away from the scatterer in the incident region~\cite{Zheng2010}. That is, for $x_{1},\cdots,x_{n}<0$,
\begin{equation}\label{eq:gn}
g^{(n)}(x) = \frac{1}{n!}\sum_{Q}h_{k_{1}}(x_{Q_{1}})\cdots h_{k_{n}}(x_{Q_{n}}),\qquad \text{with} \qquad h_{k}(x) = \frac{e^{ikx}}{\sqrt{2\pi}},
\end{equation}
where $Q=(Q_{1},\cdots,Q_{n})$ denotes the permutation of $(1,2,\cdots,n)$. Solving the Schr\"{o}dinger equation with this open boundary condition, we find the scattering eigenstates for the particular system under consideration (for a detailed derivation with a two-level system, see also the Appendix of~\cite{Zheng2010}). In the even space, the \emph{single-photon scattering eigenstate}\index{Single-photon!scattering} with energy $E=\hbar ck$ is given by the expression
\begin{equation}\label{eq:SP_SE}
\begin{aligned}
&g^{(1)}(x) \equiv  g_{k}(x)=h_{k}(x)\left[\theta(-x)+\overline{t}_{k}\theta(x)\right],\quad\quad \\
&\overline{t}_{k} = \frac{\big[ck-\epsilon_{2}+\Delta+i\Gamma_{3}/2\big]\big[ck-\epsilon_{2}+(i\Gamma_{2}-i\Gamma)/2\big]-\Omega^{2}/4 }{\big[ck-\epsilon_{2}+\Delta+i\Gamma_{3}/2\big]\big[ck-\epsilon_{2}+(i\Gamma_{2}+i\Gamma)/2\big]-\Omega^{2}/4},
\end{aligned}
\end{equation}
where $\theta(x)$ is the step function. The one-photon scattering eigenstate is exactly the same for a three-level and a four-level atom since at least two quanta are needed to excite level $|4\rangle$.

For \emph{two-photon scattering}\index{Two-photon!scattering}, we start with a propagating plane wave in the region $x_{1},x_{2}<0$, and employ the Schr\"{o}dinger equation to find the wavefunction first in the region $x_{1}<0<x_{2}$ and then for $0<x_{1},x_{2}$~\cite{Zheng2010}. Following this procedure, we obtain the following two-photon scattering eigenstate with energy $E=\hbar c(k_{1}+k_{2})$:
\begin{subequations}
\begin{eqnarray}
&&g^{(2)}(x_{1},x_{2})=\frac{1}{2!}\Big[\sum_{Q}g_{k_{1}}(x_{Q_{1}})g_{k_{2}}(x_{Q_{2}}) + \sum_{PQ}B_{k_{P_1},k_{P_2}}^{(2)}(x_{Q_{1}},x_{Q_{2}}) \theta(x_{Q_{1}}) \Big] \;,\\
&&B_{k_{P_1},k_{P_2}}^{(2)}(x_{Q_{1}},x_{Q_{2}}) = e^{iEx_{Q_{2}}} \sum_{j=1,2} C_{j}e^{-\gamma_{j}|x_{2}-x_{1}|} \theta(x_{Q_{21}}) \;,
\end{eqnarray}
\end{subequations}
where $P=(P_{1},P_{2})$ and $Q=(Q_{1},Q_{2})$ are permutations of $(1,2)$, $\theta(x_{Q_{ij}})=\theta(x_{Q_{i}}-x_{Q_{j}})$, and $B^{(2)}$ is a two-photon bound state \index{Two-photon!bound state}%
---$\text{Re}[\gamma_{1,2}]>0$. Our solution applies for the general case of arbitrary strength of the control field. The three-photon scattering eigenstate with energy $E=\hbar c(k_{1}+k_{2}+k_{3})$ follows likewise:
\begin{eqnarray}
\label{eq:g3}
&&g^{(3)}(x_{1},x_{2},x_{3}) =\frac{1}{3!}\Bigg\{
\sum_{Q}g_{k_{1}}(x_{Q_{1}})g_{k_{2}}(x_{Q_{2}})g_{k_{3}}(x_{Q_{3}}) +
\sum_{PQ}\Big[g_{k_{P_{1}}}(x_{Q_{1}})B_{k_{P_{2}},k_{P_{3}}}^{(2)}\!\!(x_{Q_{2}},x_{Q_{3}}) \;\theta(x_{Q_{2}}) \\
&&\qquad\qquad\qquad\qquad+ B_{k_{P_{1}},k_{P_{2}},k_{P_{3}}}^{(3)}\!\!(x_{Q_{1}},x_{Q_{2}},x_{Q_{3}})\; \theta(x_{Q_{1}}) \Big] \Bigg\}, \nonumber\\
&&B_{k_{P_{1}},k_{P_{2}},k_{P_{3}}}^{(3)}(x_{Q_{1}},x_{Q_{2}},x_{Q_{3}}) = e^{i\big[k_{P_{1}}x_{Q_{2}}+(k_{P_{2}}+k_{P_{3}})x_{Q_{3}}\big]}
\Big[D_{1}\;e^{-\gamma_{1}|x_{Q_{3}}-x_{Q_{1}}|}+
D_{2}\;e^{-\gamma_{2}|x_{Q_{3}}-x_{Q_{1}}|} \nonumber \\
&&\qquad\qquad\qquad\qquad\qquad\qquad +D_{3}\;e^{-\gamma_{1}|x_{Q_{3}}-x_{Q_{2}}|-\gamma_{2}|x_{Q_{2}}-x_{Q_{1}}|}+
D_{4}\;e^{-\gamma_{2}|x_{Q_{3}}-x_{Q_{2}}|-\gamma_{1}|x_{Q_{2}}-x_{Q_{1}}|} \Big] \theta(x_{Q_{32}})\theta(x_{Q_{21}}),
\end{eqnarray}
where $B^{(3)}$ is a three-photon bound state, and $P=(P_{1},P_{2},P_{3})$ and $Q=(Q_{1},Q_{2},Q_{3})$ are permutations of $(1,2,3)$. The coefficients $C_{1,2}$ and $D_{1,2,3,4}$ in the bound states depend on the system parameters and have different functional forms for different multilevel systems. From the scattering eigenstates, we construct $n$-photon ($n=1-3$) $S$-matrices using the Lippmann-Schwinger formalism, similar to the procedure followed in~\cite{Shen2007} that we delineated above. The output states are subsequently obtained by applying the $S$-matrices on the incident states~\cite{Zheng2010}.

To assess the nonclassicality of scattering, we compute the intensity correlation function of the transmitted light. For a weak incident coherent state (with a mean photon number $\overline{n}\ll1$), we consider
only the contribution of the two-photon and one-photon states in the numerator and denominator in 
\begin{equation}
 g^{(2)}(\tau)=\frac{ \langle \psi | \hat{a}^{\dagger}_R(x)\; \hat{a}^{\dagger}_R(x+c\tau)\; \hat{a}_R(x+c\tau)\; \hat{a}_R(x) |\psi \rangle }{ \langle  \psi | \hat{a}^{\dagger}_R(x)\; \hat{a}_R(x) |\psi \rangle \langle \psi| \hat{a}^{\dagger}_R(x+c\tau) \; \hat{a}_R(x+c\tau) |\psi \rangle },
\end{equation}
where $|\psi\rangle$ is the asymptotic output state. Substituting the corresponding wavefunctions for Gaussian incident wave-packets\index{Gaussian! wavepacket} with spectral amplitude 
\begin{equation}
\alpha(\omega)= \frac{1}{(2\pi \sigma^2)^{1/4}}
\exp \Big[ -\frac{(\omega-\omega_0)^2}{4\sigma^2}\Big] \;,
\end{equation}
yields the explicit expression
\begin{eqnarray}
 g^{(2)}(\tau)&=&\frac{|\int dk_1dk_2 \; \alpha(k_1) \; \alpha(k_2)\; [t_{k_1}t_{k_2}(e^{-ik_1\tau}+e^{-ik_2\tau})+B(\tau)]|^2}  {|\int dk_1dk_2\; \alpha(k_1)\; \alpha(k_2) \; t_{k_1}t_{k_2}(e^{-ik_1\tau}+e^{-ik_2\tau})|^2},\nonumber  \\ \\
 B(\tau)&=&\pi( C_1 e^{-\gamma_1c\tau}+C_2 e^{-\gamma_2c\tau} ) \;,\nonumber
\label{eq:g2}
\end{eqnarray}
with
\begin{equation}
\begin{aligned}
 &C_{1}(k_1,k_2) = \frac{\beta(k_1,k_2)-\alpha(k_1,k_2) \lambda_{2}}{\lambda_{1}-\lambda_{2}},\\
 &C_{2}(k_1,k_2)=\frac{-\beta(k_1,k_2)+\alpha(k_1,k_2) \lambda_{1}}{\lambda_{1}-\lambda_{2}},
 \end{aligned}
\end{equation}
where
\begin{equation}
 \lambda_{1,2}=\frac{\Gamma+\Gamma_2-\Gamma_3}{4}\pm\xi+i \bigg(\frac{\Delta}{2}\pm\eta \bigg),
\end{equation}
\begin{subequations}
\begin{eqnarray}
&& \xi =  \frac{\sqrt{2}}{4}
  \left(\sqrt{\chi^{2}+4\Delta^{2}\Gamma^{\prime2}}-\chi\right)^{1/2}, \qquad\qquad\quad
\eta=\frac{\sqrt{2}}{4}
  \left(\sqrt{\chi^{2}+4\Delta^{2}\Gamma^{\prime2}}+\chi \right)^{1/2} \;,\\
&& \Gamma^{\prime} = \frac{\Gamma+\Gamma_{2}-\Gamma_{3}}{2}, \qquad\qquad\qquad\qquad\qquad\quad
\chi=\Delta^{2}+\Omega^{2}-\Gamma^{\prime2},
\end{eqnarray}
\end{subequations}
\begin{equation}
 \alpha(k_1,k_2) =-\frac{(\overline{t}_{k_{1}}-1)(\overline{t}_{k_{2}}-1)}{2\pi}, \quad\quad \beta(k_1,k_2)=\frac{\Gamma\Omega^{2}}{16\pi}\left[\frac{\overline{t}_{k_1}-\nu(k_1,k_2)}{\rho_{k_2}}+\frac{\overline{t}_{k_2}-\nu(k_1,k_2)}{\rho_{k_{1}}}\right],
\end{equation}
and
\begin{equation}
 \nu(k_1,k_2)=\frac{\epsilon_{4}-E-(i\Gamma_{4}-i\Gamma)/2}{\epsilon_{4}-E-(i\Gamma_{4}+i\Gamma)/2}, \qquad\qquad\qquad
\rho_{k}  = \left(ck-\epsilon_{2}+\Delta+\frac{i\Gamma_{3}}{2}\right)\left(ck-\epsilon_{2}+\frac{i\Gamma_{2}+i\Gamma}{2}\right)-\frac{\Omega^{2}}{4}.
\end{equation}
The results for $g^{(2)}(0)$ vary as a function of the Purcell enhancement factor $P=\Gamma/\Gamma_2$ (where we assume that $\Gamma_3=0$ and $\Gamma_2=\Gamma_4$)\index{Purcell enhancement}. When $P=0$, the bound-state contribution $B$ is zero and $g^{(2)}(0)=1$. With increasing $P$, however, the bound-state term progressively cancels the plane-wave term until complete antibunching occurs. A further increase of $P$ leads to an eventual photon bunching. 

To conclude this section, let us consider a pair of two-state atoms with transition frequencies $\omega_1$ and $\omega_2$, spatial separation $L=\ell_2-\ell_1$, and dipole coupled to a 1D waveguide~\cite{Zheng2013}. As one anticipates from our earlier discussion, the Hamiltonian of this system is (we note that within the RWA, causality in photon propagation is preserved by extending the frequency integrals to $-\infty$~\cite{Milonni1995})
\begin{eqnarray}\label{eq:Hamwg2}
\hat{H}&=&\sum_{j=1,2}\hbar(\omega_j-i\Gamma_j^{\prime}/2)\hat{\sigma}_j^{+}\hat{\sigma}_j^- + \hat{H}_{wg} +\sum_{j=1,2}\sum_{\beta=R,L}\int dx\hbar V_j \delta(x-\ell_j)[\hat{a}_{\beta}^{\dagger}(x) \hat{\sigma}_j^- +{\rm H.c.}] ,\nonumber \\ \\
H_{wg}&=&\int dx\frac{\hbar v_g}{i}\Big[\hat{a}_{R}^{\dagger}(x)\frac{d}{dx}\hat{a}_{R}(x)-\hat{a}_{L}^{\dagger}(x)\frac{d}{dx}\hat{a}_{L}(x)\Big],\nonumber
\end{eqnarray}
where, as usual, $\hat{a}_{R,L}^{\dagger}(x)$ is the creation operator for a right-or left-going photon at position $x$ and $v_g$ is the group velocity\index{Group velocity} of the photon pulses in the waveguide. Also, $\hat{\sigma}_j^{+}$ and $\hat{\sigma}_j^- $ are the two-state atom raising and lowering operators, respectively. An imaginary term in the energy level is once more included to model the spontaneous emission\index{Spontaneous! emission} of the excited states at a rate $\Gamma_{1,2}^{\prime}$ to modes other than the privileged waveguide continuum~\cite{Carmichael1993}. The decay rate to that waveguide continuum is given by $\Gamma_j=2V_j^2/c$. For simplicity we assume two identical two-state atoms with  $\Gamma_1=\Gamma_2\equiv\Gamma$, $\omega_1=\omega_2\equiv\omega_0\gg\Gamma$, and $\Gamma_1^{\prime}=\Gamma_2^{\prime}\equiv\Gamma^{\prime}$.

We have already seen at that stage that quantum interference effects produce an ostensible effective optical nonlinearity on the few-photon level. The relevant physical mechanisms can be clarified through an appropriate quantum many-body approach~\cite{Longo2010}. To study the interaction effects, a Green function method allows one to calculate the full interacting scattering eigenstates and subsequently the photon-photon correlations. We then set forth with a reformulated Hamiltonian
\begin{eqnarray}\label{eq:Ham_boson}
\hat{H}&=&\hat{H}_0+\hat{V},\,\qquad \hat{V}=\sum_{j=1,2}\frac{U}{2}\hat{d}_j^{\dagger}\hat{d}_j(\hat{d}_j^{\dagger}\hat{d}_j-1), \nonumber \\ \\
\hat{H}_0&=&\sum_{j=1,2}\hbar(\omega_j-i\Gamma_j^{\prime}/2)\hat{d}_j^{\dagger}\hat{d}_j+\hat{H}_{wg} +\sum_{j=1,2}\sum_{\alpha=R,L}\int dx\hbar V_j \delta(x-a_j)[a_{\alpha}^{\dagger}(x) \hat{d}_j+{\rm H.c.}],\quad\nonumber
\end{eqnarray}
where $\hat{d}_j^{\dagger}$ and $\hat{d}_j$ are bosonic creation and annihilation operators on the two-state atom sites. The two-state atom ground and excited states correspond to zero- and one-boson states, respectively. Unphysical multiple occupation is prevented by the inclusion of a large onsite repulsive interaction term with a coefficient $U$ which is a freely varying parameter. The Hamiltonian operators in eqs.~\eqref{eq:Hamwg2} and~\eqref{eq:Ham_boson} are equivalent in the limit $U\rightarrow\infty$ (hard-core\index{Hard-core bosons} Bose-Hubbard model\index{Bose-Hubbard! model!hard-core}). The non-interacting scattering eigenstates can be obtained easily from $\hat{H}_0|\phi\rangle=E|\phi\rangle$. The full interacting scattering eigenstates $|\psi\rangle$ are linked to $|\phi\rangle$ through the Lippmann-Schwinger equation~\cite{Sakurai1994}
\begin{equation}
 |\psi\rangle = |\phi\rangle+\hat{G}^{R}(E)\hat{V}|\psi\rangle, \;\;\;
 \hat{G}^{R}(E) = \frac{1}{E-\hat{H}_0+i0^{+}}.
\label{Green_Fun}
\end{equation}
The crucial step here is the numerical evaluation of the Green functions, from which one obtains the scattering eigenstates. Once these are found, assuming a weak coherent driving field, we may calculate the intensity correlation function $g_2(\tau)$ for an arbitrary interatomic separation; this function evinces the appearance of quantum beats in both Markovian and non-Markovian regimes. 

To understand the existence of quantum beats for a small interatomic separation ($k_0L\leq\pi$), we use a master equation\index{Master equation} for the density matrix $\hat{\rho}$ of the two-state atoms in the Markov approximation\index{Master equation}. Tracing out the 1D bosonic degrees of freedom yields the equation~\cite{Ficek2002}
\begin{eqnarray}
 \frac{\partial\hat{\rho}}{\partial t}&=&\frac{i}{\hbar}[\hat{\rho},\hat{H}_{c}]-\!\!\sum_{i,j=1,2}\frac{\Gamma_{ij}}{2}(\hat{\rho}\hat{\sigma}_i^{+}\hat{\sigma}_j^- +\hat{\sigma}_i^{+}\hat{\sigma}_j^- \hat{\rho}-2\hat{\sigma}_i^- \hat{\rho}\hat{\sigma}_j^{+}),\nonumber \\ \\
 \hat{H}_{c}&=&\hbar\omega_0\sum_{i=1,2}\hat{\sigma}_i^{+}\hat{\sigma}_i^- +\hbar\Omega_{12}(\hat{\sigma}_1^{+}\hat{\sigma}_2^- +\hat{\sigma}_2^{+}\hat{\sigma}_1^- ),\nonumber
\end{eqnarray}
where $\Gamma_{ii}\equiv\Gamma+\Gamma^{\prime}$, while $\Gamma_{12}\equiv\Gamma \cos(\omega_0L/c)$ and $\Omega_{12}\equiv(\Gamma/2) \sin(\omega_0L/c)$ are the spontaneous and coherent-emission coupling rates\index{Spontaneous! emission! rate}, respectively, mediated by the vacuum field.

Transforming to the symmetric and antisymmetric states we met in sec. \ref{ssec:QEDrevisited}, $|S,A\rangle \equiv (|g_1e_2\rangle\pm|e_1g_2\rangle)/\sqrt{2}$ gives a more transparent form:
\begin{equation}
  \frac{\partial\hat{\rho}}{\partial t}=\frac{i}{\hbar}[\hat{\rho},H_{c}]-\sum_{\beta=S,A}\frac{\Gamma_{\beta}}{2}(\hat{\rho}\hat{\sigma}_{\beta}^{+}\hat{\sigma}_{\beta}^- +\hat{\sigma}_{\beta}^{+}\hat{\sigma}_{\beta}^- \hat{\rho}-2\hat{\sigma}_{\beta}^- \hat{\rho}\hat{\sigma}_{\beta}^{+}), 
  \end{equation}
  with
  \begin{equation}
 H_{c}=\sum_{\beta=S,A}\hbar\omega_{\beta}\hat{\sigma}_{\beta}^{+}\hat{\sigma}_{\beta}^- ,
\end{equation}
and where
$\hat{\sigma}_{S,A}^{+}\equiv(\hat{\sigma}_1^{+}\pm\hat{\sigma}_2^{+})/\sqrt{2}$,
$\Gamma_{S,A}\equiv\Gamma+\Gamma^{\prime}\pm\Gamma_{12}$, and
$\omega_{S,A}\equiv\omega_0\pm\Omega_{12}$. 

We note that $|S\rangle$ and $|A\rangle$ are decoupled from each other and have transition frequencies $\omega_{S,A}$ and decay rates $\Gamma_{S,A}$ oscillating as a function of $L$. When $L=0$, $\Gamma_S=2\Gamma+\Gamma^{\prime}$ and $\Gamma_A=\Gamma^{\prime}$. $|S\rangle$ is in the superradiant state, while $|A\rangle$ is
subradiant. Since the waveguide couples only to the superradiant state, the second-order photon correlation follows that for a single two-state atom, akin to the behavior depicted in Figs. \ref{fig:photonmed} (b,c). However, when $k_0L=\pi/2$, $\Gamma_S=\Gamma_A=\Gamma+\Gamma^{\prime}$, $\omega_{S,A}=\omega_0\pm\Gamma/2$, and the waveguide couples to both $|S\rangle$ and $|A\rangle$. It is then precisely the quantum interference between the transitions $|S\rangle\rightarrow|g_1g_2\rangle$ and $|A\rangle\rightarrow|g_1g_2\rangle$ that gives rise  to quantum beats at the frequency difference $\omega_S-\omega_A=\Gamma$.
\begin{figure}[t]
\includegraphics[width=10cm]{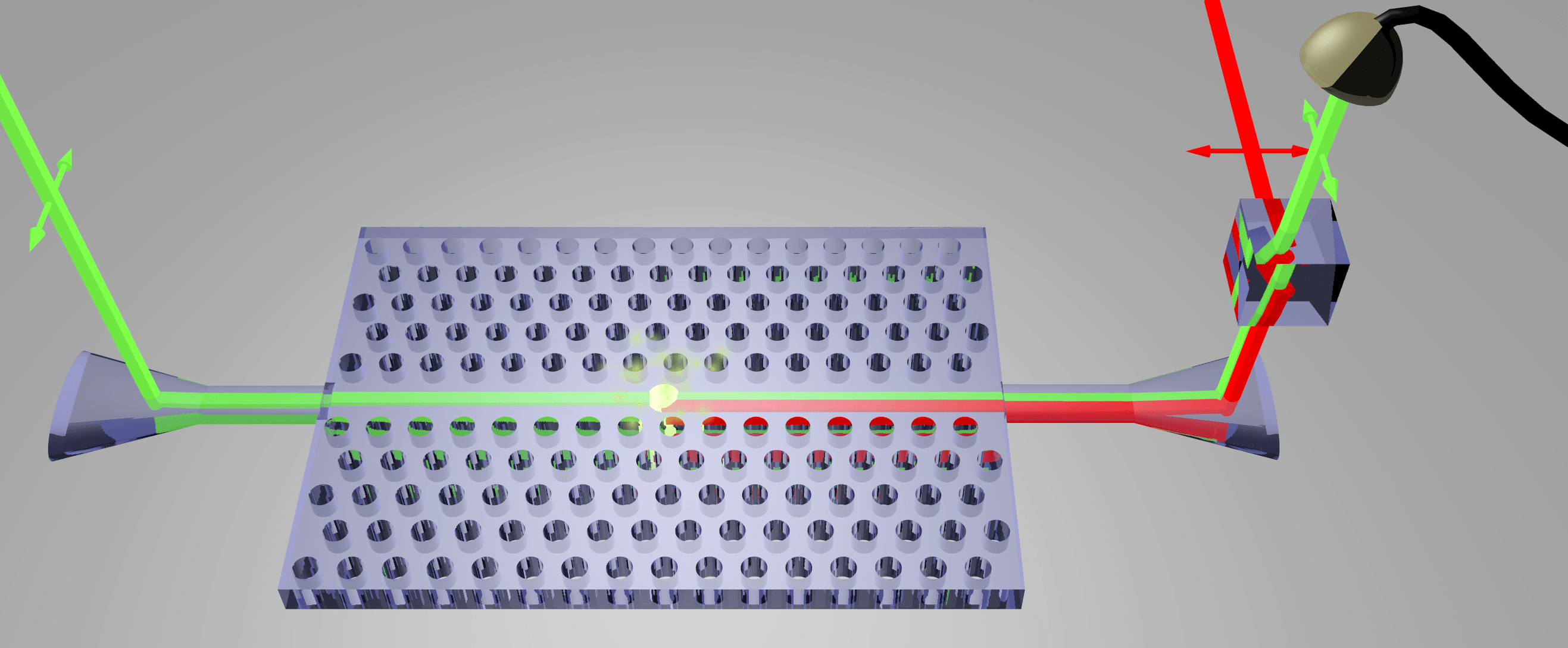} 
\vspace{3mm}
\includegraphics[width=\textwidth]{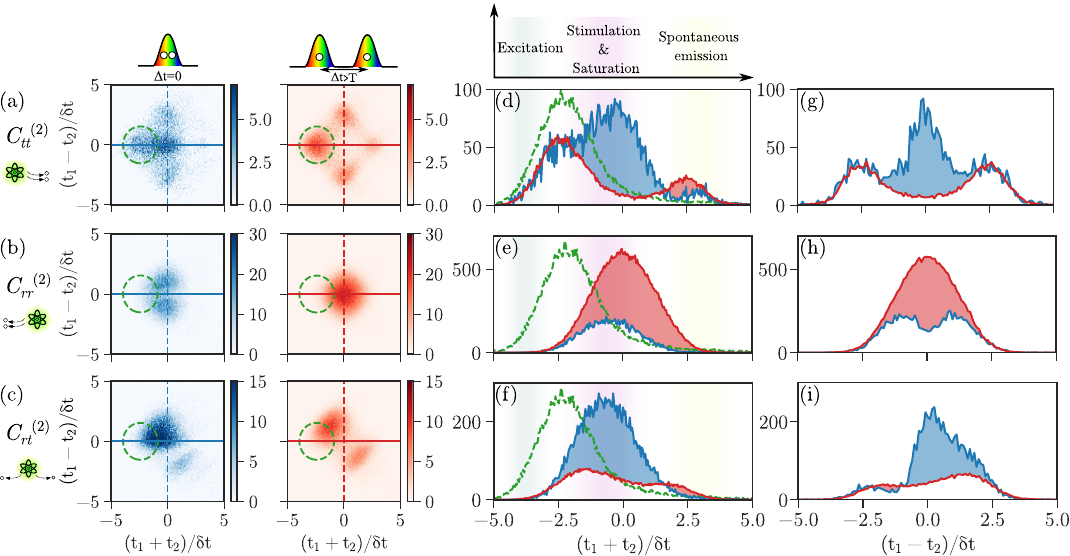}
\caption{{\bf Upper part:} Schematic representation of the experimental setup. The probe (green) and control (red) CW weak lasers of orthogonal polarizations propagate down the waveguide through opposite gratings. The transmission of the probe signal can be recorded with single-photon detectors\index{Single-photon!detector}. {\bf Lower part:} Temporal quantum correlations induced by the nonlinear interaction on a time scale determined by the pulse duration $\delta t$ and the lifetime $\tau$. Experimental measurements of the two-photon correlation\index{Two-photon!correlation} function $G^{(2)}_{xy}(t_1, t_2)$ for $\delta t/\tau=1.5$ and photons detected in different spatial modes, {\it i.e.}, both photons being transmitted {\bf (a)}, reflected {\bf (b)}, or one photon reflected and the other transmitted {\bf (c)} by the quantum dot. The two different cases correspond to two photons in the same pulse (blue data) or one photon in each subsequent pulse (red data). The green dashed line marks the position of the incident pulse exciting the quantum dot. {\bf (d)--(f)} Line cuts at $t_1=t_2$, indicated by the full line in the correlation data in frames (a)-(c), as a function of $t_1+t_2$ for the three cases. {\bf (g)--(i)} Line cuts at $t_1=-t_2$ --- indicated by the dashed line in the correlation data in frames (a)--(c) --- as a function of $t_1-t_2$ for the three cases. Source: Figs. 3 and 4 of~\cite{LeJeannicArXiv}.} 
\label{fig:2phcorr}
\end{figure}

Temporal quantum correlations induced by the interaction of two single photons via a quantum emitter in a nanophotonic waveguide were experimentally recorded in~\cite{LeJeannic2022}. In this experiment, the interaction (transmission/reflection) of a single probe photon with a quantum dot is controlled by another photon -- see the setup depicted in the upper part of fig.~\ref{fig:2phcorr}. The control photon effectively shifts the emitter resonance by a set amount that can be modified by the detuning of the control photon from the bare resonance and the control photon flux (the input intensity is about twice the saturation\index{Saturation} level of the atom). In a single experimental run it is possible to access both the correlations originating from one-photon and two-photon interactions. This is accomplished by recording the second-order intensity correlation function $G^{(2)}_{xy}(t_1, t_2)$ with two single-photon detectors\index{Single-photon!detector} in a pulsed experiment ($\Delta t$ is the separation between excitation pulses, which is much longer than the emitter lifetime). By recording two-photon detection\index{Two-photon!detection} events where $t_1 \approx t_2$ and $t_1\approx t_2 + \Delta t$, respectively, one may distinguish by post-selection two photons from the same excitation pulse or two subsequent excitation pulses interacting with the quantum dot.

Three different regimes are defined corresponding to: (1) excitation, (2) saturation\index{Saturation} and stimulated emission\index{Stimulated!  emission}, and (3) spontaneous emission\index{Spontaneous! emission} of the single quantum dot. In regime (1), the one-and two-photon dynamics is similar since the probability of absorption remains small. The rise in the excitation probability is directly revealed in the reflection measurements [frame (e) in the lower part of fig.~\ref{fig:2phcorr}], since there is no interference with the incoming pulse. When the excitation probability becomes appreciable, we enter regime (2) of stimulated emission\index{Stimulated! emission} and saturation\index{Saturation} where pronounced differences between one- and two-photon dynamics are noted. The reflection is strongly suppressed in the two-photon case [fig.~\ref{fig:2phcorr}(e)], which is a direct consequence of the fact that the emitter reflects only one photon at a time -- this leads to the dip in the time-delay plot of fig.~\ref{fig:2phcorr}(h). The single-photon response\index{Single-photon!response} is dominated by strong reflection, testifying to the efficient coupling of the quantum dot to the waveguide. This leads to a large optical extinction, confirmed by the suppression of transmission-transmission and transmission-reflection records in the frames (d) and (f) of fig.~\ref{fig:2phcorr}, respectively. In contrast, a pronounced enhancement is found for the two-photon process, since a single photon suffices for saturating the emitter enabling the transmission of a second photon. The time delay data in figs.~\ref{fig:2phcorr} (g) and (i) elucidate the dynamics of this process. The strong asymmetry in the transmission-reflection data [fig.~\ref{fig:2phcorr}(i)] reveals the temporal ordering of the scattering process, where a photon is first absorbed, then a second photon is transmitted and finally the dot re-emits a photon. In the transmission-transmission channel, the two detected photons had propagated in the same direction, giving rise to stimulated emission\index{Stimulated! emission}. One observes a pronounced preference for two-photon transmission\index{Two-photon!transmission} compared to the single-photon case\index{Single-photon!response}, see fig.~\ref{fig:2phcorr}(d). By monitoring the delay between the transmitted photons, the authors find an increased emission rate in the forward (transmission) direction by comparing the time delay data in fig.~\ref{fig:2phcorr}(g) to the transmission-reflection data in fig.~\ref{fig:2phcorr} (i). These observations are signatures of stimulated emission\index{Stimulated! emission} of a saturable emitter occurring here in the most fundamental setting of just two quanta of light and mediated by a single quantum emitter. Indeed, with the efficient coherent light-matter coupling in the photonic-crystal waveguide, even a single photon pulse is sufficient for stimulating emission. Finally, after the excitation pulse has gone through, the system enters into regime (3) where the remaining population of the emitter decays by means of spontaneous emission\index{Spontaneous! emission}. The two-photon response is suppressed relative to the one-photon response, indicating the fact that the single emitter stores only one excitation at a time (see subsec.~\ref{ssec:plasmons} and fig. 4 of~\cite{Carmichael1993}). Scattering-matrix theory has been recently applied to the study of quantum interference in multi-waveguide systems bridged by Jaynes-Cummings emitters~\cite{LiWG2022}. A recent report has investigated the minimal conditions for a superradiant burst\index{Superradiant! burst} to occur as a function of the number of emitters, the chirality of the waveguide, and the single-emitter optical depth, both for ordered and disordered ensembles. Photon bunching evinced by the second-order correlation function of ``directional superradiance''~\cite{Robicheaux2021} selects the processes in which the photon emission rate is enhanced into the waveguide coupled to an array of $N$ two-level atoms~\cite{CardenasLopez2023}. Concurrently, Liedl and collaborators have experimentally explored a wide range of regimes from weak excitation to full inversion, with ensembles comprising up to $10^3$ atoms. The Cesium emitters are trapped and optically interfaced via the evanescent field surrounding an optical nanofibre, with a guided pulse~\cite{Liedl2023}. To model the dynamics, the authors use a Lindblad master equation with the cascaded interaction\index{Cascaded! interaction} Hamiltonian $\hat{H}_{\rm casc}=-i\hbar (\gamma_f/2) \sum_{k<l} \hat{\sigma}_{l}^{\dagger}\hat{\sigma}_k + {\rm h.c.}$, where $\gamma_f$ is the collective rate of decay into the waveguide. 

Let us close this section by focusing on an extended emitter which has recently attracted considerable interest in waveguide QED. Giant atoms\index{Giant atom} are emitters which cannot be considered as pointlike, for their coupling with a guided wave extends over a significant length for which the dipole approximation no longer holds~\cite{Kockum2014}. For instance, experiments using surface acoustic waves show that a single artificial atom can be coupled to a bosonic field at several points which are located wavelengths apart~\cite{Gustafsson2014}. The multiple coupling points generate a frequency dependence in the coupling strength between the atom and its environment, as well as in the atomic Lamb shift\index{Lamb! shift}.

Santos and Bachelard have recently investigated the generation of entanglement across two giant artificial atoms in a nested configuration, interacting with a common field supported by a one-dimensional waveguide. The dynamics of the two-atom system is described by an effective Lindblad master equation\index{Master equation} with a coherent excitation exchange term and a dissipative part\index{Dissipative! part}; the latter features a cross-decay contribution on top of single-atom decay rates. The dynamical behaviour is amenable to a minimal four-level model with four distinct decay rates, characteristic to the case of giant atoms. Highly entangled states can be generated in the steady state\index{Steady state} in the case where the system is driven by a resonant classical field, while the statistics of the light emitted by the system can be used as a witness of the presence of entanglement~\cite{SantosBachelard2023}. A transmon coupling to a 1D coplanar waveguide at multiple coupling points via local capacitances is proposed in \cite{WangNori2024} to substantiate broadband chiral emission for frequency-tunable giant emitters. In this configuration, a ladder-type three-level giant atom has been shown to spontaneously emit strongly correlated photon pairs with high efficiency by optimizing the target coupling to the waveguide \cite{Gao2024}. 

The phenomenology of spontaneous emission\index{Spontaneous! emission} and two-photon scattering\index{Two-photon!scattering} has been assessed in~\cite{Alushi2023}, analyzing the cross section and the spectral features of the output channels in a waveguide QED setup where a two-state emitter is completely transparent to one-photon pulses,
while it strongly interacts with the multiphoton components of the input state\index{Multiphoton! state! component}. A single emitter, quadratically coupled to the waveguide field can realize a two-photon logic gate\index{Two-photon!logic gate} with unit fidelity, bypassing a no-go theorem formulated for conventional waveguide-QED interactions~\cite{Shapiro2006, GeaBanacloche2010, Nysteen2017}. 

\subsection{Interaction with matter in nanowire plasmons}
\label{ssec:plasmons}

We now turn to the interaction of surface plasmons\index{Surface plasmons} (SPs) with matter based on the proposal for realizing a single-photon transistor\index{Single-photon!transistor} reported in~\cite{Chang2007} concurrently with the generation of single optical plasmons in metallic nanowires coupled to quantum dots~\cite{Akimov2007}. In contrast to atoms, semiconductor quantum dots stand out due to their large optical dipole moment and engineerable emission wavelength. Other attractive properties include their fixed position and possibility of integration with cavities and waveguides using semiconductor fabrication techniques~(see e.g., \cite{Hennessy2007}). SPs are propagating electromagnetic modes confined to the surface of a conductor-dielectric interface. Their unique properties allow their confinement to sub-wavelength dimensions, which has led to waveguiding below the diffraction limit~\cite{Takahara1997}, enhanced transmission through sub-wavelength apertures~\cite{Genet2007}, and sub-wavelength imaging~\cite{Klimov2002,Smolyaninov2005,Zayats2005}. Large field enhancements associated with plasmon resonances of metallic nano-particles have also been employed for the detection of proximal single molecules via surface-enhanced Raman scattering~\cite{Kneipp1997, Nie1997}. Similar properties are directly responsible for the strong interaction between single SPs on a conducting nanowire and an individual optical emitter located nearby.

We now target more concretely a prominent experimental realization of waveguide QED, employing what we have encountered in this section and bringing in directly the Purcell enhancement of spontaneous emission\index{Spontaneous! emission! enhancement} that we met at the end of sec. \ref{ssec:lighmatter1Dcorr}\index{Purcell enhancement}. A Purcell factor of $16$ was reported in~\cite{Bleuse2011} for single quantum dots embedded in photonic nanowires. Much like what happens in a single-mode fiber, the surface plasmon (SP) modes of a conducting nanowire constitute a one-dimensional, single-mode continuum that can be labeled by the wavevectors $k$ along the direction of propagation (see \textit{e.g.},\cite{Takahara1997}). Unlike a single-mode fiber, however, the nanowire continues to display good confinement and guiding when its radius is reduced well below the optical wavelength~($R{\ll}\lambda_0$). Specifically, in this limit, the SPs exhibit strongly reduced wavelengths and small transverse mode areas relative to free-space radiation, which scale according to $\lambda_{\footnotesize\textrm{pl}}{\propto}1/k{\propto}R$ and $A_{\rm eff}{\propto}R^2$, respectively. The tight confinement results in a large interaction strength between the SP modes and any proximal emitter with a dipole-allowed transition, with a coupling constant that scales like $g\propto{1}/\sqrt{A_{\rm eff}}$. Such reduction in the group velocity\index{Group velocity} leads to an enhancement of the density of states, $D(\omega)\propto1/R$. Hence, the spontaneous emission\index{Spontaneous! emission! rate} rate into the SPs, $\Gamma^{+}{\sim}g^2(\omega)D(\omega)\propto(\lambda_0/R)^3$, can be much larger than the emission rate $\Gamma'$ into all other possible channels. A relevant figure of merit is an effective Purcell factor $P\equiv\Gamma^{+}/\Gamma'$, which can exceed $10^3$ in realistic systems. The Purcell factor plays an important role in determining the strength and fidelity of the nonlinear processes of interest\index{Purcell enhancement}.

Motivated by these considerations, we now describe in a more simplified picture a general one-dimensional model of an emitter strongly coupled to a set of travelling electromagnetic modes. Like we did before, we first consider a simple two-level configuration for the emitter, consisting of ground and excited states $\ket{-},\ket{+}$ separated by a bare frequency $\Omega$. The multimode JC Hamiltonian\index{Multimode! Hamiltonian! Jaynes-Cummings} describing this interaction in the presence of spontaneous emission\index{Spontaneous! emission} is given by
\begin{equation}\label{eq:HfullSP}
 \hat{H}= \hbar(\Omega-i\Gamma'/2)\hat{\sigma}_{ee}
+\int\,dk\,\hbar{c}|k|\hat{a}^{\dagger}_k \hat{a}_k-\hbar{g}\int\,dk\,\left(\hat{\sigma}_{+} \hat{a}_k e^{ikz_r}+\hat{\sigma}_{-} \hat{a}^{\dagger}_k e^{-ikz_r}\right),
\end{equation}
where $\hat{\sigma}_{ij} \equiv\ket{i}\bra{j}$, $\hat{a}_k$ is the annihilation operator for the mode with wavevector $k$, $g$ is the emitter-field interaction strength, and $z_r$ is the position of the emitter. We have assumed a linear dispersion relation over the relevant frequency range, $\nu_k=v_g |k|$, where $v_g$ is the group velocity\index{Group velocity} of SPs on the nanowire, and similarly that $g$ is frequency-independent. In the spirit of the theory of quantum trajectories\index{Quantum! trajectories} for an open system~\cite{BookQO2Carmichael}, we have also included a non-Hermitian term in $\hat{H}$\index{Non-Hermitian! Hamiltonian} due to the decay of state $\ket{+}$ into a reservoir of other radiative and non-radiative modes at a rate $\Gamma'$.

The propagation of SPs can be dramatically altered by the interaction with a single two-level emitter. In particular, for low incident powers, the interaction occurs with near-unit probability, and each photon can be reflected with very high efficiency. At the same time, for higher powers the emitter response rapidly saturates; the atom is not able to scatter more than one photon at a time. The low-power behavior can be easily understood by first considering the scattering of a single photon. Since we are interested only in SP modes near the optical frequency $\Omega$, we can effectively treat left- and right-propagating SPs as completely separate fields. In particular, one can define operators that annihilate a left~(right)-propagating photon at position $z$, $\hat{E}_{L(R)}(z)=(1/\sqrt{2\pi})\int\,dk\,e^{ikz}\hat{a}_{L(R),k}$, where operators acting on the left and right branches are assumed to have vanishing commutation relations with the other branch. An exact solution to the scattering from the right to left branches in the limit $P\rightarrow\infty$ was derived in~\cite{Shen2005OptLett}, and this approach can be generalized to finite $P$. In particular, it is possible to solve for the scattering eigenstates of a system containing at most one~(either atomic or photonic) excitation. The reflection coefficient for an incoming photon with wavenumber $k$ is
\begin{equation}\label{eq:rwire}
 r(\delta_k)=-\frac{1}{1+\Gamma'/\Gamma^{+}-2i\delta_k/\Gamma^{+}},
\end{equation}
where $\delta_k{\equiv}ck-\omega_{eg}$ is the photon detuning, while the transmission coefficient is related to $r$ by $t(\delta_k)=1+r(\delta_k)$. Here $\Gamma^{+}=4\pi{g^2}/c$ is the decay rate into the SPs, as obtained by application of Fermi's Golden Rule to the Hamiltonian in eq.~\eqref{eq:HfullSP}. On resonance, $r{\approx}-(1-1/P)$, and thus for large Purcell factors the emitter in state $\ket{-}$ acts as a nearly perfect mirror, which simultaneously imparts a $\pi$-phase shift upon reflection\index{Purcell enhancement}. The bandwidth $\Delta\omega$ of this process is
determined by the total spontaneous emission rate\index{Spontaneous! emission! rate}, $\Gamma=\Gamma^{+}+\Gamma'$, which can be quite large. Furthermore, the probability $\kappa$ of losing the photon to the environment is strongly suppressed for large Purcell factors, $\kappa{\equiv}1-\mathcal{R}-\mathcal{T}=2\mathcal{R}/P$, where $\mathcal{R}\,(\mathcal{T})\equiv|r|^2\,(|t|^2)$ is the reflectance~(transmittance). 

As we have demonstrated with the aforementioned theoretical results, the nonlinear response of the system can be observed when we consider the interaction of a single emitter not just with a single photon, but with multiphoton input states\index{Multiphoton! state! input}. While the reflectance and transmittance for this system are similar to those derived for a single photon at low excitation, the coincidence of several photons within the bandwidth $\Delta\omega{\sim}\Gamma$ triggers the onset of nonlinearity, saturating the atomic response. As a result, these photons cannot be efficiently reflected. We now present a mapping which allows the scattering dynamics to be accurately resolved. Without loss of generality, we assume that the incident field propagates to the right, impinging on the emitter which is initially in its ground state. Then, the initial wave function can be represented in the form
$\ket{\tilde{\psi}(t{\rightarrow}-\infty)}=\hat{D}(\{\alpha_{k}e^{-i\nu_{k}t}\})\ket{0}\ket{-}$, where the displacement operator $\hat{D}(\{\alpha_k\})\equiv\exp[\int\,dk\,(\hat{a}^{\dagger}_{R,k}\alpha_k-\alpha_k^{\ast}\hat{a}_{R,k})]$
creates a multi-mode coherent state with amplitude $\alpha_{k}e^{-i\nu_{k}t}$ when operated on the vacuum state. The definition of the displacement operator brings about a state transformation given by the expression~\cite{Mollow1975}
\begin{equation}\label{eq:disptrans}
\ket{\tilde{\psi}}=\hat{D}\left(\{\alpha_{k}e^{-i\nu_k{t}}\}\right)\ket{\psi}.
\end{equation}
The initial state is then transformed to $\ket{\psi(t{\rightarrow}-\infty)}=\ket{0}\ket{-}$. Moving now to the Heisenberg picture, the field operator transforms as $\hat{E}_{R}(z,t){\rightarrow}\hat{E}_{R}(z,t)+\hat{\mathcal{E}}_{c}(z,t)$, where the external field amplitude assumes the form $\hat{\mathcal{E}}_{c}(z,t)=(1/\sqrt{2\pi})\int\,dk\,\hat{\alpha}_{k}e^{ikz-i\nu_{k}t}$. It follows that this transformation maps the initial coherent state to a complex number in the interaction Hamiltonian, which, from a physical point of view, corresponds to a classical field with Rabi frequency  $\Omega_c=\sqrt{2\pi}g\mathcal{E}_c$. At the same time, it connects the initial photon state to the vacuum. These correspondences have the important consequence that the dynamics of the atomic emitter interacting with the field modes is amenable to the Wigner-Weisskopf approximation. Namely, the interaction with the vacuum modes gives rise to an exponential decay rate from state $\ket{+}$ to $\ket{-}$ at a rate $\Gamma$. The evolution of the atomic operators consequently reduces to the usual optical Bloch equations~\cite{BookQO1Carmichael}, something that enables the calculation of all properties of the atomic operators and the scattered field. For a resonant ($\delta_k=0$) input field of a narrow bandwidth ($\delta\omega{\ll}\Gamma$), the steady-state transmittance and reflectance evaluate as
\begin{align}
\mathcal{T} & = 
\frac{1+8(1+P)^2(\Omega_c/\Gamma)^2}{(1+P)^2(1+8(\Omega_c/\Gamma)^2)},
\\ \mathcal{R} & = 
\left(1+\frac{1}{P}\right)^{-2}\frac{1}{1+8(\Omega_c/\Gamma)^2}.
\end{align}
In the regime of low excitation~($\Omega_c/\Gamma{\ll}1$), scattering from the emitter is identical to the single-photon case\index{Single-photon!excitation}, with $\mathcal{R}{\approx}(1+1/P)^{-2},\mathcal{T}{\approx}(1+P)^{-2}$. For large Purcell enhancement factors the single emitter also acts as a perfect mirror. In the region of high drive powers~($\Omega_{c}/\Gamma{\gg}1$), however, the single emitter saturates and most of the incoming photons are simply transmitted past with no effect, $\mathcal{T}{\rightarrow}1,\mathcal{R}{\sim}\mathcal{O}((\Gamma/\Omega_c)^2)$. The significance of these results can be understood by noting that saturation\index{Saturation} is achieved at a Rabi frequency $\Omega_{c}\sim\Gamma$ that, in the limit of large $P$, corresponds to a switching energy of a single quantum~($\sim\hbar\nu$) within a pulse of duration $\sim{1}/\Gamma$.

The manifestly nonlinear response of the two-state atom at a single-photon level leads to an ostensible  modification of the photon statistics, that cannot be captured solely by steady-state average intensities. Higher-order correlations of the transmitted and reflected fields should be considered. In particular, we determine the steady-state normalized second-order correlation function for the outgoing field, $g^{(2)}_{\beta=R,L}(\tau)$, which, for a stationary process, is defined as (taking the limit $t \to \infty$)\index{Second-order correlation function}
\begin{equation}
g^{(2)}_{\beta=R,L}(z,\tau)\equiv\braket{\hat{E}^{\dagger}_{\beta}(z,t)\hat{E}^{\dagger}_{\beta}(z,t+\tau)\hat{E}_{\beta}(z,t+\tau)\hat{E}_{\beta}(z,t)}/\braket{\hat{E}^{\dagger}_{\beta}(z,t)\hat{E}_{\beta}(z,t)}^2,
\end{equation}
where $\tau$ is the difference between the two observation times $t$ and $t+\tau$. The reflected field, which is a purely scattered field, has an identical statistical behavior to the one encountered in ordinary resonance fluorescence. It follows directly that the reflected photons are strongly antibunched ($g^{(2)}(0)=0$) since the two-state atom can absorb and re-emit only one photon at a time. The transmitted field, however, has markedly different properties since it comprises a sum of the incident and scattered fields. For near-resonant weak excitation, we find
\begin{equation}\label{eq:g2R}
g^{(2)}(\tau)=e^{-\Gamma \tau}\left(P^2-e^{\Gamma \tau/2}\right)^2+\mathcal{O}(\Omega_c^2/\Gamma^2),
\end{equation}
while at high excitation $g^{(2)}(\tau)$ approaches unity for all delay times. 

The high-power behaviour of eq.~\eqref{eq:g2R} indicates that there is no appreciable change in statistics, owing to the saturation\index{Saturation} of the atomic response. The low-power behavior, on the other hand, is akin to an efficient \index{Single-photon! switch}%
\emph{single-photon switch}\index{Single-photon!switch} depending on the Purcell enhancement factor $P$\index{Purcell enhancement}. Let us trace the function of a switch of that kind. For $P{\gg}1$, individual photons have a large reflection probability, but when two photons impinge simultaneously the atomic transition saturates and the probability of transmission is considerably larger for photon pairs~(we note that when $P{\ll}1$ the statistical properties of the transmitted field are almost unaltered by the atomic emitter). This effect yields a strong bunching effect at $\tau=0$, where $g^{(2)}(0)\sim P^4$. We also encounter a later antibunching and perfect vanishing of $g^{(2)}(\tau)$ at $\tau_0=(4\ln\,P)/\Gamma$ for weak driving fields (see also the related discussion on the nonclassicality of the emission in~\cite{RiceCarmichaelIEEE}). A more complete understanding of these features can be sought from a quantum jump formalism describing the system evolution conditioned on the detection of a photon~\cite{Carmichael1991}. Unlike what happens for the reflected field, the picture for the transmitted field is less clear, since we cannot resolve the uncertainty on whether the detected photon originates from the atomic scatterer or from a direct transmission of the input field. The modification of the wave function following a photon detection is accounted for by the application of a jump operator \textemdash{in} the present case the transmitted field operator $\hat{E}_{T}=\hat{E}_{R,{\rm free}}+\hat{\mathcal{E}}_{c}+\sqrt{2\pi}ig\hat{\sigma}_{-}/c$. For a large value of $P$, the atomic component in $E_{T}$ dominates, which is responsible for the low steady-state transmittance $\mathcal{T} \approx (1+P)^{-2}$: the scattered field destructively interferes with the input field. Since multiple (co)incident photons enhance the probability of transmission, the detection of a photon increases the
conditional probability that another photon is present in the system. In the quantum-jump formalism, this translates into an instantaneous increase of the atomic coherence $\braket{\hat{\sigma}_{-}}$ by a factor of $1+P$ over its steady-state value. The destructive interference occurring between the incoming and scattered fields subsequently recedes, whence the jump causes an instantaneous increase in the field amplitude $\braket{\hat{E}_{T}}$ while it also induces a $\pi$-phase shift with respect to its steady-state value. The initial increase in the magnitude of $\braket{\hat{E}_{T}}$ generates bunching, while the $\pi$ phase shift and subsequent relaxation back to equilibrium makes $\braket{\hat{E}_{T}}$ pass through zero at some finite time $\tau_0>0$, which yields the later antibunching effect \textemdash{a} result of the cancellation of the incoming by the scattered field (see sec. 13.2.3 of~\cite{BookQO2Carmichael}). For $P=1$ the cancellation occurs precisely at $\tau=0$ giving $g^{(2)}(0)=0$.

The intensity correlation function of eq.~\eqref{eq:g2R} is to be compared with the bad-cavity limit ($2\kappa \gg g, \gamma$)\index{Bad-cavity limit} of cavity QED; in particular, the second-order correlation function\index{Second-order correlation function} of forwards scattering in the weak-excitation limit of cavity-enhanced resonance fluorescence\index{Resonance fluorescence! cavity-enhanced}~\cite{RiceCarmichaelIEEE},
\begin{equation}
g^{(2)}_{\rightarrow}(\tau) \approx [1-4C_1^2 e^{-(\gamma^{\prime}/2)\tau}]^2.
\end{equation}
where $\gamma^{\prime}=\gamma(1+2C)$ is the {\it cavity-enhanced emission rate}\index{Cavity!-enhanced emission rate}, and $C_1$ is the single-atom cooperativity (or, {\it spontaneous emission enhancement\index{Spontaneous! emission! enhancement} factor}) that we met in sec~\ref{ssec:quantumglcavQED}. This perturbative result shows that for $2C_1 \gg 1$ forwards photon scattering is highly bunched. 

Lodahl and coworkers~\cite{LodahlReview2015} provide a comprehensive review on the theoretical framework of light emission in photonic nanostructures, including photonic-crystal cavities and waveguides, dielectric nanowires, and plasmonic waveguides. Although the manuscript deals with excitons in single quantum dots embedded in photonic nanostructures, it also pertains to a broad range of quantum emitters such as molecules, nitrogen vacancy centers, or atoms. As part of a characteristic experiment conducted a few years before, Flagg {\it et al.} had reported on the development of Mollow triplet\index{Mollow! triplet} for a resonantly driven quantum dot in a microcavity~\cite{Flagg2009}. The strong light confinement in nanostructures leads to emission, scattering and absorption of photons by quantum emitters, which are dependent on the direction propagation. Locking the local polarization of light to its wavevector enables the production of non-reciprocal single-photon devices operating in a quantum superposition of two or more of their states, as well as the realization of deterministic spin–photon interfaces~\cite{Lodahl2017}. The emission properties of a quantum dot resonantly coupled to a cavity mode as well as to Majorana bound states were studied in~\cite{Ricco2022}. An alternative to electromagnetic-induced transparency\index{Transparency!
electromagnetically-induced} is proposed in Ref. \cite{AsenjoG2022}, where an electromagnetic field with periodic polarization gradient enables control over the dispersion relation of dark states that emerge in an atomic chain.

 
\section{Alternative physical systems}\label{sec:physreal}

In this section, we provide a brief account on alternative systems and platforms realizing the basic models of light-matter interaction linked to the Jaynes-Cummings (JC) model, and their associated critical behaviour that we have visited up to this point. In one of the earliest notable demonstrations of the kind, the nonperturbative dynamics of a two-state transition in a quantum box in three dimensional pillar microcavities was modelled by the multimode JC Hamiltonian\index{Multimode! Hamiltonian! Jaynes-Cummings} and the associated usual Lindblad master equation\index{Master equation} from which luminescence spectra were extracted to ascertain the presence of strong-coupling conditions~\cite{andreani1999strong}. A few years later, the interaction of cold caesium atoms with the evanescent field of a monolithic microtoroidal resonator in the experiment of~\cite{aoki2006observation} paved the way to strong coupling of single atoms with the radiation in lithographically fabricated microresonators. Particular emphasis will be laid on two systems, the first concerning the 'material', and the second the 'field' degree of freedom in the JC description. 

\subsection{Nitrogen vacancy centers}

The nitrogen-vacancy center (NVC) is a point defect with $C_{3v}$ symmetry (occurring also naturally) in a diamond lattice, comprising a substitutional nitrogen impurity adjacent to a vacancy, offering a unique combination of spin coherence at room temperature \cite{Balasubramanian2009} (with coherence times of the order of a few ns a decade ago and now in the ms range) together with efficient optical control and readout (see e.g. \cite{Buckley1212}). By virtue of its symmetry, the NVC may only have nondegenerate and two-fold degenerate energy levels. Coherent rotations and spin echoes in a single NVC spin were first reported in~\cite{Jelezko2004}. The electronic ground ($^3$A) and first excited ($^3$E) state are electron spin triplet states ($S=1$). Optical excitation is only effective between the $m_S=0$ sublevels in both states \cite{JelezkoAPL2002}. 

\begin{figure}
\includegraphics[width=8cm]{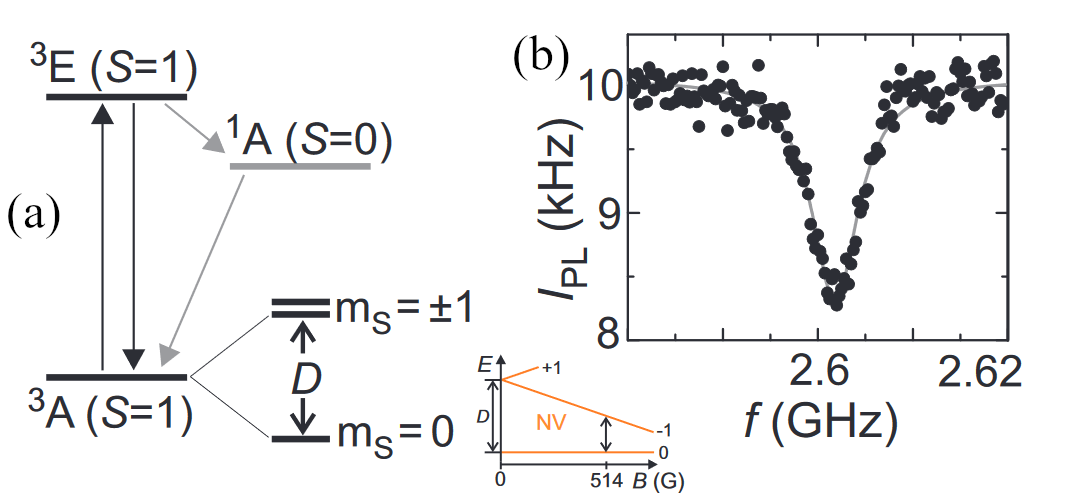} 
\includegraphics[width=8cm]{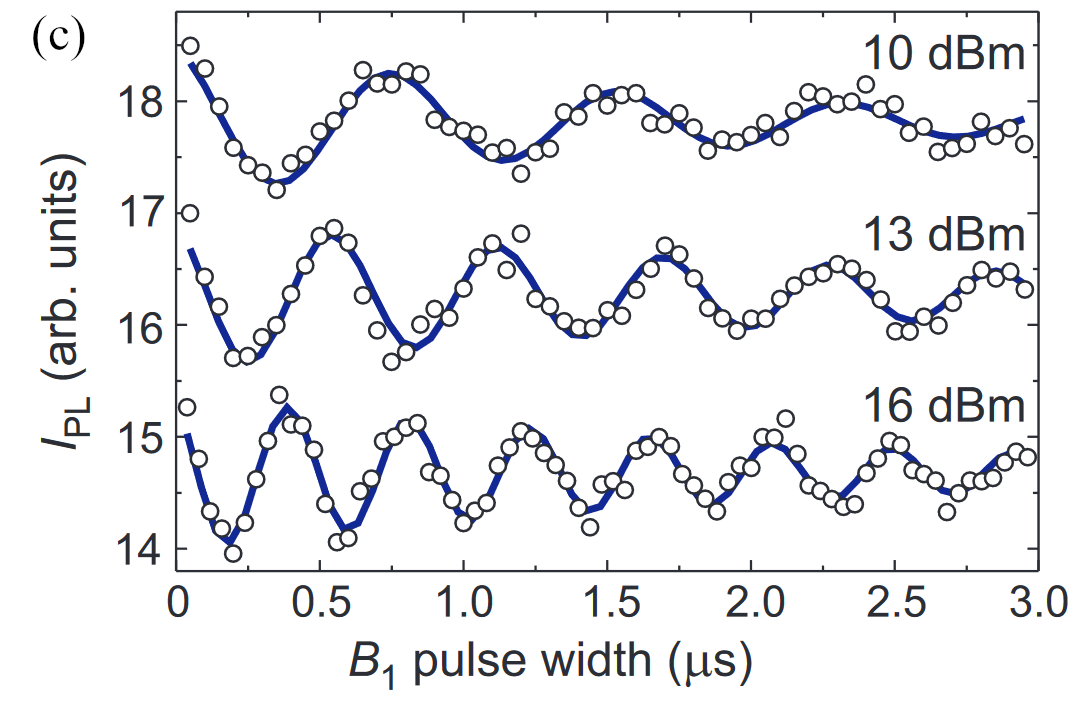}
\caption{{\bf (a)} Schematic illustration of the electronic energy level structure of the NVC. The inset on the bottom right depicts the energy levels of the NVC as a function of the external applied magnetic field $B$ along the N-V symmetry axis lifting the degeneracy of the $m_S=\pm 1$ sublevels. An energy resonance occurs around $B=514$G. {\bf (b)} Electron spin resonance, manifested as a dip in the photoluminescence intensity $I_{\rm PL}$, between the $m_S=0$ and $m_S=-1$ sublevels for a given value of the applied magnetic field. NVCs are also identified through photon antibunching measurements. {\bf (c)} Coherent Rabi oscillations for three different values of the applied RF pulse power against the pulse duration of the RF magnetic field with magnitude $B_1 \equiv B_{ \rm (RF),\, 1}$. The spin rotates between the $m_S=0$ and $m_S=-1$ sublevels, leading to the oscillation of $I_{\rm PL}$ (source: \cite{HansonPRB2006}). Reproduced with permission from the APS.}
\label{fig:NVClevelsRabi}
\end{figure}

The spin sublevels of the electronic ground state have been manipulated via magnetic resonance at room temperature in the seminal experiment of~\cite{HansonPRB2006} in 2006, as shown in fig. \ref{fig:NVClevelsRabi}. The probability to find the NVC in the $m_S=0$ state, proportional to the photoluminscence intensity $I_{\rm PL}$, reads
\begin{equation}\label{eq:RabiNVC}
P_{m_S=0}(t)=1-\frac{f_R^2}{f_R^2 + \Delta f^2}\sin^2\left(\pi \sqrt{f_R^2+\Delta f^2}\,t\right),
\end{equation}
where $f_R=(1/2)g m_B B_{\rm (RF),\, 1}/h$ is the Rabi frequency ($B_{\rm (RF),\, 1}$ is the amplitude of the AC magnetic field, $\mu_B$ is the Bohr magnetron) and $\Delta f$ is the detuning from resonance. 

Normal-mode splitting corresponding to a vacuum Rabi frequency of $g/2\pi$ for a whispering gallery mode (WGM) strongly dipole-coupled to a silica nanosphere at optical frequencies was reported in~\cite{SrongCouplingWGMNVC}. WGM cavities can support two counter-propagating traveling modes with the same frequency and profile, while the continuous total internal reflections on the boundary promote an enhanced interaction of the NVCs with the evanescent field. For a detailed analysis of the eigenmodes and the electromagnetic WGM volume see~\cite{OptimalSizesKimble}. Single NV$^{-}$ defect centers were subsequently coupled on demand to high-Q WGMs in microspheres, demonstrating the possibility of manipulating single quantum emitters probed through their modified fluorescence spectrum \cite{OnDemandNVCWGM}. The JC system Hamiltonian and the associated quantum master equation\index{Master equation} were explicitly invoked in [Appendix A of~\cite{Barclay2009}] to assess photoluminescence spectra due to coupling of a transverse magnetic (TM) WGM to NVCs mitigating the effects of electron-phonon interaction via zero-phonon (ZP) emission. Besides the ZP line, containing a small fraction of the emission, an extended phonon sideband leads to a broad emission spectrum. The transition into a Purcell-enhanced emission regime at low temperatures was reported for the coupling of an NVC to a fiber-based microcavity at room temperature~\cite{albrecht2013coupling}. Here, the coupling of the optical transitions to the cavity mode is modelled explicitly by a JC Hamiltonian and a master equation\index{Master equation} in the Markov approximation; the dynamics is however simplified since line broadening dominates at room temperature. 

The quest for high indistinguishability of the emitted photons has motivated the search for alternative active defects. Unlike the NVCs, most of the radiation emitted by negatively-charged SiV centers belongs to the ZP line, while extreme photon antibunching\index{Antibunching} (with $g^2(0)=0.04$) has been attained \cite{SiVcEnhancedEmission}. In both cases, however, the excited-state lifetime exhibits a strong dependence on temperature variations due to the presence of thermally activated nonradiative decay channels \cite{RevDiamondIQP2018}. Besides, a three-level model with an implied shelving state is found to underlie the increasing photon bunching behavior of the negatively-charged NVCs, for increasing drive power \cite{ShelvingNVCs}. 

Moving now to more sophisticated architectures, the integrated photonic network consisting of a single-mode ring resonator coupled to an optical waveguide with gratings on both ends (to enhance the collection of upward scattered photons), set a clear paradigm in single-photon generation\index{Single-photon!source} and routing  \cite{IntegratedNVCs}. At the same time, one of the most promising approaches for achieving large-scale computation relies on coupling several NVCs to optical resonators enhancing the ZP emission as opposed to the emission into phonon sidebands. Electrical control of the ZP line transition frequencies with reduced spectral diffusion was demonstrated in~\cite{ReducedDiffZPL}. For a dipole resonant with the cavity mode of quality factor $Q$ and ideally positioned with respect to the local electric field, the maximum spontaneous emission rate enhancement\index{Spontaneous! emission! rate! enhancement} (the {\it Purcell enhancement factor} we have already met several times), normalized by the fraction of the total photoluminescence emitted via the ZPL transitions (for a transition coupled to the resonator mode as opposed to the free modes in the uniform dielectric medium inside the resonator) is \cite{Faraon2011, Li20151D}
\begin{equation}\label{eq:enhancementZPL}
F_{\rm max}^{\rm ZPL} \approx \frac{3}{4\pi^2} \left(\frac{\lambda}{n}\right)^3 \frac{Q}{V},
\end{equation}
where $\lambda$ is the cavity wavelength and $V=\left[\int_{V}\varepsilon(\boldsymbol{r}) \left| \boldsymbol{E}(\boldsymbol{r})\right|^2 dV\right]/{\rm max}\left(\varepsilon(\boldsymbol{r}) \left| \boldsymbol{E}(\boldsymbol{r})\right|^2\right)$ is the optical mode volume of the resonator. Tuning two modes with quality factors $Q \sim 4,000$ over the ZP line produces an enhancement factor $F_{\rm max}^{\rm ZPL} \sim 20$~\cite{Faraon2011}. Photonic crystal cavities in one dimension have allowed the attainment of $F_{\rm max}^{\rm ZPL}=62$, with more than half of the radiated field emitted into the ZPL \cite{Li20151D}. The attainment of higher Purcell factors is currently hindered by difficulties in achieving perfect alignment between the emission dipole of NVCs and the cavity field mode. In recent experiments, NVCs in photonic structures with CVD-grown (111)-oriented diamond exhibit exceptionally good preferential orientation~\cite{CVD111}, while the ion implantation of NVCs into the mode maximum of photonic crystal nanocavities paves the way to the scalable production of optically coupled spin memories~\cite{IonImplantNVC}\index{Purcell enhancement}. 

To aid the development of scalable nanophotonic devices based on diamond, a significant reduction in the optical linewidths for NVCs was reported in~\cite{LinewidthNVC}, enabling their placement  inside a well-defined device layer, as systems with excellent optical coherence. Furthermore, entanglement between three NVCs coupled to the two counter-propagating WGMs in a microtoroidal cavity was assessed via the phenomenological master equation\index{Master equation} in~\cite{Song15}, where the diamond-induced Rayleigh scattering coupling the two modes is explicitly accounted for in the system Hamiltonian, competing with the NVC-WGM interaction of JC type. The Hamiltonian of the whole system, capturing these interactions, can be written as
\begin{equation}\label{eq:3NVCsHam}
\hat{H}=\hat{H}_S + \hat{H}_{\rm JC} + \hat{H}_{\rm sc},
\end{equation}
in which the individual terms read
\begin{equation*}
\begin{aligned}
&\hat{H}_S=\hbar \omega_0 \sum_{j=1}^{3}\hat{\sigma}_{jz} + \hbar \omega_c \hat{a}_{\rm cw}^{\dagger}\hat{a}_{\rm cw} + \hbar \omega_c \hat{a}_{\rm ccw}^{\dagger}\hat{a}_{\rm ccw}, \\
&\hat{H}_{\rm JC} = \hbar \sum_{j=1}^{3} G_{j}\left(e^{ikd_{1j}}\hat{\sigma}_{j+}\hat{a}_{\rm cw} +  e^{-ikd_{1j}}\hat{\sigma}_{j+}\hat{a}_{\rm ccw}\right) + \text{h.c.}, \\
&\hat{H}_{\rm sc} = \hbar \sum_{j=1}^{3} g_{j} \left(\hat{a}_{\rm cw}^{\dagger}\hat{a}_{\rm cw} +  \hat{a}_{\rm ccw}^{\dagger}\hat{a}_{\rm ccw}\right)+ \hbar \sum_{j=1}^{3} g_{j} \left( e^{2ikd_{1j}}\hat{a}_{\rm ccw}^{\dagger}\hat{a}_{\rm cw} + \text{h.c.} \right),
\end{aligned}
\end{equation*}
where $\hat{a}_{\rm c(c)w}$ is the annihilation operator for the (counter) propagating WGM with a common wavenumber $k$, $\hat{\sigma}_{j+}=|e_j \rangle \langle g_j|$ and $\hat{\sigma}_{jz}=|e_j \rangle \langle e_j| + |g_j \rangle \langle g_j|$ are the usual Pauli operators for the $j^{\rm th}$ NVC. The phase factors introduced in the above expressions account for the varying location of the NVCs along the toroidal cavity (with $d_{1j}$ the relative distance of the three centers with the $j=1$ placed at the origin-$d_{11}=0$). The Hamiltonian $H_S$ describes the free evolution of the system consisting of the uncoupled NVCs and the two WGMs, the Hamiltonian $H_{\rm JC}$ accounts for the NVC-WGM dipole interaction with coupling strengths $G_{j}$ and the Hamiltonian $H_{\rm sc}$ accounts for the elastic Rayleigh scattering, with a strength which is a function of the polarizability of the diamond nanocrystal. By virtue of the total angular momentum conservation, the dipole transition between the excited (a linear combination of states with $m_S=+1,-1$, one of the six allowed by the algebra of the group $C_{3v}$) and ground state (with $m_S=-1$) will be accompanied by absorbing or releasing photons with  circular polarization $\hat{\sigma}_{+}$ and frequency $\omega_0$. Formulating the dissipative dynamics\index{Dissipative! dynamics} around the quantum jumps between the eigenstates of the system Hamiltonian, a lower bound (to the convex roof) of the concurrence is computed to assess the entanglement dynamics of the tripartite system \cite{Song15}. Lastly, the possibility to realize strong collective coupling and observe cavity QED effects, such as Rabi oscillations and mode splitting, with ensembles of NV spins {\it at room temperature} has been recently assessed in~\cite{ZhangMolmer2022}. 

In summary, NVCs are one of the most promising building blocks for room-temperature manipulation in the quantum regime, amenable to a description via the JC formalism. They represent an exceptional candidate for quantum-information processing owing to extremely long electronic and nuclear spin lifetimes as well as the capability for coherent excitation in an optical fashion, such as fast initialization, high-fidelity information storage and qubit readout\index{Qubit! readout}. Instead of employing monolithic platforms, hybrid systems consisting of diamond and other materials have been investigated for the implementation of integrated quantum photonic circuits. In such a hybrid system, diamond can be used for single-photon generation\index{Single-photon!source} {\it per se} taking advantage of a wide gamut of color centers, while established photonic setups can be employed for the routing of light, thereby combining the best of both worlds.

\subsection{Strong coupling in photonic crystals}

In his pioneering proposal of 1987 that gave birth to the field of photonic crystals, Yablonovich \cite{YablonovichPRL} concluded that ``if a three-dimensionally periodic dielectric structure has an electromagnetic band gap which overlaps the electronic band edge, then spontaneous emission\index{Spontaneous! emission! forbidden} can be rigorously forbidden.'' Concurrently, prompted by Anderson localization\index{Anderson! localization}, John showed that a perturbative introduction by disorder in the dielectric constant on top of a Bravais superlattice opens up a pseudo gap of strongly localized photons in the density of states \cite{JohnPRL}. Initially, both Yablonovitch and John favored face-centered cubic (fcc) patterns which, seemingly isotropic, were supposed to block radiation in all directions. Following a report contesting the existence of a full photonic band gap in fcc structures, the successfully implemented arrangement of dielectric spheres interspersed in the diamond structure \cite{FullGap1990} demonstrated the widest gap to those days. A few years later, in 1998, a three-dimensional photonic crystal with a large bandgap and strong attenuation as well as uniform spectral response was developed with important consequences for quantum optics and quantum-optical devices \cite{Lin1998}. 

Instead of a simple exponential decay in the vacuum, spontaneous emission\index{Spontaneous! emission} displays an oscillatory behavior near the edge of a photonic band gap\index{Photonic band gap}, as was demonstrated in~\cite{SpontaneousEmission1994PBG}, while the collective spontaneous emission of $N$ two-level atoms, with an atomic resonance frequency situated at the band edge, is accompanied by symmetry breaking and the development of a phase with macroscopic polarization \cite{LocalizationofSR1995}. Let us consider $N$ identical two-level atoms with excited state $\ket{2}$, ground state $\ket{1}$, and resonance frequency $\omega_{21}$, coupled to the modes of the quantized radiation field, with $\hat{a}_{\lambda}$ and $\hat{a}_{\lambda}^{\dagger}$ the corresponding annihilation and creation operators, in a three-dimensional periodic dielectric. The Hamiltonian of this system in the interaction picture assumes the form
\begin{equation}\label{eq:PBG1}
\hat{H}=\sum_{\lambda}\hbar \Delta_{\lambda} \hat{a}_{\lambda}^{\dagger} \hat{a}_{\lambda} + i\hbar \sum_{\lambda} g_{\lambda} \left(\hat{a}_{\lambda}^{\dagger}\hat{J}_{12} - \hat{J}_{12}\hat{a}_{\lambda}\right),
\end{equation}
where $\hat{J}_{ij} \equiv \sum_{m=1}^{N} |i \rangle_{mm} \langle j|$ (with $i,j=1,2$) are the collective atomic operators, $\Delta_{\lambda} \equiv \omega_{\lambda}-\omega_{21}$ is the detuning between the field modes with frequencies $\omega_{\lambda}$ and the atomic frequency, and $g_{\lambda}=(\omega_{21} d_{21}/\hbar) \sqrt{\hbar/(2\epsilon_0 \omega_{\lambda}V)}\,\boldsymbol{e}_{\lambda}\cdot \boldsymbol{d}_{12}/d_{12}$ is the dipole coupling constant (with the dipole moment $\boldsymbol{d}_{12}$ of magnitude $d_{12}$ and $\boldsymbol{e}_{\lambda}$ the two transverse polarization unit vectors for each wavevector $\boldsymbol{k}$). Assuming that the radiation field is initially in the vacuum state, the equations of motion for the collective polarization $\braket{\hat{J}_{12}(t)}$ and inversion $\braket{\hat{J}_{3}(t)} \equiv \braket{\hat{J}_{22}(t)}-\braket{\hat{J}_{11}(t)}$ read
\begin{subequations}\label{eq:PBG2}
\begin{align}
\frac{d}{dt}\braket{\hat{J}_{12}(t)}&=\int_{0}^{t}G(t-t^{\prime}) \braket{\hat{J}_{3}(t)\hat{J}_{12}(t^{\prime})}dt^{\prime}, \\
\frac{d}{dt}\braket{\hat{J}_{3}(t)}&=-2\int_{0}^{t}G(t-t^{\prime}) \braket{\hat{J}_{21}(t)\hat{J}_{12}(t^{\prime})}dt^{\prime} + \text{c.c.},
\end{align}
\end{subequations} 
where $G(t-t^{\prime}) \equiv \sum_{\lambda} g_{\lambda}^2 e^{-i\Delta_{\lambda}(t-t^{\prime})}$ is the Green function, a kernel depending on the density of states of the dielectric medium. For an isotropic dispersion relation of the form $\omega_{\boldsymbol{k}} \approx \omega_c + A(k-k_0)^2$, after introducing a relativistic cutoff\index{Frequency cutoff!relativistic}  $\Lambda=mc/\hbar$ for the photon wavenumber, we obtain
\begin{equation}\label{eq:PBG3}
G(t-t^{\prime})=\frac{\omega_{21}^2 d_{21}^2}{6 \pi^2 \epsilon_0 \hbar} \int_{0}^{\Lambda} \frac{
k^2}{\omega_k}e^{-i(\omega_k - \omega_{21})(t-t^{\prime})}dk = \frac{\beta^{3/2} e^{-i \pi/4}}{\sqrt{\pi(t-t^{\prime})}},
\end{equation}
where $\beta^{3/2} \equiv \omega_{21}^{7/2}d_{21}^2 /(6 \pi \epsilon_0 \hbar c^3)$ and we have taken $\omega_{21}=\omega_c$ for simplicity. Assuming that the atoms in the ensemble occupy initially their excited states, with $\psi(t=0)=\prod_{m=1}^{N}\left(\sqrt{r} \ket{1}  + \sqrt{1-r}\,\ket{2}\right)$ and $r \ll 1$, neglecting quantum fluctuations by factorizing the expectation values in the system of equations (\ref{eq:PBG2}) and performing the Markov approximation by disregarding memory effects, yields
\begin{equation}\label{eq:PBG4}
\braket{\hat{J}_{3}(t)}=-N \tanh\{B[(t/\tau_{c1})^{3/2}-1]\},
\end{equation}
with $B \equiv \text{arctanh}(1-2r/N)$ and $\tau_{c1} \equiv 3^{2/3}\pi^{1/3}B^{2/3}/(2\beta N^{2/3})$. From eq. \eqref{eq:PBG4} we conclude that the collective decay rate of the inversion is proportional to $N^{2/3}$, while the superradiant intensity, being proportional to $-(d/dt) \braket{\hat{J}_{3}(t)}$ scales as $N^{5/3}$. Retaining the memory in eqs. \eqref{eq:PBG2} (but still factorizing the moments) leads to a localization of superradiant emission in the vicinity of the atomic ensemble (where $\lim_{t\to \infty} \braket{\hat{J}_{3}(t)}/N \neq -1$) as well as to the occurrence of macroscopic polarization. If, instead, one considers the anisotropic dispersion relation of the form $\omega_{\boldsymbol{k}} \approx \omega_c + A(\boldsymbol{k}-\boldsymbol{k}_0)^2$, then in the long-time limit ($\omega_c t \gg 1$) the Green function becomes
\begin{equation}\label{eq:PBG5}
G(t-t^{\prime}) \approx -\frac{1}{\sqrt{2}} \beta_3^{1/2} e^{i\pi/4}/(t-t^{\prime})^{2/3},
\end{equation} 
with $\beta_3^{1/2} \equiv \omega_{21}^2 d_{21}^2/(8 \sqrt{2} \hbar \epsilon_0 \pi^{3/2} A^{3/2} \omega_c)$, and the same procedure yields 
\begin{equation}\label{eq:PBG6}
\braket{\hat{J}_{3}(t)}=-N \tanh\{B[(t/\tau_{c2})^{1/2}-1]\},
\end{equation}
where $\tau_{c2} \equiv B^2/(4 \beta_3 N^2)$. Hence, the collective decay rate scales now as $N^2$, while the superradiance intensity is proportional to $N^3$. 

Let us now focus on the case of a single two-state atom keeping the Hamiltonian of eq. (\ref{eq:PBG1}) with $\hat{J}_{12(21)} \equiv \sigma_{12(21)}$, $\sigma_{ij}=|i \rangle \langle j|$ ($i,j=1,2$) and working with the wavefunction
\begin{equation}\label{eq:PBG7w}
\ket{\psi(t)}=c_2(t) \ket{2,\{0\}} + \sum_{\lambda}c_{1, \lambda}(t) \ket{1,\{\lambda\}}\, e^{-i\Delta_{\lambda}t}.
\end{equation} 
In the above, the state vector $\ket{2,\{0\}}$ describes the atom in its excited state in the absence of photons while $\ket{1,\{\lambda\}}$ describes the atom in its ground state with a single photon emitted into the mode $\{\lambda\}$. In that Hilbert space, the probability amplitudes obey the system of equations
\begin{subequations}\label{eq:PBG7}
\begin{align}
\frac{d}{dt}c_2(t)&=-\sum_{\lambda}g_{\lambda}c_{1, \lambda}(t)e^{-i\Delta_{\lambda} t}, \label{eq:PBG7a} \\
\frac{d}{dt} c_{1, \lambda}&=g_{\lambda} c_2(t) e^{i\Delta_{\lambda}t}. \label{eq:PBG7b}
\end{align}
\end{subequations}
Equation~\eqref{eq:PBG7a} has the formal solution
\begin{equation*}
c_{1, \lambda}(t)=g_{\lambda} \int_{0}^{t}c_{2}(t^{\prime})e^{i\Delta_{\lambda} t^{\prime}}dt^{\prime}.
\end{equation*}
yielding upon substitution to eq.~\eqref{eq:PBG7b}, 
\begin{equation}\label{eq:PBG8}
\frac{d}{dt}c_2(t)=-\sum_{\lambda} g_{\lambda}^2 \int_0^{t} c_2(t^{\prime})e^{-i\Delta_{\lambda}(t-t^{\prime})}dt^{\prime}.
\end{equation}
Taking the Laplace transform of both sides, with $\tilde{C}_2(s)=\int_0^{\infty} c_2(t) e^{-st}dt$, and assuming that $c_2(t=0)=1$ we obtain
\begin{equation}\label{eq:PBG9}
\tilde{C}_2(s)=\left[s + \sum_{\lambda} g_{\lambda}^2 \frac{1}{s + i(\omega_{\lambda}-\omega_{21})}\right]^{-1}.
\end{equation}
Adopting spherical co-ordinates in the $\boldsymbol{k}$-space and recasting the sum over transverse modes into an integral we obtain
\begin{equation}\label{eq:PBG10}
\tilde{C}_2(s)=\left[s + \frac{\omega_{21}^2 d_{21}^2}{6 \pi^2 \epsilon_0 \hbar} \int_{0}^{\Lambda} \frac{k^2 dk}{\omega_{k}[s+i(\omega_k - \omega_{21})]}\right]^{-1}.
\end{equation}
For a density of states which remains constant in the vicinity of the atomic transition, we can use the Wigner-Weisskopf approximation, substituting 
\begin{equation*}
\lim_{s \to 0^{+}} \frac{1}{s + i(\omega_k - \omega_{21})}=-i \mathcal{P}\frac{1}{\omega-\omega_{21}}+\pi \delta(\omega_k - \omega_a).
\end{equation*}
in the integrand of the right-hand side of eq.~\eqref{eq:PBG10}. We then obtain
\begin{equation}\label{eq:PBG11}
\tilde{C}_2(s)=[s+i\delta_{12} + \gamma_{21}/2],
\end{equation}
where $\delta_{21}$ and $\gamma_{21}$ are the familiar Lamb shift\index{Lamb! shift} and spontaneous emission rate\index{Spontaneous! emission! rate}, respectively. This is the exponential decay in free space. When the density of electromagnetic modes changes rapidly in the vicinity of the atomic transition, the emission profile departs from a simple exponential decay, and the exact integration in eq.~\eqref{eq:PBG10} must be performed. Selecting an isotropic dispersion relation near the band edge $\omega_c$ of the common form
\begin{equation*}
\omega_k \approx \omega_c + A(k-k_0)^2,
\end{equation*}
where $A \approx \omega_c/k_0^2$, and substituting in eq.~\eqref{eq:PBG10} yields
\begin{equation}\label{eq:PBG12}
\tilde{C}_2(s)=\frac{(s-i\delta)^{1/2}}{s(s-i\delta)^{1/2}-(i\beta)^{3/2}},
\end{equation}
in which $\delta=\omega_{21}-\omega_{c}$ and $\beta^{3/2} \equiv \omega_{21}^{7/2}d_{12}^2/(6 \pi \epsilon_0 \hbar c^3)$ as before. The inverse Laplace transform of eq.~\eqref{eq:PBG12} gives the probability amplitude
\begin{equation}\label{eq:PBG13}
c_2(t)= 2a_1 x_1 e^{\beta x_1^2 t + i\delta t} + a_2(x_2+y_2)e^{\beta x_2^2 t + i\delta t} -\sum_{j=1}^{3} a_i y_i [1-\Phi(\sqrt{\beta x_{j}^2 t})]e^{\beta x_j^2 t + i\delta t},
\end{equation}
where $x_i=f(\delta/\beta)$ and $\{a_i, y_i\}=g(x_i)$ are known constants. For $\delta=0$ we find that $\beta x_1^2=i\beta$. This means that the value of $\beta$ gives the resonant frequency splitting, which acquires the significance of the familiar vacuum Rabi splitting in cavity QED. When $\delta > \delta_{\rm crit}$ we have two dressed states occurring at frequencies $\omega_c-\beta {\rm Im}(x_1^2)$ and $\omega_c-\beta {\rm Im}(x_2^2)$, whereas the third term in the sum of eq.~\eqref{eq:PBG13} gives rise to a ``quasi-dressed'' state at the band-edge frequency $\omega_c$ with non-exponential decay. Interference between the dressed states and the ``quasi-dressed'' state generates a characteristic oscillatory behavior in the spontaneous emission\index{Spontaneous! emission} decay with no parallel in free space \cite{SpontaneousEmission1994PBG}. 

Self-induced transparency\index{Transparency! self-induced} (SIT) is a phenomenon where the propagation of a powerful ultrashort pulse of light through a resonance medium occurs without distortion and energy loss~\cite{McCallHahn1967, McCallHahn1969}. Quantum corrections in the propagation of such pulse in ordinary vacuum are negligible. This is not the case, however, in frequency gap media, such as photonic band-gap materials and Bragg reflectors, where classical linear wave propagation inside a frequency gap is blocked~\cite{SJohnRupasov1999}. The quantum Maxwell-Bloch\index{Model! Maxwell-Bloch! quantum}\index{Maxwell-Bloch! model} Hamiltonian for a collection of $N$ atoms placed inside a frequency gap material reads (setting $\hbar=c=1$)
\begin{equation}\label{eq:HamiltonianQMB}
 \hat{H}_{\rm QMB}=\omega_{21} \sum_{\alpha=1}^{N}\left(\hat{\sigma}_{\alpha}^z + \tfrac{1}{2}\right) + \int_{C}\frac{d\omega}{2\pi} \omega \hat{p}^{\dagger}(\omega)\hat{p}(\omega) -\sqrt{\gamma}\sum_{\alpha=1}^{N}\int_{C}\frac{d\omega}{2\pi} \sqrt{z(\omega)} \left[\hat{\sigma}_{\alpha}^{+} \hat{p}(\omega)e^{i k(\omega)x_{\alpha}} + \hat{p}^{\dagger}(\omega)\hat{\sigma}_{\alpha}^{-}e^{-i k(\omega)x_{\alpha}}\right],
\end{equation}
where the polariton operators\index{Polariton! operator} $\hat{p}(\omega)$ obey the commutation relation $[\hat{p}(\omega), \hat{p}^{\dagger}(\omega^{\prime})]=2\pi \delta(\omega-\omega^{\prime})$, while the Pauli operators $\boldsymbol{\hat{\sigma}}_{\alpha}=(\hat{\sigma}_{\alpha}^{x}, \hat{\sigma}_{\alpha}^{y}, \hat{\sigma}_{\alpha}^{z})$ [with $\sigma_{\alpha}^{pm}=\hat{\sigma}_{x} \pm \hat{\sigma}_{y}$] correspond to the atoms $\alpha=1,2,\ldots N$ located at the co-ordinates $x_n$. The extent of the atoms is large in comparison to the optical wavelength. The integration contour $C$ comprises two segments $C=(0,\Omega_1) + (\Omega_2, \infty)$, where $\Omega_{1,2}$ mark the boundaries of the region where states of linear propagating polariton modes\index{Polariton! mode} are forbidden. The coupling constant is $\gamma=2\pi \omega_{21} d^2/S_0$ where $d$ is the atomic dipole moment and $S_0$ is the light-beam cross-section. The atomic form factor $z(\omega)=\omega\,n^3(\omega)/\omega_{21}$ and the wavenumber $k(\omega)=\omega n(\omega)$ contain information on the frequency dispersion of the medium, here assuming the form $n(\omega)=\sqrt{\varepsilon(\omega)}$ with $\varepsilon(\omega)=(\omega^2-\Omega_2^2)/(\omega^2-\Omega_1^2)$ the dielectric permeability. 

The eigenvalues of~\eqref{eq:HamiltonianQMB}, within the $M$-particle sector of the Hilbert space for the $M$-particle Schr\"{o}dinger equation $(\hat{H}_{\rm QMB}-E)|\Psi\rangle=0$ with eigenenergy $E=\sum_{j}\omega_j$, are determined after adopting the following Bethe {\it ansatz}\index{Bethe {\it ansatz}}~\cite{SJohnRupasov1999}:
\begin{equation}\label{eq:BAE}
 \exp(i k_j L) \left( \frac{h_j - i\beta/2}{h_j + i \beta/2}\right)^N=-\prod_{l=1}^{M} \frac{h_j-h_l-i\beta}{h_j-h_l+i\beta}, \quad j=1,2,\ldots M, \quad \text{with} \quad h_j=h(\omega_j)=\frac{\omega_j-\omega_{21}}{\omega_j \, n^3(\omega_j)},
\end{equation}
where $k_j=\omega_j n(\omega_j)$ and $\beta=\gamma/\omega_{21}$. Under the above Bethe {\it ansatz}, we have placed the interacting system in a box of length $L$ imposing periodic boundary conditions at $\pm L$. The phase factor $\exp(i k_j L) $ on the LHS of eq.~\eqref{eq:BAE} is acquired by the polariton wavefunction\index{Polariton! wavefunction} through free propagation between points $\mp L/2$, while the other factor results from successive scattering from the $N$ atoms. In the course of the propagation between the points $\mp L/2$, there is also scattering from the remaining $M-1$ polaritons, accounted for by the phase factor on the RHS. The nonlinear polariton dispersion is encoded in the rapidity\index{Rapidity} $h(\omega)$. 

We will present here only the case of empty space, where the rapidity can be approximated as $h(\omega) \approx (\omega-\omega_{21})/\omega_{21}$. In the limit $L \to \infty$, eqs.~\eqref{eq:BAE} admit solutions in which the $M$ complex rapidities $h_j$ can be grouped into a number of strings, each containing $1\leq m \leq M$ rapidities. A string with $m$ rapidities is given by the expression $h_j=H + i(\beta/2)(m+1-2j)$, $j=1,2,\ldots,m$, and, due to the linear relation between rapidity, frequency and momentum, the particle frequencies and momenta also adopt similar expressions $\text{const} + (\gamma/2) (m+1-2j)$. Assuming that all particles are grouped into a single string with $M=m$, and substituting the expressions for $k_j$ and $\omega_j$ in eqs.~\eqref{eq:BAE}, we obtain 
\begin{equation}
 \exp(i Q m L)=1, \quad \text{where} \quad Q(\Omega)=K - \frac{2\rho}{m}\text{arctan}\left[\frac{m\gamma}{2(\Omega-\omega_{21})}\right]
\end{equation}
is the soliton\index{Soliton!momentum} momentum per photon, and $N=\rho L$ defines the linear atomic density. The group velocity\index{Group velocity}\index{Soliton!group velocity} of soliton propagation in the medium, $v_g=d\Omega/dQ$, is calculated as 
\begin{equation}
 \frac{1}{v_g}=1 + \frac{\gamma \rho}{(\Omega-\omega_{21})^2 + (m\gamma/2)^2}=\frac{1}{c} + \frac{2\pi}{\hbar c} \frac{\omega_{21} d^2 n_a}{(\Omega-\omega_{21})^2 + (1/\tau_s)^2},
\end{equation}
where we have reinstated the SI. In the last expression, $n_a$ is the atomic density in space and $\tau_s \sim m^{-1}$ is the duration of the soliton\index{Soliton!duration}. We conclude that only macroscopically ``long stings'' with $m \gg 1$ photons propagate without dissipation in an absorbing atomic system. Now, in the frequency gap medium there are no classical modes available for stimulated emission\index{Stimulated!  emission} and the polariton-atom coupling is highly nonlocal. This happens because an excited atom exhibits nonlocal interactions with neighbouring atoms within the classical tunneling distance, leading to coherent hopping. In contrast to the case of ordinary solitons, it is found that the particle-atom scattering speeds up a gap soliton\index{Soliton!gap}~\cite{SJohnRupasov1999}. For a review on exciton-polariton solitons\index{Soliton!exciton-polariton} in microcavities, wires, and waveguides, see~\cite{SICH2016908}. On the experimental front, Fieramosca and collaborators have shown that large single-crystal flakes of 2D layered perovskite\index{Perovskite! lattice}~\cite{Fieramosca2018} can sustain strong polariton nonlinearities at room temperature\index{Polariton! room temperature}, obviating the need to be embedded in an optical cavity with highly reflecting mirrors~\cite{Fieramosca2019}. Perovskite semiconductors\index{Perovskite! semiconductor} are excellent optical gain media typically characterised by remarkable quantum yield emission at room temperature, a bandgap which is tuneable by their composition, and a slow Auger recombination\index{Auger recombination}~\cite{Suarez2023}.   

Photonic nanocavities surrounded by a 2D photonic-crystal slab were employed in~\cite{Akahane2003} meeting the challenge of achieving Bragg reflection in multiple directions, increasing significantly the $Q$ factor in small cavities whose length is comparable to the wavelength. A year later, three dimensional mode in-plane confinement was achieved~ \cite{Yoshie2004}, demonstrating double-peaked emission in the photoluminscence spectra of a single quantum dot embedded in a photonic crystal nanocavity owing to vacuum Rabi splitting. A few years ahead, two distributed Bragg reflectors with a GaAs cavity in between were used to probe transitions up the JC excitation ladder, operating in the strong-coupling regime~\cite{Kasprzak2010}. This development followed the demonstration of photon blockade with a quantum dot strongly coupled to a photonic crystal resonator two years earlier~\cite{Faraon2008}. The non-classical nature of the generated dressed states was studied experimentally in~\cite{majumdar2012probing} using the second-order autocorrelation function\index{Second-order correlation function} at zero delay, while assessing the frequency-dependent photon statistics of the transmitted light. The dynamics of the coupled quantum dot-cavity system were explicitly modeled by the driven JC Hamiltonian. We remark that photon blockade was first demonstrated in 2005 with one atom trapped in a large Fabry-P\'{e}rot cavity\index{Fabry-P\'erot! cavity/resonator}, many times the wavelength of its resonant mode \cite{Birnbaum2005}. The ultrastrong light-matter coupling regime has been recently attained in a photonic crystal waveguide where many-body effects arise from interactions that break particle number conservation. At the same time, these interactions allow for the direct observation of multi-particle bound states with a single-excitation probe field. In that experiment, a multimode fluorescence\index{Multimode! fluorescence} measurement has been used to observe the broadband emission of entangled pairs of photons~\cite{vrajitoarea2022ultrastrong}.

Emission spectra of site-controlled quantum dots positioned at prescribed locations within a photonic crystal cavity confirmed that quantum interference of the exciton recombination paths through the cavity and free-space modes can significantly modify the radiation. The observed asymmetry in the polarization-resolved emission spectra is a function of the quantum dot position, as demonstrated in~\cite{Lyasota2022}. This leads us to the next subsection, devoted to the role of phonons for light-matter interaction in a lattice. 

\subsection{Phonon-mediated scattering}
\label{subsec:phononsc}

Phonon-induced incoherent scattering due to electron-acoustic-phonon interaction via a deformation potential is incorporated in the so-called {\it effective polaron master equation}~\cite{roy2011influence, Hughes2013}\index{Master equation!effective polaron} to assess the squeezing\index{Squeezed! state} of a cavity field strongly coupled to a single coherently driven quantum dot~\cite{Kumar2022}. 

Semiconductor quantum dots are fundamentally different from atoms, and this calls for extra care when describing the light-matter interaction. Typically, quantum dots are embedded in a solid state lattice where electron–phonon interactions, though sometimes ignored in quantum optical studies, are known to impact optical properties; they affect photoluminescence lineshapes~\cite{Besombes2001}, coherent Rabi oscillation~\cite{Ramsay2010} and the Mollow triplet\index{Mollow! triplet} spectrum of resonance fluorescence\index{Resonance fluorescence! spectrum}~\cite{Ulrich2011, roy2011phonon}; phonon-mediated scattering can cause excitation-induced dephasing~\cite{Ramsay2010, Forstner2003}, which is detrimental to the exploitation of quantum optical interactions. 

\begin{figure}
\includegraphics[width=13cm]{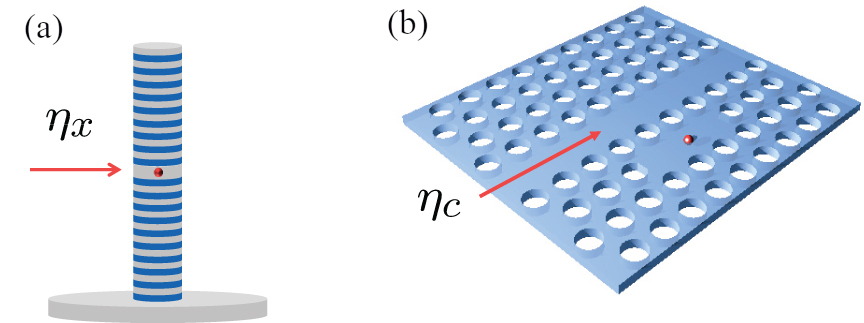} 
\vspace{4mm}
\includegraphics[width=7.2cm]{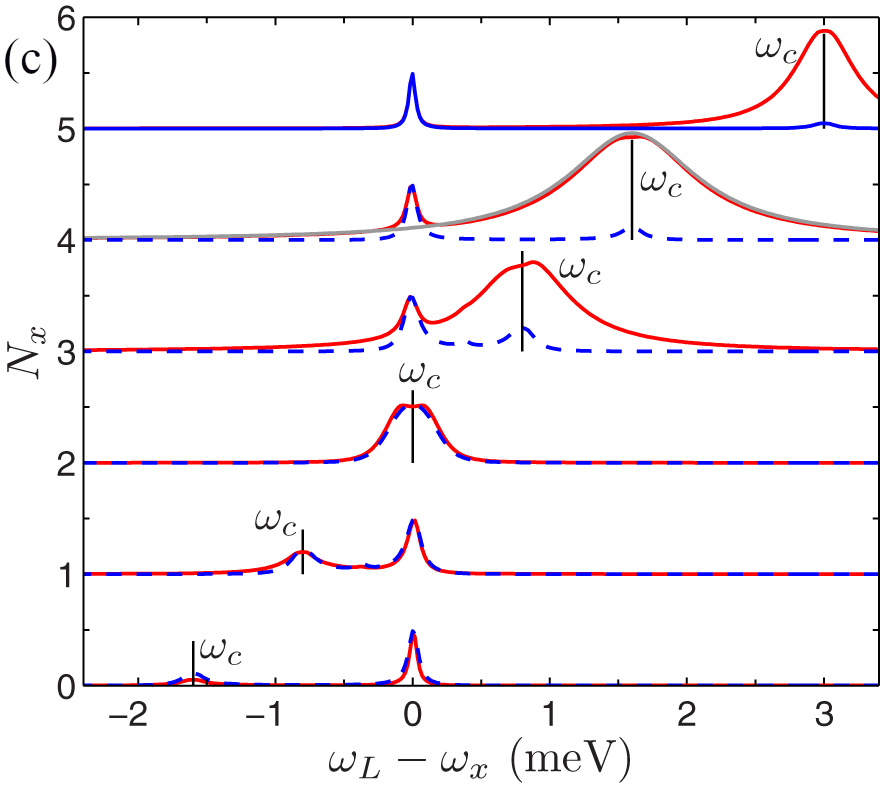} 
\includegraphics[width=8.7cm]{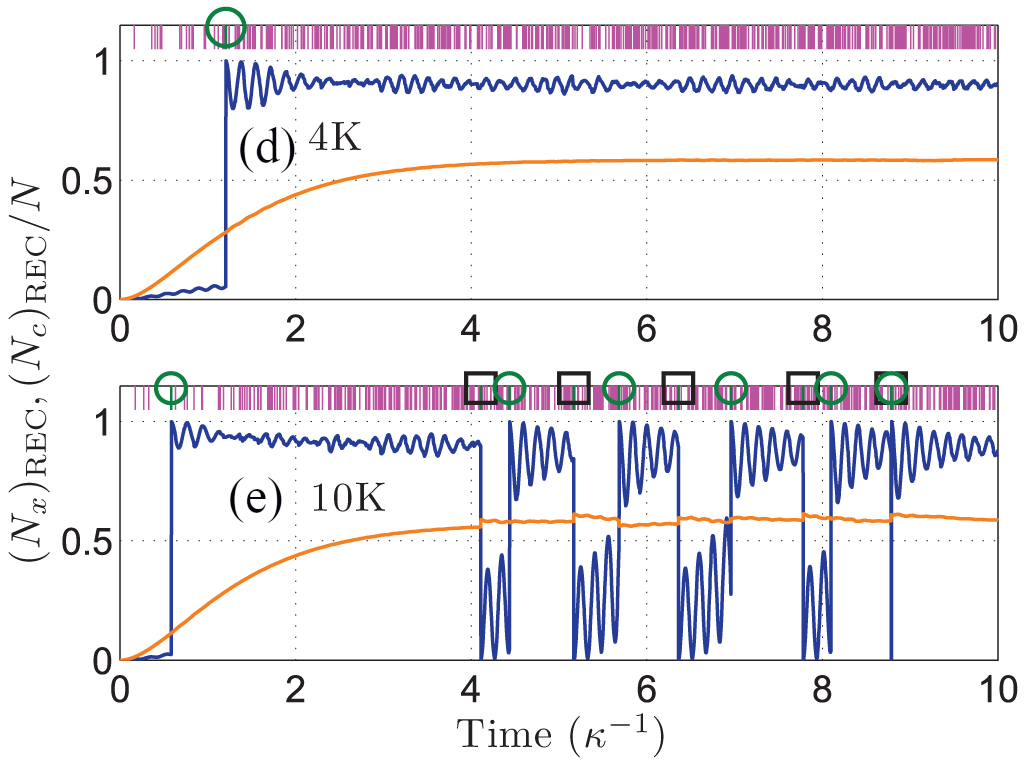} 
\caption{{\bf Schematic representation} of two possible semiconductor cavity-QED systems with coherent exciton ($\eta_x$) or cavity ($\eta_c$) driving: {\bf (a)} a micropillar system and {\bf (b)} a planar photonic crystal system. A small red sphere indicates the position of the quantum dot; {\bf (c)} Quantum-dot cavity system with cavity drive $\eta_c=0.3\mkern2mu$meV and a bath temperature of $4\mkern2mu$K. {\bf Steady-state exciton populations} are plotted as a function of laser frequency for cavity-exciton detunnings $\Delta_{cx}=-1.6$, $-0.8$, $0$, $0.8$, $1.6$, and $3.0\mkern2mu{\rm meV}$ (lower to upper). Blue dashed curves show the population without phonon scattering while the red solid curves include phonon scattering. The bare cavity resonance is indicated by the vertical black line. The grey solid curve (near $\Delta_{cx}=1.6\mkern2mu{\rm meV}$) shows results obtained with the Jaynes-Cummings term omitted from the polaron-transformed Hamiltonian; {\bf Sample quantum trajectories} at {\bf (d)} $4\mkern2mu{\rm K}$ and {\bf (e)} $10\mkern2mu{\rm K}$, for optimal detuning, $\Delta_{cx}=1.6\mkern2mu{\rm meV}$, an effective cavity-exciton coupling $g^\prime=0.1\mkern2mu{\rm meV}$, and spectral parameters\index{Spectral! parameter} $\omega_b=1\mkern2mu{\rm meV}$, $\alpha_p/(2\pi)^2=0.06\mkern2mu{\rm ps}^2$ [entering the spectral function $J(\omega)=\alpha_{p}\,\omega^{3}\exp (-{\omega^{2}}/{2\omega_{b}^{2}})$]. Orange (light) lines display the photon number expectation $(N_c)_{\rm REC}/N=\langle \hat{a}^\dagger \hat{a}\rangle_{\rm REC}/N$ ($N=60$) and blue (dark) lines the exciton expectation $(N_x)_{\rm REC}=\langle\hat{\sigma}_+\hat{\sigma}_-\rangle_{\rm REC}$. The expectations are {\em conditioned} upon the record of quantum jumps: either photon decay (magenta lines) or phonon-mediated scattering [green circles (black squares) for scattering at rate $\Gamma_{\rm ph}^{\hat{\sigma}_+\hat{a}}$ ($\Gamma_{\rm ph}^{\hat{a}^\dagger\hat{\sigma}_-}$)]. Reproduced from~\cite{Hughes2013} (figs. 1, 6 and 7 therein) with permission from the authors.}
\label{fig:phononsQD}
\end{figure}

To delineate the underpinning mechanisms we closely follow the steps of~\cite{Hughes2013} and work in a frame rotating at the laser pump frequency $\omega_L$. The simplest Hamiltonian for a single quantum dot interacting with a cavity mode and phonons, excluding the quantum dot and cavity decay, reads
\begin{equation}\label{eq:HQD}
\hat{H}=\hbar\Delta_{xL}{\hat{\sigma}}_{+}{\sigma}_{-}+\hbar\Delta_{cL}{\hat{a}}^{\dagger}{\hat{a}}+\hbar g({\hat{\sigma}}_{+}{\hat{a}}+{\hat{a}}^{\dagger}{\hat{\sigma}}_{-})+\hat{H}_{\rm{drive}}+{\hat{\sigma}}_{+}{\hat{\sigma}}_{-}\sum_{q}\hbar\lambda_{q}({\hat{b}}_{q}+{\hat{b}}_{q}^{\dagger})+\sum_{q}\hbar\omega_{q}{\hat{b}}_{q}^{\dagger}{\hat{b}}_{q},
\end{equation}
where ${\hat{b}}_{q}^{\dagger}$ and ${\hat{b}}_{q}$ are creation and annihilation operators for mode $q$ of the phonon bath, $\lambda_q$ is the (real) exciton-phonon coupling, $\hat{a}^\dagger$ and $\hat{a}$ are photon creation and annihilation operators for the cavity mode, and ${\hat{\sigma}}_+$ and ${\hat{\sigma}}_-$ are Pauli raising and lowering operators for the exciton; $\Delta_{\alpha L}\equiv \omega_\alpha-\omega_L$ ($\alpha =x,c$) designates the detuning of the exciton (frequency $\omega_{x}$) and cavity (frequency $\omega_{c}$) from the laser drive, and $\hat{H}_{\rm{drive}}=\hat{H}^c_{\rm drive}+\hat{H}^x_{\rm drive} = \hbar\eta_{c}({\hat{a}}+{\hat{a}}^{\dagger})+\hbar\eta_{x}({\hat{\sigma}_+}+\hat{\sigma}_-)$ is the drive Hamiltonian. Since we deal with quasi-resonant coherent excitation, higher lying exciton states and continuum levels in the quantum dot material are neglected. The authors adopt either non-zero $\eta_x$ or non-zero $\eta_c$.  A possible advantage of cavity over exciton driving is that it might mitigate problems with excitation-induced dephasing, a process that accompanies excitation via the exciton-phonon reservoir~\cite{Ulrich2011, roy2011phonon}; it also allows for pumping through a waveguide input channel~\cite{Kamada2004} -- see fig.~\ref{fig:phononsQD}(b) -- enabling chip-based quantum optics using semiconductor fabrication techniques.

When moving from Hamiltonian~\eqref{eq:HQD} to the effective master equation\index{Master equation}, one first transforms to a polaron frame\index{Polaron! frame} to take electron–acoustic-phonon interactions into account at a microscopic level. This formally recovers the independent boson model~\cite{wilson2002quantum, Krummheuer2002} in the appropriate limit; the independent boson model is known to capture the characteristic spectrum of an exciton coupled to a phonon bath~\cite{Besombes2001}. The derived master equation~\cite{wilson2002quantum, McCutcheon2010, roy2011phonon}\index{Master equation} treats the coherent electron-phonon interaction nonperturbatively through a mean phonon displacement,
\begin{equation}
\braket{\hat{B}}=\exp\left[-\frac{1}{2} \sum_q\left(\frac{\lambda_q}{\omega_q}\right)^2 (2\bar n_q+1) \right ]
=\exp\left[-\frac{1}{2}\int^{\infty}_{0}d\omega\frac{J(\omega)}{\omega^{2}}\coth\left (\frac{\hbar\omega} {2k_B T} \right )\right],
\end{equation}
where $\bar n_q$ is the mean phonon number (Bose-Einstein distribution\index{Distribution! Bose-Einstein} at temperature $T$), and $J(\omega) = \sum_q \lambda_q^2\delta(\omega-\omega_q)$ is the phonon spectral function. The incoherent interaction (scattering) is treated within the second-order Born approximation~\cite{roy2011influence, roy2011phonon}, although higher-order contributions are included by the polaron transform. Adding quantum dot  and cavity decay, the time-convolutionless master equation\index{Master equation!time-convolutionless} for the reduced density operator is
\begin{equation}\label{eq:MEQDconvolutionless}
\frac{d \hat{\rho}}{d t}=\frac{1}{i\hbar}[\hat{H}_{S}^{\prime},\hat{\rho}]+{\cal L}_{\rm ph}\hat{\rho}+{\cal L}\hat{\rho},
\end{equation}
with polaron-transformed Hamiltonian
\begin{equation}\label{eq:polaronHam}
\hat{H}^{\prime}_{S}= \hbar(\Delta_{xL}-\Delta_{P}){\hat{\sigma}}_{+}{\hat{\sigma}}_{-}+\hbar\Delta_{cL} {\hat{a}}^{\dagger}{\hat{a}}+\hbar {\hat{X}}_{g}+\hat{H}^c_{\rm drive},
\end{equation}
polaron shift $\Delta_P=\int^{\infty}_{0}d\omega{J(\omega)}/{\omega}$ (absorbed by $\Delta_{xL}$ below), and the phonon-scattering term
\begin{equation}\label{eq:fullratesqd}
{\cal L}_{\rm ph}\hat{\rho}=-\sum_{m=g,u}\int^{\infty}_{0} d\tau\, G_{m}(\tau)[{\hat{X}}_{m},\hat{X}_m(\tau)\hat{\rho}]
+{\rm H.c.}.
\end{equation}
where $\hat{X}_m(\tau)=\exp(-i\hat{H}_{S}^{\prime}\tau/\hbar){\hat{X}}_{m}\exp(i\hat{H}_{S}^{\prime}\tau/\hbar)$, with $(\hat{X}_g,i\hat{X}_u)=g'({\hat{a}}^{\dagger}{\hat{\sigma}}_{-}\pm{\hat{\sigma}}_{+}{\hat{a}})+\eta_x'(\hat{\sigma}_-\pm\hat{\sigma}_+)$, $g^\prime=\braket{\hat{B}}g$, $\eta_x^\prime=\braket{\hat{B}}\eta_x$.
The rescaling $g\rightarrow g'=\braket{\hat{B}}g$ was pointed out some time ago by Wilson-Rae and Imamo\ifmmode \breve{g}\else \u{g}\fi{}lu~\cite{wilson2002quantum}. It is important to note that $g^\prime$ and $\eta_x^\prime$ are temperature dependent; this dependence, however, is often ignored when fitting experiments, and compensated for by changing other parameters in an attempt to improve the fit. The response functions $G_{g}(t)=\cosh[\phi(t)]-1$ and $G_{u}(t)=\sinh[\phi(t)]$ are polaron Green functions~\cite{wilson2002quantum} obtained by assuming a separable density operator for the system and phonon-bath, and tracing over the phonon degrees of freedom. They are defined by the phonon phase term
\begin{equation}\label{eq:phase}
\phi(t)=\sum_q \left(\frac{\lambda_q}{\omega_q}\right)^2  \left [ (\bar n_q+1)e^{-i\omega_q t} + \bar n_q e^{i\omega_q t}  \right ]=\int^{\infty}_{0}d\omega\frac{J(\omega)}{\omega^{2}}\left[\coth\left(\frac{\hbar\omega}{2k_BT}\right)\cos(\omega t)-i\sin(\omega t)\right],
\end{equation}
which clearly includes contributions from multi-phonon scattering. The last term in eq.~\eqref{eq:MEQDconvolutionless} is a sum of three Lindbladians ${\cal L}=\kappa{\cal L}[\hat{a}]+(\gamma/2){\cal L}[\hat{\sigma}_-]+(\gamma^\prime/2){\cal L}[\hat{\sigma}_{ee}]$, with ${\cal L}[\hat{\xi}]\hat{\rho}=2\hat{\xi}\hat{\rho}\hat{\xi}^\dagger-\hat{\xi}^\dagger\hat{\xi}\hat{\rho}-\hat{\rho}\hat{\xi}^\dagger\hat{\xi}$ and $\hat{\sigma}_{ee} \equiv |e\rangle\langle e|$. It accounts for cavity decay at rate $2\kappa$, exciton decay at rate $\gamma$, and pure dephasing of the exciton at rate $\gamma^\prime$. These processes broaden the zero-phonon line, an essential effect not captured by the independent boson model. We adopt the established phonon spectral function, $J(\omega)=\alpha_{p}\,\omega^{3}\exp (-{\omega^{2}}/{2\omega_{b}^{2}})$, which describes the electron-acoustic-phonon interaction via a deformation potential, the dominant source of phonon scattering for InAs and GasAs quantum dots~\cite{Takagahara1999}.

The model as outlined yields an involved solution scheme with little physical insight. We turn therefore to an effective Lindblad form of the phonon scattering term, which is shown by Roy and Hughes~\cite{roy2011influence} to be in very good agreement with the predictions of eq.~\eqref{eq:MEQDconvolutionless}. For cavity excitation it makes the replacement:
\begin{equation}
{\cal L}_{\rm ph}\hat{\rho}\to\frac{\Gamma_{\rm ph}^{\hat{\sigma}_{+}\hat{a}}}{2}{\cal L}[{\hat{\sigma}}_{+}{\hat{a}}]\hat{\rho}
+\frac{\Gamma_{\rm ph}^{\hat{a}^{\dagger}\hat{\sigma}_{-}}}{2}{\cal L}[{\hat{a}}^{\dagger}{\hat{\sigma}}_{-}]\hat{\rho},
\end{equation}
with scattering rates
\begin{equation}
\label{eq:phononrates}
\Gamma_{\rm ph}^{\hat{\sigma}_{+}\hat{a}/\hat{a}^{\dagger}\hat{\sigma}_{-}} =2g^{\prime2}\,{\rm Re} \left [\int_{0}^{\infty}d\tau\,
e^{\pm i\Delta_{cx} \tau}\! \left (e^{\phi(\tau)}-1 \right )\right],
\end{equation}
where $\Delta_{cx}=\omega_c-\omega_x$ is the cavity-exciton detuning. The replacement follows by making the approximation $\hat{X}_m(\tau)\approx\exp(-i\hat{H}_{0}^{\prime}\tau/\hbar){\hat{X}}_{m}\exp(i\hat{H}_{0}^{\prime}\tau/\hbar)$, with $\hat{H}_0^{\prime} =\hbar\Delta_{xL}{\hat{\sigma}}_{+}{\hat{\sigma}}_{-}$, in~\eqref{eq:fullratesqd}. This approximation is good when $\eta_x^{-1}$  and  $g^{-1}$ are much smaller than the phonon correlation time (or when the detunings are larger than $\eta_x$ and $g$). With it we capture the dependence of the phonon scattering rates on detuning for detunings that are large enough to have a significant impact on the integral in eq.~\eqref{eq:phononrates}. In this prescription, phonon scattering amounts to a {\it one-way coupling} between the driven cavity mode and quantum dot exciton expressed through quantum jumps. There are jumps in two directions -- photon creation accompanied by exciton de-excitation (at the rate $\Gamma_{\rm ph}^{\hat{a}^\dagger\hat{\sigma}_-}$) and photon annihilation accompanied by excitation of the exciton (at the rate $\Gamma_{\rm ph}^{\hat{\sigma}_+\hat{a}}$). The expressions~\eqref{eq:phase} and~\eqref{eq:phononrates}, however, yield an asymmetry of rates. It is this asymmetry of rates that allows for phonon-mediated inversion.

For exciton excitation, the Lindblad form of phonon scattering term contains two additional contributions:
\begin{align}
\label{eq:LPh2}
{\cal L}_{\rm ph}\hat{\rho}\to & \frac{\Gamma_{\rm ph}^{\hat{\sigma}_{+}\hat{a}}}{2}{\cal L}[{\hat{\sigma}}_{+}{\hat{a}}]\hat{\rho}
+\frac{\Gamma_{\rm ph}^{\hat{a}^{\dagger}\hat{\sigma}_{-}}}{2}{\cal L}[{\hat{a}}^{\dagger}{\hat{\sigma}}_{-}]\hat{\rho} +\frac{\Gamma_{\rm ph}^{\hat{\sigma}_{+}}}{2}{\cal L}[{\hat{\sigma}}_{+}]\hat{\rho}
+\frac{\Gamma_{\rm ph}^{\hat{\sigma}_{-}}}{2}{\cal L}[{\hat{\sigma}}_{-}]\hat{\rho},
\end{align}
with additional scattering rates
\begin{equation}
\label{eq:phononrates2}
\Gamma_{\rm ph}^{\hat{\sigma}_{+}/\hat{\sigma}_{-}} =2(\eta_x^\prime)^2\,{\rm Re} \left [\int_{0}^{\infty}d\tau\,
e^{\pm i\Delta_{Lx} \tau}\! \left (e^{\phi(\tau)}-1 \right )\right],
\end{equation}
where $\Gamma_{\rm ph}^{\hat{\sigma}_+}$ (up scattering) results in incoherent excitation and $\Gamma_{\rm ph}^{\hat{\sigma}_-}$ (down scattering) causes enhanced decay. Further details appear in Ref.~\cite{roy2011influence}. The asymmetry in these rates also allows for phonon-mediated inversion.

It is widely acknowledged that any scheme to achieve population inversion must employ a multi-level manifold of material states; steady-state inversion is not possible for a driven two-level system. This is restricted, however, to a simple two-level atom coupled to classical fields in a non-engineered environment. When interacting with a \emph{quantized} radiation mode, the atom may be inverted by two-photon excitation\index{Two-photon!excitation} to higher lying levels of the  quantized \emph{matter plus field} \cite{Savage1988, HughesStInv2011, Leek2009sideband}. More recently, a two-level dipole has been set to mediate an effective interaction between a vibrational and an optical bath. This interaction leads to the occurrence of steady-state inversion for the matter system in an explicitly non-Markovian dynamical evolution which does {\it not} proceed by simply adding the individual effects of each bath~\cite{Gribben2022}.

Figure~\ref{fig:phononsQD}(c) evinces significant inversion in the presence of phonon scattering over a broad detuning range when the cavity is blue shifted (for instance, with $\Delta_{xc}=1.6$meV) for a quantum dot-cavity system and coherent excitation of the cavity mode. The quantum trajectory simulations\index{Quantum! trajectories} displayed in fig.~\ref{fig:phononsQD}(d-e) provide insight into the role of the neglected Jaynes-Cummings term. For a detuning of $\Delta_{cx}=1.6~$meV, the phonon scattering dynamic  at $4\mkern2mu{\rm K}$ and $10\mkern2mu{\rm K}$ is displayed and compared. Short magenta lines signal cavity photon jumps, while green circles and black squares identify phonon scattering jumps -- photon annihilation accompanied by exciton excitation (green circles) and photon creation accompanied by de-excitation (black squares). Phonon scattering excites the exciton within a cavity lifetime. The excitation is maintained over a relatively long time at $4\mkern2mu{\rm K}$ [frame (d)], and eventually lost primarily through radiative decay (the $\gamma$-jump is not seen in the figure). Reverse phonon scattering is the main source of de-excitation at $10\mkern2mu{\rm K}$ [frame (e)], resulting in a reduced population (from $0.9$ to $0.8$). The Jaynes-Cummings term in the polaron-transformed Hamiltonian of eq.~\eqref{eq:polaronHam} is responsible for the oscillation between photon scattering events. It hardly alters the pattern of quantum jumps but reduces somewhat the mean population.

Highly structured environments generate long-lived bath correlations characterized by complex dynamics which are very hard to simulate. These difficulties are further exacerbated when spatial correlations between different constituents of the system become important~\cite{Gribben2020}. The simultaneous coupling of quantum dots to acoustic phonons and a continuum of electromagnetic-field modes is a characteristic example of a system simultaneously coupled to two structured baths, and serves as a test case for a newly developed method termed {\it automated compression of environments} developed in~\cite{Cygorek2022}. At the heart of the method lies the explicit microscopic construction of the so-called {\it process tensor}\index{Process tensor}~\cite{Pollock2018, Jorgensen2019}, an object originally conceived as a way to interpret correlations for a general non-Markovian environment. The method is able to interpolate between infinite and short memory configurations within the same algorithm. At the same time, Fux and coworkers assessed the thermalization of individual spins in a short $XYZ$\index{Model! Heisenberg! $XYZ$} chain with strongly-coupled thermal leads. The results, obtained by introducing a general numerical tensor network simulation\index{Tensor network! simulation} to compute multitime correlations, reveal distinct effective temperatures\index{Effective! temperature} in low-, mid-, and high-frequency regimes when the Heisenberg chain is placed between a hot and a cold bath~\cite{Fux2023}. We stress that in cases of the like, correlations between the system and its environment play an important role for the correct computation of the two-time correlations, and they cannot be encoded in the reduced density matrix alone as is the case in the Born approximation\index{Born approximation}.

\begin{figure}
\includegraphics[width=0.75\textwidth]{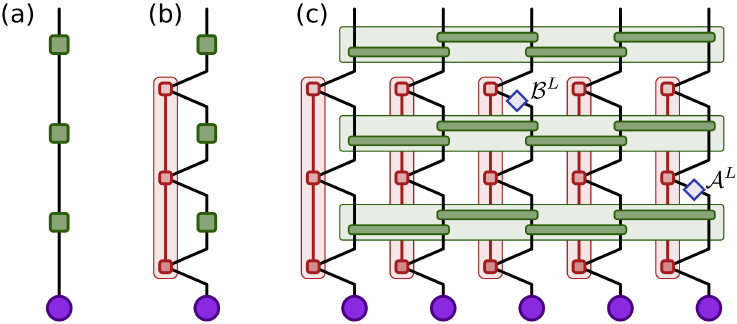} 
\caption{Tensor networks employed for the simulation of three time steps of a single closed {\bf (a)}, a single non-Markovian {\bf (b)} and a chain of non-Markovian open quantum systems {\bf (c)}. The purple circles indicate initial states, while the green squares and green rectangles represent the system propagators. The red-shaded regions enclose the process tensor in matrix-product operator form (PT-MPO), which encodes the influence of the environment onto the system. While for a Markovian open quantum system the influence of the environment is local in time, the PT-MPO couples the evolution of the system at multiple points in time and is thus capable of encoding multi-time correlations and non-Markovian effects. Frame (c) shows the proposed tensor network for the simulation of a five-site chain of system-environment pairs with uncorrelated initial states in a first-order Suzuki-Trotter splitting among the chain sites. The super-operators $\mathcal{A}^{L}$ and $\mathcal{B}^{L}$ are inserted to illustrate the calculation of the two-time correlation $\langle \hat{B}_3(2\delta t) \hat{A}_5(1 \delta t)\rangle$. Source: Fig. 1 of~\cite{Fux2023}. Reproduced with permission from the authors.} 
\label{fig:TN}   
\end{figure} 

In fig.~\ref{fig:TN}, we depict the tensor network and algorithm employed to obtain multitime correlations for 1D many-body quantum chains in the presence of strongly coupled and structured environments. The method is based on a representation of the process tensor in a matrix product operator form (PT-MPO). The PT-MPO has been developed as a general operational approach to treat non-Markovian open quantum systems, and can be obtained for a large class of different microscopic models including Gaussian and non-Gaussian, interacting and non-interacting environments formed by fermions, bosons or spins~\cite{Banuls2009, Jorgensen2019, Lerose2021, Fux2021, Sonner2021, Ye2021, Cygorek2022,Thoenniss2023}. 

The process tensor encodes the complex system-environment correlations and allows the compression of the influence of the environment to capture the most physically relevant sector of a exponentially large state space~\cite{Jorgensen2019}. Time evolution is performed with an adapted version of the so-called time evolving block decimation algorithm\index{Algorithm! block decimation}, which implements a procedure that dynamically locates the most relevant low-dimensional Liouville subspace~\cite{Daley2004, Zwolak2004}. The method presented in~\cite{Fux2023} is able to model complex situations where baths couple to every chain site, motivating the development of a versatile set of numerical tools to asses dynamics, correlations, and thermodynamic properties of many-body open quantum systems. In fact, an open source project has been very recently designed for the application of several numerical methods related to the time evolving matrix product operator (TEMPO) and the process tensor (PT) approach to open quantum systems~\cite{TEMPO}. 

\subsection{Interaction with an arbitrary electromagnetic environment}

Since mode interactions are an intrinsic feature of nanophotonic systems, a new mapping to the quantum optics formalism has been suggested in~\cite{IMedina2021} for the case where the spectral density of the electromagnetic environment cannot be written as a sum of individual Lorentzian distributions. 

In the presence of a single emitter, the electromagnetic mode basis for a continuum in macroscopic QED can be selected such that the Hamiltonian becomes~\cite{Buhmann2008, Hummer2013, SanchezBarquilla2020}
\begin{equation}\label{eq:HMQED}
    \hat{H}_{\mathrm{f}} = \hat{H}_e + \int_0^{\infty} \mathrm{d}\omega\left[\omega \hat{a}_\omega^{\dagger} \hat{a}_\omega +
    \hat{\mu}_e g(\omega) (\hat{a}_\omega + \hat{a}_\omega^{\dagger}) \right],
\end{equation}
where $\hat{H}_e$ is the bare emitter Hamiltonian and $\hat\mu_e$ is the corresponding dipole operator, while $\hat{a}_\omega$ are the bosonic annihilation operators of the electromagnetic mode at frequency $\omega$, subject to the commutation relation $[a_\omega,a_{\omega'}^\dagger] = \delta(\omega-\omega')$. The coupling between the emitter and the modes is modeled by 
\begin{equation}\label{eq:gMQED}
g^2(\omega)=\frac{\omega^2}{\pi \epsilon_0  c^2} \vec{n}\cdot
\text{Im}\{\mathbf{G}(\vec{r}_e,\vec{r}_e,\omega)\}\cdot \vec{n},
\end{equation}
where $\vec{r}_e$ is the position vector of the emitter, $\vec{n}$ indicates the orientation of
its dipole moment (assuming that all relevant transitions are identically oriented), and $\mathbf{G}(\vec{r},\vec{r}\,',\omega)$ is the classical dyadic Green function. This quantity is directly related to the spectral density\index{Spectral! density} of the environment via $J(\omega) = \mu^2 g^2(\omega)$, for a set transition dipole moment $\mu$~\cite{NovotnyHechtBook}. The environment under consideration consists of $N$ interacting electromagnetic modes with ladder operators $\hat{a}_i$, $\hat{a}_i^\dagger$, linearly coupled to the quantum emitter. Each of these modes is additionally coupled to an independent spectrally flat background bath. The resulting Hamiltonian is $\hat{\mathcal{H}}= \hat{H}_S + \hat{H}_B$, with
\begin{subequations}\label{eq:Hmodel}
\begin{align}
    \hat{H}_S &= \hat{H}_e + \sum_{i,j=1}^{N} \omega_{ij} \hat{a}^\dagger_{i} \hat{a}_{j} + \hat{\mu}_e \sum_{i=1}^N g_i (\hat{a}_i+\hat{a}_i^\dagger),\\
    \hat{H}_B &= \sum_{i=1}^{N} \int\left[\Omega \hat{b}_{i,\Omega}^\dagger \hat{b}_{i,\Omega} + \sqrt{\frac{\kappa_i}{2\pi}} (\hat{b}_{i,\Omega}^\dagger \hat{a}_i + \hat{b}_{i,\Omega} \hat{a}^\dagger_i)\right]
    \mathrm{d}\Omega.
\end{align}
\end{subequations}
\begin{figure}
 \includegraphics[width=0.34\textwidth]{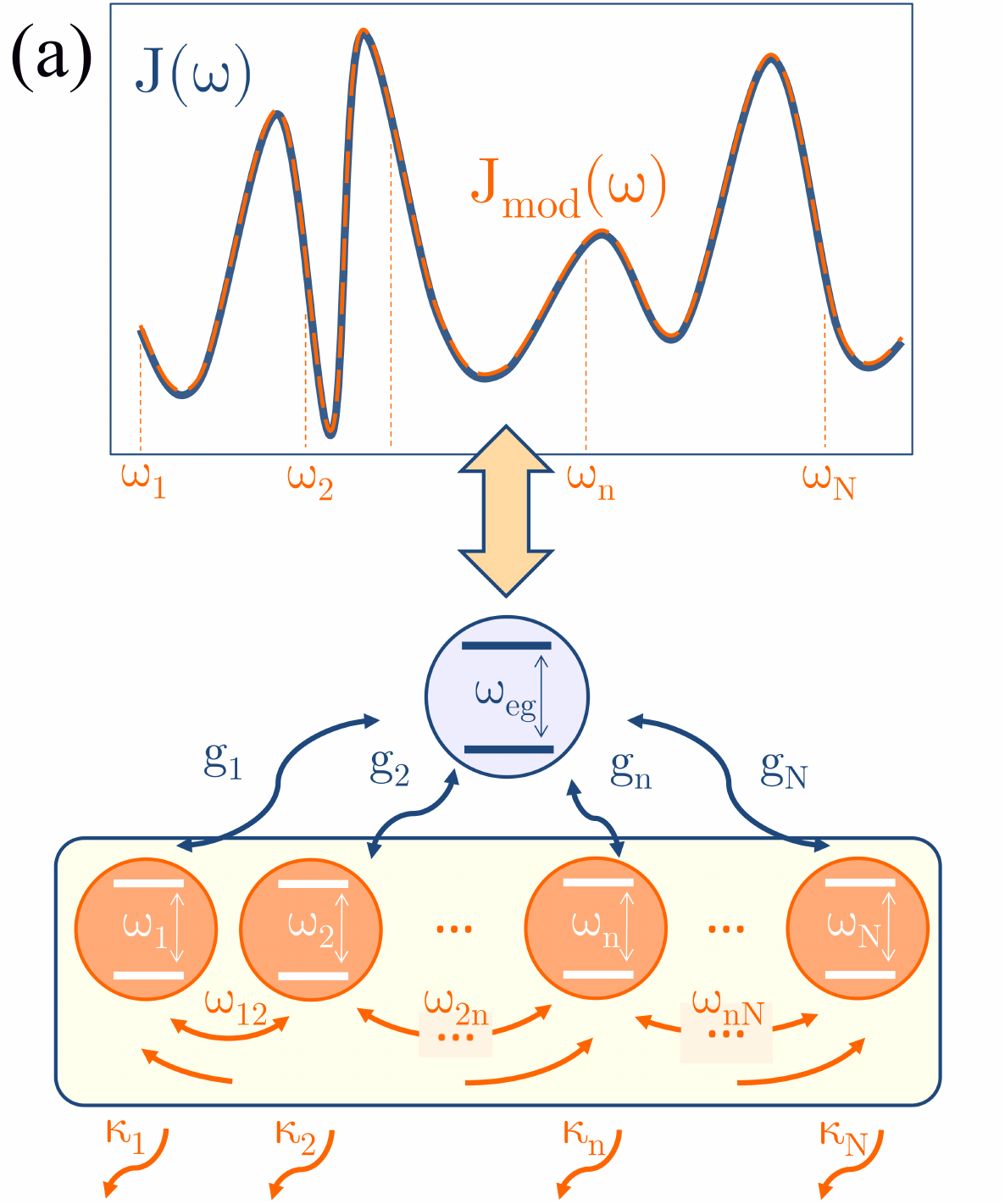}
  \includegraphics[width=0.65\textwidth]{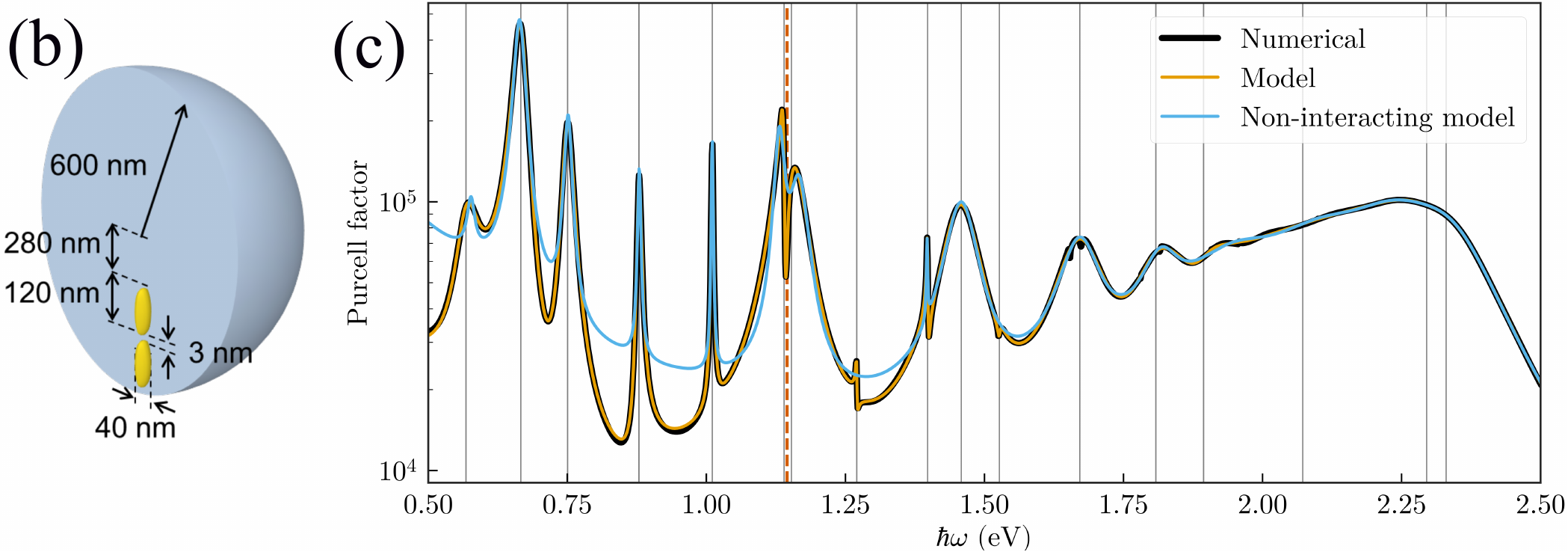}
  \caption{{\bf (a)} The original density $J(\omega)$ is fitted to a few-mode model $J_{\rm mode}(\omega)$. Equation~\eqref{eq:Jmod}, yields the modes and their coupling constants to be substituted into the Lindblad master equation~\eqref{eq:JmodME}. {\bf (b)} Sketch of a typical system comprising a silver dimer nanoantenna embedded in a dielectric microsphere, {\bf (c)} Purcell enhancement factor $J(\omega)/J_0(\omega)$ [where $J_0(\omega)=\omega^3\mu^2/(6\pi^2 \hbar \epsilon_0 c^3)$ is the spectral density in free space] for the system (solid black line), for the fitted model based on eq.~\eqref{eq:Jmod} with $20$ modes (orange line) and for a non-interacting model (where $\omega_{ij}=\omega_i \delta_{ij}$) with the same numbers of modes (dashed blue line). The thin gray lines indicate the eigenstates of $\boldsymbol{H}$ for the fitted interacting system, while the two insets depict the spatial electric field distributions for the two modes indicated by the arrows. Source: Figs. 1 and 2 of~\cite{IMedina2021}. Reproduced with permission from the authors.}
  \label{fig:Jmod}
\end{figure}
The system Hamiltonian $\hat{H}_S$ comprises the emitter and $N$ interacting electromagnetic modes, whereas Hamiltonian $\hat{H}_B$ contains both the continuous bath modes with $[\hat{b}_{j^{\prime}, \Omega^{\prime}}, \hat{b}^{\dagger}_{j,\Omega}]=\delta_{j,j^{\prime}}\delta(\Omega-\Omega^{\prime})$ and their coupling to the system modes, characterized by the rates $\kappa_i$ (see fig.~\ref{fig:Jmod}). 

The system of $N$ interacting modes and continua can be diagonalized via an adapted Fano diagonalization\index{Fano!diagonalization} process~\cite{Denning2019}, obtaining a closed expression for the few-mode spectral density\index{Spectral! density}
\begin{equation}\label{eq:Jmod}
 J_{\rm mod}(\omega) \equiv \frac{\mu^2}{\pi}{\rm Im}\left\{\boldsymbol{g}^{T}\frac{1}{\boldsymbol{H}-\omega}\boldsymbol{g}\right\},
\end{equation}
where $\boldsymbol{g}^{T}=(g_1, g_2,\ldots,g_N)$ and then $N\times N$ matrix $\boldsymbol{H}$ has entries $\boldsymbol{H}_{ij}=\omega_{ij}-\frac{i}{2}\kappa_i \delta_{ij}$. The Lamb shifts\index{Lamb! shift} due to the coupling with the baths have been absorbed in $\omega_{ii}$. We note that for non-interacting modes, with $\omega_{ij}=\omega_{i}\delta_{ij}$, the expression~\eqref{eq:Jmod} simplifies to a sum of Lorentzian distributions, 
\begin{equation}
 J_{\rm mod,\, non-interacting}(\omega)=\sum_{i}\frac{\mu^2 g_i^2}{\pi} \frac{\kappa_i/2}{[(\omega-\omega_i)^2 + \kappa_i^2/4]},
\end{equation}
reproducing a well-known relation between Lorentzian spectral densities and lossy modes~\cite{Imamoglu1994, Tudela2014, Delga2014}.

Fitting eq.~\eqref{eq:Jmod} to $J(\omega)$ for a given electromagnetic environment yields the $N(N+2)$ parameters $\omega_{ij}, \kappa_i, g_i$ which are then to be substituted in the master equation\index{Master equation}
\begin{equation}\label{eq:JmodME}
 \dot{\rho}=-i[\hat{H}_S,\rho] + \sum_{i}\kappa_i L_{a_i}[\rho], \quad \text{ with }\quad  L_{\hat{O}}[\rho]\equiv \hat{O}\rho\hat{O}^{\dagger}-\tfrac{1}{2}\{\hat{O}^{\dagger}\hat{O},\rho \},
\end{equation}
for the completely flat background baths $\hat{H}_B$ in the Markov approximation. Figure~\ref{fig:Jmod}(b) depicts the system under study -- a nanoantenna sustaining confined surface plasmons embedded in a dielectric microsphere supporting several long-lived and delocalized Mie resonances\index{Lorenz-Mie scattering}. As we can infer from fig.~\ref{fig:Jmod}(c), the non-interacting model yields a good fit for many of the peaks but overestimates the background at lower frequencies and fails to reproduce the Fano-like asymmetric profiles\index{Fano!resonance} owing to hybridization between sphere and antenna modes. The method of~\cite{IMedina2021} has also been extended to the case of ultrastrong light-matter coupling in hybrid quantum systems. Lednev and coworkers claim that these configurations can be treated using a Lindblad master equation\index{Master equation} where decay operators act only on the photonic modes by ensuring that the effective spectral density of the EM environment is sufficiently suppressed at negative frequencies~\cite{lednev2023lindblad}.

\subsection{Photon-mediated interactions in topological solid-state materials}

The non-trivial topological character of a material can be enhanced or suppressed through the coupling with a cavity mode, a property which is imprinted to the emergent polariton excitations\index{Polariton}. In our discussion below, we bring some characteristic examples up. 

\begin{figure}
 \includegraphics[width=0.7\textwidth]{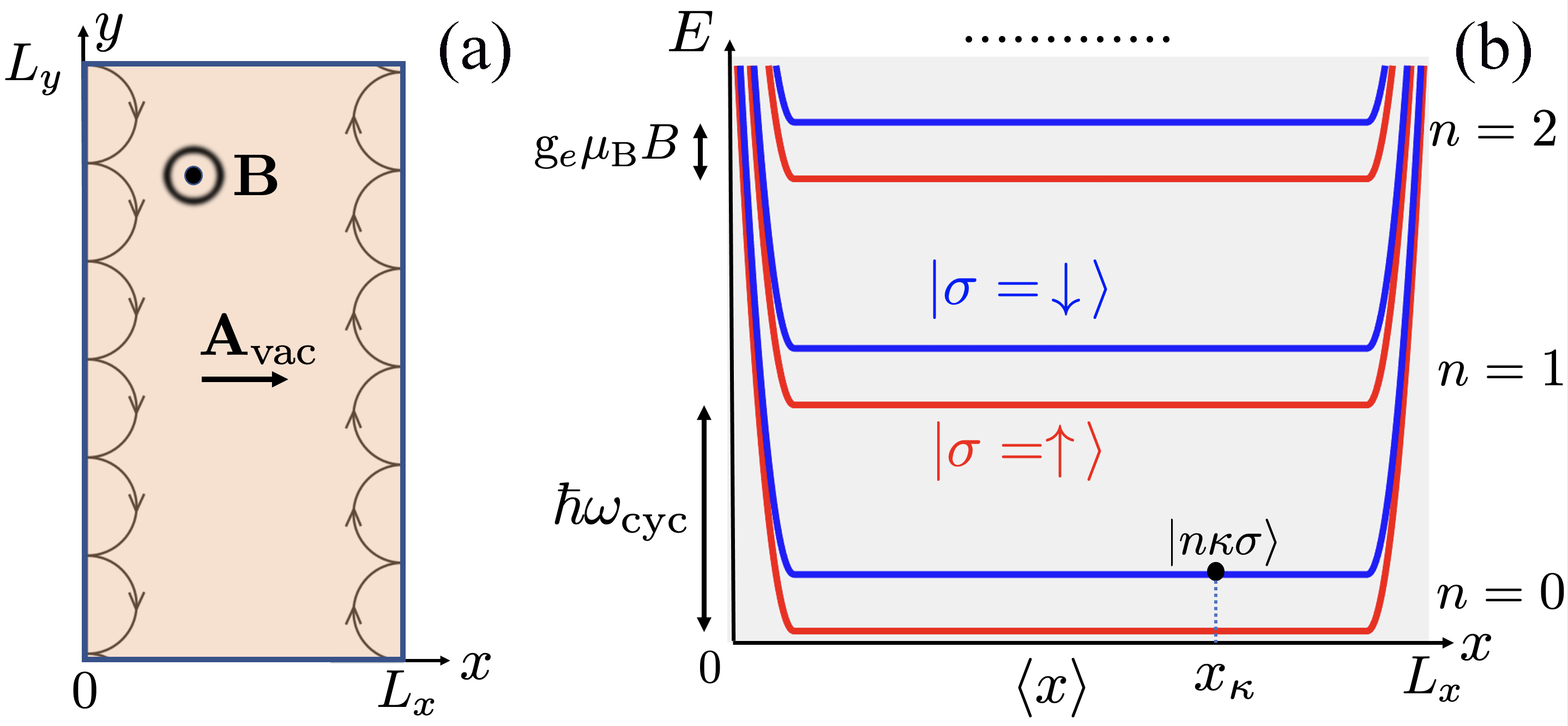}
  \includegraphics[width=0.7\textwidth]{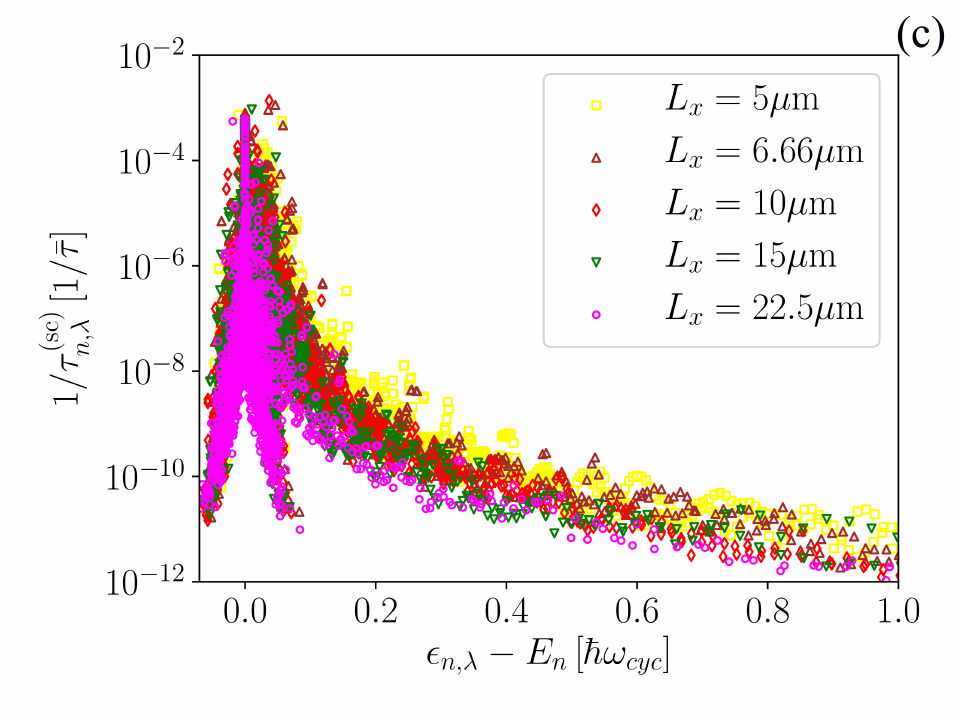}
  \caption{{\bf (a)} Pictorial representation of the 2D electron gas in a rectangle with a perpendicular static magnetic field $\boldsymbol{B}$. The gas is coupled to a cavity mode with a vacuum vector potential\index{Vector potential} $\boldsymbol{A}_{\rm vac}$. The skipping orbits represent counter-propagating Landau edge states. {\bf (b)} Schematic representation of the bare (without disorder) single-particle energy eigenstates $|n\,\kappa\,\sigma\rangle$, depicting the Zeeman splitting and a smooth wall potential at the edges along the $x$-direction. The center position of the orbit $x_{\rm \kappa}$ is proportional to the orbital quantum number $\kappa=1,2,\ldots N_{\rm deg}$, where $N_{\rm deg}$ is the Landau-level orbital degeneracy. {\bf (c)} Normalized scattering rate owing to cavity-mediated hopping against the energy difference $\epsilon_{n,\sigma}-E_n$ for five different values of the transverse length, indicated by the different colored markers in the legend. The rest of the parameters read: $N_{\rm deg} = 2400$, $n=4$, $L_x = 10 \mu$m, $L_{\rm e} = 2.5 \mu$m, $r= 0.9$,  $V_{\rm e} = 0.03 \hbar \omega_{\rm cyc}$, ${\mathcal V}^{({\rm imp})}_{\rm max}  =  1.5 \cdot 10^{-5} \hbar \omega_{\rm cyc} L_x L_y$, $N_{\rm imp} = 2000$, $B = 0.795 \,$T, $m = 0.067 m_0$ ($m_0$ is the  electron mass). For the bottom panel, $N_{\rm el} = 24000$, $g = 0.0051 \,\omega_{\rm cyc}$ and $\omega_{\rm cav} = 0.39 \, \omega_{\rm cyc}$. For these parameters, $\tilde{g} \simeq 0.61 g$. Source: Figs. 1 and 9 of~\cite{Ciuti2021}. Reproduced with permission from the APS.}
  \label{fig:QHHopping}
\end{figure}

We will first visit the case of long-range electron hopping mediated by cavity vacuum fields in disordered quantum Hall systems\index{Quantum Hall effect}, which has been investigated in~\cite{Ciuti2021} [for a general introduction to topology and the quantum Hall effect see e.g.~\cite{fradkin_2013}]. Let us consider a 2D electron gas in a rectangular geometry with a perpendicular static magnetic field, depicted in fig.~\ref{fig:QHHopping}(a). The system is coupled to a cavity mode with a linearly polarized vector potential\index{Vector potential} $\boldsymbol{A}=B \,x\,\hat{y}$. Introducing a wall potential $W(x)$ near the edges, in the second quantization the electron dynamics are governed by the Hamiltonian
\begin{equation}
 \hat{H}_{\rm el}=\sum_{n\kappa,\sigma}(E_{n,\sigma} + W_{n,\kappa})\hat{c}^{\dagger}_{n,\kappa, \sigma}\hat{c}_{n,\kappa, \sigma},
\end{equation}
where $\hat{c}^{\dagger}_{n,\kappa, \sigma}$ creates an electron with orbital quantum numbers $n=0,1,2,\ldots$, $\kappa=1,2,\ldots N_{\rm deg}$ and $\sigma=\uparrow, \downarrow$ the spin projection along the $z$ direction. The Landau energies\index{Landau! energy} are given by the expression
\begin{equation}
 E_{n,\sigma}=h\hbar\omega_{\rm cycl}- \tfrac{1}{2}\sigma g_e\mu_B B,
\end{equation}
in which $\omega_{\rm cycl}=eB/m$ is the cyclotron frequency, $g_e\mu_B B$ is the electron Zeeman splitting, $g_e$ is the effective gyromagnetic factor and $\mu_B$ is the Bohr magneton. The bare-state energies including the smooth potential are shown in fig.~\ref{fig:QHHopping}(b). Each band of Landau levels exhibits an orbital degeneracy equal to $N_{\rm deg}=L_x L_y/(2\pi l_{\rm cycl}^2)$, with $l_{\rm cycl}=\sqrt{\hbar/m\omega_{\rm cycl}}$ the cyclotron length. In the presence of the wall potential, the Landau states\index{Landau!states} have the wavefunctions
\begin{equation}
 \Psi_{n,\kappa}=\langle \boldsymbol{r}|n\,\kappa\rangle=N_n F_n\left(\frac{x-\overline{x}_k}{l_{\rm cycl}}\right)e^{i 2\pi \kappa y/L_y}, \quad \text{with} \quad N_n^2=\frac{1}{\sqrt{\pi}2^n n! l_{\rm cycl} L_y} \quad \text{and}\quad F_n(\xi)=H_n(\xi)e^{-\xi^2/2}.
\end{equation}
In the above, $H_n$ is the Hermite polynomial of order $n$. The Landau orbit center positions are $\overline{x}_{\kappa}=x_{\kappa} + \delta x_{\kappa}$, with $x_{\kappa}=2\pi (l_{\rm cycl}^2/L_y)\kappa$. The fluctuations are related to the smooth-wall potential $W_{n,\kappa}\approx W(x_{\kappa})$ as $\delta x_{\kappa} \approx -W^{\prime}(x_{\kappa})/(m\omega_{\rm cycl}^2)$. In the presence of the wall potential, the Landau states acquire a set velocity along the $y$ direction, $v_{\kappa}^{(y)}=W^{\prime}(x_{\kappa})/(m\omega_{\rm cycl})$. 

We will now assume that the system under study exhibits some moderately strong static disorder that couples Landau states with the same orbital quantum number $n$. This is reflected in the matrix elements $V_{\kappa, \kappa^{\prime}}^{(n)}$ of a sum of randomly-distributed impurity potentials, featuring in the total Hamiltonian through the term:
\begin{equation}
 \hat{H}_{\rm dis}=\sum_{n,\kappa,\kappa^{\prime}}V_{\kappa, \kappa^{\prime}}^{(n)} \hat{c}^{\dagger}_{n,\kappa, \sigma}\hat{c}_{n,\kappa, \sigma}.
\end{equation}
We also let the 2D electron gas be coupled to the field of an electromagnetic resonator mode of frequency $\omega_{\rm cav}$, having the quantized vector potential\index{Vector potential} $\hat{\boldsymbol{A}}_{\rm vac}=A_{\rm vac}(\hat{a}^{\dagger} + \hat{a})\,\hat{x}$. The bare cavity Hamiltonian is as usual $\hat{H}_{\rm cav}=\hbar \omega_{\rm cav}\hat{a}^{\dagger}\hat{a}$, while the light-matter interaction term in the Coulomb gauge reads
\begin{equation}\label{eq:Hpara}
 \hat{H}_{\rm para}=\sum_{n, \kappa, \sigma} (-i) \hbar g \sqrt{n+1} (\hat{a}^{\dagger} + \hat{a}) \hat{c}^{\dagger}_{n,\kappa, \sigma}\hat{c}_{n,\kappa, \sigma} + \text{h.c.},
\end{equation}
a coherent paramagnetic interaction with Rabi frequency
\begin{equation}
 g=\frac{eA_{\rm vac}}{\hbar} \sqrt{\frac{\hbar\omega_{\rm cycl}}{2m}}.
\end{equation}
We observe that the exchange of photons does not alter the electronic states, akin to a coupling of the form $\propto (\hat{a}^{\dagger}+\hat{a})\hat{\sigma}_z$ instead of the regular Rabi coupling involving $\hat{\sigma}_x$. The configuration we have described above following the analysis and notation of~\cite{Ciuti2021} is similar to the magneto-transport of a cavity-embedded 2D electron gas in the ultrastrong coupling regime, leading to a vacuum-induced modification of the resistivity~\cite{ParaviciniBagliani2019}.

Furthermore, there is a diamagnetic term\index{Diamagnetic term},
\begin{equation}
 \hat{H}_{\rm dia}=N_{\rm el} \frac{e^2 A_{\rm vac}^2}{2m}(\hat{a}^{\dagger} + \hat{a})^2=\frac{\hbar \Omega^2}{\omega_{\rm cycl}} (\hat{a}^{\dagger} + \hat{a})^2,
\end{equation}
featuring the collective Rabi frequency for $N_{\rm el}$ electrons, $\Omega=g\sqrt{N_{\rm el}}$. The final term which is to be added to the total Hamiltonian concerns the Coulomb interaction between electrons, which is not however required for the integer quantum Hall effect. The sum of the cavity Hamiltonian and the diamagnetic term can be diagonalized as usual via a Bogoliubov transformation\index{Bogoliubov transformation} to yield
\begin{equation}
 \hat{H}_{\rm mode}=\hat{H}_{\rm cav} + \hat{H}_{\rm dia}=\hbar \tilde{\omega}_{\rm cav}\hat{\alpha}^{\dagger}\hat{\alpha} + \text{const.}.
\end{equation}
This is a Hamiltonian attributed to a boson mode with renormalized frequency
\begin{equation}
 \tilde{\omega}_{\rm cav}=\sqrt{\omega_{\rm cav}^2 + 4\frac{\Omega^2}{\omega_{\rm cycl}}\omega_{\rm cav}}
\end{equation}
and annihilation operator
\begin{equation}
 \hat{\alpha}=\frac{1}{2\sqrt{\tilde{\omega}_{\rm cav}\omega_{\rm cav}}}[(\tilde{\omega}_{\rm cav}+\omega_{\rm cav})\hat{a} + (\tilde{\omega}_{\rm cav}-\omega_{\rm cav})\hat{a}^{\dagger}].
\end{equation}
On the other hand, the bare Landau single-particle Hamiltonian together with the disorder potential can be diagonalized as 
\begin{equation}
 \hat{H}_{\rm sp}=\hat{H}_{\rm el} + \hat{H}_{\rm dis}=\sum_{n,\kappa,\sigma}\left(\epsilon_{n,\lambda} - \tfrac{1}{2} \sigma g_e \mu_B B \right)\hat{d}^{\dagger}_{n,\lambda, \sigma}\hat{d}_{n,\lambda, \sigma},
\end{equation}
where $\epsilon_{n,\lambda}$ are the eigen-energies corresponding to the disorder eigen-states of a Landau band with a given $n$,
\begin{equation}
 |\phi_{\lambda}^{(n)}\rangle=\sum_{\kappa}\langle n\,\kappa|\phi_{\lambda}^{(n)}\rangle \,|n\,\kappa \rangle.
\end{equation}
The fermionic operators are are written as
\begin{equation}
 \hat{c}_{n,\kappa,\sigma}=\sum_{\lambda}\langle n\kappa| \phi_{\lambda}^{(n)}\rangle \hat{d}_{n,\lambda,\sigma},
\end{equation}
while the paramagnetic interaction is recast into the form~\cite{Ciuti2021}
\begin{equation}\label{eq:HparaRenm}
 \hat{H}_{\rm para}=\sum_{n,\lambda,\mu,\sigma}(-i)\hbar\, \tilde{g}_{\lambda,\mu}^{(n,n+1)}(\hat{\alpha}^{\dagger} + \hat{\alpha})\hat{d}^{\dagger}_{n+1,\mu, \sigma}\hat{d}_{n,\lambda,\sigma} + \text{h.c.},
\end{equation}
with a renormalized coupling strength:
\begin{equation}\label{eq:gRenm}
 \tilde{g}_{\lambda,\mu}^{(n,n+1)}=\tilde{g} \sqrt{n+1}\sum_{\kappa}\langle \phi_{\mu}^{(n+1)}|n+1\,\kappa \rangle \langle n\,\kappa|\phi_{\lambda}^{(n)}\rangle, \quad \text{and} \quad \tilde{g}=g\sqrt{\omega_{\rm cav}/\tilde{\omega}_{\rm cav}},
\end{equation}
depending explicitly on the disorder eigenstates. 

As we can see from eq.~\eqref{eq:Hpara}, in the absence of disorder the paramagnetic interaction preserves both spin projection $\sigma$ and the orbital quantum number $\kappa$. The spin is still conserved in the presence of disorder, however the counter-rotating terms of the renormalized paramagnetic interaction of eqs.~\eqref{eq:HparaRenm},~\eqref{eq:gRenm} can couple a general disorder eigenstate $|\phi_{\lambda}^{(n)}\rangle$ to any other disorder eigenstate $|\phi_{\lambda^{\prime}}^{(n)}\rangle$ via an intermediate virtual excited state. Indeed, in an energy non-preserving process, an electron occupying the state $|\phi_{\lambda}^{(n)}\rangle$ can be promoted to the state $\mu$ of the $(n+1)$ band, with the simultaneous creation of a photon of energy $\hbar\tilde{\omega}_{\rm cav}$. Such a virtual process has the matrix element $(-i)\tilde{g}_{\lambda,\mu}^{(n,n+1)}$ and the associated energy penalty is $\epsilon_{n,\lambda}-\epsilon_{n+1,\mu}-\hbar \tilde{\omega}_{\rm cav}$. In the conjugate process, the cavity photon can be re-absorbed and the electron will be demoted back to the $n$ band, yet into a different unoccupied disordered state $\lambda^{\prime}$; the corresponding matrix element is now $i\tilde{g}_{\lambda^{\prime},\mu}^{(n,n+1)*}$. When $\epsilon_{n,\lambda}\ll \epsilon_{n+1,\mu}+\hbar \tilde{\omega}_{\rm cav}$, the effective coupling between states $\lambda$ and $\lambda^{\prime}$ within the $n$ band can be calculated as a result of perturbation theory ~\cite{Malrieu1985, Moreira2002}:
\begin{equation}
 \tilde{\Gamma} \approx \sum_{\mu} \frac{\hbar^2 \tilde{g}_{\lambda,\mu}^{(n,n+1)}\tilde{g}_{\lambda^{\prime},\mu}^{(n,n+1)*}}{\epsilon_{n,\lambda}-\epsilon_{n+1,\mu}-\hbar \tilde{\omega}_{\rm cav}},
\end{equation}
obtained by summing over all possible intermediate states $\mu$, whose total number equals $N_{\rm deg}$. When the final state of the cavity-mediated hopping are unoccupied, the corresponding scattering rates are given by applying Fermi's golden rule:
\begin{equation}
 \frac{1}{\tau_{n,\lambda}^{\rm (sc)}}=\frac{2\pi}{\hbar}\sum_{\lambda^{\prime}\neq \lambda} |\tilde{\Gamma}^{(n)}_{\lambda, \lambda^{\prime}}|^2 \delta(\epsilon_{n,\lambda}-\epsilon_{n,\lambda^{\prime}}). 
\end{equation}
The dependence of the scattering rate on the total number of electrons and hence on the filling factor $N_{\rm el}/N_{\rm deg}$ is smooth and only enters via the renormalized cavity mode and vacuum Rabi frequencies, $\tilde{\omega}_{\rm cav}$ and $\tilde{g}$, respectively, featuring in the diamagnetic interaction. Figure~\ref{fig:QHHopping}(c) depicts scattering rates against the energy difference $\epsilon_{n,\lambda}-E_n$ (associated with the presence of disorder), normalized by a characteristic rate $\propto N_{\rm deg}$. We observe that the scattering rate for the edge states increases for decreasing confinement $L_x$.

Let us now look at an alternative manifestation of transport enhanced by the presence of disorder, in the presence of long-range hopping. To analyze the behaviour of molecular chains in optical cavities, the authors of~\cite{Chavez2021} employ the 1D Anderson model\index{Anderson! model} with all-to-all hopping,
\begin{equation}
 \hat{H}_{\rm And}=\sum_{j=1}^{N} \epsilon_j |j\rangle \langle |j| + \Omega \sum_{j=1}^{N-1} \left(|j\rangle \langle j+1| +  |j+1\rangle \langle j|\right)-\tfrac{1}{2}\gamma \sum_{i\neq j}|i\rangle \langle j|,
\end{equation}
where $|j\rangle$ is the site basis, $\epsilon_i$ are random energies uniformly distributed in the interval $[-W/2, W/2]$; $W$ is the disorder strength and $\Omega$ is the tunneling transition amplitude between nearest neighbours. It is found that disorder destroys the energy gap $N\gamma/2$, occurring for $W=0$, when its strength exceeds the threshold value $(\gamma/2)N\ln N$. To quantify the modification of the quantum transport properties by the presence of disorder, the authors model pumping and draining via the Lindblad master equation\index{Master equation} we have encountered many times so far,
\begin{equation}\label{eq:MEAnderson}
 \frac{d\hat{\rho}}{dt}=\frac{1}{i\hbar}[\hat{H}_{\rm And},\hat{\rho}]+\sum_{\eta=p,d}\mathcal{L}_{\eta}[\hat{\rho}],
\end{equation}
where the two dissipators assume the familiar forms
\begin{equation}
 \mathcal{L}_{\eta}[\hat{\rho}]=-\{\hat{L}_\eta^{\dagger}\hat{L}_{\eta},\hat{\rho}\} +2 \hat{L}_{\eta} \hat{\rho}\hat{L}_{\eta}^{\dagger}.
\end{equation}
The dissipator with $\hat{L}_{p}=\sqrt{\gamma_p/(2\hbar)}|1\rangle \langle 0|$ induces pumping on the first site at rate $\gamma_p$, while $\hat{L}_{d}=\sqrt{\gamma_d/(2\hbar)}|0\rangle \langle N|$ induces draining from the last site with rate $\gamma_d$. The steady-state solution\index{Steady state!solution} $\hat{\rho}_{\rm ss}$ of eq.~\eqref{eq:MEAnderson} determines the stationary current
\begin{equation}
 I=\frac{\gamma_d}{\hbar}\langle N|\hat{\rho}_{\rm ss}|N\rangle.
\end{equation}
Now, a chain of emitters coupled to the same cavity mode can be described in terms of an effective long-range hopping model. For a molecular chain on resonance with the cavity mode, one may write the Hamiltonian as
\begin{equation}
 \hat{H}_{\rm mol.ch.}=\sum_{j=1}^{N} \epsilon_j |j\rangle \langle |j| + \Omega \sum_{j=1}^{N-1} \left(|j\rangle \langle j+1| +  |j+1\rangle \langle j|\right) + g\sum_{j=1}^{N}(|j\rangle \langle c| + |c\rangle \langle j|),
\end{equation}
where $|c\rangle$ represents a single excitation in the cavity mode with no excitation in the chain. The emitters are coupled to the optical mode with a strength given by the familiar expression
\begin{equation}
 g=\sqrt{\frac{2\pi \mu^2 \hbar \omega_{\rm cav}}{V_c}},
\end{equation}
where $\mu$ is the molecular transition dipole, $\omega_{\rm cav}$ is the cavity mode frequency and $V_c$ is the mode volume. Furthermore, a recent report argues that global and periodic interactions between a quantum gas of bosons and a standing-wave field of a high-finesse optical cavity, which are incommensurate with the lattice spacing, give rise to many-body localization\index{Many-body! localization} (MBL) for a finite-size system. Bypassing any additional random or quasi-periodic onsite disorder, as was the case in the cavity-mediated hopping we discussed in connection to the quantum Hall effect, this configuration challenges the standard MBL mechanism characterized by the existence of a complete set of local integrals of motion~\cite{Kubala2021}.  

The coupling of a topological superconductor to a cavity mode has been recently discussed in~\cite{dmytruk2023hybrid}. The authors employ either a Kitaev chain\index{Kitaev chain} or a tight-binding Hamiltonian for the `material part' to calculate the cavity-photon spectral function with these two models, highlighting the emergence of polariton excitations\index{Polariton} and their topological signatures.   

\subsection{Hybrid systems: from nanomechanics to atomic ensembles}
\label{ssec:hybridsys}

We will now focus on the coherent coupling of light to the bosonic modes describing mechanical oscillators and magnets, before visiting some notable recent realizations of the quantum Rabi, Dicke and Tavis-Cummings models in mesoscopic systems. The largest part of this section is devoted to cavity optomechanics\index{Cavity! optomechanics}. As the levitation and control of microscopic objects in vacuum is progressively attaining the quantum regime, several interesting prospects open up in different fields, including ground-state cooling, the generation of nonclassical states alongside macroscopic quantum superpositions, motional entanglement, quantum metrology and the study of quantum many-particle systems~\cite{GonzalezBallestero}. For example, polarization fluctuations of a laser beam transmitted through an optical nanofiber have been used to assess the fundamental torsional mechanical mode of the nanofiber waist in the recent experiment of Tebbenjohanns and coworkers~\cite{Tebbenjohanns2023}. Equally, quantum control has been sought over mechanical rotors using measurement-based parametric feedback~\cite{vanderLaan2021} in the so-called field of {\it rotational} optomechanics\index{Rotational! optomechanics}. 

At the same time, there exists a vast number of optomechanical devices and ways to realize optomechanical interactions. Among the ones that have stood out during the development of the field since the early 1990s~\cite{Abramovici1992} are: suspended micro- and macro-scopic mirrors --  allowing the study of the center of mass motion of truly macroscopic test masses, coated cantilevers, micropillars, optical microresonators where light is guided in whispering gallery modes, clouds of ultra-cold atoms, on-chip waveguides and photonic crystal cavities. For the latter, and further to our discussion in sec.~\ref{sec:physreal}, we mention that the idea of generating a bandgap by imposing periodic boundary conditions can be extended to the modes of a mechanical beam. Introducing a surrounding periodic structure matched to the phonon wavelength results in a 1D photonic crystal cavity with co-localized photonic and phononic modes. Representative experimental parameters for some of these setups are tabulated in Table II of~\cite{AspelmeyerRev2014}. 

In the following subsection, we will only present a limited subset of the above developments, and we will make use of the simplest possible model in cavity optomechanics, focusing on those pats of the phenomenology which are akin to the predictions of the Jaynes-Cummings model and its extensions. 

\subsubsection{Cavity optomechanics I: Radiation pressure force and optomechanical coupling}

\begin{figure}
 \includegraphics[width=\textwidth]{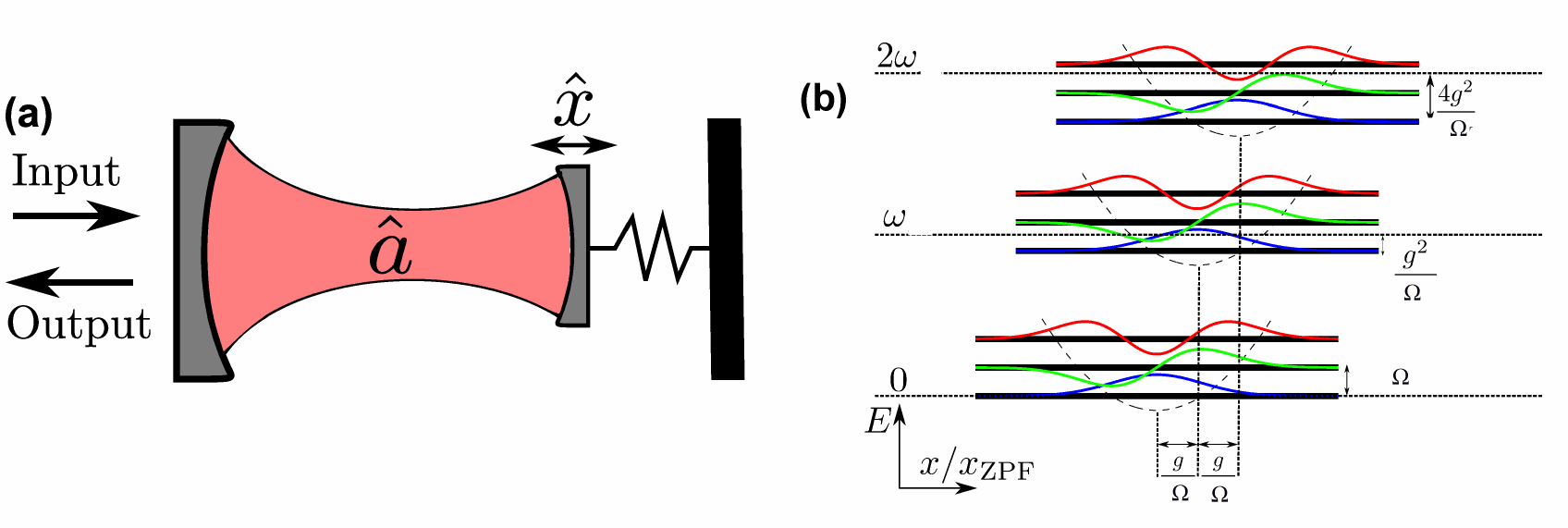}
 \caption{{\bf (a)} Standard optomechanical setup, with the position $\hat{x} \propto (\hat{b} + \hat{b}^{\dagger})$ parametrically coupled to the driven cavity mode $\hat{a}$. {\bf (b)} Spectrum and eigenfunctions of the optomechanical Hamiltonian $\hat{H}=\hbar \omega \hat{a}^{\dagger}\hat{a} + \hbar \Omega \hat{b}^{\dagger}\hat{b} - \hbar g_0 \hat{a}^{\dagger}\hat{a}(\hat{b}^{\dagger}+\hat{b})$. The parabolas (dashed curves) indicate the displaced harmonic oscillator potentials for $n=0,1,2$ photons. Source: Fig. 1 of ~\cite{Nunnenkamp2011}. Reproduced with permission from the APS.}
 \label{fig:OptfigSetup}
\end{figure}

The standard optomechanical setup we will be dealing with is depicted in fig.~\ref{fig:OptfigSetup} (a). The momentum transfer of photons is the fundamental mechanism that couples the intracavity field to the motion of a mechanical oscillator~\cite{AspelmeyerRev2014}. Since a single photon of wavelength $\lambda$ transfers momentum $|\Delta p|=\hbar 2\pi/\lambda$, the average radiation pressure force exerted on the mirrors of a Fabry-P\'{e}rot resonator\index{Fabry-P\'erot! cavity/resonator} is
\begin{equation}\label{eq:Fopt}
 \braket{\hat{F}}=2\hbar k \frac{\braket{\hat{a}^{\dagger}\hat{a}}}{\tau_c}=\hbar  \frac{\omega}{L} \braket{\hat{a}^{\dagger}\hat{a}},
\end{equation}
where $\tau_c=2L/c$ is the cavity round trip time. From eq.~\eqref{eq:Fopt} we immediately see that the quantity $\hbar\omega/L$ gives the radiation-pressure force exerted by a single intracavity photon. The parameter $G\equiv \omega/L$ featuring in this formula indicates the change of the cavity resonance frequency with respect to position. Formalizing our description, we consider the Hamiltonian describing the coupling of a radiation (referred to as optical) mode to a vibrational (mechanical) mode modelled as two harmonic oscillators,
\begin{equation}
 \hat{H}_{0}=\hbar\omega \hat{a}^{\dagger}\hat{a} + \hbar \Omega \hat{b}^{\dagger}\hat{b}.
\end{equation}
For a cavity with a moveable mirror on the one end, the interaction between the two modes is {\it parametric}\index{Parametric!coupling}, which means that the cavity resonance frequency is modulated by the mechanical amplitude in view of the expansion
\begin{equation}
 \omega(x) \approx \omega + x\,\frac{\partial \omega}{\partial x} + \ldots,
\end{equation}
where the first-order term defines the optical frequency shift per displacement, $G \equiv - \partial \omega/\partial x$ (see also the early derivation of Law in ref.~\cite{Law1995}). For a simple cavity of length $L$, we obtain $G=\omega/L$. Expanding to leading order in the displacement, which is sufficient in most experimental realizations, we get
\begin{equation}
 \hbar \omega(x)\hat{a}^{\dagger}\hat{a} \approx \hbar (\omega-G\hat{x})\hat{a}^{\dagger}\hat{a}, \quad \text{where}\quad \hat{x} \equiv x_{ZPF}(\hat{b} + \hat{b}^{\dagger}),
\end{equation}
and $x_{ZPF}\equiv \sqrt{\hbar/(2 m_{\rm eff}\Omega)}$ is the zero-point fluctuation amplitude\index{Zero-point! fluctuations! amplitude} of the mechanical oscillator with effective mass\index{Effective! mass} $m_{\rm eff}$: $\langle 0 |\hat{x}^2|0\rangle=x_{\rm ZPF}^2$. Hence, the interaction part of the Hamiltonian can be expressed as
\begin{equation}\label{eq:optHam}
 \hat{H}_{\rm int}=-\hbar g_0 \hat{a}^{\dagger}\hat{a}(\hat{b} + \hat{b}^{\dagger}), \quad \text{with} \quad g_0=G x_{\rm ZPF}.
\end{equation}
Here, $g_0$ is the so-called vacuum optomechanical coupling strength written in units of angular frequency. Having now written down the interaction part of the Hamiltonian, the radiation pressure force is simply given by its derivative with respect to displacement: 
\begin{equation}
 \hat{F}=-\frac{d\hat{H}_{\rm int}}{d\hat{x}}=\hbar G \hat{a}^{\dagger}\hat{a}=\hbar \frac{g_0}{x_{\rm ZPF}} \hat{a}^{\dagger}\hat{a}.
\end{equation}
To tackle the full problem which includes dissipation, thermal fluctuations and an external coherent drive, it proves advantageous to work in a frame rotating at the frequency $\omega_L$ of the drive (laser) field coupled to the cavity mode. The new Hamiltonian in this frame reads
\begin{equation}
 \hat{H}=-\hbar \Delta \hat{a}^{\dagger}\hat{a} + \hbar \Omega \hat{b}^{\dagger}\hat{b} - \hbar g_0 \hat{a}^{\dagger}\hat{a}(\hat{b}^{\dagger}+\hat{b})+\ldots,
\end{equation}
where $\Delta \equiv \omega_L-\omega$ is the cavity-drive detuning. Having made this step, we can now introduce the so-called ``linearized'' approximation of cavity optomechanics\index{Linearized! approximation in optomechanics} by separating the cavity field into an average coherent amplitude $\langle \hat{a} \rangle=\overline{\alpha}$ and a fluctuating term:
\begin{equation}
 \hat{a}=\overline{\alpha} + \delta\hat{a}.
\end{equation}
Then the interaction part of the Hamiltonian, now reading
\begin{equation}
 \hat{H}_{\rm int}=-\hbar g_0 (\overline{\alpha} + \delta \hat{a})^{\dagger}(\overline{\alpha} + \delta \hat{a})(\hat{b} + \hat{b}^{\dagger}),
\end{equation}
can be expanded in powers of $\delta \hat{a}$. The zero-order term, $-\hbar g_0 |\overline{\alpha}|^2(\hat{b} + \hat{b}^{\dagger})$ corresponds to an average radiation pressure force $\overline{F}=\hbar G |\overline{\alpha}|^2$, which may be cancelled by an appropriate shift in the origin of the displacement by $\delta \overline{x} \equiv \overline{F}/(m_{\rm eff} \Omega^2)$. The linear term in the fluctuations is the one of importance, reading
\begin{equation}
 -\hbar g_0 (\overline{\alpha}^{*}\,\delta\hat{a} + \overline{\alpha}\,\delta\hat{a}^{\dagger}) (\hat{b} + \hat{b}^{\dagger}),
\end{equation}
while we neglect terms which are second-order in fluctuations. Assuming a real $\overline{\alpha}=\sqrt{\overline{n}_{\rm cav}}$ (where $\overline{n}_{\rm cav}$ is the average intracavity photon number) and putting the pieces together we obtain [see also the linearization procedure discussed in sec.~\ref{sssec:dia} for the Dicke model in the thermodynamic limit\index{Thermodynamic limit}]
\begin{equation}
 \hat{H} \approx -\hbar \Delta  \hat{a}^{\dagger}\hat{a} + \hbar \Omega_m \hat{b}^{\dagger}\hat{b} - \hbar g_0\sqrt{\overline{n}_{\rm cav}}(\delta\hat{a}+\delta\hat{a}^{\dagger})(\hat{b} + \hat{b}^{\dagger}).
\end{equation}
The resulting Heisenberg equations of motion will be linear under the action of the above Hamiltonian coupling the two degrees of freedom. This is a quadratic form than can be solved via a Bogoliubov transformation\index{Bogoliubov transformation}. The product $g=g_0\sqrt{\overline{n}_{\rm cav}}$ is referred to as the {\it optomechanical coupling strength}, amplifying the single-photon coupling by the intracavity amplitude. The linearized description may be adequate even if the intracavity photon number is not so large, provided that the dissipation at rate $\kappa$ is appreciable, such that the mechanical system is not capable of resolving individual photons. On the other end, the resolved-sideband\index{Resolved-sideband regime} regime requires $\kappa \ll \Omega$.

Three different regimes of interaction can be distinguished depending on the detuning in the sideband-resolved regime. For $\Delta \approx -\Omega$, the two harmonic oscillators -- the mechanical oscillator and the driven cavity mode --  exchange quanta. The exchange is captured in the RWA\index{Rotating-wave approximation} by a so-called beam-splitter interaction\index{Beam-splitter! interaction} of the form
\begin{equation}
 -\hbar g (\delta \hat{a}^{\dagger} \hat{b} + \delta \hat{a}\,\hat{b}^{\dagger}).
\end{equation}
This case is encountered in cavity cooling~\cite{Genes2008, Marquardt2007, WilsonRae2007}, where thermal phonons are transferred into the ``cold'' cavity mode in the weak-coupling regime ($g \ll \kappa$). Photons impinging at a frequency which is red-detuned with respect to the cavity resonance will preferentially scatter upwards in energy -- in an anti-Stokes process\index{Stokes and anti-Stokes process} -- in order to attain the cavity resonance frequency. Consequently they will carry a phonon away and will be reflected away blue-shifted by $\Omega$. More precisely, the transition rate from the phonon state $n$ to $n-1$ in a lase-driven cavity reads~\cite{AspelmeyerRev2014}
\begin{equation}
 \Gamma_{n\to n-1}=nA^{-}\equiv ng_0^2 S_{NN}(\omega=-\Omega),\quad \text{where} \quad  S_{NN}(\omega)=\overline{n}_{\rm cav} \frac{\kappa}{\kappa^2/4 + (\Delta + \omega)^2}
\end{equation}
is the photon-number noise spectrum\index{Noise spectrum} originating from the radiation pressure force spectrum. Conversely, the Stokes-process, in which photons are reflected red-shifted leaving being an extra phonon occurs at a smaller rate
\begin{equation}
 A^{+}\equiv ng_0^2 S_{NN}(\omega=+\Omega),
\end{equation}
due to the suppression in the final density of photon states if the laser field is red-detuned. The difference between the two rates, $\Gamma_{\rm opt}=A^{-}-A^{+}$ defines the {\it optomechanical damping rate}\index{Damping! rate, optomechanical}. From the corresponding rate equations, the minimum phonon number in steady state\index{Steady state} is 
\begin{equation}
 \overline{n}_{\rm min}=\left(\frac{A^{-}}{A^{+}} -1 \right)^{-1}=\left[\frac{(\kappa/2)^2 + (\Delta-\Omega)^2}{(\kappa/2)^2 + (\Delta+\Omega)^2} -1\right]^{-1}.
\end{equation}
This quantity can be minimized by varying the laser detuning $\Delta$. In the resolved sideband regime, with $\kappa\ll \Omega$, this leads to $\overline{n}_{\rm min}=[\kappa/(4\Omega)]^2\ll 1$, approaching ground-state cooling.

The red-detuned case $\Delta \approx -\Omega$ is also very relevant for quantum state transfer between the light mode and the mechanical oscillator be reliably converting an optical pulse into a mechanical excitation [see also sec.~\ref{sec:ion} for the fundamental transitions involved in a trapped-ion configuration]. The coupling $g=g(t)$ is now explicitly time dependent through the laser intensity, remaining on for the right amount of time to allow for a complete state swap between the oscillators $\delta \hat{a}$ and $\hat{b}$. A second ``signal'' pulse excites the oscillator $\delta \hat{a}$ into the desirable target state which is subsequently written onto the mechanical part. For a detailed discussion on the swapping protocols see ref.~\cite{WangClerk2012}.   

In the blue-detuned case now, the dominant terms in the RWA read
\begin{equation}
 -\hbar g (\delta\hat{a}^{\dagger}\,\hat{b}^{\dagger} + \delta\hat{a}\,\hat{b}),
\end{equation}
representing a two-mode squeezing\index{Squeezing! two-mode} interaction, having a close affinity to parametric amplification. For this reason, it may lead to strong quantum correlations between the vibrational and the driven optical mode~\cite{ClerkDevoret2010}. Finally, when $\Delta=0$, the interaction is captured by the terms $-\hbar g (\delta\hat{a}^{\dagger}+\delta\hat{a})(\hat{b} + \hat{b}^{\dagger})$. This means that the mechanical oscillator through its position induces a phase shift to the mode field, which can be used in displacement detection. It can also be used for a quantum non-demolition measurement\index{Non-demolition measurement} of the quadrature $\delta\hat{a}^{\dagger}+\delta\hat{a}$, which commutes with the Hamiltonian~\cite{Braginsky1996}. 

In 2011, Rabl reported on the occurrence of photon blockade\index{Photon! blockade! optomechanics} (see sec.~\ref{sssec:JCzerodim} for a discussion in terms of the JC dynamics), ascertained from photon antibunching\index{Antibunching} in coincidence measurements on the output of a weakly-driven optomechanical system~\cite{Rabl2011}. Concurrently, Nunenkamp and coworkers focused on the consequences of the single-photon strong-coupling regime $g_0 \geq \Omega$, allowing radiation pressure of a single photon to displace the mechanical oscillator by more than its zero-point uncertainty\index{Zero-point! uncertainty}. They showed that in this regime the power spectrum has multiple sidebands and that the cavity response has several resonances in the resolved-sideband limit~\cite{Nunnenkamp2011}. The nonlinear quantum Langevin\index{Heisenberg-Langevin equations} equations are solved in the weak-drive limit after diagonalizing the optomechanical Hamiltonian~\eqref{eq:optHam} via the polaron transformation\index{Polaron! transformation} $\hat{U}=e^{-\hat{S}}$ with $\hat{S}\equiv(g_0/\Omega)\hat{a}^{\dagger}\hat{a}(\hat{b}^{\dagger}-\hat{b})$ [see also eq.~\eqref{polaron2}]. The mean intracavity photon number in the steady state\index{Steady state} evinces a series of resonance peaks spaced apart by the mechanical frequency $\Omega$:
\begin{equation}
 \frac{\langle\hat{a}^{\dagger}\hat{a}\rangle}{n_0}=\sum_{n=0}^{\infty} \frac{(g_0/\Omega)^{2n}}{(2n)!} \sum_{k=0}^{n} \begin{pmatrix}
  n \\ k                                                                                                                                                                                                 \end{pmatrix} (n_{\rm th}+1)^{n-k}n_{\rm th}^{k}\frac{\kappa(\kappa+n\gamma)\,e^{-(g_0/\Omega)^2(2n_{\rm th}+1)}}{(\frac{\kappa+n\gamma}{2})^2 + [\Delta-(n-2k)\Omega]^2},
\end{equation}
where $n_{\rm th}=[\exp(\hbar \Omega/k_B T)-1]^{-1}$ is the thermal occupation of the mechanical bath, $\kappa, \gamma$ are the dissipation rates for the optical mode and the mechanical system, respectively, and $n_0=4|\alpha_{\rm in}|^2$ is the mean photon number on resonance in the cavity (with $\alpha_{\rm in}$ the coherent amplitude of the input light). For $n_{\rm th}=0$ only terms with $k=0$ contribute, and we find that the resonance peaks are weighted by a Poisson distribution\index{Poisson! distribution} of variance $(g_0/\Omega)^2$ and have widths $\kappa + n \gamma$. The resonances occur if the laser frequency $\omega_L$ matches the frequency difference between the vacuum state $|0,0\rangle$ and the manifold of the one-photon eigenstates $|1,m \rangle$, satisfying
$\hbar \omega_L=\hbar (E_{1m}-E_{00})=\hbar(\omega-g_0^2/\Omega-m\Omega)$, while the Poisson weights\index{Poisson! weights} originate from the Franck-Condon factors\index{Franck-Condon! factors} [see also the Hamiltonian spectrum depicted in fig.~\ref{fig:OptfigSetup}(b)]
\begin{equation}
 C_m \equiv |\langle m |e^{\hat{X}}|0 \rangle|^2=\left|\int dx\, \phi_m^{*}(x-x_0)\phi_0(x) \right|^2=(g_0/\Omega)^{2m} \frac{e^{-(g_0/\Omega)^2}}{m!}.
\end{equation}
Moreover, for $\Delta=-n g^2/\Omega$, multiphoton transitions\index{Multiphoton! transition} occur between the vacuum state $|0,0\rangle$ and the lowest-energy $n$-photon state $|n,0\rangle$ (since $n\omega_L=E_{n0}-E_{00}$) and the system is in the Franck-Condon blockade\index{Franck-Condon! blockade} regime with all intermediate transitions being off-resonant. In 2015, a design of cavity optomechanics in the microwave frequency regime was proposed, involving a Josephson junction\index{Josephson! junction} qubit; the radiation-pressure interaction was enhanced by six orders of magnitude, accessing the strong-coupling regime~\cite{Pirkkalainen2015}. 

Elste et al.~\cite{Elste2009} proposed a novel kind of optomechanics, where the mechanical motion modulates the linewidth of the cavity\index{Linewidth! cavity}. In this case, the radiation pressure force spectrum features a Fano resonance\index{Fano!resonance}, substantially modifying the interaction between light and mechanical degrees of freedom. The potential of dissipative optomechanics~\index{Dissipative! optomechanics} as a source for producing squeezed light and correlated photons was explored in~\cite{Kilda2016}. 

We should also add here a short comment on {\it dispersive optomechanics}\index{Dispersive! optomechanics}, focusing on a so-called `membrane in the middle' of a cavity configuration~\cite{Jayich2008}. Such an optomechanical device can be operated in a regime where the cavity frequency depends directly on $x^2$--the position squared of the macroscopic mechanical oscillator. As a consequence, the experimenter can perform a direct measurement to determine the energy of the oscillator, $E=\hbar \Omega (n+\frac{1}{2})$, where $n$ is the number of phonons in the state of the membrane~\cite{Thompson2008}. If the cavity is driven on resonance, the phase of the transmitted beam is proportional to the energy $E$. Such a measurement allows the detection of energy quantization, something which is not possible with a cavity whose frequency directly depends on $x$. Such a linear-position detection, like the ones we have discussed above in this section, is limited by quantum backaction\index{Backaction}~\cite{Braginsky1996}.   

Controlling the external degree of freedom of trapped molecular ions to eliminate systematic effects such as Doppler shifts is paramount for their applications to spectroscopy, precision measurements of fundamental constants, and quantum information technology. The experiment conducted by Qi and collaborators~\cite{QiLu2023} has recently attained near ground-state cooling of the axial motional modes of a calcium mono-oxide ion via sympathetic sideband cooling\index{Sympathetic cooling} with a co-trapped calcium ion~\cite{Kielpinski2000} [see also the ion-cooling experiments discussed in sec.~\ref{ssec:ionfur}]. Remarkably, the phonon state of the axial out-of-phase mode of the ion chain has been preserved as the mode frequency is adiabatically varied. In the sub-field of {\it active optomechanics}, where the energy source is the intracavity active medium, the motion of a mechanical oscillator may be also detected by measuring the laser intensity, spectral broadening, and laser frequency stability~\cite{YuVollmer2022}. 

The ability to achieve coherent quantum control over the center of mass motion of massive mechanical objects provides a powerful scheme to conduct experiments to assess the foundations of quantum theory~\cite{Kippenberg2008, AspelmeyerRev2014}. Optomechanical systems are also very good candidates for detecting quantum signatures of gravity. The models of~\cite{Karolyhazy1966, Diosi1984, Penrose1996} identify gravity as the dominant agent of the state-vector collapse. A notable example in a different direction is provided in~\cite{Pikovski2012}, where a protocol is proposed using quantum optical control and readout of a mechanical system to probe possible deviations from the quantum commutation relation even at the Planck scale\index{Planck!scale}. 

\subsubsection{Cavity optomechanics II: Photon-phonon interaction formulated as a scattering problem}

Quantum optics experiments are scattering experiments, where the experimenter controls the applied inputs and measures the generated outputs. We will now formalize further the photon-phonon interaction as a scattering problem, following closely the analysis of~\cite{maurer2022quantum} where the motional quantum dynamics of a levitated dielectric sphere interacting with the quantum electromagnetic field are discussed beyond the point-dipole approximation. The Hamiltonian is formulated under the assumption that the dielectric sphere of mass $M$ primarily interacts only with electromagnetic field modes in a sufficiently narrow frequency window, so that its electromagnetic response can be described by a single scalar and real relative permittivity $\epsilon$. For sufficiently small center-of-mass displacements, meaning that the standard deviation $\Delta r_\mu \equiv \braket{\hat r_\mu^2- \braket{\hat r_\mu}^2}^{1/2}$ is smaller than any relevant length scale associated with the variation of the electromagnetic field, the total Hamiltonian describing the interaction between the center-of-mass motion and the electromagnetic field -- whose deterministic part generates the conservative confining potential $V(\hat{\boldsymbol{r}})$ -- reads
\begin{equation}\label{eq:QEDHamiltonian}
\hat H = \frac{\hat{\boldsymbol{p}}^2}{2M} + V(\hat{\boldsymbol{r}}) + \hat{H}_\text{em} -\hat{\mathcal{F}} \cdot \hat{\boldsymbol{r}},
\end{equation}
where $\hat{\mathcal{F}}$ is the radiation-pressure force operator, which does not commute with the Hamiltonian $\hat{H}_\text{em}$ describing the free evolution of the electromagnetic field. From the Hamiltonian~\eqref{eq:QEDHamiltonian}, the Heisenberg equation of motion for the $\mu$-component of the center-of-mass position vector is
\begin{equation}\label{eq:eomD}
 \frac{d^2\hat{r}_{\mu}}{dt^2} + \Omega_{\mu}^2 \hat{r}_{\mu}=\hat{\mathcal{F}}_{\mu}/M,
\end{equation}
where $\Omega_{\mu}$ is the corresponding frequency of the (anisotropic) harmonic trap. The right-hand side of eq.~\eqref{eq:eomD} is the radiation pressure exerted by self-consistent electromagnetic fields, expressed as a surface integral of the Maxwell stress tensor\index{Maxwell stress tensor} in the far field of the dielectric sphere and defined as
\begin{equation}
 \hat{\mathcal{F}}_{\mu}=-\frac{1}{2}\epsilon_0 \lim_{r \to \infty}r^2 \int d\Omega (\hat{r}\cdot \hat{\mu})[\hat{\mathbf{E}}^2(\boldsymbol{r}) + c^2 \hat{\mathbf{B}}^2(\boldsymbol{r})].
\end{equation}
The electric and magnetic field operators can be expanded in terms of the normalized eigenmodes for a scattering problem where the sphere is motionless and placed at the origin of coordinates,
\begin{subequations}
 \begin{align}
 \hat{\mathbf{E}}(\boldsymbol{r})&=i\sum_{\kappa}\sqrt{\frac{\hbar \omega_{\kappa}}{2\epsilon_0}}[\boldsymbol{F}_{\kappa}(\boldsymbol{r})\hat{a}_{\kappa}-\text{h.c.}], \\
 \hat{\mathbf{B}}(\boldsymbol{r})&=\sum_{\kappa}\sqrt{\frac{\hbar}{2\epsilon_0 \omega_{\kappa}}}\{[\nabla \times \boldsymbol{F}_{\kappa}(\boldsymbol{r})]\hat{a}_{\kappa}-\text{h.c.}\},
 \end{align}
\end{subequations}
where the normalized scattering eigenmodes are labeled by the multi-index $\kappa \equiv (g,\boldsymbol{k})$, with $g=1,2$ the polarization index and $\mathbf{k}$ the wavevector. They comprise a plane-wave contribution alongside an elastically-scattered field off the dielectric sphere. All the elastic scattering processes arise as solutions of the Lorenz-Mie problem~\cite{Mie1908, maurer2021quantum}\index{Lorenz-Mie scattering}. Now, expressing the Hamiltonian~\eqref{eq:QEDHamiltonian} in terms of photons interacting with center-of-mass phonons, we may write
\begin{equation}\label{eq:StokesAntiStokes}
 \hat{H}=\sum_{\mu}\hbar \Omega_{\mu}\hat{b}_{\mu}^{\dagger}\hat{b}_{\mu} + \sum_{\kappa}\hbar \omega_{\kappa} \hat{a}_{\kappa}^{\dagger}\hat{a}_{\kappa} + \hbar\sum_{\kappa \kappa^{\prime}\mu}g_{\kappa \kappa^{\prime}\mu} \hat{a}_{\kappa}^{\dagger}\hat{a}_{\kappa^{\prime}}(\hat{b}^{\dagger}_{\mu} + \hat{b}_{\mu}).
\end{equation}
The third term accounts for the two fundamental processes that make the optomechanical light-matter interaction: (i) Stokes processes\index{Stokes process} where a photon in mode $\kappa^{\prime}$ is inelastically scattered into a lower-frequency photon in mode $\kappa$ by generating a phonon in mode $\mu$, and (ii) anti-Stokes processes\index{anti-Stokes process} where a photon in mode $\kappa^{\prime}$ is inelastically scattered into a higher-frequency photon in mode $\kappa$ by absorbing a phonon in mode $\mu$. The photon-phonon interaction is captured by the coupling rates $g_{\kappa \kappa^{\prime}\mu}$, which can be written as a surface integral of a far-field quantity depending on the normalized scattering eigenmodes:
\begin{equation}
 g_{\kappa \kappa^{\prime}\mu}=\sqrt{\frac{\hbar}{2M\Omega_{\mu}}}\frac{c\sqrt{k k^{\prime}}}{2}\lim_{r\to \infty}r^2 \int d\Omega (\hat{r}\cdot \hat{\mu})\left\{\boldsymbol{F}_{\kappa}^{*}(\boldsymbol{r}) \cdot \boldsymbol{F}_{\kappa^{\prime}}(\boldsymbol{r}) + \frac{1}{k k^{\prime}} [\nabla \times \boldsymbol{F}_{\kappa}^{*}(\boldsymbol{r})] \cdot [\nabla \times \boldsymbol{F}_{\kappa^{\prime}}(\boldsymbol{r})]\right\},
\end{equation}
containing products of vector spherical harmonics\index{Vector spherical harmonics}~\cite{RanhaNeves19} and expressed as a sum of discrete angular momentum states with $l=1,2,\ldots,$ $m=-l, -l+1, \ldots, l-1, l$ and polarization $p$ indicating TE or TM modes.   

We will now specialize to the case where some of the electromagnetic field modes are in a coherent state, given by
\begin{equation}
 \ket{\Psi_{\rm coh}}=\exp\left[\sum_{\kappa}(\alpha_{\kappa}\hat{a}^{\dagger}_{\kappa}-\text{h.c.})\right]\ket{0},
\end{equation}
with an associated electric field
\begin{equation}
 \mathbf{E}_{\rm coh}(\boldsymbol{r},t)=i\sum_{\kappa}\sqrt{\frac{\hbar\omega_{\kappa}}{2\epsilon_0}}[\boldsymbol{F}_{\kappa}(\boldsymbol{r})\alpha_{\kappa}e^{-i\omega_{\kappa}t}-\text{c.c.}].
\end{equation}
Assuming that all modes found in the coherent state generated by a monochromatic laser have the same frequency $\omega_0$, the {\it linearized Hamiltonian} accounting for the fluctuations of the electromagnetic field about $\ket{\Psi_{\rm coh}}$ in a frame rotating with $\omega_0$ is given by
\begin{equation}\label{eq:linHamOPT}
 \hat{H}_{\rm lin}=\sum_{\kappa}\hbar \Delta_{\kappa} \hat{a}_{\kappa}^{\dagger}\hat{a}_{\kappa} + \sum_{\mu}\hbar \Omega_{\mu}\hat{b}_{\mu}^{\dagger}\hat{b}_{\mu} + \hbar \sum_{\kappa \mu}(G_{\kappa\mu}\hat{a}_{\kappa}^{\dagger} + G_{\kappa \mu}^{*}\hat{a}_{\kappa})(\hat{b}_{\mu}^{\dagger} + \hat{b}_{\mu}),
\end{equation}
where $\Delta_{\kappa} \equiv \omega_{\kappa}-\omega_0$ and $G_{\kappa\mu}=\sum_{\kappa^{\prime}}\alpha_{\kappa^{\prime}}g_{\kappa\kappa^{\prime} \mu}$. For example, in the case of a single Gaussian beam\index{Gaussian! beam} propagating along the positive $z$-axis with a focus at the origin of co-ordinates, the corresponding coherent amplitudes for each polarization $g=1, 2$ are
\begin{equation}
 \begin{pmatrix}
  \alpha_1(k, \theta_k, \phi_k) \\
  \alpha_2(k, \theta_k, \phi_k)
 \end{pmatrix}=\sqrt{\frac{4\pi k_0 P |\cos\theta_k|}{\hbar c^2 \sin^2\theta_{\rm NA}}} \, \exp\left(-\frac{\sin^2\theta_k}{f_0^2 \sin^2\theta_{\rm NA}}\right) \frac{\delta(k-k_0)}{k_0^2}  \begin{pmatrix}
  i\sin\phi_k \\
  \cos\phi_k
 \end{pmatrix},
\end{equation}
where $P$ is the optical power of the focused field, ${\rm NA}=\sin\theta_{\rm NA}$ is the numerical aperture of the lens, and $f_0\gg 1$ is the filling factor over the lens, guaranteeing maximum focusing of the beam~\cite{NovotnyHechtBook}. 
\begin{figure}
 \includegraphics[width=0.65\textwidth]{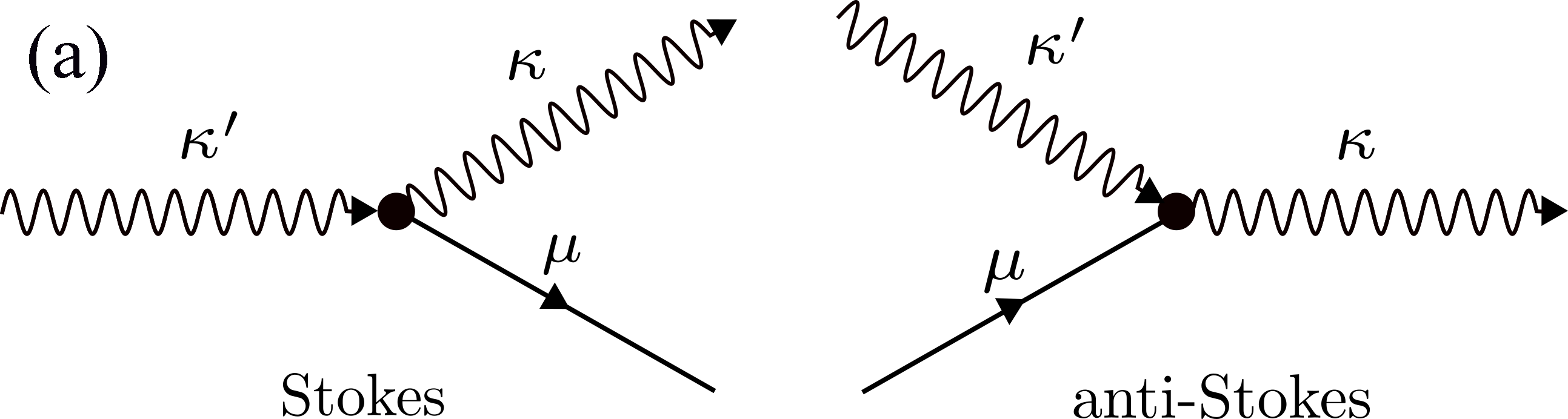}
  \includegraphics[width=0.55\textwidth]{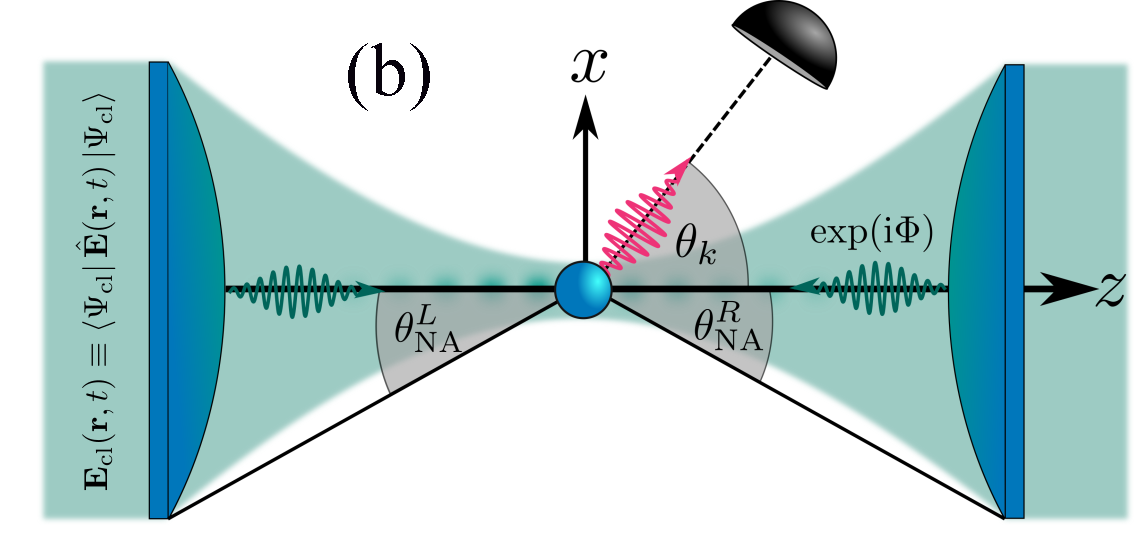}
   \includegraphics[width=0.85\textwidth]{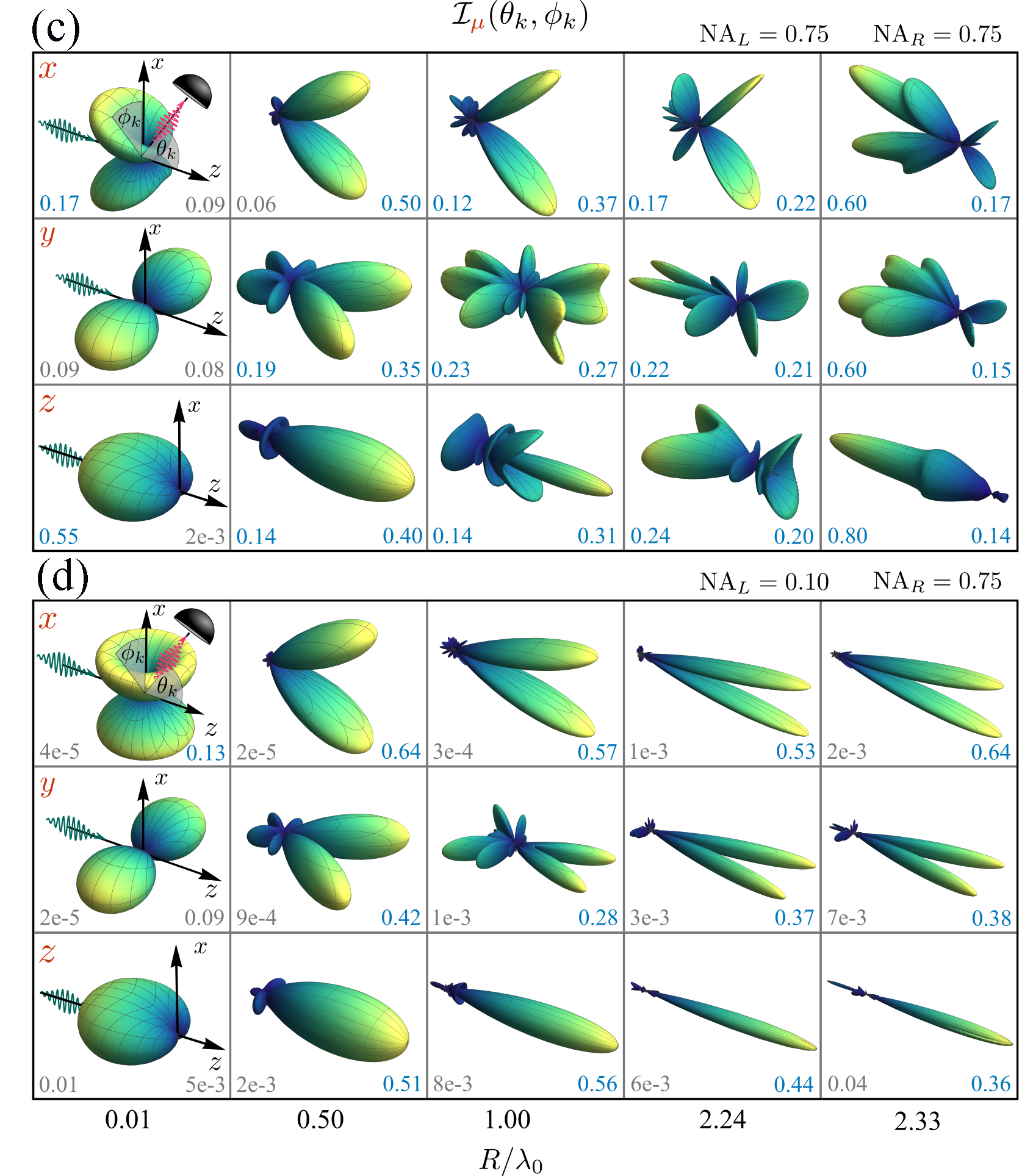}
\end{figure}

\begin{figure} [t!]
  \caption{{\it Scattered light probing the center-of-mass motion.} {\bf (a)} Feynman diagrams of the Stokes and anti-Stokes processes described by the third term of eq.~\eqref{eq:StokesAntiStokes}. {\bf (b)} Schematic representation of the scattering experiment. Either one or two $x$-polarized monochromatic (frequency $\omega_0$) Gaussian beams propagate along the direction of $\hat{z}$ and $-\hat{z}$. The beams are strongly focused by an aplanatic lens to the location where a dielectric sphere of radius $R$ is placed. For the two-beam configuration, the relative phase between the two counter-propagating beams is $\Phi$. The information radiation patterns\index{Information radiation pattern (IRP)} (IRPs) $\mathcal{I}_{\mu}^{p}(\theta_k, \phi_k)=\mathcal{I}_{\mu}(\theta_k, \phi_k)$ (in the broadband coupling regime) are defined with respect to the same co-ordinate system, and $\theta_k$ is the polar angle. {\bf (c, d)} IRPs of a silica sphere irradiated by a single focused Gaussian beam\index{Gaussian! beam} polarized along the $x$-axis and propagating along the $z$-axis. The values of the IRP are encoded both in the radial distance from the origin and in the color scale. The focusing lens has numerical aperture NA$_L=0.75, 0.15$ in panels (c, d) respectively, while the collection lens has numerical aperture NA$_L=0.75$ in both panels. The detection efficiency of the focusing and collection lenses is displayed in each frame, and the values exceeding $\eta_{\mu}^{\rm det}=1/9$ [see eq.~\eqref{eq:efflens}] are highlighted in blue. The ratio $R/\lambda_0$ is the same across each column. Source: Figs. 1, 2 and 4 of~\cite{maurer2022quantum}. Reproduced with permission from the APS.}
  \label{fig:FigscattOpt}
\end{figure}

The linearized Hamiltonian of eq.~\eqref{eq:linHamOPT}, modeling both Stokes and anti-Stokes processes as depicted in fig.~\ref{fig:FigscattOpt}(a), describes two important phenomena: (a) the laser-light recoil heating (dissipation of the center-of-mass motion) and (b) the scattering of light from the dielectric sphere, yielding information on the position fluctuation of the center of mass. The angular distribution of the scattered light is termed {\it information radiation pattern} (IRP)\index{Information radiation pattern (IRP)}, which will be determined below through the scattering amplitudes.

The transition amplitudes associated with these processes are evaluated with respect to the input and output states as
\begin{equation}
 \tau_{\kappa\mu}^{p} \equiv \braket{\Psi_{\rm out}^p|\hat{U}(T/2,-T/2)|\Psi_{\rm in}}.
\end{equation}
The input is formed by the product state $\ket{\Psi_{\rm in}}=\ket{n_x, n_y, n_z} \otimes \ket{\Psi_{\rm coh}}$ (the first component refers to the center-of-mass degrees of freedom with $n_{\mu}>0$), while the output state is $\ket{\Psi_{\rm out}^{\rm p=S}}=\hat{a}_{\kappa}^{\dagger}\hat{b}_{\mu}^{\dagger}\ket{\Psi_{\rm in}}$ and $\ket{\Psi_{\rm out}^{\rm p=aS}}=\hat{a}_{\kappa}^{\dagger}\hat{b}_{\mu}\ket{\Psi_{\rm in}}$ for the Stokes and anti-Stokes transitions, respectively. The time-evolution operator is given as $\hat{U}(t,t^{\prime})=\exp[-i\hat{H}(t-t^{\prime})/\hbar]$, where $\hat{H}$ is the Hamiltonian~\eqref{eq:QEDHamiltonian}. For $\tau_{\kappa\mu}^{p}$ to be nonzero, the incoming photon must belong to a mode excited by the classical electromagnetic field, whereas the photon scattered into the mode $\kappa$ may belong to any of the available electromagnetic field modes.

Using Fermi's golden rule, one may calculate the transition probability rates for the two processes under consideration in the asymptotic limit as
\begin{subequations}
\begin{align}
  \lim_{T\to \infty} \frac{\sum_{\kappa}|\tau_{\kappa\mu}^{\rm S}|^2}{T}=(n_{\mu}+1)\Gamma_{\mu}^{+},\\
  \lim_{T\to \infty} \frac{\sum_{\kappa}|\tau_{\kappa\mu}^{\rm aS}|^2}{T}=(n_{\mu}+1)\Gamma_{\mu}^{-},
\end{align}
\end{subequations}
where
\begin{equation}
 \Gamma_{\mu}^{\pm} \equiv 2\pi \sum_{\kappa}|G_{\kappa\mu}|^2 \delta(\omega_{\kappa}-\omega_0 \pm \Omega_{\mu}).
\end{equation}
Since the linearized couplings extend over a broad frequency range set by $\Omega_{\mu}$, we may consider%
\begin{equation}
 \Gamma_{\mu} \approx \Gamma_{\mu}^{+} \approx \Gamma_{\mu}^{-} = 2\pi \sum_{\kappa}|G_{\kappa\mu}|^2 \delta(\omega_{\kappa}-\omega_0).
\end{equation}
The above equality does not hold in environments that modify the mode structure of the electromagnetic field.  

The optical configuration is displayed in fig.~\ref{fig:FigscattOpt}(b). Photon scattering via Stokes ($p={\rm S}$) or anti-Stokes ($p={\rm aS}$) processes yields information about the center-of-mass motion about a particular axis $\mu$ via the information radiation pattern (IRP)\index{Information radiation pattern (IRP)}
\begin{equation}
 \mathcal{I}_{\mu}^{p}(\theta_k,\phi_k) \equiv \frac{\lim_{T\to \infty}\int_{0}^{\infty}dk\,k^2 \sum_g |\tau_{\kappa\mu}^p|^2}{\lim_{T\to\infty}\sum_{\kappa}|\tau_{\kappa\mu}^p|^2},
\end{equation}
a normalized quantity providing the solid-angle probability distribution of a photon scattered either through a Stokes or an anti-Stokes process. In the broadband coupling regime, the IRP is the same for both Stokes and anti-Stokes processes, which means that these can be distinguished by means of the frequency of a scattered photon but not its angular distribution. 

The determination of the laser recoil heating rate $\Gamma_{\mu}$ and the IRP $\mathcal{I}_{\mu}^{p}(\theta_k,\phi_k)$ -- both of which can be derived from the coupling rates $g_{\kappa \kappa^{\prime} \mu}$ and the coherent-state amplitudes $\alpha_{\kappa}$ associated with a classical electromagnetic field in a linearized treatment -- is indispensable for the implementation of active feedback cooling through optical detection~\cite{Magrini2021, Tebbenjohanns2021, Kamba22}. In the experiment of~\cite{Tebbenjohanns2021}, in particular, a silica nanoparticle is optically levitated in a cryogenic environment. The light scattered off the particle in the backwards direction is split between heterodyne and homodyne receivers\index{Heterodyne detection}. A linear filter then processes the homodyne signal and feeds it back to the trap in the form of a Coulomb force exerted to the charged particle in order to cool its centre-of-mass motion along the optical axis. The minimum attainable phonon-occupation number is given by the formula
\begin{equation}
 \overline{n}_{\mu} \equiv \frac{1}{2}\left(\frac{1}{\sqrt{\eta_{\mu}}}-1\right),
\end{equation}
where $\eta_{\mu}$ denotes the total efficiency of the cooling. It is formed by the product of three individual efficiencies: the {\it detection efficiency}
\begin{equation}\label{eq:efflens}
 \eta_{\rm \mu}^{\rm det} \equiv \int_{S_d}d\theta_k d\phi_k \sin\theta_k \mathcal{I}_{\mu}^{p}(\theta_k,\phi_k),
\end{equation}
evaluated over the solid angle $S_d$ subtended by the collection lens [see figs.~\ref{fig:FigscattOpt}(c, d)], the efficiency associated with information loss to the environment,
\begin{equation}
 \eta_{\rm \mu}^{\rm env} \equiv \frac{\Gamma_{\mu}}{\Gamma_{\mu} + \Gamma_{\mu}^{g}}
\end{equation}
where $\Gamma_{\mu}^{g} \propto p\sqrt{k_B T}$ is the heating rate due to the presence of the environmental gas at pressure $p$ and temperature $T$. The last term in the product making the total efficiency $\eta_\mu=\eta_{\mu}^{\rm det} \eta_{\mu}^{\rm env} \eta_{\mu}^0$ is an efficiency coefficient $\eta_{\mu}^0$ amassing the contribution from all other information channels in the active feedback scheme. Scattering of a two-photon state off a dielectric sphere leads to a vanishing simultaneous joint detection probability of the two scattered photons along diagonal stripes in the second-order correlation function\index{Second-order correlation function} due to the Hong-Ou-Mandel interference\index{Hong-Ou-Mandel! interference}~\cite{maurer2021quantum}. Tuneable interactions between different mechanical modes mediated by photons scattered by spatially separated particles have been recently demonstrated in the experiment of~\cite{vijayan2023cavitymediated}, where measured spectrograms reveal the normal mode splittings arising from cavity mediated particle-particle interactions for different values of the cavity detuning. 

\subsubsection{Coherent interaction of a magnon and a qubit}

The collective bosonic spin excitations of magnets are called {\it magnons}\index{Magnon}. An anisotropic ferromagnet can host magnons with a controllable degree of intrinsic squeezing in equilibrium, while exchange-coupling a single qubit with this ferromagnet\index{Ferromagnet} can realize the quantum Rabi model. In the limit of zero anisotropy, the mapping reduces to the JC Hamiltonian.

Ferromagnetic films often exhibit strong shape anisotropy on top of magnetocrystalline anisotropies~\cite{KamraPRL2016}. These effects are modeled by including a single-ion anisotropy contribution to the isotropic terms. We consider a spin model (where the spin operator at lattice site $i$ is denoted by $\hat{\boldsymbol{S}}_i$) with nearest-neighbor interactions (the sum over which is indicated by $\langle i,j \rangle$). The total spin Hamiltonian in a ferromagnet of $N_F$ lattice sites includes the Zeeman energy\index{Zeeman energy} induced by an external magnetic field (taken along the direction of the $z$-axis), the exchange interaction between nearest neighbors and a possible (generalized) anisotropy term:
\begin{equation}\label{eq:ferromagneticHam}
 \hat{H}_f=|\gamma|\mu_0 H_0 \sum_{i}\hat{S}_{iz} - J \sum_{\langle i,j \rangle} \hat{\boldsymbol{S}}_i \cdot \hat{\boldsymbol{S}}_j + \sum_{i} [K_x(\hat{S}_{ix})^2 + K_y(\hat{S}_{iy})^2 + K_z(\hat{S}_{iz})^2],
\end{equation}
where $\gamma<0$ is the gyromagnetic ratio\index{Gyromagnetic ratio}, $H_0$ is the applied external field in the direction of the $z$-axis and $J$ is the exchange coupling strength $J$. The coefficients $K_x$, $K_y$ and $K_z$ characterize the magnetic anisotropy~\cite{Skogvoll2021, Kamra2016, romling2023resolving}. Assuming only small deviations from the ground state, where the spins point in the $-\hat{z}$ direction owing to a dominant Zeeman-energy contribution, we can operate under the spin wave approximation\index{Spin wave approximation}~\cite{Auerbach1994} using the Holstein-Primakoff transformations\index{Holstein-Primakoff! transformation}~\cite{holstein1940field} 
\begin{equation}\label{eq:magnonHP}
 \hat{S}_{i+}=\sqrt{2S}\,\hat{a}_i^{\dagger}, \quad \quad \hat{S}_{i-}=\sqrt{2S}\,\hat{a}_i, \quad \quad \hat{S}_{iz}=-S + \hat{a}_i^{\dagger}\hat{a}_i,
\end{equation}
where $\hat{S}_{i\pm}=\hat{S}_{ix} \pm i \hat{S}_{iy}$ and $S$ stands for the magnitude of the spin (see also the application of these transformations to derived the 'bosonised Dicke Hamiltonian'~\eqref{hbdicke} in sec.~\ref{sssec:dicke}). The bosonic magnon operator $\hat{a}_i^{\dagger}$ -- defined in the context of the above transformation -- creates a ``spin flip'' at lattice site $i$. Taking the corresponding Fourier transform we write
\begin{equation}
 \hat{a}_i=\sqrt{\frac{1}{N_F}}\sum_{\boldsymbol{k}} \hat{a}_{\boldsymbol{k}} e^{-i\boldsymbol{k}\cdot \boldsymbol{r}_i}.
\end{equation}
Using the Holstein-Primakoff relations~\eqref{eq:magnonHP} and the Fourier decomposition, and retaining only the magnon mode with $\boldsymbol{k}=\boldsymbol{0}$ by focusing on coherent resonant interactions, it can be shown~\cite{Skogvoll2021} that the ferromagnetic Hamiltonian~\eqref{eq:ferromagneticHam}\index{Ferromagnetic! Hamiltonian} is recast into the form
\begin{equation}\label{eq:Hfsqmagnon}
 \hat{H}_f=A \hat{a}^{\dagger}\hat{a} + B \hat{a}^2 + B^{*}\hat{a}^{\dagger 2},
\end{equation}
where $\hat{a} \equiv \hat{a}_{\boldsymbol{k}=\boldsymbol{0}}$. The constant $A=|\gamma|\mu_0 H_0 + (K_x + K_y-2K_z)$ reflects the impact of the external magnetic field, while the constant $B$ directly reflects the transverse anisotropy as $B=S(K_x-K_y)/2$. The Hamiltonian~\eqref{eq:Hfsqmagnon} can be diagonalized via a Bogoliubov transformation\index{Bogoliubov transformation}, defining a so-called {\it bare squeezed magnon}\index{Magnon!bare squeezed}:
\begin{equation}
 \hat{\alpha}=\hat{a}\cosh r + \hat{a}^{\dagger}e^{i\theta} \sinh r, \quad \text{with} \quad \tanh(2r)=\frac{2|B|}{A} \quad \text{and} \quad e^{i\theta}=\frac{B^{*}}{|B|},
\end{equation}
and the single-mode squeeze operator\index{Squeezing! operator}\index{Operator! squeezing} $\hat{S}(\xi)$, defined in eq.~\eqref{sqop}, employed in the transformation
\begin{equation}
 \hat{\alpha}^{\dagger}=\hat{S}(\xi)\hat{a}^{\dagger}\hat{S}^{\dagger}(\xi), \quad \text{with} \quad \hat{S}(\xi) \equiv \exp\left[\frac{\xi^*}{2}\hat{a}^2 - \frac{\xi}{2} (\hat{a}^{\dagger})^2\right]. 
\end{equation}
The diagonalized Hamiltonian then reads
\begin{equation}
 \hat{H}_F=\omega_{\alpha}\hat{\alpha}^{\dagger}\hat{\alpha} + \text{const.}, \quad \text{with} \quad \omega_{\alpha}=\sqrt{A^2-4|B|^2},
\end{equation}
which sets $|A| \geq 2|B|$ and $(A+\omega_\alpha)^2>4|B|^2$ as requirements for stability. A uniformly ordered ground state can be maintained for a sufficiently large applied magnetic field $H_0$. In typical experiments, it is shown that $\sinh r \sim 1$~\cite{KamraPRL2016, Kamra2016}. 

Having noted the link between anisotropy and squeezing\index{Squeezing}, let us now see how the dipolar interaction of a single qubit with the ferromagnet\index{Ferromagnet} can lead to coherent dynamics described by a Jaynes-Cummings Hamiltonian~\cite{Skogvoll2021, Trifunovic2013}. The interfacial exchange interaction is parametrized via the operator
\begin{equation}
 \hat{H}_{\rm int}=J_{\rm int} \sum_{l}\hat{\boldsymbol{S}}_l \cdot \hat{\boldsymbol{s}}_l,
\end{equation}
where $l$ labels the interfacial site. The ferromagnetic spin\index{Ferromagnetic! spin} represented by the operator $\hat{\boldsymbol{S}}$ is coupled to the electronic states comprising the qubit of spin $\hat{\boldsymbol{s}}$, with the interfacial coupling strength $J_{\rm int}$. The qubit spin can be expressed in terms of the Pauli matrices as $\hat{\boldsymbol{s}}_l=|\psi_l|^2/2\,\hat{\boldsymbol{\sigma}}$, where $\psi_l$ is the spatial wavefunction amplitude of the qubit orbital at position $l$, and $\hat{\boldsymbol{\sigma}}=(\hat{\sigma}_x, \hat{\sigma}_y, \hat{\sigma}_z)$. Writing the interfacial interaction in the form of a central spin model\index{Central spin model}\index{Model! central spin} (see sec.~\ref{sssec:poormodel})
\begin{equation}\label{eq:HintSs}
  \hat{H}_{\rm int}=J_{\rm int} \sum_{l}\left[\hat{S}_{lz}\hat{s}_{lz} + \tfrac{1}{2}(\hat{S}_{l+}\hat{s}_{l-}+\hat{S}_{l-}\hat{s}_{l+}) \right],
\end{equation}
using the Fourier decomposition and the Holstein-Primakoff relations, and retaining only the uniform magnon mode ($\boldsymbol{k}=\boldsymbol{0}$), we find that the term in parentheses describes a coherent interaction between the magnon and the qubit, which can be expressed as
\begin{equation}
 \hat{H}_{\rm coh}=g (\hat{a}^{\dagger}\hat{\sigma}_{-} + \hat{a}\hat{\sigma}_{+})\quad \text{with} \quad g\equiv J_{\rm int}N_{\rm int} |\psi|^2 \sqrt{\frac{S}{2N_F}}
\end{equation}
In the above formula, $|\psi|^2=\sum_{l}|\psi_l|^2/N_{\rm int}$ is the  wavefunction averaged over the $N_{\rm int}$ interfacial sites. Breaking the symmetry in the plane transverse to the equilibrium spin order generates squeezing, while a uniaxial anisotropy does not. For this configuration one obtains~\cite{Skogvoll2021}
\begin{equation}
 \hat{H}_{\rm coh}=g \cosh r (\hat{a}^{\dagger}\hat{\sigma}_{-} + \hat{a}\hat{\sigma}_{+}) + g \sinh r (\hat{a}\hat{\sigma}_{-} + \hat{a}^{\dagger}\hat{\sigma}_{+}), 
\end{equation}
showing that the squeeze parameter $r$ controls the relative importance of the rotating and counter-rotating terms. The coupling strengths can also reach the deep-strong regime. The composite nature of the squeezed magnon enables a simultaneous excitation of three spin qubits coupled to the same ferromagnet\index{Ferromagnet}. This allows for the robust generation of three-qubit Greenberger-Horne-Zeilinger (GHZ)\index{State! GHZ} and related states~\cite{Skogvoll2021} needed for implementing Shor's quantum error-correction code~\cite{Shor1995, Neeley2010, DiCarlo2010, Reed2012}\index{Shor's algorithm}.

Finally, summing over non-uniform modes, the first term in eq.~\eqref{eq:HintSs} can be recast in the form
\begin{equation}
 \hat{H}_{\rm int, z}=-\frac{S J_{\rm int} N_{\rm int}|\psi|^2}{2} \hat{\sigma}_z + \frac{J_{\rm int}}{2N_F} \sum_{l}|\psi_l|^2 \sum_{\boldsymbol{k}, \boldsymbol{k}^{\prime}} \hat{a}_{\boldsymbol{k}}^{\dagger}\hat{a}_{\boldsymbol{k}} e^{-i(\boldsymbol{k}-\boldsymbol{k}^{\prime})\cdot \boldsymbol{r}_l}\,\hat{\sigma}_z.
\end{equation}
The first term amounts to a mere renormalization of the qubit frequency, and assuming $|\psi_l|^2$ to be spatially independent we may sum over the interfacial sites $l$ to obtain
\begin{equation}
 \hat{H}_{\rm int (disp), z}=\frac{J_{\rm int}}{2N_F} |\psi|^2 \sum_{\boldsymbol{k}, \boldsymbol{k}^{\prime}} \hat{a}_{\boldsymbol{k}}^{\dagger}\hat{a}_{\boldsymbol{k}} \delta_{\boldsymbol{k},\boldsymbol{k}^{\prime}}\,\hat{\sigma}_z=\frac{J_{\rm int}}{2N_F} |\psi|^2 \sum_{\boldsymbol{k}, \boldsymbol{k}^{\prime}} \hat{a}_{\boldsymbol{k}}^{\dagger}\hat{a}_{\boldsymbol{k}}\,\hat{\sigma}_z.
\end{equation}
Limiting once again our attention to the uniform magnon mode, we end up in an effective dispersive interaction\index{Dispersive! Hamiltonian} of the form $\hat{H}_{\rm int (disp), z}=\chi \hat{a}^{\dagger}\hat{a}\hat{\sigma}_z$~\cite{romling2023resolving}, which we have already met in sec.~\ref{ssec:JCspectrcQED}. 

Following up our discussion in secs.~\ref{ssec:rabi} and~\ref{ssec:exjc}, we note that a hybrid quantum system, consisting of a single-mode cavity simultaneously coupled to both a two-level system and an yttrium-iron-garnet (YIG) sphere supporting magnons with Kerr nonlinearity has recently been proposed to restore the supperradiant phase transition against the $\boldsymbol{A}^2$ term~\cite{LiuXiong2023}. It is found that the Kerr-magnon-induced phase transition can occur in both cases of ignoring and including the intrinsic diamagnetic term\index{Diamagnetic term}. Regarding now extended systems, a central element in an array of YIG spheres coupled to identical lossy cavity modes is driven by an external quantum squeezed drive in a magnon-magnon entanglement generation scheme proposed to generate bipartite and tripartite entanglement between distant magnon modes across the entire array~\cite{ullah2023macroscopic}. 

\subsubsection{Further hybrid setups}

According to the proposal of~\cite{palyi2012spin}, a quantum dot with an odd number of electrons can be modelled by the Jaynes-Cummings Hamiltonian in the strong-coupling regime. A quantized flexural mode of the suspended tube plays the role of the optical mode while two distinct two-level subspaces -- for small and large magnetic fields -- can be modelled by two-state atoms in this nanomechanical setup. The authors start from the Hamiltonian 
\begin{equation}
\hat{H} = \frac{\Delta_{\rm so}}{2} \hat{\tau}_z ({\bf s}\cdot \hat{{\bf t}}) + \Delta_{KK'} \hat{\tau}_x
- \mu_{\rm orb} \hat{\tau}_z  ({\bf B}\cdot \hat{{\bf t}})  + \mu_B (\hat{\vec s} \cdot \vec B),
\label{eq:HnP}
\end{equation}
where $\Delta_{\rm so}$ and $ \Delta_{KK'}$ are the spin-orbit and intervalley couplings,
$\hat{\tau}_i$ and $\hat{s}_i$ ($i=x,y,z$) are the Pauli matrices for the valley and spin degrees of freedom, ${\bf t}$ is the tangent vector along the carbon nanotube (CNT) axis, 
and ${\bf B}$ denotes the applied magnetic field. 

A generic deformation of the CNT with deflection $u(z)$ makes the tangent vector ${\bf t}(z)$ coordinate-dependent. Expanding ${\bf t}(z)$ for small deflections, we rewrite the coupling terms in Hamiltonian \eqref{eq:HnP}) as 
${\bf s}\cdot {\bf t}  \simeq s_z + (du/dz) s_x$ and ${\bf B}\cdot {\bf t}  \simeq B_z + (du/dz) B_x$.
The deflection $u(z)$ is then expressed in terms of the creation and annihilation operators $\hat{a}^\dagger$ and $\hat{a}$ for a quantized flexural phonon mode, $u(z) = f(z) \frac{\ell_0}{\sqrt 2} (\hat{a}+\hat{a}^\dag)$, where
$f(z)$ and $\ell_0$ are the waveform and zero-point amplitude\index{Zero-point! amplitude} of the phonon mode. Then, each of the three qubit types (spin, spin-valley qubits) acquires a coupling to the oscillator mode which can be modeled as 
\begin{equation}
\frac{\hat{H}}{\hbar} = 
	\frac{\omega_{q}}{2} \hat{\sigma}_z +
	 g  (\hat{a}+\hat{a}^\dag) \hat{\sigma}_x +
	\omega_p \hat{a}^\dag \hat{a} +
	2 \lambda (\hat{a}+\hat{a}^\dag) \cos\omega t.
	\label{eq:jcmP}
\end{equation}
In the above Hamiltonian, a term describing external driving of the
oscillator with frequency $\omega$ and coupling strength $\lambda$ has been included, which can be achieved by coupling to the ac electric field of a nearby antenna. The coupling strength $g$ is directly proportional to $\Delta_{\rm so}$ as well as to the derivative of the waveform corresponding to the phonon mode averaged against the electron
density profile in the quantum dot. Upon effecting the RWA, the authors employ the standard master equation\index{Master equation} in Lindblad form to study strong-coupling conditions at finite temperatures.  

An optomechanical setup consisting of a laser-pumped cavity partitioned by a flexible membrane which is partially transmissive to light, and thereby subject to radiation pressure, has been proposed as a comparison to the critical behaviour encountered in the Dicke model~\cite{mumford2015dicke}. This proposal follows the consideration of a Fabry-P\'{e}rot cavity\index{Fabry-P\'erot! cavity/resonator} with a moveable mirror to realize the Dicke–Hepp–Lieb phase transition\index{Dicke! phase transition}\index{Phase transition! Dicke} in~\cite{bhattacherjee2013dicke}. The authors report on a new type of superradiance\index{Superradiance} effect in this system for a very low value of the coupling between the optical and mechanical mode.

The simplest many-body description for bosons in a double-well potential is the single-band two-site Bose-Hubbard model\index{Bose-Hubbard! model!single-band, two-site}. For a small overlap between the two condensates the ground state is a macroscopic Schr\"{o}dinger cat state~\cite{cirac1998quantum}. The ``size of the cat'' is inversely proportional to the overlap. With the addition of an atomic impurity, the Hamiltonian of the system reads
\begin{equation}
\hat{H} =  - J \hat{B} - J^a \hat{A} +
\frac{W}{2} \Delta \hat{N} \Delta \hat{M} + \frac{\Delta \epsilon}{2} \Delta \hat{N} + \frac{\Delta
  \epsilon^a}{2} \Delta \hat{M} \, .
\label{eq:mbham}
\end{equation}
Here, $\Delta \hat{N} \equiv \hat{b}^{\dagger}_R
\hat{b}_R - \hat{b}^{\dagger}_L \hat{b}_L$ is the number difference
operator between the two wells confining the bosons, and $\hat{B} \equiv
\hat{b}^{\dagger}_L \hat{b}_R + \hat{b}^{\dagger}_R \hat{b}_L$ is the
boson hopping operator responsible for the coherence between the two wells. Likewise, $\hat{M} \equiv \hat{a}^{\dagger}_R
\hat{a}_R - \hat{a}^{\dagger}_L \hat{a}_L$ and $\hat{A} \equiv
\hat{a}^{\dagger}_L \hat{a}_R + \hat{a}^{\dagger}_R \hat{a}_L$ are the equivalent operators for the impurity. We assume that both the boson and impurity creation/annihilation operators obey the standard bosonic commutation relations, {\it i.e.}, $[\hat{b}_\alpha,\hat{b}_\alpha^\dagger]=[\hat{a}_\alpha,\hat{a}_\alpha^\dagger]=1$ with $\alpha=L,\,R$ and all the remaining commutators are identically zero. $W$ parametrizes the boson-impurity interaction, while $J$ and $J^a$ are the hopping amplitudes for the bosons and the impurity, respectively. Using similar notation, $\Delta \epsilon$ and $\Delta\epsilon^a$ are the respective differences between the
zero-point energies\index{Zero-point! energy} of the two wells, {\it i.e.}, the tilt for the bosons and the impurity.

The Hamiltonian of eq.~\eqref{eq:mbham} can be recast in a simpler form via a spin notation which makes use of the symmetric/antisymmetric (S/AS) modes instead of the left/right (L/R) modes as a basis. The S/AS modes are the eigenmodes of the single-particle problem, {\it i.e.}, in the absence of interactions. Therefore, in the limit $W \rightarrow 0$, the ground state corresponds to all the particles belonging to the S mode, since it has a lower energy. Using a simple Hadamard rotation\index{Hadamard! transformation} of the L/R creation (annihilation) operators we have,
\begin{eqnarray}
\hat{b}_L &=& \frac{1}{\sqrt{2}} \left ( \hat{b}_{S} + \hat{b}_{AS} \right ), \\
\hat{b}_R &=& \frac{1}{\sqrt{2}} \left ( \hat{b}_{S} - \hat{b}_{AS} \right) 
\end{eqnarray}
and similar expressions hold for the impurity operators.
In the new basis, and for vanishing tilts $\Delta\epsilon=\Delta\epsilon^a=0$, eq.~(\ref{eq:mbham}) takes the form 
\begin{equation}
\hat{H}_{\mathrm{S,AS}} = 2J \hat{S}_z + 2J^a \hat{S}^a_z + 2 W \hat{S}_x \hat{S}^a_x, 
\label{eq:smallham}
\end{equation}
where we have used the Schwinger spin representation\index{Schwinger spin-boson mapping}~\cite{Sakurai1994}, {\it i.e.}, $\hat{S}_z \equiv ( \hat{b}^{\dagger}_{AS} \hat{b}_{AS} -
  \hat{b}^{\dagger}_{S} \hat{b}_{S} )/2 =-\hat{B}/2$ and
$\hat{S}_x \equiv  ( \hat{b}^{\dagger}_{AS} \hat{b}_{S} +
  \hat{b}^{\dagger}_{S} \hat{b}_{AS} )/2 =-\Delta \hat{N}/2$. 
Apart from the trivial $U(1)$ symmetry related to particle conservation, we note that the Hamiltonian supports a $Z_2$ parity symmetry under the transformation $\hat{S}_x\rightarrow-\hat{S}_x$, $\hat{S}_y\rightarrow-\hat{S}_y$, $\hat{S}_z\rightarrow\hat{S}_z$, and likewise for the impurity spin operators. This spin rotation preserves the $SU(2)$ angular momentum commutation relations. Note that in the original L/R-basis this symmetry is nothing but a reflection of the double-well about the origin. It follows that a non-zero tilt $\Delta\epsilon \neq 0$ or $\Delta\epsilon^a \neq 0$ breaks this symmetry.   

The model Hamiltonian written in the form of eq.~(\ref{eq:smallham}) shows some resemblance to the Dicke Hamiltonian
\begin{equation}
\hat{H}_{\mathrm{D}} = \omega_B \hat{S}_z + \omega_A \hat{c}^{\dagger}\hat{c} +
2 g (\hat{c} + \hat{c}^{\dagger})\hat{S}_x \, .
\label{eq:dickeImp}
\end{equation}
Here, $\hat{c}^{\dagger} (\hat{c})$ are photon, {\it i.e.}, bosonic creation (annihilation) operators and $\hat{S}_z$ and $\hat{S}_x$ are spin operators. eq.~(\ref{eq:dickeImp}) describes a spin-$N/2$ system coupled to the position coordinate of a harmonic oscillator.  The frequencies $\omega_B$ and $\omega_A$ are the spin precession and harmonic oscillator frequencies, respectively, and $g$ is the coupling strength. As we have seen many times so far, this system experiences a QPT at a certain critical value $g_c$~\cite{mumford2014impurity}. Meanwhile, the semiclassical dynamical behaviour of a fermionic condensate interacting with a single bosonic mode is studied via a generalization of the Tavis-Cummings model\index{Tavis-Cummings model} in~\cite{yuzbashyan2005integrable}. 
  
Collapses and revivals typical of the JC dynamics can be observed in the regime of perfect Rydberg blockade where the mesoscopic atomic ensemble is effectively a two-level system with two levels represented by collective Dicke states, say $\ket{G}$ and $\ket{R}$. The `atom' is initially prepared in the state $\ket{G}$, and the field is in state $\sum\limits_{n=0}^{\infty}{C\left( n
\right) \ket{n}}$, where the probability to find $n$ photons in the field mode is $p(n)=|C(n)|^2$. The number of photons in a coherent state of the electromagnetic field (with average photon number $\overline{n}$) follows the Poissonian distribution\index{Distribution! Poisson}~\cite{Loudon1983},
\begin{equation}
p\left({n} \right) = \left( {\bar {n}} \right)^{n}\mathrm{exp}\left( { - \bar {n}} \right)/\left( {n!} \right).
\end{equation}
Following up our discussion from sec. \ref{sssec:crsubsec}, we note that the solution of Schr\"odinger's equation yields the probability $P_e$ to find the atom in the excited state~\cite{Banacloche1990},
\begin{equation}\label{eq:PeJC}
P_e = \sum\limits_{n=1}^{\infty} {p\left( {n} \right)\mathrm{sin}^{2}\left( {g t\sqrt {n}}
\right)} .
\end{equation}
The familiar $\sqrt{n}$ dependence in eq. \eqref{eq:PeJC} arises from matrix elements $\left\langle {n} \right|\hat{a}^{ +} \left|{n-1} \right\rangle =
\left\langle {n-1} \right|\hat{a}\left| {n} \right\rangle = \sqrt {n}$ of creation and annihilation operators.
Similarly, due to the dependence of the Rabi frequency\index{Rabi! frequency} of single-atom excitation in Rydberg-blockaded ensembles on the number of atoms,
the probability of single-atom excitation in an ensemble with a random number of atoms is given by
\begin{equation}\label{eq:P1R}
P_1 = \sum\limits_{N} {p\left( {N} \right)\mathrm{sin}^{2}\left( {\sqrt{N}\Omega t/2}
\right)},
\end{equation}
where $p(N)$ is the probability to have $N$ atoms in the ensemble. Given the similarity between eqs.~\eqref{eq:PeJC} and \eqref{eq:P1R}, one can introduce a new quantum number for the mesoscospic ensemble of atoms which is equivalent to the number of photons $n$ and is described by similar annihilation and creation operators~\cite{beterov2014jaynes}. Moving on to a hybrid system of ultracold atoms in interaction with the field of waveguide resonators, coherent coupling of a Rydberg transition of atoms trapped inside an integrated superconducting chip to the microwave field of an on-chip coplanar waveguide resonator has been very recently reported in~\cite{kaiser2021cavity}. The setup supports cavity-driven Rabi oscillations between a pair of Rydberg states of atoms in an inhomogeneous electric field near the atom chip surface.

The photonic realization of the quantum Rabi model in a Hilbert space, based on a semi-infinite curved binary waveguide array, was analyzed in~\cite{crespi2012photonic}. The experiment and the accompanying simulations were performed in the deep strong-coupling regime ($g \simeq \omega$). In the limit where the qubit frequency tends to zero, strictly periodic temporal dynamics of photon number distribution were observed, with the periodicity dictated by the frequency of the quantized field. Alternatively, for a cloud of ultracold atoms subjected to two laser-induced potentials, a periodic lattice and a harmonic trap, when the system state which is fully contained in the first Brillouin zone\index{Brillouin zone} the system Hamiltonian can be mapped to the quantum Rabi model in terms of the momentum operator and in a rotated spin basis~\cite{felicetti2017quantum}. Similarly, the anisotropic 1D spin-orbit coupling realized in a trapped cold atomic gas has been mapped to the quantum Rabi model with spin-momentum coupling in~\cite{hu2013mapping}. The energy spectrum is extracted from the zeros of a transcendental equation in the spirit of~\cite{braak2011integrability}. 

Reporting from the experimental front once again, a hybrid quantum system that combines two heterogeneous collective-excitation  modes -- a narrow-linewidth\index{Linewidth} magnetostatic mode and a superconducting qubit -- was modeled by a JC-type Hamiltonian to interpret the normal-mode splitting arising due to strong coupling between the qubit and the magnon~\cite{tabuchi2015coherent}. A couple of years later, magnon number states were resolved by employing a qubit dispersively coupled to a harmonic oscillator for the experiment reported in~\cite{lachance2017resolving}. At the same time, two-level hybrid-spin qubits can be defined using magnetic dipole coupling of an ensemble of spins to microstrip cavities with an embedded nonlinear subsystem such as a transmon qubit~\cite{imamouglu2009cavity}.

We conclude this section with a brief reference to solid-state devices and models that have been of paramount importance throughout the last two decades up to our days. The cyclotron motion in graphene monolayers positioned  perpendicular to a magnetic field is described by a Hamiltonian which is formally equivalent to the JC model~\cite{schliemann2008cyclotron}. Within that correspondence, the solution for the time-dependent position operators in the Heisenberg picture is considerably simplified if one assumes that the electron energy is larger than the characteristic energy scale of the Hamiltonian, set by the magnetic length and the Fermi velocity in graphene. In connection to graphene and its relativistic Landau levels, and anti-JC term arises when modelling the quantum Hall effect in square lattice subjected to a non-Abelian gauge potential\index{Gauge! potential! non-Abelian}~\cite{goldman2009non}. Fermionic degrees of freedom are also involved in the edge dynamics of a spin-Hall insulator interacting with circularly polarized quantized electromagnetic field through a Zeeman term were considered in~\cite{gulacsi2015floquet}. The pertinent Hamiltonian reads
\begin{equation}
\hat{H}/\hbar=\omega \hat{a}^{\dagger}\hat{a}+v\sum_p p \hat{S}^z_p+\frac{g}{\sqrt{L}}\sum_p\left(\hat{a}^{\dagger}\hat{S}^{-}_p+\hat{a}\hat{S}^{+}_p\right).
\end{equation}
Here, $v$ is the Fermi velocity of the edge state,
$\omega$ and $\hat{a}$ denote the resonance frequency and annihilation operator of the cavity mode, while $\hat{c}_{p,\sigma}$ annihilates
an electron with momentum $p$ and spin $\sigma$. 
The last term describes the Zeeman interaction with the spin operators expressed as
$\hat{S}^{+}_p=\hat{c}^+_{p,+}\hat{c}_{p,-}$ and $\hat{S}^z_p=\sum_{\sigma=\pm}\sigma \hat{c}^+_{p,\sigma}\hat{c}_{p,\sigma}$. The Zeeman coupling depends on the photon frequency as $g=\sqrt{ \tilde g|\omega|}$. Considering a single momentum $p$, the Hamiltonian reduces to the Jaynes-Cummings model. Hence, the spectrum consists of the energy doublet
\begin{equation}
  \epsilon_{\pm}(p,m)=\omega\left(m-\frac{1}{2}\right)\pm\sqrt{\left(vp-\frac{\omega}{2}\right)^2+\frac{g^2}{L}m}, 
\end{equation}
with $m$ being a positive integer, plus a single mode $\epsilon_{-}(p,0)=-v p$. The Purcell enhancement\index{Purcell enhancement} factor and the excited-state population dynamics for a quantum emitter in the vicinity of a graphene nanodisk were calculated in~\cite{Thanopoulos2022}, and distinct signatures of non-Markovianity were identified.

A very recent notable proposal in solid-state quantum technology concerns the use of nuclear spin states of magnetic ions. The hyperfine interaction with an electronic spin mediates the coupling of nuclear spins to to electromagnetic radiation, enhancing the nuclear Rabi frequencies with respect to those of isolated nuclei~\cite{Hussain2018}. This enhancement can be of use when exploring the coupling of on-chip superconducting resonators to the nuclear spins of magnetic molecules~\cite{Gaita2019}. Avoided crossings have been experimentally observed when the level splittings of $^{171}$Yb and $^{173}$Yb are brought into resonance with the cavity photons~\cite{Rollano2022}. 

Furthermore, we mention that Sommerfeld's theory~\cite{Sommerfeld1928}\index{Sommerfeld theory} of the free electron gas\index{Free electron gas} has been revisited in the framework of cavity quantum electrodynamics. Rokaj and coworkers find that the electron-photon ground state is a Fermi liquid containing virtual photons. In contrast to models of finite systems, no ground state exists if the diamagentic $\boldsymbol{A}^2$ term\index{Diamagnetic term} is omitted. Furthermore, linear response theory shows that the cavity field induces plasmon-polariton excitations\index{Polariton}, and alters the optical and the DC conductivity of the electron gas~\cite{Rokaj2022}. Furthermore, The Tavis-Cummings Hamiltonian\index{Tavis-Cummings model} has been used in~\cite{Lei2023} to model an inhomogeneously broadened ensemble of solid-state emitters coupled with high cooperativity to a nanophotonic resonator under strong excitation. A collectively induced transparency in the cavity reflection spectrum results from quantum interference as a collective effect, while the coherent excitation of the system produces a nonlinear response spanning a broad range from subradiance\index{Subradiance} to superradiance\index{Superradiance}.

Opening a new path, Karnieli and coworkers have developed a unified theory to account for the quantum three-particle interaction of a free electron, a bound electron, and a photon in a single-mode resonator~\cite{Karnieli2023}. Modeling the bound-electron system by a two-state atom, they are able to reduce the problem to the JC interaction and the free-electron bound-electron resonant interaction  models for the coupling to the cavity field and the free electron, respectively. The coupling between the free electron and the cavity field is generally captured by the quantum photon–induced near-field electron microscopy model~\index{Model! Near-field electron microscopy}. This simplified formulation can be extended to include a wide range of models for the emitter-cavity system, including the quantum Rabi model for ultrastrong light-matter coupling, the Dicke model for an ensemble of emitters, or the Hopfield model for collective matter excitations. The latter may replace the JC model, e.g, in dielectric micro- and nanocavities, accounting for a larger Rabi splitting.


\section{Extensions to many-body configurations and additional degrees of freedom}\label{sec:ext}

Some years after the millennium, we saw the first realization of the Mott-insulator phase transition\index{Mott!--insulator phase transition}\index{Phase transition! Mott-insulating phase} in an ultracold atomic system~\cite{greiner2002quantum}. This was a breakthrough demonstrating how truly strongly interacting many-body quantum systems could be studied in a quantum optical setting. These systems had a clear advantage over many other systems in the condensed matter community due to their flexibility and controllability. This spurred the quest for realizing {\it quantum simulators}\index{Quantum! simulator}, {\it i.e.}, controllable quantum systems mimicking other quantum systems which are not subject to the same control and are too complicated to be simulated on a classical computer. Apart from being capable of realizing known many-body systems from (mainly) the condensed-matter physics community, these atomic setups also enable the study of novel systems lacking counterparts in other communities. An avalanche of experimental and theoretical activities set off, and it did not take long before new proposals of different systems implementing quantum simulators popped up~\cite{carusotto2013quantum,georgescu2014quantum}. In 2006, two notable works suggested how one may simulate interacting many-body systems with photons~\cite{hartmann2006strongly,greentree2006quantum}. Following in particular the footsteps of the development of circuit QED (see section~\ref{sec:cirQED}), it was explained how to couple Jaynes-Cummings (JC) sub-systems into a lattice---the {\it Jaynes-Cummings-Hubbard model}: each JC sub-system, comprising a qubit and a photon mode, becomes connected by allowing for photons to tunnel between neighboring JC systems. The JC nonlinearity\index{Jaynes-Cummings! nonlinearity} induces the effective interaction, and it was argued how the strongly-correlated regime could be reached making truly `many-body' systems beyond the perturbative regime. Another feature of interest, not as important for ultracold atoms in optical lattices, is the nonequilibrium character of the system; typically one has in mind driven-dissipative resonators\index{Dissipative! resonator}. These types of systems will be discussed in sec.~\ref{ssec:jch} below. Roughly at the same time, people started to think what new physics will emerge by coupling condensed atoms to a single mode of an optical resonator~\cite{brennecke2007cavity,colombe2007strong}. We have already seen two phenomena where the atom number plays an important role; in sec.~\ref{sssec:optbis} we discussed optical bistability and how the coupling of thermal atomic gases to the cavity field may cause\index{Optical! bistability} a bistable behaviour, and in sec~\ref{sssec:dicke} the superradiance\index{Superradiance} was considered. By lowering the temperature of the atomic gas such that the atoms condense, a new regime is reached where the typical atomic kinetic energy, set by the recoil energy $E_\mathrm{R}$, becomes comparable to other energy scales, see sec.~\ref{sssec:qatmo}. In a condensate, contrary to a thermal gas, we expect {\it long-range order} meaning that the atomic gas shows quantum coherence over large spatial distances, and it makes sense to talk about collective excitation modes of the condensate rather than individual atomic excitations. An important ingredient in these systems, which are discussed in sec.~\ref{ssec:mbcQED}, that makes them different from other ultracold atomic setups, is the {\it infinite-range interaction} between the atoms stemming from the common coupling to the quantized light field in the cavity. This leads to, as we will show, a set of novel phenomena. Finally, in this section we also discuss a relatively young field where one lets a gas of molecules, rather than atoms, be coherently coupled to optical resonators. While not condensed, the molecular cavity QED setting still adds a complexity in terms of the internal molecular structures. In a Born-Oppenheimer picture, the electronic transition frequencies are set by the internuclear distances, which of course are dynamical degrees of freedom. Thus, apart from the energy exchange between electronic transitions and photons, we also derive results concerning transitions in the molecular vibrations. One idea is that `dressing the molecules' with the cavity photons can alter chemical reactions, and thereby control them to some extent. 

\subsection{Jaynes-Cummings-Hubbard models}\label{ssec:jch}
The Jaynes-Cummings Hubbard (JCH) Hamiltonian was formulated to describe an optical system displaying strong correlations on a mesoscopic scale; in other words, we are dealing with a model imitating equilibrium predictions in situations where the photon number is not conserved. The original setup, proposed in 2006~\cite{Greentree2006}, comprised a two-dimensional array of connected cavities coupled to two-level atoms such that photon blockade occurs, which means that absorption of one photon and the formation of a vacuum Rabi resonance ``blocks'' the absorption of a second\index{Photon! blockade}. This behavior is generalized to multiphoton resonances\index{Multiphoton! resonance} which, however, do not present an absolute barrier to the absorption of additional photons since, in a nonequilibrium configuration\index{Nonequilibrium! configuration}, the coherent drive strength broadens the energy levels~\cite{CarmichaelPRX}. Onsite JC interactions with photon hopping between cavities and an appropriate chemical-potential term comprise the JCH Hamiltonian,
\begin{equation}\label{eq:JCHHam}
\hat{H}=\sum_{i}\hat{H}_{i}^{\rm JC}-\sum_{\braket{i,j}}t_{ij}\hat{a}_{i}^{\dagger}\hat{a}_{j}-\sum_{i}\mu_{i}\hat{N}_{i},
\end{equation} 
where $t_{ij}$ are the hopping coefficients, $\hat{N}_{i}$ is the total (photonic and atomic) excitation number operator at site $i$ (where $\sum_{i} \hat{N}_{i}$ commutes with $\hat{H}$), associated with the chemical potential $\mu_i$. Assuming zero disorder, introducing a superfluid order parameter\index{Superfluid! order parameter} $\psi \equiv \braket{\hat{a}_i}$, and applying the decoupling transformation $\hat{a}_{i}^{\dagger}\hat{a}_{j}=\braket{\hat{a}_i^{\dagger}}\hat{a}_j + \braket{\hat{a}_j}\hat{a}_i^{\dagger}-\braket{\hat{a}_i^{\dagger}}\braket{\hat{a}_j}$, results in the mean-field single-site Hamiltonian
\begin{equation}\label{eq:singlesiteJCH}
\hat{H}=(\omega_0-\mu)\left(\hat{a}^{\dagger}\hat{a}+\hat{\sigma}_{+}\hat{\sigma}_{-}\right) + g\left(\hat{a}\hat{\sigma}_{+} + \hat{a}^{\dagger}\hat{\sigma}_{-}\right)
-zt\psi\left(\hat{a}^{\dagger}+\hat{a}\right)+zt |\psi|^2,
\end{equation}
where $z$ is the number of nearest neighbors, between which inter-cavity hopping occurs with frequency $t$. When the order parameter is zero ($\psi=0$), one encounters a Mott phase with a fixed number of excitations per site in the absence of fluctuations; the presence of a finite order parameter, $\psi \neq 0$, on the other hand, indicates a superfluid phase\index{Superfluid! phase}. 

\begin{figure*}
\begin{center}
\includegraphics[width=9cm]{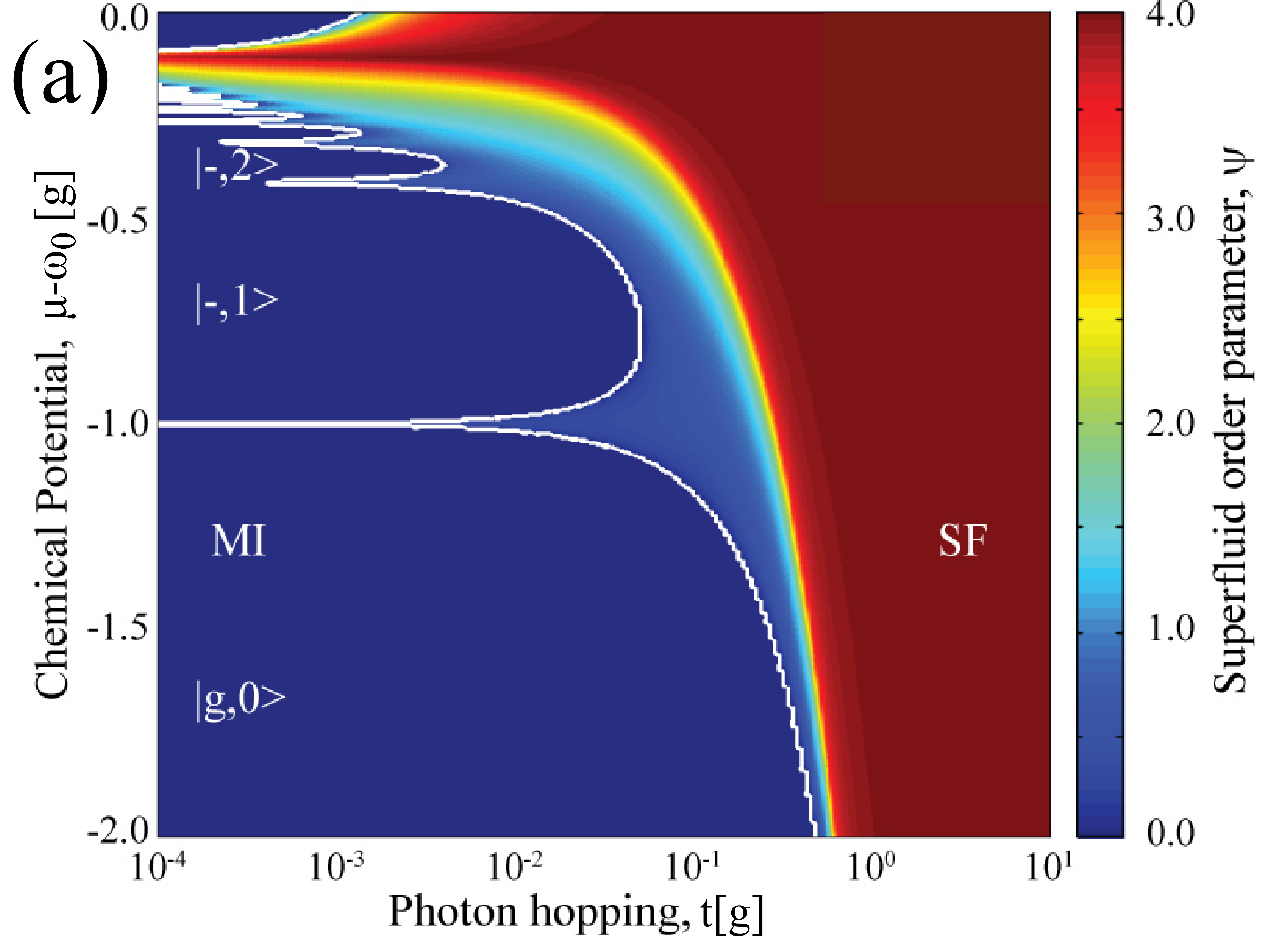} 
\includegraphics[width=8cm]{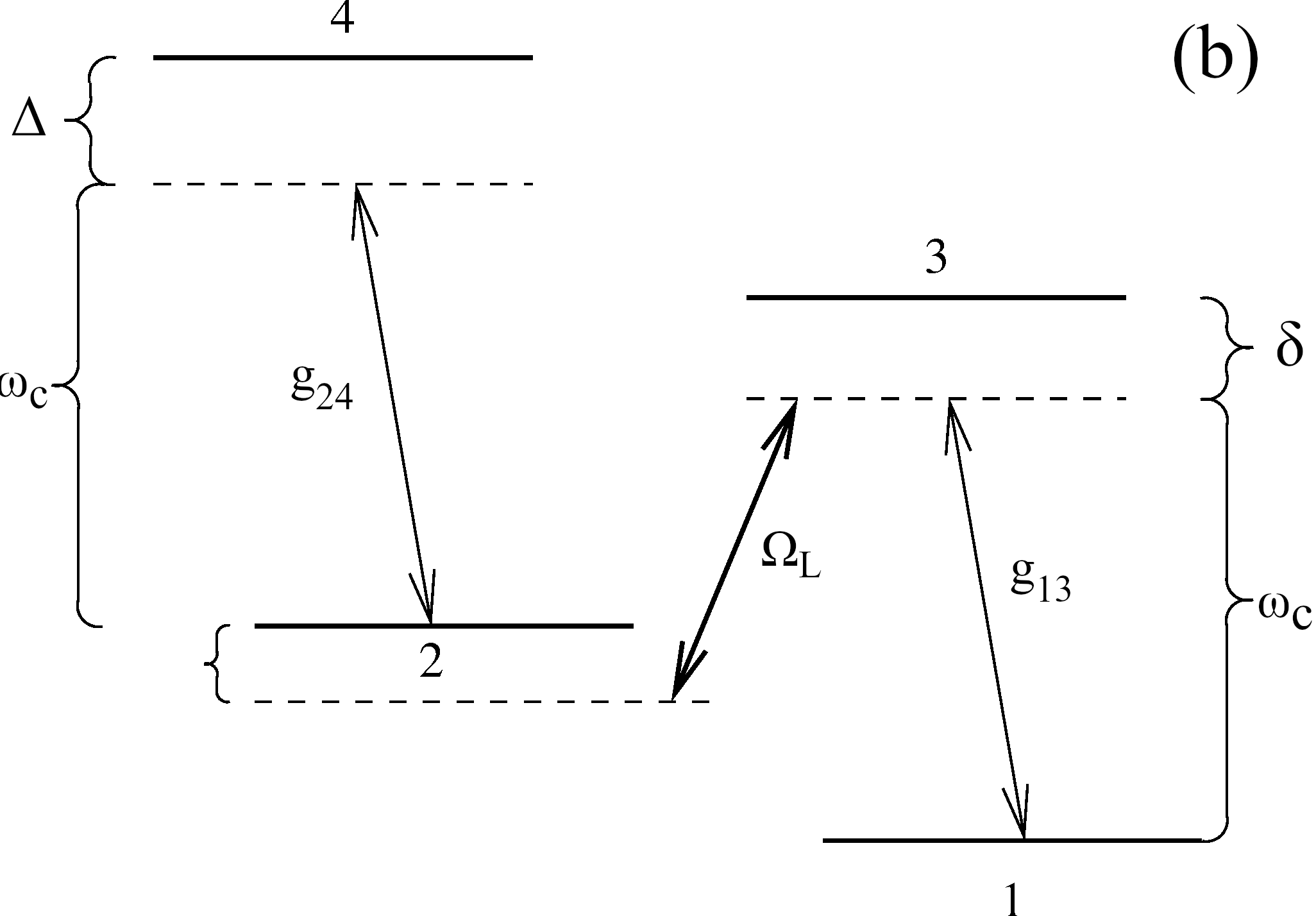} 
\end{center}
\caption{{\bf (a)} Slices depicting the superfluid order parameter $\psi$ as a function of the photon hopping frequency $t$ (divided by the onsite light-matter coupling strength $g$) and the (shifted) chemical potential $\mu-\omega_0$ (also normalized to $g$) for zero detuning ($\Delta=0$). The diagrams show the lowest-order Mott-insulator lobes, indicated by the regions of $\psi=0$. On the left-hand side (where photonic repulsion dominates over hopping) we find the Mott insulator phase\index{Mott! insulator phase} (denoted by MI), while the superfluid phase (denoted by SF) is located on the right-hand side. The white contour delineates the region where $\psi$ vanishes, hence delineating the quantum phase transition\index{Phase transition! quantum}. [Adapted from fig. 4(a) of~\cite{Greentree2006}] {\bf (b)} The energy-level structure of electromagnetically induced transparency\index{Transparency! electromagnetically-induced} depicting externally-driven atoms with a corresponding Rabi frequency $\Omega_L$, dipole-coupled to the intracavity mode with strengths $g_{13}$ and $g_{24}$. In this scheme employing cavity polaritons, $\delta, \Delta$ and $\varepsilon$ are appropriately selected detunings featuring in the interaction Hamiltonian. [Adapted from fig. 2 of~\cite{Hartmann2006} and fig. 1 of~\cite{EITWerner1999}]}
\label{fig:FigJCH}
\end{figure*}

The extent of the characteristic Mott-insulator lobes is determined by an appropriate boundary set by the critical chemical potential; such a potential corresponds to the increment of the excitation per site by one, yielding
\begin{equation}\label{eq:mucritical}
\mu_c=\omega_0-[\Omega_{n+1}-\Omega_{n}],
\end{equation} 
where now we have taken into account a detuning between the onsite coupled cavity mode and two-level atom $\Delta$, featuring in the familiar Rabi frequencies $\Omega_{n}\equiv \sqrt{(1/4)\Delta^2+g^2 n}$\index{Rabi! frequency}. The phase diagram is depicted in fig. \ref{fig:FigJCH} (a). 
The number of excitations per site, $\rho$, in each lobe of the Mott-insulator phase has been computed post-minimization for the ground state, following the expression $\rho=-\partial E_g (\psi=\psi_{\rm min})/\partial \mu$ as a function of $t$ and $\mu$ in a grand-canonical ensemble formulation. The produced plateaus point to regimes of constant density and incompressibility, attributed to the gapped Mott-insulator phase. In \cite{Greentree2006}, negatively-charged NV centers, placed inside each cavity of the ensemble, have been proposed as a promising candidate for realizing this equilibrium quantum phase transition\index{Phase transition! quantum!equilibrium} of light. 

The Bose-Hubbard Hamiltonian and the corresponding two phases of matter has been associated with hopping of cavity polaritons\index{Polariton} due to the mode-profile overlap of adjacent resonators. Atoms coupled to the intracavity modes are driven by external detuned coherent fields \cite{Hartmann2006}. To generate repulsion between polaritons, four-level atoms are driven in such a fashion to emulate Electromagnetically Induced Transparency \cite{EITWerner1999}. With reference to the level-structure of fig. \ref{fig:FigJCH}, in a frame rotating with the Hamiltonian
\begin{equation*}
\hat{H}_0= \hbar \omega_C \left(\hat{a}^{\dagger}\hat{a} + \frac{1}{2}\right) + \sum_{j=1}^{N}\left(\omega_C \hat{\sigma}_{22}^{j} + \omega_C \hat{\sigma}_{33}^{j} + 2\omega_C \hat{\sigma}_{44}^{j}\right),
\end{equation*} 
the interaction Hamiltonian of the atoms coupled to the cavity mode reads
\begin{equation}\label{eq:HIpolaritons}
\hat{H}_I=\sum_{j=1}^{N}\left(\varepsilon \hat{\sigma}_{22}^{j} + \delta \hat{\sigma}_{33}^{j} + (\Delta + \varepsilon)\hat{\sigma}_{44}^{j}\right) + \left(\Omega_L \hat{\sigma}_{23}^{j} + g_{13}\hat{\sigma}_{13}^{j}\hat{a}^{\dagger} + g_{24}\hat{\sigma}_{24}^{j}\hat{a}^{\dagger} + {\rm H.c.}\right),
\end{equation}
where the operator $\hat{\sigma}_{kl}^{j} \equiv |k_j \rangle \langle l_j|$ projects the level $l$ to the level $k$ of the $j^{\rm th}$ atom, $\omega_C$ is the cavity resonance frequency, $\Omega_L$ is the Rabi frequency determined by the laser driving, and $g_{13}, g_{24}$ are the dipole coupling strengths of the cavity mode to the respective atomic transitions. 

We are interested in the situations where there is on average one excitation per cavity; hence, we consider states with at most two excitations per cavity, suppressed by a Kerr nonlinearity\index{Kerr! nonlinearity}. Neglecting two-photon detuned transitions\index{Two-photon!transition} and the coupling to the atomic level 4 \cite{polariton4thlvl}, as illustrated in fig. \ref{fig:FigJCH} (b), the Hamiltonian of eq.~\eqref{eq:HIpolaritons} can be reduced to include interactions between three species of polaritons when the number of atoms is large ($N \gg 1$). These polaritons, {\it i.e.}, quasi particles formed by an atomic and a photonic contribution, are created by three operators, denoted by $\hat{p}_0^{\dagger}, \hat{p}_{\pm}^{\dagger}$, in the subspace of at most two excitations. After defining collective spin operators $\hat{S}_{12}^{\dagger}=(1/\sqrt{N})\sum_{j=1}^{N}\hat{\sigma}_{21}^{j}$ and $\hat{S}_{13}^{\dagger}=(1/\sqrt{N})\sum_{j=1}^{N}\hat{\sigma}_{31}^{j}$ alongside the renormalized coupling rates $g\equiv \sqrt{N}\, g_{13}$, $B=\sqrt{g^2 + \Omega_L^2}$ and $A \equiv \sqrt{4B^2 + \delta^2}$, the aforementioned creation operators assume the form
\begin{equation}\label{eq:polaritonicops}
\hat{p}_0^{\dagger}=\frac{1}{B}\left(g \hat{S}_{12}^{\dagger} - \Omega_L \hat{a}^{\dagger} \right),\qquad 
\hat{p}_{\pm}^{\dagger} = \sqrt{\frac{2}{A(A\pm \delta)}}\, \left(\Omega_L \hat{S}_{12}^{\dagger} + \hat{a}^{\dagger} \pm \frac{A \pm \delta}{2}\hat{S}_{13}^{\dagger} \right).
\end{equation}
These operators satisfy the familiar bosonic commutation relations $[\hat{p}_{\eta}, \hat{p}_{\lambda}]=0$, $[\hat{p}_{\eta}, \hat{p}_{\lambda}^{\dagger}]=\delta_{\eta\lambda}$, with $\eta, \lambda=0, \pm$. For $g_{24}=0$ and $\varepsilon=0$, the Hamiltonian of eq.~\eqref{eq:HIpolaritons} is then reduced to
\begin{equation}\label{eq:HIpolaritonsRed}
\hat{H}_{I, (g_{24}=\varepsilon=0)}=\mu_0 \hat{p}_0^{\dagger}\hat{p}_0 + \mu_{+} \hat{p}_{+}^{\dagger}\hat{p}_{+} + \mu_{-} \hat{p}_{-}^{\dagger}\hat{p}_{-},
\end{equation}
with new frequencies $\mu_0=0$, $\mu_{\pm}=(\delta \pm A)/2$. 
We now need to write the full interaction Hamiltonian in the polariton basis\index{Polariton! basis}, expressing the operators $\sum_{j=1}^N \hat{\sigma}_{22}^{j}$ and $a^{\dagger}\sum_{j=1}^N \hat{\sigma}_{24}^{j}$ in terms of $\hat{p}_0^{\dagger}$, $\hat{p}_{\pm}^{\dagger}$ and decoupling the fourth level from the dynamics. In the two-excitation manifold and the rotating-wave approximation, we express the coupling of the polaritons\index{Polariton} to the level in question as
\begin{equation*}
g_{24} \left(\sum_{j=1}^{N} \hat{\sigma}_{42}^{j}\hat{a} + {\rm H.c.}\right) \approx -g_{24}\frac{g\Omega_L}{B^2} \left(\hat{S}_{14}^{\dagger} \hat{p}_0^2 + {\rm H.c.}\right),
\end{equation*}   
where $\hat{S}_{14}^{\dagger} \equiv (1/\sqrt{N}) \sum_{j=1}^{N}\hat{\sigma}_{41}^{j}$. In a frame rotating with respect to the Hamiltonian of eq.~\eqref{eq:HIpolaritonsRed}, the operator $\hat{S}_{14}^{\dagger}$ rotates at the frequency $2\mu_0$. In the hierarchy of timescales set by
\begin{equation}\label{eq:hierarchyts}
|g_{24}|, |\varepsilon|, |\Delta| \ll |\mu_{\pm}|,
\end{equation}
we neglect all terms rotating with the frequencies $2\mu_0-(\mu_{+} \pm \mu_{-})$, $\mu_0 \pm \mu_{\pm}$, which amounts to eliminating the interactions coupling $\hat{p}_0^{\dagger}$ and $\hat{S}_{14}^{\dagger}$ to the remaining polaritons. If we further assume that $g_{24} \ll |\Delta|$, the presence of the fourth level results only in an energy shift of $2\tilde{g}$, where (see eq. 5 of~\cite{EITWerner1999})
\begin{equation}\label{eq:effectiveg}
\tilde{g}=-\frac{g_{24}^2}{\Delta}\frac{N g_{13}^2 \Omega_L^2}{(N g_{13}^2 + \Omega_L^2)^2}
\end{equation}
A state containing one excitation of the operator $\hat{S}_{14}^{\dagger}$ comes with a probability $-2\tilde{g}/\Delta$, which determines an effective decay rate for the polariton species\index{Polariton! species} $p_0^{\dagger}$ via the route of spontaneous emission\index{Spontaneous! emission} from the fourth level. Similarly, the two-photon detuning\index{Two-photon!detuning} $\varepsilon$ is responsible for an energy shift $\varepsilon g^2/B^2$, playing the role of a chemical potential in the resulting effective Hamiltonian in the perturbative regime,
\begin{equation}\label{eq:Hp0}
\hat{H}_{\rm eff}=\tilde{g} \left(\hat{p}_0^{\dagger}\right)^2 (\hat{p}_0)^2 + \varepsilon \frac{g^2}{B^2}\hat{p}_0^{\dagger} \hat{p}_0,
\end{equation}
which is the normally-ordered Hamiltonian of the nonlinear polarizability model \cite{DrummondWallsDuffing} \textemdash{the} second term accounting for the Kerr nonlinearity\index{Kerr! nonlinearity} and the associated repulsion. Photon loss at a rate $2\kappa$ and spontaneous emission\index{Spontaneous! emission} from the fourth atomic level at a rate $\gamma_4$ lead to the following decoherence rate
\begin{equation}\label{eq:decohratesmall}
\tilde{\gamma}=\frac{2\Omega_L^2}{N g_{13}^2 + \Omega_L^2} \left(\kappa + \frac{g_{24}^2 N g_{13}^2}{\Delta^2 (N g_{13}^2 + \Omega_L^2)} \gamma_4\right).
\end{equation}
To close this subsection out, we mention that a trapped-ion quantum simulator with up to 32 ions has been used to study the transition from Markovian to non-Markovian spin dynamics in different regimes of the JCH model, tuning the spin frequencies to different locations of the phonon band. In addition, collapse and revival signals provide evidence for the occurrence of localization~\cite{LiMei2022}. The ground state quantum phase transition\index{Phase transition! quantum! ground state} in the Rabi-Hubbard model\index{Rabi-Hubbard model} has been experimentally observed via a slow quench of the coupling strength once again in a system of trapped ions, while the dynamical evolution for average spin inversion, phonon number and entanglement entropy is displayed for a set of parameter regimes in~\cite{LiMeiRH2022}.

An array of coupled cavities in the photon blockade regime can be mapped to the transverse-field anisotropic $XY$ model~\cite{BardynIm2012} (see also subsec.~\ref{ssec:ionfur}). Two-time correlations leading to the absorption and fluorescence spectra for such an array were calculated in~\cite{Kilda2019, KildaPhD}. The structure of the distribution function, calculated by the applying the fluctuation dissipation theorem~\cite{breuer2002theory, Sieberer2016}\index{Fluctuation-dissipation theorem}, and the associated thermalization can be understood by using a spin-wave theory, while the Fourier-transform of fluorescence in momentum space reveals the nature of {\it quasi}particle excitations in the {\it quasi}thermal state. 

\subsection{Many-body cavity QED}\label{ssec:mbcQED}
In sec.~\ref{sssec:qatmo}, we saw that as the atomic kinetic energy is lowered, the motional states\index{State! motional} of the atoms couple to the other degrees of freedom via absorption/emission of cavity photons. This implies the alteration of their kinetic energy by one recoil-energy unit, $E_\mathrm{r}=\hbar^2k^2/2m\equiv\hbar\omega_\mathrm{r}$\index{Recoil energy}, introduced in that subsection (here we also define the recoil frequency $\omega_\mathrm{r}$). By ``many-body cavity QED''\index{Many-body! cavity QED} we envisage a large number of ultracold atoms coherently coupled to the cavity mode~\cite{ritsch2013cold,mivehvar2021cavity}. Thus, in this section the motional degrees of freedom\index{Motional degrees of freedom} are considered, while they were not taken into account when discussing the TC and Dicke models in sec.~\ref{sssec:dicke}. Instead of coupling the individual atomic kinetic energies to the light field, it is often more convenient to think in terms of collective effects where low-energy phonon modes of the atomic cloud interact with the photons~\cite{ritsch2013cold,mivehvar2021cavity}. These collective modes become relevant once the atoms are condensed\index{Bose-Einstein! condensate} and coherences extend over the full atomic cloud~\cite{pethick2008bose}. At these low temperatures an additional effect that may arise is that of atom-atom interaction, which may indeed qualitatively modify the picture~\cite{Lewenstein2007a,Bloch2008a}. 

With the development of experimental techniques in laser cooling and trapping of neutral atoms it became feasible to also couple a gas of ultracold, or even condensed, atoms to the cavity field. In these first pioneering experiments, the collective ultrastrong coupling regime of a large number of two-level atoms was demonstrated in two back-to-back papers by the ETH group of T. Esslinger~\cite{brennecke2007cavity} and the ENS group of J. Reichel~\cite{colombe2007strong}. The square-root dependence of the collective coupling on the number of atoms, $g_N=g\sqrt{N}$, was verified in both experiments. The reason why the coupling scales as $N^{1/2}$ and not simply as $N$ can be found in the construction of the Dicke states\index{Dicke! state}\index{State! Dicke}~(\ref{dickestates}). This is similar to the enhancement of collective spontaneous emission\index{Spontaneous! emission! collective} -- the superradiance we discussed in sec.~\ref{sssec:dicke}. In a strict sense, for the results of these works it was not crucial that the atoms were condensed. Nevertheless, at least in the ETH experiment, condensation was confirmed. After these experiments, a couple of independent experimental works on optical bistability followed, in Fabry-P\'erot cavities\index{Fabry-P\'erot! cavity/resonator}~\cite{gupta2007cavity,brennecke2008cavity} and in a ring cavity~\cite{schmidt2014dynamical}. It turned out that these systems were excellent candidates for studying optomechanics~\cite{brooks2012non}, {\it i.e.} the coupling between light and mechanical modes we discuss in sec.~\ref{ssec:hybridsys}. The last decade has seen tremendous progress in this field, both theoretically and experimentally, which we will briefly discuss in this section. The interest is two-fold; the system can be tailored in order to study known exotic models and phenomena, falling in the realm of quantum simulators\index{Quantum! simulator}, but also new phases of matter may emerge with no known counterparts in other systems. An important feature is that these systems are open by nature since photons are inevitably lost, and one ends up with nonequilibrium configurations\index{Nonequilibrium! configuration}.

Schmit and collaborators have developed a quantum kinetic theory for describing the dynamics of atoms interacting with a multimode high-finesse cavity\index{Multimode! cavity} or multiple cavities. The recent study of atomic self-organization is based on the formulation of a quantum master equation\index{Master equation} for many-body QED, which is obtained after adiabatic elimination of the cavity degree of freedom via a Schrieffer-Wolff\index{Schrieffer-Wolff transformation} transformation~\cite{JagerSchmit2022}.

\subsubsection{Mean-field explorations}\label{sssec:mfmQED}
Let us for now focus on bosonic atoms. Until nowadays, they are the ones considered in experiments. When exploring many-body systems composed of a large number of identical particles, it is convenient to employ the method of second quantization\index{Second quantization}. In this monograph we do not go through the ideas behind the method, but instead refer the reader to books like~\cite{altland2010condensed,nagaosa2013quantum}. We introduce the atomic field operators $\hat\psi^\dagger({\bf x})$ and $\hat\psi({\bf x})$\index{Field! operators}, that create respectively annihilate an atom at position $x$. For bosons they obey the commutation relations $\left[\hat\psi({\bf x}),\hat\psi^\dagger({\bf x}')\right]=\delta({\bf x}-{\bf x}')$ and $\left[\hat\psi({\bf x}),\hat\psi({\bf x}')\right]=0$. For fermions we need to replace the commutators by anti-commutators. We note that for a complete orthonormal set of functions $\{\phi_n({\bf x})\}$, the field operators can be expanded as
\begin{equation}\label{fieldop}
\hat\psi({\bf x})=\sum_n\hat c_n\phi_n({\bf x}),
\end{equation}
where the $\hat c_n$'s are bosonic operators\index{Operator! bosonic}, {\it i.e.} $\left[\hat c_n,\hat c_m^\dagger\right]=\delta_{nm}$ and so on. Thus, the operator $\hat c_n^\dagger$ ($\hat c_n$) creates (annihilates) an atom in the single-particle state, which is often called {\it orbital}, $\phi_n(x)$. A Fock state $|n_1,n_2,\dots\rangle$ represents a state with $n_1$ atoms in the orbital $\phi_1(x)$, $n_2$ atoms in the orbital $\phi_2(x)$, and so on. Hence, we may think of second quantization as a way to determine the occupation of particles in different orbitals, to be compared with first quantization where we instead look for the particular wave-function $\psi(x_1,x_2,\dots)$ attributed to the particles.

Expanding the field operators as in~(\ref{fieldop}) is usually practical when considering various approximations~\cite{Lewenstein2007a,Bloch2008a}, often restricting the states to low-energy modes -- projecting the Hilbert space to a low-energy sector. There are two special sets of functions $\phi_n$ used for the expansion~(\ref{fieldop}):
\begin{enumerate}
\item[1)] When the particles experience a relatively strong periodic potential $V(x)$, it is customary to use the orthonormal {\it Wannier fuctions}\index{Wannier functions}, $w_{\nu,{\bf R}_i}({\bf x})$. These are exponentially localized functions centered around the position ${\bf R}_i$ of site $i$ in the periodic potential~\cite{kohn1959analytic}. The quantum number $\nu=1,\,2,\,\dots$ is the eigenenergy band index. The eigenfunctions of a single-particle periodic Hamiltonian are the (unnormalized) Bloch functions\index{Bloch! function} $\psi_{{\bf k},\nu}({\bf x})$, and the Wannier functions are given in terms of these as
 \begin{equation}\label{wanf}
 w_{\nu,{\bf R}_i}({\bf x})=\frac{1}{V}\int_\mathrm{BZ}e^{-i{\bf k\cdot R}_i}\psi_{{\bf k},\nu}({\bf x})\,d^3{\bf k},
\end{equation}
where the integration is over the first Brillouin zone\index{Brillouin zone} (This is not a unique definition of the Wannier functions, but the most common one.). For ultracold atoms, and relatively deep lattice potentials, it is often justified to restrict the expansion~(\ref{fieldop}) to Wannier functions of the lowest energy band, $\nu=1$, -- the so-called {\it single-band approximation}~\cite{Lewenstein2007a,Bloch2008a}\index{Single-band approximation}. 

\item[2)] The expansion in terms of Wannier functions is particularly suited for strongly-interacting systems, {\it e.g.} for describing the {\it Mott insulating} physics\index{Mott! insulator} of the Bose-Hubbard model\index{Bose-Hubbard! model}\index{Model! Bose-Hubbard}~\cite{Jaksch1998a}. For weakly-interacting systems, on the other hand, the relevant degrees of freedom might not be those tied to the sites, but rather the collective ones. For a condensate these are the phonon vibrational modes\index{Phonon mode} characterized by some momentum ${\bf k}$. Here it might be convenient to expand $\hat\psi({\bf x})$ in plane-wave Fourier components instead.
\end{enumerate}
We will consider both choices in what follows. 

Almost exclusively, we will furthermore assume that the internal atomic hyperfine level $|e\rangle$ has been adiabatically eliminated\index{Adiabatic! elimination}, see Subsec.~\ref{ssec:JCm}. In this respect, we do not need to specify the internal atomic state for the field operators $\hat\psi({\bf x})$. In case we consider both internal states we would need two field operators $\hat\psi_{g,e}({\bf x})$. For bosonic atoms the many-body Hamiltonian becomes (in one dimension for simplicity)~\cite{ritsch2013cold,mivehvar2021cavity} 
\begin{equation}\label{mbham}
\hat{\mathcal H}=\displaystyle{\hbar\delta\hat a^\dagger\hat a+\int dx\,\hat\psi^\dagger(x)\Bigg[-\frac{\hbar^2}{2m}\frac{d^2}{dx^2}+V(x)}\displaystyle{+\frac{U_0}{2}\hat\psi^\dagger(x)\hat\psi(x)\Bigg]\hat\psi(x)},
\end{equation}
where $m$ is the atomic mass, $\delta=\omega-\omega_\mathrm{L}$ the cavity-pump detuning ({\it i.e.} we imagine a pumped system with $\omega_\mathrm{L}$ the pump frequency), and $U_0$ is the effective interaction strength (to a good approximation being proportional to the $s$-wave scattering length $a_s$; $U_0=4\pi a_s\hbar^2/m$~\cite{Jaksch1998a}), The `potential' $V(x)$ includes the Stark shifts\index{Stark shift} felt by the atoms and depends on the particular situation under consideration. The two most common setups are {\it transverse} or {\it longitudinal} (also called {\it on-axial}) pumping\index{Transverse pumping}\index{Longitudinal pumping}, see Fig~\ref{fig18}. In the first, a laser illuminates the atomic cloud such that photons can be scattered by the atoms into the cavity via a two-photon process\index{Two-photon!process}. This process takes place in two steps: ($i$) a photon is absorbed from the laser and emitted into the cavity or vice versa, and ($ii$) a cavity photon is absorbed and emitted back into the cavity (the process involving two laser photons is left out as it constitutes an overall constant energy shift within our one-dimensional treatment). With the mode function of the cavity $\propto\cos(kx)$, the effective potential for transverse pumping takes the form~\cite{domokos2002collective,asboth2005self,vukics2007microscopic,maschler2008ultracold,nagy2008self,fernandez2010quantum,nagy2010dicke,ritsch2013cold,mivehvar2021cavity}
\begin{equation}\label{trpot}
V_\mathrm{tr}(x)=\frac{g_0^2}{\Delta_\mathrm{a}}\hat a^\dagger\hat a\cos^2(kx)+i\frac{g_0\eta(z)}{\Delta_\mathrm{a}}\left(\hat a^\dagger-\hat a\right)\cos(kx),
\end{equation}
with $g_0$ the vacuum Rabi frequency, $\eta(x,z)$ the pump with a possible transverse spatial dependence ($z$ is taken as the transverse direction), and $\Delta_\mathrm{a}=\Omega-\omega_\mathrm{L}$ the atom-pump detuning. Since the effective potential is generated from the quantized cavity field, this type of optical lattice is sometimes referred to as {\it quantum optical lattice}~\cite{maschler2008ultracold,ritsch2013cold,caballero2015quantum,caballero2016quantum}\index{Optical! lattice}. For the case of longitudinal pumping, the cavity is directly excited via a one-end mirror, and the second term of (\ref{trpot}) is strictly zero, leaving us with
\begin{equation}\label{onpot}
V_\mathrm{oa}(x)=\frac{g_0^2}{\Delta_\mathrm{a}}\hat a^\dagger\hat a\cos^2(kx).
\end{equation}
The pump term (see sec.~\ref{sssec:drivejc})
\begin{equation}
i\eta\left(\hat a^\dagger-\hat a\right),
\end{equation}
in turn, must be added to the full Hamiltonian~(\ref{mbham}). The Hamiltonian~(\ref{mbham}), together with the explicit potentials (\ref{trpot}) and (\ref{onpot}), form the starting point for deriving an effective low-energy many-body Hamiltonian. The procedure goes on with expanding the field operators $\hat \psi(x)$ in the appropriate basis. However, before discussing the full many-body problem we pause to consider the mean-field approximation.

For an atomic condensate containing thousands of atoms, a series of phenomena experimentally explored to date can be qualitatively (and sometimes quantitatively) understood from a mean-field treatment~\cite{ritsch2013cold,pethick2002bose}. Introducing an optical lattice is a means to enhance the role of atom-atom interactions and thereby access a strongly-correlated regime in which the mean-field treatment breaks down. When the mean-field approximation is valid, the field operators are replaced by their expectations $\hat\psi(x)\rightarrow\psi(x)=\langle\hat\psi(x)\rangle$\index{Mean-field approximation}. The function $\psi(x)$ serves as the condensate order parameter, {\it i.e.} it typically assumes non-zero values when the atoms condense. The mean-field version of (\ref{mbham}) is a modified {\it Gross-Pitaevskii equation}~\cite{nagy2008self}\index{Gross-Pitaevskii equation}
\begin{equation}\label{gpeq}
i\hbar\frac{\partial}{\partial t}\psi(x,t)=\left[\frac{p^2}{2m}+V(x)+U_0N|\psi(x,t)|^2\right]\psi(x,t)
\end{equation}
accompanied by the (mean-field) equation for the photon field amplitude
\begin{equation}
i\frac{\partial}{\partial t}\alpha=\left(-\Delta+N\langle W_1\rangle-i\kappa\right)\alpha+\eta\langle W_2\rangle,
\end{equation}
where $N$ is the number of atoms such that we have normalized the order parameter $\int dx|\psi(x,t)|^2=1$, $\kappa$ is the photon decay rate and we have assumed a zero-temperature photon bath ({\it i.e.} the Langevin noise\index{Langevin noise} terms vanish, see eq.~(\ref{langeq}) and corresponding discussion), and
\begin{equation}
\langle W_1\rangle=\frac{g_0^2}{\Delta_\mathrm{a}}\int dx|\psi(x,t)|^2\cos^2(kx)
\end{equation}
gives the single-atom dispersive frequency shift of the cavity resonance, and finally
\begin{equation}
\langle W_2\rangle=\left\{
\begin{array}{lll}
\displaystyle{\frac{Ng_0}{\Delta_\mathrm{a}}\int dx\,|\psi(x,t)|^2\cos(kx),} &\quad & \mathrm{Transverse\,\,pump},\\ \\
1, & \quad& \mathrm{On-axis\,\,pump}.
\end{array}\right.
\end{equation}
The two quantities $\Theta\equiv\langle W_1\rangle$ and $\mathcal{B}\equiv\langle W_2\rangle$ have been referred to as {\it order parameter} and {\it bunching parameter}\index{Bunching parameter} respectively~\cite{nagy2008self}. When the quantity $|-\Delta-i\kappa|$ sets the fast time scale we may adiabatically eliminate the photon degrees of freedom~\cite{fernandez2010quantum} [see eqs.~(\ref{adel0}) and~(\ref{adel1})], {\it i.e.}, solve for the steady state\index{Steady state}
\begin{equation}\label{ass}
\alpha_\mathrm{ss}=\frac{i\eta\langle W_2\rangle}{\kappa+i(-\Delta+N\langle W_1\rangle)}
\end{equation}
and insert into the Gross-Pitaevskii equation~(\ref{gpeq}). Even in the absence of interatomic interactions ($U_0=0$), the resulting equation is nonlinear since $\langle W_1\rangle$, and possibly also $\langle W_2\rangle$, depend on $\psi(x,t)$. It is clear that the nonlinearity is not local~\cite{venkatesh2011band}, which is a manifestation of the cavity-mediated infinite-range interaction between the atoms~\cite{bhaseen2012dynamics,landig2016quantum,klinder2015observation}. When working in the full quantum picture, care has to be taken when eliminating the photon degrees of freedom since operators may not commute~\cite{larson2008mott,larson2008quantum,larson2008cold,fernandez2010quantum}. Another approximation deriving from eliminating the cavity degrees of freedom amounts to neglecting any quantum correlations between the matter and light subsystems. In particular, the evolution of the matter fields $\psi(x,t)$ will be fully coherent, while, in a more complete picture, its effective evolution would be non-unitary. This question has only been
raised in a preliminary fashion in~\cite{szirmai2009excess}, where the effect of the fluctuations induced upon the condensate due to the coupling to the quantized photon field was analyzed. Such noise tends to deplete the condensate, meaning that the condensate fraction decreases.  

\begin{figure}
\includegraphics[width=10cm]{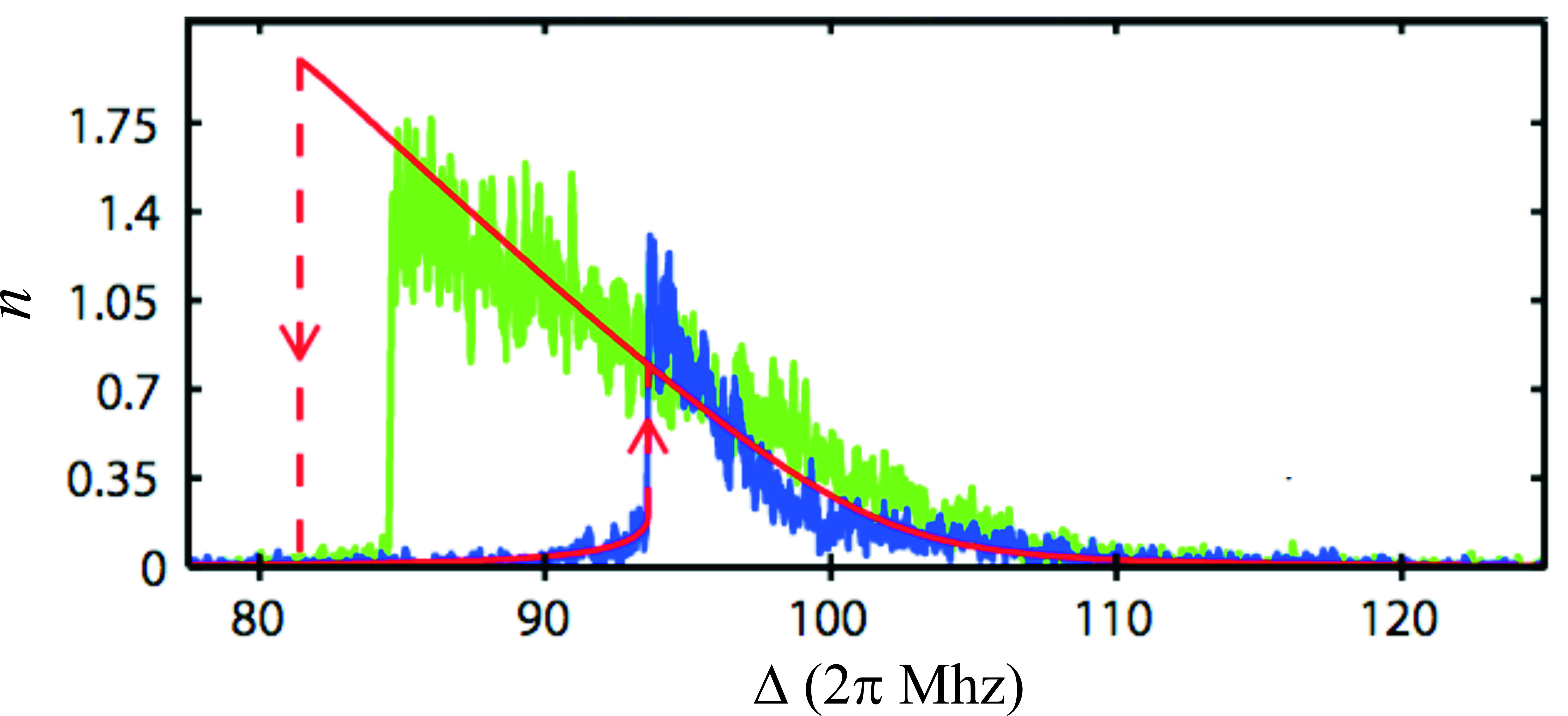} 
\caption{The measured intracavity field intensity $n$, see eq.~(\ref{ssphoton}), as a function of the cavity-pump detuning~\cite{ritter2009dynamical}. For slowly ramping up or down the detuning the intensity displays characteristic jumps due to the bistability\index{Optical! bistability}. The jump occurs for different detunings whether it is raised or lowered, see fig.~\ref{bistability}. In this experiment a condensate of $10^5$ $^{87}$Rb atoms were harmonically trapped in an optical Fabry-P\'erot cavity\index{Fabry-P\'erot! cavity/resonator} with $g_0=2\pi\times14.1$ MHz and $\kappa=2\pi\times1.3$ MHz. Reproduced with permission from Springer Nature.}
\label{bistfig}
\end{figure}

Since the order parameter\index{Order parameter} $\psi(x,t)$ depends on the photon amplitude $\alpha$, the nonlinearity naturally also shows up in the light intensity,
\begin{equation}\label{ssphoton}
n=\frac{\eta^2f_2^2(n)}{\kappa^2+(-\Delta+Nf_1(n))^2},
\end{equation}
where we introduced the functions $f_i(n)=\langle W_i(x)\rangle$ for $i=1,\,2$. The physical picture is that the atoms constitute an index of refraction which in turn shifts the effective mode frequency; the atomic state is then to be determined by the light field, which yields the backaction interaction\index{Backaction! interaction} between the two subfields. Going back to the expressions for the effective potentials, eqs.~(\ref{trpot}) and (\ref{onpot}), we notice that bistable behavior in this configuration already emerges when the photonic part of the Hamiltonian is quadratic. This is a new type of optical bistability compared to that discussed in sec.~\ref{sssec:optbis} which requires a Kerr-type nonlinearity\index{Kerr! nonlinearity}. The present type of bistability, encapsulated in the functions $f_i(n)$ on the mean-field level, results from the quantization of the atomic degrees of freedom. This has been verified from a set of experiments for on-axis pumping by varying the drive frequency $\omega_\mathrm{L}$ and detecting the output field intensity (which is proportional to $n$ of eq.~(\ref{ssphoton}))~\cite{{gupta2007cavity,brennecke2008cavity,ritter2009dynamical,purdy2010tunable}}, see fig.~\ref{bistfig} for one of the experimental results. We note that optical bistability was also demonstrated earlier for a thermal gas in a ring cavity~\cite{nagorny2003collective,elsasser2004optical}.

\begin{figure}
\includegraphics[width=8cm]{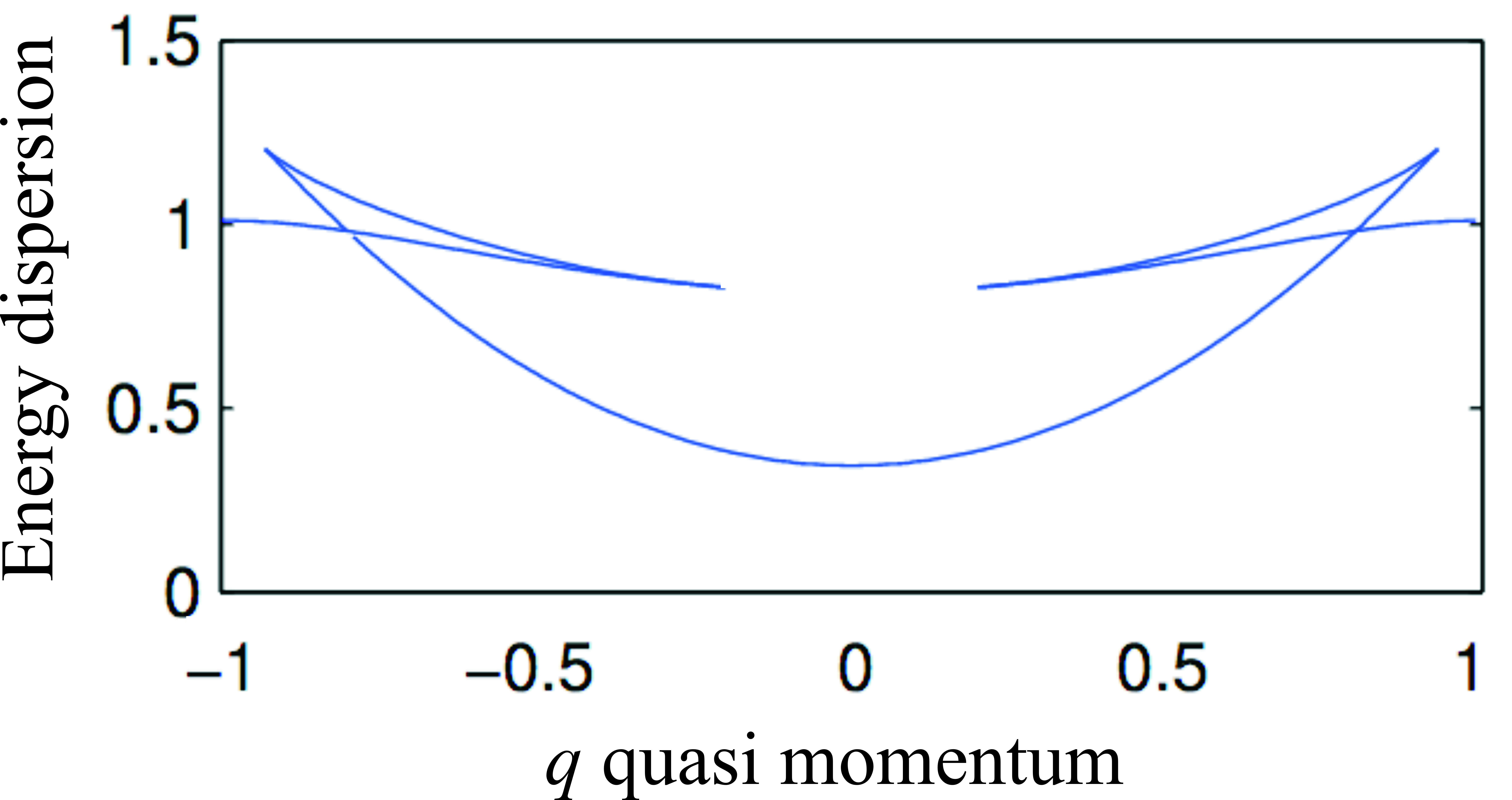} 
\caption{The lowest energy dispersion of the on-axis pumped Gross-Pitaevskii equation~(\ref{gpeq}) as a function of the quasi momentum. The `swallow-tail' loops formed close to the edges of the Brillouin zone\index{Brillouin zone} derive from the inherent nonlinearity of the system. Such loops imply among other things that Bloch oscillations\index{Bloch! oscillations} rapidly would decay due to non-adiabatic excitations~\cite{venkatesh2009atomic,venkatesh2013bloch}. The figure is taken from ref.~\cite{venkatesh2011band}, and the parameters for this example are $N=10^4$, $\Delta=2100\omega_\mathrm{r}$, $\kappa=350\omega_\mathrm{r}$, $\eta=980\omega_\mathrm{r}$, and $g_0^2/\Delta_\mathrm{a}=\omega_\mathrm{r}$.}
\label{Swallowfig}
\end{figure}

The multistability deriving from eq.~(\ref{mfbistability}) is limited to bistability -- at most three solutions for the mean-field photon number $n$ exist; two of them are stable and the other is unstable to fluctuations. To achieve genuine multistability,\index{Optical! multistability} higher-order terms in the Kerr Hamiltonian\index{Kerr! Hamiltonian} (\ref{kerrham}) need to be included. Such terms are, however, typically very small. The nonlinearity of (\ref{ssphoton}) can in principle be strong enough to induce multistability~\cite{venkatesh2011band}. The bistability shown in Figs.~\ref{bistability} and \ref{bistfig} appears in the intracavity photon number $n$, but it is clear that it might manifest in other quantities as well. Going back to the Gross-Pitaevskii equation (\ref{gpeq}) and letting $U_0=0$, we note that we are actually dealing with a periodic (nonlinear) Schr\"odinger equation, for which we may expect a band spectrum. For the regular Gross-Pitaevskii equation, in which the nonlinearity stems from atom-atom interaction ($U_0\neq0$), such energy dispersion has been studied in the past~\cite{diakonov2002loop,machholm2003band}. Hysteresis\index{Hysteresis} appears here in what has been called `swallow-tail loops'. The cavity-induced nonlinearity is in a sense more complex, and so are also the resulting swallow-tail loops~\cite{venkatesh2011band,zheng2012controllable}. One example can be seen in fig.~\ref{Swallowfig}, where, contrary to the ordinary case, the loops are formed around quasi-momenta different from 0 or $\pm 1$. In this example, two additional solutions of the nonlinear equations emerge, but for other parameters it is possible to identify cases with more solutions, {\it i.e.}, genuine multistability. 

Paradigmatic systems displaying hysteresis are the ones encountered in {\it cavity optomechanics}\index{Cavity! optomechanics}~\cite{aspelmeyer2014cavity} and discussed in sec.~\ref{ssec:hybridsys}. In these, light modes of the cavity are coherently coupled to mechanical oscillators, typically an end mirror~\cite{dorsel1983optical} of a Fabry-P\'erot cavity\index{Fabry-P\'erot! cavity/resonator} or some membrane~\cite{thompson2008strong} inserted into the resonator. The light fields induce effective forces felt by the oscillators, analogous to that discussed in sec.~\ref{ssec:lightforce} for atomic cooling. The ultimate aim is to cool down macroscopic objects to the quantum regime, {\it e.g.} prepare macroscopic entangled or superposition states or use these for metrology. For such optomechanical purposes, the atomic BEC is an attractive candidate. In this setting, the vibrational (phonon) modes $\hat{b}_k$ characterizing the BEC properties, are coherently coupled to the cavity mode $\hat a$~\cite{brennecke2008cavity}. Hamiltonians emulating the exemplary optomechanical systems can then be realized~\cite{nagy2009nonlinear}. In the many-body cavity QED setting, cooling the oscillator is much simpler than cooling a mirror or membrane, and several experiments have reached the deep quantum regime~\cite{brennecke2008cavity,kanamoto2010optomechanics,purdy2010tunable,brooks2012non}. The possibility to reach this regime relies of course on the fact that the atoms can be efficiently laser-cooled before being coupled to the cavity mode. As for other optomechanical systems, it has been recently investigated whether quantum entanglement could be sustained in two spatially separated condensates interacting with the same optical mode~\cite{joshi2015cavity}. Even when taking photon losses into account it was found that these condensates may exhibit steady-state entanglement.

For transverse pumping, see eq.~(\ref{trpot}), a phenomenon called {\it self-organization}\index{Self-!organization} occurs above a critical effective pump amplitude $\tilde\eta_c=g\eta/\Delta_\mathrm{a}$~\cite{domokos2002collective,asboth2005self,nagy2008self}. After the photon degree of freedom is eliminated, the effective potential felt by the atoms has the form of a {\it superlattice} $V(x)=U_1\cos(kx)+U_2\cos^2(kx)$\index{Superlattice}. Such a superimposed lattice potential comprises either a double-well structure whose minima are separated by alternating large or low potential barriers, or potential minima with alternating depths. The actual situation we are dealing with is determined by the signs of the amplitudes $U_1$ and $U_2$ (which depend nonlinearly on the condensate state $\psi(x)$). In particular, if $\Delta<-Ng_0^2/\Delta_\mathrm{a}$ the second case occurs~\cite{nagy2008self}. As a consequence, the atoms will tend to populate the deep potential sites more than the shallow ones. However, there is a trade-off between the involved terms; forming such modulations in the BEC incurs a cost in kinetic energy which, for weak potentials, doesn't prove favorable. By increasing the pump amplitude, the system will not be able to sustain a small potential, but photons will scatter into the cavity and build up a field of amplitude $\alpha_\mathrm{ss}$ given by eq.~(\ref{ass}) simultaneously with the repopulation of atoms dictated by the potential profile. This occurs for a critical pump amplitude: below threshold, $\tilde\eta<\tilde\eta_c$, the condensate is uniformly distributed, while above threshold it forms a periodic structure with period $\lambda=2\pi/k$. As this happens, the self-organization order parameter\index{Order parameter} $\Theta$ turns from zero to a finite value; the mean-field steady state\index{Steady state!mean-field} solutions undergo a pitchfork bifurcation identical to the one we encounter in the Dicke model, depicted in fig.~\ref{bifurcationDicke}. As we will see in the next section, the accompanying phase transition can be identified as the Dicke PT\index{Phase transition! Dicke}, and the spontaneously-broken\index{Spontaneous! symmetry breaking} $\mathbb{Z}_2$-parity symmetry\index{$\mathbb{Z}_2$ symmetry} reflects the fact that the condensate populates either even or odd sites in the lattice.

Already from the very first experiments on atomic condensation it was realized that these systems provide clean and controllable platforms in which macroscopic quantum phenomena can be studied. {\it Superfluidity} and {\it superconductivity} are among the most spectacular examples where quantum mechanics manifests at a scale visible to the bare eye. For many applications, it would be desirable to couple the neutrally-charged atomic BEC to some analog of an external magnetic field. In the early days of the field, this was achieved by rotating the condensates and identifying the emerging Coriolis force with an effective Lorentz force~\cite{pethick2008bose}. A drawback of such method of realizing {\it synthetic magnetic fields} is that not very strong fields can be sustained. Strong fields entail such a rapid rotation that the condensate does not stay trapped and flies apart. As a result, new methods with laser-induced {\it synthetic gauge fields} were developed~\cite{dalibard2011colloquium,larson2020conical}. These ideas can be extended to gauge fields induced by cavity fields~\cite{dong2014cavity,mivehvar2014synthetic,deng2014bose,padhi2014spin,dong2015photon}.

This identification has already been introduced in sec.~\ref{ssec:rabi} when we considered the quantum Rabi model. We recall the quantum Rabi Hamiltonian~(\ref{rabiham2}), expressed in the quadrature representation, has the general form 
\begin{equation}
\hat H=\frac{\hat p^2}{2}+V(\hat x),
\end{equation}
where the ``potential'' is $\hat x$-dependent and has a $2\times2$ matrix structure. Upon diagonalizing the potential-matrix $V(\hat x)$, a synthetic gauge potential\index{Gauge! potential! synthetic} $\hat A(\hat x)$ with the same matrix dimension as $V(\hat x)$ arises. The size of the matrix will depend on the number of internal atomic states, {\it e.g.} for the quantum Rabi model we obtain
\begin{equation}
\hat A_x(\hat x)=i\frac{\sqrt{2}g\Omega}{\Omega^2+8g^2\hat x^2}.
\end{equation}
In higher dimensions, the gauge potentials becomes a vector field: $\hat A(\hat{\bf x})=\left(\hat A_x(\hat x,\hat y,\hat z),\hat A_y(\hat x,\hat y,\hat z),\hat A_z(\hat x,\hat y,\hat z)\right)$ in three dimensions. Since the $\hat A_\alpha(\hat{\bf x})$'s are matrices, they do not necessarily commute. For non-commuting components, the vector field is termed {\it non-Abelian}, as already discussed in sec.~\ref{sssec:multi}, in connection to the bimodal quantum Rabi model~(\ref{jt2}). 

In the above example of the quantum Rabi model, the operator $\hat x$ represents a quadrature of the field, while for synthetic gauge-field experiments with cold atoms it represents the center-of-mass position of the atom or condensate much like the emergence of gauge fields in eq.~(\ref{jcgaugepot}) of sec.~\ref{sssec:qatmo}. Hence, there is a conceptual difference between the synthetic gauge fields discussed in secs.~\ref{ssec:rabi} and~\ref{ssec:exjc} for cavity QED models, and those for ultracold atoms in spatially dependent fields as in sec.~\ref{sssec:qatmo}. In the latter, the potential $V(\hat x)$ derives from dressing the condensate with classical laser fields with a spatial $\hat x$-dependence, {\it e.g.} with standing-wave or traveling-wave fields~\cite{dalibard2011colloquium,larson2020conical}. The emerging gauge fields will represent classical synthetic electromagnetic fields -- classical in the sense that they will not be quantized. To mimic gauge theories akin to those arising in high energy physics, it is, however, desirable to come up with schemes generating {\it dynamical gauge fields}, where $\hat A(\hat{\bf x})$ comprises dynamical degrees of freedom. This is the main motivation behind introducing ``cavity-induced gauge potentials''~\cite{dong2015photon}\index{Gauge! potential! cavity-induced}.  

Let us visualize the idea by considering an atomic $\Lambda$ configuration (see fig.~\ref{fig12}) coupled to two degenerate quantized cavity modes $a$ and $b$ in a Raman setup similar to ref.~\cite{mivehvar2014synthetic}. In fact, most works to date have in mind the $\Lambda$ setup, but with one classical and one quantized field rather than both fields quantized. We assume now that the atom is confined to move on a 2D plane, and that the cavity modes have some profiles $g_{1,2}(x,y)$ within this plane. Furthermore, we consider longitudinal pumping for the two modes, and we adiabatically eliminate the excited state of the atomic $\Lambda$ system\index{$\Lambda$ system} such that we are left with a two-photon transition\index{Two-photon!transition} between the two lower atomic states. We find the effective Hamiltonian
\begin{equation}\label{ham1}
\begin{array}{lll}
\hat{H} & = & \displaystyle{\int dxdy\,\hat{\psi}(x,y)\left[\hat{p}_x^2+\hat{p}_y^2+|\Omega_1|^2\hat{a}^\dagger\hat{a}\hat{\sigma}_{11}+|\Omega_2|^2\hat{b}^\dagger\hat{b}\hat{\sigma}_{11}+\left(\Omega_1\Omega_2^*\hat{a}^\dagger\hat{b}\hat{\sigma}_{12}+\Omega_1^*\Omega_2\hat{b}^\dagger\hat{a}\hat{\sigma}_{21}\right)\right]\hat{\psi}(x,y)}\\ \\
& & +\Delta_c\left(\hat{a}^\dagger\hat{a}+\hat{b}^\dagger\hat{b}\right)-i\eta\left(\hat{a}+\hat{b}-\hat{a}^\dagger-\hat{b}^\dagger\right).
\end{array}
\end{equation}
Here, we use the same notation as in eq.~(\ref{lambda}), {\it i.e.} $\hat{\sigma}_{ij}=|i\rangle\langle j|$, and the atomic field operator is written as $\hat{\psi}(x,y)=\hat{\psi}_1(x,y)|1\rangle+\hat{\psi}_2(x,y)|2\rangle$, while $\Omega_{1,2}\propto g_{1,2}(x,y)$. Let us now consider the mean-field case as in eq.~(\ref{gpeq}), and adiabatically eliminate the photon fields. Their steady-state solutions\index{Steady state!solution} read
\begin{equation}\label{gss}
\begin{array}{l}
\displaystyle{\alpha=\frac{\eta\left(\kappa-i\left(\Delta_c-N_{11}\right)\right)-i\eta N_{12}}{D}},\\ \\
\displaystyle{\beta=\frac{\eta\left(\kappa-i\left(\Delta_c-N_{22}\right)\right)-i\eta N_{21}}{D}},
\end{array}
\end{equation}
where
\begin{equation}
D=\left(\kappa-i\left(\Delta_c-N_{11}\right)\right)\left(\kappa-i\left(\Delta_c-N_{22}\right)\right)+|N_{12}|^2,
\end{equation}
and
\begin{equation}
N_{ij}=\int dxdy\,\Omega_i^*(x,y)\Omega_j(x,y)\psi_i^*(x,y)\psi_j(x,y)
\end{equation}

Within the mean-field approach, and neglecting atom-atom interaction, the equation of motion for the atomic order parameters becomes
\begin{equation}\label{mfGP}
i\frac{\partial}{\partial t}\psi(x,y)=\left\{-\frac{\partial^2}{\partial x^2}-\frac{\partial^2}{\partial y^2}+\left[\begin{array}{cc}
|\Omega_1(x,y)|^2|\alpha|^2 & \Omega_1(x,y)\Omega_2^*(x,y)\alpha^*\beta\\ \\
\Omega_1^*(x,y)\Omega_2(x,y)\beta^*\alpha & |\Omega_2(x,y)|^2|\beta|^2\end{array}\right]\right\}\psi(x,y),
\end{equation}
with the coherent field amplitudes $\alpha$ and $\beta$ given in eq.~(\ref{gss}). In the Born-Oppenheimer approximation (BOA) we move to an adiabatic basis which diagonalizes the matrix $\hat V(x,y)$ in~(\ref{mfGP}), {\it i.e.} $\hat D(x,y)=\hat U(x,y)\hat V(x,y)\hat U^{-1}(x,y)$, and the resulting gauge potential\index{Gauge! potential} is given by~(\ref{syngauge}). Alternatively, denoting the adiabatic states by $|\chi_{1,2}(x,y)\rangle$ (the unitary operator is $\hat U(x,y)=[|\chi_1(x,y)\rangle\,\,|\chi_{2}(x,y)\rangle]$), we can write the components of the vector potential\index{Vector potential} as~\cite{ruseckas2005non,dalibard2011colloquium,larson2020conical}
\begin{equation}
\left(A_\alpha\right)_{nm}=i\langle\chi_n|\partial_{\alpha}\chi_m\rangle,
\end{equation}
while the scalar potential is given by
\begin{equation}
\Phi_{nm}=\langle\nabla\chi_n|\nabla\chi_m\rangle+\sum_{k=1}^2\langle\chi_n|\nabla\chi_k\rangle\langle\chi_k|\nabla\chi_m\rangle.
\end{equation}
The corresponding magnetic field\index{Synthetic! magnetic field} is
\begin{equation}
\hat{B}_i=\frac{1}{2}\varepsilon_{ijk}\hat{F}_{kl},\hspace{2cm}\hat{F}_{kl}=\partial_k \hat{A}_l-\partial_l \hat{A}_k-i[\hat{A}_k,\hat{A}_l],
\end{equation}
where $\hat{A}_i$ is the $i$'th component of the vector potential\index{Vector potential}, and the last term of the field tensor $\hat{F}_{kl}$ stems from the non-Abelian property of the vector potential~\cite{wong1970field}. These expressions for the synthetic potentials are, of course, general and do not only apply to the studied model. For a real $\hat V(x,y)$, the adiabatic states are given by (\ref{adbas}). Our potential is in general complex, and in polar coordinates $\Omega_1\Omega_2^*\alpha^*\beta=\varrho e^{i\phi}$, we instead have
\begin{equation}
|\xi_1(x,y)\rangle=\left[
\begin{array}{c}
\cos\theta\, e^{-i\phi/2}\\
\sin\theta\, e^{i\phi/2}
\end{array}\right],\hspace{1cm}|\xi_2(x,y)\rangle=\left[
\begin{array}{c}
\sin\theta\, e^{-i\phi/2}\\
-\cos \theta\, e^{i\phi/2}
\end{array}\right],
\end{equation}
where the angles $\theta(x,y)$ and $\phi(x,y)$ are in general spatially dependent. The vector potential\index{Vector potential} takes the form
\begin{equation}
\hat{A}_\alpha=\frac{\sin\theta}{2}\partial_\alpha\phi\,\hat{\sigma}_x-\frac{1}{2}\partial_\alpha\theta\,\hat{\sigma}_y+\frac{\cos\theta}{2}\partial_\alpha\phi\,\hat{\sigma}_z.
\end{equation}

Let us look at one example of two traveling waves~\cite{gunter2009practical}, which could be implemented with two ring cavities overlapping in the region where the condensate forms. Here,
\begin{equation}\label{travpu}
\Omega_1(x,y)=\Omega_0e^{ix-\frac{(y-x_0)^2}{w^2}},\hspace{1cm}\Omega_2(x,y)=\Omega_0e^{i y-\frac{(x+x_0)^2}{w^2}},
\end{equation}
in dimensionless variables, where $x_0$ is the displacement and $w$ the mode waist. With this, the vector potential\index{Vector potential} becomes
\begin{equation}
\begin{array}{l}
\hat{A}_x=\frac{\sin\theta}{2}\hat{\sigma}_x-\frac{1}{2}\partial_x\theta\,\hat{\sigma}_y+\frac{\cos\theta}{2}\hat{\sigma}_z, \\ \\
\hat{A}_y=-\frac{\sin\theta}{2}\hat{\sigma}_x-\frac{1}{2}\partial_y\theta\,\hat{\sigma}_y-\frac{\cos\theta}{2}\hat{\sigma}_z,
\end{array}
\end{equation}
and the synthetic magnetic field, which points in the perpendicular $z$-direction, assumes the form
\begin{equation}\label{smag}
\hat{B}_z = \frac{3}{4}(\partial_x\theta+\partial_y\theta)\left(e^{i\theta}\hat{\sigma}_{+}+e^{-i\theta}\hat{\sigma}_{-}\right)
\end{equation}

\begin{figure}
\includegraphics[width=13cm]{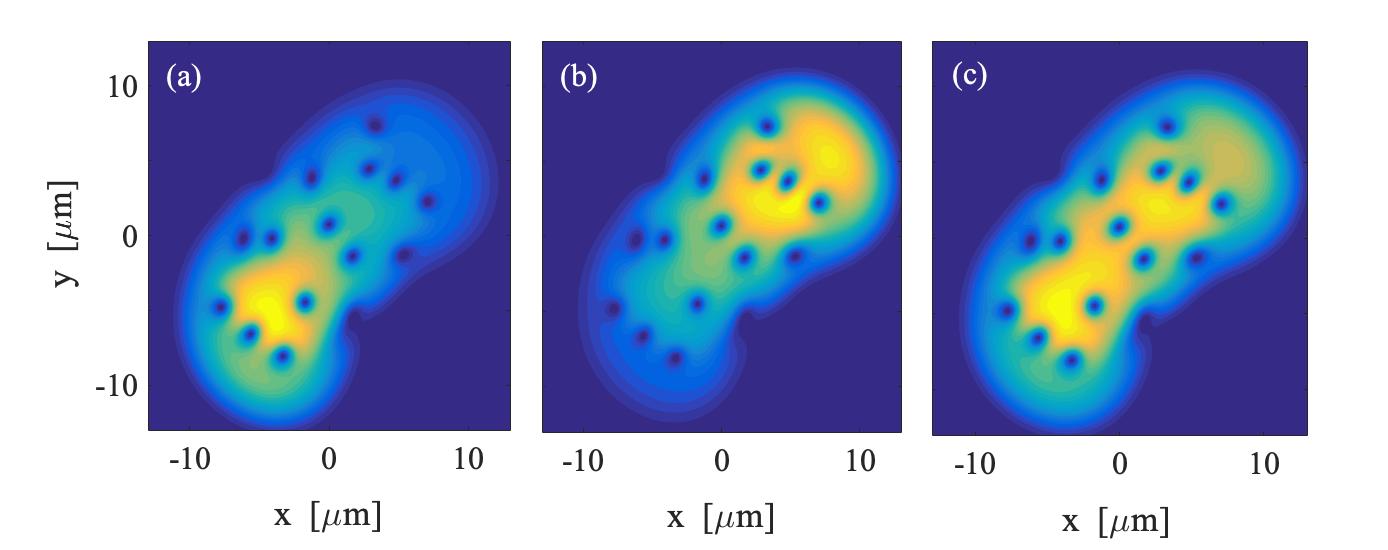} 
\caption{The ground-state atomic densities $|\psi_1(x,y)|^2$ {\bf (a)}, $|\psi_2(x,y)|^2$ {\bf (b)}, and $|\psi_1(x,y)|^2+|\psi_1(x,y)|^2$ {\bf (c)} for a realistic configuration with a condensate of $10^4$ $^{87}$Rb atoms in the Hamiltonian~(\ref{mfGP}) including a harmonic trap with $\omega_\mathrm{tr}=40$ Hz. With a characteristic trap length $a_0=\sqrt{\hbar/m\omega_\mathrm{tr}}$, for the modes we used $x_0=w=24a_0$, and the amplitudes $\Omega_0/\hbar=10^4$s$^{-1}$. Further, the pump amplitude $\eta=2.7\times10^9$ $s^{-1}$, the pump-cavity detuning $\Delta_c=7\times10^9$ s$^{-1}$, the photon decay rate $\kappa=6\times10^5$ $s^{-1}$, and the wave length of the two modes $\lambda=795$ nm. The Rb $s$-wave scattering length $a_s=5.77$ nm, and the condensate is assumed tightly confined in the perpendicular $z$-direction with a width $w_z=20$ $\mu m$. The synthetic magnetic field~(\ref{smag}) manifest as vortices in the condensate, despite the rather weak field amplitudes $|\alpha|^2\approx|\beta|^2\approx1$. We also not how the two components tend to repel each other even though the scattering lengths have been taken the same for every scattering process. }
\label{gdens}
\end{figure}

In fig.~\ref{gdens}, we give an example of how the synthetic magnetic field creates vortices in the ground state. We propagate the Hamiltonian~(\ref{mfGP}) in imaginary time such that the initial state `relaxes' towards the system ground state. We have propagated the state long enough to reach convergence, and we have also taken atom-atom scattering into account while including a harmonic trap for the atoms. The parameters used in the figure have been chosen to represent realistic experimental values (see caption). It is evident that the presence of the magnetic field creates vortices forming in the superfluid condensate, similar to what is found in a superconductor~\cite{goodstein2014states}. The same type of patterns have been found in atomic condensates dressed with spatially-dependent classical laser fields~\cite{lin2009synthetic}. Here, the emerging magnetic field~(\ref{smag}) is dynamical in the sense that it depends on the condensate order parameter, and the ground state has to be solved self-consistently. However, with a single cavity mode the gauge field is still acting globally, making it impossible to simulate something like the {\it Meissner effect}\index{Meissner effect} in which the magnetic field is expelled from the condensate/superconductor. To overcome this problem of locality, Ballentinne {\it et al.} extended the problem of a cavity-induced gauge field to a multi-mode cavity, and realized a close analogue of the Meissner effect~\cite{ballantine2017meissner}.

In their initial work on cavity-induced synthetic magnetic fields, the authors of~\cite{dong2014cavity} considered a generalization of the scheme used for implementing spin-orbit couplings\index{Spin-orbit coupling} in bosonic condensates~\cite{lin2011spin} [it should be remembered that spin-orbit couplings have an underlying gauge structure, as we pointed out in sec.~\ref{sssec:multi}, see eq.~(\ref{jt2})]. The 1D spin-orbit coupling in the cold-atom community has the structure of a quantum Rabi interaction, {\it i.e.} $\hat V_\mathrm{SOC}=\lambda\hat p_x\hat\sigma_y$, and, interestingly, in the pioneering work~\cite{lin2011spin} the corresponding normal-superradiant PT was experimentally demonstrated for a Rb$^{87}$ atomic condensate coupled to a laser field. The spin-orbit coupling becomes especially simple if the two transitions in the $\Lambda$ system are driven by counter-propagating plane waves~\cite{zhu2016effects}. After eliminating the excited atomic level in the dispersive regime, the Hamiltonian becomes
\begin{equation}
\hat H_\Lambda=\frac{\hat p^2}{2m}+\Omega\left(\hat a^\dagger\hat b\hat{\sigma}_{-}e^{2ikx}+\hat b^\dagger\hat a\hat{\sigma}_{+}e^{-2ikx}\right).
\end{equation}
Going to an interaction picture with $\hat U=e^{-ikx}\hat\sigma_{11}+e^{ikx}\hat\sigma_{22}$ removes the $x$-dependence,
\begin{equation}\label{gham}
\hat H_\Lambda'=\hat U\hat H_\Lambda\hat U^{-1}=\frac{\left(\hat p-k\hat\sigma_z\right)^2}{2m}+\Omega\left(\hat a^\dagger\hat b\hat{\sigma}_{-}+\hat b^\dagger\hat a\hat{\sigma}_{+}\right).
\end{equation}
Hence, we are left with a Hamiltonian which is diagonal in the momentum representation, and the effect of the spin-orbit coupling is to shift the momentum by one wave number $\pm k$ depending on the state inversion. In ref.~\cite{dong2014cavity} the Raman transition is driven by one classical laser field and one cavity field. It was shown that, on the mean-field level, as the cavity field is adiabatically eliminated, the dispersion builds up loops like those depicted in fig.~\ref{Swallowfig}. Thus, we are again dealing with the case where the intrinsic atom-light effective nonlinearity induces a hysteresis behavior. The zero-temperature mean-field phase diagram for a related setup was first discussed in ref.~\cite{deng2014bose}. It was found that in the superradiant phase, as the energy dispersion builds up a double-well structure [which is evident from the kinetic term of $\hat H_\Lambda'$ in eq.~(\ref{gham})], the condensate ground state is formed from counter-propagating phonon modes, and the density follows a striped pattern -- the {\it striped phase} (see also~\cite{mivehvar2015enhanced}). The striped phase is also found in the spin-orbit coupled BEC using classical lasers~\cite{li2012quantum,zhang2012mean}. However, due to the self-organization occurring in these systems (see next subsection), an additional {\it checkerboard phase} has been identified. In this phase, the two condensate components $\psi_{1,2}(x,y)$ arrange themselves in a lattice structure such that they avoid one another, while on top of this arrangement a vortex/anti-vortex lattice forms with alternating clockwise/anti-clockwise vortices. For the bimodal case, with two parallel traveling waves similar to those given in eq.~(\ref{travpu}), plus two classical drives of the $\Lambda$ system, the phase diagram was discussed in ref.~\cite{mivehvar2014synthetic}. What is different in these works is that they do not consider the mean-field approximation, but they instead look at the full Hamiltonian, like in sec.~\ref{sssec:qatmo} where we discussed the JC model with quantized atomic motion included. The adiabatic states are then the well- known dressed states introduced in~\ref{sssec:qatmo}, {\it i.e.} the JC dressed states~(\ref{dstate}) complemented with the atomic motional degrees of freedom. The bimodal JC-like Hamiltonian, in the bare basis, has a potential term $\hat V(x)$ in matrix form as we saw for example in eq.~(\ref{qm}). This large matrix structure, deriving from the photon degrees of freedom, results in a whole set of energy dispersion curves, which, however, split into two branches due to the internal two-level structure of the atom. The spin-orbit coupling, giving rise to the synthetic magnetic field, implies that the lower branch forms the well-known double-well shape within the mean-field treatment brought up in our discussion above. Since the photonic field is not eliminated in this approach, the emerging synthetic magnetic field will depend on the bosonic degrees of freedom rather than indirectly on the degrees of freedom for the atom. One reason which allows the full problem to be handled in~\cite{mivehvar2014synthetic} is that photon losses are not being considered. 

A `dynamical' spin-orbit coupling has been experimentally realized in the group of B. Lev~\cite{kroeze2019dynamical}. In particular, in~\cite{kroeze2019dynamical} the Raman transition within a $^{87}$Rb atomic condensate was considered with two classical drives and one standing-wave mode supported by a Fabry-P\'erot cavity\index{Fabry-P\'erot! cavity/resonator}. As in eq.~(\ref{gham}), this coupling causes a state-dependent shift of the atomic momentum, and the lower dispersion solution $\varepsilon_-(p)$ may form a double-well structure above a critical coupling, controlled by the classical pump amplitudes. Like for the self-organization Dicke PT discussed below, above the critical pumping the photon scattering from the classical pump fields into the cavity mode is enhanced thanks to constructive interference, and the condensate builds up a coherent spin texture. In such a superradiant phase, the condensate spontaneously arranges\index{Spontaneous! arrangement} itself around either momentum $\pm k$ following a $\mathbb{Z}_2$ symmetry breaking. Spin textures of a dipolar $^{87}$Rb condensate coupled to a cavity were also demonstrated in~\cite{landini2018formation}, for a slightly different coupling scheme. Dreon and collaborators realized a synthetic crystal by a $^{87}$Rb BEC self-organizing in an optical cavity~\cite{Dreon2022}. The atoms were illuminated by a pair of counter-propagating laser beams, with different Rabi frequencies, intersecting the cavity mode at an angle $\pi/3$. Photons scatter in the cavity from this pair of imbalanced transverse beams and leak out of the cavity, and the field is being recorded by a heterodyne detection setup\index{Heterodyne detection}. A finite photon occupation in the real (Q) or imaginary (P) quadrature of the cavity field breaks a $\mathbb{Z}_2$ symmetry. In the absence of dissipation, increasing the coupling strength results in a structural phase transition between the two symmetry-broken states. Dissipation comes in to mix the optical quadratures, and a third regime appears between these states, in which the cavity field undergoes persistent oscillations.  

Raman transitions have been suggested to control the tunneling rates of atoms in optical lattices, and in~\cite{zheng2016superradiance} this idea was generalized to the construction of a Raman-transition scheme employing one classical field and one cavity field. Thus, an atom at site $i$ can transfer to site $i\pm1$ via an absorption/emission of one cavity and one laser photon. By considering a lossy cavity it is possible to get a directed motion in the lattice; for motion towards the right a laser photon is absorbed and emitted into the cavity, while for moving to the left a cavity photon is absorbed from the cavity and emitted into the laser beam. However, if the cavity is kept in the vacuum state, one finds a net flow to the right. The steady state of this setup, for a finite system, accumulates particles at the right edge of the lattice, similar to what in modern days has been termed the {\it non-Hermitian skin effect}\index{Non-Hermitian! skin effect}. The same idea of using the irreversibility of photon losses to bring about directed motion was also suggested in ref.~\cite{larson2011diode} in order to realize a `diode' for an atomic condensate.

The Peierls substitution\index{Peierls! substitution} is a standard approach when describing the dynamics of a charged particle in a periodic potential and in the presence of a magnetic field. Given a rather smooth magnetic field, when we start from a minimal coupling Hamiltonian, and introduce the second quantization to the model in site-localized states, {\it i.e.} bringing the Wannier functions~(\ref{wanf}) in, the tunneling amplitudes will acquire complex phases, $t_{ij}\rightarrow t_{ij}e^{i\phi_{ij}}$~\cite{larson2020conical}. Summing these phases around a loop in the lattice gives the magnetic flux $\varphi$ through the loop. The flux $\varphi$ is a gauge invariant quantity, while the tunneling phases $\phi_{ij}$ are not. Usually one talks about the magnetic flux penetrating a single plaquette in the lattice. Hence, for example, when dealing with cold atoms in optical lattices or photons in photonic lattices, the aim is to control the phases of the tunneling amplitudes~\cite{dalibard2011colloquium}. For cold atoms, this is typically achieved via laser-assisted Raman transitions between neighboring sites, or {\it lattice shaking}. An exemplary model is that of a single particle in a square tight-binding lattice with a constant flux $\varphi$ through each plaquette, the so-called {\it Harper model}\index{Harper model}~\cite{harper1955single,hofstadter1976energy}. The flux typically breaks the periodicity of the lattice, except for integer values $\varphi=p/q$, for which the unit cell becomes enlarged and includes many lattice sites. As a result, the spectrum, called the {\it Hofstadter butterfly}\index{Hofstadter butterfly}, consists of a single energy band for $\varphi=2\pi n$, and forms several fractal sub-bands whenever $\varphi=p/q$, while if $\varphi$ is irrational it becomes point-like. In refs.~\cite{sheikhan2016cavity,colella2019hofstadter}, such a Harper/Hofstadter model was analyzed in the situation where the gauge field was induced partly by a dynamical cavity field. In particular, as the atoms enter into the superradiant phase (see section below), a dynamical version of the Harper model is realized within the tight-binding approximation. Furthermore, the normal-superradiant PT may turn into a first-order PT in this model and, in the classical limit of very large pumping, the dynamical model can be approximated with the static Harper model as expected.

In a 1D lattice, any phase factors $\exp(i\phi_{ij})$ of the tunneling amplitudes can be gauged away, unless one considers periodic boundary conditions~\cite{nunnenkamp2011synthetic}. This follows since there are no loops that can be formed in 1D. The {\it ladder lattices}\index{Ladder! lattice} consist of connected 1D chains, with hopping along the rungs. Here a synthetic magnetic field is non-trivial since it is possible to form loops. The simplest ladder is that with two legs, and for a constant magnetic field through the plaquettes it can be seen as a quasi 1D Harper model. Kollath {\it et al.} considered spin-less fermions in two-leg ladder system in the presence of a cavity-induced synthetic magnetic field~\cite{kollath2016ultracold,sheikhan2016cavity}. The magnetic field breaks time-reversal symmetry and it is possible to construct {\it chiral liquids} in which the atoms move upwards along one leg and downwards along the other leg, as observed experimentally for a cold atomic bose gas in an optical lattice~\cite{atala2014observation}. Using the cavity for such realizations opens up two nice possibilities: the chiral currents can be easily switched and the output cavity field can work as a probe for the atomic state. We will talk about such cavity-photon probes in more detail below. The above ladder system was also numerically analyzed for interacting bosons, {\it i.e.} the Bose-Hubbard ladder~\cite{halati2017cavity}. Once again, chiral liquid states were found, together with vortex phases.

The construction of low-energy tight-binding Hamiltonians for condensed-matter systems with a strong coupling to the quantized electromagnetic field are customarily obtained by projecting the continuum theory onto a given set of Wannier orbitals. However, different representations of the continuum theory lead to different low-energy formulations since they may transform orbitals into light-matter hybrid states before the projection. The Coulomb-gauge\index{Coulomb gauge} Hamiltonian is contrasted to the dipolar-gauge\index{Dipole! gauge} Hamiltonian in~\cite{Jiajun2020} when studying the coupling of a 1D solid to a quantized cavity mode, which renormalizes the band-structure into electron-polariton bands. A model of a 1D chain of interacting spinless fermions coupled to the first transmittance resonance of a cavity in the dipole approximation, fixing the wave vector perpendicular to the direction of the chain, has been very recently considered in~\cite{Passetti2022}. The authors find that quantum fluctuations of the current operator (material system) play a pivotal role in achieving light-matter entanglement; in turn, such an entanglement is the key to understand the modification of the electronic properties by the cavity field. 

\subsubsection{Critical phenomena I -- bosons}
Cavity systems are inevitably open due to photon losses. As a result, in order to study the physical behavior governed by Hermitian Hamiltonians, the experiments have to be performed on times much shorter than the inverse photon decay rate. Such finite time scales limit what can be explored in the lab. However, we have mentioned repeatedly in this monograph that critical phenomena may also appear for open quantum systems in which losses and driving balance. Such nonequilibrium driven-dissipative PTs\index{Driven! dissipative phase transition}\index{Nonequilibrium! phase transition} have gained enormous attention within the last decade~\cite{sieberer2013dynamical,sieberer2016keldysh}. For the high-$Q$ optical cavities we have considered here, photon losses can be accurately captured with the Lindblad equation~(\ref{lindblad}). Since there is no ground state for the open system, one is interested in the steady state\index{Steady state}, $\hat{\mathcal{L}}\hat\rho_\mathrm{ss}=0$\index{Steady state}. Criticality is then manifested as non-analyticity in $\hat\rho_\mathrm{ss}=0$ for some critical parameter $\lambda_c$ when the thermodynamic limit\index{Thermodynamic limit} is taken. Criticality in these systems arises from different mechanisms~\cite{hedvall2017dynamics}, delineated below:

\begin{itemize}
\item[(i)] The system Hamiltonian $\hat H$ is already critical, and fluctuations induced by the dissipation are not sufficient to disrupt the PT and the order in the symmetry-broken phase. In this case, the phases may carry on, but they can possibly also change, and the universality can be different. The open Dicke and critical quantum Rabi model exemplify this case. The two models may remain critical even when photon losses are included in the dynamics, and we find a normal and a superradiant phase~\cite{larson2017some}. The universality is altered by the open nature of the system, since the critical exponents are not the same for the open and closed versions of the model~\cite{nagy2010dicke,nagy2011critical,oztop2012excitations,nagy2015nonequilibrium,hwang2018dissipative}. The TC and critical JC models, on the other hand, are not critical as photon losses are taken into account; the normal phase extends for all parameters.  

\item[(ii)] The interplay between unitary and dissipative evolution\index{Dissipative! evolution} underlies the displayed criticality. For a single Lindblad jump operator\index{Lindblad! jump operator} $\hat L$ we then have $\left[\hat L,\hat H\right]\neq0$, and the Hamiltonian ground state $|\psi_0\rangle$ cannot be dark, {\it i.e.} $\hat L|\psi_0\rangle\neq0$. The two terms favour different steady states\index{Steady state}. An example related to the present section on many-body cavity QED is given by $\hat H=\omega\hat S_x$ and $\hat L=\hat S^-$ with the collective spin-$S$ operators describing a set of $N=2S$ two-level atoms ~\cite{drummond1978volterra,puri1979exact,schneider2002entanglement}. Separately, both the Hamiltonian and the dissipation are trivial; the Hamiltonian ground state simply describes all atomic dipoles pointing in the $\sigma_x$-direction -- the {\it paramagnetic phase}, while the dissipator pushes the atoms into their lower state $|-\rangle$ -- the {\it magnetic phase}. There is a continuous PT between these two phases in the thermodynamic limit\index{Thermodynamic limit} $S\rightarrow\infty$, but interestingly it is not related to any sort of symmetry breaking~\cite{hannukainen2018dissipation}, which demonstrates how these types of PTs can be conceptually different from QPTs. Note though that for open systems {\it Noether's theorem}\index{Noether's theorem} does not apply, whence symmetries and conserved quantities cease to be connected in a profound way~\cite{albert2014symmetries}.

\item[(iii)] Finally we can imagine a situation in which criticality emerges solely from the action of the dissipators. In the simplest case $\hat H=0$, and we have two jump operators $\hat L_1$ and $\hat L_2$, such that $\left[\hat L_1,\hat L_2\right]\neq0$. In fact, given an initial multi-qubit state $|\psi_0\rangle$, it can be shown that any target state $|\psi_f\rangle$ can be reached utilizing only dissipative evolution\index{Dissipative! evolution} generated by some set of Lindblad jump operators $\hat L_k$~\cite{verstraete2009quantum}. 
\end{itemize}

It is often argued that the onset of laser action was the first example in which nonequilibrium driven-dissipative PTs\index{Nonequilibrium! phase transition} were discussed for a quantum system involving light-matter interaction~\cite{graham1970laserlight,haken1975cooperative}. In a certain class of these systems, a classical EM field drives an active medium confined inside a resonator, and, above some critical pump amplitude $\eta_c$, intense scattering of photons into the cavity sets off and a large-amplitude coherent cavity field is established. The transition for the onset of lasing can be envisaged as a continuous PT accompanied with a spontaneous symmetry breaking\index{Spontaneous! symmetry breaking} of the coherent phase. In practice, the prescribed definite phase of the electric field above threshold diffuses due to fluctuations, giving the laser a nonzero linewidth\index{Linewidth} -- the well-known Schawlow-Townes linewidth~\index{Schawlow-Townes! linewidth}\index{Linewidth! Schawlow-Townes}. Photon losses limit the steady-state field amplitude; the turning on of the laser is caused by an explosion of stimulated emission\index{Stimulated! emission}, an explosion which is held in check below threshold by the cavity loss~\cite{Rice1994laser}. The lasing PT shows many similarities with the Dicke PT, as we will discuss in some more detail below, which complements the more general discussion on the Dicke PT presented in sec.~\ref{sssec:dicke}. 

Coherence for an ideal laser beam can be construed as the number of photons emitted consecutively into the beam with the same phase. The bound on this number has been recently reappraised and established as being on the order of the fourth power of the number of photons in the laser cavity~\cite{BakerWiseman2021}. The Heisenberg limit\index{Heisenberg!limit}, which is the ultimate quantum limit, fares quadratically better than the standard Schawlow-Townes limit\index{Schawlow-Townes!limit}. Besides, using atoms with an ultranarrow clock transition (on the scale of mHz), Liu and collaborators have predicted that a laser linewidth\index{Linewidth! laser} (the reciprocal of the coherence time) below $100\mu$Hz can in principle be attained~\cite{LiuWiseman2021}. 

As already discussed in subsecs.~\ref{sssec:dicke} and~\ref{sssec:dia}, reaching the normal-superradiant critical point\index{Critical! point} in a closed quantum system is at the present stage experimentally unrealistic, and the light-matter coupling must be instead indirectly controlled via an external pump. In the Raman setting for the $\Lambda$ atoms, with one transition driven by a classical laser field and one transition coupled via the cavity field, according to eqs.~(\ref{lambdaeff}) and~(\ref{ramnd}), the effective light-matter coupling, after eliminating the excited atomic state, becomes $\tilde g=\eta g/\Delta$ with $\eta$ proportional to laser pump amplitude and $\Delta$ the detuning of the excited atomic level from the pump and cavity frequencies. Since the diamagnetic self-energy does not depend on $\eta$, the light-matter coupling can become sufficiently large while keeping the diamagnetic term\index{Diamagnetic term} negligible~\cite{dimer2007proposed}. The emergence of superradiance in such a setup was experimentally demonstrated in a thermal gas of a few hundreds of thousands of $^{87}$Rb atoms Raman-coupled to a high finesse cavity mode~\cite{baden2014realization}. The two internal atomic levels were taken as two hyperfine levels [$(F,m)=(2,2)$ and $(F,m)=(1,1)$] of the rubidium atoms. Depending on the specific Raman coupling considered it is possible to engineer the anisotropic Dicke model~(\ref{anRabi})\index{Anisotropic! Dicke model} (for an analysis of the dynamics of the anisotropic Dicke model see refs.~\cite{keeling2010collective,bhaseen2012dynamics}), and both limiting cases of the symmetric Dicke and the TC models were considered in the experiment. A similar coupling scheme was also employed in ref.~\cite{ferri2021emerging}. Here, Zeeman levels of a $^{87}$Rb condensate were Raman-coupled with two external fields and the cavity mode, and by tuning the drive amplitudes it was possible to explore the influence of photon losses on the phase diagram. At strong anisotropy, the superradiant phase became unstable and a {\it dissipation-stabilized normal phase} and multistability was found, first predicted in~\cite{soriente2018dissipation,stitely2020nonlinear}. We recall from sec.~\ref{sssec:dicke} that the Dicke PT survives moderate photon losses, while the TC PT does not, which is a result of the lower symmetry of the Dicke model, a discrete $\mathbb{Z}_2$, compared to the TC model which has a continuous $U(1)$ symmetry. When the anisotropy of the model is increased, the gap in the Higgs mode\index{Higgs mode} gradually decreases, giving rise to a gapless Goldstone mode\index{Goldstone mode}, while the dissipation-induced fluctuations destabilize the ordered superradiant phase. Multistability was experimentally explored by tuning the system through the critical point\index{Critical! point}, and hysteresis was observed~\cite{ferri2021emerging}. In addition to the ground-state quantum phase transition from the normal to the superradiant phase, the anisotropic Dicke\index{Anisotropic! Dicke model} model also exhibits other transitions, namely the excited state quantum phase transition, the ergodic to nonergodic transition, and the temperature-dependent phase transition [see also sec.~\ref{sssec:dicke}]. These transitions have been recently assessed with recourse to eigenvector quantities, such as the von Neumann entanglement entropy\index{Entanglement! von Neumann entropy}\index{von Neumann! entropy} and the participation ratio~\cite{DasP2023}. 

Let us spend some time to understand how the Dicke model can emerge through the coupling of photon and phonon modes in systems where the atomic BEC interacts with a single mode supported by a high-finesse cavity. With a transverse standing-wave pump, {\it i.e.} when an optical lattice arises from the action of an external laser field, according to eqs.~(\ref{mbham}) and (\ref{trpot}) the (atomic) single-particle Hamiltonian reads
\begin{equation}\label{1stbec}
\hat H=\frac{p_x^2}{2m}+\frac{\hat p_z^2}{2m}+\frac{g_0^2}{\Delta_\mathrm{a}}\hat a^\dagger\hat a\cos^2(kx)+\frac{g_0\eta_0}{\Delta_\mathrm{a}}(\hat a^\dagger+\hat a)\cos(kx)\cos(kz),
\end{equation}
where $\eta_0$ is the classical pump amplitude, the pump frequency is taken the same as that of the cavity mode, and we have replaced the bosonic field operators by their steady states~(\ref{ass})\index{Steady state}. The steady-state value of the cavity-field amplitude $\alpha$ determines the shape of the effective potential felt by the atom. For a non-zero $\alpha$ the lattice is a square superlattice meaning that shallow and deep sites alternate, whence a unit cell contains two lattice sites. The phase of $\alpha$ determines which sites will be the deep or shallow, and, as we will see, this appearance of these two possibilities is a manifestation of the $\mathbb{Z}_2$ parity symmetry breaking in the Dicke model.

As explained above, in turning from Hamiltonian~(\ref{1stbec}) in the first quantization to a many-body Hamiltonian, via the process of second quantization, we need to identify the appropriate low-energy expansion of the atomic field operators~(\ref{fieldop}). At this stage we are not interested in the strong-coupling regime, but in superfluid phases whose properties are assumed to be mainly governed by the single-particle Hamiltonian. For $\alpha=0$, the ground state is a flat distribution $\psi_0(x,y)=N/L^2$ with $N$ the number of atoms and $L^2$ the size of the square lattice. This is clearly a zero-momentum state minimizing the kinetic energy (it also minimizes the repulsive-interaction energy as it is the most extended state). The emerging lattice will couple the $|k_x,k_y\rangle=|0,0\rangle$ state to momentum modes $|k_x,k_y\rangle=|n_x,n_y\rangle$ for integers $n_{x,y}$. It is apparent that the kinetic energy grows with $|n_{x,y}|$, and, as we get close to the onset of the superradiant transition when $\alpha$ becomes non-zero, we may restrict our analysis to the momentum eigenstates $|0,0\rangle$ and $|\pm1,\pm1\rangle$. The momentum modes with $k_{x,y}\neq0$ represent the first-excited condensate phonon modes. Populating these modes causes a modulation of the condensate profile, which will no longer be completely uniform. One can see this as if the condensate adjusts to the potential and tries to avoid the potential heights. When expanding the field operators~(\ref{fieldop}) we include the zero-momentum mode and the symmetric combination of the other four modes (which should have the lowest energy)~\cite{nagy2010dicke,baumann2010dicke}
\begin{equation}\label{dickeexpan}
\hat\psi(x,z)=\hat c_0+\hat c_1\cos(kx)\cos(kz).
\end{equation}
This is of course an approximation; as we go deeper into the superradiant phase the photon number increases and higher modes will get populated. However, this low-energy model is adequate for our purposes, since it captures criticality.   

Having identified the relevant modes we now proceed to derive the many-body Hamiltonian by plugging~(\ref{dickeexpan}) into eq.~(\ref{mbham}), resulting in
\begin{equation}
\hat H_\mathrm{cD}=\omega_0\hat a^\dagger\hat a+\frac{\Omega_0}{2}\hat S_z+\frac{g}{\sqrt{N}}\left(\hat a^\dagger+\hat a\right)\hat S_x+\frac{U_1}{4}\hat a^\dagger\hat a\hat c_1^\dagger\hat c_1,
\end{equation}
where we have used the Schwinger-boson mapping\index{Schwinger-boson mapping}~\cite{sakurai1995modern,assa1994interacting}
\begin{equation}
\hat S^+=\left(\hat S^-\right)^\dagger=\hat c_1^\dagger\hat c_0,\hspace{1cm}\hat S_z=\frac{1}{2}\left(\hat c_1^\dagger\hat c_1-\hat c_0^\dagger\hat c_0\right).
\end{equation}
The parameters involved in this expression are: the shifted photon resonance $\omega_0=-\Delta_c-NU_0/2$, the effective atomic transition frequency $\Omega_0=4E_\mathrm{r}$ equal to four recoil energies, $U_1=g_0^2/\Delta_\mathrm{a}$, and the coupling $g=g_0\eta_0\sqrt{N}/\Delta_\mathrm{a}$. Apart from the last term, the Hamiltonian $\hat H_\mathrm{cD}$ identifies the Dicke Hamiltonian~(\ref{dickeham}). This last term scales as $\sim g_0^2$ to be compared against the other interaction term scaling as $\sim g_0\eta_0$; since $|\eta_0|\gg|g_0|$ we may neglect it~\cite{bhaseen2012dynamics}. Note that this term does not break the $\mathbb{Z}_2$ parity symmetry of the Dicke model. Atom-atom interactions could be effectively described by a term proportional to $\hat S_z^2$, which again would not alter the parity symmetry~\cite{dalafi2014effect}, yet it shifts the value of the critical coupling. 

It is important to appreciate that the effective model described by the Hamiltonian~(\ref{dickeexpan}) emerges in a driven system -- the atomic condensate is pumped transversely by a classical laser field. This is indirectly reflected in the Raman-induced coupling strength $g$ that is proportional to the pump amplitude $\eta_0$. Further, the Hamiltonian is given in an interaction picture rotating with the laser frequency. The ground state of $\hat H_\mathrm{cD}$ should therefore be regarded as a nonequilibrium state\index{Nonequilibrium! state}. As mentioned several times already (see sec.~\ref{sssec:dicke}), owing to this configuration, neither the Dicke no-go theorem\index{Dicke! no-go theorem}~\cite{bialynicki1979no} nor the Thomas-Reiche-Kuhn sum rule\index{Sum rule!Thomas-Reiche-Kuhn}~\cite{rzazewski1975phase} apply, and it is therefore possible to ramp up the light-matter coupling $g$ beyond the critical value~(\ref{critcoupnew}). Physically, the mechanism behind the normal-superradiant phase transition\index{Normal-superradiant phase transition}\index{Phase transition! normal-superradiant} can be explained by the scattering of laser photons into the cavity. In the normal phase, $\langle\hat c_0^\dagger\hat c_0\rangle=\mathcal{O}(N)$ and $\langle\hat c_1^\dagger\hat c_1\rangle=\mathcal{O}(1)$, such that the condensate density is uniform, and destructive interference prohibits photons to be scattered into the cavity -- as a result, $\langle\hat a^\dagger\hat a\rangle\approx0$. On the other hand, when higher-phonon modes of the condensate start to get populated, the scattering becomes coherent, leading to constructive interference and the photon field rapidly builds up. 

The phase diagram, including the normal and the superradiant phases, was experimentally mapped out in the $\eta_0-\Delta_\mathrm{c}$ plane by measuring the output cavity field in ref.~\cite{baumann2010dicke}. Note that in order for the expansion of the atomic field operator in terms of the collective phonon modes to make sense, we need to assume a condensate and not a thermal gas of atoms. The coherent coupling of an atomic condensate to the quantized EM cavity field had been demonstrated~\cite{brennecke2007cavity}, but it was the experimental realization of the nonequilibrium Dicke PT\index{Nonequilibrium! phase transition! Dicke} that really caused a resurgence of interest in the community. Under such control, several known models could be emulated.  The same group pushed the explorations of the Dicke PT further in a couple of follow-up works. In ref.~\cite{baumann2011exploring}, the focus was on breaking the $\mathbb{Z}_2$-parity symmetry in the superradiant phase. Recall that the parity symmetry manifests as an invariance under the simultaneous change $\hat a\leftrightarrow-\hat a$ and $\hat S_x\leftrightarrow-\hat S_x$, such that in the symmetry-broken superradiant phase, the phase of the field $\mathrm{angle}(\alpha)=0,\,\pi$. Hence, upon entering this phase, the order parameter (in the rotating frame) will be real and either positive or negative, determined in a random fashion. This was experimentally shown in~\cite{baumann2011exploring}, as well as in studies on the conservation of coherence in the system. The dynamical quench response across the critical point was explored in the Hamburg setup~\cite{klinder2015dynamical}. The difference with this experiment compared to the above ETH one is the much lower decay rate $\kappa$. The most interesting result reported therein was the observation of hysteresis across the critical point\index{Critical! point}; by ramping the parameters from the normal up to the superradiant phase the transition occurred in a sudden jump, and upon quenching back into the normal state no such jump was found. Furthermore, the amount of excitations followed a power-law behaviour with respect to the ramp velocity, as predicted by the {\it Kibble-Zurek mechanism}\index{Kibble-Zurek mechanism}~\cite{zurek2005dynamics}. Finally, numerical simulations showed that the hysteresis area decreased as $\kappa$ was increased. The aspect studied in~\cite{brennecke2013real} was how the reservoir-induced fluctuations affect criticality. In particular, the dynamical critical exponent defined in. eq.~(\ref{dcrit}) is either $\beta=1/2$ or $\beta=1$, depending on whether the system is closed or open respectively~\cite{nagy2010dicke,nagy2011critical,oztop2012excitations,nagy2015nonequilibrium,hwang2018dissipative}. In the experiment they found indeed that the openness due to photon losses cannot be disregarded; the extracted exponent $\beta=0.9\pm0.1$ is more in line with the predicted exponent for an open system.

At the critical point\index{Critical! point}, when the atoms start to coherently scatter photons into the cavity, the $\hat c_1$ mode gets macroscopically populated and the condensate density changes accordingly. As pointed out above, the effective potential formed when $\alpha\neq0$ will be a square lattice with alternating deep and shallow sites, and the polaritonic condensate (the atomic condensate gets dressed by the cavity photons) will predominantly populate the deep sites in a `checkerboard' manner. This is reminiscent of a {\it supersolid phase}\index{Supersolid phase}, in which a superfluid builds up a spatial structure (in this case the structure derives from the cavity induced dynamical lattice)~\cite{leonard2017supersolid,leonard2017monitoring}. Actually, before phrasing things in terms of a Dicke PT, we note that the transition in question had been studied in the realm of what is called {\it self-organization}\index{Self-!organization}~\cite{domokos2002collective,asboth2005self,nagy2008self}. The self-ordering, or equivalently the Dicke transition, has also been studied for a spinor Bose-Einstein condensate~\index{Spinor Bose-Einstein condensate}~\cite{kroeze2018spinor}, {\it i.e.} several Zeeman levels of the condensate atoms play a role. In particular, two classical transverse drive fields were considered and two detuned Raman transitions (one for each drive field together with the cavity field) coupling two hyperfine levels, say $|\uparrow\rangle$ and $|\downarrow\rangle$, of $^{87}$Rb atoms. The detunings were chosen such that the $|\uparrow\rangle$ and $|\downarrow\rangle$ atoms felt opposite effective potentials in the superradiant phase; the alternating deep/shallow sites were shifted between the two hyperfine levels. As a result, the condensate populates different sites depending on the internal state. As we see in eqs.~(\ref{lmg}) and~(\ref{effis}), the photon field induces an effective spin-spin coupling among the atoms. This has been experimentally explored in a thermal gas of $^{87}$Rb atoms, in which three internal hyperfine levels, $m=-1,0,+1$, were coupled via Raman transitions~\cite{davis2019photon}. Initializing the atoms in the $m=0$ states a coherent population of the $m=\pm1$ states was observed, which can be understood from the spin-exchange interactions. Collective tunneling has been recently implemented via cavity-assisted Raman scattering of photons by a spinor BEC into an optical cavity~\cite{RosaMedina2022}. The field responsible for the tunneling processes suffers from dissipation, giving rise to an effective directional dynamics in a non-Hermitian formulation.  

For a set of two-level atoms identically coupled to the light mode, the adiabatic elimination of the cavity mode results in the effective spin Hamiltonian of Lipkin-Meshkov-Glick\index{Lipkin-Meshkov-Glick! Hamiltonian}~(\ref{lmg})~\cite{morrison2008dynamical,larson2010circuit}. In ref.~\cite{muniz2020exploring} one version of the LMG model was realized in a gas of roughly a million $^{88}$Sr thermal atoms. The focus of this work was on the exploration of the dynamical phase diagram. By a {\it dynamical phase transition}\index{Dynamical phase transition}\index{Phase transition! dynamical} we mean that criticality is found in the time averages of physical quantities as the system parameters are varied. To understand the dynamical phase diagram it is instructive to return to the Dicke model. In the superradiant phase, the ground state is a Schr\"odinger cat\index{Schr\"odinger cat! states}, which in the thermodynamic limit\index{Thermodynamic limit} becomes~(\ref{dickecat}). It can be understood from the Born-Oppenheimer approximation\index{Born-Oppenheimer approximation}, see eq.~(\ref{mfdicke}), in which the lowest adiabatic potential~(\ref{adpot2}) takes the form of a double-well potential; see also fig~\ref{fig9} depicting the adiabatic potential for the quantum Rabi model. From the mean-field Dicke Hamiltonian~(\ref{mfdicke}), the semiclassical steady-state solutions\index{Steady state!solution} display the typical pitchfork bifurcation portrayed in fig.~\ref{bifurcationDicke}; the Dicke PT occurs as one goes from a single to two solutions. A state initialized in one of the potential minima will eventually tunnel through the barrier to the other minimum~\cite{irish2013oscillator}. The tunneling time diverges in the thermodynamic limit\index{Thermodynamic limit} as the potential barrier becomes infinite. Now, if we initialize the state in one minimum and then perturb it slightly it will start to oscillate around the steady state solution\index{Steady state}. The Bloch vector will approximately precess around a given axis on the Bloch sphere. Upon time averaging, $\hat S_x$, for instance, will show a non-zero value. If the light-matter coupling strength $g$ is lowered, the potential barrier also decreases and the system may start to oscillate between the two potential minima; as a consequence, the time average of $\hat S_x$ vanishes. The  time average as a function of $g$ shows resemblances of a continuous phase transition\index{Phase transition! continuous} which was demonstrated in~\cite{muniz2020exploring}. The authors therein compare this behavior to the bosonic Josephson effect\index{Josephson! effect}, which can be realized with an atomic condensate in a double-well potential. This system may as well, in certain approximations, be mapped onto the LMG model~\cite{cirac1998quantum}, and for strong enough atom-atom interaction it enters into a {\it self-trapped} state\index{Self-!trapping}. In the present setup this represents the phases in which the time-averaged $\hat S_x$ is non-zero. A recent connection between the {\it unbalanced Dicke model}\index{Dicke! model! unbalanced}, with unequal coupling strengths for the co- and counter-rotating interaction terms, and the LMG model is explored in~\cite{Stitely2022}.    
 
\begin{figure}[ht]
\begin{center}
\includegraphics[width=10cm]{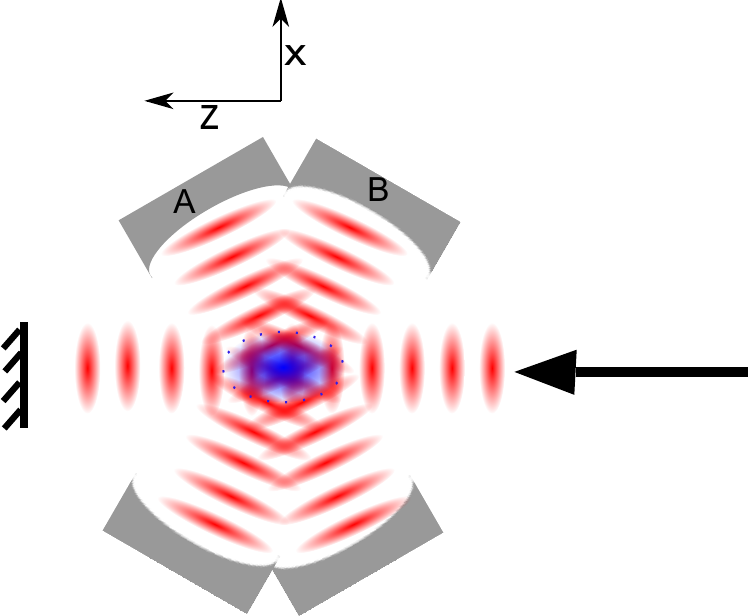}
\caption{Schematic picture of the setup for the experimental realization of the $SU(3)$ Dicke model~(\ref{su3dicke}). The two cavities together with the pump laser forms an effective 2D triangular lattice in the center. Photons from the external pump (marked by the thick black arrow) can be scattered into either of the two cavities. If the setup is fully symmetric and isotropic there is no preferred cavity to scatter into and as a result a continuous $U(1)$ symmetry emerges. In the superradiant phase, spontaneously breaking\index{Spontaneous! symmetry breaking} of this symmetry implies that the two cavities get populated unevenly.  } \label{fig:setup}
\end{center}
\end{figure}

In sec.~\ref{sssec:dicke} we mentioned the $SU(3)$ Dicke model of eq.~(\ref{su3dicke}), which couples two boson modes to $N$ three-level atoms\index{Three-level atom} according to the $\Lambda$-configuration shown in fig.~\ref{fig12} (b). As for the regular Dicke model, its $SU(3)$ version can also be effectively studied with an atomic condensate coupled this time to two cavity modes. The setup is presented in fig.~\ref{fig:setup}: two Fabry-Perot cavities\index{Fabry-P\'erot! cavity/resonator} are crossed together with an external standing wave laser. They are forming an angle of 60 degrees with each other, such that the three standing-wave fields form a perfect triangular lattice in the region of overlap where also the condensate is held fixed by a harmonic trap. This system was experimentally realized in the ETH group~\cite{leonard2017monitoring,leonard2017supersolid}. The qualitatively new physics of the $SU(3)$ Dicke model\index{$SU(3)$ Dicke model} is the presence of a continuous $U(1)$ symmetry, as discussed in more depth in sec.~\ref{sssec:dicke}. This symmetry emerges in the isotropic case when $\omega_A=\omega_B$ and $g_A=g_B$. In this case, there is no {\it a priori} preferred cavity into which the photons will be scattered. If the isotropy condition is lifted, either the $A$ or the $B$ cavity will get populated once the pumping amplitude exceeds the critical value. Hence, there exist four phases as depicted in fig.~\ref{fig:phasediag}: a normal phase N, two superradiant phases -- SR-A and SR-B --  with either of the two cavities populated, or a superradiant phase SR-$\vartheta$. In the SR-$\vartheta$ phase the $U(1)$ symmetry has been spontaneously broken\index{Spontaneous! symmetry breaking}. Furthermore, this is a gapless phase, as demonstrated in fig.~\ref{excfig} which gives plots of the different excitation modes. The setup can be changed so that the experiment operates on either isotropy point  or such that it avoids both, and the excitations can be subsequently probed. In this way it is possible to demonstrate how the massive Higgs mode turns into a massless Goldstone mode~\cite{leonard2017monitoring}, while the spontaneous breaking of the continuous symmetry was explored in some detail in~\cite{leonard2017supersolid} by, e.g., measuring the field amplitudes of the two cavities. The above enhanced symmetry created when going from one to two crossed cavities can be expanded to more cavities, and for three crossed cavities it is possible to realize an effective low-energy model supporting an $SO(3)$ rotational symmetry\index{Rotational! symmetry} [rather than the $U(1)$ symmetry], and thereby two massless Goldstone modes~\cite{chiacchio2018emergence}. A continuous $U(1)$ symmetry also emerges in the ring-cavity setup. Here, instead of a standing-wave cavity mode, the condensate couples to two counter-propagating travelling-wave modes. The self-organization transition has been recently demonstrated in such a system~\cite{schuster2020supersolid}.

So far we have discussed mainly a realization in which the role of quantum fluctuations is not prominent. The systems are well described by mean-field theories and hence, they don't truly belong to the class of `quantum many-body models'. In the realm of quantum simulators~\cite{buluta2009quantum,hauke2012can} or the study of quantum phases of matter, the importance of quantum fluctuations should be extensive in the system size. Optical lattices\index{Optical! lattice} have been utilized for this purpose when it comes to ultracold atoms~\cite{lewenstein2007ultracold,bloch2008many}; for sufficiently strong lattices the atoms get tightly confined within the light-induced potential minima, something which greatly enhances the atom-atom interaction. It was theoretically suggested~\cite{jaksch1998cold}, and experimentally realized~\cite{greiner2002quantum}, that ultracold bosonic atoms can realize the paradigmatic Mott-superfluid QPT\index{Mott!--superfluid phase transition}\index{Phase transition! Mott-superfluid}. The insulating Mott phase cannot be correctly described with an effective few-mode model, even though in 2D and 3D mean-field methods like the {\it Gutzwiller approximation} provide a qualitatively accurate picture. 

The many-body Hamiltonian for the ultracold atoms confined within the cavity was given by eq.~(\ref{mbham}), together with the equations for the effective potentials, (\ref{trpot}) and (\ref{onpot}) for transverse and on-axial pumping respectively. We recall here the approximations imposed in order to arrive at this Hamiltonian: we assume a JC type dipole interaction between every atom and the quantized mode; we then turn to a rotating frame and adiabatically eliminate the excited atomic level in the large detuning regime. The same approximations also lie behind the models describing ultracold atoms in laser-induced optical lattices, and indeed we find such models by replacing the bosonic operators by $c$-numbers, {\it e.g.} $\hat a^\dagger\hat a$ is proportional to the laser amplitude. What is new here is that the periodic potentials thus generated, given by eqs. (\ref{trpot}) or (\ref{onpot}), are dynamical. One early idea was actually to employ the dynamical structure of the lattices in order to build quantum simulators that could realize `phonon-like physics'~\cite{lewenstein2006travelling}, {\it i.e.} the particles would exchange momentum with the underlying lattice via scattering processes involving the vibrational modes. However, in typical model Hamiltonians the matter field couples to a continuum of phonon modes, while in the single-mode approximation of cavity QED there is only a single momentum component involved and the interaction is completely non-local. Including many cavity modes is a way to, on the one hand realize purely quantum many-body states, and, on the other hand construct models with a local atom-light interaction, as mentioned above in terms of the proposal for realizing the Meissner effect~\cite{ballantine2017meissner}. This direction was mainly initiated by the group of Lev in 2009~\cite{gopalakrishnan2009emergent}, and it has since then been greatly developed, both theoretically~\cite{gopalakrishnan2010atom,gopalakrishnan2011frustration,strack2011dicke,muller2012quantum,buchhold2013dicke,rylands2020photon,kollar2017supermode,guo2021optical,erba2021self} and experimentally~\cite{kollar2015adjustable,kollar2017supermode,vaidya2018tunable,guo2019sign,guo2019emergent}. 
Considering a Fabry-P\'erot cavity\index{Fabry-P\'erot! cavity/resonator}, the modes can be arranged into the family of {\it Hermite-Gaussian modes}\index{Hermite-Gaussian mode}, labeled TEM$_{nm}$ ({\it Transverse Electromagnetic})\index{TEM mode}. The subscripts $n$ and $m$ denote the number of nodes in the two transverse directions $x$ and $y$. The {\it fundamental mode}, TEM$_{00}$ has a simple Gaussian transverse mode shape\index{Gaussian! mode! profile} as mentioned in sec.~\ref{sssec:qatmo}, see eq.~(\ref{fpprof}). Typically it is this mode considered in the cavity QED experiments since it is more stable -- the higher-order modes have a larger photon decay rate. The actual shape of the cavity can be designed in order to control the photon frequencies; for a confocal cavity the TEM-modes are almost degenerate, while a more planar cavity lifts this degeneracy and individual modes can be singled out~\cite{kollar2015adjustable,mivehvar2021cavity}. As an example, for the cavity used at ETH for the observation of the Dicke PT, the frequency difference between two consecutive modes, TEM$_{n\,m}$ and TEM$_{n\pm 1\,m}$, is $\Delta_\mathrm{TEM}=18.5$ GHz~\cite{baumann2010dicke}, while for the multimode cavity\index{Multimode! cavity} used at Stanford it can be adjusted in the range $\Delta_\mathrm{TEM}=0-100$ MHz~\cite{kollar2015adjustable}. For further parameter values and other details of the specific cavities we refer the reader to the review~\cite{mivehvar2021cavity}. The atom-light coupling strengths depend on the subscripts $n$ and $m$, and they are now denoted by $g_{nm}(x,y,z)$; the multi-mode JC coupling term~(\ref{mmcoup}) can be written as
\begin{equation}
\hat V=\sum_{i=1}^N\sum_{n,m}\left[g_{nm}(x_i,y_i,z_i)\hat a^\dagger_{nm}\hat{\sigma}_{-i}+H.c.\right],
\end{equation}
where the first sum runs over the $N$ atoms, and the second over the cavity modes. Combining many of the TEM-modes gives rise to a complex potential landscape felt by the atoms. The normal-superradiant transition in this scenario implies scattering of photons into a set of distinct cavity modes, a {\it supermode}~\cite{kollar2017supermode}. Due to the specific mode patterns, the self-organized condensate will in this case display a complicated structure with, for example, matter-wave phase slips. In addition, a strong transverse pump forms an optical lattice in this direction and the atoms move on effective 2D planes. Each plane has a somewhat different effective light-induced potential. The atoms can thereby self-organize differently on the various planes. This is sometimes called a {\it Brazovskii transition}\index{Brazovskii transition}, and, since there are many possible configurations in which the atoms can build up density modulations with almost the same energy, the system becomes {\it frustrated}\index{Frustrated state}. This was suggested in ref.~\cite{gopalakrishnan2009emergent} by deriving an effective low-energy action for the motional states of the atoms (after integrating out the cavity and internal atomic degrees of freedom). 

The complicated mode structure emerging when many modes are taken into account lends itself to the realization of various disordered models. We mentioned that upon eliminating the cavity field one derives an effective model for the atoms which -- to the lowest order -- comprises an Ising-type interaction, see eq.~(\ref{effis}). The spin-spin coupling between spins $i$ and $j$ has the form $J_{ij}\propto\sum_{n,m}g_{nm}(x_iy_i,z_i)g_{nm}^*(x_j,y_j,z_j)$~\cite{gopalakrishnan2011frustration,strack2011dicke}. In a ring cavity, where travelling waves are sustained, we simply have $g(x_iy_i,z_i)g^*(x_j,y_j,z_j)\sim\cos[k(x_i-x_j)]$, and such spin models can be used in order to study a type of interaction called RKKY (Ruderman-Kittel-Kasuya-Yosida)\index{RKKY interaction}, which is thought to be able to explain some unexpected observations, like the line broadening of some metals. For the more involved TEM-modes, the {\it quasi}disordered coupling may result in a {\it spin glass phase}\index{Spin! glass} provided that sufficiently many cavity modes contribute to the interaction. The number of modes can effectively be thought of as controlling the disorder strength. As the number of modes is increased, the self-organized state (or the normal phase, depending on the strength of the light-matter coupling) will go through a transition into the quantum spin glass phase. In fact, a set of different models can be realized by simply adding more modes~\cite{gopalakrishnan2011frustration}. In the language of ``hard-core bosons'' one can map the above model to a disordered Bose-Hubbard Hamiltonian\index{Bose-Hubbard! model}, where the spin states represent empty or occupied sites. The above references did not provide a detailed study on whether the glassy phases would actually survive fluctuations stemming from photon losses, {\it e.g.} can the decay rate $\kappa$ act as some effective temperature\index{Effective! temperature}, and how will universality be affected by the fluctuations? These questions were addressed in~\cite{buchhold2013dicke} by implementing the {\it Keldysh path integral}\index{Keldysh! method} method to open systems~\cite{dalla2013keldysh,sieberer2016keldysh}. It was found that the glassiness may indeed survive photon losses, but the universality is altered. The glassy phases have also been explored by starting from a Hubbard model of spinless fermions (obtained from considering ultracold fermionic atoms in an optical lattice) and then inducing a long-range interaction among the fermions via coupling to a cavity. Again, in the multimode configuration, this interaction will be {\it quasi}disordered, and as a result various types of glassy phases are possible~\cite{muller2012quantum}. A confocal cavity QED spin system is employed in~\cite{marsh2023entanglement} to show that quantum trajectories\index{Quantum! trajectories} avoid a semiclassical energy barrier severely inhibiting passage to a low-energy manifold of states in a dissipative spin glass\index{Dissipative! spin glass}, where a replica symmetry breaking occurs\index{Symmetry breaking! replica}. 

A pump is a transport mechanism, which can have topological origin, where a cyclic evolution of the potential generates direct currents. An emergent mechanism of the kind for a quantum gas coupled to an optical resonator has been realized without recourse to an external drive~\cite{Dreon2022}. Besides, a topological Thouless pump\index{Thouless pump} with fully tuneable Hubbard interactions in an optical lattice has been recently realized in the experiment by A.-S. Walter and coworkers~\cite{Walter2023}. They observed regimes with robust pumping, as well as a breakdown of transport for strong repulsive interactions.  A dynamically tuneable superlattice is attained by superimposing phase-controlled standing waves with an additional running-wave component and the experiment assesses the topological charge pumping in the periodically driven, interacting Rice-Mele model\index{Model! Rice-Mele}.  

As we have already pointed out, one of the initial motivations for studying these types of systems was to find pure quantum many-body lattice systems, in which the lattices are somehow dynamical. What forms the background is, of course, the inspiration drawn from real metals with the electrons coupled to phonon modes, {\it e.g.} {\it Fr\"ohlich-type models}\index{Fr\"ohlich model}~\cite{mahan2013many}. The single-mode case would be a small step in this direction, but it would not lead alone to some Fr\"ohlich Hamiltonian. In this respect, it is the cavity mode that provides the periodic potential for the atoms. This was first studied in ref.~\cite{maschler2005cold}, and later elaborated in refs.~\cite{larson2008mott,larson2008quantum}. Below we will also study the many-body case in which an external classical potential forms the lattice, and the cavity field is only additionally coupled to the atoms. Furthermore, we assume that the cavity is pumped longitudinally (on-axis), such that the light-induced potential has the form~(\ref{onpot}). Thus, the dynamics of the lattice potential depend on its amplitude $\hat a^\dagger\hat a$, and are therefore global. Turning to the strongly coupled regime in which atom-atom interactions play a major role, it is no longer practical to expand the atomic field operators~(\ref{fieldop}) in terms of Fourier modes, but rather use the localized Wannier functions $w_{\nu,{\bf R}_i}({\bf x})$\index{Wannier functions}, see eq.~(\ref{wanf}). Recall that these are typically exponentially localized around the potential minima at $\mathbf{R}_i$, and form an $ON$-basis. To derive a tractable low-energy model we impose the single-band approximation, {\it i.e.}, we limit the expansion to Wannier functions of the first band with $\nu=1$. Furthermore, for sufficiently deep lattices we can impose the tight-binding approximation\index{Tight-binding approximation}. This means that the overlap integrals satisfy
\begin{equation}\label{ovlap}
J_{ij}=\int d^3x\,w_{1,{\bf R}_i}^*({\bf x})\hat Hw_{1,{\bf R}_j}({\bf x})=\left\{
\begin{array}{lll}
E_\mathrm{os}, &\hspace{0.5cm}& i=j,\\
J, &\hspace{0.5cm}& i=j\pm1,\\
0, &\hspace{0.5cm}& i\neq j,j\pm1
\end{array}\right.
\end{equation}
Here, $E_\mathrm{os}$ is the onsite energy and $J$ is the tunneling amplitude/rate. In addition to the above, the tight-binding approximation usually also entails limiting the atom-atom interaction to occur only onsite, and not between neighbouring sites. From the equation~(\ref{mbham}), we then find an interaction term in the language of second quantization as
\begin{equation}
\frac{U}{2}\sum_i\hat c_i^\dagger\hat c_i^\dagger\hat c_i\hat c_i=\frac{U}{2}\sum_i\hat n_i(\hat n_i-1),\hspace{2cm}U=U_0\int d^3x\,|w_{1,\mathbf{R}_i}(\mathbf{x})|^4.
\end{equation}
One could easily generalize~(\ref{ovlap}) to more complicated lattices such that $E$ and $J$ have a spatial dependence, but for the simple potential (\ref{onpot}) we do not have such dependence. One important example of such a situation occurs when one takes a harmonic trapping potential into account. Assuming that this potential is smooth on the scale of the lattice wavelength, it is then justified to assume that the potential only shifts the onsite energies locally, while it does not alter the tunneling amplitudes -- the so-called {\it local density approximation}\index{Local! density approximation}~\cite{lewenstein2007ultracold,bloch2008many}. This procedure of expanding in a Wannier basis and imposing the single-band and tight-binding approximations is identical to the standard procedure for describing strongly-interacting ultracold atoms in optical lattices~\cite{jaksch1998cold}. What is new here is that the lattice amplitude of~(\ref{onpot}), $g_0^2\hat n/\Delta_\mathrm{a}$ depends explicitly on the photon number $\hat n$. Since the Wannier functions depend on the potential amplitude, they become `operator-valued', $w_{i,\hat n}(\mathbf{x})$, where we simplified the notation by omitting the band index $\nu=1$ and in 1D we label the site by the index $i$. This, of course, implies that the coefficients $E_\mathrm{os}=E_\mathrm{os}(\hat n)$ and $J=J(\hat n)$ also become operator-valued. This qualification was overlooked in the original work~\cite{maschler2005cold}, but was handled with care in~\cite{larson2008mott,larson2008quantum}. It is especially when the cavity field is eliminated, under the assumption that $|\kappa-i\Delta|$ sets the fast timescale, that this may cause some technical problems. The problem arises from non-commuting atomic operators. Once expressed in terms of the atomic operators, one may expand the results as a function of the total atomic number $\hat N^{-1}$ ($\hat N=\sum_i\hat n_i$), and the resulting Hamiltonian has a Bose-Hubbard structure\index{Bose-Hubbard! model}~\cite{larson2008quantum}:
\begin{equation}\label{cavityBH}
\hat H_\mathrm{cBH1}=-J(\hat N)\sum_i\left(\hat c_i^\dagger\hat c_{i+1}+\text{h.c.}\right)+\frac{U(\hat N)}{2}\sum_i\hat n_i(\hat n_i-1)-\mu(\hat N)\hat N,
\end{equation}
with density-dependent coefficients $J(\hat N)$ and $U(\hat N)$, and we have included a chemical potential $\mu(\hat N)$ (also depending on $\hat N$) which determines the number of atoms [the onsite energies proportional to $E_\mathrm{os}(\hat N)$ are included into the last term]. The $\hat N$-dependence of the coefficients is understood as a result of the cavity-induced infinite range atom-atom interaction. 
 
\begin{figure}
\includegraphics[width=9cm]{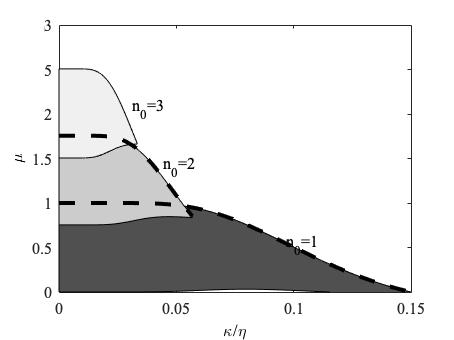} 
\caption{An example of the phase diagram of the generalized Bose-Hubbard model of eq.~(\ref{cavityBH}), in the plane of the scaled decay rate $\kappa/\eta$ ($\eta$ is the pump amplitude) and the chemical potential $\mu$. The three gray shaded regions mark the Mott insulating phases with $n_0=1,2,3$ atoms per site, and hence the white region gives the superfluid phase. The spectacular feature of the phase diagram is the regions in which the insulating phases overlap. The true ground state is the one solution with the lowest energy. The parameters of the figure is taken to represent a gas of $^{87}$Rb atoms with the pump-cavity detuning $\Delta_\mathrm{c}=\kappa$ and $g_0^2/\Delta_\mathrm{a}=2\kappa$, and the number of sites was taken as 50. }
\label{cdp}
\end{figure}

Bose-Hubbard models with density-dependent coefficients may result in exotic phases~\cite{maik2013density}, like {\it charge density waves} (see more below). However, such phases were not found in~\cite{larson2008mott,larson2008quantum}, which looked at the Mott insulators with an integer number of atoms per sites and the superfluid phase\index{Superfluid! phase}. As we are dealing with a 1D model, pertubative methods yield more accurate results than mean-field explorations. The resulting phase diagram, obtained using the {\it strong coupling expansion} method~\cite{freericks1996strong} is shown in fig.~\ref{cdp} for one set of physically relevant parameters. The ``Mott insulating lobes''\index{Mott! insulator phase} show similar features as for the regular Bose-Hubbard model~\cite{freericks1996strong,jaksch1998cold}, with the exception that, given a chemical potential, there might exist multiple solutions evidenced by the ``overlapping'' insulating phases. This is a manifestation of the light-matter backaction\index{Backaction! light-matter}, discussed earlier at the mean-field level as hysteresis and multistability. Here it appears at the quantum level, and the true ground state in the overlapping regime is the one with the lowest energy. The example of the figure is given for a positive pump-atom detuning $\Delta_\mathrm{a}$, while for a negative $\Delta_\mathrm{a}$ (red vs. blue detuning) the Mott lobes may be even more strangely shaped: disconnected, showing re-entrant behaviour, and subject to a first-order PT~\cite{larson2008quantum}.  

Another type of cavity Bose-Hubbard Hamiltonian is obtained by forming the actual lattice by means of an external optical lattice and confining it within the resonator, as realized in the experiment~\cite{landig2016quantum}. This means expanding the setup used for the realization of the Dicke PT~\cite{baumann2010dicke,klinder2015dynamical} into the strongly-correlated regime where atom-atom interactions get strong enough to localize the atoms. More precisely, in this scenario a 2D square optical lattice was formed and the cavity field was aligned along one of the axes. Generalizing eq.~(\ref{1stbec}), this potential takes the form
\begin{equation}\label{bhpot}
V(x,z)=V_{2D}\left[\cos^2(kx)+\cos^2(kz)\right]+\frac{g_0\eta_0}{\Delta_\mathrm{a}}\left(\hat a^\dagger+\hat a\right)\cos(kx)\cos(kz)+\frac{g_0^2}{\Delta_\mathrm{a}}\cos^2(kx)\hat a^\dagger\hat a.
\end{equation}
A reasonable assumption is that the classical optical lattice forms the main lattice felt by the atoms, and we can thereby expand in its corresponding Wannier functions. Hence, the light-matter backaction\index{Backaction! light-matter} is not taken into account at this level, in contrast to what we discussed above, and the overlap integrals~(\ref{ovlap}) will not depend on the cavity properties. Within this regime one derives the 2D tight-binding model~\cite{li2013lattice,maschler2008ultracold,bakhtiari2015nonequilibrium,landig2016quantum}
\begin{equation}
\hat H_\mathrm{cBH2}=-J\sum_{\langle ij\rangle}\left(\hat c_i^\dagger\hat c_j+h.c\right)+\frac{U}{2}\sum_i\hat n_i(\hat n_i-1)+\frac{g_0\eta_0}{\Delta_\mathrm{a}}M_0\left(\hat a^\dagger+\hat a\right)\left(\sum_{i\in\mathcal{O}}\hat n_i-\sum_{i\in\mathcal{E}}\hat n_i\right)-\left(\Delta_\mathrm{c}-\frac{g_0^2}{\Delta_\mathrm{a}M_1}N\right)\hat a^\dagger\hat a-\mu\sum_i\hat n_i,
\end{equation}
with the overlap integrals
\begin{equation}
\begin{array}{l}
\displaystyle{M_0=\int dxdz\,w_i^*(x,z)\cos(kx)\cos(kz)w_i(x,z)},\\ \\
\displaystyle{M_1=\int dxdz\,w_i^*(x,z)\cos^2(kx))w_i(x,z)}.
\end{array}
\end{equation}
After eliminating the cavity field one obtains the effective model for the atoms~\cite{landig2016quantum}
\begin{equation}\label{cbh2}
\hat H_\mathrm{cBH2}=-J\sum_{\langle ij\rangle}\left(\hat c_i^\dagger\hat c_j+h.c\right)+\frac{U}{2}\sum_i\hat n_i(\hat n_i-1)-\frac{U_1}{K}\left(\sum_{i\in\mathcal{O}}\hat n_i-\sum_{i\in\mathcal{E}}\hat n_i\right)^2-\mu\sum_i\hat n_i,
\end{equation}
where $K$ is the number of sites, and $\mathcal{E}/\mathcal{O}$ denotes even/odd sites. The separation into even/odd sites derives from the superlattice structure induced by the second term in~(\ref{bhpot}), while the cavity-induced infinite-range interaction is captured by the term proportional to $U_1$; by squaring this term one finds all combinations of density-density interactions $\hat n_i\hat n_j$. Whenever $i,j\in\mathcal{O}$ or $i,j\in\mathcal{E}$, this interaction term is attractive/repulsive for $U_1$ positive/negative. The sign of $U_1$ exhibits a particular dependence on the detuning $\Delta_\mathrm{c}$~\cite{landig2016quantum}. The attractiveness/repulsiveness is reversed for $i\in\mathcal{O}$ and $j\in\mathcal{E}$ (or vice versa). It is this combination of attractive and repulsive interaction at infinite length scales that makes the model qualitatively different from a regular Bose-Hubbard model -- an interplay between short and infinite range interactions emerges. The phase diagram was studied theoretically in~\cite{bakhtiari2015nonequilibrium,lin2021mott} and in~\cite{chen2020extended} for finite temperature, and experimentally in~\cite{klinder2015observation,landig2016quantum,lin2021mott}. We already know from our discussion above that in the superfluid regime, where the tunneling dominates over the onsite interaction (set by $U$), the system is either in a regular superfluid or in a state reminiscent of a supersolid\index{Supersolid phase}\index{Superfluid}. These two phases correspond to the normal and superradiant phase respectively (In ref.~\cite{klinder2015observation} they are actually referred to as {\it normal superfluid} and {\it self-organized superfluid}). A natural question that arises is whether the insulating phase also displays some sort of self-ordering transition, just like the superfluid. This happens indeed, and one finds four phases characterized according to table~\ref{ordertab}.
\begin{table}[h!]
  \centering
  \begin{tabular}{ | c | c | c |}
    \hline
    Phase & $U(1)$ & $\mathbb{Z}_2$ \\ \hline \hline
    Superfluid SF & Broken & Not broken \\ \hline
    Supersolid SS & Broken & Broken \\ \hline
    Mott insulator MI &\hspace{0.5cm}Not broken\hspace{0.5cm} & \hspace{0.5cm}Not broken\hspace{0.5cm}  \\ \hline
    \hspace{0.2cm}Charge density wave CDW \hspace{0.2cm}& Not broken & Broken\\
    \hline
  \end{tabular}
 \caption{The different phases of the extended Bose-Hubbard model of eq.~(\ref{cbh2}). For strong interaction the system is insulating, with either an isotropic Mott insulator or in a charge density wave\index{Charge density wave}. In the latter, the long-range interaction causes the number of atoms per site to vary from even to odd sites, and hence the translational invariance ($\mathbb{Z}_2$ symmetry) is spontaneously broken\index{Spontaneous! symmetry breaking}. In the insulating phases the continuous $U(1)$ symmetry corresponding to particle conservation is not broken. In the superfluid phases, SF and SS, condensation of the atoms breaks the $U(1)$, and beyond the superradiant transition also the $\mathbb{Z}_2$ is broken. }
 \label{ordertab}
\end{table}
The charge density wave state is a phase in which there is a population imbalance between even and odd sites, but no superfluid order. Thus, this is the self-organized insulating phase. A word of caution is warranted here, since the correctness of the phrasing ``supersolid'' and ``charge density wave'' is a matter of ongoing debate. Normally such phases are connected to a breaking of translational invariance of the underlying lattice, {\it e.g.} if there is an imbalance between consecutive sites, the period of the state is double that of the underlying lattice. However, in the present setup where the atoms self-organize, the profile of the light-induced potential has the same period as that of the atomic state. In both experiments~\cite{klinder2015observation,landig2016quantum} the phase diagram was mapped out. The atomic condensate coherence shows up in a {\it time-of-flight measurement}\index{Time-of-flight measurement} which is a standard experimental technique in cold-atom experiments~\cite{bloch2008many}. That is, at some set time, the optical lattices together with the trapping potential are suddenly switched off and the atoms freely expand as they fall in the gravity field and are being detected by a CCD camera. The expansion time is long enough, ensuring that what is detected by the camera is actually an image of the atomic momentum distribution. In the insulating phase, there is no long-range order coherence of the atoms, and the time-of-flight image lacks interference structures. In the superfluid phase, as the atoms are condensed, long-range order is established and the detection reveals coherences in terms of {\it Bragg peaks}. For the present experiments, such measurements are capable of detecting the transitions between the insulating and the superfluid phase, {\it i.e.} the breaking of the $U(1)$ symmetry. The breaking of the $\mathbb{Z}_2$ symmetry is reflected in the build-up of the intra cavity field. We may note that both the theoretical~\cite{bakhtiari2015nonequilibrium} and the experimental~\cite{klinder2015observation} works seem to have overlooked the presence of a regular Mott insulating phase with preserved $\mathbb{Z}_2$ symmetry. The Bose-Hubbard model~(\ref{cbh2}) with infinite-range interaction can be mapped onto a spin-1/2 chain in the hard-core boson limit\index{Hard-core bosons}~\cite{li2020fast,li2021long}. In this limit the repulsive onsite interaction is taken so large that no two atoms can populate the same site, and the number of atoms on each site is then limited 0 or 1. This can then be described by a spin-1/2 particle per site, and one finds a 1D Heisenberg $X\!X\!Z$ model\index{Heisenberg! model} accompanied by an infinite spin-exchange interaction term akin to those of the Lipkin-Meshkov-Glick models\index{Lipkin-Meshkov-Glick! model}, see eq.~(\ref{lmg}). 

\begin{figure}
\includegraphics[width=9cm]{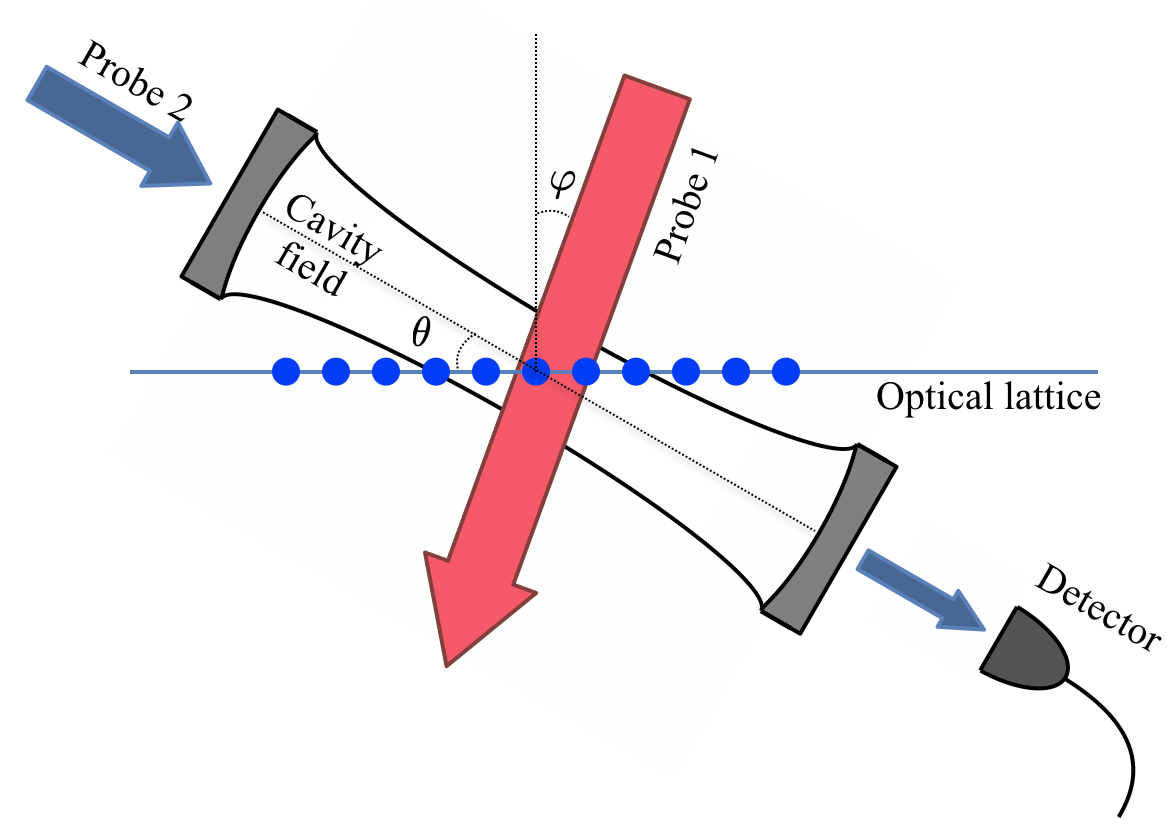} 
\caption{The setup for the cavity mediated measurement of many-body states~\cite{mekhov2007probing}. Interacting ultracold atoms are confined to a horizontal optical lattice (blue balls). The field of a cavity is overlapping with the atomic gas, and it forms an angle $\theta$ relative the optical lattice. Furthermore, the cavity may be longitudinally pumped by a classical field (dark blue arrow) or from the side by another classical field (red arrow). These two act as probes, while the smaller dark blue arrow represent the out-put field from the cavity that will be recorded by a detector. Due to the (cavity) light-matter interaction, the photons escaping the cavity and being detected with contain information about the atomic many-body state.}
\label{mekfig}
\end{figure}

One essential aspect of many-body cavity QED is the openness arising due to photon losses. On the one hand, losses induce noise and bring about decoherence which is felt by the atoms, but on the other hand they form an output channel carrying with them information about the atomic state. We have met examples of this a few times already in section~\ref{sec:cavQED} for single atoms, as for example the `atomic microscope' where the out-put cavity field revealed the presence or absence of an atom inside the cavity~\cite{hood2000atom,pinkse2000trapping}. This idea relies on the frequency shift of the light field induced by the single atom in the cavity. Already back in 1997 this was extended to atomic many-body states in terms of a condensate in a double-well potential~\cite{corney1998homodyne}. The atomic condensate trapped in a double-well potential has been used to realize a bosonic analogue of the {\it Josephson effect}\index{Josephson! effect}~\cite{albiez2005direct}, and the theoretical proposal of ref.~\cite{corney1998homodyne} suggests using a cavity to non-destructively measure the coherent Josephson oscillations\index{Josephson! oscillations}. The bosonic double-well problem can be seen as a two-site Bose-Hubbard model, and when only two localized modes are considered in the expansion~(\ref{fieldop}) one derives a Lipkin-Meshkov-Glick model\index{Lipkin-Meshkov-Glick! model}, see eq.~(\ref{lmg}). The homodyne detection\index{Homodyne detection} scheme of~\cite{corney1998homodyne} measures essentially the population of bosons in either potential well. This has not been demonstrated experimentally, but instead a cavity QED non-demolition measurement of an atomic condensate undergoing Bloch oscillations\index{Bloch! oscillations} has been performed~\cite{kessler2016situ,georges2017bloch}. The second-order coherence of the two bosonic modes alongside the emergence of limit cycles in the phase-space profile for the {\it Bose-Hubbard dimer}\index{Bose-Hubbard! dimer} were studied in~\cite{LledoBHD}. Of course, probing more involved quantum many-body state requires that information beyond the atomic number/density can be extracted. In the following analysis we will assume that the properties of the system to be probed are primarily determined by cold atoms loaded in an optical lattice; the cavity acts as a probe and does not appreciably affect the atomic dynamics~\cite{mekhov2007probing,mekhov2007cavity,mekhov2007light,mekhov2009quantum,mekhov2012quantum}. For ultracold bosonic atoms in a 1D optical lattice, and for moderate interaction and lattice strengths, the physical response can be accurately captured by a Bose-Hubbard Hamiltonian\index{Bose-Hubbard! model}~(\ref{cavityBH}) (with $\hat N$-independent $J$ and $U$) and one is often interested in its possible ground states; {\it i.e.} the Mott insulating phase\index{Mott! insulator} and the superfluid phase\index{Superfluid! phase}~\cite{jaksch1998cold}. In the former, the superfluid order parameter is $\psi=\langle\hat c_i\rangle=0$, with $\hat c_i$ given in eq.~(\ref{fieldop}), where we assume an expansion in terms of Wannnier functions~(\ref{wanf}). To lowest order in the expansion in powers of the ratio $J/U$, we may approximate the quantum many-body states by
\begin{equation}
|\Psi_\mathrm{Mott}\rangle=\prod_{i=1}^K|n_i\rangle_i,\hspace{1.3cm}|\Psi_\mathrm{SF}\rangle=\frac{1}{\sqrt{K^NN!}}\!\left(\sum_{i=1}^K\hat c_i^\dagger\right)^N\![0\rangle,
\end{equation}
where $K$ is the number of sites and $N$ the number of atoms. The system setup for the non-demolition detection is sketched in fig.~\ref{mekfig}; the atoms are trapped in a horizontal optical lattice, surrounded by a cavity whose symmetry axis forms an angle $\theta$ to the lattice potential. We can consider either transverse pumping of the atoms or longitudinal pumping of the cavity, and in either case the photons leaking through the cavity are detected. As we have already mentioned, the optical lattice, formed by a classical standing-wave laser field, mainly dictates the potential felt by the atoms, which motivates the expansion of the atomic operators in terms of the Wannier functions\index{Wannier functions}, while imposing the single-band and tight-binding approximations. If we consider transverse pumping, the many-body Hamiltonian has a similar form to that of equation~(\ref{mbham}), but we have to take into account more than one dimensions, {\it i.e.} the single-particle Hamiltonian in the present case reads 
\begin{equation}
\hat H_0=\frac{{\mathbf{p}}^2}{2m}+V_\mathrm{cl}(\mathbf{x})+\frac{g_0^2}{\Delta_\mathrm{a}}\hat a^\dagger\hat a \,u_c^2(\mathbf{x})+\frac{g_0\eta_0}{\Delta_\mathrm{a}}\left(\hat a^\dagger+\hat a\right)u_c(\mathbf{x})u_p(\mathbf{x}),
\end{equation}
where $V_\mathrm{cl}(\mathbf{x})$ provides the classical optical lattice potential, and $u_{c,p}(\mathbf{x})$ are the mode profiles for the cavity/probe fields (taken to be real). We assume that the two last terms are smooth and weak compared to the classical field $V_\mathrm{cl}(\mathbf{x})$ such that we can neglect any tunneling induced by them. These terms will, however, induce onsite energy shifts $\hat h_i$ in the Bose-Hubbard model that describes the atomic motion. That is, we should include additional terms 
\begin{equation}
\hat h_i=\left[\mu_i\hat a^\dagger\hat a+\nu_i\left(\hat a^\dagger+\hat a\right)\right]\hat c_i^\dagger\hat c_i,
\end{equation}
with amplitudes
\begin{equation}
\mu_i=\frac{g_0^2}{\Delta_\mathrm{a}}\int d{\bf x}\,|w_{{\bf R}_i}({\bf x})|^2u_c^2({\bf x}),\hspace{1.23cm}\mu_i=\frac{g_0\eta_0}{\Delta_\mathrm{a}}\int d{\bf x}\,|w_{{\bf R}_i}({\bf x})|^2u_c({\bf x})u_p({\bf x}).
\end{equation}
If the optical lattice strongly localizes the atomic onsite states $w_{{\bf R}_i}({\bf x})$ on the scale of the mode profiles $u_{c,p}(\mathbf{x})$ we can replace the integrals by $u_{c,p}^i\equiv u_{c,p}({\bf R}_i)$. We can now follow the scheme above by assuming that the cavity field reaches steady state on a short time scale and, similar to eq.~(\ref{ass}), we can write down the steady state\index{Steady state} solutions of the field operators, which will now be slaved to the atomic operators, {\it i.e} $\hat a_\mathrm{ss}=\hat a_\mathrm{ss}(\hat b_i,\hat b_i^\dagger)$. Since the output cavity field contains the information of the field operators $\hat a_\mathrm{ss}$ it carries over information about the atomic state, and in particular about the coherence of the atoms; {\it i.e.} whether they belong to an insulating or a superfluid state. Upon varying the detuning, an insulating atomic state will typically display a single broadened  resonance peak in the field intensity, while a superfluid displays several peaks that reflect the number fluctuations~\cite{mekhov2007probing}.

As mentioned, the loss of photons causes an indirect decoherence of the atomic states~\cite{szirmai2009excess}. Similarly, by recording photons escaping the cavity we extract information about the atomic state, which must be updated accordingly. This is a realization of a quantum non-demolition measurement\index{Non-demolition measurement} first discussed in sec.~\ref{ssec:JCdyn}. It turns out that this measurement-induced backaction\index{Backaction! measurement-induced} may lead to drastic effects~\cite{mazzucchi2016quantum,mekhov2009quantum}. Utilizing such backactions in order to prepare non-classical atomic states, like number squeezed\index{Squeezed! state} and Schr\"odinger cat states\index{Schr\"odinger cat! states} was first suggested in~\cite{mekhov2009quantum} in a quantum-trajectory approach~\cite{BookQO2Carmichael} where detection of a leaking photon updates the state as $|\Psi(t)\rangle\rightarrow\hat a|\Psi(t)\rangle$ up to a normalization constant. Between such jumps the system evolves according to the non-Hermitian Hamiltonian\index{Non-Hermitian! Hamiltonian} $\hat H-i\kappa\hat a^\dagger\hat a$. The photon detections are stochastic processes, and for a {\it quantum trajectory} labelled by $n$ counts, the state $|\Psi_n(t)\rangle$ is pure and characterized by the applications of $\hat a$ at times $t_1,\,t_2,\dots$ -- see the photon counting records for a decaying even-cat state in subsec.~\ref{sssec:openjc}\index{Quantum! trajectories}. In fig.~\ref{fig2} we explained how the shift of the atomic Bloch vector depends on the field intensity (number of photons); the larger the field amplitude the larger the displacement of the Bloch vector towards the equator of the Bloch sphere. Alternatively, in fig.~\ref{fig5} we turned things around and thought of the dispersive interaction from the perspective of the photon field instead. If the latter is initially in a coherent state, the atom-induced shift will effectively act as a coherent displacement operation. The same idea can be extended to many atoms -- naturally the displacement will get enhanced with increased atom number. Nevertheless, in the quadratic approximation, an initial coherent field state will remain coherent even in the presence of photon loss~\cite{gardiner2004quantum}. Within this regime, it is possible to find an analytic expression for the evolved state $|\Psi_n(t)\rangle$, given, say, $m$ detection events. Assuming the existence of the superfluid phase, the atom-number fluctuations at a given site approximately follow a Poissonian distribution, but consecutive photon detections narrow down the atomic distribution to become sub-Poissonian\index{Sub-Poissonian}, much in the same way as the {\it filter functions}~\cite{larson2003photon} discussed in relation to eq.~(\ref{filter}). In secs.~\ref{ssec:JCdyn} and.~\ref{sssec:micro} we explained a similar idea behind trapping states\index{Trapping states}~\cite{meystre1988very,slosser1989harmonic} for the photon field of the cavity. Hence, this localization (squeezing) bears similarities to the trapping states but now for the atomic field instead. The measurement-induced backaction\index{Backaction! measurement-induced} was further analyzed in refs.~\cite{mazzucchi2016quantum,ivanov2020feedback}, extending the earlier works to dynamical studies. It was shown, for example, how initially localized states undergo large-amplitude oscillations within the lattice, and how the information feedback into the system may lead to a critical response. We conclude this subsection by mentioning that a continuous time crystal\index{Time crystal! continuous} -- in the form of a temporally robust limit cycle --  has been very recently reported for a transversely pumped atom-cavity system with an optical pump lattice which is blue
-detuned with respect to an atomic transition~\cite{Kongkhambut2022}. Ergodicity breaking\index{Ergodicity!breaking} holds the key to discrete time crystals, while the delaying of ergodicity is the source of several phenomena that share many of the properties of discrete time crystals, including the ac Josephson effect, coupled map lattices, and Faraday waves~\cite{Zaletel2023}.  

\subsubsection{Critical phenomena II -- fermions}
Fermions, contrary to bosons, obey the Pauli exclusion principle\index{Pauli! principle}, {\it i.e.} any two fermions cannot occupy the same quantum state. This, of course, has far-reaching consequences for several kinds of physical phenomena, and we may ask how things change as we consider fermions instead of bosons, with reference to our previous section. Will, for example, the superradiant PT occur in a system of ultracold fermionic atoms? That is, instead of populating predominantly the $k=0$ momentum mode at zero temperature, we now fill up a {\it Fermi sea}\index{Fermi! sea} with a range of different momenta. For a thermal gas, in which both bosons and fermions obey a Maxwell-Boltzmann distribution\index{Distribution! Maxwell-Boltzmann}\index{Maxwell-Boltzmann distribution} a superradiant transition has been observed~\cite{black2003observation}, which is not, however, guaranteed to survive at lower temperatures in the fermionic case. 

Considering spin-polarized fermions and the transverse pumping setup, the many-body Hamiltonian is $\hat{\mathcal H}=\displaystyle{\hbar\delta\hat a^\dagger\hat a+\int dx\,\hat\psi^\dagger(x)\hat H\hat\psi(x)}$, with the single-particle Hamiltonian $\hat H$ given by~(\ref{1stbec}). The atomic-field operators $\hat\psi(x)$ obey now the fermionic anti-commutation relations, {\it e.g.} $\left\{\hat\psi(x),\hat\psi^\dagger(x')\right\}=\delta(x-x')$. For spin-full fermions we need to add a subscript to the atomic operators in order to refer to their corresponding spin, and we may then have to include also interaction terms between atoms with different spins. Since the potential is periodic, we expect a band spectrum, but it must be remembered that, just like for bosons, the same backaction\index{Backaction} between the cavity field and the atoms exists; the resulting equations become nonlinear and must be solved self-consistently. In fact, this nonlinearity was the topic of one of the very first studies on many-body cavity QED with fermions~\cite{larson2008cold}. The cavity hysteresis was studied, and the topology of the Fermi sea was also explored numerically; one would imagine that the light-matter nonlinear interaction could give rise to exotic Fermi surfaces including those which are not simply connected. This was, however, not found. The cavity hysteresis was further considered in~\cite{kanamoto2010optomechanics} in the realm of optomechanics. The light-induced potential, being expressed in terms of a few trigonometric functions,~(\ref{1stbec}), will couple different momentum modes in a rather simple manner, especially in 1D (if we impose a tight transverse confinement). We will see that this will have consequences for the phase diagram. In a sense, the fermionic problem becomes more complex since we must include more momentum modes, {\it i.e.} the low-energy bosonic expansion~(\ref{dickeexpan}) of the field operators is not justified for fermions. Let us expand the field operator in plane waves
\begin{equation}
\hat\psi(x)=\sum_{\bf q}\hat c_{\bf q}e^{-i{\bf qx}},
\end{equation}
where the fermionic operators obey $\left\{\hat c_{\bf q},\hat c_{\bf q'}^\dagger\right\}=\delta_{\bf qq'}$. The Hamiltonian in the second quantization (unscaled units) becomes
\begin{equation}\label{ferham}
\hat H=\hbar\delta\hat a^\dagger\hat a+\sum_{\bf q}\left[\frac{\hbar^2{\bf q}^2}{2m}\hat c_{\bf q}^\dagger\hat c_{\bf q}+\frac{g_0^2}{4\Delta_\mathrm{a}}\hat a^\dagger\hat a\sum_n\hat c_{\bf q}^\dagger\hat c_{{\bf q}+2n{\bf k}_x}+\frac{g_0\eta_0}{4\Delta_\mathrm{a}}\left(\hat a^\dagger+\hat a\right)\sum_{n,n'}\hat c_{\bf q}^\dagger\hat c_{{\bf q}+n{\bf k}_x+n'{\bf k}_z}\right].
\end{equation}
In the ground state the Fermi sea is filled, {\it i.e.} $\hat c_{\bf q}^\dagger\hat c_{\bf q}=1$ for the occupied momentum modes and $0$ for the unoccupied modes. The {\it Fermi momentum} $k_F$\index{Fermi! momentum} marks the surface of the Fermi sea, which for the regular geometry indicates the maximum momentum among the occupied modes. It should be clear that the Fermi momentum $k_F$ relative to the lattice momentum $k$ will play an important role for the possible phases, {\it e.g.} whether self-organization can occur for incommensurate momenta. In the 1D setup, where the atoms are tightly confined in the transverse direction, the last sum runs only over $n$. If the wave number of the lattice is exactly twice the Fermi momentum, $k=2k_F$, we encounter a perfect {\it Fermi surface nesting}~\cite{chen2014superradiance}, meaning that the shape of the Fermi sea is recovered upon a shift by $k$. Alternatively, states at the Fermi surface undergo {\it Umklapp scattering processes}\index{Umklapp processes} as momentum is exchanged with the optical lattice~\cite{piazza2014umklapp}. A result of this is that in 1D, and for a particular filling such that $k=2k_F$, the normal phase is unstable and any pumping, no matter how weak, causes the state of the system to become superradiant. This was demonstrated in refs.~\cite{chen2014superradiance,piazza2014umklapp} by analyzing the free energy in the thermodynamic limit\index{Thermodynamic limit} with the mean-field description of the cavity field, similar to the original analysis by Wang and Hioe for the Dicke-model phase transition~\cite{wang1973phase}. Photon losses will, however, stabilize the normal phase for small pump amplitudes~\cite{piazza2014umklapp,keeling2014fermionic}. In 2D, perfect nesting is not obtained, yet the larger the nesting, the weaker the critical pump amplitude $\eta_c$ for attaining superradiance~\cite{chen2014superradiance}. Thus, the counterpart of the normal-superradiant transition found for bosons survives at $T=0$ for the fermion case as well, and the Pauli exclusion even enhances superradiance. In refs.~\cite{keeling2014fermionic}, the same system was analyzed also in terms of its free energy, and it was argued that the normal-superradiant transition can become first order for certain cavity-pump detunings. This reference also considered fermionic fillings beyond the first band, such that the second band gets partly occupied. The onsite orbitals on the second band are $p_x$ and $p_z$ and, for an isotropic square lattice, these are degenerate. However, in the general case they do not need to be degenerate, and it was found that there is a crossing between them leading to a first-order transition connecting two possible phases called {\it high-field} and {\it low-field} superradiant phase, where high/low refers to the cavity-field amplitude.

The above three references,~\cite{keeling2014fermionic,chen2014superradiance,piazza2014umklapp}, appeared back-to-back and were the first to demonstrate a fermionic normal-superradiant phase transition\index{Phase transition! normal-superradiant}. The following year, it was shown that the superradiant phase can be topologically non-trivial if the system is extended to include spin-full fermions~\cite{pan2015topological}. Let us recall the SSH \index{SSH model} (Su-Schrieffer-Heeger) model~\cite{su1979solitons} of sec.~\ref{ssec:focklattice}, represented by the 1D tight-binding lattice model
\begin{equation}\label{ssh2}
\hat H_\mathrm{SSH}=\sum_{i=1}^N\left(v\hat c_{i,1}^\dagger\hat c_{i,2}+w\hat c_{i,2}^\dagger\hat c_{i+1,1}+\text{h.c.}\right).
\end{equation}
This is a dimerized model with the unit cell comprising two lattice sites; $v$ is the intra-cell tunneling rate and $w$ the inter-cell tunneling rate. It is straightforward to diagonalize the Hamiltonian by turning to the momentum representation, as was done in sec.~\ref{sssec:strans}. The Bloch Hamiltonian\index{Bloch! Hamiltonian} is a $2\times2$ Hamiltonian and its eigenstates $|\theta_\pm(k)\rangle$ can be represented on the Bloch sphere. As a function of the quasi momentum $k$, these states map out a curve in the $xy$-plane; for $w>v$ the origin lies within this curve, but not for $w<v$. One finds that the Zak phase\index{Zak phase}
\begin{equation}
\gamma=i\oint\langle\theta_\pm(k)|\nabla_k|\theta_\pm(k)\rangle dk
\end{equation}
equals either $\pm\pi$ or $0$ in the two cases. In the first case, one says that the state is topologically non-trivial\index{Topological! state}, and, according to the bulk-boundary correspondence~\cite{hasan2010colloquium}\index{Bulk-boundary correspondence} we know that in a finite system there are two $E=0$ energy eigenstates localized to the edges of the chain, the so-called {\it edge states}\index{Edge state}. To realize the SSH model as an effective low-energy model, we need a superlattice potential $V(x)$ with two sites within every unit cell. This is what happens in the superradiant phase with the self-contained potentials given by eq.~(\ref{trpot}). We remember that the $\mathbb{Z}_2$ symmetry breaking is reflected in the sign $\pm$ of the amplitude $\alpha=\langle\hat a\rangle$, leading to the fact that even sites are either deeper or shallower than odd ones. Such a superlattice gives a dimerized lattice model when we expand in the Wannier basis~(\ref{wanf}) and impose the tight-binding approximation, but it will not be of the SSH form due to the energy offset between the sites. However, this can be adjusted by flipping the sign of the atom-pump detuning $\Delta_\mathrm{a}$, {\it i.e.} consider {\it blue} instead of {\it red detuning}\index{Blue! detuning}\index{Red! detuning}. In this case, the entire potential changes sign and is flipped upside-down. We thus obtain a chain of double-well potentials, which is exactly what is required for realizing the SSH model. Such a model still supports a $\mathbb{Z}_2$ symmetry, and any breaking of it would manifest in the double-wells being formed between even-odd sites or between odd-even sites. We note that the SSH model may serve as a simple model demonstrating the {\it Peierls transition}\index{Peierls! transition}~\cite{voit1995one}. For a 1D system at half filling, the {\it Peierls instability} occurs when the atoms in a crystal deform such that the period is doubled, while a gap opens up in the lowest band at the Fermi momentum $k_F$. By causing this deformation (every second atom is slightly closer to its right than to its left neighbour), and the cavity-induced lattice serves precisely that role. The deformation will result in tunneling amplitudes alternating between the neighbouring sites, which is the characteristic of the SSH model. For the Peierls transition one needs a dynamical lattice (or possibly long range interaction), but that we have in the present case with the cavity induced lattice.

In ref.~\cite{pan2015topological}, the setup is pretty much the same as for the transversely-pumped configuration, but it is modified to a Raman transition between two lower atomic Zeeman levels, such that the fermions possess two internal states which are flipped upon scattering cavity photons. This coupling therefore realizes a spin-orbit type of coupling, similar to that discussed in terms of synthetic gauge fields in sec.~\ref{sssec:mfmQED}. By again analyzing the free energy we find a critical pump amplitude for the normal-superradiant transition, and furthermore that the superradiant phase may either be topological or topologically trivial, {\it i.e.} the corresponding Zak phase is non-vanishing or zero, and there exist zero-energy edge states. Such localized edge states are described by {\it Majorana fermions}\index{Majorana fermion}, {\it i.e.} they are their own anti-particles. The mean-field phase diagram was mapped out in the pump-Zeeman field plane, with the Zeeman field marking the energy offset between the two atomic Zeeman levels. Apart from the two superradiant phases, the normal phases split in two, one gapless metallic and one insulating. 

It was later shown that in order to realize a topological superradiant phase it is not necessary to consider fermions with internal degrees of freedom; it may also be realized with polarized fermions by using the aforementioned blue detuning~\cite{mivehvar2017superradiant}. In this situation, as soon as the system is subject to a spontaneous breaking of the $\mathbb{Z}_2$ symmetry\index{Spontaneous! symmetry breaking} and becomes superradiant, the emerging effective potential will have the desired double-well structure akin to the SSH model, for a Zak phase $\gamma=\pm\pi$. As we have mentioned above, the Peierls instability depends on the filling, such that the Fermi momentum $k_F$ becomes a relevant parameter in determining the phase diagram. For half filling, when $k_F$ lies at the center of the Brillouin zone\index{Brillouin zone}, a Peierls transition occurs as the pump amplitude is increased beyond a critical value and the system transitions from a metallic Fermi gas to a superradiant insulator which can either be trivial or topological. For other fillings, the superradiant phase is instead metallic, {\it i.e.} the bulk is gapless. 

\begin{figure}
\includegraphics[width=13cm]{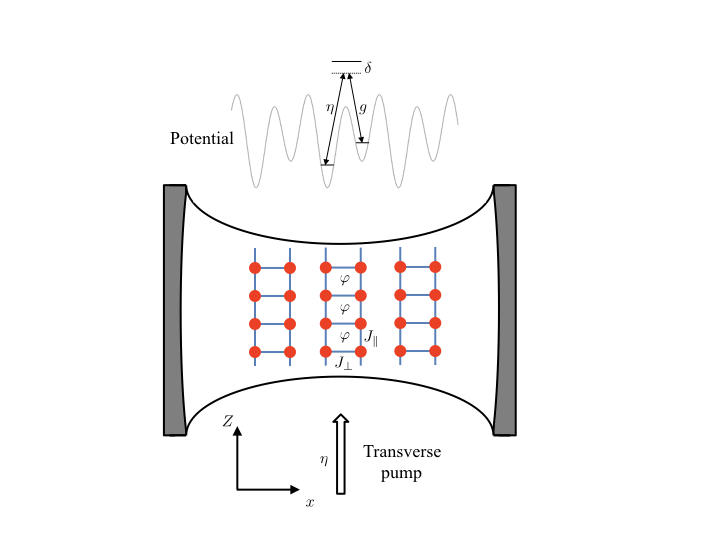} 
\caption{The system setup of refs.~\cite{kollath2016ultracold,sheikhan2016cavity}. Two classical standing-wave potentials form a square lattice in the $xz$-plane, with the sites marked by red dots. An additional optical lattice along the $x$-direction creates an energy difference between neighbouring sites along this direction. Tunneling in the transverse direction is real with a rate $J_\parallel$, while the tunneling in the longitudinal direction is approximately zero. A transverse classical pump, together with the standing-wave cavity mode, realize a Raman transition between neighbouring sites in the $x$-direction (upper sketch), and by adjusting the phase of the pump laser, the tunneling amplitudes $J_\perp$ can be made complex mimicking a synthetic flux $\varphi$ through every plaquette of the emerging ladder lattices.  
}
\label{cfig}
\end{figure}

Topological states in fermionic many-body cavity QED systems were also studied in~\cite{kollath2016ultracold,sheikhan2016cavity}. A {\it topological insulator}\index{Topological! insulator} is characterized by a gapped bulk and conducting edges, by which we mean that in a finite (but large) system with open boundary conditions one finds eigenstates localized in the bulk organized with a clear band gap separating them, while in addition there are gapless states localized to the boundary. This does not contradict the Bloch theorem\index{Bloch! theorem}, as the system is finite and translational symmetry has been broken. For a trivial, or non-topological, insulator there will be no conducting edge states. The {\it Haldane model}\index{Haldane model}\index{Model! Haldane} is an example of a 2D topological model~\cite{larson2020conical}. It describes a hexagonal lattice with nearest and next nearest neighbouring tunneling. For nearest neighbour tunneling, {\it i.e.} the tight-binding limit, one recovers a gapless spectrum where the two points meet in points of conical intersections\index{Conical intersection} or, as they also are called, in the {\it Dirac points}\index{Dirac! point}. While the spectrum is not gapped as the Fermi surface, this is actually not a true metal since the density of states at the Fermi surface vanishes (the Fermi surface reduces to single points for unit filling). It is not an insulator either. To achieve an insulator one must open up a gap, which can be achieved by inducing an energy offset between the neighbouring sites in the hexagonal lattice. This offset breaks the inversion symmetry of the model, but, as it turns out, it is not enough for establishing a topological insulator. With real nearest neighbouring tunneling rates, the model also possess a chiral symmetry\index{Chiral symmetry}, and this can be broken by adding complex next nearest neighbouring tunneling rates, which is what happens in the Haldane model. Hence, breaking a chiral symmetry can be accomplished by introducing a gauge field that makes the tunneling rates complex (this follows from the {\it Peierls substitution}~\cite{larson2020conical}). In refs.~\cite{kollath2016ultracold,sheikhan2016cavity} the gauge field is induced by the cavity field, following similar ideas to those discussed above. The polarized fermions sit on a 2D lattice created by three standing-wave laser fields; in the transverse direction it is a regular $cos^2(kz)$ potential, while in the longitudinal direction two lasers form a superlattice; see fig.~\ref{cfig} for a sketch of the setup. As shown in the figure, the system comprises a set of ladders. Each ladder has two legs and a large number of rungs. The sites along the two legs have an energy offset, and, employing a transverse classical pump, a Raman transition is driven from sites along one leg to sites along the other leg. The two-photon Raman transition\index{Two-photon!transition!Raman} involves the pump and the cavity mode, and with a plane-wave pump field one achieves a synthetic gauge field representing a constant flux $\varphi$ through every plaquette along the ladders. The synthetic flux translates into complex tunneling rates $J_\perp$ along the rungs, and also a broken chiral symmetry. This implies the formation of edge currents, rotating clockwise or anti-clockwise, along the legs of the ladders and for certain fillings. Chiral states appear for non-zero fluxes $\varphi$, which are in turn non-zero only after the onset of self-organization when the cavity field builds up. This reflects the dynamic nature of the gauge field. The edge currents carry a non-zero momentum along the $z$-direction, reminiscent of the shifted minima of the dispersion to non-zero momenta (see fig.~\ref{fig11} for an example where the zero momentum is no longer at the origin). Depending on the filling, the superradiant chiral phase will either be an insulator or a liquid, {\it e.g.} half filling results in an insulator given the flux $\varphi=\pi/2$, while a quarter filling gives a liquid. The ladder formation was extended to interacting spin-$1/2$ fermions in~\cite{sheikhan2019cavity}. The focus was not on the synthetic flux and chirality, but the appearance of a charge density wave\index{Charge density wave} state with a wavevenumber equal to $4k_F$. To treat an interacting many-body system of that sort, more sophisticated numerical methods were employed, namely DMRG (density matrix renormalization group).

Our basic understanding of superconductivity relies on the {\it BCS theory} (Bardeen-Cooper-Schrieffer)\index{BCS theory}. The coupling of electrons to nuclei (phonons) causes an effective positive electron-electron interaction between electrons in a small momentum window around the Fermi surface. A small attractive electron-electron interaction gives rise to what is called the {\it Cooper instability}; electrons with opposite spins, and momenta, ${\bf k}$ and ${\bf -k}$, bind to form a Cooper pair\index{Cooper pair}. Such Cooper pairs are bosonic and at low temperatures they may condense. ~\cite{altland2010condensed}. In its simplest form, the BCS Hamiltonian is given by
 \begin{equation}
 \hat H=\sum_{{\bf k},\sigma}\epsilon({\bf k})\hat c_{{\bf k}\sigma}^\dagger\hat c_{{\bf k}\sigma}-\sum_{\bf k,k'}V_{\bf kk'}\hat c_{{\bf k}'\uparrow}^\dagger\hat c_{-{\bf k}'\downarrow}^\dagger\hat c_{{\bf k}\uparrow}\hat c_{-{\bf k}\downarrow}.
 \end{equation}
 At the mean-field level, the quartic interaction is handled by introducing the order parameter\index{Order parameter}
 \begin{equation}
\Delta_{\bf k}=\sum_{\bf k'}V_{\bf kk'}\langle\psi_0|\hat c_{{\bf k}'\uparrow}^\dagger\hat c_{-{\bf k}'\downarrow}^\dagger|\psi_0\rangle,
\end{equation}
with $|\psi_0\rangle$ the ground state.  The resulting {\it Bogoliubov-de Gennes Hamiltonian}\index{Bogoliubov-de Gennes Hamiltonian} 
\begin{equation}
\hat H_\mathrm{BdG}=\sum_{\bf k}\left[\epsilon({\bf k})\hat c_{{\bf k}\sigma}^\dagger\hat c_{{\bf k}\sigma}-\left(\Delta_{\bf k}\hat c_{{\bf k}\uparrow}\hat c_{-{\bf k}\downarrow}+\Delta_{\bf k}^*\hat c_{-{\bf k}\downarrow}^\dagger\hat c_{{\bf k}\uparrow}^\dagger\right)\right]
\end{equation} 
is then quadratic and analytically solvable. Such mean-field treatment is capable of identifying a critical temperature at which the Cooper instability\index{Cooper instability} occurs, {\it i.e.} when the order parameter $\Delta$ becomes non-zero~\cite{altland2010condensed}. Thus, a weak attractive interaction among ultracold fermions may cause the formation of Cooper pairs. The properties of the state are determined by the ${\bf k}$-dependence of the order parameter. Examples of a $p$-wave superconductor have
\begin{equation}\label{porder}
\Delta_{\bf k}=\left\{\begin{array}{l}
\alpha(k_x+ik_y),\\
\alpha(k_x-k_y),
\end{array}\right.
\end{equation}
where the first example preserves chiral symmetry but instead breaks time-reversal symmetry, and vice versa for the second example. Only the first example gives rise to a topological superconductor.
Examples of a $d$-wave superconductor include
\begin{equation}\label{dorder}
\Delta_{\bf k}=\left\{\begin{array}{l}
\alpha\left(k_x^2-k_y^2\right),\\
\alpha\left(k_x^2-k_y^2+ik_xk_y\right).
\end{array}\right.
\end{equation}
If the interaction between the fermions becomes strongly attractive, the BCS state formed by Cooper pairs\index{Cooper pair} will not survive and instead fermions bind in pairs in real space as bosonic molecules, which can form a Bose-Einstein condensate. This instance is called the {\it BCS-BEC crossover}\index{BCS-BEC crossover}~\cite{bloch2008many}.

Ultracold atomic experiments are ideal for studying this crossover since the interaction between the fermionic atoms can be largely controlled via the so-called {\it Feshbach resonances}\index{Feshbach resonance}~\cite{bloch2008many}. Neutral ultracold atoms scatter predominantly via $s$-waves (with no angular dependence), and the interaction strength is determined by a single quantity, the $s$-wave scattering length $a_s$\index{Scattering length}. In the presence of magnetic fields, the atomic Zeeman levels can be Stark shifted\index{Stark shift}, and one can tune $a_s$ across these resonances in order to move from weak to strong interaction, and from attractive to repulsive. The quantity $-1/k_Fa_s$ is usually taken as the knob controlling the BCS-BEC crossover; in the BCS limit $-1/k_Fa_s\rightarrow+\infty$, while in the BEC limit $-1/k_Fa_s\rightarrow-\infty$. In these two limits, the gas obeys mainly fermionic or bosonic statistics, respectively. It is interesting to explore how the cavity-induced long-range fermion-fermion interaction affects this behavior, which was the topic of refs.~\cite{colella2018quantum,schlawin2019cavity,chen2015superradiant}. For example, we have so far studied the bosonic and fermionic versions of self-organization and the transition to a superradiant phase, but at the Feshbach resonance, when $a_s$ diverges, we have $-1/k_Fa_s\rightarrow0$ and the system cannot correctly be described by neither fermionic nor bosonic statistics. In general, the critical pump amplitude at which the system becomes superradiant can be expressed as~\cite{baumann2010dicke,chen2014superradiance,chen2015superradiant}
\begin{equation}
\eta_c=\frac{1}{2}\sqrt{\frac{\omega_0^2+\kappa^2}{-\omega_0\chi}},
\end{equation}
where $\omega_0=-\Delta_c-\frac{g_0^2}{\Delta_\mathrm{a}}\int d{\bf x}\,\langle n({\bf x})\rangle\cos^2(kx)$ is the shifted cavity detuning, and $\chi$ is a {\it density-wave susceptibility}. Normally we have $|\chi_F|>|\chi_B|$, where $\chi_F$ and $\chi_B$ are the fermion and boson susceptibilities respectively, and the superrariant transition occurs for weaker pump amplitudes in the fermionic case. In~\cite{chen2015superradiant}, it was shown that the susceptibility can be decomposed into a fermionic and a bosonic part, $\chi=\chi_F+\chi_B$. On the BCS side, $\chi\approx\chi_F$, on the BEC side $\chi\approx\chi_B$, while at the Feshbach resonance both terms contribute. 

In ref.~\cite{schlawin2019cavity} a different setup was considered (see also~\cite{camacho2017quantum}); two perpendicularly crossed ring cavities in the $xy$-plane. The atomic cloud is located at the overlapping region of the two cavities and experiences a square lattice potential; in addition it is pumped by a classical field. The cavity-induced long-range interaction causes exotic pairing of the fermions, {\it i.e.} the order parameter $\Delta_{\bf k}$ attains an interesting momentum dependence. For a cubic lattice, it was shown that $s$-, $p$-, and $d$-wave pairings are quasi degenerate. However, this degeneracy is lifted by symmetry-breaking perturbations, like for example a contact interaction induced by Feshbach resonances, or a spin imbalance. Such perturbations will determine the state of the system, and they may thereby also serve as control parameters for switching between different phases. The chiral $p$-wave state is topological, contrary to the topologically trivial $s$-wave state. For a spin-balanced gas, with positive local interaction the ground state is a $p+id$-wave superconductor which is topological and hosts zero-energy Majorana edge states. These states are localized at the corners of the 2D lattice in the $xy$-plane. 

In the setup of a transversely pumped cavity, the mean-field Hamiltonian (corresponding to the Bogoliubov-de Gennes Hamiltonian) becomes~\cite{colella2018quantum}
\begin{equation}\label{cbdg}
\hat H_\mathrm{cBdG}=\sum_{k}\left[\epsilon(k)\hat c_{k\sigma}^\dagger\hat c_{k\sigma}-\left(\Delta_{\bf k}\hat c_{{\bf k}\uparrow}\hat c_{-{\bf k}\downarrow}+\Delta_{\bf k}^*\hat c_{-{\bf k}\downarrow}^\dagger\hat c_{{\bf k}\uparrow}^\dagger\right)\right]+\sum_{k,k'=k_L}S(k')\hat c_{k\downarrow}^\dagger\hat c_{k+k'\uparrow},
\end{equation} 
where $k_L$ is the optical-lattice wave vector. Here, $\epsilon(k)=\hbar^2 k^2/(2m)$ is the free-energy dispersion, and $S(k')=-\frac{g\eta}{\Delta_\mathrm{a}}\sum_k\langle\hat c_{k\downarrow}^\dagger\hat c_{k+k'\uparrow}\rangle$ is a {\it spin density-wave} order parameter. Hence, the long-range interaction brings in an additional term that will compete with the Cooper pairing\index{Cooper pair}; in particular, this term favours a spin density-wave. The mean-field Hamiltonian must be solved self-consistently, since both order parameters depend on the system ground state. This can typically be done iteratively. In ref.~\cite{colella2018quantum} the analysis of the Hamiltonian~(\ref{cbdg}) was complemented with exact diagonalization in a bosonization procedure. The system was found to be either a superfluid or a spin density-wave, depending on the sign of the effective interaction strength.  
 On the experimental side, the first experiment achieving a coherent coupling between a fermionic gas and an optical cavity was reported in~\cite{braverman2019near}. In this experiment, a thermal gas of about 1000 $^{171}$Yb atoms, with effectively two involved Zeeman levels, interacted with a single-mode cavity. The single-atom cooperativity,~(\ref{coop}), in the experiment is $C\approx1.8$, indicating coherent evolution\index{Cooperativity}. This work revealed a very high degree of spin squeezing, see sec.~\ref{sssec:squeez}, which, as an entanglement witness, signifies the presence of entanglement between the two-level atoms forming the thermal gas. The first demonstration of coherent coupling with a {\it degenerate unitary Fermi gas}, {\it i.e.} one where the states up to the Fermi energy are occupied, appeared one year later, in 2020~\cite{roux2020strongly} (see ref.~\cite{roux2021cavity} for technical details on the experiment). This work benchmarks the system, in the same way as refs.~\cite{brennecke2007cavity,colombe2007strong} did for the bosonic case. The transition between two internal Zeeman levels of gas of $^6$Li atoms was driven by a cavity field. Like in the squeezing experiment, the cooperativity exceeded unity, $C\approx2.02$. About $2\times10^5$ atoms were coupled to the cavity mode, and the population imbalance between the two levels was tuned, such that both the polarized and balanced cases could be studied. The transmission spectroscopy, {\it i.e.} pumping the cavity longitudinally and detecting the output cavity-field intensity as a function of the atom-pump detuning, was used in order to demonstrate the JC-like avoided crossing between the resonances (dressed states). That is, upon varying the detuning, different bare states become degenerate, but due to the light-matter coupling the crossing becomes an avoided crossing in the dressed basis. It was found that other modes, besides the fundamental $\text{TEM}_{00}$, could play a role, and their corresponding resonances were identified. The multi-mode contribution effectively scales the light-matter coupling, as we discussed in sec.~\ref{sssec:smapp}, which was further confirmed by {\it ab initio} numerical calculations. In the balanced case, a few additional avoided crossings appeared, an effect that could be due to the transitions between different Zeeman levels.   
 
In a recent proposal to realize the Sachdev-Ye-Kitaev (SYK) model\index{Sachdev-Ye-Kitaev model}, a quasi-two dimensional cloud of fermionic atoms is to be trapped at an antinode of a longitudinal mode supported by a multimode optical cavity\index{Multimode! cavity}, while a random phase mask (speckle) is imprinted on a light-shift beam focused onto the atomic cloud to create a disordered intensity distribution~\cite{uhrich2023cavity}. Driving the cloud and subjecting it to a spatially disordered AC-Stark shift\index{Stark shift} emulates the physical behaviour of the SYK model characterized by random all-to-all interactions and fast scrambling\index{Scrambling}. Data for the OTOC and the spectral form factor have been obtained in~\cite{uhrich2023cavity} for a system of 10 and 14 fermionc nodes at half filling, averaged over $10^3$ realizations governed by the SYK Hamiltonian with Cauchy-distributed interactions.


\subsection{Polaritonic chemistry}\label{ssec:chem}
Rydberg atoms (section~\ref{sec:cavQED}) or superconducting q-dots (section~\ref{sec:cirQED}), in conjunction with a high-$Q$ cavity/resonator, are ideal building blocks for realizing the JC model. The discrete energy level structure of such atoms/q-dots is simple enough such that single transitions can be isolated and the two-level approximation, see sec.~\ref{sssec:tla}, is applicable. The internal energies of molecules, comprising {\it electronic}, {\it vibrational}, and {\it rotational} contributions\index{Rotational! contribution}, are typically much more complex (as briefly mentioned in sec.~\ref{sssec:qatmo} when we discussed cavity cooling) and one would expect the light-matter interaction to be too complicated to allow drawing any simple conclusions. In this section we will show that it is in fact still possible to say much from rather simple models. We will focus on simple molecules, like diatomic and triatomic ones, but experiments have demonstrated that many phenomena seen for very complex polyatomic organic molecules coupled to quantized cavity fields can be explained with basic cavity QED knowledge, for example by invoking the concept of {\it Rabi splitting}. Currently, the field has attracted much attention in both theory and experiment. 

In the next subsection we provide a very short introduction to the Born-Oppenheimer approach as applied to molecular physics. This is used once the molecules are coupled to the light field, as we did in the preceding subsection. We will recognize many steps in the derivations from earlier sections like~\ref{ssec:rabi} and~\ref{sssec:qatmo}. We will, for example, see how, in certain approximations, the resulting Hamiltonian shares the same shape as the one for an ultracold two-level atom in a cavity, eq.~(\ref{qm}). 

\subsubsection{Born-Oppenheimer theory}\label{sssec:boa}
On several occasions in the monograph, we have talked about the Born-Oppenheimer approximation\index{Born-Oppenheimer approximation} (BOA), see {\it e.g.}, sec.~\ref{ssec:rabi}. In the JC setting it naturally emerges in the quadrature representation~(\ref{quad}), in which the boson degree of freedom is expressed in terms of a `position' and `momentum' operator rather than the creation/annihilation operators. We then diagonalized the Hamiltonian within the spin subspace under the assumption that the field evolves on a much slower time-scale than the spin. The origin of the BOA comes from molecular/chemical physics, where such separation of time-scales arises naturally due to the large mass difference between the electrons and nuclei forming the molecule. We will sketch the general method here, not because we will need the actual approximation so much, but since it gives a representative idea about the structure of molecules.

We may assign coordinates $\hat{\bf R}_i$\index{Nuclear! coordinates} to the nuclei and coordinates $\hat{\bf r}_j$ to the electrons, with corresponding masses $m_{ni}$ and $m_e$ and conjugate momenta $\hat{\bf P}_i$ and $\hat{\bf p}_j$ (hence, $\left[{\hat\bf R}_n,\hat{\bf P}_m\right]=\delta_{nm}$ and so on). The Hilbert space for the full system can be written as $\mathcal{H}=\mathcal{H}_\mathrm{slow}\otimes\mathcal{H}_\mathrm{fast}$, with the `slow' and `fast' parts comprising the nuclear and electron degrees of freedom $\hat{\bf R}_i$ and $\hat{\bf r}_j$ respectively. We do not need to care about the spin for this demonstration, nor about the actual shape of the potential $V(\hat{\bf R}_i,\hat{\bf r}_j)$ capturing the interaction among all electrons and nuclei. The Hamiltonian becomes
\begin{equation}\label{molham}
\hat H_\mathrm{mol}=\sum_i\frac{\hat{\bf P}_i^2}{2m_{ni}}+\hat h_{\bf R}=\sum_i\frac{\hat{\bf P}_i^2}{2m_{ni}}+\sum_j\frac{\hat{\bf p}_j^2}{2m_{e}}+V(\hat{\bf R}_i,\hat{\bf r}_j),
\end{equation}
where we defined the {\it electronic Hamiltonian} $\hat h_{\bf R}$ which is parametrized with respect to the coordinates of the nuclei $\hat{\bf R}_i$. We assume that $\hat h_{\bf R}$ behaves well as ${\bf R}$ is varied (we omit the index $i$ for brevity), {\it i.e.} we can parametrize its eigenstates and eigenvalues with ${\bf R}$; 
\begin{equation}
\hat h_{\bf R}|\phi_n({\bf R}\rangle=\varepsilon_n({\bf R})|\phi_n({\bf R})\rangle,
\end{equation}
with $\langle\phi_n({\bf R})|\phi_{n'}({\bf R})\rangle=\delta_{nn'}$. Note that the eigenstates and eigenvalues are truly many-body wavefunctions and energies (for frozen nuclear coordinates ${\bf R}$). The eigenenergies $\varepsilon_n({\bf R})$ are the {\it adiabatic potential surfaces}\index{Adiabatic! potential}, to be compared with, for example, eq.~(\ref{adpot}) which derives from diagonalizing the light-matter interaction term of the quantum Rabi model, while here the corresponding quantities are eigenvalues of the electronic Hamiltonian $\hat h_{\bf R}$. 

We are seeking the solution to the full eigenvalue problem 
\begin{equation}\label{mols}
\hat H|\psi_m\rangle=E_m|\psi_m\rangle,
\end{equation}
with the corresponding wavefunction $\psi_m({\bf R},{\bf r})=\langle{\bf R},{\bf r}|\psi_m\rangle$. In the Born-Oppenheimer expansion the full wavefunction is expanded as
\begin{equation}
|\psi({\bf R})\rangle=\sum_n\varphi_n({\bf R})|\phi_n({\bf R})\rangle,
\end{equation}
such that the $\varphi_n^m({\bf R})$ are ${\bf R}$-dependent coefficients. Likewise, any eigenstate $|\psi_m\rangle$ of the full Hamiltonian can also be expanded as
\begin{equation}
|\psi_m({\bf R})\rangle=\sum_n\varphi_n^m({\bf R})|\phi_n({\bf R})\rangle.
\end{equation}
Inserting this {\it ansatz} into (\ref{mols}), and multiplying from the left with $\langle\phi_l({\bf R})|$, we derive~\cite{baer2006beyond,larson2020conical,bohm2013geometric}
\begin{equation}
\sum_l\left[\sum_n\sum_i\frac{1}{2m_{ni}}\left(\delta_{mn}\hat{\bf P}_i-\mathbf{A}^{mn}\right)\cdot\left(\delta_{nl}\hat{\bf P}_i-\mathbf{A}^{nl}\right)+\varepsilon_m({\bf R})\delta_{ml}\right]\varphi_l({\bf R})=E_m\varphi_m({\bf R}),
\end{equation}
where the matrix
\begin{equation}\label{gc}
{\bf A}^{nl}=i\langle\phi_n({\bf R})|\nabla_{\bf R}|\phi_l({\bf R})\rangle
\end{equation}
is the {\it synthetic gauge connection}\index{Gauge! connection}, which comprises the {\it non-adiabatic coupling terms}\index{Non-adiabatic! coupling term}.  It is in general non-diagonal, which causes transitions between the various electronic levels $\varepsilon_m({\bf R})$. By introducing $\varphi({\bf R})=[\varphi_1({\bf R}),\varphi_2({\bf R}),\dots]^T$ , $\varepsilon({\bf R})=\mathrm{diag}[\varepsilon_m({\bf R})]$ we may write the above vibrational Schr\"odinger equation as
\begin{equation}\label{vibh}
i\partial_t\varphi({\bf R})=\left[\sum_i\frac{1}{2m_{ni}}\left(\hat{\bf P}_i-\mathbf{A}\right)\cdot\left(\hat{\bf P}_i-\mathbf{A}\right)+\varepsilon({\bf R})\right]\varphi({\bf R}).
\end{equation}
With the large mass difference between the electrons and the nuclei, we expect the electronic many-body eigenstates $|\phi_n(R)\rangle$ to vary very smoothly with respect to the nuclear coordinates ${\bf R}$, and the gauge connection~(\ref{gc}) can be neglected; in other words, we obtain the equation for the nuclear motion, 
\begin{equation}\label{boah}
i\partial_t\varphi({\bf R})=\left[\sum_i\frac{\hat{\bf P}_i^2}{2m_{ni}}+\varepsilon({\bf R})\right]\varphi({\bf R}),
\end{equation}
which is diagonal in terms of the $\varphi_m({\bf R})$'s. This is the molecular Born-Oppenheimer approximation (BOA)\index{Born-Oppenheimer approximation} that was discussed first in sec.~\ref{ssec:rabi} for the quantum Rabi model, and later in, for example, sec.~\ref{sssec:qatmo} for the ultracold cavity QED setting. The validity of this approximation relies on the separation of characteristic time-scales. It is important to note that despite the large mass difference, (or other differences, like for the quantum Rabi model), it might break down for certain nuclear distances $R$, typically when the energy separation between the neighbouring electronic energies $\varepsilon_n({\bf R})$ becomes close to zero. In sec.~\ref{sssec:qatmo}, the gauge connection is explicitly written out for the JC model with quantized atomic motion and, provided that the light-matter coupling $g(z)$ is not varying too rapidly, we see that ${\bf A}\rightarrow 0$ in the large-detuning limit $|\Delta|\rightarrow\infty$, {\it i.e.} when the bare/dressed energies are well separated. 

Within the BOA, ${\bf A}\equiv0$, according to eq.~(\ref{boah}) we find that only the eigenstates of $\hat h_{\bf R}$ are needed in order to solve for the nuclear wavefunction $\varphi({\bf R})$. In practice, well-developed quantum chemistry codes are typically used for solving for the eigenvalues $\varepsilon_n({\bf R})$, and these codes need not be efficient in order to extract the eigenstates $|\phi_n({\bf R})\rangle$ required for calculating ${\bf A}({\bf R})$. Hence, in quantum chemistry numerics, it is often the adiabatic Schr\"odinger equation~(\ref{boah}) that is given, while in quantum optical systems one often has access to the diabatic Hamiltonian, {\it e.g.} the JC one. If we are to solve the full problem beyond the BOA, we must somehow also determine ${\bf A}({\bf R})$.

\begin{figure}
\includegraphics[width=8cm]{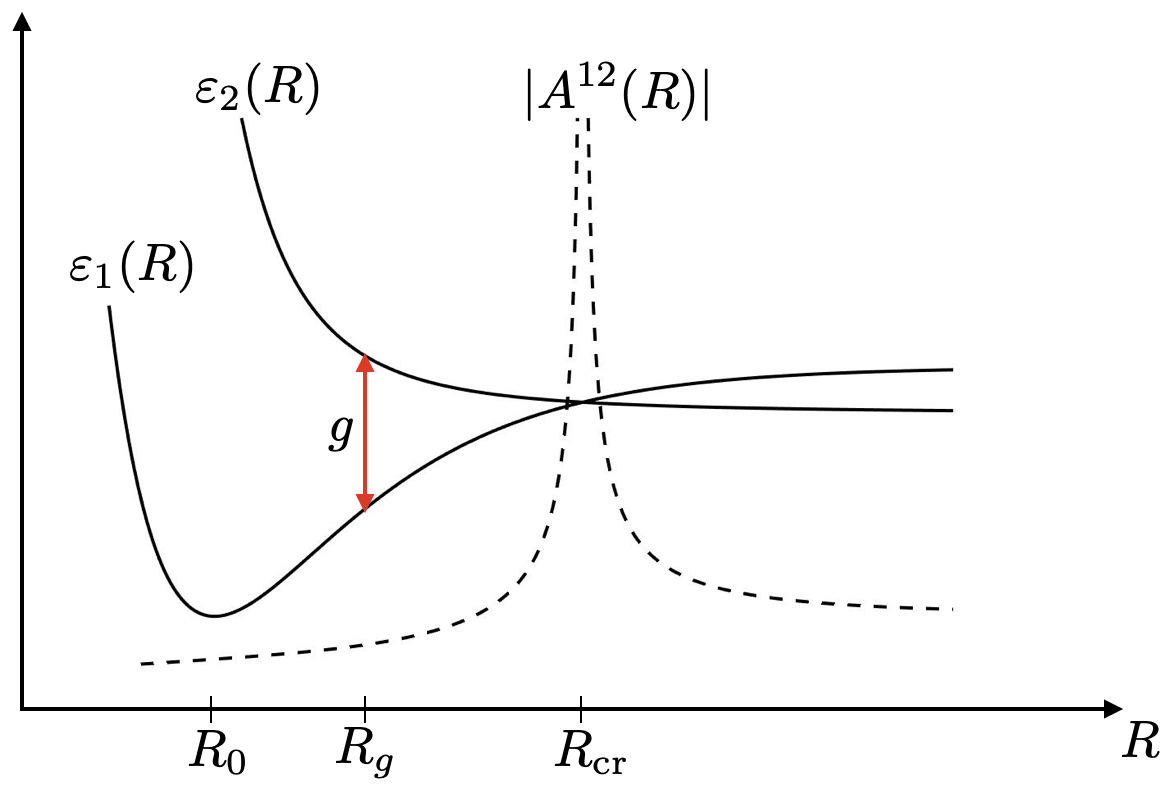} 
\caption{Schematic of the adiabatic potential curves $\varepsilon_{1,2}(R)$ and the non-adiabatic coupling $|A^{12}(R)|$ for a diatomic molecule where $R$ gives the distance between the two nuclei. The adiabatic curve $\varepsilon_{1}(R)$ shows how bound molecular vibrational states may exist; given a not too high energy, the molecule will vibrate around the internuclear distance $R_0$. The molecule eventually dissociates with growing excitation strength. A molecule in the electronic state corresponding to $\varepsilon_{2}(R)$ instead will dissociate; there are no bound vibrational states. At $R_\mathrm{cr}$ the two potentials cross (become degenerate), {\it i.e.} their energy difference vanishes and here the non-adiabatic coupling $|A^{12}(R)|$ becomes non-negligible. In the vicinity of the crossing the BOA breaks down and transitions between the electronic states take place. The vertical red arrow shows some internuclear distance $R_g$ for which the electronic transition becomes degenerate with the cavity mode, $\varepsilon_{2}(R_g)-\varepsilon_{1}(R_g)=\omega$.
}
\label{molpol}
\end{figure}

To build a better understanding of the molecular system, as presented in the Born-Oppenheimer approach, let us consider the most simple configuration, namely a diatomic molecule where we truncate the electronic configuration space to only two states $|\phi_1(R)\rangle$ and $|\phi_2(R)\rangle$. Hence, we invoke the correspondence to the two-level approximation discussed in sec.~\ref{sssec:tla}. Turning to spherical coordinates (assuming Coulomb interactions) we simplify the nuclear problem with a single coordinate $R$ labeling the distance between the two nuclei, while $m_{ni}$ becomes now the reduced mass $\mu$. The rotation and central motion of the molecule are constants of motion and we leave them aside. We thereby obtain
\begin{equation}\label{vibham}
i\partial_t\left[
\begin{array}{c}\varphi_1(R)\\
\varphi_2(R)
\end{array}\right]=\left[\frac{1}{2\mu}\left(-i\nabla_R-\left[
\begin{array}{cc}
0 & A^{12}(R)\\
A^{12*}(R) & 0
\end{array}\right]\right)^2+\left[
\begin{array}{cc}
\varepsilon_1(R) & 0\\
0 & \varepsilon_2(R)
\end{array}\right]\right]\left[
\begin{array}{c}\varphi_1(R)\\
\varphi_2(R)
\end{array}\right],
\end{equation}
with $A^{12}(R)=i\langle\phi_1(R)|\frac{d}{dR}|\phi_2(R)\rangle$, and the centrifugal term has been absorbed into the adiabatic potentials $\varepsilon_{1,2}(R)$. In fig.~\ref{molpol}, we give an example on how the adiabatic potential curves and the amplitude of the non-adiabatic coupling may look like as a function of the internuclear distance $R$. Loosely speaking, the Hamiltonian in eq.~(\ref{vibham}) is our molecular counterpart of the bare atomic term $\Omega\hat\sigma_z/2$ in eq.~(\ref{inth}) giving the definition of the JC Hamiltonian. If $R$ is considered to be frozen, the kinetic term vanishes, and so do the off-diagonal terms $A^{12}(R)$; the system thus obtained is equivalent to the atomic two-level case. Remember, though, that the electronic potential curves $\varepsilon_{1,2}(R)$ and the states $|\phi_{1,2}(R)\rangle$ are true many-body states, while for a Rydberg atom the two electronic states can (approximately) be thought of as single-particle states. The fact that the nuclei are not static causes the bare system to become dynamic, and it is this change that gives rise to non-adiabatic transitions between the adiabatic states with energies $\varepsilon_{1,2}(R)$. In the figure we picture two curves; one which supports bound vibrational states and one purely dissociative. In the former, when localized discrete states are present, the molecule can be bound and oscillate around the potential minima $R_0$. In the latter, the molecule is dissociative and the distance $R$ between the two nuclei will grow as time evolves. However, we also imagine that the two curves cross at $R_\mathrm{cr}$; in the vicinity of that point, the amplitude of the non-adiabatic coupling term becomes non-negligible. For a localized wavepacket $\psi(R)$, when we approach the crossing region around $R_\mathrm{cr}$ there will be a population transfer between the levels~\cite{baer2006beyond}.

\subsubsection{Molecular JC Hamiltonian}
Next we wish to couple the molecules to the quantized radiation of an optical resonator, whence we are dealing with some Hamiltonian of the form
\begin{equation}
\hat H_\mathrm{mc}=\hat H_\mathrm{mol}+\sum_k\omega_k\hat a_k^\dagger\hat a_k+\hat H_\mathrm{int},
\end{equation}
with the bare molecular Hamiltonian $\hat H_\mathrm{mol}$ of eq.~(\ref{molham}), and $\hat H_\mathrm{int}$ capturing the light-matter interaction. Like for the atoms, the molecule will be dressed by the photons to form polaritons\index{Polariton}. Let us for simplicity impose the single-mode approximation, see sec.~\ref{sssec:smapp} (its validity is, of course, more questionable in this more complex molecular setting). In the quadrature representation, eq~(\ref{quad})\index{Quadrature representation},
\begin{equation}\label{quad2}
\begin{array}{l}
\hat{x}=\displaystyle{\frac{1}{\sqrt{2}}\left(\hat{a}^\dagger+\hat{a}\right)},\\ \\
\hat{p}=\displaystyle{\frac{i}{\sqrt{2}}\left(\hat{a}^\dagger-\hat{a}\right)}
\end{array}
\end{equation}
and it becomes clear that the photon field introduces an additional continuous degree of freedom. Following the Born-Oppenheimer approach of the previous section we now split the system into the subsystems formed by the vibrational, electronic and photonic degrees of freedom. Refs.~\cite{flick2017atoms,flick2017cavity} discuss the Born-Oppenheimer treatment in the presence of a quantized photon field, labeled {\it cavity Born-Oppenheimer approximation}\index{Cavity! Born-Oppenheimer approximation}. The idea is to treat the quadrature photon degrees of freedom like the coordinates $\hat{\bf P}_i$ and $\hat{\bf R}_i$ for the nuclei, while the electronic Hamiltonian $\hat h_{\bf R,\bf x}$ will be parametrized by ${\bf R}$ and the $x$-quadratures\index{$x$-quadrature} of the boson fields (it may also depend on the $p$-quadratures\index{$p$-quadrature}, {\it e.g.} in a RWA). The interaction term in the dipole approximation (see sec.~\ref{eda}) becomes
\begin{equation}
\hat H_\mathrm{int}=\sum_{k,i}\omega_k\lambda_{k,i}\cdot{\bf r}_i+\sum_{k,j}\omega_k\lambda_{k,j}\cdot{\bf R}_j,
\end{equation}
where the $\lambda_{k,i}$'s give the coupling strengths as discussed in section~\ref{ssec:approx}. Here we have not included the diamagnetic self-energy term~(\ref{set})\index{Diamagnetic term}. Following our discussion in secs.~\ref{sssec:dicke} and~\ref{sssec:dia}, this term is expected to conceptually alter the system properties beyond the ultrastrong coupling regime, see tab.~\ref{regimetable}. This was also numerically verified in~\cite{flick2017atoms} by diagonalizing the full Hamiltonian $\hat H_\mathrm{mc}$ for a diatomic system with two electrons and a single photon mode. It has also been argued that this term may be important for the complete understanding of some experimental results~\cite{george2016multiple}. Within the cavity Born-Oppenheimer treatment, the dimension of the adiabatic potential surfaces obtained from diagonalizing the electronic Hamiltonian $\hat h_{\bf R,\bf x}$ will now be increased to include the photon degrees of freedom, whence we write the eigenvalues as $\varepsilon_n({\bf R},{\bf x})$~\cite{flick2017atoms,flick2017cavity}. In molecular physics, the coordinates ${\bf R}_i$ are cleverly chosen in order to decouple as many of them as possible, {\it i.e.} block-diagonalize the vibrational Hamiltonian. Such coordinates describe the {\it normal modes}; for a diatomic molecule there is only a single vibrational mode, but already for a triatomic molecule one finds more vibrational modes. The cavity Born-Oppenheimer approach shows how the normal modes will couple the vibrational excitations (phonons) of the molecule to the photons. Like a classical field, the coupling to a quantized field may break some symmetries and therefore qualitatively affect the system. As an example, triatomic molecules $X_3$ (all three atoms identical) support a $2\pi/3$ rotational planar symmetry\index{Rotational! symmetry! planar}, while it is common to find conical intersections\index{Conical intersection} among the lower adiabatic potential surfaces in such systems~\cite{larson2020conical}. These are point degeneracies of two adiabatic potential surfaces at some nuclear configuration ${\bf R}^*$, $\varepsilon_n({\bf R}^*)=\varepsilon_n({\bf R}^*)$. We have already discussed conical intersection in some depth in sec.~\ref{sssec:multi} for the $E\times\varepsilon$ Jahn-Teller Hamiltonian\index{$E\times\varepsilon$ Jahn-Teller Hamiltonian}~(\ref{jt}), with the corresponding adiabatic potential surfaces depicted in fig.~\ref{fig11}. In that setting it described a two-level atom coupled to two boson modes. For a triatomic molecule, the $E\times\varepsilon$ model may derive from expanding the vibrational Hamiltonian to linear order in ${\bf R}_i$. Note that if one is interested in the breakdown of the BOA, such a linear expansion may be sufficient, and, furthermore, for the topological properties only the presence or absence of the conical intersection matters; a conical intersection always hosts a synthetic magnetic $\pi$-flux penetrating it. There exists also {\it glancing intersections}, or {\it Renner-Teller models}\index{Glancing intersection}\index{Renner-Teller model}, where the linear term vanishes while the quadratic term is non-zero~\cite{larson2020conical}. Here the two surfaces touch, but not intersect, and in this case the geometric phase\index{Geometric phase} related to the sysnthetic magnetic flux is trivial, {\it i.e.} a multiple of $2\pi$. In the triatomic molecular system the potential surfaces are actually 3D, but one direction, representing a symmetric stretch vibrational mode, is trivial -- for any plane perpendicular to this one finds a conical intersection. The remaining two coordinates are called {\it reaction coordinates}. Now, if such a triatomic molecule supporting a conical intersection is coupled to a single cavity mode, the field polarization will determine whether the conical intersection survives or not~\cite{flick2017atoms,galego2015cavity}. As such, lifting the intersection off implies a decrease in the non-adiabatic couplings, which can, in principle, restore the BOA. 

Many works do not approach the problem from the cavity Born-Oppenheimer point of view, but instead add the coupling to the photon field once the standard molecular separation has been performed. Hence, the bare Hamiltonian for the molecule in this case is the vibrational term of eq.~(\ref{vibh}). If, furthermore, one imposes the two-level, single-mode, and dipole approximations together with the molecular BOA, one finds the full Hamiltonian~\cite{kowalewski2016cavity,kowalewski2016non,csehi2019ultrafast}
\begin{equation}\label{mc2}
\hat H_\mathrm{mc}=-\frac{1}{2\mu}\frac{\partial^2}{\partial R^2}+\omega\hat a^\dagger\hat a+\frac{\Omega(R)}{2}\hat\sigma_z+V(R)+g(R)\left(\hat a^\dagger+\hat a\right)\hat\sigma_x,
\end{equation}
where $\Omega(R)=\varepsilon_1(R)-\varepsilon_2(R)$ and $V(R)=(\varepsilon_1(R)+\varepsilon_2(R))/2$. By including the non-adiabatic coupling terms ({\it i.e.}, by giving up the BOA) we get an $R$-dependent term proportional to $\hat\sigma_{x,y}$, which -- in our language of sec.~\ref{ssec:drjc} -- represents a drive of the two-level system. Apart from the constant potential term $V(R)$, note how the Hamiltonian has the same form as the JC model with quantized atomic motion~(\ref{qm}). We have not performed the RWA in the above equation though. The physical meaning is different for the two cases: in (\ref{qm}) the additional degree of freedom corresponds to the atomic motion, while in the cavity-molecular Hamiltonian~(\ref{mc2}) it describes the molecular vibrations. Furthermore, the main spatial dependence is in the transition frequency $\Omega(R)$ and not in the light-matter coupling amplitude $g(R)$ as is the case for the moving atoms. In sec.~\ref{sssec:qatmo} we introduced the concept of well-dressed states\index{Well-dressed states}, and the same could be considered here. The two-level system is then dressed by both photons (polariton)\index{Polariton} and phonons. In the above Hamiltonian, we consider a coupling between the electronic states corresponding to the adiabatic potentials $\varepsilon_{1,2}(R)$. We could also imagine a situation in which the lower adiabatic potential supports a set of molecular bound states separated by some characteristic energy scale $\nu$. If the photon frequency is such that $\omega\sim\nu$, and at the same time the first excited electronic state is far off in energy, one can adiabatically eliminate\index{Adiabatic! elimination} the latter, as we have done for the JC model in sec.~\ref{ssec:JCm}, and we are then left with an effective model coupling phonons to photons. However, let us instead consider Hamiltonian (\ref{mc2}), and assume that there exists some internuclear distance $R_g$ such that $\Omega(R_g)=\omega$. In fig.~\ref{molpol} we mark such distance with the green arrow. For this distance $R_g$, the transition becomes resonant with the photon energy. 

Returning to the example of fig.~\ref{molpol}, as we have already mentioned, the breakdown of the molecular BOA implies population transfer between the adiabatic states. The vibrational motion of the molecule can thereby become very complex if the potentials $\varepsilon_n(R)$ display many crossings; interference between the different paths will occur, and for unbound dissociative states the molecule might break apart. Understanding such evolution may be necessary in order to correctly describe chemical reactions. The coupling to the photon field will modify this energy landscape and will affect the reactions~\cite{kowalewski2016cavity,kowalewski2016non,csehi2019ultrafast,galego2015cavity,herrera2016cavity,hutchison2012modifying,ebbesen2016hybrid,ribeiro2018polariton}. We could in principle perform an adiabatic diagonalization of (\ref{mc2}), and we would then find that the corresponding non-adiabatic coupling terms will be non-vanishing around $R_g$ meaning that the main population transfer between the two internal states occurs at that point. However, depending on all the involved timescales, the picture might well be more involved than such simple arguments, {\it e.g.} if the light-matter coupling $g$ is small compared to the characteristic vibrational frequency, the cavity has a small influence over one oscillation period. In ref.~\cite{kowalewski2016cavity}, the authors numerically solved for the evolution generated by the Hamiltonian~(\ref{mc2}). Computationally it appears favorable to propagate the state in the quadrature representation~(\ref{quad2}) rather than expanding in photon Fock states. They considered the diatomic molecule NaI, and obtained the adiabatic potential curves $\varepsilon_{1,2}(R)$ from quantum chemistry calculations. The actual potentials $\varepsilon_{1,2}(R)$ are similar in shape to those of fig.~\ref{molpol} depicting the crossing between a bound and a dissociative potential. The analysis focused on the survival probability of the molecule; we recall that the molecule breaks apart as it oscillates and populates the dissociative level. It was found that the coupling of the cavity mode to the material system could stabilize the molecule. Similar potential curves have also been found for the LiF molecule~\cite{csehi2019ultrafast}. By analyzing the adiabatic potential surfaces when the photon degree of freedom is taken into account, giving rise to $\varepsilon_{1,2}(R,x)$, the crossing of the curves $\varepsilon_{1,2}(R)$ becomes a proper conical intersection. Again the influence of the cavity field on the dissociation rate was considered, and like for the NaI molecule it was found that the survival time could be controlled and enhanced by the dressing of the photon field. Another exploration into how the chemical reactions are modified due to the cavity field was presented in~\cite{herrera2016cavity}. This study did not involve any {\it ab initio} chemical calculations, but the arguments were based on a toy model in which the adiabatic potentials were given by displaced harmonic oscillators, coupled via the cavity field. In particular, $N$ molecules identically coupled to the cavity mode were considered. It was shown, using the polaron transfomration\index{Polaron! transformation}~(\ref{polaron}), similar to what was employed for a mean-field decoupling in refs.~\cite{kurcz2014hybrid,gagge2020superradiance} for the Jaynes-Cummings-Hubbard models, that in the ultrastrong coupling regime the system decouples into a symmetric $\mathcal{P}$ and non-symmetric $\mathcal{Q}$ manifold. The symmetric manifold is spanned by the product state with all molecules in the lower electronic level, $|g_1,g_2,\dots,g_N\rangle$, and the symmetric state with one molecule with an excited electronic state, {\it i.e.} the first excited Dicke state~(\ref{dickestates}). In this regime where the Rabi frequency dominates over the molecular vibrational frequencies, the phonon degrees of freedom decouple from the electronic ones, and the effective model becomes a simple JC model, that parametrically depends on the phonon degrees of freedom such that a self-consistent solution is called upon. It was argued that this decoupling will actually influence chemical processes within the molecular cloud.

As we have seen, vibrational degrees of freedom are crucial for the proper description of organic molecular cavity QED systems. This introduces additional phonon modes and qubit–phonon couplings in the Tavis-Cummings Hamiltonian\index{Tavis-Cummings model}, yielding the Holstein-Tavis-Cummings (HTC) model\index{Holstein-Tavis-Cummings model} describing phonon sidebands in absorption or emission spectra. At the same time, for molecules placed inside a cavity, inhomogeneities in the material, different levels of aggregation, or solvent fluctuations are sources of disorder in the molecular energy levels. Using the {\it disordered Tavis-Cummings model}\index{Tavis-Cummings model!disordered}, Gera and Sebastian report on a disorder-induced width of the polaritonic peak\index{Polariton! peak}, presenting results for several densities of states and the absorption spectrum. The authors conclude that the variation of the width of the polaritonic peaks as a function of the Rabi splitting indicates the distribution of molecular energy levels~\cite{Gera2022}. Moreover, Sun and collaborators report on the dynamics of the TC model and the HTC models in the presence of diagonal disorder and cavity-qubit coupling disorder~\cite{Sun2022HTC}.

The current experiments do not operate at the level of individual molecules, contrary to what has been achieved with atoms in cavity QED, see section~\ref{sec:cavQED}. Instead, ensembles (or thin films) of polyatomic organic molecules are coupled to low-$Q$ optical cavities formed for example by metallic plates. Both the strong~\cite{lidzey1998strong,lidzey2000photon,tischler2005strong,kena2008strong,hutchison2012modifying,kena2010room} and ultrastrong coupling~\cite{george2015ultra,schwartz2011reversible,kena2013ultrastrongly,thomas2016ground} regimes have been demonstrated. It should be clear that for such complex molecules the model Hamiltonian is not as simple as~(\ref{mc2}), yet coherent Rabi oscillations have been verified~\cite{schwartz2011reversible} alongside the modification of the dissociation and other reaction rates~\cite{hutchison2012modifying,thomas2016ground}. It's worth remarking here that the Franck-Condon principle\index{Franck-Condon! principle} has been recently incorporated into a nonadditive master equation\index{Master Equation!nonadditive}~\cite{Maguire2019} -- see subec.~\ref{subsubsec:oldquantum}.

Another interesting aspect related to molecular cavity QED is the photonic Bose-Einstein condensate\index{Bose-Einstein! condensate}. Photons are massless and their chemical potential which fixes their number vanishes. As a result, the number of photons is not preserved and, as the temperature is lowered or the density is increased, the photons should not go through a critical point\index{Critical! point} and condense. Nevertheless, by using a cavity filled with dye molecules, an effective mass\index{Effective! mass} can be ascribed to the photons and condensation becomes possible. This has been experimentally verified in the group of Martin Weitz~\cite{klaers2010bose}. Their results sparked numerous discussions since these are manifestly nonequilibrium systems\index{Nonequilibrium! system} and the fluctuations due to photon losses could prohibit condensation ~\cite{kirton2013nonequilibrium,sieberer2013dynamical,sieberer2016keldysh}. 


\section{Conclusions -- a projection for the coming decades}

Upon reading the historical note by Cummings on how it all started and how the JC model came about~\cite{cummings2013reminiscing}, it is evident that the impact of the model was not anticipated at that stage. The present monograph, celebrating the model and exploring its numerous ramifications, attempts to provide an account on what has emerged in the wake of the 1963 report whose principal aim was ``to clarify the relationship between the quantum theory of radiation, where the electromagnetic field-expansion coefficients satisfy commutation relations, and the semiclassical theory, where the electromagnetic field is considered as a definite function of time rather than as an operator''~\cite{jaynes1963comparison}. Like any process, it is hard to predict how things would have turned out differently if the setting had been changed, ``What if...''? Hence, we cannot with certainty pinpoint the significance of the JC model in deterministically shaping the development of quantum optics and related fields. Nevertheless, it should stand clear that it has been one of the most paradigmatic models in the development of modern quantum mechanics---one that initiated a progressive yet visible shift from the laser (maser) it was supposed to describe---and that it will carry on serving that purpose over the coming half-century as well. It is also clear that the initial focus has been {\it de facto} reappraised since 1963, as the JC report laid more emphasis on the successes of semiclassical theory rather than on the need for a treatment which is fully quantum~\cite{CarmichaelPRX}. 

Quantum optics as a research field is rather young compared to other branches of quantum physics. The interaction between light and matter is as old as other fields, but the systematic studies on the importance of the quantum nature of the electromagnetic field were established primarily in the course of the last 50-60 years. Glauber, Klauder, Sudarshan, Eberly and others started exploring the quantum nature of the light fields in the first half of the 1960s, but the emphasis was not so much laid on the importance of such a quantization in conjunction with single atoms in their idealized two-state form -- the JC model. Even if Jaynes and Cummings did not foresee the potential of their model and its exemplary nonlinearity to reveal quantum coherence, it actually became quickly the theoretical workhorse in a subfield of quantum optics where tailored processes were devised. Today, the study of multiphoton quantum-nonlinear optics\index{Multiphoton! quantum optics} is a topic that has been subject of extensive research due to the exquisite control acquired over cavity and circuit quantum electrodynamics architectures; some of them have been extensively visited in the corresponding sections. In addition, other quantum optical systems like ultracold atoms and optical lattices (not {\it ab initio} in need of a quantized field theory) are key players in the study of coherent quantum processes. Nevertheless, in the late 1980s and early 1990s, cavity QED advanced to such degree such that the JC model and its predictions became most relevant, as discussed in section~\ref{sec:cavQED}. In some sense, this takes us back to the 1920s and the heydays of quantum physics. The thought experiments of the pioneers of quantum mechanics with, at best, {\it quasi}monochromatic blackbody radiation\index{Blackbody radiation} at their disposal, were no longer an unattainable vision. Isolated quantum processes, working deep in the quantum regime where quantum fluctuations cannot be overlooked nor reduced to classical noise, were demonstrated. This marks the departure from the theory developed for the laser in the late 1960s and the representation of two-level atoms by means of a characteristic function and a corresponding distribution by Haken and others~\cite{Hakenbook, BookQO1Carmichael}. In addition, sustainable coherence was achieved in order to demonstrate the first quantum logic gates. These demonstrations signal the breakthrough of reaching the strong-coupling regime of cavity QED. About the same time, in the quest for quantum information processing (QIP), the trapped ion physics also demonstrated isolated coherent evolution at the single particle(s) level, see section~\ref{sec:ion}. In fact, as we explained in this section, the first two-qubit logic gate was demonstrated in a trapped ion system relying on JC physics. The discovery that classically driven harmonically trapped ions could realize JC physics was also a first example in which the model goes beyond light-matter interactions; matter coherently coupled to phonons. In this monograph we have seen other examples of this, specifically in sec.~\ref{sec:cirQED} which demonstrates how the JC model can be realized in a solid-state setup comprised of superconducting quantum dots coherently coupled to transmission lines. Here, the two-level systems emerge from collective electronic states of quantum dots. 

As Wiseman and Milburn remark in their book~\cite{wiseman_milburn_2009}, projective measurements on their own are insufficient for describing real measurements since an experimenter never measures directly the system of interest. Rather, the system under study, say a two-level atom or a laser, interacts with its environment which is a continuum of electromagnetic field modes. To substantiate this interaction, P. Rice, in his pedagogical account on quantum optics, defines the photon as ``click in an absorptive detector...It is not a particle, or a wave, but an excitation of a bosonic field. So, how [does one] tell whether the light is coming from an incandescent bulb or a He-Ne laser? Glauber~\cite{glauber1963coherent} taught us that the difference is in photon correlations. One must look past expectation values and examine two-time correlations at the least....''
 
When collecting data to determine these correlations, the experimenter, in fact, observes the effect of the system on the radiated field or rather the field interacting with the photodetector, with an observable sequence of stimuli leading to the decision on the result of measurement. This chain of systems is known as a von Neumann chain\index{von Neumann! chain}~\cite{vN1932}. What is essential to note is that, at some stage before reaching the mind of the observer, one needs to cut the chain by applying the projection postulate. This cut is termed the Heisenberg's cut\index{Heisenberg! cut}, the instance at which one considers that the measurement has been effected~\cite{Heisenberg1930}. A direct application of the projection postulate directly to the system would yield wrong predictions. However, assuming a projective measurement on the field will generate results which are negligibly different to those obtained by performing a projective measurement at any later stage. This is due to the rapid decoherence of macroscopic objects such as photodetectors. It is therefore sufficient to consider the field to be measured projectively. Since the field has interacted with the system, their quantum states are correlated. The projective measurement on the field is then effectively a measurement on the system. This type of measurement, however, is not projective, whence we are in need of a more general formalism to account for it. 

We have already mentioned that the JC physics, expressed in the guise of cavity QED, trapped ions, and later circuit QED, enabled the study of `textbook quantum mechanics' in the lab at a level not envisaged so far. But it also went further, to traverse a long path with exploring the meaning of {\it quantum measurements} and the appearance of a `classical world' within quantum mechanics. Especially the seminal work on the {\it in situ} measurement of the progressive decoherence of the Schr\"odinger cat state as discussed in sec.~\ref{sssec:subscat}, took the field into a new direction. This was followed by many theoretical and experimental works, including {\it quantum jumps} and {\it quantum trajectories}, {\it interaction-free} and {\it quantum non-demolition measurements}, {\it quantum Zeno measurements}, {\it weak measurements}, decoherence and the quantum-to-classical transition. In short, the field has given us glimpses into what is usually known as the ``measurement problem'' or the ``state collapse''. In more recent years, the inherently open character of quantum systems has raised a plethora of new research problems, like the manifestation of {\it criticality in driven-dissipative configurations} and the role of a {\it strong-coupling ``thermodynamic limit''}\index{Thermodynamic limit! strong-coupling} in zero dimensions, as well as the requirement for a formalism with a new conceptual outlook. Photon antibunching advocates the discrete nature of light (particles), in contrast to amplitude squeezing, which speaks of the continuous (waves). One hundred years after Planck stated his law for the blackbody radiation\index{Blackbody radiation} for the first time in public, the tension between particles and waves was found at the focus of experiments which combined the measurement strategies used to observe these nonclassical behaviors of light~\cite{CarmichaelTalk2000}. Moreover, the quantum trajectory theory accounts for many possible unravelings of an open system evolution, each one matched to the nonlocal action of the system in question upon a specific environment -- the crux of Bohr's {\it complementarity}~\cite{BookQO2Carmichael, CarmichaelQJRev}\index{Complementarity/contextuality}\index{Copenhagen interpretation}. These aspects are briefly discussed in secs.~\ref{ssec:drjc} and \ref{ssec:quantumglcavQED}.   

Another fundamental topic, brought up in sec.~\ref{ssec:approx} and concerning a more fundamental aspect, concerns the {\it gauge invariance} of the JC model and the truncation of the material system to two levels. Keeping in mind the importance of the model, it is surprising that it took more than half a century to thoroughly appreciate this issue. Naturally, it is only in the advent of circuit QED that the approximations stemming from truncation of the Hilbert spaces were cast into doubt. It was the development of this new field since the early 2000s that drove the interest in the facets of gauge invariance. The last word has not been said on this topic, since there is no consensus on the problem, {\it i.e.}, which is the most accurate low-energy effective model within some two-level and single/few-mode approximations?

On the theory side, we must also alight to the work of Braak on the integrability/solvability of the quantum Rabi model, see sec.~\ref{sssec:rabiint}. His report not only initiated a debate on the completeness of its solution, but an avalanche of studies on generalizations of the solution to other systems. The work has gained much attention in the community, but it has also made researchers interested in integrable models to open their eyes up for fully- connected quantum optical models. Braak's results are not so much of practical interest, as they are important from a fundamental point of view. The meaning of `quantum integrability' is debated, and the solvability of the driven quantum Rabi model, which lacks any avoided energy crossings, ``..signifies that integrability and solvability are not equivalent in the realm of quantum physics''. 

Let us try to summarize the developments related to the JC model during the last 50-60 years in some general terms. The field has been driven much by technological advancements in cavity QED and trapped ions, and later in other systems. Early on, QIP was a key factor motivating this progress, but, as we have already mentioned, other aspects emerging from the investigation of isolated quantum systems constituted an additional drive. In later years, we have witnessed a clear tendency towards larger system sizes, primarily by increasing the number of degrees of freedom. Many trapped-ion experiments go towards {\it quantum simulators}, where Coulomb crystals of trapped ions are prepared, focusing primarily on their evolution. In cavity QED, single two-level atoms have been replaced by atomic Bose-Einstein condensates, and, once again, collective phenomena rather than an evolution at the level of individual quanta are considered. One motivation with circuit QED, apart from achieving desirable system parameters, was the scalability of these systems. It is much easier to increase the number of qubits, and keep control and access to individual ones, in such systems than in, say, cavity QED. This is also what is envisioned by commercial companies, like {\it Google} and {\it IBM}, in their striving towards {\it quantum supremacy} and building reliable {\it quantum computers}. We have also seen how new physical systems are entering the game in sec.~\ref{sec:physreal}. In general, on the experimental side, much has been about ``pushing boundaries'', with the goal of tackling either questions of fundamental quantum nature, or finding applications in QIP or quantum simulations. 

One can only speculate on what the coming decades will bring in the footsteps of the JC physics. One trend lately is going towards {\it hybrid systems}. The idea is to `make the best out of several worlds' and combine them; {\it e.g.} photons can transmit information between remote places, and matter can store information. The JC-type models will keep showing up in other fields. Already today, the boundaries between different communities have become fuzzier. We have seen numerous examples of this in our monograph, especially as JC physics moves in the direction of many-body problems which have a much longer history in condensed-matter physics than in quantum optics. Techniques and knowledge are bound to become more universal than community based. Technological advancements are still partly driven by QIP - faster gates, higher fidelities, implementing error correction codes, larger system sizes and the like. Yet, after three decades of QIP, we see a broadening of perspective from universal digital QIP to analog systems like quantum simulators. Phenomena like {\it quantum thermalization} and {\it scrambling}\index{Scrambling}, {\it quantum chaos}, {\it many-body localization}, and {\it quantum scarring} have become of interest in the community during the last years, intertwined with critical nonequilibrium phenomena\index{Nonequilibrium! phenomenon}. While we already have experimental results at hand along these lines, one can question whether a quantum simulator in its proper sense has been realized. The idea of a quantum simulator is to provide answers to open questions or give insight beyond present understanding. We might not be there at this stage, but it should only be a question of time until we are. 

Other young fields gathering much momentum at the moment include {\it quantum thermodynamics}, {\it quantum machine learning}, and {\it quantum feedback control}. We have met some examples of these fields in our monograph, but we expect many more to appear in the near future. The JC systems, being highly and precisely controllable as well as versatile, are natural candidates for implementing diverse schemes in these disciplines. Such configurations typically need to include a very large number of degrees of freedom, but most theoretical proposals are dealing with small systems at the moment. There are still many gaps in our fundamental understanding, and we do not yet know all possible applications that may emerge. Related to our partial understanding is the ever-unresolved problem of {\it quantum measurement} that we touched upon before. The JC physics, lying at the core of low-energy light-matter interaction, has proved so far a very suitable framework for addressing this problem, and we may expect new insights particularly when investigating the nature of quantum jumps. 

\bibliography{draft_rev_II}
\clearpage
\printindex

\end{document}